\documentclass[a4paper]{article}

\usepackage{graphicx,bm}
\usepackage{color}
\usepackage{amssymb}
\usepackage{amsmath}

\definecolor{lightgreen}{rgb}{0.2,1.0,0.5}
\definecolor{lightblue}{rgb}{0.5,0.5,1.0}
\definecolor{gray}{rgb}{0.5,0.5,0.5}
\definecolor{brown}{rgb}{0.55,0.3,0.3}   
\definecolor{darkgreen}{rgb}{0.1,0.5,0.4}
\definecolor{color1}{rgb}{0.2,0.5,0.5}

\newcommand{\ie}{{\it i.e.}}
\newcommand{\eg}{{\it e.g.}}
\newcommand{\ds}{\displaystyle}
\newcommand{\ra}{\rightarrow}
\newcommand{\la}{\leftarrow}
\newcommand{\Ra}{\Rightarrow}
\newcommand{\Lra}{\Leftrightarrow}

\newcommand{\rlas}{\rightleftarrows}
\newcommand{\sgn}{{\rm sgn}}
\newcommand{\ev}{{\bm e}}
\newcommand{\bev}{\bar{\bm e}}
\newcommand{\nablav}{{\bm\nabla}}
\newcommand{\tnablav}{\tilde{\bm\nabla}}

\newcommand{\rd}{\partial}
\newcommand{\brd}{\bar{\partial}}

\newcommand{\dg}{\dagger}

\newcommand{\bsgn}{\bar{\sgn}}
\newcommand{\cd}{\cdot}

\newcommand{\Av}{{\bm A}}
\newcommand{\tAv}{\tilde{\bm A}}
\newcommand{\tA}{\tilde{A}}
\newcommand{\bA}{\bar{A}}
\newcommand{\hA}{\hat{A}}
\newcommand{\cA}{{\cal A}}
\newcommand{\bcA}{\bar{\cal A}}
\newcommand{\tcA}{\tilde{\cal A}}
\newcommand{\mfA}{{\mathfrak A}}
\newcommand{\bmfA}{\bar{\mathfrak A}}
\newcommand{\abar}{\bar{a}}
\newcommand{\ha}{\hat{a}}
\newcommand{\mfa}{{\mathfrak a}}

\newcommand{\ta}{\tilde{a}}
\newcommand{\tb}{\tilde{b}}
\newcommand{\tc}{\tilde{c}}
\newcommand{\td}{\tilde{d}}

\newcommand{\bB}{\bar{B}}
\newcommand{\cB}{{\cal B}}
\newcommand{\bcB}{\bar{\cal B}}
\newcommand{\tcB}{\tilde{\cal B}}
\newcommand{\cBv}{\vec{\cal B}}
\newcommand{\tB}{\tilde{B}}
\newcommand{\hB}{\hat{B}}
\newcommand{\mfB}{{\mathfrak B}}

\newcommand{\Bv}{{\bm B}}
\newcommand{\tBv}{\tilde{\bm B}}

\newcommand{\bbar}{\bar{b}}
\newcommand{\hb}{\hat{b}}
\newcommand{\mfb}{{\mathfrak b}}

\newcommand{\bC}{\bar{C}}
\newcommand{\hC}{\hat{C}}
\newcommand{\cC}{{\cal C}}
\newcommand{\bcC}{\bar{\cal C}}
\newcommand{\tcC}{\tilde{\cal C}}
\newcommand{\bbC}{\mathbb{C}}

\newcommand{\hc}{\hat{c}}
\newcommand{\mfc}{{\mathfrak c}}

\newcommand{\hD}{\hat{D}}
\newcommand{\Dv}{{\bm D}}
\newcommand{\tDv}{\tilde{\bm D}}
\newcommand{\bD}{\bar{D}}
\newcommand{\cD}{{\cal D}}
\newcommand{\bcD}{\bar{\cal D}}
\newcommand{\tcD}{\tilde{\cal D}}
\newcommand{\hcD}{\hat{\cal D}}
\newcommand{\tD}{\tilde{D}}
\newcommand{\mfD}{{\mathfrak D}}
\newcommand{\bmfD}{\bar{\mathfrak D}}
\newcommand{\hd}{\hat{d}}
\newcommand{\mfd}{{\mathfrak d}}

\newcommand{\bE}{\bar{E}}
\newcommand{\brE}{\breve{E}}
\newcommand{\chE}{\check{E}}
\newcommand{\cE}{{\cal E}}
\newcommand{\bcE}{\bar{\cal E}}
\newcommand{\ccE}{\check{\cal E}}
\newcommand{\hcE}{\hat{\cal E}}
\newcommand{\brcE}{\breve{\cal E}}
\newcommand{\cEv}{\vec{\cal E}}
\newcommand{\hE}{\hat{E}}
\newcommand{\tE}{\tilde{E}}
\newcommand{\mfE}{{\mathfrak E}}

\newcommand{\Ev}{{\bm E}}
\newcommand{\tEv}{\tilde{\bm E}}
\newcommand{\hEv}{\hat{\bm E}}
\newcommand{\Fv}{{\bm F}}
\newcommand{\bF}{{\bar F}}
\newcommand{\cF}{{\cal F}}
\newcommand{\tF}{{\tilde F}}
\newcommand{\hF}{{\hat F}}

\newcommand{\mfF}{{\mathfrak F}}
\newcommand{\mff}{{\mathfrak f}}
\newcommand{\bmff}{\bar{\mathfrak f}}
\newcommand{\bmfF}{\bar{\mathfrak F}}

\newcommand{\barf}{\bar{f}}

\newcommand{\cG}{{\cal G}}

\newcommand{\hG}{{\hat G}}

\newcommand{\mfg}{{\mathfrak g}}
\newcommand{\mfG}{{\mathfrak G}}
\newcommand{\bmfg}{\bar{\mathfrak g}}
\newcommand{\bmfG}{\bar{\mathfrak G}}
\newcommand{\bg}{\bar{g}}

\newcommand{\bH}{\bar{H}}
\newcommand{\cH}{{\cal H}}
\newcommand{\mfH}{{\mathfrak H}}

\newcommand{\bI}{\bar{I}}
\newcommand{\tI}{\tilde{I}}
\newcommand{\hI}{\hat{I}}
\newcommand{\cI}{{\cal I}}
\newcommand{\bcI}{\bar{\cal I}}
\newcommand{\cJ}{{\cal J}}
\newcommand{\bcJ}{\bar{\cal J}}

\newcommand{\tJ}{{\tilde J}}
\newcommand{\hJ}{{\hat J}}
\newcommand{\mfJ}{{\mathfrak J}}
\newcommand{\Jv}{{\bm J}}
\newcommand{\tJv}{\tilde{\bm J}}
\newcommand{\hJv}{\hat{\bm J}}
\newcommand{\bK}{\bar{K}}
\newcommand{\cK}{{\cal K}}
\newcommand{\bcK}{\bar{\cal K}}

\newcommand{\hK}{\hat{K}}
\newcommand{\mfK}{{\mathfrak K}}
\newcommand{\bk}{\bar{k}}
\newcommand{\hk}{\hat{k}}

\newcommand{\bL}{\bar{L}}
\newcommand{\hL}{\hat{L}}
\newcommand{\cL}{{\cal L}}
\newcommand{\bcL}{\bar{\cal L}}

\newcommand{\brL}{\breve{L}}
\newcommand{\mfL}{{\mathfrak L}}

\newcommand{\hM}{\hat{M}}
\newcommand{\hMv}{\hat{\bm M}}
\newcommand{\tMv}{\tilde{\bm M}}
\newcommand{\cM}{{\cal M}}
\newcommand{\bcM}{\bar{\cal M}}
\newcommand{\hcM}{\hat{\cal M}}
\newcommand{\tM}{\tilde{M}}

\newcommand{\mfm}{\mathfrak{m}}
\newcommand{\bmfm}{\bar{\mathfrak{m}}}
\newcommand{\hmfm}{\hat{\mathfrak{m}}}

\newcommand{\bN}{\bar{N}}
\newcommand{\hN}{\hat{N}}
\newcommand{\hNv}{\hat{\bm N}}
\newcommand{\tNv}{\tilde{\bm N}}
\newcommand{\cN}{{\cal N}}
\newcommand{\bcN}{\bar{\cal N}}
\newcommand{\hcN}{\hat{\cal N}}
\newcommand{\tN}{\tilde{N}}
\newcommand{\bbN}{\mathbb{N}}

\newcommand{\mfn}{\mathfrak{n}}
\newcommand{\bmfn}{\bar{\mathfrak{n}}}
\newcommand{\hmfn}{\hat{\mathfrak{n}}}

\newcommand{\hP}{\hat{P}}

\newcommand{\hPv}{\hat{\bm P}}
\newcommand{\tPv}{\tilde{\bm P}}
\newcommand{\cP}{{\cal P}}
\newcommand{\hcP}{\hat{\cal P}}
\newcommand{\tP}{\tilde{P}}

\newcommand{\mfp}{\mathfrak{p}}
\newcommand{\bmfp}{\bar{\mathfrak{p}}}
\newcommand{\hmfp}{\hat{\mathfrak{p}}}

\newcommand{\hQ}{\hat{Q}}
\newcommand{\hQv}{\hat{\bm Q}}
\newcommand{\tQv}{\tilde{\bm Q}}
\newcommand{\cQ}{{\cal Q}}
\newcommand{\hcQ}{\hat{\cal Q}}
\newcommand{\tQ}{\tilde{Q}}

\newcommand{\mfq}{\mathfrak{q}}
\newcommand{\bmfq}{\bar{\mathfrak{q}}}
\newcommand{\hmfq}{\hat{\mathfrak{q}}}

\newcommand{\cR}{{\cal R}}
\newcommand{\bbR}{\mathbb{R}}

\newcommand{\hR}{\hat{R}}
\newcommand{\hr}{\hat{r}}

\newcommand{\br}{\bar{r}}
\newcommand{\brr}{\bar{r}}
\newcommand{\mfr}{\mathfrak{r}}

\newcommand{\Sv}{{\bm S}}
\newcommand{\tSv}{\tilde{\bm S}}
\newcommand{\cSv}{\vec{\cal S}}
\newcommand{\bS}{\bar{S}}
\newcommand{\tS}{{\tilde S}}
\newcommand{\cS}{{\cal S}}
\newcommand{\bcS}{\bar{\cal S}}
\newcommand{\hcS}{\hat{\cal S}}
\newcommand{\mfS}{{\mathfrak S}}

\newcommand{\mfs}{\mathfrak{s}}
\newcommand{\hS}{\hat{S}}

\newcommand{\Tv}{{\bm T}}
\newcommand{\tTv}{\tilde{\bm T}}
\newcommand{\cTv}{\vec{\cal T}}
\newcommand{\tT}{{\tilde T}}
\newcommand{\bT}{\bar{T}}
\newcommand{\bt}{\bar{t}}

\newcommand{\cT}{{\cal T}}
\newcommand{\mfT}{{\mathfrak T}}

\newcommand{\tUv}{\tilde{\bm U}}
\newcommand{\tVv}{\tilde{\bm V}}
\newcommand{\tU}{\tilde{U}}
\newcommand{\tV}{\tilde{V}}
\newcommand{\bU}{\bar{U}}
\newcommand{\hU}{\hat{U}}
\newcommand{\hu}{\hat{u}}
\newcommand{\cU}{{\cal U}}
\newcommand{\bcU}{\bar{\cal U}}
\newcommand{\mfU}{{\mathfrak U}}

\newcommand{\hUv}{\hat{\bm U}}
\newcommand{\hVv}{\hat{\bm V}}

\newcommand{\bV}{\bar{V}}
\newcommand{\hV}{\hat{V}}
\newcommand{\hv}{\hat{v}}
\newcommand{\cV}{{\cal V}}
\newcommand{\bcV}{\bar{\cal V}}
\newcommand{\mfV}{{\mathfrak V}}

\newcommand{\Uv}{{\bm U}}
\newcommand{\Vv}{{\bm V}}
\newcommand{\bu}{\bar{u}}
\newcommand{\bv}{\bar{v}}
\newcommand{\chu}{\check{u}}
\newcommand{\chv}{\check{v}}

\newcommand{\bW}{\bar{W}}
\newcommand{\brW}{\breve{W}}
\newcommand{\cW}{{\cal W}}
\newcommand{\bcW}{\bar{\cal W}}
\newcommand{\mfW}{{\mathfrak W}}

\newcommand{\bw}{\bar{w}}

\newcommand{\hx}{\hat{x}}
\newcommand{\xv}{{\bm x}}

\newcommand{\bX}{\bar{X}}
\newcommand{\cX}{{\cal X}}
\newcommand{\mfX}{{\mathfrak X}}
\newcommand{\cY}{{\cal Y}}
\newcommand{\bcY}{\bar{\cal Y}}
\newcommand{\hcY}{\hat{\cal Y}}
\newcommand{\bZ}{\bar{Z}}
\newcommand{\cZ}{{\cal Z}}

\newcommand{\bz}{\bar{z}}

\newcommand{\bzeta}{\bar{\zeta}}
\newcommand{\veps}{\varepsilon}
\newcommand{\hveps}{\hat{\varepsilon}}
\newcommand{\bveps}{\bar{\varepsilon}}
\newcommand{\vrho}{\varrho}
\newcommand{\vphi}{\varphi}
\newcommand{\bphi}{\bar{\phi}}
\newcommand{\Gam}{\Gamma}
\newcommand{\bGam}{\bar{\Gamma}}

\newcommand{\Gamv}{{\boldsymbol \Gamma}}
\newcommand{\gam}{\gamma}

\newcommand{\alp}{\alpha}
\newcommand{\balp}{\bar{\alpha}}
\newcommand{\bbeta}{\bar{\beta}}
\newcommand{\bdelta}{\bar{\delta}}
\newcommand{\del}{\delta}
\newcommand{\Del}{\Delta}
\newcommand{\eps}{\epsilon}

\newcommand{\kap}{\kappa}
\newcommand{\bkap}{\bar{\kappa}}
\newcommand{\vkap}{\varkappa}
\newcommand{\vbkap}{\bar{\varkappa}}
\newcommand{\Lam}{\Lambda}
\newcommand{\bLam}{\bar{\Lambda}}

\newcommand{\Lamv}{{\boldsymbol \Lambda}}
\newcommand{\lm}{\lambda}
\newcommand{\lam}{\lambda}
\newcommand{\blam}{\bar{\lambda}}
\newcommand{\hlam}{\hat{\lambda}}
\newcommand{\tlam}{\tilde{\lambda}}

\newcommand{\hrho}{\hat{\rho}}
\newcommand{\bmu}{{\bar\mu}}
\newcommand{\cmu}{\check{\mu}}
\newcommand{\hmu}{\hat{\mu}}
\newcommand{\bnu}{{\bar\nu}}
\newcommand{\hnu}{\hat{\nu}}
\newcommand{\cnu}{\check{\nu}}

\newcommand{\bpi}{\bar{\pi}}
\newcommand{\vpi}{\varpi}
\newcommand{\Sig}{\Sigma}
\newcommand{\bSig}{\bar{\Sigma}}
\newcommand{\sig}{\sigma}
\newcommand{\bsig}{\bar{\sigma}}
\newcommand{\vsig}{\varsigma}

\newcommand{\Phiv}{{\boldsymbol \Phi}}

\newcommand{\Psiv}{{\boldsymbol \Psi}}

\newcommand{\tpsi}{\tilde{\psi}}
\newcommand{\bpsi}{\bar{\psi}}
\newcommand{\Th}{\Theta}
\newcommand{\bTh}{\bar{\Theta}}
\newcommand{\hTh}{\hat{\Theta}}
\newcommand{\tht}{\theta}
\newcommand{\vth}{\vartheta}
\newcommand{\btheta}{\bar{\theta}}

\newcommand{\Ups}{\Upsilon}

\newcommand{\omg}{\omega}
\newcommand{\Omg}{\Omega}

\newcommand{\ub}{\underbrace}
\newcommand{\ob}{\overbrace}
\newcommand{\os}{\overset}
\newcommand{\us}{\underset}

\newcommand{\Ai}{{\rm Ai}}
\newcommand{\Bi}{{\rm Bi}}

\begin{document}

\title{Transient fields of coherent synchrotron radiation\\ in a rectangular pipe}
\author{{\sc Tomonori Agoh}}
\date{\today}
\maketitle


\begin{abstract}
We found an exact analytical solution of the wave equation for a transient electromagnetic 
field of synchrotron radiation in the frequency domain.
The exact solution represents the field
which consists of the coherent and incoherent components of synchrotron radiation
and the space charge field of the particle beam moving in a bending magnet.
The field in the time domain is gotten by numerically Fourier transforming
the values of the field calculated using the exact solution.
The beam has an arbitrary charge density and current density
which satisfy the equation of continuity.
The beam is moving in a perfectly conducting rectangular pipe
which is uniformly curved in a semi-infinite bending magnet.
The exact solution is not self-consistent, \ie,
this is an exact expression of the field for a given beam current.
We do not solve the equation of motion of the beam in the present paper.
On the basis of the exact expression of the field found in the present study,
we discuss the applicability and accuracy of the paraxial approximation
which is sometimes used to calculate the field and spectrum of coherent or incoherent 
synchrotron radiation.
\end{abstract}

\tableofcontents
\listoffigures

\clearpage


\section{Introduction}
\label{sec:introduction}

We discuss a classical electromagnetic field of synchrotron radiation
which is emitted by relativistic particles in a bending magnet of a particle accelerator.
We would like first to introduce synchrotron radiation to explain
coherent synchrotron radiation which is the main subject in the present paper.
In an accelerator, a particle beam is bent by a dipole magnet and moves
 nearly on an arc of circle within it.
As it is generally known, if a moving charged particle receives a force perpendicular to
its velocity, it emits an electromagnetic wave called synchrotron radiation
forwards and backwards in general.
It has a continuous and broad spectrum up to the critical frequency which depends on $\gam$
\cite{schwinger} as will be shown in Eq.(\ref{eq:kc}),
where $\gam$ is the Lorentz factor given by Eq.(\ref{eq:z}).  
At a high energy as $\gam\gg1$, it is emitted almost within an angle of $\gam^{-1}$
in the forward direction.

\subsection{Coherent and incoherent synchrotron radiation}
\label{sec:csr_isr}

In this paragraph,
let us consider synchrotron radiation without limiting the discussion to accelerators
in order to understand the fundamental relation of synchrotron radiation to
the distribution of the source charge.
We suppose an annular electric current which is a continuous charge distribution
like a fluid, circulating around an axis.
If this annular current keeps a toroidal distribution which has an axial symmetry,
it does not emit synchrotron radiation even though the charge is circulating,
because, if it did, it would follow that the electromagnetic wave breaks
the axial symmetry of this system.
In terms of Fourier transform, an annular current, which has a uniform distribution in
the azimuthal direction, has the spectrum of a $\delta$-function at  zero frequency.
As seen from this example, when a current is bent, synchrotron radiation is emitted
if the current has a nonuniform charge distribution in the moving direction
such as a point charge, a Gaussian distribution and so on.
If we assume a point charge or a bunch of particles,
which has infinitely large  frequency components  in its Fourier series,
the synchrotron radiation has a broad spectrum up to the critical frequency
as described above.
As the source charge of the field if we assume, for example, a current which is
a continuous distribution of charge having a smooth Gaussian distribution along
a circular orbit,
then the synchrotron radiation has a spectrum up to a frequency which corresponds to
either the period of revolution or the length of the Gaussian distribution,
because this Gaussian current does not have higher frequency components
unlike a bunch of point charges or a rectangular continuous   distribution  in
the longitudinal direction.
Thus, in general, the longitudinal distribution of the current determines
the spectrum of synchrotron radiation.

A beam in an accelerator consists of a number of charged particles
which are bunched in the direction of motion.
Synchrotron radiation in an accelerator is the sum of the electromagnetic waves
emitted by the point charges forming the bunch.
In the field of light source which is devoted to the application of synchrotron light
mainly to material science, the phrase {\it synchrotron radiation} may usually refer to
the high frequency components around the critical frequency.
The radiation is incoherent if the wavenumber of the field, $k$, is much larger than
the inverse of $z_c$ which denotes a characteristic length of the bunch
such as the bunch length $\sig_z$ or the structure length in general
if the bunch has a density fluctuation in the moving direction.
The crests of the density fluctuation are equivalent to short bunches.
On the contrary, if the synchrotron radiation by a bunch has a frequency component
whose wavenumber $k$ is comparable or smaller than $z_c^{-1}$,
then the electromagnetic waves, emitted by the particles forming the bunch, are nearly on 
the same phase and hence are amplified by the number of  particles engaging in
the coherence.
Coherent synchrotron radiation (CSR) refers to  the low frequency components of
synchrotron radiation such that
\begin{align}
  kz_c\lesssim1 .
   \label{eq:kz0}
\end{align}
In the present paper we use the phrase {\it synchrotron radiation} in the sense of
the entire spectrum of the emitted wave by a bent current regardless of
the coherence and frequency band.

CSR is a collective effect of a bunch of charged particles in a bending magnet,
which is inseparable from the space charge effect of the particles in their vicinity.
Accordingly, in the present paper the term {\it field} means an electromagnetic field
which consists of coherent and incoherent synchrotron radiation
and space charge of the bunch.
In terms of mathematics we can distinguish the radiation and space charge 
as the terms of the oscillatory mode and damped mode with respect to
the direction of motion of the bunch.
Also, there are cases where we can distinguish them through a specific consideration such as
the asymptotic expression of the frequency domain field in the low frequency limit.
Since the spectrum of synchrotron radiation depends on
the longitudinal distribution of the bunch,
a higher frequency region will have coherence for a smaller $z_c$.
A shorter bunch consisting of more charges is more influenced by its own field of CSR.
It has the potential to cause problems to the bunch such as an nonuniform energy variation
which can give rise to a longitudinal microwave instability or a deterioration in
the quality of the beam, depending on the type of accelerator and its parameters.
At the moment, it is not clear to us how  particles are affected by
the horizontal Lorentz force of the field of CSR on a curved trajectory
and how it differs from that of the space charge field in a straight section.

It is true that the longitudinal distribution of the charge determines
the coherence of the field of synchrotron radiation.
Assuming a rigid bunch, however, we can describe the radiation as a quantity
which does not depend on the charge distribution of the bunch.
This is true because we can normalize the field in the frequency domain by the bunch spectrum
which denotes the distribution of the frequency components of the bunch.
This is the impedance which depends only on the geometry of the beamline and $\gam$.
In short, it is the Green function in the frequency domain,
which is called the propagator in the parlance of particle theory.
The impedance is a more fundamental quantity than the field itself in the sense that
it has no information on the longitudinal charge distribution of the bunch and
coherence of the field.
This is one of the reasons we define the phrase {\it synchrotron radiation} as
the entire spectrum of the emitted wave regardless of the coherence and range of frequency.

\subsection{Properties of coherent synchrotron radiation}
\label{sec:properties}

In terms of the geometric configuration, a beamline of an accelerator consists of
straight sections and bending sections.
In classical electromagnetism each particle forming a bunch constantly emits 
incoherent synchrotron radiation in a bend.
Figuratively speaking in this regard, the incoherent radiation is likened to a searchlight 
which is turned on or off when the particle is moving inside or outside the bend.
On the other hand, the field of CSR gradually forms as the bunch travels in the bend,
because it takes time so that the radiation, emitted by the particles
around the tail of the bunch, reaches the head part
by propagating on a shorter path than the arc-shaped trajectory of the bunch in the bend.
Then the field finally reaches a steady state with respect to the bunch if the bend is long 
enough.
Saldin, Schneidmiller and Yurkov discussed a transient field of CSR in free space by 
taking a bend of a finite length into account in a two-particle model
\cite{saldin_schneidmiller_yurkov}.
They derived an approximate expression of the energy change rate of the particles due to
CSR in the time domain.
CSR having a wavenumber $k$ has a rough criterion on the formation length of
the field, $s_{\rm f}$, along the trajectory of the beam in the bend,
which will be shown later in Fig.\ref{fig:asymp_csr} (p.\pageref{fig:asymp_csr}),
\begin{align}
  s_{\rm f}
  \simeq \bigg(\frac{24\rho^2}{k}\bigg)^{1/3} .
   \label{eq:formation}
\end{align}
$\rho$ is the bending radius of the beam trajectory which is an arc of circle.
If $k\rho\gg1$, the formation length $s_{\rm f}$ is much longer than $k^{-1}$, \ie,
the field of CSR is usually formed much slower than its wave period.

In addition to transient state, shielding of the field by a metallic vacuum chamber
is also an important effect in considering CSR, in particular,
if the bunch length is relatively long such as in a storage ring.
A field of steady CSR having a wavenumber $k$ has a transverse spread $\ell_{\perp}$
which is roughly given as
\begin{align}
  \ell_{\perp}
  \simeq \left(\frac{\rho}{2k^2}\right)^{1/3}
   \quad\Lra\quad
  k_{\rm th}
  \simeq \bigg(\frac{\rho}{d_{\perp}^3}\bigg)^{1/2} .
  \label{eq:shield}
\end{align}
Conversely, as shown by the second equation of (\ref{eq:shield}),
the wall of the vacuum chamber suppresses the emission of CSR in the range
below the wavenumber $k_{\rm th}$ called the shielding threshold \cite{warnock}.
$d_{\perp}$ denotes the transverse dimension of the vacuum chamber, \eg,
the width $w$ or height $h$ of the cross section if it is a rectangular pipe.
$k_{\rm th}$ differs from the so called cutoff frequency below
which electromagnetic waves cannot propagate in a waveguide.
The cutoff frequency depends only on the cross section of the waveguide
unlike $k_{\rm th}$ which depends also on $\rho$.
To be  precise, Eqs.(\ref{eq:shield}) each have a coefficient which depends on
the definition: we define $k_{\rm th}$ by $\Re Z\propto\exp[-(k/k_{\rm th})^2]$ in
\cite{agoh} and Eq.(\ref{eq:kth}) of the present paper,
which differs from Eqs.(1.2-1.3) in \cite{warnock} by $(2\pi/3)^{1/2}$,
where $Z$ denotes the longitudinal impedance of synchrotron radiation.
Also, the shielding effect depends on the angle of the wall of the vacuum chamber
with respect to the plane on which the beam trajectory lies, \ie,
parallel or perpendicular to the horizontal plane.
We can interpret the shielding effect by a vacuum chamber as a partial cancelation of
the waves which have nearly the opposite phase to each other
because of the fixed-end reflection on the wall of the vacuum chamber.
Assuming a pair of parallel plates, we can interpret the effect as the cancelation by
the radiation from the mirror charges which move together with the true charges.
The sidewalls of the beam pipe also have a shielding effect for the field of CSR,
while the effect is weaker than that by the upper-lower walls \cite{agoh}.

The shielding effect is negligible for the high frequency components of CSR
such as $k\gg k_{\rm th}$.
But if the beam channel has sidewalls, since the field can resonate on them,
the steady field has a discrete spectrum unlike the one in the absence of sidewalls.
Therefore we should take into account a pipe-shaped chamber like a waveguide
in calculating the spectrum of CSR even if the shielding effect is weak.
Warnock and Morton studied a field of CSR emitted by a beam which is circulating in
a rectangular toroid \cite{warnock_morton}.
They derived the exact expressions of the field and longitudinal impedance of a steady CSR.
In their model the field resonates also in the azimuthal direction,
because their model of the system has a periodicity along the circular orbit of the beam.

If $\gam\gg1$, as described in section \ref{sec:csr_isr}, the synchrotron radiation is 
emitted almost in the forward  direction of motion of the bunch.
In the bend, the field of CSR overtakes the bunch which moves along an arc of circle.
Therefore there is no field of CSR behind the bunch in the absence of vacuum chamber, \ie, 
in free space, only the space charge field of the bunch spreads behind it
within a short range which is roughly given by the inverse of Eq.(129) in \cite{agoh}.
On the other hand, in the presence of a vacuum chamber,
the field can lag behind the bunch, \ie, it has a tail,
because the group velocity of the field in a vacuum chamber is smaller than
the speed of the bunch along the trajectory.
We can interpret this lag as the effect that
the wave, reflected on the wall, propagates along a longer path than
the arc-shaped trajectory of the bunch in the bend.
For the sake of clarity, we can add that the lag of the field behind the bunch due to
the effect of the vacuum chamber is not propagation backward (toward the upstream) in
the beamline.
If $\gam\gg1$, the field of synchrotron radiation propagates almost forward
while reflecting on the walls of the vacuum chamber.

After being emitted in a bend, the field of synchrotron radiation goes out of it
together with the bunch through the exit of the bend and will propagate in
the next straight section.
In terms of the dynamics of collective effect, the energy change of the particles
by their own field of CSR in the next straight section can be larger than that in the bend, 
depending on the parameters of the bunch and geometry of the beam channel.
When the field of CSR traverses the common area between the bend and straight section,
a part of the field can, in theory, reflect and will propagate backward in 
the bend since the curvature of the beam pipe abruptly changes on the common area.
However, the reflection of the field around
the common area is usually much smaller than the transmitted field.
In quality, this is similar to the reflection and transmission of light on a boundary
between two media which have different indices of refraction.

\subsection{Theories and calculations of the field of CSR}
\label{sec:theories}

There are various ways to formulate and calculate the electromagnetic field due to CSR.
Ideally  we wish we had an expression of the field
taking a more realistic model into consideration, 
but in reality it is not so easy to formulate the field of CSR
since the causality of the field and particles is complicated in a  curved beam channel.
In order to  solve Maxwell equations analytically,
we have to reconcile ourselves to employing a relatively simple model,
otherwise we have to rely on numerical calculations to solve it.
In order to solve the differential equation for the field by a numerical method such as
 finite difference, the grid must, in general, be fine enough to resolve the field, \ie,
the spacing of the grid points must be much smaller than $k^{-1}$, where
$k$ is the wavenumber of the field.
The numerical calculation of CSR in an accelerator requires a large number of grid points 
since the dimension of the bend or beamline is of the order of one meter to tens of meters
which is very much larger than $k^{-1}$.
This large difference between the scale lengths of the electromagnetic wave and bend is
one of the difficulties in performing a numerical calculation of the field of CSR
using a grid in an accelerator  beamline.

Using a paraxial approximation,  we derived a parabolic wave equation for
a field in the frequency domain in order to calculate a transient field of 
CSR which is shielded by a beam pipe \cite{agoh_yokoya}.
Then we numerically solved it using a finite difference on a grid in a rectangular pipe
which has a finite or infinite conductivity.
Since the boundary condition on a resistive wall is formulated in the frequency domain,
it is relatively easier to take the resistivity of the walls into consideration
in the numerical calculation of the field in the frequency domain
rather than the time domain.
In addition to the field in a bend, we calculated the field which propagates in
the  straight section following the bend.
Stupakov and Kotelnikov  solved analytically the parabolic equation for the field in
a perfectly conducting pipe \cite{stupakov_kotelnikov}.
In general, analytical study can provide physical pictures and instances of 
the phenomenon in terms of mathematics.
Analytical solutions have the additional merit that they provide the value of the field,
usually, much faster than the numerical calculation using  finite differences,
in particular, for the field at higher frequency which requires a finer grid.

The theories in \cite{warnock,agoh,warnock_morton,agoh_yokoya,stupakov_kotelnikov} are 
formulated in (or through) the frequency domain.
On the other hand, the theory in \cite{saldin_schneidmiller_yurkov} is formulated
throughout in the time domain.
The latter requires somewhat complicated geometrical considerations about the positions of
the particles and field across the edges of the bend, because we must take into account
the advancement and retardation of the field in the beamline which consists of
straight and bending sections.
Furthermore, if one wants to take a beam pipe into account in the beam channel,
it must be difficult to formulate the field of CSR throughout in the time domain
since the causality of the field is very complicated
due to the reflection on the walls of the curved pipe.
In this regard, it is relatively easier to formulate the field of CSR in the frequency 
domain, because the causality of the field and particles is decomposed into
the frequency components.
Therefore, in calculating the field in the frequency domain, we do not have to consider
the complicated geometry which involves the positions of the particles and field at each time.
It is enough to match the value of the field in frequency domain
on the common area of the straight and bending sections.
On the other hand, in the calculations of the field in the frequency domain,
since the current is also Fourier transformed into the frequency domain,
we cannot straightforwardly take into account the dynamic redistribution of the current
by its own field.
Thus the method of calculation in the frequency domain hardly has
a self-consistent formalism or algorithm.
In this regard, it is easier to take into account the dynamic variation of
the current distribution by its own field in the time domain.

Let us review the theories on synchrotron radiation from a different point of view.
The papers \cite{warnock,warnock_morton} present exact theories which are formulated
in terms of the Bessel functions.
On the other hand, the theories in
\cite{schwinger, saldin_schneidmiller_yurkov, agoh, agoh_yokoya,stupakov_kotelnikov}
are formulated on the basis of the paraxial approximation
which is valid at high frequency such as $k\rho\gg 1$ in free space or $kd_{\perp}\gg\pi$ in
the presence of a vacuum chamber, where $d_{\perp}$ denotes the transverse dimension of
the chamber as described below Eq.(\ref{eq:shield}).
Therefore, as shown by Eq.(D13) in \cite{stupakov_kotelnikov_0} and Eq.(208) in \cite{agoh},
it is possible to derive the formula for the power spectrum of synchrotron radiation
respectively in different ways from the one in \cite{schwinger}.
The theories on the basis of the paraxial approximation have a simpler formalism than
the exact ones.
The paraxial approximation have an accuracy often good enough to calculate the field at
high frequency from a practical point of view.
So far, the paraxial approximation has actually contributed to the calculations of
the field and spectrum of synchrotron radiation at high frequency.
In addition, it is important also in understanding the basic properties of CSR through 
the asymptotic expressions of the field and impedance.
Conversely, the paraxial approximation has a limit of applicability, \ie,
it produces an error which is not negligible at low frequencies such as
$k\rho\lesssim1$ in free space or $kd_{\perp}\lesssim\pi$ in the presence of
a vacuum chamber.

There are various ways to calculate the field of CSR as described above.
Each method can be characterized by exact/approximate method,
the analytical/numerical method and the explicit/implicit method.
We also have other choices in calculating the field such as the time/frequency domain,
the electromagnetic field or its potential with the aid of the method of image charges
if one wants to take into account the shielding effect by infinite parallel plates.
In addition, there are several kinds of models of the system to calculate the field of CSR:
free space/parallel plates/pipe, a circular orbit or a finite trajectory in a single bend.
Depending on the model and parameters of the calculation,
these various methods  have both  advantages and disadvantages such as
solvability, extensibility, tractability, simplicity, self-consistency,
clarification of the physical picture and spectrum,
efficiency and speed of calculation, accuracy and range of applicability, etc.
As far as we know, no method covers all these points, \ie,
these various methods are often complementary to each other.

In general, numerical methods have higher solvability than analytical ones.
But numerical calculation of the field is usually more time consuming than
that which uses an analytical solution.
The parabolic equation in the paraxial approximation allows us to calculate the field of 
CSR much faster than the exact method
since the factors of the fast oscillation in the time domain field are removed,
similar to a wave function in Schr\"odinger equation.
However, we cannot use the paraxial approximation if the shielding by the vacuum chamber is
very strong.
We need an exact solution to precisely calculate the field of CSR at low frequency
such as $kd_{\perp}\lesssim \pi$.
If we have the exact solution, we can examine the error of a calculation
in the paraxial approximation.
Also, with this method, it is difficult to calculate the horizontal Lorentz force
which the particles receive from the field of CSR,
because the space charge field in the bend is also involved.
That is, the grid to calculate the field in the pipe is not fine enough to
resolve the horizontal charge distribution of the bunch.
These defects of the paraxial approximation and numerical calculation motivate us to find
an exact and explicit expression of the transient field of CSR emitted in a curved beam pipe.
We think that the theory presented in the present paper has some worth in improving or 
designing an accelerator in which the influence of CSR would be a concern
because of a short bunch and/or a large current.
But this is just one of our motivations to engage in this study.
As a theorist, we want to  pursue  a more precise or exact theory in
the field of synchrotron radiation regardless of its utility and application.
This motivated us to start and complete this work.
In science, it is not sound to attach too much importance to utilities and applications.

\subsection{Outline of the present paper}
\label{sec:outline}

This paper presents the classical theory of an electromagnetic field of
synchrotron radiation which has coherent and incoherent components.
As described in section \ref{sec:csr_isr}, the field includes also the space charge field 
of the source current.
The present goal of  this  study is to get an explicit and exact expression of
the transient field of synchrotron radiation in the frequency domain,
taking into account the shielding by a rectangular pipe like a waveguide.
In classical electromagnetism, the Green function often means an expression of
a field which is created by a point charge instead of a charge distribution.
In other words, the Green function is the solution of Maxwell equations
in which we replace the charge distribution by a $\delta$-function of
the spatial coordinate(s).
Therefore, in this usage, the impedance denotes the Green function in the frequency domain.
On the other hand, throughout the present paper, we use the term {\it Green function}
in the sense of the expression of the field, which is gotten by solving
the wave equation whose driving term is replaced by a $\delta$-function of
the spatial coordinate(s).

We outline the present paper which consists of 10 sections and 21 appendices.
In section \ref{sec:frame} we introduce a curvilinear coordinate system and
model of the system to calculate the field.
We first employ the curvilinear coordinate system as a frame of reference to derive
the wave equation for an arbitrary field at any position in a beamline
regardless of the curvature of the beam trajectory.
We need this general wave equation using the curvilinear coordinate system
to understand the  singular behavior of the field on the common area
between a straight section and neighboring bending section as discussed 
later in section \ref{sec:disconti}.
In solving the wave equation for the field in each section,
after removing the singularity of the field on the common area,
we rewrite the general wave equation in the curvilinear coordinates
of the straight and bending sections
respectively using the Cartesian and cylindrical coordinate systems
as shown in appendix \ref{sec:travel_wave} and section \ref{sec:we_fc}.

In section \ref{sec:we} we begin the formulation with the wave equation for
the field in the time domain.
Similar to the way in \cite{warnock_morton},
we first Fourier transform the field and current in the frequency domain.
However, our model differs from \cite{warnock_morton},
because the authors in \cite{warnock_morton} 
consider a periodic field along the circular orbit of a beam,
which is resonating in a rectangular toroid.
As described later in section \ref{sec:assumption},
our model of the vacuum chamber is not a toroid but a uniformly curved pipe
which is connected to a straight pipe at the entrance of the bend.
In section \ref{sec:disconti} we discuss the singularity of the field in the common area
between the straight and bending sections.
In section \ref{sec:ymode} we expand the field and current in the frequency domain into
the modes of the vertical oscillation between the upper-lower walls of the rectangular pipe.
In order to solve the wave equation for the Fourier coefficients of the transient field in
the frequency domain, we Laplace transform the Fourier coefficients of the field and current
with respect to the reference axis along the beamline.
By this transform, we will have the wave equation for the Fourier coefficients of
the frequency domain field in the Laplace domain.
The Laplace transform requires the initial values of the field and its derivative with
respect to the reference axis at the entrance of the bend.
In section \ref{sec:LD_field} we solve the wave equation for the Laplace domain field
under the boundary condition of the sidewalls of the curved pipe in the bend.
Then we get the analytical solution of the wave equation for the field in the Laplace domain.

In order to get the expressions of the Fourier coefficients of the field through
the inverse Laplace transform, section \ref{sec:pole} examines the singularity of
the field in the Laplace domain.
This analysis shows that it is possible to calculate the Bromwich integral
using the residue theorem in the Laplace plane.
In section \ref{sec:ILT} we transform the field in the Laplace domain to
its Fourier coefficients.
In section \ref{sec:expression}
we summarize the various  expressions of the field.  
The expression of the field gotten in section \ref{sec:ILT} is useful in
calculating the field value on a computer
since it does not involve the radial and longitudinal $\delta$-functions.
In section \ref{sec:normal_form} we rewrite the Fourier coefficients of the field
into expressions which are useful in the analytical calculation.
At last, in section \ref{sec:FD_field} we will have the expression of the field in
the frequency domain from the Fourier coefficients.
Thus we get the explicit expression of the frequency domain field in the bend,
which represents the exact relation between the field in the bend and its initial values
at the entrance of the bend.
Eq.(\ref{eq:transforms}) gives a summary on the transforms and expansion of the field
in the present derivation.
In terms of the exact solution of the field gotten in the present paper,
section \ref{sec:discussion} discusses the validity of the calculation of the CSR field
on the basis of the paraxial approximation.
In section \ref{sec:conclusion} we make some remarks on the present study.
Also, we would like to describe our perspective for further developments of
the theory and calculation on CSR.

The present paper has 24 appendices where we summarize the mathematical considerations 
including particular functions which we need in formulating the field of
synchrotron radiation.
In addition, we describe the asymptotic limit of the exact expression of the field in
the bending magnet to that in the straight section in order to verify
the exact expression analytically.
Appendix \ref{sec:coordinates} describes the curvilinear coordinate system and
differential operations on the basis of a planar reference trajectory.
The general wave equation using the curvilinear coordinates has a singular term
due to the discontinuity of the field derivative with respect to
the reference axis at the edge (common area) of the bend
since we assume a hard edge bend as described in section \ref{sec:assumption}.
In appendix \ref{sec:discontinuity} we show that we can remove the singularity 
of the field on the common area while keeping the wave equation exact, \ie,
without making any approximation in the wave equation.
In the intermediate calculation to find the exact solution of the field,
we use Maxwell equations and the equation of continuity for the beam current
in order to rearrange equations into elegant forms depending on the context.
In appendices \ref{sec:boundary} and \ref{sec:Maxeqs_LD} we write down Maxwell equations
for the Fourier coefficients of the field and those in the Laplace domain respectively.
Appendix \ref{sec:eqcontin} shows the equations of continuity for the beam current in
all the four domains of the present derivation:
time domain, frequency domain, vertical Fourier mode and Laplace domain.

As described in the last paragraph of section \ref{sec:properties},
the field of synchrotron radiation is emitted in the forward direction out of a bend
and propagates in the next straight section together with the bunch.
The particles forming a bunch are affected by the field of CSR also
in the straight section following the bend.
In order to evaluate the influence of the field on  the bunch in the straight section,
as shown in appendix \ref{sec:travel_wave}, we formulate an arbitrary electromagnetic field 
which propagates in a straight rectangular pipe together with a current.
Thus we derive the exact expression of a transient field propagating in
a straight pipe in a similar way to deriving the solution of the field in the bend.
Assuming a beam moving in a straight section, we can define the space charge field of
the beam unlike the one in a bend in which
the space charge field is mixed up with the radiation field.
In section \ref{sec:space_charge},
using the final value theorem of the Laplace transform, we derive the expression of
the space charge field which clings to a rigid bunch moving in a straight rectangular pipe.
Then in section \ref{sec:discussion} we use the space charge field in
the straight pipe as the approximate initial value of the field at the entrance of the bend
to  calculate  numerically  the field of CSR.
However, this is an approximate use, because the field partially reflects at
the entrance of the bend due to the abrupt change in curvature of the trajectory.
Since a real straight section ends at the next bending section, to be precise,
the partial reflection of the field can occur also at the exit of the straight section.
In the present paper we do not have a rigorous discussion on the reflection and 
transmission of the electromagnetic field in the common areas of the beamline.

As shown later by Eqs.(\ref{eq:EBy_trans}-\ref{eq:EBxs_trans}) in section \ref{sec:ILT},
we have three ways to get the expressions of the radial and longitudinal components of
the fields.
In section \ref{sec:sol_mfEB_xs} we derive them from the vertical components of
the fields through Maxwell equations in the Laplace domain,
which are given in appendix \ref{sec:Maxeqs_LD}.
On the other hand, as shown in appendix \ref{sec:we_xs}, we can get them also by 
straightforwardly solving the wave equations for the radial and longitudinal components of 
fields in the Laplace domain using the eigenfunctions of the operator of the wave equations.
Furthermore, we can derive them from the Fourier coefficients of the vertical fields
as shown in appendix \ref{sec:HL_field_sdom}.
We show all these three ways on purpose to verify all the components of the fields
analytically.

We use the cross products of the Bessel functions in describing the field of
synchrotron radiation since we deal with the exact solution of the field in
a uniformly curved pipe as the vacuum chamber in the bending section.
In appendix \ref{sec:bessel} we summarize useful identities of these cross products
in deriving and verifying the expressions of the field.
We deal also with the Bessel functions of purely imaginary order, because the field has 
also the imaginary poles in the Laplace plane as shown in section \ref{sec:pole}.
We can calculate these functions as well as the modified Bessel functions of purely 
imaginary order using Dunster's expressions of the Bessel functions \cite{dunster}.
In appendix \ref{sec:dunster} we summarize the useful formulae involving them.

To understand the pole structure of the field in the Laplace domain in
section \ref{sec:pole}, we use Cochran's study on the zeros of the cross products
with respect to the order of the Bessel functions \cite{cochran}.
We briefly introduce his work in appendix \ref{sec:cochran}.
Appendix \ref{sec:analytic} shows our interpretation and treatment for the pole of
the second order which the field can have at the origin of the Laplace plane.
This may be important in the inverse Fourier transform of the field to the time domain.
The exact expression of the field gotten in the present study is explicit except for
the expressions of the poles of the field in the Laplace domain and those of
the cutoff wavenumbers of the curved pipe.
That is, we cannot find the explicit and exact expressions of the zeros of
the cross products of the Bessel functions with respect to their order.
As appendix \ref{sec:poles} shows, however, we can find the asymptotic expressions of
the zeros using the uniform asymptotic expansions of the Bessel functions
under some assumptions.
In appendix \ref{sec:orthogonality} we find the orthogonality relations of
the cross products of the Bessel functions with respect to their orders,
which we need to verify the solution of the field.
Then appendix \ref{sec:verify} demonstrates that the solution of the field actually
satisfies the exact wave equations and Maxwell equations in the frequency domain.

As shown later by Eqs.(\ref{eq:4express}), we have four expressions of the field of 
synchrotron radiation in the bend.
We explicitly write out one of the expression in appendix \ref{sec:opera_express},
which is suited to the numerical calculation of the field on a computer.
In appendix \ref{sec:steady_field} we find the expressions of a steady field and 
longitudinal impedance of synchrotron radiation by using the final value theorem of
the Laplace transform under the assumption of a rigid and thin bunch.
We will show that this impedance is equivalent to that given in \cite{warnock_morton} for
\begin{align}
  k\rho\in\mathbb{Z}
  ~\to~ k\rho\in \mathbb{R}
   \label{eq:krho}
\end{align}
which was discussed around Eq.(A6) in appendix A of \cite{agoh}.
In short, this is the limit corresponding to the removal of the longitudinal periodicity of
the field along the beam trajectory.
In appendix \ref{sec:impedance} we derive the expressions of a transient field and 
impedance of synchrotron radiation under the assumption of a rigid and thin bunch.
These expressions may be useful in a simple calculation of the field of CSR.
In the present study we use the Fourier transform with respect to time,
which differs from the one in \cite{agoh}.

\clearpage

\section{Framework of the theory}
\label{sec:frame}

\subsection{Symbols and notation}

We use the following symbols and notation in the present paper.

\vspace{2mm}

\noindent
Physical constants in the SI Units
\begin{itemize}
\setlength{\itemsep}{0mm}
\item
  $c$, $\eps_0$, $\mu_0$:
  Speed of light in vacuum, permittivity and permeability of vacuum; $\eps_0\mu_0=c^{-2}$.
\item
  $Z_0=c\mu_0 \,[\Omg]$: Impedance of vacuum.
\end{itemize}

\noindent
Curvilinear coordinate system and time
\begin{itemize}
\setlength{\itemsep}{0mm}
\item
  Space $\xv=(x,y,s)$ and time $t$.
  Transverse coordinates $\xv_{\perp}=(x,y)$ and longitudinal coordinate $s$.
\item
  $z$: Advance/delay from the reference particle
  in terms of length along the $s$-axis instead of $t$; Eq.(\ref{eq:z}).
\item
  $r$: Radial variable in a constant bend;
  Eq.(\ref{eq:r}).
\item
  $\omg$, $k$: Frequency and total wavenumber;
  Eqs.(\ref{eq:Fourier_trans}-\ref{eq:omg}).
\item
  $\nu$: Laplace variable (rotated clockwise by $\pi/2$);
  Eq.(\ref{eq:u_nu}).
\end{itemize}

\noindent
Mathematical symbols and notation
\begin{itemize}
\setlength{\itemsep}{0mm}
\item 
      Kronecker delta: $\delta_{j}^{\ell}\in\{0,1\}$ ; Eq.(\ref{eq:lprm}).
\item 
      Reversals of the Kronecker delta:
      $\bdelta_{0}^{\ell},\bdelta_{01}^{\ell}\in\{0,1\}$;
      Eqs.(\ref{eq:d01_d0}).
\item 
      Heaviside step function and Dirac $\delta$-function of $s$:
      $\theta(s)\in\{0,1\}$ and $\delta(s)=\rd_s\theta(s)$;
      Eqs.(\ref{eq:step}-\ref{eq:delta_sgn}).
\item 
      Extended step function and $\delta$-function of $\vsig=\{s,s-s'\}$:
      $\btheta(\vsig)$ and $\bdelta(\vsig)=\rd_{\vsig}\btheta(\vsig)$;
      Eqs.(\ref{eq:bstep}-\ref{eq:bdelta}).
\item
      Window functions of the wavenumber:
      $\Th_{\pm}^{n\ell}\in\{0,1\}$; Eqs.(\ref{eq:Th_def}-\ref{eq:sum_Th}).
\item
      Floor function: $\lfloor x\rfloor$ denotes
      the largest integer less than $x\in\mathbb{R}_0^{+}$.
\item 
      Infinitesimals: $\pm0=\pm\eps$ and $\pm i0=\pm i\eps$ ($\eps\to+0\in\mathbb{R}^{+}$).
\item
      The double signs ($\pm$) correspond in each equation unless describing unrelated.
\item 
      The double subscripts correspond as $g_{a,b}=r_{a,b}/\rho$ represents
      $g_a=r_a/\rho$ and $g_b=r_b/\rho$.
\item 
      The quantities in the parentheses (or braces) correspond as
      $(g_a,g_b)=(r_a,r_b)/\rho=1+(x_a,x_b)/\rho$.
\item
  For brevity, we write the derivative of a single variable function $f(s)$
  using the symbol of the partial derivative $\rd_s=\rd /\rd s$ as $df/ds=\rd_sf$,
  \eg, $\delta(s)=d\theta(s)/ds=\rd_s\theta(s)$.
  If $f$ is a multivariable function $f(x,s)$, $\rd_sf$ denotes
  the partial derivative of $f$ with respect to $s$ as usual, \ie, $\rd_sf=\rd f/\rd s$.
\item
  In introducing a symbol
  which is used in another paper in a different meaning from the present paper,
  we use the symbol with double quotation marks in order to avoid confusion.
  For example, ``$k$'' denotes the radial wavenumber of the field in \cite{horvat_prosen},
  which differs from $k$ in the present paper.
\end{itemize}

\noindent
Sets of numbers
\begin{itemize}
\setlength{\itemsep}{0mm}
\item[] $\mathbb{N}=\{1,2,3,\cdots\}$: All natural numbers excluding 0.
\item[] $\mathbb{Z}=\{0,\pm n|n\in\mathbb{N}\}$: All integers.
\item[] $\mathbb{Z}_0^{+}=\{0,n|n\in\mathbb{N}\}=\mathbb{N}+\{0\}$:
       All nonnegative integers.
\item[] $\mathbb{Z}^{-}=\{-n|n\in\mathbb{N}\}=\mathbb{Z}-\mathbb{Z}_0^{+}$:
       All negative natural numbers.
\item[] $\mathbb{Z}_0^{-}=\{-n|n\in\mathbb{Z}_0^{+}\}=\mathbb{Z}^{-}+\{0\}$:
       All negative integers including 0.
\item[] $\mathbb{R}=$ Set of all real numbers.
\item[] $\mathbb{R}^{+}=\{x|x\in\mathbb{R},\, x>0\}$:
       All positive real numbers.
\item[] $\mathbb{R}_0^{+}=\{x|x\in\mathbb{R},\, x\geq0\}=\mathbb{R}^{+}+\{0\}$:
       All nonnegative real numbers.
\item[] $i\mathbb{R}=\{ix|x\in\mathbb{R}\}$:
       All purely imaginary numbers including 0.
\item[] $i\mathbb{R}^{+}=\{ix|x\in\mathbb{R}^{+}\}$:
       All purely imaginary numbers on the upper half-plane excluding 0.
\item[] $\mathbb{A}=\{x,ix|x\in\mathbb{R}\}=\{\mathbb{R},i\mathbb{R}\}$:
       All real and purely imaginary numbers (\ie, on the axes in $\mathbb{C}$).
\item[] $\mathbb{A}_0^{+}=\{x,ix|x\in\mathbb{R}_0^{+}\}=\{\mathbb{R}_0^{+},i\mathbb{R}^{+}\}$:
       All numbers on the positive real and imaginary axes in $\mathbb{C}$.
\item[] $\mathbb{C}=\{x+iy|x,y\in\mathbb{R}\}$: All complex numbers.
\end{itemize}

\clearpage

\subsection{Coordinate system}
\label{sec:curv_coordinates}

We define a curvilinear coordinate system on the basis of a planar reference trajectory
in order to express a beamline which has no vertical bend.
This is the planar version ($\vkap_y=0$) of the coordinate system given in
appendix A of \cite{yokoya} in which the vertical curvature of the reference axis
$\vkap_y$ is taken into consideration in a different way from
the Frenet-Serret formula which uses the torsion of the reference axis
instead of $\vkap_y$ to express a three dimensional curve.
Assuming a planar reference trajectory, however, the latter is equivalent to the former
since the reference trajectory has no torsion.

In defining the curvilinear coordinate system, 
we first consider a reference particle which moves smoothly at
a constant speed on a plane in a three dimensional Euclidean space.
Then we define $s$ as the length along the trajectory of the reference particle toward
the downstream of the particle from a fixed point defined as the origin $(s=0)$.
This is the reference axis ($s$-axis) which defines the longitudinal direction of
the beamline.
We define $\ev_s$ as the unit tangential vector of the $s$-axis,
which depends on $s$ in general as shown in Eq.(\ref{eq:coordinates}).
$\ev_s$ is differentiable with respect to $s$ since we assume that
the reference trajectory is smooth.
In addition to $\ev_s$, we define $\ev_x$ and $\ev_y$ respectively as the horizontal and 
vertical unit vectors such that the triad $(\ev_x,\ev_y,\ev_s)$ forms a right-handed 
orthonormal basis of the curvilinear coordinates $\xv=(x,y,s)$ in this order.
Thus we define the transverse coordinates $\xv_\perp=(x,y)$ where
$x$ and $y$ represent the horizontal and vertical deviations from the $s$-axis.
In the present paper, since we assume that the reference axis is planar,
$\ev_y$ does not depend on $s$,
\begin{align}
  \frac{d}{ds}
  \Bigg(\!\begin{array}{l} \ev_x \\ \ev_y \\ \ev_s \end{array}\!\Bigg)
  =\vkap \Bigg(\!\begin{array}{c} \ev_s \\ 0 \\ -\ev_x \end{array}\!\Bigg) ,
    \qquad
  \vkap
  =\frac{1}{\vrho} ,
    \qquad
  g=1+\vkap x .
  \label{eq:coordinates}
\end{align}
$\vkap$ and $\vrho$ are the local curvature and radius of the $s$-axis,
which depend on $s$ in general since the geometry of the $s$-axis is arbitrary
if it is a differentiable curve lying on the horizontal plane $y=0$ perpendicular to $\ev_y$.
When $\ev_x$ points to the center of curvature,
$\vkap$ has a negative value and vice versa.
$g$ is the geometric factor which denotes the ratio of $\vrho+x$ to $\vrho$ at a certain $s$.
$g$ is a function of $x$ and $s$ in general.
To be precise, we define $x$ in $x>-|\vrho|$ so that
any point in this system is uniquely specified in reference to $s$.
When $\vkap=0$, $\xv=(x,y,s)$ is the local Cartesian coordinate system
which means a straight section of the beamline.

We define the nabla $\nablav$ using $d{\bm\ell}$
which is the line element vector in the curvilinear coordinate system,
\begin{align}
  d{\bm\ell}
  =\ev_xdx+\ev_ydy+\ev_sgds ,
    \qquad
  df
  \equiv  d{\bm\ell}\cdot\nablav f
   \quad~~\Lra~~\quad
  \nablav=\ev_x\rd_x+\ev_y\rd_y+\ev_s\brd_s,
   \label{eq:nabla}
\end{align}
where $f$ is an arbitrary function which is differentiable with respect to $x,y$ and $s$.
In the present paper we use the following differential operators with respect to
the space $\xv=(x,y,s)$ and time $t$ for simpler expression,
\begin{align}
  \brd_x=\rd_x+\frac{\vkap}{g}
   ,\qquad
  \brd_s=\frac{\rd_s}{g}
   ,\qquad
  (\rd_x,\rd_y,\rd_s)
  =\bigg(\frac{\rd}{\rd x},\frac{\rd}{\rd y},\frac{\rd}{\rd s}\bigg)
   ,\qquad
  \rd_t
  =\frac{\rd}{\rd t}
  =-v\rd_z .
  \label{eq:brdxs}
\end{align}
$\brd_x$ and $\brd_s$ are the horizontal and longitudinal differential operators
which come out of the differential operations in the curvilinear coordinates
as seen from Eqs.(\ref{eq:curlA}-\ref{eq:divA}).
$\brd_{x,s}$ is the curvilinear version of $\rd_{x,s}$ in the sense that
$\brd_{x,s}$ goes to $\rd_{x,s}$ in the limit of $\vkap\to0$.
$\brd_{x,s}$ does not commute with $\rd_{x,s}$ in general.
Taking the advantage of the incommutability between $\brd_x$ and $\rd_x$,
we can briefly write the terms which involve the derivatives with respect to $x$
in the Laplacian $\nablav^2$ as shown in Eqs.(\ref{eq:laplacian}-\ref{eq:nablav2A}).
If $\vrho=\rho$ (constant), the curvilinear coordinates are the local cylindrical 
coordinates denoting a constant bend as shown later in Eq.(\ref{eq:r}).
See appendix \ref{sec:coordinates} for the details on the differential 
operations in the curvilinear coordinates.

The reference particle is not a particle forming a beam
but an imaginary particle that is used to define the frame of reference.
Let the reference particle pass the origin ($s=0$) at time $t=0$
while keeping a constant speed $v$.
We define $z$ using $v$ as
\begin{align}
  z=s-vt
   ,\qquad
  dz=-vdt
   ,\qquad
  v=c\beta
   =\text{const.}
   \quad
  (0<\beta<1)
   ,\qquad
  \gamma=(1-\beta^2)^{-1/2} .
   \label{eq:z}
\end{align}
$\beta$ is the relative speed of the reference particle to the speed of light $c$ in vacuum.
$\gamma$ is the Lorentz factor.
We do not necessarily assume $\gamma\gg 1$ in formulating the present theory.
In the Eulerian specification, an electric current and electromagnetic field are expressed
in terms of $\xv$ and $t$ which are independent of each other
unlike the case of the Lagrangian specification.
Instead of $(\xv,t)$, we can describe the current and field using $(\xv,z)$.
In section \ref{sec:discussion} we use $z$ instead of $t$ in showing
transient fields of CSR in the time domain.
As a special case, if the current and field do not depend on $s$ explicitly,
\ie, if they depend on $s$ and $t$ only through $z$ in the time domain, then we refer to
the beam as ``rigid'' and to the field as ``steady'' with respect to the beam.
In other words, rigid beam means that the longitudinal charge distribution of the beam 
does not change while moving in the beamline regardless of the straight or bending section
as described later in section \ref{sec:discussion}.
In the present paper, rigid beam does not mean that the beam is not bent by
an external field.

\clearpage

\subsection{Assumptions}
\label{sec:assumption}

\begin{figure} [h]
  \begin{center}
    \includegraphics[scale=0.6,clip]{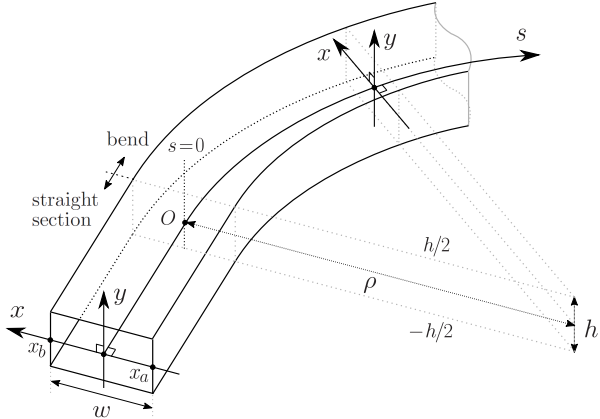}
    \caption[Beamline and coordinate system]
       {\small
       Beamline and coordinate system.
       The reference axis ($s$-axis) lies on the median horizontal plane ($y=0$).
       $O$ is the origin $\xv=0$.
       The bend has an entrance at $s=0$.
       The $s$-axis has a curvature radius $\vrho=\rho$ in $s\geq 0$ (bend)
       and $\vrho=\infty$ in $s<0$ (straight section).
       The rectangular pipe represents the vacuum chamber which extends along the $s$-axis.
       $w$ and $h$ are the full width and full height of the beam pipe.
       $x_a$ and $x_b$ are the horizontal positions of
       the inner and outer walls of the beam pipe.
       The upper and lower walls of the beam pipe are located at $y=\pm h/2$.
        }
    \label{fig:pipe3D}
  \end{center}
\end{figure}

We will solve the exact wave equation for the electromagnetic field created by
a beam current which is going into a dipole magnetic field from a drift space.
Fig.\ref{fig:pipe3D} shows the model of the system to find the exact expression of
a transient field of (coherent) synchrotron radiation in the frequency domain.
We assume the following in formulating the present theory:
\begin{itemize}
\item[(a)]
  Beamline
  \\
  We assume that the beamline consists of a straight section and a bending section
  which are connected at $s=0$.
  The bending section is a uniform dipole magnet (constant bend)
  which has a hard edge at $s=0$ (no fringe field).
  This bend has no exit, \ie, it extends semi-infinitely in the longitudinal direction
  from $s=0$ to $\infty$ without overlapping.
  However, this beamline is not a ring having a periodicity with respect to $s$.
  The $s$-axis has a curvature radius $\vrho=\infty$ in $s<0$ and
  $\vrho=\rho$ (constant) in $s\geq0$.
  In other words, the curvature of the $s$-axis changes only at $s=0$ discontinuously
  from 0 to a constant $\rho^{-1}$ like a step function.
  We assume that there is no external electromagnetic field in the beamline
  except for the uniform dipole magnetic field in $s\geq0$.
\item[(b)]
  Beam pipe (vacuum chamber)
  \\
  We assume that the vacuum chamber is a perfectly conducting pipe.
  In accordance with the $s$-axis assumed in (a),
  the beam pipe consists of a straight section in $s<0$ and
  a uniformly curved section (bending section) in $s\geq0$.
  There is a uniform dipole magnetic field inside the curved pipe.
  Along the $s$-axis, the pipe has a uniform cross section
  which is a rectangle of a full width $w$ and full height $h$.
  We define $x_a$ and $x_b$ as the horizontal position of the inner and outer walls
  (sidewalls) of the rectangular beam pipe.
  They are constants which satisfy
   \begin{align}
     w=x_b-x_a
      \quad\text{and}\quad
    -\rho<x_a<0<x_b .
   \end{align}
  For simplicity, we assume that the upper and lower walls of
  the rectangular pipe are located at $y=\pm h/2$ symmetrically with respect to
  the median horizontal plane $(y=0)$ on which the $s$-axis lies.
  We assume that the walls of the beam pipe have a smooth surface (no roughness).
\item[(c)]
  Beam
  \\
  We consider a beam which is an electric current flowing around the $s$-axis
  toward the positive direction in the beam pipe.
  We assume that the beam is bunched in the transverse and longitudinal directions.
  There is no electric charge on the walls of the beam pipe and at $\pm\infty$.
  The charge and current densities of the bunch are arbitrary
  if they satisfy the equation of continuity in the beam pipe.
  Also, the motion of the particles forming the bunch is arbitrary
  unless the particles hit a wall of the pipe.
\item[(d)]
  Absence of self-interaction of the beam
  \\
  We assume that the bunch is not affected by its own field
  which consists of the radiation field and the space charge field.
  In other words, we do not take into account the dynamic redistribution of the particles, 
  which is caused by the interaction with their own field (self-interaction).
  In the present paper we do not solve the equations of motion of the particles
  forming the bunch.
  We will merely solve the wave equations for the electromagnetic field
  created by a bunch given under the assumptions (a,b,c).
  Therefore the present theory is not self-consistent which means that
  the current and field satisfy the equations of motion and
  Maxwell equations simultaneously.
  But it does not necessarily mean that we assume a rigid bunch.
  That is, the charge distribution of the bunch can arbitrarily vary
  while moving in the bending magnet
  if we know the variation of the charge distribution in advance
  as the solution of the equations of motion in the absence of self-interaction.
  For example, we can consider a bunch compression in the absence of self-interaction
  in the bend, which is made by giving a particular initial distribution to
  the bunch in the longitudinal phase space.
  On the other hand, when we consider a steady field in appendices \ref{sec:space_charge} 
  and \ref{sec:steady_field} as a special case,
  we assume a rigid bunch as a matter of course.
\item[(e)]
  Initial field at the entrance of the bend
  \\
  We give the value of the field and its derivative with respect to $s$ at $s=0$
  as the initial condition.
  In the present paper, the initial field means the field on the transverse plane at $s=0$,
  which depends on $(x,y)$.
  The initial field created by the beam is arbitrary
  if it satisfies Maxwell equations and the transverse boundary condition
  on the walls of the beam pipe under the assumptions (b,c).
  The initial field can be either steady or transient.
  We must take care that the longitudinal derivatives of all the components of
  the electromagnetic field are discontinuous at $s=0$
  regardless whether the field is steady or transient,
  because we assume that the bend has a hard edge at $s=0$ as described in (a).
  In section \ref{sec:disconti} we will discuss the discontinuity of
  the longitudinal derivative of the field at the edge of the bend.
\end{itemize}

The beam pipe plays the role like a waveguide to the electromagnetic field
created by the beam.
The beam pipe imposes the transverse boundary condition to the field.
On the other hand, we impose the longitudinal boundary condition by giving
the values of the field and its longitudinal derivative on
the transverse plane at $s=0$ as described in (e).
We will solve the wave equation as an initial value problem using the Laplace transform 
with respect to $s$.
Then the solution of the wave equation is uniquely determined as a function of $(x,y,s)$
in the frequency domain.
The solution has an explicit expression which represents the exact relation of
the field in $s>0$ to the initial field at $s=0$.
That is, the analytical solution is the exact map of the field from $s=0$ to
an arbitrary $s$ in the bending magnet, which involves the beam current moving in the pipe
unlike a usual waveguide which only transfers an electromagnetic wave.

We may need to explain assumption (a) in detail
since our model of the bending magnet is unrealistic in the following sense.
That is, in reality, it is impossible that a uniformly curved pipe extends semi-infinitely
from $s=0$ to $\infty$ without overlapping.
In the present paper we do not assume that the beam is circulating in a toroidal chamber.
Our model of the system differs from that of \cite{warnock_morton} in which
the field and current satisfy the periodic boundary condition with respect to $s$
in a toroidal chamber.
Instead of the periodicity along the $s$-axis, we fix the initial value of the field
and its longitudinal derivative at $s=0$ as described in assumption (e).
Although our model of the bend is unrealistic, the formulation is valid in theory
on finding the solution of transient fields of (coherent) synchrotron radiation
in the absence of longitudinal periodicity.
In this regard we had a similar discussion in appendix A of \cite{agoh},
though it is about a steady field of CSR in the paraxial approximation.
The expression of the field, gotten under assumption (a), is more realistic than
the one which has the longitudinal periodicity as in \cite{warnock_morton}
in the sense that we take into account the transient state of the field in a single bend
which is assumed to be a part of a beamline in an accelerator.
In addition, the frequency of the field is not discretized in our model as described in
Eq.(\ref{eq:krho}).

As described in assumption (a), our model does not take into account the exit of
the bending section to the next straight section toward the downstream of the beamline.
Needless to say, however, a bending magnet in an actual accelerator has an exit to
the next straight section.
Here we mention the difference which occurs, depending on
whether the bend has an exit or not.
That is, assuming that the bend has a finite length,
we briefly examine the effect of the common area between the bend and
the downstream straight section following the bend.
If the bend has an exit as usual,
a small partial reflection of the field may occur at the exit of the bend,
because the curvature of the trajectory and the beam pipe changes abruptly at the exit.
As the result, the reflected wave propagates backward in the bend.
This phenomenon is similar to the reflection and transmission of light on the boundary
between two media which have different indices of refraction.
Since the bend in our model has no exit,
the field in our model may be slightly different from that in which
the partial reflection at the exit is taken into account
(if we could get such a solution of the field).
The problem is how small the reflected field is, compared to the transmitted field.
The field, reflected at the exit of the bend, may be of a relative order of $x/\rho$ to
the transmitted field, according to the estimation in section \ref{sec:reflection}.
On the other hand, in the calculation of the field using the paraxial approximation as in
\cite{agoh_yokoya, stupakov_kotelnikov}, we do not have to consider the reflection of
the field at the exit of the bending magnet since all the backward waves are neglected in
the paraxial approximation because of neglecting $\rd_s^2$ in the wave equation.

\clearpage

\section{Wave equations}
\label{sec:we}

We derive the wave equations for the fields in
the frequency domain using the curvilinear coordinate system in section \ref{sec:we_TDFD}.
In section \ref{sec:disconti} we remove the singular terms from the wave equations
with no approximation.
Expanding the fields and current in the Fourier series with respect to $y$ in
section \ref{sec:ymode}, we further Laplace transform the Fourier coefficients of
the fields with respect to $s$ in section \ref{sec:we_in_LD}.
Then we will derive Eqs.(\ref{eq:wenu}) and (\ref{we_mfEBx}-\ref{we_mfEBs}) which are
the exact wave equations for the components of the fields in the Laplace domain.

\subsection{Fourier transform of the field with respect to time}
\label{sec:we_TDFD}

We define $\Ev(\xv,t)$ and $\Bv(\xv,t)$ as the electric and magnetic fields in
the time domain,
\begin{align}
  \Ev=E_x\ev_x+E_y\ev_y+E_s\ev_s ,
   \qquad
  \Bv=B_x\ev_x+B_y\ev_y+B_s\ev_s .
\end{align}
From Maxwell equations in vacuum, we get the wave equations for the fields in
the time domain,
\begin{align}
  \bigg(\nablav^2-\frac{\rd_t^2}{c^2}\bigg)\bigg({ \Ev \atop c\Bv }\bigg)
  =Z_0\bigg({ \Sv \atop \Tv }\bigg) ,
    \qquad
  \Sv
  =\nablav J_0+\frac{\rd_t}{c}\Jv ,
    \qquad
  \Tv
  =-\nablav\times\Jv .
  \label{eq:wez}
\end{align}
$Z_0$ and $\mu_0$ are the impedance and permeability of vacuum.
The source terms $\Sv(\xv,t)$ and $\Tv(\xv,t)$ consist of
\begin{align}
  J_0/c
   \quad \text{and}\quad
  \Jv=J_x\ev_x+J_y\ev_y+J_s\ev_s
\end{align}
which are the charge density and current density of the source current of
the fields in the time domain.
Hereinafter, we call $J(\xv,t)$ $[=(J_0,\Jv)]$ simply a current.
In order to solve Eq.(\ref{eq:wez}) under the assumptions listed in
section \ref{sec:assumption}, we Fourier transform the fields and current with
respect to $t$,
\begin{align}
  \tF(\xv,\omg)=\int_{-\infty}^{\infty}F(\xv,t)e^{i\omg t}dt ,
    \qquad
  F(\xv,t)=\int_{-\infty}^{\infty}\frac{d\omg}{2\pi}\tF(\xv,\omg)e^{-i\omg t} .
  \label{eq:Fourier_trans}
\end{align}
$F$ represents the quantities in the time domain, \ie, $F=(\Ev,\Bv,J,\Sv,\Tv)$.
$\omg$ is the frequency of $\tF$ which is the Fourier transform of $F$ with respect to $t$:
$\tF=(\tEv,\tBv,\tJ,\tSv,\tTv)$.
Using $v$ which is the speed of the reference particle as described in Eq.(\ref{eq:z}),
we define the wavenumber $k$ as
\begin{align}
  \omg=kv .
   \label{eq:omg}
\end{align}
$k$ is equivalent to $\omg$ in practice
since Eq.(\ref{eq:omg}) is just a scaling by $v$ which is a positive constant.
Therefore we often call the wavenumber $k$ frequency, \eg,
high/low frequency means large/small $|k|$.
In the present paper
we use $k$ instead of $\omg$, because it is convenient to keep the reference particle
in mind in considering the fields and current which move along the $s$-axis.
Besides, since $k$ has the dimension of inverse length,
it is easy to compare the scale of $k$ with the other quantities
having the dimension of length or inverse length.
For example, we can readily compare $k$ with the radial and vertical wavenumbers
given by Eq.(\ref{eq:asymp_lim}),
the bunch length and the geometric parameters of the beamline as we used in
Eqs.(\ref{eq:kz0}-\ref{eq:shield}) in describing the basic properties of CSR.
$v$ is not the flowing speed of the source current of the field
unless we consider a rigid bunch moving at the speed of $v$.
If we consider only a steady field created by a rigid bunch, then instead of
Eqs.(\ref{eq:Fourier_trans}), we should use Eqs.(\ref{eq:FT_zk_9})
which are the Fourier transform with respect to $z$ and $k$.
Also, when we consider transient fields in the paraxial approximation as described later in
section \ref{sec:parax_approx},
we should use Eqs.(\ref{eq:FT_zk_9}) instead of Eqs.(\ref{eq:Fourier_trans})
since we do not need to consider backward propagating waves.
If we take into account an infinitesimal damping of the field in the frequency domain,
we must put it as $k=k-i\eps$ ($\eps=+0$) in Eq.(\ref{eq:omg}) as shown in
Eq.(\ref{eq:torus_kmn_resona}),
unlike $k=k+i\eps$ given by Eqs.(136-137) in \cite{agoh}.

Fourier transforming the fields and current through Eq.(\ref{eq:Fourier_trans}),
the fields in the frequency domain, $\tEv(\xv,k)$ and $\tBv(\xv,k)$,
satisfy the following wave equations,
\begin{alignat}{2}
  &
  \{\nablav^2+(k\beta)^2\}\tEv=Z_0\tSv ,
    \qquad &&
  \tEv=\tE_x\ev_x+\tE_y\ev_y+\tE_s\ev_s ,
  \label{eq:wek_E}
   \\
  &
  \{\nablav^2+(k\beta)^2\}c\tBv=Z_0\tTv ,
   \qquad &&
  \tBv=\tB_x\ev_x+\tB_y\ev_y+\tB_s\ev_s .
  \label{eq:wek_B}
\end{alignat}
$\tSv(\xv,k)$, $\tTv(\xv,k)$ and $\tJ(\xv,k)$ $[=(\tJ_0,\tJv)]$ 
are the source terms and current in the frequency domain,
\begin{alignat}{2}
  \tSv
  &=\tS_x\ev_x+\tS_y\ev_y+\tS_s\ev_s
  &&=\nablav\tJ_0-ik\beta\tJv,
  \label{eq:tSv}
   \\
  \tTv
  &=\tT_x\ev_x+\tT_y\ev_y+\tT_s\ev_s
  &&=-\nablav\times\tJv,
  \label{eq:tTv}
   \\
  \tJv
  &=\tJ_x\ev_x+\tJ_y\ev_y+\tJ_s\ev_s.
\end{alignat}
Using the curvilinear coordinates defined in section \ref{sec:curv_coordinates},
we decompose the wave equations (\ref{eq:wek_E}-\ref{eq:wek_B}) into
the components of $x$, $s$ and $y$,
\begin{align}
  &(\nablav_{\vdash}^2-\alp x\brd_s)\bigg({\tE_x \atop c\tB_x}\bigg)
   -\frac{2\vkap\bar{\rd}_s+\alp}{g}
    \bigg({\tE_s \atop c\tB_s}\bigg)
  =Z_0\bigg({\tS_x \atop \tT_x}\bigg) ,
   \label{eq:we_tExBx}
   \\
  &(\nablav_{\vdash}^2-\alp x\brd_s)\bigg({\tE_s\atop c\tB_s}\bigg)
  +\frac{2\vkap\bar{\rd}_s+\alp}{g}
   \bigg({\tE_x \atop c\tB_x}\bigg)  
  =Z_0\bigg({\tS_s \atop \tT_s}\bigg) ,
   \label{eq:we_tEsBs}
   \\
  &(\nablav_{\rm v}^2-\alp x\brd_s)\bigg({\tE_y \atop c\tB_y}\bigg)
  =Z_0\bigg({\tS_y \atop \tT_y}\bigg) ,
    \qquad
  \alp
  =\frac{\rd_s\vkap}{g^2} .
    \label{eq:wey0}
\end{align}
Eqs.(\ref{eq:we_tExBx}-\ref{eq:wey0}) are the general wave equations in
the frequency domain,
which hold regardless of the assumptions listed in section \ref{sec:assumption}.
That is, Eqs.(\ref{eq:we_tExBx}-\ref{eq:wey0}) hold anywhere in the beamline
regardless of the straight or bending section including their boundary area.
$\nablav_{\vdash}^2$ and $\nablav_{\rm v}^2$ are symbols which denote the differential 
operators of the wave equations
[excluding $\alp x\brd_s$ and $\pm(2\vkap\bar{\rd}_s+\alp)/g$]
for the $(x,s)$ and $y$ components of the fields in the frequency domain
($\nablav_{\vdash}$ and $\nablav_{\rm v}$ do not exist unlike $\nablav$),
\begin{align}
  \bigg({\nablav_\vdash^2 \atop \nablav_{\rm v}^2 }\bigg)
  =\bigg({\rd_x\brd_x \atop \brd_x\rd_x }\bigg)
   +\rd_y^2+(k\beta)^2 +\frac{\rd_s^2}{g^2} .
  \label{eq:operator}
\end{align}
$\vkap$ is the curvature of the $s$-axis, given by Eq.(\ref{eq:coordinates}).
$\alp$ is a function of $x$ and $s$ in general.
Since $\alp$ originates in the change of the curvature of the $s$-axis,
$\alp=0$ in a space where $\vkap$ does not depend on $s$ like in
a constant bend or a straight section.
Under assumption (a) in section \ref{sec:assumption}, $\alp\ne0$ only at $s=0$
which is the entrance of the bending magnet.
In Eqs.(13-14) of \cite{agoh_yokoya}, which are the wave equations for
the transverse components of the electric field in the frequency domain,
we neglected the terms involving $\alp$ (hereinafter called the $\alp$-term) in order
to derive Eq.(21) in \cite{agoh_yokoya}, which is the parabolic wave equation
in the paraxial approximation.
On the other hand, in the present study we do not neglect the $\alp$-terms in
Eqs.(\ref{eq:we_tExBx}-\ref{eq:wey0}) since our purpose is to find
the exact expression of the fields.
Since this is a problem which requires a somewhat intricate consideration due to
a singular behavior of the field at the edge of the bend, we will discuss it
in section \ref{sec:disconti} in detail.

If Eq.(\ref{eq:wey0}) does not have the $\alp$-term
under the assumptions listed in section \ref{sec:assumption},
it is not so difficult to solve it without making an approximation,
because $\tE_y$ and $\tB_y$ do not couple with the other components of the fields in
Eq.(\ref{eq:wey0}) and the transverse boundary conditions on the walls of the beam pipe.
Although $\tE_{x,s}$ and $\tB_{x,s}$ couple in Eqs.(\ref{eq:we_tExBx}-\ref{eq:we_tEsBs}), 
if they do not have the $\alp$-terms, we can solve
Eqs.(\ref{eq:we_tExBx}-\ref{eq:we_tEsBs}) exactly as shown in appendix \ref{sec:we_xs}.
There is another way to find the expressions of $\tE_{x,s}$ and $\tB_{x,s}$,
because the electromagnetic field, $\tEv$ and $\tBv$, has only two degrees of freedom.
That is, if arbitrary two components of the electromagnetic field are given out of
the six components $(\tE_x,\tE_y,\tE_s)$ and $(\tB_x,\tB_y,\tB_s)$,
then the rest four are uniquely determined by the former two through Maxwell equations.
The same is true in terms of the potentials $\phi$ and $\Av$, \ie,
the field has two degrees of freedom out of the four $(\phi;A_x,A_y,A_s)$ in total,
excluding the choice of the gauge
which is an utterly different matter from what we are describing here.
As shown in appendix \ref{sec:boundary}, we can get the Fourier coefficients of
the horizontal and longitudinal components of the fields from the vertical ones through
Eqs.(\ref{eq:cBx}-\ref{eq:cBs}) which are derived from Maxwell equations for
the Fourier coefficients of the fields.
Accordingly, in order to find the expressions of $\tE_{x,s}$ and $\tB_{x,s}$,
it is enough to solve Eq.(\ref{eq:wey0}) with respect to $\tE_y$ and $\tB_y$
under the initial and boundary conditions.
Nevertheless, as described later in section \ref{sec:domains},
we will derive the expressions of the horizontal and longitudinal fields in
three different ways in order to ensure the correctness of the expressions of
all the components of the fields in an analytical way.

\subsection{Longitudinal derivative of the field at the edge of the bend}
\label{sec:disconti}

We examine the $\alp$-term given by Eq.(\ref{eq:wey0}) in order to solve
Eqs.(\ref{eq:we_tExBx}-\ref{eq:wey0}) exactly under the assumptions listed in
section \ref{sec:assumption}.
Assuming that the bend has a hard edge at the entrance,
the electromagnetic field is singular at $s=0$.
In order to clarify the singular behavior of the field on the edge of the bend,
we consider the edge in a somewhat more general form than assumption (a) in
section \ref{sec:assumption}.
That is, we consider an edge such that $\vrho$ changes only at $s=0$
abruptly from $\rho_{-}$ to $\rho_{+}$ as follows,
\begin{align}
  \frac{1}{\vrho}
  =\frac{\theta(s)}{\rho_{+}}+\frac{\theta(-s)}{\rho_{-}} ,
    \qquad
  \rd_s\vkap
  =\delta(s)\bigg(\frac{1}{\rho_{+}}-\frac{1}{\rho_{-}}\bigg) ,
    \qquad
  \theta(s)
  =\left\{\begin{array}{ll} 1 &(s\geq0) \\ 0 &(s<0) \end{array}\right. .
  \label{eq:curv}
\end{align}
$\theta$ and $\delta$ are the unit step function and the $\delta$-function.
$\rho_{+}$ and $\rho_{-}$ are constants which denote the curvature radii of the $s$-axis in 
$s\geq0$ and $s<0$ respectively.
For example, if the origin $(s=0)$ is set to the entrance of the semi-infinite bend as 
in assumption (a) of section \ref{sec:assumption},
they are given as $\rho_{-}=\infty$ and $\rho_{+}=\rho$.
On the contrary, if the bend has exit at $s=0$ and has no entrance,
then they reverse, \ie, $\rho_{-}=\rho$ and $\rho_{+}=\infty$.

The exact wave equations (\ref{eq:we_tExBx}-\ref{eq:wey0}) are
partial differential equations of the second order with respect to $s$.
$\alp$ is proportional to $\delta(s)$ if $\vkap$ is given by Eq.(\ref{eq:curv}).
It follows that the first and second derivatives of the field with respect to $s$ behave as
$\theta(s)$ and $\delta(s)$ respectively.
The field itself is continuous at $s=0$ while having a kink at $s=0$, and hence
$\rd_s\tE_{x,y,s}$ and $\rd_s\tB_{x,y,s}$ are discontinuous at $s=0$.
In the present study we do not neglect the $\alp$-terms involved in
Eqs.(\ref{eq:we_tExBx}-\ref{eq:wey0}) unlike \cite{agoh_yokoya}
which is formulated on the basis of the paraxial approximation.
We will solve Eqs.(\ref{eq:we_tExBx}-\ref{eq:wey0}) exactly, taking into account
the discontinuity of the longitudinal derivative of the field at $s=0$.
In discussing the singularity of the field at $s=0$ for Eq.(\ref{eq:curv}),
we concentrate on the vertical electric field $\tE_y$
since the current discussion on the singularity at $s=0$ is basically common to
all the components of the fields.
We separate $\tE_y$ into those in $s<0$ and $s\geq0$ using the step function
$\theta$ given by Eq.(\ref{eq:curv}),
\begin{align}
  \tE_y(s)=\theta(-s)\tE_y(s)+\theta(s)\tE_y(s),
   \qquad
  \tE_y(+0)=\tE_y(-0) .
  \label{eq:tEytJi_sep}
\end{align}
$\tE_y$ is continuous at $s=0$ while having a kink.
In section \ref{sec:disconti} we do not indicate the transverse arguments $(x,y)$ of
all the functions for clarity to show the longitudinal argument $s$.
From Eqs.(\ref{eq:tEytJi_sep}), we get the first and second derivatives of $\tE_y$ with
respect to $s$,
\begin{align}
  \rd_s\tE_y(s)=\theta(-s)\rd_s\tE_y(s)+\theta(s)\rd_s\tE_y(s) ,
   \qquad
  \rd_s^2\tE_y(s)
  =\theta(s)\rd_s^2\tE_y(s)+\theta(-s)\rd_s^2\tE_y(s)+\delta(s)\Delta\tE_y' .
  \label{eq:rds2_tEy_pm}
\end{align}
$\Delta\tE_y'$ is the discontinuity of $\rd_s\tE_y$ at $s=0$,
\begin{align}
  \Delta\tE_y'=\rd_s\tE_y(+0)-\rd_s\tE_y(-0)
   \qquad\text{where}\quad
  \rd_s\tE_y(\pm0)=[\rd_s\tE_y(s)]_{s=\pm0} .
  \label{eq:Del_tEy}
\end{align}
Substituting Eqs.(\ref{eq:rds2_tEy_pm}) into Eq.(\ref{eq:wey0}), we have
\begin{align}
  \theta(s)\nablav_{\rm v}^2\tE_y+\theta(-s)\nablav_{\rm v}^2\tE_y
  +\frac{\delta(s)}{g^2}\Delta\tE_y'-\alp x\brd_s\tE_y
  =Z_0\tS_y .
  \label{eq:we_sep}
\end{align}
We examine the discontinuity of $\rd_s\tE_y$ at $s=0$ for Eq.(\ref{eq:curv}).
We consider the horizontal component of Faraday's law in the frequency domain,
\begin{align}
  \rd_y\tE_s-\brd_s\tE_y-ik\beta c\tB_x=0.
   \label{eq:Faraday_x_FD}
\end{align}
For Eq.(\ref{eq:curv}), $\rd_y\tE_s$ and $\tB_x$ are both continuous at $s=0$.
From Eq.(\ref{eq:Faraday_x_FD}), we get the relation between the values of $\rd_s\tE_y$
at $s=+0$ and $s=-0$,
\begin{align}
  \frac{\rd_s\tE_y(+0)}{g_{+}}=\frac{\rd_s\tE_y(-0)}{g_{-}} ,
    \qquad
  g_{\pm}=1+\frac{x}{\rho_{\pm}} .
  \label{eq:Frdy_x}
\end{align}
$g_{+}$ and $g_{-}$ are the geometric factors in $s\geq0$ and $s<0$ respectively.
From Eq.(\ref{eq:Frdy_x}), we get the following two expressions of $\Delta\tE_y'$,
depending on taking the value of $\rd_s\tE_y$ at either $s=+0$ or $-0$,
\begin{align}
  \Delta\tE_y'
  =\bigg(1-\frac{g_{-}}{g_{+}}\bigg)\rd_s\tE_y(+0)
  =\bigg(\frac{g_{+}}{g_{-}}-1\bigg)\rd_s\tE_y(-0) .
  \label{eq:rds_tEy_jump}
\end{align}
There is another way to derive Eqs.(\ref{eq:rds_tEy_jump}), because $\rd_s^2\tE_y$ in
Eq.(\ref{eq:we_sep}) is responsible for the singular term $\alp x\brd_s\tE_y$ in
Eqs.(\ref{eq:wey0}), which is not 0 only at $s=0$.
That is, we can get Eqs.(\ref{eq:rds_tEy_jump}) by integrating $\alp x\brd_s\tE_y$
in Eqs.(\ref{eq:wey0}) in the vicinity of the common area $s=\pm\eps$
where $\eps$ denotes a positive infinitesimal ($\eps\to+0$),
\begin{align}
  \Delta \tE_y'&=\int_{-\eps}^{\eps}\rd_s^2\tE_yds
  =\int_{-\eps}^{\eps}g^2\alp x\brd_s\tE_yds
  =\int_{-\eps}^{\eps}ds\frac{g_{+}-g_{-}}{g}\delta(s)\rd_s\tE_y
  =\bigg[\frac{g_{+}-g_{-}}{g}\rd_s\tE_y\bigg]_{s=\pm0} .
  \label{eq:tEy_jump}
\end{align}
Eq.(\ref{eq:tEy_jump}) agrees with Eqs.(\ref{eq:rds_tEy_jump}).
From Eq.(\ref{eq:Frdy_x}) and Eq.(\ref{eq:curv}), we get the relation of
$\Delta\tE_y'$ and $\alp$,
\begin{align}
  \frac{g_{+}-g_{-}}{g}\delta(s)\rd_s\tE_y
  =\frac{x}{g}(\rd_s\vkap)\rd_s\tE_y
   \qquad\Lra\qquad
  \frac{\delta(s)}{g^2}\Delta\tE_y'=\alp x\brd_s\tE_y .
  \label{eq:del_DelEy}
\end{align}
Eq.(\ref{eq:del_DelEy}) implies that the third and fourth terms in Eq.(\ref{eq:we_sep}) 
cancel out and vanish.
It follows that, while keeping the formalism exact,
we can remove the singularity of the field at $s=0$ from the wave equation
by separating the component of the field into those in $s<0$ and $s\geq0$.
Similar to $\tE_y$ given by Eq.(\ref{eq:tEytJi_sep}),
we separate the current $\tJ$ $[=(\tJ_0,\tJv)]$ into those for $s<0$ and $s\geq0$, \ie,
\begin{align}
  \tJ(s)=\theta(-s)\tJ(s)+\theta(s)\tJ(s),
   \qquad
  \rd_s\tJ(s)=\theta(-s)\rd_s\tJ(s)+\theta(s)\rd_s\tJ(s) .
\end{align}
By this, we can separate the wave equation (\ref{eq:we_sep}) for $\tE_y$ into
those for $s<0$ and $s\geq0$ as follows,
\begin{align}
  \theta(-s)\big[\nablav_{\rm v}^2\tE_y-Z_0\tS_y\big]
  +\theta(s)\big[\nablav_{\rm v}^2\tE_y-Z_0\tS_y\big]
  =0
   \quad~~\Lra~~\quad
  \nablav_{\rm v}^2\tE_y=Z_0\tS_y
    \quad
  (\forall s\in\mathbb{R}) .
  \label{eq:we_tEy_pm}
\end{align}
The second equation of (\ref{eq:we_tEy_pm}) is exact,
even though it no longer has the singular term involving $\alp$.
Thus we can remove the singularity of the field from the wave equation
without neglecting any term.

As can be seen from Eq.(\ref{eq:rds_tEy_jump}), $\Delta\tE_y'$ is of relative order of
$(g_{+}-g_{-})/g_{\pm}$ to the magnitude of $\rd_s\tE_y$ at $s=\pm0$.
If the $s$-axis changes at $s=0$ from a straight line to an arc of circle
having a radius $\rho$, as assumed in (a) of section \ref{sec:assumption},
$\rd_s\tE_y$ jumps at $s=0$ by a relative order of $x/\rho$ as shown below,
\begin{align}
  \frac{\Delta\tE_y'}{\rd_s\tE_y(-0)}
  =\frac{g_{+}}{g_{-}}-1
  =\frac{x}{\rho} .
  \label{eq:relative_error}
\end{align}
Therefore if we neglect the small terms of $O(x/\rho)$ in the wave equations
as we did from Eqs.(13-14) to Eq.(21) in \cite{agoh_yokoya},
then it is nonsense to keep the $\alp$-term in the paraxial approximation.
In this regard we must mention our wrong discussion in evaluating the edge effect
in Eqs.(13-14) of \cite{agoh_yokoya}.
Our statement ``$E_{x,y}$ itself can be step function like'' in \cite{agoh_yokoya}
is wrong, \ie, the field itself is not discontinuous at $s=0$
as in Eq.(\ref{eq:tEytJi_sep}), while it has a kink.
The first derivative of the field with respect 
to $s$ is discontinuous at $s=0$ since the field has a kink at $s=0$
because of the abrupt change of the curvature of the $s$-axis.
The second derivative of the field with respect to $s$ has a term involving
$\delta(s)$, which is canceled exactly with the $\alp$-term in the exact wave equation
as shown in Eq.(\ref{eq:del_DelEy}).
Accordingly, it is nonsense to keep the $\alp$-term in the parabolic wave equation,
assuming that the bending magnet has a fringe field around the entrance.
It is also irrelevant to compare the bunch length with the length of
the fringe field of the bend, which we described in p.3 of \cite{agoh_yokoya}
to justify the neglect of the $\alp$-term.
Since the term of the second derivative of the field with respect to $s$ is fully 
responsible for the $\alp$-term in the wave equation,
if we neglect $\rd_s^2$ for $ik\rd_s$ in order to derive the parabolic equation with 
respect to $s$, then we should also neglect the $\alp$-term to keep the consistency of
the theoretical framework in the paraxial approximation.
Otherwise the field itself will be discontinuous at $s=0$, which is unphysical.

Similar to $\rd_s\tE_y$, we examine the discontinuity of $\rd_s\tB_y$ at $s=0$
using Eq.(\ref{eq:Ampere_x_FD2}) which is the horizontal component of Ampere's law in
the frequency domain.
The singular term $\alp x\brd_s\tB_y$ in Eq.(\ref{eq:wey0}) is exactly canceled with 
the term having $\delta(s)$ which comes out of $\rd_s^2\tB_y$.
Therefore we can separate Eq.(\ref{eq:wey0}) for $\tB_y$ into the component in
the straight section and the bending section without making an approximation,
\begin{align}
  \theta(-s)\big[\nablav_{\rm v}^2c\tB_y-Z_0\tT_y\big]
  +\theta(s)\big[\nablav_{\rm v}^2c\tB_y-Z_0\tT_y\big]
  =0
   \quad~~\Lra~~\quad
  \nablav_{\rm v}^2c\tB_y=Z_0\tT_y
    \quad
  (\forall s\in\mathbb{R}) .
  \label{eq:we_tBy_pm}
\end{align}
Similar to the second equation of (\ref{eq:tEytJi_sep}) and Eq.(\ref{eq:Frdy_x}),
$\tB_y$ and $\rd_s\tB_y$ at $s=0$ satisfy Eqs.(\ref{eq:cont}) and
(\ref{eq:rds_tBy_jump}).
Similar to $\tE_y$ and $\tB_y$, the $\alp$-term in
Eqs.(\ref{eq:we_tExBx}-\ref{eq:we_tEsBs}) is canceled exactly with the one which
comes out of $\rd_s^2\tE_{x,s}$ and $\rd_s^2\tB_{x,s}$ as shown in
appendix \ref{sec:discontinuity}.
Then we get the exact wave equations for the horizontal and longitudinal components 
of the fields in either $s<0$ or $s\geq0$,
\begin{align}
  \nablav_{\vdash}^2\bigg({\tE_x\atop c\tB_x}\bigg)
   -\frac{2\brd_s}{g\vrho}\bigg({\tE_s\atop c\tB_s}\bigg)
  =Z_0\bigg({\tS_x \atop \tT_x}\bigg) ,
    \qquad
  \nablav_{\vdash}^2\bigg({\tE_s\atop c\tB_s}\bigg)
  +\frac{2\brd_s}{g\vrho}\bigg({\tE_x\atop c\tB_x}\bigg)  
  =Z_0\bigg({\tS_s \atop \tT_s}\bigg) .
  \label{eq:we_tExsBxs}
\end{align}
Eqs.(\ref{eq:we_tExsBxs}) have no singular terms at $s=0$, similar to
the second equations of (\ref{eq:we_tEy_pm}) and (\ref{eq:we_tBy_pm}).
According to the above considerations, even if we assume a bending magnet
having a hard edge,
by separating the components of the fields into those in each section $s<0$ and $s\geq0$,
we can derive the exact wave equations which have no singular term.
Then we can solve them exactly.
We have three ways to find the expressions of $\tE_{x,s}$ and $\tB_{x,s}$
as summarized in section \ref{sec:domains}.
We can straightforwardly solve Eqs.(\ref{eq:we_tExsBxs}) as shown in
appendix \ref{sec:we_xs}.
However, this solution requires a somewhat complicated treatment to decouple
the field variables using the eigenfunctions of the operator of the wave equations.
Instead of this way, we can get their Fourier coefficients with respect to $y$
from the vertical components as shown in appendix \ref{sec:HL_field_sdom}.
In addition to these two ways, we can get the horizontal and longitudinal fields
in the Laplace domain from the vertical ones as shown in section \ref{sec:sol_mfEB_xs}.
Then we can transform the Laplace domain field to their Fourier coefficients
as shown later in section \ref{sec:XS_field_sdom}.

\subsection{Field in a bend of a finite length}
\label{sec:reflection}

\begin{figure} [h]
  \begin{center}
    \includegraphics[scale=0.65,clip]{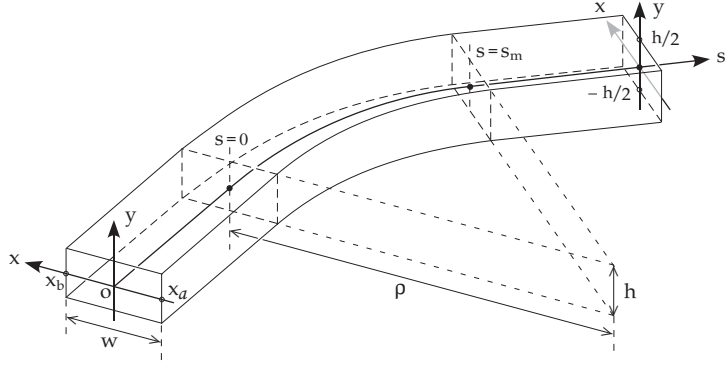}
    \caption[Bend of finite length]
       {\small
       Beamline and coordinate system for a bend of finite length.
       The reference axis ($s$-axis) lies on the median horizontal plane ($y=0$).
       $O$ is the origin $\xv_{\perp}=(x,y)=0$ of the transverse plane.
       The bend has an entrance at $s=0$.
       The $s$-axis has a curvature radius $\vrho=\rho$ in $0\leq s< s_{\rm m}$ (bend)
       and $\vrho=\infty$ in $s<0$ and $s\geq s_{\rm m}$ (straight sections).
       Similar to Fig.\ref{fig:pipe3D},
       the rectangular pipe represents the beam pipe which extends along the $s$-axis.
       $w$ $(=x_b-x_a)$ and $h$ are respectively the full width and
       full height of the pipe, where $x_a$ and $x_b$ are the horizontal positions of
       the inner and outer walls of the beam pipe.
       The upper and lower walls of the pipe are located at $y=\pm h/2$.
        }
    \label{fig:pipe_exit}
  \end{center}
\end{figure}

Let us consider the wave equation for $\tE_y$ in a bend
which has a finite length $s_{\rm m}$ along the $s$-axis instead of assumption (a) in
section \ref{sec:assumption}.
Only in section \ref{sec:reflection} we assume that semi-infinite straight sections connect
the entrance and exit of the bending section as shown in Fig.\ref{fig:pipe_exit}.
Assuming that the $s$-axis has a constant radius $\rho$ in the bend, the curvature of
the $s$-axis and its longitudinal derivative in this beamline are given as
\begin{align}
  \frac{1}{\vrho}
  &=\frac{1}{\rho}\{\theta(s)-\theta(s-s_{\rm m})\} ,
    \qquad
  \rd_s\vkap
  =\frac{1}{\rho}\{\delta(s)-\delta(s-s_{\rm m})\} .
  \label{eq:curv_bend}
\end{align}

We separate $\tE_y$ into those in the three sections: the upstream straight section ($s<0$),
the bending section ($0\leq s<s_{\rm m}$) and the downstream straight section
($s\geq s_{\rm m}$),
\begin{align}
  \tE_y(s)
  &=\theta(-s)\tE_y(s)+\{\theta(s)-\theta(s-s_{\rm m})\}\tE_y(s)
    +\theta(s-s_{\rm m})\tE_y(s) .
  \label{eq:tEy_dmd}
\end{align}
$\tE_y$ is continuous on the edges of the bending magnet, \ie,
\begin{align}
  \tE_y(-0)=\tE_y(+0)
   ,\qquad
  \tE_y(s_{\rm m}-0)=\tE_y(s_{\rm m}+0) .
\end{align}
From Eq.(\ref{eq:tEy_dmd}), the first and second derivatives of $\tE_y$ with respect to $s$ 
are given as
\begin{align}
  \rd_s\tE_y(s)
  &=\theta(-s)\rd_s\tE_y(s)
    +\{\theta(s)-\theta(s-s_{\rm m})\}\rd_s\tE_y(s)
    +\theta(s-s_{\rm m})\rd_s\tE_y(s) ,
   \\
  \rd_s^2\tE_y(s)
  &=\theta(-s)\rd_s^2\tE_y(s)
    +\{\theta(s)-\theta(s-s_{\rm m})\}\rd_s^2\tE_y(s)
    +\theta(s-s_{\rm m})\rd_s^2\tE_y(s)
    +\Delta_y(s) .
  \label{eq:rds2_tEy_mag}
\end{align}
$\Delta_y$ has a singular term which originates at the edges of the bend.
We rewrite $\Delta_y$ as follows,
\begin{align}
  \Delta_y(s)
  &=\delta(s)\{\rd_s\tE_y(+0)-\rd_s\tE_y(-0)\}
    +\delta(s-s_{\rm m})\{\rd_s\tE_y(s_{\rm m}+0)-\rd_s\tE_y(s_{\rm m}-0)\}
  \\
  &=\frac{x}{g\rho}\{\delta(s)-\delta(s-s_{\rm m})\}\rd_s\tE_y(s)
   =g^2\alp x\brd_s\tE_y(s) .
   \label{eq:Dely}
\end{align}
$\alp$ is the function defined in Eq.(\ref{eq:wey0}).
Since we assume that the bending magnet has a length $s_{\rm m}$,
\begin{align}
  \alp
  =\frac{1}{g^2\rho}\{\delta(s)-\delta(s-s_{\rm m})\} .
\end{align}
Eq.(\ref{eq:Dely}) implies that the terms $\Delta_y/g^2$ and $\alp x\brd_s\tE_y$ cancel out
in the wave equation (\ref{eq:wey0}),
\begin{align}
  \theta(-s)W_y(s)
  +\{\theta(s)-\theta(s-s_{\rm m})\}W_y(s)
  +\theta(s-s_{\rm m})W_y(s)
  =0 ,
\end{align}
where
\begin{align}
  W_y(s)=\nablav_{\rm v}^2\tE_y(s)-Z_0\tS_y(s) .
  \label{eq:we_tEy_sm}
\end{align}
Thus, from Eq.(\ref{eq:wey0}), we can derive the wave equation $W_y(s)=0$ in each section
$s<0$, $0\leq s<s_{\rm m}$ and $s\geq s_{\rm m}$ without making approximation.
This discussion holds not only for $\tE_y$ but also for all the components of
the electromagnetic field.
If the beamline has several bending magnets having hard edges,
we can separate the wave equations for $\tE_{x,y,s}$ and $\tB_{x,y,s}$ into 
those in each section without approximation.
If we neglect $\Delta_y$ in Eq.(\ref{eq:rds2_tEy_mag}),
it produces a relative error of the order of $x/\rho$ in the field,
similar to Eq.(\ref{eq:relative_error}).
Accordingly, in calculating the field of synchrotron radiation on the basis of
the paraxial approximation as in \cite{agoh_yokoya}, we should neglect the edge effect
at both the entrance and exit of the bending section.

\subsection{Fourier expansion of the field in the vertical eigenmodes}
\label{sec:ymode}

We go back to the assumptions listed in section \ref{sec:assumption},
\ie, we assume a semi-infinite bend in what follows.
The exact wave equations for $\tE_y$ and $\tB_y$ are given by
the second equations of (\ref{eq:we_tEy_pm}) and (\ref{eq:we_tBy_pm}).
In order to solve them, we expand the fields and current in the frequency domain
$\tF=(\tEv,\tBv;\tJ)$ in a Fourier series with respect to $y$.
Since the upper-lower walls of the rectangular pipe are perpendicular to $\ev_y$,
we can expand $\tF$ as
\begin{align}
  \tF(\xv)=\sum_{n=-\infty}^{\infty}c_y^n(x,s)e^{ik_y^ny}
   ,\qquad
  k_y^n=\frac{n\pi}{h}
    \quad
  (n\in\mathbb{Z}) .
   \label{eq:kyn}
\end{align}
$k_y^n$ is the vertical wavenumber of the $n$th oscillatory mode
between the upper-lower walls of the vacuum chamber.
$n$ is the index to denote the mode number of the oscillation in the vertical direction.
$c_y^n$ is the $n$th Fourier coefficient of $\tF$ with respect to $y$.
Throughout the present paper, we indicate the vertical mode index $n$ with the superscript as
$c_y^n$ and $k_y^n$ except $(-1)^n\in\{\pm1\}$ which means $-1$ to the power of $n$ as usual.
The boundary conditions of the fields on the upper-lower walls of
the rectangular pipe are given as
\begin{align}
  \tE_x=\tE_s=c\tB_y=0
   ,\qquad
  \rd_yc\tB_x=\rd_yc\tB_s=\rd_y\tE_y=0
   \qquad\text{at}~~y=\pm h/2 .
  \label{eq:t_BC_UL}
\end{align}
According to Eqs.(\ref{eq:t_BC_UL}), we classify the components of the fields and current in
the pipe into the following two kinds $\tF=\tF_{\pm}$,
depending on the positive or negative parity with respect to $y$,
\begin{align}
  \tF_{+}&=(\tB_y,\tE_{x,s};\tJ_0,\tJ_{x,s}),
  \qquad
   \tF_{-}=(\tE_y,\tB_{x,s};\tJ_y) .
   \label{eq:compo_Apm}
\end{align}
Imposing Eqs.(\ref{eq:t_BC_UL}) to Eq.(\ref{eq:kyn}), $\tF_{\pm}$ is given as
\begin{align}
  \tF_{+}(\xv)
  &=\sum_{p=1}^{\infty}
    \big[\hat{\cF}_{+}^{2p-1}(x,s)\cos(k_y^{2p-1}y)
        +\check{\cF}_{+}^{2p}(x,s)\sin(k_y^{2p}y)
    \big] ,
  \label{eq:Four_Exp_plus}
  \\
  \tF_{-}(\xv)
  &=\sum_{p=0}^{\infty}
    \bigg[
        \frac{\check{\cF}_{-}^{2p}(x,s)}{1+\delta_0^p}\cos(k_y^{2p}y)
       +\hat{\cF}_{-}^{2p+1}(x,s)\sin(k_y^{2p+1}y)
    \bigg] .
  \label{eq:Four_Exp_minus}
\end{align}
$\delta_0^p\in\{0,1\}$ is the Kronecker delta.
$p\in\mathbb{Z}_0^{+}$ represents the vertical mode number $n=\{2p,2p+1\}$ instead of $n$.
$\cF_\pm^n=(\hat{\cF}_\pm^{2p\mp 1},\check{\cF}_\pm^{2p})$ denotes
the $n$th Fourier coefficient of $\tF_\pm$,
\begin{align}
  \cF_{+}^n(x,s)
  &=(\cB_y^n,\cE_{x,s}^n;\cJ_0^n,\cJ_{x,s}^n)
   ,\qquad
  \cF_{-}^n(x,s)
  =(\cE_y^n,\cB_{x,s}^n;\cJ_y^n) .
   \label{eq:coeff_FJ}
\end{align}
$\vec{\cE}^n$, $\vec{\cB}^n$ and $\cJ^n$ are the $n$th Fourier coefficients of
$\tEv$, $\tBv$ and $\tJ$,
\begin{alignat}{2}
  \vec{\cE}^n
  &=\cE_x^n\ev_x+\cE_y^n\ev_y+\cE_s^n\ev_s,
    \qquad&
  \cJ^n
  &=(\cJ_0^n,\vec{\cJ}^n),
   \\
  \vec{\cB}^n
  &=\cB_x^n\ev_x+\cB_y^n\ev_y+\cB_s^n\ev_s,
    \qquad&
  \vec{\cJ}^n
  &=\cJ_x^n\ev_x+\cJ_y^n\ev_y+\cJ_s^n\ev_s .
\end{alignat}
In general, $\cF_\pm^n$ has both the odd modes $\hat{\cF}_\pm^{2p\mp1}$ and even modes
$\check{\cF}_\pm^{2p}$,
\begin{alignat}{3}
  \cF_{+}^n(x,s)
  &=\Bigg\{{\hat{\cF}_{+}^{2p-1} \atop \check{\cF}_{+}^{2p} }\Bigg\}
  &&=\int_{-h}^{h}\frac{dy}{h}\tF_{+}(\xv)
  \Bigg\{{\cos(k_y^{2p-1}y) \atop \sin(k_y^{2p}y)}\Bigg\}
  \label{eq:cFpn}
   \\
  &&
  &=\int_{-h/2}^{h/2}\frac{dy}{h/2}\tF_{+}(\xv)
  \Bigg\{{\cos(k_y^{2p-1}y) \atop \sin(k_y^{2p}y)}\Bigg\}
   \quad~~&&(n,p\in\mathbb{N}) ,
  \label{eq:Fp_coeff}
   \\
  \cF_{-}^n(x,s)
  &=\Bigg\{{\hat{\cF}_{-}^{2p+1} \atop \check{\cF}_{-}^{2p} }\Bigg\}
  &&=\int_{-h}^{h}\frac{dy}{h}\tF_{-}(\xv)
  \Bigg\{{\sin(k_y^{2p+1}y) \atop \cos(k_y^{2p}y)}\Bigg\}
  \label{eq:cFmn}
   \\
  &&
  &=\int_{-h/2}^{h/2}\frac{dy}{h/2}\tF_{-}(\xv)
  \Bigg\{{\sin(k_y^{2p+1}y) \atop \cos(k_y^{2p}y)}\Bigg\}
   \quad~~&&(n,p\in\mathbb{Z}_0^{+}) .
  \label{eq:Fm_coeff}
\end{alignat}
We can rewrite Eqs.(\ref{eq:Four_Exp_plus}-\ref{eq:Four_Exp_minus}) into the following
expressions, similar to Eqs.(2.2-2.3) in \cite{warnock_morton},
\begin{alignat}{2}
  \tF_{+}(\xv)
  &=\sum_{n=1}^{\infty}f_{+}^n(x,s)\sin[k_y^n(y-h/2)]
  ,\qquad &
  f_{+}^n(x,s)
  &=\int_{-h/2}^{h/2}\frac{dy}{h/2}\tF_{+}(\xv)\sin[k_y^n(y-h/2)] ,
   \\
  \tF_{-}(\xv)
  &=\sum_{n=0}^{\infty}\frac{f_{-}^n(x,s)}{1+\delta_0^n}\cos[k_y^n(y-h/2)]
  ,\qquad &
  f_{-}^n(x,s)
  &=\int_{-h/2}^{h/2}\frac{dy}{h/2}\tF_{-}(\xv)\cos[k_y^n(y-h/2)] .
\end{alignat}
$f_{\pm}^n$ are the Fourier coefficients which are related to $\cF_{\pm}^n$ as follows,
\begin{align}
  \cF_{\pm}^{2p}=(-1)^pf_{\pm}^{2p} ,
   \qquad
  \cF_{\pm}^{2p+1}=(-1)^pf_{\pm}^{2p+1} .
\end{align}
Instead of $\cF_{\pm}^n$, one can use $f_{\pm}^n$ to describe the Fourier coefficients of
the fields and current, which is a matter of taste.
We use $\cF_{\pm}^n$ in the present paper.
It follows that the Fourier expansions,
given by Eqs.(\ref{eq:Four_Exp_plus}-\ref{eq:Four_Exp_minus}),
are formally equivalent to replacing
the operator $\rd_y$ in the differential equation for $\tF_{\pm}$
by the factor $\pm(-1)^nk_y^n$ in the equation for $\cF_{\pm}^n$.

In order to explain the meaning of the vertical parity which is indicated by the sign
$(\pm)$ of $\tF_{\pm}$ and $\cF_{\pm}^n$, only in this paragraph let us consider a current
which is symmetric with respect to the median horizontal plane ($y=0$) as
$\tJ_{0,x,s}(-y)=\tJ_{0,x,s}(y)$ and $\tJ_y(-y)=-\tJ_y(y)$.
This symmetric current creates an electric and magnetic fields whose components have
the positive and negative parities as $\tF_{\pm}(-y)=\pm\tF_{\pm}(y)$
since the entire system has the mirror symmetry with respect to the median horizontal plane.
$\check{\cF}_{\pm}^{2p}=0$ in this system, \ie,
the field does not have the even modes with respect to $y$.
For an arbitrary current which is asymmetric with respect to $y=0$ in general,
the fields have no vertical symmetry in the pipe as a matter of course.
Nevertheless, we use the sign of the vertical parity as $\tF_{\pm}$ and $\cF_{\pm}^n$
for convenience in classifying the components of the fields and current as
Eqs.(\ref{eq:compo_Apm}) and (\ref{eq:coeff_FJ}) in a systematic manner.

\subsection{Wave equations for the Fourier coefficients of the field}
\label{sec:we_fc}

We discuss the fields which are created by an arbitrary current
under the assumptions listed in section \ref{sec:assumption}.
According to Eqs.(\ref{eq:we_tEy_pm}) and (\ref{eq:we_tBy_pm}-\ref{eq:we_tExsBxs}),
the $n$th Fourier coefficients of the components of the fields in the straight section or 
the bending section satisfy the following wave equations,
\begin{align}
  &
  \rd_{\vdash}^2\bigg({\cE_x^n \atop c\cB_x^n }\bigg)
  -\frac{2\brd_s}{g\vrho}\bigg({\cE_s^n \atop  c\cB_s^n}\bigg)
  =Z_0\bigg({\cS_x^n \atop \cT_x^n}\bigg) ,
    \qquad
  \rd_{\vdash}^2\bigg({\cE_s^n \atop c\cB_s^n}\bigg)
  +\frac{2\brd_s}{g\vrho}\bigg({\cE_x^n \atop c\cB_x^n}\bigg)
  =Z_0\bigg({\cS_s^n \atop \cT_s^n}\bigg) ,
  \label{eq:wen_xs}
   \\
  &
  \rd_{\rm v}^2\bigg({\cE_y^n \atop c\cB_y^n }\bigg)
  =Z_0\bigg({\cS_y^n \atop \cT_y^n }\bigg) ,
    \qquad\text{where}\qquad
  \bigg({\rd_{\vdash}^2 \atop \rd_{\rm v}^2 }\bigg)
  =\bigg({\rd_x\brd_x \atop \brd_x\rd_x }\bigg)
   +(k\beta)^2-(k_y^n)^2+\brd_s^2 .
  \label{eq:wen}
\end{align}
The operators $\brd_x$ and $\brd_s$ are given by Eqs.(\ref{eq:brdxs}).
$\rd_{\vdash}^2$ and $\rd_{\rm v}^2$ are the operators of the wave equations for
the $n$th Fourier coefficients of the fields, which correspond to
$\nablav_\vdash^2$ and $\nablav_{\rm v}^2$ given by Eq.(\ref{eq:operator}).
$\cSv^n$ and $\cTv^n$ are the $n$th Fourier coefficients of the source terms
$\tSv$ and $\tTv$ given by Eqs.(\ref{eq:tSv}-\ref{eq:tTv}),
\begin{align}
  \cSv^n=\cS_x^n\ev_x+\cS_y^n\ev_y+\cS_s^n\ev_s ,
     \qquad
  \cTv^n=\cT_x^n\ev_x+\cT_y^n\ev_y+\cT_s^n\ev_s .
   \label{eq:cSv_cTv}
\end{align}
Eqs.(\ref{eq:wen_xs}-\ref{eq:wen}) hold in either a straight section or a constant bend
where the $s$-axis has a constant curvature.
In the limit of $\vrho\to\infty$, Eqs.(\ref{eq:wen_xs}-\ref{eq:wen}) go to
Eqs.(\ref{eq:we_xprim_drift}) in appendix \ref{sec:travel_wave}.
On the other hand, in describing the field in a constant bend in which the $s$-axis has
a constant radius $\vrho=\rho$, we use the cylindrical coordinate system $\xv=(r,y,s)$
where $r$ is the radial variable to describe the fields and current in the constant bend,
\begin{align}
  r
  =\rho+x
  =g\rho
   ,\qquad
  r_a=\rho+x_a
   ,\qquad
  r_b=\rho+x_b .
   \label{eq:r}
\end{align}
$r_a$ and $r_b$ are the radii of the inner and outer walls of the curved rectangular pipe 
in the bending section.
Although one can use the angle $\vphi$ $(=s/\rho)$ instead of $s$
as the longitudinal variable to describe the fields and current in the constant bend,
we use $s$ throughout the present paper, because we can take the limit of $\rho\to\infty$ 
for the expression of the field in considering the asymptotic expression of the field in
the straight pipe in which $\xv=(x,y,s)$ denotes the local Cartesian coordinates.

In solving Eqs.(\ref{eq:wen_xs}-\ref{eq:wen}) for the fields in the constant bend,
we rewrite them using $r$ instead of $x$,
\begin{align}
  &
  \rd_{\vdash}^2\bigg({\cE_x^n \atop \cE_s^n}\bigg)
  +\frac{2}{r}\brd_s\bigg({-\cE_s^n \atop +\cE_x^n}\bigg)
  =Z_0\bigg({\cS_x^n \atop \cS_s^n}\bigg) ,
    \qquad
  \rd_{\vdash}^2\bigg({c\cB_x^n \atop c\cB_s^n}\bigg)
  +\frac{2}{r}\brd_s\bigg({-c\cB_s^n \atop +c\cB_x^n}\bigg)
  =Z_0\bigg({\cT_x^n \atop \cT_s^n}\bigg) ,
  \label{eq:we_cFxs}
  \\
  &
  \rd_{\rm v}^2
  \bigg({\cE_y^n \atop c\cB_y^n}\bigg)
  =Z_0\bigg({\cS_y^n \atop \cT_y^n}\bigg) ,
    \qquad\text{where}\qquad
  \bigg({ \rd_{\vdash}^2 \atop \rd_{\rm v}^2 }\bigg)
  =\bigg({ \rd_r\brd_r \atop \brd_r\rd_r}\bigg) +(k_r^n)^2+\brd_s^2 .
  \label{eq:wen_bend}
\end{align}
Using $r$, the components of Eqs.(\ref{eq:cSv_cTv}) are given as follows,
\begin{alignat}{2}
  \cS_x^n(r,s)
  &=\rd_r\cJ_0^n(r,s)-ik\beta\cJ_x^n(r,s) ,
     \qquad&
  \cT_x^n(r,s)
  &=\brd_s\cJ_y^n(r,s)-(-1)^nk_y^n\cJ_s^n(r,s) ,
  \label{eq:cSx_cTx}
  \\
  \cS_y^n(r,s)
  &=(-1)^nk_y^n\cJ_0^n(r,s)-ik\beta\cJ_y^n(r,s) ,
     \qquad&
  \cT_y^n(r,s)
  &=\brd_r\cJ_s^n(r,s)-\brd_s\cJ_x^n(r,s) ,
  \label{eq:cSy_cTy}
   \\
  \cS_s^n(r,s)
  &=\brd_s\cJ_0^n(r,s)-ik\beta\cJ_s^n(r,s) ,
     \qquad&
  \cT_s^n(r,s)
  &=(-1)^nk_y^n\cJ_x^n(r,s)-\rd_r\cJ_y^n(r,s) .
  \label{eq:cSs_cTs}
\end{alignat}
The differential operators in Eqs.(\ref{eq:we_cFxs}-\ref{eq:cSs_cTs})
are given by Eqs.(\ref{eq:rdv2}),
\begin{align}
  \brd_r=\rd_{r}+\frac{1}{r} ,
     \qquad
  \brd_s=\frac{\rho}{r}\rd_s .
  \label{eq:rs_opera}
\end{align}
$\brd_r$ does not commute with $\rd_r$.
On the other hand, $\brd_s$ commutes with $\rd_s$ in the constant bend.
$k_r^n$ is the radial wavenumber which is real or purely imaginary, \ie,
$k_r^n\in\mathbb{A}=\{\mathbb{R},i\mathbb{R}\}$, depending on $\beta$, $k$ and $k_y^n$,
\begin{align}
  (k_r^n)^2=(k\beta)^2-(k_y^n)^2
   ,\qquad
  \hr=k_r^nr
   ,\qquad
  \hr_{a,b}=k_r^nr_{a,b}
   ,\qquad
  \hrho=k_r^n\rho .
  \label{eq:krn}
\end{align}
In considering the fields in the constant bend, we often use
the dimensionless radial variable $\hr\in\mathbb{A}$ which is normalized by $k_r^n$.
Since $k_r^n$ depends on the vertical mode number $n$,
although $\hr$ should have the index as $\hr^n$ in precise,
we omit the index $n$ of $\hr^n$ for brevity.

As shown in appendix \ref{sec:we_xs},
we can straightforwardly solve Eqs.(\ref{eq:we_cFxs}) for $\cE_{x,s}^n$ and $\cB_{x,s}^n$
using the variation of parameters under the boundary conditions of the sidewalls of
the curved rectangular pipe.
Instead of this way, we can get the expressions of $\cE_{x,s}^n$ and $\cB_{x,s}^n$
from the solutions of $\cE_y^n$ and $\cB_y^n$ through Maxwell equations for
the Fourier coefficients of the field, given by
Eqs.(\ref{eq:cBx}-\ref{eq:cBs}) in appendix \ref{sec:boundary}.
In the present paper we demonstrate both ways to verify the solutions analytically.
As shown later in section \ref{sec:sol_mfEB_xs},
we have the third way to get $\cE_{x,s}^n$ and $\cB_{x,s}^n$
through the vertical components of the fields in the Laplace domain.
Eqs.(\ref{eq:EBy_trans}-\ref{eq:EBxs_trans}) summarize the sections and appendices
which show these three ways to find the expressions of $\cE_{x,s}^n$ and $\cB_{x,s}^n$.

By the way, we would like to mention the mode expansion method with respect to
the radial oscillation.
If Eq.(\ref{eq:wen_bend}) is homogeneous as $\rd_{\rm v}^2\cF=0$,
the general solution may be given as
\begin{align}
  \cF(r,s)
  \propto C_\nu(\hr)e^{\pm i(\nu/\rho)s} ,
    \qquad
  C_\nu
  =J_\nu,Y_\nu,H_\nu^{(1,2)}~\text{etc}.
  \label{eq:gnrl_sol_hmg}
\end{align}
$C_\nu$ is the Bessel function of the order $\nu$.
$\nu/\rho$ denotes the longitudinal wavenumber of the field in the bend.
Giving two indices $m$ and $n$ to the order $\nu$ as $\nu_m^n$, if we expand $\cF$
using a linearly independent pair of $C_\nu$ for $\nu=\nu_m^n$ as the basis of
the $m$th radial oscillation mode, then it may be possible to solve Eqs.(\ref{eq:wen_bend})
under the boundary conditions of the sidewalls, though we do not do it in the present paper.
We solve them using the Laplace transform with respect to $s$ and
the variation of parameters with respect to $r$ as shown in section \ref{sec:we_in_LD}.

\subsection{Laplace transform with respect to the reference axis}
\label{sec:we_in_LD}

\begin{figure}[h]
  \begin{center}
    \includegraphics[scale=0.32,clip]{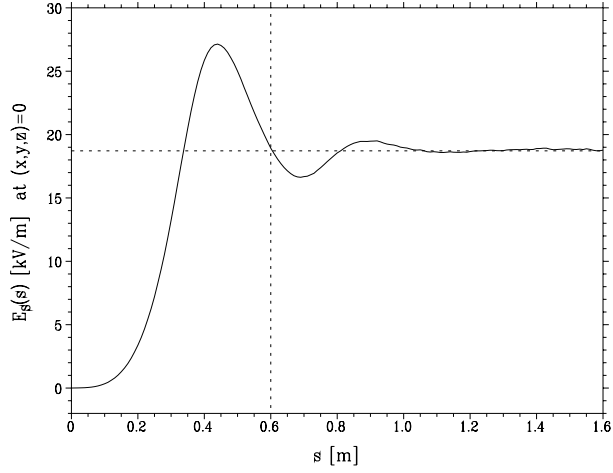}
    \caption[Longitudinal electric field of transient CSR in a bend]
      {\small
       Longitudinal electric field of transient CSR in a bend in the time domain,
       which is shielded by a curved rectangular pipe.
       The horizontal axis is the longitudinal distance $s$ from the entrance of the bend
       ($s=0$).
       The vertical axis is $E_s(x,y,s;z)$ at the center of the bunch $(x,y,z)=0$.
       We assume a rigid Gaussian bunch which has no transverse spread.
       We calculated $E_s$ using Eq.(\ref{eq:PE_Es}) by numerically solving
       the parabolic wave equation given by Eq.(\ref{eq:PE_Exy}).
       The horizontal dotted line shows the steady $E_s$ at $z=0$ in the pipe,
       given by Eq.(158) in \cite{agoh}.
       The parameters are given as
       $\rho=3{\rm m}$, $w=8{\rm cm}$, $h=4{\rm cm}$,
       $\sig_z=1{\rm mm}$ and $q=1{\rm nC}$.
       From Eq.(\ref{eq:formation}) for $k\sim\sig_z^{-1}$, the criterion of
       the formation length of the field in the bend is about $s_{\rm f}\approx 0.6{\rm m}$
       which is shown by the vertical dotted line.
       }
    \label{fig:asymp_csr}
  \end{center}
\end{figure}

As described in section \ref{sec:assumption}, we assume a semi-infinite bending magnet
in which the reference axis ($s$-axis) has a constant radius $\rho$.
However, the beam pipe is not a rectangular toroid as the one employed in
\cite{warnock_morton}, \ie,
we do not impose the periodic boundary condition to the field with respect to $s$.
We use the Laplace transform in order to solve Eqs.(\ref{eq:we_cFxs}-\ref{eq:wen_bend})
as the initial value problem with respect to $s$.
Giving the values of the field and its derivative with respect to $s$ on the plane
at the entrance of the bend ($s=+0$: common area between the straight and bending sections),
we will find the explicit and exact expression of the field in $s>0$.
Fig.\ref{fig:asymp_csr} shows a transient behavior of $E_s$ in a long bend
as a function of $s$.
$E_s$ is the longitudinal electric field of CSR in the time domain.
The field tends to be steady as the bunch travels in the bend
if the bend is much longer than $s_{\rm f}$ as given by Eq.(\ref{eq:formation})
which is the formation length of the field of CSR in the bend.

We Laplace transform $\cF^n(s)$ with respect to $s$,
\begin{align}
  \mfF^n(\kap)
 &=\cL[\cF^n(s)]
  =\int_0^\infty\frac{ds}{\rho}\cF^n(s)e^{-\kap s/\rho} ,
   \qquad
  \cF^n(s)=\cL^{-1}[\mfF^n(\kap)]
  =\int_{\vpi-i\infty}^{\vpi+i\infty}\frac{d\kap}{2\pi i}\mfF^n(\kap)e^{\kap s/\rho} .
  \label{eq:Laplace}
\end{align}
$\cL$ denotes the Laplace transform;
$\kap\in\mathbb{C}$ is the Laplace variable.
$\mfF^n$ is the Laplace transform of $\cF^n$ $(=\cF^n_{\pm})$ which represents
the $n$th Fourier coefficients of the components of the fields and current in
the frequency domain as in Eqs.(\ref{eq:coeff_FJ}),
\begin{align}
  \mfE_{x,y,s}^n
  =\cL[\cE_{x,y,s}^n] ,
    \qquad
  \mfB_{x,y,s}^n
  =\cL[\cB_{x,y,s}^n] ,
    \qquad
  \mfJ^n
  =(\mfJ_0^n,\mfJ_{x,y,s}^n)
  =\cL[\cJ^n]  .
   \label{eq:J_LD}
\end{align}
$\vpi\in\mathbb{R}$ is the abscissa of convergence of
the inverse Laplace transform ($\cL^{-1}$).
The field of CSR in the semi-infinite bend tends to be steady in the limit of $s\to\infty$ 
as shown in Fig.\ref{fig:asymp_csr}.
Since the field does not grow exponentially for $s\to\infty$,
it is enough to set $\vpi=+0$.
For convenience in describing the fields in the Laplace domain,
we change the Laplace variable from $\kap$ to $\nu$ as follows,
\begin{align}
  \kap=i\nu
    \quad
  (\nu\in\mathbb{C}) .
   \label{eq:u_nu}
\end{align}
This means that we rotate the Laplace plane clockwise by $\pi/2$.
The new Laplace variable $\nu$ corresponds to the order of
the Bessel functions as it can be seen from Eq.(\ref{eq:wenu}).
Using $\nu$ instead of $\kap$, we rewrite the Bromwich integral
given by the second equation of (\ref{eq:Laplace}) for $\vpi=+0$,
\begin{align}
  \cF^n(s)
  =\cL^{-1}[\mfF^n(\nu)]
  =\int_{-\infty-i0}^{\infty-i0}\frac{d\nu}{2\pi}
   \mfF^n(\nu)e^{i\nu s/\rho} .
  \label{eq:ILT_nu}
\end{align}
It follows that the contour of the Bromwich integral goes under the real axis of
the $\nu$-plane by an infinitesimal amount in any case as it will be shown later in
Fig.\ref{fig:contour} (p.\pageref{fig:contour}).
Thus the causality of the field is determined in the Laplace transform.
Since Eq.(\ref{eq:ILT_nu}) is equivalent to the Fourier transform with respect to $\nu$,
$\nu/\rho$ denotes the longitudinal wavenumber of the field in the bend,
which is complex in general.
As already done in Eq.(\ref{eq:ILT_nu}), we redefine the argument of all the functions in
the Laplace domain $\mfF^n(\kap)=\mfF^n(i\nu)$ as $\mfF^n(\nu)$ omitting $i$ for brevity.
We do not use $\kap$ hereafter in the present paper.
Also, we call $\mfF^n(\nu)$ the field and current in the Laplace domain
(Laplace plane, $\nu$-plane/domain, order domain/plane).

According to Eq.(\ref{eq:wen_bend}), the vertical components of the fields in
the Laplace domain satisfy
\begin{align}
  \bigg(\brd_{\hr}\rd_{\hr}+1-\frac{\nu^2}{\hr^2}\bigg)
  \bigg\{{\mfE_y^n(r,\nu) \atop c\mfB_y^n(r,\nu)}\bigg\}
  =\frac{1}{k_r^n\hr}\bigg\{{\bmfD_y^n(r,\nu) \atop \bmfA_y^n(r,\nu)}\bigg\} .
  \label{eq:wenu}
\end{align}
$\brd_{\hr}$ is the radial differential operator which is the dimensionless version of
$\brd_r$ given by Eq.(\ref{eq:rs_opera}),
\begin{alignat}{2}
  \brd_{\hr}
  =\rd_{\hr}+\frac{1}{\hr}
  =\frac{\brd_{r}}{k_r^n}
   ,\qquad
  \brd_{\hr'}
  =\rd_{\hr'}+\frac{1}{\hr'}
  =\frac{\brd_{r'}}{k_r^n}
   ;\qquad
  \brd_{r}
  =\rd_{r}+\frac{1}{r}
   ,\qquad
  \brd_{r'}
  =\rd_{r'}+\frac{1}{r'} .
  \label{eq:rd_hr}
\end{alignat}
$r'$ and $\hr'$ are dummy variables for $r$ and $\hr$ given by Eq.(\ref{eq:krn}),
which we will use in the following sections,
\begin{align}
  r'
  =\rho+x'
  \in\mathbb{R}^{+} ,
   \qquad
  \hr'
  =k_r^nr'
  \in\mathbb{A} .
  \label{eq:hrp}
\end{align}
$\bmfD_y^n$ and $\bmfA_y^n$ are the driving terms in the wave equations for
the vertical fields in the Laplace domain,
\begin{alignat}{2}
  \bmfD_y^n(r,\nu)
  &=\mfD_y^n(r,\nu)+Z_0r\mfS_y^n(r,\nu) ,
   \qquad
  &\mfD_y^n(r,\nu)&
   =\frac{1}{r}[(i\nu+\rho\rd_{s})\cE_y^n(r,s)]_{s=+0} ,
  \label{eq:bmfDy}
  \\
  \bmfA_y^n(r,\nu)
  &=\mfA_y^n(r,\nu)+Z_0r\mfT_y^n(r,\nu) ,
   \qquad
  &\mfA_y^n(r,\nu)&
   =\frac{1}{r}[(i\nu+\rho\rd_{s})c\cB_y^n(r,s)]_{s=+0} .
  \label{eq:bmfAy}
\end{alignat}
$\mfD_y^n$ and $\mfA_y^n$ are the parts of the driving terms,
which contain the initial values of $\cE_y^n$ and $\cB_y^n$ at the entrance of
the bend ($s=+0$).
$\mfS_y^n$ and $\mfT_y^n$ are the Laplace transforms of the Fourier coefficients of
the source terms $\cS_y^n$ and $\cT_y^n$ given by Eqs.(\ref{eq:cSy_cTy}),
\begin{alignat}{2}
  \mfS_y^n(r,\nu)
  &=\cL[\cS_y^n(r,s)]
  &&=(-1)^nk_y^n\mfJ_0^n(r,\nu)-ik\beta\mfJ_y^n(r,\nu) ,
   \label{eq:mfSy}
   \\
  \mfT_y^n(r,\nu)
  &=\cL[\cT_y^n(r,s)]
  &&=\brd_r\mfJ_s^n(r,\nu)-\frac{1}{r}\{i\nu\mfJ_x^n(r,\nu)-\cJ_x^n(r,0)\} ,
   \label{eq:mfTy}
\end{alignat}
where $\cJ_x^n(r,0)$ is the initial value of $\cJ_x^n(r,s)$ at $s=0$.

The Fourier coefficients of the radial and longitudinal components of the fields are 
coupled in the wave equations (\ref{eq:we_cFxs}) since the $s$-axis is curved in the bend.
We Laplace transform $\cE_{x,s}^n$ and $\cB_{x,s}^n$ with respect to $s$
through Eq.(\ref{eq:Laplace}) as shown in Eqs.(\ref{eq:J_LD}).
The wave equations for the radial and longitudinal components of the fields
in the Laplace domain, $\mfE_{x,s}^n$ and $\mfB_{x,s}^n$, are given as
\begin{align}
  &
  \bigg(\rd_{\hr}\brd_{\hr}+1-\frac{\nu^2}{\hr^2}\bigg)
  \bigg\{{\mfE_x^n(r,\nu) \atop c\mfB_x^n(r,\nu) }\bigg\}
  -\frac{2i\nu}{\hr^2}\bigg\{{\mfE_s^n(r,\nu) \atop c\mfB_s^n(r,\nu) }\bigg\}
  =\frac{1}{k_r^n\hr}\bigg\{{\bmfD_x^n(r,\nu) \atop \bmfA_x^n(r,\nu) }\bigg\} ,
   \label{we_mfEBx}
  \\
  &
  \bigg(\rd_{\hr}\brd_{\hr}+1-\frac{\nu^2}{\hr^2}\bigg)
  \bigg\{{\mfE_s^n(r,\nu) \atop c\mfB_s^n(r,\nu)}\bigg\}
  +\frac{2i\nu}{\hr^2}\bigg\{{\mfE_x^n(r,\nu) \atop c\mfB_x^n(r,\nu) }\bigg\}
  =\frac{1}{k_r^n\hr}\bigg\{{\bmfD_s^n(r,\nu) \atop \bmfA_s^n(r,\nu) }\bigg\} ,
   \label{we_mfEBs}
\end{align}
where the radial differential operator $\rd_{\hr}\brd_{\hr}$ differs from
$\brd_{\hr}\rd_{\hr}$ which is involved in Eq.(\ref{eq:wenu}) as shown in
Eqs.(\ref{eq:rdx_brdx}).
$\bmfD_{x,s}^n$ and $\bmfA_{x,s}^n$ are the driving terms of
Eqs.(\ref{we_mfEBx}-\ref{we_mfEBs}),
\begin{align}
  \bigg\{{\bmfD_x^n(r,\nu) \atop \bmfA_x^n(r,\nu)}\bigg\}
  &=\bigg\{{\mfD_x^n(r,\nu) \atop \mfA_x^n(r,\nu)}\bigg\}
   +Z_0r\bigg\{{\mfS_x^n(r,\nu) \atop \mfT_x^n(r,\nu)}\bigg\} ,
   \label{eq:drv_x}
   \\
  \bigg\{{\bmfD_s^n(r,\nu) \atop \bmfA_s^n(r,\nu)}\bigg\}
  &=\bigg\{{\mfD_s^n(r,\nu) \atop \mfA_s^n(r,\nu)}\bigg\}
   +Z_0r\bigg\{{\mfS_s^n(r,\nu) \atop \mfT_s^n(r,\nu)}\bigg\} .
   \label{eq:drv_s}
\end{align}
$\mfD_{x,s}^n$ and $\mfA_{x,s}^n$ are the parts of the driving terms,
which contain the initial values of $(\cE_{x,s}^n,\cB_{x,s}^n)$
and their longitudinal derivatives at $s=+0$,
\begin{align}
  \bigg\{{\mfD_x^n(r,\nu) \atop \mfA_x^n(r,\nu)}\bigg\}
  &=\frac{1}{r}
   \bigg[
      (i\nu+\rho\rd_{s})\bigg\{{\cE_x^n(r,s) \atop c\cB_x^n(r,s) }\bigg\}
     -2\bigg\{{\cE_s^n(r,s) \atop c\cB_s^n(r,s) }\bigg\}
   \bigg]_{s=+0} ,
   \\
  \bigg\{{\mfD_s^n(r,\nu) \atop \mfA_s^n(r,\nu)}\bigg\}
  &=\frac{1}{r}
   \bigg[
      (i\nu+\rho\rd_{s})\bigg\{{\cE_s^n(r,s) \atop c\cB_s^n(r,s)}\bigg\}
     +2\bigg\{{\cE_x^n(r,s) \atop c\cB_x^n(r,s)}\bigg\}
   \bigg]_{s=+0} .
\end{align}
$\mfS_{x,s}^n$ and $\mfT_{x,s}^n$ are the Laplace transform of the radial and longitudinal
components of the source terms $\cS_{x,s}^n$ and $\cT_{x,s}^n$
given by Eqs.(\ref{eq:cSx_cTx}) and (\ref{eq:cSs_cTs}),
\begin{alignat}{2}
  \mfS_x^n(r,\nu)
  &=\cL[\cS_x^n(r,s)]
  &&=\rd_{r}\mfJ_0^n(r,\nu)-ik\beta\mfJ_x^n(r,\nu) ,
   \label{eq:mfSx}
   \\
  \mfS_s^n(r,\nu)
  &=\cL[\cS_s^n(r,s)]
  &&=\frac{1}{r}\{i\nu\mfJ_0^n(r,\nu)-\cJ_0^n(r,0)\}
   -ik\beta\mfJ_s^n(r,\nu) ,
   \\
  \mfT_x^n(r,\nu)
  &=\cL[\cT_x^n(r,s)]
  &&=\frac{1}{r}\{i\nu\mfJ_y^n(r,\nu)-\cJ_y^n(r,0)\}
   -(-1)^nk_y^n\mfJ_s^n(r,\nu) ,
   \\
  \mfT_s^n(r,\nu)
  &=\cL[\cT_s^n(r,s)]
  &&=(-1)^nk_y^n\mfJ_x^n(r,\nu)-\rd_{r}\mfJ_y^n(r,\nu) ,
   \label{eq:mfTs}
\end{alignat}
where $\cJ_{0,y}^n(r,0)$ is the initial value of $\cJ_{0,y}^n(r,s)$ at $s=0$.

We will solve Eqs.(\ref{we_mfEBx}-\ref{we_mfEBs}) in appendix \ref{sec:we_xs}
using the eigenfunctions for the operator of the wave equations,
\begin{align}
  \mfE_{\pm}^n
   =\frac{\mfE_x^n\pm i\mfE_s^n}{2}
     \qquad\text{and}\qquad
  \mfB_{\pm}^n
   =\frac{\mfB_x^n\pm i\mfB_s^n}{2} ,
   \label{eq:mfEB_pm}
\end{align}
in order to disentangle the radial and longitudinal components of the fields
as shown in Eq.(\ref{eq:we_xs_LD}).
Instead of this straightforward solution in appendix \ref{sec:we_xs},
we can get the expressions of $\mfE_{x,s}^n$ and $\mfB_{x,s}^n$ from the solutions of
$\mfE_y^n$ and $\mfB_y^n$ through Eqs.(\ref{eq:mfBx_EyBy}-\ref{eq:mfBs_EyBy})
as shown later in section \ref{sec:sol_mfEB_xs}.
We find the expressions of the radial and longitudinal components of the fields 
in three different ways in order to ensure them analytically.

\clearpage

\section{Fields in the Laplace domain}
\label{sec:LD_field}

We will derive Eqs.(\ref{eq:mfEBy_solution}) and (\ref{eq:mfBsEx}-\ref{eq:mfBxEs})
which are the components of the fields in the Laplace domain.

\subsection{Solutions of the vertical fields in the Laplace domain}
\label{sec:mfEBy}

The wave equation for the vertical components of the fields in the Laplace domain is
given by Eq.(\ref{eq:wenu}).
In order to solve it using the variation of parameters, 
we first consider its homogeneous version,
\begin{align}
  O_{\nu}(\hr)C_{\nu}(\hr)=0
   \qquad\text{where}\quad
  O_{\nu}(\hr)
  =\brd_{\hr}\rd_{\hr}+1-\frac{\nu^2}{\hr^2}
   \quad
  (\nu\in\mathbb{C},~ \hr\in\mathbb{A}) .
  \label{eq:BDE}
\end{align}
$O_{\nu}(\hr)$ is the operator of the Bessel differential equation with respect to $\hr$.
$C_{\nu}$ represents a Bessel function of order $\nu$.
The general solution of Eq.(\ref{eq:BDE}) is given by an arbitrary pair of
linearly independent Bessel functions.
Choosing $J_\nu$ and $Y_\nu$ as the pair of the fundamental solutions,
the general solution of Eq.(\ref{eq:wenu}) is given as
\begin{align}
  \mfE_y^n(r)
  =\mfa_{+}(r)J_\nu(\hr)+\mfb_{+}(r)Y_\nu(\hr) ,
    \qquad
  c\mfB_y^n(r)
  =\mfa_{-}(r)J_\nu(\hr)+\mfb_{-}(r)Y_\nu(\hr) .
  \label{eq:mfEBy_JY}
\end{align}
$\mfa_{\pm}$ and $\mfb_{\pm}$ are unknown functions.
Substituting Eqs.(\ref{eq:mfEBy_JY}) into Eq.(\ref{eq:wenu}),
we integrate them with respect to $r$,
\begin{align}
  \bigg\{{\mfa_{+}(r) \atop \mfa_{-}(r)}\bigg\}
  &=\bigg\{{\mfa_{+}^0 \atop \mfa_{-}^0}\bigg\}
   +\frac{\pi}{2}\int_{r}^{r_b}dr'
    \bigg\{{\bmfD_y^n(r') \atop \bmfA_y^n(r')}\bigg\}Y_\nu(\hr') ,
  \label{eq:a_pm}
   \\
  \bigg\{{\mfb_{+}(r) \atop \mfb_{-}(r)}\bigg\}
  &=\bigg\{{\mfb_{+}^0 \atop \mfb_{-}^0}\bigg\}
   +\frac{\pi}{2}\int_{r_a}^{r}dr'
    \bigg\{{\bmfD_y^n(r') \atop \bmfA_y^n(r')}\bigg\}J_\nu(\hr') .
  \label{eq:b_pm}
\end{align}
$r'$ and $\hr'$ are given by Eqs.(\ref{eq:hrp}).
$\mfa_{\pm}^0$ and $\mfb_{\pm}^0$ are undetermined constants of integration.
They are determined in order that $\mfE_y^n$ and $\mfB_y^n$ satisfy the boundary conditions 
on the sidewalls of the curved pipe,
\begin{align}
  \mfE_y^n=0
   ,\qquad
  \rd_r\mfB_y^n=0
   \quad
  \text{at}~~ r=r_a, r_b .
  \label{eq:BC_side}
\end{align}
Eqs.(\ref{eq:BC_side}) fix the constants $\mfa_{\pm}^0$ and $\mfb_{\pm}^0$ as follows,
\begin{align}
  \bigg\{{\mfa_{+}^0 \atop \mfb_{+}^0}\bigg\}
  &=\frac{\pi}{2}\int_{r_a}^{r_b}dr'\bmfD_y^n(r')
    \bigg[\frac{Y_\nu(\hr_b)J_\nu(\hr_a)}{p_\nu(\hr_b,\hr_a)}
           \bigg\{{Y_\nu(\hr')\atop J_\nu(\hr')}\bigg\}
         -\frac{J_\nu(\hr_b)Y_\nu(\hr_a)}{p_\nu(\hr_b,\hr_a)}
           \bigg\{{J_\nu(\hr')\atop Y_\nu(\hr')}\bigg\}
    \bigg] ,
  \label{eq:const_e}
  \\
  \bigg\{{\mfa_{-}^0 \atop \mfb_{-}^0}\bigg\}
  &=\frac{\pi}{2}\int_{r_a}^{r_b}dr'\bmfA_y^n(r')
    \bigg[\frac{Y_\nu'(\hr_b)J_\nu'(\hr_a)}{s_\nu(\hr_b,\hr_a)}
           \bigg\{{Y_\nu(\hr')\atop J_\nu(\hr')}\bigg\}
         -\frac{J_\nu'(\hr_b)Y_\nu'(\hr_a)}{s_\nu(\hr_b,\hr_a)}
           \bigg\{{J_\nu(\hr')\atop Y_\nu(\hr')}\bigg\}
    \bigg] .
  \label{eq:const_b}
\end{align}
$p_\nu$ and $s_\nu$ are the cross products of the Bessel functions,
defined in 9.1.32 of \cite{abramo_stegun},
\begin{alignat}{2}
  p_\nu(b,a)
  &=J_\nu(b)Y_\nu(a)-Y_\nu(b)J_\nu(a) ,
     \qquad&
  q_\nu(b,a)
  &=J_\nu(b)Y_\nu'(a)-Y_\nu(b)J_\nu'(a)
   =\rd_ap_\nu(b,a) ,
  \label{eq:CP_pq}
  \\
  s_\nu(b,a)
  &=J_\nu'(b)Y_\nu'(a)-Y_\nu'(b)J_\nu'(a) ,
    \qquad&
  r_\nu(b,a)
  &=J_\nu'(b)Y_\nu(a)-Y_\nu'(b)J_\nu(a)
   =\rd_bp_\nu(b,a) ,
  \label{eq:CP_sr}
\end{alignat}
where $a,b\in\mathbb{C}$ in general, however, $a,b\in\mathbb{A}$ in the present study
since $k_r^n\in\mathbb{A}$ and $r\in\mathbb{R}^{+}$.
The prime $(')$ denotes the derivative with respect to the argument.
The Wronskian of $J_\nu$ and $Y_\nu$ is given by 9.1.16 in \cite{abramo_stegun},
\begin{align}
   W[J_\nu(\hr),Y_\nu(\hr)]=q_\nu(\hr,\hr)=\frac{2}{\pi\hr} .
\end{align}
From Eqs.(\ref{eq:mfEBy_JY}-\ref{eq:b_pm}) and (\ref{eq:const_e}-\ref{eq:const_b}),
we get the vertical components of the fields in the Laplace domain,
\begin{align}
  \mfE_y^n(r,\nu)=\int_{r_a}^{r_b}\frac{dr'}{\rho}\bmfD_y^n(r',\nu)\mfG_{+}^n(r,r',\nu) ,
    \qquad
  c\mfB_y^n(r,\nu)=\int_{r_a}^{r_b}\frac{dr'}{\rho}\bmfA_y^n(r',\nu)\mfG_{-}^n(r,r',\nu) .
  \label{eq:mfEBy_solution}
\end{align}
$\mfG_{+}^n$ and $\mfG_{-}^n$ are the Green functions of the wave equations (\ref{eq:wenu})
for $\mfE_y^n$ and $\mfB_y^n$ which satisfy Eqs.(\ref{eq:BC_side}),
\begin{align}
  \mfG_{+}^n(r,r',\nu)
  &=\frac{\pi}{2}\rho
  \bigg\{\theta(r-r')\frac{p_\nu(\hr_b,\hr)p_\nu(\hr',\hr_a)}{p_\nu(\hr_b,\hr_a)}
        +\theta(r'-r)\frac{p_\nu(\hr_b,\hr')p_\nu(\hr,\hr_a)}{p_\nu(\hr_b,\hr_a)}
  \bigg\} ,
  \label{eq:mfGe}
   \\
  \mfG_{-}^n(r,r',\nu)
  &=\frac{\pi}{2}\rho
  \bigg\{\theta(r-r')\frac{r_\nu(\hr_b,\hr)q_\nu(\hr',\hr_a)}{s_\nu(\hr_b,\hr_a)}
        +\theta(r'-r)\frac{r_\nu(\hr_b,\hr')q_\nu(\hr,\hr_a)}{s_\nu(\hr_b,\hr_a)}
  \bigg\}
  \label{eq:mfGb}
   \\
  &=\frac{\pi}{2}\rho
    \bigg\{
      \theta(r-r')
      \frac{\hr_br_\nu(\hr_b,\hr)\hr_aq_\nu(\hr',\hr_a)}{\hr_b\hr_as_\nu(\hr_b,\hr_a)}
     +\theta(r'-r)
      \frac{\hr_br_\nu(\hr_b,\hr')\hr_aq_\nu(\hr,\hr_a)}{\hr_b\hr_as_\nu(\hr_b,\hr_a)}
  \bigg\} .
  \label{eq:mfGm_ba}
\end{align}
$r$ and $r'$ denote the radial positions of the observation point and
the source particle of the field respectively.
$\theta$ is the Heaviside step function similar to the one in Eq.(\ref{eq:curv}).
In considering $\mfG_{-}^n$ in the limit of $k_r^n\to0$,
we use Eq.(\ref{eq:mfGm_ba}) instead of Eq.(\ref{eq:mfGb}),
because the cross products $q_{\nu}$, $r_{\nu}$ and $s_{\nu}$ for $\nu\ne0$ diverge in
the limit of $k_r^n\to0$ unlike $p_{\nu}$ as seen from Eqs.(\ref{lim_kr0_aq_br_bas}).
Therefore we use Eq.(\ref{eq:mfGm_ba}) in discussing the pole structure of
$\mfG_{-}^n$ in the $\nu$-domain for $\forall k\in\mathbb{R}$ and
$\forall n\in\mathbb{Z}_0^{+}$.

In electromagnetism, the Green function often means the solution of
Maxwell equations for a field which is created by a point charge.
Unlike this usage, the Green function in the present paper means the solution of the wave 
equation whose driving term is a $\delta$-function of the coordinate variable.
That is, the Green function $\mfG_{\pm}^n$ satisfies the wave equation (\ref{eq:wenu})
whose driving term is replaced by the radial $\delta$-function,
\begin{align}
  O_{\nu}(\hr)\mfG_{\pm}^n(r,r',\nu)
  &=\frac{\rho}{\hr}\delta(\hr-\hr') ,
    \qquad
  \mfG_{\pm}^n(r,r',\nu)
  =\mfG_{\pm}^n(r',r,\nu) .
   \label{eq:BDE_mfGpm}
\end{align}
$O_{\nu}(\hr)$ is the operator given by Eq.(\ref{eq:BDE}).
The first equation of (\ref{eq:BDE_mfGpm}) holds also for $O_{\nu}(\hr')$
since $\mfG_{\pm}^n$ is symmetric with respect to the exchange of $r$ and $r'$
as shown in the second equation of (\ref{eq:BDE_mfGpm}).
$\mfG_{\pm}^n$ is continuous at $r'=r$ while having a kink at $r'=r$.
Therefore $\rd_{\hr}\mfG_{\pm}^n$ and $\rd_{\hr}^2\mfG_{\pm}^n$ have
$\theta(\hr-\hr')$ and $\delta(\hr-\hr')$ respectively.
If the sidewalls are symmetric with respect to the $s$-axis, \ie, if $x_b=-x_a$,
the sign of $\mfG_{\pm}^n$ denotes the horizontal parity of the field in the limit of
$\rho\to\infty$ (straight section) as shown by Eq.(\ref{eq:lim_mfGpm}).
Although the fields in the bending section are not symmetric with respect to $r=\rho$
since the trajectory is curved, we use the sign $(\pm)$ of $\mfG_{\pm}^n$ for convenience 
to describe the fields in the curved pipe.

Since $\nu/\rho\in\mathbb{C}$ denotes the longitudinal wavenumber of the field in
the constant bend as seen from Eq.(\ref{eq:ILT_nu}), $\nu/\rho$ goes to $k_s$ in
the limit of $\rho\to\infty$, \ie,
\begin{align}
  \lim_{\rho\to\infty}\frac{\nu}{\rho}
  =k_s ,
    \qquad
  \lim_{\rho\to\infty}\kap_x
  =k_x ,
   \qquad
  \kap_x^2
  =(k_r^n)^2-\frac{\nu^2}{r^2}
    \quad
  (k_{x,s},\kap_x\in\mathbb{C}) .
  \label{eq:lim_nu}
\end{align}
$k_s$ and $k_x$ are respectively the longitudinal and horizontal wavenumbers of
the field in a straight section, which are related through the second equation of
(\ref{eq:def_kx}).
$k_s$ and $k_x$ are complex variables unlike $k_s^{mn}\in\mathbb{A}$ and
$k_x^m\in\mathbb{R}_0^{+}$ which are involved in Eq.(\ref{eq:asymp_lim}).
$\kap_x^2$ is a complex variable which corresponds to $1-\nu^2/\hr^2$ in $O_{\nu}(\hr)$
given by Eq.(\ref{eq:BDE}).
$\kap_x$ goes to $k_x$ in the limit of $\rho\to\infty$.
As described in appendix \ref{sec:coordinates}, the radial operator in the bend $\brd_r$ 
goes to $\rd_x$ in the limit of $\rho\to\infty$.
Therefore, for $\rho\to\infty$, the Bessel differential equation (\ref{eq:BDE}) tends to be 
the differential equation of the harmonic oscillation in the horizontal direction,
\begin{align}
  \bigg[\brd_r\rd_r+(k_r^n)^2-\frac{\nu^2}{r^2}\bigg]
  \bigg\{{J_{\nu}(k_r^nr) \atop Y_{\nu}(k_r^nr)}\bigg\}
  =0
  \quad~~\os{\frac{}{}(\rho\to\infty)\frac{}{}}{\longrightarrow}\quad~~
  (\rd_x^2+k_x^2)\bigg\{{\cos(k_xx) \atop \sin(k_xx)}\bigg\}
  =0 .
  \label{eq:BDE_asympt}
\end{align}
Accordingly, the first equation of (\ref{eq:BDE_mfGpm}) goes to Eq.(\ref{eq:we_mfG_str}) in
the limit of $\rho\to\infty$,
\begin{align}
  (\rd_x^2+k_x^2)\mfG_{\pm}^n(x,x',k_s)
  =\delta(x-x') ,
    \qquad
  \mfG_{\pm}^n(x,x',k_s)
  =\lim_{\rho\to\infty}\mfG_{\pm}^n(r,r',\nu) .
   \label{eq:we_mfGpm_str}
\end{align}
$\mfG_{\pm}^n(x,x',k_s)$ denotes the Green functions of the Laplace domain fields in
the straight pipe, given by Eq.(\ref{eq:mfGbe_strt}).

\subsection{Solutions of the radial and longitudinal fields in the Laplace domain}
\label{sec:sol_mfEB_xs}

We can get the expressions of $\mfE_{x,s}^n$ and $\mfB_{x,s}^n$
from the solutions of $\mfE_y^n$ and $\mfB_y^n$ given by Eqs.(\ref{eq:mfEBy_solution}) 
through Eqs.(\ref{eq:mfBx_EyBy}-\ref{eq:mfBs_EyBy}).
Multiplying $ik\beta$ or $(-1)^nk_y^n$ with Eqs.(\ref{eq:mfEBy_solution}),
we rewrite them in order that $\mfE_y^n$ and $\mfB_y^n$ each have the initial values of
only the electric field $\cE_{x,s}^n$ or only the magnetic field $\cB_{x,s}^n$ as follows,
\begin{align}
  -\bigg\{{(-1)^nk_y^n \atop ik\beta}\bigg\}\mfE_y^n(r,\nu)
  &=\int_{r_a}^{r_b}\frac{dr'}{\rho}
    \bigg\{{ \mfU_{+}^n(r,r',\nu) \atop \mfU_{-}^n(r,r',\nu) }\bigg\}
   +\frac{1}{r}\bigg\{{\cE_s^n(r,0) \atop -c\cB_x^n(r,0)}\bigg\}
   +Z_0\bigg\{{\mfJ_0^n(r,\nu) \atop  -\mfJ_y^n(r,\nu)}\bigg\} ,
  \label{eq:mfEy_rwrtn}
  \\
  \bigg\{{(-1)^nk_y^n \atop ik\beta}\bigg\}c\mfB_y^n(r,\nu)
  &=\int_{r_a}^{r_b}\frac{dr'}{\rho}
    \bigg\{{ \mfV_{+}^n(r,r',\nu) \atop \mfV_{-}^n(r,r',\nu) }\bigg\}
   +\frac{1}{r}\bigg\{{c\cB_s^n(r,0)\atop -\cE_x^n(r,0)}\bigg\} ,
  \label{eq:mfBy_rwrtn}
\end{align}
where
\begin{align}
  \bigg\{{ \mfU_{+}^n(r,r',\nu) \atop \mfU_{-}^n(r,r',\nu) }\bigg\}
  &=
      \bigg\{{\bmfD_x^n(r',\nu) \atop \bmfA_s^n(r',\nu) }\bigg\}
      \rd_{r'}\mfG_{+}^n(r,r',\nu)
     +\bigg\{{-\bmfD_s^n(r',\nu) \atop \bmfA_x^n(r',\nu) }\bigg\}
      \frac{i\nu}{r'}\mfG_{+}^n(r,r',\nu) ,
   \\
  \bigg\{{ \mfV_{+}^n(r,r',\nu) \atop \mfV_{-}^n(r,r',\nu) }\bigg\}
  &=
      \bigg\{{\bmfA_x^n(r',\nu) \atop \bmfD_s^n(r',\nu)}\bigg\}
      \rd_{r'}\mfG_{-}^n(r,r',\nu)
     +\bigg\{{-\bmfA_s^n(r',\nu) \atop \bmfD_x^n(r',\nu)}\bigg\}
      \frac{i\nu}{r'}\mfG_{-}^n(r,r',\nu) .
\end{align}
$\bmfD_{x,s}^n$ and $\bmfA_{x,s}^n$ are the driving terms of the wave equations
(\ref{we_mfEBx}-\ref{we_mfEBs}) for $\mfE_{x,s}^n$ and $\mfB_{x,s}^n$,
given by Eqs.(\ref{eq:drv_x}-\ref{eq:drv_s}),
\begin{align}
  \bmfD_{x,s}^n(r',\nu)=\mfD_{x,s}^n(r',\nu)+Z_0r'\mfS_{x,s}^n(r',\nu) ,
    \qquad
  \bmfA_{x,s}^n(r',\nu)=\mfA_{x,s}^n(r',\nu)+Z_0r'\mfT_{x,s}^n(r',\nu) ,
  \label{eq:bmfDA_xs}
\end{align}
where the subscripts, $x$ and $s$, correspond in this order.
$\mfD_{x,s}^n$ and $\mfA_{x,s}^n$ are the parts of the driving terms
which involve the initial values of the Fourier coefficients of
the radial and longitudinal fields at $s=+0$,
\begin{align}
  \bigg\{{\mfD_x^n(r',\nu) \atop \mfA_x^n(r',\nu)}\bigg\}
  &=\frac{1}{r'}
    \bigg[
      (i\nu+\rho\rd_{s'})
      \bigg\{{\cE_{x}^n(r',s') \atop c\cB_x^n(r',s')}\bigg\}
     -2\bigg\{{\cE_{s}^n(r',s') \atop c\cB_s^n(r',s')}\bigg\}
    \bigg]_{s'=+0} ,
  \label{eq:mfDx_mfAx}
  \\
  \bigg\{{\mfD_s^n(r',\nu) \atop \mfA_s^n(r',\nu) }\bigg\}
  &=\frac{1}{r'}
    \bigg[
      (i\nu+\rho\rd_{s'})
      \bigg\{{\cE_{s}^n(r',s') \atop c\cB_s^n(r',s')}\bigg\}
     +2\bigg\{{\cE_{x}^n(r',s') \atop c\cB_x^n(r',s')}\bigg\}
    \bigg]_{s'=+0} .
  \label{eq:mfDs_mfAs}
\end{align}
$s'$ is a dummy variable for $s$.
$\mfS_{x,s}^n$ and $\mfT_{x,s}^n$ are the source terms given by
Eqs.(\ref{eq:mfSx}-\ref{eq:mfTs}), which are involved in the wave equations
(\ref{we_mfEBx}-\ref{we_mfEBs}) for $\mfE_{x,s}^n$ and $\mfB_{x,s}^n$
through the driving terms given by Eqs.(\ref{eq:bmfDA_xs}),
\begin{alignat}{2}
  \mfS_x^n(r',\nu)
  &=\rd_{r'}\mfJ_0^n(r',\nu)-ik\beta\mfJ_x^n(r',\nu) ,
   \\
  \mfS_s^n(r',\nu)
  &=\frac{1}{r'}\{i\nu\mfJ_0^n(r',\nu)-\cJ_0^n(r',0)\}
   -ik\beta\mfJ_s^n(r',\nu) ,
   \\
  \mfT_x^n(r',\nu)
  &=\frac{1}{r'}\{i\nu\mfJ_y^n(r',\nu)-\cJ_y^n(r',0)\}
   -(-1)^nk_y^n\mfJ_s^n(r',\nu) ,
   \\
  \mfT_s^n(r',\nu)
  &=(-1)^nk_y^n\mfJ_x^n(r',\nu)-\rd_{r'}\mfJ_y^n(r',\nu) .
\end{alignat}
As defined in Eqs.(\ref{eq:J_LD}), $\mfJ^n=(\mfJ_0^n,\mfJ_{x,y,s}^n)$ is
the current in the Laplace domain.

From Eqs.(\ref{eq:mfEy_rwrtn}-\ref{eq:mfBy_rwrtn}) and
(\ref{eq:mfBx_EyBy}-\ref{eq:mfBs_EyBy}),
we get the expressions of the radial and longitudinal components of the fields
in the Laplace domain,
\begin{align}
  \bigg\{{\mfE_x^n(r,\nu) \atop c\mfB_s^n(r,\nu)}\bigg\}
  &=\int_{r_a}^{r_b}\frac{dr'}{\rho}
    \bigg[
      \bigg\{{\bmfD_x^n(r',\nu) \atop \bmfA_s^n(r',\nu)}\bigg\}\mfH_{-}^n(r,r',\nu)
     +\bigg\{{-\bmfD_s^n(r',\nu) \atop \bmfA_x^n(r',\nu)}\bigg\}\mfK_{-}^n(r,r',\nu)
    \bigg] ,
  \label{eq:mfBsEx}
  \\
  \bigg\{{c\mfB_x^n(r,\nu) \atop \mfE_s^n(r,\nu)}\bigg\}
  &=\int_{r_a}^{r_b}\frac{dr'}{\rho}
    \bigg[
      \bigg\{{\bmfA_x^n(r',\nu) \atop \bmfD_s^n(r',\nu)}\bigg\}\mfH_{+}^n(r,r',\nu)
     +\bigg\{{-\bmfA_s^n(r',\nu) \atop \bmfD_x^n(r',\nu)}\bigg\}\mfK_{+}^n(r,r',\nu)
    \bigg] .
  \label{eq:mfBxEs}
\end{align}
$\mfH_{\pm}^n$ and $\mfK_{\pm}^n$ are respectively the Green functions and
the coupling Green functions of the radial and longitudinal components of the fields in
the Laplace domain,
\begin{align}
  \mfH_{\pm}^n
  &=\frac{\hr}{\hr'}(\brd_{\hr}\rd_{\hr}+1)\mfG_{\pm}^n
   +\rd_{\hr}\rd_{\hr'}\mfG_{\mp}^n
   =\frac{\nu^2}{\hr\hr'}\mfG_{\pm}^n
   +\bmfG_{\mp}^n ,
   \label{eq:mfHpm}
   \\
  \mfK_{\pm}^n
  &=i\nu
   \bigg(
     \frac{\rd_{\hr'}}{\hr}\mfG_{\pm}^n
    +\frac{\rd_{\hr}}{\hr'}\mfG_{\mp}^n
   \bigg) ,
   \label{eq:mfKpm}
   \\
  \bmfG_{\pm}^n
  &=\rd_{\hr}\rd_{\hr'}\mfG_{\pm}^n+\frac{\rho}{\hr}\delta(\hr-\hr') .
  \label{eq:bmfG_pm_delta}
\end{align}
$\brd_{\hr}$ is given by Eq.(\ref{eq:rd_hr}).
Eqs.(\ref{eq:mfHpm}-\ref{eq:mfKpm}) are related to the following identities
involving the radial $\delta$-function,
\begin{align}
  &
  \Big\{\frac{\hr}{\hr'}(\brd_{\hr}\rd_{\hr}+1)+\rd_{\hr}\rd_{\hr'}\Big\}
  [\hr\delta(\hr-\hr')]
  =\hr\delta(\hr-\hr')
    \label{eq:rdelr}
   ,\qquad
  (\hr\rd_{\hr}+\hr'\rd_{\hr'})\delta(\hr-\hr')
  = -\delta(\hr-\hr') .
\end{align}
$\rd_{\hr}\rd_{\hr'}\mfG_{\pm}^n$ includes $\delta(\hr-\hr')$ since $\mfG_{\pm}^n$ has
a kink at $r=r'$ as seen from Eqs.(\ref{eq:mfGe}-\ref{eq:mfGb}).
From Eq.(\ref{eq:bmfG_pm_delta}),
\begin{align}
  \bmfG_{+}^n(r,r',\nu)
  &=\frac{\pi}{2}\rho
    \bigg\{\theta(r-r')\frac{q_{\nu}(\hr_b,\hr)r_{\nu}(\hr',\hr_a)}{p_{\nu}(\hr_b,\hr_a)}
          +\theta(r'-r)\frac{q_{\nu}(\hr_b,\hr')r_{\nu}(\hr,\hr_a)}{p_{\nu}(\hr_b,\hr_a)}
    \bigg\} ,
  \label{eq:bmfG_pls}
   \\
  \bmfG_{-}^n(r,r',\nu)
  &=\frac{\pi}{2}\rho
    \bigg\{
       \theta(r-r')\frac{s_{\nu}(\hr_b,\hr)s_{\nu}(\hr',\hr_a)}{s_{\nu}(\hr_b,\hr_a)}
      +\theta(r'-r)\frac{s_{\nu}(\hr_b,\hr')s_{\nu}(\hr,\hr_a)}{s_{\nu}(\hr_b,\hr_a)}
    \bigg\} .
  \label{eq:bmfG_mns}
\end{align}
$\bmfG_{\pm}^n$ does not include $\delta(\hr-\hr')$
since $\delta(\hr-\hr')$ on the R.H.S. of Eq.(\ref{eq:bmfG_pm_delta}) 
is canceled with the one coming out of $\rd_{\hr}\rd_{\hr'}\mfG_{\pm}^n$.
$\bmfG_{\pm}^n$ is finite and continuous at $r'=r$ while it has a kink at $r'=r$.
Therefore $\rd_{\hr}\bmfG_{\pm}^n$ and $\rd_{\hr'}\bmfG_{\pm}^n$ are discontinuous at
$r'=r$.
According to the second equation of (\ref{eq:BDE_mfGpm}), the Green functions have
the following symmetries with respect to the exchange of $r$ and $r'$,
\begin{align}
  \mfH_{\pm}^n(r,r',\nu)=\mfH_{\pm}^n(r',r,\nu)
   ,\qquad
  \mfK_{\pm}^n(r,r',\nu)=\mfK_{\mp}^n(r',r,\nu)
   ,\qquad
  \bmfG_{\pm}^n(r,r',\nu)=\bmfG_{\pm}^n(r',r,\nu) .
   \label{eq:mfHK_symm}
\end{align}
According to the first equation of (\ref{eq:BDE_mfGpm}),
$\mfH_{\pm}^n$ and $\mfK_{\pm}^n$ satisfy the following wave equations,
\begin{align}
  \bigg(\rd_{\hr}\brd_{\hr}+1-\frac{\nu^2}{\hr^2}\bigg)
  \mfH_{\pm}^n(r,r',\nu)
  -\frac{2i\nu}{\hr^2}
   \mfK_{\mp}^n(r,r',\nu)
  &=\frac{\rho}{\hr}\delta(\hr-\hr') ,
  \label{we_mfH}
   \\
  \bigg(\rd_{\hr}\brd_{\hr}+1-\frac{\nu^2}{\hr^2}\bigg)\mfK_{\pm}^n(r,r',\nu)
  +\frac{2i\nu}{\hr^2}\mfH_{\mp}^n(r,r',\nu)
  &=0 .
   \label{we_mfK}
\end{align}
The operator of Eq.(\ref{we_mfH}) differs from $O_{\nu}$ given by Eq.(\ref{eq:BDE}) 
which is for $\mfG_{\pm}^n$ as shown in Eq.(\ref{eq:BDE_mfGpm}).
Comparing Eq.(\ref{we_mfH}) with Eqs.(\ref{we_mfEBx}-\ref{we_mfEBs}),
we can tell that $\mfH_{\pm}^n$ and $\mfK_{\pm}^n$ are respectively
the Green functions and the coupling Green functions of the wave equations for
the radial and longitudinal components of the fields in the Laplace domain.
According to Eqs.(\ref{eq:lim_nu}) and the second equation of (\ref{eq:we_mfGpm_str}),
\begin{align}
  \lim_{\rho\to\infty}\mfH_{\pm}^n(r,r',\nu)
  =\mfG_{\pm}^n(x,x',k_s) ,
     \qquad
  \lim_{\rho\to\infty}\bmfG_{\pm}^n(r,r',\nu)
  =\frac{k_x^2 \mfG_{\mp}^n(x,x',k_s)}{(k\beta)^2-(k_y^n)^2} .
  \label{eq:lim_mfH_bmfG}
\end{align}
$\mfG_{\pm}^n(x,x',k_s)$ is given by Eq.(\ref{eq:mfGbe_strt})
which is the Green function of the wave equations for the fields in the straight section.
$k_s\in\mathbb{C}$ is the Laplace variable of the fields in the straight section.
The coupling Green functions $\mfK_{\pm}^n$ vanish in the limit of $\rho\to\infty$,
\begin{align}
  \lim_{\rho\to\infty}\mfK_{\pm}^n(r,r',\nu)=0 .
  \label{eq:lim_mfKpm}
\end{align}
%

\subsection{Domains of the fields in the linear transforms}
\label{sec:domains}

We found the explicit and exact expressions of the fields in the Laplace domain
by solving the wave equations through the following domains,
\begin{align}
  \us{\text{\small Time domain}}{\frac{}{}E(x,y,s,t)\frac{}{}}
  \os{\frac{}{}\cF_\text{t}(t,k)\frac{}{}}{\ra}
  \us{\text{\small Frequency domain}}{\frac{}{}\tE(x,y,s,k)\frac{}{}}
  \us{\cF_\text{e}^{-1}}{\os{~\cF_\text{e}(y,n)}{\rlas}}
  \us{\text{\small $y$-Fourier mode}}{\frac{}{}\cE^n(x,s,k)\frac{}{}}
  \us{\cL^{-1}}{\os{~\cL(s,\nu)}{\rlas}}
  \us{\text{\small Laplace domain}}{\frac{}{}\mfE^n(x,\nu,k)\frac{}{}}
  \label{eq:transforms}
\end{align}
where $E$ and $\tE$ represent the components of the electromagnetic field in
the time and frequency domains.
$\cE^n$ is the $n$th Fourier coefficient of $\tE$ in the Fourier series with respect to $y$
($y$-Fourier mode).
$\mfE^n$ is the Laplace transform of $\cE^n$ with respect to $s$.
$\cF_\text{t}$, $\cF_\text{e}$ and $\cL$ denote the following linear transforms,
\begin{alignat}{2}
  \cF_\text{t}(t,k)
  &=\text{Fourier transform from $t$ to $k$ $(=\omg/v \in\mathbb{R})$},
   \qquad&&
  \text{Eqs.(\ref{eq:Fourier_trans})} ,
  \nonumber
   \\
  \cF_\text{e}(y,n)
  &=\text{Fourier expansion from $y$ to $n\in\mathbb{Z}$},
   \qquad&&
  \text{Eqs.(\ref{eq:Fp_coeff}) and (\ref{eq:Fm_coeff})} ,
  \nonumber
   \\
  \cL(s,\nu)
  &=\text{Laplace transform from $s$ to $\nu$ $(=-i\kap\in\mathbb{C})$},
   \qquad&&
  \text{Eqs.(\ref{eq:Laplace}) and (\ref{eq:ILT_nu})} .
  \nonumber
\end{alignat}
In section \ref{sec:ILT} we will transform the fields in the Laplace domain to the Fourier 
coefficients through the inverse Laplace transform $(\cL^{-1})$ given by
Eq.(\ref{eq:ILT_nu}).
In section \ref{sec:FD_field} we will find the expressions of the fields in
the frequency domain $(\cF_\text{e}^{-1})$ using the Fourier coefficients through
Eqs.(\ref{eq:Four_Exp_plus}-\ref{eq:Four_Exp_minus}).
Although we attempted the inverse Fourier transform ($\cF_\text{t}^{-1}$) of the fields
to the time domain by creating the poles in the $k$-plane on purpose,
it is not successful at present since we cannot find the proper contours.

We summarize the sections and appendices in which
we find the expressions of the fields in each domain,
\begin{alignat}{5}
  &\text{\small Time domain}
  &
  &\text{\small Freq.\,domain} ~~
  &
  &\text{\small $y$-Fourier mode}
  &
  &\text{\small Laplace domain}
   \nonumber
   \\
  &\us{ }
      { \os{\text{\small  Sec.\,\ref{sec:we_TDFD}}}{(E_y,B_y)} }
  &\us{~}{\os{~\cF_{\rm t}}{\ra}}~~
  &\us{\text{\small  Sec.\,\ref{sec:FD_field}}}
      { \os{\text{\small  Sec.\,\ref{sec:we_TDFD}}}{(\tE_y,\tB_y)} }
  &\us{~\cF_\text{e}^{-1}}{\os{~\cF_{\rm e}}{\rlas}}~~
  &\us{\text{\small  Sec.\,\ref{sec:ILT_vertical}}}
      { \os{\text{\small  Sec.\,\ref{sec:ymode}}}{(\cE_y^n,\cB_y^n)} }
  &\us{~\cL^{-1}}{\os{\cL}{\rlas}}~~
  &\os{\text{\small  Sec.\,\ref{sec:mfEBy}}}{(\mfE_y^n,\mfB_y^n)}
   \label{eq:EBy_trans}
   \\
  &
  &
  &
  &
  &\downarrow \text{\small  App.\,\ref{sec:HL_field_sdom}}
  &
  &\downarrow \text{\small  Sec.\,\ref{sec:sol_mfEB_xs}}
   \nonumber
   \\
  &\us{\text{\small  Sec.\,\ref{sec:we_TDFD}}}{(E_{x,s},B_{x,s})}
  &\os{~\cF_{\rm t}}{\ra}~~
  &\us{\text{\small  Sec.\,\ref{sec:FD_field}}}{(\tE_{x,s},\tB_{x,s})}
  &\us{~\cF_{\rm e}^{-1}\!\!}{\la}~~
  &\us{\text{\small  Sec.\,\ref{sec:XS_field_sdom}}}{(\cE_{x,s}^n,\cB_{x,s}^n)}
  &\us{\cL^{-1}\!\!}{\la}~~~
  &\us{\text{\small  App.\,\ref{sec:we_xs}}}{(\mfE_{x,s}^n,\mfB_{x,s}^n)}
   \label{eq:EBxs_trans}
\end{alignat}
As already mentioned,
there are three ways to get the expressions of the radial and longitudinal components of
the fields, which are shown in section \ref{sec:sol_mfEB_xs},
appendices \ref{sec:HL_field_sdom} and \ref{sec:we_xs}.
As shown in Eqs.(\ref{eq:mfBsEx}-\ref{eq:mfBxEs}), we derived the expressions of
$\mfE_{x,s}^n$ and $\mfB_{x,s}^n$ from $\mfE_y^n$ and $\mfB_y^n$
given by Eqs.(\ref{eq:mfEBy_solution}).
But it is burdensome to rewrite Eqs.(\ref{eq:mfEBy_solution}) into
Eqs.(\ref{eq:mfEy_rwrtn}-\ref{eq:mfBy_rwrtn}) using
Eqs.(\ref{eq:Gauss_cF}-\ref{eq:Ampere_s_cF})
which are Maxwell equations for the Fourier coefficients of the fields.
Instead of this way, as shown in appendix \ref{sec:we_xs},
we can get Eqs.(\ref{eq:mfBsEx}-\ref{eq:mfBxEs}) by straightforwardly solving
the wave equations (\ref{we_mfEBx}-\ref{we_mfEBs}) for $\mfE_{x,s}^n$ and $\mfB_{x,s}^n$
using the eigenfunctions $\mfE_{\pm}^n$ and $\mfB_{\pm}^n$ given by Eqs.(\ref{eq:mfEB_pm}).
These solutions gotten in the different ways exactly agree as a matter of course.
We did both ways on purpose to ensure the expressions of $\mfE_{x,s}^n$ and $\mfB_{x,s}^n$
in an analytical way.
However, the way shown in appendix \ref{sec:we_xs} also requires elaborate calculations,
because we must disentangle the coupling between the radial and longitudinal components of
the field in their wave equations by means of the eigenfunctions of the operator of
the wave equations, given by Eqs.(\ref{eq:mfEpm}-\ref{eq:mfBpm}).
In addition, this disentanglement in the wave equations conversely gives rise to
another entanglement of the eigenfunctions in the boundary conditions of
the sidewalls of the curved pipe.
Thus, in order to derive the exact expressions of the radial and longitudinal components of 
the transient fields in the curved pipe, it seems that we cannot avoid
complicated calculations as far as we know.

\clearpage

\section{Pole structure of the fields in the Laplace domain}
\label{sec:nupole}

The expressions of the fields in the Laplace domain, $\mfE^n$ and $\mfB^n$, are given by
Eqs.(\ref{eq:mfEBy_solution}) and (\ref{eq:mfBsEx}-\ref{eq:mfBxEs}).
They involve the Green functions $\mfG_{\pm}^n$ given by
Eqs.(\ref{eq:mfGe}-\ref{eq:mfGm_ba}).
In section \ref{sec:ILT} we will find the expressions of the Green functions of
the Fourier coefficients of the fields $\cG_{\pm}^n(s)$ from $\mfG_{\pm}^n(\nu)$
through Eq.(\ref{eq:ILT_nu}) which is the inverse Laplace transform with respect to $\nu$.
In order to calculate the Bromwich integral which will be given by Eq.(\ref{eq:cGeb_def}),
we investigate the singularity of $\mfG_{\pm}^n$ in the $\nu$-plane.
For given $k$ and $k_y^n$, $\mfG_{+}^n$ and $\mfG_{-}^n$ each have poles in
the $\nu$-plane at the points such that
\begin{align}
  p_\nu(\hr_b,\hr_a)=0
   \quad\text{for}~~\mfG_{+}^n
   ,\qquad\quad
  \hr_b\hr_as_\nu(\hr_b,\hr_a)=0
   \quad\text{for}~~\mfG_{-}^n .
  \label{eq:ps_nu_0}
\end{align}
$\nu/\rho$ denotes the longitudinal wavenumber of the field in the bend of radius $\rho$.
The solutions of Eqs.(\ref{eq:ps_nu_0}) with respect to $\nu$ are
the dispersion relations of $\tE_y$ and $\tB_y$ in the curved pipe,
which we call the pole structure of the fields in the Laplace domain.
The pole structure will be shown in Fig.\ref{fig:poles}.
We use the term ``pole'' (of $\mfG_{\pm}^n$ in the $\nu$-plane) as a synonym of 
``zero'' (of $p_\nu$ and $\hr_b\hr_as_\nu$ with respect to $\nu$) except
$\mfG_{+}^n$ in the limit of $k_r^n\to0$ for which
both the numerator and denominator of $\mfG_{+}^n$ become zero at $\nu=0$
as seen from Eqs.(\ref{lim_kr0_mfGpm}-\ref{eq:mfLp}).
$\hr_a$ and $\hr_b$ are the dimensionless radii of the inner and outer walls of
the curved pipe, normalized by the radial wavenumber
$k_r^n\in\mathbb{A}$ given by Eq.(\ref{eq:krn}).
$k\in\mathbb{R}$ in Eqs.(\ref{eq:ps_nu_0}) since we are calculating $\tEv$ and $\tBv$ 
which are involved in the Fourier integral given by Eq.(\ref{eq:Fourier_trans}).
We regard $k$ as a complex variable only in considering the integral of
the inverse Fourier transform with respect to $k$.

\subsection{Dispersion relations through the cross products of the Bessel functions}
\label{sec:pole}

Although we cannot find the explicit and exact solutions of Eqs.(\ref{eq:ps_nu_0}) 
with respect to $\nu$,
we can understand the pole structure of $\mfG_{\pm}^n$ in the $\nu$-plane,
owing to Cochran's work \cite{cochran} which is reviewed in
appendix \ref{sec:cochran}.
Also, we can find the asymptotic expressions of the poles for large radius ($\rho\gg w$)
as shown in appendix \ref{sec:poles}.
See appendices \ref{sec:bessel}, \ref{sec:dunster}, \ref{sec:cochran} and
\ref{sec:poles} for the details on the solutions of Eqs.(\ref{eq:ps_nu_0}) with
respect to $\nu$.
In the present section we describe the zeros of $p_{\nu}$ and $s_{\nu}$
while quoting the important points in these appendices.
For brevity, $t_{\nu}$ represents the four cross products $(p_\nu,q_\nu,r_\nu,s_\nu)$
defined in Eqs.(\ref{eq:CP_pq}-\ref{eq:CP_sr}).
$t_{\nu}$ is even with respect to $\nu$ as shown in Eqs.(\ref{eq:cp_symm}).
In addition, $t_{\nu}$ has mirror symmetry in the $\nu$-plane, \ie,
\begin{align}
  t_{-\nu}(b,a)=t_\nu(b,a) ,
    \qquad
  t_{\nu^{\ast}}(b,a)=t_\nu^{\ast}(b,a) ,
   \qquad\text{where}\quad
  t_\nu
  =(p_\nu,q_\nu,r_\nu,s_\nu) .
  \label{eq:even_symm_ps}
\end{align}
$a,b\in\mathbb{C}$ in general, however, $a,b\in\mathbb{A}$ in the present study.
The asterisk $({}^{\ast})$ denotes the complex conjugate.
According to Eqs.(\ref{eq:even_symm_ps}), it is enough to consider only the first quadrant 
of the $\nu$-plane including the positive axes
$\mathbb{A}_0^{+}=\{\mathbb{R}_0^{+},i\mathbb{R}^{+}\}$
in order to understand the pole structure of $\mfG_{\pm}^n$.
$t_\nu$ is a meromorphic function with respect to $\nu$, \ie,
$\mfG_{\pm}^n$ does not have a branch point or an essential singular point in
the $\nu$-plane.
In short, $\mfG_{\pm}^n$ has only poles in the $\nu$-plane.
$t_\nu$ is meromorphic also in the complex $k$-plane.
In addition, $t_\nu$ is even with respect to $k$ from the definition of $k_r^n$.

Although the fields in the curved pipe have a somewhat complicated pole structure due to
the curvature, they have similar features to those in a straight pipe
except the whispering gallery modes which are particular to the curved pipe.
Therefore, in order to understand the pole structure of the fields in the curved pipe,
it is often instructive to refer to the fields in a straight pipe,
which is formulated in appendix \ref{sec:travel_wave} in a similar way to the one in
the curved pipe on purpose.
Eqs.(\ref{eq:ps_nu_0}) correspond to the first equation of (\ref{eq:kxm}) in
a straight section.
In a straight pipe which has the same rectangular cross section as that of the curved pipe, 
the dispersion relation of all the components of the fields is given by
Eq.(\ref{eq:ks_mn}),
\begin{align}
  (k_r^n)^2
  =(k\beta)^2-(k_y^n)^2
  =(k_s^{mn})^2+(k_x^m)^2 ;
   \qquad
  k\beta
  =\frac{\omg}{c} ,
    \qquad
  k_x^m
  =\frac{m\pi}{w} ,
    \qquad
  k_y^n
  =\frac{n\pi}{h} .
  \label{eq:asymp_lim}
\end{align}
The indices $m$ and $n$ denote respectively the horizontal and vertical mode numbers 
of the field in the pipe which has a rectangular cross section as shown in
Fig.\ref{fig:pipe3D}.
$k_x^m$, $k_y^n$ and $k_s^{mn}$ are the horizontal, vertical and
longitudinal wavenumbers of the field in the straight rectangular pipe.
$k_r^n$ can also be given using $k_s^{mn}$ and $k_x^m$
which are the variables of the field in the straight pipe.
In terms of mathematics, $k_s^{mn}$ denotes the pole of the field in the straight pipe
in the complex $k_s$-plane (Laplace domain of the field in the straight pipe)
as shown in Eq.(\ref{eq:kxm}).
One must distinguish $k_s^{mn}\in\mathbb{A}$ from $k_s\in\mathbb{C}$ in Eq.(\ref{eq:lim_nu}).

According to appendix \ref{sec:cochran}, $p_{\nu}$ and $\hr_b\hr_as_{\nu}$ have zeros
only on the real and imaginary axes of the $\nu$-plane,
similar to $k_s^{mn}$ in the $k_s$-plane,
\begin{alignat}{4}
  p_\nu(\hr_b,\hr_a)&=0
   \qquad&\Lra&\qquad&
  \nu&=\pm\nu_m^n \in\mathbb{A}
   \quad
  &&(m\in\mathbb{N}) ,
  \label{eq:pnu_poles}
  \\
  \hr_b\hr_as_\nu(\hr_b,\hr_a)&=0
   \qquad&\Lra&\qquad&
  \nu&=\pm\mu_m^n \in\mathbb{A}
   \quad
  &&(m\in\mathbb{Z}_0^+) .
  \label{eq:snu_poles}
\end{alignat}
$\mfG_{+}^n$ and $\mfG_{-}^n$ each have a finite number of poles on the real axis
and an infinite number of poles on the imaginary axis of the $\nu$-plane,
\begin{align}
  \nu_m^n
  =
  \left\{
  \begin{array}{ll}
    \hnu_m^n\in\bbR_0^{+} & (m\leq m_{+})
    \\
    i\cnu_m^n\in i\bbR^{+} & (m>m_{+})
  \end{array}
  \right. ,
    \qquad
  \mu_m^n
  =
  \left\{
  \begin{array}{ll}
    \hmu_m^n\in\bbR_0^{+} & (m\leq m_{-})
    \\
    i\cmu_m^n\in i\bbR^{+} & (m>m_{-})
  \end{array}
  \right. .
  \label{eq:ps_poles}
\end{align}
The poles $(\nu_m^n,\mu_m^n)$ are functions of $\hr_b$ and $\hr_a$.
The exact expressions of the poles must be symmetric with respect to
the exchange of $r_b$ and $r_a$ since Eqs.(\ref{eq:ps_nu_0}) are so.
Similarly, Eq.(\ref{eq:asymp_lim}) is symmetric with respect to
the exchange of $x_b$ and $x_a$ which are involved in $k_x^m$ through $w=x_b-x_a$.
The radial (=\,horizontal) mode number is defined as $m=(\mathbb{N},\mathbb{Z}_{0}^{+})$ 
for $\nu=(\nu_m^n,\mu_m^n)$, \ie, $m$ begins with 1 and 0 respectively for
$\nu_m^n$ and $\mu_m^n$ in accordance with the one in Eqs.(\ref{eq:asymp_lim}).
$\mfG_{-}^n$ has the poles of the zeroth radial mode $\nu=\pm\mu_0^n$
which are somewhat different from the higher order modes ($m\in\mathbb{N}$) as
Eqs.(\ref{eq:lim_nurho_ksmn}) and (\ref{eq:cXm}) imply.

The radial mode indices $m_{\pm}\in\mathbb{Z}_0^{+}$ correspond to the numbers of
the real poles of $\mfG_{\pm}^n$ for given $k$ and $k_y^n$.
That is, the numbers of the zeros of $p_\nu$ and $\hr_b\hr_as_\nu$ in $\mathbb{R}_0^{+}$
(positive real $\nu$-axis including the origin) are $m_{+}$ and $m_{-}+1$ respectively.
To be precise, only when $m_{\pm}=0$, the upper expressions in
Eqs.(\ref{eq:ps_poles}) are incorrect,
because $\mfG_{\pm}^n$ do not have the real poles $\hnu_m^n$ and $\hmu_m^n$ in this case.
In short, $m_{\pm}=0$ means that all the poles are purely imaginary
$\nu=(\nu_m^n,\mu_m^n)\in i\mathbb{R}$.
Although we cannot find the exact expressions of $m_{\pm}$, the asymptotic expression of
$m_{\pm}$ for large radius is given by Eq.(\ref{eq:m_pm}),
\begin{align}
  (m_{+},m_{-})
  =\bigg\lfloor
   \frac{k_r^n}{k_x^1}
   \bigg\{1+\frac{(1,-3)}{2^3\hr_b\hr_a}
      +\frac{(-25,63)}{2^7(\hr_b\hr_a)^2}
      +O(\hr_{a,b}^{-6})
   \bigg\}
   \bigg\rfloor
   \qquad
  ( |k\beta|>k_y^n,~ w/\rho \ll 1 \ll k_r^n\rho) ,
  \label{eq:Re_poles}
\end{align}
where the quantities in the parentheses correspond.
$\lfloor\chi\rfloor$ is the floor function which denotes the largest integer
less than $\chi\in\mathbb{R}_0^{+}$.
$k_x^1$ $(=\pi/w)$ is the fundamental mode of $k_x^m$.
$m_{+}$ given by Eq.(\ref{eq:Re_poles}) agrees with Eq.(19) in \cite{horvat_prosen} up to
$O(\hr_{a,b}^{-2})$.
$\mfG_{\pm}^n$ has no real poles when $|k\beta|<k_y^n$.
In the limit of $\rho\to\infty$, $m_{\pm}$ goes to $m_0$ given by Eq.(\ref{eq:Re_poles_st}),
which is the number of the positive real poles of the fields in the straight pipe.
The asymptotic expression of $m_{\pm}$, given by Eq.(\ref{eq:Re_poles}), is equal to $m_0$ 
in most cases under the assumptions $w/\rho\ll1\ll\hr_{a,b}$
since these assumptions mean that the bend is not so sharp to the fields.

\subsection{Cutoff wavenumbers of the curved pipe}
\label{sec:cutoff}

\begin{figure}[h]
  \begin{center}
    \includegraphics[scale=0.32,clip]{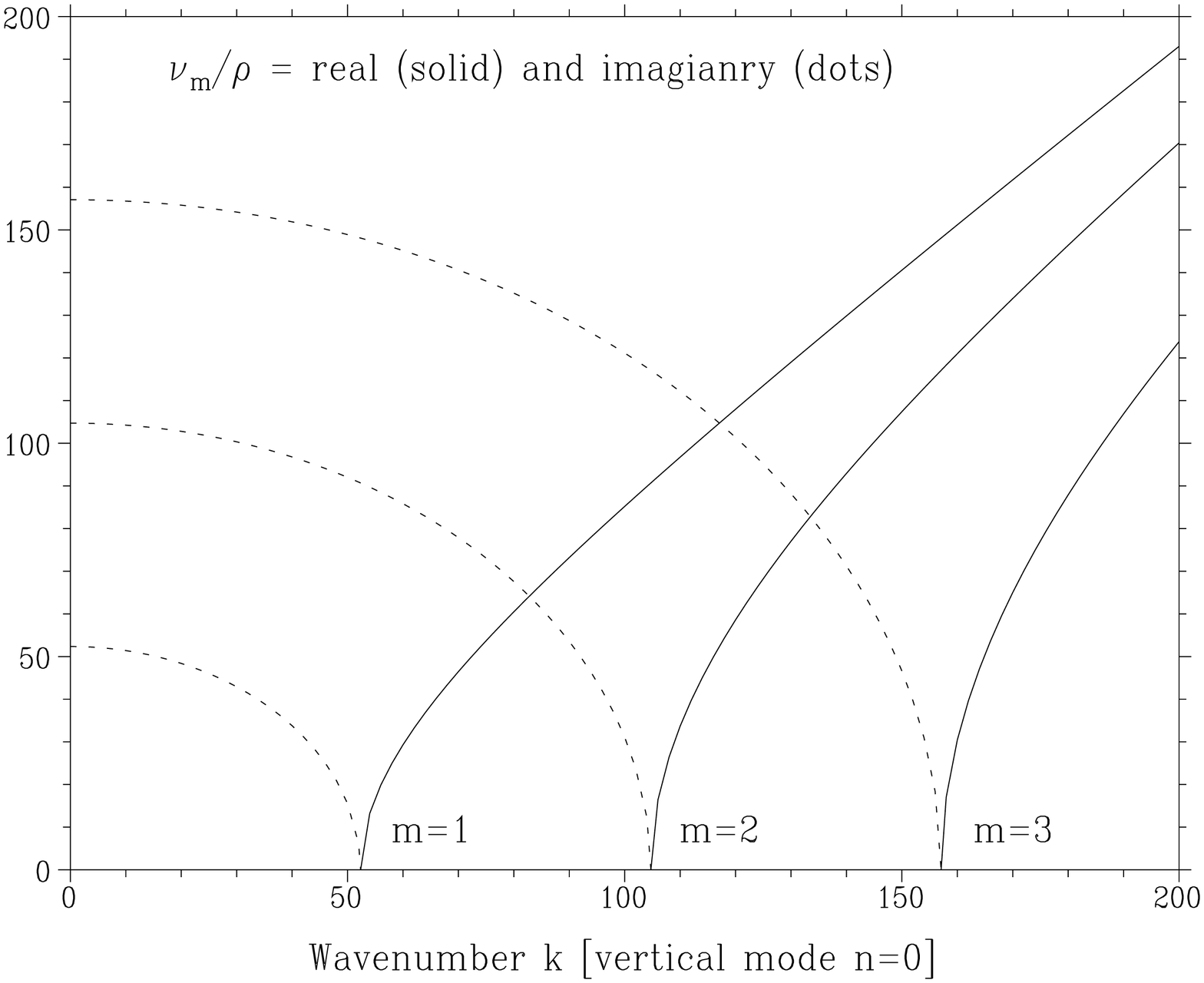} ~~
    \includegraphics[scale=0.32,clip]{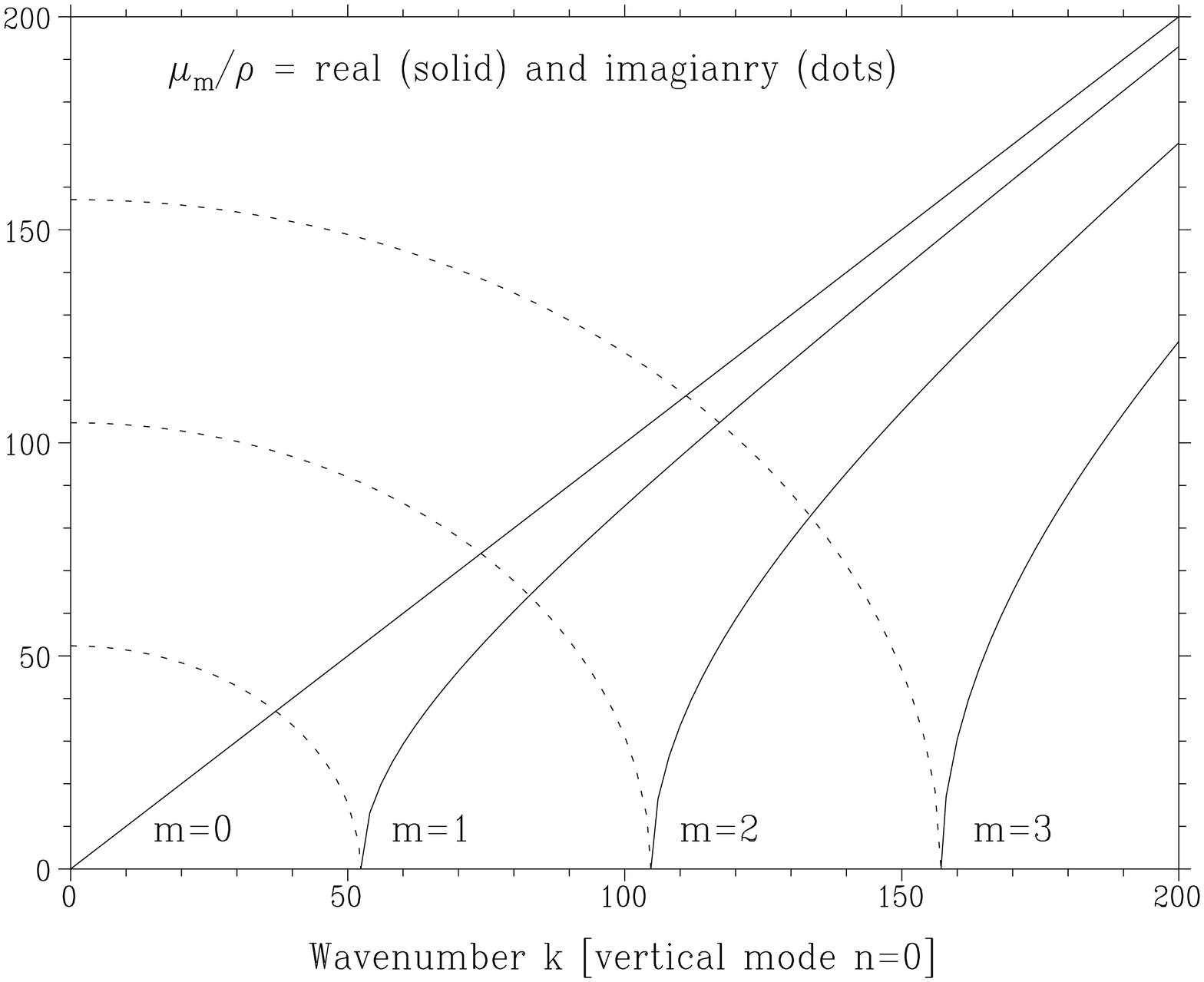}
      \vspace{3mm}
      \\
    \includegraphics[scale=0.32,clip]{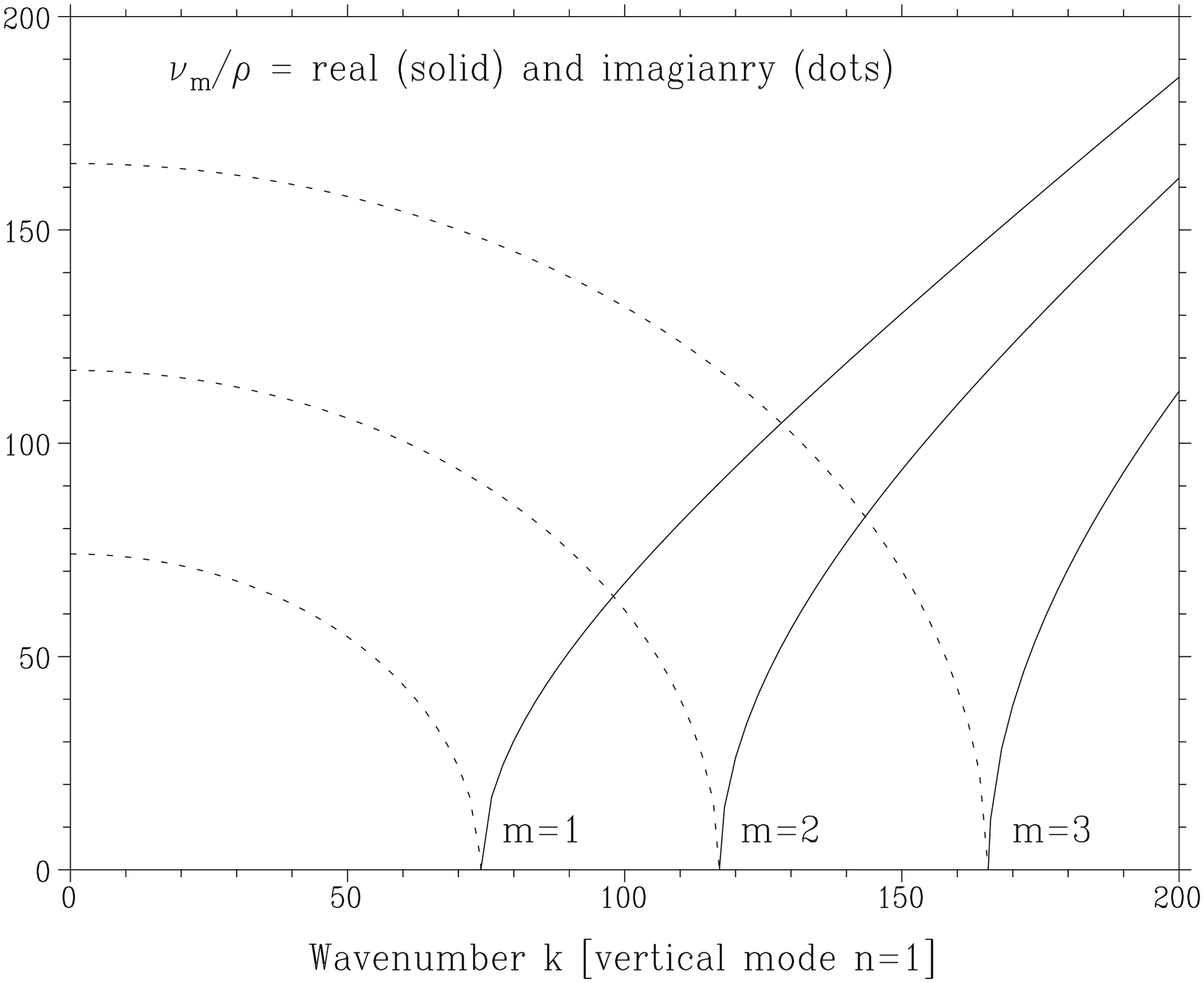} ~~
    \includegraphics[scale=0.32,clip]{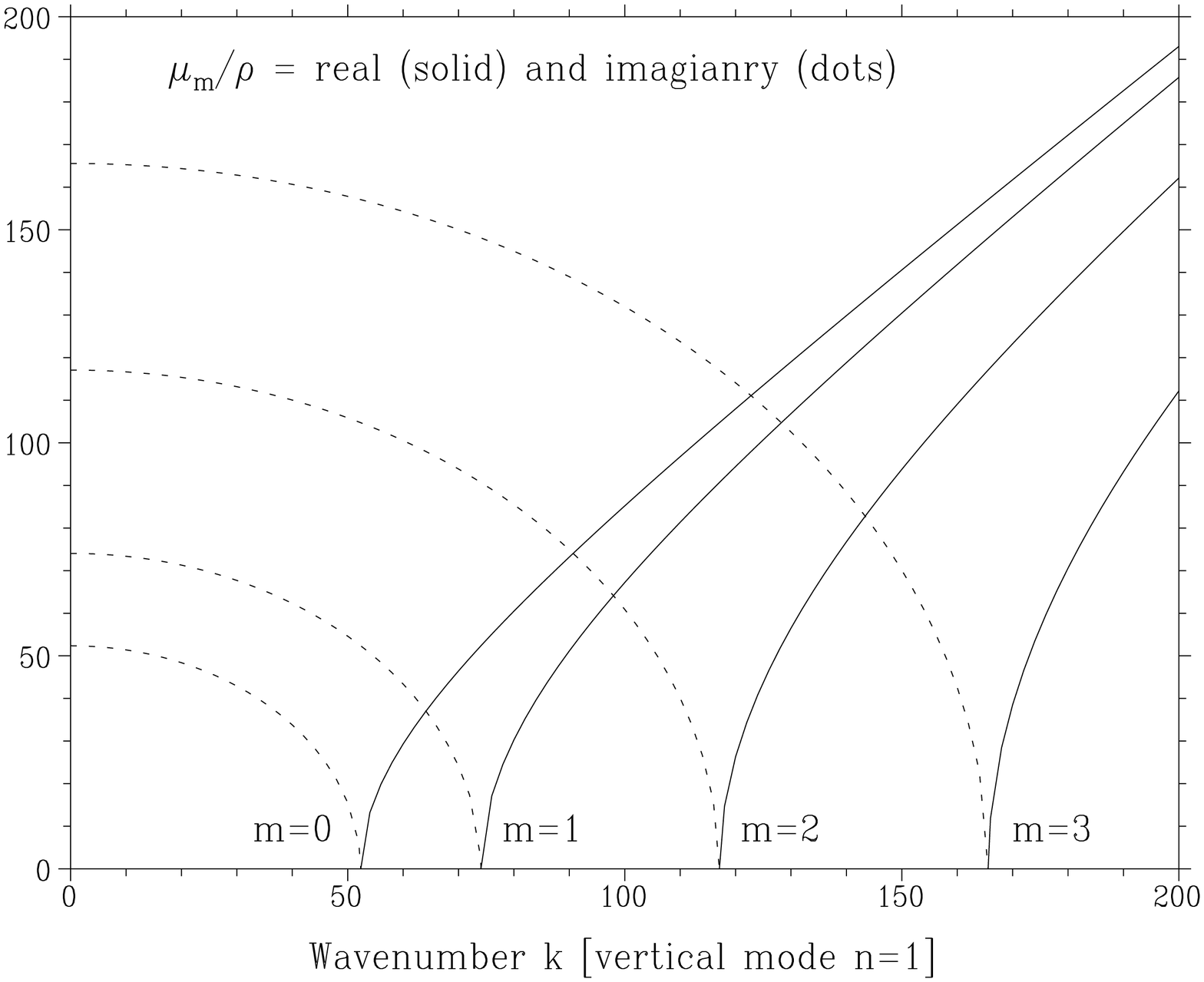}
    \caption[Pole structure of the transient field in a curved pipe]{
      \small
       Poles $\nu_m^n/\rho$ (left) and $\mu_m^n/\rho$ (right)
       as functions of the wavenumber $k\,[{\rm m}^{-1}]$.
       The vertical mode number is $n=0$ (upper) and $n=1$ (lower).
       The solid and dotted curves show respectively the real poles
       ($\nu\in\mathbb{R}$) and imaginary poles ($\nu=i\bnu\in i\mathbb{R}$).
       The curves of $\nu_m^n/\rho$ and $\mu_m^n/\rho$ represent
       $\mathbb{H}_{+}$ and $\mathbb{H}_{-}$ given by
       Eqs.(\ref{eq:H_pls}-\ref{eq:H_mns}).
       We indicate only the radial mode number ($m$) of the poles,
       omitting the superscript $n$ as $\nu_m^n=\nu_m$ and $\mu_m^n=\mu_m$ for brevity.
       The parameters are given as $\rho=10{\rm m}$, $x_b=-x_a=3{\rm cm}$
       ($w=6{\rm cm}$) and $h=6{\rm cm}$.
       }
    \label{fig:poles}
  \end{center}
\end{figure}

The number of the real poles of $\mfG_{\pm}^n$ depend on $k$ through $k_r^n$.
The origin of the Laplace plane ($\nu=0$) is the transition point that
the longitudinal wavenumber $\nu/\rho$ interchanges between $\nu\in\mathbb{R}$ 
(oscillation) and $\nu\in i\mathbb{R}$ (exponential damping)
as shown in Fig.\ref{fig:poles} (p.\pageref{fig:poles}).
For a pair of the transverse mode numbers $(m,n)$, we define the cutoff wavenumbers of
the curved pipe, $k_m^n$ and $\bk_m^n$, as $k$ such that $p_{\nu}(\hr_b,\hr_a)$ and
$\hr_b\hr_as_{\nu}(\hr_b,\hr_a)$ each become zero of the second order at $\nu=0$, \ie,
\begin{alignat}{4}
  \nu_m^n
  &=0
   \qquad&\Lra&\qquad&
  k&=\pm k_m^n
   \quad&&
  (m\in\mathbb{N}) ,
  \label{eq:kmn}
   \\
  \mu_m^n
  &=0
   \qquad&\Lra&\qquad&
  k&=\pm\bk_m^n
   \quad&&
  (m\in\mathbb{Z}_0^+) .
  \label{eq:bkmn}
\end{alignat}
That is, when $k=(k_m^n,\bk_m^n)$, $\mfG_{\pm}^n$ has a pole of the second order at
$\nu=0$ as described in appendix \ref{sec:orgn_odr}.
$k_m^n$ and $\bk_m^n$ are symmetric with respect to the exchange of $x_a$ and $x_b$
since $\nu_m^n$ and $\mu_m^n$ are so.
From Eqs.(\ref{eq:asymp_lim}) and (\ref{eq:nu_krn0}),
we get $\bk_0^n=k_y^n/\beta$ since $k_r^n=0$ at $k=k_y^n/\beta$.
When $k_r^n=0$, $\hr_b\hr_a s_{\nu}(\hr_b,\hr_a)$ has a zero of the second order at $\nu=0$, 
because $\hr_b\hr_a\propto (k_r^n)^2\to0^2$ and $s_0(\hr_b,\hr_a)=p_1(\hr_b,\hr_a)$
which is not zero in the limit of $k_r^n\to0$ as shown later in Eq.(\ref{eq:lim_krn_s0}).
Although $k_m^n$ is defined for $m\in\mathbb{N}$, we define $k_0^n$ as the one equal to
$\bk_0^n$ just for convenience later in defining the rectangular window function
$\Th_{+}^{n\ell}$ in Eq.(\ref{eq:Th_def}),
\begin{align}
  k_0^n
  \equiv\bk_0^n
  =\hk_0^n
  =\frac{k_y^n}{\beta} ,
    \qquad
  \lim_{\rho\to\infty}(k_m^n,\bk_m^n)
  &=\hk_m^n
   =\frac{1}{\beta}\{(k_x^m)^2+(k_y^n)^2\}^{1/2} .
   \label{eq:lim_kmn}
\end{align}
$\hk_m^n$ is given by Eq.(\ref{eq:hkmn}) which is the cutoff wavenumber of
the straight rectangular pipe, \ie, $\hk_m^n$ is $k$ such that $k_s^{mn}=0$.
Both $k_m^n$ and $\bk_m^n$ tend to be $\hk_m^n$ in the limit of $\rho\to\infty$ as
shown in the second equation of (\ref{eq:lim_kmn}).
As already described, we use $k_0^n$ just for convenience to express $\cG_{+}^n$ in
a similar manner to the expression of $\cG_{-}^n$ after the inverse Laplace transform.
We defined $k_m^n$ and $\bk_m^n$ implicitly by Eqs.(\ref{eq:kmn}-\ref{eq:bkmn})
since we cannot find their explicit and exact expressions except $\bk_0^n$.
Assuming $\eps\equiv w/\rho\ll1$, the asymptotic expressions of the square of
the cutoff wavenumbers are given by Eq.(\ref{eq:cutoff}),
\begin{alignat}{2}
  \bigg\{\frac{(k_m^n,\bk_m^n)}{\hk_m^n}\bigg\}^2
  &\simeq
     1
    -\frac{(1,-3)}{4r_br_a(\hk_m^n\beta)^2}
     \bigg\{1-\frac{3}{2r_br_a(k_x^m)^2}\bigg\}
    +O(\eps^6)
   \qquad&&(m\in\mathbb{N}) ,
  \label{eq:kcutoff}
\end{alignat}
where the coefficients $(1,-3)$ correspond to $(k_m^n,\bk_m^n)$.

In terms of geometry, Eqs.(\ref{eq:pnu_poles}-\ref{eq:snu_poles}) denote the sets
$\mathbb{H}_{\pm}$ which are subsets of the following space $\mathbb{S}$,
\begin{align}
  \mathbb{S}
  &=\{(k,\nu,n)|k\in\mathbb{R},~\nu\in\mathbb{C},~n\in\mathbb{Z}_0^{+}\} ,
  \label{eq:hypersurface}
   \\
  \mathbb{H}_{+}
  &=\{(k,\nu)|k\in\mathbb{R},~\nu\in\mathbb{C},~p_\nu(\hr_b,\hr_a)=0,~n\in\mathbb{Z}_0^{+}\}
   \subset\mathbb{S} ,
  \label{eq:H_pls}
   \\
  \mathbb{H}_{-}
  &=\{(k,\nu)|k\in\mathbb{R},~\nu\in\mathbb{C},~\hr_b\hr_as_\nu(\hr_b,\hr_a)=0,
      ~n\in\mathbb{Z}_0^{+}\}
   \subset\mathbb{S} .
  \label{eq:H_mns}
\end{align}
As shown in Fig.\ref{fig:poles}, $\mathbb{H}_{\pm}$ consists of the infinite number of
curves like parabolas and ellipses which are lying on
the $(\Re\nu,k)$-plane and the $(\Im\nu,k)$-plane respectively.
The parabola-like curves and ellipse-like curves tend to be parabolas and ellipses 
rigorously in the limit of $\rho\to\infty$ as seen from Eqs.(\ref{eq:ks_mn}) and
Fig.\ref{fig:poles_str} in p.\pageref{fig:poles_str} except the line for $(m,n)=(0,0)$.
A parabola-like curve on the $(\Re\nu,k)$-plane contacts an ellipse-like curve on
the $(\Im\nu,k)$-plane at the point $k=k_m^n$ or $\bk_m^n$ on the $\nu=0$ line.
For given $k=k'$ and $n=n'$, the poles of $\mfG_{\pm}^n$ in the $\nu$-plane denote
the cross points of $\mathbb{H}_{\pm}$ with the $k=k'$ plane in $\mathbb{S}$.
For a given $n$, we consider the cross points of $\mathbb{H}_{\pm}$ for various $k'$, 
\ie, varying $k$ with $n$ fixed as shown in Fig.\ref{fig:poles}.
Since the cutoff wavenumbers $k_m^n$ and $\bk_m^n$ are the contacts between
the parabola-like curves and the ellipse-like curves on the $\nu=0$ line, 
they each form the following sets $\mathbb{K}_{+}$ and $\mathbb{K}_{-}$
which are the subsets of $\mathbb{H}_{+}$ and $\mathbb{H}_{-}$ respectively,
\begin{align}
  \mathbb{K}_{+}
  &=\{k|k\in\mathbb{R},~p_{\nu}(\hr_b,\hr_a)=0,~\nu_m^n=0,
    ~m\in\mathbb{N},~n\in\mathbb{Z}_0^{+}\}
   \subset\mathbb{H}_{+} ,
   \\
  \mathbb{K}_{-}
  &=\{k|k\in\mathbb{R},~\hr_b\hr_as_{\nu}(\hr_b,\hr_a)=0,~\mu_m^n=0,
      ~m\in\mathbb{Z}_0^{+},~n\in\mathbb{Z}_0^{+}\}
   \subset\mathbb{H}_{-} .
\end{align}
In what follows, we often use the index $\ell$ instead of $m_{\pm}$ for convenience,
\begin{align}
  \ell=m_{\pm}+1 .
   \label{eq:ell_m}
\end{align}
When $|k|<k_1^n$ for a given $n$, $p_\nu$ has no zeros on the real $\nu$-axis,
\ie, all the zeros of $p_\nu$ for $m\in\mathbb{N}$ are purely imaginary.
When $k_{\ell}^n<|k|<k_{\ell+1}^n$, $p_\nu$ has $\ell$ zeros on the positive real $\nu$-axis.
When $|k|=k_{\ell}^n$, $p_\nu$ has $\ell$ zeros on the positive real axis including
the zero of the second order at the origin $\nu=\nu_{\ell}^n=0$.
Similarly, when $|k|<\bk_0^n$, $\hr_b\hr_as_\nu$ has no zeros on the real $\nu$-axis.
When $\bk_{\ell}^n<|k|<\bk_{\ell+1}^n$, $\hr_b\hr_as_\nu$ has $\ell+1$ zeros on
the positive real axis including the zeroth pole $\nu=\mu_0^n$.
When $|k|=\bk_{\ell}^n$, $\hr_b\hr_as_\nu$ has $\ell+1$ zeros on the positive real axis
including the zero of the second order at the origin $\nu=\mu_{\ell}^n=0$.
Thus, the numbers of the zeros of $p_\nu$ and $\hr_b\hr_as_\nu$ on the real $\nu$-axis 
intermittently increase as $|k|$ increases.
These patterns of the zeros for $n=0$ and 1 are shown in
Fig.\ref{fig:poles} and Eqs.(\ref{eq:pole_struct_pls}-\ref{eq:pole_struct_mns}).
For a given $n$, in order to fix $\ell$ which depends on $k$,
we will partition $k$ using the rectangular window functions $\Th_{\pm}^{n\ell}$ defined in
Eqs.(\ref{eq:Th_def}-\ref{eq:bTh_def}).

Although $\ell$ intermittently increases as $|k|$ increases, the real and imaginary poles 
are complementary to each other as seen from Fig.\ref{fig:poles}.
We illustrate it using the following function $\phi$ having a real parameter $\delta$,
\begin{align}
  \phi(\nu)=\nu^2-\delta
   \qquad
  (\nu\in\mathbb{C},~\delta\in\mathbb{R}) .
  \label{eq:phi_simple}
\end{align}
$\phi$ has zeros in the $\nu$-plane, which depend on $\delta$ as follows,
\begin{align}
  \phi(\nu)=0
   \quad~~\Lra~~\quad
  \nu
  =
  \left\{\begin{array}{ll}
    \pm \delta^{1/2} & (\delta>0)\\
    0 & (\delta=0) \\
    \pm i(-\delta)^{1/2} & (\delta<0)
  \end{array}\right. .
\end{align}
If $\delta\ne0$, $\phi$ has a pair of simple zeros which are symmetrically located on
the real or imaginary axis of the $\nu$-plane.
On the contrary, if $\delta=0$, $\phi$ has a zero of the second order at
the origin of the $\nu$-plane.
We can regard $\nu=0$ as the consequence that the pair of the real zeros coupled at
the origin as $\nu=\pm\delta^{1/2}\to\pm0$ in the limit of $\delta\to+0$.
That is, the zero of the second order $\phi=0^2$ can be regarded as a special case of
the pair of the simple zeros on the real axis for $\delta\to+0$.
This idea fits our definition of the step function given by Eq.(\ref{eq:curv}),
\ie, $\theta(\vsig)=1$ at $\vsig=+0$.
Also, the idea that the pair of the real poles couples at $\nu=0$ fits the contours of
the Bromwich integral in the $\nu$-plane as shown in Fig.\ref{fig:contour}
(p.\pageref{fig:contour}), because the pair of the real poles is
either the inside or the outside of the closed semicircular contour.
Although we can regard $\phi=0^2$ as the special case that the pair of the imaginary poles 
couples at the origin for $\delta\to-0$, this idea does not fit the contours of
the Bromwich integral, because the pair of the imaginary poles are split into
the inside and outside of the closed contour.
For $p_\nu(\hr_b,\hr_a)$ and $\hr_b\hr_as_\nu(\hr_b,\hr_a)$, the cutoff wavenumbers,
$k=k_m^n$ and $\bk_m^n$ defined in Eqs.(\ref{eq:kmn}-\ref{eq:bkmn}),
correspond to the case $\delta=0$ in Eq.(\ref{eq:phi_simple}).
That is, when $k=k_m^n$ and $\bk_m^n$, $p_\nu(\hr_b,\hr_a)$ and
$\hr_b\hr_as_\nu(\hr_b,\hr_a)$ each have a zero of the second order at the origin of
the Laplace plane ($\nu=0$) as shown in appendix \ref{sec:orgn_odr}.

\subsection{Whispering gallery mode}
\label{sec:wgm}

As shown in Eqs.(\ref{eq:pnu_poles}-\ref{eq:snu_poles}),
$(\nu_m^n,\mu_m^n)\in\mathbb{A}=\{\mathbb{R},i\mathbb{R}\}$
similar to $k_s^{mn}\in\mathbb{A}$.
$k_s^{mn}$ and $k_x^m$ are respectively different from $k_s$ and $k_x$
which are involved in Eqs.(\ref{eq:lim_nu}-\ref{eq:we_mfGpm_str}).
The latter are the complex wavenumbers in a straight section
before discretizing the field into the eigenmodes of the pipe.
For $\rho\to\infty$, as shown in Eq.(\ref{eq:lim_nu}),
the complex longitudinal wavenumber in the bend, $\nu/\rho\in\mathbb{C}$,
goes to $k_s\in\mathbb{C}$ in the straight section.
On the other hand, not all $\nu_m^n/\rho$ and $\mu_m^n/\rho$ for
$\forall m\in(\mathbb{N},\mathbb{Z}_0^{+})$ go to $k_s^{mn}$ in the limit of $\rho\to\infty$,
because the curved pipe has the whispering gallery modes
which the straight pipe does not.
The inner and outer walls of the curved pipe are asymmetric to the field in it 
(unless $\rho=\infty$).
That is, the field in the curved pipe is not symmetric with respect to the exchange of
$r_b$ and $r_a$ regardless of the presence or absence of the electric current in the pipe.
We can classify the modes of the field in the curved pipe into the following two kinds:
those which go to the ones in the straight pipe for $\rho\to\infty$,
and those which do not.
The former is the normal mode which is symmetric with respect the exchange of
the inner and outer walls of the curved pipe.
The latter is called the whispering gallery mode
which propagates in the curved pipe by reflecting only on the outer wall of the pipe,
similar to the whispering gallery of light on a round or circular dielectric boundary.
According to the discussion below Eq.(\ref{eq:z_uae}), the real poles
$\nu=(\nu_m^n,\mu_m^n)\in\mathbb{R}^{+}$ satisfying $\hr_a<\nu<\hr_b$ are
the whispering gallery modes of the fields.
In general, $\hr_b>\nu=(\nu_m^n,\mu_m^n)\in\mathbb{R}^{+}$ holds for
$m=(\mathbb{N},\mathbb{Z}_0^{+})$ and $k_r^n\in\mathbb{R}^{+}$.
We define $m_a^{+}$ and $m_a^{-}$ as the largest radial mode numbers among
the whispering gallery modes for $\nu_m^n$ and $\mu_m^n$,
\begin{align}
  \nu_{m+1}^n
  <\hr_a
  \leq\nu_m^n
   \quad
  (m=m_a^{+}\in\mathbb{N}) ,
    \qquad
  \mu_{m+1}^n
  <\hr_a
  \leq\mu_m^n
   \quad
  (m=m_a^{-}\in\mathbb{Z}_0^{+}) .
   \label{eq:nu_mt}
\end{align}
That is, when $m\leq m_a^{\pm}$,
$C_{\nu}(\hr_a)$ and $C_{\nu}'(\hr_a)$ for $k_r^n\in\mathbb{R}^{+}$ behave as 
exponential functions at the real poles $\nu=(\nu_m^n,\mu_m^n)\in\mathbb{R}^{+}$,
where $C_{\nu}=(J_\nu,Y_\nu)$ and $C_{\nu}'=(J_\nu',Y_\nu')$.
Conversely, $C_{\nu}(\hr_a)$ and $C_{\nu}'(\hr_a)$ are oscillatory at
$\nu=(\nu_m^n,\mu_m^n)\in\mathbb{R}^{+}$ for $m>m_a^{\pm}$, similar to
$C_{\nu}(\hr_b)$ and $C_{\nu}'(\hr_b)$ at $\nu=(\nu_m^n,\mu_m^n)\in\mathbb{R}^{+}$ for
$m\in(\mathbb{N},\mathbb{Z}_0^{+})$.
Since $\nu_m^n\ne\mu_m^n$ unless $\rho=\infty$, $m_a^{+}$ can differ from $m_a^{-}$.
Excluding $m_a^{-}=0$, however, $m_a^{+}=m_a^{-}$ in most cases
under the assumptions $w/\rho\ll 1 \ll k_r^n\rho$ and $|k\beta|>k_y^n$,
because the difference between $\nu_m^n$ and $\mu_m^n$ is of $O[(k_r^n\rho)^{-2}]$
as shown in Eqs.(\ref{eq:nu_eps}) and (\ref{eq:nu_rho2}).
In what follows, we write $m_a^{\pm}$ simply as $m_a$ 
omitting the sign ($\pm$) unless $m_a^{+}\ne m_a^{-}$.

As defined in Eqs.(\ref{eq:nu_mt}), the radial modes in $m\leq m_a$ are the whispering
gallery of $\nu_m^n$ and $\mu_m^n$ except $\mu_0^n$ which is $\mu_m^n$ for $m=0$.
Among all the poles, $\mu_0^n$ is exceptional, similar to the fact that
$k_s^{0n}$ is equal to $k_r^n$ as seen from Eq.(\ref{eq:asymp_lim}).
In addition to $\mu_0^n$, the radial modes such that $m>m_a$ go to the ones in
the straight pipe for $\rho\to\infty$, which are not the whispering gallery modes
but the normal modes of the curved pipe,
\begin{align}
  \lim_{\rho\to\infty}\frac{\mu_0^n}{\rho}
  =k_s^{0n}
  =k_r^n
   \quad
  (m=0) ,
    \qquad
  \lim_{\rho\to\infty}\frac{(\nu_m^n,\mu_m^n)}{\rho}
  =k_s^{mn}
   \quad
  (m\geq m_a+1) .
   \label{eq:lim_nurho_ksmn}
\end{align}
From Eqs.(\ref{eq:nu_mt}) and (\ref{eq:bnu_sols}),
we get the asymptotic expression of $m_a$ for large radius,
\begin{align}
  m_a
  &=\bigg\lfloor \frac{k_r^n}{k_x^1}\gam_a^{1/2} \bigg\rfloor
   \leq m_{\pm}
   \qquad
  (|k\beta|>k_y^n,~ \eps=w/\rho \ll 1 \ll k_r^n\rho ) ,
   \label{eq:mt}
   \\
  \gam_a
  &=\frac{g_l}{2}\{f_a+(f_a^2+f_{\pm})^{1/2}\}
   \simeq
   \eps\Big\{\frac{1}{2}+\frac{1}{6^{1/2}}+O[\eps,(k_r^nw)^{-2}]\Big\} .
   \label{eq:gamt}
\end{align}
$\lfloor~\rfloor$ is the floor function as also used in Eq.(\ref{eq:Re_poles}).
The functions $f_{\pm}=(f_{+},f_{-})$ correspond to $\nu=(\nu_m^n,\mu_m^n)$ and
$m_a=(m_a^{+},m_a^{-})$.
$f_{\pm}$ is the term which can give a difference between $m_a^{+}$ and $m_a^{-}$,
depending on the parameters of the field and the geometry of the pipe,
\begin{align}
  f_a
  =g_1-g_lg_a^2
   ,\qquad
  f_{\pm}
  =g_1\bigg\{(g_1-g_lg_2)\mp\frac{4g_l}{(k_r^n\rho)^2}\bigg\}
   ,\qquad
  g_1-g_lg_2
  \simeq  -\frac{\eps^2}{3}+O(\eps^3) .
   \label{eq:f0_fpm}
\end{align}
If $(k_r^nw)^2\gg1$, $m_a^{+}=m_a^{-}$ since $4g_l/(k_r^n\rho)^2$ is 
negligible for $(g_1-g_lg_2)$ in the braces of $f_{\pm}$.
The geometric factors are given as
\begin{align}
  g_{b,a}
  =\frac{r_{b,a}}{\rho}
    ,\qquad
  g_1
  =\frac{g_b+g_a}{2}
    ,\qquad
  g_2
  =\frac{g_b^2+g_a^2}{2}
    ,\qquad
  g_l
  =\frac{\rho}{w}\log\bigg(\frac{g_b}{g_a}\bigg) .
   \label{eq:gl_g12}
\end{align}
$g_{b,a}$ denotes $g$ for $r=r_{b,a}$, given by Eqs.(\ref{eq:r}).
If $\eps=w/\rho\ll 1$, $g_l$ is expanded as
\begin{align}
  g_l
  \simeq
      1
      -\frac{x_b+x_a}{2\rho}
      +\frac{x_b^2+x_bx_a+x_a^2}{3\rho^2}
      -\frac{(x_b+x_a)(x_b^2+x_a^2)}{4\rho^3}
      +O(\eps^4) .
\end{align}
$g_{1,2}$ and $g_l$ go to 1 in the limit of $\rho\to\infty$ similar to $g_{a,b}$.
$m_a$ goes to zero in this limit, \ie,
\begin{align}
  \lim_{\rho\to\infty}(g_{b,a},g_{1,2},g_l)=1
   ,\qquad
   \lim_{\rho\to\infty}(f_a,f_{\pm})=0
   ,\qquad
  \lim_{\rho\to\infty}m_a=0 .
   \label{eq:lim_gl_g12}
\end{align}
The straight pipe does not have the whispering gallery mode
except the zeroth radial mode ($m=0$).
$\mu_0^n$ is the only pole that satisfies Eq.(\ref{eq:lim_nurho_ksmn}) among
$\nu_m^n$ and $\mu_m^n$ for $m\leq m_a$.

\subsection{Imaginary poles for imaginary radial wavenumber}

In section \ref{sec:wgm} we assumed $|k\beta|>k_y^n$ ($k_r^n\in\mathbb{R}$) to explain
the whispering gallery mode of the curved pipe.
On the contrary, when $|k\beta|<k_y^n$, since $k_r^n=i\bk_r^n\in i\mathbb{R}$,
Eqs.(\ref{eq:ps_nu_0}) are rewritten as Eqs.(2.15) and (4.2) in \cite{warnock_morton},
\begin{align}
  P_\nu(\brr_b,\brr_a)=0
   \quad\text{for}~~\mfG_{+}^n
   , \qquad\quad
  \brr_b\brr_aS_\nu(\brr_b,\brr_a)=0
   \quad\text{for}~~\mfG_{-}^n .
   \label{eq:PS_nu_zero}
\end{align}
$\brr_{a,b}$ denotes the dimensionless radii which are real when $|k\beta|<k_y^n$,
opposite to $\hr_{a,b}$ given by Eq.(\ref{eq:krn}),
\begin{align}
  \brr_{a,b}
  =\bk_r^nr_{a,b}
  \in\mathbb{R}
  \quad (\bk_r^n\in\mathbb{R})
   ,\qquad
  (\bk_r^n)^2
  =-(k_r^n)^2
  =(k_y^n)^2-(k\beta)^2
  >0 .
   \label{eq:bkrn}
\end{align}
$P_\nu(\brr_b,\brr_a)$ and $S_\nu(\brr_b,\brr_a)$ are the cross products of
the modified Bessel functions $I_{\nu}(\brr)$ and $K_{\nu}(\brr)$,
\begin{alignat}{3}
  P_\nu(\bbar,\abar)
  &=\bigg|
    \begin{array}{cc}
      I_{\nu}(\bbar) & K_{\nu}(\bbar) \\
      I_{\nu}(\abar) & K_{\nu}(\abar)
    \end{array}
    \bigg|
  &&=\frac{\sin(\pi\nu)}{i\pi}
    \bigg|
    \begin{array}{cc}
      L_{\nu}(\bbar) & K_{\nu}(\bbar) \\
      L_{\nu}(\abar) & K_{\nu}(\abar)
    \end{array}
    \bigg|
  &&=-\frac{\pi}{2}p_\nu(i\bbar,i\abar) ,
  \label{eq:CP_P_mod}
   \\
  Q_\nu(\bbar,\abar)
  &=\bigg|
    \begin{array}{cc}
      I_{\nu}(\bbar) & K_{\nu}(\bbar) \\
      I_{\nu}'(\abar) & K_{\nu}'(\abar)
    \end{array}
    \bigg|
  &&=\frac{\sin(\pi\nu)}{i\pi}
    \bigg|
    \begin{array}{cc}
      L_{\nu}(\bbar) & K_{\nu}(\bbar) \\
      L_{\nu}'(\abar) & K_{\nu}'(\abar)
    \end{array}
    \bigg|
  &&=-i\frac{\pi}{2}q_\nu(i\bbar,i\abar) ,
  \label{eq:CP_Q_mod}
   \\
  R_\nu(\bbar,\abar)
  &=\bigg|
    \begin{array}{cc}
      I_{\nu}'(\bbar) & K_{\nu}'(\bbar) \\
      I_{\nu}(\abar) & K_{\nu}(\abar)
    \end{array}
    \bigg|
  &&=\frac{\sin(\pi\nu)}{i\pi}
    \bigg|
    \begin{array}{cc}
      L_{\nu}'(\bbar) & K_{\nu}'(\bbar) \\
      L_{\nu}(\abar) & K_{\nu}(\abar)
    \end{array}
    \bigg|
  &&=-i\frac{\pi}{2}r_\nu(i\bbar,i\abar) ,
  \label{eq:CP_R_mod}
  \\
  S_\nu(\bbar,\abar)
  &=\bigg|
    \begin{array}{cc}
      I_{\nu}'(\bbar) & K_{\nu}'(\bbar) \\
      I_{\nu}'(\abar) & K_{\nu}'(\abar)
    \end{array}
    \bigg|
  &&=\frac{\sin(\pi\nu)}{i\pi}
    \bigg|
    \begin{array}{cc}
      L_{\nu}'(\bbar) & K_{\nu}'(\bbar) \\
      L_{\nu}'(\abar) & K_{\nu}'(\abar)
    \end{array}
    \bigg|
  &&=\frac{\pi}{2}s_\nu(i\bbar,i\abar) ,
  \label{eq:CP_S_mod}
\end{alignat}
where the arguments $\abar$ and $\bbar$ can be complex in general,
similar to $a$ and $b$ in Eqs.(\ref{eq:CP_pq}-\ref{eq:CP_sr}).
But since we assume Eqs.(\ref{eq:bkrn}), we limit $\abar$ and $\bbar$ to real variables.
$R_\nu(z,z)=1/z$ is the Wronskian of $K_\nu(z)$ and $I_\nu(z)$,
given by 9.6.15 in \cite{abramo_stegun} and Eq.(\ref{eq:W_KI}).
When $\nu\in i\mathbb{R}$ in Eqs.(\ref{eq:CP_P_mod}-\ref{eq:CP_S_mod}), instead of $I_\nu$,
it is convenient to use $L_\nu$ which is the modified Bessel function
defined in Eq.(\ref{eq:def_KL_nu}) excluding the origin of the $\nu$-plane,
\begin{align}
  L_\nu(z)=i\pi\frac{I_{-\nu}(z)+I_\nu(z)}{2\sin(\pi\nu)}
   \qquad
  (\nu\in\mathbb{C},~\nu\ne0,~z\in\mathbb{C}) .
\end{align}
The Wronskian of $K_{\nu}$ and $L_{\nu}$ is given by Eq.(\ref{eq:W_KL}).
$K_{\nu}$ and $L_{\nu}$ for $\nu\in i\mathbb{R}$ have the power series representations
(\ref{eq:LK_ibnu}).
Their uniform asymptotic expansions are given by
Eqs.(\ref{eq:Kibnu_uae}-\ref{eq:L_prm_uae}).
Although $L_{\nu}$ is undefined at $\nu=0$,
Eqs.(\ref{eq:CP_P_mod}-\ref{eq:CP_S_mod}) are well defined $\forall\nu\in\mathbb{C}$
since $L_{\nu}\sin(\pi\nu)$ is defined at $\nu=0$.
When $\bk_r^n\in\mathbb{R}$, Eqs.(\ref{eq:PS_nu_zero}) have no solution on the real axis
($\nu\in\mathbb{R}$).
They have zeros only on the imaginary axis ($\nu\in i\mathbb{R}$) as seen from
Eq.(\ref{eq:p_ibnu_ixy}) and (\ref{eq:Kibnu_zero}-\ref{eq:Kp_ibnu_zero}).
The asymptotic expressions of the imaginary poles for $|\nu|\gg k_x^m\rho$ or
$|\nu|=O(k_x^m\rho)$ are given by Eq.(\ref{eq:iwgm}) or (\ref{eq:bnu_mn}).
When $|k|=k_y^n/\beta$, since $k_r^n=i\bk_r^n=0$, all the arguments of the cross products
($\hr=k_r^nr$ for $\forall r\in[r_a,r_b]$), which are involved in $\mfG_{\pm}^n$, become zero.
$P_\nu(\brr_b,\brr_a)$ and $S_\nu(\brr_b,\brr_a)$ are not identically zero in the limit of
$\bk_r^n\to0$, however, they have zeros with respect to $\nu\in i\mathbb{R}$.

We consider the poles of $\mfG_{\pm}^n$ in the limit of $|k|\to k_y^n/\beta$.
The asymptotic limit of the cross products for $k_r^n\to0$ with $\nu$ fixed
is equivalent to that for $|\nu|\to\infty$ with $\hr$ fixed as seen from
Eqs.(\ref{lim_kr0_aq_br_bas}), (\ref{eq:pnu_asymp})
and (\ref{eq:p_ibnu_ixy}-\ref{eq:s_ibnu_ixy}).
From Eqs.(\ref{lim_kr0_aq_br_bas}), the asymptotic limit of $\mfG_{\pm}^n$ for
$k_r^n\to0$ is given as
\begin{align}
  \lim_{k_r^n\to0}\mfG_{\pm}^n(r,r',\nu)
  =-\frac{\mfL_{\pm}(r,r',\nu)+\mfL_{\pm}(r',r,\nu)}{(\nu/\rho)\sinh[\nu\log(r_b/r_a)]} ,
  \label{lim_kr0_mfGpm}
\end{align}
where
\begin{align}
  \mfL_{+}(r,r',\nu)
  &=\theta(r-r')\sinh[\nu\log(r_b/r)]\sinh[\nu\log(r'/r_a)] ,
   \label{eq:mfLp}
   \\
  \mfL_{-}(r,r',\nu)
  &=\theta(r-r')\cosh[\nu\log(r_b/r)]\cosh[\nu\log(r'/r_a)] .
\end{align}
Accordingly, $\mfG_{\pm}^n$ in the limit of $k_r^n\to0$ has poles in $i\mathbb{R}$
as follows,
\begin{align}
  \lim_{k_r^n\to0}\frac{(\nu_m^n,\mu_m^n)}{\rho}
  =i\frac{k_x^m}{g_l}
  \in i\mathbb{R}
   ,\qquad
  m
  \in(\mathbb{N},\mathbb{Z}_0^{+}) .
   \label{eq:nu_krn0}
\end{align}
We can also derive Eq.(\ref{eq:nu_krn0}) from
Eqs.(\ref{eq:CP_pq_appendix}-\ref{eq:CP_sr_appendix}) through the leading order terms of
Eqs.(\ref{eq:Fimu_x}-\ref{eq:Gimu_x}) for small arguments $\hr_{b,a}\to0$.
$k_x^m$ is the horizontal wavenumber of the straight pipe,
given by Eq.(\ref{eq:asymp_lim}).
$g_l$ is the geometric factor given by Eq.(\ref{eq:gl_g12}),
which is of the order of 1 under the assumption $\eps\ll1$,
\begin{align}
  \frac{1}{g_l}
  \simeq g_1-\frac{\eps^2}{12}+O(\eps^3) ,
    \qquad
  \eps
  =\frac{w}{\rho} .
   \label{eq:gl_eps}
\end{align}
In the limit of $k_r^n\to0$, $\mfG_{-}^n$ has a pole also at $\nu=0$
unlike $\mfG_{+}^n$ since $\mfL_{+}=0^2$ and $\mfL_{-}\ne0$ at $\nu=0$.
As described in Eq.(\ref{eq:snu_poles}),
$\hr_b\hr_as_\nu$ has the zeros of the zeroth radial mode $\nu=\pm\mu_0^n$
which are simple unless $k_r^n=0$.
In the limit of $k_r^n\to0$, the pair of the zeroth poles $\nu=\pm\mu_0^n$ couple at
the origin of the $\nu$-plane and become a pole of the second order,
because $\mu_0^n\to0$ in the limit of $k_r^n\to0$ from Eq.(\ref{eq:nu_krn0}) for $m=0$
regardless of the curvature of the pipe.
This is similar to $k_s^{mn}$ for $m=0$, \ie,
$k_s^{0n}\to0$ in the limit of $k_r^n\to0$ since $k_s^{0n}=k_r^n$.
That is, when $k_r^n=0$, $\hr_b\hr_a s_\nu(\hr_b,\hr_a)$ is zero of the second order at 
$\nu=\mu_0^n=0$, because $\hr_b\hr_a\propto(k_r^n)^2$,
and $s_0(\hr_b,\hr_a)$ cannot be zero in the limit of $k_r^n\to0$
unless $\eps=w/\rho=0$ as shown in Eq.(\ref{eq:lim_kr0_p0s0}),
\begin{align}
  \lim_{k_r^n\to0}s_0(\hr_b,\hr_a)
  =-\frac{2}{\pi}\sinh(g_l\eps)
  \ne0 .
  \label{eq:lim_krn_s0}
\end{align}
%

\subsection{Imaginary whispering gallery mode}
\label{sec:iwgm}

We consider the Green functions $\mfG_{\pm}^n$, given by Eqs.(\ref{eq:mfGe}-\ref{eq:mfGb}),
for $\nu=(\nu_m^n,\mu_m^n)\in i\mathbb{R}^{+}$ and $k_r^n\in i\mathbb{R}^{+}$, \ie,
\begin{align}
  \nu
  =i\bnu ,
   \qquad
  k_r^n
  =i\bk_r^n
   \quad
  (\hr=i\brr,~\brr=\bk_r^nr) ,
    \quad\text{where}\quad
  \bnu
  =(\bnu_m^n,\bmu_m^n)
  \in\mathbb{R}^{+} ,
    \quad
  \bk_r^n
  \in\mathbb{R}^{+} .
    \label{eq:bnu_bk}
\end{align}
In Eqs.(\ref{eq:nu_mt}) we defined the whispering gallery mode of the field
for $k_r^n\in\mathbb{R}$ in the curved pipe.
It is a mode of the real pole which hardly depends on the inner wall $r_a$.
The real pole of the whispering gallery mode is asymmetric with respect to $r_b$ and $r_a$.
Similar to Eqs.(\ref{eq:nu_mt}), we define the imaginary whispering gallery mode
which hardly depends on the outer wall $r_b$ under the conditions given by
Eqs.(\ref{eq:bnu_bk}),
\begin{align}
  \bnu_m^n
  \leq\brr_b
  <\bnu_{m+1}^n
   \quad
  (m=m_b^{+}) ,
    \qquad
  \bmu_m^n
  \leq\brr_b
  <\bmu_{m+1}^n
   \quad
  (m=m_b^{-}) .
    \label{eq:mb_def}
\end{align}
That is, the imaginary whispering gallery denotes the following radial modes,
\begin{align}
  m \leq m_b^{+}
   \quad\text{for}~~ \bnu_m^n ,
    \qquad
  m \leq m_b^{-}
   \quad\text{for}~~ \bmu_m^n .
    \label{eq:mb_rng}
\end{align}
Eqs.(\ref{eq:mb_rng}) are the conditions such that
$|L_{i\bnu}(\brr_b)|\gg |K_{i\bnu}(\brr_b)|$ and
$|L_{i\bnu}'(\brr_b)|\gg |K_{i\bnu}'(\brr_b)|$
respectively at the imaginary poles $\bnu=\bnu_m^n$ and $\bmu_m^n$.
And therefore the zeros of $P_{i\bnu}(\brr_b,\brr_a)$ and $S_{i\bnu}(\brr_b,\brr_a)$
with respect to $\bnu$ are approximately equal to the zeros of $K_{i\bnu}(\brr_a)$ and
$K_{i\bnu}'(\brr_a)$ as described in Eqs.(\ref{eq:Kibnu_zero}-\ref{eq:Kp_ibnu_zero}).
It follows that $\bnu_m^n$ and $\bmu_m^n$ do not depend on
the outer wall radius $r_b$, opposite to the real whispering gallery mode.
The asymptotic expressions of the poles of the imaginary whispering gallery modes are
given by Eq.(\ref{eq:iwgm}).

In what follows we write $m_b^{\pm}$ simply as $m_b$ for brevity
unless we need to distinguish $m_b^{+}$ and $m_b^{-}$.
The poles for $m\geq m_b+1$ satisfy Eq.(\ref{eq:lim_nurho_ksmn})
when $k_s^{mn}\in i\mathbb{R}^{+}$.
Similar to Eq.(\ref{eq:mt}), the asymptotic expression of $m_b$ for large radius is
gotten from Eq.(\ref{eq:bnu_sols}),
\begin{align}
  m_b^{\pm}
  =\bigg\lfloor\frac{\bk_r^n}{k_x^1}\gam_b^{1/2}\bigg\rfloor
   \qquad
  (|k\beta|<k_y^n,~ \eps=w/\rho \ll 1 \ll \bk_r^n\rho ) .
    \label{eq:mb}
\end{align}
$\lfloor~\rfloor$ denotes the floor function similar to the one in Eq.(\ref{eq:mt}).
$\gam_b$ is given as
\begin{align}
  \gam_b
  &=\frac{g_l}{2}\{\barf_b+(\barf_b^2+\barf_{\pm})^{1/2}\}
   \simeq
    \eps
    \Big\{
      \frac{1}{2}
     +\frac{1}{6^{1/2}}
     +O[\eps,(\bk_r^nw)^{-2}]
    \Big\} ,
   \\
  \barf_b
  &=g_lg_b^2-g_1 ,
    \qquad
  \barf_{\pm}
  =g_1\bigg\{(g_1-g_lg_2) \pm\frac{4g_l}{(\bk_r^n\rho)^2}\bigg\} .
\end{align}
The geometric factors $g_{1,2}$ and $g_l$ are given by Eqs.(\ref{eq:gl_g12}).
The double sign of $\barf_{\pm}$ corresponds to that of $m_b^{\pm}$ in Eqs.(\ref{eq:mb_rng}).
Similar to Eqs.(\ref{eq:lim_gl_g12}), $m_b$ goes to zero in the limit of $\rho\to\infty$, 
\ie, the straight pipe does not have the imaginary whispering gallery mode,
\begin{align}
   \lim_{\rho\to\infty}(\barf_b,\barf_{\pm})=0
   ,\qquad
  \lim_{\rho\to\infty}m_b=0 .
   \label{eq:lim_mb}
\end{align}
%

\subsection{Asymptotic expressions of the poles for large radius}
\label{sec:asymptotic_poles}

Using the wavenumbers given by Eqs.(\ref{eq:asymp_lim}),
we define $\eps_{x,r,s}$ as follows,
\begin{align}
  \eps_x
  =\bigg(\frac{k_x^m}{k_r^n}\bigg)^2 ,
    \qquad
  \eps_r
  =\bigg(\frac{k_r^n}{k_x^m}\bigg)^2 ,
    \qquad
  \eps_s
  =\bigg(\frac{k_s^{mn}}{k_r^n}\bigg)^2 ,
    \qquad
  \eps
  =\frac{w}{\rho} .
  \label{eq:eps_rsx}
\end{align}
In appendix \ref{sec:poles} we derive the asymptotic expressions of the poles
$\nu=(\nu_m^n,\mu_m^n)$ for large radius ($\eps\ll1$).
In addition to $\eps$, we assume that one of $\eps_{x,r,s}$ is small
in deriving the asymptotic expressions of the poles:
\begin{alignat}{5}
  |\eps_x|
  &\ll1,
    \qquad&
  k_s^{mn}
  &=k_r^n(1-\eps_x)^{1/2}
   \quad&~\Lra~&\quad&
  &|\nu/\rho|\gg k_x^m,
    \quad&
  &\text{App.\,\ref{sec:whisper},~~Eqs.(\ref{eq:rwgm}-\ref{eq:iwgm})} ,
   \label{eq:cond3}
   \\
  |\eps_r|&\ll1,
    \qquad&
  k_s^{mn}
  &=ik_x^m(1-\eps_r)^{1/2}
   \quad&~\Lra~&\quad&
  &|\nu/\rho|= O(k_x^m),
    \quad&
  &\text{App.\,\ref{sec:asympt_numu},~~Eqs.(\ref{eq:nu_ks_body}-\ref{eq:bnu_mn})} ,
   \label{eq:cond1}
   \\
  |\eps_s|&\ll1,
    \qquad&
  k_x^m
  &=k_r^n(1-\eps_s)^{1/2}
   \quad&~\Lra~&\quad&
  &|\nu/\rho|\ll k_x^m,
    \quad&
  &\text{App.\,\ref{sec:HP2},~~Eq.(\ref{eq:asympt_sol_ks0})} .
   \label{eq:cond2}
\end{alignat}
$m\in\mathbb{N}$ in Eqs.(\ref{eq:cond1}-\ref{eq:cond2}).
In order to derive the asymptotic expression of $\mu_0^n$ which is symmetric
with respect to the exchange of $r_b$ and $r_a$, we assume that
\begin{align}
  \nu^{-1}
  =O(\eps)
   ,\qquad
  (k_r^n\rho)^{-1}
  =O(\eps)
   \quad~~\Lra~~\quad
  \nu/\rho
  =O(k_r^n)
   ,\qquad
  \text{App.\,\ref{sec:0th_0},~~Eq.(\ref{eq:mu0n_2nd})} .
   \label{eq:mu0n_cond}
\end{align}

Also, the expression of $\mu_0^n$ is gotten for Eq.(\ref{eq:cond3}).
Assuming Eq.(\ref{eq:cond3}), the asymptotic solutions of Eqs.(\ref{eq:ps_nu_0}) with
respect to $\nu\in\mathbb{R}$ and $\nu=i\bnu\in i\mathbb{R}$ are given by
Eqs.(\ref{cochran_sol}) and (\ref{eq:imag_whisper}),
\begin{alignat}{2}
  \frac{\{\nu_m^n,\mu_m^n\}}{\hr_b}
  &\simeq
   1+\sum_{j=1}^{\infty}\frac{\{A_m^{(j)},B_m^{(j)}\}}{\hr_b^{2j/3}}
   \qquad&&
  (\nu\in\mathbb{R},~ k_r^n\in\mathbb{R},~ m\leq m_a) ,
   \label{eq:rwgm}
   \\
  \frac{\{\bnu_m^n,\bmu_m^n\}}{\brr_a}
  &\simeq
   1+\sum_{j=1}^{\infty}\frac{\{A_m^{(j)},B_m^{(j)}\}}{(-1)^j\brr_a^{2j/3}}
  \qquad&&
   (\bnu\in\mathbb{R},~k_r^n\in i\mathbb{R},~ m\leq m_b) .
   \label{eq:iwgm}
\end{alignat}
$\hr_b$ and $\brr_a$ are given by Eqs.(\ref{eq:krn}) and (\ref{eq:bkrn}).
The coefficients $A_m^{(j)}$ and $B_m^{(j)}$ are given by Eqs.(\ref{eq:Ajm}-\ref{eq:Bjm}).
Eqs.(\ref{eq:rwgm}) and (\ref{eq:iwgm}) do not depend on $r_a$ and $r_b$ respectively
since they are the poles of the real and imaginary whispering 
gallery modes described in sections \ref{sec:wgm} and \ref{sec:iwgm}.
The pole of the zeroth radial mode, $\mu_0^n$ and $\bmu_0^n$ given by
Eqs.(\ref{eq:rwgm}-\ref{eq:iwgm}) for $m=0$, satisfies Eq.(\ref{eq:lim_nurho_ksmn})
since $k_s^{0n}$ does not depend on $w$ unlike $k_s^{mn}$ for $m\in\mathbb{N}$.
On the other hand, under the assumptions $k_r^nw=O(1)$ and Eqs.(\ref{eq:mu0n_cond}),
we get the asymptotic expression of $\mu_0^n$ which is symmetric with 
respect to the exchange of $r_b$ and $r_a$ as shown in Eq.(\ref{eq:tau0_2nd}),
\begin{align}
  \bigg(\frac{\mu_0^n}{k_r^n\rho}\bigg)^2
  =
    \frac{r_br_a}{\rho^2}
   +\frac{w^2}{6\rho^2}\bigg\{1+\frac{(k_r^nw)^2}{5}\bigg\}
   +O(\eps^3) .
  \label{eq:mu0n_2nd}
\end{align}

\begin{figure}[h]
  \begin{center}
    \includegraphics[scale=0.32,clip]{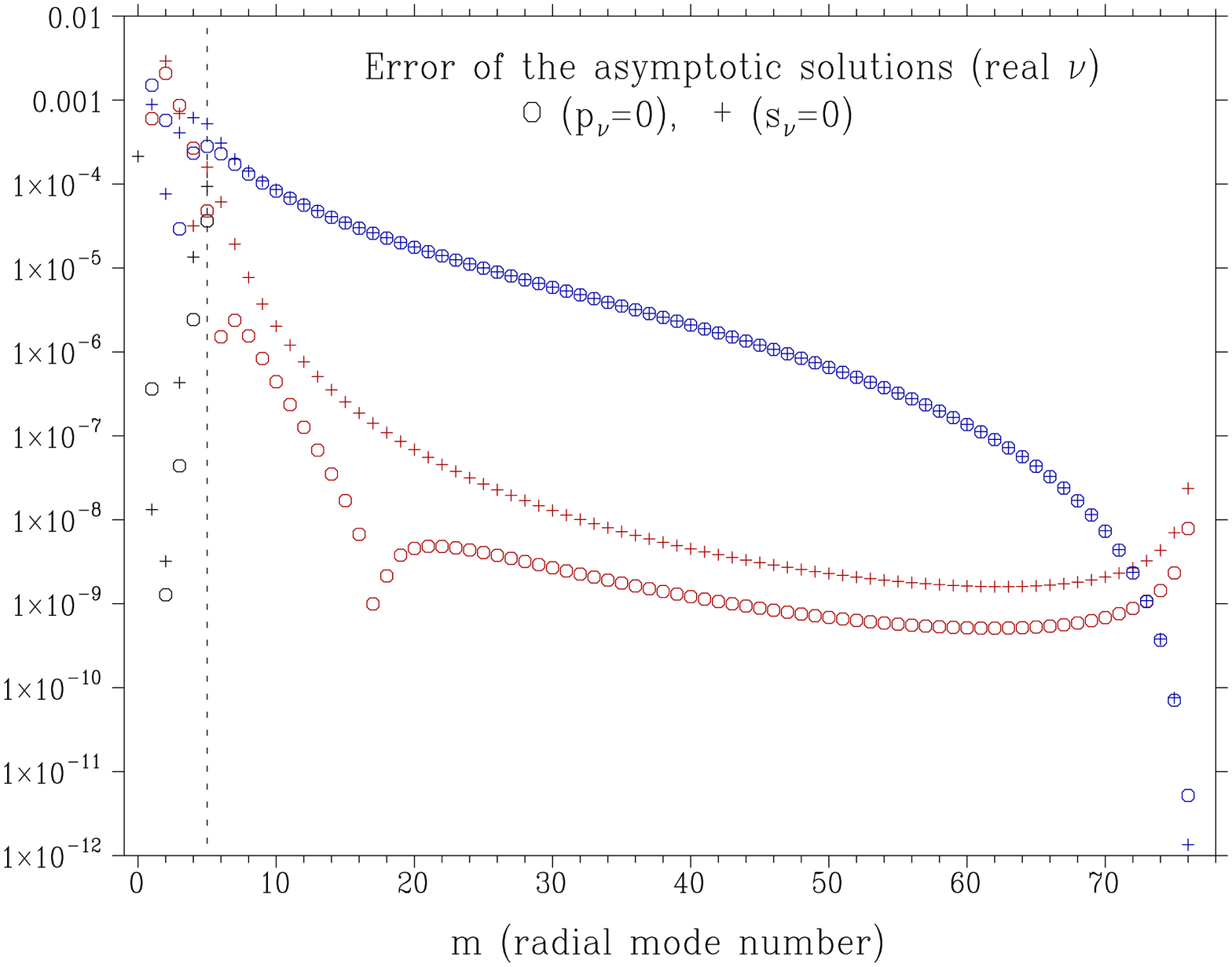}
     \hspace{1mm}
    \includegraphics[scale=0.32,clip]{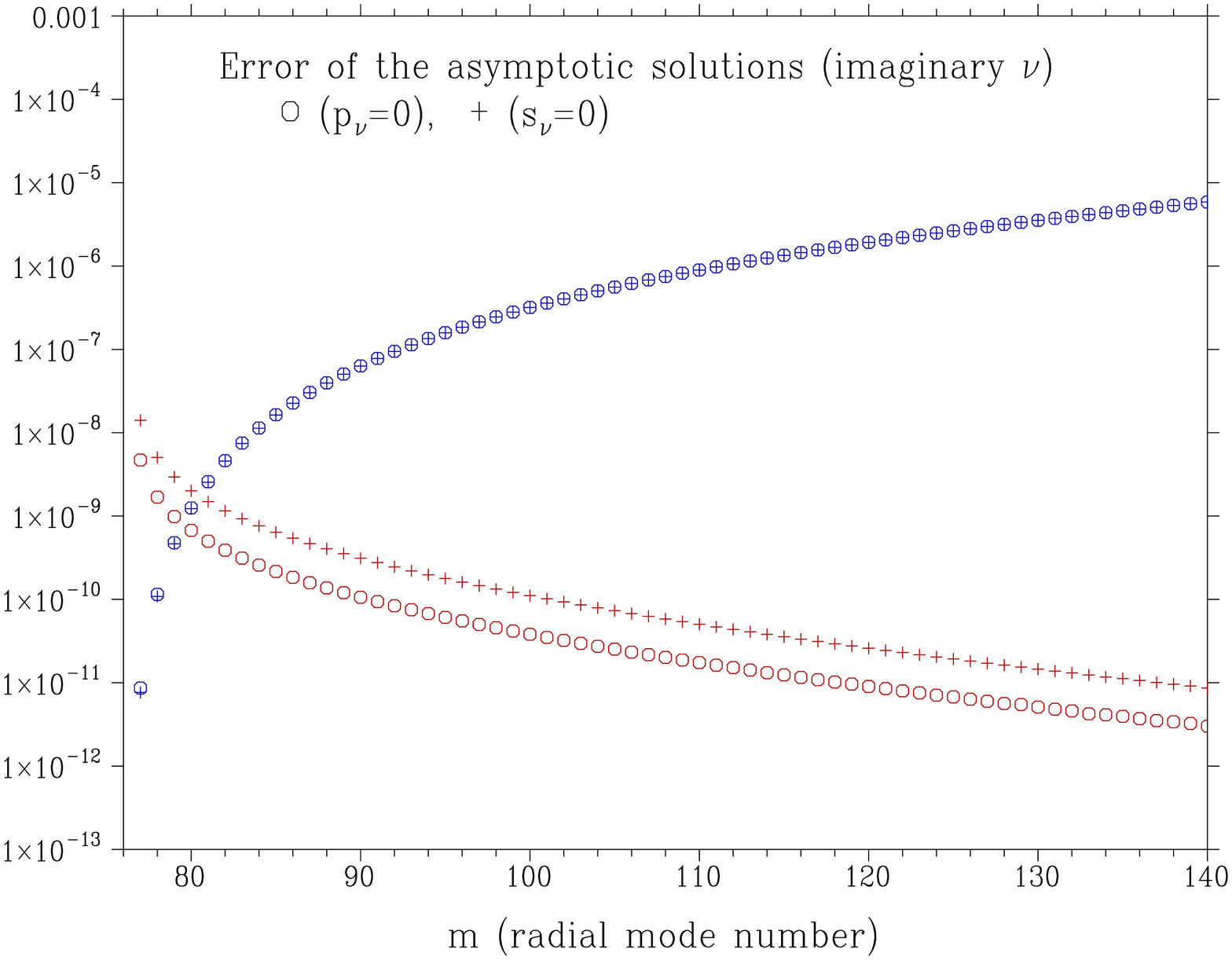}
    \caption[Error of the asymptotic expressions of the poles for real radial wavenumber]
     {\small 
     Absolute value of the relative error of the asymptotic expressions of
     the real poles (left) and the imaginary poles (right) for a real radial wavenumber
     ($k_r^n\in\mathbb{R}$).
     The horizontal axis is the radial mode number
     $m\in(\mathbb{N},\mathbb{Z}_0^{+})$ for $(\nu_m^n,\mu_m^n)$.
     The circles and crosses $(\circ,+)$ correspond to $(\nu_m^n,\mu_m^n)$.
     The asymptotic solutions are given by Eqs.(\ref{eq:rwgm}), (\ref{eq:numu_sol}) and
     (\ref{nu2_6th}) which are plotted respectively with black ($|\nu|\gg\nu_x$),
     red [$|\nu|=O(\nu_x)$] and blue ($|\nu|\ll\nu_x$) where $\nu_x=k_x^m\rho$.
     We got the numerical solutions of Eqs.(\ref{eq:ps_nu_0}) using Newton's method.
     The vertical dashed line in the left figure denotes $m_a=5$ which is
     the number of the real whispering gallery modes given by Eq.(\ref{eq:mt}).
     The number of the real poles is $m_{\pm}=76$ (right/left end of the left/right figure)
     for the following parameters:
     $\rho=10{\rm m}$, $x_b=-x_a=3{\rm cm}$ ($w=6{\rm cm}$), $h=6{\rm cm}$
     and $E=1\,\text{GeV}$.
     }
    \label{fig:nu_err_4K}
  \end{center}
    \vspace{1mm}
  \begin{center}
    \includegraphics[scale=0.32,clip]{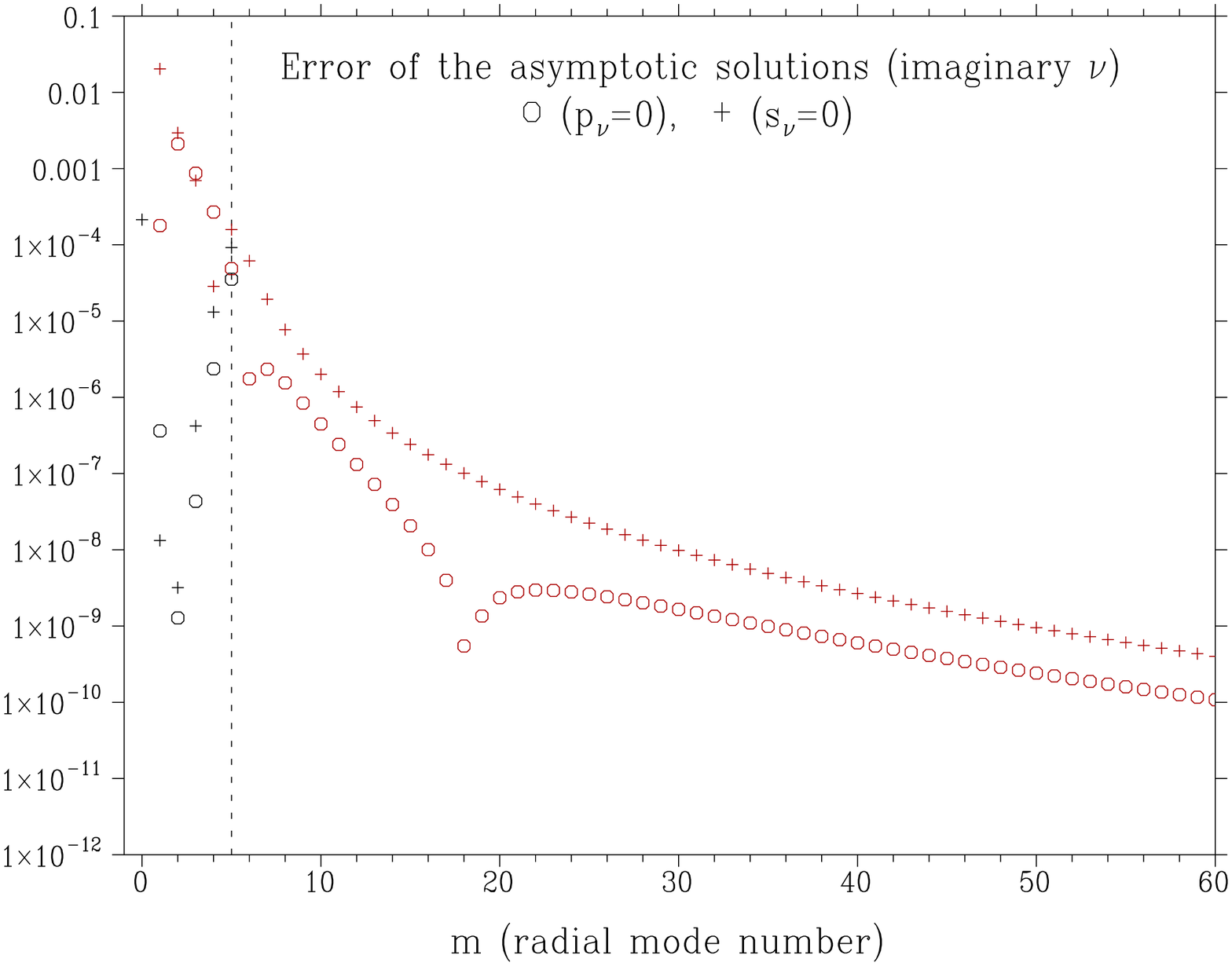}
    \caption[Error of the asymptotic expressions of the poles for
      imaginary radial wavenumber]
     {\small 
     Absolute value of the relative error of the asymptotic expressions of
     the imaginary poles $\nu=i\cnu\in i\mathbb{R}$ for an imaginary radial wavenumber
     ($k_r^n=i\bk_r^n\in i\mathbb{R}$).
     The horizontal axis is the radial mode number
     $m\in(\mathbb{N},\mathbb{Z}_0^{+})$ for $(\cnu_m^n,\cmu_m^n)$.
     The circles and crosses $(\circ,+)$ correspond to $(\cnu_m^n,\cmu_m^n)$.
     The asymptotic solutions are given by Eqs.(\ref{eq:rwgm}) and (\ref{eq:bnu_mn})
     which are plotted respectively with black ($\cnu\gg\nu_x$) and
     red [$\cnu=O(\nu_x)$] where $\nu_x=k_x^m\rho$.
     We got the numerical solutions of Eqs.(\ref{eq:ps_nu_0}) for
     $\nu=i\cnu$ and $\hr=i\bk_r^nr$ using Newton's method.
     The vertical dashed line denotes $m_b=5$ which is the number of
     the imaginary whispering gallery modes given by Eq.(\ref{eq:mb}).
     The parameters are given as  $\rho=10\,{\rm m}$,
     $x_b=-x_a=3\,{\rm cm}$ ($w=6{\rm cm}$), $h=6\,{\rm cm}$ and $E=1\,\text{GeV}$.
     }
    \label{fig:bnu_err_bkr}
  \end{center}
\end{figure}

Assuming Eq.(\ref{eq:cond1}), the asymptotic expression of the square of the poles of
the normal mode ($m>m_{a,b}$) is given by Eq.(\ref{eq:nu_ks}),
\begin{align}
  \bigg\{\frac{(\nu_m^n,\mu_m^n)}{k_s^{mn}\rho}\bigg\}^2
  &\simeq
    \frac{1}{g_l^2}
   +\frac{1-g_1g_l}{g_l^2}\bigg(\frac{k_r^n}{k_x^m}\bigg)^2
   +\bigg\{
      \frac{1-g_1g_l}{g_l^2}+g_1\frac{g_2g_l-g_1}{4}\pm\frac{g_1g_l}{(k_r^n\rho)^2}
    \bigg\}
    \bigg(\frac{k_r^n}{k_x^m}\bigg)^4
   +O(\eps_r^3)
  \label{eq:nu_ks_body}
   \\
  &\simeq
    \frac{r_br_a}{\rho^2}
   -\frac{w^2}{12\rho^2}\bigg(\frac{k_s^{mn}}{k_x^m}\bigg)^2
   \pm\frac{1}{(k_x^m\rho)^2}\bigg(1+\frac{w^2}{12\rho^2}\bigg)
    \bigg(\frac{k_r^n}{k_x^m}\bigg)^2
   +O(\eps_r^3,\eps^3) ,
  \label{eq:numu_sol}
\end{align}
where the double sign $(+,-)$ corresponds to $(\nu_m^n,\mu_m^n)$.
$g_{1,2}$ and $g_l$ are the geometric factors given by Eqs.(\ref{eq:gl_g12})
which are expanded with respect to $x_{b,a}/\rho$ as shown in Eqs.(\ref{eq:g0_gl}).
Under the assumption (\ref{eq:cond1}), the square of the poles of
the normal mode on the imaginary axis is given by Eq.(\ref{eq:bnu_sols}),
\begin{align}
  \bigg\{\frac{(\bnu_m^n,\bmu_m^n)}{k_x^m\rho}\bigg\}^2
  &\simeq
    \frac{1}{g_l^2}
   -\frac{g_1}{g_l}\bigg(\frac{k_r^n}{k_x^m}\bigg)^2
   +g_1\bigg\{\frac{g_2g_l-g_1}{4}\pm\frac{g_l}{(k_r^n\rho)^2}\bigg\}
    \bigg(\frac{k_r^n}{k_x^m}\bigg)^4
   +O(\eps_r^3) ,
   \label{eq:bnu_mn}
\end{align}
where $k_r^n\in\mathbb{A}$ in general.
If $k_r^n\in\mathbb{R}$, Eq.(\ref{eq:bnu_mn}) holds also for the real poles.
Actually, we got Eq.(\ref{eq:nu_ks_body}) from Eq.(\ref{eq:bnu_mn}) for $\nu=i\bnu$.
On the other hand, assuming Eq.(\ref{eq:cond2}), the asymptotic expression of the square of
the poles of the normal mode for $k_r^n\in\mathbb{R}$ around $m=m_{\pm}$
(\ie, around $\nu=0$) is given by Eq.(\ref{eq:nu_rho2}),
\begin{align}
  (\nu_m^n,\mu_m^n)^2
  \simeq
    r_br_a(k_s^{mn})^2
   +\frac{(1,-3)}{4}
   +O(\eps_s,\eps^2)
   ,\qquad
  |m-m_{\pm}|=O(1) ,
   \label{eq:asympt_sol_ks0}
\end{align}
where the coefficient $(1,-3)$  corresponds to $(\nu_m^n,\mu_m^n)$.
In Fig.\ref{fig:nu_err_4K} ($k_r^n\in\mathbb{R}$) and Fig.\ref{fig:bnu_err_bkr}
($k_r^n\in i\mathbb{R}$) we plot the relative error of the asymptotic expressions of
the poles, which are given by Eqs.(\ref{eq:rwgm}-\ref{eq:iwgm}) and
Eqs.(\ref{eq:numu_sol}-\ref{eq:asympt_sol_ks0}),
for the numerical solutions of Eqs.(\ref{eq:ps_nu_0}) using Newton's method.
We used these asymptotic expressions of the poles as the initial values of Newton's method.

Eqs.(\ref{eq:mu0n_2nd}-\ref{eq:asympt_sol_ks0}) are symmetric with respect to
the exchange of $r_b$ and $r_a$ unlike Eqs.(\ref{eq:rwgm}-\ref{eq:iwgm}).
Eqs.(\ref{eq:mu0n_2nd}-\ref{eq:asympt_sol_ks0}) satisfy the asymptotic condition
given by Eq.(\ref{eq:lim_nurho_ksmn})
since they are not the whispering gallery modes ($m\leq m_{a,b}$)
but the normal modes ($m> m_{a,b}$).
Although we derived Eqs.(\ref{eq:mu0n_2nd}-\ref{eq:asympt_sol_ks0}) under the different 
conditions as listed in Eqs.(\ref{eq:cond3}-\ref{eq:cond2}),
Eqs.(\ref{eq:mu0n_2nd}-\ref{eq:asympt_sol_ks0}) have common terms up to
the first order with respect to $\eps$,
\begin{align}
  \bigg\{\frac{(\nu_m^n,\mu_m^n)}{k_s^{mn}\rho}\bigg\}^2
  \simeq 
  1+\frac{x_b+x_a}{\rho}+O(\eps^2)
   \qquad (m> m_{a,b},~\eps=w/\rho\ll1) .
   \label{nu_rho_ks}
\end{align}
If the $s$-axis is located at the center of the sidewalls of the rectangular pipe,
Eq.(\ref{nu_rho_ks}) has no first order term with respect to $\eps$.
In this case, $\nu_m^n$ differs from $\mu_m^n$ in the terms of $O(\eps^2)$.

%

\subsection{Residue of the integrand at the second order pole}
\label{sec:analytic_continuation}

In section \ref{sec:ILT} we will find the expressions of the Fourier coefficients of
the Green functions $\cG_{\pm}^n$ from $\mfG_{\pm}^n$ through Eq.(\ref{eq:cGeb_def})
which is the Bromwich integral with respect to $\nu$.
$\mfG_{\pm}^n$ is the Green functions in the Laplace domain,
given by Eqs.(\ref{eq:mfGe}-\ref{eq:mfGb}).
We consider the residue of $\mfG_{\pm}^n$ in the Bromwich integral at the origin of
the $\nu$-plane ($\nu=0$) which is a pole of the second order.
It happens at the cutoff wavenumbers
$k=k_m^n$ for $\mfG_{+}^n$ and $k=\bk_m^n$ for $\mfG_{-}^n$,
which are implicitly defined in Eqs.(\ref{eq:kmn}-\ref{eq:bkmn}).
See appendix \ref{sec:analytic} for the details on the coupling of the simple poles at
$\nu=0$.
Here we show the treatment of the residue of the Bromwich integral at the second order pole.
One can skip section \ref{sec:analytic_continuation}
since it is possible to calculate the Bromwich integral analytically
without taking into account the second order pole at $\nu=0$, which can happen by accident 
depending on the parameters of the field and the geometry of the curved pipe.
In short, we need no special treatment for the pole at $\nu=0$ in 
calculating the Bromwich integral using the residue theorem,
because the residue of the integrand at $\nu=0$ is equal to the sum of the residues at
the corresponding pair of the simple poles on the real axis.
In other words, we can regard the pole of the second order as a pair of
the simple real poles which coupled at $\nu=0$.
This is because $p_{\nu}$ and $s_{\nu}$ are analytic and even
with respect to $\nu$ for $\forall k$ and $\forall k_y^n$.
The Green functions in the Laplace domain $\mfG_{\pm}^n$ have the following properties:
(i) they have a fractional form,
(ii) their numerator and denominator are both analytic with respect to $\nu$,
(iii) their denominator is even with respect to $\nu$.
In what follows in section \ref{sec:analytic_continuation},
we discuss the poles of $\mfG_{\pm}^n$ and the residue of the integrand,
which is involved in the Bromwich integral (\ref{eq:cGeb_def}),
using a more general form than Eqs.(\ref{eq:mfGe}-\ref{eq:mfGb}).

We consider the following fraction function $F$ of $\nu\in\bbC$,
\begin{align}
  F(\nu)=\frac{f(\nu)}{\phi(\nu)}
   ,\qquad
  \phi(0)=0^2
   \qquad
  ( f\ne0 ~~\text{when}~~ \phi=0 ) .
  \label{eq:frac_func}
\end{align}
We assume that $f$ and $\phi$ are both analytic for $\forall\nu\in\bbC$.
Let $\phi$ have a zero of the second order at $\nu=0$,
similar to the simple example given by Eq.(\ref{eq:phi_simple}) for $\delta=0$.
We assume that $f$ is not zero at $\nu$ such that $\phi=0$.
We define $R(0)$ as the residue of $F$ at $\nu=0$ which is the pole of the second order,
\ie, $\phi(0)=0^2$.
Under these assumptions, $R(0)$ is given as
\begin{align}
  R(0)
  =\bigg[\frac{2}{\rd_\nu^2\phi}(\rd_{\nu}-u)f\bigg]_{\nu=0}
    \quad\text{where}\quad
  u
  =\frac{\rd_\nu^3\phi}{3\rd_\nu^2\phi} ,
   \qquad
  \rd_\nu^j\phi
  =\frac{d^j\phi}{d\nu^j}
   ,\qquad
  \rd_\nu=\frac{d}{d\nu} .
  \label{eq:Res2}
\end{align}
$\rd_\nu^j\phi$ denotes the $j$th derivative of $\phi(\nu)$ with respect to $\nu$.
If $\phi$ is even with respect to $\nu$, $u=0$ since $\rd_\nu^3\phi=0$ at $\nu=0$.

We consider the Bromwich integral given by Eq.(\ref{eq:cGeb_def})
which is the inverse Laplace transform of $\mfG_{\pm}^n$.
For convenience in describing the residue at the second order pole,
we rewrite the integrands, involved in Eq.(\ref{eq:cGeb_def}),
into the form of fraction function as follows,
\begin{align}
  \mfG_{+}^n(\nu,k)\frac{e^{i\nu s/\rho}}{2\pi\rho}
  =\frac{f_{+}^n(\nu,k,s)}{p_\nu(\hr_b,\hr_a)},
   \qquad
  \mfG_{-}^n(\nu,k)\frac{e^{i\nu s/\rho}}{2\pi\rho}
  =\frac{f_{-}^n(\nu,k,s)}{\hr_b\hr_a s_\nu(\hr_b,\hr_a)} ,
  \label{eq:mfGeb_frac}
\end{align}
where we omit the radial arguments $(r,r')$ of $\mfG_{\pm}^n$ and $f_{\pm}^n$ for brevity.
Instead, we indicate the argument $k$ which is the wavenumber of the field,
defined in Eq.(\ref{eq:omg}).
$f_{\pm}^n$ are the numerators of the integrands which are involved in
the Bromwich integral (\ref{eq:cGeb_def}).
$p_\nu$ and $s_\nu$ are even with respect to $\nu$ as shown in Eq.(\ref{eq:even_symm_ps}).
As shown in Fig.\ref{fig:poles}, when $k=k_m^n$ or $\bk_m^n$ which is the cutoff wavenumber
of the curved pipe, the first or second equation of (\ref{eq:mfGeb_frac}) has
a pole of the second order at $\nu=0$.
The residues of Eqs.(\ref{eq:mfGeb_frac}) at $\nu=0$ are gotten
from Eq.(\ref{eq:Res2}) for $u=0$,
\begin{align}
  R_{+}(0)
  =\bigg[\frac{2\rd_{\nu}f_{+}^n(\nu,k_m^n,s)}{\rd_\nu^2p_\nu(\hr_b,\hr_a)}
   \bigg]_{\nu=0} ,
   \qquad
  R_{-}(0)
  =\bigg[\frac{2\rd_{\nu}f_{-}^n(\nu,\bk_m^n,s)}{\hr_b\hr_a \rd_\nu^2s_\nu(\hr_b,\hr_a)}
   \bigg]_{\nu=0} .
  \label{eq:R2}
\end{align}
The first and second equations of (\ref{eq:R2}) do not hold simultaneously
unless $\rho=\infty$ since $k_m^n\ne\bk_m^n$ in the curved pipe.
As shown in appendix \ref{sec:analytic}, we can find Eqs.(\ref{eq:R2}) in another way.
That is, in the limit that the pair of the real poles $\nu=\pm\hnu_m^n$ or $\pm\hmu_m^n$ 
couples at $\nu=0$, the sum of their residues at the simple poles is given as
\begin{alignat}{3}
  R_{+}(0)
  &=\lim_{k\to k_m^n}\{R_{+}(\hnu_m^n)+R_{+}(-\hnu_m^n)\}
   ,\qquad&
  \lim_{k\to k_m^n}\hnu_m^n
  &=0
   ,\qquad&
  R_{+}(\nu)
  &=\frac{f_{+}^n(\nu,k,s)}{\rd_\nu p_\nu(\hr_b,\hr_a)} ,
  \label{eq:R2p_lim}
   \\
  R_{-}(0)
  &=\lim_{k\to \bk_m^n}\{R_{-}(\hmu_m^n)+R_{-}(-\hmu_m^n)\}
   ,\qquad&
  \lim_{k\to \bk_m^n}\hmu_m^n
  &=0
   ,\qquad&
  R_{-}(\nu)
  &=\frac{f_{-}^n(\nu,k,s)}{\hr_b\hr_a \rd_\nu s_\nu(\hr_b,\hr_a)} .
  \label{eq:R2m_lim}
\end{alignat}
The poles on the real axis $\hnu_m^n$ and $\hmu_m^n$ go to zero for
$k\to k_m^n$ and $k\to \bk_m^n$ respectively.
As shown by the first equations of (\ref{eq:R2p_lim}-\ref{eq:R2m_lim}),
the sum of the residues at $\nu=\pm\hnu_m^n$ and $\pm\hmu_m^n$ tends to be the one at
the pole of the second order in the limit of $\nu\to0$.
Therefore, in the calculation of the Bromwich integral using the residue theorem,
we need no special treatment for the residue of the pole at $\nu=0$.
That is, we can suppose that $\mfG_{\pm}$ has only pairs of the simple poles
on the real and imaginary axes of the Laplace plane.
If $k$ is accidentally right on the cutoff wavenumber $k_m^n$ or $\bk_m^n$ in
the numerical calculation of the fields, it is enough to replace the sum of
the residues of the corresponding pair of the simple real poles by
that of the pole of the second order given by Eqs.(\ref{eq:R2}).
The cutoff wavenumbers may be important in the complex wavenumber plane ($k$-domain)
in deriving the expressions of the fields in the time domain through
the inverse Fourier transform given by Eq.(\ref{eq:Fourier_trans}).

\clearpage

\section{Inverse Laplace transform}
\label{sec:ILT}

From the fields in the Laplace domain $\mfE^n$ and $\mfB^n$
through the Bromwich integral with respect to $\nu$,
we will derive Eqs.(\ref{eq:cEBy_sol}) and (\ref{eq:cExs_cBxs_sprt}) which are
the $n$th Fourier coefficients of the fields $\cE^n$ and $\cB^n$ with respect to $y$.

\subsection{Fourier coefficients of the vertical fields}
\label{sec:ILT_vertical}

From Eq.(\ref{eq:ILT_nu}) and Eqs.(\ref{eq:mfEBy_solution}),
we get the Fourier coefficients of the vertical components of the fields in $s>0$,
\begin{align}
  \cE_y^n(r,s)
  &=\int_{-\infty-i0}^{\infty-i0}\frac{d\nu}{2\pi} e^{i\nu s/\rho}
    \int_{r_a}^{r_b}\frac{dr'}{\rho}
    \{\mfD_y^n(r',\nu)+Z_0r'\mfS_y^n(r',\nu)\}\mfG_{+}^n(r,r',\nu) ,
  \label{eq:ILT_cEy}
  \\
  c\cB_y^n(r,s)
  &=\int_{-\infty-i0}^{\infty-i0}\frac{d\nu}{2\pi} e^{i\nu s/\rho}
    \int_{r_a}^{r_b}\frac{dr'}{\rho}
    \{\mfA_y^n(r',\nu)+Z_0r'\mfT_y^n(r',\nu)\}\mfG_{-}^n(r,r',\nu) .
  \label{eq:ILT_cBy}
\end{align}
Using Eqs.(\ref{eq:cSy_cTy}) which are the Fourier coefficients of the source terms
$(\cS_y^n,\cT_y^n)$, we rewrite the source terms in the Laplace domain $(\mfS_y^n,\mfT_y^n)$
through the Laplace integral given by the first equation of (\ref{eq:Laplace}) and
(\ref{eq:u_nu}),
\begin{align}
  \bigg\{{\mfS_y^n(r',\nu) \atop \mfT_y^n(r',\nu)}\bigg\}
  =\int_0^\infty\frac{ds'}{\rho}
   \bigg\{{\cS_y^n(r',s') \atop \cT_y^n(r',s')}\bigg\}e^{-i\nu s'/\rho} .
  \label{eq:mfJbe_LT_dum}
\end{align}
$s'$ is a dummy variable for $s$.
We use the variables $(r',s')$ as the radial and longitudinal positions of
a source particle contained in the bunch,
or as the arguments of the initial fields at $s=0$.
On the other hand, $(r,s)$ denotes the observation point of the fields.
Substituting Eq.(\ref{eq:mfJbe_LT_dum}) into Eqs.(\ref{eq:ILT_cEy}-\ref{eq:ILT_cBy}),
we rearrange them as
\begin{align}
  \bigg\{{\cE_y^n(r,s) \atop c\cB_y^n(r,s)}\bigg\}
  &=\int_{r_a}^{r_b}dr'
    \bigg[
        \bigg\{{\cP_y^n(r,r',s) \atop \cQ_y^n(r,r',s)}\bigg\}
       +Z_0\int_0^{\infty}ds'
        \bigg\{{\cM_y^n(r,r',s-s',s') \atop \cN_y^n(r,r',s-s',s')}\bigg\}
    \bigg] ,
  \label{eq:cEBy_sol}
\end{align}
where the first and second integrands involve $(\cD_y^n,\cA_y^n)$ and $(\cS_y^n,\cT_y^n)$ 
respectively,
\begin{alignat}{2}
  \cP_y^n(r,r',s)
  &=\cD_y^n(r')\cG_{+}^n(r,r',s) ,
    \qquad&
  \cM_y^n(r,r',\vsig,s')
  &=\frac{r'}{\rho}\cS_y^n(r',s')\cG_{+}^n(r,r',\vsig) ,
   \label{eq:cPy_cMy}
   \\
  \cQ_y^n(r,r',s)
  &=\cA_y^n(r')\cG_{-}^n(r,r',s) ,
    \qquad&
  \cN_y^n(r,r',\vsig,s')
  &=\frac{r'}{\rho}\cT_y^n(r',s')\cG_{-}^n(r,r',\vsig) .
  \label{eq:cQy_cNy}
\end{alignat}
$\cD_y^n$ and $\cA_y^n$ are operators with respect to $s$ (not $s'$),
which have the initial values of the fields at $s=+0$,
\begin{align}
  \cD_y^n(r')
  =\frac{\rho}{r'}
   [\rd_{s'}\cE_y^n(r',s')+\cE_y^n(r',s')\rd_s]_{s'=+0} ,
    \qquad
  \cA_y^n(r')
  =\frac{\rho}{r'}
   [\rd_{s'}c\cB_y^n(r',s')+c\cB_y^n(r',s')\rd_s]_{s'=+0} .
  \label{eq:cDy_cAy}
\end{align}
Eqs.(\ref{eq:cDy_cAy}) have the longitudinal operator $\rd_s$ which acts on
$\cG_{\pm}^n(s)$ in the first equations of (\ref{eq:cPy_cMy}-\ref{eq:cQy_cNy}).
$\cG_{+}^n$ and $\cG_{-}^n$ are the $n$th Fourier coefficients of the Green functions of
the wave equations for $\cE_y^n$ and $\cB_y^n$ respectively,
\begin{align}
  \cG_{\pm}^n(r,r',\vsig)
  =\int_{-\infty-i0}^{\infty-i0}\frac{d\nu}{2\pi\rho}
   \mfG_{\pm}^n(r,r',\nu)e^{i\nu\vsig/\rho} ,
    \qquad
  \vsig=
  \bigg\{
  \begin{array}{l}
    s-s'\in\mathbb{R}  \\
    s\in\mathbb{R}^{+}
  \end{array}
  .
  \label{eq:cGeb_def}
\end{align}
$\vsig$ is the longitudinal variable which represents either $s-s'$ or $s$ as
$\cG_{\pm}^n(\vsig)$ represents $\cG_{\pm}^n(s-s')$ or $\cG_{\pm}^n(s)$ in
Eqs.(\ref{eq:cPy_cMy}-\ref{eq:cQy_cNy}).
As shown in section \ref{sec:pole}, $\mfG_{\pm}^n$ has the poles on
the real and imaginary axes of the $\nu$-plane.
Since $\mfG_{\pm}^n$ has no other singularities in the $\nu$-plane such as
a branch point or an essential singularity, we can calculate the Bromwich integral
(\ref{eq:cGeb_def}) analytically using the residue theorem.
$s$ is greater than 0 since we are using the Laplace transform.
We give the initial values of the fields at $s=+0$,
which are contained in Eqs.(\ref{eq:cDy_cAy}).
On the other hand, $s-s'$ can be positive or negative
depending on the relation of the observer and the source particle.
Therefore we regard $\vsig=\{s-s',s\}$ in Eq.(\ref{eq:cGeb_def}) as
a real parameter in general.
After we got the expression of $\cG_{\pm}^n(\vsig)$, however, we must replace
the step functions $\theta(\pm s)$ as $\theta(s)\to1$ and $\theta(-s)\to0$
in the expression of $\cG_{\pm}^n(s)$,
because $\theta(s)$ is equivalent to 1 in the Laplace transform defined in $s>0$.

The wavenumber $k$ is a real parameter in Eq.(\ref{eq:cGeb_def})
since $\tE_y$ and $\tB_y$ are involved in the Fourier integral
given by Eq.(\ref{eq:Fourier_trans}).
In counting $m_{\pm}$ which corresponds to the numbers of the real poles of
$\mfG_{\pm}^n$ as shown in Eqs.(\ref{eq:ps_poles}),
we need to partition $k$ since $m_{\pm}$ depends on $k$ through $k_r^n$.
It is enough to consider the range $k\geq0$ since $k_r^n$ is even with respect to $k$.
For a given $n$, in order to fix the number of the real poles of $\mfG_{\pm}^n$
using the partition of $k$, we define $\Th_{\pm}^{n\ell}(k)\in\{0,1\}$ as
the following rectangular window functions,
\begin{alignat}{2}
  \Th_{+}^{n\ell}(k)
  &=\theta(|k|-k_{\ell-1}^n)-\theta(|k|-k_{\ell}^n)
    \qquad
  (\ell\in\mathbb{N}) ,
  \label{eq:Th_def}
  \\
  \Th_{-}^{n\ell}(k)
  &=\theta(|k|-\bk_{\ell-1}^n)-\theta(|k|-\bk_{\ell}^n)
    \qquad
  (\ell\in\mathbb{N}) .
  \label{eq:bTh_def}
\end{alignat}
For a given $n$, we define the zeroth window functions $\Th_{\pm}^{n0}$ so that
the sum of $\Th_{\pm}^{n\ell}$ over all $\ell\in\mathbb{Z}_0^{+}$ becomes unity
to cover the entire range of $k\in\mathbb{R}_0^{+}$,
\begin{align}
  \sum_{\ell=0}^{\infty}\Th_{\pm}^{n\ell}(k)
  =1 .
   \label{eq:sum_Th}
\end{align}
$\ell$ is the index which is related to $m_{\pm}$ as in Eq.(\ref{eq:ell_m}) except for
$\ell=0$.
$k_m^n$ and $\bk_m^n$ are the cutoff wavenumbers of the curved pipe for
$\tE_y$ and $\tB_y$, which are defined implicitly in Eqs.(\ref{eq:kmn}-\ref{eq:bkmn}).
$\Th_{\pm}^{n\ell}$ partitions the full range of $k\in\mathbb{R}$ into the following 
infinite number of windows by $k_m^n$ and $\bk_m^n$ for $m\in\mathbb{Z}_0^{+}$,
\begin{alignat}{3}
  \Th_{+}^{n\ell}(k)&=
  \left\{
  \begin{array}{ll}
    1 & (k_{\ell-1}^n\leq |k|<k_{\ell}^n) \\
    0 & (\text{elsewhere})
  \end{array}
  \right.
  \quad (\ell\in\mathbb{N}),
  \qquad
  &&\Th_{+}^{n0}(k)=
  \left\{
  \begin{array}{ll}
    1 & (|k|<k_0^n) \\
    0 & (|k|\geq k_0^n)
  \end{array}
  \right. ,
  \label{eq:part_e}
   \\
  \Th_{-}^{n\ell}(k)&=
  \left\{
  \begin{array}{ll}
    1 & (\bk_{\ell-1}^n\leq |k|<\bk_{\ell}^n) \\
    0 & (\text{elsewhere})
  \end{array}
  \right.
  \quad (\ell\in\mathbb{N}),
  \qquad
  &&\Th_{-}^{n0}(k)=
  \left\{
  \begin{array}{ll}
    1 & (|k|<\bk_0^n) \\
    0 & (|k|\geq\bk_0^n)
  \end{array}
  \right. .
  \label{eq:part_b}
\end{alignat}
Eqs.(\ref{eq:pole_struct_pls}-\ref{eq:pole_struct_mns}) show the partitions of
the range $k\in\mathbb{R}_0^{+}$ by $\Th_{\pm}^{n\ell}$.
When $|k|<k_1^n$ which corresponds to $\ell=0$ and 1, $\mfG_{+}^n$ has no real poles, \ie,
$\mfG_{+}^n$ has only the imaginary poles in the $\nu$-plane.
Similarly, when $|k|<\bk_0^n$ which is for $\ell=0$, $\mfG_{-}^n$ has no real poles.
For $k$ such that $\Th_{+}^{n\ell}=1$ $(\ell\geq2)$,
$\mfG_{+}^n$ has $2(\ell-1)$ poles on the real axis at $\nu=\pm\hnu_m^n$
as shown in Eq.(\ref{eq:ps_poles}),
where $m$ is the radial mode number $m=\{1,2,\cdots,\ell-1\}$,
\begin{align}
  \begin{array}{cccccccccccc}
  k=0&  &k_0^n&   &k_1^n&  &k_2^n&  &k_3^n&  &k_3^n&
  \\
  \big|&
  \quad~~\Th_{+}^{n0}
  &\big|& \hspace{-2mm} \Th_{+}^{n1} &\big|& \Th_{+}^{n2}
  &\big|& \Th_{+}^{n3} &\big|& \Th_{+}^{n4} &\big|&
  \\
   \Big|&&\hspace{-5mm}(\text{no real poles})\hspace{-5mm} &  
  &\Big|&\hspace{-1.5mm} m=1\hspace{-1.5mm}
  &\Big|& \hspace{-2mm}m=1,2\hspace{-2mm}
  &\Big|& \hspace{-2mm}m=1\sim 3\hspace{-2mm}
  &\Big|&
  \\
  \Big|&& (\nu_m^n)^2<0 &
  &\hspace{-2mm}\nu_1^n=0\hspace{-2mm}&
  &\hspace{-2mm}\nu_2^n=0\hspace{-2mm}&
  &\hspace{-2mm}\nu_3^n=0\hspace{-2mm}&
  &\hspace{-2mm}\nu_4^n=0\hspace{-2mm}&
  \end{array}
  \cdots
   \label{eq:pole_struct_pls}
\end{align}
For $k$ such that $\Th_{-}^{n\ell}=1$ $(\ell\geq1)$,
$\mfG_{-}^n$ has $2\ell$ poles on the real axis at $\nu=\pm\hmu_m^n$
for $m=\{0,1,2,\cdots,\ell-1\}$,
\begin{align}
  \begin{array}{cccccccccccc}
  k=0&  &\bk_0^n&   &\bk_1^n&  &\bk_2^n&  &\bk_3^n&  &\bk_4^n&
  \\
  \big|&
  \quad\Th_{-}^{n0}
  &\big|& \Th_{-}^{n1} &\big|& \Th_{-}^{n2}
  &\big|& \Th_{-}^{n3} &\big|& \Th_{-}^{n4} &\big|&
  \\
   \Big|&\hspace{-3mm}(\text{no real poles})\hspace{-3mm}
  &\Big|& \hspace{-3mm}m=0\hspace{-3mm}
  &\Big|& \hspace{-3mm}m=0,1\hspace{-3mm}
  &\Big|& \hspace{-3mm}m=0,1,2\hspace{-3mm}
  &\Big|& \hspace{-3mm}m=0\sim 3\hspace{-3mm}
  &\Big|&
  \\
  \Big|& \hspace{-4mm}(\mu_m^n)^2<0
  &\mu_0^n=0\hspace{-2mm}&
  &\hspace{-2mm}\mu_1^n=0\hspace{-2mm}&
  &\hspace{-2mm}\mu_2^n=0\hspace{-2mm}&
  &\hspace{-2mm}\mu_3^n=0\hspace{-2mm}&
  &\hspace{-2mm}\mu_4^n=0\hspace{-2mm}&
  \end{array}
  \cdots
   \label{eq:pole_struct_mns}
\end{align}

The cutoff wavenumbers $k=k_m^n~(m\in\mathbb{N})$ and
$\bk_m^n~(m\in\mathbb{Z}_0^{+})$ correspond to the poles of the second order
at $\nu_m^n=0$ and $\mu_m^n=0$ which are the origin of the $\nu$-plane.
We defined $k_0^n$ in Eq.(\ref{eq:lim_kmn}) to define $\Th_{+}^{n\ell}$ for
$\ell\in\mathbb{Z}_0^{+}$ similar to $\Th_{-}^{n\ell}$.
Instead of Eqs.(\ref{eq:part_e}), we can define $\Th_{+}^{n\ell}$ ($\ell\in\mathbb{N}$) 
without $k_0^n$ since $k_0^n$ has no physical meaning; it is a matter of taste.
According to Eqs.(\ref{eq:lim_kmn}), $\Th_{\pm}^{n\ell}$ goes to $\bTh_{\ell}^n$ in
the limit of $\rho\to\infty$,
\begin{align}
  \lim_{\rho\to\infty}\Th_{\pm}^{n\ell}(k)
  =\bTh_{\ell}^n(k)
   \qquad
  (\ell\in\mathbb{Z}_0^{+}) .
   \label{eq:lim_Th}
\end{align}
$\bTh_{\ell}^n$ is the rectangular window function for the fields in the straight pipe,
given by Eq.(\ref{eq:def_Th_ell}).

We consider a wavenumber $k$ which satisfies either Eq.(\ref{eq:kapm_k}) or
(\ref{eq:bkapm_k}) in the range $|k\beta|\geq k_y^n$ ($k_r^n\in\mathbb{R}$),
\begin{alignat}{4}
  k_m^n
  &\leq |k|
  <k_{m+1}^n
    \qquad&\Lra&\qquad&
  \kap_r^m
  &\leq k_r^n
  <\kap_r^{m+1} ,
     \qquad&
  (\kap_r^m)^2
  &=(k_m^n\beta)^2-(k_y^n)^2
   >0 ,
    \label{eq:kapm_k}
   \\
  \bk_m^n
  &\leq |k|
  <\bk_{m+1}^n
    \qquad&\Lra&\qquad&
  \bkap_r^m
  &\leq k_r^n
  <\bkap_r^{m+1} ,
     \qquad&
  (\bkap_r^m)^2
  &=(\bk_m^n\beta)^2-(k_y^n)^2
   >0 .
    \label{eq:bkapm_k}
\end{alignat}
$(\kap_r^m,\bkap_r^m)$ are the radial wavenumbers $k_r^n$ at the cutoff wavenumbers
$k=(k_m^n,\bk_m^n)$.
From Eq.(\ref{eq:kaprm2_p}) which is the asymptotic expression of $(\kap_r^m,\bkap_r^m)$
for $w\ll\rho$, we get the cutoff radial mode number $m_c^{\pm}$ which satisfies
Eq.(\ref{eq:kapm_k}) and (\ref{eq:bkapm_k}) for $m=m_c^{+}$ and $m_c^{-}$ respectively,
\begin{align}
  &
  m_c^{\pm}
  =\bigg\lfloor\frac{k_r^n}{k_x^1}T_n^{1/2}\bigg\rfloor
   ,\qquad
  T_n
  =\frac{\hat{c}+(\hat{c}^2-4\hat{a}\hat{b})^{1/2}}{2}
   \qquad
  (k_r^n\in\mathbb{R}) ,
   \\
  &
  \hat{a}
  =\frac{(1,-3)}{4\hr_b\hr_a}
   ,\qquad
  \hat{b}
  =\frac{3}{2\hr_b\hr_a}
   ,\qquad
  \hat{c}
  =1+\hat{a}
   \qquad
  (w\ll\rho) ,
\end{align}
where $\hat{a}$'s coefficient $(1,-3)$ corresponds to $(m_c^{+},m_c^{-})$.
$\lfloor~\rfloor$ is the floor function as used in Eq.(\ref{eq:Re_poles}).
$k_x^1$ $(=\pi/w)$ is the fundamental mode of $k_x^m$ given by Eq.(\ref{eq:asymp_lim}).
Since $T_n\to1$ in the limit of $\rho\to\infty$, $m=\lfloor k_r^n/k_x^1\rfloor$
satisfies $\hk_m^n\leq k<\hk_{m+1}^n$ where $\hk_m^n$ is the cutoff wavenumber of
the straight pipe, given by Eq.(\ref{eq:lim_kmn}).

\subsection{Convergence of the Bromwich integral}
\label{sec:cGpm}

In section \ref{sec:rad_eigen} we will calculate the Bromwich integral (\ref{eq:cGeb_def}) 
analytically using the residue theorem.
In order to fix the number of the real poles of $\mfG_{\pm}^n$ for a given $n$,
we partition $\cG_{\pm}^n$ with respect to $k$ using Eq.(\ref{eq:sum_Th}),
\begin{align}
  \cG_{\pm}^n(r,r',\vsig)
  =\sum_{\ell=0}^{\infty}I_{\pm}^{n\ell}(r,r',\vsig) ,
    \qquad
  I_{\pm}^{n\ell}(r,r',\vsig)
  =\Th_{\pm}^{n\ell}
   \int_{-\infty-i0}^{\infty-i0}\frac{d\nu}{2\pi\rho}
   \mfG_{\pm}^n(r,r',\nu)e^{i\nu\vsig/\rho} .
  \label{eq:cGe_Th}
\end{align}
$\mfG_{\pm}^n$ is given by Eqs.(\ref{eq:mfGe}-\ref{eq:mfGb}).
$I_{\pm}^{n\ell}$ is the Bromwich integral for a given $k$ whose range is limited by
$\Th_{\pm}^{n\ell}$ as shown in Eqs.(\ref{eq:Th_def}-\ref{eq:bTh_def}).
By this constraint, the number of the real poles of $\mfG_{\pm}^n$ 
does not depend on $k$ in the integrand of $I_{\pm}^{n\ell}$.
$\vsig$ represents either $s$ or $s-s'$ as shown in Eq.(\ref{eq:cGeb_def}).
Unlike $s\in\mathbb{R}^{+}$, $s-s'$ can be either positive or negative,
depending on the longitudinal positions of the observation point $(s)$ and
the source particle $(s')$ of the field.
$r$ and $r'$ denote their radial positions similar to $s$ and $s'$.
In the integrand of $I_{\pm}^{n\ell}$ we regard the radial and longitudinal variables,
$(r,r')$ and $\vsig$, as real parameters independent of $\nu$.

In order to investigate the convergence of the integral $I_{\pm}^{n\ell}$,
we examine the asymptotic behavior of the integrand $\mfG_{\pm}^n(\nu)e^{i\nu\vsig/\rho}$ 
for $|\nu|\to\infty$.
It is enough to examine the behavior of $\mfG_{\pm}^n$ only in the first quadrant of
the $\nu$-plane, $\Re\nu\geq0$ and $\Im\nu\geq0$,
since the cross products of the Bessel functions
$t_{\nu}=(p_{\nu},q_{\nu},r_{\nu},s_{\nu})$ have symmetries with respect to $\nu$ as
shown in Eqs.(\ref{eq:even_symm_ps}).
We can get the asymptotic expression of $\mfG_{\pm}^n$ for $|\nu|\to\infty$ from
Eqs.(\ref{eq:pnu_asymp}) and (\ref{eq:pimu_asymp_mu_inf}-\ref{eq:simu_asymp_mu_inf})
which are the asymptotic expressions of $t_{\nu}$ respectively for
$\Re\nu\to\infty$ and $\nu=i\bnu\to i\infty$ with the other variables fixed,
\begin{alignat}{2}
  \mfG_{\pm}^n(r,r',\nu)
  &\simeq
   -\{\theta(r-r')U_{\pm}(r,r',\nu)+\theta(r'-r)U_{\pm}(r',r,\nu)\}
   \qquad&&(\Re\nu\to\infty) ,
   \label{eq:mfGp_asym}
  \\
  \mfG_{\pm}^n(r,r',i\bnu)
  &\simeq
   \mp\{\theta(r-r')\bU_{\pm}(r,r',\bnu)+\theta(r'-r)\bU_{\pm}(r',r,\bnu)\}
   \qquad&&(\bnu\to\infty) .
  \label{eq:mfGe_imu_inf}
\end{alignat}
$\bnu$ is the variable defined in Eq.(\ref{eq:bnu_bk}) for convenience while 
considering the behavior of $\mfG_{\pm}^n$ on the imaginary axis of the $\nu$-plane.
$U_{\pm}$ and $\bU_{\pm}$ are given as
\begin{alignat}{2}
  U_{+}(r,r',\nu)
  &=\frac{\sinh[\nu\log(r_b/r)]\sinh[\nu\log(r'/r_a)]}{(\nu/\rho)\sinh[\nu\log(r_b/r_a)]}
   ,\qquad
  &\bU_{+}(r,r',\bnu)
  &=\frac{\sin[\bnu\log(r_b/r)]\sin[\bnu\log(r'/r_a)]}{(\bnu/\rho)\sin[\bnu\log(r_b/r_a)]} ,
   \label{eq:Up_bUp}
   \\
  U_{-}(r,r',\nu)
  &=\frac{\cosh[\nu\log(r_b/r)]\cosh[\nu\log(r'/r_a)]}{(\nu/\rho)\sinh[\nu\log(r_b/r_a)]}
   ,\qquad
  &\bU_{-}(r,r',\bnu)
  &=\frac{\cos[\bnu\log(r_b/r)]\cos[\bnu\log(r'/r_a)]}{(\bnu/\rho)\sin[\bnu\log(r_b/r_a)]} .
   \label{eq:Um_bUm}
\end{alignat}
They have the following relations,
\begin{align}
  U_{\pm}(r,r',i\bnu)
  &=\pm\bU_{\pm}(r,r',\bnu) .
\end{align}
Eqs.(\ref{eq:mfGp_asym}-\ref{eq:mfGe_imu_inf}) do not depend on $k_r^n$,
because they are equivalent to Eq.(\ref{lim_kr0_mfGpm})
which is the asymptotic limit of $\mfG_{\pm}^n$ for $k_r^n\to0$.
Therefore $\mfG_{\pm}^n$ has imaginary poles in the limit of $k_r^n\to0$ as
shown in Eq.(\ref{eq:nu_krn0}).
Neglecting the small terms in Eq.(\ref{eq:mfGp_asym}),
$\mfG_{\pm}^n$ tends to behave for $\Re\nu\to\infty$ eventually as follows,
\begin{align}
  \mfG_{\pm}^n(r,r',\nu)
  \sim -\frac{\rho}{2\nu}
  \{\theta(r-r')(r'/r)^{\nu}+\theta(r'-r)(r/r')^{\nu}\}
   \qquad
  (\Re\nu\to\infty) .
  \label{eq:mfGeb_asymp_nuinf}
\end{align}
Accordingly, when $r\ne r'$, $\mfG_{\pm}^n$ goes to 0 exponentially for $\Re\nu\to\infty$.
Conversely, when $r=r'$, $\mfG_{\pm}^n$ goes to 0 algebraically as
$\mfG_{\pm}^n\sim-\rho/2\nu$ for $\Re\nu\to\infty$.
This means that the integral $I_{\pm}^{n\ell}$ diverges only when $r=r'$ and $\vsig=0$.
This divergence is correct, because $I_{\pm}^{n\ell}$ involves the Green functions
$\mfG_{\pm}^n$ which satisfy Eq.(\ref{eq:BDE_mfGpm}), \ie, $\mfG_{\pm}^n$ is
the solution of the wave equation whose driving term is the radial $\delta$-function.
In other words, since $I_{\pm}^{n\ell}$ includes $\delta(r-r')$ as shown later in
Eqs.(\ref{eq:rdel}),
it is right that $I_{\pm}^{n\ell}=\infty$ only at $r=r'$ and $\vsig=0$ as
the $\delta$-function.
$I_{\pm}^{n\ell}$ is involved in the integral with respect to $r'$ as in
Eq.(\ref{eq:cEBy_sol}) through $\cG_{\pm}^n$ given by Eq.(\ref{eq:cGe_Th}).

We got Eq.(\ref{eq:mfGe_imu_inf}) using Eqs.(\ref{eq:Fimu_x}-\ref{eq:Gimu_x}) which are
the asymptotic expansions of the Bessel functions for large purely imaginary order
($\nu\to i\infty$) with the argument fixed.
Besides, we can also get Eq.(\ref{eq:mfGe_imu_inf}) by simply substituting $\nu=i\bnu$
into Eq.(\ref{eq:mfGp_asym}), though this is a formal derivation without proper proof.
Eq.(\ref{eq:mfGe_imu_inf}) shows the fact that $\mfG_{\pm}^n$ algebraically tend to be zero
while oscillating for $\bnu\to\infty$.
$I_{\pm}^{n\ell}$ converges if $e^{i\nu\vsig/\rho}$ decays to zero exponentially
for $\nu\to\pm i\infty$ in accordance with the sign of $\vsig$ as shown in
Fig.\ref{fig:contour} (p.\pageref{fig:contour}).
Therefore $I_{\pm}^{n\ell}$ converges unless $r=r'$ and $\vsig=0$ (\ie, $s=s'$ since $s>0$)
which imply $\delta(r-r')$ and $\delta(s-s')$ respectively.

We describe the order $\nu$ of the Bessel functions before and after
the Bromwich integral (\ref{eq:cGe_Th}).
The Laplace variable $\nu$ before the integration differs from the poles
$(\nu_m^n,\mu_m^n)$ which are the orders of the Bessel functions after the integration.
We must distinguish $\nu\in\mathbb{C}$ in Eq.(\ref{eq:cGe_Th}) from
$(\nu_m^n,\mu_m^n)\in\mathbb{A}$ which are the solutions of
Eqs.(\ref{eq:pnu_poles}-\ref{eq:snu_poles}) with respect to $\nu$.
The poles $\nu_m^n$ and $\mu_m^n$ are functions of $\hr_{b,a}\in\mathbb{A}$
unlike $\nu$ in Eq.(\ref{eq:cGe_Th}) where
$\nu\in\mathbb{C}$ is independent of the other variables and parameters.
The relation between $\nu\in\mathbb{C}$ and $(\nu_m^n,\mu_m^n)\in\mathbb{A}$ is
similar to that between $k_s\in\mathbb{C}$ and $k_s^{mn}\in\mathbb{A}$
which are described below Eqs.(\ref{eq:asymp_lim}).

\subsection{Radial eigenfunctions of the curved pipe}
\label{sec:rad_eigen}

\begin{figure}[h]
  \begin{center}
    \includegraphics[scale=0.4,clip]{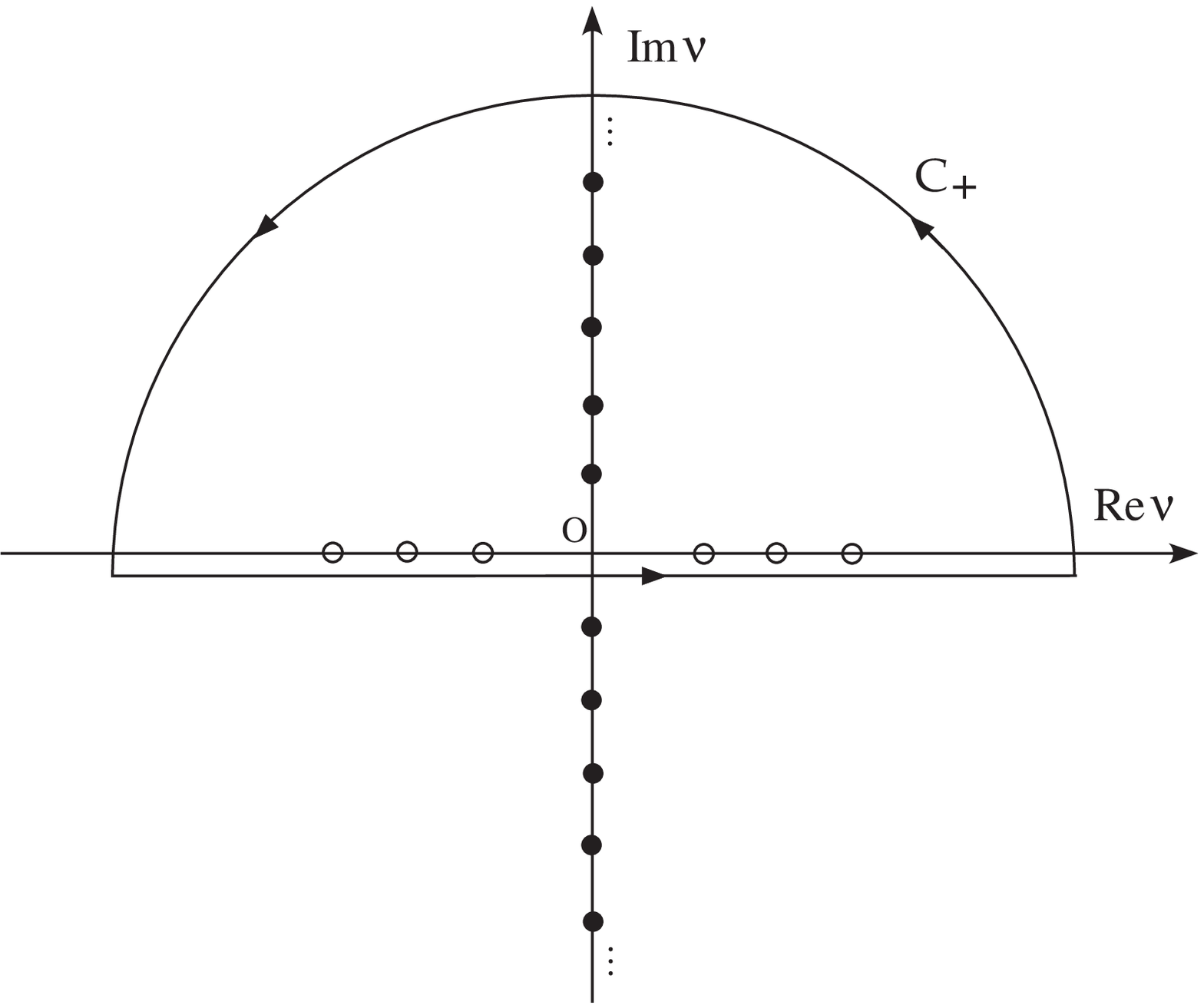} \qquad
    \includegraphics[scale=0.4,clip]{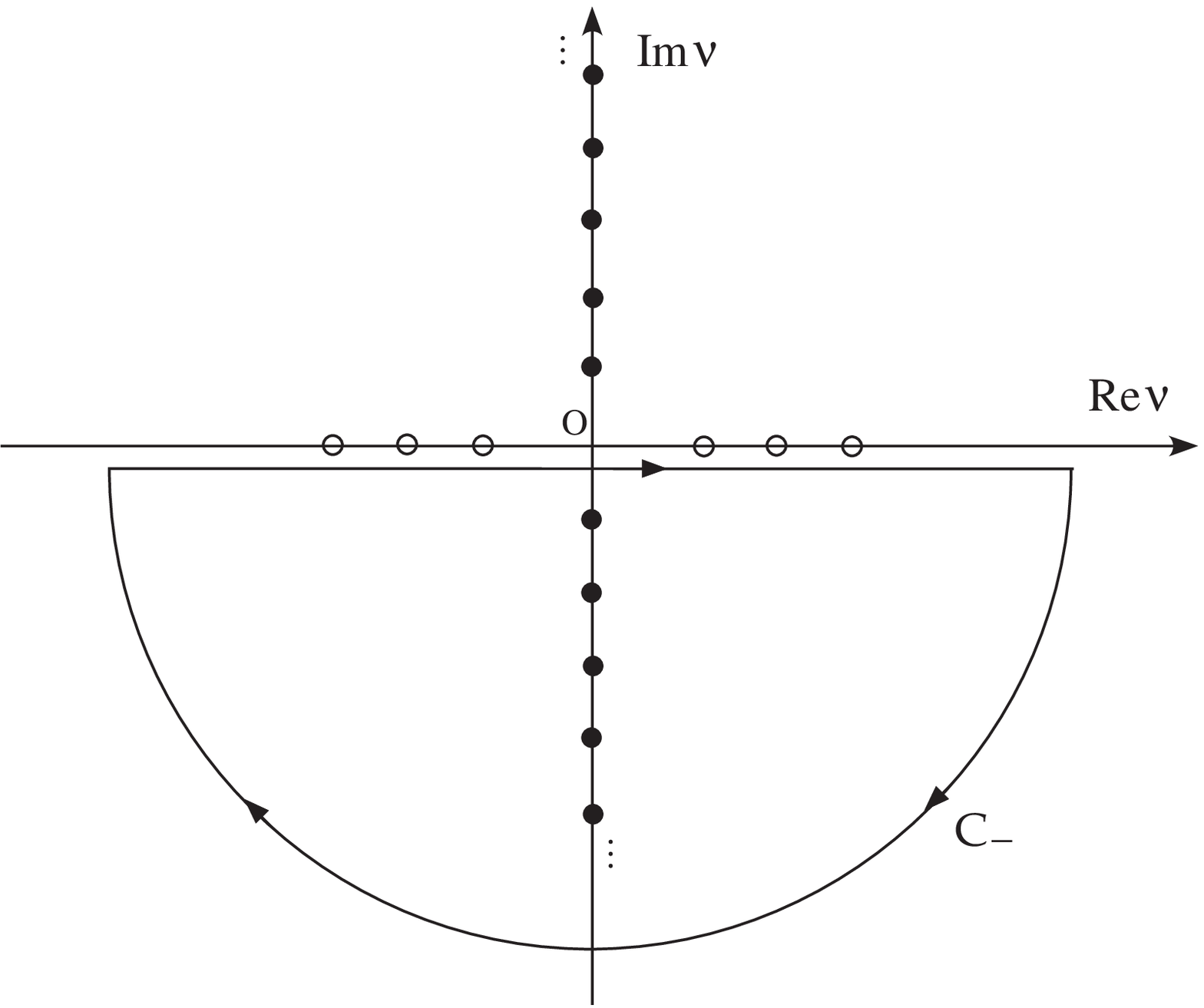}
    \caption[Contours of the Bromwich integral in the Laplace domain]{
     \small
     Contours of the Bromwich integral $I_{\pm}^{n\ell}$ in the Laplace domain.
     $C_{+}$ (left) and $C_{-}$ (right) are the contours respectively for
     $\vsig\geq0$ and $\vsig<0$.
     The white and black circles on the axes denote respectively 
     the real poles $(\hnu_m^n,\hmu_m^n)$ and the imaginary poles $(\cnu_m^n,\cmu_m^n)$ of
     the fields in the Laplace domain, which are shown in Eqs.(\ref{eq:ps_poles}).
     $C_{\pm}$ goes through the real axis infinitesimally under the real poles
     since we rotated the original Laplace variable $\kap$ clockwise by $\pi/2$
     as in Eq.(\ref{eq:u_nu}) for convenience in describing the fields
     in terms of the Bessel functions.
     }
    \label{fig:contour}
  \end{center}
\end{figure}

We consider the contour of the integral $I_{\pm}^{n\ell}$ given by Eq.(\ref{eq:cGe_Th}).
According to the limits of the integration,
the contour along the real $\nu$-axis goes through infinitesimally under it.
Since $e^{i\nu\vsig/\rho}$ must go to zero in the limit of $|\Im\nu|\to\infty$,
the sign of $\Im\nu$ on the contour at infinity must be the same as that of $\vsig$.
Therefore we must separate $\cG_{\pm}^n(\vsig)$ into those for $\vsig\geq0$ and $\vsig<0$
in accordance with $\Im\nu\geq0$ and $\Im\nu<0$.
It follows that $I_{\pm}^{n\ell}$ converges
if we choose the contours as shown in Fig.\ref{fig:contour}.

$I_{\pm}^{n\ell}$ has the window $\Th_{\pm}^{n\ell}$ in which the number of
the real poles does not depend on $k$ as shown in
Eqs.(\ref{eq:pole_struct_pls}-\ref{eq:pole_struct_mns}).
As shown by the left figure in Fig.\ref{fig:contour}, when $\vsig\geq 0$,
the positive imaginary poles ($\Im\nu>0$) and all the real poles including the origin are 
inside the closed contour $C_{+}$ on the upper half-plane.
On the contrary, when $\vsig<0$,
only the negative imaginary poles ($\Im\nu<0$) are inside the closed contour $C_{-}$
on the lower half-plane as the right figure in Fig.\ref{fig:contour} shows.
When $k$ is exactly at the cutoff wavenumber $k_m^n$ or $\bk_m^n$
which is defined in Eqs.(\ref{eq:kmn}-\ref{eq:bkmn}),
$\mfG_{\pm}^n$ has a pole of the second order at $\nu=0$
as described in section \ref{sec:analytic_continuation}.
As shown in Eqs.(\ref{eq:R2p_lim}-\ref{eq:R2m_lim}), however,
the residue at $\nu=0$ can be represented 
by the sum of the residues at the corresponding pair of the simple poles on the real axis
$\nu=\pm\hnu_m^n$ in the limit of $\hnu_m^n\to0$ ($k\to k_m^n$).
It means that the integral with respect to $\nu$ commutes with the limit of
$k\to(k_m^n,\bk_m^n)$.
Therefore, as in Fig.\ref{fig:contour},
we can suppose that $\mfG_{\pm}^n$ has only the pairs of the simple poles in 
applying the residue theorem to $I_{\pm}^{n\ell}$
without caring the pole of the second order which happens at $k=(k_m^n,\bk_m^n)$.
According to the above considerations, we get
\begin{align}
  I_{+}^{n\ell}(\vsig)
  &=\frac{\Th_{+}^{n\ell}}{2i}
    \bigg[\bdelta_{01}^{\ell}\theta(\vsig)\sum_{m=1}^{\ell-1}
      \{R_{+}(\hnu_m^n)+R_{+}(-\hnu_m^n)\}
      +\sum_{m=\ell'}^{\infty}
       \{\theta(\vsig)R_{+}(i\cnu_m^n)-\theta(-\vsig)R_{+}(-i\cnu_m^n)\}
    \bigg] ,
  \label{eq:Imn_done}
   \\
  I_{-}^{n\ell}(\vsig)
  &=\frac{\Th_{-}^{n\ell}}{2i}
    \bigg[
      \bdelta_{0}^{\ell}\theta(\vsig)
      \sum_{m=0}^{\ell-1}\{R_{-}(\hmu_m^n)+R_{-}(-\hmu_m^n)\}
     +\sum_{m=\ell}^{\infty}
      \{\theta(\vsig)R_{-}(i\cmu_m^n)-\theta(-\vsig)R_{-}(-i\cmu_m^n)\}
    \bigg] .
  \label{eq:bImn_done}
\end{align}
The lower limit of the second sum $\ell'$ in Eq.(\ref{eq:Imn_done}) is given using
the Kronecker delta $\delta_{j}^\ell$ as follows,
\begin{align}
  \ell'
  =\ell+\delta_0^{\ell}
  =1,1,2,3,\cdots
   \quad\text{for}~~
  \ell
  =0,1,2,3,\cdots  ,
    \qquad
  \delta_{j}^{\ell}
  =\left\{
    \begin{array}{ll}
       1 & (\ell=j) \\
       0 & (\ell\ne j)
    \end{array}
    \right. .
   \label{eq:lprm}
\end{align}
$\bdelta_{0}^{\ell}$ is the reversal of $\delta_{0}^{\ell}$ as given below.
$\bdelta_{01}^{\ell}$ is the reversal of $\delta_{j}^{\ell}$ for $j=0$ and 1, \ie,
\begin{align}
  \bdelta_{0}^{\ell}
  =1-\delta_{0}^\ell
  =\left\{
    \begin{array}{ll}
       0 & (\ell=0) \\
       1 & (\ell\ne 0)
    \end{array}
    \right. ,
    \qquad
  \bdelta_{01}^{\ell}
  =1-(\delta_0^\ell+\delta_1^\ell)
  =\left\{
    \begin{array}{ll}
       0 & (\ell=0~\text{or}~1) \\
       1 & (\ell\ne 0~\text{or}~1)
    \end{array}
    \right. .
  \label{eq:d01_d0}
\end{align}
$R_{+}(\nu)$ and $R_{-}(\nu)$ in Eqs.(\ref{eq:Imn_done}-\ref{eq:bImn_done}) are
the residues of the integrands in the second equation of (\ref{eq:cGe_Th}) at the poles
$\nu=\nu_m^n$ and $\mu_m^n$ given by Eqs.(\ref{eq:pnu_poles}-\ref{eq:snu_poles}),
\begin{align}
  R_{+}(\nu_m^n)=\frac{\rho}{\nu_m^n}\cR_{+}^{mn}(r,r')e^{i\nu_m^n\vsig/\rho}
   ,\qquad
  R_{-}(\mu_m^n)=\frac{\rho}{\mu_m^n}\cR_{-}^{mn}(r,r')e^{i\mu_m^n\vsig/\rho} .
  \label{eq:residues}
\end{align}
$\cR_{+}^{mn}$ and $\cR_{-}^{mn}$ are the radial eigenfunctions of $\tE_y$ and $\tB_y$ in
the uniformly curved rectangular pipe,
\begin{align}
  \cR_{+}^{mn}(r,r')
  &=-\pi\frac{\nu_m^n}{\rho}
  \bigg[
     \theta(r-r')
     \frac{p_\nu(\hr_b,\hr)p_\nu(\hr',\hr_a)}{\rd_\nu p_\nu(\hr_b,\hr_a)}
    +\theta(r'-r)
     \frac{p_\nu(\hr_b,\hr')p_\nu(\hr,\hr_a)}{\rd_\nu p_\nu(\hr_b,\hr_a)}
  \bigg]_{\nu=\nu_m^n}
  \label{eq:Ven}
   \\
  &=-\pi
    \bigg[\frac{\nu}{\rho}\cd
          \frac{p_\nu(\hr_b,\hr)p_\nu(\hr',\hr_a)}{\rd_\nu p_\nu(\hr_b,\hr_a)}
    \bigg]_{\nu=\nu_m^n}
   =-\pi
    \bigg[\frac{\nu}{\rho}\cd
          \frac{p_\nu(\hr_b,\hr')p_\nu(\hr,\hr_a)}{\rd_\nu p_\nu(\hr_b,\hr_a)}
    \bigg]_{\nu=\nu_m^n} ,
  \label{eq:mfRp}
   \\
  \cR_{-}^{mn}(r,r')
  &=-\pi\frac{\mu_m^n}{\rho}
  \bigg[
     \theta(r-r')
     \frac{r_\nu(\hr_b,\hr)q_\nu(\hr',\hr_a)}{\rd_\nu s_\nu(\hr_b,\hr_a)}
    +\theta(r'-r)
     \frac{r_\nu(\hr_b,\hr')q_\nu(\hr,\hr_a)}{\rd_\nu s_\nu(\hr_b,\hr_a)}
  \bigg]_{\nu=\mu_m^n}
  \label{eq:Vbn}
  \\
  &=-\pi
    \bigg[\frac{\nu}{\rho}\cd
          \frac{r_\nu(\hr_b,\hr)q_\nu(\hr',\hr_a)}{\rd_\nu s_\nu(\hr_b,\hr_a)}
    \bigg]_{\nu=\mu_m^n}
   =-\pi
    \bigg[\frac{\nu}{\rho}\cd
          \frac{r_\nu(\hr_b,\hr')q_\nu(\hr,\hr_a)}{\rd_\nu s_\nu(\hr_b,\hr_a)}
    \bigg]_{\nu=\mu_m^n} .
  \label{eq:mfRm}
\end{align}
Using Eqs.(\ref{eq:lim_kr0_rd_pnu}-\ref{eq:lim_kr0_rd_snu}),
we get $\cR_{\pm}^{mn}$ in the limit of $k_r^n\to 0$,
\begin{alignat}{2}
  \lim_{k_r^n\to0}\cR_{+}^{mn}(r,r')
  &=\bigg[
      \frac{2(\nu/\rho)\sinh[\nu\log(r_b/r)]\sinh[\nu\log(r'/r_a)]}
           {\nu\log(r_b/r_a)\cosh[\nu\log(r_b/r_a)]-\sinh[\nu\log(r_b/r_a)]}
    \bigg]_{\nu=\nu_m^n(0)} ,
  \label{lim_kr0_cRp}
   \\
  \lim_{k_r^n\to0}\cR_{-}^{mn}(r,r')
  &=\bigg[
      \frac{2(\nu/\rho)\cosh[\nu\log(r_b/r)]\cosh[\nu\log(r'/r_a)]}
           {\nu\log(r_b/r_a)\cosh[\nu\log(r_b/r_a)]+\sinh[\nu\log(r_b/r_a)]}
    \bigg]_{\nu=\mu_m^n(0)} ,
  \label{lim_kr0_cRm}
\end{alignat}
where $\nu_m^n(0)$ and $\mu_m^n(0)$ denote the poles $\nu_m^n(k_r^n)$ and $\mu_m^n(k_r^n)$ 
in the limit of $k_r^n\to0$ ($|k|\to\bk_0^n$).

According to Eqs.(\ref{eq:R2}),
when $k=k_m^n$ or $\bk_m^n$, the radial eigenfunctions are given as
\begin{align}
  \cR_{+}^{mn}(r,r')
  &=-\frac{\pi}{\rho}
  \bigg[
     \theta(r-r')\frac{p_\nu(\hr_b,\hr)p_\nu(\hr',\hr_a)}{\rd_\nu^2 p_\nu(\hr_b,\hr_a)}
    +\theta(r'-r)\frac{p_\nu(\hr_b,\hr')p_\nu(\hr,\hr_a)}{\rd_\nu^2 p_\nu(\hr_b,\hr_a)}
  \bigg]_{\nu=0}
  \label{eq:cRp_nu0}
  \\
  &=-\frac{\pi}{\rho}
    \bigg[\frac{p_\nu(\hr_b,\hr)p_\nu(\hr',\hr_a)}{\rd_\nu^2 p_\nu(\hr_b,\hr_a)}
    \bigg]_{\nu=0}
   =-\frac{\pi}{\rho}
    \bigg[\frac{p_\nu(\hr_b,\hr')p_\nu(\hr,\hr_a)}{\rd_\nu^2 p_\nu(\hr_b,\hr_a)}
    \bigg]_{\nu=0}
   \quad
  (k=k_m^n) ,
  \label{eq:cRp_nu0_0}
   \\ 
  \cR_{-}^{mn}(r,r')
  &=-\frac{\pi}{\rho}
  \bigg[
     \theta(r-r')\frac{r_\nu(\hr_b,\hr)q_\nu(\hr',\hr_a)}{\rd_\nu^2s_\nu(\hr_b,\hr_a)}
    +\theta(r'-r)\frac{r_\nu(\hr_b,\hr')q_\nu(\hr,\hr_a)}{\rd_\nu^2s_\nu(\hr_b,\hr_a)}
  \bigg]_{\nu=0}
  \label{eq:cRm_nu0}
  \\
  &=-\frac{\pi}{\rho}
   \bigg[\frac{r_\nu(\hr_b,\hr)q_\nu(\hr',\hr_a)}{\rd_{\nu}^2s_\nu(\hr_b,\hr_a)}
   \bigg]_{\nu=0}
   =-\frac{\pi}{\rho}
   \bigg[\frac{r_\nu(\hr_b,\hr')q_\nu(\hr,\hr_a)}{\rd_{\nu}^2s_\nu(\hr_b,\hr_a)}
   \bigg]_{\nu=0}
   \quad
  (k=\bk_m^n) .
  \label{eq:cRm_nu0_0}
\end{align}
Eqs.(\ref{eq:cRp_nu0}) and (\ref{eq:cRm_nu0}) do not hold simultaneously in a curved pipe
since $k_m^n\ne\bk_m^n$ ($m\in\mathbb{N}$) unless $\rho=\infty$.
Substituting Eqs.(\ref{eq:lim_kr0_aq0_br0}) and (\ref{eq:lim_kr0_rd2_snu}) into
the first expression of (\ref{eq:cRm_nu0_0}) for $k_r^n=0$,
we get $\cR_{-}^{0n}$ for $k_r^n\to 0$,
\begin{align}
  \lim_{k_r^n\to0}\cR_{-}^{0n}
  &=-\frac{\pi}{\rho}\lim_{k_r^n\to0}
   \bigg[\frac{\hr_br_\nu(\hr_b,\hr)\hr_aq_\nu(\hr',\hr_a)}
              {\rd_{\nu}^2\{\hr_b\hr_as_\nu(\hr_b,\hr_a)\}}
   \bigg]_{\nu=0}
   =\frac{1}{wg_l} ,
     \qquad
  g_l
  =\frac{\rho}{w}\log\Big(\frac{r_b}{r_a}\Big) .
  \label{eq:cRm_0n}
\end{align}
$g_l$ is the geometric factor given by Eq.(\ref{eq:gl_g12}),
which goes to 1 in the limit of $\rho\to\infty$.
We can also get Eq.(\ref{eq:cRm_0n}) from Eq.(\ref{lim_kr0_cRm})
since Eq.(\ref{eq:cRm_nu0_0}) is the asymptotic limit of Eq.(\ref{lim_kr0_cRm})
for $k\to\bk_0^n~(k_r^n\to0)$ for which the pair of the zeroth poles
$\nu=\pm\mu_0^n$ couples at $\nu=0$ and becomes a pole of the second order.
That is, Eq.(\ref{lim_kr0_cRm}) tends to be Eq.(\ref{eq:cRm_0n}) in the limit of $\nu\to0$, 
\begin{align}
  \lim_{k_r^n\to0}\cR_{-}^{0n}
  =\lim_{\nu\to0}\big\{\text{Eq.(\ref{lim_kr0_cRm})}\big\}
  &=\frac{1}{wg_l} ,
    \qquad
  \lim_{\rho\to\infty}\lim_{k_r^n\to0}\cR_{-}^{0n}
  =\frac{1}{w}
  =\cX_{-}^0 .
  \label{lim_kr0_cRm_m0}
\end{align}
$\cX_{-}^0$ is given by Eq.(\ref{eq:cXm}) which is the zeroth horizontal eigenfunction of 
negative parity in the straight rectangular pipe.
In general, $\cR_{-}^{0n}$ for $\forall k_r^n$ goes to $\cX_{-}^0$ in
the limit of $\rho\to\infty$.
See appendices \ref{sec:cRcX} and \ref{sec:R0n_AE} for the details on Eq.(\ref{eq:lim_RX})
which is the asymptotic limit of $\cR_{\pm}^{mn}$ for $\forall k_r^n$
in the limit of $\rho\to\infty$.

\subsection{Properties of the radial eigenfunctions}

We describe the properties of the radial eigenfunctions $\cR_{\pm}^{mn}$.
According to Eqs.(\ref{eq:nu_mn_rels}-\ref{eq:mu_mn_rels}), $\cR_{\pm}^{mn}$
is symmetric with respect to the exchange of $r$ and $r'$, \ie,
\begin{align}
  \cR_{\pm}^{mn}(r,r')=\cR_{\pm}^{mn}(r',r) .
   \label{eq:cRpm_symm}
\end{align}
$\cR_{+}^{mn}$ and $\cR_{-}^{mn}$ satisfy the homogeneous Bessel differential equation
(\ref{eq:BDE}) respectively for $\nu=\nu_m^n$ and $\mu_m^n$,
\begin{align}
  O_{\nu}(\hr)\cR_{+}^{mn}(r,r')
  &=O_{\nu}(\hr')\cR_{+}^{mn}(r,r')
   =0
   \qquad
  (\nu=\nu_m^n) ,
  \label{eq:BDE_cRp}
   \\
  O_{\nu}(\hr)\cR_{-}^{mn}(r,r')
  &=O_{\nu}(\hr')\cR_{-}^{mn}(r,r')
  =0
   \qquad
  (\nu=\mu_m^n) .
  \label{eq:BDE_cRm}
\end{align}
The sum of $\cR_{\pm}^{mn}$ over all the radial modes forms the radial $\delta$-function
as shown in Eqs.(\ref{eq:delta_Ve}) and (\ref{eq:delta_Vb}),
\begin{align}
  \sum_{m=1}^{\infty}\cR_{+}^{mn}(r,r')
  =\sum_{m=0}^{\infty}\cR_{-}^{mn}(r,r')
  =\frac{r}{\rho}\delta(r-r') ,
    \qquad
  \int_{r_a}^{r_b}dr\frac{\rho}{r}\cR_{\pm}^{mn}(r,r)
  =1 .
   \label{eq:rdel}
\end{align}
According to Eqs.(\ref{eq:ortho_pnu_pnu}) and (\ref{eq:ortho_rmu_qmu}),
$(\rho/r)\cR_{\pm}^{mn}(r,r)$ is normalized to 1 between $r_a$ and $r_b$ as shown in
the second equation of (\ref{eq:rdel}).
Eqs.(\ref{eq:rdel}) go to Eqs.(\ref{eq:delta_Tbe}) in the limit of $\rho\to\infty$.
$\cR_{\pm}^{mn}$ and their radial derivatives satisfy the following boundary conditions
on the sidewalls of the curved pipe,
\begin{align}
  &\cR_{+}^{mn}(r_a,r')=\cR_{+}^{mn}(r_b,r')
  =\cR_{+}^{mn}(r,r_a)=\cR_{+}^{mn}(r,r_b)
  =0 ,
  \label{eq:BC_mfRp}
   \\
  &[\rd_{\hr}\cR_{-}^{mn}(r,r')]_{r=r_a,r_b}
  =[\rd_{\hr'}\cR_{-}^{mn}(r,r')]_{r'=r_a,r_b}
  =0 ,
  \label{eq:BC_mfRm}
   \\
  &[\rd_{\hr}\cR_{+}^{mn}(r,r')]_{r'=r_a,r_b}
  =[\rd_{\hr'}\cR_{+}^{mn}(r,r')]_{r=r_a,r_b}
  =0 ,
   \\
  &[\rd_{\hr}\rd_{\hr'}\cR_{-}^{mn}(r,r')]_{r=r_a,r_b}
  =[\rd_{\hr}\rd_{\hr'}\cR_{-}^{mn}(r,r')]_{r'=r_a,r_b}
  =0 .
\end{align}
Eqs.(\ref{eq:BC_mfRp}-\ref{eq:BC_mfRm}) go to Eqs.(\ref{eq:cX_BC})
in the limit of $\rho\to\infty$.

$\cR_{\pm}^{mn}$ has $\rd_{\nu}p_{\nu}(\hr_b,\hr_a)$ and $\rd_{\nu}s_{\nu}(\hr_b,\hr_a)$ 
for $\nu=\nu_m^n$ and $\mu_m^n$.
We can rewrite them using the implicit function theorem.
That is, differentiating $p_{\nu_m^n}(\hr_b,\hr_a)=0$ and $s_{\mu_m^n}(\hr_b,\hr_a)=0$
with respect to $k_r^n$, we get
\begin{align}
  &
  [\rd_{\nu}p_{\nu}(\hr_b,\hr_a)]_{\nu=\nu_m^n}
  =\bigg[\frac{\alp_{+}^{-1}-\alp_{+}}{d_{+}}\bigg]_{\nu=\nu_m^n}
   ,\qquad
  [\rd_{\nu}s_{\nu}(\hr_b,\hr_a)]_{\nu=\mu_m^n}
  =\bigg[\frac{e_b\alp_{-}^{-1}-e_a\alp_{-}}{d_{-}}\bigg]_{\nu=\mu_m^n} ,
  \label{eq:rdnu_psnu}
   \\
  &
  (d_{+},d_{-})
  =\frac{\pi}{2}k_r^n\frac{d(\nu_m^n,\mu_m^n)}{dk_r^n}
   ,\qquad
  \alpha_{+}
  =\frac{J_{\nu}(\hr_b)}{J_{\nu}(\hr_a)}
   ,\qquad
  \alpha_{-}
  =\frac{J_{\nu}'(\hr_b)}{J_{\nu}'(\hr_a)}
   ,\qquad
  e_{b,a}
  =1-\frac{\nu^2}{\hr_{b,a}^2} .
  \label{eq:alpha}
\end{align}
We can rewrite $\alp_{\pm}$ for $\nu=(\nu_m^n,\mu_m^n)$
as shown in Eqs.(\ref{eq:qr_p0}-\ref{eq:qr_s0}).
%
According to Eqs.(\ref{eq:int_pnu_pnu}) and (\ref{eq:int_qnu_rnu}),
$\rd_\nu p_\nu(\hr_b,\hr_a)$ and $\rd_\nu s_\nu(\hr_b,\hr_a)$ at the poles
$\nu=(\nu_m^n,\mu_m^n)$ have the following integral representations,
\begin{alignat}{2}
  [\rd_\nu p_\nu(\hr_b,\hr_a)]_{\nu=\nu_m^n}
  &=-\pi\nu_m^n\int_{r_a}^{r_b}\frac{dr}{r}
    [p_\nu(\hr_b,\hr)p_\nu(\hr,\hr_a)]_{\nu=\nu_m^n}
   \qquad&&(m\in\mathbb{N}) ,
  \label{eq:int_pp}
   \\
  [\rd_{\nu}s_\nu(\hr_b,\hr_a)]_{\nu=\mu_m^n}
  &=-\pi\mu_m^n\int_{r_a}^{r_b}\frac{dr}{r}
     [r_\nu(\hr_b,\hr)q_\nu(\hr,\hr_a)]_{\nu=\mu_m^n}
   \qquad&&(m\in\mathbb{Z}_0^{+}) .
  \label{eq:int_qr}
\end{alignat}
We can use Eqs.(\ref{eq:int_pp}-\ref{eq:int_qr}) to confirm
Eqs.(\ref{lim_kr0_cRp}-\ref{lim_kr0_cRm})
since the integral with respect to $r$ commutes with the limit of $k_r^n\to0$.
From Eqs.(\ref{eq:int_pnu_pnu_nu0}) and (\ref{eq:int_rnu_qnu_nu0}), we can get
the integral representations of the second derivatives of $p_{\nu}(\hr_b,\hr_a)$ and
$s_{\nu}(\hr_b,\hr_a)$ with respect to $\nu$ at $\nu=0$
which is the pole of the second order,
\begin{alignat}{2}
  [\rd_\nu^2 p_\nu(\hr_b,\hr_a)]_{\nu=0}
  &=-\pi\int_{r_a}^{r_b}\frac{dr}{r}p_0(\hr_b,\hr)p_0(\hr,\hr_a)
   \qquad&&
  (k=k_m^n,~m\in\mathbb{N}) ,
   \\
  [\rd_\nu^2 s_\nu(\hr_b,\hr_a)]_{\nu=0}
  &=-\pi\int_{r_a}^{r_b}\frac{dr}{r}r_0(\hr_b,\hr)q_0(\hr,\hr_a)
   \qquad&&
  (k=\bk_m^n,~m\in\mathbb{Z}_0^{+}) .
  \label{eq:int_r0q0}
\end{alignat}
We can use Eq.(\ref{eq:int_r0q0}) to confirm Eq.(\ref{eq:cRm_0n})
since $\bk_0^n=k_y^n/\beta$ at which $k_r^n=0$ as shown in Eq.(\ref{eq:lim_kmn}).

We rewrite Eqs.(\ref{eq:Ven}-\ref{eq:mfRm}) using Eqs.(\ref{eq:rdnu_psnu}) 
and (\ref{eq:Cnu_pnu_ba}-\ref{eq:Cnu_snu_ba}),
\begin{alignat}{2}
  \frac{\cR_{+}^{mn}(r,r')}{\rho\vpi_{+}^{mn}}
  &=\bigg[\frac{p_{\nu}(\hr_b,\hr)p_{\nu}(\hr',\hr_a)}{\alp_{+}-\alp_{+}^{-1}}
    \bigg]_{\nu=\nu_m^n},
   \qquad&
  \vpi_{+}^{mn}
  &=k_r^n\frac{d}{dk_r^n}\Big(\frac{\pi\nu_m^n}{2\rho}\Big)^2 ,
  \label{eq:cRp_ba}
   \\
  \frac{\cR_{-}^{mn}(r,r')}{\rho\vpi_{-}^{mn}}
  &=\bigg[
      \frac{r_{\nu}(\hr_b,\hr)q_{\nu}(\hr',\hr_a)}{e_a\alpha_{-}-e_b\alpha_{-}^{-1}}
    \bigg]_{\nu=\mu_m^n},
   \qquad&
  \vpi_{-}^{mn}
  &=k_r^n\frac{d}{dk_r^n}\Big(\frac{\pi\mu_m^n}{2\rho}\Big)^2 .
  \label{eq:cRm_ba}
\end{alignat}
We can further rewrite Eqs.(\ref{eq:cRp_ba}-\ref{eq:cRm_ba}) as follows,
\begin{align}
  \frac{\cR_{+}^{mn}(r,r')}{\rho\vpi_{+}^{mn}}
  &=\bigg[
      \frac{p_{\nu}(\hr_b,\hr)p_{\nu}(\hr_b,\hr')}{1-\alpha_{+}^2}
    \bigg]_{\nu=\nu_m^n}
   =\bigg[
      \frac{p_{\nu}(\hr,\hr_a)p_{\nu}(\hr',\hr_a)}{\alpha_{+}^{-2}-1}
    \bigg]_{\nu=\nu_m^n} ,
   \\
  \frac{\cR_{-}^{mn}(r,r')}{\rho\vpi_{-}^{mn}}
  &=\bigg[
     \frac{r_{\nu}(\hr_b,\hr)r_{\nu}(\hr_b,\hr')}{e_b-e_a\alpha_{-}^2}
    \bigg]_{\nu=\mu_m^n}
   =\bigg[
     \frac{q_{\nu}(\hr,\hr_a)q_{\nu}(\hr',\hr_a)}{e_b\alpha_{-}^{-2}-e_a}
    \bigg]_{\nu=\mu_m^n} .
\end{align}
We use Eqs.(\ref{eq:cRp_ba}-\ref{eq:cRm_ba}) in appendices \ref{sec:cRcX} and
\ref{sec:R0n_AE} to verify
\begin{align}
  \lim_{\rho\to\infty}\cR_{\pm}^{mn}(r,r')
  &=\cX_{\pm}^m(x,x') .
  \label{eq:lim_RX}
\end{align}
$\cX_{\pm}^m$ is given by Eqs.(\ref{eq:mfXp_new}-\ref{eq:mfXm_new})
which are the horizontal eigenfunctions of the straight pipe.

\subsection{Fourier coefficients of the Green functions of the separated form}
\label{sec:cGpm_we}

We rearrange the expressions of the Green functions $\cG_{\pm}^n(\vsig)$ which are
defined in Eq.(\ref{eq:cGeb_def}) and involved in Eq.(\ref{eq:cEBy_sol}) for
$\vsig=s-s'$ and $\vsig=s$ through Eqs.(\ref{eq:cPy_cMy}-\ref{eq:cQy_cNy}).
From Eqs.(\ref{eq:cGe_Th}), (\ref{eq:Imn_done}-\ref{eq:bImn_done}) and (\ref{eq:residues}),
we get
\begin{align}
  \cG_{+}^n(r,r',\vsig)
  &=\sum_{\ell=0}^{\infty}\Th_{+}^{n\ell}
   \bigg[
    \bdelta_{01}^{\ell}\btheta(\vsig)
    \sum_{m=1}^{\ell-1}\cR_{+}^{mn}(r,r')
     \frac{\rho}{\hnu_m^n}\sin\!\bigg(\frac{\hnu_m^n\vsig}{\rho}\bigg)
    -\sum_{m=\ell'}^{\infty}\cR_{+}^{mn}(r,r')
     \frac{\rho}{2\cnu_m^n}e^{-\cnu_m^n\bar{|\vsig|}/\rho}
   \bigg] ,
  \label{eq:cGp_sep}
  \\
  \cG_{-}^n(r,r',\vsig)
  &=\sum_{\ell=0}^{\infty}\Th_{-}^{n\ell}
   \bigg[
      \bdelta_{0}^{\ell}\btheta(\vsig)
      \sum_{m=0}^{\ell-1}\cR_{-}^{mn}(r,r')
      \frac{\rho}{\hmu_m^n}\sin\!\bigg(\frac{\hmu_m^n\vsig}{\rho}\bigg)
     -\sum_{m=\ell}^{\infty}\cR_{-}^{mn}(r,r')
      \frac{\rho}{2\cmu_m^n}e^{-\cmu_m^n\bar{|\vsig|}/\rho}
   \bigg] .
  \label{eq:cGm_sep}
\end{align}
The window functions $\Th_{\pm}^{n\ell}$ are given by Eqs.(\ref{eq:Th_def}-\ref{eq:sum_Th}).
$\ell'$, $\bdelta_{01}^{\ell}$ and $\bdelta_{0}^{\ell}$ are given by
Eqs.(\ref{eq:lprm}-\ref{eq:d01_d0}).
$\btheta(\vsig)$ and $\bar{|\vsig|}$ each represent two functions of $\vsig$,
corresponding to the two longitudinal variables $\vsig=\{s-s',s\}$.
For $\vsig=s\in\mathbb{R}^{+}$, \ie, in the expressions of $\cG_{\pm}^n(s)$,
we must replace the Heaviside step function of $s$ as $\theta(s)=1$ and $\theta(-s)=0$,
which is the manner of the Laplace transform.
The step function of $\vsig\in\mathbb{R}$ is given as
\begin{align}
   \theta(\vsig)
   =
    \bigg\{
    \begin{array}{ll}
       1 & (\vsig\geq0) \\
       0 & (\vsig<0)
    \end{array} ,
    \qquad
  \theta(\vsig)+\theta(-\vsig)
  =1.
   \label{eq:step}
\end{align}
The Dirac $\delta$-function and the sign function of $\vsig$ are gotten from
$\theta(\vsig)$ as follows,
\begin{align}
  \delta(\vsig)
  =\rd_{\vsig}\theta(\vsig)
  =\frac{1}{2}\rd_{\vsig}\sgn(\vsig),
    \qquad
  \sgn(\vsig)
  =\frac{\vsig}{|\vsig|}
  =\theta(\vsig)-\theta(-\vsig) .
   \label{eq:delta_sgn}
\end{align}
$\cG_{\pm}^n(s)$ has the same expression as $\cG_{\pm}^n(s-s')$
except $\theta(\vsig)$ and $|\vsig|$.
For a concise and convenient expression, we want to put together the expressions
of $\cG_{\pm}^n(s)$ and $\cG_{\pm}^n(s-s')$ into the single one
as already shown in Eqs.(\ref{eq:cGp_sep}-\ref{eq:cGm_sep}).
So, by extending $\theta(\vsig)$,
we define the step function $\btheta(\vsig)$ of $\vsig=\{s-s',s\}$ as follows,
\begin{align}
  &
  \btheta(\vsig)+\btheta(-\vsig)
  =1,
   \qquad
  \btheta(\vsig)
  =\bigg\{{\theta(\vsig) \atop 1}\bigg\} ,
   \qquad
  \btheta(-\vsig)
  =\bigg\{{\theta(-\vsig) \atop 0}\bigg\} ,
   \qquad
  \vsig
  =\bigg\{
   \begin{array}{ll}
      s-s' \in\mathbb{R}\\
      s \in\mathbb{R}^{+}
   \end{array} .
  \label{eq:bstep}
\end{align}
The upper and lower quantities in the braces of Eqs.(\ref{eq:bstep}-\ref{eq:babsol}) 
correspond.
Using $\btheta(\vsig)$, we define $\bdelta(\vsig)$ and $\bsgn(\vsig)$ in order that
they follow the same relations as Eqs.(\ref{eq:delta_sgn}) which hold among
$\theta(\vsig)$, $\delta(\vsig)$ and $\sgn(\vsig)$, \ie,
\begin{align}
  &
  \bdelta(\vsig)
  =\rd_{\vsig}\btheta(\vsig)
  =\bigg\{{\delta(\vsig) \atop 0}\bigg\}
   ,\qquad
  \bsgn(\vsig)
  =\btheta(\vsig)-\btheta(-\vsig)
  =\bigg\{{\sgn(\vsig) \atop 1}\bigg\} ,
  \label{eq:bdelta}
   \\
  &
  \bar{|\vsig|}
  =\vsig\,\bar{\sgn}(\vsig)
  =\bigg\{{|\vsig| \atop \vsig}\bigg\}
  =\bigg\{{|s-s'| \atop s}\bigg\}
  \qquad\text{for}\quad
  \vsig
  =\bigg\{\begin{array}{ll} s-s' \\ s \end{array} .
  \label{eq:babsol}
\end{align}
$\bar{|\vsig|}$ represents the absolute values of $s-s'$ and $s$.
Since $\theta(s)$ and $\theta(-s)$ are respectively equivalent to 1 and 0
in the Laplace transform defined in Eqs.(\ref{eq:Laplace}),
we must replace $\delta(s)$ by 0 in the present theory,
otherwise the second derivative of the field with respect to $s$ will create
a bare $\delta(s)$ which is not involved in the integral with respect to $s$.
This replacement for $\vsig=s$ is necessary, because the Laplace transform
is a technique to solve a differential equation as an initial value problem
without taking into account the value of the field in $s<0$.
On the other hand, $\theta(s-s')$ has a meaning in $\cG_{\pm}^n(s-s')$,
because $s-s'$ denotes the longitudinal relation of
the observation point and the source particle of the fields.
Thus, corresponding to the longitudinal variables $\vsig=\{s-s',s\}$,
$\cG_{\pm}^n(\vsig)$ represents the two expressions of the Green functions:
$\cG_{\pm}^n(s-s')$ for the source terms in $s>0$ and
$\cG_{\pm}^n(s)$ for the initial values of the fields at $s=+0$.
In the analytical calculation of the fields, we can deal with the extended special 
functions [$\btheta$, $\bdelta$, $\bar{|\vsig|}$, $\bsgn(\vsig)$]
in the same manner as the usual ones [$\theta$, $\delta$, $|\vsig|$, $\sgn(\vsig)$]
since we defined the former using the same relations as the latter.
Therefore, in what follows, we do not have to pay particular attention to
the mathematical treatments for these extended special functions.

According to Eqs.(\ref{eq:BDE_cRp}-\ref{eq:rdel}) and (\ref{eq:sum_Xmn}),
$\cG_{\pm}^n$ satisfies the following wave equation and symmetry with respect to
the exchange of $r$ and $r'$,
\begin{align}
  \rd_{\rm v}^2\cG_{\pm}^n(r,r',\vsig)
  =\frac{\rho}{r}\delta(r-r')\bdelta(\vsig),
    \qquad
  \cG_{\pm}^n(r,r',\vsig)
  =\cG_{\pm}^n(r',r,\vsig) .
  \label{eq:BessDE_delta}
\end{align}
$\rd_{\rm v}^2$ is given by Eq.(\ref{eq:wen_bend}) which is the operator of
the wave equations for $\cE_y^n$ and $\cB_y^n$.
Eqs.(\ref{eq:BessDE_delta}) correspond to Eqs.(\ref{eq:BDE_mfGpm})
in the Laplace domain.
The first equation of (\ref{eq:BessDE_delta}) has the radial and longitudinal
$\delta$-functions in the driving term instead of $\cS_y^n$ and $\cT_y^n$
which are the source terms of Eq.(\ref{eq:wen_bend}).
According to Eq.(\ref{eq:cRpm_symm}), since $\cG_{\pm}^n$ is symmetric
with respect to the exchange of $r$ and $r'$ as shown in the second equation of
(\ref{eq:BessDE_delta}), the first equation of (\ref{eq:BessDE_delta}) holds also for
the operator $\rd_{\rm v}^2$ with respect to $r'$ instead of $r$.

In the present paper, the term {\it Green function} does not mean
a field created by a point charge.
$\cG_{\pm}^n(\vsig)$ denotes the Green functions of Eq.(\ref{eq:wen_bend})
in which we replace the driving term by the radial and longitudinal $\delta$-functions
as shown in the first equation of (\ref{eq:BessDE_delta}).
Since $\cG_{\pm}^n(s)$ satisfies the homogeneous wave equation unlike $\cG_{\pm}^n(s-s')$,
rigorously speaking, $\cG_{\pm}^n(s)$ is the eigenfunction of $\rd_{\rm v}^2$.
Nevertheless, we call $\cG_{\pm}^n(\vsig)$ the Fourier coefficient of
the Green functions regardless of $\vsig=s-s'$ or $s$.
According to the first equation of (\ref{eq:BessDE_delta}), the integrands of
$\cE_y^n$ and $\cB_y^n$, involved in Eq.(\ref{eq:cEBy_sol}) through
Eqs.(\ref{eq:cPy_cMy}-\ref{eq:cQy_cNy}), satisfy the following wave equations,
\begin{align}
  \rd_{\rm v}^2\bigg\{{ \cP_y^n(s) \atop \cQ_y^n(s) }\bigg\}
  =0 ,
     \qquad
  \rd_{\rm v}^2\bigg\{{ \cM_y^n(\vsig,s') \atop \cN_y^n(\vsig,s') }\bigg\}
  =\bigg\{{ \cS_y^n(s') \atop \cT_y^n(s') }\bigg\}
   \delta(r-r')\delta(\vsig) ,
   \label{eq:we_cPQy_cMNy}
\end{align}
where, for clarity to show the longitudinal arguments, we do not indicate
the radial arguments $(r,r')$ of all the functions excluding $\delta(r-r')$.
From Eqs.(\ref{eq:lim_nurho_ksmn}), (\ref{eq:lim_Th}) and (\ref{eq:lim_RX}),
we can show that
\begin{align}
  \lim_{\rho\to\infty}\cG_{\pm}^n(r,r',\vsig)
  =\cG_{\pm}^n(x,x',\vsig) .
   \label{eq:lim_cGpm}
\end{align}
$\cG_{\pm}^n(x,x',\vsig)$ denotes the Fourier coefficients of the Green functions in
the straight pipe, given by Eqs.(\ref{eq:cGp_st}-\ref{eq:cGm_st}).

As seen from Eq.(\ref{eq:cGp_sep}), $\cG_{+}^n$ has the oscillatory modes
$\sin(\hnu_m^n\vsig/\rho)$ and the damped modes $e^{-\cnu_m^n|\vsig|/\rho}$ with respect to
$\vsig$, where $\hnu_m^n\in\mathbb{R}_0^{+}$ and $\cnu_m^n\in\mathbb{R}^{+}$ are
the poles of the field on the real and imaginary axes of the $\nu$-plane
as shown in Eq.(\ref{eq:ps_poles}).
$\cG_{-}^n$ is similar to $\cG_{+}^n$ in this regard.
Since these two kinds of modes are separated into the two sums with respect to
the radial mode number $m$ in Eqs.(\ref{eq:cGp_sep}-\ref{eq:cGm_sep}), 
$\cG_{\pm}^n$ is referred to as the Green functions of {\it the separated form}.
$\cG_{\pm}^n$ is suited for using in the numerical calculation of the fields
since $\cG_{\pm}^n$ has no terms which diverge for $s\to\infty$.
But $\cG_{\pm}^n$ is somewhat unwieldy in the analytical calculation,
because it is hard to deal with the separated sums which involve
the window functions $\Th_{\pm}^{n\ell}$ as it can be seen later from
Eqs.(\ref{eq:sum_Xmn}).
In short, $\cG_{\pm}^n$ has no complete sum of the radial eigenfunctions
$\cR_{\pm}^{mn}$ over all $m$ from the lowest mode to infinity.
In section \ref{sec:normal_form}
we will derive the Green functions of {\it the complete form} $\cC_{\pm}^n$
which has the complete sum over all the radial eigenmodes of the curved pipe.
We can describe the fields using $\cC_{\pm}^n$ instead of $\cG_{\pm}^n$
as shown later in Eq.(\ref{eq:cEy_cBy_cmpl}).
We can derive the first equation of (\ref{eq:BessDE_delta}) from
Eq.(\ref{eq:cGpm_cCpm_bcCpm}) and Eqs.(\ref{eq:we_cCpm})
which are the wave equations for $\cC_{\pm}^n$.
We need the Green functions of the complete form $\cC_{\pm}^n$ in
the analytical verification of the fact that the solutions of the fields satisfy
Maxwell equations and the initial conditions at $s=0$.
On the other hand, $\cC_{\pm}^n$ is not suited to the numerical calculation
since it has terms which diverge exponentially for $s\to\infty$,
while the divergent terms of the opposite sign cancel out and vanish in
the expressions of the fields given by Eq.(\ref{eq:cEy_cBy_cmpl}).
In other words, the separated form $\cG_{\pm}^n$ is the expression of the Green functions 
from which we removed the divergent terms with respect to $s$ from
the complete form $\cC_{\pm}^n$.

\subsection{Fourier coefficients of the radial and longitudinal fields}
\label{sec:XS_field_sdom}

We find the expressions of the Fourier coefficients of the radial and longitudinal 
components $\cE_{x,s}^n$ and $\cB_{x,s}^n$ by
inverting the Laplace transform $\mfE_{x,s}^n$ and $\mfB_{x,s}^n$.
Substituting Eqs.(\ref{eq:mfBsEx}-\ref{eq:mfBxEs}) into Eq.(\ref{eq:ILT_nu}),
we rewrite them as
\begin{align}
  \bigg\{{\cE_{x,s}^n(r,s) \atop c\cB_{x,s}^n(r,s)}\bigg\}
  &=\int_{r_a}^{r_b}dr'
    \bigg[
        \bigg\{{\cP_{x,s}^n(r,r',s) \atop \cQ_{x,s}^n(r,r',s)}\bigg\}
       +Z_0\int_0^{\infty}ds'
        \bigg\{{\cM_{x,s}^n(r,r',s-s',s') \atop \cN_{x,s}^n(r,r',s-s',s')}\bigg\}
    \bigg] .
  \label{eq:cExs_cBxs_sprt}
\end{align}
$\cP_{x,s}^n$ and $\cQ_{x,s}^n$ are the Fourier coefficients
which have the initial values of $\cE_{x,s}^n$ and $\cB_{x,s}^n$ at $s=0$,
\begin{alignat}{2}
  \cP_x^n(s)
  &=\cD_x^n\cH_{-}^n(s)-\cD_s^n\cK_{-}^n(s),
    \qquad
  &\cM_x^n(\vsig,s')
  &=\frac{r'}{\rho}\{\cS_{x}^n(s')\cH_{-}^n(\vsig)-\cS_{s}^n(s')\cK_{-}^n(\vsig)\} ,
   \label{eq:cPxn}
   \\
  \cP_s^n(s)
  &=\cD_s^n\cH_{+}^n(s)+\cD_x^n\cK_{+}^n(s),
    \qquad
  &\cM_s^n(\vsig,s')
  &=\frac{r'}{\rho}\{\cS_{s}^n(s')\cH_{+}^n(\vsig)+\cS_{x}^n(s')\cK_{+}^n(\vsig)\} ,
   \label{eq:cPsn}
   \\
  \cQ_x^n(s)
  &=\cA_x^n\cH_{+}^n(s)-\cA_s^n\cK_{+}^n(s),
    \qquad
  &\cN_x^n(\vsig,s')
  &=\frac{r'}{\rho}\{\cT_{x}^n(s')\cH_{+}^n(\vsig)-\cT_{s}^n(s')\cK_{+}^n(\vsig)\} ,
   \label{eq:cQxn}
   \\
  \cQ_s^n(s)
  &=\cA_s^n\cH_{-}^n(s)+\cA_x^n\cK_{-}^n(s),
    \qquad
  &\cN_s^n(\vsig,s')
  &=\frac{r'}{\rho}\{\cT_{s}^n(s')\cH_{-}^n(\vsig)+\cT_{x}^n(s')\cK_{-}^n(\vsig)\} ,
   \label{eq:cQsn}
\end{alignat}
where we omit the radial arguments $(r,r')$ of the functions.
$\cM_{x,s}^n$ and $\cN_{x,s}^n$ are the Fourier coefficients
which have the source terms of $\cE_{x,s}^n$ and $\cB_{x,s}^n$ in $s>0$.
$\cS_{x,s}^n$ and $\cT_{x,s}^n$ are the Fourier coefficients of the source terms
given by Eqs.(\ref{eq:cSx_cTx}) and (\ref{eq:cSs_cTs}).
We rewrite them using the dummy variables $(r',s')$ for $(r,s)$,
\begin{alignat}{2}
  \cS_x^n(r',s')
  &=\rd_{r'}\cJ_0^n(r',s')-ik\beta\cJ_x^n(r',s')
   ,\qquad&
  \cT_x^n(r',s')
  &=\frac{\rho}{r'}\rd_{s'}\cJ_y^n(r',s')-(-1)^nk_y^n\cJ_s^n(r',s') ,
   \label{eq:cSx_cTx_prm}
  \\
  \cS_s^n(r',s')
  &=\frac{\rho}{r'}\rd_{s'}\cJ_0^n(r',s')-ik\beta\cJ_s^n(r',s')
   ,\qquad&
  \cT_s^n(r',s')
  &=(-1)^nk_y^n\cJ_x^n(r',s')-\rd_{r'}\cJ_y^n(r',s') .
   \label{eq:cSs_cTs_prm}
\end{alignat}
$\cD_{x,s}^n$ and $\cA_{x,s}^n$ are the inverse Laplace transform of
$\mfD_{x,s}^n$ and $\mfA_{x,s}^n$ given by Eqs.(\ref{eq:mfDx_mfAx}-\ref{eq:mfDs_mfAs}).
Multiplying $i\nu/\rho$ with a quantity in the Laplace domain is equivalent to acting
the longitudinal operator $\rd_s$ on the corresponding Fourier coefficient
which is a function of $s$.
$\cD_{x,s}^n$ and $\cA_{x,s}^n$ are the operators with respect to $s$,
which have the initial values of $\cE_{x,s}^n$ and $\cB_{x,s}^n$ at the entrance of
the bending magnet,
\begin{align}
  \cD_x^n(r')
  &=\frac{\rho}{r'}
    \Big[\rd_{s'}\cE_{x}^n(r',s')+\cE_x^n(r',s')\rd_s-\frac{2}{\rho}\cE_s^n(r',s')
    \Big]_{s'=+0} ,
  \label{eq:bcEx_init}
  \\
  \cD_s^n(r')
  &=\frac{\rho}{r'}
    \Big[\rd_{s'}\cE_{s}^n(r',s')+\cE_s^n(r',s')\rd_s+\frac{2}{\rho}\cE_x^n(r',s')
    \Big]_{s'=+0} ,
  \label{eq:bcEs_init}
  \\
  \cA_x^n(r')
  &=\frac{\rho}{r'}
    \Big[\rd_{s'}c\cB_x^n(r',s')+c\cB_x^n(r',s')\rd_s-\frac{2}{\rho}c\cB_s^n(r',s')
    \Big]_{s'=+0} ,
  \label{eq:bcBx_init}
  \\
  \cA_s^n(r')
  &=\frac{\rho}{r'}
    \Big[\rd_{s'}c\cB_s^n(r',s')+c\cB_s^n(r',s')\rd_s+\frac{2}{\rho}c\cB_x^n(r',s')
    \Big]_{s'=+0} .
  \label{eq:bcBs_init}
\end{align}

Eqs.(\ref{eq:bcEx_init}-\ref{eq:bcBs_init}) have the operator $\rd_s$
which acts on $\cH_{\pm}^n(s)$ and $\cK_{\pm}^n(s)$ in the first equations of
(\ref{eq:cPxn}-\ref{eq:cQsn}).
$\cH_{\pm}^n(\vsig)$ and $\cK_{\pm}^n(\vsig)$ are the Fourier coefficients of
the Green functions and the coupling Green functions of the separated form.
They are the inverse Laplace transform of $\mfH_{\pm}^n(\nu)$ and $\mfK_{\pm}^n(\nu)$
given by Eqs.(\ref{eq:mfHpm}-\ref{eq:mfKpm}),
\begin{align}
  \bigg\{{ \cH_{\pm}^n(r,r',\vsig) \atop \cK_{\pm}^n(r,r',\vsig) }\bigg\}
  &=\int_{-\infty-i0}^{\infty-i0}\frac{d\nu}{2\pi\rho}
    \bigg\{{ \mfH_{\pm}^n(r,r',\nu) \atop \mfK_{\pm}^n(r,r',\nu) }\bigg\}
    e^{i\nu\vsig/\rho} ,
   \label{eq:cHK}
\end{align}
where $\vsig=\{s-s',s\}$.
Using the residue theorem for the contours shown in Fig.\ref{fig:contour}
(p.\pageref{fig:contour}), we calculate the Bromwich integral (\ref{eq:cHK})
in a similar way to the one involving $\mfG_{\pm}^n(\nu)$ given by
Eq.(\ref{eq:cGeb_def}) and (\ref{eq:cGe_Th}),
\begin{align}
  \cH_{\pm}^n
  &=\frac{\hr}{\hr'}(\brd_{\hr}\rd_{\hr}+1)\cG_{\pm}^n
   +\rd_{\hr}\rd_{\hr'}\cG_{\mp}^n ,
    \qquad
  \cK_{\pm}^n
  =\rho\rd_s
    \bigg(\frac{\rd_{\hr'}}{\hr}\cG_{\pm}^n
          +\frac{\rd_{\hr}}{\hr'}\cG_{\mp}^n
    \bigg) .
  \label{eq:cHpm}
\end{align}
$\brd_{\hr}$ is given by Eq.(\ref{eq:rd_hr}).
Substituting Eqs.(\ref{eq:cGp_sep}-\ref{eq:cGm_sep}) into the first equation of
(\ref{eq:cHpm}), we get
\begin{align}
  \cH_{+}^n(\vsig)
  &=\sum_{\ell=0}^{\infty}\Th_{+}^{n\ell}
    \Bigg[
       \bdelta_{01}^{\ell}\btheta(\vsig)\sum_{m=1}^{\ell-1}
       \frac{\rho\hnu_m^n}{\hr\hr'}\cR_{+}^{mn}
       \sin\!\bigg(\frac{\hnu_m^n\vsig}{\rho}\bigg)
      +\sum_{m=\ell'}^{\infty}
       \frac{\rho\cnu_m^n}{\hr\hr'}\cR_{+}^{mn}\frac{e^{-\cnu_m^n\bar{|\vsig|}/\rho}}{2}
    \Bigg]
   \nonumber\\&\quad
   +\sum_{\ell=0}^{\infty}\Th_{-}^{n\ell}
    \Bigg[
      \bdelta_{0}^{\ell}\btheta(\vsig)\sum_{m=0}^{\ell-1}
      \frac{\rd_{\hr}\rd_{\hr'}\cR_{-}^{mn}}{\hmu_m^n/\rho}
      \sin\!\bigg(\frac{\hmu_m^n\vsig}{\rho}\bigg)
     -\sum_{m=\ell}^{\infty}
      \frac{\rd_{\hr}\rd_{\hr'}\cR_{-}^{mn}}{\cmu_m^n/\rho}
      \cd\frac{e^{-\cmu_m^n\bar{|\vsig|}/\rho}}{2}
    \Bigg] ,
   \label{eq:cHp}
   \\
  \cH_{-}^n(\vsig)
  &=\sum_{\ell=0}^{\infty}\Th_{-}^{n\ell}
    \Bigg[
       \bdelta_{0}^{\ell}\btheta(\vsig)\sum_{m=0}^{\ell-1}
       \frac{\rho\hmu_m^n}{\hr\hr'}\cR_{-}^{mn}
       \sin\!\bigg(\frac{\hmu_m^n\vsig}{\rho}\bigg)
      +\sum_{m=\ell}^{\infty}\frac{\rho\cmu_m^n}{\hr\hr'}\cR_{-}^{mn}
       \frac{e^{-\cmu_m^n\bar{|\vsig|}/\rho}}{2}
    \Bigg]
   \nonumber\\&\quad
   +\sum_{\ell=0}^{\infty}\Th_{+}^{n\ell}
    \Bigg[
       \bdelta_{01}^{\ell}\btheta(\vsig)\sum_{m=1}^{\ell-1}
       \frac{\rd_{\hr}\rd_{\hr'}\cR_{+}^{mn}}{\hnu_m^n/\rho}
       \sin\!\bigg(\frac{\hnu_m^n\vsig}{\rho}\bigg)
      -\sum_{m=\ell'}^{\infty}
       \frac{\rd_{\hr}\rd_{\hr'}\cR_{+}^{mn}}{\cnu_m^n/\rho}
       \cd\frac{e^{-\cnu_m^n\bar{|\vsig|}/\rho}}{2}
   \Bigg] .
   \label{eq:cHm}
\end{align}
$\ell'$ is given by Eq.(\ref{eq:lprm}).
From Eqs.(\ref{eq:cGp_sep}-\ref{eq:cGm_sep}) and (\ref{eq:cHpm}),
we get $\cK_{\pm}^n$ which denotes the Fourier coefficients of the coupling Green functions 
between the radial and longitudinal components of the fields,
\begin{align}
  \cK_{+}^n(\vsig)
  &=\sum_{\ell=0}^{\infty}\Th_{+}^{n\ell}
    \Bigg[
       \bdelta_{01}^{\ell}\btheta(\vsig)\sum_{m=1}^{\ell-1}
       \frac{\rho}{\hr}\rd_{\hr'}\cR_{+}^{mn}\cos\!\bigg(\frac{\hnu_m^n\vsig}{\rho}\bigg)
      +\bar{\sgn}(\vsig)\sum_{m=\ell'}^{\infty}
       \frac{\rho}{\hr}\rd_{\hr'}\cR_{+}^{mn}\frac{e^{-\cnu_m^n\bar{|\vsig|}/\rho}}{2}
   \Bigg]
    \nonumber\\&\quad
   +\sum_{\ell=0}^{\infty}\Th_{-}^{n\ell}
    \Bigg[
       \bdelta_{0}^{\ell}\btheta(\vsig)\sum_{m=0}^{\ell-1}
       \frac{\rho}{\hr'}\rd_{\hr}\cR_{-}^{mn}\cos\!\bigg(\frac{\hmu_m^n\vsig}{\rho}\bigg)
      +\bar{\sgn}(\vsig)\sum_{m=\ell}^{\infty}\frac{\rho}{\hr'}\rd_{\hr}\cR_{-}^{mn}
       \frac{e^{-\cmu_m^n\bar{|\vsig|}/\rho}}{2}
    \Bigg] ,
   \label{eq:cKp}
  \\
  \cK_{-}^n(\vsig)
  &=\sum_{\ell=0}^{\infty}\Th_{-}^{n\ell}
    \Bigg[
       \bdelta_{0}^{\ell}\btheta(\vsig)\sum_{m=0}^{\ell-1}
       \frac{\rho}{\hr}\rd_{\hr'}\cR_{-}^{mn}
       \cos\!\bigg(\frac{\hmu_m^n\vsig}{\rho}\bigg)
      +\bar{\sgn}(\vsig)\sum_{m=\ell}^{\infty}\frac{\rho}{\hr}\rd_{\hr'}\cR_{-}^{mn}
       \frac{e^{-\cmu_m^n\bar{|\vsig|}/\rho}}{2}
    \Bigg]
    \nonumber\\&\quad
   +\sum_{\ell=0}^{\infty}\Th_{+}^{n\ell}
    \Bigg[
       \bdelta_{01}^{\ell}\btheta(\vsig)\sum_{m=1}^{\ell-1}
       \frac{\rho}{\hr'}\rd_{\hr}\cR_{+}^{mn}
       \cos\!\bigg(\frac{\hnu_m^n\vsig}{\rho}\bigg)
      +\bar{\sgn}(\vsig)\sum_{m=\ell'}^{\infty}
       \frac{\rho}{\hr'}\rd_{\hr}\cR_{+}^{mn}
       \frac{e^{-\cnu_m^n\bar{|\vsig|}/\rho}}{2}
    \Bigg] .
   \label{eq:cKm}
\end{align}
In Eqs.(\ref{eq:cHp}-\ref{eq:cKm}) we omit the radial arguments
$(r,r')$ of $\cR_{\pm}^{mn}$, $\cH_{\pm}^n$ and $\cK_{\pm}^n$ for brevity.

We get the radial derivatives of the radial eigenfunctions
from Eqs.(\ref{eq:Ven}) and (\ref{eq:Vbn}),
\begin{align}
  \rd_{\hr}\cR_{+}^{mn}(r,r')
  &=-\pi\frac{\nu_m^n}{\rho}
    \bigg[
     \theta(\hr-\hr')
     \frac{q_\nu(\hr_b,\hr)p_\nu(\hr',\hr_a)}{\rd_\nu p_\nu(\hr_b,\hr_a)}
    +\theta(\hr'-\hr)
     \frac{p_\nu(\hr_b,\hr')r_\nu(\hr,\hr_a)}{\rd_\nu p_\nu(\hr_b,\hr_a)}
    \bigg]_{\nu=\nu_m^n} ,
  \label{eq:Dr_cRp}
   \\
  \rd_{\hr}\cR_{-}^{mn}(r,r')
  &=-\pi\frac{\mu_m^n}{\rho}
    \bigg[
     \theta(\hr-\hr')
     \frac{s_\nu(\hr_b,\hr)q_\nu(\hr',\hr_a)}{\rd_\nu s_\nu(\hr_b,\hr_a)}
    +\theta(\hr'-\hr)
     \frac{r_\nu(\hr_b,\hr')s_\nu(\hr,\hr_a)}{\rd_\nu s_\nu(\hr_b,\hr_a)}
    \bigg]_{\nu=\mu_m^n} ,
   \\
  \rd_{\hr'}\cR_{+}^{mn}(r,r')
  &=-\pi\frac{\nu_m^n}{\rho}
    \bigg[
     \theta(\hr-\hr')
     \frac{p_\nu(\hr_b,\hr)r_\nu(\hr',\hr_a)}{\rd_\nu p_\nu(\hr_b,\hr_a)}
    +\theta(\hr'-\hr)
     \frac{q_\nu(\hr_b,\hr')p_\nu(\hr,\hr_a)}{\rd_\nu p_\nu(\hr_b,\hr_a)}
    \bigg]_{\nu=\nu_m^n} ,
   \\
  \rd_{\hr'}\cR_{-}^{mn}(r,r')
  &=-\pi\frac{\mu_m^n}{\rho}
    \bigg[
     \theta(\hr-\hr')
     \frac{r_\nu(\hr_b,\hr)s_\nu(\hr',\hr_a)}{\rd_\nu s_\nu(\hr_b,\hr_a)}
    +\theta(\hr'-\hr)
     \frac{s_\nu(\hr_b,\hr')q_\nu(\hr,\hr_a)}{\rd_\nu s_\nu(\hr_b,\hr_a)}
    \bigg]_{\nu=\mu_m^n} .
  \label{eq:Drp_cRm}
\end{align}
Using Eqs.(\ref{eq:pppp_rqrq}-\ref{eq:qrqr_ssss}),
we can rewrite Eqs.(\ref{eq:Dr_cRp}-\ref{eq:Drp_cRm}) as follows,
\begin{align}
  \rd_{\hr}\cR_{+}^{mn}(r,r')
  &=-\pi
    \bigg[\frac{\nu}{\rho}\cd
          \frac{q_\nu(\hr_b,\hr)p_\nu(\hr',\hr_a)}{\rd_\nu p_\nu(\hr_b,\hr_a)}
    \bigg]_{\nu=\nu_m^n}
   =-\pi
    \bigg[\frac{\nu}{\rho}\cd
          \frac{p_\nu(\hr_b,\hr')r_\nu(\hr,\hr_a)}{\rd_\nu p_\nu(\hr_b,\hr_a)}
    \bigg]_{\nu=\nu_m^n} ,
   \\
  \rd_{\hr}\cR_{-}^{mn}(r,r')
  &=-\pi
    \bigg[\frac{\nu}{\rho}\cd
          \frac{s_\nu(\hr_b,\hr)q_\nu(\hr',\hr_a)}{\rd_\nu s_\nu(\hr_b,\hr_a)}
    \bigg]_{\nu=\mu_m^n}
   =-\pi
    \bigg[\frac{\nu}{\rho}\cd
          \frac{r_\nu(\hr_b,\hr')s_\nu(\hr,\hr_a)}{\rd_\nu s_\nu(\hr_b,\hr_a)}
    \bigg]_{\nu=\mu_m^n} ,
   \\
  \rd_{\hr'}\cR_{+}^{mn}(r,r')
  &=-\pi
    \bigg[\frac{\nu}{\rho}\cd
          \frac{p_\nu(\hr_b,\hr)r_\nu(\hr',\hr_a)}{\rd_\nu p_\nu(\hr_b,\hr_a)}
    \bigg]_{\nu=\nu_m^n}
   =-\pi
    \bigg[\frac{\nu}{\rho}\cd
          \frac{q_\nu(\hr_b,\hr')p_\nu(\hr,\hr_a)}{\rd_\nu p_\nu(\hr_b,\hr_a)}
    \bigg]_{\nu=\nu_m^n} ,
   \\
  \rd_{\hr'}\cR_{-}^{mn}(r,r')
  &=-\pi
    \bigg[\frac{\nu}{\rho}\cd
          \frac{r_\nu(\hr_b,\hr)s_\nu(\hr',\hr_a)}{\rd_\nu s_\nu(\hr_b,\hr_a)}
    \bigg]_{\nu=\mu_m^n}
   =-\pi
    \bigg[\frac{\nu}{\rho}\cd
          \frac{s_\nu(\hr_b,\hr')q_\nu(\hr,\hr_a)}{\rd_\nu s_\nu(\hr_b,\hr_a)}
    \bigg]_{\nu=\mu_m^n} .
  \label{eq:D2_mfRm}
\end{align}
Also, we get $\rd_{\hr}\rd_{\hr'}\cR_{\pm}^{mn}$ which appears in
Eqs.(\ref{eq:cHp}-\ref{eq:cHm}),
\begin{align}
  \rd_{\hr}\rd_{\hr'}\cR_{+}^{mn}(r,r')
  &=-\pi
    \bigg[\theta(\hr-\hr')\frac{\nu}{\rho}\cd
           \frac{q_{\nu}(\hr_b,\hr)r_{\nu}(\hr',\hr_a)}{\rd_{\nu}p_{\nu}(\hr_b,\hr_a)}
          +\theta(\hr'-\hr)\frac{\nu}{\rho}\cd
           \frac{q_{\nu}(\hr_b,\hr')r_{\nu}(\hr,\hr_a)}{\rd_{\nu}p_{\nu}(\hr_b,\hr_a)}
    \bigg]_{\nu=\nu_m^n}
  \label{eq:bU_pls}
  \\
  &=-\pi
    \bigg[\frac{\nu}{\rho}\cd
          \frac{q_{\nu}(\hr_b,\hr)r_{\nu}(\hr',\hr_a)}{\rd_{\nu}p_{\nu}(\hr_b,\hr_a)}
    \bigg]_{\nu=\nu_m^n}
   =-\pi
    \bigg[\frac{\nu}{\rho}\cd
          \frac{q_{\nu}(\hr_b,\hr')r_{\nu}(\hr,\hr_a)}{\rd_{\nu}p_{\nu}(\hr_b,\hr_a)}
    \bigg]_{\nu=\nu_m^n} ,
  \\
  \rd_{\hr}\rd_{\hr'}\cR_{-}^{mn}(r,r')
  &=-\pi
    \bigg[
       \theta(\hr-\hr')\frac{\nu}{\rho}\cd
       \frac{s_{\nu}(\hr_b,\hr)s_{\nu}(\hr',\hr_a)}{\rd_{\nu}s_{\nu}(\hr_b,\hr_a)}
      +\theta(\hr'-\hr)\frac{\nu}{\rho}\cd
       \frac{s_{\nu}(\hr_b,\hr')s_{\nu}(\hr,\hr_a)}{\rd_{\nu}s_{\nu}(\hr_b,\hr_a)}
    \bigg]_{\nu=\mu_m^n}
  \label{eq:bU_mns}
  \\
  &=-\pi
    \bigg[\frac{\nu}{\rho}\cd
          \frac{s_{\nu}(\hr_b,\hr)s_{\nu}(\hr',\hr_a)}{\rd_{\nu}s_{\nu}(\hr_b,\hr_a)}
    \bigg]_{\nu=\mu_m^n}
   =-\pi
    \bigg[\frac{\nu}{\rho}\cd
          \frac{s_{\nu}(\hr_b,\hr')s_{\nu}(\hr,\hr_a)}{\rd_{\nu}s_{\nu}(\hr_b,\hr_a)}
    \bigg]_{\nu=\mu_m^n} .
  \label{eq:rdr_rdrp_cRm_2}
\end{align}
Similar to Eqs.(\ref{eq:cRp_ba}-\ref{eq:cRm_ba}),
we rewrite Eqs.(\ref{eq:bU_pls}-\ref{eq:rdr_rdrp_cRm_2}) using
Eqs.(\ref{eq:Cnu_prm_pnu_ba}-\ref{eq:Cnu_prm_snu_ba}),
\begin{align}
  \frac{\rd_{\hr}\rd_{\hr'}\cR_{+}^{mn}(r,r')}{\rho\vpi_{+}^{mn}}
  &=\bigg[\frac{q_{\nu}(\hr_b,\hr)r_{\nu}(\hr',\hr_a)}{\alpha_{+}-\alpha_{+}^{-1}}
    \bigg]_{\nu=\nu_m^n}
   \\
  &=\bigg[\frac{q_{\nu}(\hr_b,\hr)q_{\nu}(\hr_b,\hr')}{1-\alpha_{+}^2}\bigg]_{\nu=\nu_m^n}
   =\bigg[\frac{r_{\nu}(\hr,\hr_a)r_{\nu}(\hr',\hr_a)}{\alpha_{+}^{-2}-1}
    \bigg]_{\nu=\nu_m^n},
   \\
  \frac{\rd_{\hr}\rd_{\hr'}\cR_{-}^{mn}(r,r')}{\rho\vpi_{-}^{mn}}
  &=
    \bigg[\frac{s_{\nu}(\hr_b,\hr)s_{\nu}(\hr',\hr_a)}{e_a\alpha_{-}-e_b\alpha_{-}^{-1}}
    \bigg]_{\nu=\mu_m^n}
  \\
  &=
    \bigg[\frac{s_{\nu}(\hr_b,\hr)s_{\nu}(\hr_b,\hr')}{e_b-e_a\alpha_{-}^2}
    \bigg]_{\nu=\mu_m^n}
   =
    \bigg[\frac{s_{\nu}(\hr,\hr_a)s_{\nu}(\hr',\hr_a)}{e_b\alpha_{-}^{-2}-e_a}
    \bigg]_{\nu=\mu_m^n} .
\end{align}
$\alpha_{\pm}$, $e_{a,b}$ and $\vpi_{\pm}^{mn}$ are given by
Eqs.(\ref{eq:alpha}) and (\ref{eq:cRp_ba}-\ref{eq:cRm_ba}).
The Green functions $\cG_{\pm}^n$ and $\cH_{\pm}^n$ are continuous and 
smooth at $r=r'$ in the range $[r_a,r_b]$, because
Eqs.(\ref{eq:nu_mn_rels}-\ref{eq:mu_mn_rels}) hold at the pole values
$\nu=(\nu_m^n,\mu_m^n)$.
$\cG_{\pm}^n(s-s')$ and $\cH_{\pm}^n(s-s')$ are continuous at $s=s'$
while having a kink at $s=s'$.

\clearpage

\subsection{Properties of the Fourier coefficients of the Green functions}

We describe the properties of $\cH_{\pm}^n$ and $\cK_{\pm}^n$ given by Eqs.(\ref{eq:cHpm}).
According to the second equation of (\ref{eq:BessDE_delta}), $\cH_{\pm}^n$ and $\cK_{\pm}^n$
each have the following symmetries with respect to the exchange of $r$ and $r'$,
\begin{align}
  \cH_{\pm}^n(r,r',\vsig)=\cH_{\pm}^n(r',r,\vsig) ,
    \qquad
  \cK_{\pm}^n(r,r',\vsig)=\cK_{\mp}^n(r',r,\vsig) .
  \label{eq:cHcK_rrp}
\end{align}
Eqs.(\ref{eq:cHcK_rrp}) correspond to Eqs.(\ref{eq:mfHK_symm}) in the Laplace domain.
From the first equation of (\ref{eq:BessDE_delta}) and Eqs.(\ref{eq:cHpm}),
we get the wave equations for $\cH_{\pm}^n$ and $\cK_{\pm}^n$,
\begin{align}
  \rd_{\,\vdash}^2\cH_{\pm}^n(r,r',\vsig)
  -\frac{2}{r}\brd_s \cK_{\mp}^n(r,r',\vsig)
  &=\frac{\rho}{r}\delta(r-r')\bdelta(\vsig) ,
   \label{eq:Grn_cHpm_cKpm}
   \\
  \rd_{\,\vdash}^2\cK_{\pm}^n(r,r',\vsig)
  +\frac{2}{r}\brd_s\cH_{\mp}^n(r,r',\vsig)
  &=0 .
   \label{eq:Grn_cKpm_cHpm}
\end{align}
$\rd_{\,\vdash}^2$ and $\brd_s$ are the operators given by Eqs.(\ref{eq:wen_bend}) and
(\ref{eq:rs_opera}).
$\bdelta(\vsig)$ is the extended $\delta$-function defined in Eq.(\ref{eq:bdelta}).
Eqs.(\ref{eq:Grn_cHpm_cKpm}-\ref{eq:Grn_cKpm_cHpm}) correspond to
Eqs.(\ref{we_mfH}-\ref{we_mfK}) in the Laplace domain.
As seen from Eqs.(\ref{eq:we_cFxs}) and (\ref{eq:Grn_cHpm_cKpm}),
$\cH_{\pm}^n$ and $\cK_{\pm}^n$ are respectively the Green functions and
the coupling Green functions of the wave equations for $\cE_{x,s}^n$ and $\cB_{x,s}^n$.
Since $\bdelta(s)=0$, $\cH_{\pm}^n(s)$ satisfies the homogeneous wave equation,
similar to $\cG_{\pm}^n(s)$ in Eq.(\ref{eq:BessDE_delta}).
Also, $\cK_{\pm}^n(\vsig)$ for $\vsig=\{s-s',s\}$ satisfies the homogeneous wave equation 
as shown in Eq.(\ref{eq:Grn_cKpm_cHpm}).
On the other hand, $\cH_{\pm}^n(s-s')$ and $\cG_{\pm}^n(s-s')$ satisfy
the inhomogeneous wave equations.

According to Eq.(\ref{eq:Grn_cHpm_cKpm}), the integrands of
$\cE_{x,s}^n$ and $\cB_{x,s}^n$, given by Eqs.(\ref{eq:cPxn}-\ref{eq:cQsn}) and
involved in Eq.(\ref{eq:cExs_cBxs_sprt}), satisfy the following wave equations,
\begin{align}
  \rd_{\vdash}^2\bigg\{{ \cP_x^n(s) \atop \cQ_x^n(s) }\bigg\}
  -\frac{2}{r}\brd_s\bigg\{{ \cP_s^n(s) \atop \cQ_s^n(s) }\bigg\}
  =0 ,
    \qquad
  \rd_{\vdash}^2\bigg\{{ \cP_s^n(s) \atop \cQ_s^n(s) }\bigg\}
  +\frac{2}{r}\brd_s\bigg\{{ \cP_x^n(s) \atop \cQ_x^n(s) }\bigg\}
  =0 ,
   \label{eq:we_cPxs_cQxs}
\end{align}
\begin{align}
  \rd_{\vdash}^2\bigg\{{ \cM_x^n(\vsig,s') \atop \cN_x^n(\vsig,s') }\bigg\}
  -\frac{2}{r}\brd_s\bigg\{{ \cM_s^n(\vsig,s') \atop \cN_s^n(\vsig,s') }\bigg\}
  &=\bigg\{{ \cS_x^n(s') \atop \cT_x^n(s') }\bigg\}
    \delta(r-r')\delta(\vsig) ,
   \label{eq:we_cMx_cNx}
   \\
  \rd_{\vdash}^2\bigg\{{ \cM_s^n(\vsig,s') \atop \cN_s^n(\vsig,s') }\bigg\}
  +\frac{2}{r}\brd_s\bigg\{{ \cM_x^n(\vsig,s') \atop \cN_x^n(\vsig,s') }\bigg\}
  &=\bigg\{{ \cS_s^n(s') \atop \cT_s^n(s') }\bigg\}
    \delta(r-r')\delta(\vsig) ,
   \label{eq:we_cMs_cNs}
\end{align}
where, for clarity to show the longitudinal arguments,
we omit the radial arguments $(r,r')$ of the functions excluding $\delta(r-r')$.
According to Eqs.(\ref{eq:we_cPxs_cQxs}-\ref{eq:we_cMs_cNs}),
$\cE_{x,s}^n$ and $\cB_{x,s}^n$, given by Eq.(\ref{eq:cExs_cBxs_sprt}),
satisfy the wave equations (\ref{eq:we_cFxs}).
The Green functions $\cG_{\pm}^n$, $\cH_{\pm}^n$ and $\cK_{\pm}^n$ have
the following relations,
\begin{alignat}{2}
  \frac{\rho}{\hr}\rd_s\cH_{\pm}^n(\vsig)-\brd_{\hr}\cK_{\mp}^n(\vsig)
  &=\frac{\rho}{\hr'}\rd_s\cG_{\pm}^n(\vsig) ,
    \qquad&
  \brd_{\hr}\cH_{\mp}^n(\vsig)+\frac{\rho}{\hr}\rd_s\cK_{\pm}^n(\vsig)
  &=-\rd_{\hr'}\cG_{\pm}^n(\vsig) ,
   \label{eq:cHcK_rel}
   \\
  \frac{\rho}{\hr'}\rd_s\cH_{\pm}^n(\vsig)-\brd_{\hr'}\cK_{\pm}^n(\vsig)
  &=\frac{\rho}{\hr}\rd_s\cG_{\pm}^n(\vsig) ,
    \qquad&
  \brd_{\hr'}\cH_{\mp}^n(\vsig)+\frac{\rho}{\hr'}\rd_s\cK_{\mp}^n(\vsig)
  &=-\rd_{\hr}\cG_{\pm}^n(\vsig) .
   \label{eq:cKcH_rel}
\end{alignat}
The dimensionless radial operators $\brd_{\hr}$ and $\brd_{\hr'}$ are given by
Eqs.(\ref{eq:rd_hr}).
Eqs.(\ref{eq:cHcK_rel}-\ref{eq:cKcH_rel}) are the relations which connect
the radial eigenfunctions $\cR_{+}^{mn}$ and $\cR_{-}^{mn}$ given by
Eqs.(\ref{eq:Ven}-\ref{eq:mfRm}).
The horizontal eigenfunctions of the straight pipe, $\cX_{+}^m$ and $\cX_{-}^m$, have
the simple relation given by Eqs.(\ref{eq:we_cX}).
On the other hand, we cannot find a simple relation between $\cR_{+}^{mn}$ and $\cR_{-}^{mn}$
since their pole values are different ($\nu_m^n\ne\mu_m^n$) in the curved pipe.
We have no idea whether $\cR_{+}^{mn}$ and $\cR_{-}^{mn}$ obey a simple relation
similar to the one between $\cX_{+}^m$ and $\cX_{-}^m$.

We consider the asymptotic limit of the Green functions for large radius $\rho\to\infty$
keeping the width of the pipe constant.
In this limit, both $\cG_{\pm}^n(r,r',\vsig)$ and $\cH_{\pm}^n(r,r',\vsig)$ go to
$\cG_{\pm}^n(x,x',\vsig)$ given by Eqs.(\ref{eq:cGp_st}-\ref{eq:cGm_st}),
\begin{align}
  \lim_{\rho\to\infty}\cG_{\pm}^n(r,r',\vsig)
  =\lim_{\rho\to\infty}\cH_{\pm}^n(r,r',\vsig)
  =\cG_{\pm}^n(x,x',\vsig) ,
    \qquad
  \lim_{\rho\to\infty}\cK_{\pm}^n(r,r',\vsig)
  =0 .
   \label{eq:lim_cH_cK}
\end{align}
$\cG_{\pm}^n(x,x',\vsig)$ is the Green function of the wave equation for
the Fourier coefficients of the fields in the straight pipe.
The coupling Green functions $\cK_{\pm}^n$ vanish in the limit of $\rho\to\infty$
as shown in the second equation of (\ref{eq:lim_cH_cK}).
Eqs.(\ref{eq:lim_cH_cK}) correspond to the second of (\ref{eq:we_mfGpm_str}) and
Eq.(\ref{eq:lim_mfH_bmfG}-\ref{eq:lim_mfKpm}) in the Laplace domain.
According to Eqs.(\ref{eq:lim_cH_cK}), the second equations of
(\ref{eq:cHcK_rel}-\ref{eq:cKcH_rel}) go to Eq.(\ref{eq:rdxGp_rdxp_cGm})
in the limit of $\rho\to\infty$.
According to Eqs.(\ref{eq:BC_mfRp}-\ref{eq:BC_mfRm}), $\cH_{\pm}^n$ and $\cK_{\pm}^n$
each satisfy the following boundary conditions on the sidewalls of the curved pipe,
\begin{alignat}{6}
  &\cH_{+}^{n}(r_{a,b},r',\vsig)
  &&=\cH_{+}^{n}(r,r_{a,b},\vsig)
  &&=0 ,
    \qquad
  &&[\brd_{\hr}\cH_{-}^{n}(r,r',\vsig)]_{r=r_{a,b}}
  &&=[\brd_{\hr'}\cH_{-}^{n}(r,r',\vsig)]_{r'=r_{a,b}}
  &&=0 ,
   \label{eq:BC_cHpm}
  \\
  &\cK_{+}^{n}(r_{a,b},r',\vsig)
  &&=\cK_{-}^{n}(r,r_{a,b},\vsig)
  &&=0 ,
    \qquad
  &&[\brd_{\hr}\cK_{-}^{n}(r,r',\vsig)]_{r=r_{a,b}}
  &&=[\brd_{\hr'}\cK_{+}^{n}(r,r',\vsig)]_{r'=r_{a,b}}
  &&=0 .
   \label{eq:BC_cKpm}
\end{alignat}
The asymptotic limits of Eqs.(\ref{eq:BC_cHpm}) for $\rho\to\infty$ are
given by Eqs.(\ref{eq:cGpm_st_BC}).

\clearpage

\section{Complete form of the fields}
\label{sec:normal_form}

The Fourier coefficients of the fields $\cE^n$ and $\cB^n$ are given by
Eqs.(\ref{eq:cEBy_sol}) and (\ref{eq:cExs_cBxs_sprt}).
They are given using the Green functions of {\it the separated form}
($\cG_{\pm}^n$, $\cH_{\pm}^n$, $\cK_{\pm}^n$) given by
Eqs.(\ref{eq:cGp_sep}-\ref{eq:cGm_sep}) and (\ref{eq:cHpm}).
They consist of the two kinds of the modes with respect to $\vsig$:
the oscillatory modes $\sin(\hnu\vsig/\rho)$ and
the damped modes $e^{-\cnu\bar{|\vsig|}/\rho}$ where
$\hnu=(\hnu_m^n,\hmu_m^n)\in\mathbb{R}_0^{+}$ and
$\cnu=(\cnu_m^n,\cmu_m^n)\in\mathbb{R}^{+}$
as shown in Eqs.(\ref{eq:ps_poles}).
In section \ref{sec:normal_form} we will derive the Green functions of
{\it the complete form} ($\cC_{\pm}^n$, $\cU_{\pm}^n$, $\cV_{\pm}^n$)
which are useful in the analytical calculation of the fields unlike
the separated form functions which are usable in the numerical calculation.

\subsection{Complete form of the Fourier coefficients of the vertical fields}
\label{sec:cCpm}

We first derive the Green functions of the complete form $\cC_{\pm}^n$ for
$\cE_y^n$ and $\cB_y^n$ by rewriting the separated form $\cG_{\pm}^n$
so that $\cG_{\pm}^n$ has the complete sum over all the radial modes.
In the brackets of Eqs.(\ref{eq:cGp_sep}-\ref{eq:cGm_sep}), adding and subtracting
$\btheta(\vsig)e^{\cnu\vsig/\rho}$ for $\cnu=(\cnu_m^n,\cmu_m^n)\in\mathbb{R}^{+}$,
which diverges exponentially for $\vsig\to\infty$,
we rewrite $\cG_{\pm}^n$ as
\begin{align}
  \cG_{\pm}^n(r,r',\vsig)
  &=\cC_{\pm}^n(r,r',\vsig)-\bcC_{\pm}^n(r,r',\vsig) .
  \label{eq:cGpm_cCpm_bcCpm}
\end{align}
$\bcC_{\pm}^n$ denotes the following functions which diverge exponentially for
$\vsig\to\infty$,
\begin{align}
  \bcC_{+}^n(r,r',\vsig)
  &=\frac{1}{2}\sum_{\ell=0}^{\infty}\Th_{+}^{n\ell}\sum_{m=\ell'}^{\infty}
     \cR_{+}^{mn}(r,r')\frac{\rho}{\cnu_m^n}e^{\cnu_m^n\vsig/\rho} ,
   \label{eq:bcCp}
   \\
  \bcC_{-}^n(r,r',\vsig)
  &=\frac{1}{2}\sum_{\ell=0}^{\infty}\Th_{-}^{n\ell}\sum_{m=\ell}^{\infty}
     \cR_{-}^{mn}(r,r')\frac{\rho}{\cmu_m^n}e^{\cmu_m^n\vsig/\rho} .
   \label{eq:bcCm}
\end{align}
$\Th_{\pm}^{n\ell}$ and $\ell'$ are given by Eqs.(\ref{eq:Th_def}-\ref{eq:bTh_def})
and (\ref{eq:lprm}).
$\bcC_{\pm}^n$ will vanish as shown later in Eq.(\ref{eq:cEy_cBy_cmpl}).
$\cC_{\pm}^n$ is given as
\begin{align}
  \cC_{+}^n(r,r',\vsig)
  &=\btheta(\vsig)
    \sum_{\ell=0}^{\infty}\Th_{+}^{n\ell}
     \bigg[\bdelta_{01}^{\ell}\sum_{m=1}^{\ell-1}\frac{\cR_{+}^{mn}(r,r')}{\hnu_m^n/\rho}
           \sin\!\bigg(\frac{\hnu_m^n\vsig}{\rho}\bigg)
          +\sum_{m=\ell'}^{\infty}\frac{\cR_{+}^{mn}(r,r')}{i\cnu_m^n/\rho}
           \sin\!\bigg(\frac{i\cnu_m^n\vsig}{\rho}\bigg)
     \bigg] ,
  \label{eq:cGp_sin_sinh}
  \\
  \cC_{-}^n(r,r',\vsig)
  &=\btheta(\vsig)
    \sum_{\ell=0}^{\infty}\Th_{-}^{n\ell}
     \bigg[\bdelta_{0}^{\ell}\sum_{m=0}^{\ell-1}\frac{\cR_{-}^{mn}(r,r')}{\hmu_m^n/\rho}
           \sin\!\bigg(\frac{\hmu_m^n\vsig}{\rho}\bigg)
          +\sum_{m=\ell}^{\infty}\frac{\cR_{-}^{mn}(r,r')}{i\cmu_m^n/\rho}
           \sin\!\bigg(\frac{i\cmu_m^n\vsig}{\rho}\bigg)
     \bigg] .
  \label{eq:cGm_sin_sinh}
\end{align}
$\bdelta_{01}^{\ell}$ and $\bdelta_{0}^{\ell}$ are given by Eqs.(\ref{eq:d01_d0}).
Using the following identities which hold for an arbitrary summand $l_m$,
\begin{align}
  \sum_{m=1}^{\infty}l_m
  &=\sum_{\ell=0}^{\infty}\Th_{+}^{n\ell}
    \bigg\{
      \bdelta_{01}^{\ell}\sum_{m=1}^{\ell-1}l_m
     +\sum_{m=\ell'}^{\infty}\! l_m
    \bigg\} ,
    \qquad
  \sum_{m=0}^{\infty}l_m
  =\sum_{\ell=0}^{\infty}\Th_{-}^{n\ell}
    \bigg\{
       \bdelta_{0}^{\ell}\sum_{m=0}^{\ell-1}l_m
      +\sum_{m=\ell}^{\infty}l_m
    \bigg\} ,
  \label{eq:sum_Xmn}
\end{align}
we rewrite Eqs.(\ref{eq:cGp_sin_sinh}-\ref{eq:cGm_sin_sinh}) so that they have
the complete sum with respect to $m$ from the lowest mode to $\infty$,
which is not separated into the real poles $(\hnu_m^n,\hmu_m^n)$ and
the imaginary poles $(\cnu_m^n,\cmu_m^n)$.
Then we get the Green functions of the complete form $\cC_{+}^n$ and $\cC_{-}^n$
for the vertical components of the fields $\cE_y^n$ and $\cB_y^n$,
\begin{align}
  \cC_{+}^n(r,r',\vsig)
  &=\btheta(\vsig)
    \sum_{m=1}^{\infty}\cR_{+}^{mn}(r,r')
    \frac{\rho}{\nu_m^n}\sin\!\bigg(\frac{\nu_m^n\vsig}{\rho}\bigg) ,
  \label{eq:cCp}
   \\
  \cC_{-}^n(r,r',\vsig)
  &=\btheta(\vsig)
    \sum_{m=0}^{\infty}\cR_{-}^{mn}(r,r')
    \frac{\rho}{\mu_m^n}\sin\!\bigg(\frac{\mu_m^n\vsig}{\rho}\bigg) .
  \label{eq:cCm}
\end{align}
$\cC_{\pm}^n$ also diverges exponentially for $\vsig\to\infty$
since $\nu=(\nu_m^n,\mu_m^n)\in i\mathbb{R}$ when $m\geq m_{\pm}+1$ as shown in
Eqs.(\ref{eq:ps_poles}).
Therefore the complete form $\cC_{\pm}^n$ is not suited to
the numerical calculation of the fields.

We review the derivation of Eq.(\ref{eq:cGpm_cCpm_bcCpm}).
Let $R^{\pm}$ be the sums of the residues of the integrand in
Eqs.(\ref{eq:cGe_Th}) at the real poles on the positive and negative sides, \ie,
the sign $(\pm)$ of $R^{\pm}$ denotes the sides of the real poles:
$\Re\nu>0$ and $\Re\nu<0$.
Similarly, let $I^{\pm}$ be the sums of the residues at
the positive and negative imaginary poles: $\Im\nu>0$ and $\Im\nu<0$.
The sign of $\cG_{\pm}^n$ has nothing to do with that of $R^{\pm}$ and $I^{\pm}$.
Using the residue theorem to the Bromwich integral (\ref{eq:cGe_Th}),
we got $\cG_{\pm}^n$ as shown in Eq.(\ref{eq:cGcC})
which is the separated form given by rearranging
Eqs.(\ref{eq:Imn_done}-\ref{eq:bImn_done}).
Adding and subtracting $\btheta(\vsig)I^{-}$ in Eq.(\ref{eq:cGcC}),
we created $\cC_{\pm}^n$ which is the complete sum of the residues at
all the poles in the $\nu$-plane as shown in Eq.(\ref{eq:cG_cC}).
$\bcC_{\pm}^n$ is the remainder $I^{-}$.
\begin{alignat}{2}
  \cG_{\pm}^n(\vsig)
  &=\btheta(\vsig)\ob{ (R^{+}+R^{-}) }^{\sin(\hnu\vsig/\rho)}
   +\ob{ \btheta(\vsig)I^{+}-\btheta(-\vsig)I^{-} }^{e^{-\cnu\bar{|\vsig|}/\rho}}
    \qquad&&
   \text{: Eqs.(\ref{eq:cGp_sep}-\ref{eq:cGm_sep}) $=$ separated form}
   \label{eq:cGcC}
   \\
  &=\ub{\btheta(\vsig)(R^{+}+R^{-}+I^{+}+I^{-})}_{\cC_{\pm}^n}
   -\ub{I^{-}_{} }_{\bcC_{\pm}^n}
     \qquad&&
    \text{: Eq.(\ref{eq:cGpm_cCpm_bcCpm}) $=$ complete form $-$ $\bcC_{\pm}^n$} .
   \label{eq:cG_cC}
\end{alignat}
Thus, by making the divergent term $\bcC_{\pm}^n$ in $\cG_{\pm}^n$,
we got the Green functions of the complete form $\cC_{\pm}^n$.
Therefore both $\cC_{\pm}^n$ and $\bcC_{\pm}^n$ diverge exponentially for $\vsig\to\infty$.
$\cG_{\pm}^n$ does not diverge for $\vsig\to\infty$
as seen from Eqs.(\ref{eq:cGp_sep}-\ref{eq:cGm_sep}).
Substituting Eq.(\ref{eq:cGpm_cCpm_bcCpm}) into Eq.(\ref{eq:cEBy_sol}),
$\bcC_{\pm}^n(s)$ and $\bcC_{\pm}^n(s-s')$ cancel out
and vanish through Eq.(\ref{eq:wen_bend}).
Then we get the Fourier coefficients of the vertical components of the fields in
the complete form,
\begin{align}
  \bigg\{{ \cE_y^n(r,s) \atop c\cB_y^n(r,s) }\bigg\}
  &=\int_{r_a}^{r_b}dr'
    \bigg[
       \bigg\{{ \hcP_y^n(r,r',s) \atop \hcQ_y^n(r,r',s) }\bigg\}
      +Z_0\int_0^{\infty}ds'
       \bigg\{{ \hcM_y^n(r,r',s-s',s') \atop \hcN_y^n(r,r',s-s',s') }\bigg\}
    \bigg] .
  \label{eq:cEy_cBy_cmpl}
\end{align}
Eq.(\ref{eq:cEy_cBy_cmpl}) does not involve $\bcC_{\pm}^n$.
The first and second integrands of Eq.(\ref{eq:cEy_cBy_cmpl}) are given as follows,
\begin{alignat}{2}
  \hcP_y^n(r,r',s)
  &=\cD_y^n(r')\cC_{+}^n(r,r',s) ,
    \qquad&
  \hcM_y^n(r,r',\vsig,s')
  &=\frac{r'}{\rho}\cS_y^n(r',s')\cC_{+}^n(r,r',\vsig) ,
   \label{eq:hcPy_hcMy}
   \\
  \hcQ_y^n(r,r',s)
  &=\cA_y^n(r')\cC_{-}^n(r,r',s) ,
    \qquad&
  \hcN_y^n(r,r',\vsig,s')
  &=\frac{r'}{\rho}\cT_y^n(r',s')\cC_{-}^n(r,r',\vsig) .
   \label{eq:hcQy_hcNy}
\end{alignat}
Although $\cC_{\pm}^n$ has terms which diverge exponentially for $s\to\infty$,
Eq.(\ref{eq:cEy_cBy_cmpl}) does not diverge for $s\to\infty$,
because the divergent terms, included in $\cC_{\pm}^n(s)$ and $\cC_{\pm}^n(s-s')$,
cancel out and vanish through Eq.(\ref{eq:wen_bend}).
On the other hand, the vertical fields in the separated form (\ref{eq:cEBy_sol}) have
no terms which diverge for $s\to\infty$.

\subsection{Properties of the Green functions of the complete form}

According to Eqs.(\ref{eq:BC_mfRp}-\ref{eq:BC_mfRm}), $\cC_{\pm}^n$ satisfies
the following boundary conditions,
\begin{align}
  &
  \cC_{+}^n(r_{b,a},r',\vsig)
  =\cC_{+}^n(r,r_{b,a},\vsig)
  =0,
    \qquad
  [\rd_{\hr}\cC_{-}^n(r,r',\vsig)]_{r=r_{b,a}}
  =[\rd_{\hr'}\cC_{-}^n(r,r',\vsig)]_{r'=r_{b,a}}
  =0 .
   \label{eq:cCpm_BCr}
\end{align}
From Eqs.(\ref{eq:cCp}-\ref{eq:cCm}), we get the derivative of $\cC_{\pm}^n$
with respect to $s$,
\begin{align}
  \rd_s\cC_{+}^n(r,r',\vsig)
  &=\btheta(\vsig)\sum_{m=1}^{\infty}\cR_{+}^{mn}(r,r')
    \cos\!\bigg(\frac{\nu_m^n\vsig}{\rho}\bigg) ,
   \\
  \rd_s\cC_{-}^n(r,r',\vsig)
  &=\btheta(\vsig)\sum_{m=0}^{\infty}\cR_{-}^{mn}(r,r')
    \cos\!\bigg(\frac{\mu_m^n\vsig}{\rho}\bigg) .
\end{align}
Using the first equation of (\ref{eq:rdel}), we get the second derivative of
$\cC_{\pm}^n$ with respect to $s$,
\begin{align}
  \rd_s^2\cC_{+}^n(r,r',\vsig)
  &=\frac{r}{\rho}\delta(r-r')\bdelta(\vsig)
   -\btheta(\vsig)\sum_{m=1}^{\infty}\cR_{+}^{mn}(r,r')
    \frac{\nu_m^n}{\rho}\sin\!\bigg(\frac{\nu_m^n\vsig}{\rho}\bigg) ,
   \\
  \rd_s^2\cC_{-}^n(r,r',\vsig)
  &=\frac{r}{\rho}\delta(r-r')\bdelta(\vsig)
   -\btheta(\vsig)\sum_{m=0}^{\infty}\cR_{-}^{mn}(r,r')
    \frac{\mu_m^n}{\rho}\sin\!\bigg(\frac{\mu_m^n\vsig}{\rho}\bigg) .
\end{align}
$\btheta(\vsig)$ and $\bdelta(\vsig)$ are given by Eqs.(\ref{eq:bstep}-\ref{eq:bdelta}).
$\cC_{\pm}^n$ and their longitudinal derivatives at $\vsig=0$ are gotten as
\begin{align}
  [\cC_{\pm}^n(r,r',\vsig)]_{\vsig=0}=0
   ,\qquad
  [\rd_s\cC_{\pm}^n(r,r',\vsig)]_{\vsig=0}
  =\frac{r}{\rho}\delta(r-r')
   ,\qquad
  [\rd_s^2\cC_{\pm}^n(r,r',s)]_{s=0}
  =0 .
  \label{eq:rds_cCpm_vs0}
\end{align}
$\cC_{\pm}^n$ and $\bcC_{\pm}^n$ satisfy the wave equations (\ref{eq:wen_bend}) in which
the source terms are replaced by the $\delta$-function or 0,
\begin{align}
  \rd_{\rm v}^2\cC_{\pm}^n(r,r',\vsig)=\frac{\rho}{r}\delta(r-r')\bdelta(\vsig) ,
    \qquad
  \rd_{\rm v}^2\bcC_{\pm}^n(r,r',\vsig)=0 .
  \label{eq:we_cCpm}
\end{align}
$\rd_{\rm v}^2$ is the operator given by Eq.(\ref{eq:wen_bend}).
Since $\bdelta(s)=0$, 
$\cC_{\pm}^n(s)$ satisfies the homogeneous wave equation,
similar to $\cG_{\pm}^n(s)$ in Eq.(\ref{eq:BessDE_delta}).
According to Eqs.(\ref{eq:BDE_cRp}-\ref{eq:BDE_cRm}), Eqs.(\ref{eq:we_cCpm}) hold
also for the operator with respect to $r'$ instead of $r$ in $\rd_{\rm v}^2$.
According to Eqs.(\ref{eq:we_cCpm}), the integrands
$(\hcP_y^n,\hcQ_y^n)$ and $(\hcM_y^n,\hcN_y^n)$,
given by Eqs.(\ref{eq:hcPy_hcMy}-\ref{eq:hcQy_hcNy}),
satisfy the following wave equations, similar to Eqs.(\ref{eq:we_cPQy_cMNy}),
\begin{align}
  \rd_{\rm v}^2\bigg\{{ \hcP_y^n(s) \atop \hcQ_y^n(s) }\bigg\}
  =0 ,
     \qquad
  \rd_{\rm v}^2\bigg\{{ \hcM_y^n(\vsig,s') \atop \hcN_y^n(\vsig,s') }\bigg\}
  =\bigg\{{ \cS_y^n(s') \atop \cT_y^n(s') }\bigg\}
   \delta(r-r')\delta(\vsig) ,
   \label{eq:we_hcPQy_hcMNy}
\end{align}
where we omit the radial arguments $(r,r')$ of the functions except for $\delta(r-r')$.
According to Eqs.(\ref{eq:we_hcPQy_hcMNy}), Eq.(\ref{eq:cEy_cBy_cmpl}) satisfies
Eq.(\ref{eq:wen_bend}) which is the wave equation for $\cE_y^n$ and $\cB_y^n$.
According to Eqs.(\ref{eq:lim_nurho_ksmn}) and (\ref{eq:lim_RX}),
\begin{align}
  \lim_{\rho\to\infty}\cC_{\pm}^n(r,r',\vsig)
  =\cC_{\pm}^n(x,x',\vsig) .
\end{align}
Thus, in the limit of $\rho\to\infty$, $\cC_{\pm}^n(r,r',\vsig)$ goes to
$\cC_{\pm}^n(x,x',\vsig)$ given by Eqs.(\ref{eq:cCp_str}-\ref{eq:cCm_str})
which are the Fourier coefficients of the complete form of the Green functions of
the straight rectangular pipe.

\subsection{Complete form of the Fourier coefficients of the radial and longitudinal fields}

We derive the complete form of $\cE_{x,s}^n$ and $\cB_{x,s}^n$ from
Eq.(\ref{eq:cExs_cBxs_sprt}), similar to deriving Eq.(\ref{eq:cEy_cBy_cmpl}).
We first rewrite Eqs.(\ref{eq:cHpm})
so that they have the complete sums with respect to $m$
in a similar way to Eqs.(\ref{eq:cGpm_cCpm_bcCpm}) and (\ref{eq:cG_cC}),
\begin{align}
  \cH_{\pm}^n(r,r',\vsig)=\cU_{\pm}^n(r,r',\vsig)-\bcU_{\pm}^n(r,r',\vsig) ,
    \qquad
  \cK_{\pm}^n(r,r',\vsig)=\cV_{\pm}^n(r,r',\vsig)-\bcV_{\pm}^n(r,r',\vsig) .
  \label{eq:cHK_cUV_bcUV}
\end{align}
$\cU_{+}^n$ and $\cU_{-}^n$ are the Green functions of the complete form for
$(\cE_s^n,\cB_x^n)$ and $(\cE_x^n,\cB_s^n)$ respectively.
$\cV_{\pm}^n$ denotes the coupling Green functions of the complete form
between the $(r,s)$-components of the fields in the bend,
\begin{align}
  \bigg\{{\cU_{\pm}^n \atop \bcU_{\pm}^n }\bigg\}
  &=\frac{\hr}{\hr'}(\brd_{\hr}\rd_{\hr}+1)
    \bigg\{{\cC_{\pm}^n \atop \bcC_{\pm}^n }\bigg\}
   +\rd_{\hr}\rd_{\hr'}\bigg\{{\cC_{\mp}^n \atop \bcC_{\mp}^n }\bigg\} ,
  \label{eq:cUpm}
   \\
  \bigg\{{\cV_{\pm}^n \atop \bcV_{\pm}^n }\bigg\}
  &=\rho\rd_s
    \bigg[
       \frac{\rd_{\hr'}}{\hr}\bigg\{{\cC_{\pm}^n \atop \bcC_{\pm}^n }\bigg\}
      +\frac{\rd_{\hr}}{\hr'}\bigg\{{\cC_{\mp}^n \atop \bcC_{\mp}^n }\bigg\}
    \bigg] .
  \label{eq:cVpm}
\end{align}
$\cC_{\pm}^n$ and $\bcC_{\pm}^n$ are given by Eqs.(\ref{eq:cCp}-\ref{eq:cCm})
and (\ref{eq:bcCp}-\ref{eq:bcCm}).
$\cU_{\pm}^n$ and $\cV_{\pm}^n$ have the complete sums with respect to $m$,
\begin{align}
  \cU_{+}^n(r,r',\vsig)
  &=\btheta(\vsig)
    \bigg\{
      \sum_{m=0}^{\infty}\frac{\rho}{\mu_m^n}\rd_{\hr}\rd_{\hr'}\cR_{-}^{mn}(r,r')
      \sin\!\bigg(\frac{\mu_m^n\vsig}{\rho}\bigg)
     +\sum_{m=1}^{\infty}\frac{\rho\nu_m^n}{\hr\hr'}\cR_{+}^{mn}(r,r')
      \sin\!\bigg(\frac{\nu_m^n\vsig}{\rho}\bigg)
    \bigg\} ,
   \label{eq:cUp}
  \\
  \cU_{-}^n(r,r',\vsig)
  &=\btheta(\vsig)
    \bigg\{
      \sum_{m=1}^{\infty}\frac{\rho}{\nu_m^n}\rd_{\hr}\rd_{\hr'}\cR_{+}^{mn}(r,r')
      \sin\!\bigg(\frac{\nu_m^n\vsig}{\rho}\bigg)
     +\sum_{m=0}^{\infty}\frac{\rho\mu_m^n}{\hr\hr'}\cR_{-}^{mn}(r,r')
      \sin\!\bigg(\frac{\mu_m^n\vsig}{\rho}\bigg)
    \bigg\} ,
   \label{eq:cUm}
   \\
  \cV_{+}^n(r,r',\vsig)
  &=\btheta(\vsig)
    \bigg\{
      \sum_{m=1}^{\infty}\frac{\rho}{\hr}\rd_{\hr'}\cR_{+}^{mn}(r,r')
      \cos\!\bigg(\frac{\nu_m^n\vsig}{\rho}\bigg)
     +\sum_{m=0}^{\infty}\frac{\rho}{\hr'}\rd_{\hr}\cR_{-}^{mn}(r,r')
      \cos\!\bigg(\frac{\mu_m^n\vsig}{\rho}\bigg)
    \bigg\} ,
   \label{eq:cVp}
  \\
  \cV_{-}^n(r,r',\vsig)
  &=\btheta(\vsig)
    \bigg\{
      \sum_{m=0}^{\infty}\frac{\rho}{\hr}\rd_{\hr'}\cR_{-}^{mn}(r,r')
      \cos\!\bigg(\frac{\mu_m^n\vsig}{\rho}\bigg)
     +\sum_{m=1}^{\infty}\frac{\rho}{\hr'}\rd_{\hr}\cR_{+}^{mn}(r,r')
      \cos\!\bigg(\frac{\nu_m^n\vsig}{\rho}\bigg)
    \bigg\} .
   \label{eq:cVm}
\end{align}
$\bcU_{\pm}^n$ and $\bcV_{\pm}^n$ have sums which are truncated at $m=\ell$ or 
$\ell'$ $(=\ell+\delta_0^\ell)$ given by Eq.(\ref{eq:lprm}),
\begin{align}
  \bcU_{+}^n(r,r',\vsig)
  &=    
    \frac{1}{2}\sum_{\ell=0}^{\infty}
    \bigg\{
      \Th_{-}^{n\ell}\sum_{m=\ell}^{\infty}
      \frac{\rho}{\cmu_m^n}\rd_{\hr}\rd_{\hr'}\cR_{-}^{mn}(r,r')e^{\cmu_m^n\vsig/\rho}
     -\Th_{+}^{n\ell}\sum_{m=\ell'}^{\infty}
      \frac{\rho\cnu_m^n}{\hr\hr'}\cR_{+}^{mn}(r,r')e^{\cnu_m^n\vsig/\rho}
    \bigg\} ,
  \label{eq:bcUp}
   \\
  \bcU_{-}^n(r,r',\vsig)
  &=
    \frac{1}{2}\sum_{\ell=0}^{\infty}
    \bigg\{
      \Th_{+}^{n\ell}\sum_{m=\ell'}^{\infty}
      \frac{\rho}{\cnu_m^n}\rd_{\hr}\rd_{\hr'}\cR_{+}^{mn}(r,r')e^{\cnu_m^n\vsig/\rho}
     -\Th_{-}^{n\ell}\sum_{m=\ell}^{\infty}
      \frac{\rho\cmu_m^n}{\hr\hr'}\cR_{-}^{mn}(r,r')e^{\cmu_m^n\vsig/\rho}
    \bigg\} ,
   \\
  \bcV_{+}^n(r,r',\vsig)
  &=\frac{1}{2}\sum_{\ell=0}^{\infty}
    \bigg\{
       \Th_{+}^{n\ell}\sum_{m=\ell'}^{\infty}
       \frac{\rho}{\hr}\rd_{\hr'}\cR_{+}^{mn}(r,r')e^{\cnu_m^n\vsig/\rho}
      +\Th_{-}^{n\ell}\sum_{m=\ell}^{\infty}
       \frac{\rho}{\hr'}\rd_{\hr}\cR_{-}^{mn}(r,r')e^{\cmu_m^n\vsig/\rho}
    \bigg\} ,
   \\
  \bcV_{-}^n(r,r',\vsig)
  &=\frac{1}{2}\sum_{\ell=0}^{\infty}
    \bigg\{
      \Th_{-}^{n\ell}\sum_{m=\ell}^{\infty}
      \frac{\rho}{\hr}\rd_{\hr'}\cR_{-}^{mn}(r,r')e^{\cmu_m^n\vsig/\rho}
     +\Th_{+}^{n\ell}\sum_{m=\ell'}^{\infty}
      \frac{\rho}{\hr'}\rd_{\hr}\cR_{+}^{mn}(r,r')e^{\cnu_m^n\vsig/\rho}
    \bigg\} .
  \label{eq:bcVm}
\end{align}
According to Eq.(\ref{eq:cRpm_symm}), Eqs.(\ref{eq:cUpm}-\ref{eq:cVpm}) have
the following symmetry with respect to the exchange of $r$ and $r'$,
\begin{alignat}{2}
  \cU_{\pm}^n(r,r',\vsig)&=\cU_{\pm}^n(r',r,\vsig) ,
    \qquad&
  \cV_{\pm}^n(r,r',\vsig)&=\cV_{\mp}^n(r',r,\vsig) ,
    \\
  \bcU_{\pm}^n(r,r',\vsig)&=\bcU_{\pm}^n(r',r,\vsig) ,
    \qquad&
  \bcV_{\pm}^n(r,r',\vsig)&=\bcV_{\mp}^n(r',r,\vsig) .
\end{alignat}
Similar to Eq.(\ref{eq:Grn_cHpm_cKpm}),
Eqs.(\ref{eq:cUpm}-\ref{eq:cVpm}) satisfy the following wave equations,
\begin{align}
  \rd_{\,\vdash}^2
  \bigg\{{\cU_{\pm}^n(\vsig) \atop \bcU_{\pm}^n(\vsig)}\bigg\}
  -\frac{2}{r}\brd_s
  \bigg\{{ \cV_{\mp}^n(\vsig) \atop \bcV_{\mp}^n(\vsig)}\bigg\}
  &=\bigg\{{1 \atop 0}\bigg\}
   \frac{\rho}{r}\delta(r-r')\bdelta(\vsig) ,
  \label{eq:we_cUpm}
   \\
  \rd_{\,\vdash}^2\bigg\{{\cV_{\pm}^n(\vsig) \atop \bcV_{\pm}^n(\vsig)}\bigg\}
  +\frac{2}{r}\brd_s
   \bigg\{{ \cU_{\mp}^n(\vsig) \atop \bcU_{\mp}^n(\vsig)}\bigg\}
  &=0 .
  \label{eq:we_cVpm}
\end{align}
$\bcU_{\pm}^n$ and $\bcV_{\pm}^n$ satisfy the homogeneous wave equations
since they do not have $\theta(\vsig)$.
In the 16 Green functions given by Eqs.(\ref{eq:cUpm}-\ref{eq:cVpm}) for $\vsig=\{s-s',s\}$,
only $\cU_{\pm}^n(s-s')$ satisfies the inhomogeneous wave equation.
$\cU_{\pm}^n(s)$ satisfies the homogeneous wave equation since $\bdelta(s)=0$ as in
Eq.(\ref{eq:bdelta}).
Similar to Eqs.(\ref{eq:cHcK_rel}-\ref{eq:cKcH_rel}),
$\cU_{\pm}^n$ and $\cV_{\pm}^n$ have the following relations for
both $\vsig=s-s'$ and $s$,
\begin{alignat}{2}
  \frac{\rho}{\hr}\rd_s\cU_{\pm}^n(\vsig)-\brd_{\hr}\cV_{\mp}^{n}(\vsig)
  &=\frac{\rho}{\hr'}\rd_s\cC_{\pm}^n(\vsig) ,
    \qquad&
  \brd_{\hr}\cU_{\mp}^{n}(\vsig)+\frac{\rho}{\hr}\rd_s\cV_{\pm}^{n}(\vsig)
  &=-\rd_{\hr'}\cC_{\pm}^n(\vsig) ,
   \label{eq:cUVC}
   \\
  \frac{\rho}{\hr'}\rd_s\cU_{\pm}^n(\vsig)-\brd_{\hr'}\cV_{\pm}^n(\vsig)
  &=\frac{\rho}{\hr}\rd_s\cC_{\pm}^n(\vsig) ,
    \qquad&
  \brd_{\hr'}\cU_{\mp}^n(\vsig)+\frac{\rho}{\hr'}\rd_s\cV_{\mp}^n(\vsig)
  &=-\rd_{\hr}\cC_{\pm}^n(\vsig) .
   \label{eq:cUcVcC}
\end{alignat}
$\cU_{\pm}^n$ and $\cV_{\pm}^n$ satisfy the following boundary conditions on
the sidewalls of the curved pipe,
\begin{alignat}{2}
  \cU_{+}^{n}(r_{a,b},r',\vsig)
  &=\cU_{+}^{n}(r,r_{a,b},\vsig)
  =0 ,
   \qquad&
  [\brd_{\hr}\cU_{-}^{n}(r,r',\vsig)]_{r=r_{a,b}}
  &=[\brd_{\hr'}\cU_{-}^{n}(r,r',\vsig)]_{r'=r_{a,b}}
  =0 ,
  \label{eq:BC_cUp_r}
  \\
  \cV_{+}^{n}(r_{a,b},r',\vsig)
  &=\cV_{-}^{n}(r,r_{a,b},\vsig)
  =0 ,
    \qquad&
  [\brd_{\hr}\cV_{-}^{n}(r,r',\vsig)]_{r=r_{a,b}}
  &=[\brd_{\hr'}\cV_{+}^{n}(r,r',\vsig)]_{r'=r_{a,b}}
  =0 .
  \label{eq:BC_cVpm_r}
\end{alignat}
From Eqs.(\ref{eq:rds_cCpm_vs0}) and (\ref{eq:we_cUpm}), $\cU_{\pm}^n$, $\cV_{\pm}^n$
and their longitudinal derivatives at $\vsig=0$ are gotten as follows,
\begin{alignat}{3}
  \cU_{\pm}^n(r,r',0)&=0
   ,\qquad&
  [\rd_s\cU_{\pm}^n(r,r',\vsig)]_{\vsig=0}
  &=\frac{r}{\rho}\delta(r-r')
   ,\qquad&
  [\rd_s^2\cU_{\pm}^n(r,r',s)]_{s=0}
  &=0 ,
  \label{eq:cUVpm_s0}
   \\
  \cV_{\pm}^n(r,r',0)&=0
   ,\qquad&
  [\rd_s\cV_{\pm}^n(r,r',\vsig)]_{\vsig=0}
  &=0
   ,\qquad&
  [\rd_s^2\cV_{\pm}^n(r,r',\vsig)]_{\vsig=0}
  &=-\frac{2r}{\rho^2}\delta(r-r') ,
  \label{eq:rds_cUVpm_s0}
\end{alignat}
where $\vsig=\{s-s',s\}$.
Similar to the behavior of $\rd_s^2\cC_{\pm}^n(s-s')$ at $s=s'$,
$\rd_s^2\cU_{\pm}^n(s-s')$ diverges at $s=s'$ in theory
since $\rd_s^2\cU_{\pm}^n(\vsig)$ has $\bdelta(\vsig)$ given by Eq.(\ref{eq:bdelta}).
Similar to Eqs.(\ref{eq:lim_cH_cK}), $\cU_{\pm}^n$ and $\cV_{\pm}^n$ go to
the following limits for $\rho\to\infty$,
\begin{align}
  \lim_{\rho\to\infty}\cU_{\pm}^n(r,r',\vsig)
  =\cC_{\pm}^n(x,x',\vsig) ,
    \qquad
  \lim_{\rho\to\infty}\cV_{\pm}^n(r,r',\vsig)
  =0 .
   \label{eq:lim_Cpm}
\end{align}
$\cC_{\pm}^n(x,x',\vsig)$ is given by Eqs.(\ref{eq:cCp_str}-\ref{eq:cCm_str})
which are the complete form of the Green functions in the straight pipe.
The coupling Green functions $\cV_{\pm}^n$ vanish in the limit of $\rho\to\infty$.

We derive the complete forms of $\cE_{x,s}^n$ and $\cB_{x,s}^n$
from their separated forms given by Eq.(\ref{eq:cExs_cBxs_sprt}).
Substituting Eqs.(\ref{eq:cHK_cUV_bcUV}) into Eq.(\ref{eq:cExs_cBxs_sprt}),
since all the terms involving $\bcU_{\pm}^n$ and $\bcV_{\pm}^n$ vanish through
Eqs.(\ref{eq:we_cFxs}), we get
\begin{align}
  \bigg\{{\cE_{x,s}^n(r,s) \atop c\cB_{x,s}^n(r,s)}\bigg\}
  &=\int_{r_a}^{r_b}dr'
    \bigg[
      \bigg\{{\hcP_{x,s}^n(r,r',s) \atop \hcQ_{x,s}^n(r,r',s)}\bigg\}
     +Z_0\int_0^{\infty}ds'
      \bigg\{{\hcM_{x,s}^n(r,r',s-s',s') \atop \hcN_{x,s}^n(r,r',s-s',s')}\bigg\}
    \bigg] .
  \label{eq:cExs_cBxs_cmpl}
\end{align}
Similar to Eq.(\ref{eq:cEy_cBy_cmpl}) which does not have $\bcC_{\pm}^n$,
Eq.(\ref{eq:cExs_cBxs_cmpl}) does not have $\bcU_{\pm}^n$ and $\bcV_{\pm}^n$
since they are canceled out between the first and second terms in the bracket of
Eq.(\ref{eq:cExs_cBxs_sprt}) through Eqs.(\ref{eq:we_cFxs})
after substituting Eqs.(\ref{eq:cHK_cUV_bcUV}).
The integrands $(\hcP_{x,s}^n,\hcQ_{x,s}^n)$ and $(\hcM_{x,s}^n,\hcN_{x,s}^n)$
in Eq.(\ref{eq:cExs_cBxs_cmpl}) involve respectively the initial values of
$(\cE_{x,s}^n,\cB_{x,s}^n)$ at $s=0$ and
the source terms $(\cS_{x,s}^n,\cT_{x,s}^n)$ in $s>0$,
\begin{alignat}{2}
  \hcP_x^n(s)
  &=\cD_x^n\cU_{-}^n(s)-\cD_s^n\cV_{-}^n(s) ,
    \qquad
  &\hcM_x^n(\vsig,s')
  &=\frac{r'}{\rho}\{\cS_{x}^n(s')\cU_{-}^n(\vsig)-\cS_{s}^n(s')\cV_{-}^n(\vsig)\} ,
   \label{eq:hcPx_hcMx}
   \\
  \hcP_s^n(s)
  &=\cD_s^n\cU_{+}^n(s)+\cD_x^n\cV_{+}^n(s) ,
    \qquad
  &\hcM_s^n(\vsig,s')
  &=\frac{r'}{\rho}\{\cS_{s}^n(s')\cU_{+}^n(\vsig)+\cS_{x}^n(s')\cV_{+}^n(\vsig)\} ,
   \label{eq:hcPs_hcMs}
   \\
  \hcQ_x^n(s)
  &=\cA_x^n\cU_{+}^n(s)-\cA_s^n\cV_{+}^n(s) ,
    \qquad
  &\hcN_x^n(\vsig,s')
  &=\frac{r'}{\rho}\{\cT_{x}^n(s')\cU_{+}^n(\vsig)-\cT_{s}^n(s')\cV_{+}^n(\vsig)\} ,
   \label{eq:hcQx_hcNx}
   \\
  \hcQ_s^n(s)
  &=\cA_s^n\cU_{-}^n(s)+\cA_x^n\cV_{-}^n(s) ,
    \qquad
  &\hcN_s^n(\vsig,s')
  &=\frac{r'}{\rho}\{\cT_{s}^n(s')\cU_{-}^n(\vsig)+\cT_{x}^n(s')\cV_{-}^n(\vsig)\} .
   \label{eq:hcQs_hcNs}
\end{alignat}
In Eqs.(\ref{eq:hcPx_hcMx}-\ref{eq:we_hcMs_hcNs}) we omit the radial arguments $(r,r')$ of
the functions for clarity to show the longitudinal arguments.
$\vsig=s-s'$ ($\vsig\ne s$) in Eqs.(\ref{eq:hcPx_hcMx}-\ref{eq:we_hcMs_hcNs}).
$\cD_{x,s}^n(r')$ and $\cA_{x,s}^n(r')$ are the longitudinal operators given by
Eqs.(\ref{eq:bcEx_init}-\ref{eq:bcBs_init}).
$\cS_{x,s}^n(r',s')$ and $\cT_{x,s}^n(r',s')$, given by
Eqs.(\ref{eq:cSx_cTx_prm}-\ref{eq:cSs_cTs_prm}), are the source terms in
the wave equations (\ref{eq:we_cFxs}) for $\cE_{x,s}^n$ and $\cB_{x,s}^n$.
Eq.(\ref{eq:cExs_cBxs_cmpl}) is the expression of the fields in $s>0$,
which is related to the initial fields at $s=0$ through
Eqs.(\ref{eq:cUVpm_s0}-\ref{eq:rds_cUVpm_s0}).
That is, the complete forms of $\cE^n$ and $\cB^n$, given by
Eqs.(\ref{eq:cEy_cBy_cmpl}) and (\ref{eq:cExs_cBxs_cmpl}),
produce their initial values at $s=0$ through $\rd_s\cC_{\pm}^n(s)$ and
$\rd_s\cU_{\pm}^n(s)$ which go to $(r/\rho)\delta(r-r')$ for $s\to+0$
as shown in appendix \ref{sec:initial_field}.
This is one of the reasons that the complete form is useful in the analytical calculation 
of the fields, though it is not usable in the numerical calculation
since $\cC_{\pm}^n$ and $\cU_{\pm}^n$ diverge exponentially for $s\to\infty$.
According to Eq.(\ref{eq:we_cUpm}), Eqs.(\ref{eq:hcPx_hcMx}-\ref{eq:hcQs_hcNs}) satisfy 
the following wave equations, similar to Eqs.(\ref{eq:we_cPxs_cQxs}-\ref{eq:we_cMs_cNs}),
\begin{align}
  &
  \rd_{\vdash}^2\bigg\{{ \hcP_x^n(s) \atop \hcQ_x^n(s) }\bigg\}
  -\frac{2}{r}\brd_s\bigg\{{ \hcP_s^n(s) \atop \hcQ_s^n(s) }\bigg\}
  =0 ,
    \qquad
  \rd_{\vdash}^2\bigg\{{ \hcP_s^n(s) \atop \hcQ_s^n(s) }\bigg\}
  +\frac{2}{r}\brd_s\bigg\{{ \hcP_x^n(s) \atop \hcQ_x^n(s) }\bigg\}
  =0 ,
   \label{eq:we_hcPxs_hcQxs}
   \\
  &
  \rd_{\vdash}^2\bigg\{{ \hcM_x^n(\vsig,s') \atop \hcN_x^n(\vsig,s') }\bigg\}
  -\frac{2}{r}\brd_s\bigg\{{ \hcM_s^n(\vsig,s') \atop \hcN_s^n(\vsig,s') }\bigg\}
  =\bigg\{{ \cS_x^n(s') \atop \cT_x^n(s') }\bigg\}
    \delta(r-r')\delta(\vsig) ,
   \\
  &
  \rd_{\vdash}^2\bigg\{{ \hcM_s^n(\vsig,s') \atop \hcN_s^n(\vsig,s') }\bigg\}
  +\frac{2}{r}\brd_s\bigg\{{ \hcM_x^n(\vsig,s') \atop \hcN_x^n(\vsig,s') }\bigg\}
  =\bigg\{{ \cS_s^n(s') \atop \cT_s^n(s') }\bigg\}
    \delta(r-r')\delta(\vsig) .
   \label{eq:we_hcMs_hcNs}
\end{align}
According to Eqs.(\ref{eq:we_hcPxs_hcQxs}-\ref{eq:we_hcMs_hcNs}), 
$\cE_{x,s}^n$ and $\cB_{x,s}^n$ given by Eq.(\ref{eq:cExs_cBxs_cmpl}) satisfy
the wave equations (\ref{eq:we_cFxs}).

\subsection{Fourier coefficients of the Green functions and eigenfunctions}
\label{sec:Green}

We summarize the Green functions and the coupling Green functions of the wave equations
for $\cE^n$ and $\cB^n$.
Also, we mention the transverse eigenfunctions and the integrands
involved in the expressions of $\cE^n$ and $\cB^n$.
\begin{alignat}{2}
  \text{Separated form:}& \quad\text{(converging for $s\to\infty$)}
  && \text{Eqs.(\ref{eq:cEBy_sol}) and (\ref{eq:cExs_cBxs_sprt})}
  \nonumber
   \\
  \cG_{+}^n ~\text{and}~ \cG_{-}^n
  &=\text{Green functions of $\cE_y^n$ and $\cB_y^n$ }
    \qquad&& \text{Eqs.(\ref{eq:cGp_sep}-\ref{eq:cGm_sep})}
  \nonumber
   \\
  \cH_{+}^n ~\text{and}~ \cH_{-}^n
  &=\text{Green functions of $(\cE_s^n,\cB_x^n)$ and $(\cE_x^n,\cB_s^n)$ }
    \qquad&& \text{Eqs.(\ref{eq:cHpm}) and (\ref{eq:cHp}-\ref{eq:cHm})}
  \nonumber
   \\
  \cK_{+}^n ~\text{and}~ \cK_{-}^n
  &=\text{Coupling functions of $(\cE_s^n,\cB_x^n)$ and $(\cE_x^n,\cB_s^n)$ }
    \qquad&& \text{Eqs.(\ref{eq:cHpm}) and (\ref{eq:cKp}-\ref{eq:cKm})}
  \nonumber
   \\
  \text{Complete form:}& \quad\text{(diverging for $s\to\infty$)}
  && \text{Eqs.(\ref{eq:cEy_cBy_cmpl}) and (\ref{eq:cExs_cBxs_cmpl})}
  \nonumber
   \\
  \cC_{+}^n ~\text{and}~ \cC_{-}^n
  &=\text{Green functions of $\cE_y^n$ and $\cB_y^n$ }
    \qquad&& \text{Eqs.(\ref{eq:cCp}-\ref{eq:cCm})}
  \nonumber
   \\
  \cU_{+}^n ~\text{and}~ \cU_{-}^n
  &=\text{Green functions of $(\cE_s^n,\cB_x^n)$ and $(\cE_x^n,\cB_s^n)$ }
    \qquad&& \text{Eqs.(\ref{eq:cUpm}) and (\ref{eq:cUp}-\ref{eq:cUm})}
  \nonumber
   \\
  \cV_{+}^n ~\text{and}~ \cV_{-}^n
  &=\text{Coupling functions of $(\cE_s^n,\cB_x^n)$ and $(\cE_x^n,\cB_s^n)$ }
    \qquad&& \text{Eqs.(\ref{eq:cVpm}) and (\ref{eq:cVp}-\ref{eq:cVm})}
  \nonumber
   \\
  \text{Their difference:}
  &\quad \text{(vanishing functions in the complete form)}
  \nonumber
   \\
  \bcC_{\pm}^n
  &=\cC_{\pm}^n-\cG_{\pm}^n
    \qquad&& \text{Eqs.(\ref{eq:bcCp}-\ref{eq:bcCm})}
  \nonumber
   \\
  \bcU_{\pm}^n
  &=\cU_{\pm}^n-\cH_{\pm}^n ,
    \qquad
  \bcV_{\pm}^n
  =\cV_{\pm}^n-\cK_{\pm}^n
    \qquad&& \text{Eqs.(\ref{eq:cUpm}-\ref{eq:cVpm}) and (\ref{eq:bcUp}-\ref{eq:bcVm})}
   \nonumber
   \\
   \nonumber
   \\
  \text{First integrand:}& \quad\text{(initial fields at $s=0$)}
  \nonumber
   \\
  \cP_{x,y,s}^n
  &=\text{Separated form having $\cE_{x,y,s}^n(0)$}
    \qquad&& \text{Eqs.(\ref{eq:cPy_cMy}) and (\ref{eq:cPxn}-\ref{eq:cPsn})}
   \nonumber
   \\
  \cQ_{x,y,s}^n
  &=\text{Separated form having $\cB_{x,y,s}^n(0)$}
    \qquad&& \text{Eqs.(\ref{eq:cQy_cNy}) and (\ref{eq:cQxn}-\ref{eq:cQsn})}
   \nonumber
   \\
  \hcP_{x,y,s}^n
  &=\text{Complete form having $\cE_{x,y,s}^n(0)$}
    \qquad&& \text{Eqs.(\ref{eq:hcPy_hcMy}) and (\ref{eq:hcPx_hcMx}-\ref{eq:hcPs_hcMs})}
   \nonumber
   \\
  \hcQ_{x,y,s}^n
  &=\text{Complete form having $\cB_{x,y,s}^n(0)$}
    \qquad&& \text{Eqs.(\ref{eq:hcQy_hcNy}) and (\ref{eq:hcQx_hcNx}-\ref{eq:hcQs_hcNs})}
   \nonumber
   \\
  \text{Second integrand:}& \quad\text{(source terms in $s>0$)}
  \nonumber
   \\
  \cM_{x,y,s}^n
  &=\text{Separated form having $\cS_{x,y,s}^n$}
    \qquad&& \text{Eqs.(\ref{eq:cPy_cMy}) and (\ref{eq:cPxn}-\ref{eq:cPsn})}
   \nonumber
   \\
  \cN_{x,y,s}^n
  &=\text{Separated form having $\cT_{x,y,s}^n$}
    \qquad&& \text{Eqs.(\ref{eq:cQy_cNy}) and (\ref{eq:cQxn}-\ref{eq:cQsn})}
   \nonumber
   \\
  \hcM_{x,y,s}^n
  &=\text{Complete form having $\cS_{x,y,s}^n$}
    \qquad&& \text{Eqs.(\ref{eq:hcPy_hcMy}) and (\ref{eq:hcPx_hcMx}-\ref{eq:hcPs_hcMs})}
   \nonumber
   \\
  \hcN_{x,y,s}^n
  &=\text{Complete form having $\cT_{x,y,s}^n$}
    \qquad&& \text{Eqs.(\ref{eq:hcQy_hcNy}) and (\ref{eq:hcQx_hcNx}-\ref{eq:hcQs_hcNs})}
   \nonumber
   \\
   \nonumber
   \\
  \text{Eigenfunctions:}& \quad\text{(rectangular pipe)}
  \nonumber
   \\
  \cR_{\pm}^{mn}(r,r')
  &=\text{Radial eigenfunctions of the curved pipe}
    \qquad&& \text{Eqs.(\ref{eq:Ven}-\ref{eq:mfRm}) and (\ref{eq:cRp_ba}-\ref{eq:cRm_ba})}
  \nonumber
   \\
  \cX_{\pm}^m(x,x')
  &=\text{Horizontal eigenfunctions of the straight pipe}
    \qquad&& \text{Eqs.(\ref{eq:mfXp_new}-\ref{eq:cXm}) and (\ref{eq:lim_RX})}
  \nonumber
   \\
  \cY_{\pm}^n(y,y')
  &=\text{Vertical eigenfunctions of the pipe}
    \qquad&& \text{Eqs.(\ref{eq:cYp}) and (\ref{eq:cYm})}
  \nonumber
\end{alignat}
In section \ref{sec:FD_field} we will define $\cY_{\pm}^n$ in deriving the expressions of 
the fields in the frequency domain $(\tEv,\tBv)$ from $(\cE^n,\cB^n)$.
We will define $\cX_{\pm}^n$ in appendix \ref{sec:travel_wave} where we discuss
a field created by an arbitrary beam moving in a straight rectangular pipe.
In the following table we summarize the symbols which represent the physical quantities
and the differential operators of the wave equations (W.E.) in each domain.
The operators $(\nablav_{\vdash}^2,\nablav_{\rm v}^2)$ and $(\rd_{\vdash}^2,\rd_{\rm v}^2)$
are given by Eqs.(\ref{eq:operator}) and (\ref{eq:wen_bend}).
$\hr$ and $\brd_{\hr}$ are defined in Eqs.(\ref{eq:krn}) and (\ref{eq:rd_hr}).

\begin{center}
\begin{tabular}{l||l|l|l|l}
  \hline
    & Time domain & Frequency domain & $y$-mode & Laplace domain $\ds\rule{0pt}{10pt}$ 
   \\ & ~~$(x,y,s,t)$ & ~~$(x,y,s,k)$ & ~~$(r,n,s,k)$ & ~~$(r,n,\nu,k)$
   \\ \hline\hline
  Fields  $\us{}{\rule{0pt}{12pt}}$
  & $E_{x,y,s}, B_{x,y,s}$
  & $\tE_{x,y,s}, \tB_{x,y,s}$
  & $\cE_{x,y,s}^n, \cB_{x,y,s}^n$
  & $\mfE_{x,y,s}^n, \mfB_{x,y,s}^n$
   \\ \hline
  Current  $\us{}{\rule{0pt}{12pt}}$
  & $J_0, J_{x,y,s}$
  & $\tJ_0, \tJ_{x,y,s}$
  & $\cJ_0^n, \cJ_{x,y,s}^n$
  & $\mfJ_0^n, \mfJ_{x,y,s}^n$
   \\ \hline
  Source terms  $\us{}{\rule{0pt}{12pt}}$
  & $S_{x,y,s}, T_{x,y,s}$
  & $\tS_{x,y,s}, \tT_{x,y,s}; \tS_{x,y,s}^{\dg}, \tT_{x,y,s}^{\dg}$
  & $\cS_{x,y,s}^n, \cT_{x,y,s}^n$
  & $\mfS_{x,y,s}^n, \mfT_{x,y,s}^n$
   \\ \hline
  Initial fields  $\us{}{\rule{0pt}{12pt}}$
  & $D_{x,y,s}, A_{x,y,s}$
  & $\tD_{x,y,s}, \tA_{x,y,s}; \tD_{x,y,s}^{\dg}, \tA_{x,y,s}^{\dg}$
  & $\cD_{x,y,s}^n, \cA_{x,y,s}^n$
  & $\mfD_{x,y,s}^n, \mfA_{x,y,s}^n$
   \\ \hline
  Green functions
  &  Not found  $\rule{0pt}{11pt}$ 
  & $\Phi_{x,y,s}, \Psi_{x,y,s}, \Phi_{x,s}^{\ast}, \Psi_{x,s}^{\ast}$
  & $\cG_{\pm}^n, \cH_{\pm}^n, \cK_{\pm}^n$
  & $\mfG_{\pm}^n, \mfH_{\pm}^n, \mfK_{\pm}^n$
   \\
  (complete form)
  &   $\us{}{\rule{0pt}{10pt}}$
  & ($\Gam_{x,y,s}, \Lam_{x,y,s}, \Gam_{x,s}^{\ast}, \Lam_{x,s}^{\ast}$)
  & ($\cC_{\pm}^n, \cU_{\pm}^n, \cV_{\pm}^n$)
  & (none)
   \\ \hline
  W.E. operators
  & $\ds \nablav^2-\frac{\rd_t^2}{c^2}$
  & $\nablav_{\vdash}^2, \nablav_{\rm v}^2$
  & $\rd_{\vdash}^2, \rd_{\rm v}^2$  $\us{\rule{0pt}{9pt}}{\rule{0pt}{18pt}}$
  & $\ds \bigg( {\rd_{\hr}\brd_{\hr} \atop \brd_{\hr}\rd_{\hr}} \bigg)
     +1-\frac{\nu^2}{\hr^2}$ 
   \\ \hline
\end{tabular}
\end{center}

\clearpage

\section{Fields in the frequency domain}
\label{sec:FD_field}

We find the expressions of the fields in the frequency domain $\tEv$ and $\tBv$
from their Fourier coefficients $\cE^n$ and $\cB^n$ through
Eqs.(\ref{eq:Four_Exp_plus}-\ref{eq:Four_Exp_minus}).
The vertical dimension of the system is limited to $|y|\leq h/2$
by the upper-lower walls of the pipe as shown in Fig.\ref{fig:pipe3D}.
But we have considered the extended range $|y|\leq h$ in order to expand the fields
and current in the Fourier modes of the period $2h$ with respect to $y$.
We first rewrite the Fourier coefficients of the source terms and the initial values of
the fields into those in the frequency domain through
Eqs.(\ref{eq:cFpn}) and (\ref{eq:cFmn}) using a dummy variable $y'$ for $y$.
The fields and source terms (\ref{eq:Four_Exp_plus}-\ref{eq:Four_Exp_minus}) have
the following symmetries with respect to $y=\pm h/2$, depending on their vertical parities,
\begin{alignat}{2}
  \tF_{+}(\pm h-y)
  &=-\tF_{+}(y) ,
   \qquad&
  \tF_{+}
  &=\{\tE_{x,s},\tB_y;\,\tS_{x,s},\tT_y\} ,
  \label{eq:fp_ysymm}
  \\
  \tF_{-}(\pm h-y)
  &=+\tF_{-}(y) ,
   \qquad&
  \tF_{-}
  &=\{\tB_{x,s},\tE_y;\,\tT_{x,s},\tS_y\} .
  \label{eq:fm_ysymm}
\end{alignat}
Therefore we can fold the $y'$-integrals outside the upper-lower walls
$h/2\leq|y'|\leq h$ into 
the inside $|y'|\leq h/2$ as shown in Eqs.(\ref{eq:Fp_coeff}) and (\ref{eq:Fm_coeff}).
Then we describe the fields and current using only the quantities in the pipe.

\subsection{Scalar expression of the fields in the separated form}

We rewrite the Fourier coefficients of the initial fields and current
in Eqs.(\ref{eq:cEBy_sol}) and (\ref{eq:cExs_cBxs_sprt}) using
Eqs.(\ref{eq:Fp_coeff}) and (\ref{eq:Fm_coeff}) with respect to $y'$.
From Eqs.(\ref{eq:Four_Exp_plus}-\ref{eq:Four_Exp_minus}),
we get the separated form of the fields in the frequency domain,
\begin{align}
  \bigg\{{\tEv(\xv) \atop c\tBv(\xv)}\bigg\}
  &=\int_{r_a}^{r_b}dr'\int_{-h/2}^{h/2}dy'
    \bigg[
       \bigg\{{\tPv(\xv,\xv_\perp') \atop \tQv(\xv,\xv_\perp')}\bigg\}
      +Z_0\int_{0}^{\infty}ds'
       \bigg\{{\tMv(\xv_{\perp},s-s',\xv') \atop \tNv(\xv_{\perp},s-s',\xv')}\bigg\}
    \bigg] .
  \label{eq:tEv_tBv}
\end{align}
$\xv$ and $\xv'$ are respectively the observation point and
the position of the source particle of the fields in the bend,
\begin{align}
  \xv=(\xv_\perp,s)=(r,y,s) ,
   \qquad
  \xv_\perp=(r,y) ;
   \qquad
  \xv'=(\xv_\perp',s')=(r',y',s') ,
   \qquad
  \xv_\perp'=(r',y') .
\end{align}
$\tPv$ and $\tQv$ have the operators $\tDv$ and $\tAv$
which involve the initial values of $\tEv$ and $\tBv$ at $s=0$.
On the other hand,
$\tMv$ and $\tNv$ involve the components of the source terms $\tSv$ and $\tTv$ in $s>0$,
\begin{alignat}{2}
  \tPv
  &=\tP_x\ev_x+\tP_y\ev_y+\tP_s\ev_s ,
    \qquad&
  \tMv
  &=\tM_x\ev_x+\tM_y\ev_y+\tM_s\ev_s ,
   \label{eq:tPv}
   \\
  \tQv
  &=\tQ_x\ev_x+\tQ_y\ev_y+\tQ_s\ev_s ,
    \qquad&
  \tNv
  &=\tN_x\ev_x+\tN_y\ev_y+\tN_s\ev_s .
   \label{eq:tQv}
\end{alignat}
$\ev_x$ and $\ev_s$ depend on $s$ as shown in Eq.(\ref{eq:coordinates}).
The components of Eqs.(\ref{eq:tPv}-\ref{eq:tQv}) are given as follows,
\begin{alignat}{2}
  \tP_x(s)
  &=\tD_x\Phi_x(s)-\tD_s\Phi_x^{\ast}(s) ,
    \qquad&
  \tM_x(\vsig,s')
  &=\frac{r'}{\rho}\{\tS_x(s')\Phi_x(\vsig)-\tS_s(s')\Phi_x^{\ast}(\vsig)\} ,
   \label{eq:tPx}
   \\
  \tP_y(s)
  &=\tD_y\Phi_y(s) ,
    \qquad&
  \tM_y(\vsig,s')
  &=\frac{r'}{\rho}\tS_y(s')\Phi_y(\vsig) ,
   \label{eq:tPy}
   \\
  \tP_s(s)
  &=\tD_s\Phi_s(s)+\tD_x\Phi_s^{\ast}(s) ,
    \qquad&
  \tM_s(\vsig,s')
  &=\frac{r'}{\rho}\{\tS_s(s')\Phi_s(\vsig)+\tS_x(s')\Phi_s^{\ast}(\vsig)\} ,
   \\
  \tQ_x(s)
  &=\tA_x\Psi_x(s)-\tA_s\Psi_x^{\ast}(s) ,
    \qquad&
  \tN_x(\vsig,s')
  &=\frac{r'}{\rho}\{\tT_x(s')\Psi_x(\vsig)-\tT_s(s')\Psi_x^{\ast}(\vsig)\} ,
   \\
  \tQ_y(s)
  &=\tA_y\Psi_y(s) ,
    \qquad&
  \tN_y(\vsig,s')
  &=\frac{r'}{\rho}\tT_y(s')\Psi_y(\vsig) ,
   \label{eq:tQy}
   \\
  \tQ_s(s)
  &=\tA_s\Psi_s(s)+\tA_x\Psi_s^{\ast}(s) ,
    \qquad&
  \tN_s(\vsig,s')
  &=\frac{r'}{\rho}\{\tT_s(s')\Psi_s(\vsig)+\tT_x(s')\Psi_s^{\ast}(\vsig)\} ,
   \label{eq:tQs}
\end{alignat}
where the transverse arguments $(\xv_\perp,\xv_\perp')$ of the functions are omitted
for clarity to show the longitudinal arguments.
$\vsig=s-s'$ ($\vsig\ne s$) in Eqs.(\ref{eq:tPx}-\ref{eq:tQs}).
Using $\xv'$, we write the source terms $\tSv$ and $\tTv$, given by
Eqs.(\ref{eq:tSv}-\ref{eq:tTv}), which have the derivatives of
the current components $\tJ=(\tJ_0,\tJv)$,
\begin{alignat}{2}
  \tS_x(\xv')
  &=\rd_{r'}\tJ_0(\xv')-ik\beta\tJ_x(\xv') ,
    \qquad&
  \tT_x(\xv')
  &=\frac{\rho}{r'}\rd_{s'}\tJ_y(\xv')-\rd_{y'}\tJ_s(\xv') ,
   \label{eq:tSx_tTx}
  \\
  \tS_y(\xv')
  &=\rd_{y'}\tJ_0(\xv')-ik\beta\tJ_y(\xv') ,
    \qquad&
  \tT_y(\xv')
  &=\brd_{r'}\tJ_s(\xv')-\frac{\rho}{r'}\rd_{s'}\tJ_x(\xv') ,
  \\
  \tS_s(\xv')
  &=\frac{\rho}{r'}\rd_{s'}\tJ_0(\xv')-ik\beta\tJ_s(\xv') ,
    \qquad&
  \tT_s(\xv')
  &=\rd_{y'}\tJ_x(\xv')-\rd_{r'}\tJ_y(\xv') .
   \label{eq:tSs_tTs}
\end{alignat}
The radial operator $\brd_{r'}$ is given by Eq.(\ref{eq:rd_hr}).

\noindent
$\tDv$ and $\tAv$ are the longitudinal operators which have $\rd_s$ and
the initial fields at the entrance of the bend,
\begin{alignat}{2}
  \tDv(\xv_\perp')
  &=\tD_x\ev_x+\tD_y\ev_y+\tD_s\ev_s
   \label{eq:tDv}
   \\
  &=\frac{\rho}{r'}
    \Big[
      \rd_{s'}\tEv(\xv')+\tEv(\xv')\rd_s
     +\frac{1}{\rho}\{\tE_x(\xv')\ev_s-\tE_s(\xv')\ev_x\}
    \Big]_{s'=+0} ,
   \\
  \tAv(\xv_\perp')
  &=\tA_x\ev_x+\tA_y\ev_y+\tA_s\ev_s
   \\
  &=\frac{\rho}{r'}
    \Big[
      \rd_{s'}c\tBv(\xv')+c\tBv(\xv')\rd_s
     +\frac{1}{\rho}\{c\tB_x(\xv')\ev_s-c\tB_s(\xv')\ev_x\}
    \Big]_{s'=+0} .
   \label{eq:tAv}
\end{alignat}
Namely,
\begin{align}
  \tD_x(\xv_\perp')
  &=\frac{\rho}{r'}
    \Big[\rd_{s'}\tE_x(\xv')+\tE_x(\xv')\rd_s-\frac{2}{\rho}\tE_s(\xv')\Big]_{s'=+0} ,
  \label{eq:bEx}
   \\
  \tD_y(\xv_\perp')
  &=\frac{\rho}{r'}\big[\rd_{s'}\tE_y(\xv')+\tE_y(\xv')\rd_s\big]_{s'=+0} ,
   \label{eq:tDy}
   \\
  \tD_s(\xv_\perp')
  &=\frac{\rho}{r'}
    \Big[\rd_{s'}\tE_s(\xv')+\tE_s(\xv')\rd_s+\frac{2}{\rho}\tE_x(\xv')\Big]_{s'=+0} ,
  \label{eq:bEs}
   \\
  \tA_x(\xv_\perp')
  &=\frac{\rho}{r'}
    \Big[\rd_{s'}c\tB_x(\xv')+c\tB_x(\xv')\rd_s-\frac{2}{\rho}c\tB_s(\xv')\Big]_{s'=+0} ,
  \label{eq:bBx}
   \\
  \tA_y(\xv_\perp')
  &=\frac{\rho}{r'}\big[\rd_{s'}c\tB_y(\xv')+c\tB_y(\xv')\rd_s\big]_{s'=+0} ,
   \label{eq:tAy}
   \\
  \tA_s(\xv_\perp')
  &=\frac{\rho}{r'}
    \Big[\rd_{s'}c\tB_s(\xv')+c\tB_s(\xv')\rd_s+\frac{2}{\rho}c\tB_x(\xv')\Big]_{s'=+0} .
  \label{eq:bBs}
\end{align}
As discussed in section \ref{sec:disconti},
$\rd_{s'}\tE_{x,y,s}(s')$ and $\rd_{s'}\tB_{x,y,s}(s')$ are discontinuous at $s'=0$.

Eqs.(\ref{eq:bEx}-\ref{eq:bBs}) have $\rd_s$ which acts on
the Green functions of the separated form in the frequency domain,
\begin{alignat}{2}
  \Phiv
  =(\Phi_x,\Phi_y,\Phi_s)
   ,\qquad
  \Phiv^{\ast}
  =(\Phi_x^{\ast},\Phi_s^{\ast})
   ,\qquad
  \Psiv
  =(\Psi_x,\Psi_y,\Psi_s)
   ,\qquad
  \Psiv^{\ast}
  =(\Psi_x^{\ast},\Psi_s^{\ast}) .
   \label{eq:tPhi_tPsi}
\end{alignat}
They consist of the Fourier coefficients $\cG_{\pm}^n$, $\cH_{\pm}^n$ and $\cK_{\pm}^n$ 
given by Eqs.(\ref{eq:cGp_sep}-\ref{eq:cGm_sep}) and (\ref{eq:cHpm}).
$\Phiv^{\ast}$ and $\Psiv^{\ast}$ are the coupling Green functions between 
the radial and longitudinal components of the fields.
The Green functions of the separated form for $\tE_y$ and $\tB_y$ are given as
\begin{align}
  \Phi_y(\xv_\perp,\xv_\perp',\vsig)
  =\sum_{n=0}^{\infty}\cG_{+}^n(r,r',\vsig)\cY_{-}^n(y,y')
  ,\qquad
  \Psi_y(\xv_\perp,\xv_\perp',\vsig)
  =\sum_{n=1}^{\infty}\cG_{-}^n(r,r',\vsig)\cY_{+}^n(y,y') .
   \label{eq:tPhiy_tPsiy}
\end{align}
In Eqs.(\ref{eq:tPhiy_tPsiy}-\ref{eq:tPsi_xs})
$\vsig$ represents $s-s'$ and $s$ as in Eq.(\ref{eq:cGeb_def}).
$\cG_{\pm}^n$ denotes the Green functions for $\cE_y^n$ and $\cB_y^n$,
given by Eqs.(\ref{eq:cGp_sep}-\ref{eq:cGm_sep}).
The Green functions of the separated form for $\tE_{x,s}$ and $\tB_{x,s}$ are
given as follows,
\begin{align}
  \bigg[
  \begin{array}{cc}
    \Phi_x(\xv_\perp,\xv_\perp',\vsig) & \Phi_x^{\ast}(\xv_\perp,\xv_\perp',\vsig) \\
    \Phi_s(\xv_\perp,\xv_\perp',\vsig) & \Phi_s^{\ast}(\xv_\perp,\xv_\perp',\vsig)
  \end{array}
  \bigg]
  &=\sum_{n=1}^{\infty}
  \bigg[
  \begin{array}{cc}
      \cH_{-}^{n}(r,r',\vsig) & \cK_{-}^{n}(r,r',\vsig)
       \\
      \cH_{+}^{n}(r,r',\vsig) & \cK_{+}^{n}(r,r',\vsig)
  \end{array}
  \bigg]
    \cY_{+}^n(y,y') ,
  \label{eq:tPhi_xs}
   \\
  \bigg[
  \begin{array}{cc}
     \Psi_x(\xv_\perp,\xv_\perp',\vsig) & \Psi_x^{\ast}(\xv_\perp,\xv_\perp',\vsig) \\
     \Psi_s(\xv_\perp,\xv_\perp',\vsig) & \Psi_s^{\ast}(\xv_\perp,\xv_\perp',\vsig)
  \end{array}
  \bigg]
  &=\sum_{n=0}^{\infty}
    \bigg[
    \begin{array}{cc}
      \cH_{+}^{n}(r,r',\vsig) & \cK_{+}^{n}(r,r',\vsig)
       \\
      \cH_{-}^{n}(r,r',\vsig) & \cK_{-}^{n}(r,r',\vsig)
    \end{array}
    \bigg]
    \cY_{-}^n(y,y') ,
  \label{eq:tPsi_xs}
\end{align}
where the functions, arrayed like matrices, correspond between
the left and right hand sides; \eg, $\Phi_x$ and $\Psi_s$ correspond to $\cH_{-}^n$.
$\cY_{\pm}^n$ denotes the vertical eigenfunctions of the positive and negative 
parities between the upper-lower walls of the rectangular pipe as mentioned in
section \ref{sec:Green},
\begin{align}
  \cY_{+}^n(y,y')
  &=\frac{1}{h}\big\{\cos[k_y^n(y-y')]-(-1)^n\cos[k_y^n(y+y')]\big\}
   \label{eq:cYp_0}
  \\
  &=\frac{2}{h}\sin[k_y^n(y\pm h/2)]\sin[k_y^n(y'\pm h/2)]
    \qquad (n\in\mathbb{N}) ,
   \label{eq:cYp}
  \\
  \cY_{-}^n(y,y')
  &=\frac{1}{h(1+\delta_0^n)}
    \big\{\cos[k_y^n(y-y')]+(-1)^n\cos[k_y^n(y+y')]\big\}
   \\
  &=\frac{2}{h(1+\delta_0^n)}\cos[k_y^n(y\pm h/2)]\cos[k_y^n(y'\pm h/2)]
    \qquad (n\in\mathbb{Z}_0^{+}) .
   \label{eq:cYm}
\end{align}
$\delta_0^n$ is the Kronecker delta defined in Eq.(\ref{eq:lprm}).
Similar to Eqs.(\ref{eq:cX_symm}-\ref{eq:we_cX}), $\cY_{\pm}^n$ satisfies
\begin{align}
  \{\rd_y^2+(k_y^n)^2\}\cY_{\pm}^n
  =\{\rd_{y'}^2+(k_y^n)^2\}\cY_{\pm}^n=0 ,
    \qquad
  \rd_y\cY_{\pm}^n+\rd_{y'}\cY_{\mp}^n=0 .
  \label{eq:cYpm_relations}
\end{align}

We rewrite the operators given by Eqs.(\ref{eq:brdxs}) and (\ref{eq:operator})
using $r$ instead of $x$,
\begin{align}
  \bigg({ \nablav_{\vdash}^2 \atop \nablav_{\rm v}^2 }\bigg)
   =\bigg({\rd_r\brd_r \atop  \brd_r\rd_r}\bigg)
    +(k\beta)^2+\rd_y^2+\brd_s^2 ,
   \qquad
  \brd_r=\rd_r+\frac{1}{r} ,
    \qquad
  \brd_s=\frac{\rho}{r}\rd_s .
  \label{eq:vopera}
\end{align}
Using $\cY_{\pm}^n$ defined for $|y|\leq h/2$, the vertical $\delta$-function is 
represented as follows,
\begin{align}
  \delta(y-y')
  =\sum_{n=1}^{\infty}\cY_{+}^n(y,y')
  =\sum_{n=0}^{\infty}\cY_{-}^n(y,y')
  \qquad
    (|y|\leq h/2,~ |y'|\leq h/2) .
  \label{eq:delta_yyp}
\end{align}
According to Eqs.(\ref{eq:delta_yyp}), the vertical Green functions
$\Phi_y$ and $\Psi_y$ satisfy the following wave equation,
\begin{align}
  \nablav_{\rm v}^2
  \bigg\{{\Phi_y(\vsig) \atop \Psi_y(\vsig)}\bigg\}
  =\bdelta^3(\xv_\perp-\xv_\perp',\vsig)
   \qquad
    (|y|\leq h/2, ~|y'|\leq h/2) ,
   \label{eq:we_Phiy_Psiy}
\end{align}
where
\begin{align}
  \bdelta^3(\xv_\perp-\xv_\perp',\vsig)
  =\frac{\rho}{r}\delta(r-r')\delta(y-y')\bdelta(\vsig)
   ,\qquad
  \vsig
  =\{s-s',s\} .
   \label{eq:bdelta3}
\end{align}
$\bdelta(\vsig)$ is the $\delta$-function defined in Eq.(\ref{eq:bdelta}).
In Eqs.(\ref{eq:we_Phiy_Psiy}-\ref{eq:we_tMNs}) we omit the transverse arguments
$(\xv_\perp,\xv_\perp')$ of the functions for brevity excluding the $\delta$-functions.
According to Eqs.(\ref{eq:Grn_cHpm_cKpm}-\ref{eq:Grn_cKpm_cHpm}) and (\ref{eq:delta_yyp}),
the Green functions for the radial and longitudinal components of
the fields satisfy the following wave equations,
\begin{alignat}{2}
  \nablav_{\vdash}^2\bigg\{{\Phi_{x,s}(\vsig) \atop \Psi_{x,s}(\vsig)}\bigg\}
 -\frac{2}{r}\brd_s
  \bigg\{{ \Phi_{s,x}^{\ast}(\vsig) \atop \Psi_{s,x}^{\ast}(\vsig) }\bigg\}
  &=\bdelta^3(\xv_\perp-\xv_\perp',\vsig) ,
   \label{eq:we_Phis_Psix}
   \\
  \nablav_{\vdash}^2
  \bigg\{{ \Phi_{x,s}^{\ast}(\vsig) \atop \Psi_{x,s}^{\ast}(\vsig) }\bigg\}
  +\frac{2}{r}\brd_s\bigg\{{\Phi_{s,x}(\vsig) \atop \Psi_{s,x}(\vsig)}\bigg\}
  &=0 .
   \label{eq:we_Phis_Psix_ast}
\end{alignat}
According to Eq.(\ref{eq:we_Phiy_Psiy}), Eqs.(\ref{eq:tPy}) and (\ref{eq:tQy}) satisfy
the following wave equations,
\begin{align}
  \nablav_{\rm v}^2\bigg\{{ \tP_y(s) \atop \tQ_y(s) }\bigg\}
  =0 ,
    \qquad
  \nablav_{\rm v}^2\bigg\{{ \tM_y(\vsig,s') \atop \tN_y(\vsig,s') }\bigg\}
  =\bigg\{{ \tS_y(s') \atop \tT_y(s') }\bigg\}
   \delta(r-r')\delta(y-y')\delta(\vsig) .
   \label{eq:we_tPQy}
\end{align}
According to Eqs.(\ref{eq:we_Phis_Psix}-\ref{eq:we_Phis_Psix_ast}),
the radial and longitudinal components of Eqs.(\ref{eq:tPv}-\ref{eq:tQv}) satisfy
\begin{align}
  &
  \nablav_{\vdash}^2\bigg\{{ \tP_x(s) \atop \tQ_x(s) }\bigg\}
  -\frac{2}{r}\brd_s\bigg\{{ \tP_s(s) \atop \tQ_s(s) }\bigg\}
  =0 ,
    \qquad
  \nablav_{\vdash}^2\bigg\{{ \tP_s(s) \atop \tQ_s(s) }\bigg\}
  +\frac{2}{r}\brd_s\bigg\{{ \tP_x(s) \atop \tQ_x(s) }\bigg\}
  =0 ,
   \label{eq:we_tPQxs}
   \\
  &
  \nablav_{\vdash}^2\bigg\{{ \tM_x(\vsig,s') \atop \tN_x(\vsig,s') }\bigg\}
  -\frac{2}{r}\brd_s\bigg\{{ \tM_s(\vsig,s') \atop \tN_s(\vsig,s') }\bigg\}
  =\bigg\{{ \tS_x(s') \atop \tT_x(s') }\bigg\}
    \delta(r-r')\delta(y-y')\delta(\vsig) ,
   \\
  &
  \nablav_{\vdash}^2\bigg\{{ \tM_s(\vsig,s') \atop \tN_s(\vsig,s') }\bigg\}
  +\frac{2}{r}\brd_s\bigg\{{ \tM_x(\vsig,s') \atop \tN_x(\vsig,s') }\bigg\}
  =\bigg\{{ \tS_s(s') \atop \tT_s(s') }\bigg\}
    \delta(r-r')\delta(y-y')\delta(\vsig) .
   \label{eq:we_tMNs}
\end{align}
In the limit of $\rho\to\infty$, the Green functions of the separated form go to
the following limits,
\begin{align}
  \lim_{\rho\to\infty}\Phi_x
  &=\lim_{\rho\to\infty}\Psi_y
  =\Phi_{-}^{+}
   ,\qquad
  \lim_{\rho\to\infty}\Phi_y
  =\lim_{\rho\to\infty}\Psi_x
  =\Phi_{+}^{-} ,
   \\
  \lim_{\rho\to\infty}\Phi_s
  &=\Phi_{+}^{+}
   ,\qquad
  \lim_{\rho\to\infty}\Psi_s
  =\Phi_{-}^{-}
   ,\qquad
  \lim_{\rho\to\infty}\Phi_{x,s}^{\ast}
  =\lim_{\rho\to\infty}\Psi_{x,s}^{\ast}
  =0 .
   \label{eq:lim_Phi_ast}
\end{align}
$\Phi_{\pm}^{\pm}$ is given by Eqs.(\ref{eq:Phi})
which are the Green functions of the straight pipe in the frequency domain.
Let us classify the components of the electromagnetic field in the curved pipe.
They are summarized to the following table with respect to the transverse parities
in the limit of $\rho\to\infty$,
\begin{alignat}{3}
  &
  &&\cY_{+}^n\,(n\in\mathbb{N}), \qquad &&\cY_{-}^n\,(n\in\mathbb{Z}_0^{+})
    \nonumber\\
  &\cR_{+}^{mn}\,(m\in\mathbb{N}):       \quad &&\quad \tE_s,       &&\quad \tB_x,\tE_y
    \nonumber\\
  &\cR_{-}^{mn}\,(m\in\mathbb{Z}_0^{+}): \quad &&\quad \tE_x,\tB_y, && \quad \tB_s
    \nonumber
\end{alignat}
Needless to say, the fields in the curved pipe are not symmetric with respect to $r=\rho$.
The horizontal parity, which is indicated by the sign of $\cR_{\pm}^{mn}$, is correct
if $x_b=-x_a$ and in the limit of $\rho\to\infty$ as shown in Eq.(\ref{eq:lim_RX}).

\subsection{Differential expression of the fields in the separated form}

The source terms $\tSv$ and $\tTv$ given by Eqs.(\ref{eq:tSx_tTx}-\ref{eq:tSs_tTs})
are involved in Eq.(\ref{eq:tEv_tBv}) through Eqs.(\ref{eq:tPx}-\ref{eq:tQs}).
They have the derivatives of the current components $\tJ=(\tJ_0,\tJv)$.
Integrating the derivatives of $\tJ(\xv')$ by parts with respect to $r'$, $y'$ or $s'$ 
in Eq.(\ref{eq:tEv_tBv}), we get the differential expressions of the fields in
the separated form,
\begin{align}
  \bigg\{{\tEv(\xv) \atop c\tBv(\xv)}\bigg\}
  &=\int_{r_a}^{r_b}dr'\int_{-h/2}^{h/2}dy'
    \bigg[
       \bigg\{{\tPv^{\dg}(\xv,\xv_\perp') \atop
               \tQv^{\dg}(\xv,\xv_\perp')}
       \bigg\}
      -Z_0\int_{0}^{\infty}ds'
       \bigg\{{\tMv^{\dg}(\xv_{\perp},s-s',\xv') \atop \tNv^{\dg}(\xv_{\perp},s-s',\xv')}
       \bigg\}
    \bigg] ,
  \label{eq:tEBv_dg}
\end{align}
where
\begin{alignat}{2}
  \tPv^{\dg}
  &=\tP_x^{\dg}\ev_x+\tP_y^{\dg}\ev_y+\tP_s^{\dg}\ev_s ,
     \qquad&
  \tMv^{\dg}
  &=\tM_x^{\dg}\ev_x+\tM_y^{\dg}\ev_y+\tM_s^{\dg}\ev_s ,
   \label{eq:tPv_dg}
   \\
  \tQv^{\dg}
  &=\tQ_x^{\dg}\ev_x+\tQ_y^{\dg}\ev_y+\tQ_s^{\dg}\ev_s ,
     \qquad&
  \tNv^{\dg}
  &=\tN_x^{\dg}\ev_x+\tN_y^{\dg}\ev_y+\tN_s^{\dg}\ev_s .
   \label{eq:tQv_dg}
\end{alignat}
Omitting the transverse arguments $(\xv_\perp,\xv_\perp')$,
these components are given as follows,
\begin{alignat}{2}
  \tP_x^{\dg}(s)
  &=\tD_x^{\dg}\Phi_x(s)-\tD_s^{\dg}\Phi_x^{\ast}(s) ,
    \qquad&
  \tM_x^{\dg}(\vsig,s')
  &=\frac{r'}{\rho}\{\tS_x^{\dg}(s')\Phi_x(\vsig)-\tS_s^{\dg}(s')\Phi_x^{\ast}(\vsig)\} ,
  \label{eq:tPx_dg}
  \\
  \tP_y^{\dg}(s)
  &=\tD_y^{\dg}\Phi_y(s) ,
    \qquad&
  \tM_y^{\dg}(\vsig,s')
  &=\frac{r'}{\rho}\tS_y^{\dg}(s')\Phi_y(\vsig) ,
  \label{eq:tPy_tMy_dg}
  \\
  \tP_s^{\dg}(s)
  &=\tD_s^{\dg}\Phi_s(s)+\tD_x^{\dg}\Phi_s^{\ast}(s) ,
    \qquad&
  \tM_s^{\dg}(\vsig,s')
  &=\frac{r'}{\rho}\{\tS_s^{\dg}(s')\Phi_s(\vsig)+\tS_x^{\dg}(s')\Phi_s^{\ast}(\vsig)\} ,
  \\
  \tQ_x^{\dg}(s)
  &=\tA_x^{\dg}\Psi_x(s)-\tA_s^{\dg}\Psi_x^{\ast}(s) ,
    \qquad&
  \tN_x^{\dg}(\vsig,s')
  &=\frac{r'}{\rho}\{\tT_x^{\dg}(s')\Psi_x(\vsig)-\tT_s^{\dg}(s')\Psi_x^{\ast}(\vsig)\} ,
  \\
  \tQ_y^{\dg}(s)
  &=\tA_y^{\dg}\Psi_y(s) ,
    \qquad&
  \tN_y^{\dg}(\vsig,s')
  &=\frac{r'}{\rho}\tT_y^{\dg}(s')\Psi_y(\vsig) ,
  \label{eq:tQy_tNy_dg}
  \\
  \tQ_s^{\dg}(s)
  &=\tA_s^{\dg}\Psi_s(s)+\tA_x^{\dg}\Psi_s^{\ast}(s) ,
    \qquad&
  \tN_s^{\dg}(\vsig,s')
  &=\frac{r'}{\rho}\{\tT_s^{\dg}(s')\Psi_s(\vsig)+\tT_x^{\dg}(s')\Psi_s^{\ast}(\vsig)\} .
  \label{eq:tQs_dg}
\end{alignat}
$(\tDv^{\dg},\tAv^{\dg})$ and $(\tSv^{\dg},\tTv^{\dg})$ are operators
which act on the Green functions given by Eqs.(\ref{eq:tPhi_tPsi}),
\begin{alignat}{2}
  \tDv^{\dg}(\xv_\perp')
  &=\tD_x^{\dg}\ev_x+\tD_y^{\dg}\ev_y+\tD_s^{\dg}\ev_s ,
     \qquad&
  \tSv^{\dg}(\xv')
  &=\tJv(\xv')ik\beta+\tJ_0(\xv')\Big(\nablav'+\frac{\ev_x}{r'}\Big) ,
   \\
  \tAv^{\dg}(\xv_\perp')
  &=\tA_x^{\dg}\ev_x+\tA_y^{\dg}\ev_y+\tA_s^{\dg}\ev_s ,
     \qquad&
  \tTv^{\dg}(\xv')
  &=\tJv(\xv')\times\nablav'-\tJ_y(\xv')\frac{\ev_s}{r'} .
\end{alignat}
$\nablav'$ is the nabla with respect to $\xv'=(r',y',s')$
instead of $\xv=(r,y,s)$ given by Eq.(\ref{eq:nabla}).
$\nablav'$ acts on the Green functions of $\vsig=s-s'$
in the second equations of (\ref{eq:tPx_dg}-\ref{eq:tQs_dg}).
Since $\rd_{s'}f(s-s')=-\rd_sf(s-s')$ for an arbitrary differentiable function $f$,
we can replace $\rd_{s'}$, involved in $\nablav'$, by $-\rd_s$ as follows,
\begin{align}
  \nablav'
  =\ev_x\rd_{r'}+\ev_y\rd_{y'}-\ev_s\brd_s'
   ,\qquad
  \brd_s'
  =\frac{\rho}{r'}\rd_s .
   \label{eq:nabla_prm}
\end{align}
The operators $\tDv^{\dg}$ and $\tAv^{\dg}$ involve the initial fields at
the entrance of the bending magnet,
\begin{align}
  \tD_x^{\dg}(\xv_\perp')
  &=[
     \tE_{x}(\xv')\brd_s'
    -\tE_s(\xv')\brd_{r'}
    +c\tB_y(\xv')ik\beta
    ]_{s'=0} ,
  \label{eq:tDx_dg}
  \\
  \tD_y^{\dg}(\xv_\perp')
  &=[
      \tE_y(\xv')\brd_s'
     -\tE_s(\xv')\rd_{y'}
     -c\tB_x(\xv')ik\beta
    ]_{s'=0} ,
  \label{eq:tDy_dg}
  \\
  \tD_s^{\dg}(\xv_\perp')
  &=[
      \tE_s(\xv')\brd_s'
     +\tE_x(\xv')\brd_{r'}
     +\tE_y(\xv')\rd_{y'}
    ]_{s'=0} ,
  \\
  \tA_x^{\dg}(\xv_\perp')
  &=[
       c\tB_x(\xv')\brd_s'
      -c\tB_s(\xv')\brd_{r'}
      -\tE_y(\xv')ik\beta
    ]_{s'=0} ,
  \\
  \tA_y^{\dg}(\xv_\perp')
  &=[
     c\tB_y(\xv')\brd_s'
    -c\tB_s(\xv')\rd_{y'}
    +\tE_x(\xv')ik\beta
    ]_{s'=0} ,
  \label{eq:tAy_dg}
  \\
  \tA_s^{\dg}(\xv_\perp')
  &=[
      c\tB_s(\xv')\brd_s'
     +c\tB_x(\xv')\brd_{r'}
     +c\tB_y(\xv')\rd_{y'}
    ]_{s'=0} .
  \label{eq:tAs_dg}
\end{align}
$\tSv^{\dg}$ and $\tTv^{\dg}$ are the operators which involve the current components,
\begin{alignat}{2}
  \tS_x^{\dg}(\xv')
  &=\tJ_0(\xv')\brd_{r'}+\tJ_x(\xv')ik\beta ,
    \qquad
  &\tT_x^{\dg}(\xv')
  &=\tJ_y(\xv')\frac{\rho}{r'}\rd_{s'}-\tJ_s(\xv')\rd_{y'} ,
  \label{eq:tSxs_dg}
   \\
  \tS_y^{\dg}(\xv')
  &=\tJ_0(\xv')\rd_{y'}+\tJ_y(\xv')ik\beta ,
    \qquad
  &\tT_y^{\dg}(\xv')
  &=\tJ_s(\xv')\rd_{r'}-\tJ_x(\xv')\frac{\rho}{r'}\rd_{s'} ,
   \\
  \tS_s^{\dg}(\xv')
  &=\tJ_0(\xv')\frac{\rho}{r'}\rd_{s'}+\tJ_s(\xv')ik\beta ,
    \qquad
  &\tT_s^{\dg}(\xv')
  &=\tJ_x(\xv')\rd_{y'}-\tJ_y(\xv')\brd_{r'} .
  \label{eq:tTxs_dg}
\end{alignat}
We can replace $\rd_{s'}$ in Eqs.(\ref{eq:tSxs_dg}-\ref{eq:tTxs_dg}) by $-\rd_s$
since it acts on the Green functions of $\vsig=s-s'$
in the second equations of (\ref{eq:tPx_dg}-\ref{eq:tQs_dg}).
Eqs.(\ref{eq:tSxs_dg}-\ref{eq:tTxs_dg}) do not have the derivative of
the current components $(\tJ_0,\tJ_{x,y,s})$ unlike $\tSv$ and $\tTv$
given by Eqs.(\ref{eq:tSx_tTx}-\ref{eq:tSs_tTs}).
Instead of it, $\tSv^{\dg}$ and $\tTv^{\dg}$ are operators
which act on the Green functions in Eqs.(\ref{eq:tPx_dg}-\ref{eq:tQs_dg}).
Since the Green functions are differentiated by Eqs.(\ref{eq:tSxs_dg}-\ref{eq:tTxs_dg}),
Eq.(\ref{eq:tEBv_dg}) is referred to as {\it the differential expression} of the fields.
On the other hand, $\tSv$ and $\tTv$ are scalar functions in
Eq.(\ref{eq:tEv_tBv}) which is {\it the scalar expression}.
We substitute Eqs.(\ref{eq:tPhiy_tPsiy}-\ref{eq:tPsi_xs}) and
(\ref{eq:tDx_dg}-\ref{eq:tAs_dg}) into the first equations of
(\ref{eq:tPx_dg}-\ref{eq:tQs_dg}),
\begin{alignat}{2}
  \tP_y^{\dg}(s)
  &=
    \Big\{\tE_y(0)\frac{\rho}{r'}\rd_s-\tE_s(0)\rd_{y'}-c\tB_x(0)ik\beta\Big\}
    \sum_{n=0}^{\infty}
    \cG_{+}^n(s)\cY_{-}^n
   ,\qquad&
  \tP_{x,s}^{\dg}(s)
  &=\sum_{n=1}^{\infty} \tilde{p}_{x,s}^{\dg}(s)\cY_{+}^n ,
   \label{eq:tPy_dg_cG}
   \\
  \tQ_y^{\dg}(s)
  &=
    \Big\{c\tB_y(0)\frac{\rho}{r'}\rd_s-c\tB_s(0)\rd_{y'}+\tE_x(0)ik\beta\Big\}
    \sum_{n=1}^{\infty}
    \cG_{-}^n(s)\cY_{+}^n
   ,\qquad&
  \tQ_{x,s}^{\dg}(s)
  &=\sum_{n=0}^{\infty} \tilde{q}_{x,s}^{\dg}(s)\cY_{-}^n .
   \label{eq:tQy_dg_cG}
\end{alignat}
We omit the transverse arguments $(\xv_{\perp},\xv_{\perp}')$ in
Eqs.(\ref{eq:tPy_dg_cG}-\ref{eq:DtFv_std}).
$\tEv(0)$ and $\tBv(0)$ are the initial fields at $s'=0$,
\begin{align}
  \tEv(0)
  =\tEv(\xv_{\perp}',0)
   ,\qquad
  \tBv(0)
  =\tBv(\xv_{\perp}',0) .
   \label{tEv_tBv_sp0}
\end{align}
$\tilde{p}_{x,s}^{\dg}$ and $\tilde{q}_{x,s}^{\dg}$ are the following operators
with respect to $y'$, which act on $\cY_{\pm}^n$,
\begin{align}
  \tilde{p}_{x}^{\dg}(s)
  &=
    \{\tE_x(0)\brd_s\cG_{-}^n(s)+\tE_s(0)\rd_{r}\cG_{+}^n(s)\}
   +\{c\tB_y(0)\cH_{-}^{n}(s)ik\beta-\tE_y(0)\cK_{-}^{n}(s)\rd_{y'}\} ,
   \label{eq:tPx_dg_cG_0}
   \\
  \tilde{p}_{s}^{\dg}(s)
  &=
    \{\tE_s(0)\brd_s\cG_{+}^n(s)-\tE_x(0)\rd_{r}\cG_{-}^n(s)\}
   +\{\tE_y(0)\cH_{+}^n(s)\rd_{y'}+c\tB_y(0)\cK_{+}^n(s)ik\beta\} ,
   \label{eq:tPs_dg_cG_0}
   \\
  \tilde{q}_{x}^{\dg}(s)
  &=
    \{c\tB_x(0)\brd_s\cG_{+}^n(s)+c\tB_s(0)\rd_{r}\cG_{-}^n(s)\}
   -\{\tE_y(0)\cH_{+}^n(s)ik\beta+c\tB_y(0)\cK_{+}^n(s)\rd_{y'}\} ,
   \label{eq:tQx_dg_cG_0}
   \\
  \tilde{q}_{s}^{\dg}(s)
  &=
    \{c\tB_s(0)\brd_s\cG_{-}^n(s)-c\tB_x(0)\rd_{r}\cG_{+}^n(s)\}
   +\{c\tB_y(0)\cH_{-}^n(s)\rd_{y'}-\tE_y(0)\cK_{-}^n(s)ik\beta\} .
   \label{eq:tQs_dg_cG_0}
\end{align}
$\brd_s$ is given by Eq.(\ref{eq:vopera}).
We substitute Eqs.(\ref{eq:tPhiy_tPsiy}-\ref{eq:tPsi_xs}) and
(\ref{eq:tSxs_dg}-\ref{eq:tTxs_dg}) into the second equations of
(\ref{eq:tPx_dg}-\ref{eq:tQs_dg}),
\begin{alignat}{2}
  \tM_y^{\dg}(\vsig,s')
  &=\frac{r'}{\rho}\Big\{\tJ_0(s')\rd_{y'}+\tJ_y(s')ik\beta\Big\}
    \sum_{n=0}^{\infty}\cG_{+}^n(\vsig)\cY_{-}^n
   ,\qquad&
  \tM_{x,s}^{\dg}(\vsig,s')
  &=\frac{r'}{\rho}\sum_{n=1}^{\infty} \tilde{m}_{x,s}^{\dg}(\vsig,s')\cY_{+}^n ,
  \label{eq:tMy_dg_rw}
   \\
  \tN_y^{\dg}(\vsig,s')
  &=\frac{r'}{\rho}\Big\{\tJ_s(s')\rd_{r'}+\tJ_x(s')\frac{\rho}{r'}\rd_s\Big\}
    \sum_{n=1}^{\infty}\cG_{-}^n(\vsig)\cY_{+}^n
   ,\qquad&
  \tN_{x,s}^{\dg}(\vsig,s')
  &=\frac{r'}{\rho}\sum_{n=0}^{\infty} \tilde{n}_{x,s}^{\dg}(\vsig,s')\cY_{-}^n .
  \label{eq:tNy_dg_rw}
\end{alignat}
$\vsig=s-s'$ in Eqs.(\ref{eq:tMy_dg_rw}-\ref{eq:tNs_dg_rw}).
$\tilde{m}_{x}^{\dg}$ and $\tilde{m}_{s}^{\dg}$ are not operators but the following functions.
$\tilde{n}_{x,s}^{\dg}$ denotes the following operators which have $\rd_{y'}$,
\begin{align}
  \tilde{m}_x^{\dg}(\vsig,s')
  &=ik\beta\{\tJ_x(s')\cH_{-}^n(\vsig)-\tJ_s(s')\cK_{-}^n(\vsig)\}
    -\tJ_0(s')\rd_{r}\cG_{+}^n(\vsig) ,
  \label{eq:tMx_dg}
   \\
  \tilde{m}_s^{\dg}(\vsig,s')
  &=ik\beta\{\tJ_s(s')\cH_{+}^n(\vsig)+\tJ_x(s')\cK_{+}^n(\vsig)\}
      -\tJ_0(s')\brd_s\cG_{+}^n(\vsig) ,
  \label{eq:tMs_dg}
   \\
  \tilde{n}_x^{\dg}(\vsig,s')
  &=-\{\tJ_s(s')\cH_{+}^n(\vsig)+\tJ_x(s')\cK_{+}^n(\vsig)\}\rd_{y'}
    -\tJ_y(s')\brd_s\cG_{+}^n(\vsig) ,
  \label{eq:tNx_dg_rw}
   \\
  \tilde{n}_s^{\dg}(\vsig,s')
  &=\{\tJ_x(s')\cH_{-}^n(\vsig)-\tJ_s(s')\cK_{-}^n(\vsig)\}\rd_{y'}
    +\tJ_y(s')\rd_{r}\cG_{+}^n(\vsig) .
  \label{eq:tNs_dg_rw}
\end{align}
$(\tPv^{\dg},\tQv^{\dg})$, $(\tMv^{\dg},\tNv^{\dg})$ and $(\tSv^{\dg},\tTv^{\dg})$ satisfy 
the wave equations (\ref{eq:we_tPQy}-\ref{eq:we_tMNs}), similar to the combination of
$(\tPv,\tQv)$, $(\tMv,\tNv)$ and $(\tSv,\tTv)$.

Eq.(\ref{eq:tEBv_dg}) is the differential expression of the field in the semi-infinite bend.
We will get a better expression if we assume a rigid bunch as the source current of
the field.
Under the assumption of a rigid bunch, we should subtract the steady field in the limit of
$s\to\infty$ from Eq.(\ref{eq:tEBv_dg})
since it includes the space charge field on the curved trajectory.
However, we cannot subtract $(\tEv,\tBv)$ for $s\to\infty$ directly from
Eq.(\ref{eq:tEBv_dg}), because the expression of the steady field in the frequency domain,
which does not depend on $s$ in the limit of $s\to\infty$, is not $(\tEv,\tBv)$
but $(\tEv,\tBv)e^{-iks}$ in the Fourier transform with respect to $t$.
Accordingly, if we subtract the steady field from Eq.(\ref{eq:tEBv_dg}),
we must do it in the following manner,
\begin{align}
  \bigg\{{\Delta\tEv(s) \atop \Delta\tBv(s)}\bigg\}
  =\bigg\{{\tEv(s) \atop \tBv(s)}\bigg\}
   -e^{iks}\lim_{s\to\infty}\bigg\{{\tEv(s) \atop \tBv(s)}\bigg\}e^{-iks} .
  \label{eq:DtFv_std}
\end{align}
$\Delta\tEv$ and $\Delta\tBv$ are the differences of the transient fields
from the steady fields for $s\to\infty$ in the semi-infinite bending magnet.
We use Eq.(\ref{eq:DtFv_std}) in appendices \ref{sec:steady_field} and \ref{sec:impedance}
in calculating the fields created by a rigid bunch.
If we use the Fourier transform with respect to $z$ as in Eqs.(\ref{eq:FT_zk_9}),
we can simply subtract the steady field from the transient field in the frequency domain 
instead of Eq.(\ref{eq:DtFv_std}).

\subsection{Scalar expression of the fields in the complete form}
\label{sec:scalar_complete}

We found two expressions of the electromagnetic field in the frequency domain
as shown in Eqs.(\ref{eq:tEv_tBv}) and (\ref{eq:tEBv_dg}).
They are referred to as the scalar expression and the differential expression of the field 
in the separated form since they consist of the Green functions of the separated form
$(\Phiv,\Phiv^{\ast})$ and $(\Psiv,\Psiv^{\ast})$ given by Eq.(\ref{eq:tPhi_tPsi}).
Besides, we derived the expressions of $\cC_{\pm}^n$, $\cU_{\pm}^n$ and $\cV_{\pm}^n$ in
section \ref{sec:normal_form}, which are the Fourier coefficients of the Green functions
in the complete form as summarized in section \ref{sec:Green}.
Using them, we will derive the complete forms of the field in the frequency domain.
It follows that we eventually have four expressions of the field in total as
the combinations of the separate/complete forms and the scalar/differential expressions
as reviewed in section \ref{sec:expression}.

From Eqs.(\ref{eq:cEy_cBy_cmpl}) and (\ref{eq:cExs_cBxs_cmpl}),
we get the scalar expression of the electromagnetic field in the complete form,
\begin{align}
  \bigg\{{\tEv(\xv) \atop c\tBv(\xv)}\bigg\}
  &=\int_{r_a}^{r_b}dr'\int_{-h/2}^{h/2}dy'
    \bigg[
       \bigg\{{\hPv(\xv,\xv_\perp') \atop
               \hQv(\xv,\xv_\perp')}
       \bigg\}
      +Z_0\int_{0}^{\infty}ds'
       \bigg\{{\hMv(\xv_{\perp},s-s',\xv') \atop \hNv(\xv_{\perp},s-s',\xv')}\bigg\}
    \bigg] .
  \label{eq:tEBv_cmpl}
\end{align}
$(\hPv,\hQv)$ and $(\hMv,\hNv)$ involve respectively
the initial fields at $s=0$ and the source terms in $s>0$,
\begin{alignat}{2}
  \hPv
  &=\hP_x\ev_x+\hP_y\ev_y+\hP_s\ev_s ,
     \qquad&
  \hMv
  &=\hM_x\ev_x+\hM_y\ev_y+\hM_s\ev_s ,
   \label{eq:hPv_hMv}
   \\
  \hQv
  &=\hQ_x\ev_x+\hQ_y\ev_y+\hQ_s\ev_s ,
     \qquad&
  \hNv
  &=\hN_x\ev_x+\hN_y\ev_y+\hN_s\ev_s .
   \label{eq:hQv_hNv}
\end{alignat}
$\ev_x$ and $\ev_s$ depend on $s$ as shown in Eq.(\ref{eq:coordinates}).
The components of Eqs.(\ref{eq:hPv_hMv}-\ref{eq:hQv_hNv}) are given as follows,
\begin{alignat}{2}
  \hP_x(s)&=\tD_x\Gam_x(s)-\tD_s\Gam_x^{\ast}(s) ,
    \qquad&
  \hM_x(\vsig,s')
  &=\frac{r'}{\rho}\{\tS_x(s')\Gam_x(\vsig)-\tS_s(s')\Gam_x^{\ast}(\vsig)\} ,
   \label{eq:hPx}
   \\
  \hP_y(s)&=\tD_y\Gam_y(s) ,
    \qquad&
  \hM_y(\vsig,s')
  &=\frac{r'}{\rho}\tS_y(s')\Gam_y(\vsig) ,
   \label{eq:hPy}
   \\
  \hP_s(s)&=\tD_s\Gam_s(s)+\tD_x\Gam_s^{\ast}(s) ,
    \qquad&
  \hM_s(\vsig,s')
  &=\frac{r'}{\rho}\{\tS_s(s')\Gam_s(\vsig)+\tS_x(s')\Gam_s^{\ast}(\vsig)\} ,
   \label{eq:hPs}
   \\
  \hQ_x(s)&=\tA_x\Lam_x(s)-\tA_s\Lam_x^{\ast}(s) ,
    \qquad&
  \hN_x(\vsig,s')
  &=\frac{r'}{\rho}\{\tT_x(s')\Lam_x(\vsig)-\tT_s(s')\Lam_x^{\ast}(\vsig)\} ,
   \\
  \hQ_y(s)&=\tA_y\Lam_y(s) ,
    \qquad&
  \hN_y(\vsig,s')
  &=\frac{r'}{\rho}\tT_y(s')\Lam_y(\vsig) ,
   \label{eq:hQy}
   \\
  \hQ_s(s)&=\tA_s\Lam_s(s)+\tA_x\Lam_s^{\ast}(s) ,
    \qquad&
  \hN_s(\vsig,s')
  &=\frac{r'}{\rho}\{\tT_s(s')\Lam_s(\vsig)+\tT_x(s')\Lam_s^{\ast}(\vsig)\} .
   \label{eq:hQs}
\end{alignat}
In Eqs.(\ref{eq:hPx}-\ref{eq:we_hMNs})
we omit the transverse arguments $(\xv_\perp,\xv_\perp')$ of the functions
except the $\delta$-functions for clarity to show the longitudinal arguments.
$\vsig=s-s'$ (\ie, $\vsig\ne s$) in Eqs.(\ref{eq:hPx}-\ref{eq:we_hMNs}).
$(\tDv,\tAv)$ and $(\tSv,\tTv)$ are given by
Eqs.(\ref{eq:tDv}-\ref{eq:tAv}) and (\ref{eq:tSx_tTx}-\ref{eq:tSs_tTs}).
Similar to Eqs.(\ref{eq:we_tPQy}), the vertical components of
Eqs.(\ref{eq:hPv_hMv}-\ref{eq:hQv_hNv}) satisfy the following wave equations,
\begin{align}
  \nablav_{\rm v}^2\bigg\{{\hP_y(s) \atop \hQ_y(s)}\bigg\}
  =0
     ,\qquad
  \nablav_{\rm v}^2
  \bigg\{{\hM_y(\vsig,s') \atop \hN_y(\vsig,s')}\bigg\}
  =\bigg\{{\tS_y(s') \atop \tT_y(s')}\bigg\}
   \delta(r-r')\delta(y-y')\delta(\vsig) .
   \label{eq:we_hPQMN_y}
\end{align}
The radial and longitudinal components of Eqs.(\ref{eq:hPv_hMv}-\ref{eq:hQv_hNv}) satisfy
\begin{align}
  &
  \nablav_{\vdash}^2\bigg\{{ \hP_x(s) \atop \hQ_x(s) }\bigg\}
  -\frac{2}{r}\brd_s\bigg\{{ \hP_s(s) \atop \hQ_s(s) }\bigg\}
  =0
   ,\qquad
  \nablav_{\vdash}^2\bigg\{{ \hP_s(s) \atop \hQ_s(s) }\bigg\}
  +\frac{2}{r}\brd_s\bigg\{{ \hP_x(s) \atop \hQ_x(s) }\bigg\}
  =0 ,
   \label{eq:we_hPxs}
   \\
  &
  \nablav_{\vdash}^2\bigg\{{ \hM_x(\vsig,s') \atop \hN_x(\vsig,s') }\bigg\}
  -\frac{2}{r}\brd_s\bigg\{{ \hM_s(\vsig,s') \atop \hN_s(\vsig,s') }\bigg\}
  =\bigg\{{ \tS_x(s') \atop \tS_s(s') }\bigg\}
    \delta(r-r')\delta(y-y')\delta(\vsig) ,
   \label{eq:we_hMNx}
   \\
  &
  \nablav_{\vdash}^2\bigg\{{ \hM_s(\vsig,s') \atop \hN_s(\vsig,s') }\bigg\}
  +\frac{2}{r}\brd_s\bigg\{{ \hM_x(\vsig,s') \atop \hN_x(\vsig,s') }\bigg\}
  =\bigg\{{\tT_x(s')  \atop \tT_s(s') }\bigg\}
    \delta(r-r')\delta(y-y')\delta(\vsig) .
   \label{eq:we_hMNs}
\end{align}
The operators $(\nablav_{\rm v}^2,\nablav_\vdash^2)$ and $\brd_s$ are
given by Eqs.(\ref{eq:vopera}).
Eqs.(\ref{eq:we_hPQMN_y}-\ref{eq:we_hMNs}) correspond to Eqs.(\ref{eq:we_hcPQy_hcMNy})
and (\ref{eq:we_hcPxs_hcQxs}-\ref{eq:we_hcMs_hcNs})
which are the wave equations for the Fourier coefficients of the integrands.
According to Eqs.(\ref{eq:we_hPQMN_y}-\ref{eq:we_hMNs}), the components of $\tEv$ and $\tBv$,
given by Eq.(\ref{eq:tEBv_cmpl}), satisfy the wave equations
(\ref{eq:we_tEy_pm}) and (\ref{eq:we_tBy_pm}-\ref{eq:we_tExsBxs}).

$\Gamv$ and $\Lamv$ are the Green functions of the complete form for the fields
in the frequency domain,
\begin{align}
  \Gamv
  =(\Gam_x,\Gam_y,\Gam_s)
    ,\qquad
  \Lamv
  =(\Lam_x,\Lam_y,\Lam_s)
    ,\qquad
  \Gamv^{\ast}
  =(\Gam_x^{\ast},\Gam_s^{\ast})
    ,\qquad
  \Lamv^{\ast}
  =(\Lam_x^{\ast},\Lam_s^{\ast}) .
   \label{eq:tGam_tLam}
\end{align}
$\Gamv$ and $\Lamv$ consist of the Fourier coefficients of the Green functions of
the complete form $\cC_{\pm}^n$ and $\cU_{\pm}^n$ given by
Eqs.(\ref{eq:cCp}-\ref{eq:cCm}) and (\ref{eq:cUp}-\ref{eq:cUm}).
$\Gamv^{\ast}$ and $\Lamv^{\ast}$ are the coupling Green functions of 
the complete form for the radial and longitudinal components of the electric and
magnetic fields $\tE_{x,s}$ and $\tB_{x,s}$ respectively.
$\Gamv^{\ast}$ and $\Lamv^{\ast}$ consist of the Fourier coefficients $\cV_{\pm}^n$
given by Eqs.(\ref{eq:cVp}-\ref{eq:cVm}).
$\Gam_y$ and $\Lam_y$ are the Green functions of the complete form for
the vertical components of the fields $\tE_y$ and $\tB_y$ respectively,
\begin{align}
  \Gam_y(\xv_\perp,\xv_\perp',\vsig)
  =\sum_{n=0}^{\infty}\cC_{+}^n(r,r',\vsig)\cY_{-}^n(y,y') ,
   \qquad
  \Lam_y(\xv_\perp,\xv_\perp',\vsig)
  =\sum_{n=1}^{\infty}\cC_{-}^n(r,r',\vsig)\cY_{+}^n(y,y') .
   \label{eq:Gamy_Lamy}
\end{align}
$\vsig=\{s-s',s\}$ for all the Green functions given by Eqs.(\ref{eq:tGam_tLam}) as
those involved in Eqs.(\ref{eq:hPx}-\ref{eq:hQs}).
$\cY_{\pm}^n$ is the vertical eigenfunctions given by Eqs.(\ref{eq:cYp_0}-\ref{eq:cYm}).
$\Gam_{x,s}$ and $\Lam_{x,s}$ are the Green functions of the complete form
for $\tE_{x,s}$ and $\tB_{x,s}$ respectively,
\begin{align}
  \bigg[
  \begin{array}{cc}
    \Gam_x(\xv_\perp,\xv_\perp',\vsig) & \Gam_x^{\ast}(\xv_\perp,\xv_\perp',\vsig)
     \\
    \Gam_s(\xv_\perp,\xv_\perp',\vsig) & \Gam_s^{\ast}(\xv_\perp,\xv_\perp',\vsig)
  \end{array}
  \bigg]
  &=\sum_{n=1}^{\infty}
  \bigg[
  \begin{array}{cc}
      \cU_{-}^{n}(r,r',\vsig) & \cV_{-}^{n}(r,r',\vsig)
       \\
      \cU_{+}^{n}(r,r',\vsig) & \cV_{+}^{n}(r,r',\vsig)
  \end{array}
  \bigg]
    \cY_{+}^n(y,y') ,
  \label{eq:tGam_xs}
   \\
  \bigg[
  \begin{array}{cc}
     \Lam_x(\xv_\perp,\xv_\perp',\vsig) & \Lam_x^{\ast}(\xv_\perp,\xv_\perp',\vsig)
       \\
     \Lam_s(\xv_\perp,\xv_\perp',\vsig) & \Lam_s^{\ast}(\xv_\perp,\xv_\perp',\vsig)
  \end{array}
  \bigg]
  &=\sum_{n=0}^{\infty}
    \bigg[
    \begin{array}{cc}
      \cU_{+}^{n}(r,r',\vsig) & \cV_{+}^{n}(r,r',\vsig)
       \\
      \cU_{-}^{n}(r,r',\vsig) & \cV_{-}^{n}(r,r',\vsig)
    \end{array}
    \bigg]
    \cY_{-}^n(y,y') ,
  \label{eq:tLam_xs}
\end{align}
where the functions in the matrices on the left and right hand sides
behave as in Eqs.(\ref{eq:tPhi_xs}-\ref{eq:tPsi_xs}).

According to Eqs.(\ref{eq:cCpm_BCr}), (\ref{eq:cYp}) and (\ref{eq:cYm}),
$\Gam_y$ and $\Lam_y$ satisfy the following boundary conditions on the walls of
the curved rectangular pipe,
\begin{alignat}{4}
  \Gam_y
  &=0
   \quad&\text{at}&~~&
  r
  &=r_{b,a} ~~\text{and}~~ r'=r_{b,a} ,
    \qquad&&
  [\rd_{\hr}\Lam_y]_{r=r_{b,a}}
  =[\rd_{\hr'}\Lam_y]_{r'=r_{b,a}}
  =0 ,
   \label{eq:BC_Gamy_Lamy}
   \\
  \Lam_y
  &=0
   \quad&\text{at}&~~&
  y
  &=\pm h/2 ~~\text{and}~~ y'=\pm h/2 ,
    \qquad&&
  [\rd_y\Gam_y]_{y=\pm h/2}
  =[\rd_{y'}\Gam_y]_{y'=\pm h/2}
  =0 .
\end{alignat}
$\Gam_y(\vsig)$, $\Lam_y(\vsig)$ and their derivatives with respect to $\vsig=\{s-s',s\}$ 
at $\vsig=0$ are given as
\begin{align}
  [\Gam_y(\vsig)]_{\vsig=0}=[\Lam_y(\vsig)]_{\vsig=0}=0 ,
    \qquad
  [\rd_s\Gam_y(\vsig)]_{\vsig=0}
  =[\rd_s\Lam_y(\vsig)]_{\vsig=0}
  =\frac{r}{\rho}\delta(r-r')\delta(y-y') .
  \label{eq:tGamy_tLamy_vsig0}
\end{align}
The second derivatives of $\Gam_y(s)$ and $\Lam_y(s)$ with respect to $s$ are
zero at $s=0$, \ie,
\begin{align}
  [\rd_s^2\Gam_y(s)]_{s=0}
  =[\rd_s^2\Lam_y(s)]_{s=0}
  &=0 .
  \label{eq:rds2_Gamy_s0}
\end{align}
We omit the transverse arguments $(\xv_\perp,\xv_\perp')$ of
the Green functions in Eqs.(\ref{eq:tGamy_tLamy_vsig0}-\ref{eq:lim_Gam_s}).
According to Eqs.(\ref{eq:we_cCpm}) and (\ref{eq:delta_yyp}),
$\Gam_y$ and $\Lam_y$ satisfy the following wave equation,
\begin{align}
  \nablav_{\rm v}^2
  \bigg\{{\Gam_y(\vsig) \atop \Lam_y(\vsig)}\bigg\}
  =\bdelta^3(\xv_\perp-\xv_\perp',\vsig) .
  \label{eq:we_Gamy_Lamy}
\end{align}
$\bdelta^3(\vsig)$ is given by Eq.(\ref{eq:bdelta3}).
According to Eqs.(\ref{eq:we_cUpm}-\ref{eq:we_cVpm}) and (\ref{eq:delta_yyp}),
the Green functions given by Eqs.(\ref{eq:tGam_xs}-\ref{eq:tLam_xs}) satisfy
the following wave equations,
\begin{alignat}{2}
  \nablav_\vdash^2\bigg\{{\Gam_{x,s}(\vsig) \atop \Lam_{x,s}(\vsig)}\bigg\}
  -\frac{2}{r}\brd_s
   \bigg\{{ \Gam_{s,x}^{\ast}(\vsig) \atop \Lam_{s,x}^{\ast}(\vsig) }\bigg\}
  &=\bdelta^3(\xv_\perp-\xv_\perp',\vsig) ,
   \label{eq:we_GL}
   \\
  \nablav_\vdash^2
  \bigg\{{ \Gam_{x,s}^{\ast}(\vsig) \atop \Lam_{x,s}^{\ast}(\vsig) }\bigg\}
  +\frac{2}{r}\brd_s\bigg\{{\Gam_{s,x}(\vsig) \atop \Lam_{s,x}(\vsig)}\bigg\}
  &=0 .
   \label{eq:we_GL_ast}
\end{alignat}
Eqs.(\ref{eq:we_GL}-\ref{eq:we_GL_ast}) correspond to Eqs.(\ref{eq:we_tExsBxs})
which are the wave equations for $\tE_{x,s}$ and $\tB_{x,s}$.
Eqs.(\ref{eq:we_hPQMN_y}-\ref{eq:we_hMNs}) are gotten from
Eqs.(\ref{eq:we_Gamy_Lamy}-\ref{eq:we_GL_ast}).

According to Eqs.(\ref{eq:BC_cUp_r}-\ref{eq:BC_cVpm_r}),
the Green functions of the wave equations for $\tE_{x,s}$ and $\tB_{x,s}$ satisfy
the following boundary conditions on the sidewalls of the curved pipe,
\begin{alignat}{4}
  \Gam_s
  &=\Lam_x=\Gam_s^{\ast}=\Lam_x^{\ast}=0,
    \qquad&
  \brd_{\hr}\Gam_x
  &=\brd_{\hr}\Lam_s=\brd_{\hr}\Gam_x^{\ast}=\brd_{\hr}\Lam_s^{\ast}=0
    \qquad&\text{at}&~~
  r
  &=r_{b,a} ,
    \\
  \Gam_s
  &=\Lam_x=\Gam_x^{\ast}=\Lam_s^{\ast}=0,
    \qquad&
  \brd_{\hr'}\Gam_x
  &=\brd_{\hr'}\Lam_s=\brd_{\hr'}\Gam_s^{\ast}=\brd_{\hr'}\Lam_x^{\ast}=0
    \qquad&\text{at}&~~
  r'
  &=r_{b,a} .
\end{alignat}
According to Eqs.(\ref{eq:cYp}) and (\ref{eq:cYm}), the Green functions of the complete 
form satisfy the boundary conditions on the upper and lower walls of the pipe as follows,
\begin{align}
  \Gam_{x,s}=\Gam_{x,s}^{\ast}=0
   ,\qquad
  \rd_y\Lam_{x,s}=\rd_y\Lam_{x,s}^{\ast}=0
   \qquad\text{at}~~
  y=\pm h/2 .
\end{align}
From Eqs.(\ref{eq:cUVpm_s0}-\ref{eq:rds_cUVpm_s0}), we get the Green functions
and their longitudinal derivatives at $\vsig=0$,
\begin{alignat}{3}
  \bigg[{\Gam_{x,s}(\vsig)\atop \Lam_{x,s}(\vsig)}\bigg]_{\vsig=0}
  &=0 ,
    \qquad&
  \bigg[\rd_s\bigg\{{\Gam_{x,s}(\vsig) \atop \Lam_{x,s}(\vsig)}\bigg\}\bigg]_{\vsig=0}
  &=\frac{r}{\rho}\delta(r-r')\delta(y-y'),
   \label{eq:GamLam_vsig0}
   \\
  \bigg[{\Gam_{x,s}^{\ast}(\vsig)\atop \Lam_{x,s}^{\ast}(\vsig)}\bigg]_{\vsig=0}
  &=0 ,
   \qquad&
  \bigg[\rd_s\bigg\{{\Gam_{x,s}^{\ast}(\vsig) \atop \Lam_{x,s}^{\ast}(\vsig)}\bigg\}
  \bigg]_{\vsig=0}
  &=0 .
   \label{eq:rds_GamLam_vsig0}
\end{alignat}
From the third equations of (\ref{eq:cUVpm_s0}-\ref{eq:rds_cUVpm_s0}),
we get their second derivatives at $s=0$ or $\vsig=0$,
\begin{align}
  &
  \bigg[\rd_s^2\bigg\{{\Gam_{x,s}(s)\atop \Lam_{x,s}(s)}\bigg\}\bigg]_{s=0}
  =0
   ,\qquad
  \bigg[
    \rd_s^2
    \bigg\{{ \Gam_{x,s}^{\ast}(\vsig) \atop \Lam_{x,s}^{\ast}(\vsig) }\bigg\}
  \bigg]_{\vsig=0}
  =-\frac{2r}{\rho^2}\delta(r-r')\delta(y-y') .
   \label{eq:rds2_GL}
\end{align}
In the limit of $\rho\to\infty$, the Green functions of the complete form go to
the following limits,
\begin{align}
  \lim_{\rho\to\infty}\Gam_x
  &=\lim_{\rho\to\infty}\Lam_y
  =\Gam_{-}^{+}
   ,\qquad
  \lim_{\rho\to\infty}\Gam_y
  =\lim_{\rho\to\infty}\Lam_x
  =\Gam_{+}^{-} ,
   \label{eq:lim_Gam_xy}
   \\
  \lim_{\rho\to\infty}\Gam_s
  &=\Gam_{+}^{+}
   ,\qquad
  \lim_{\rho\to\infty}\Lam_s
  =\Gam_{-}^{-}
   ,\qquad
  \lim_{\rho\to\infty}\Gam_{x,s}^{\ast}
  =\lim_{\rho\to\infty}\Lam_{x,s}^{\ast}
  =0 .
   \label{eq:lim_Gam_s}
\end{align}
$\Gam_{\pm}^{\pm}$ is given by Eqs.(\ref{eq:Gam_pm_pm}) which are
the Green functions of the complete form in the straight rectangular pipe.
The coupling Green functions $\Gamv^{\ast}$ and $\Lamv^{\ast}$ vanish in
the limit of $\rho\to\infty$, similar to $\Phiv^{\ast}$ and $\Psiv^{\ast}$ in
Eqs.(\ref{eq:lim_Phi_ast}).

\subsection{Differential expression of the fields in the complete form}

We find the differential expressions of the complete form of the fields,
which are the complete form version of Eq.(\ref{eq:tEBv_dg}).
We rewrite Eq.(\ref{eq:tEBv_cmpl}) by integrating by parts the derivatives of
the current components, which are involved in the source terms
$\tilde{\bm S}$ and $\tilde{\bm T}$, with respect to $r'$, $y'$ or $s'$,
\begin{align}
  \bigg\{{\tEv(\xv) \atop c\tBv(\xv)}\bigg\}
  &=\int_{r_a}^{r_b}dr'\int_{-h/2}^{h/2}dy'
    \bigg[
       \bigg\{{\hPv^{\dg}(\xv,\xv_\perp') \atop
               \hQv^{\dg}(\xv,\xv_\perp')}
       \bigg\}
      -Z_0\int_{0}^{\infty}ds'
       \bigg\{{\hMv^{\dg}(\xv_{\perp},s-s',\xv') \atop \hNv^{\dg}(\xv_{\perp},s-s',\xv')}
       \bigg\}
    \bigg] ,
  \label{eq:tEBv_cmpl_dg}
\end{align}
where
\begin{alignat}{2}
  \hPv^{\dg}
  &=\hP_x^{\dg}\ev_x+\hP_y^{\dg}\ev_y+\hP_s^{\dg}\ev_s ,
     \qquad&
  \hMv^{\dg}
  &=\hM_x^{\dg}\ev_x+\hM_y^{\dg}\ev_y+\hM_s^{\dg}\ev_s ,
   \label{eq:hPv_hMv_dg}
   \\
  \hQv^{\dg}
  &=\hQ_x^{\dg}\ev_x+\hQ_y^{\dg}\ev_y+\hQ_s^{\dg}\ev_s ,
     \qquad&
  \hNv^{\dg}
  &=\hN_x^{\dg}\ev_x+\hN_y^{\dg}\ev_y+\hN_s^{\dg}\ev_s .
   \label{eq:hQv_hNv_dg}
\end{alignat}
The components of Eqs.(\ref{eq:hPv_hMv_dg}-\ref{eq:hQv_hNv_dg}) are given as follows,
\begin{alignat}{2}
  \hP_x^{\dg}(s)
  &=\tD_x^{\dg}\Gam_x(s)-\tD_s^{\dg}\Gam_x^{\ast}(s) ,
    \qquad&
  \hM_x^{\dg}(\vsig,s')
  &=\frac{r'}{\rho}\{\tS_x^{\dg}(s')\Gam_x(\vsig)-\tS_s^{\dg}(s')\Gam_x^{\ast}(\vsig)\} ,
  \label{eq:hPx_dg}
   \\
  \hP_y^{\dg}(s)
  &=\tD_y^{\dg}\Gam_y(s) ,
    \qquad&
  \hM_y^{\dg}(\vsig,s')
  &=\frac{r'}{\rho}\tS_y^{\dg}(s')\Gam_y(\vsig) ,
   \\
  \hP_s^{\dg}(s)
  &=\tD_s^{\dg}\Gam_s(s)+\tD_x^{\dg}\Gam_s^{\ast}(s) ,
    \qquad&
  \hM_s^{\dg}(\vsig,s')
  &=\frac{r'}{\rho}\{\tS_s^{\dg}(s')\Gam_s(\vsig)+\tS_x^{\dg}(s')\Gam_s^{\ast}(\vsig)\} ,
  \label{eq:hPs_dg}
   \\
  \hQ_x^{\dg}(s)
  &=\tA_x^{\dg}\Lam_x(s)-\tA_s^{\dg}\Lam_x^{\ast}(s) ,
    \qquad&
  \hN_x^{\dg}(\vsig,s')
  &=\frac{r'}{\rho}\{\tT_x^{\dg}(s')\Lam_x(\vsig)-\tT_s^{\dg}(s')\Lam_x^{\ast}(\vsig)\} ,
  \label{eq:hQx_dg}
   \\
  \hQ_y^{\dg}(s)
  &=\tA_y^{\dg}\Lam_y(s) ,
    \qquad&
  \hN_y^{\dg}(\vsig,s')
  &=\frac{r'}{\rho}\tT_y^{\dg}(s')\Lam_y(\vsig) ,
   \\
  \hQ_s^{\dg}(s)
  &=\tA_s^{\dg}\Lam_s(s)+\tA_x^{\dg}\Lam_s^{\ast}(s) ,
    \qquad&
  \hN_s^{\dg}(\vsig,s')
  &=\frac{r'}{\rho}\{\tT_s^{\dg}(s')\Lam_s(\vsig)+\tT_x^{\dg}(s')\Lam_s^{\ast}(\vsig)\} .
  \label{eq:hQs_dg}
\end{alignat}
We omit the transverse arguments $(\xv_{\perp},\xv_{\perp}')$ in
Eqs.(\ref{eq:hPx_dg}-\ref{eq:hNs_cCp}) for clarity to show the longitudinal arguments.
$\vsig=s-s'$ ($\vsig\ne s$) in Eqs.(\ref{eq:hPx_dg}-\ref{eq:hNs_cCp}).
$(\tilde{\bm D}^{\dg},\tilde{\bm A}^{\dg})$ and
$(\tilde{\bm S}^{\dg},\tilde{\bm T}^{\dg})$ are the operators given by
Eqs.(\ref{eq:tDx_dg}-\ref{eq:tAs_dg}) and (\ref{eq:tSxs_dg}-\ref{eq:tTxs_dg}).
We rearrange the components of $\hPv^{\dg}$ and $\hQv^{\dg}$ as follows,
\begin{alignat}{2}
  \hP_y^{\dg}(s)
  &=\Big\{\tE_y(0)\frac{\rho}{r'}\rd_s-\tE_s(0)\rd_{y'}-c\tB_x(0)ik\beta\Big\}
    \sum_{n=0}^{\infty}\cC_{+}^n(s)\cY_{-}^n ,
     \qquad&
  \hP_{x,s}^{\dg}(s)
  &=\sum_{n=1}^{\infty}\hat{p}_{x,s}^{\dg}(s)\cY_{+}^n ,
   \label{eq:hPy_dg_cCp}
   \\
  \hQ_y^{\dg}(s)
  &=\Big\{c\tB_y(0)\frac{\rho}{r'}\rd_s-c\tB_s(0)\rd_{y'}+\tE_x(0)ik\beta\Big\}
    \sum_{n=1}^{\infty}\cC_{-}^n(s)\cY_{+}^n ,
     \qquad&
  \hQ_{x,s}^{\dg}(s)
  &=\sum_{n=0}^{\infty}\hat{q}_{x,s}^{\dg}(s)\cY_{-}^n
   \label{eq:hQy_dg_cCm} .
\end{alignat}
$\tEv(0)$ and $\tBv(0)$ denote the initial fields at the entrance of the bend as shown in
Eqs.(\ref{tEv_tBv_sp0}).
The components of the fields $\tE_{x,y,s}(s')$ and $\tB_{x,y,s}(s')$ are continuous at
$s'=0$ unlike their longitudinal derivatives $\rd_{s'}\tE_{x,y,s}(s')$ and
$\rd_{s'}\tB_{x,y,s}(s')$ which are discontinuous at $s'=0$ as discussed in
section \ref{sec:disconti}.
$\hat{p}_{x,s}^{\dg}$ and $\hat{q}_{x,s}^{\dg}$ are operators having $\rd_{y'}$
which act on $\cY_{\pm}^n$ given by Eqs.(\ref{eq:cYp_0}-\ref{eq:cYm}),
\begin{align}
  \hat{p}_{x}^{\dg}(s)
  &=
    \{\tE_x(0)\brd_s\cC_{-}^n(s)+\tE_s(0)\rd_{r}\cC_{+}^n(s)\}
   +\{c\tB_y(0)\cU_{-}^n(s)ik\beta -\tE_y(0)\cV_{-}^n(s)\rd_{y'}\} ,
   \\
  \hat{p}_{s}^{\dg}(s)
  &=
    \{\tE_s(0)\brd_s\cC_{+}^n(s)-\tE_x(0)\rd_{r}\cC_{-}^n(s)\}
   +\{\tE_y(0)\cU_{+}^n(s)\rd_{y'}+c\tB_y(0)\cV_{+}^n(s)ik\beta\} ,
   \\
  \hat{q}_{x}^{\dg}(s)
  &=
    \{c\tB_x(0)\brd_s\cC_{+}^n(s)+c\tB_s(0)\rd_{r}\cC_{-}^n(s)\}
   -\{\tE_y(0)\cU_{+}^n(s)ik\beta+c\tB_y(0)\cV_{+}^n(s)\rd_{y'}\} ,
   \\
  \hat{q}_{s}^{\dg}(s)
  &=
    \{c\tB_s(0)\brd_s\cC_{-}^n(s)-c\tB_x(0)\rd_{r}\cC_{+}^n(s)\}
   +\{c\tB_y(0)\cU_{-}^n(s)\rd_{y'}-\tE_y(0)\cV_{-}^n(s)ik\beta\} ,
   \label{eq:hqs_dg}
\end{align}
where $\brd_s=(\rho/r)\rd_s$ as in Eq.(\ref{eq:vopera}).
We rearrange the components of $\hMv^{\dg}$ and $\hNv^{\dg}$ as follows,
\begin{alignat}{2}
  \hM_y^{\dg}(\vsig,s')
  &=\frac{r'}{\rho}\big\{\tJ_0(s')\rd_{y'}+\tJ_y(s')ik\beta\big\}
    \sum_{n=0}^{\infty}\cC_{+}^n(\vsig)\cY_{-}^n
    ,\qquad&
  \hM_{x,s}^{\dg}(\vsig,s')
  &=\frac{r'}{\rho}\sum_{n=1}^{\infty}
    \hat{m}_{x,s}^{\dg}(\vsig,s')\cY_{+}^n ,
   \label{eq:hMy_cCp}
  \\
  \hN_y^{\dg}(\vsig,s')
  &=\frac{r'}{\rho}\Big\{\tJ_s(s')\rd_{r'}+\tJ_x(s')\frac{\rho}{r'}\rd_s\Big\}
    \sum_{n=1}^{\infty}\cC_{-}^n(\vsig)\cY_{+}^n
    ,\qquad&
  \hN_{x,s}^{\dg}(\vsig,s')
  &=\frac{r'}{\rho}\sum_{n=0}^{\infty}
    \hat{n}_{x,s}^{\dg}(\vsig,s')\cY_{-}^n .
   \label{eq:hNy_cCm}
\end{alignat}
$\hat{m}_{x}^{\dg}$ and $\hat{m}_{s}^{\dg}$ are the functions given below.
On the other hand,
$\hat{n}_{x}^{\dg}$ and $\hat{n}_{s}^{\dg}$ are operators having $\rd_{y'}$,
\begin{align}
  \hat{m}_x^{\dg}(\vsig,s')
  &=
    ik\beta\{\tJ_x(s')\cU_{-}^n(\vsig)-\tJ_s(s')\cV_{-}^n(\vsig)\}
   -\tJ_0(s')\rd_{r}\cC_{+}^n(\vsig) ,
   \label{eq:hMx_cCp}
   \\
  \hat{m}_s^{\dg}(\vsig,s')
  &=
    ik\beta\{\tJ_s(s')\cU_{+}^n(\vsig)+\tJ_x(s')\cV_{+}^n(\vsig)\}
   -\tJ_0(s')\brd_s\cC_{+}^n(\vsig) ,
   \\
  \hat{n}_x^{\dg}(\vsig,s')
  &=
   -\{\tJ_s(s')\cU_{+}^n(\vsig)+\tJ_x(s')\cV_{+}^n(\vsig)\}\rd_{y'}
   -\tJ_y(s')\brd_s\cC_{+}^n(\vsig) ,
   \\
  \hat{n}_s^{\dg}(\vsig,s')
  &=
    \{\tJ_x(s')\cU_{-}^n(\vsig)-\tJ_s(s')\cV_{-}^n(\vsig)\}\rd_{y'}
   +\tJ_y(s')\rd_{r}\cC_{+}^n(\vsig) .
   \label{eq:hNs_cCp}
\end{align}
%

\subsection{Expressions of the fields and their utilities}
\label{sec:expression}

We found four expressions of the transient electromagnetic field of synchrotron radiation 
in the frequency domain.
They are the combination of the separated/complete forms and the scalar/differential 
expressions, which involve the following integrands,
\begin{align}
  \begin{array}{rll}
                           & \text{Scalar expressions},
                           & \text{Differential expressions}
    \\
    \text{Separated forms}:& (\tPv,\tQv,\tMv,\tNv)~\text{in Eq.(\ref{eq:tEv_tBv})},\quad
                           & (\tPv^{\dg},\tQv^{\dg},\tMv^{\dg},\tNv^{\dg})~
                             \text{in Eq.(\ref{eq:tEBv_dg})} ,
    \\
    \text{Complete forms}: & (\hPv,\hQv,\hMv,\hNv)~\text{in Eq.(\ref{eq:tEBv_cmpl})},\quad
                           & (\hPv^{\dg},\hQv^{\dg},\hMv^{\dg},\hNv^{\dg})~
                             \text{in Eq.(\ref{eq:tEBv_cmpl_dg})} .
  \end{array}
  \label{eq:4express}
\end{align}
In appendix \ref{sec:opera_express} for later use in appendices
\ref{sec:verify} and \ref{sec:impedance}, we will explicitly write out
the differential expressions given by Eqs.(\ref{eq:tEBv_dg}) and (\ref{eq:tEBv_cmpl_dg}).
Since the explanations about these four expressions of the field are dispersed over
sections \ref{sec:ILT}, \ref{sec:normal_form} and \ref{sec:FD_field},
we review their characteristics and utilities in the rest of this section.

We first describe the difference of the separated form and the complete form of the field.
They consist of the following Green functions in the Fourier mode with respect to
the vertical variable (see section \ref{sec:Green} in p.\pageref{sec:Green}) and those in 
the frequency domain,
\begin{align}
  \begin{array}{rll}
                           & \text{$y$-Fourier mode,}~~
                           & \text{Frequency domain}
    \\
    \text{Separated form}: & \cG_{\pm}^n,\cH_{\pm}^n,\cK_{\pm}^n,
                           & (\Phiv,\Psiv)~\text{and}~(\Phiv^{\ast},\Psiv^{\ast})
                             ~\text{in Eqs.(\ref{eq:tPhi_tPsi})} ,
    \\
    \text{Complete form}:  & \cC_{\pm}^n,\cU_{\pm}^n,\cV_{\pm}^n,
                           & (\Gamv,\Lamv)~\text{and}~(\Gamv^{\ast},\Lamv^{\ast})
                             ~\text{in Eqs.(\ref{eq:tGam_tLam})} .
  \end{array}
  \label{eq:sep_cmp}
\end{align}
The separated form and the complete form are related through the identities
(\ref{eq:sum_Xmn}) as shown in Eqs.(\ref{eq:cGcC}-\ref{eq:cG_cC}).
The separated and complete forms are suited for use respectively in
the numerical and analytical calculations,
because the separated form of the field consists of a finite number of
the oscillatory modes and infinite number of the damped modes with respect to
$\vsig=\{s-s',s\}$, which correspond respectively to the real and imaginary poles of
the field in the Laplace domain as shown in Eqs.(\ref{eq:ps_poles}).
Therefore none of the terms in the separated form diverges for $s\to\infty$.
The oscillatory and damped modes are separated into the sums with respect to the radial mode
number $m$ as in Eqs.(\ref{eq:cGp_sep}-\ref{eq:cGm_sep}) and (\ref{eq:cHp}-\ref{eq:cKm}).
\ie, the sum with respect to the radial mode is truncated at $m=m_{\pm}$
as shown in Eqs.(\ref{eq:ps_poles}-\ref{eq:Re_poles}).
The separated sums with respect to $m$ each do not produce the radial $\delta$-function
$\delta(r-r')$ which is hard to deal within the numerical calculation.
On the contrary, it is harder to deal with the separated sums in the analytical 
calculation than the complete sum, because the separated sums are partitioned by
the window functions $\Th_{\pm}^{n\ell}$,
given by Eqs.(\ref{eq:Th_def}-\ref{eq:bTh_def}),
with respect to the cutoff wavenumbers $k=(k_m^n,\bk_m^n)$ which are defined implicitly by 
Eqs.(\ref{eq:kmn}-\ref{eq:bkmn}) for the second order poles at the origin of
the Laplace plane.

The complete form of the field has the complete sum over all the radial modes from
the lowest mode to infinity as shown in Eqs.(\ref{eq:cCp}-\ref{eq:cCm}) and
(\ref{eq:cUp}-\ref{eq:cVm}).
The complete sum of the radial eigenfunctions $\cR_{\pm}^{mn}$ produces
the radial $\delta$-function $\delta(r-r')$ as shown in Eqs.(\ref{eq:rdel})
which are gotten from the orthogonality relations of the cross products of
the Bessel functions as described in appendix \ref{sec:orthogonality}.
Using Eqs.(\ref{eq:rdel}), we can show that the Green functions of the complete form
$(\Gamv,\Lamv)$ produce $\delta(r-r')$ in Eqs.(\ref{eq:tGamy_tLamy_vsig0}),
(\ref{eq:GamLam_vsig0}) and (\ref{eq:rds2_GL}).
It is true that the Fourier coefficients of the Green functions of the separated form,
$\cG_{\pm}^n$ and ($\cH_{\pm}^n,\cK_{\pm}^n$), satisfy the wave equations
which have the $\delta$-function or 0 in the driving term as 
shown in Eqs.(\ref{eq:BessDE_delta}) and (\ref{eq:Grn_cHpm_cKpm}).
But we derived these wave equations from Eqs.(\ref{eq:we_cCpm}) and (\ref{eq:we_cUpm})
which are the wave equations for the complete form Green functions
$\cC_{\pm}^n$ and ($\cU_{\pm}^n,\cV_{\pm}^n$).
Thus, the complete form is suited to using in the analytical calculation
since it can produce the radial and longitudinal $\delta$-functions.
In particular, it is necessary in the analytical verification of the solution of the field,
shown in appendix \ref{sec:verify}.
On the other hand, the complete form is not suited to using in the numerical 
calculation since the complete sum involves the hyperbolic functions of $\vsig=\{s-s',s\}$,
which diverge exponentially for $s\to\infty$.
Since the divergent terms for $s\to\infty$ in the complete form cancel out and vanish in
the expression of the field through the wave equations with respect to $\xv'$,
the field itself does not diverge for $s\to\infty$.
As described at the end of section \ref{sec:cGpm_we}, the separated form is the expression 
of the field that we removed the divergent terms for $s\to\infty$ from the complete form of
the field.

Next, we describe the characteristics of the scalar and differential expressions of
the field.
To put it simply, their relation is merely the integration by parts with respect to
$\xv'=(r',y',s')$.
That is, we derived the differential expressions from the scalar expressions by integrating
the derivatives of the current components which are convoluted with the Green functions in
the expression of the field.
We dubbed these expressions, {\it the scalar} and {\it differential} 
expressions, in terms of the Green functions which are involved in the expressions of
the field (not in terms of the current).
That is, in the scalar expression of the field, the components of the source terms
$(\tSv,\tTv)$, which are scalar functions (not operators),
are multiplied to the Green functions as shown in
Eqs.(\ref{eq:tPx}-\ref{eq:tQs}) and (\ref{eq:hPx}-\ref{eq:hQs}).
On the other hand, in the differential expression of the field,
the components of $(\tSv^{\dg},\tTv^{\dg})$,
which are differential operators (not scalar functions), act on the Green functions
as shown in Eqs.(\ref{eq:tPx_dg}-\ref{eq:tQs_dg}) and (\ref{eq:hPx_dg}-\ref{eq:hQs_dg}).
Conversely, in the scalar expression of the field, the differential operators
$(\rd_{r'},\rd_{y'},\rd_{s'})$ act on the current $\tJ(\xv')$
instead of acting on the Green functions.

The differential expression is useful if the beam current has a transverse distribution
much thinner than the transverse dimension of the beam pipe $d_{\perp}=(w,h)$
since the current components are not differentiated with respect to the coordinate variables
$\xv'$ instead that the Green functions are differentiated.
We need the differential expression if the current is infinitely thin, \ie,
if the beam has the transverse distribution as $\delta(r-r')$ or $\delta(y-y')$ or both
since the differential expression does not involve the derivative of the transverse
$\delta$-functions $\rd_{r'}\delta(r-r')$ and $\rd_{y'}\delta(y-y')$.
In addition, we can calculate the integrals analytically with respect to $r'$ and $y'$
for an infinitely thin current if the transverse $\delta$-functions are not differentiated 
with respect to $r'$ and $y'$.
Thus, the differential expression is useful in calculating a field created by
a beam current whose transverse size $\sig_{x,y}$ is zero or very small compared to
the transverse dimension of the beam pipe.
In numerically calculating the longitudinal electric field in section \ref{sec:discussion},
we use the differential expression of the separated form which is rewritten using
the steady radiation field for $s\to\infty$ in the curved pipe as shown in
appendix \ref{sec:Es_bend_std}.

The scalar expression of the complete form is useful in the analytical verification
where the solution of the field components satisfies the wave equations
(\ref{eq:we_tEy_pm}) and (\ref{eq:we_tBy_pm}-\ref{eq:we_tExsBxs})
since the driving term of the wave equations involve the derivatives of
the current components as seen from Eqs.(\ref{eq:tSv}-\ref{eq:tTv}).
But since usually $\sig_{x,y}\ll d_{\perp}$ in modern accelerators,
the scalar expressions of the field are not useful in the numerical 
calculation of the field of synchrotron radiation emitted by a thin beam.
The scalar expression of the field is the intermediate expression in
deriving the differential expression.
We think that the difference between the scalar and differential expressions is
less important than the difference between the separated and complete forms
since the former implies merely the integration by parts with respect to $\xv'$
as already described.
Nevertheless, we showed both the scalar and differential expressions of the field in
the present paper since it is somewhat complicated to do the integration by parts and
to rearrange the expressions through Maxwell equations.

\clearpage

\section{Numerical results and discussion}
\label{sec:discussion}

We verify the exact solution of the longitudinal electric field of CSR by comparing with
the numerical results in other theories.
Fig.\ref{fig:Esz} shows $E_s$ in the time domain, which is the longitudinal electric field
of transient CSR shielded by a curved pipe in a bend.
In general, $E_s$ of CSR hardly depends on the transverse beam size $\sig_{x,y}$
if $\sig_{x,y}\ll \ell_{\perp}$ which is the typical transverse extent of
the steady field of CSR as described in Eq.(\ref{eq:shield}).
In this section we assume a rigid and thin electron bunch
which has no horizontal extent as $\psi_x=\delta(x)$ and no correlation among the coordinates $\xv=(x,y,s)$ for simplicity.
Conversely, if the horizontal charge distribution $\psi_x$ has an extent, \ie,
$\psi_x\ne\delta(x)$, since the charge distribution of
the bunch $J_0/c$ will have a correlation between $x$ and $s$ in the horizontal bend,
it contradicts the assumption that the bunch is rigid in the bend
(if it is a straight section, we can assume a rigid bunch having a horizontal extent).
Assuming a rigid bunch, besides the simplicity,
we can define the impedance of the beamline as described later in Eq.(\ref{eq:Z_bend}).

Assuming that a rigid and thin bunch is moving along the $s$-axis at a constant speed $v$
in a straight or bending section, the electric current of the beam in the frequency domain
(\ref{eq:Fourier_trans}-\ref{eq:omg}) is given as
\begin{align}
  \tJ_0(\xv,k)
  =cq\tlam(k,s)\delta(x)\psi_y(y)
    ,\qquad
  \tJv(\xv,k)
  =\beta\tJ_0(\xv,k)\ev_s
    \qquad
  (v=c\beta,~ 0<\beta<1) .
   \label{eq:thin_bunch}
\end{align}
$q$ is the bunch charge.
$\tJ_0/c$ and $\tJv$ are the charge and current densities of the bunch in
the frequency domain.
In this section we assume that the vertical charge distribution $\psi_y$ has
a Gaussian distribution of spread $\sig_y$ as in Eq.(\ref{eq:tpsi_gaussian})
in order to reduce the computing time, \ie,
the sum with respect to the vertical mode number converges faster than the bunch that 
assumes $\psi_y=\delta(y)$ owing to the decrease of the Fourier coefficients $\tpsi_y^n$
in computing the values of $\tE_s$ using the exact solution (\ref{eq:tEs_std_sep}).
In this section we assume that the longitudinal charge distribution $\lam$ has
a smooth Gaussian distribution of bunch length $\sig_z$, \ie,
\begin{align}
  \lam(z)
  =\frac{e^{-(z/\sig_z)^2/2}}{\sig_z(2\pi)^{1/2}} ,
     \qquad
  \hlam(k)
  =\tlam(k,s)e^{-iks}
  =\frac{e^{-(k\sig_z)^2/2}}{v} .
   \label{eq:rigid_bunch}
\end{align}
$z$ is defined in Eq.(\ref{eq:z}) on the basis of the reference particle.
$\tlam$ is the  Fourier transform of $\lam$ with respect to $\{t,\omg\}$
defined in Eqs.(\ref{eq:Fourier_trans}-\ref{eq:omg}).
$\hlam$ is the bunch spectrum scaled by $v$ in accordance with
Eqs.(\ref{eq:Fourier_trans}-\ref{eq:omg}).
As described below Eq.(\ref{eq:z}), we define a ``rigid'' beam (bunch) as the one
whose $\lambda$ does not depend on $s$ explicitly, \ie, $\hlam$ does not depend on $s$
regardless of the curvature of the $s$-axis.
For example, a single point charge which is moving on the $s$-axis at
a constant speed $v$ similar to the reference particle, \ie,
$\lambda=\delta(z)$ and $\psi_{x,y}=\delta(\xv_{\perp})$, is also a rigid beam.
To be rigorous in terms of beam dynamics, a rigid beam is an approximate model
as the source current to calculate the radiation field
since any rigid beam loses no energy by definition.

Assuming Eqs.(\ref{eq:thin_bunch}), the exact expression of the transient field $\tE_s$
shielded by the curved pipe in the bend is given by Eq.(\ref{eq:tEs_std_sep}).
As shown in Fig.\ref{fig:Esz}, we calculate the field in the time domain $E_s$ as
a function of $(\xv,z)$ instead of $(\xv,t)$ by Fourier transforming
$\tE_s(\xv,k)e^{-iks}$ $(=\hE_s)$ numerically with respect to $\{z,k\}$,
\begin{align}
  F(\xv,z)
  &=\frac{v}{2\pi}\int_{-\infty}^{\infty}\hF(\xv,k)e^{ikz}dk ,
    \qquad
  \hF(\xv,k)
  =\int_{-\infty}^{\infty}\frac{dz}{v}F(\xv,z)e^{-ikz} ,
   \label{eq:FT_zk_9}
\end{align}
where $F(z)$ and $\hF(k)$ represent the quantities in the time and frequency domains.
$e^{ikz}$ contains $e^{iks}$ which is factored out of the field and current in
the frequency domain (\ref{eq:FT_zk_9}) unlike $\tF(k)$ defined in
Eqs.(\ref{eq:Fourier_trans}-\ref{eq:omg}),
\begin{align}
  \tF(\xv,k)
  =\hF(\xv,k) e^{iks} .
  \label{eq:tF_hF_exp}
\end{align}
%

\subsection{Comparison of the exact solution with the paraxial approximation}
\label{sec:parax_approx}

We first verify Eq.(\ref{eq:tEs_std_sep}), which is the exact solution (ES) of $\tE_s$ in
a bend, by comparing with the numerical solution of the parabolic wave equation
with respect to $\xv_{\perp}$ and $s$.
It also implies that we examine the accuracy of the calculation of the field
on the basis of the paraxial approximation (PA).
We use the following parabolic wave equation \cite{agoh_ws} for the transverse electric field
$\hEv_{\perp}$ in the frequency domain (\ref{eq:FT_zk_9}),
\begin{align}
  &
   \Big\{
     \nablav_{\!\perp}^2
    +k^2\Big(\beta^2-\frac{1}{g^2}\Big)
    +\frac{2ik}{g^2}\rd_s
   \Big\}
   \hEv_{\perp}
  -\frac{2ik}{g^2\vrho}\hE_s\bev_x
  \simeq
   Z_0\nablav_{\!\perp} \hJ_0 ,
   \label{eq:PE_Exy}
  \\
  &
  \hE_s
  \simeq \frac{g}{ik}(Z_0\hJ_0-\nablav_{\!\perp}\cdot\hEv_\perp) ,
    \qquad
  \hEv_{\perp}
  =\hE_x\bev_x+\hE_y\bev_y .
   \label{eq:PE_Es}
\end{align}
In general, $g$ and $\vrho$ depend on $s$ as in Eqs.(\ref{eq:coordinates}).
$\hEv=(\hEv_{\perp},\hE_s)$ and $\hJ_0/c$ are the electric field and the charge density of 
the electric current in the frequency domain (\ref{eq:FT_zk_9}).
$\bev_i$ is a two dimensional basis to represent the horizontal and vertical directions
$(i=x,y)$ in the transverse plane regardless of $s$,
\begin{align}
  \bev_i\cd\bev_j
  =\delta_j^i
   ,\qquad
  \rd_{x,y,s}\bev_i
  =0
   ,\qquad
  \nablav_{\!\perp}
  =\bev_x\rd_x+\bev_y\rd_y .
\end{align}
$\delta_j^i$ is the Kronecker delta extended for $(x,y)$.
$\hEv$ and $\hJ_0$ depend on $(\xv,k)$ in general.
Assuming a rigid bunch, however, $\hJ_0$ does not depend on $s$,
\begin{align}
  \hJ_0(\xv_\perp,k)
  =cq\hlam(k)\delta(x)\psi_y(y) ,
    \qquad
  \hJv(\xv_\perp,k)
  =\beta\hJ_0(\xv_\perp,k)\ev_s .
\end{align}
$\hJv$ is the current density in the frequency domain (\ref{eq:FT_zk_9}).
Unless $\vkap=0$, $\ev_s$ depends on $s$ as shown in Eq.(\ref{eq:coordinates}).

The values of the field gotten by solving Eqs.(\ref{eq:PE_Exy}-\ref{eq:PE_Es}) are more 
accurate than that of Eqs.(8) and (21) of \cite{agoh_yokoya} since we take into account
the entanglement of the horizontal and longitudinal components of the field in the bend.
A lower frequency component of the field tends to be less accurate in
the paraxial approximation since we neglected $\rd_s^2$ for $2ik\rd_s$ in the braces of
Eq.(\ref{eq:PE_Exy}) under the assumption that the field in the bend evolves much slower
than the change of $e^{iks}$ with respect to $s$, \ie,
\begin{align}
  \rd_s
  =O(s_{\rm f}^{-1})
  \ll k .
   \label{eq:rds_ll_k}
\end{align}
$s_{\rm f}$ is given by Eq.(\ref{eq:formation}) which denotes
the formation length of the field of unshielded CSR in a bend of radius $\rho$.
In Fig.\ref{fig:asymp_csr} (p.\pageref{fig:asymp_csr}) we can see the speed of
the transient behavior of $E_s$ of shielded CSR in the bend.
In Eq.(\ref{eq:PE_Exy}), if the magnitude of $2ik\rd_s$ is much smaller than
$\nablav_{\!\perp}^2$, it means that the field is nearly steady.
The criterion of whether the field reaches the steady state is roughly estimated as in
Eq.(\ref{eq:formation}) by dimensional analysis,
\begin{align}
   |2ik\rd_s|
   \sim |\nablav_{\perp}^2|
   \qquad\Lra\qquad
  \rd_s
  \sim s_{\rm f}^{-1}
   ,\qquad
  \rd_{x,y}
  \sim \ell_{\perp}^{-1} .
\end{align}
$\ell_{\perp}$ is given by Eq.(\ref{eq:shield}).
As discussed in Eq.(199) of \cite{agoh}, 
the paraxial approximation is applicable to the high frequency waves such that
\begin{align}
  k\rho
  \gg 1
   \quad\text{and}\quad
  kd_{\perp}
  \gg \pi
   \qquad\text{where}\quad
  d_{\perp}={\rm min}[w,h] .
   \label{eq:PA_cond}
\end{align}
$d_{\perp}$ represents the transverse dimension of the beam pipe.
The first equation of (\ref{eq:PA_cond}) is equivalent to Eq.(\ref{eq:rds_ll_k}).
Since usually $\rho\gg d_{\perp}$ and $w\geq h$, the condition that 
Eq.(\ref{eq:PE_Exy}) is applicable to calculating the field of CSR shielded by
a rectangular pipe or a pair of superimposed parallel plates is given as follows in 
practice,
\begin{align}
  k
  \gg  \frac{\pi}{h} .
   \label{eq:PA_condh}
\end{align}

We solve Eq.(\ref{eq:PE_Exy}) numerically by discretizing it with imposing
the transverse boundary condition of a perfectly conducting rectangular pipe as
described in section III of \cite{agoh_yokoya}.
Besides the perfectly conducting pipe, we can take into account a finite conductivity
$\sig_c$ (resistive wall) of the beam pipe in the numerical solution of
Eqs.(\ref{eq:PE_Exy}-\ref{eq:PE_Es}) \cite{agoh_yokoya1}.
In this section, in calculating the transient field of CSR emitted in the curved beam pipe,
for simplicity,
we assume that the initial field at the entrance of the bend ($s=0$) is the steady field in 
the straight pipe, \ie, the space charge field or resistive wall wakefield respectively in 
the perfectly conducting pipe or resistive pipe which has the same cross section as
the curved pipe.
Assuming a steady field at the entrance of the bend,
the initial value of the field is gotten by numerically solving
Eqs.(\ref{eq:PE_Exy}-\ref{eq:PE_Es}) for $\rd_s=0$ and $\vrho=0$ in
the transverse plane $s=0$.

In calculating $E_s$ in the bend using the exact solution (\ref{eq:tEs_std_sep}),
we use Eq.(\ref{eq:tEs_init}) as the initial value which is the exact expression of
the steady field in the perfectly conducting straight rectangular pipe.
If we assume that the initial field at $s=0$ is not steady but transient, to be exact,
we must take into account the discontinuity of the first derivative of the field with 
respect to $s$ as shown in Eqs.(\ref{eq:cEx_jump_plus_sum}-\ref{eq:cEs_jump_plus_sum}) and
(\ref{eq:cBx_jump_plus_sum}-\ref{eq:cBs_jump_plus_sum}).
In the present paper we do not assume the ultrarelativistic limit in computing
the exact solution of the field, 
\ie, we take into account a finite Lorentz factor $\gamma$ given by Eq.(\ref{eq:z})
which represents the beam energy $E=\gamma mc^2$ under the assumption of a rigid bunch,
where $m$ is the rest mass of the particles forming the beam.
We assume an electron beam ($mc^2\approx 0.511{\rm MeV}$)
as the source current of the field in section 9, however,
all the results in the present paper hold for arbitrary charged particles such as
muon or proton if we use $\gamma$ instead of $E$.

\begin{figure}[h]
  \begin{center}
    \includegraphics[scale=0.32,clip]{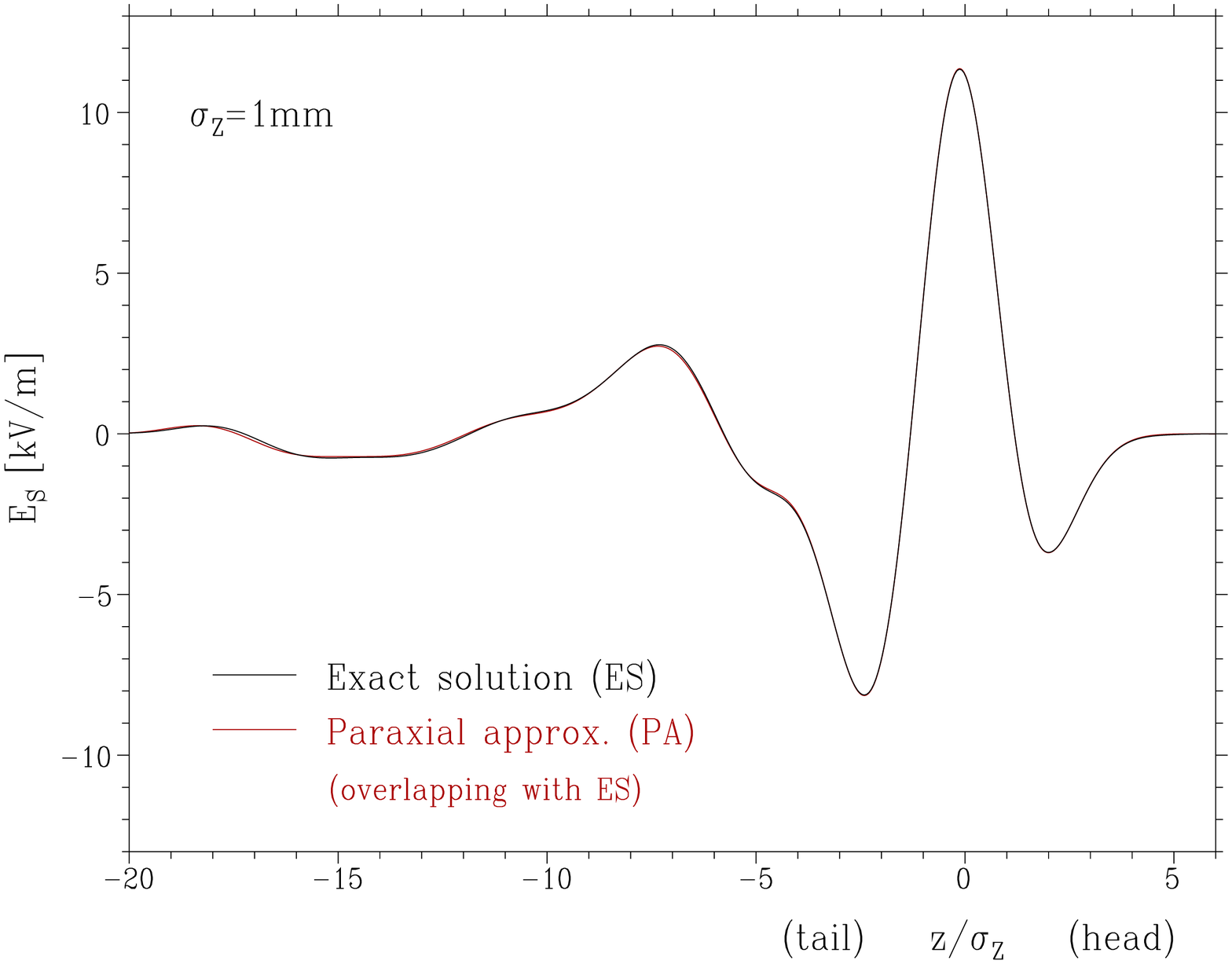}~~
    \includegraphics[scale=0.32,clip]{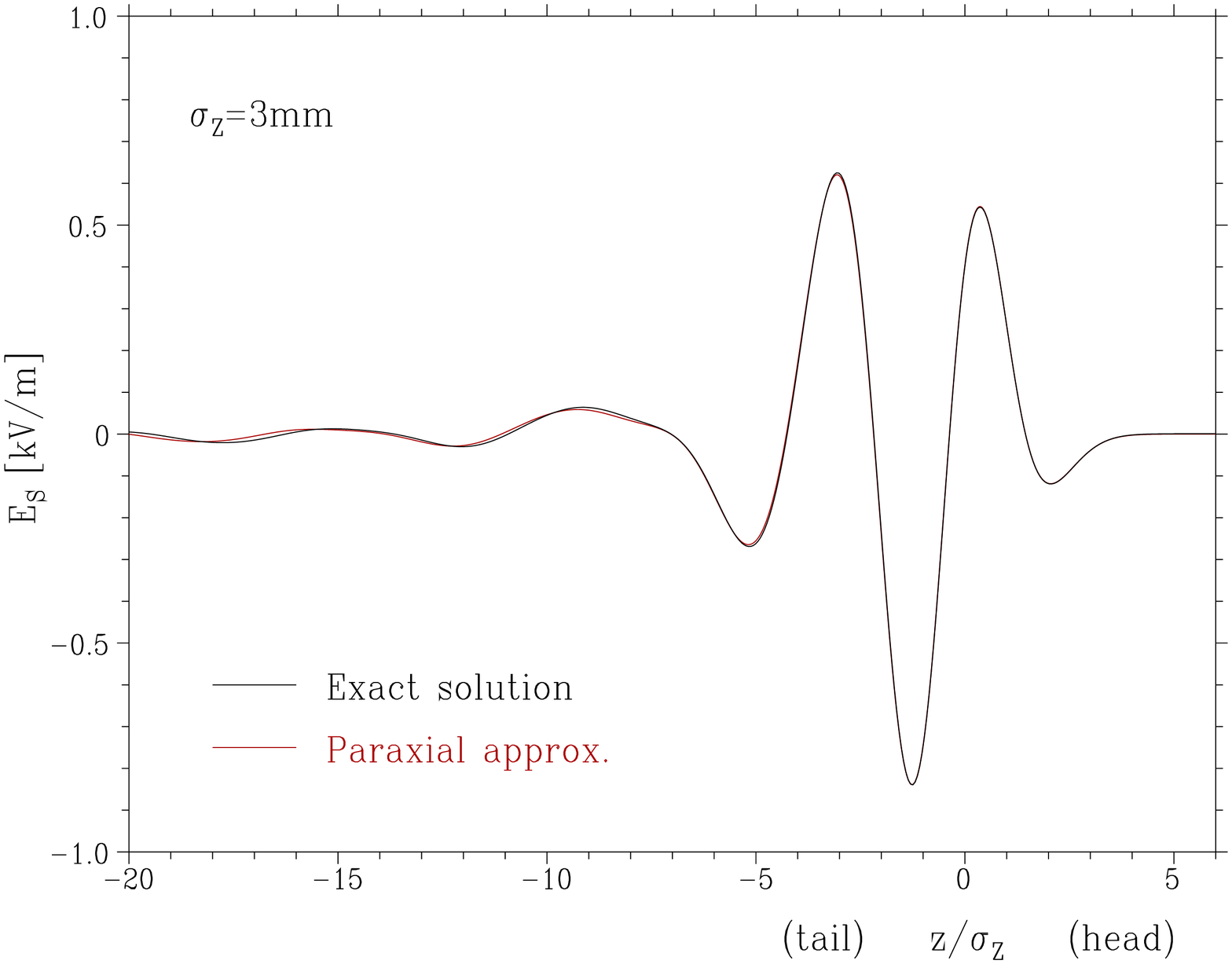}
      \vspace{1mm}
      \\
    \includegraphics[scale=0.32,clip]{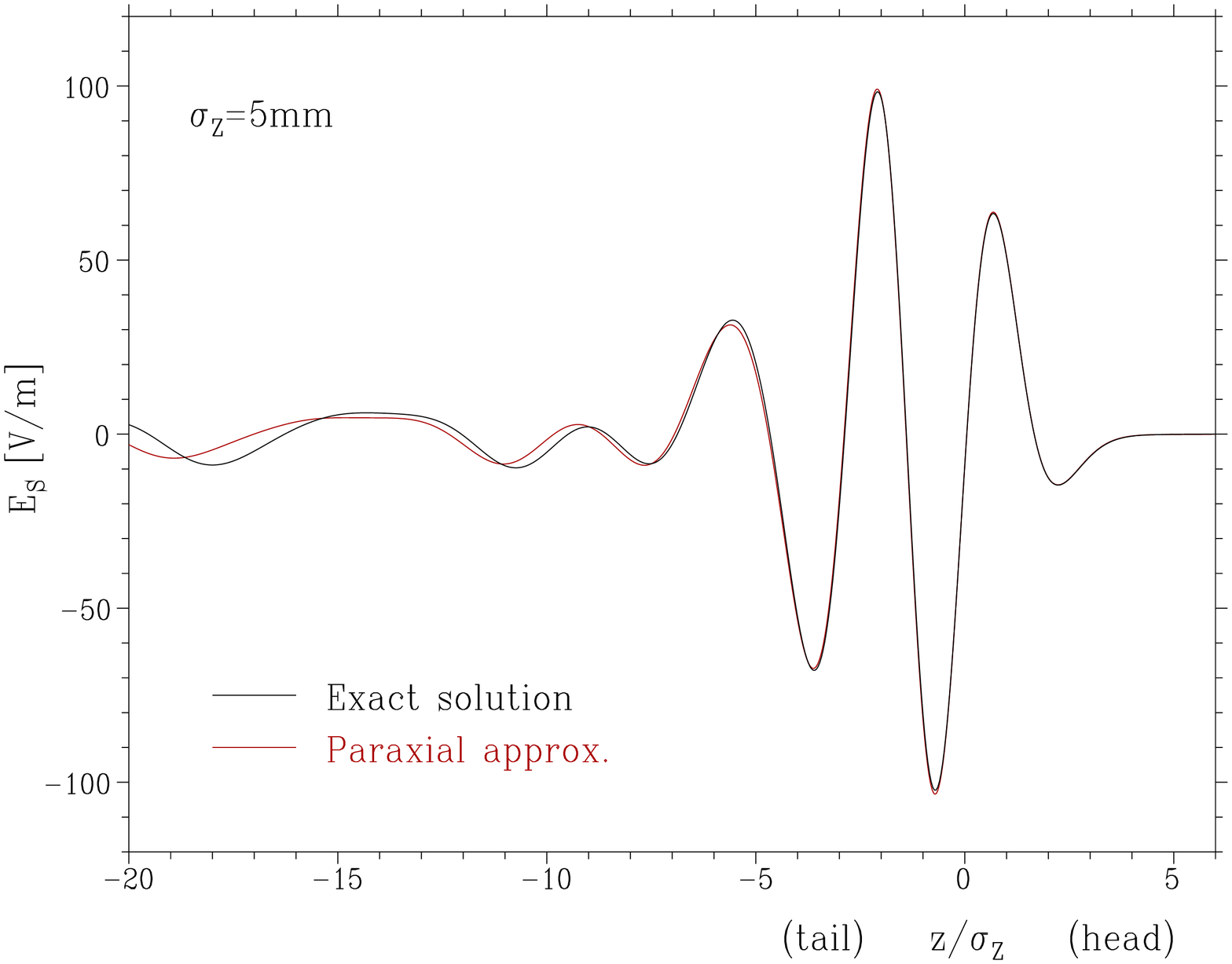}~~
    \includegraphics[scale=0.32,clip]{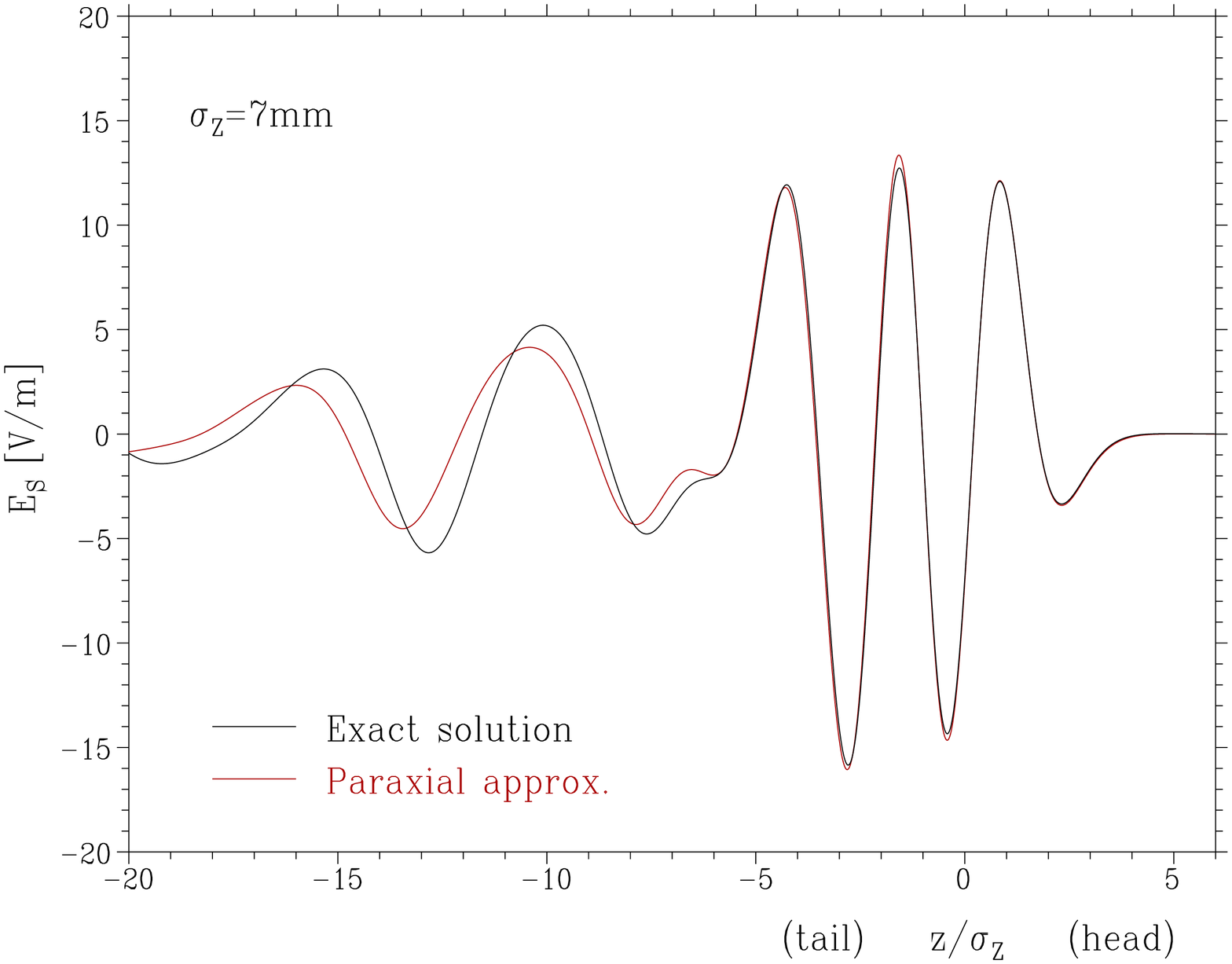}
      \vspace{1mm}
      \\
    \includegraphics[scale=0.32,clip]{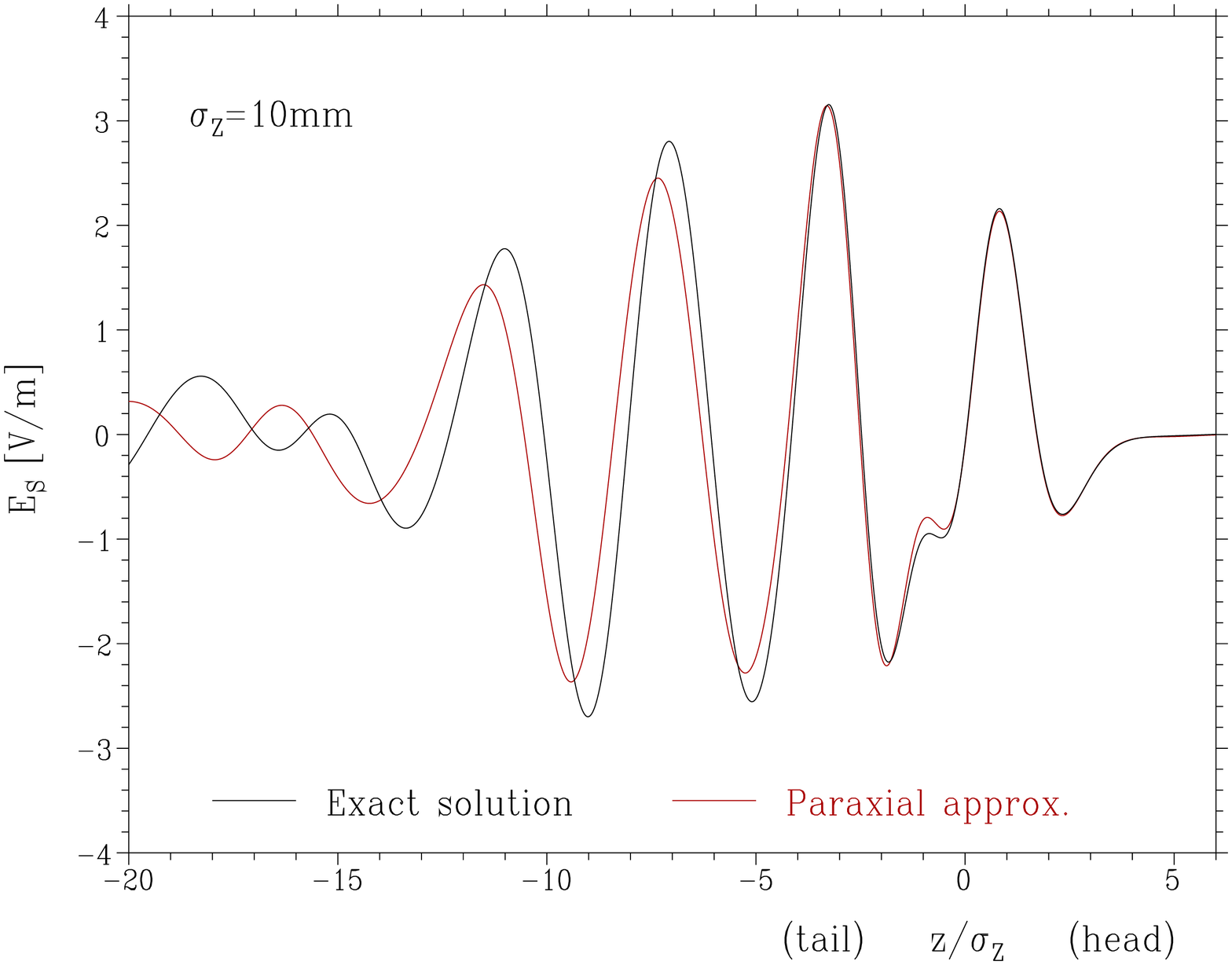}~~
    \includegraphics[scale=0.32,clip]{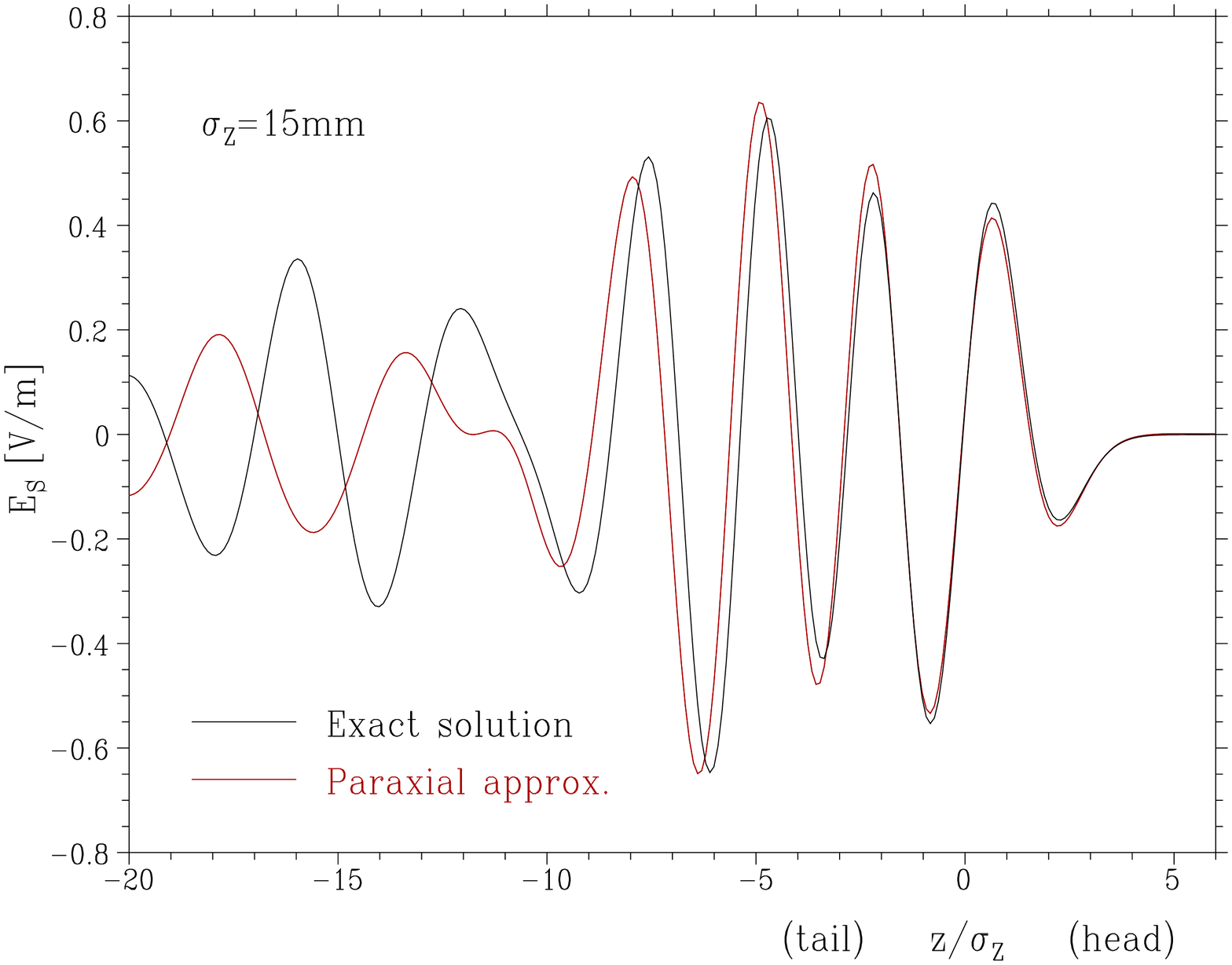}
    \caption[Longitudinal electric field of transient CSR in the time domain]{\small 
     Longitudinal electric field of transient CSR in the time domain $E_s(z)$
     at $\xv_{\perp}=0$ and $s=1{\rm m}$ in a bend.
     We assume that the electron bunch (\ref{eq:thin_bunch}-\ref{eq:rigid_bunch}) is
     moving at the center of a perfectly conducting square pipe.
     The horizontal axis is $z/\sig_z$; $z=0$ at the longitudinal center of the bunch,
     positive/negative toward the head/tail of the bunch.
     The black curve (ES) shows $E_s$ computed using the exact solution
     (\ref{eq:tEs_std_sep}).
     The red curve (PA) shows the numerical solution of $E_s$ using
     Eqs.(\ref{eq:PE_Exy}-\ref{eq:PE_Es}) on the basis of the paraxial approximation.
     The bunch length is $\sig_z=1$, 3, 5, 7, 10, 15mm.
     The other parameters are given as follows:
     $\rho=10{\rm m}$, $x_b=-x_a=3{\rm cm}$, $w=h=6{\rm cm}$,
     $\sig_y=20\mu{\rm m}$, $E=1{\rm GeV}$ and $q=-1{\rm nC}$.
     When $\sig_z=1{\rm mm}$ and 3mm, the curve of PA (red) almost overlaps with ES (black) 
     in the figure.
     }
    \label{fig:Esz}
  \end{center}
\end{figure}

As shown in Fig.\ref{fig:Esz}, the longitudinal electric field of CSR in the time domain
$E_s$, which is gotten by solving the parabolic wave equation
(\ref{eq:PE_Exy}-\ref{eq:PE_Es}) [PA: paraxial approximation, red curve], tends to
agree with the exact solution [ES, black curve] given by Eq.(\ref{eq:tEs_std_sep}) for
a shorter bunch as expected in Eq.(\ref{eq:PA_cond}).
Conversely, Eqs.(\ref{eq:PE_Exy}-\ref{eq:PE_Es}) are not applicable to calculating
the field if the bunch length $\sig_z$ is comparable to $d_{\perp}$ as described in
Eq.(\ref{eq:PA_condh}).
But even if the bunch length is short as $\sig_z\ll d_{\perp}$ in using
Eqs.(\ref{eq:PE_Exy}-\ref{eq:PE_Es}), the value of $E_s$ in the long range
$|z|\geq O(d_{\perp})$ is less accurate than that in the short range $|z|\ll d_{\perp}$
since a large $|z|$ implies a small $|k|$ by the uncertainty principle of
the Fourier transform.
Accordingly, assuming the presence of the vacuum chamber
(beam pipe or superimposed parallel plates) in the bend,
Eq.(\ref{eq:PE_Exy}) is not applicable to calculating the field in the long range
$|z|\geq O(d_{\perp})$ even if the bunch length is short enough as $\sig_z\ll d_{\perp}$.
If we consider a field in a bend in the absence of vacuum chamber,
Eqs.(\ref{eq:PE_Exy}-\ref{eq:PE_Es}) are applicable to the field in
the range $|z|\ll \rho$.
We must take care of the applicable range of Eq.(\ref{eq:PE_Exy}) with respect to $z$
if we calculate the tail of a CSR field at the position of the succeeding bunch moving 
behind the source bunch of CSR.

\clearpage

\subsection{Longitudinal impedances of CSR and resistive wall}
\label{sec:Zk}

Under the assumption of a rigid beam moving in a straight or bending section of
the beamline, we define the longitudinal impedance $Z\,[\Omg]$ for a length $s_0$ on
the $s$-axis by Ohm's law (with the opposite sign convention) for
the longitudinal voltage $\hV$ and current $\hI$ in the frequency domain (\ref{eq:FT_zk_9}),
\begin{align}
  \hV(k)
  =-\hI(k)Z(k)
  =\int_0^{s_0}ds [\hE_s(\xv,k)]_{\xv_\perp=0} ,
    \qquad
  \hI
  =\int_{x_a}^{x_b}\int_{-h/2}^{h/2}\hJ_s dxdy
  =qv\hlam .
  \label{eq:Z_bend}
\end{align}
$\hE_s$ is related to $\tE_s$ through Eq.(\ref{eq:tF_hF_exp}).
We often refer to $Z$ as simply ``impedance'' in what follows
since we do not discuss the transverse impedance in the present paper.
In considering the impedance of synchrotron radiation (SR), since it is irrelevant
whether the field is coherent or not, $Z$ can be referred to as ``impedance of CSR''
instead of saying ``impedance of SR''; this is merely a matter of expression.
That is, substituting $v\hlam=1$ into $\hV$ is equivalent to dividing $\hV$ by
an arbitrary nonzero $v\hlam$ which determines the coherence and spectrum of
the field as described in section \ref{sec:csr_isr}.
$Z\in\mathbb{C}$ is a kind of Green's function in the frequency domain,
which is a more fundamental quantity than $\hV$ in the sense that
$Z$ does not depend on the beam current $\hI$.

In this section we calculated the impedance of the transient field of CSR by integrating
$\hE_s$ numerically with respect to $s$.
Although we derived three expressions of $Z$ using the exact solution (\ref{eq:tEs_std_sep})
as shown in Eqs.(\ref{eq:Zk}) and (\ref{zetab_app}),
none of them can be used to compute the values of $Z$,
because they have infinite series which do not converge with respect to 
the radial mode number in the numerical calculation.

Fig.\ref{fig:logZ} shows the longitudinal impedances of CSR
which we calculated in several different models or ways.
For brevity, we use the acronyms in the following list to denote the impedances of CSR 
emitted in the bend of radius $\rho$ and length $s$ in the absence of longitudinal 
periodicity, \ie, $k\rho\in\mathbb{R}$ as described in section \ref{sec:assumption}.
\begin{center}
  \begin{tabular}{l|l|l|l}
   \hline
    CSR (curves in Fig.\ref{fig:logZ}) & state & vacuum chamber & calculation
     \\ \hline
    ES (black) & transient & perfectly conducting pipe & Eq.(\ref{eq:tEs_std_sep})
     \\
    PA (red) & transient & perfectly conducting pipe & Eqs.(\ref{eq:PE_Exy}-\ref{eq:PE_Es})
     \\
    PA-RW (blue) & transient & resistive pipe (copper) &
    Eqs.(\ref{eq:PE_Exy}-\ref{eq:PE_Es})
     \\
    TF (cyan) & transient & none (free space) &
        Eq.(87) in \cite{saldin_schneidmiller_yurkov}
     \\
    SF (dotted blue) & steady & none (free space) & Eq.(A10) in \cite{agoh_yokoya}
     \\
    SP (dotted cyan) & steady & infinite parallel plates & Eq.(A1) in \cite{agoh_yokoya}
     \\
    AE (dotted green)& steady ($k\to 0$) & perfectly conducting pipe & Eq.(\ref{eq:cZ_AE})
     \\
     \hline
  \end{tabular}
\end{center}
In Fig.\ref{fig:logZ} we assume a curved square pipe with a side length $w=h$ as
the beam pipe for ES (exact solution), PA (paraxial approximation),
PA-RW (paraxial approximation, resistive wall) and
AE (asymptotic expression of $\Im Z$ for $k\to 0$).
We plotted TF (transient, free space), SF (steady, free space) and SP (steady, plates) in
Fig.\ref{fig:logZ} in order to show the criteria of the transient effect in free space and 
the shielding effect by a pair of perfectly conducting infinite parallel plates,
which correspond to $k_{\rm f}$ and $k_{\rm th}$ given by Eqs.(\ref{eq:kth}).

For $k\to0$, the asymptotic expression (AE) of the impedance of the steady field of 
CSR in a perfectly conducting rectangular pipe is given by Eq.(87) in \cite{agoh}.
Let us rearrange it using our present notation,
\begin{align}
  \frac{\cZ(k)}{Z_0}
  &=\frac{ik}{2\beta}
    \sum_{p=0}^{\infty}\frac{\tpsi_y^{2p+1}}{k_y^{2p+1}}(\mfU_0^{2p+1}+\mfU_1^{2p+1}k^2)
     \quad [{\rm m}^{-1}] .
    \label{eq:cZ_AE}
\end{align}
In the present paper we use the symbol $\cZ\,[\Omg/{\rm m}]$ in general to represent
a longitudinal impedance per unit length along the $s$-axis as also used in 
Eqs.(\ref{eq:kc}) and (\ref{eq:cZ_thin}).
$\tpsi_y^n$ is given by Eq.(\ref{eq:tpsi}) which is the Fourier coefficient of
the vertical charge distribution $\psi_y$ of the bunch (\ref{eq:thin_bunch}).
Eq.(\ref{eq:cZ_AE}) is the truncated power series of the asymptotic expression of $\cZ$
for $k\to0$ to $O(k^3)$, where
$\mfU_j^n$ is the coefficient of the $(2j+1)$th order term excluding $\tpsi_y^n/k_y^n$.
For better accuracy, we modify Eq.(87) in \cite{agoh} using $S$ and $T$
 instead of $\bS$ and $\bT$, respectively given by Eqs.(85-86) and (88) in \cite{agoh}, 
although this is not important,
\begin{alignat}{2}
  \mfU_0^n
  &=\bigg\{\frac{1}{\gam^2}+\frac{1}{2\rho^2(k_y^n)^2}\bigg\}T(k_y^nw) ,
    \qquad&
  T(u)
  &=\frac{\sinh u}{\cosh u+1} ,
   \\
  \mfU_1^n
  &=-\frac{3}{2\rho^2 (k_y^n)^4}S(k_y^nw) ,
    \qquad&
  S(u)
  &=\frac{\sinh u-u}{\cosh u+1} .
    \label{eq:mfU1}
\end{alignat}
Eq.(\ref{eq:cZ_AE}) is purely imaginary since the real impedance of the steady field in
the perfectly conducting pipe is completely discretized as in Eq.(139) of \cite{agoh}.
The radiation is represented by $\Re Z$ which means the energy loss.
The bunch as a whole does not lose energy through the imaginary part $\Im Z$.
The particles forming the bunch interact with each other through $\Im Z$.
In a broad sense, $\Im Z$ in a perfectly conducting vacuum chamber or free space denotes
a kind of space charge effect of the bunch moving on the curved trajectory in the bend.

Fig.\ref{fig:logZ} has the following three criteria (VW, TH, FW) with respect to
the wavenumber $k$:
\begin{align}
  k_y^1
  =\frac{\pi}{h} ,
    \qquad
  k_{\rm th}
  =\Big\{\frac{2\rho}{3}(k_y^1)^3\Big\}^{1/2} ,
    \qquad
  k_{\rm f}
  =\frac{24\rho^2}{s^3} .
   \label{eq:kth}
\end{align}
$k_y^1$ is given by Eq.(\ref{eq:kyn}) for $n=1$, which is the fundamental vertical 
wavenumber (VW) between the upper and lower walls of the vacuum chamber
(rectangular pipe or infinite parallel plates).
As described in Eq.(\ref{eq:PA_condh}), $k_y^1$ is the criterion of the wavenumber
if the paraxial approximation is applicable to calculating the field
which is shielded by the upper and lower walls.
$k_{\rm th}$ is the shielding threshold (TH) by a pair of infinite parallel 
plates located at $y=\pm h/2$ with a spacing of $h$.
$k_{\rm th}$ is the criterion of the wavenumber below which
$\Re Z$ of SP deviates significantly from SF.
$k_y^1$ and $k_{\rm th}$ are shown in Fig.\ref{fig:logZ} with
the black vertical dotted line (VW) and dash-dotted line (TH) respectively.
$k_{\rm f}$ is gotten from Eq.(\ref{eq:formation}), which is
the formation wavenumber (FW) of the wave in the bend of length $s$.
$k_{\rm f}$ is plotted with the gray vertical dotted line in Fig.\ref{fig:logZ}.
$k_{\rm f}$ is the wavenumber above which the field of unshielded CSR is in
the steady state at a longitudinal position $s$ in the bend.
That is, $k_{\rm f}$ is the criterion of $k$ below which TF deviates significantly from 
SF in Figs.\ref{fig:logZ}, \ref{fig:logZ_s2m} and \ref{fig:logZ_w12}.
\begin{center}
  \begin{tabular}{l|l|l}
    \hline
    VW: $k_y^1$ (dotted black) & Fundamental vertical wavenumber & Eq.(\ref{eq:kyn})
     \\
    TH: $k_{\rm th}$ (dash-dotted black) & Shielding threshold (parallel plates) &
    Eq.(179) in \cite{agoh}
     \\
    FW: $k_{\rm f}$ (dotted gray) & Formation wavenumber (free space)
     & Eq.(\ref{eq:kth})
     \\ \hline
    RW: $\beta<1$ (yellow-brown) & Resistive wall $Z$ (straight square pipe) &
    Eqs.(\ref{eq:PE_Exy}-\ref{eq:PE_Es})
     \\
    RW-RP: $\beta=1$ (magenta dots) & Resistive wall $Z$ (straight round pipe)
     & Eq.(4.75) in \cite{yokoya_rw}
     \\ \hline
  \end{tabular}
\end{center}
In addition to CSR,
we plotted the resistive wall impedances (RW and RW-RP) in Fig.\ref{fig:logZ} in order 
to confirm that PA-RW behaves correctly for $k\to0$.
The resistive wall impedances also show the order of magnitude of $Z$.
The impedances of RW (resistive wall, rectangular/square pipe, $\beta<1$) and RW-RP
(resistive wall, round pipe, $\beta=1$) for a length $s$ are shown in Fig.\ref{fig:logZ}
respectively with the yellow-brown curve and magenta dots.
We calculated RW-RP using Eq.(4.75) in \cite{yokoya_rw} in order to confirm the correctness
of RW which we computed by numerically solving Eqs.(\ref{eq:PE_Exy}-\ref{eq:PE_Es}) for
$\rd_s=0$ and $\vrho=\infty$, taking into account a finite conductivity $\sig_c$ in
the transverse boundary condition of the beam pipe.
According to Fig.8 in \cite{yokoya_rw}, in the ultrarelativistic limit,
the longitudinal resistive wall wakefield $E_s$ at
the center of a square pipe equals that at the center of the round pipe
which is inscribed in the square.
$Z$ of RW includes the space charge field of the beam unlike RW-RP
which is derived assuming $\beta=1$.
The longitudinal wakefield of RW and RW-RP is plotted in Fig.\ref{fig:Esz_rw_pcw}.

In Fig.\ref{fig:logZ} the difference between PA (paraxial approximation) and
ES (exact solution) denotes the error of PA, which depends on $k$.
PA (red) tends to agree with ES (black) at high frequencies such that $k\gg k_y^1$
(roughly $k\gtrsim 10k_y^1 \approx 524{\rm m}^{-1}$ in Fig.\ref{fig:logZ})
as expected in Eq.(\ref{eq:PA_condh}).
On the contrary, PA deviates from ES at low frequencies 
when $k$ is comparable to $k_y^1$ or smaller than $k_y^1$.
They differ especially when $k<k_y^1$, \ie, $k_r^n\in i\mathbb{R}$ for
$\forall n\in\mathbb{N}$ which means the imaginary whispering gallery mode
discussed in section \ref{sec:iwgm}.
Since the field does not oscillate in the low frequency limit ($k\to0$),
the imaginary impedance $\Im Z$ of ES for $k\to0$ approaches the green dashed curve
(AE: asymptotic expression) given by Eq.(\ref{eq:cZ_AE}).
In section \ref{sec:Z_energy} we will discuss the small disagreement between
AE and ES for $k\to0$ in Figs.\ref{fig:logZ} and \ref{fig:logZ_s2m}.
If the width of the pipe $w$ is larger than the height $h$,
AE agrees well with ES for $k\to0$ as shown in Fig.\ref{fig:logZ_w12}
which is for $w=2h$.

When $k\ll k_{\rm th}$, $|\Im Z|$ of ES is larger than $|\Re Z|$
since the radiation is suppressed by the shielding effect of the pipe.
Therefore, when $k\ll k_{\rm th}$, the difference of PA from ES in $\Re Z$ is less important 
than their difference in $\Im Z$ which represents the space charge field of the bunch as
described below Eq.(\ref{eq:mfU1}).
That is, when the shielding effect by the perfectly conducting pipe is strong,
the influence through $\Im Z$ is more important than $\Re Z$ for the beam
especially when the beam energy is lower as discussed later in section \ref{sec:Z_energy}.
When $E=1{\rm GeV}$, since the relative difference of PA from ES in $\Im Z$ is not so large
(compared to their difference in $\Re Z$) as seen from the lower figure of
Fig.\ref{fig:logZ}, the solution of the parabolic wave equation
(\ref{eq:PE_Exy}) is fairly accurate around the center of the bunch
such that $|z|\ll d_{\perp}$ as seen from Fig.\ref{fig:Esz}.
On the other hand, $\Im Z$ computed using the solution of the low accuracy parabolic 
equation, given by Eq.(21) in \cite{agoh_yokoya}, tends to follow Eq.(84) in
\cite{agoh} for $k\to0$ as shown with the green solid line in Fig.6 of \cite{agoh}.
This is the difference of Eq.(21) in \cite{agoh_yokoya} from
Eq.(\ref{eq:PE_Exy}) which is the high accuracy parabolic equation
taking into account the entanglement between $\hE_x$ and $\hE_s$ in the bend.

In Fig.\ref{fig:logZ} the impedance of the transient field of CSR in
the perfectly conducting beam pipe (ES and PA) is comparable to or smaller than
the resistive wall impedance (RW) when $k\leq O(k_y^1)$ for which
the paraxial approximation breaks down as described in Eq.(\ref{eq:PA_condh}).
$Z$ of the transient field of CSR emitted in the resistive pipe is shown with
the blue solid curve (PA-RW: paraxial approximation, resistive wall) in Fig.\ref{fig:logZ}.
PA-RW approaches RW for $k\to0$ and PA for $k\to\infty$.
From a practical point of view, if the actual beam pipe is made of a metal having a finite 
conductivity $\sig_c$ such as copper or aluminum, then the assumption that the beam pipe is 
perfectly conducting is not true as seen from Figs.\ref{fig:logZ} and \ref{fig:Esz_rw}.
If the order of magnitude of a CSR field is comparable to or smaller than
the resistive wall wakefield, however, CSR may not cause a serious problem to the particles
forming the bunch unless its longitudinal charge distribution fluctuates for some reason.

Here we would like to explain the reason why we discuss the impedance
assuming a perfectly conducting beam pipe which is unusual.
If we take into account the resistivity in the pipe in the field calculation,
the error by the paraxial approximation is largely covered up and buried under
the resistive wall effect as seen from the low frequency range
$k<O(k_y^1)$ of Figs.\ref{fig:logZ} and \ref{fig:logZ_s2m}.
From a practical point of view in calculating the field of CSR by
taking a more realistic situation into account,
one would think that it is irrelevant or unimportant to discuss the impedance in
the low frequency range in detail under the assumption of a perfectly conducting pipe.
From a scientific point of view, however,
it is important to discuss $Z$ of CSR in a perfectly conducting pipe,
similar to discussing a motion of an object in the absence of friction.
In physics, primarily, it is important to find the fundamental law of the phenomenon
in the absence of friction or resistance such as
the equation of motion under the gravity in the absence of air resistance.
Then we extend the fundamental model, which may not be necessarily realistic,
to other models by taking more practical circumstances into consideration.
In this regard, the resistance is an additional effect to extend the model in
considering the motion of the object in accordance with reality after
understanding the fundamental law of the phenomenon which is prescribed by
a certain mathematical framework.
This is the reason for discussing the impedance assuming a perfectly conducting pipe
even though we can take into account the resistivity in the pipe in calculating the field
of CSR.
We can examine the difference of Eq.(\ref{eq:PE_Exy}) from the exact solution
or Eq.(7) in \cite{agoh} by assuming a perfectly conducting pipe.
Assuming a perfectly conducting pipe, we can tell how large or small
the field of CSR is, compared to the resistive wall wakefield.
In addition, we can understand the fundamental features of synchrotron radiation emitted in
the pipe such as the shielding effect and the resonance by calculating $Z$ of CSR in
a perfectly conducting pipe.
The impedances of CSR in the simpler models listed in p.\pageref{eq:Z_bend} also have
meanings in understanding the fundamental features of synchrotron radiation in
classical electrodynamics regardless whether they are applicable or not to
designing or improving the beamlines of an accelerator.
This is the basic way of thinking in physics which is a field of science
unlike engineering and technology oriented toward the application and practicality 
for some specific purpose.

\subsection{Resonance of the fields in a curved pipe}

\begin{figure}[h]
\begin{center}
  \includegraphics[scale=0.46,clip]{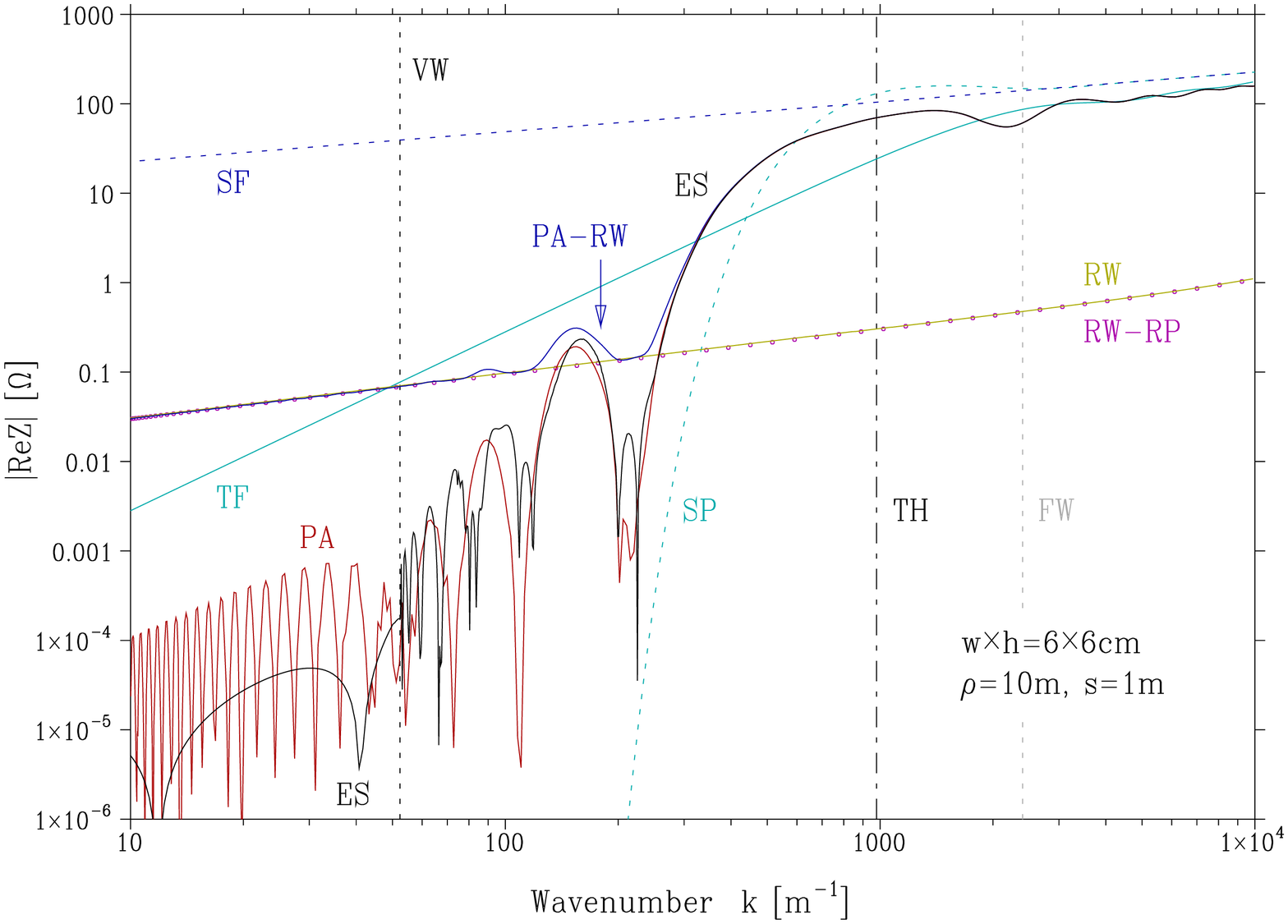}
   \vspace{2mm}
   \\
  \includegraphics[scale=0.46,clip]{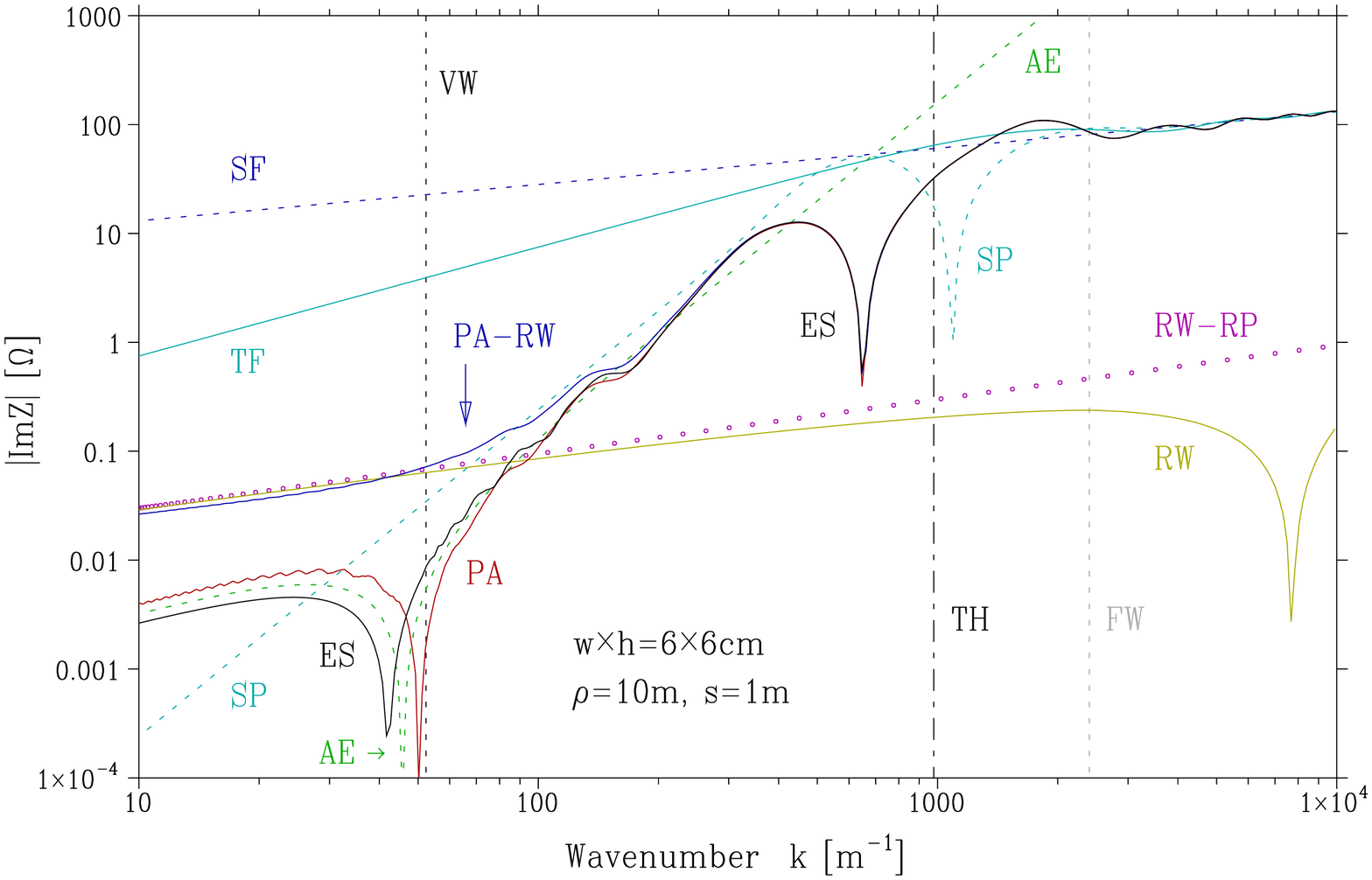}
  \caption[Longitudinal impedances of CSR and resistive wall in a square pipe
    ($s=1{\rm m}$)]
    { \small
    Absolute value of the real part $\Re Z$ (upper) and imaginary part $\Im Z$ (lower) of
    $Z\,[\Omg]$ which represents the longitudinal impedances of CSR and
    resistive wall (RW) for a length $s=1{\rm m}$.
    The horizontal axis is the wavenumber $k$ given by Eq.(\ref{eq:omg}).
    The black vertical dotted line (VW) is the fundamental vertical wavenumber of
    the beam pipe $k_y^1$.
    The vertical dash-dotted line (TH) is the shielding threshold $k_{\rm th}$.
    The gray vertical dotted line (FW) shows $k_{\rm f}$ given by Eq.(\ref{eq:kth})
    which is the formation wavenumber of CSR in the bend.
    The black and red curves (ES and PA) show $Z$ of a transient field of CSR
    emitted in a perfectly conducting square pipe.
    The black curve (ES: exact solution) shows $Z$ computed by integrating
    the exact solution of $\tE_s$ given by Eq.(\ref{eq:tEs_std_sep}).
    The red curve (PA: paraxial approximation) shows $Z$ computed using the numerical 
    solution of Eqs.(\ref{eq:PE_Exy}-\ref{eq:PE_Es}).
    The blue curve (PA-RW: paraxial approximation, resistive wall) shows $Z$ of
    the transient CSR in a copper pipe, computed by solving
    Eqs.(\ref{eq:PE_Exy}-\ref{eq:PE_Es}) imposing the resistive boundary condition.
    The yellow-brown curve (RW, $E=1{\rm GeV}$) shows the resistive wall impedance of
    a straight copper pipe which has the same cross section and length along the $s$-axis as
    the curved square pipe.
    The magenta dots (RW-RP: resistive wall, round pipe) show $Z$ computed using
    Eq.(4.75) in \cite{yokoya_rw}, which is the resistive wall impedance of
    a straight round pipe of diameter $h$ in the ultrarelativistic limit ($\beta=1$).
    The cyan curve (TF: transient, free space) shows $Z$ of the transient field of CSR
    in free space, computed using Eq.(87) in \cite{saldin_schneidmiller_yurkov}.
    The cyan dotted curve (SP: steady, plates) shows $Z$ of the steady field of CSR
    between perfectly conducting infinite parallel plates located at $y=\pm h/2$.
    We computed SP using Eq.(A1) in \cite{agoh_yokoya},
    which tends to be SF for $k\to\infty$.
    The blue dotted line (SF: steady, free space) shows Eq.(A10) in \cite{agoh_yokoya},
    which is $Z$ of the steady field of CSR in free space.
    The green dotted curve (AE) in the lower figure shows
    $\Im Z=s\Im\cZ$ given by Eq.(\ref{eq:cZ_AE}) which is the asymptotic expression of
    $\cZ$ of the steady field of CSR in the perfectly conducting pipe for $k\to0$.
    The parameters are common to Fig.\ref{fig:Esz}:
    $\rho=10{\rm m}$, $x_b=-x_a=3{\rm cm}$, $w=h=6{\rm cm}$, $s=1{\rm m}$,
    $\sig_y=20\mu{\rm m}$ and $E=1{\rm GeV}$.
    The electrical conductivity of copper is $\sig_c=6\times 10^7/\Omg {\rm m}$.
    }
  \label{fig:logZ}
\end{center}
\end{figure}

\begin{figure}[h]
  \begin{center}
    \includegraphics[scale=0.46,clip]{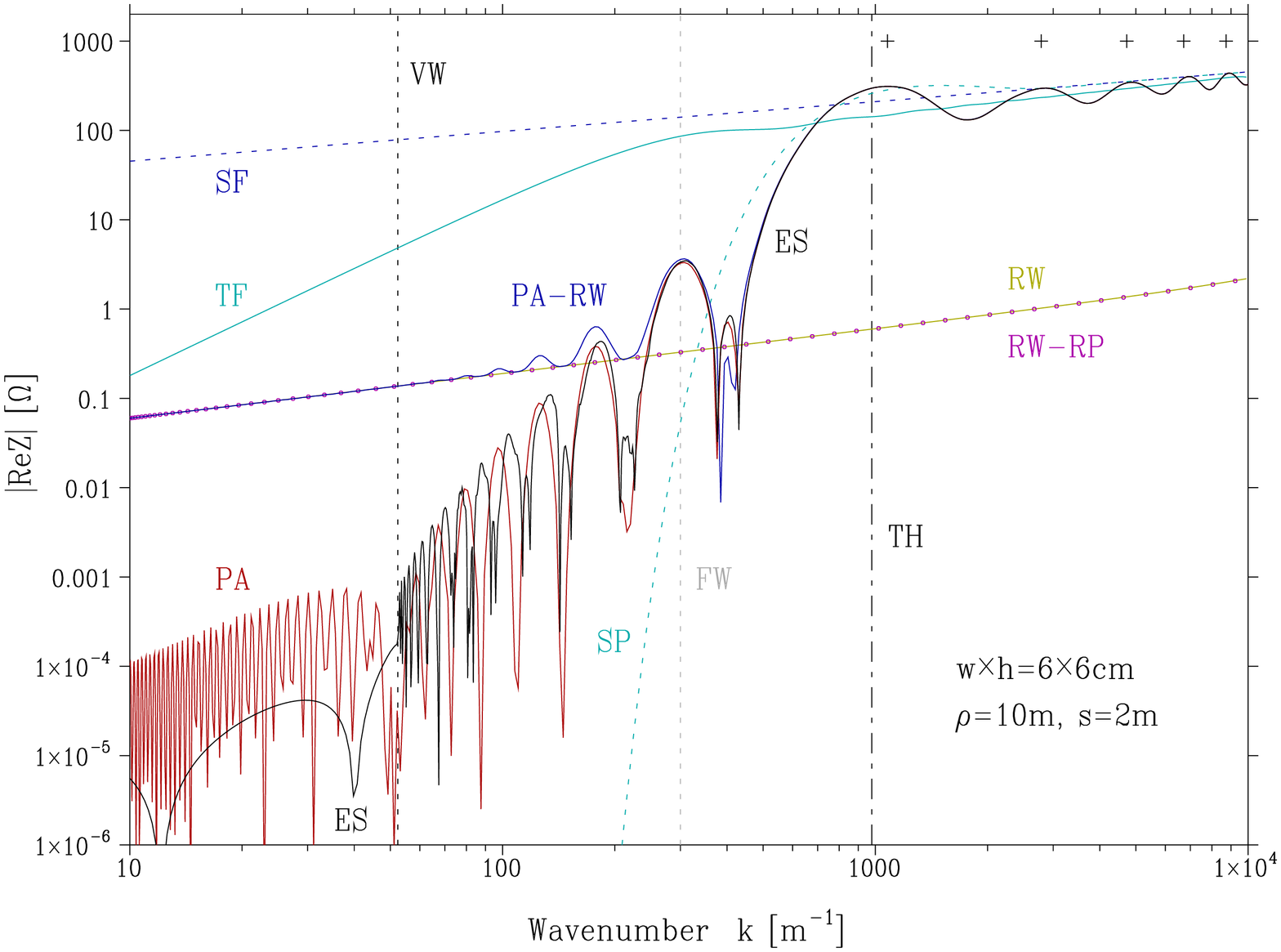}
      \vspace{2mm}
      \\
    \includegraphics[scale=0.46,clip]{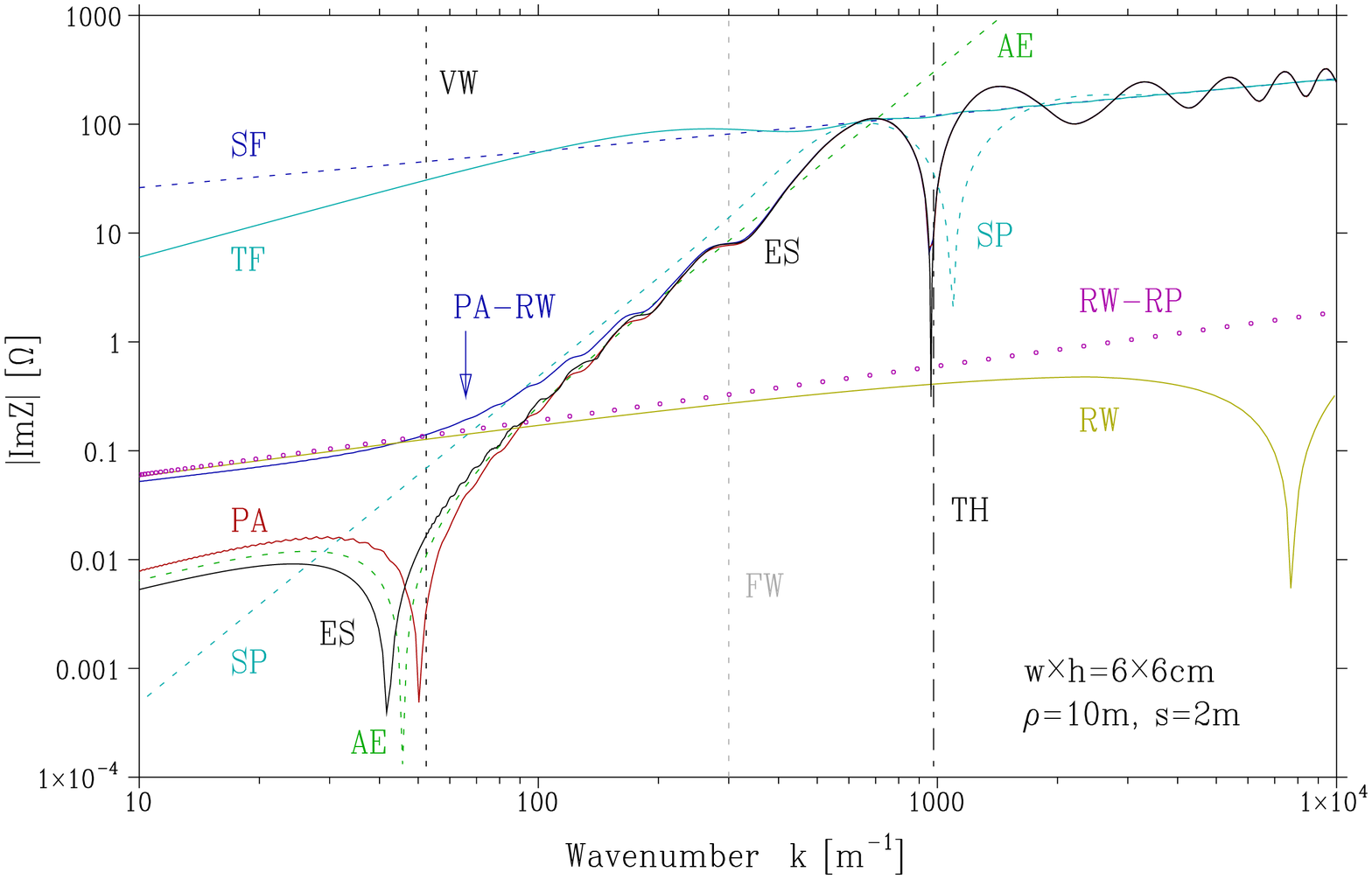}
    \caption[Longitudinal impedances of CSR and resistive wall in a square pipe
    ($s=2{\rm m}$)]
    {\small 
    Absolute value of the real part $\Re Z$ (upper) and imaginary part $\Im Z$ (lower) of
    $Z\,[\Omg]$ which represents the longitudinal impedances of CSR and 
    resistive wall (RW) for a length $s=2{\rm m}$.
    The horizontal axis is the wavenumber $k$.
    The parameters are common to Fig.\ref{fig:logZ} excluding
    the length of the bend.
    The nomenclature is also common to Fig.\ref{fig:logZ}:
    VW (fundamental vertical wavenumber) and TH (shielding threshold) are 
    given by Eqs.(\ref{eq:kth}).
    The formation wavenumber (FW) is $k_{\rm f}=300{\rm m}^{-1}$ whose value differs from
    that in Fig.\ref{fig:logZ} ($s=1{\rm m}$).
    RW is the longitudinal resistive wall impedance of the straight copper square pipe.
    The other curves show the impedances of CSR listed in p.\pageref{eq:Z_bend}.
    ES (exact solution) and PA (paraxial approximation) show $Z$ of the transient field of 
    CSR in the perfectly conducting square pipe,
    which are shown respectively with the black and red curves.
    The blue curve (PA-RW: paraxial approximation, resistive wall) shows $Z$ of
    the transient field of CSR in the copper square pipe.
    The cyan curve (TF: transient, free space) and blue dotted line
    (SF: steady, free space) show $Z$ of the transient and steady fields of CSR in
    free space.
    The cyan dotted curve (SP: steady, plates) shows $Z$ of the steady field of CSR on
    the mid-plane between the infinite parallel plates of spacing $h$.
    The green dotted curve (AE) in the lower figure shows $\Im Z=s\Im\cZ$ given by
    Eq.(\ref{eq:cZ_AE}) which is the asymptotic expression of $\cZ$ of
    the steady field of CSR in the perfectly conducting pipe for $k\to0$.
    When $k>k_{\rm th}$, the crests of the ripples of $\Re Z$ of ES correspond to
    the resonant wavenumbers $\vbkap_m^n$ given by Eq.(\ref{eq:vkapm_approx})
    for $m\in\mathbb{Z}_0^{+}$ and $n=1$.
    As for the lowest five radial modes ($m=0\sim4$), we plotted $k=\vbkap_m^1$ at
    the vertical position $\Re Z=10^3$ with the crosses ($+$) in the upper figure.
     }
    \label{fig:logZ_s2m}
  \end{center}
\end{figure}

\begin{figure}[h]
\begin{center}
  \includegraphics[scale=0.46,clip]{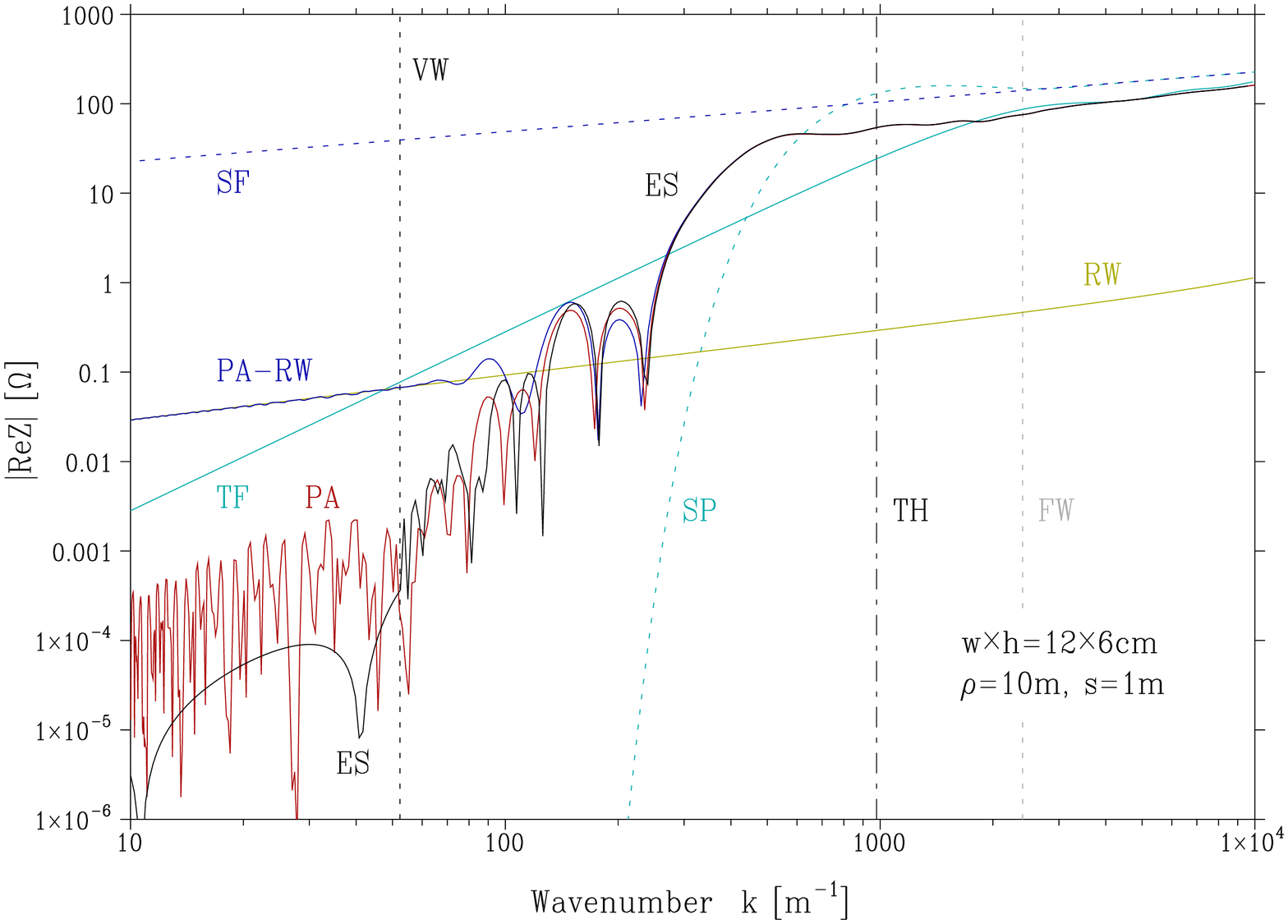}
   \vspace{2mm}
   \\
  \includegraphics[scale=0.46,clip]{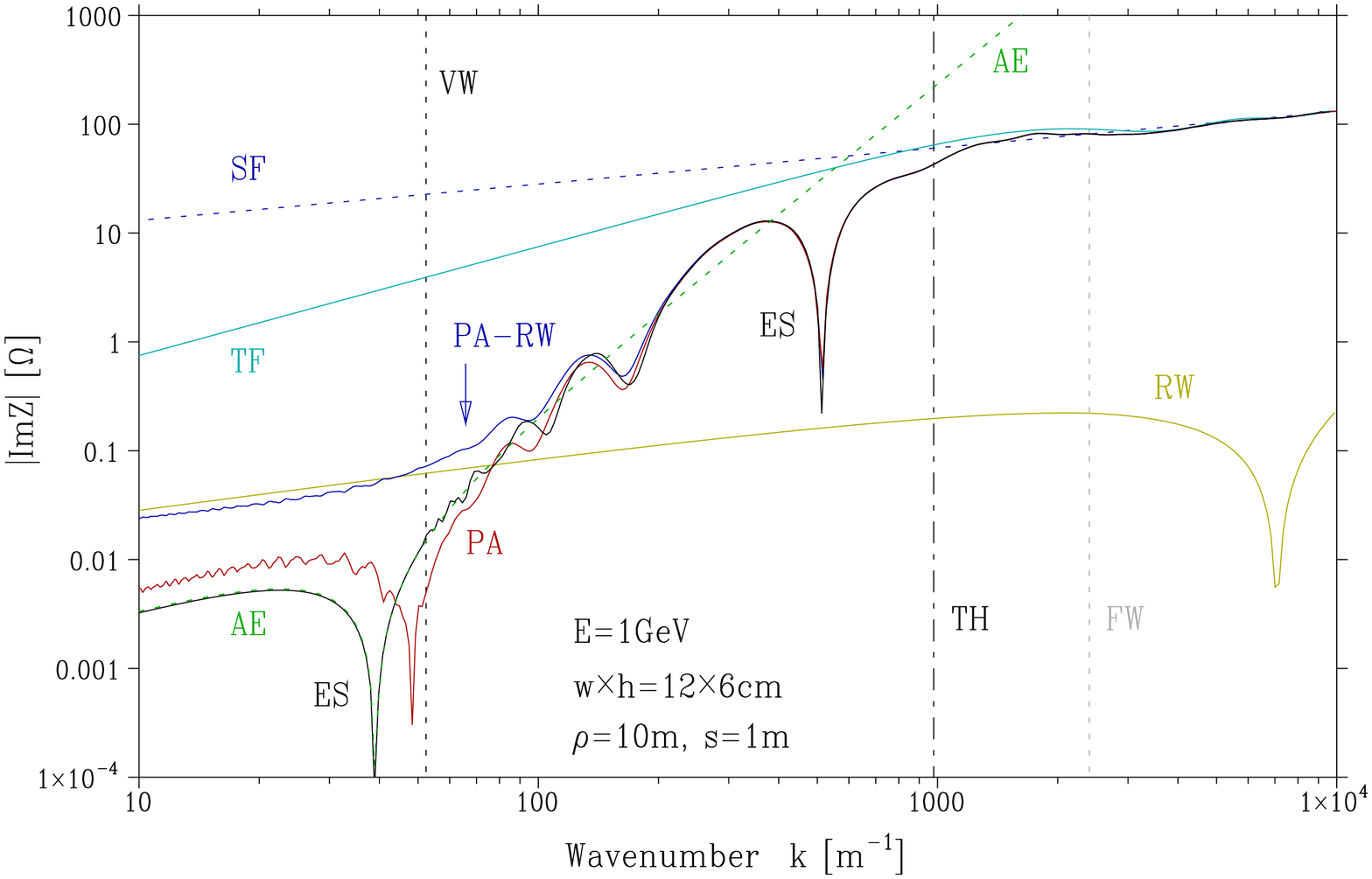}
  \caption[Longitudinal impedances of CSR and resistive wall in a rectangular pipe]
    { \small
    Absolute value of the real part $\Re Z$ (upper) and imaginary part $\Im Z$ (lower) of
    $Z\,[\Omg]$ which represents the longitudinal impedances of CSR and resistive wall (RW) 
    in a rectangular pipe $w=2h=12{\rm cm}$ for a length $s=1{\rm m}$.
    The horizontal axis is the wavenumber $k$.
    Excluding $x_{a,b}$ and $w$, the parameters are common to Fig.\ref{fig:logZ}.
    The nomenclature is also common to Fig.\ref{fig:logZ}.
    The black vertical dotted line (VW) is the fundamental vertical wavenumber of
    the pipe $k_y^1$.
    The vertical dash-dotted line (TH) is the shielding threshold $k_{\rm th}$.
    The gray vertical dotted line (FW) shows $k_{\rm f}$ given by Eq.(\ref{eq:kth})
    which is the formation wavenumber of CSR in the bend.
    The black and red curves (ES and PA) show $Z$ of a transient field of CSR
    emitted in the perfectly conducting rectangular pipe.
    The black curve (ES) shows $Z$ computed by integrating
    the exact solution of $\tE_s$ given by Eq.(\ref{eq:tEs_std_sep}).
    The red curve (PA) shows $Z$ computed using the numerical 
    solution of Eqs.(\ref{eq:PE_Exy}-\ref{eq:PE_Es}).
    The blue curve (PA-RW) shows $Z$ of
    the transient CSR in a copper pipe, computed by solving
    Eqs.(\ref{eq:PE_Exy}-\ref{eq:PE_Es}) imposing the resistive boundary condition.
    The yellow-brown curve (RW) shows the resistive wall impedance of a straight copper pipe
    which has the same cross section and length along the $s$-axis as
    the curved rectangular pipe.
    The cyan curve (TF) shows $Z$ of the transient field of CSR
    in free space, computed using Eq.(87) in \cite{saldin_schneidmiller_yurkov}.
    The cyan dotted curve (SP) shows $Z$ of the steady field of CSR emitted
    between perfectly conducting infinite parallel plates located at $y=\pm h/2$.
    We computed $\Re Z$ of SP using Eq.(A1) in \cite{agoh_yokoya}.
    The blue dotted line (SF) shows Eq.(A10) in \cite{agoh_yokoya},
    which is $Z$ of the steady field of CSR in free space.
    The green dotted curve (AE) in the lower figure shows $\Im Z$ ($=s\Im\cZ$) given by
    Eq.(\ref{eq:cZ_AE}), which almost overlaps with the ES curve (black) for
    $k\leq O(k_y^1)$.
    }
  \label{fig:logZ_w12}
\end{center}
\end{figure}

Let us see the longitudinal impedance in the high frequency range above
the shielding threshold $k_{\rm th}$ (TH) in Fig.\ref{fig:logZ}.
For $k\to\infty$, the impedance of the transient field of CSR emitted in the beam pipe
(ES, PA, PA-RW) tends to agree with TF (transient, free space)
which is shown with the cyan curve in Fig.\ref{fig:logZ}.
It implies that we can assume free space in calculating the transient field of CSR in
the time domain if the field consists of the frequency components mostly in
the range $k\gg k_{\rm th}$.
But if the field of CSR in the curved pipe is reaching the steady state in a long bend,
then the spectrum in the high frequency range $k\gtrsim k_{\rm th}$ tends to ripple
as in Fig.\ref{fig:logZ_s2m} since the field resonates in the pipe.
Reaching the steady state in the perfectly conducting curved pipe, the spectrum of
the field is completely discretized, because the energy of the electromagnetic field is not 
dissipated in this system as discussed in section IV of \cite{agoh}.
For the set of parameters used in Fig.\ref{fig:logZ} ($s=1{\rm m}$),
since the transient field of CSR in the curved pipe has not yet resonated in the bend,
$Z$ of ES (black) has a continuous spectrum
even though it is emitted in the perfectly conducting curved beam pipe.

Assuming a longer bend, \eg, $s=2{\rm m}$, as shown in Fig.\ref{fig:logZ_s2m},
$\Re Z$ of ES has ripples in the high frequency range ($k\gtrsim k_{\rm th}$)
because of the resonance by the reflection on the outer wall of the curved pipe.
When $k\gtrsim k_{\rm th}$ in the upper figure of Fig.\ref{fig:logZ_s2m},
the wavenumbers of the crests of the ripples of $\Re Z$ correspond to $\vbkap_m^n$
($m\in\mathbb{Z}_0^{+}$) for the fundamental vertical mode ($n=1$),
which are shown with the crosses ($+$) in the figure,
\begin{align}
  &
  (\vkap_m^n,\bar{\vkap}_m^n)
  \simeq\frac{k_y^n}{\beta} \bigg\{\frac{1}{\veps_bf_r^3(\xi_m^n)}+1\bigg\}^{1/2} ,
    \qquad
  \xi_m^n
  =\pi(m\mp 1/4)\frac{\beta^3g_b^2}{\veps_bk_y^n\rho} ,
    \label{eq:vkapm_approx}
   \\
  &
  \veps_b
  =\frac{(\beta g_b)^2-1}{2}
  >0 ,
    \qquad
  m\in(\mathbb{N},\mathbb{Z}_0^{+}) ,
    \qquad
  n=2p-1
   \quad
  (p\in\mathbb{N}) ,
    \label{eq:vepsb}
\end{align}
where
\begin{align}
  f_r(\xi)
  =\{(1+\xi^2/3)^{1/2}+1\}^{1/3}-\{(1+\xi^2/3)^{1/2}-1\}^{1/3} .
\end{align}
The sign $(-,+)$ of $m \mp 1/4$ corresponds to $(\vkap_m^n,\bar{\vkap}_m^n)$ and
$m\in(\mathbb{N},\mathbb{Z}_0^{+})$ in Eqs.(\ref{eq:vkapm_approx}-\ref{eq:vepsb}) and
(\ref{eq:ivkapm_approx}-\ref{eq:vepsa}).
$\vkap_m^n$ and $\vbkap_m^n$ are respectively the resonant wavenumbers of
the $\cG_{+}^n$ and $\cG_{-}^n$ modes.
In the upper figure of Fig.\ref{fig:logZ_s2m}
the wavenumbers of the crests of the ripples of $\Re Z$ roughly agree with $\vbkap_m^1$
since the $\vbkap_m^1$ mode dominates over the others.
Also in Figs.\ref{fig:logZ_wh20} and \ref{fig:logZ_r5}, $\vbkap_m^1$ corresponds to
the wavenumbers of the crests of the ripples of $\Re Z$ above $k_{\rm th}$.
But since the ripples for $k>k_{\rm th}$ still depend on $s$ a little bit in
Figs.\ref{fig:logZ_wh20} and \ref{fig:logZ_r5}, the wavenumbers of the crests differ
a little bit from $\vbkap_m^1$
which is for the steady radiation field resonating in the curved pipe.

The condition to emit the steady synchrotron radiation in the curved pipe is given by
Eq.(1.3) of \cite{warnock_morton}, 
\begin{align}
  \beta g_b
  >1 .
   \label{eq:sr_cond}
\end{align}
$g_b$ is the geometric factor given by Eq.(\ref{eq:gl_g12}).
Assuming a steady field in the absence of entanglement between the radial and longitudinal 
components in a constant bend, Eq.(\ref{eq:PE_Exy}) is rewritten as
\begin{align}
  (\nablav_{\!\perp}^2 +\bkap^2)\hEv_{\perp}
  \simeq
   Z_0\nablav_{\!\perp} \hJ_0 ,
    \qquad
  \bkap^2(x,\beta)
  =k^2\Big(\beta^2-\frac{1}{g^2}\Big) ,
    \qquad
  g
  =1+\frac{x}{\rho} .
   \label{eq:PE_Exy_std}
\end{align}
We can also tell Eq.(\ref{eq:sr_cond}) from the sign of $\bkap^2$ for $k\in\mathbb{R}$,
\ie, $\hEv_{\perp}$ is oscillatory with respect to $\xv_{\perp}$ if $\bkap^2>0$.
$\delta x$ ($=\rho/2\gam^2$) given by Eq.(95) in \cite{agoh} is the approximate expression
of $\bkap^2(x+\delta x,\beta)=\bkap^2(x,1)$ for $|x/\rho|\ll 1$.
It means that the effect of taking into account a finite $\gam$ for the ultrarelativistic 
limit is equivalent to the radial displacement by $\delta x$ in theory.
For a given $k\in\mathbb{R}$, if $\gam$ is large such as $\delta x\ll \ell_{\perp}$
which is the typical transverse spread of the radiation field given by Eq.(\ref{eq:shield}), 
the field hardly depends on $\gam$, \ie,
it is nearly converging to the ultrarelativistic limit.
The criterion of whether the radiation field depends on $\gam$ is roughly estimated as
$\delta x\sim \ell_{\perp}$ $\Lra$ $k\sim(4/3)k_c^0$ which roughly agrees with
the critical wavenumber (\ref{eq:kc}).

According to the definition of the Fourier transform (\ref{eq:Fourier_trans}-\ref{eq:omg}) 
in the present paper, we have assumed $k\in\mathbb{R}$ and so are
the resonant wavenumbers $k=(\vkap_m^n,\bar{\vkap}_m^n)\in\mathbb{R}$.
In general, however, the resonant wavenumbers of the fields in the uniformly curved pipe are
the implicit solutions of Eqs.(\ref{eq:p_krho}-\ref{eq:s_krho}) with respect to $k$, \ie,
\begin{align}
  p_{k\rho}(\hr_b,\hr_a)=0
   \quad\Lra\quad
  k=\vkap_m^n ,
    \qquad\quad
  s_{k\rho}(\hr_b,\hr_a)=0
   \quad\Lra\quad
  k=\vbkap_m^n .
   \label{eq:p_krho_vkap}
\end{align}
Accordingly, $k=(\vkap_m^n,\bar{\vkap}_m^n)$ satisfies
\begin{align}
  k\rho
  =(\nu_m^n,\mu_m^n)
  \in\mathbb{A} .
   \label{eq:kreson}
\end{align}
The precise values of $\vkap_m^n$ and $\bar{\vkap}_m^n$ are gotten by solving
Eqs.(\ref{eq:p_krho_vkap}) using Newton's method if necessary.
To be exact, $\vkap_m^n$ and $\bar{\vkap}_m^n$ depend on both $x_a$ and $x_b$
which are the horizontal positions of the inner and outer walls of the curved pipe.
On the other hand, $\vkap_m^n$ and $\bar{\vkap}_m^n$ given by
Eq.(\ref{eq:vkapm_approx}) do not depend on $x_a$ since they are the asymptotic solutions of 
Eq.(\ref{eq:kreson}) for large $k\in\mathbb{R}$ such that the uniform asymptotic expansion 
is applied to calculating the cross products $p_{\nu}$ and $s_{\nu}$.
That is, Eq.(\ref{eq:vkapm_approx}) is gotten by solving Eq.(\ref{eq:kreson}) using
Eq.(\ref{eq:rwgm}) to $O(\hr_b^{-2/3})$, which is the asymptotic expression of
the real poles of the whispering gallery mode as described in section \ref{sec:wgm}.
Eq.(98) in \cite{agoh} is the approximate expression of Eq.(\ref{eq:vkapm_approx}) for
$x_b\ll\rho$ and $\gam\to\infty$.

Similar to Eq.(\ref{eq:vkapm_approx}), the asymptotic expression of
the imaginary resonant wavenumbers ($k=i\bk\in i\mathbb{R}$) of the steady fields in
the curved pipe are gotten by solving Eq.(\ref{eq:kreson}) using Eq.(\ref{eq:iwgm}) as
$\nu=i\bnu\in i\mathbb{R}$ which is the asymptotic expression of the poles of
the imaginary whispering gallery mode as described in section \ref{sec:iwgm},
\begin{align}
   &
  (\vkap_m^n,\bar{\vkap}_m^n)
  \simeq  i\frac{k_y^n}{\beta}\bigg\{\frac{1}{\veps_af_r^3(\eta_m^n)}-1\bigg\}^{1/2}
   ,\qquad
  \eta_m^n
  =\pi(m\mp1/4)\frac{\beta^3g_a^2}{\veps_ak_y^n\rho} ,
    \label{eq:ivkapm_approx}
   \\
   &
  \veps_a
  =\frac{1-(\beta g_a)^2}{2}
  >0
   ,\qquad
  m\in(\mathbb{N},\mathbb{Z}_0^{+})
   ,\qquad
  n=2p-1
   \quad
  (p\in\mathbb{N}) .
    \label{eq:vepsa}
\end{align}
The geometric factor of the inner wall $g_a$ is given by Eq.(\ref{eq:gl_g12}).
Eq.(129) in \cite{agoh} is the approximate expression of Eq.(\ref{eq:ivkapm_approx}) for
$-x_a\ll\rho$ and $\gam\to\infty$.
Contrary to Eq.(\ref{eq:vkapm_approx}), $\vkap_m^n$ and $\bar{\vkap}_m^n$ given by
Eq.(\ref{eq:ivkapm_approx}) do not depend on the outer wall $x_b$
since they are the asymptotic expressions for $k\in i\mathbb{R}$.
When $k\in i\mathbb{R}$ and $\beta g<1$, $\bkap^2$ involved in Eq.(\ref{eq:PE_Exy_std}) is 
positive, similar to that for $k\in\mathbb{R}$ and $\beta g>1$.

In calculating a field of CSR,
if we do not impose the periodic boundary condition with respect to $s$ to the field as
in the present paper, the impedance of CSR emitted in free space or between parallel plates 
has a continuous spectrum regardless whether the field is steady or transient 
since the energy of the electromagnetic field is dissipated
unlike CSR emitted in a perfectly conducting pipe in
which the energy of the field is completely confined and conserved.
When the field resonates in a curved pipe, the spectrum of
the steady or quasi-steady field of CSR differs from that in free space
(or between parallel plates)
even if their fields in the time domain look similar around the bunch
(\eg, $|z|\lesssim 5\sig_z$)
when the bunch length is very short as $\sig_z\ll k_{\rm th}^{-1}$.
Therefore we cannot necessarily assume free space (or infinite parallel plates) in 
investigating the spectrum of CSR emitted in a beam pipe
even if the frequency components of the field are mostly in the range $k>k_{\rm th}$.
That is, we must take into account the presence of the beam pipe in investigating
the spectrum of CSR (SR) not only in the low frequency range $k<k_{\rm th}$
but also in the high frequency range $k>k_{\rm th}$ if the resonance is not negligible.
But in practice, our assumption that the beam pipe has no roughness 
on the inner surface may not be correct for extremely high frequency waves.
That is, when $k^{-1}$ is comparable to or smaller than the scale of the microscopic 
surface roughness of the beam pipe in the order of microns or less for example,
the resonance of the field in the pipe may not happen clearly
due to the diffuse reflection on the rough surface.
We think that it would be better to assume free space in calculating the field of CSR
in such an extremely high frequency range
that the diffuse reflection happens on the rough surface of the beam pipe.

\clearpage

\begin{figure}[h]
\begin{center}
  %
  \includegraphics[scale=0.46,clip]{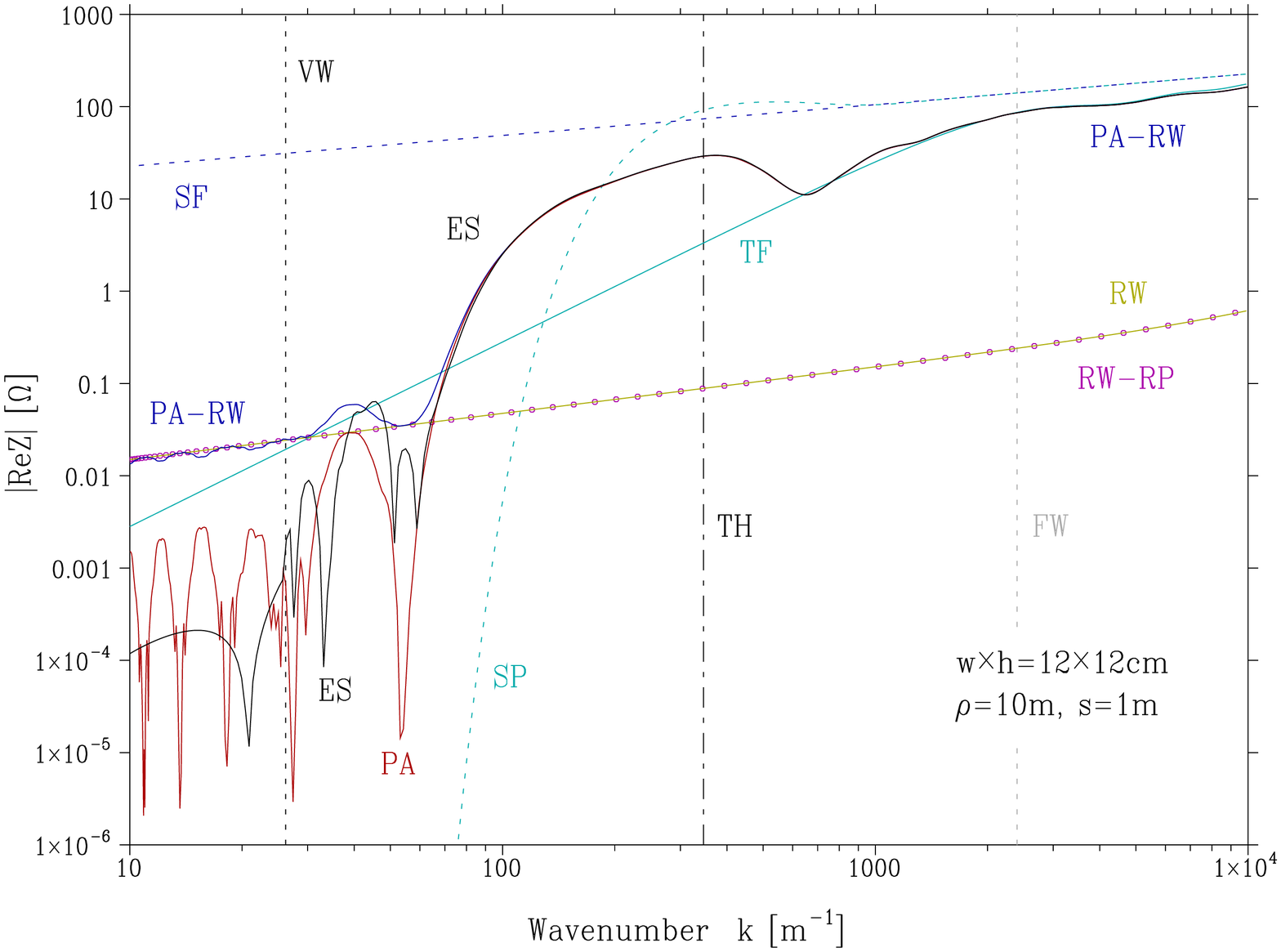}
   \vspace{2mm}
   \\
  \includegraphics[scale=0.46,clip]{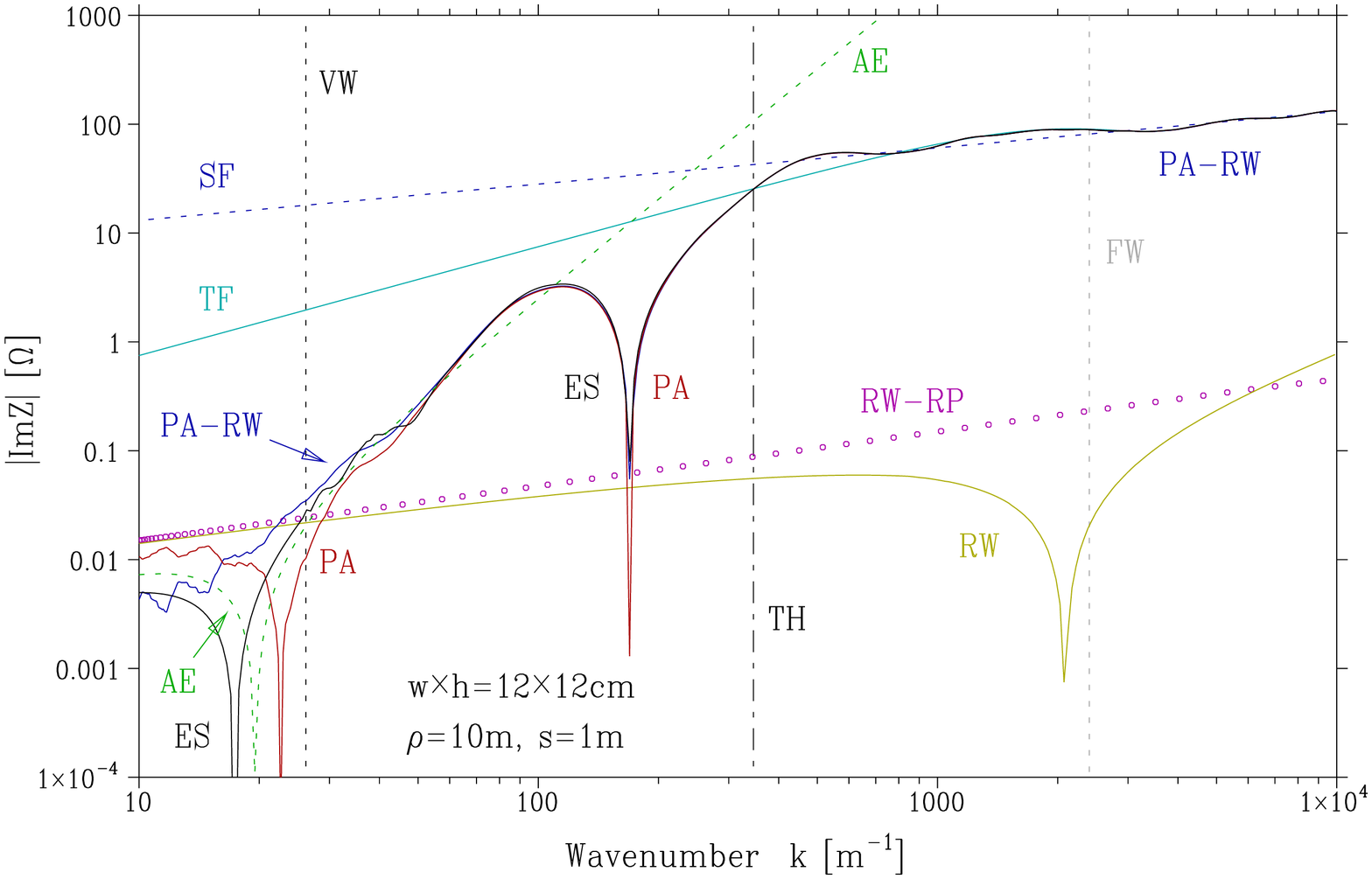}
  \caption[Longitudinal impedances of CSR and resistive wall in a large square pipe]
    { \small
    Absolute value of the real part $\Re Z$ (upper) and imaginary part $\Im Z$ (lower) of
    $Z\,[\Omg]$ which represents the longitudinal impedances of CSR and resistive wall (RW) 
    in a square pipe $w=h=12{\rm cm}$ for a length $s=1{\rm m}$.
    The horizontal axis is the wavenumber $k$.
    Excluding $w$ and $h$, the parameters are common to Fig.\ref{fig:logZ}.
    The nomenclature is also common to Fig.\ref{fig:logZ}.
    The black vertical dotted line (VW) is the fundamental vertical wavenumber of
    the pipe $k_y^1$.
    The vertical dash-dotted line (TH) is the shielding threshold $k_{\rm th}$.
    The gray vertical dotted line (FW) shows $k_{\rm f}$ given by Eq.(\ref{eq:kth})
    which is the formation wavenumber of CSR in the bend.
    The black and red curves (ES and PA) show $Z$ of a transient field of CSR
    emitted in the perfectly conducting rectangular pipe.
    The black curve (ES) shows $Z$ computed by integrating
    the exact solution of $\tE_s$ given by Eq.(\ref{eq:tEs_std_sep}).
    The red curve (PA) shows $Z$ computed using the numerical 
    solution of Eqs.(\ref{eq:PE_Exy}-\ref{eq:PE_Es}).
    The blue curve (PA-RW) shows $Z$ of
    the transient CSR in a copper pipe, computed by solving
    Eqs.(\ref{eq:PE_Exy}-\ref{eq:PE_Es}) imposing the resistive boundary condition.
    The yellow-brown curve (RW) shows the resistive wall impedance of a straight copper pipe
    ($E=1{\rm GeV}$) which has the same cross section and length along the $s$-axis as
    the curved square pipe.
    The magenta dots (RW-RP) show $Z$ computed using
    Eq.(4.75) in \cite{yokoya_rw}, which is the resistive wall impedance of
    a straight round pipe of diameter $h$ for $\beta=1$.
    The cyan curve (TF) shows $Z$ of the transient field of CSR
    in free space, computed using Eq.(87) in \cite{saldin_schneidmiller_yurkov}.
    The cyan dotted curve (SP) shows $Z$ of the steady field of CSR emitted
    between perfectly conducting infinite parallel plates located at $y=\pm h/2$.
    We computed $\Re Z$ of SP using Eq.(A1) in \cite{agoh_yokoya}.
    The blue dotted line (SF) shows Eq.(A10) in \cite{agoh_yokoya},
    which is $Z$ of the steady field of CSR in free space.
    The green dotted curve (AE) in the lower figure shows $\Im Z$ ($=s\Im\cZ$) given by
    Eq.(\ref{eq:cZ_AE}).
  }
  \label{fig:logZ_wh12}
\end{center}
\end{figure}

\clearpage

\subsection{Power spectrum of synchrotron radiation emitted in a beam pipe}
\label{sec:SR_Z}

\begin{figure}[h]
  \begin{center}
    \includegraphics[scale=0.32,clip]{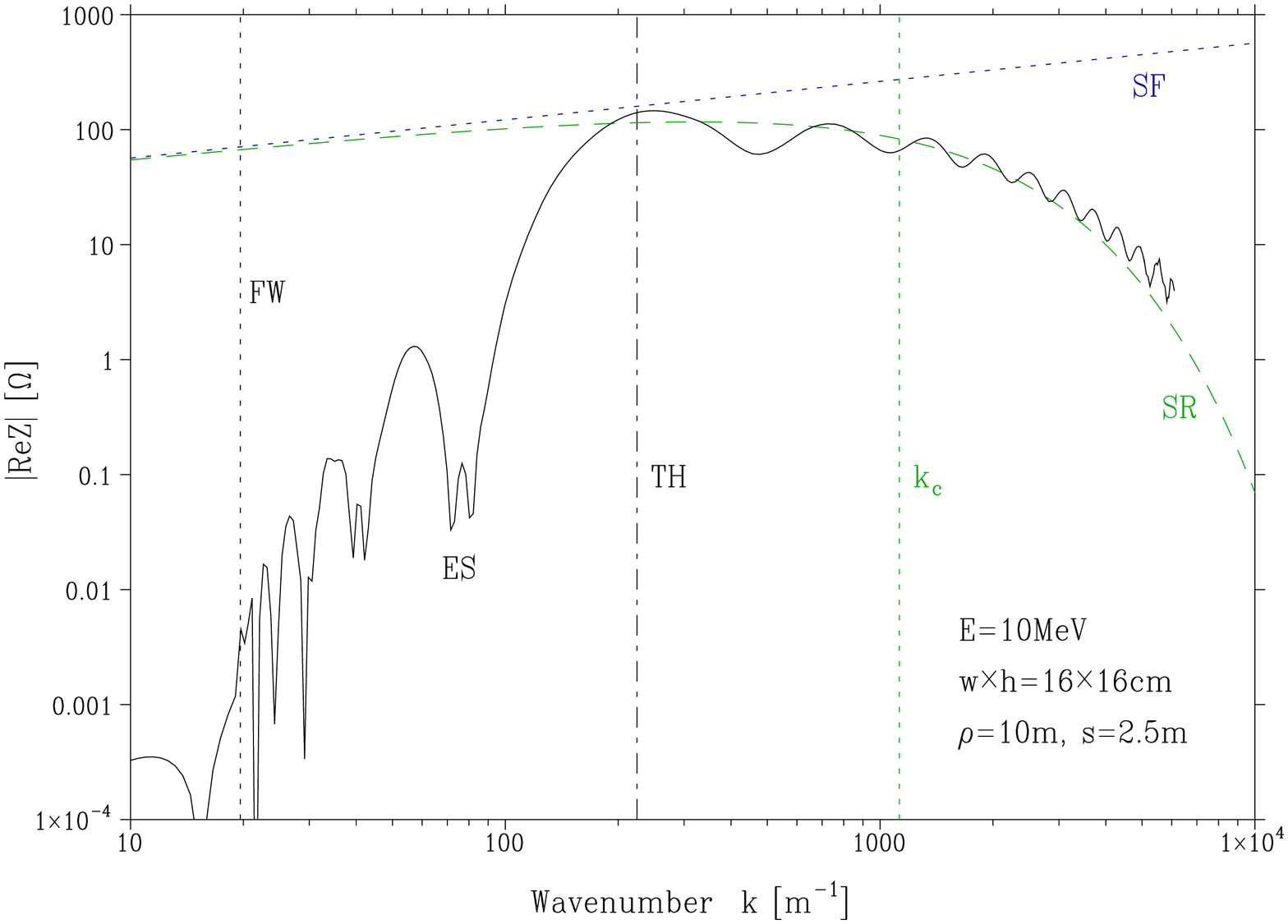}
    \includegraphics[scale=0.32,clip]{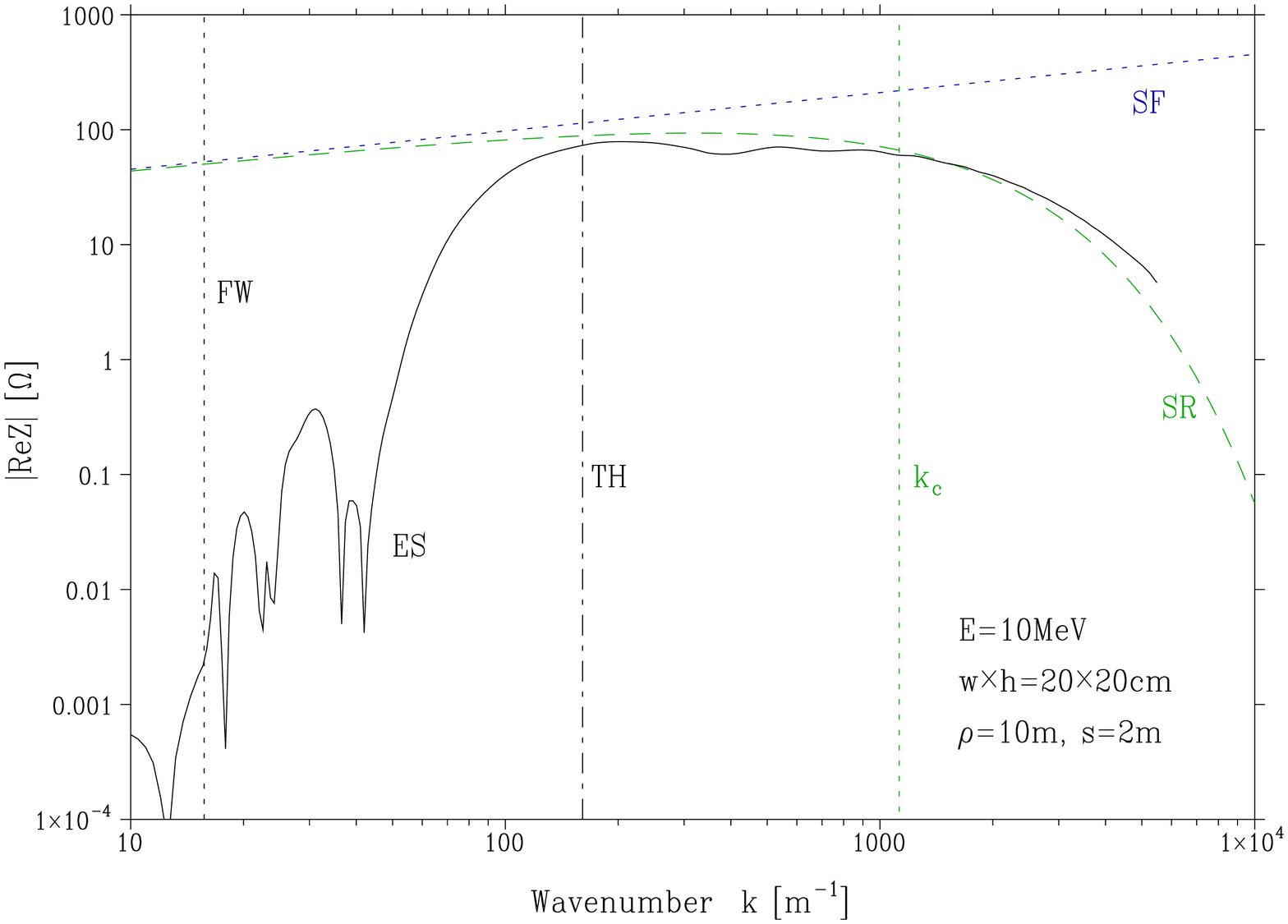}
      \vspace{1mm}
      \\
    \includegraphics[scale=0.32,clip]{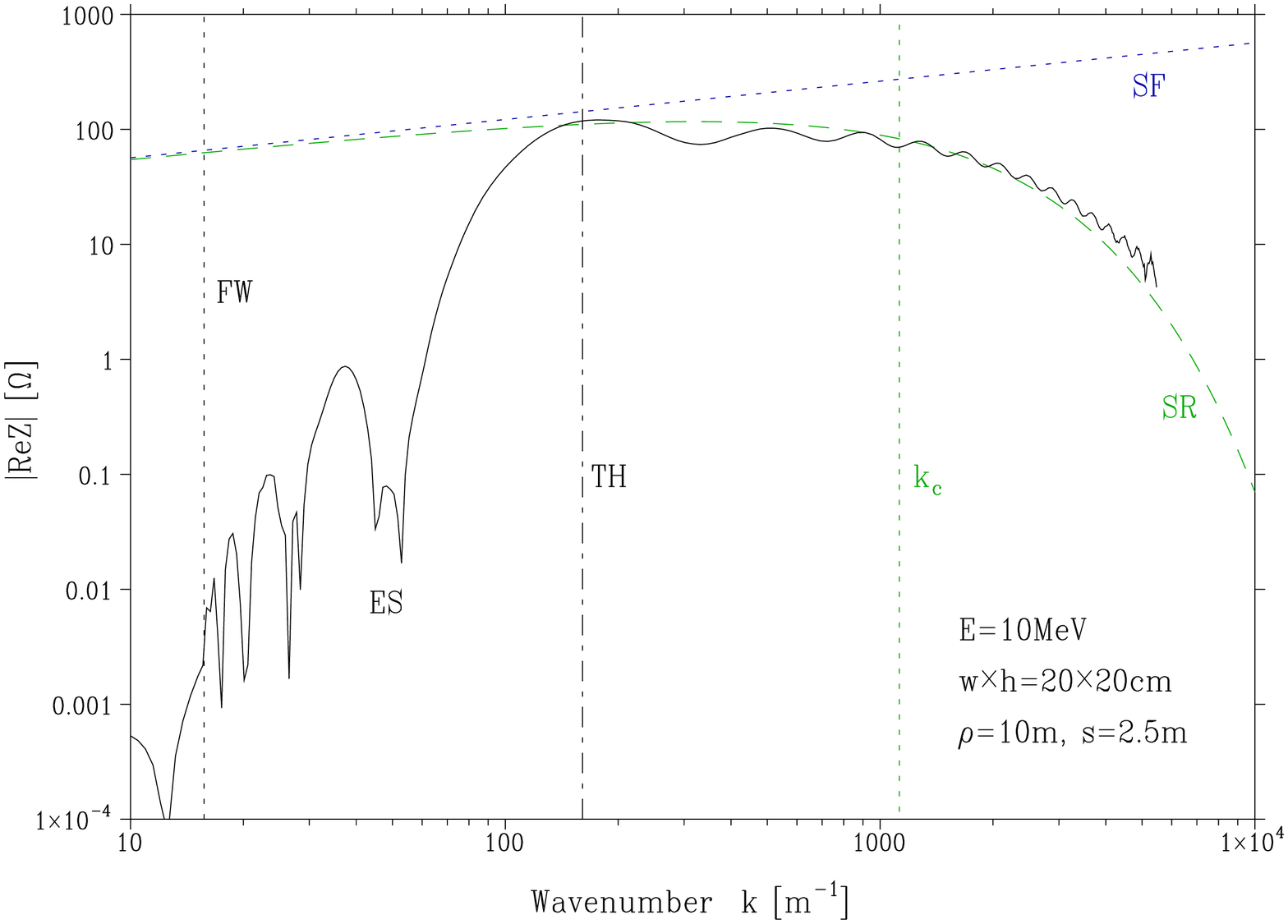}
    \includegraphics[scale=0.32,clip]{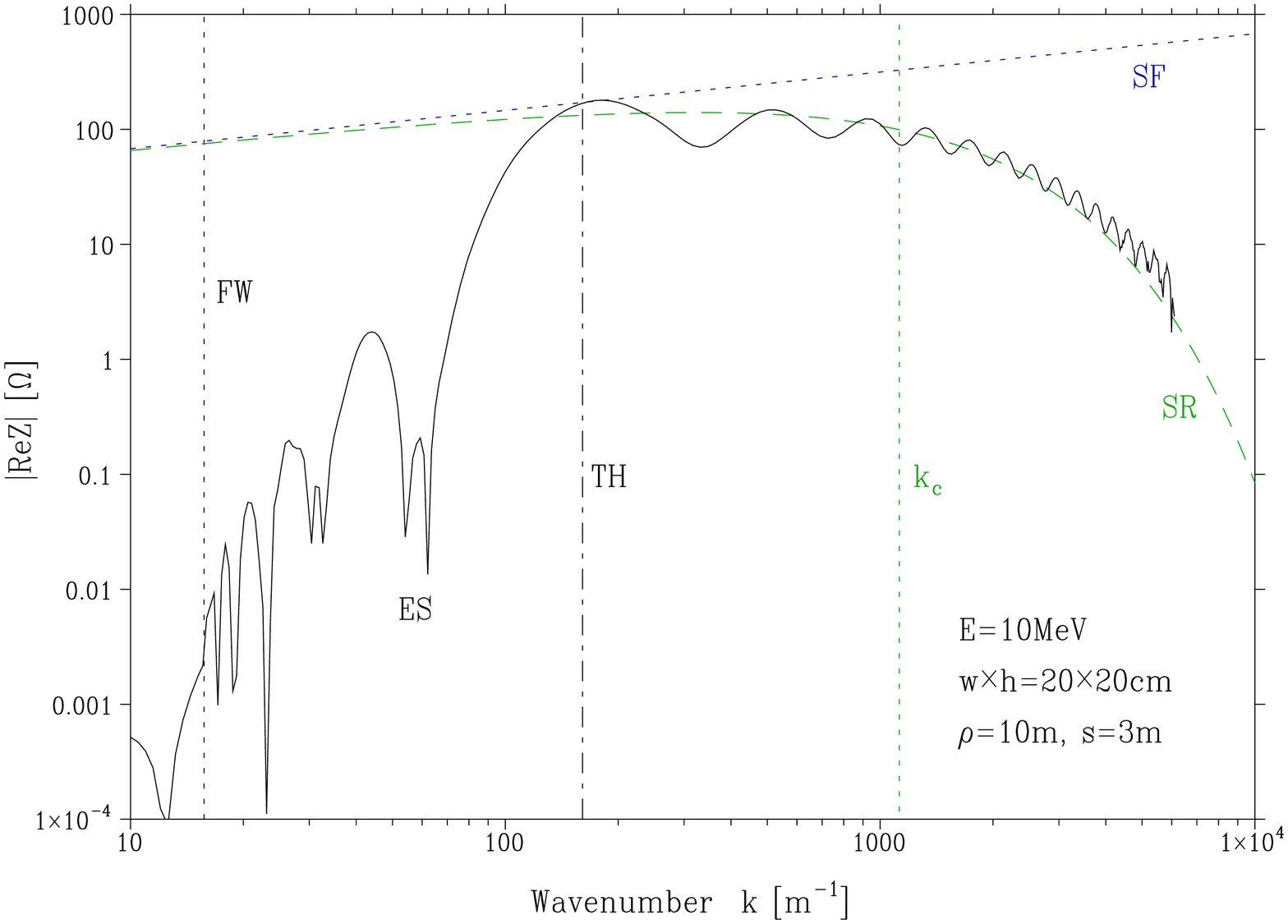}
    \caption[Power spectrum of shielded synchrotron radiation ($\rho=10{\rm m}$)]
    {\small 
    Absolute value of the real part of the longitudinal impedance of CSR
    as a function of the wavenumber $k$.
    The vertical black dotted line (VW) is the fundamental vertical wavenumber $k_y^1$.
    The vertical dash-dotted line (TH) is the shielding threshold $k_{\rm th}$
    given by Eq.(\ref{eq:kth}).
    The black curve (ES) shows $|\Re Z|$ computed using the exact solution of
    the transient field of CSR in the perfectly conducting pipe.
    The green dashed curve (SR) shows $\Re Z$ $(=s\Re\cZ)$ of the steady field of SR in
    free space, given by Eq.(\ref{eq:kc}).
    The vertical green dotted line is the critical wavenumber $k_c$ for
    $\rho=10{\rm m}$ and $E=10{\rm MeV}$ (electron).
    In order to confirm the correct behavior of SR for $k/k_c^0\to0$, we plotted
    the blue dotted line (SF) which shows $\Re Z$ of the steady field of CSR in free space
    for $\gamma=\infty$, given by Eq.(A10) in \cite{agoh_yokoya}.
    We assume $w=h=16{\rm cm}$ and $s=2.5{\rm m}$ in the upper left figure.
    Assuming a larger pipe $w=h=20{\rm cm}$, we calculated $\Re Z$ for
    $s=2{\rm m}$ (upper right), $2.5{\rm m}$ (lower left) and $3{\rm m}$ (lower right)
    to reproduce SR for $k>k_{\rm th}$ using ES.
     }
    \label{fig:logZ_wh20}
  \end{center}
\end{figure}

\begin{figure}[h]
  \begin{center}
    \includegraphics[scale=0.32,clip]{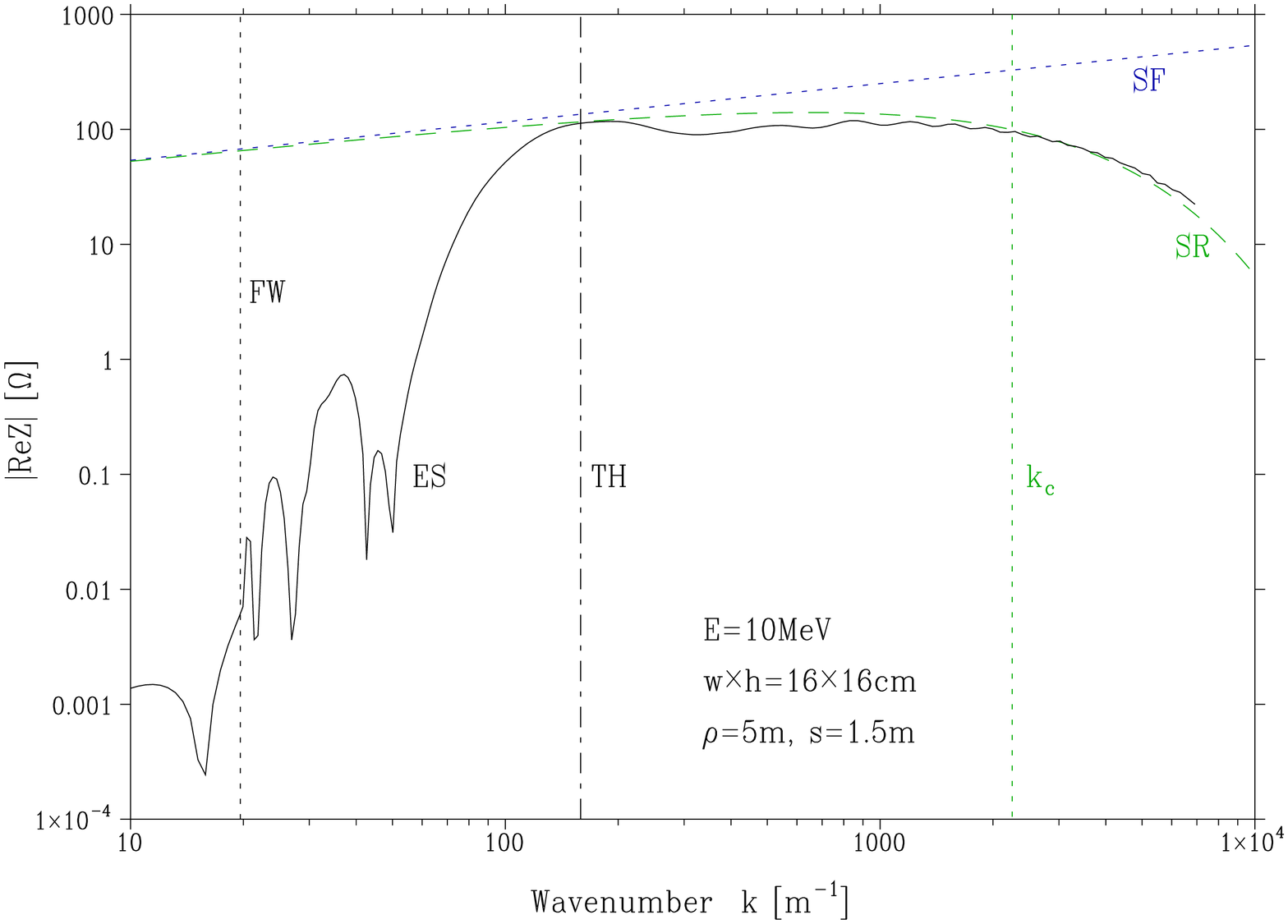}
    \includegraphics[scale=0.32,clip]{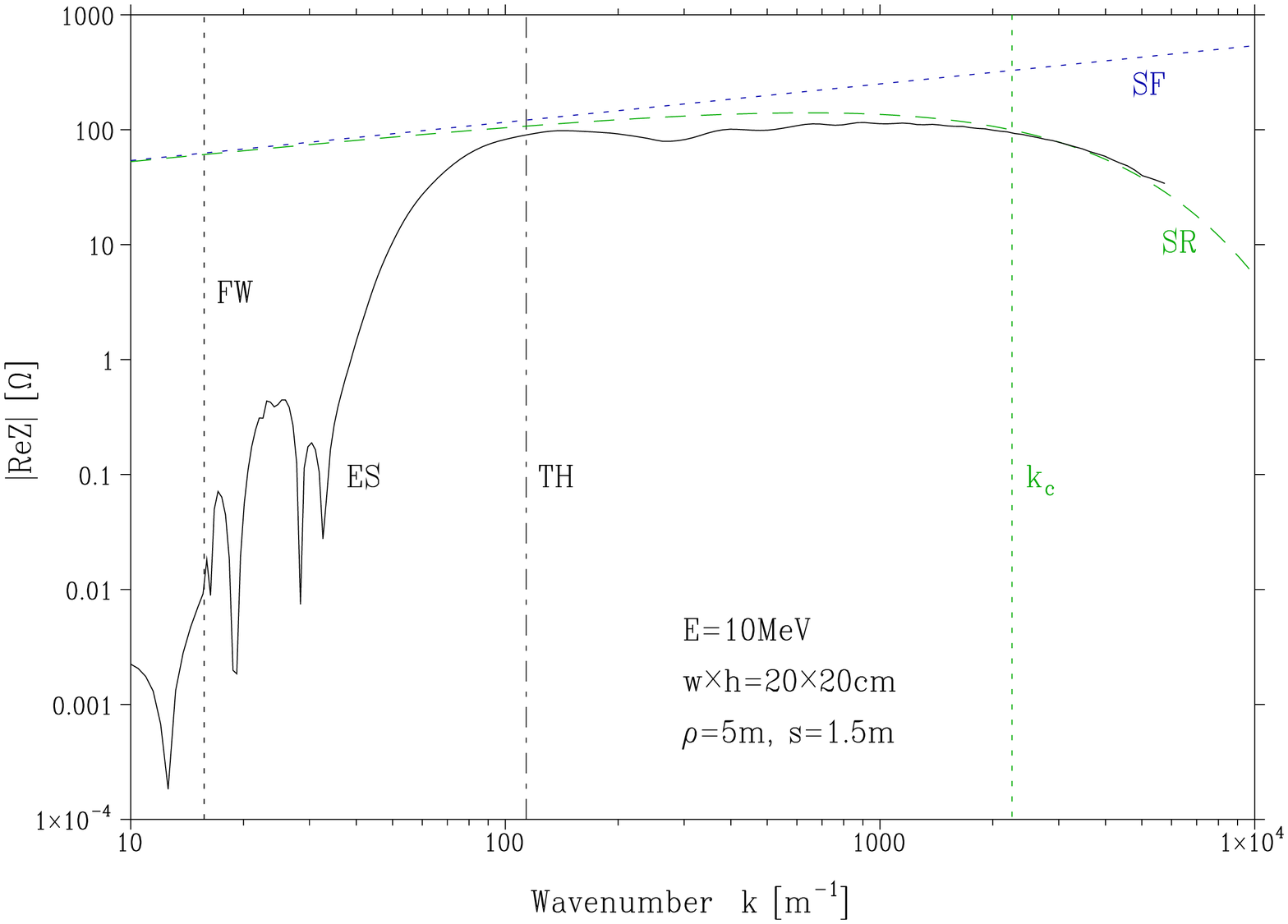}
      \vspace{1mm}
      \\
    \includegraphics[scale=0.32,clip]{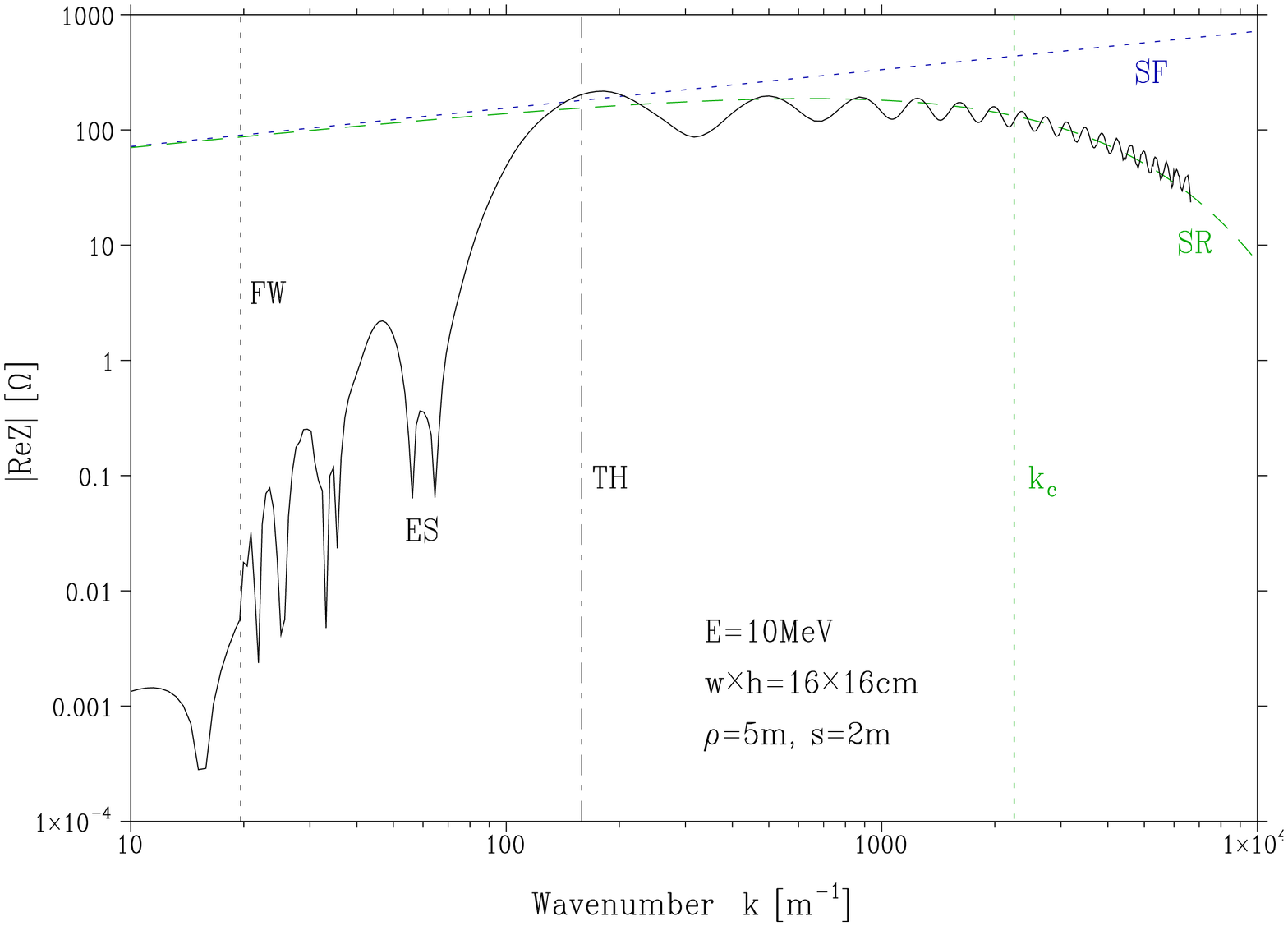}
    \includegraphics[scale=0.32,clip]{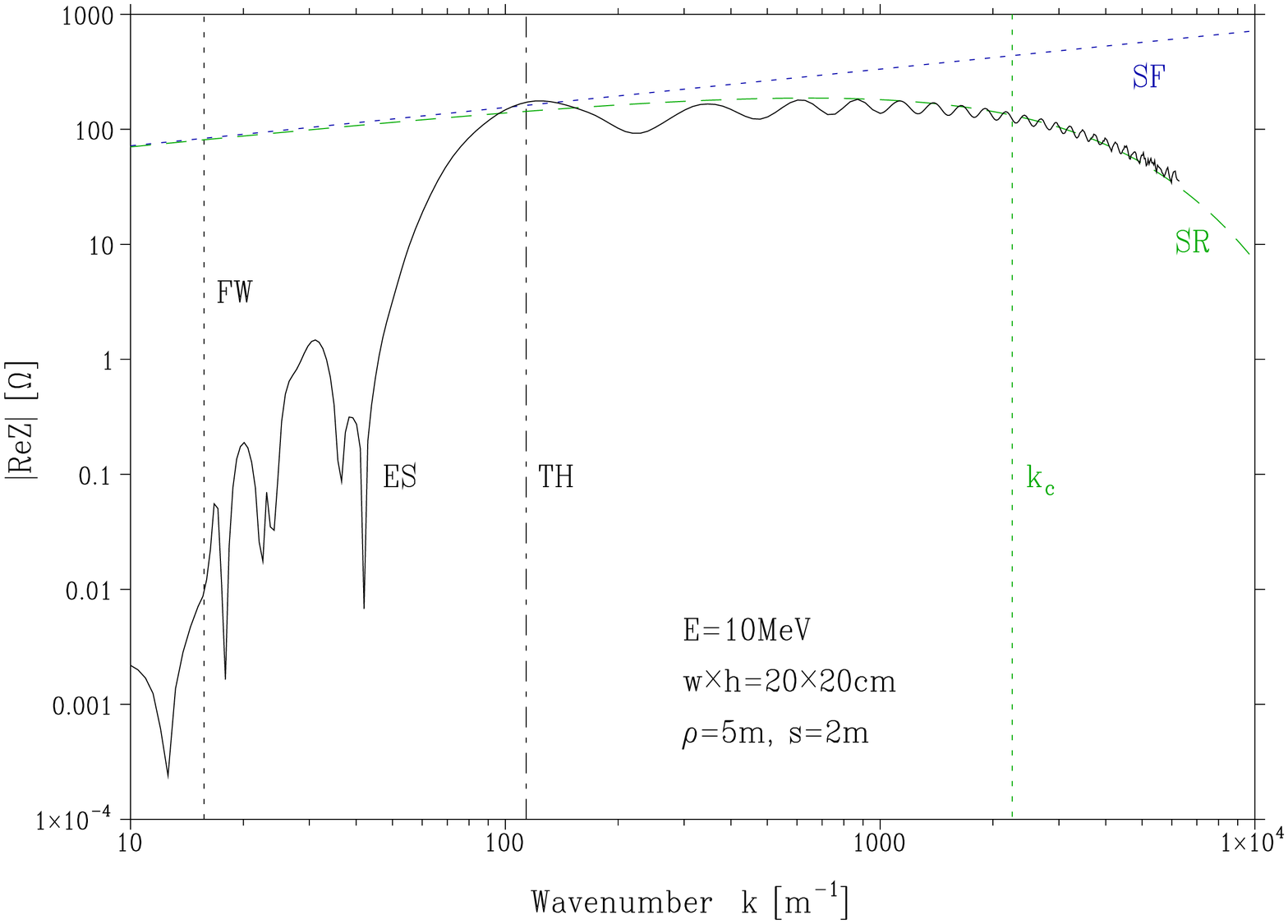}
      \vspace{1mm}
      \\
    \includegraphics[scale=0.32,clip]{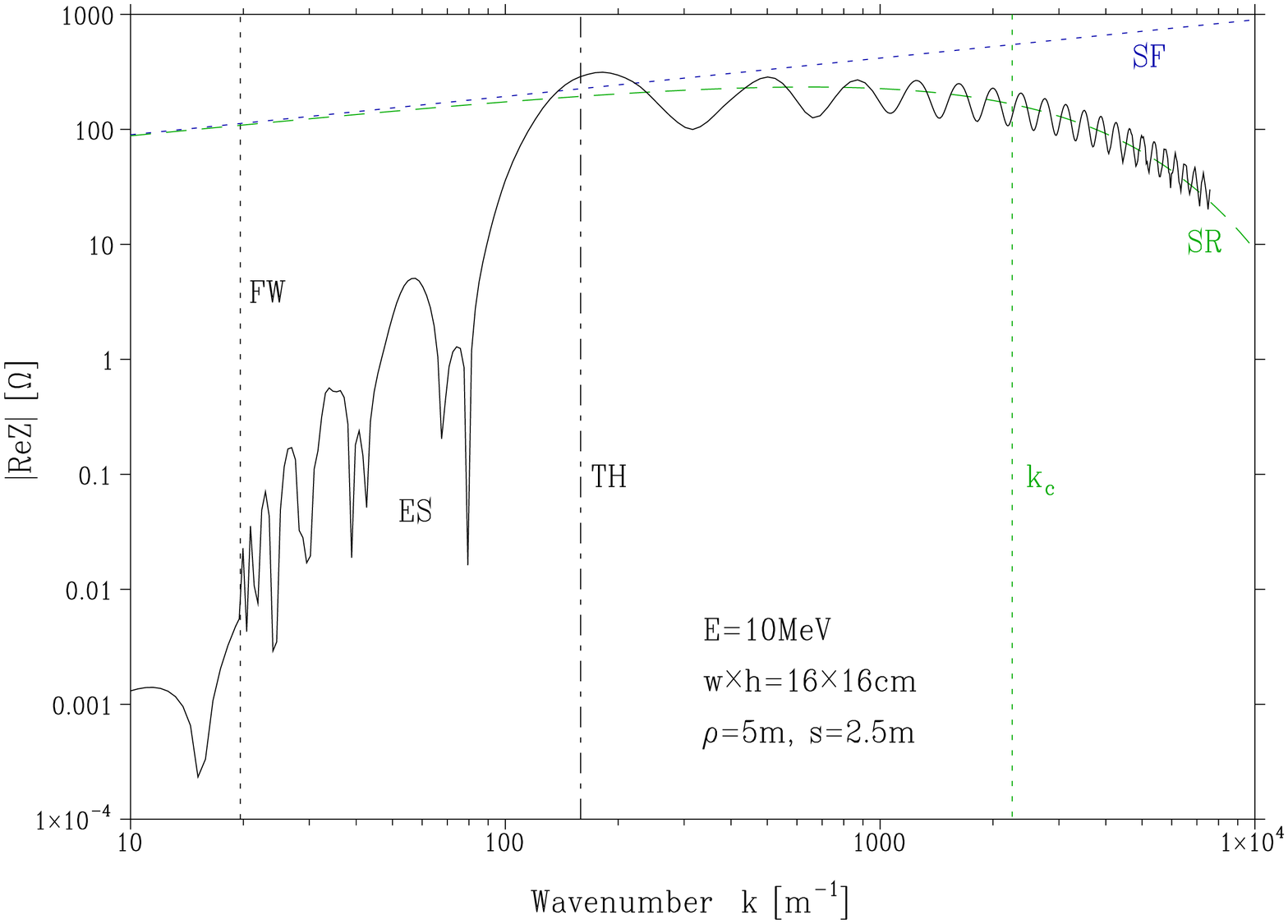}
    \includegraphics[scale=0.32,clip]{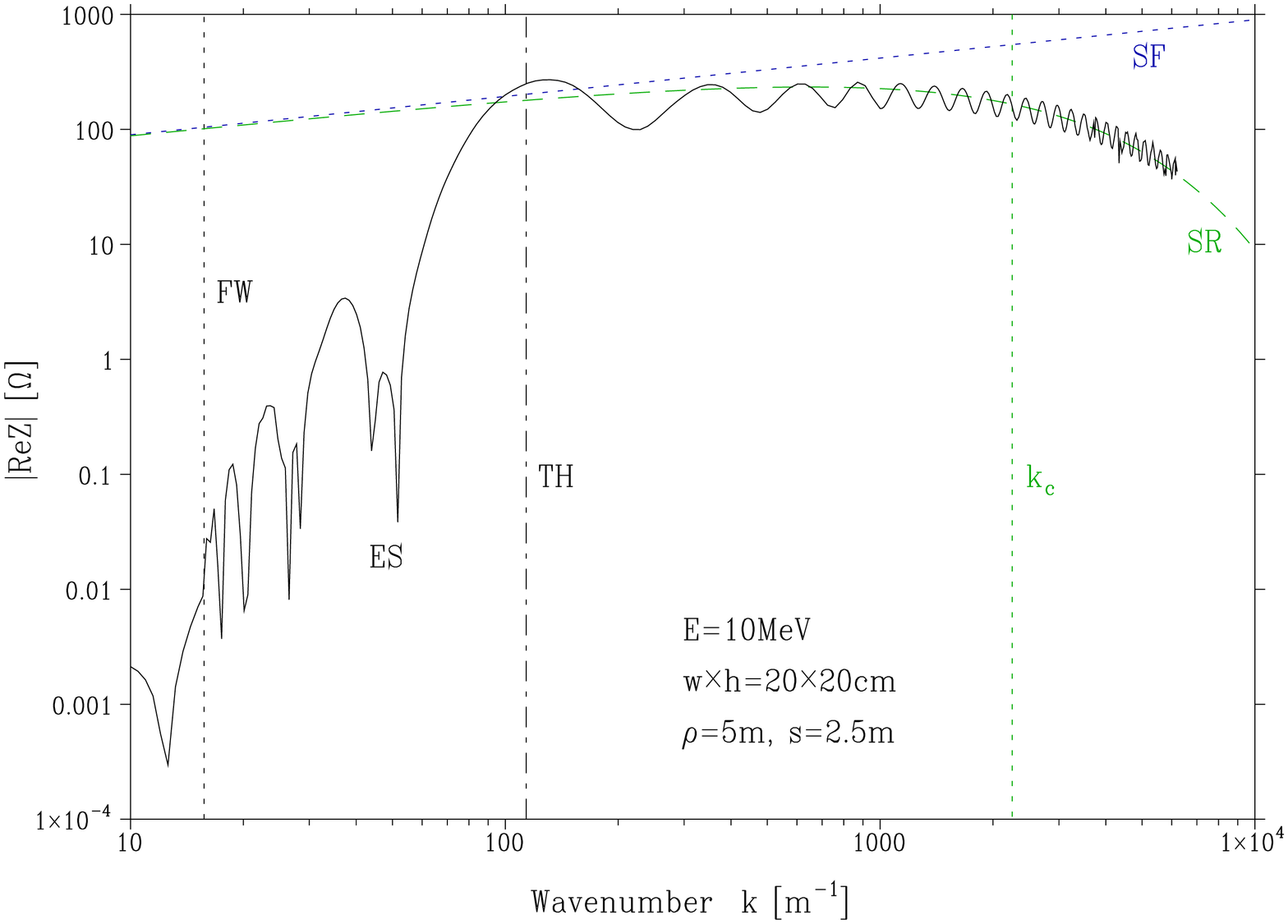}
    \caption[Power spectrum of shielded synchrotron radiation ($\rho=5{\rm m}$)]
    {\small 
    Absolute value of the real part of the longitudinal impedance of CSR
    as a function of the wavenumber $k$.
    The vertical black dotted line (VW) is the fundamental vertical wavenumber $k_y^1$.
    The vertical dash-dotted line (TH) is the shielding threshold $k_{\rm th}$
    given by Eq.(\ref{eq:kth}).
    The black curve (ES) shows $|\Re Z|$ computed using the exact solution of
    the transient field of CSR in the perfectly conducting pipe.
    The blue dotted line (SF: steady, free space, $\gam\to\infty$) shows $\Re Z$ of
    the steady field of CSR in free space, given by Eq.(A10) in \cite{agoh_yokoya}.
    We plotted SF in order to confirm the correct behavior of SR for $k/k_c^0\to0$.
    The green dashed curve (SR) shows $\Re Z$ $(=s\Re\cZ)$ of the steady field of SR
    in free space, computed using Eq.(\ref{eq:kc}).
    The vertical green dotted line is the critical wavenumber $k_c$ for
    $\rho=5{\rm m}$ and $E=10{\rm MeV}$ (electron).
    We assume a perfectly conducting square pipe
    $w=h=16{\rm cm}$ (left) and $20{\rm cm}$ (right) in computing ES.
    The length of the bend is $s=1.5{\rm m}$ (upper),
    $2{\rm m}$ (middle) and $2.5{\rm m}$ (lower).
    When $k>k_{\rm th}$ for $s=2{\rm m}$ and $2.5{\rm m}$, $\Re Z$ of ES has the ripples
    whose crests correspond to the resonant wavenumbers $\vbkap_m^n$ ($m\in\mathbb{Z}_0^{+}$)
    given by Eq.(\ref{eq:vkapm_approx}) for the fundamental vertical mode ($n=1$).
     }
    \label{fig:logZ_r5}
  \end{center}
\end{figure}

As mentioned at the beginning of introduction, synchrotron radiation (SR) emitted by
a single point charge has a power spectrum up to the critical frequency $\omg_c$ $(=k_cv)$
\cite{schwinger},
\begin{align}
  \frac{c\eps_0}{2e^2\rho\beta}\cd\frac{dP}{d\omega}
  =\frac{3^{1/2}}{8\pi\rho}\cd\frac{k\gam}{k_c^0\beta}
   \int_{k/k_c^0}^{\infty}K_{5/3}(x)dx
  =\frac{\Re\cZ(k)}{Z_0}
   \quad[{\rm m}^{-1}] ,
    \qquad
  k_c^0
  =k_c\beta
  =\frac{3\gamma^3}{2\rho} .
   \label{eq:kc}
\end{align}
$\gam$ $(=E/mc^2)$ is the Lorentz factor.
$E$, $m$ and $e$ are the energy, mass and charge of the particle.
Eq.(\ref{eq:kc}) can be gotten from Eq.(\ref{eq:PE_Exy_std})
as shown in Eq.(208) of \cite{agoh} and Eq.(D13) in \cite{stupakov_kotelnikov_0}.
In the present paper we define the critical wavenumber $k_c$ as
it is consistent with $\omg_c$ through Eq.(\ref{eq:omg}).
$\cZ\,[\Omg/{\rm m}]$ is the longitudinal impedance of the steady field of unshielded SR
per unit length along the $s$-axis, which tends to be SF for $k/k_c^0\to0$,
given by Eq.(A10) in \cite{agoh_yokoya}.
Schwinger's power spectrum formula (\ref{eq:kc}) is plotted in Figs.\ref{fig:logZ_wh20} and
\ref{fig:logZ_r5} with the green dashed curve (SR) as the real part of
the impedance $Z=s\cZ\,[\Omg]$ for a given trajectory length $s$ through Eq.(\ref{eq:dP_dw}).

In theory, $\Re Z$ computed using the exact solution (ES) must agree with $s\Re\cZ$ 
given by Eq.(\ref{eq:kc}) in the limit of large pipe ($w,h\to\infty$: free space)
and long bend ($s\to\infty$: steady state).
In order to confirm that ES and its numerical code are correct around $k_c$,
we tried to reproduce Eq.(\ref{eq:kc}) using ES as it is on purpose
(\ie, without using the asymptotic expression for $w,h,s\to\infty$ and $k\to\infty$)
simply by entering large values $(w,h,s)$ into the ES code.
As shown in Fig.\ref{fig:logZ_wh20}, when $k>k_{\rm th}$,
$\Re Z$ of ES tends to agree with SR for a larger pipe in a longer bend.
When $k<k_{\rm th}$, ES differs from SR due to the shielding effect by
the beam pipe as expected.
Examining $\Im Z$ in section \ref{sec:Z_energy}, we will confirm that
ES is also correct in the low frequency range $k\leq O(k_y^1)$.


\clearpage

Although the curve of the exact solution (ES, black) does not overlap perfectly with SR
(dashed green) for $k>k_{\rm th}$ in Fig.\ref{fig:logZ_wh20} ($\rho=10{\rm m}$),
it does not mean that ES is incorrect.
It is enough to observe the reasonable asymptotic behavior of ES for
$w,h,s\to\infty$ and $k\to\infty$ in Figs.\ref{fig:logZ_wh20} and \ref{fig:logZ_r5}.
As shown in Fig.\ref{fig:logZ_r5} ($\rho=5{\rm m}$), assuming a smaller bending radius, 
it is easier to compute the impedance using ES so that the values of $\Re Z$ of ES get 
close to SR since $k_{\rm th}$ and $k_{\rm f}$ become smaller as seen from
Eqs.(\ref{eq:kth}).
But also in Fig.\ref{fig:logZ_r5} the curve of ES does not overlap perfectly with SR
for $k>k_{\rm th}$.
In practice, it is not easy to reproduce Eq.(\ref{eq:kc}) using ES due to
the numerical limits: computing time and accuracy in calculating $Z$ especially
above the critical wavenumber $k_c$ for which $\Re Z$ decays exponentially.
In order to reproduce Eq.(\ref{eq:kc}) using ES for $k>k_{\rm th}$ in the numerical 
calculation, the dimension of the beam pipe $(w,h,s)$ must be large enough to lessen
both the transient effect and the influence of the pipe.
If $s$ is not long enough, since the field does not reach the steady state in the bend,
$\Re Z$ of ES deviates from SR as shown in the upper right figure of Fig.\ref{fig:logZ_wh20}.
Conversely, if $s$ is long, since the radiation field hits the outer wall of the curved pipe 
and reflects, $\Re Z$ of ES tends to ripple in $k>k_{\rm th}$ due to the resonance as
seen from the lower right figure in Fig.\ref{fig:logZ_wh20}.
Similar to the upper figure in Fig.\ref{fig:logZ_s2m}, the crests of the ripples of $\Re Z$ 
for $k>k_{\rm th}$ correspond to $\vbkap_m^1$ given by Eq.(\ref{eq:vkapm_approx}) 
which are the resonant wavenumbers of the whispering gallery mode in the curved pipe.

Assuming a beam pipe which has a larger cross section to avoid the shielding and resonance,
it takes a very long time to compute the values of $Z$ using ES
since the field consists of a larger number of the transverse modes $(m,n)$ for
higher frequencies.
Using ES, it takes time to compute $Z$ not only in the high frequency range $k>k_{\rm th}$
but also in the very low frequency range $k<k_y^1$ which corresponds to
the imaginary whispering gallery mode described in section \ref{sec:iwgm}.
In addition, the numerical accuracy tends to be reduced for $w\to\infty$ and $k\to\infty$ in 
computing the impedance using the ES code.
Therefore, assuming a very large pipe such as $w=h=16{\rm cm}$ or 20cm,
it is not easy to compute the values of $Z$ accurately using ES for $k>k_c$ in particular.
That is why we could not calculate the values of $\Re Z$ using ES to 
the right end ($k=10^4/{\rm m}$) of Figs.\ref{fig:logZ_wh20} and \ref{fig:logZ_r5}.
Because of these numerical constraints, it is difficult to reproduce Eq.(\ref{eq:kc}) 
perfectly for $k>k_{\rm th}$
using the exact solution of the transient field of CSR emitted in the pipe
even though it is possible in theory.
As described above, we can explain the reasons why $\Re Z$ computed using
the exact solution (ES) has a small difference (\ie, deviation or ripples) from
Schwinger's power spectrum formula (\ref{eq:kc}) for $k>k_{\rm th}$ in the figure.
Since we assumed the very large beam pipe on purpose to demonstrate the fact that 
ES tends to agree with SR for $k>k_{\rm th}$ in the numerical calculation as shown in
Figs.\ref{fig:logZ_wh20} and \ref{fig:logZ_r5}, it is not important that,
due to the numerical limits of our computer and code, we cannot get a numerical result
such that the curve of ES overlaps perfectly with SR for $k>k_{\rm th}$ in the figure.
For faster and more accurate calculations, we should derive the asymptotic expressions of 
the field and impedance for both $k\to0$ and $k\to\infty$ by expanding the exact solution.
This is one of the work that remains to be done in the current study.

\subsection{Energy dependence of the longitudinal impedance of CSR}
\label{sec:Z_energy}

\begin{figure}[h]
  \begin{center}
    \includegraphics[scale=0.32,clip]{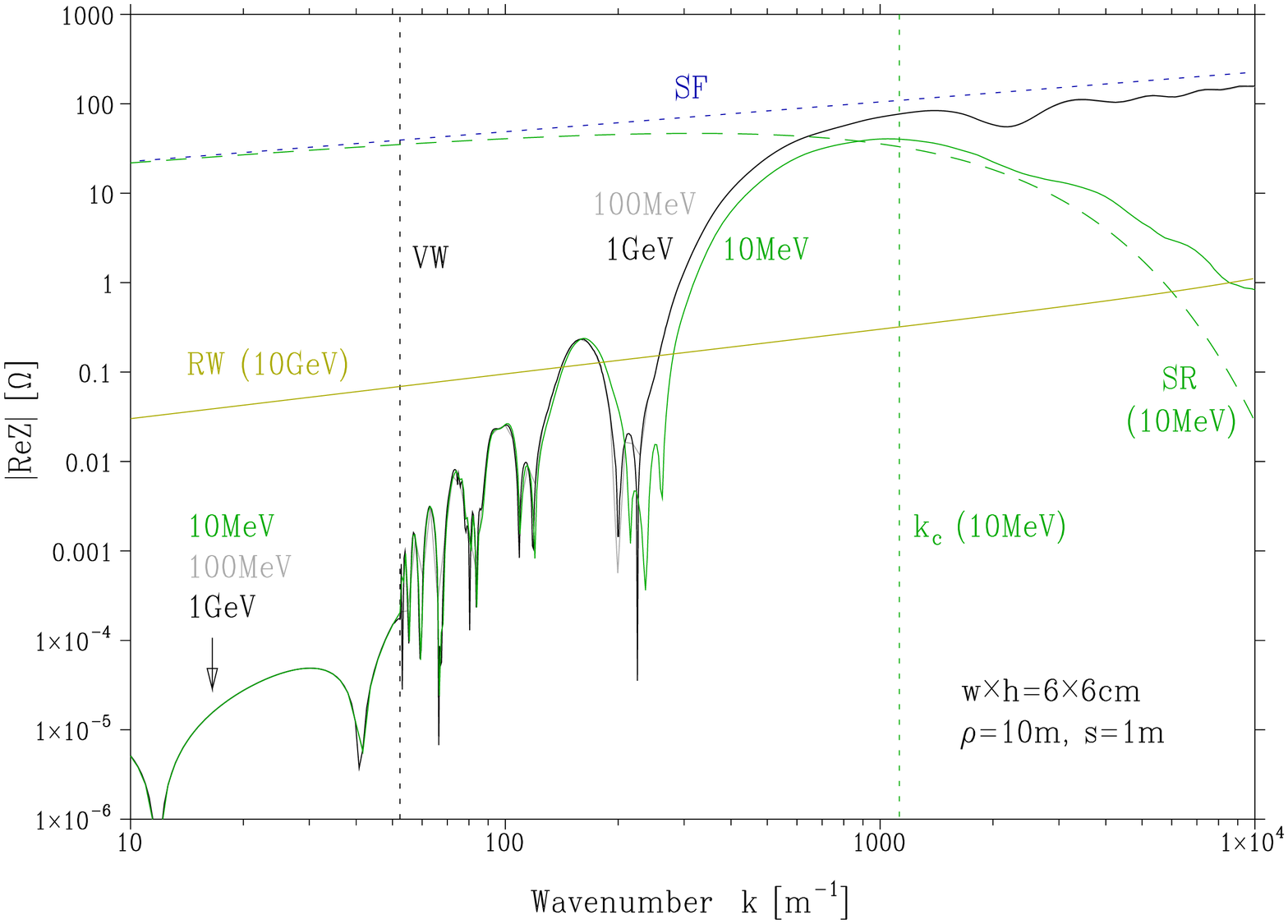}
    \includegraphics[scale=0.32,clip]{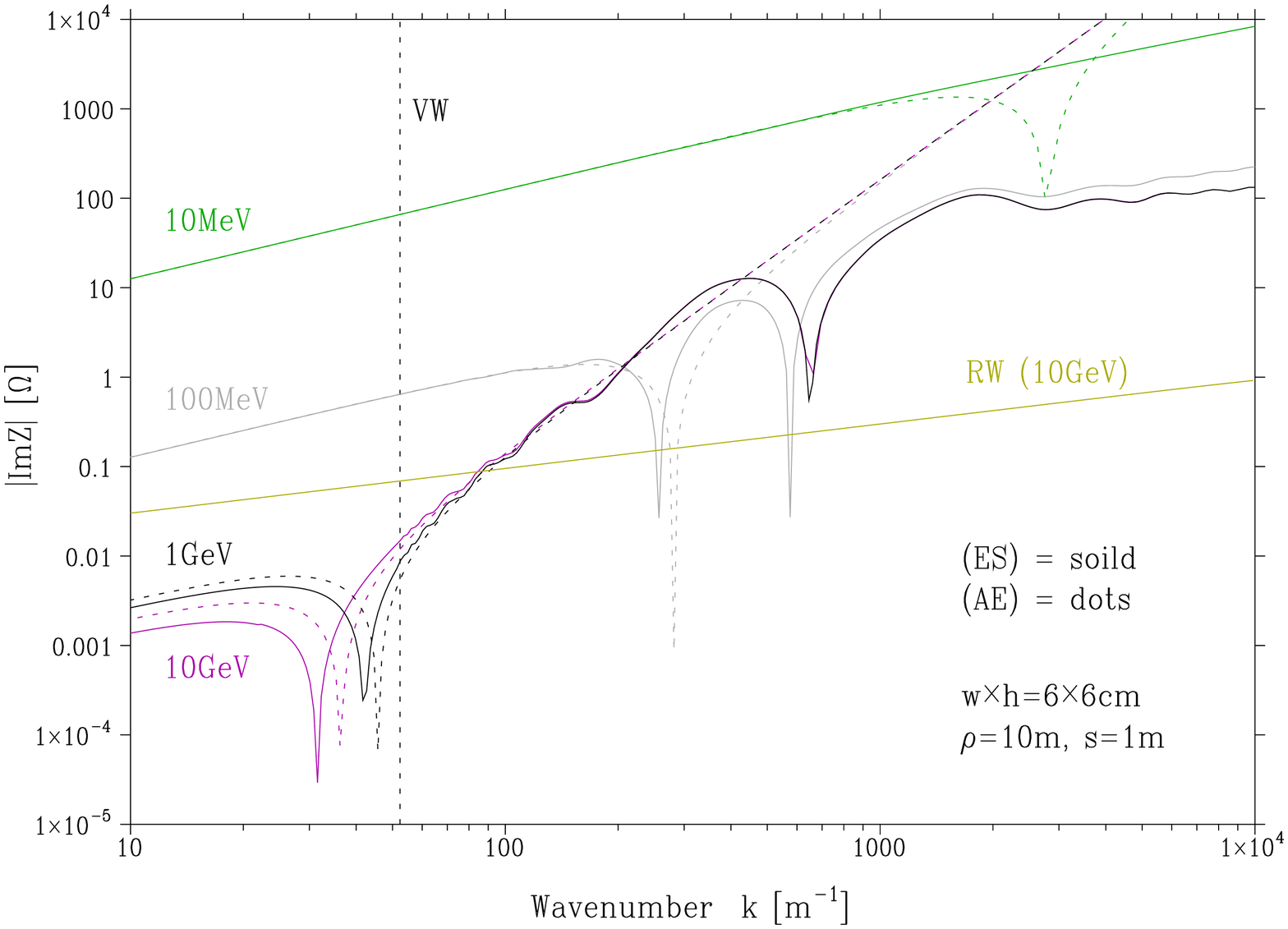}
    \caption[Energy dependence of the longitudinal impedance of CSR]
    {\small 
    Absolute value of the real part (left) and imaginary part (right) of $Z$
    which is the longitudinal impedance of CSR (SR) for various beam energies.
    The horizontal axis is the wavenumber $k$.
    The vertical black dotted line (VW) is the fundamental vertical wavenumber $k_y^1$ 
    given by Eq.(\ref{eq:kth}).
    The solid curves show $|\Re Z|$ and $|\Im Z|$ computed using the exact solution (ES) of
    the transient field of CSR in the perfectly conducting pipe.
    The blue dotted line (SF: steady, free space; $\gam=\infty$) in the left figure is
    the same as that in Fig.\ref{fig:logZ}.
    The green dashed curve (SR) in the left figure shows $\Re Z=s\Re\cZ$ using
    Eq.(\ref{eq:kc}) for $E=10{\rm MeV}$, which is for the steady field of
    synchrotron radiation in free space.
    The vertical green dotted line in the left figure is the critical wavenumber
    $k_c$ for $E=10{\rm MeV}$.
    The dotted curves (AE) in the right figure show $\Im Z=s\Im\cZ$ using
    Eq.(\ref{eq:cZ_AE}) which is the asymptotic expression of $\cZ$ of
    the steady field of CSR in the perfectly conducting pipe for $k\to0$.
    The parameters are the same as Fig.\ref{fig:logZ}:
    $\rho=10{\rm m}$, $w=h=6{\rm cm}$ and $s=1$ except for
    the beam energy (electron): $E=10{\rm MeV}$ (green), $100{\rm MeV}$ (gray),
    $1{\rm GeV}$ (black) and $10{\rm GeV}$ (magenta, only $\Im Z$).
    The solid gray (100MeV) and black (1GeV) curves of $\Re Z$ almost overlap with each other
    over the full range of $k$ in the left figure; the gray $\Re Z$ curve is invisible.
    The solid green (10MeV) curve of $\Re Z$ deviates from the black curve
    for $k\gtrsim 100{\rm m}^{-1}$ in the left figure.
    The solid black (1GeV) and magenta (10GeV) curves of $\Im Z$ almost overlap with
    each other for $k\gtrsim 200{\rm m}^{-1}$ in the right figure.
    When $E\gtrsim 5{\rm GeV}$, $\Im Z$ hardly depends on $E$ over the full range of $k$ for
    this set of parameters.
    The yellow-brown curve (RW) is the resistive wall impedance of
    the straight square pipe having the same size $(w,h,s)$ as the curved pipe,
    which is plotted to show the order of magnitude of $Z$.
    We assumed $E=10{\rm GeV}$ and $\sig_c=6\times 10^7/\Omg {\rm m}$ (copper) in
    computing RW by numerically solving Eq.(\ref{eq:PE_Exy_std}) for $\rho=\infty$.
     }
    \label{fig:logZ_Energy}
  \end{center}
\end{figure}

We examine how the impedance of CSR depends on the beam energy $E~(=\gamma mc^2)$
in order to confirm that the exact solution (ES) is correct in the low frequency range
$k\leq O(k_y^1)$ in which the paraxial approximation is not applicable 
as described in Eq.(\ref{eq:PA_condh}).
The solid curves in Fig.\ref{fig:logZ_Energy} show the real and imaginary parts of $Z$
for various $E$ computed using ES, which is the impedance of the transient field of CSR 
emitted in a perfectly conducting pipe.
For $E=10{\rm MeV}$, Schwinger's formula (\ref{eq:kc}) for a length
$s=1{\rm m}$ is plotted with the green dashed curve (SR) in the left figure of
Fig.\ref{fig:logZ_Energy}.
When $k\ll k_c$, the real part $\Re Z$ of CSR (SR) does not depend on $\gamma$ as expected.
Due to this, the curve of $\Re Z$ for $E=1{\rm GeV}$ (black) overlaps perfectly with
that for $E=100{\rm MeV}$ (gray, invisible) in the left figure of Fig.\ref{fig:logZ_Energy}.
On the other hand, unless
\begin{align}
  \gamma
  \gg k_y^1\rho ,
  \label{eq:gam_cond}
\end{align}
the imaginary part $\Im Z$ depends on $\gamma$ since $\Im Z$ represents
the space charge field of the beam moving on the curved trajectory in the bend.
When $\gamma\gtrsim 10^4$ ($E\gtrsim 5{\rm GeV}$) for the set of parameters used in
Fig.\ref{fig:logZ_Energy}, $\Im Z$ hardly depends on $\gamma$ over the full range of $k$.
Eq.(\ref{eq:gam_cond}) is gotten from the leading order term of Eq.(\ref{eq:cZ_AE})
which is plotted with the dotted curves (AE: asymptotic expression of $Z$ for $k\to0$)
in the right figure of Fig.\ref{fig:logZ_Energy}.

In section \ref{sec:SR_Z} we showed that $\Re Z$
computed using the exact solution (ES) tends to agree with Eq.(\ref{eq:kc}) for 
the same length $s$ as $\Re Z=s\Re\cZ$ in the limit of $k\to\infty$ and $s\to\infty$.
On the other hand, in the limit of $k\to0$ and $s\to\infty$, $\Im Z$ computed using ES must
agree with AE given by Eq.(\ref{eq:cZ_AE}) for the same length $s$ as $\Im Z=s\Im \cZ$ 
since the low frequency limit ($k\to0$) means that
the field does not oscillate, which is equivalent to steady state.
As shown in the right figure of Fig.\ref{fig:logZ_Energy},
when $E=10{\rm MeV}$ (green) and $100{\rm MeV}$ (gray), $\Im Z$ of ES (solid curves) agrees 
with AE (dotted curves) in the low frequency range $k\leq O(k_y^1)$.
These numerical results show that the exact solution (ES) is also correct in the range
$k\leq O(k_y^1)$.
But, assuming a higher energy beam $E=1{\rm GeV}$ (black) or $10{\rm GeV}$ (magenta)
moving in the square pipe $w=h=6{\rm cm}$, $\Im Z$ of ES differs somewhat from AE in
$k\leq O(k_y^1)$.

In general, $\Im Z$ of ES differs from AE for a nonzero $k$ and a finite $s$
since ES and AE are respectively the impedances of the transient and steady fields of CSR.
Although we examined $\Im Z$ of ES in the square pipe ($w=h=6{\rm cm}$) for a longer bend
in order that the field gets almost steady,
still ES does not agree perfectly with AE in the range $k\leq O(k_y^1)$
when the beam energy is high as $E\gtrsim 1{\rm GeV}$.
We guess that this small disagreement between ES and AE in $\Im Z$ for $k\leq O(k_y^1)$ and 
$E\gtrsim 1{\rm GeV}$ is probably caused by the truncation error of the higher order terms
$O(k^5)$ in AE given by Eq.(\ref{eq:cZ_AE}),
which may not be negligible for $\gamma\to\infty$.
Conversely, if the beam energy is low such as $\gamma\ll k_y^1\rho$,
the term which is proportional to $k/\gamma^2$ dominates over
the other terms in Eq.(\ref{eq:cZ_AE}) when $k$ is small.
That is, when $\gamma\leq O(10^2)$ and $k\leq O(k_y^1)$ for the set of parameters
used in Fig.\ref{fig:logZ_Energy}, the higher order terms $O(k^5)$ may be negligible for 
the $k/\gamma^2$ term in the asymptotic expression of $\cZ$ of the steady field for $k\to0$.
This may be the reason that $s\Im\cZ$ of AE agrees with $\Im Z$ of ES
when the beam energy is low such as $E=10{\rm MeV}$ and $100{\rm MeV}$ as in
the right figure of Fig.\ref{fig:logZ_Energy}.

\begin{figure}[h]
\begin{center}
  \includegraphics[scale=0.32,clip]{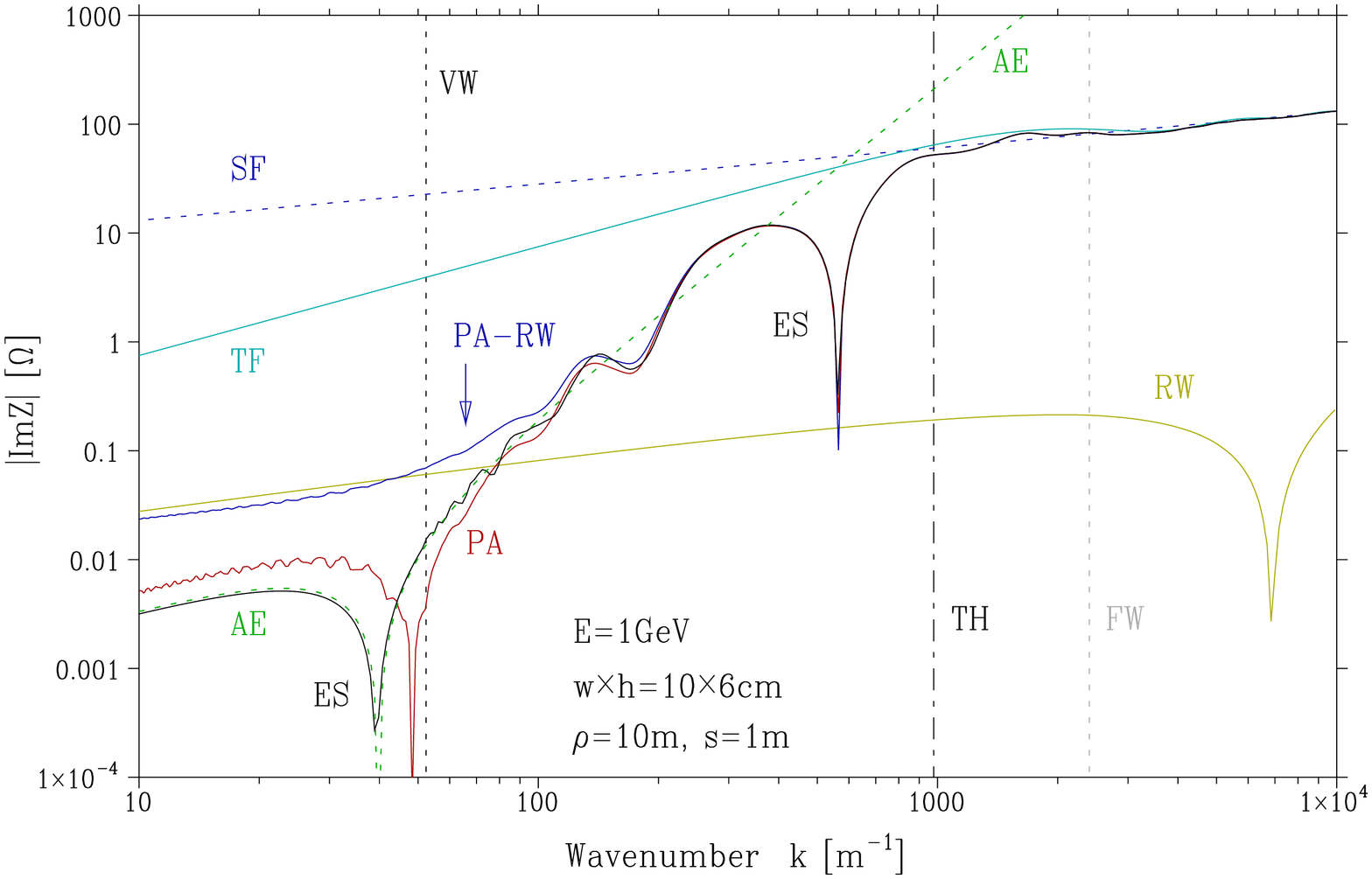}
  \includegraphics[scale=0.32,clip]{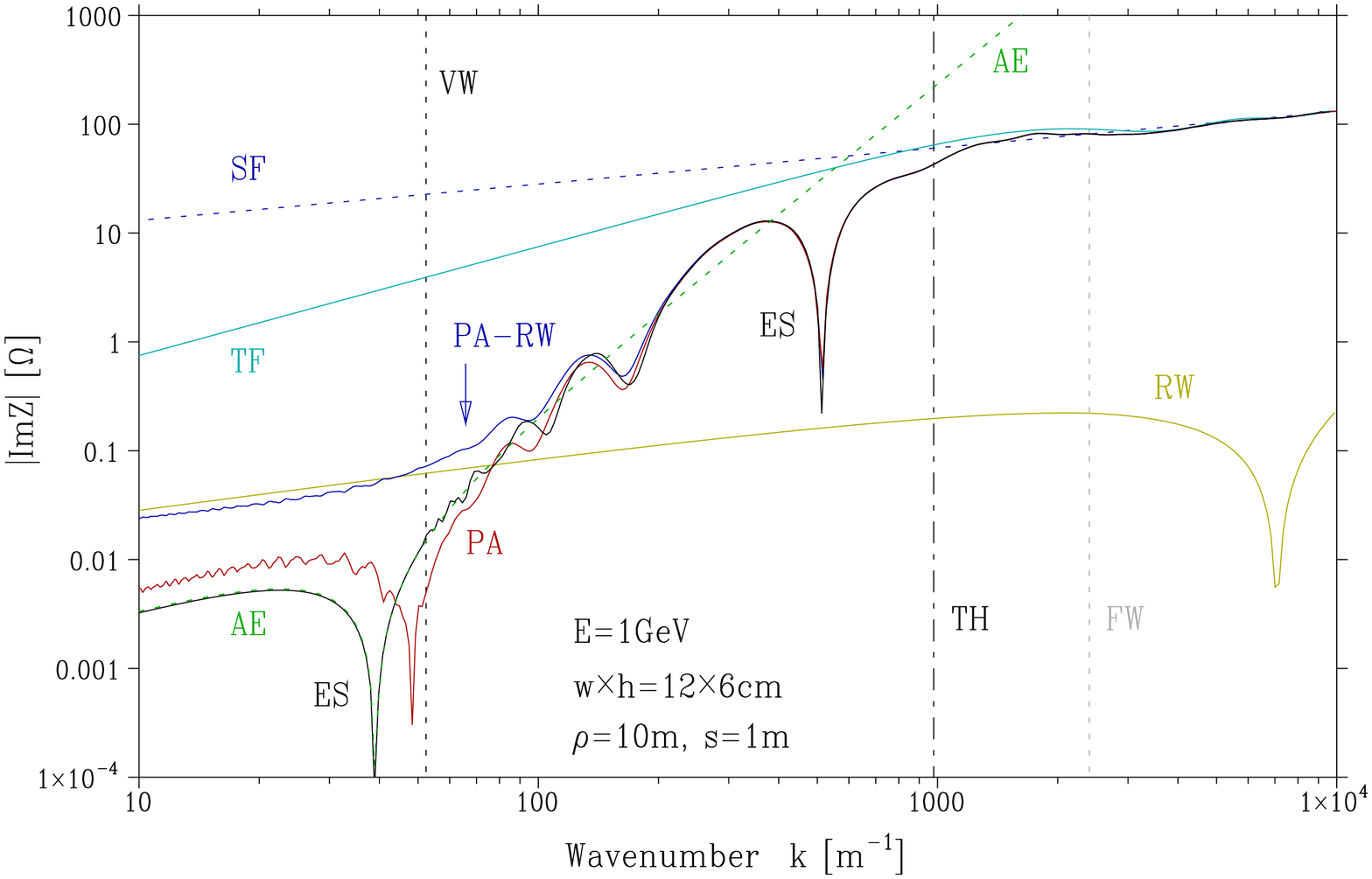}
  \caption[Imaginary impedance of CSR in a rectangular pipe]{ \small
    Absolute value of the imaginary part of $Z\,[\Omg]$
    which represents the longitudinal impedances of CSR and RW (resistive wall).
    The horizontal axis is the wavenumber $k$.
    The parameters are common to Fig.\ref{fig:logZ}
    excluding the width of the pipe $w=10{\rm cm}$ (left) and $12{\rm cm}$ (right)
    in calculating ES, AE, PA, PA-RW and RW listed in p.\pageref{eq:Z_bend}-\pageref{eq:kth}.
    The black vertical dotted line (VW) is the fundamental vertical wavenumber $k_y^1$.
    The vertical dash-dotted line (TH) is the shielding threshold $k_{\rm th}$.
    The gray vertical dotted line (FW) shows $k_{\rm f}$ given by Eq.(\ref{eq:kth}),
    which is the formation wavenumber of CSR in the bend.
    The black and red curves (ES and PA) show $\Im Z$ of the transient field of CSR
    emitted in the perfectly conducting rectangular pipe.
    The black curve (ES) shows $\Im Z$ computed using the exact solution.
    The red curve (PA) shows $\Im Z$ computed using the numerical solution of
    Eqs.(\ref{eq:PE_Exy}-\ref{eq:PE_Es}).
    The blue curve (PA-RW) shows $\Im Z$ of
    the transient field of CSR in a copper pipe, computed by solving
    Eqs.(\ref{eq:PE_Exy}-\ref{eq:PE_Es}) imposing the resistive boundary condition.
    The yellow-brown curve (RW) shows the resistive wall impedance of the straight copper 
    pipe which has the same dimension $(w,h,s)$ as the curved rectangular pipe.
    The cyan curve (TF) and blue dotted line (SF) show $\Im Z$ of
    the transient and steady fields of CSR in free space.
    The green dotted curve (AE) shows Eq.(\ref{eq:cZ_AE})
    which is the asymptotic expression of $\cZ$ of the steady field of CSR in
    the perfectly conducting rectangular pipe for $k\to0$.
    When $k<k_y^1$, AE almost overlaps with ES.
    The parameters are given as follows:
    $\rho=10{\rm m}$, $w=(10,12)\,{\rm cm}$, $h=6{\rm cm}$, $s=1{\rm m}$,
    $\sig_y=20\mu{\rm m}$, $E=1{\rm GeV}$ and $\sig_c=6\times 10^7/\Omg {\rm m}$ (copper).
     }
  \label{fig:logZ_w12cm}
\end{center}
\end{figure}

Assuming a wider pipe $w=10{\rm cm}$ or 12cm ($>h=6{\rm cm}$) as shown in
Fig.\ref{fig:logZ_w12cm}, when $E=1{\rm GeV}$,
$\Im Z$ of ES (black) tends to agree very well with AE (green dots) for $k\to0$
despite the fact that ES has a small difference from AE under the assumption of
the square pipe ($w=h=6{\rm cm}$) as shown in the right figure of
Fig.\ref{fig:logZ_Energy} with the black solid and dotted curves ($E=1{\rm GeV}$).
As far as we know at present, no evidence shows that the exact solution (ES) is incorrect 
for $k\leq O(k_y^1)$.
As described in the previous paragraph, we think that AE given by Eq.(\ref{eq:cZ_AE}) may 
not be accurate enough in using for the detailed comparison with ES not only in
the very low frequency range $k\rho<1$ but also in a wider range $k\leq O(k_y^1)$
for an arbitrary beam energy including the ultrarelativistic limit.
As explained above Eq.(87) in \cite{agoh}, Eq.(\ref{eq:cZ_AE}) is the asymptotic expression 
of $\cZ$ which we got by rearranging Eq.(4.23) in \cite{ng_warnock},
neglecting the small terms under the assumption $x_b=-x_a=w/2$.
In order to compare $\Im\cZ$ of AE accurately with ES in the low frequency range
$k\leq O(k_y^1)$, we need to review Eq.(\ref{eq:cZ_AE}) by deriving
the asymptotic expression of Eq.(\ref{eq:cZ_thin}) for $k\to0$,
taking into account the higher order terms $O(k^{\ell\geq 5})$.
In deriving the asymptotic expression of Eq.(\ref{eq:cZ_thin}) for $k\to0$,
we need to figure out the correct way to expand the Bessel functions, \ie,
Debye's asymptotic expansion or the uniform asymptotic expansion or something else.
Since this problem deviates somewhat from the main theme of the present paper,
we would like to discuss it in detail somewhere else in the future.

\clearpage

%

\subsection{Applicability of the paraxial approximation for low energy particles}
\label{sec:PA_applicability}

\begin{figure}[h]
  \begin{center}
    \includegraphics[scale=0.32,clip]{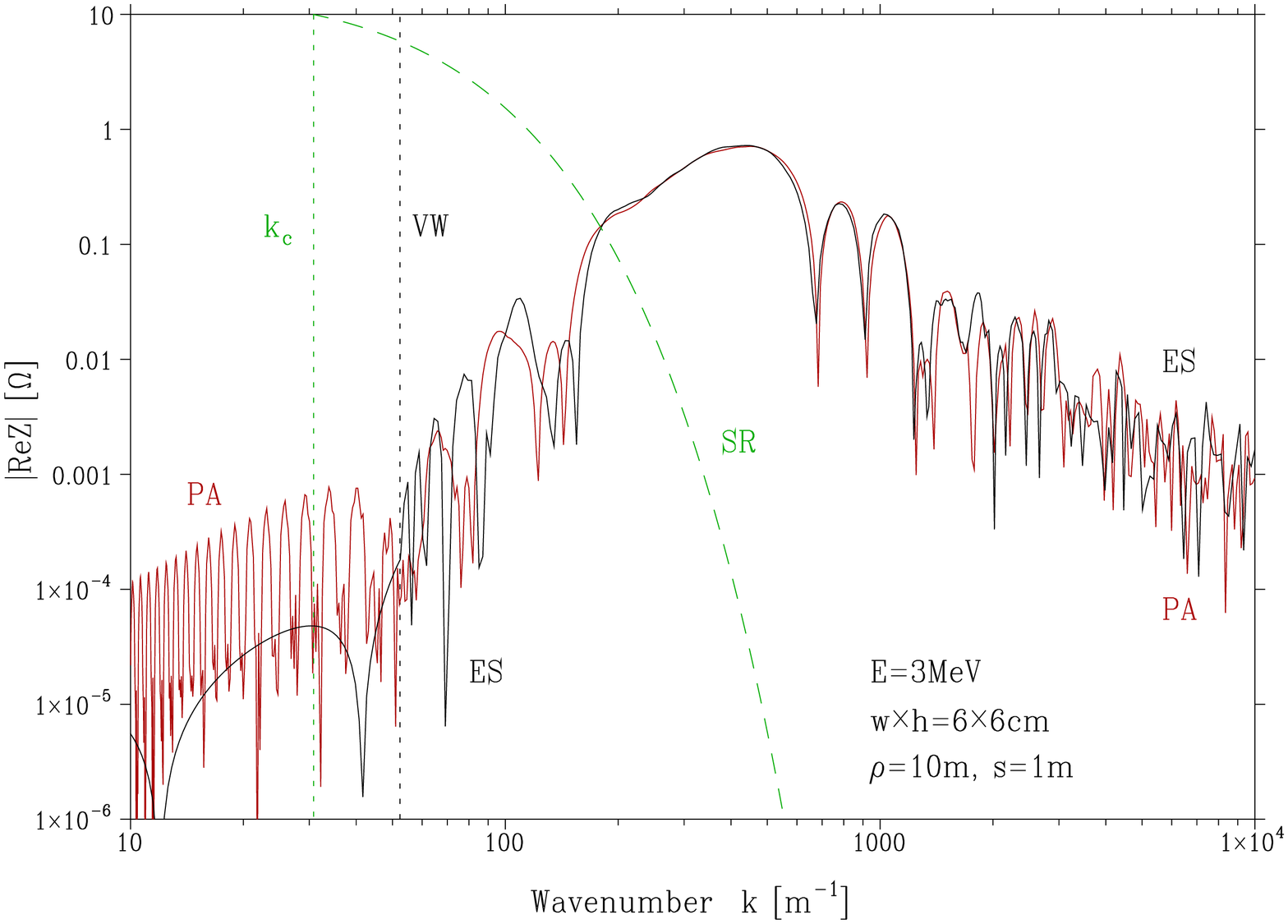}
    \includegraphics[scale=0.32,clip]{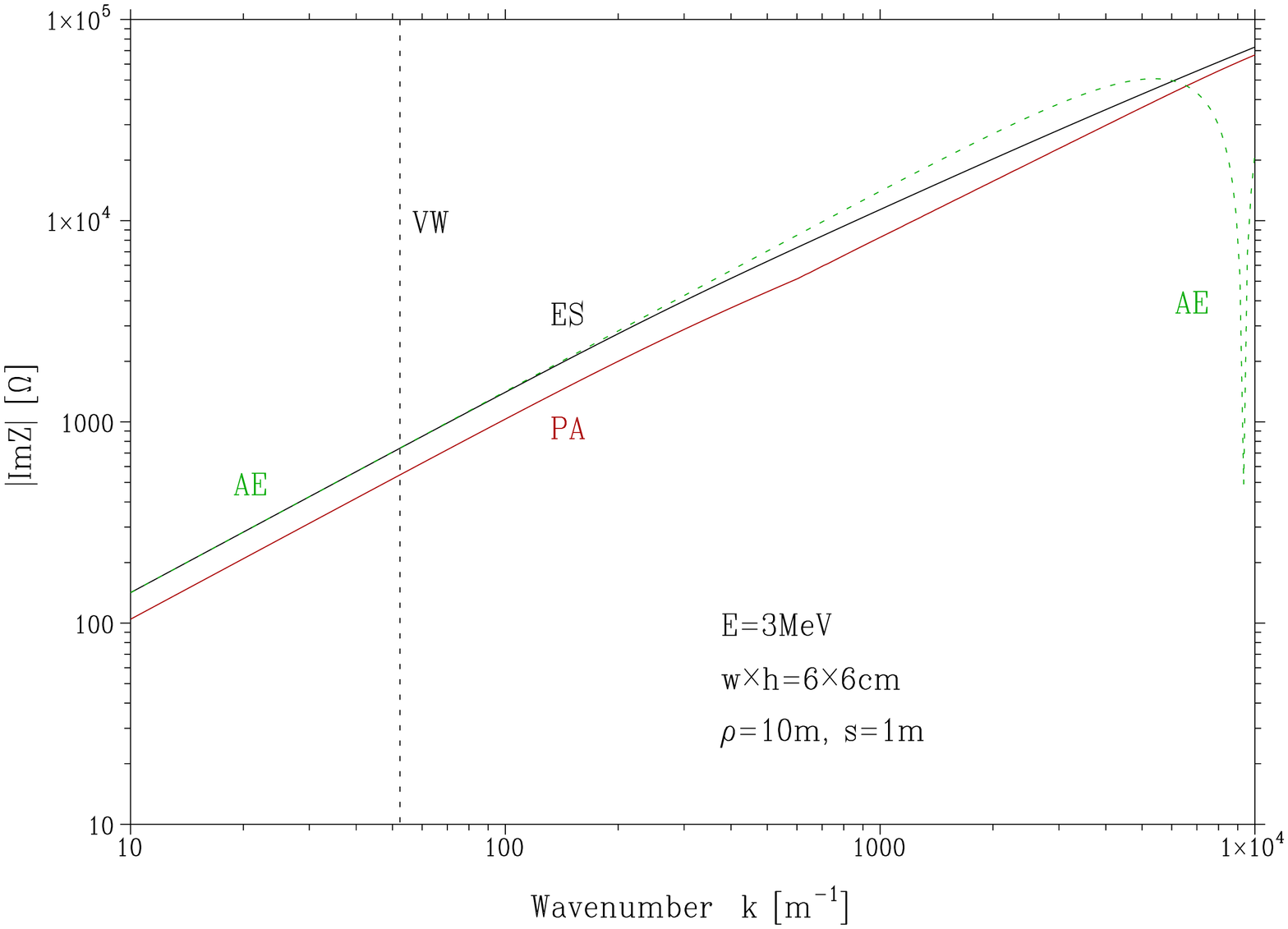}
      \vspace{1mm}
      \\
    \includegraphics[scale=0.32,clip]{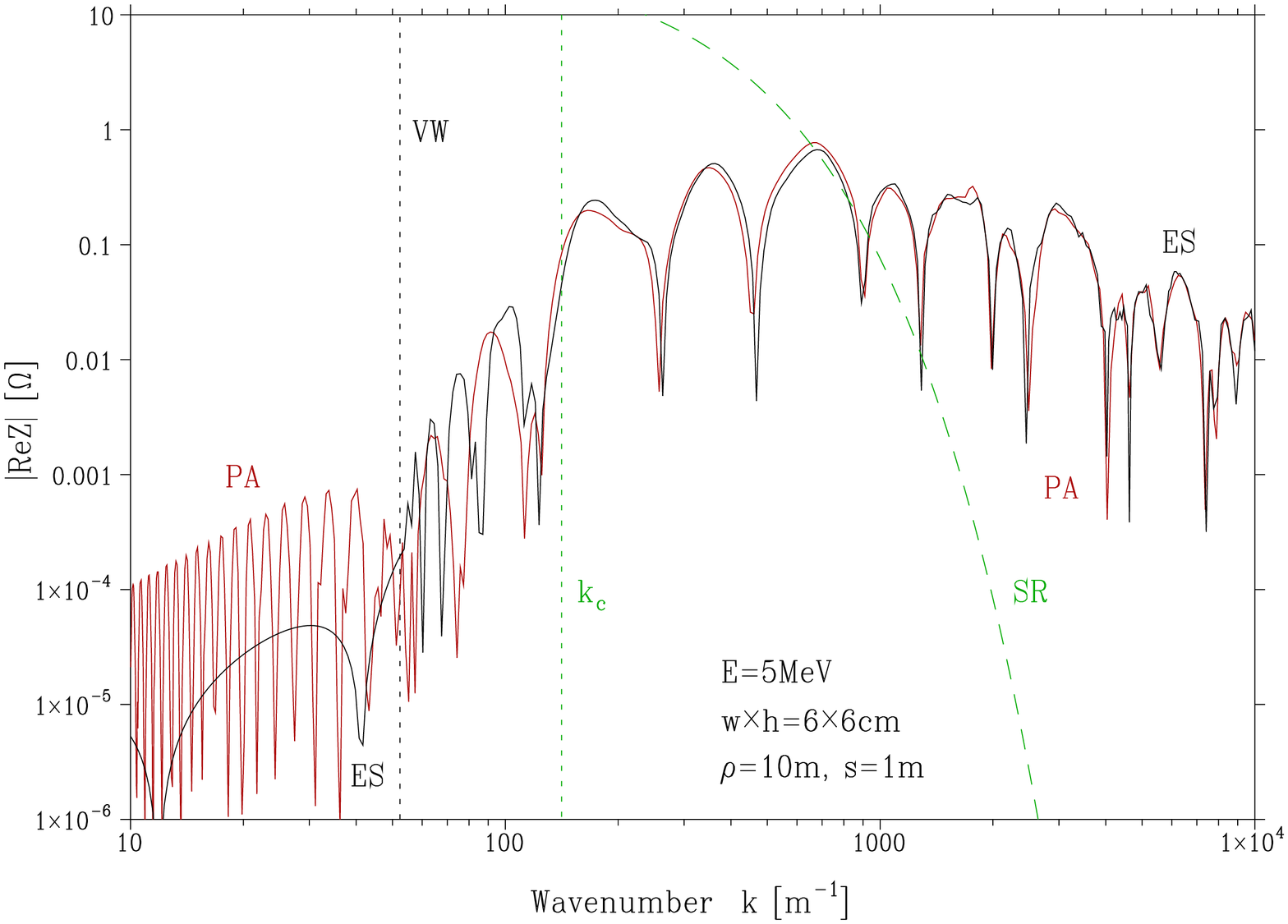}
    \includegraphics[scale=0.32,clip]{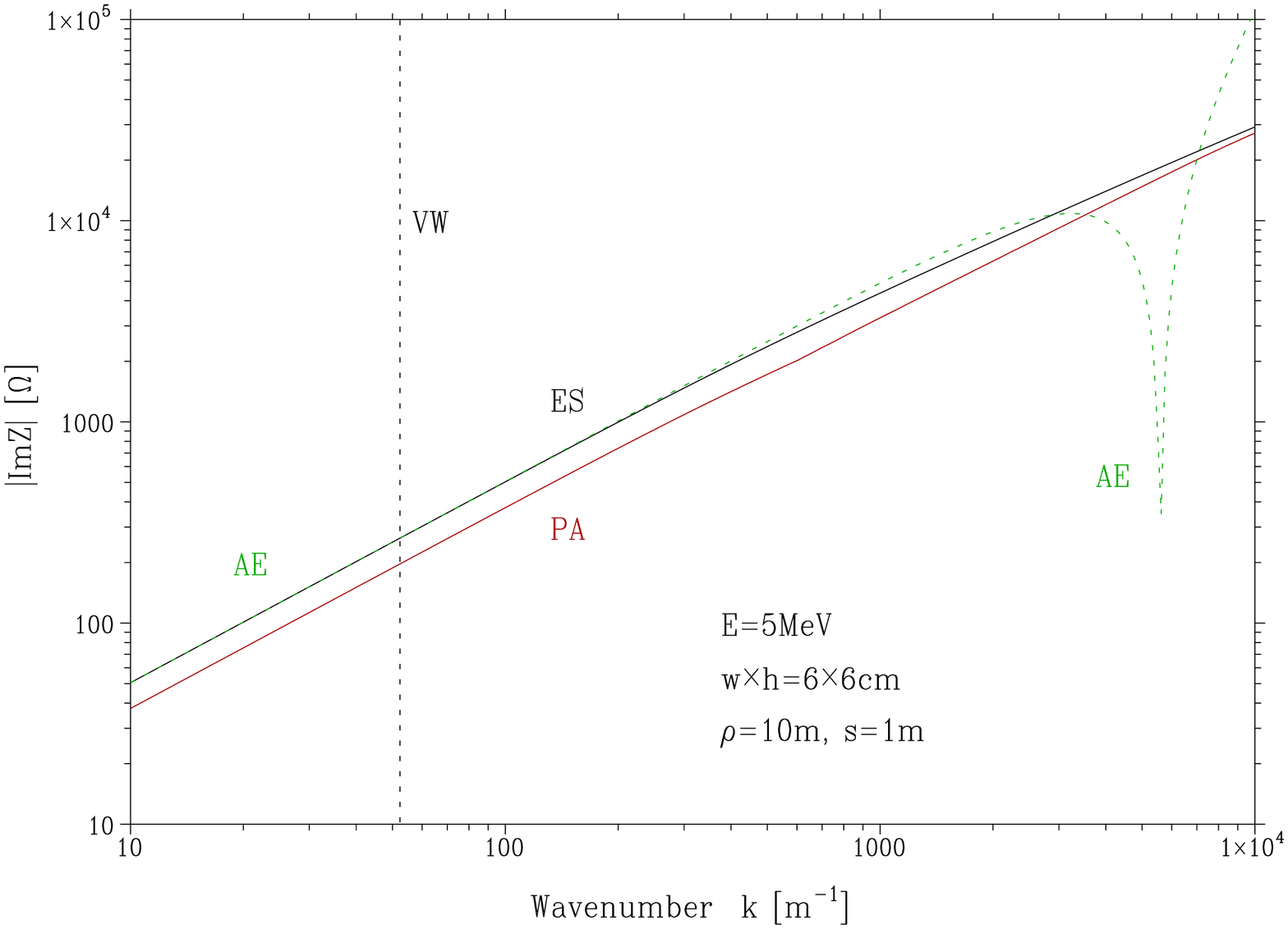}
      \vspace{1mm}
      \\
    \includegraphics[scale=0.32,clip]{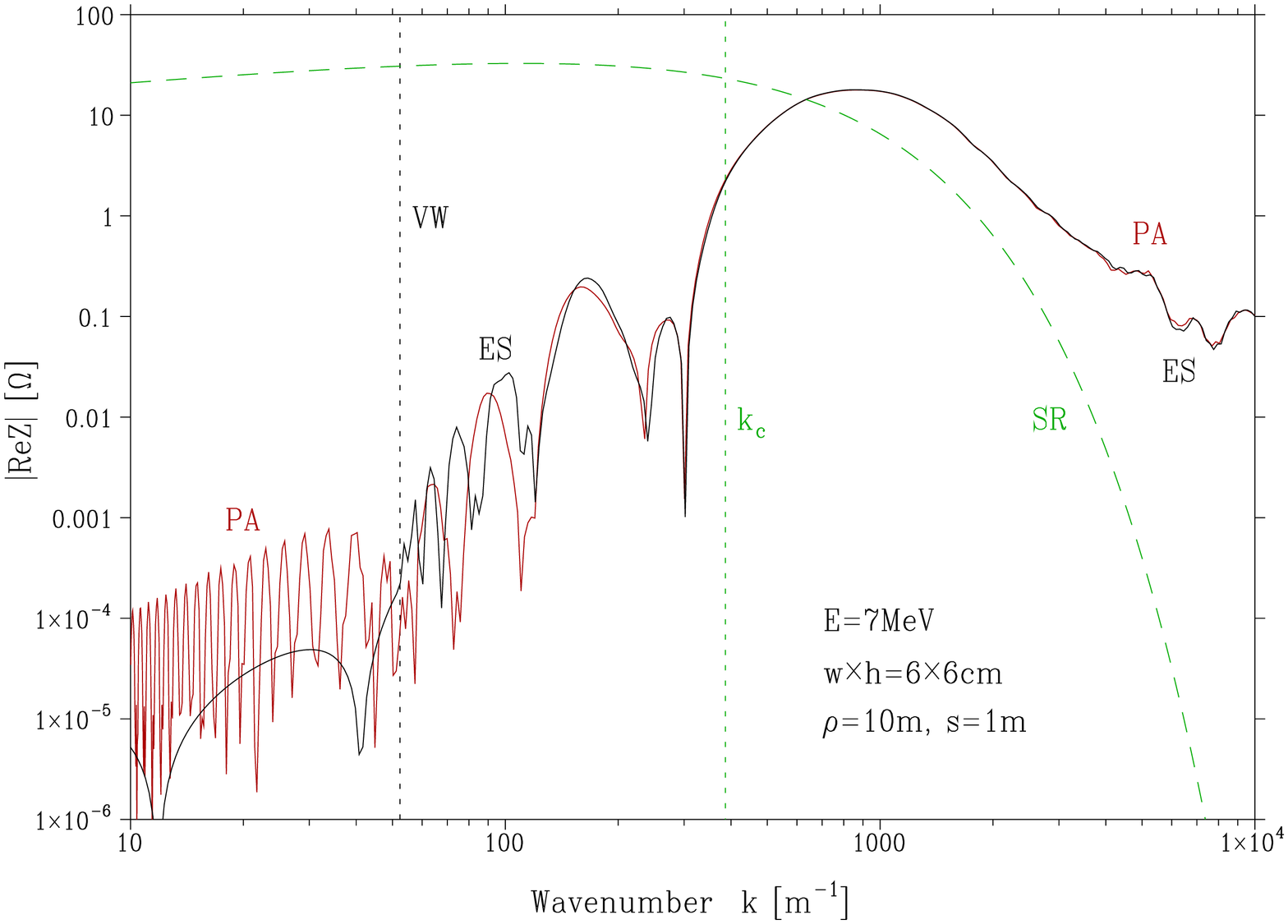}
    \includegraphics[scale=0.32,clip]{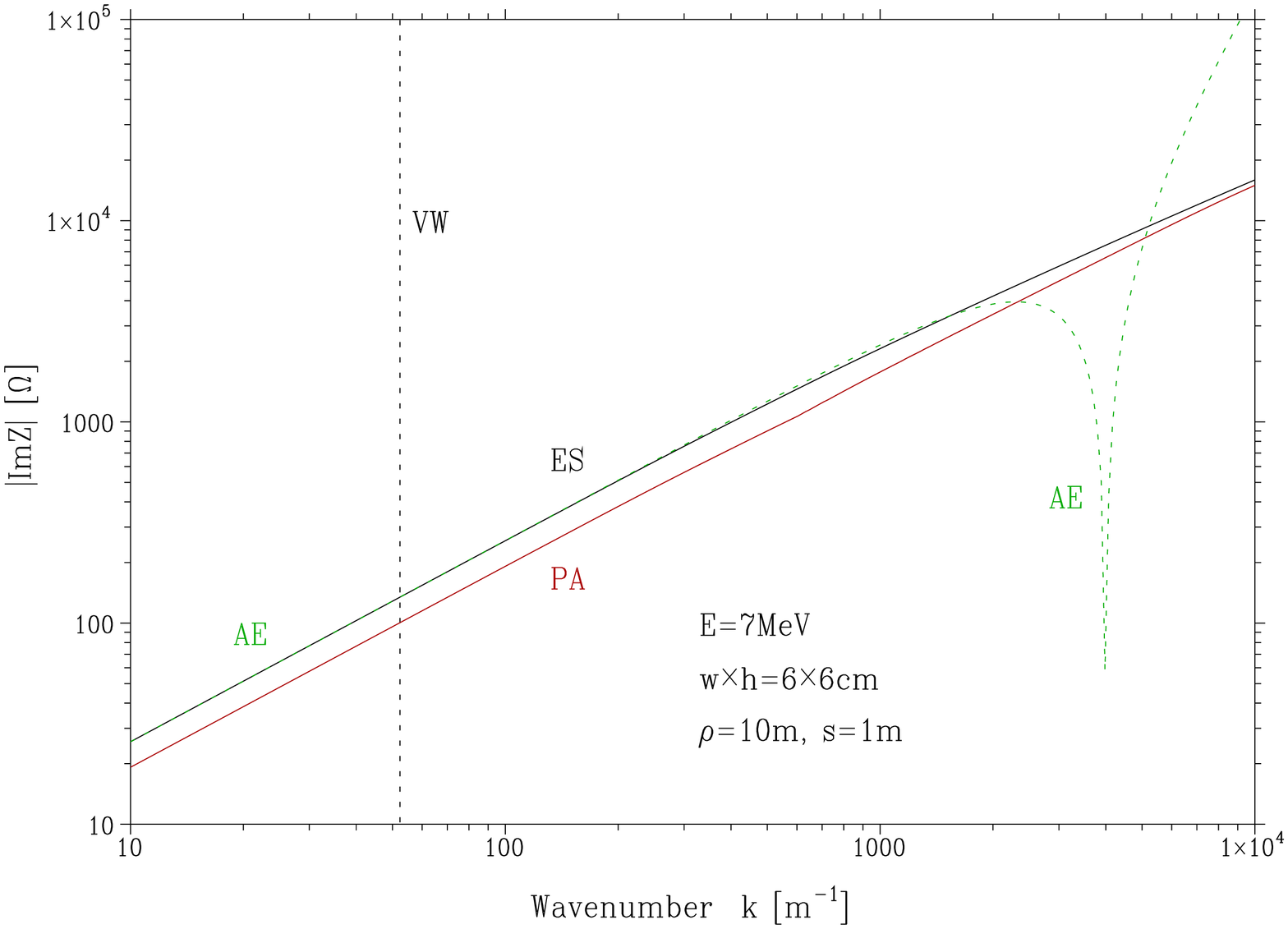}
    \caption[Longitudinal impedance of CSR ($E=3, 5, 7{\rm MeV}$)]
    {\small 
    Absolute value of the real part (left) and imaginary part (right) of $Z$
    which is the longitudinal impedance of a transient field in
    a perfectly conducting pipe.
    The horizontal axis is the wavenumber $k$.
    We assume an electron beam (\ref{eq:thin_bunch}) having an energy
    $E=3{\rm MeV}$ (upper), $5{\rm MeV}$ (middle) and $7{\rm MeV}$ (lower).
    The nomenclature and colors of $Z$ are common to Fig.\ref{fig:logZ}:
    ES (exact solution, black), PA (paraxial approximation, red) and
    AE (asymptotic expression of $\Im Z$ for $k\to0$, green dots).
    ES and PA denote $Z$ of the transient field in the bend,
    which is shielded by a perfectly conducting square curved pipe.
    AE is gotten using Eq.(\ref{eq:cZ_AE})
    which tends to agree with ES for $k\to0$ in the right figures.
    The vertical black dotted line (VW) is the fundamental vertical wavenumber $k_y^1$ 
    given by Eq.(\ref{eq:kth}).
    The green dashed curve (SR) in the left figures shows $\Re Z$ ($=s\Re\cZ$) using
    Eq.(\ref{eq:kc})
    which is the power spectrum of the steady field of synchrotron radiation in free space.
    The vertical green dotted line in the left figures is the critical wavenumber $k_c$.
    Excluding the beam energy $E$, the parameters are common to Fig.\ref{fig:logZ}:
    $\rho=10{\rm m}$, $x_b=-x_a=3{\rm cm}$, $w=h=6{\rm cm}$ and $s=1{\rm m}$.
    For this set of parameters, the condition to emit steady synchrotron radiation in
    the curved pipe is $E>E_0=6.61{\rm MeV}$ ($\gamma>12.9$) according to
    Eq.(\ref{eq:sr_cond}).
    The energy range $E=3\sim 7{\rm MeV}$ means $E\lesssim E_0$, 
    \ie, $E$ does not satisfy Eq.(\ref{eq:sr_cond}) or barely does.
    $\Re Z$ of ES (black) is fluctuating due to a numerical error in the left figures
    when $k$ is large such as about $k\gtrsim 1000{\rm m}^{-1}$ (upper),
    $k\gtrsim 2000{\rm m}^{-1}$ (middle) and $k\gtrsim 4000{\rm m}^{-1}$ (lower).
     }
    \label{fig:logZ_7MeV}
  \end{center}
\end{figure}

\begin{figure}[h]
  \begin{center}
    \includegraphics[scale=0.32,clip]{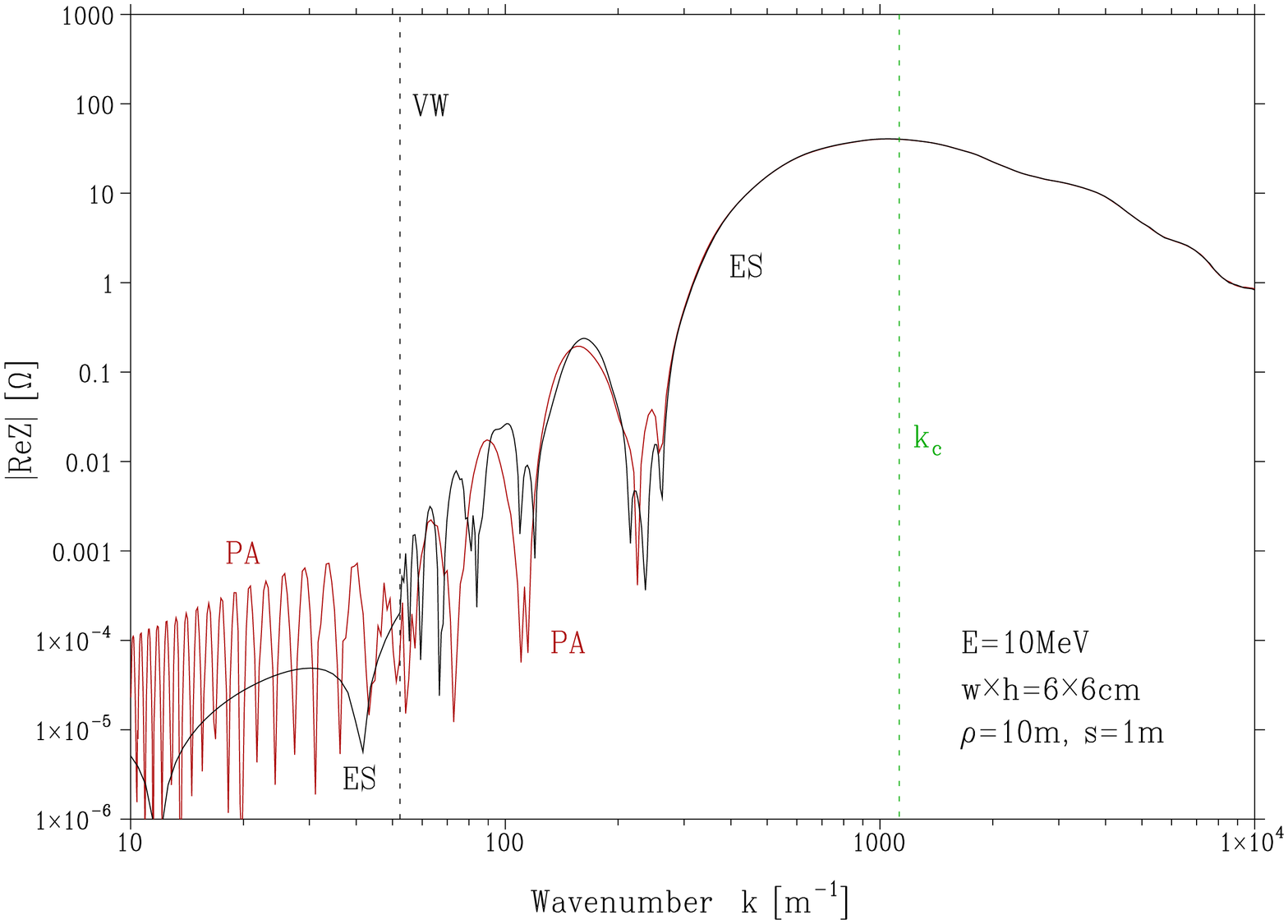}
    \includegraphics[scale=0.32,clip]{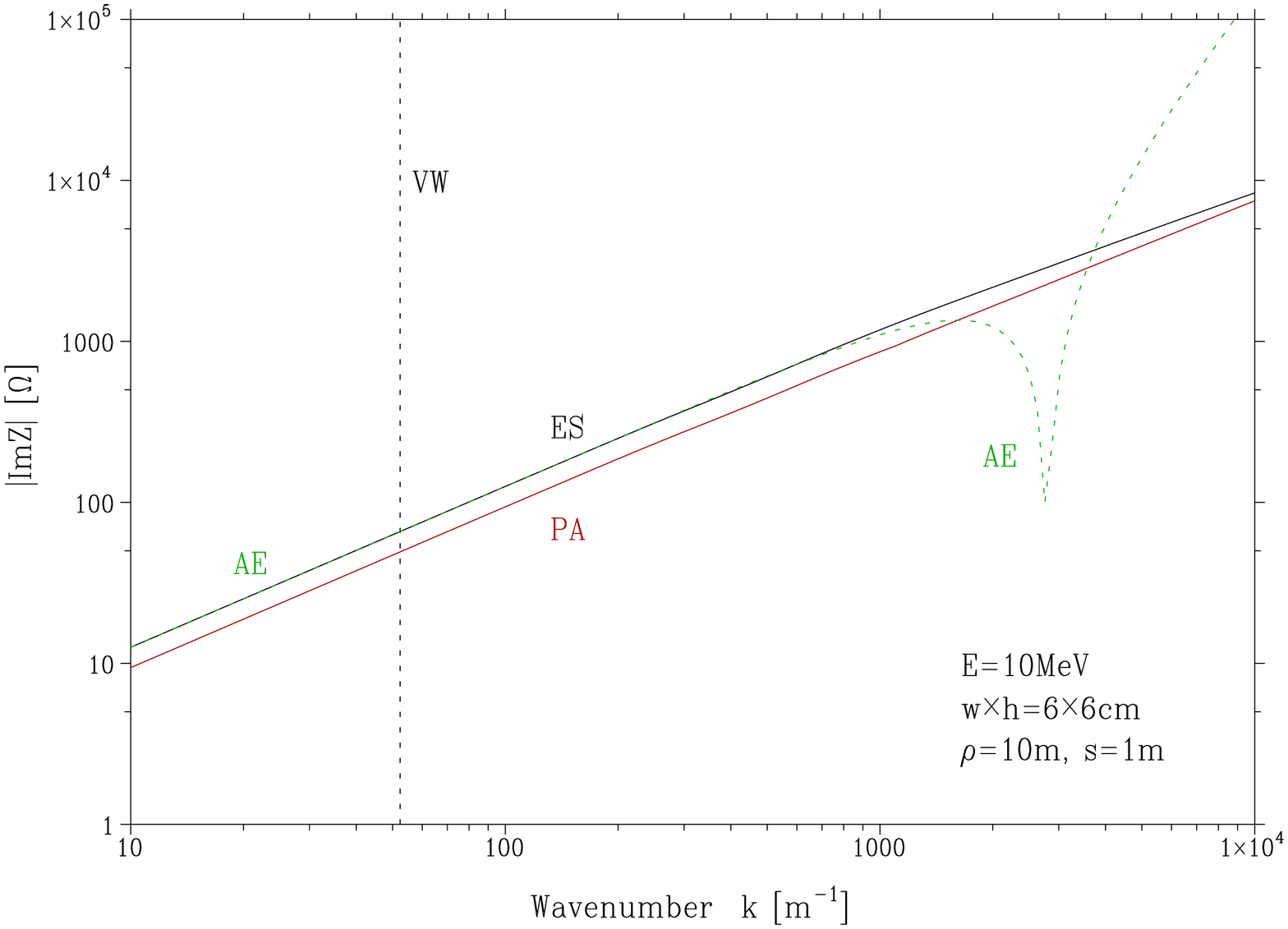}
      \vspace{1mm}
      \\
    \includegraphics[scale=0.32,clip]{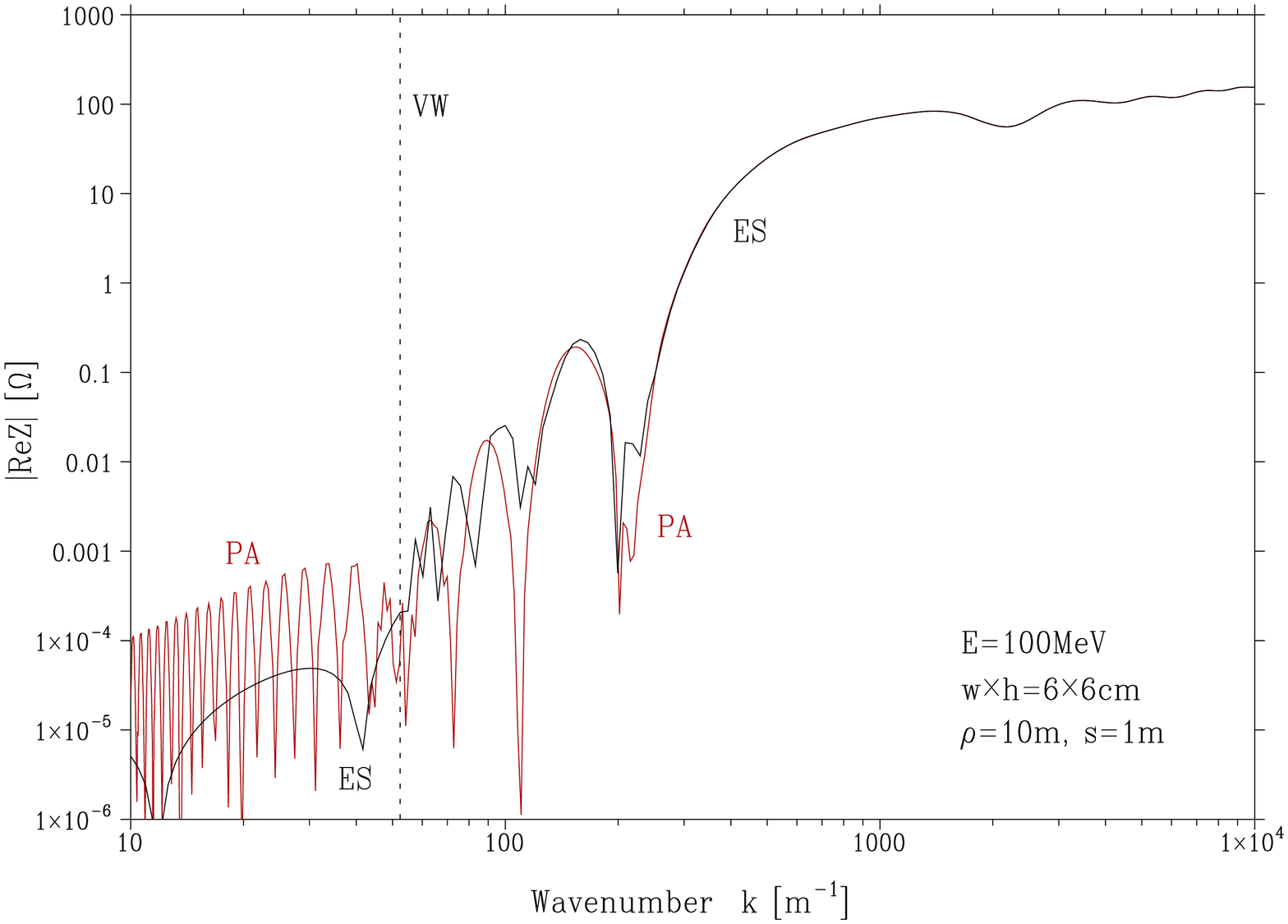}
    \includegraphics[scale=0.32,clip]{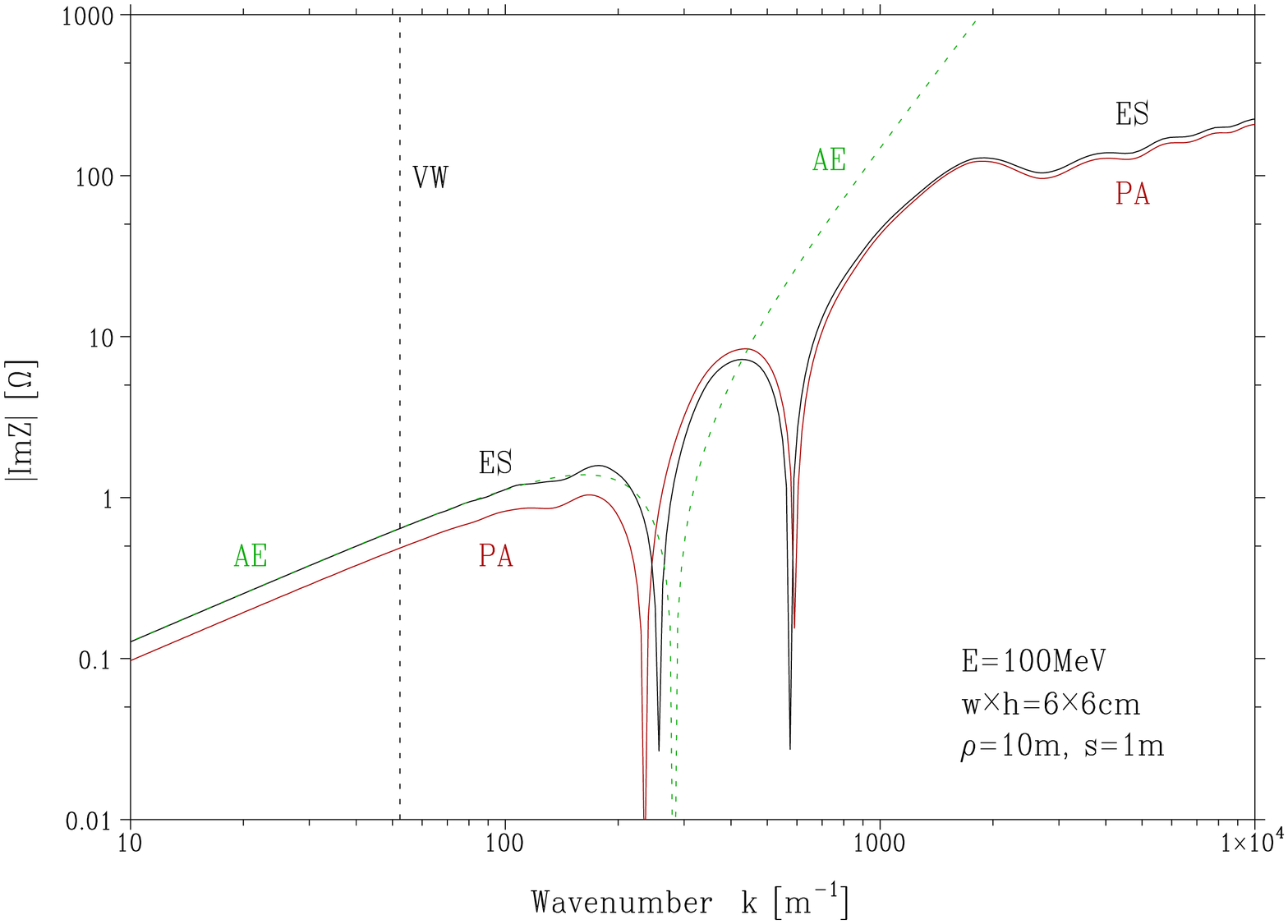}
      \vspace{1mm}
      \\
    \includegraphics[scale=0.32,clip]{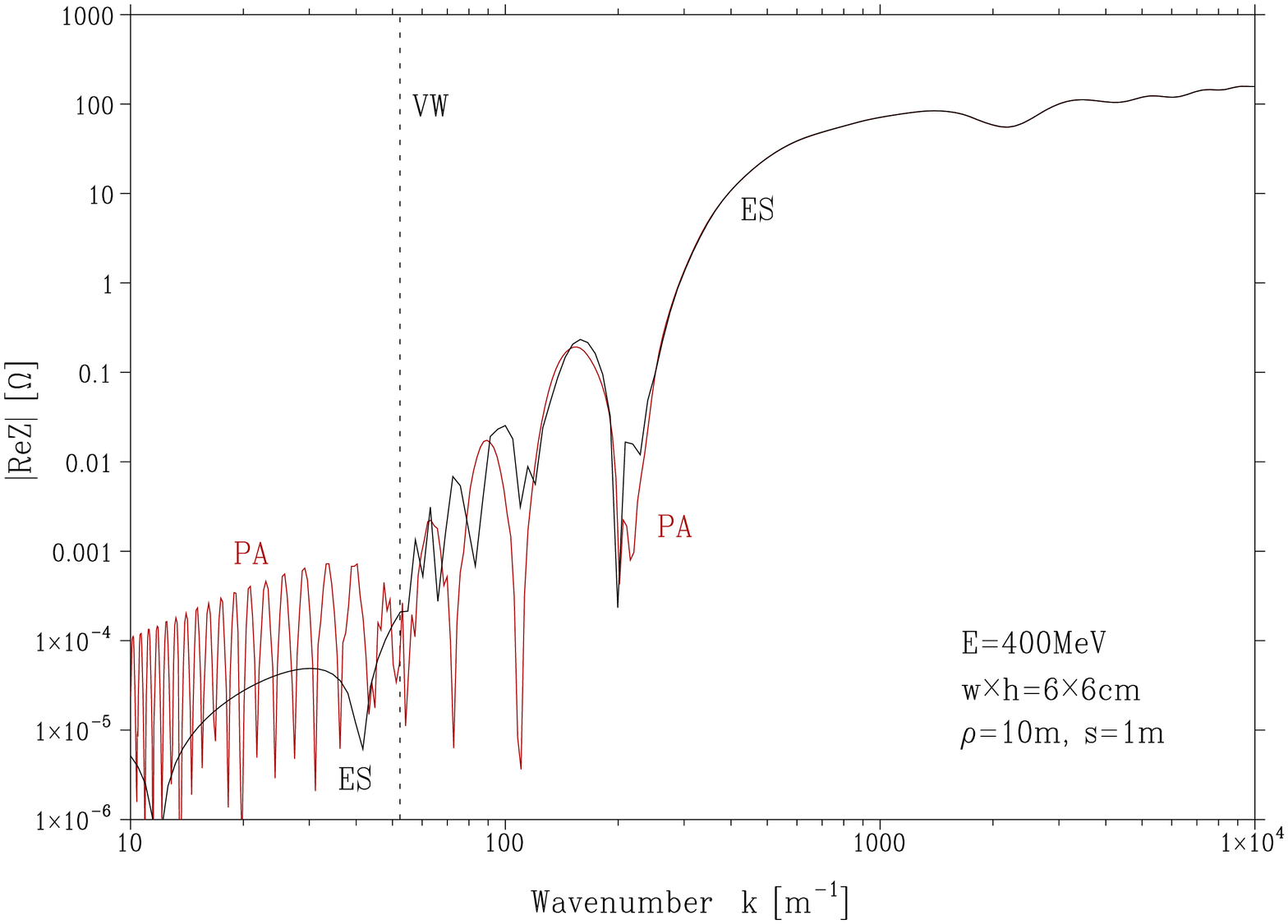}
    \includegraphics[scale=0.32,clip]{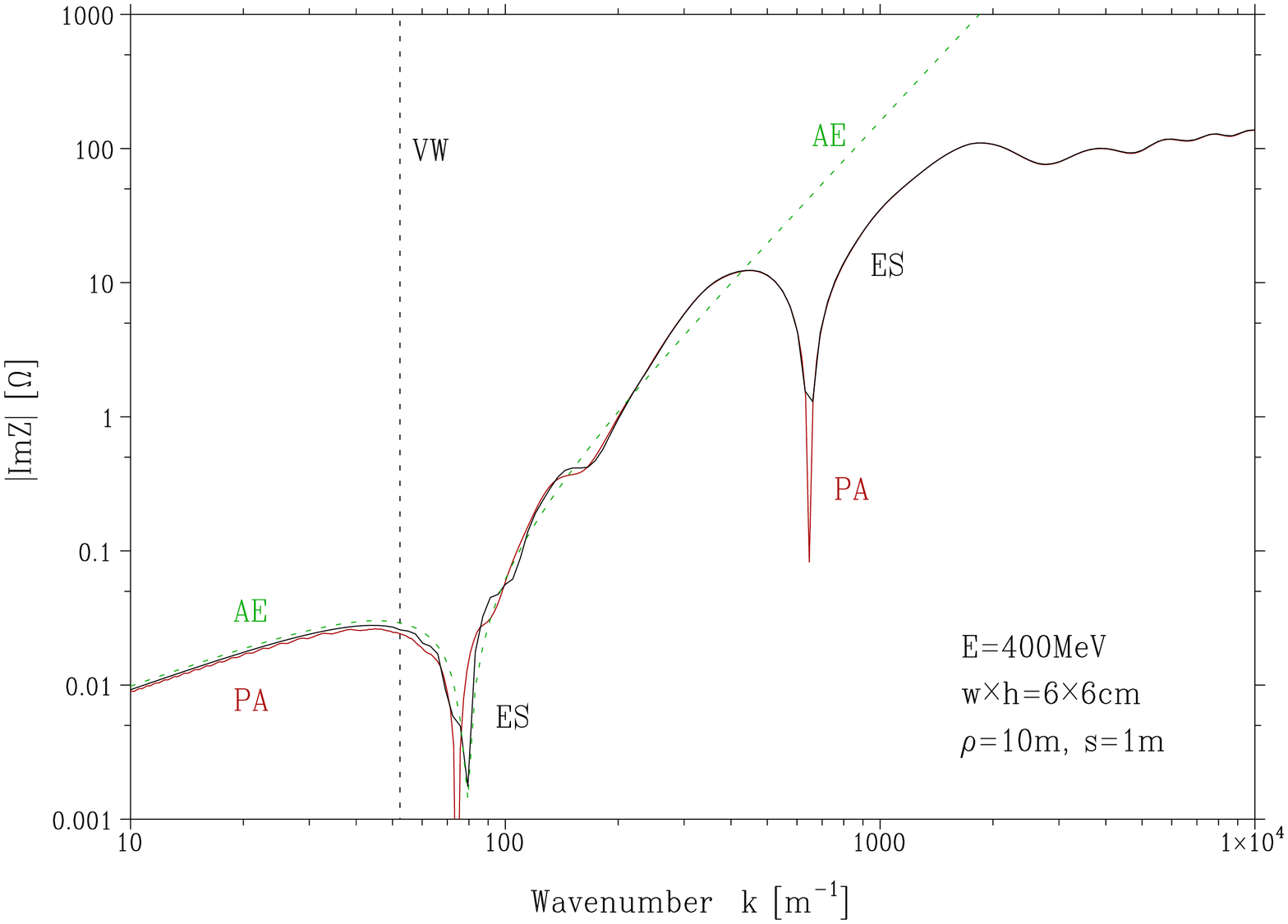}
    \caption[Longitudinal impedance of CSR ($E=10, 100, 400{\rm MeV}$)]
    {\small 
    Absolute value of the real part (left) and imaginary part (right) of $Z$
    which is the longitudinal impedance of a transient field of CSR emitted in
    a perfectly conducting pipe.
    The horizontal axis is the wavenumber $k$.
    We assume an electron beam having an energy
    $E=10{\rm MeV}$ (upper), $100{\rm MeV}$ (middle) and $400{\rm MeV}$ (lower).
    The nomenclature and colors of $Z$ of CSR are common to those in Fig.\ref{fig:logZ}:
    ES (exact solution, black), PA (paraxial approximation, red) and
    AE (asymptotic expression of $\Im Z$ for $k\to0$, green dots).
    The green dotted curve (AE) in the right figures shows $|s\Im\cZ|$ computed using
    Eq.(\ref{eq:cZ_AE}) which is the asymptotic expression of $\cZ$ of
    the steady field of CSR in the perfectly conducting pipe for $k\to0$.
    AE tends to agree with ES for $k\to0$ in the right figures.
    The vertical black dotted line (VW) is the fundamental vertical wavenumber $k_y^1$ 
    given by Eq.(\ref{eq:kth}).
    The vertical green dotted line in the upper left figure is the critical wavenumber
    $k_c$ for $E=10{\rm MeV}$, given by Eq.(\ref{eq:kc}).
    The geometric parameters are the same as Fig.\ref{fig:logZ}: $\rho=10{\rm m}$,
    $w=h=6{\rm cm}$ and $s=1{\rm m}$.
     }
    \label{fig:logZ_ES_PA}
  \end{center}
\end{figure}

We discuss the range of applicability of Eqs.(\ref{eq:PE_Exy}-\ref{eq:PE_Es}) with
respect to the beam energy $E$.
To conclude first, Eqs.(\ref{eq:PE_Exy}-\ref{eq:PE_Es}) can be used to calculate
the longitudinal electric field at least under the condition (\ref{eq:sr_cond}).
But if $E$ is small, the algorithm to solve Eqs.(\ref{eq:PE_Exy}-\ref{eq:PE_Es}) 
using a grid does not work well in calculating the space charge field.
Figs.\ref{fig:logZ_7MeV} and \ref{fig:logZ_ES_PA} show the longitudinal impedance of
a transient field created by an electron beam moving in a bend,
calculated using the exact solution (ES, black) and the paraxial approximation (PA, red).
The real part $\Re Z$ of PA agrees with ES under the conditions
(\ref{eq:PA_cond}) and (\ref{eq:sr_cond}) at least.
[Eqs.(\ref{eq:PE_Exy}-\ref{eq:PE_Es}) themselves may be applicable for an arbitrary $\gam$.
But we cannot show it numerically with precision since
$\Re Z$ of ES (black) is fluctuating due to a numerical error
when $k$ is large in the left figures of Fig.\ref{fig:logZ_7MeV}.]

If the beam energy is low such as $E\leq 100{\rm MeV}$, however,
the imaginary part $\Im Z$ of PA does not agree with ES over the full range 
of $k$ as in the right figures of Figs.\ref{fig:logZ_7MeV} and \ref{fig:logZ_ES_PA}.
This is because the values of $\Im Z$ of PA do not converge with respect to
the transverse grid size if $E$ is small, \ie,
the space charge field of the low energy beam is not resolved well
due to using a coarse grid in the numerical solution of Eq.(\ref{eq:PE_Exy}).
In short, the advantage that a coarse grid can be used in calculating $\Re Z$ by solving 
Eq.(\ref{eq:PE_Exy}) has the disadvantage that $\Im Z$ cannot be resolved sufficiently
at the transverse beam position if $E$ is small.
This is a numerical error caused by the defect of the algorithm to solve
Eq.(\ref{eq:PE_Exy}) using a coarse grid, and hence the paraxial approximation itself is
not responsible for the error of $\Im Z$ for small $E$.
If the beam energy is higher, \eg, $E=400{\rm MeV}$, since the space charge effect,
which is proportional to $\gamma^{-2}$, is less important,
$\Im Z$ of PA tends to agree with ES for $k\gg k_y^1$ as shown in
the lower right figure of Fig.\ref{fig:logZ_ES_PA}.
Thus, if $E\gtrsim 400{\rm MeV}$, the values of $\Im Z$ converge with respect to
the transverse grid size in solving Eq.(\ref{eq:PE_Exy}) numerically
using a coarse grid.

Since it takes time to compute the values of the field using the exact solution (ES)
as discussed later in section \ref{sec:computing_time}, when $k\gg k_y^1$,
instead of ES, we should use the solution of Eqs.(\ref{eq:PE_Exy}-\ref{eq:PE_Es})
taking the space charge field properly into account.
One of the solutions is to find the analytical solution of Eq.(\ref{eq:PE_Exy}),
which may be possible in a similar way to finding the exact solution outlined in
Eq.(\ref{eq:transforms}), approximating $\beta^2-g^{-2}$ as $2x/\rho-\gam^{-2}$, \ie,
Eq.(\ref{eq:PE_Exy}) is approximated as Eq.(7) in \cite{agoh}
which can be rewritten into the Airy differential equation 
with respect to $x$ by the Laplace transform with respect to $s$.
Besides this way, it may be possible to get the analytical solution of
Eqs.(\ref{eq:PE_Exy}-\ref{eq:PE_Es}) by expanding the exact solution in
the uniform asymptotic series.
The analytical solutions of Eqs.(\ref{eq:PE_Exy}-\ref{eq:PE_Es}), found in these two ways, 
must agree with each other at least to first order with respect to
$x/\rho$ and $(k\rho)^{-1}$.


\clearpage

\subsection{CSR shielded by a resistive pipe}
\label{sec:resistivity}

\begin{figure}[h]
  \begin{center}
    \includegraphics[scale=0.32,clip]{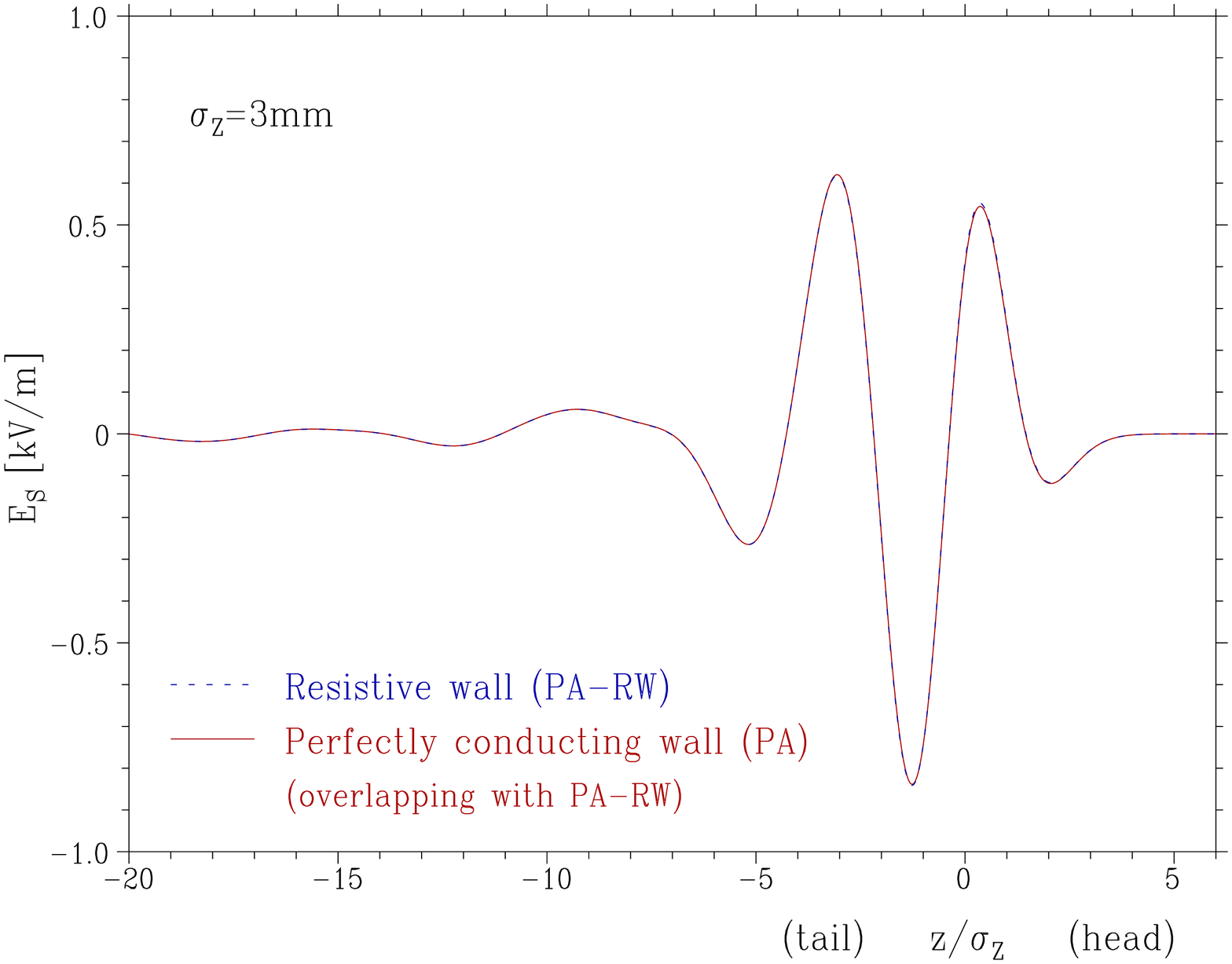}~~
    \includegraphics[scale=0.32,clip]{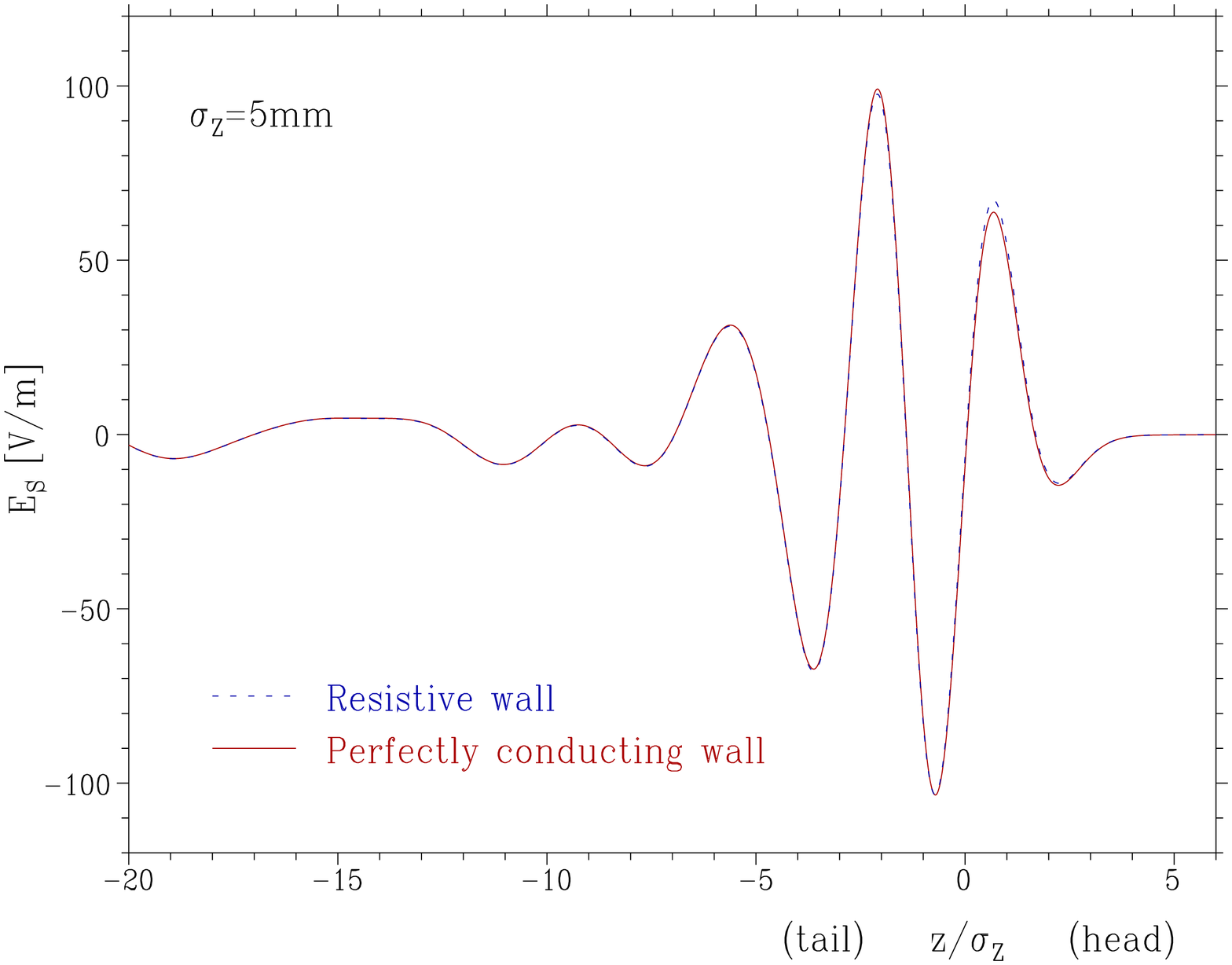}
      \vspace{1mm}
      \\
    \includegraphics[scale=0.32,clip]{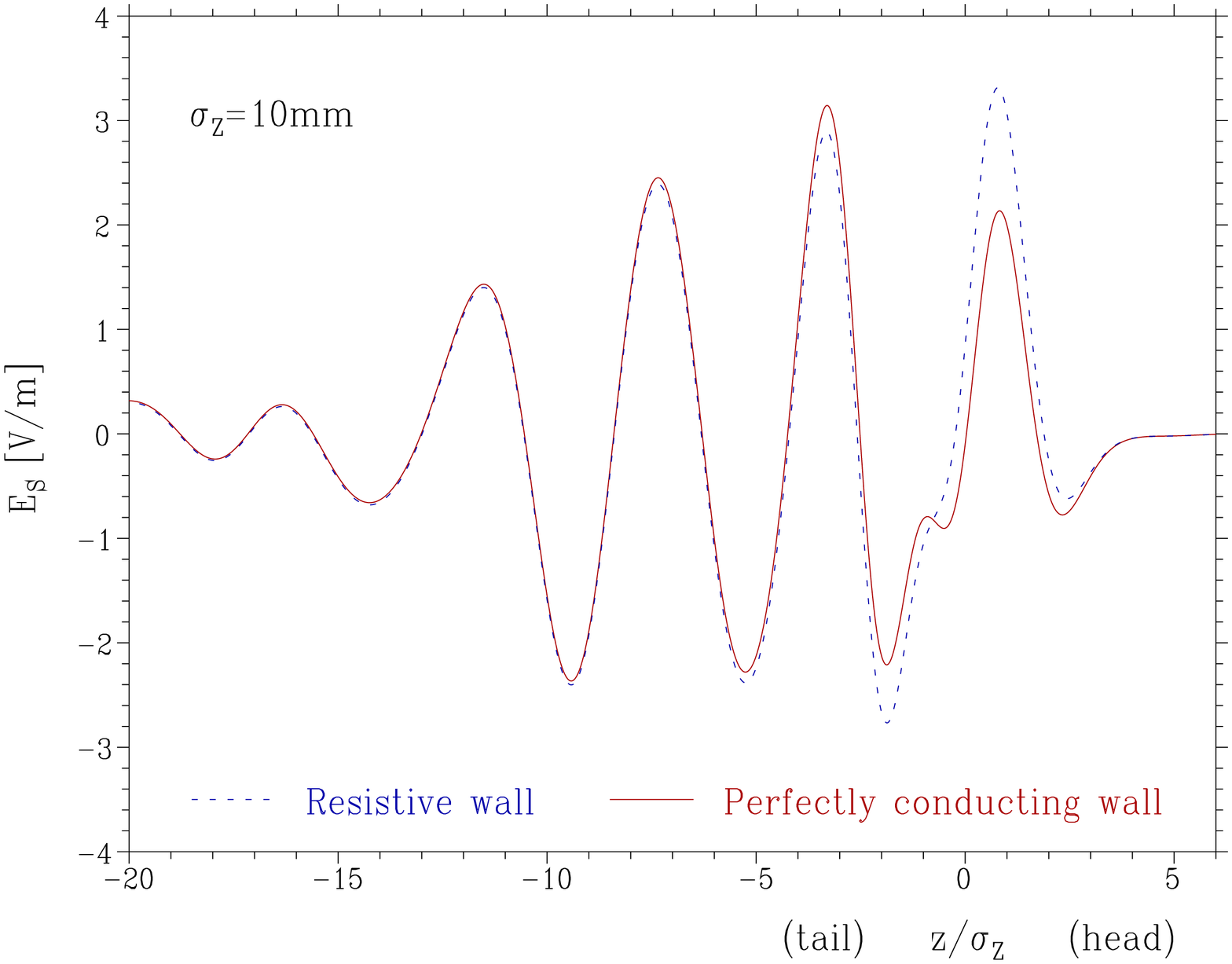}~~
    \includegraphics[scale=0.32,clip]{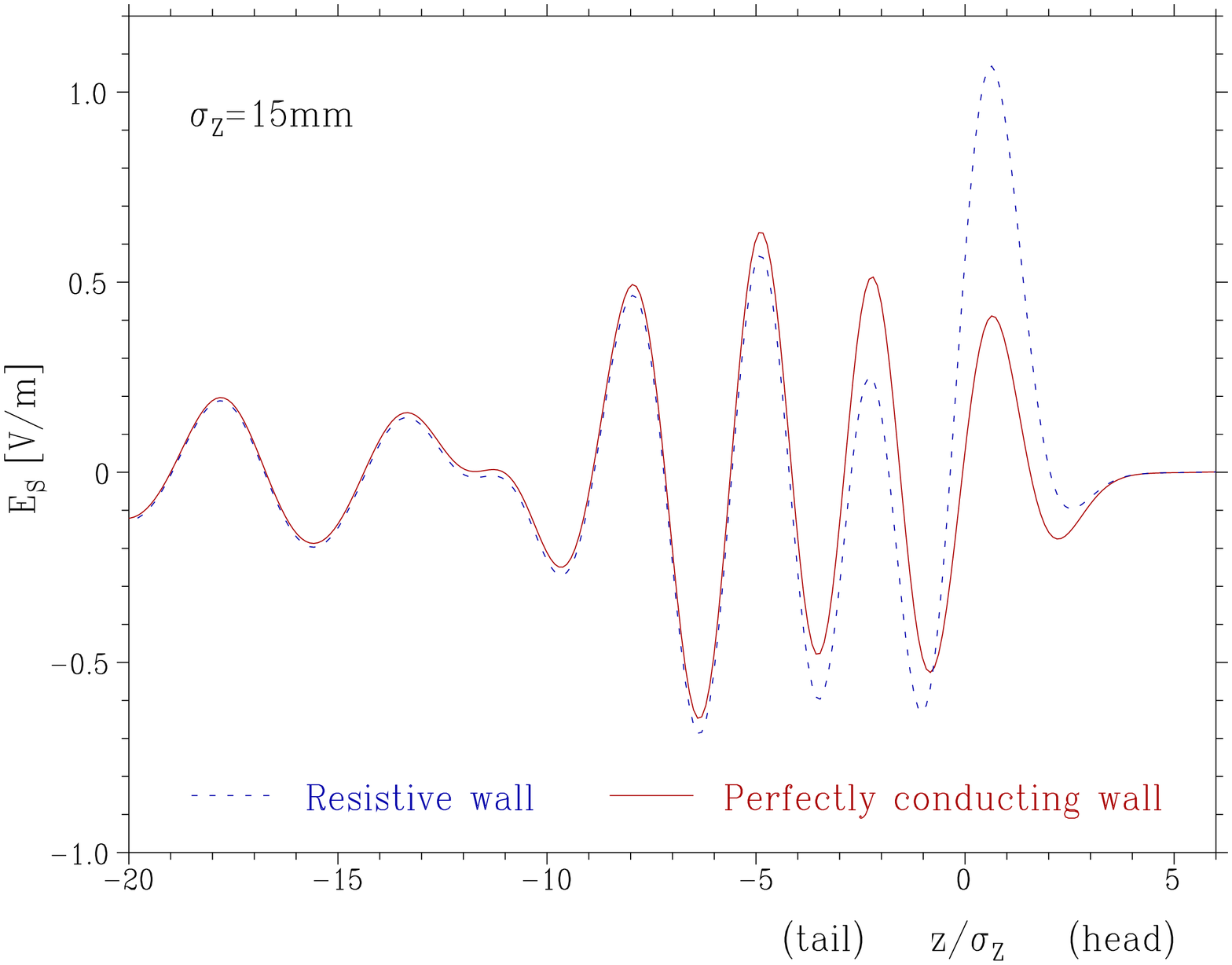}
    \caption
    [Longitudinal electric field of CSR in a resistive pipe using the paraxial approximation]
     {\small 
     Longitudinal electric field of CSR in the time domain $E_s(z)$ at
     $\xv_{\perp}=0$ and $s=1{\rm m}$ in a bend.
     We assume the electron bunch (\ref{eq:thin_bunch}-\ref{eq:rigid_bunch})
     which has a length $\sig_z=(3,5,10,15)\,{\rm mm}$.
     The horizontal axis is $z/\sig_z$ similar to Fig.\ref{fig:Esz}.
     We computed $E_s$ by solving Eqs.(\ref{eq:PE_Exy}-\ref{eq:PE_Es}).
     The blue dots (PA-RW) and red curves (PA) show $E_s$ of CSR emitted in
     a copper pipe (blue dots) and perfectly conducting pipe (red) respectively.
     The red curves and blue dots correspond to PA and PA-RW in Fig.\ref{fig:logZ}.
     The electrical conductivity of copper is $\sig_c=6\times 10^7/\Omg {\rm m}$.
     The other parameters are the same as Figs.\ref{fig:Esz} and \ref{fig:logZ}:
     $\rho=10\,{\rm m}$, $x_b=-x_a=3\,{\rm cm}$, $w=h=6\,{\rm cm}$,
     $\sig_y=20\,\mu{\rm m}$, $q=-1\,{\rm nC}$ and $E=1\,{\rm GeV}$.
      }
    \label{fig:Esz_rw}
  \end{center}
\end{figure}

As described in section VI.\,C of \cite{agoh}, to be precise from a practical point of view,
it is not correct to assume a perfectly conducting pipe in calculating the field of CSR in 
the low frequency range $k\leq O(k_y^1)$ since the value of the field of CSR for
$k\leq O(k_y^1)$ is comparable to or smaller than the resistive wall wakefield
as seen from Fig.\ref{fig:logZ}.
Let us see the longitudinal electric field of CSR in the time domain $E_s$,
taking into account the resistivity to the pipe in the numerical solution of
Eqs.(\ref{eq:PE_Exy}-\ref{eq:PE_Es}).
Fig.\ref{fig:Esz_rw} shows $E_s$ of CSR in a copper pipe (blue dots) and
in a perfectly conducting pipe (red curve), which correspond respectively to $Z$ of
the blue curve (PA-RW) and the red curve (PA) in Fig.\ref{fig:logZ}.
If $\sig_z\ll d_{\perp}\,(=w~\text{or}~h)$, \eg, $\sig_z=3{\rm mm}$ and
$d_{\perp}=6{\rm cm}$, the effect of the resistive wall is almost negligible to
$E_s$ of CSR for this set of parameters.
But the resistive wall effect is not negligible if $\sig_z$ is comparable to
$d_{\perp}$ as seen from the lower figures in Fig.\ref{fig:Esz_rw}.

\begin{figure}[h]
  \begin{center}
    \includegraphics[scale=0.32,clip]{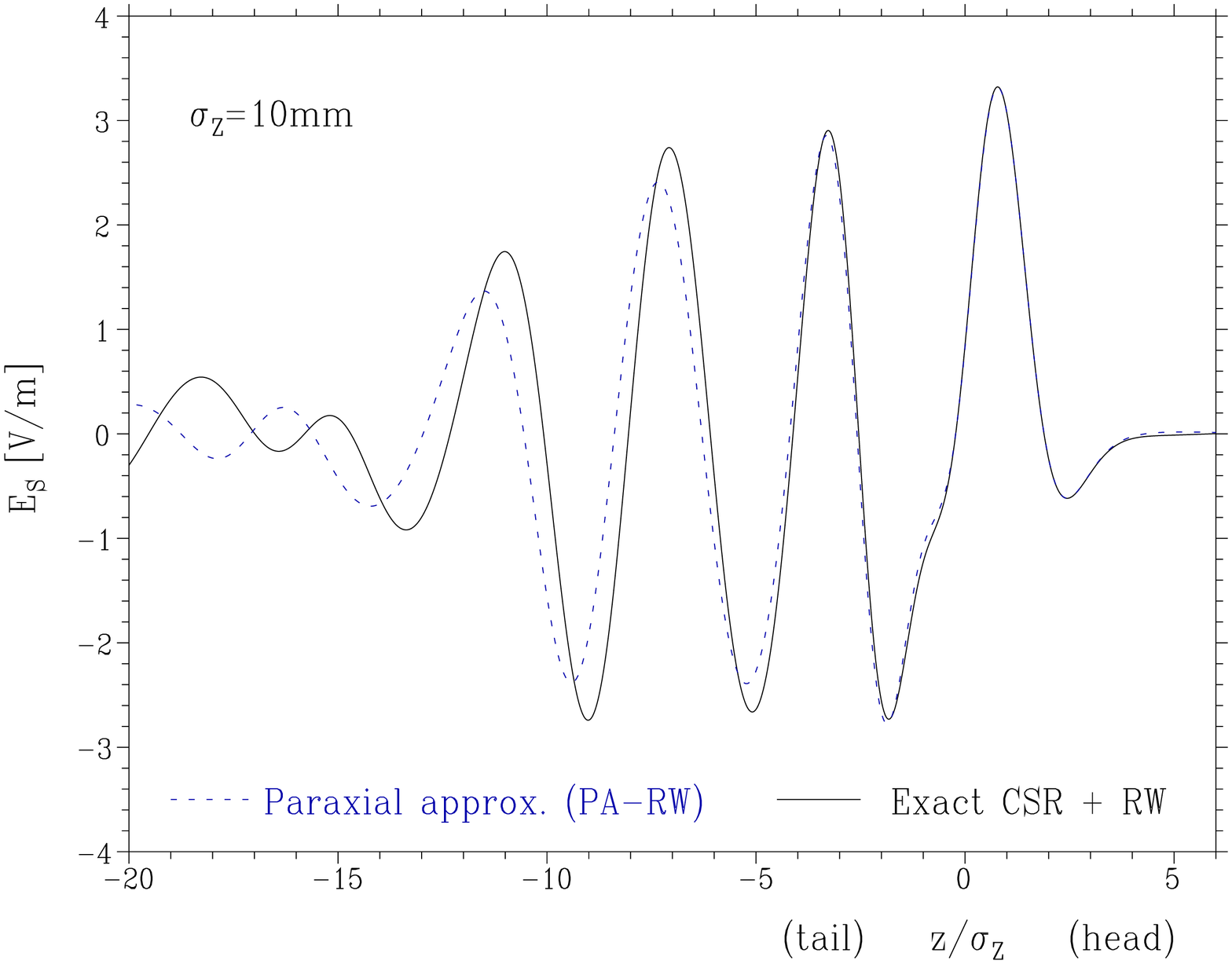}~~
    \includegraphics[scale=0.32,clip]{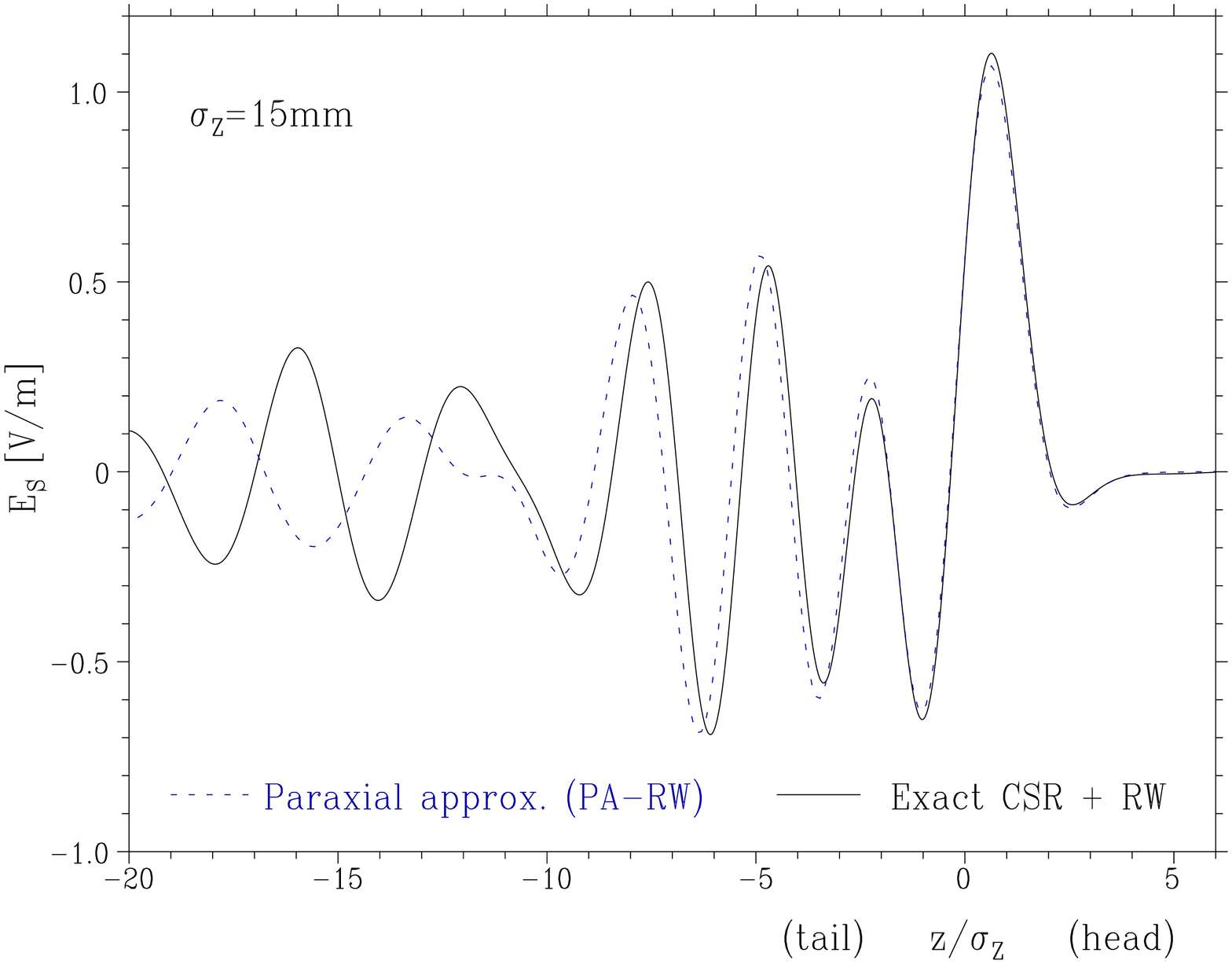}
    \caption
     [Longitudinal electric field of CSR in a resistive pipe using the exact solution]
     {\small 
     Longitudinal electric field of CSR in the time domain $E_s(z)$ at
     $\xv_{\perp}=0$ and $s=1{\rm m}$ in a copper square pipe in a bend.
     The horizontal axis is $z/\sig_z$ similar to Fig.\ref{fig:Esz}
     ($z=0$ is the longitudinal center of the bunch).
     We assume the electron bunch (\ref{eq:thin_bunch}-\ref{eq:rigid_bunch})
     which has a length $\sig_z=10{\rm mm}$ (left) and 15mm (right).
     The blue dots (PA-RW) are the same as those in the lower figures of
     Fig.\ref{fig:Esz_rw}, \ie, we computed PA-RW by solving
     Eqs.(\ref{eq:PE_Exy}-\ref{eq:PE_Es}) imposing the resistive boundary condition of
     the beam pipe.
     We calculated the black curve (Exact CSR + RW) by adding the resistive wall wakefield
     $E_s$ in the copper square pipe (RW: yellow-brown curve in Fig.\ref{fig:Esz_rw_pcw})
     to the exact solution of $E_s$ of CSR emitted in the perfectly conducting square pipe
     (ES: black curve in Fig.\ref{fig:Esz}).
     The parameters are common to Figs.\ref{fig:Esz}, \ref{fig:logZ} and \ref{fig:Esz_rw}: 
     $\rho=10{\rm m}$, $x_b=-x_a=3{\rm cm}$, $w=h=6{\rm cm}$, $q=-1{\rm nC}$, $E=1{\rm GeV}$,
     excluding the conductivity of the beam pipe $\sig_c=6\times 10^7/\Omg {\rm m}$ (copper).
     }
    \label{fig:Esz_rw_es}
  \end{center}
   \vspace{1mm}
  \begin{center}
    \includegraphics[scale=0.32,clip]{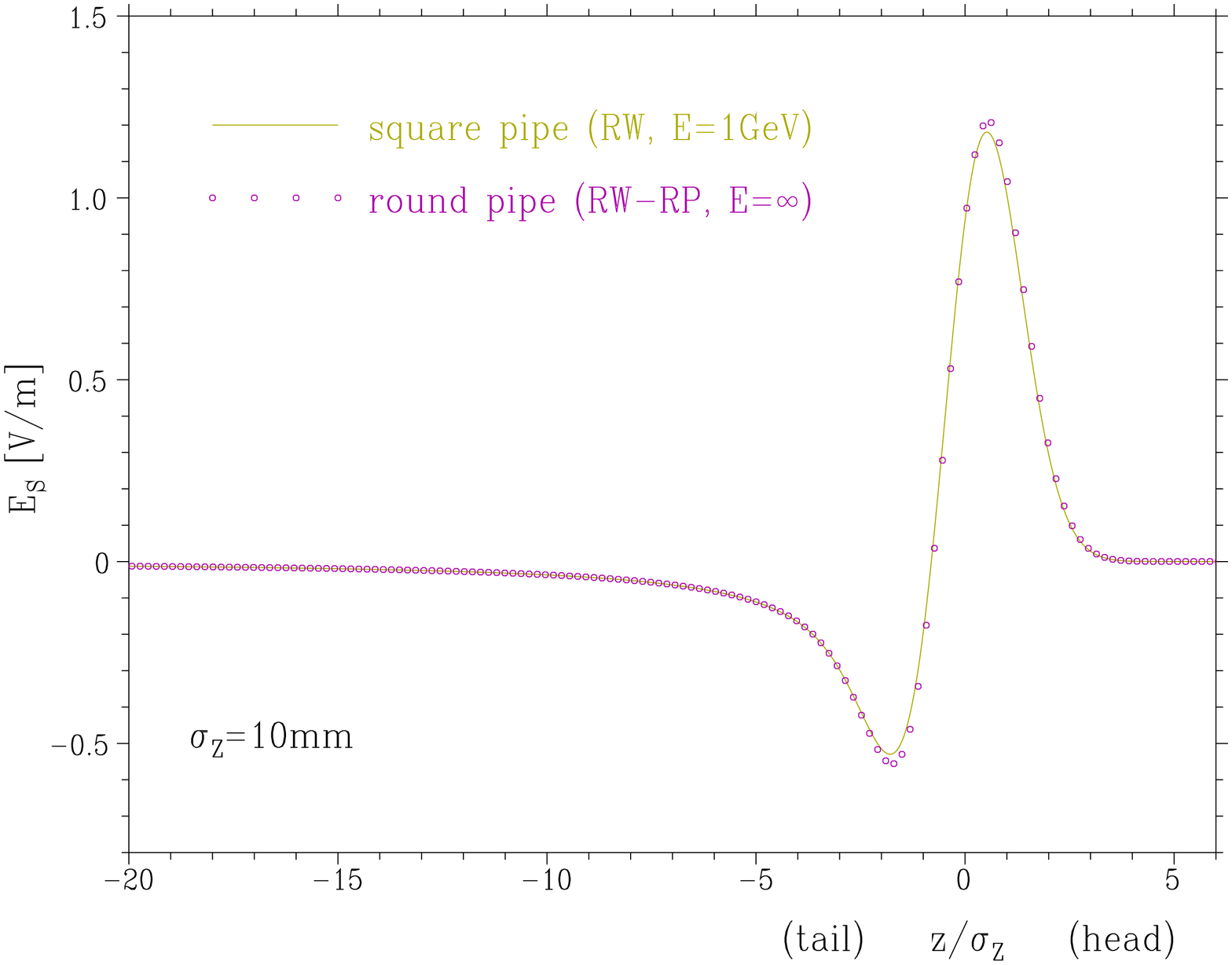}~~
    \includegraphics[scale=0.32,clip]{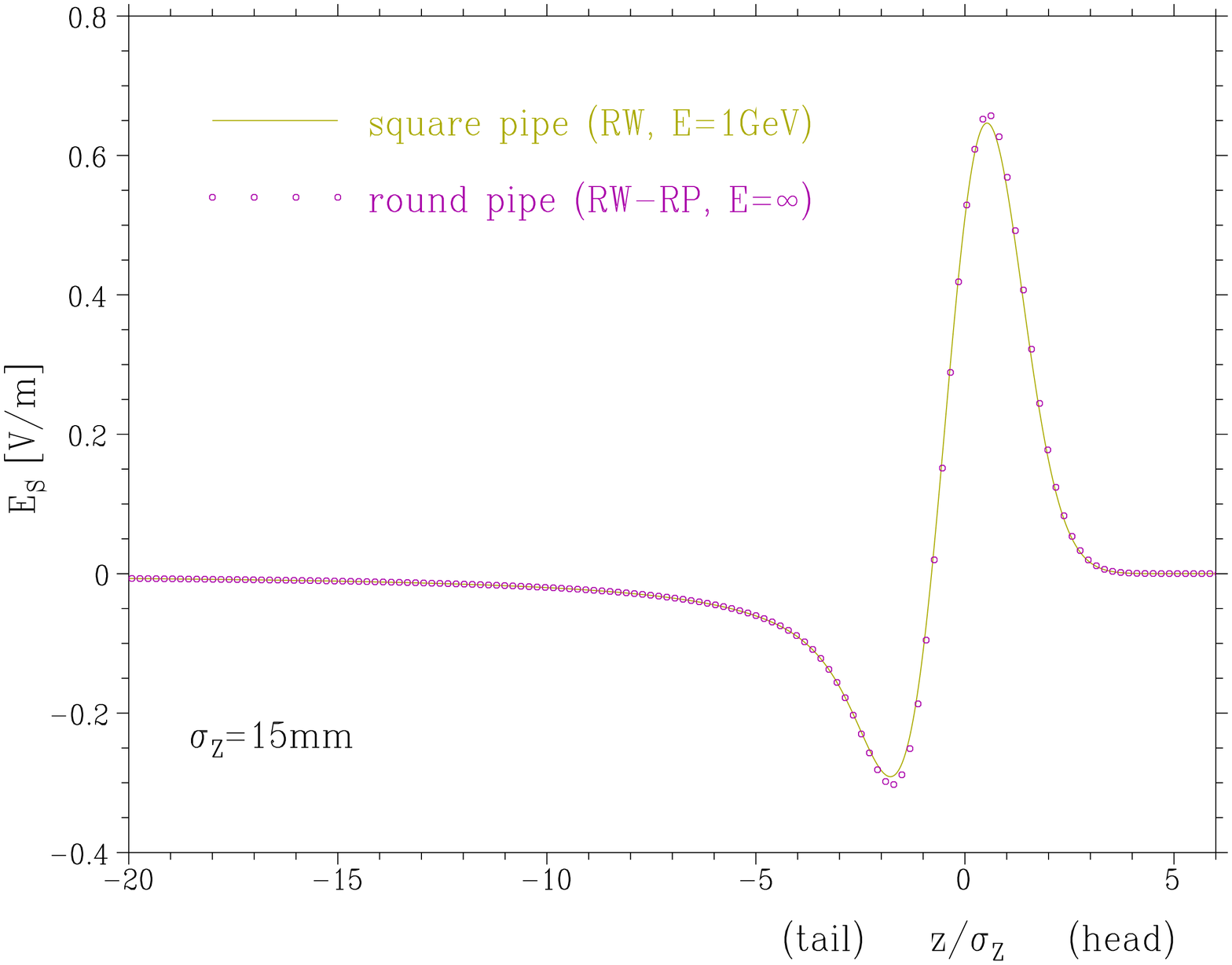}
    \caption[Resistive wall wakefield in a curved/straight resistive pipe]
     {\small
     Longitudinal resistive wall wakefield $E_s(z)$
     at the center ($\xv_{\perp}=0$) of a straight beam pipe for $s=1{\rm m}$.
     The horizontal axis is $z/\sig_z$ similar to Fig.\ref{fig:Esz_rw_es}.
     We assume the electron bunch (\ref{eq:thin_bunch}-\ref{eq:rigid_bunch}) moving at
     the center of a straight copper pipe which has a square or round cross section.
     The bunch length is $\sig_z=10{\rm mm}$ (left) and 15mm (right).
     We computed the resistive wall $E_s$ in a square pipe
     (RW, $E=1{\rm GeV}$: yellow-brown) and round pipe (RW-RP, $E=\infty$: magenta dots).
     We computed $E_s$ in the straight square pipe (RW) by numerically
     solving Eqs.(\ref{eq:PE_Exy}-\ref{eq:PE_Es}) for the steady field in the straight pipe,
     imposing the resistive boundary condition of the beam pipe.
     We computed $E_s$ at the center of the straight round pipe (RW-RP) using Eq.(4.75) in 
     \cite{yokoya_rw} which is derived assuming an ultrarelativistic beam ($\beta=1$).
     The yellow-brown curve and magenta dots correspond to the resistive wall impedances
     respectively shown with the yellow-brown curve (RW) and magenta dots (RW-RP) in
     Fig.\ref{fig:logZ}.
     The parameters are common to Figs.\ref{fig:Esz_rw} and \ref{fig:Esz_rw_es},
     excluding the beam energy of RW-RP and the bending radius $\vrho=\infty$.
      }
    \label{fig:Esz_rw_pcw}
  \end{center}
\end{figure}

When $k\leq O(k_y^1)$, we should take into account the resistive wall effect in
calculating the field of CSR emitted in a pipe having a finite conductivity.
As shown in section 4.3 (p.70) of \cite{agoh_phD},
if the bend is not so sharp and the pipe is relatively thin,
there is an approximate way to take into account the resistive wall effect in
the calculation of the field of CSR which is emitted in a perfectly conducting beam pipe.
That is, if $\rho\gg w$, we can approximately add the longitudinal resistive wall wakefield 
of a straight (or curved) resistive pipe to that of CSR in the perfectly conducting pipe
which are calculated separately.
As shown in Fig.\ref{fig:Esz_rw_es}, we can get the approximate value of $E_s$
shielded by the resistive pipe in the bend by adding the resistive wall $E_s$
(RW in Fig.\ref{fig:Esz_rw_pcw}) to $E_s$ of CSR in the perfectly conducting pipe
which is computed using the exact solution (ES in Fig.\ref{fig:Esz}).
Thus, even if the resistive wall effect is not negligible in calculating
the field of CSR, we can utilize the values of the field of CSR emitted in
a perfectly conducting pipe under the condition $\rho\gg w$.


\clearpage

\subsection{Computing time}
\label{sec:computing_time}

The exact solution (\ref{eq:tEs_std_sep}) takes much more time to compute
the value of the longitudinal electric field $E_s$ created in a bend, compared to 
that of solving the parabolic wave equation (\ref{eq:PE_Exy}-\ref{eq:PE_Es}) numerically
(paraxial approximation).
The following is the time to compute the values of $E_s$ for various bunch lengths $\sig_z$ 
shown in Fig.\ref{fig:Esz}, using an ordinary personal computer
Dell Precision 490 (Intel Xeon 5160, 3GHz, 2007).
\begin{center}
  \begin{tabular}{c||l|l}
   \hline
    $\sig_z$\,{\small [mm]} & {\small Exact solution} & {\small Paraxial approx.}
     \\ \hline
    1 & 23\,hr 6\,min & 53\,$\sim$\,60\,min
     \\
    3 & 60\,min & 5\,$\sim$\,7\,min
     \\
    5 & 21\,min & 2\,$\sim$\,3\,min
     \\
    7 & 12\,min & 1\,$\sim$\,3\,min
     \\
    10 & 9\,min & 1\,$\sim$\,3\,min
     \\
    15 & 5\,min & 1\,$\sim$\,3\,min
     \\
     \hline
  \end{tabular}
\end{center}
We rounded off the unit of second in this table since
it is sufficient to know the approximate values for the current discussion.
The computing time to solve Eq.(\ref{eq:PE_Exy}) depends on the grid size 
to resolve the field, which determines the accuracy and condition of stable iteration.
The computing time for $\sig_z=(7,10,15)\,\text{mm}$ in
the paraxial approximation is about a few minutes, which depends on $\sig_z$ weakly,
because the grid size of the space $(x,y,s)$ must satisfy Eq.(52) in
\cite{agoh_yokoya}, which is the condition of stable iteration in evolving the field
with respect to $s$ using the discretized parabolic wave equation.
Also, our numerical code to calculate the exact solution in Fortran has
several numerical parameters which determine the accuracy and computing time.

In computing the values of the frequency domain field $\tE_s$ using the exact solution 
(\ref{eq:tEs_std_sep}), one of the most time consuming process is to calculate the integral 
with respect to the horizontal variable $x'$ in Eq.(\ref{eq:Del_brcEs}).
We used a sixth order Runge-Kutta method to calculate the integral
which involves $K_{\ell}^n$ given by Eq.(\ref{eq:Kln}).
The numerical integration takes time, because a small spacing with respect to $x'$ is needed
to resolve the radial eigenfunctions $\cR_{\pm}^{mn}$ which are given by
Eqs.(\ref{eq:Ven}) and (\ref{eq:Vbn}) and involved in $K_{\ell}^n$ through
Eqs.(\ref{eq:kap_p}-\ref{eq:kap_m}) and (\ref{eq:bkap_Delbkap_p}-\ref{eq:bkap_Delbkap_m}).
It takes time especially for higher order radial modes $m$ which denote
the number of nodes between the inner and outer walls of the curved pipe $x=[x_a,x_b]$.
As seen from Fig.\ref{fig:poles} (p.\pageref{fig:poles}) and
Eqs.(\ref{eq:pole_struct_pls}-\ref{eq:pole_struct_mns}),
since the field tends to have real poles of the higher order 
radial mode $m$ for a larger $k$, it takes more time to compute
the numerical values of the exact solution for shorter bunch lengths.

We neglected $\rd_s^2$ in the braces of Eq.(\ref{eq:PE_Exy}), which is
the core of the paraxial approximation in computing the transient field of CSR.
It follows that we neglect the backward waves $e^{-ik(s+vt)}$ of the field.
The high frequency wave factor $e^{iks}$, which is contained in $e^{ikz}$, is removed from 
the time domain field through the Fourier transform given by Eq.(\ref{eq:FT_zk_9}).
Therefore the frequency domain field $\hE_s$ in the paraxial approximation does not have
so many nodes in the space $(x,y,s)$.
Owing to the removal of $e^{iks}$ from the time domain field,
we can use a coarse grid $(\Del x, \Del y, \Del s)$ to resolve
the frequency domain field in the paraxial approximation.
If the beam energy is high such that the space charge field is much smaller than
the radiation field, the condition of stable iteration for Eq.(\ref{eq:PE_Exy}) is not so
severe unless the field contains a very low frequency wave such as $k<10\,\text{m}^{-1}$.
Therefore this condition is not a big problem in practice in solving
Eqs.(\ref{eq:PE_Exy}-\ref{eq:PE_Es}) numerically unless the beam energy is low
as discussed in section \ref{sec:PA_applicability}.
Thus, in computing the values of the field of CSR for $k$ satisfying Eqs.(\ref{eq:PA_cond}),
it is much more efficient to solve Eqs.(\ref{eq:PE_Exy}-\ref{eq:PE_Es}) than computing
the exact solution which contains the high frequency factor $e^{iks}$ also in
the frequency domain (\ref{eq:Fourier_trans}-\ref{eq:omg}).

As seen from Eqs.(\ref{eq:Kln}) and (\ref{eq:bkap_Delbkap_p}-\ref{eq:bkap_Delbkap_m}),
it takes more time to compute the field using the exact solution for a smaller $s$
(shorter distance from the entrance of the bend),
because we must compute the higher order radial modes in the infinite sums which 
involve $\bkap_{\pm}^{mn}\propto e^{-(\bnu/\rho)s}$ with respect to $m$ in $K_{\ell}^n$,
where $\bnu=(\cnu_m^n,\cmu_m^n)$ represents the imaginary poles of the field in
the Laplace domain.
This trait is opposite to the numerical solution of the parabolic wave equation
by which the field evolves with respect to $s$.

The radial eigenfunctions $\cR_{\pm}^{mn}$ consist of the cross products of
the Bessel functions and their derivatives with respect to the order $\nu$
which is either real or purely imaginary.
We must calculate quadruple sums with respect to the indices
$(k,s,\ell,j)\in\mathbb{Z}^4$ in calculating the cross products using
the uniform asymptotic expansion of the Bessel functions as described in
appendices \ref{sec:uae_coeff}-\ref{sec:JY_uae},
where $k\in\mathbb{Z}$ and $s\in\mathbb{Z}$ are not the wavenumber $k\in\mathbb{R}$ and
the longitudinal coordinate $s\in\mathbb{R}^{+}$ but the indices of the series given by
Eqs.(\ref{eq:uv_app}-\ref{eq:tcj}).
By this, it takes time to compute the values of the cross products of
the Bessel functions using the uniform asymptotic expansion.
In order to calculate $\tE_s$ using the exact solution, moreover,
we must compute $\cR_{\pm}^{mn}$ for various $m$, $n$, $k$ and $x'$,
where $k\in\mathbb{R}$ is the wavenumber of the field.
It follows that we must compute an 8-fold sums in total
in computing $\tE_s$ using the exact solution (\ref{eq:tEs_std_sep}).
We wonder if we could not reformulate the uniform asymptotic expansion of
the cross products of the Bessel functions into a simpler formalism than
those we summarized in appendices \ref{sec:uae_coeff}-\ref{sec:uae_KL}.
It would be good if we could reformulate it in general for $\forall\nu\in\mathbb{C}$ without 
distinguishing $\nu\in\mathbb{R}$ and $i\mathbb{R}$.
Instead, we should find an approximate numerical method to compute the cross products of
the Bessel functions faster.

\clearpage
\section{Conclusions}
\label{sec:conclusion}

Under the assumptions listed in section \ref{sec:assumption},
we solved the exact wave equations for the electromagnetic field in
the analytical way shown in section \ref{sec:domains}.
Then we got the exact expression of the transient field of (coherent) synchrotron radiation
which is shielded by a perfectly conducting rectangular pipe.
We found the expression of all the components of the electric and magnetic fields in
the frequency domain $\tEv(\xv,k)=(\tE_x,\tE_y,\tE_s)$ and $\tBv(\xv,k)=(\tB_x,\tB_y,\tB_s)$
as functions of all the variables $\xv=(x,y,s)$ and $k$.
This is an explicit solution with respect to the field which includes not only
the coherent and incoherent components of synchrotron radiation but also
the space charge field of an arbitrary beam current moving in
a uniformly curved rectangular pipe.
The solution represents the exact relation of the field in the sector bend ($s>0$) 
as a function of its initial value on the plane at the entrance ($s=0$) of the bend.
In the derivation of the exact solution, as described in section \ref{sec:disconti},
we also took into account the discontinuity of the longitudinal derivative of
the field at $s=0$,
which is caused due to the abrupt change of the curvature of the reference trajectory.

In section \ref{sec:discussion} we demonstrated the fact that we can compute the values of
the longitudinal electric field of CSR in a bend using the exact solution
(\ref{eq:tEs_std_sep}).
As described in sections \ref{sec:parax_approx} and \ref{sec:Zk},
we need the exact solution in calculating the field in the low frequency range
which is equivalent to the long range with respect to $z$.
Using the exact solution, however, it takes much more time to compute the values of
the field than the numerical solution of the parabolic wave equation
(\ref{eq:PE_Exy}) on the basis of the paraxial approximation
as shown in section \ref{sec:computing_time}.
Therefore the exact solution is not so useful from a practical point of view in computing
the values of the field in the high frequency range
especially above the shielding threshold of the vacuum chamber.
Nevertheless, the exact solution is worth as it can be a reference to evaluate the validity 
and accuracy of an approximate calculation of the field of synchrotron radiation.

\subsection{Expressions of the field}

As described in section \ref{sec:expression}, we found four expressions of
the field created by a beam moving in a bend, which are the combination of
the scalar/differential expressions and separated/complete forms.
We should use the one which fits the purpose of calculating the field.
The separated form and complete form are suited respectively to the numerical
and analytical calculations of the field.
The four expressions of the field are given by the components of $(x,y,s)$
since we can find no elegant expression in terms of the vector formalism.

The exact expression of the field is given using $(\nu_m^n,\mu_m^n)$
defined in Eqs.(\ref{eq:pnu_poles}-\ref{eq:snu_poles})
which are the implicit solutions of the poles of the field in the Laplace domain.
Since we cannot find the explicit and exact expression of the poles
which denote the dispersion relation of the field in the curved pipe, to be rigorous,
the exact solution of the field, found in the present study, is not fully explicit.
We need to use Newton's method to compute the pole values precisely.
In appendix \ref{sec:poles} we found the asymptotic expressions of the poles under 
several different conditions as described in section \ref{sec:asymptotic_poles}.
Their accuracy is shown in Figs.\ref{fig:nu_err_4K} and \ref{fig:bnu_err_bkr}.
We use the asymptotic expressions of the poles as the initial value of Newton's method.

\subsection{Verification of the exact solution of the field}
\label{sec:verification}

We verified the exact solution of the field analytically and numerically in several ways.
Using the scalar and differential expressions of the complete form of the field in
appendix \ref{sec:verify}, we analytically verified that the solution of the field
satisfies the exact wave equation and Maxwell equations.
In addition, we analytically verified that the solution of the field satisfies
the initial condition at the entrance of the bend and
the boundary condition of the beam pipe.
Moreover, in order to check the expressions of the radial and longitudinal 
components of the field, we derived them in three different ways as shown in
section \ref{sec:sol_mfEB_xs}, appendices \ref{sec:we_xs} and \ref{sec:HL_field_sdom}.
Then we confirmed that all these expressions gotten by the three different ways
agree exactly.
Also, using the final value theorem of the Laplace transform, we derived
Eq.(\ref{eq:cZ_thin}) which is equivalent to the expression of the longitudinal impedance
of steady CSR in a rectangular torus, given by Eq.(3.9) in \cite{warnock_morton}.
In appendix \ref{sec:AL_GR} we analytically showed that the Green functions in
the Laplace domain and the radial eigenfunctions of the normal mode of the curved pipe 
converge to those of the straight pipe in the limit of $\rho\to\infty$.

We verified the exact solution of the longitudinal electric 
field in numerical ways using the differential expression of the separated form.
We compared the numerical values of the exact solution with the numerical solution of
Eq.(\ref{eq:PE_Exy}) which is the parabolic wave equation
on the basis of the paraxial approximation.
In section \ref{sec:discussion},
assuming a rigid and thin bunch in the absence of radial extent,
we showed that the longitudinal electric field of a transient CSR,
which is computed using the exact solution, agrees with the numerical solution of 
Eqs.(\ref{eq:PE_Exy}-\ref{eq:PE_Es}) for higher frequencies ($k\to\infty$) as in
Figs.\ref{fig:Esz}, \ref{fig:logZ} and \ref{fig:logZ_s2m}.
Also, as shown in Figs.\ref{fig:logZ_wh20} and \ref{fig:logZ_r5}, we confirmed that
the exact solution tends to agree with Schwinger's power spectrum formula (\ref{eq:kc}) 
around the critical wavenumber $k_c$ above the shielding threshold $k_{\rm th}$.
In addition, as shown in Figs.\ref{fig:logZ_Energy} and \ref{fig:logZ_w12cm},
the longitudinal impedance of CSR in the low frequency range $k\leq O(k_y^1)$,
computed using the exact solution, tends to agree with Eq.(\ref{eq:cZ_AE})
which is the asymptotic expression of the impedance for $k\to0$.

\subsection{Accuracy and speed of the calculation in the paraxial approximation}

In section \ref{sec:discussion} we examined the accuracy of the paraxial approximation
by comparing with the exact solution.
As shown in Figs.\ref{fig:logZ} and \ref{fig:logZ_s2m}, the longitudinal impedance in
the paraxial approximation tends to have a larger error for a lower frequency
($k\to0$) as expected in Eqs.(\ref{eq:PA_cond}).
By this, the field of CSR in the long range as $|z|\gtrsim d_{\perp}$ ($=w~\text{or}~h$) is 
incorrect in the paraxial approximation as seen from Fig.\ref{fig:Esz}.
Conversely, the error of the field in the paraxial approximation is relatively small in
the short range as $|z|\ll d_{\perp}$ unless the wavenumber $k$ is comparable to or 
smaller than the fundamental vertical wavenumber $k_y^1=\pi/h$
between the upper and lower walls of the beam pipe.
In such a low frequency range $k\leq O(k_y^1)$, however, the field of CSR is comparable to 
or smaller than the resistive wall wakefield as seen from
Figs.\ref{fig:logZ} and \ref{fig:logZ_s2m}.
The numerical solution has the advantage that we can easily take into account
the resistive wall effect by the pipe having a finite conductivity.
In this regard, we can approximately take into account the resistive wall effect by 
superimposing the resistive wall wakefield of the straight pipe on the exact solution of 
the field in the perfectly conducting pipe as shown in section \ref{sec:resistivity}.

When $k\gg k_y^1$ as in Eq.(\ref{eq:PA_condh}),
the numerical solution of Eqs.(\ref{eq:PE_Exy}-\ref{eq:PE_Es}) using a grid is 
correct for a high energy beam such as $\gamma\gtrsim 10^3$ as discussed in
section \ref{sec:PA_applicability}.
Conversely, if the beam energy is low such as $\gamma\ll 10^3$, as shown in
Fig.\ref{fig:logZ_ES_PA}, the numerical solution of Eqs.(\ref{eq:PE_Exy}-\ref{eq:PE_Es}) 
using a grid produces a significant error in the imaginary impedance $\Im Z$
since it is difficult to resolve the space charge field at the transverse beam position
using a coarse grid in the transverse plane.
We need to improve the algorithm to solve Eq.(\ref{eq:PE_Exy}) numerically.
On the other hand, if the beam energy satisfies Eq.(\ref{eq:sr_cond}),
we can get the correct values of the real impedance $\Re Z$ using
a coarse grid in numerically solving Eqs.(\ref{eq:PE_Exy}-\ref{eq:PE_Es})
unless $k\leq O(k_y^1)$.

Although  Eqs.(\ref{eq:PE_Exy}-\ref{eq:PE_Es}) have the range of applicability
limited to the high frequencies as in Eqs.(\ref{eq:PA_cond}),
the paraxial approximation has the advantage that
it takes much less time to compute the values of the field,
compared to computing the exact solution as described in section \ref{sec:computing_time}.
Therefore, if the error of the field value in the paraxial approximation is negligible or
acceptable, we should use Eqs.(\ref{eq:PE_Exy}-\ref{eq:PE_Es})
instead of using the exact solution in order to reduce the computing time.

\subsection{Further developments of the theories on CSR}

We still have many things to work out in the theoretical study on
(coherent) synchrotron radiation.
We would like to make some remarks for further developments of
the formulation and calculation of the field of CSR and its influence to the particles, 
including on the technical details in the mathematical and numerical methods.

Although we found the exact expression of the electromagnetic field of CSR,
it does not necessarily mean that we can compute the numerical values of
all the components of the field using the exact expression.
That is, it is not certain whether the exact expression can be used in any case under
the assumptions listed in section \ref{sec:assumption} to compute the value of the field on
a personal computer within a reasonable computing time, \eg, several hours.
One of the problems is that the field includes the space charge field
created by the beam.
Assuming a rigid and thin bunch in appendix \ref{sec:Es_bend_std},
we found the expression of the longitudinal electric field $\tE_s$ in a transient state,
which can be used in the numerical calculation
since we separated the steady field which includes the space charge field in the curved pipe.
For an arbitrary beam current which is not necessarily rigid, however,
we have not derived the expression of $\tE_s$ usable in the numerical calculation.
Also, we have not yet computed the other components of the fields except $E_s$.
In appendix \ref{sec:impedance} we showed three expressions of
the longitudinal impedance $Z$ using the exact solution of $\tE_s$.
But all these expressions of $Z$ cannot be used in the numerical calculation
since they have series which do not converge with respect to the radial mode number $m$.
The series include the space charge field of the beam current as the radial $\delta$-function
$\delta(r-r')$ which must be separated from the radiation field
as we did for $\tE_s$ in appendix \ref{sec:Es_bend_std}.
We want to find the expression of $Z$ usable to compute its value
since it takes time to integrate $\tE_s$ numerically with respect to $s$,
particularly, when $s$ is small.
This is because the higher order imaginary poles are required for the convergence of
the series with respect to the radial mode in computing $\tE_s$ for smaller $s$.

In appendix \ref{sec:travel_wave} we derived the exact expression of the field
propagating in a straight rectangular pipe.
But we have not yet written a numerical code to compute the field of CSR which gets out of
the bend and propagates in the straight pipe following the bend.
We should also examine the field propagating in the straight pipe
since the effect of CSR to the beam current in the downstream pipe is
usually comparable to or larger than that in the bend.
If we need only the total energy change by CSR in the straight pipe from
the exit of the bend to $s=\infty$ in the perfectly conducting straight pipe,
then we can compute it in the semi-analytical way shown in \cite{agoh_yokoya1,agoh_phD}
under the assumption of ultrarelativistic particles ($\gamma\to\infty$), \ie,
excluding the space charge field.
If necessary, we can add the space charge effect for a finite length to
the semi-analytical solution.
We should implement it into our numerical code using the exact solution.

It is difficult to find the explicit and exact expression of $\nu=(\nu_m^n,\mu_m^n)$
which are the poles of the field in the Laplace domain.
Also, we could not find the exact expression of the cutoff wavenumbers
$k=(k_m^n,\bk_m^n)$ in the curved pipe,
which are defined in Eqs.(\ref{eq:kmn}-\ref{eq:bkmn}) implicitly.
We derived their asymptotic expressions in appendix \ref{sec:poles}.
As discussed in appendix \ref{sec:uae_problem}, each Bessel function has
a different mathematical structure from that of the cross products of
the Bessel functions in the sense that the latter has symmetry with respect to
the order and argument as shown in Eqs.(\ref{eq:cp_symm}-\ref{eq:cp_symm2}).
Therefore we want to reformulate the uniform asymptotic expansion of
the cross products of the Bessel functions, not that of each Bessel function.
If we made it, we might be able to find the explicit and exact series representation of
$(\nu_m^n,\mu_m^n)$ and $(k_m^n,\bk_m^n)$.
Besides, since there are cancelations in the cross products, it is meaningful to
reformulate the uniform asymptotic expansion of the cross products also from
a computational point of view to avoid digit loss and overflow.
Similar to Dunster's uniform asymptotic series given by
Eqs.(\ref{eq:Fimu_muz}-\ref{eq:Gimu_muz}), it would be better to expand the cross products 
directly in the trigonometric or exponential functions not through the Airy functions.

Although we found the exact analytical solution of the field of (coherent) synchrotron 
radiation including the space charge field, it takes time to compute the values of
the field since the exact expression involves the Bessel functions
which are computed using several kinds of series representations.
One of the ways of improving the speed and accuracy of the exact solution code is
to develop a numerical method to compute the values of the Bessel functions faster.
Instead of this way, it might be possible to get the numerical solution of the field in
a semi-analytical way without using the Bessel functions.
That is, it might be possible to solve Eq.(\ref{eq:we_xs_LD}) numerically by
the polynomial expansion, which is the wave equation (Bessel differential equation) 
for the field in the Laplace domain $(\mfE^n,\mfB^n)$.
If we get the numerical solution of $(\mfE^n,\mfB^n)$,
we may be able to get the frequency domain field $(\tEv,\tBv)$ through
the inverse Laplace transform (\ref{eq:ILT_nu}) and the Fourier expansion
(\ref{eq:Four_Exp_plus}-\ref{eq:Four_Exp_minus}) in the numerical calculation.

Using the model given in section \ref{sec:assumption}, we may be able to find
the analytical solution of Eq.(\ref{eq:PE_Exy}) which is the parabolic wave equation on
the basis of the paraxial approximation taking into account the entanglement of
the radial and longitudinal components of the field in the bend.
We may be able to find it in the following two ways:
(i) solving Eq.(\ref{eq:PE_Exy}) using the Laplace transform with 
respect to $s$ similar to what we did in the present paper,
(ii) approximating the exact solution, found in the present study,
using the uniform asymptotic expansion of the Bessel functions.
These two ways can verify each other analytically as described in
section \ref{sec:PA_applicability}.
If we found the analytical solution of Eqs.(\ref{eq:PE_Exy}-\ref{eq:PE_Es}),
perhaps it could help compute the value of the field faster than computing
the exact solution of the field, taking the space charge field correctly into account.
In addition, we should also find the asymptotic expression of the field for $k\to0$ by
expanding the exact solution as described in section \ref{sec:Z_energy}.
Furthermore, we would like to consider whether we can derive the analytical solution of
the field in a bend, taking into account the resistive wall effect of the curved beam pipe.

Using the exact solution, we have not yet investigated the transverse Lorentz force
$\Fv_{\perp}=(F_x,F_y)$ which the particles receive from their own field on
a curved trajectory in a bend.
Assuming a rigid and thin bunch, as shown in section 4.4 of \cite{agoh_phD},
it may not be difficult to calculate the vertical Lorentz force $F_y$ using
the exact solution since $F_y$ does not depend on the transverse beam size
if $\sig_{x,y}\ll \ell_{\perp}$ given by Eq.(\ref{eq:shield}),
similar to the longitudinal electric field $E_s$ discussed in section \ref{sec:discussion}.
On the other hand, it may be difficult to calculate the horizontal Lorentz force $F_x$ on
the curved trajectory since it may depend on the horizontal beam size $\sig_x$ unlike $F_y$.
In a horizontal bend, $F_x$ has a different trait from $F_y$ and $F_s$ as far as we know.
The horizontal force may diverge at the transverse beam position
if the bunch has no horizontal extent.
As described in section \ref{sec:discussion}, we cannot assume a rigid bunch moving in
a bending magnet if it has a horizontal extent.
Therefore, in order to calculate $F_x$, it is necessary to take into account
the horizontal and longitudinal motion of the particles in the bend.
In the future, taking into account the motion of the source current,
we would like to discuss whether the horizontal force in a bend is negligible,
compared to the effect of the longitudinal force which gives rise to
a non-uniform energy variation to the particles forming the bunch.
We hope to establish a method to calculate the motion of the charged particles,
taking into account the total effects of the field of (coherent) synchrotron radiation.
A goal of ours in this research is to establish a theory and method to find
the simultaneous solution of Maxwell equations and equations of motion of
the source particles.


\clearpage

\appendix

\section{Differential operations in the curvilinear coordinate system}
\label{sec:coordinates}

We describe the differential operations in the curvilinear coordinates
on the basis of a planar reference axis.
The basis $(\ev_x,\ev_y,\ev_s)$ consists of the horizontal, vertical and
longitudinal unit vectors as shown in Eq.(\ref{eq:coordinates}).
The reference axis ($s$-axis) lies on a horizontal plane perpendicular to $\ev_y$.
The nabla $\nablav$ is given by Eq.(\ref{eq:nabla}),
\begin{align}
  \nablav=\ev_x\rd_x+\ev_y\rd_y+\ev_s\brd_s
   ,\qquad
  g=1+\vkap x
   ,\qquad
  \vkap\vrho
  =1.
  \label{eq:xys_basis}
\end{align}
$\vkap$ and $\vrho$ are the local curvature and radius of the $s$-axis,
which depend on $s$ in general.
$g$ is the geometric factor which denotes the relative radius of $\vrho+x$ to $\vrho$.
The curl and divergence of a vector $\Av$ are given as
\begin{align}
  &
  \nablav\times\Av
  =\ev_x(\rd_yA_s-\brd_sA_y)
   +\ev_y(\brd_sA_x-\brd_xA_s)
   +\ev_s(\rd_xA_y-\rd_yA_x),
  \label{eq:curlA}
  \\
  &
  \nablav\cdot\Av
  =\brd_xA_x+\rd_yA_y+\brd_sA_s ,
    \qquad
  \Av=A_x\ev_x+A_y\ev_y+A_s\ev_s.
  \label{eq:divA}
\end{align}
$\brd_s$ and $\brd_x$ are the longitudinal and horizontal differential operators
which go to $\rd_s$ and $\rd_x$ for $\vkap\to 0$,
\begin{align}
  \brd_s=\frac{\rd_s}{g}
   ,\qquad
  \brd_x=\rd_x+\frac{\vkap}{g}
   \quad\Lra\quad
  \brd_xf=\frac{1}{g}\rd_x(gf)
   ;\qquad
  \lim_{\vkap\to 0}\brd_{s,x}
  =\rd_{s,x}.
   \label{eq:brd}
\end{align}
$f$ is an arbitrary differentiable function in this appendix.
$\brd_s$ does not commute with $\rd_s$ unless $\vkap$ is a constant,
\begin{align}
  g(\brd_s\rd_s-\rd_s\brd_s)
  =\rd_s^2-g^2\brd_s^2
  =x(\rd_s\vkap)\brd_s .
   \label{eq:rds_brds}
\end{align}
$\brd_x$ does not commute with $\rd_x$ unless $\vkap=0$.
Therefore, acting $\rd_x$ and $\brd_x$ on $f$ in the opposite order, we get
\begin{align}
  \brd_x\rd_xf
  =\frac{1}{g}\rd_x(g\rd_xf)
  =\Big(\rd_x^2+\frac{\vkap}{g}\rd_x\Big)f ,
    \qquad
  \rd_x\brd_xf
  =\rd_x\bigg\{\frac{1}{g}\rd_x(gf)\bigg\}
  =\bigg(\brd_x\rd_x-\frac{\vkap^2}{g^2}\bigg)f .
   \label{eq:rdx_brdx}
\end{align}
Eqs.(\ref{eq:rdx_brdx}) are the terms which involve the derivatives
with respect to $x$ in the Laplacian $\nablav^2$.
In the curvilinear coordinates,
the Laplacians for a function $f$ and a vector $\Av$ are given as
\begin{align}
  \nablav^2f
  =\nablav\cd(\nablav f)
  =(\brd_x\rd_x+\rd_y^2+\brd_s^2)f ,
    \qquad
  \nablav^2\Av
  =\nablav(\nablav\cdot\Av)-\nablav\times(\nablav\times\Av).
   \label{eq:laplacian}
\end{align}
$\brd_s^2$ is gotten from Eq.(\ref{eq:rds_brds}).
Since we assume that the $s$-axis has no vertical curvature,
$[\nablav^2\Av]_y=\nablav^2A_y$ similar to $\nablav^2f$.
On the other hand, $[\nablav^2\Av]_{x,s}\ne\nablav^2A_{x,s}$ unless $\vkap=0$
which means a straight section,
\begin{align}
  \nablav^2\Av
  &=\ev_y
    \bigg[
      \bigg(\brd_x\rd_x+\rd_y^2+\frac{\rd_s^2}{g^2}\bigg)A_y
     -\frac{(\rd_s\vkap)}{g^2}
      x\brd_sA_y
    \bigg]
  \nonumber\\&\quad
   +\ev_x
    \bigg[
        \bigg(\rd_x\brd_x+\rd_y^2+\frac{\rd_s^2}{g^2}\bigg)A_x
       -\frac{(\rd_s\vkap)}{g^2}\bigg(x\brd_sA_x+\frac{A_s}{g}\bigg)
       -\frac{2\brd_s}{g\vrho}A_s
    \bigg]
  \nonumber\\&\quad
   +\ev_s
    \bigg[
      \bigg(\rd_x\brd_x+\rd_y^2+\frac{\rd_s^2}{g^2}\bigg)A_s
     -\frac{(\rd_s\vkap)}{g^2}\bigg(x\brd_sA_s-\frac{A_x}{g}\bigg)
     +\frac{2\brd_s}{g\vrho}A_x
    \bigg] .
  \label{eq:nablav2A}
\end{align}
From Eq.(\ref{eq:nablav2A}), we get Eqs.(\ref{eq:we_tExBx}-\ref{eq:wey0}) which are
the $(x,s)$ and $y$ components of the wave equations (\ref{eq:wek_E}-\ref{eq:wek_B}) for
the fields in the frequency domain.

$\nablav_\vdash^2$ is the operator of the wave equations
(\ref{eq:we_tExBx}-\ref{eq:we_tEsBs}) for $\tE_{x,s}$ and $\tB_{x,s}$,
excluding the coupling term and the term which involves $\rd_s\vkap$.
$\nablav_{\rm v}^2$ is the operator of the wave equations (\ref{eq:wey0}) for
$\tE_y$ and $\tB_y$,
\begin{align}
  \bigg({ \nablav_\vdash^2 \atop \nablav_{\rm v}^2}\bigg)
  &=\bigg({ \rd_x\brd_x \atop \brd_x\rd_x}\bigg)
    +(k\beta)^2+\rd_y^2+\frac{\rd_s^2}{g^2} ,
    \qquad
  \bigg({ \rd_{\,\vdash}^2 \atop \rd_{\rm v}^2}\bigg)
  =\bigg({ \rd_x\brd_x \atop \brd_x\rd_x }\bigg)
    +(k\beta)^2-(k_y^n)^2+\frac{\rd_s^2}{g^2}.
  \label{eq:Nb_vdash2}
\end{align}
As in Eqs.(\ref{eq:we_cFxs}-\ref{eq:wen_bend}), 
$\rd_{\,\vdash}^2$ and $\rd_{\rm v}^2$ are the operators of the wave equations
for $(\cE_{x,s}^n,\cB_{x,s}^n)$ and $(\cE_y^n,\cB_y^n)$
which are the Fourier coefficients of $(\tE_{x,s},\tB_{x,s})$ and $(\tE_y,\tB_y)$
with respect to $y$.
For $\vrho\to\infty$, both $\rd_{\,\vdash}^2$ and $\rd_{\rm v}^2$ go to
$\vec{\rd}^2+(k\beta)^2$ which is the d'Alembertian for the field in
a straight section as in Eq.(\ref{eq:we_xprim_drift}).
When $\vrho=\rho$ (constant), we rewrite $\rd_{\,\vdash}^2$ and $\rd_{\rm v}^2$ using
the radial variable $r$ $(=\rho+x=g\rho)$ and $k_r^n$ given by
Eqs.(\ref{eq:r}) and (\ref{eq:krn}),
\begin{align}
  \bigg({ \rd_{\,\vdash}^2 \atop \rd_{\rm v}^2 }\bigg)
  &=\bigg({ \rd_r\brd_r \atop \brd_r\rd_r }\bigg)
   +(k_r^n)^2+\brd_s^2 ,
    \qquad
  \brd_s=\frac{\rho}{r}\rd_s ,
    \qquad
  \brd_r
  =\rd_r+\frac{1}{r}
   \quad\Lra\quad
  \brd_rf
  =\frac{1}{r}\rd_r(rf).
  \label{eq:rdv2}
\end{align}
$\brd_r\rd_r$ is the radial operator in the Bessel differential equation (\ref{eq:BDE}).
In describing the fields in a constant bend,
if we use the angle variable $\phi$ $(=s/\rho)$ instead of $s$,
we cannot take the limit of $\rho\to\infty$ for $\phi$.

\section{Discontinuity of the field derivative at the edge of a bend}
\label{sec:discontinuity}

$\rd_s\tE_y$ is discontinuous at the entrance of a hard-edge bend $(s=0)$ as shown in
Eq.(\ref{eq:rds_tEy_jump}).
In this appendix we describe the discontinuity of $\rd_s\tE_{x,s}$ and $\rd_s\tB_{x,s}$ at
$s=0$, which we need in deriving the wave equations (\ref{eq:we_tExsBxs}).
We first write out all the components of Maxwell equations in the frequency domain
(\ref{eq:Fourier_trans}) using the curvilinear coordinates $\xv=(x,y,s)$ 
defined in section \ref{sec:curv_coordinates} and appendix \ref{sec:coordinates},
\begin{alignat}{2}
  \nablav\cdot\Ev=c\mu_0J_0:
  &&\qquad&
  \brd_x\tE_x+\brd_s\tE_s+\rd_y\tE_y=Z_0\tJ_0 ,
  \label{eq:Gauss_law_FD2}
  \\
  \nablav\cdot\Bv=0:
  &&\qquad&
  \brd_xc\tB_x+\brd_sc\tB_s+\rd_yc\tB_y=0 ,
  \label{eq:No_monopole_FD2}
   \\
  \nablav\times\Ev+\rd_t\Bv=0:
  &&\qquad&
  \rd_y\tE_s-\brd_s\tE_y-ik\beta c\tB_x=0 ,
    \label{eq:Faraday_x_FD2}
  \\
  &&\qquad&
  \brd_s\tE_x-\brd_x\tE_s-ik\beta c\tB_y=0 ,
    \label{eq:Faraday_y_FD2}
  \\
  &&\qquad&
  \rd_x\tE_y-\rd_y\tE_x-ik\beta c\tB_s=0 ,
    \label{eq:Faraday_s_FD2}
   \\
  \nablav\times\Bv-c^{-2}\rd_t\Ev=\mu_0\Jv:
  &&\qquad&
  \rd_yc\tB_s-\brd_sc\tB_y+ik\beta\tE_x=Z_0\tJ_x ,
    \label{eq:Ampere_x_FD2}
  \\
  &&\qquad&
   \brd_sc\tB_x-\brd_xc\tB_s+ik\beta \tE_y=Z_0\tJ_y ,
    \label{eq:Ampere_y_FD2}
  \\
  &&\qquad&
  \rd_xc\tB_y-\rd_yc\tB_x+ik\beta \tE_s=Z_0\tJ_s .
    \label{eq:Ampere_s_FD2}
\end{alignat}
The fields $(\tEv,\tBv)$ and current $\tJ=(\tJ_0,\tJv)$ in the frequency domain have
the arguments $(\xv,k)$.
The boundary conditions of the fields on the walls of the rectangular pipe are
given as follows,
\begin{alignat}{4}
  \tE_y&=\tE_s=c\tB_x=0 ,
    \qquad&
  \brd_x\tE_x&=\brd_xc\tB_s=\rd_xc\tB_y=0
    \qquad&&\text{at}~~&
  x&=x_a,x_b ,
   \\
  \tE_x&=\tE_s=c\tB_y=0 ,
    \qquad&
  \rd_y\tE_y&=\rd_yc\tB_x=\rd_yc\tB_s=0
    \qquad&&\text{at}~~&
  y&=\pm h/2 .
   \label{eq:BCy}
\end{alignat}

We assume that the curvature of the $s$-axis changes abruptly from
$\kap_{-}$ to $\kap_{+}$ at $s=0$ as shown in Eq.(\ref{eq:curv}).
Assuming Eq.(\ref{eq:curv}), the geometric factor $g$ is given as
\begin{align}
  g=\theta(s)g_{+} +\theta(-s)g_{-}
   ,\qquad
  g_\pm=1+\kap_{\pm}x
    \quad
   (\kap_{\pm}=\rho_{\pm}^{-1}=\text{constant}).
   \label{eq:g_pm}
\end{align}
In this appendix we omit the transverse arguments $\xv_{\perp}=(x,y)$ of
the functions for clarity with regard to the longitudinal argument $s$.
The fields and current are continuous at $s=0$, \ie,
\begin{align}
  \tEv(+0)=\tEv(-0),
   \qquad
  \tBv(+0)=\tBv(-0),
   \qquad
  \tJ_0(+0)=\tJ_0(-0),
   \qquad
  \tJv(+0)=\tJv(-0) .
   \label{eq:cont}
\end{align}
However, the fields have a kink at $s=0$ due to the abrupt change of the curvature.
The first derivatives of the components of the fields with respect to $s$ are
discontinuous at $s=0$.
From Eqs.(\ref{eq:Gauss_law_FD2}) and (\ref{eq:Faraday_y_FD2}),
\begin{alignat}{2}
  \frac{\rd_s\tE_s(+0)+\kap_{+}\tE_x(0)}{g_{+}}
  &=\frac{\rd_s\tE_s(-0)+\kap_{-}\tE_x(0)}{g_{-}} ,
    \qquad&
  \rd_s\tE_s(\pm0)
  &=[\rd_s\tE_s(s)]_{s=\pm0} ,
  \label{eq:tExs_cont}
   \\
  \frac{\rd_s\tE_x(+0)-\kap_{+}\tE_s(0)}{g_{+}}
  &=\frac{\rd_s\tE_x(-0)-\kap_{-}\tE_s(0)}{g_{-}} ,
    \qquad&
  \rd_s\tE_x(\pm0)
  &=[\rd_s\tE_x(s)]_{s=\pm0} .
  \label{eq:tEsx_cont}
\end{alignat}
Eqs.(\ref{eq:tExs_cont}-\ref{eq:tEsx_cont}) correspond to Eq.(\ref{eq:Frdy_x})
which is the relation of the values of $\rd_s\tE_y$ at $s=\pm0$.
From Eq.(\ref{eq:Frdy_x}) and Eqs.(\ref{eq:tExs_cont}-\ref{eq:tEsx_cont}),
we get the discontinuity of $\rd_s\tE_{x,y,s}$ at $s=0$,
\begin{align}
  \rd_s\tE_x(+0)-\rd_s\tE_x(-0)
  &=u_{\pm}\{x\rd_s\tE_x(\pm0)+\tE_s(0)\} ,
  \label{eq:cEx_jump_plus_sum}
   \\
  \rd_s\tE_y(+0)-\rd_s\tE_y(-0)
  &=u_{\pm}\{x\rd_s\tE_y(\pm0)\} ,
  \label{eq:cEy_jump_plus_sum}
   \\
  \rd_s\tE_s(+0)-\rd_s\tE_s(-0)
  &=u_{\pm}\{x\rd_s\tE_s(\pm0)-\tE_x(0)\} ,
  \label{eq:cEs_jump_plus_sum}
\end{align}
where
\begin{align}
  u_{\pm}
  &=\frac{\kap_{+}-\kap_{-}}{g_\pm} .
\end{align}
Similar to deriving Eq.(\ref{eq:tEy_jump}) which is the discontinuity of
$\rd_s\tE_y$ at $s=0$, we can also get Eqs.(\ref{eq:cEx_jump_plus_sum}) and
(\ref{eq:cEs_jump_plus_sum}) by integrating the terms $\rd_s^2\tE_{x,s}$ in
Eqs.(\ref{eq:we_tExBx}-\ref{eq:we_tEsBs}) infinitesimally in the vicinity of $s=0$.
Substituting $\kap_{-}=0$ and $\kap_{+}=\rho^{-1}$ into Eqs.(\ref{eq:cEs_jump_plus_sum}),
and rewriting the $\alp$-terms of Eq.(\ref{eq:we_tEsBs}) for $\tE_s$, we get
\begin{align}
  \bigg[\frac{(\rd_s\vkap)}{g}\{x\rd_s\tE_s(s)-\tE_x(s)\}\bigg]_{s=\pm0}
  &=\delta(s)\{\rd_s\tE_s(+0)-\rd_s\tE_s(-0)\} .
  \label{eq:Es_dels_pm0}
\end{align}
Eq.(\ref{eq:Es_dels_pm0}) cancels the term involving $\delta(s)$
which comes out of $\rd_s^2\tE_s$ in Eq.(\ref{eq:we_tEsBs}).
Thus, by removing the singular term which is involved in Eq.(\ref{eq:we_tEsBs}) for
$\tE_s$ at $s=0$, we get the wave equation (\ref{eq:we_tExsBxs}) which holds for
$s>0$ and $s<0$.
$\tE_x$ is similar to $\tE_s$ in this regard.

Similar to the electric field, we get the discontinuity of $\rd_s\tB_{x,s}$ at $s=0$ from
Eqs.(\ref{eq:No_monopole_FD2}) and (\ref{eq:Ampere_y_FD2}),
\begin{align}
  \rd_s\tB_x(+0)-\rd_s\tB_x(-0)
  &=u_{\pm}\{x\rd_s\tB_x(\pm0)+\tB_s(0)\} ,
  \label{eq:cBx_jump_plus_sum}
   \\
  \rd_s\tB_y(+0)-\rd_s\tB_y(-0)
  &=u_{\pm}\{x\rd_s\tB_y(\pm0)\} ,
  \label{eq:rds_tBy_jump}
   \\
  \rd_s\tB_s(+0)-\rd_s\tB_s(-0)
  &=u_{\pm}\{x\rd_s\tB_s(\pm0)-\tB_x(0)\} .
  \label{eq:cBs_jump_plus_sum}
\end{align}
If the beamline consists of several straight sections and constant bends which have
hard edges (no fringe field) on their entrance and exit as discussed in
section \ref{sec:reflection},
we can separate the wave equations for the components of the field
into those in each bend and straight section as shown in
Eqs.(\ref{eq:we_tEy_pm}) and (\ref{eq:we_tBy_pm}-\ref{eq:we_tExsBxs}).
Then the wave equations in each section no longer involve $\rd_s\vkap$ 
which has the $\delta$-functions with respect to $s$.

\section{Maxwell equations for the Fourier coefficients of the fields}
\label{sec:boundary}

In appendix \ref{sec:HL_field_sdom} we will use Maxwell equations for
the Fourier coefficients of the fields in the frequency domain in order to get
the Fourier coefficients of the horizontal and longitudinal components of the fields
$(\cE_{x,s}^n,\cB_{x,s}^n)$ from the vertical components $(\cE_y^n,\cB_y^n)$.
Through Eqs.(\ref{eq:Four_Exp_plus}-\ref{eq:Four_Exp_minus}), we expand the fields and 
current in the vertical oscillation modes between the upper and lower walls of
the rectangular pipe which has a full height $h$,
\begin{align}
  \tF_{\pm}(y) \to \cF_{\pm}^n ,
    \qquad
  \rd_y\tF_{\pm}(y) \to \pm(-1)^nk_y^n\cF_{\pm}^n ,
    \qquad
  \rd_y^2\tF_{\pm}(y) \to  -(k_y^n)^2\cF_{\pm}^n ,
\end{align}
where the double sign ($\pm$) denotes the vertical parities of the components of
the fields and current as shown in Eqs.(\ref{eq:compo_Apm}) and (\ref{eq:coeff_FJ}).
$k_y^n$ is the $n$th vertical wavenumber of the fields between the upper and lower walls of
the vacuum chamber, given by Eq.(\ref{eq:kyn}).
We expanded the fields in the vertical Fourier modes so that the components of
the fields satisfy Eqs.(\ref{eq:BCy}) which are the boundary conditions of the fields on
the upper and lower walls of the rectangular pipe.

The following are Maxwell equations for the Fourier coefficients of the fields in
the frequency domain,
\begin{alignat}{2}
  \nablav\cdot\Ev=c\mu_0J_0:
  &&\qquad&
   \brd_x\cE_x^n+\brd_s\cE_s^n-(-1)^nk_y^n\cE_y^n=Z_0\cJ_0^n ,
  \label{eq:Gauss_cF}
  \\
  \nablav\cdot\Bv=0:
  &&\qquad&
   \brd_x\cB_x^n+\brd_s\cB_s^n+(-1)^nk_y^n\cB_y^n=0 ,
  \label{eq:No_monopole_cF}
  \\
  \nablav\times\Ev+\rd_t\Bv=0:
  &&\qquad&
  (-1)^nk_y^n\cE_s^n-\brd_s\cE_y^n-ik\beta c\cB_x^n=0 ,
    \label{eq:Faraday_x_cF}
  \\
  &&\qquad&
  \brd_s\cE_x^n-\brd_x\cE_s^n-ik\beta c\cB_y^n=0 ,
    \label{eq:Faraday_y_cF}
  \\
  &&\qquad&
  \rd_x\cE_y^n-(-1)^nk_y^n\cE_x^n-ik\beta c\cB_s^n=0 ,
    \label{eq:Faraday_s_cF}
  \\
  \nablav\times\Bv-c^{-2}\rd_t\Ev=\mu_0\Jv:
  &&\qquad&
   -(-1)^nk_y^nc\cB_s^n-\brd_sc\cB_y^n+ik\beta\cE_x^n=Z_0 \cJ_x^n ,
    \label{eq:Ampere_x_cF}
  \\
  &&\qquad&
  \brd_sc\cB_x^n-\brd_xc\cB_s^n+ik\beta\cE_y^n=Z_0\cJ_y^n ,
    \label{eq:Ampere_y_cF}
  \\
  &&\qquad&
  \rd_xc\cB_y^n+(-1)^nk_y^nc\cB_x^n+ik\beta \cE_s^n=Z_0 \cJ_s^n .
    \label{eq:Ampere_s_cF}
\end{alignat}
The Fourier coefficients of the fields $(\vec{\cE}^n,\vec{\cB}^n)$ and current
$(\cJ_0^n,\vec{\cJ}^n)$ have the arguments $(x,s,k)$.
$\beta$ $(=v/c)$ is the relative speed of the reference particle,
which is a nonzero positive constant as in Eq.(\ref{eq:z}).
From Eq.(\ref{eq:Gauss_cF}) and Eqs.(\ref{eq:Ampere_y_cF}-\ref{eq:Ampere_s_cF}),
we get the boundary conditions of the fields on the sidewalls of the rectangular pipe,
\begin{align}
  \cE_y^n=\cE_s^n=c\cB_x^n=0 ,
    \qquad
  \brd_x\cE_x^n=\brd_xc\cB_s^n=\rd_xc\cB_y^n=0
  \qquad\text{at}~~
  x=x_a,x_b .
  \label{eq:BC_side_cF}
\end{align}

As shown in appendix \ref{sec:HL_field_sdom}, we can get the expressions of
$\cE_{x,s}^n$ and $\cB_{x,s}^n$ from those of $\cE_y^n$ and $\cB_y^n$ given by
Eq.(\ref{eq:cEBy_sol}) or (\ref{eq:cEy_cBy_cmpl}).
From Eqs.(\ref{eq:Faraday_x_cF}) and (\ref{eq:Ampere_s_cF}),
we get the expressions of $\cB_x^n$ and $\cE_s^n$ using $\cE_y^n$ and $\cB_y^n$.
Similarly, we get the expressions of $\cE_x^n$ and $\cB_s^n$ from
Eqs.(\ref{eq:Faraday_s_cF}-\ref{eq:Ampere_x_cF}),
\begin{align}
   \bigg\{{ c\cB_x^n \atop \cE_s^n }\bigg\}
  &=\frac{1}{(k_r^n)^2}
    \bigg[
       \bigg\{{ (-1)^nk_y^n \atop ik\beta }\bigg\}
      (\rd_xc\cB_y^n-Z_0\cJ_s^n)
     + \bigg\{{ ik\beta \atop -(-1)^nk_y^n }\bigg\}
      \brd_s\cE_y^n
    \bigg] ,
  \label{eq:cBx}
  \\
   \bigg\{{ \cE_x^n \atop c\cB_s^n }\bigg\}
   &=\frac{-1}{(k_r^n)^2}
     \bigg[
         \bigg\{{ (-1)^nk_y^n \atop ik\beta }\bigg\}
        \rd_x\cE_y^n
       + \bigg\{{ ik\beta \atop -(-1)^nk_y^n }\bigg\}
        (\brd_sc\cB_y^n+Z_0\cJ_x^n)
     \bigg] .
  \label{eq:cBs}
\end{align}
$k_r^n$ is the radial wavenumber given by Eq.(\ref{eq:krn}).


\section{Maxwell equations for the fields in the Laplace domain}
\label{sec:Maxeqs_LD}

Laplace transforming the Fourier coefficients of the fields $(\cE^n,\cB^n)$
and current $\cJ^n$ with respect to $s$
through Eqs.(\ref{eq:Laplace}) and (\ref{eq:u_nu}),
Maxwell equations for the Laplace domain fields in the bend are given as follows,
\begin{alignat}{2}
  \nablav\cdot\Ev=c\mu_0J_0:
  &&\qquad&
   \brd_r\mfE_x^n-(-1)^nk_y^n\mfE_y^n+\frac{1}{r}\{i\nu\mfE_s^n-\cE_s^n(0)\}
   =Z_0\mfJ_0^n ,
  \label{eq:LD_Gauss}
  \\
  \nablav\cdot\Bv=0:
  &&\qquad&
   \brd_r\mfB_x^n+(-1)^nk_y^n\mfB_y^n+\frac{1}{r}\{i\nu\mfB_s^n-\cB_s^n(0)\}=0 ,
  \label{eq:LD_No_monopole}
  \\
  \nablav\times\Ev+\rd_t\Bv=0:
  &&\qquad&
  (-1)^nk_y^n\mfE_s^n-\frac{1}{r}\{i\nu\mfE_y^n-\cE_y^n(0)\}-ik\beta c\mfB_x^n=0 ,
    \label{eq:LD_Faraday_x}
  \\
  &&\qquad&
  \frac{1}{r}\{i\nu\mfE_x^n-\cE_x^n(0)\}-\brd_r\mfE_s^n-ik\beta c\mfB_y^n=0 ,
    \label{eq:LD_Faraday_y}
  \\
  &&\qquad&
  \rd_r\mfE_y^n-(-1)^nk_y^n\mfE_x^n-ik\beta c\mfB_s^n=0 ,
    \label{eq:LD_Faraday_s}
  \\
  \nablav\times\Bv-c^{-2}\rd_t\Ev=\mu_0\Jv:
  &&\qquad&
   \!-(-1)^nk_y^nc\mfB_s^n-\frac{1}{r}\{i\nu c\mfB_y^n-c\cB_y^n(0)\}
   +ik\beta \mfE_x^n=Z_0\mfJ_x^n ,
    \label{eq:LD_Ampere_x}
  \\
  &&\qquad&
  \frac{1}{r}\{i\nu c\mfB_x^n-c\cB_x^n(0)\}-\brd_rc\mfB_s^n
  +ik\beta\mfE_y^n=Z_0\mfJ_y^n ,
    \label{eq:LD_Ampere_y}
  \\
  &&\qquad&
  \rd_rc\mfB_y^n+(-1)^nk_y^nc\mfB_x^n+ik\beta\mfE_s^n=Z_0\mfJ_s^n .
    \label{eq:LD_Ampere_s}
\end{alignat}
$\cE^n(0)=\cE^n(r,0)$ and $\cB^n(0)=\cB^n(r,0)$ are
the initial values of $\cE^n$ and $\cB^n$ at $s=0$.
As defined in Eqs.(\ref{eq:J_LD}),
$(\mfE^n,\mfB^n)$ and $\mfJ^n$ are the fields and current 
in the Laplace domain, which depend on $(r,\nu,k)$ in general.
Using Eqs.(\ref{eq:LD_Gauss}-\ref{eq:LD_Ampere_s}), the $(x,s)$-components of the fields in 
the Laplace domain are gotten from the vertical ones,
\begin{align}
   \bigg\{{ c\mfB_x^n \atop \mfE_s^n }\bigg\}
  &=\frac{1}{(k_r^n)^2}
    \bigg[
      \bigg\{{ (-1)^nk_y^n \atop ik\beta }\bigg\}
      (\rd_rc\mfB_y^n-Z_0\mfJ_s^n)
     + \bigg\{{ ik\beta \atop -(-1)^nk_y^n }\bigg\}
      \frac{i\nu\mfE_y^n-\cE_y^n(0)}{r}
    \bigg] ,
  \label{eq:mfBx_EyBy}
  \\
   \bigg\{{ \mfE_x^n \atop c\mfB_s^n }\bigg\}
   &=\frac{-1}{(k_r^n)^2}
     \bigg[
        \bigg\{{ (-1)^nk_y^n \atop ik\beta }\bigg\}
        \rd_r\mfE_y^n
       +\bigg\{{ ik\beta \atop -(-1)^nk_y^n }\bigg\}
        \bigg(\frac{i\nu c\mfB_y^n-c\cB_y^n(0)}{r}+Z_0\mfJ_x^n\bigg)
     \bigg] .
  \label{eq:mfBs_EyBy}
\end{align}
We use Eqs.(\ref{eq:mfBx_EyBy}-\ref{eq:mfBs_EyBy}) in deriving
Eqs.(\ref{eq:mfBsEx}-\ref{eq:mfBxEs}) from Eqs.(\ref{eq:mfEy_rwrtn}-\ref{eq:mfBy_rwrtn}) 
without solving Eqs.(\ref{we_mfEBx}-\ref{we_mfEBs}) which are the wave equations for
$\mfE_{x,s}^n$ and $\mfB_{x,s}^n$.
On the other hand, appendix \ref{sec:we_xs} shows the straightforward solution of
Eqs.(\ref{we_mfEBx}-\ref{we_mfEBs}) using Eqs.(\ref{eq:mfEB_pm})
which are the eigenfunctions of the operator of the wave equations.

\section{Equation of continuity for the electric current}
\label{sec:eqcontin}

Using $\nablav$ given by Eq.(\ref{eq:xys_basis}), the equation of continuity for
the current in the time domain is given as
\begin{align}
  \rd_t(J_0/c)+\nablav\cdot\Jv=0:
   \qquad
  \rd_t(J_0/c)+\brd_xJ_x+\brd_sJ_s+\rd_yJ_y=0 .
  \label{eq:eq_cont_td}
\end{align}
The continuity equation for the current in the frequency domain $\tJ=(\tJ_0,\tJv)$ is 
given through Eq.(\ref{eq:Fourier_trans}),
\begin{align}
  -ik\beta\tJ_0+\brd_x\tJ_x+\brd_s\tJ_s+\rd_y\tJ_y=0 .
  \label{eq:eq_cont_fd}
\end{align}
Expanding $\tJ$ in the vertical Fourier modes through
Eqs.(\ref{eq:Four_Exp_plus}-\ref{eq:Four_Exp_minus}),
the $n$th Fourier coefficients satisfy
\begin{align}
  -ik\beta\cJ_0^n+\brd_x\cJ_x^n+\brd_s\cJ_s^n-(-1)^nk_y^n\cJ_y^n=0 .
  \label{eq:eq_cont_vmode}
\end{align}
Laplace transforming $\cJ^n=(\cJ_0^n,\cJ_{x,y,s}^n)$ with respect to $s$ in
a straight section or a constant bend through Eq.(\ref{eq:Lap_trans_kap}) or
Eq.(\ref{eq:Laplace}), the equation of continuity for the current in
the Laplace domain is given as follows,
\begin{alignat}{2}
  &
  -ik\beta\mfJ_0^n+\rd_x\mfJ_x^n+\{ik_s\mfJ_s^n-\cJ_s^n(0)\}
  -(-1)^nk_y^n\mfJ_y^n=0
  &&\qquad\text{(straight section)} ,
   \label{eq:Eq_cont_LD_st}
   \\
  &
  -ik\beta\mfJ_0^n+\brd_r\mfJ_x^n+\frac{1}{r}\{i\nu\mfJ_s^n-\cJ_s^n(0)\}
  -(-1)^nk_y^n\mfJ_y^n=0
  &&\qquad\text{(bending section)} .
   \label{eq:Eq_cont_LD}
\end{alignat}
$\cJ_s^n(0)$ is the initial value of $\cJ_s^n$ at $s=0$.
We use Eq.(\ref{eq:Eq_cont_LD}) in rewriting $\mfE_y^n$ and $\mfB_y^n$ from
Eqs.(\ref{eq:mfEBy_solution}) into Eqs.(\ref{eq:mfEy_rwrtn}-\ref{eq:mfBy_rwrtn}).
Since we define the Laplace transform in the constant bend, given by Eqs.(\ref{eq:Laplace}),
by dividing by $\rho$, $\mfJ^n$ in Eq.(\ref{eq:Eq_cont_LD_st}) corresponds to $\rho\mfJ^n$ 
in Eq.(\ref{eq:Eq_cont_LD}), \ie,
\begin{align}
  \mfJ^n(\text{straight})
  =\lim_{\rho\to\infty}\rho\mfJ^n(\text{bend}) .
   \label{eq:LD_st_bend}
\end{align}
Also, the Laplace domain fields in the straight and bending sections are related
similar to Eq.(\ref{eq:LD_st_bend}).

\clearpage

\section{Transient fields in a straight rectangular pipe}
\label{sec:travel_wave}

We will find an expression of an electromagnetic field which propagates in a straight 
rectangular pipe together with an arbitrary electric current flowing in
the positive $s$-direction.
There are several ways to find the expression of the field propagating in a straight pipe.
We will do it in a similar way to finding the solution of the field in the bending section 
on purpose to explain how we did it in sections \ref{sec:we}-\ref{sec:FD_field}.
In formulating the field in the straight section,
we use the same symbols as those in the bend:
$(\tEv,\tBv,\tJ)$, $(\vec{\cE}^n,\vec{\cB}^n,\cJ^n)$, $(\cG_{\pm}^n,\cC_{\pm}^n)$, etc.
Excluding the whispering gallery modes which we discussed in
sections \ref{sec:wgm} and \ref{sec:iwgm}, all the quantities in the bend must agree with 
those in the straight section in the limit of $\rho\to\infty$ as in
Eqs.(\ref{eq:lim_mfGpm}) and (\ref{eq:lim_cRcX}).
But sometimes we use different symbols, \eg, the window functions in the bend
($\Th_{\pm}^{n\ell}$) and the one in the straight section ($\bTh_{\ell}^{n}$), which are 
respectively given by Eqs.(\ref{eq:Th_def}-\ref{eq:bTh_def}) and (\ref{eq:def_Th_ell}).

In this appendix we define $s=0$ as
the longitudinal position of the entrance of the straight section.
We assume a perfectly conducting straight rectangular pipe
which is semi-infinite in the absence of exit, \ie,
the beam pipe extends semi-infinitely along the straight $s$-axis from $s=0$ to $\infty$.
It follows that we do not take into account the reflection of the field
at the exit of the straight section, which connects to the next bending section.
Similar to assumption (e) in section \ref{sec:assumption},
the initial field at the entrance of the straight pipe $(s=0)$ is arbitrary
if it satisfies Maxwell equations and the boundary condition of
the straight rectangular pipe.
The solution of the field represents the exact relation of the field at an arbitrary 
position in the straight pipe ($s>0$) to the initial field at $s=+0$.
The solution of the field in the straight pipe, which we will find in
this appendix, does not necessarily represent the radiation field emitted in
the preceding bending section.

\subsection{Solution of the fields in the Laplace domain}
\label{sec:LD_field_str}

We use the Cartesian coordinates $\xv=(x,y,s)$
where the basis $(\ev_x,\ev_y,\ev_s)$ does not depend on $s$,
\begin{align}
  \nablav
  =\ev_x\rd_x+\ev_y\rd_y+\ev_s\rd_s
   ,\qquad
  \nablav^2
  =\rd_x^2+\rd_y^2+\rd_s^2 .
\end{align}
The wave equations for the electric and magnetic fields in the frequency domain
(\ref{eq:Fourier_trans}) are given as
\begin{align}
  \{\nablav^2+(k\beta)^2\}\tEv=Z_0\tSv ,
    \qquad
  \{\nablav^2+(k\beta)^2\}c\tBv=Z_0\tTv .
   \label{eq:we_EB_str}
\end{align}
$Z_0$ is the impedance of vacuum.
$\tSv$ and $\tTv$ are the source terms of the fields in the frequency domain,
\begin{alignat}{3}
  \tEv(\xv)
  &=\tE_x\ev_x+\tE_y\ev_y+\tE_s\ev_s ,
    \qquad&
  \tSv(\xv)
  &=\tS_x\ev_x+\tS_y\ev_y+\tS_s\ev_s
  &&=\nablav\tJ_0-ik\beta\tJv ,
   \label{eq:src_E_str}
   \\
  \tBv(\xv)
  &=\tB_x\ev_x+\tB_y\ev_y+\tB_s\ev_s ,
    \qquad&
  \tTv(\xv)
  &=\tT_x\ev_x+\tT_y\ev_y+\tT_s\ev_s
  &&=-\nablav\times\tJv .
   \label{eq:src_B_str}
\end{alignat}
$\tJ_0/c$ and $\tJv$ are the charge and current densities of
the source current in the frequency domain,
\begin{alignat}{2}
  \tS_x(\xv)
  &=\rd_x\tJ_0(\xv)-ik\beta\tJ_x(\xv) ,
    \qquad&
  \tT_x(\xv)
  &=\rd_s\tJ_y(\xv)-\rd_y\tJ_s(\xv) ,
   \label{eq:tSTx_str}
  \\
  \tS_y(\xv)&=\rd_y\tJ_0(\xv)-ik\beta\tJ_y(\xv) ,
    \qquad&
  \tT_y(\xv)
  &=\rd_x\tJ_s(\xv)-\rd_s\tJ_x(\xv) ,
  \\
  \tS_s(\xv)
  &=\rd_s\tJ_0(\xv)-ik\beta\tJ_s(\xv) ,
    \qquad&
  \tT_s(\xv)
  &=\rd_y\tJ_x(\xv)-\rd_x\tJ_y(\xv) .
   \label{eq:tTs_str}
\end{alignat}
We expand the fields and current in the vertical oscillation modes
between the upper-lower walls
through Eqs.(\ref{eq:Four_Exp_plus}-\ref{eq:Four_Exp_minus}).
The $n$th Fourier coefficients of the fields and source terms in the frequency domain are 
given as
\begin{alignat}{2}
  \vec{\cE}^n(x,s)
  &=\cE_x^n\ev_x+\cE_y^n\ev_y+\cE_s^n\ev_s ,
    \qquad&
  \vec{\cS}^n(x,s)
  &=\cS_x^n\ev_x+\cS_y^n\ev_y+\cS_s^n\ev_s ,
   \\
  \vec{\cB}^n(x,s)
  &=\cB_x^n\ev_x+\cB_y^n\ev_y+\cB_s^n\ev_s ,
    \qquad&
  \vec{\cT}^n(x,s)
  &=\cT_x^n\ev_x+\cT_y^n\ev_y+\cT_s^n\ev_s .
\end{alignat}
$\vec{\cE}^n$ and $\vec{\cB}^n$ satisfy the following wave equations,
\begin{align}
  \{\vec{\rd}^2+(k\beta)^2\}\cEv^n=Z_0\cSv^n ,
   \qquad
  \{\vec{\rd}^2+(k\beta)^2\}c\cBv^n=Z_0\cTv^n ,
   \qquad
  \vec{\rd}^2
  =\rd_x^2-(k_y^n)^2+\rd_s^2 .
  \label{eq:we_xprim_drift}
\end{align}
The vertical wavenumber $k_y^n\,(=\pi n/h)$ is given by Eq.(\ref{eq:kyn}).
$\vec{\cS}^n$ and $\vec{\cT}^n$ are the $n$th Fourier coefficients of the source terms in
the wave equations for $\vec{\cE}^n$ and $\vec{\cB}^n$,
\begin{alignat}{2}
  \cS_x^n(x,s)
  &=\rd_x\cJ_0^n-ik\beta\cJ_x^n ,
   \qquad&
  \cT_x^n(x,s)
  &=\rd_s\cJ_y^n-(-1)^nk_y^n\cJ_s^n ,
   \label{eq:cSx_cTx_st}
  \\
  \cS_y^n(x,s)
  &=(-1)^nk_y^n\cJ_0^n-ik\beta\cJ_y^n ,
   \qquad&
  \cT_y^n(x,s)
  &=\rd_x\cJ_s^n-\rd_s\cJ_x^n ,
   \label{eq:cSy_cTy_st}
  \\
  \cS_s^n(x,s)
  &=\rd_s\cJ_0^n-ik\beta\cJ_s^n ,
   \qquad&
  \cT_s^n(x,s)
  &=(-1)^nk_y^n\cJ_x^n-\rd_x\cJ_y^n .
   \label{eq:cSs_cTs_st}
\end{alignat}
$\cJ^n=(\cJ_0^n,\vec{\cJ}^n)$ is the $n$th Fourier coefficient of
the current in the frequency domain $\tJ=(\tJ_0,\tJv)$.

In order to solve Eqs.(\ref{eq:we_xprim_drift}) as an initial value problem
with respect to $s$, we Laplace transform the fields and current with respect to $s$,
\begin{align}
  \mfF^n(\kap)
  =\cL[\cF^n]
  =\int_0^\infty \cF^n(s)e^{-\kap s}ds ,
    \qquad
  \cF^n(s)
  =\int_{\vpi-i\infty}^{\vpi+i\infty}\frac{d\kap}{2\pi i}\mfF^n(\kap)e^{\kap s}
  =\int_{-\infty-i\vpi}^{\infty-i\vpi}\frac{dk_s}{2\pi}\mfF^n(k_s)e^{ik_ss} .
  \label{eq:Lap_trans_kap}
\end{align}
$\cL$ denotes the Laplace transform with respect to $s$.
$\cF^n$ represents the Fourier coefficients of the fields and current.
$\mfF^n$ represents those in the Laplace domain.
$\kap\in\mathbb{C}$ and $\vpi\in\mathbb{R}^{+}$ are
the Laplace variable and the abscissa of convergence.
For convenience, similar to Eq.(\ref{eq:u_nu}),
we change the Laplace variable from $\kap$ to $k_s$,
\begin{align}
  k_s
  =-i\kap
  \in\mathbb{C} ,
    \qquad
  k_x^2=(k\beta)^2-(k_y^n)^2-k_s^2 .
  \label{eq:def_kx}
\end{align}
$k_x\in\mathbb{C}$ is the complex horizontal wavenumber
which is defined using the new Laplace variable $k_s$.
Using $k_s$ instead of $\kap$, the wave equations for the components of the fields in
the Laplace domain are given as
\begin{alignat}{2}
  (\rd_x^2+k_x^2)\mfE_{x,y,s}^n(x,k_s)
  &=Z_0\mfS_{x,y,s}^n(x,k_s) +[(\rd_{s'}+ik_s)\cE_{x,y,s}^n(x,s')]_{s'=+0} ,
   \label{eq:we_mfEy_st_ks}
   \\
  (\rd_x^2+k_x^2)c\mfB_{x,y,s}^n(x,k_s)
  &=Z_0\mfT_{x,y,s}^n(x,k_s) +[(\rd_{s'}+ik_s)c\cB_{x,y,s}^n(x,s')]_{s'=+0} ,
   \label{eq:we_mfBy_st_ks}
\end{alignat}
where $(\mfE^n,\mfB^n)$ and $(\mfS^n,\mfT^n)$ are the Laplace transform of 
the fields $(\cE^n$, $\cB^n)$ and source terms $(\cS^n,\cT^n)$, \ie,
\begin{alignat}{2}
  (\mfE_{x,y,s}^n,\mfB_{x,y,s}^n)
  &=\cL[(\cE_{x,y,s}^n,\cB_{x,y,s}^n)] ,
    \qquad
  (\mfS_{x,y,s}^n,\mfT_{x,y,s}^n)
  =\cL[(\cS_{x,y,s}^n,\cT_{x,y,s}^n)] .
\end{alignat}
The vertical components of the source terms in the Laplace domain are given as
\begin{align}
  \mfS_y^n(x,k_s)
  &=(-1)^nk_y^n\mfJ_0^n(x,k_s)-ik\beta\mfJ_y^n(x,k_s) ,
   \\
  \mfT_y^n(x,k_s)
  &=\rd_x\mfJ_s^n(x,k_s)-ik_s\mfJ_x^n(x,k_s)+\cJ_x^n(x,0) .
\end{align}
As described in section \ref{sec:disconti} and appendix \ref{sec:discontinuity},
the longitudinal derivatives of the fields are discontinuous on the transverse plane that
the curvature of the $s$-axis abruptly changes like a step function.
We can get the initial values of $\rd_{s'}\cE_y^n(s')$ and $\rd_{s'}\cB_y^n(s')$ at $s'=+0$
from those at $s'=-0$ through Eqs.(\ref{eq:cEy_jump_plus_sum}) and (\ref{eq:rds_tBy_jump}).

The fields in the Laplace domain satisfy the following boundary conditions on
the sidewalls of the pipe,
\begin{align}
  \mfE_y^n=\mfE_s^n=c\mfB_x^n=0,
   \qquad
  \rd_xc\mfB_y^n=\rd_xc\mfB_s^n=\rd_x\mfE_x^n=0
  \qquad  \text{at}~~ x=x_{a,b} .
  \label{eq:BC_hx_straight}
\end{align}
According to Eqs.(\ref{eq:BC_hx_straight}), we classify the components of the fields and 
source terms into the following two kinds of groups with respect to
their horizontal parities $(\pm)$,
\begin{alignat}{3}
  \cF_{(+)}^n&=(\cE_y^n,\cE_s^n,c\cB_x^n) ,
   \qquad&
  \mfF_{(+)}^n&=(\mfE_y^n,\mfE_s^n,c\mfB_x^n) ,
   \qquad&
  \mfS_{(+)}^n&=(\mfS_y^n,\mfS_s^n,\mfT_x^n) ,
   \label{eq:cFp_mfFp}
   \\
  \cF_{(-)}^n&=(c\cB_y^n,c\cB_s^n,\cE_x^n) ,
   \qquad&
  \mfF_{(-)}^n&=(c\mfB_y^n,c\mfB_s^n,\mfE_x^n) ,
   \qquad&
  \mfS_{(-)}^n&=(\mfT_y^n,\mfT_s^n,\mfS_x^n) .
   \label{eq:cFm_mfFm}
\end{alignat}
$\cF_{(\pm)}^n$ denotes the Fourier coefficients of the components of the fields
which have the positive/negative parity with respect to $x$ if $x_b=-x_a$.
$\mfF_{(\pm)}^n$ and $\mfS_{(\pm)}^n$ are the components of the fields and the source terms
in the Laplace domain, which have positive/negative parity with respect to $x$.
Solving Eqs.(\ref{eq:we_mfEy_st_ks}-\ref{eq:we_mfBy_st_ks}) with respect to $x$
under the boundary conditions (\ref{eq:BC_hx_straight}),
we get the solutions of the fields in the Laplace domain,
\begin{align}
  \mfF_{(\pm)}^n(x,k_s)
  &=\int_{x_a}^{x_b}dx'\mfG_{\pm}^n(x,x',k_s)
    \big\{
      Z_0\mfS_{(\pm)}^n(x',k_s)
     +[(\rd_{s'}+ik_s)\cF_{(\pm)}^n(x',s')]_{s'=+0}
    \big\} .
  \label{eq:mfEy_strt}
\end{align}
$\mfG_{\pm}^n$ denotes the Green functions of the straight rectangular pipe in
the Laplace domain,
\begin{align}
  \mfG_{\pm}^n(x,x',k_s)
  =\pm\frac{\Xi_{\pm}(x,x',k_x)+\Xi_{\pm}(x',x,k_x)}{k_x p(\hx_b,\hx_a)} ,
  \label{eq:mfGbe_strt}
\end{align}
where
\begin{alignat}{2}
  \Xi_{+}(x,x',k_x)
  &=\theta(x-x')\sin(\hx_b-\hx)\sin(\hx'-\hx_a)
  &&=+\theta(x-x')p(\hx_b,\hx)p(\hx',\hx_a) ,
  \label{eq:Te_strt}
   \\
  \Xi_{-}(x,x',k_x)
  &=\theta(x-x')\cos(\hx_b-\hx)\cos(\hx'-\hx_a)
  &&=-\theta(x-x')r(\hx_b,\hx)q(\hx',\hx_a) .
  \label{eq:Tb_strt}
\end{alignat}
$\hx$, $\hx'$ and $\hx_{a,b}$ are the dimensionless horizontal variables
which are $x$, $x'$ and $x_{a,b}$ normalized by $k_x$,
\begin{align}
  \hx=k_xx ,
    \qquad
  \hx'=k_xx' ,
    \qquad
  \hx_{a,b}=k_xx_{a,b} .
\end{align}
$(p,q,r,s)$ in Eqs.(\ref{eq:mfGbe_strt}-\ref{eq:Tb_strt}) are the cross products of
$\cos\hx$ and $\sin\hx$, which correspond to Eqs.(\ref{eq:CP_pq}-\ref{eq:CP_sr}),
\begin{alignat}{2}
  p(\hx,\hx')
  &=\cos\hx\sin\hx'-\sin\hx\cos\hx'
   =-\sin(\hx-\hx') ,
    \qquad &&
  q(\hx,\hx')
  =\rd_{\hx'}p(\hx,\hx')
  =\cos(\hx-\hx') ,
  \label{eq:CP_p_st}
  \\
  s(\hx,\hx')
  &=\rd_{\hx}\rd_{\hx'}p(\hx,\hx')
   =p(\hx,\hx')
   =-\sin(\hx-\hx') ,
    \qquad &&
  r(\hx,\hx')
  =\rd_{\hx}p(\hx,\hx')
  =-\cos(\hx-\hx') .
  \label{eq:CP_s_st}
\end{alignat}
We rewrite Eqs.(\ref{eq:mfGbe_strt}) using Eqs.(\ref{eq:CP_p_st}-\ref{eq:CP_s_st}), 
\begin{align}
  \mfG_{+}^n(x,x',k_s)
  &=\theta(x-x')\frac{p(\hx_b,\hx)p(\hx',\hx_a)}{k_x p(\hx_b,\hx_a)}
   +\theta(x'-x)\frac{p(\hx_b,\hx')p(\hx,\hx_a)}{k_x p(\hx_b,\hx_a)} ,
  \label{eq:mfGp_st}
  \\
  \mfG_{-}^n(x,x',k_s)
  &=\theta(x-x')\frac{r(\hx_b,\hx)q(\hx',\hx_a)}{k_x s(\hx_b,\hx_a)}
   +\theta(x'-x)\frac{r(\hx_b,\hx')q(\hx,\hx_a)}{k_x s(\hx_b,\hx_a)} .
  \label{eq:mfGm_st}
\end{align}
Eqs.(\ref{eq:mfGp_st}-\ref{eq:mfGm_st}) correspond to Eqs.(\ref{eq:mfGe}-\ref{eq:mfGb})
which are the Green functions of the curved pipe.
$\mfG_{\pm}^n$ satisfies
\begin{align}
  &
  (\rd_x^2+k_x^2)\mfG_{\pm}^n(x,x',k_s)
  =(\rd_{x'}^2+k_x^2)\mfG_{\pm}^n(x,x',k_s)
  =\delta(x-x') ,
   \label{eq:we_mfG_str}
   \\
  &
  \mfG_{\pm}^n(x,x',k_s)
  =\mfG_{\pm}^n(x',x,k_s) ,
    \qquad
  \rd_{x'}\mfG_{\pm}^n(x,x',k_s)
  +\rd_{x}\mfG_{\mp}^n(x,x',k_s)
  =0 .
\end{align}
%


\subsection{Space charge fields in a straight rectangular pipe}
\label{sec:space_charge}

Assuming a rigid bunch which moves at a constant speed $v$ $(=c\beta)$ in
a straight rectangular pipe, we consider the space charge field
in the pipe which is perfectly conducting.
It is gotten as the steady field created by the bunch in the limit of $s\to\infty$.
In section \ref{sec:discussion} we use it as the initial field at the entrance of the bend.
We consider $\mfS_{(\pm)}^n$ given by Eqs.(\ref{eq:cFp_mfFp}-\ref{eq:cFm_mfFm})
which is the Laplace transform of the Fourier coefficients of the source terms
$\cS_{(\pm)}^n$ through Eqs.(\ref{eq:Lap_trans_kap}-\ref{eq:def_kx}).
Assuming a rigid bunch, we find $\mfS_{(\pm)}^n$ in terms of $(x',s')$,
\begin{align}
  \mfS_{(\pm)}^n(x',k_s)
  =\cL[\cS_{(\pm)}^n(x',s')]
   =\frac{\cS_{(\pm)}^n(x',s')}{i(k_s-k)}e^{-iks'} .
\end{align}
$\cS_{(\pm)}^n(x',s')e^{-iks'}$ does not depend on $s'$.
Similarly, $\cF_{(\pm)}^n(x,s)e^{-iks}$ does not depend on $s$ if the field is steady.
We get the Fourier coefficients of the steady field for $s\to\infty$
using the final value theorem,
\begin{align}
  \lim_{s\to\infty}\cF_{(\pm)}^n(x,s)e^{-iks}
  &=\lim_{k_s\to k}i(k_s-k)\mfF_{(\pm)}^n(x,k_s)
   =Z_0\int_{x_a}^{x_b}dx'\mfG_{\pm}^n(x,x',k)\cS_{(\pm)}^n(x',s')e^{-iks'} .
  \label{eq:stdy_strt_FVT}
\end{align}
From Eq.(\ref{eq:mfGbe_strt}), we get the Green functions in the Laplace domain
$\mfG_{\pm}^n(k_s)$ for $k_s=k$,
\begin{align}
  \mfG_{\pm}^n(x,x',k)
  &=-\frac{H_{\pm}^n(x,x')+H_{\pm}^n(x',x)}{\bk_x\sinh(\bk_x^nw)} ,
   \label{eq:mfGpm_k_str}
\end{align}
where
\begin{align}
  H_{+}^n(x,x')
  &=\theta(x-x')\sinh[\bk_x^n(x_b-x)]\sinh[\bk_x^n(x'-x_a)] ,
   \\
  H_{-}^n(x,x')
  &=\theta(x-x')\cosh[\bk_x^n(x_b-x)]\cosh[\bk_x^n(x'-x_a)] .
\end{align}
$i\bk_x^n$ $(=k_x)$ denotes the horizontal wavenumber of the steady field in
the straight rectangular pipe,
\begin{align}
  (\bk_x^n)^2
  =(k_y^n)^2+(k/\gamma)^2
  =-k_x^2
  >0 ,
    \qquad
  \gamma
  =(1-\beta^2)^{-1/2} .
  \label{eq:bkx}
\end{align}
$\bk_x^n$ tends to be the vertical wavenumber $k_y^n$ in the limit of $\gamma\to\infty$.
Considering the fields $\tEv$ and $\tBv$ in the limit of $s\to\infty$,
Eq.(\ref{eq:mfGpm_k_str}) is rewritten as follows,
\begin{align}
  \mfG_{+}^n(x,x',k)
  &=\sum_{m=1}^{\infty}\frac{\cX_{+}^m(x,x')}{(k_s^{mn})^2-k^2} ,
    \qquad
  \mfG_{-}^n(x,x',k)
   =\sum_{m=0}^{\infty}\frac{\cX_{-}^m(x,x')}{(k_s^{mn})^2-k^2} .
  \label{eq:mfGpm_cXpm}
\end{align}
$\cX_{\pm}^m$ denotes the horizontal eigenfunctions of the straight rectangular pipe,
given by Eqs.(\ref{eq:mfXp_new}-\ref{eq:cXm}).
Eqs.(\ref{eq:mfGpm_cXpm}) correspond to Eqs.(\ref{eq:mfGpm_mfRpm})
which are the expressions of $\mfG_{\pm}^n$ in the curved rectangular pipe.
We can rewrite the denominator of Eqs.(\ref{eq:mfGpm_cXpm}) using $\bk_x^n$ as follows,
\begin{align}
  k^2-(k_s^{mn})^2
  &=(k_x^m)^2+(\bk_x^n)^2 .
\end{align}
From Eqs.(\ref{eq:stdy_strt_FVT}), we get the steady fields in the straight rectangular pipe
in the frequency domain,
\begin{align}
  \lim_{s\to\infty}
  \bigg\{{ \tE_y(\xv) \atop c\tB_x(\xv) }\bigg\}
  e^{-iks}
  &=Z_0\int_{x_a}^{x_b}dx'\int_{-h/2}^{h/2}dy'
    \bigg\{{ \tS_y(\xv') \atop  \tT_x(\xv')}\bigg\}
    e^{-iks'}\Ups_{+}^{-}(\xv_\perp,\xv_\perp') ,
   \label{eq:tEy_tBx_str}
   \\
  \lim_{s\to\infty}
  \bigg\{{ c\tB_y(\xv) \atop \tE_x(\xv) }\bigg\}
  e^{-iks}
  &=Z_0\int_{x_a}^{x_b}dx'\int_{-h/2}^{h/2}dy'
    \bigg\{{ \tT_y(\xv') \atop  \tS_x(\xv')}\bigg\}
    e^{-iks'}\Ups_{-}^{+}(\xv_\perp,\xv_\perp') ,
\end{align}
\begin{align}
  \lim_{s\to\infty}\tE_s(\xv)e^{-iks}
  &=Z_0\int_{x_a}^{x_b}dx'\int_{-h/2}^{h/2}dy'
    \tS_s(\xv')e^{-iks'}\Ups_{+}^{+}(\xv_\perp,\xv_\perp') ,
   \\
  \lim_{s\to\infty}c\tB_s(\xv)e^{-iks}
  &=Z_0\int_{x_a}^{x_b}dx'\int_{-h/2}^{h/2}dy'
    \tT_s(\xv')e^{-iks'}\Ups_{-}^{-}(\xv_\perp,\xv_\perp') .
   \label{eq:tBs_str}
\end{align}
$\Ups$ represents the Green functions of the fields in the straight rectangular pipe
$(\Ups_{+}^{+},\Ups_{+}^{-},\Ups_{-}^{+},\Ups_{-}^{-})$:
\begin{align}
  \Ups_{\pm}^{+}(\xv_\perp,\xv_\perp')
  =\sum_{n=1}^{\infty}\mfG_{\pm}^n(x,x',k)\cY_{+}^n(y,y') ,
    \qquad
  \Ups_{\pm}^{-}(\xv_\perp,\xv_\perp')
  =\sum_{n=0}^{\infty}\mfG_{\pm}^n(x,x',k)\cY_{-}^n(y,y') .
\end{align}
The signs of $\Ups_{\pm}$ (subscript) and $\Ups^{\pm}$ (superscript) correspond to
those of $\mfG_{\pm}^n$ and $\cY_{\pm}^n$ respectively.
$\cY_{\pm}^n$ is given by Eqs.(\ref{eq:cYp_0}-\ref{eq:cYm}) which are
the vertical eigenfunctions of the upper-lower walls of the pipe.
The source terms $\tSv$ and $\tTv$ have the derivatives of the current components
as in Eqs.(\ref{eq:tSTx_str}-\ref{eq:tTs_str}).
Since we assume a rigid bunch in appendix \ref{sec:space_charge}, \ie,
$\rd_{s'}(\tJ e^{-iks'})=0$,
we can replace $\rd_{s'}$ for $\tSv(\xv')$ and $\tTv(\xv')$ by $ik$.
In Eqs.(\ref{eq:tEy_tBx_str}-\ref{eq:tBs_str}), integrating the derivatives of
the current components by parts with respect to $x'$ or $y'$, we get
\begin{align}
  \lim_{s\to\infty}\bigg\{{ \tE_y(\xv) \atop c\tB_x(\xv) }\bigg\}e^{-iks}
  &=-Z_0\int_{x_a}^{x_b}dx'\int_{-h/2}^{h/2}dy'
    \bigg\{{ \tS_y^{\dg}(\xv') \atop  \tT_x^{\dg}(\xv')}\bigg\}
    e^{-iks'}\Ups_{+}^{-}(\xv_\perp,\xv_\perp') ,
  \label{eq:tEy_tBx_str_std_dg}
   \\
  \lim_{s\to\infty}\bigg\{{ c\tB_y(\xv) \atop \tE_x(\xv) }\bigg\}e^{-iks}
  &=-Z_0\int_{x_a}^{x_b}dx'\int_{-h/2}^{h/2}dy'
    \bigg\{{ \tT_y^{\dg}(\xv') \atop  \tS_x^{\dg}(\xv')}\bigg\}
    e^{-iks'}\Ups_{-}^{+}(\xv_\perp,\xv_\perp') ,
  \label{eq:tBy_tEx_str_std_dg}
\end{align}
and
\begin{align}
  \lim_{s\to\infty}\tE_s(\xv)e^{-iks}
  &=-Z_0\int_{x_a}^{x_b}dx'\int_{-h/2}^{h/2}dy'
    \tS_s^{\dg}(\xv')e^{-iks'}\Ups_{+}^{+}(\xv_\perp,\xv_\perp') ,
  \label{eq:tEs_str_std_dg}
   \\
  \lim_{s\to\infty}c\tB_s(\xv)e^{-iks}
  &=-Z_0\int_{x_a}^{x_b}dx'\int_{-h/2}^{h/2}dy'
    \tT_s^{\dg}(\xv')e^{-iks'}\Ups_{-}^{-}(\xv_\perp,\xv_\perp') .
  \label{eq:tBs_str_std_dg}
\end{align}
Eqs.(\ref{eq:tEy_tBx_str_std_dg}-\ref{eq:tBs_str_std_dg}) are the differential expressions 
of the steady field in the straight pipe.
$\tSv^{\dg}$ and $\tTv^{\dg}$ are operators
which have the current components as the coefficients,
\begin{align}
  \tSv^{\dg}(\xv')
  =\tJ_0(\xv')\tnablav'+ik\beta\tJv(\xv') ,
    \qquad
  \tTv^{\dg}(\xv')
  =\tJv(\xv')\times\tnablav' ,
    \qquad
  \tnablav'
  =\ev_x\rd_{x'}+\ev_y\rd_{y'}-\ev_sik ,
\end{align}
namely,
\begin{alignat}{3}
  \tS_x^{\dg}
  &=ik\beta\tJ_x+\tJ_0\rd_{x'} ,
   \qquad&
  \tS_y^{\dg}
  &=ik\beta\tJ_y+\tJ_0\rd_{y'} ,
   \qquad&
  \tS_s^{\dg}
  &=ik(\beta\tJ_s-\tJ_0) ,
  \\
  \tT_x^{\dg}
  &=-(ik\tJ_y+\tJ_s\rd_{y'}) ,
   \qquad&
  \tT_y^{\dg}
  &=\tJ_s\rd_{x'}+ik\tJ_x ,
   \qquad&
  \tT_s^{\dg}
  &=\tJ_x\rd_{y'}-\tJ_y\rd_{x'} .
\end{alignat}
From Eq.(\ref{eq:mfGpm_k_str}), $\rd_{x'}\mfG_{-}^n(x,x',k)$ is gotten as follows,
\begin{align}
  \rd_{x'}\mfG_{-}^n(x,x',k)
  &=\theta(x-x')\mfW_{-}^n(x_b,x_a)
   +\theta(x'-x)\mfW_{-}^n(x_a,x_b) ,
   \\
  \mfW_{-}^n(x_1,x_2)
  &=\frac{\cosh[\bk_x^n(x_1-x)]\sinh[\bk_x^n(x_2-x')]}{\sinh(\bk_x^nw)} .
\end{align}

We consider a rigid bunch in the absence of horizontal extent.
Assuming that the thin bunch is moving along the $s$-axis at a constant speed $v$
$(=c\beta)$, the current in the frequency domain is given by Eqs.(\ref{eq:thin_bunch}),
\begin{align}
  \tJ_0(\xv)
  &=cq\delta(x)\psi_y(y)\tlam(s)
   =cq\delta(x)\psi_y(y)\tlam_0e^{iks} ,
   \label{eq:ribbon_str}
    \\
  \tJv(\xv)
  &=\beta\tJ_0(\xv)\ev_s ,
    \qquad
  \tlam_0
  =\tlam(0)
  =\tlam(s)e^{-iks} .
\end{align}
$\tlam_0$ is the bunch spectrum which does not depend on $s$
under the assumption of a rigid bunch.
For Eqs.(\ref{eq:ribbon_str}), the steady electric field in the straight pipe is given as
\begin{align}
  \lim_{s\to\infty}\tEv(\xv)e^{-iks}
  &=\frac{q\tlam_0}{\eps_0h}\sum_{n=1}^{\infty}
    2
    \bigg\{
      \ev_x\mfX_{-}^n(x)
     -\frac{\mfX_{+}^n(x)}{\bk_x^n}\bigg(\ev_y\rd_y+\ev_s\frac{ik}{\gamma^2}\bigg)
    \bigg\}
    \hcY_{+}^n(y) .
   \label{eq:tEs_init}
\end{align}
Similarly, for Eqs.(\ref{eq:ribbon_str}), the steady magnetic field in
the straight pipe is given using Eq.(\ref{eq:tEs_init}) as
\begin{align}
  \lim_{s\to\infty}(\tB_x,\tB_y,\tB_s)ce^{-iks}
  &=\lim_{s\to\infty}(-\tE_y,\tE_x,0)\beta e^{-iks} .
   \label{eq:tBv_ini}
\end{align}
$\hcY_{+}^n$ is the following dimensionless function of $y$,
\begin{align}
  \hcY_{+}^n(y)
  &=\frac{h}{2}\int_{-h/2}^{h/2}dy'\psi_y(y')\cY_{+}^n(y,y')
   =\frac{h}{2}\tpsi_y^n
    \bigg\{{ \cos(k_y^ny) \atop \sin(k_y^ny)}\bigg\} ,
   \qquad
  n
  =\bigg\{{ 2p-1 \atop 2p }\bigg\}
   \in\mathbb{N}
     \quad
   (p\in\mathbb{N}) .
  \label{eq:hcYp_straight}
\end{align}
$\tpsi_y^n$ is the $n$th Fourier coefficient of $\psi_y$ in the vertical Fourier series,
given by Eq.(\ref{eq:tpsi}).
If the beam has no vertical extent as $\psi_y(y)=\delta(y)$,
then $\tpsi_y^{2p-1}=2/h$ and $\tpsi_y^{2p}=0$
as shown in Eq.(\ref{eq:psiy_delta}).
$\mfX_{\pm}^n$ is given as
\begin{align}
  \mfX_{+}^n(x)
  =-\bk_x^n\mfG_{+}^n(x,0,k) ,
    \qquad
  \mfX_{-}^n(x)
  =-[\rd_{x'}\mfG_{-}^n(x,x',k)]_{x'=0} .
   \label{eq:mfHp_mfGp}
\end{align}
Substituting Eq.(\ref{eq:mfGpm_k_str}) into Eqs.(\ref{eq:mfHp_mfGp}), we get
\begin{align}
  \mfX_{+}^n(x)
  &=\theta(x)\frac{\sinh[\bk_x^n(x_b-x)]\sinh[\bk_x^n(-x_a)]}{\sinh(\bk_x^nw)}
   +\theta(-x)\frac{\sinh(\bk_x^nx_b)\sinh[\bk_x^n(x-x_a)]}{\sinh(\bk_x^nw)} ,
   \label{eq:mfHp}
   \\
  \mfX_{-}^n(x)
  &=\theta(x)\frac{\cosh[\bk_x^n(x_b-x)]\sinh[\bk_x^n(-x_a)]}{\sinh(\bk_x^nw)}
   -\theta(-x)\frac{\sinh(\bk_x^nx_b)\cosh[\bk_x^n(x-x_a)]}{\sinh(\bk_x^nw)} .
   \label{eq:mfHm}
\end{align}
$\bk_x^n$ is given by Eq.(\ref{eq:bkx}).
$\mfX_{\pm}^n$ satisfies the following equations,
\begin{alignat}{2}
  \rd_x\mfX_{+}^n(x)+\bk_x^n\mfX_{-}^n(x)
  &=0,
   \qquad&
  \{\rd_x^2-(\bk_x^n)^2\}\mfX_{+}^n(x)
  &=-\bk_x^n\delta(x) ,
   \\
  \rd_x\mfX_{-}^n(x)+\bk_x^n\mfX_{+}^n(x)
  &=\delta(x),
   \qquad&
  \{\rd_x^2-(\bk_x^n)^2\}\mfX_{-}^n(x)
  &=\rd_x\delta(x) .
\end{alignat}
From the longitudinal component of Eq.(\ref{eq:tEs_init}), we get $\cZ_0\,[\Omega/{\rm m}]$ 
which is the longitudinal impedance of the space charge field per unit length in
the straight rectangular pipe,
\begin{align}
  \cZ_0(\xv_{\perp},k)
  &=-\frac{Z_0}{h\beta}
    \sum_{n=1}^{\infty}\ccE_s^n(x,0)\hcY_{+}^n(y) ,
   \qquad
  \ccE_s^n(x,0)
  =\ccE_s^n(x,s)e^{-iks}
  =-\frac{2ik}{\gamma^2}\cd\frac{\mfX_{+}^n(x)}{\bk_x^n} .
  \label{eq:cZ0_straight}
\end{align}
$\cZ_0$ is purely imaginary since we assume that the beam pipe is perfectly conducting, \ie,
the energy of the electromagnetic field is confined in the pipe
(not dissipated to the outside of the system).

\subsection{Fourier coefficients of the fields in a straight pipe}

We go back to the discussion in appendix \ref{sec:LD_field_str} where we derived
the transient fields in the Laplace domain, which are created by an arbitrary beam
moving in the straight pipe.
Inverting Eq.(\ref{eq:mfEy_strt}) through the second equation of (\ref{eq:Lap_trans_kap}),
we get the Fourier coefficients of the vertical components of the fields,
\begin{align}
  \cE_y^n(x,s)
  &=\int_{x_a}^{x_b}dx'
    \Big[\cD_y^n(x')\cG_{+}^n(x,x',s)+Z_0\int_0^\infty ds'\cS_y^n(x',s')\cG_{+}^n(x,x',s-s')
    \Big] ,
  \label{eq:cEy_strt}
   \\
  c\cB_y^n(x,s)
  &=\int_{x_a}^{x_b}dx'
    \Big[\cA_y^n(x')\cG_{-}^n(x,x',s)+Z_0\int_0^\infty ds'\cT_y^n(x',s')\cG_{-}^n(x,x',s-s')
    \Big] .
  \label{eq:cBy_strt}
\end{align}
$\cD_y^n$ and $\cA_y^n$ are the following operators which have the initial fields at
the entrance of the straight pipe,
\begin{align}
  \cD_y^n(x')
  =[\rd_{s'}\cE_y^n(s')+\cE_y^n(s')\rd_s]_{s'=+0} ,
    \qquad
  \cA_y^n(x')
  =[\rd_{s'}c\cB_y^n(s')+c\cB_y^n(s')\rd_s]_{s'=+0} .
\end{align}
They have the longitudinal operator $\rd_s$ which acts on the Green functions
$\cG_{\pm}^n$ in the straight pipe,
\begin{align}
  \cG_{\pm}^n(x,x',\vsig)
  &=\int_{-\infty-i\vpi}^{\infty-i\vpi}\frac{dk_s}{2\pi}
    \mfG_{\pm}^n(x,x',k_s)e^{ik_s\vsig},
   \qquad
  \vsig=
  \left\{
  \begin{array}{l}
    s-s'\in\mathbb{R}  \\
    s\in\mathbb{R}^{+}
  \end{array}
  \right. .
  \label{eq:cGbe_strt}
\end{align}
The longitudinal variable $\vsig$ represents $s-s'$ and $s$,
similar to $\vsig$ in Eq.(\ref{eq:cGeb_def}).
$s-s'$ can be either positive or negative depending on the positions of
the observation point $(s)$ and the source particle $(s')$ of the fields.

\begin{figure}[h]
  \begin{center}
    \includegraphics[scale=0.30,clip]{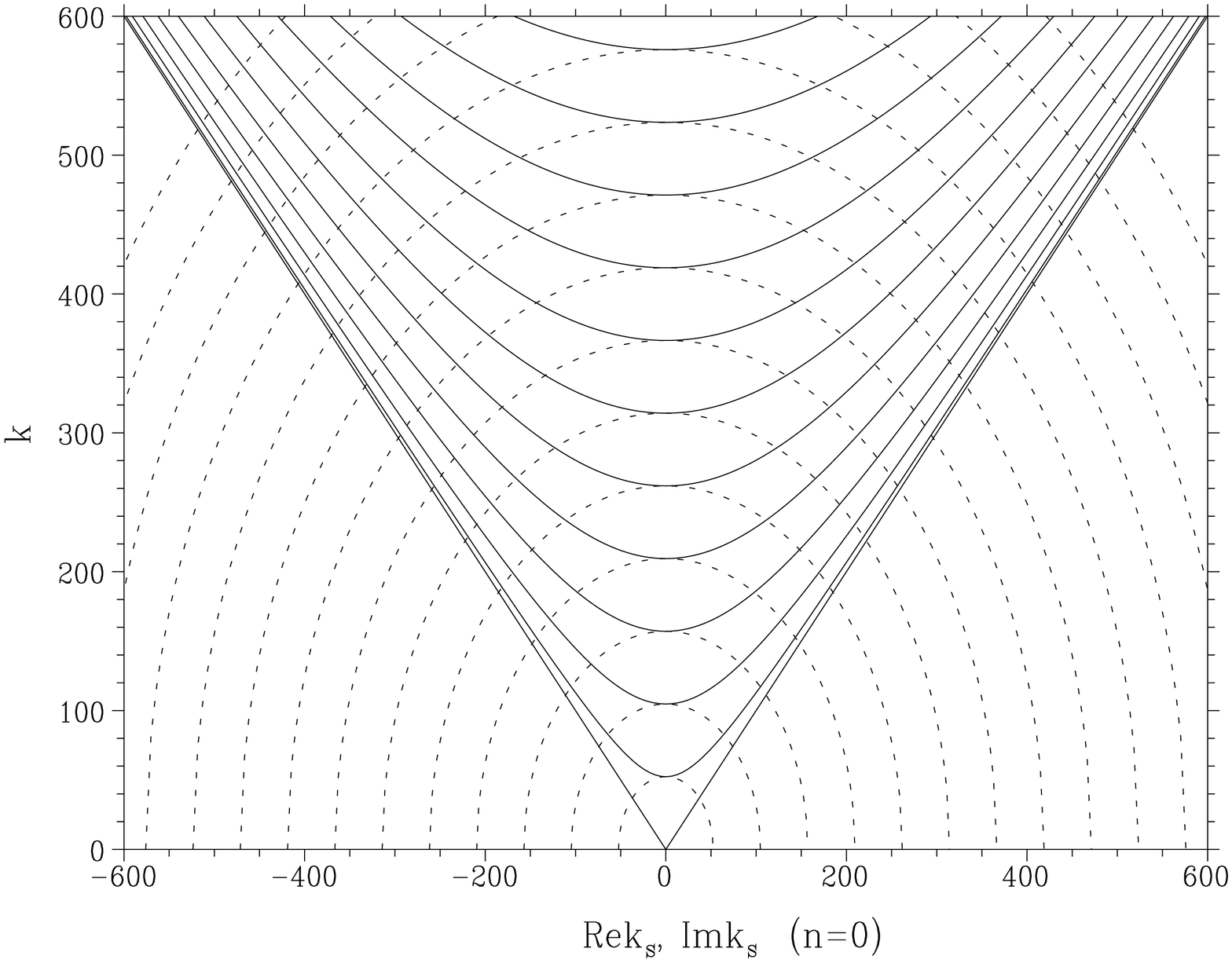}~~
    \includegraphics[scale=0.30,clip]{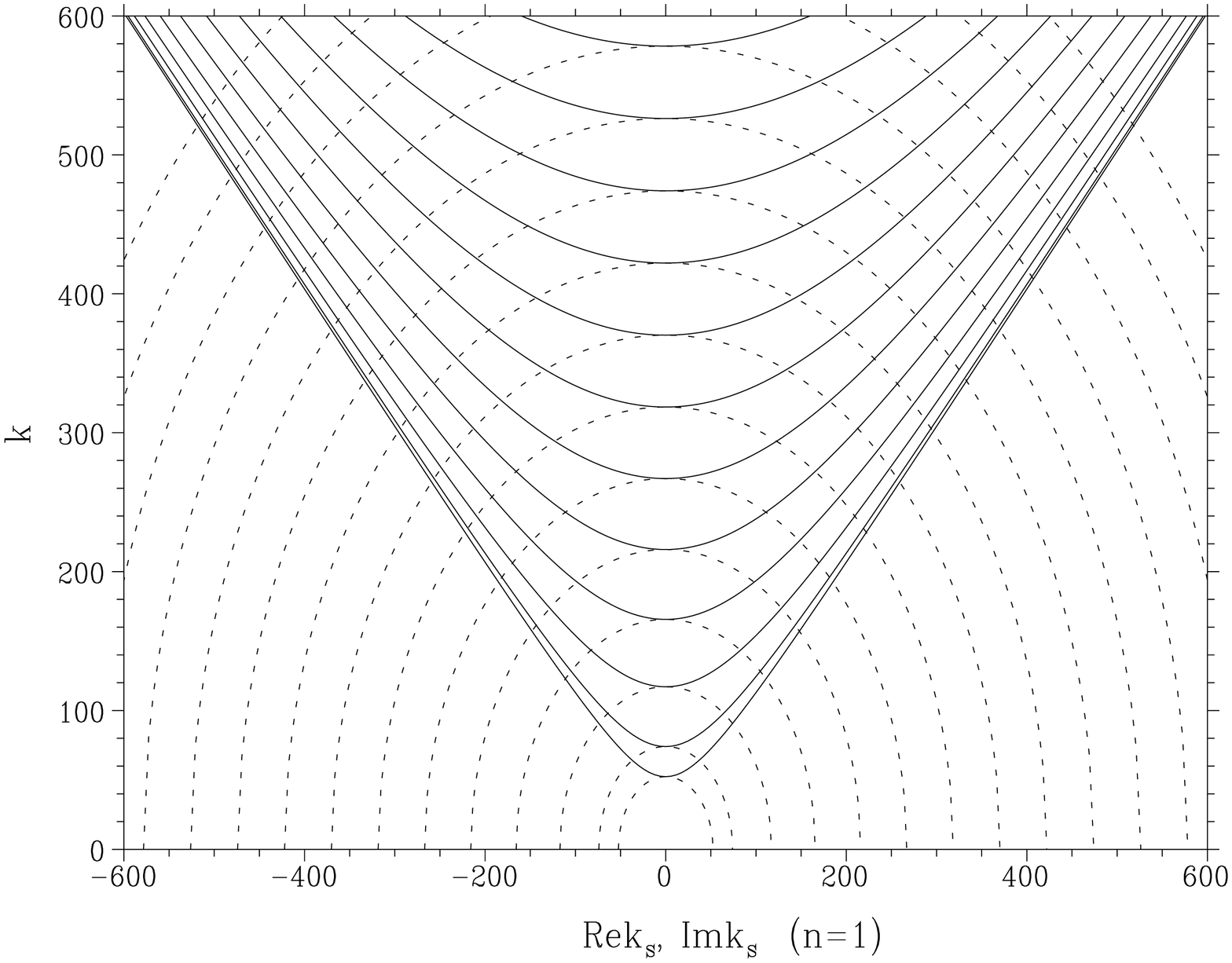}
     \\
    \vspace{2mm}
    \includegraphics[scale=0.30,clip]{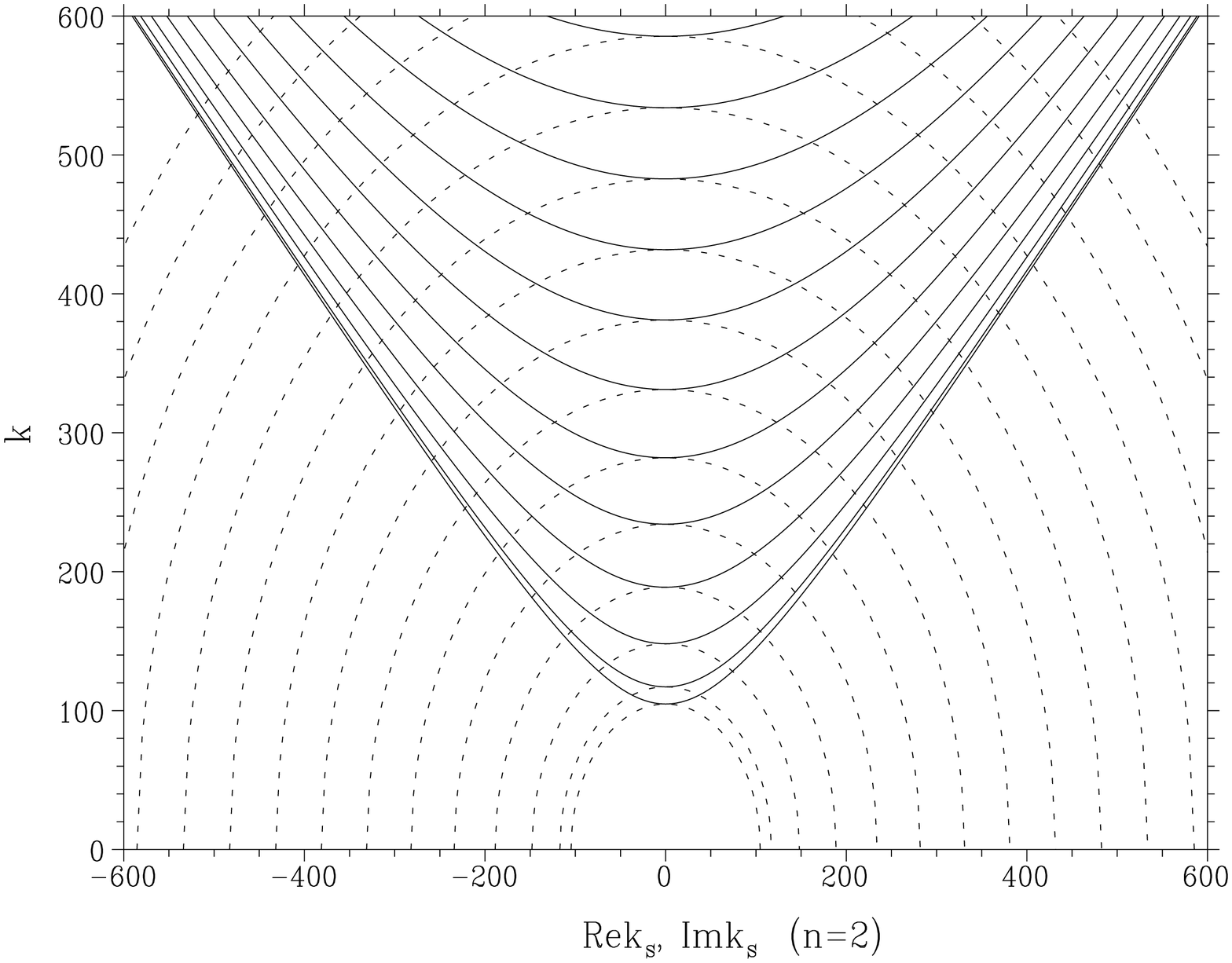}~~
    \includegraphics[scale=0.30,clip]{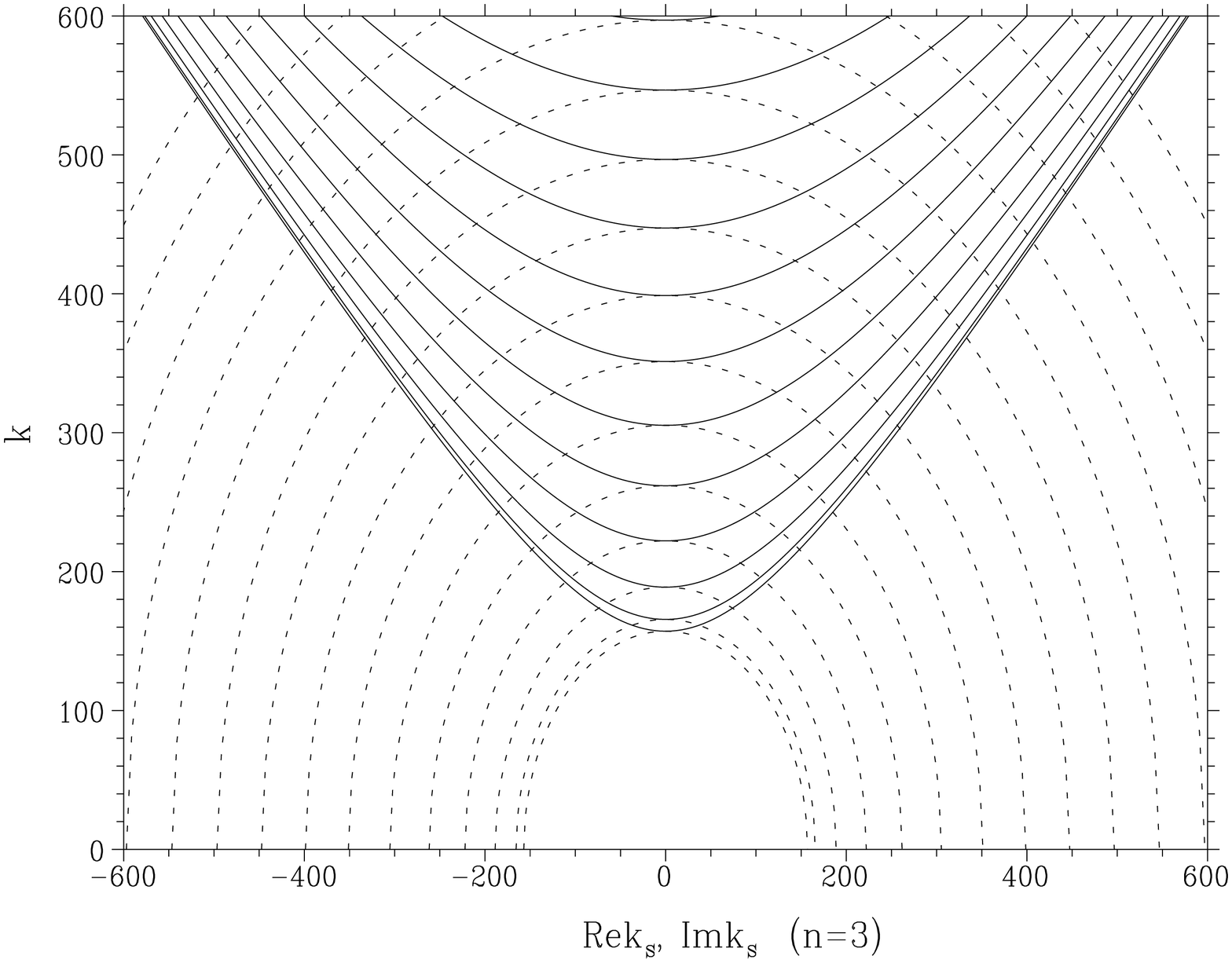}
    \caption[Pole structure of the transient field in a straight pipe]
       {\small
       Pole structure of the field in a straight rectangular pipe
       for $m\in\mathbb{Z}_0^{+}$ and $n\in\{0,1,2,3\}$.
       The vertical axis is the wavenumber $k\,[{\rm m}^{-1}]$.
       The horizontal axis represents $\Re k_s$ (solid curves) and
       $\Im k_s$ (dotted curves) in units of $[{\rm m}^{-1}]$.
       The solid and dotted curves show respectively the real and imaginary poles $k_s^{mn}$
       in the $k_s$-plane for various $k$ and $m$ with $n$ fixed in each figure.
       The values of $k$ at the bottom of the solid curves (which are also the vertices of 
       the dotted curves) are the cutoff wavenumbers of the straight pipe,
       given by Eq.(\ref{eq:hkmn}).
       The solid curves tend to appear toward the upper region (larger $k$)
       for larger $m$; the lowest solid curve in each figure is the zeroth pole ($m=0$).
       The parameters are given as $w=h=6\,{\rm cm}$ and $\beta=1$.
       }
    \label{fig:poles_str}
  \end{center}
\end{figure}

In order to calculate the Bromwich integral (\ref{eq:cGbe_strt}) analytically,
we examine the singularity of $\mfG_{\pm}^n$ in the $k_s$-plane.
At first, $\mfG_{\pm}^n$ has no branch point in the $k_s$-plane.
The points such that $k_x=0$, \ie,
\begin{align}
  k_s=\pm\{(k\beta)^2-(k_y^n)^2\}^{1/2}
  \label{eq:ks_0n}
\end{align}
are not the branch points of $\mfG_{\pm}$ in the $k_s$-plane
since $\mfG_{\pm}$ has $k_x\sin(k_xw)$ in the denominator.
In addition, the points given by Eq.(\ref{eq:ks_0n}) are not the poles of $\mfG_{+}$
since its numerator (\ref{eq:Te_strt}) also becomes zero at these points unlike
Eq.(\ref{eq:Tb_strt}).
In the $k_s$-plane, $\mfG_{+}$ has poles at points such that
\begin{align}
  \sin(k_xw)=0
   \quad\text{and}\quad
  k_x\ne0
   \qquad\Lra\qquad
  k_x=\frac{m\pi}{w}\equiv k_x^m
   \qquad
  (m\in\mathbb{N}~\,\text{for}~\mfG_{+}^n) .
  \label{eq:kxm}
\end{align}
$m$ is the mode number of the horizontal oscillation in the straight rectangular pipe.
$k_x^m$ is the horizontal eigen-wavenumber of the pipe.
The points given by Eq.(\ref{eq:ks_0n}) are the poles of $\mfG_{-}^n$
in the $k_s$-plane.
That is, $\mfG_{-}^n$ has the zeroth horizontal mode unlike $\mfG_{+}^n$.
Accordingly, for a given $n$, $\mfG_{\pm}^n$ has poles at $k_s=k_s^{mn}$, where
\begin{align}
  (k_s^{mn})^2
  =(k\beta)^2-(k_x^m)^2-(k_y^n)^2
  \in\mathbb{R} ,
    \qquad
  m\in(\mathbb{N},\mathbb{Z}_0^{+})
   ~~\text{for}~~
  (\mfG_{+}^n,\mfG_{-}^n) .
 \label{eq:ks_mn}
\end{align}
Eq.(\ref{eq:ks_mn}) is plotted in Fig.\ref{fig:poles_str} (p.\pageref{fig:poles_str})
which shows the dispersion relation of the field in the straight rectangular pipe.
$k_s^{mn}$ is the longitudinal wavenumber of the field in the straight rectangular pipe,
which corresponds to $\nu_m^n/\rho$ and $\mu_m^n/\rho$ in the curved pipe.
$\mfG_{\pm}^n$ has poles only on the axes of the $k_s$-plane, \ie,
$k_s^{mn}\in\mathbb{A}=\{\mathbb{R},i\mathbb{R}\}$.
$\mfG_{-}^n$ has the poles at the same points as $\mfG_{+}^n$ except
the zeroth poles at the points given by Eq.(\ref{eq:ks_0n})
which corresponds to $m=0$ in Eq.(\ref{eq:ks_mn}).
The poles $k_s^{mn}$ are symmetrically located with respect to the origin of
the $k_s$-plane.
A pair of the symmetrically located simple poles couples at the origin and
becomes a pole of the second order when $k=\hk_m^n$ which is the cutoff wavenumber of
the straight rectangular pipe,
\begin{align}
  \hk_m^n
  =\frac{1}{\beta}\{(k_x^m)^2+(k_y^n)^2\}^{1/2} ,
    \qquad
  k_x^m
  =\frac{m\pi}{w} ,
    \qquad
  k_y^n
  =\frac{n\pi}{h}
  \qquad
  (m,n\in\mathbb{Z}_0^{+}) .
  \label{eq:hkmn}
\end{align}
$\hk_m^n$ is the transition wavenumber at which the $(m,n)$th longitudinal mode
interchanges from a damped mode ($k_s^{mn}\in i\mathbb{R}$) to an oscillatory mode
($k_s^{mn}\in\mathbb{R}$) through the origin ($k_s=0$).
As shown in appendix \ref{sec:analytic}, we can treat the pole of the second order
at $k_s=0$ in a similar manner to a pair of the simple poles on the real axis.
In the straight pipe, $\cG_{+}^n$ and $\cG_{-}^n$ have the common cutoff wavenumber
$\hk_m^n$ unlike $k=(k_m^n,\bk_m^n)$ given by Eqs.(\ref{eq:kmn}-\ref{eq:bkmn})
which are the cutoff wavenumbers respectively for $\cG_{+}^n$ and $\cG_{-}^n$ in
the curved pipe.

$k_s^{mn}$ is either real or purely imaginary, depending on $k$, $m$ and $n$.
When $|k\beta|<k_y^n$, all the poles $k_s^{mn}$ for $\forall m\in\mathbb{Z}_0^+$ are
purely imaginary; the field does not propagate in the pipe.
Conversely, when $|k\beta|>k_y^n$, the field in the straight pipe can have
oscillatory modes, depending on $m$.
We define $m_0$ as the largest $m$ such that $k_s^{mn}\in\mathbb{R}$,
which corresponds to the number of the poles on the positive real axis in the $k_s$-plane,
\begin{align}
  m_0
  =\bigg\lfloor \frac{k_r^n}{k_x^1}\bigg\rfloor
  =\Big\lfloor \frac{w}{\pi}\{(k\beta)^2-(k_y^n)^2\}^{1/2}\Big\rfloor
   \qquad
  (|k\beta|>k_y^n) .
   \label{eq:Re_poles_st}
\end{align}
${\lfloor\xi\rfloor}$ is the floor function which denotes the greatest integer
less than $\xi\in\mathbb{R}_0^{+}$.
The condition $|k\beta|>k_y^n$ is not sufficient but necessary so that
$k_s^{mn}\in\mathbb{R}$.
Eq.(\ref{eq:Re_poles_st}) is the asymptotic limit of Eqs.(\ref{eq:Re_poles}) for
$\rho\to\infty$.

We consider the contours of the Bromwich integral (\ref{eq:cGbe_strt}) in the $k_s$-plane.
Since the fields do not diverge exponentially for $s\to\infty$,
it is enough to set $\vpi=+0$ in Eqs.(\ref{eq:cGbe_strt}), \ie,
the contour goes along the real axis infinitesimally under it.
For the integral involving $e^{ik_ss}$, the contour must close as
an infinite semicircle in the upper half-plane since $s>0$.
On the other hand, for the integral involving $e^{ik_s(s-s')}$, the contour must close
either in the upper or lower half plane, depending on the sign of $s-s'$.
Therefore, for $\vsig=s-s'$, we must separate Eq.(\ref{eq:cGbe_strt}) into the cases for
$\vsig>0$ and $\vsig<0$, similar to Fig.\ref{fig:contour} (p.\pageref{fig:contour}).
As seen from Eq.(\ref{eq:ks_mn}), the number of the real poles of $\mfG_{\pm}^n$ in
the $k_s$-plane depends on $k$ and $n$.
In order to fix the numbers of the real poles for a given $n$, we define
the rectangular window functions $\bTh_\ell^n$ for the straight pipe,
\begin{align}
  \bTh_{\ell}^n(k)=\theta(|k|-\hk_{\ell-1}^n)-\theta(|k|-\hk_{\ell}^n)
  \qquad(\ell\in\mathbb{N}),
    \qquad
  \bTh_0^n(k)=1-\sum_{\ell=1}^\infty\bTh_\ell^n(k) .
  \label{eq:def_Th_ell}
\end{align}
%
%
$\bTh_\ell^n$ is common to all the components of the fields in the straight pipe
unlike $\Th_{\pm}^{n\ell}$ given by Eqs.(\ref{eq:Th_def}-\ref{eq:bTh_def}).
%
As shown in Eq.(\ref{eq:lim_Th}),
$\Th_{\pm}^{n\ell}$ goes to $\bTh_\ell^n$ in the limit of $\rho\to\infty$
since $(k_m^n,\bk_m^n)\to \hk_m^n$ for $\rho\to\infty$.

By partitioning $k$ using Eq.(\ref{eq:def_Th_ell}), we fix and count the number of
the real poles of $\mfG_{\pm}^n$, similar to what we did in section \ref{sec:ILT_vertical}.
Then $\cG_{\pm}^n$ is gotten as the sum of the residues at
the real and imaginary poles.
Using the residue theorem for the contours similar to Fig.\ref{fig:contour}
(p.\pageref{fig:contour}), we calculate the Bromwich integral (\ref{eq:cGbe_strt}),
\begin{align}
  \cG_{+}^n(x,x',\vsig)
  &=\sum_{\ell=0}^{\infty}\bTh_{\ell}^n
     \bigg[
       \bdelta_{01}^{\ell}\btheta(\vsig)
       \sum_{m=1}^{\ell-1}\frac{\cX_{+}^m(x,x')}{k_s^{mn}}\sin(k_s^{mn}\vsig)
      +\sum_{m=\ell'}^{\infty}\frac{\cX_{+}^m(x,x')}{2ik_s^{mn}}
       e^{ik_s^{mn}\bar{|\vsig|}}
     \bigg] ,
  \label{eq:cGp_st}
  \\
  \cG_{-}^n(x,x',\vsig)
  &=\sum_{\ell=0}^\infty \bTh_{\ell}^n
    \bigg[
      \bdelta_{0}^{\ell}\btheta(\vsig)
      \sum_{m=0}^{\ell-1}\frac{\cX_{-}^m(x,x')}{k_s^{mn}}\sin(k_s^{mn}\vsig)
      +\sum_{m=\ell}^{\infty}\frac{\cX_{-}^m(x,x')}{2ik_s^{mn}}
       e^{ik_s^{mn}\bar{|\vsig|}}
    \bigg] .
  \label{eq:cGm_st}
\end{align}
$\ell'$, $\bdelta_{01}^\ell$ and $\bdelta_0^\ell$ are given by
Eqs.(\ref{eq:lprm}-\ref{eq:d01_d0}).
As defined in Eqs.(\ref{eq:bstep}-\ref{eq:babsol}),
$\btheta(\vsig)$ and $\bar{|\vsig|}$ are the extended step function and absolute value 
of $\vsig$ which represents $s-s'$ and $s$ in the Green functions $\cG_{\pm}^n(\vsig)$.
$\cX_{\pm}^m$ denotes the horizontal eigenfunctions of the straight rectangular pipe,
\begin{align}
  \cX_{+}^m(x,x')
  &=\frac{1}{w}\big\{\cos[k_x^m(x-x')]-(-1)^m\cos[k_x^m(x+x'-x_b-x_a)]\big\} ,
   \label{eq:mfXp_new}
  \\
  \cX_{-}^m(x,x')
  &=\frac{1}{w(1+\delta_0^m)}
    \big\{\cos[k_x^m(x-x')]+(-1)^m\cos[k_x^m(x+x'-x_b-x_a)]\big\} .
  \label{eq:mfXm_new}
\end{align}
Eqs.(\ref{eq:mfXp_new}-\ref{eq:mfXm_new}) are rewritten as follows,
\begin{align}
  \cX_{+}^m(x,x')
  &=-\frac{2(-1)^m}{w}\sin[k_x^m(x_b-x)]\sin[k_x^m(x'-x_a)]
    \qquad~~(m\in\mathbb{N}) ,
  \label{eq:cXp}
  \\
  \cX_{-}^m(x,x')
  &=\frac{2(-1)^m}{w(1+\delta_0^m)}\cos[k_x^m(x_b-x)]\cos[k_x^m(x'-x_a)]
    \qquad(m\in\mathbb{Z}_0^{+}) .
   \label{eq:cXm}
\end{align}
$\cX_{\pm}^m$ satisfies the following equations,
\begin{align}
  &
  \cX_{\pm}^m(x',x)
  =\cX_{\pm}^m(x,x') ,
    \qquad
  \{\rd_x^2+(k_x^m)^2\}\cX_{\pm}^m(x,x')
  =\{\rd_{x'}^2+(k_x^m)^2\}\cX_{\pm}^m(x,x')
  =0 ,
  \label{eq:cX_symm}
   \\
  &
  \rd_{x}\cX_{\pm}^m(x,x')+\rd_{x'}\cX_{\mp}^m(x,x')
  =0,
    \qquad
  \rd_{x}\rd_{x'}\cX_{\pm}^m(x,x')
  =(k_x^m)^2\cX_{\mp}^m(x,x') .
  \label{eq:we_cX}
\end{align}
$\cX_{+}^m$ and $\cX_{-}^m$ satisfy the following boundary conditions on the sidewalls of
the straight pipe,
\begin{align}
  \cX_{+}^m(x_{a,b},x')
  =\cX_{+}^m(x,x_{a,b})
  =0
   ,\qquad
  [\rd_{x}\cX_{-}^m(x,x')]_{x=x_{a,b}}
  =[\rd_{x'}\cX_{-}^m(x,x')]_{x'=x_{a,b}}
  =0 .
   \label{eq:cX_BC}
\end{align}

In appendix \ref{sec:orthogonality} we discuss the orthogonality of
the cross products of the Bessel functions with respect to the order $\nu$.
Similar to this, the trigonometric functions have the following orthogonality relations,
\begin{alignat}{2}
  \int_{x_a}^{x_b}dx\sin[k_x^m(x_b-x)]\sin[k_x^{m'}(x-x_a)]
  &=-\frac{w}{2}\delta_{m'}^m(-1)^m 
  &\qquad&(m,m'\in\mathbb{N}) ,
  \label{eq:sin_ortho_strt}
   \\
  \int_{x_a}^{x_b}dx\cos[k_x^m(x_b-x)]\cos[k_x^{m'}(x-x_a)]
  &=\frac{w}{2}\delta_{m'}^m \{(-1)^m+\delta_0^m\}
  &\qquad&(m,m'\in\mathbb{Z}_0^+) .
  \label{eq:cos_ortho_strt}
\end{alignat}
$\delta_{m'}^m$ is the Kronecker delta.
From Eqs.(\ref{eq:sin_ortho_strt}-\ref{eq:cos_ortho_strt}),
the horizontal $\delta$-function is given as
\begin{align}
  \delta(x-x')
  =\sum_{m=1}^{\infty}\cX_{+}^m(x,x')
  =\sum_{m=0}^{\infty}\cX_{-}^m(x,x') ,
    \qquad
  \int_{x_a}^{x_b}\cX_{\pm}^m(x,x)dx
  =1 .
  \label{eq:delta_Tbe}
\end{align}
$\cG_{\pm}^n$ satisfies the following wave equation
which is the asymptotic limit of Eq.(\ref{eq:BessDE_delta}) for $\rho\to\infty$,
\begin{align}
  \{\vec{\rd}^2+(k\beta)^2\}\cG_{\pm}^n(x,x',\vsig)
  =\delta(x-x')\bdelta(\vsig) .
  \label{eq:bNb_cGpm_st}
\end{align}
$\vec{\rd}^2$ is given by Eq.(\ref{eq:we_xprim_drift}) which is the Laplacian for
the Fourier coefficients of the fields in the straight pipe.
$\bdelta(\vsig)$ is the extended $\delta$-function
defined for $\vsig=\{s-s',s\}$ in Eq.(\ref{eq:bdelta}).
As seen from Eqs.(\ref{eq:we_xprim_drift}) and (\ref{eq:bNb_cGpm_st}),
$\cG_{\pm}^n$ denotes the Green functions of the wave equations for
the Fourier coefficients of the fields.
According to the second equation of (\ref{eq:we_cX}), $\cG_{\pm}^n$ satisfies
\begin{align}
  \rd_{x}\cG_{\pm}^n(x,x',\vsig)
  +\rd_{x'}\cG_{\mp}^n(x,x',\vsig)
  =0 .
   \label{eq:rdxGp_rdxp_cGm}
\end{align}
According to Eqs.(\ref{eq:cX_BC}), $\cG_{\pm}^n$ satisfies
the following boundary conditions on the sidewalls of the pipe,
\begin{align}
  \cG_{+}^n(x_{a,b},x')
  =\cG_{+}^n(x,x_{a,b})
  =0
   ,\qquad
  [\rd_{x}\cG_{-}^n(x,x')]_{x=x_{a,b}}
  =[\rd_{x'}\cG_{-}^n(x,x')]_{x'=x_{a,b}}
  =0 .
   \label{eq:cGpm_st_BC}
\end{align}
Eqs.(\ref{eq:cGpm_st_BC}) are the asymptotic limits of Eqs.(\ref{eq:BC_cHpm}) for
$\rho\to\infty$.

From $\cE_y^n$ and $\cB_y^n$ given by Eqs.(\ref{eq:cEy_strt}-\ref{eq:cBy_strt})
through Eqs.(\ref{eq:cBx}-\ref{eq:cBs}) for $\vrho=\infty$,
we can get $\cE_{x,s}^n$ and $\cB_{x,s}^n$ which are the Fourier coefficients of
the horizontal and longitudinal components of the fields.
But it is somewhat troublesome to calculate them in this way, because we must rewrite
the initial fields involved in Eqs.(\ref{eq:cEy_strt}-\ref{eq:cBy_strt})
as shown in appendix \ref{sec:HL_field_sdom}.
In order to derive the expressions of $\cE_{x,s}^n$ and $\cB_{x,s}^n$ in
the straight section,
it may be easier to solve their wave equations under the boundary conditions
(\ref{eq:BC_side_cF}) since the horizontal and longitudinal 
components of the fields are not entangled in the wave equations,
\begin{align}
  \bigg\{{c\cB_x^n(x,s) \atop  \cE_s^n(x,s)}\bigg\}
  &=
   \int_{x_a}^{x_b}dx'
   \bigg[
     \bigg\{{\cA_x^n(x') \atop \cD_s^n(x')}\bigg\}\cG_{+}^n(x,x',s)
    +Z_0\int_0^\infty ds'
     \bigg\{{\cT_x^n(x',s') \atop \cS_s^n(x',s')}\bigg\}\cG_{+}^n(x,x',s-s')
   \bigg] ,
  \label{eq:sols_even_strt}
  \\
  \bigg\{{\cE_x^n(x,s) \atop c\cB_s^n(x,s)}\bigg\}
  &=
   \int_{x_a}^{x_b}dx'
   \bigg[
     \bigg\{{\cD_x^n(x') \atop \cA_s^n(x')}\bigg\}\cG_{-}^n(x,x',s)
    +Z_0\int_0^\infty ds'
     \bigg\{{\cS_x^n(x',s') \atop \cT_s^n(x',s')}\bigg\}\cG_{-}^n(x,x',s-s')
   \bigg] .
  \label{eq:sols_odd_strt}
\end{align}
$\cS_{x,s}^n$ and $\cT_{x,s}^n$ are the Fourier coefficients of the source terms,
given by Eqs.(\ref{eq:cSx_cTx_st}-\ref{eq:cSs_cTs_st}).
$\cD_{x,s}^n$ and $\cA_{x,s}^n$ are the following operators which involve
the initial values of $\cE_{x,s}^n$ and $\cB_{x,s}^n$ at the entrance of the straight pipe,
\begin{align}
  \cD_{x,s}^n(x')
  &=[\rd_{s'}\cE_{x,s}^n(x',s')+\cE_{x,s}^n(x',s')\rd_s]_{s'=+0} ,
   \label{eq:cDxs_str}
   \\
  \cA_{x,s}^n(x')
  &=[\rd_{s'}c\cB_{x,s}^n(x',s')+c\cB_{x,s}^n(x',s')\rd_s]_{s'=+0} .
   \label{eq:cAxs_str}
\end{align}
Eqs.(\ref{eq:cDxs_str}-\ref{eq:cAxs_str}) have the longitudinal operator $\rd_s$
which acts on $\cG_{\pm}^n(s)$ in Eqs.(\ref{eq:sols_even_strt}-\ref{eq:sols_odd_strt}).

\subsection{Separated form of the fields in a straight pipe}

We find the expressions of the frequency domain fields
which propagate together with the beam moving in the straight rectangular pipe.
Using Eqs.(\ref{eq:Fp_coeff}-\ref{eq:Fm_coeff}), we first rewrite the Fourier coefficients
$(\cD^n,\cA^n)$ and $(\cS^n,\cT^n)$ which are involved in
Eqs.(\ref{eq:cEy_strt}-\ref{eq:cBy_strt}) and
(\ref{eq:sols_even_strt}-\ref{eq:sols_odd_strt}).
Substituting them into Eqs.(\ref{eq:Four_Exp_plus}-\ref{eq:Four_Exp_minus}), we get
\begin{alignat}{2}
  \bigg\{{\tEv(\xv) \atop c\tBv(\xv)}\bigg\}
  &=\int_{x_a}^{x_b}dx'\int_{-h/2}^{h/2}dy'
    \bigg\{{\tUv(\xv,\xv_\perp') \atop \tVv(\xv,\xv_\perp')}\bigg\}
     \qquad &&\text{(scalar expr.)}
   \label{eq:EB_str_sep_scl}
   \\
  &=\int_{x_a}^{x_b}dx'\int_{-h/2}^{h/2}dy'
    \bigg\{{\tUv^{\dg}(\xv,\xv_\perp') \atop \tVv^{\dg}(\xv,\xv_\perp')}\bigg\}
     \qquad &&\text{(differential expr.)} .
   \label{eq:EB_str_sep}
\end{alignat}
The arguments of the functions are given as
\begin{align}
  \xv=(x,y,s),
   \qquad
  \xv_\perp=(x,y);
   \qquad
  \xv'=(x',y',s'),
   \qquad
  \xv_\perp'=(x',y') .
\end{align}
$\tUv$ and $\tVv$ are the integrands of the scalar expression of the separated form
of the fields in the straight pipe,
\begin{alignat}{2}
  \tUv
  &=\tU_x\ev_x+\tU_y\ev_y+\tU_s\ev_s ,
    \qquad&
  \tUv^{\dg}
  &=\tU_x^{\dg}\ev_x+\tU_y^{\dg}\ev_y+\tU_s^{\dg}\ev_s ,
   \label{eq:tUv_tVv}
   \\
  \tVv
  &=\tV_x\ev_x+\tV_y\ev_y+\tV_s\ev_s ,
    \qquad&
  \tVv^{\dg}
  &=\tV_x^{\dg}\ev_x+\tV_y^{\dg}\ev_y+\tV_s^{\dg}\ev_s .
   \label{eq:tUv_tVv_dg}
\end{alignat}
The components of $\tUv$ and $\tVv$ are given as follows,
\begin{align}
  \bigg\{{\tU_x(\xv,\xv_\perp') \atop \tV_y(\xv,\xv_\perp')}\bigg\}
  &=
       \bigg\{{\tD_x(\xv_\perp') \atop \tA_y(\xv_\perp')}\bigg\}
       \Phi_{-}^{+}(\xv_\perp,\xv_\perp',s)
      +Z_0\int_0^{\infty}ds'\bigg\{{\tS_x(\xv') \atop \tT_y(\xv')}\bigg\}
       \Phi_{-}^{+}(\xv_\perp,\xv_\perp',s-s') ,
  \label{eq:tEx_tBy_st}
   \\
  \bigg\{{\tU_y(\xv,\xv_\perp') \atop \tV_x(\xv,\xv_\perp')}\bigg\}
  &=
       \bigg\{{\tD_y(\xv_\perp') \atop \tA_x(\xv_\perp')}\bigg\}
       \Phi_{+}^{-}(\xv_\perp,\xv_\perp',s)
      +Z_0\int_0^{\infty}ds'\bigg\{{\tS_y(\xv') \atop \tT_x(\xv')}\bigg\}
       \Phi_{+}^{-}(\xv_\perp,\xv_\perp',s-s') ,
  \label{eq:tEy_tBx_st}
   \\
  \tU_s(\xv,\xv_\perp')
  &=
       \tD_s(\xv_\perp')\Phi_{+}^{+}(\xv_\perp,\xv_\perp',s)
      +Z_0\int_0^{\infty}ds'\tS_s(\xv')\Phi_{+}^{+}(\xv_\perp,\xv_\perp',s-s')  ,
  \label{eq:tEs_st}
   \\
  \tV_s(\xv,\xv_\perp')
  &=
       \tA_s(\xv_\perp')\Phi_{-}^{-}(\xv_\perp,\xv_\perp',s)
      +Z_0\int_0^{\infty}ds'\tT_s(\xv')\Phi_{-}^{-}(\xv_\perp,\xv_\perp',s-s') .
  \label{eq:tBs_st}
\end{align}
The source terms $\tSv$ and $\tTv$, given by Eqs.(\ref{eq:src_E_str}-\ref{eq:src_B_str}),
have the derivatives of the current components $(\tJ_0,\tJv)$
with respect to $r'$, $y'$ or $s'$.
Integrating them by parts in Eq.(\ref{eq:EB_str_sep_scl}),
we get Eq.(\ref{eq:EB_str_sep}) which is the differential expression of the separated form 
of the fields in the straight rectangular pipe.
$\tUv^{\dg}$ and $\tVv^{\dg}$ are the integrands of the differential expression of
the separated form of the fields in the straight pipe,
\begin{align}
  \bigg\{{\tU_x^{\dg}(\xv,\xv_\perp') \atop \tV_y^{\dg}(\xv,\xv_\perp')}\bigg\}
  &=
       \bigg\{{\tD_x^{\dg}(\xv_\perp') \atop \tA_y^{\dg}(\xv_\perp')}\bigg\}
       \Phi_{-}^{+}(\xv_\perp,\xv_\perp',s)
      -Z_0\int_0^{\infty}ds'\bigg\{{\tS_x^{\dg}(\xv') \atop \tT_y^{\dg}(\xv')}\bigg\}
       \Phi_{-}^{+}(\xv_\perp,\xv_\perp',s-s') ,
  \label{eq:tEx_st_DS}
   \\
  \bigg\{{\tU_y^{\dg}(\xv,\xv_\perp') \atop \tV_x^{\dg}(\xv,\xv_\perp')}\bigg\}
  &=
       \bigg\{{\tD_y^{\dg}(\xv_\perp') \atop \tA_x^{\dg}(\xv_\perp')}\bigg\}
       \Phi_{+}^{-}(\xv_\perp,\xv_\perp',s)
      -Z_0\int_0^{\infty}ds'\bigg\{{\tS_y^{\dg}(\xv') \atop \tT_x^{\dg}(\xv')}\bigg\}
       \Phi_{+}^{-}(\xv_\perp,\xv_\perp',s-s') ,
  \label{eq:tEy_st_DS}
   \\
  \tU_s^{\dg}(\xv,\xv_\perp')
  &=
       \tD_s^{\dg}(\xv_\perp')\Phi_{+}^{+}(\xv_\perp,\xv_\perp',s)
      -Z_0\int_0^{\infty}ds'\tS_s^{\dg}(\xv')\Phi_{+}^{+}(\xv_\perp,\xv_\perp',s-s') ,
  \label{eq:tEs_st_DS}
  \\
  \tV_s^{\dg}(\xv,\xv_\perp')
  &=
       \tA_s^{\dg}(\xv_\perp')\Phi_{-}^{-}(\xv_\perp,\xv_\perp',s)
      -Z_0\int_0^{\infty}ds'\tT_s^{\dg}(\xv')\Phi_{-}^{-}(\xv_\perp,\xv_\perp',s-s') .
  \label{eq:tBs_st_AT}
\end{align}
$\Phi$ $[=(\Phi_{+}^{+},\Phi_{-}^{+},\Phi_{+}^{-},\Phi_{-}^{-})]$ represents
the Green functions of the frequency domain fields in the straight pipe,
\begin{align}
  \Phi_{\pm}^{+}(\xv_\perp,\xv_\perp',\vsig)
  &=\sum_{n=1}^{\infty}\cG_{\pm}^n(x,x',\vsig)\cY_{+}^n(y,y') ,
    \qquad
  \Phi_{\pm}^{-}(\xv_\perp,\xv_\perp',\vsig)
  =\sum_{n=0}^{\infty}\cG_{\pm}^n(x,x',\vsig)\cY_{-}^n(y,y') .
   \label{eq:Phi}
\end{align}
The sign of the subscript of $\Phi_{\pm}$ denotes the horizontal parity of the field,
which corresponds to that of $\cG_{\pm}^n$ as in Eqs.(\ref{eq:Phi}).
On the other hand, the sign of the superscript of $\Phi^{\pm}$ denotes the vertical parity 
of the field, which corresponds to that of the vertical eigenfunctions $\cY_{\pm}^n$
given by Eqs.(\ref{eq:cYp_0}-\ref{eq:cYm}).
$\tilde{\bm D}$ and $\tilde{\Av}$ are longitudinal operators which have $\rd_s$ and
the initial values of the fields at the entrance of the straight section,
\begin{align}
  \tilde{\bm D}(\xv_\perp')
  =[\rd_{s'}\tEv(\xv')+\tEv(\xv')\rd_s]_{s'=+0} ,
    \qquad
  \tilde{\Av}(\xv_\perp')
  =[\rd_{s'}c\tBv(\xv')+c\tBv(\xv')\rd_s]_{s'=+0} .
   \label{eq:tDv_tDA_str}
\end{align}
$\rd_s$ in Eqs.(\ref{eq:tDv_tDA_str}) act on $\Phi(s)$ in the first terms of
Eqs.(\ref{eq:tEx_tBy_st}-\ref{eq:tBs_st}).
$\Phi(\vsig)$ satisfies the following wave equation,
\begin{align}
  \{\nablav^2+(k\beta)^2\}\Phi(\xv_\perp,\xv_\perp',\vsig)
  &=\delta(x-x')\delta(y-y')\bdelta(\vsig) ,
   \qquad
  \vsig=\{s-s',s\} .
\end{align}
$\bdelta(\vsig)$ is given by Eq.(\ref{eq:bdelta}).
$\tDv^{\dg}$ and $\tAv^{\dg}$ are operators
which have the initial values of the fields at $s=0$,
\begin{alignat}{2}
  \tDv^{\dg}
  &=\tD_x^{\dg}\ev_x+\tD_y^{\dg}\ev_y+\tD_s^{\dg}\ev_s ,
    \qquad&
  \tAv^{\dg}
  &=\tA_x^{\dg}\ev_x+\tA_y^{\dg}\ev_y+\tA_s^{\dg}\ev_s ,
   \\
  \tD_x^{\dg}
  &=\tE_x(0)\rd_s-\tE_s(0)\rd_{x'}+c\tB_y(0)ik\beta ,
   \label{eq:tDx_dg_st}
    \qquad&
  \tA_x^{\dg}
  &=c\tB_x(0)\rd_s-c\tB_s(0)\rd_{x'}-\tE_y(0)ik\beta ,
  \\
  \tD_y^{\dg}
  &=\tE_y(0)\rd_s-\tE_s(0)\rd_{y'}-c\tB_x(0)ik\beta ,
    \qquad&
  \tA_y^{\dg}
  &=c\tB_y(0)\rd_s-c\tB_s(0)\rd_{y'}+\tE_x(0)ik\beta ,
  \\
  \tD_s^{\dg}
  &=\tE_s(0)\rd_s+\tE_x(0)\rd_{x'}+\tE_y(0)\rd_{y'} ,
   \label{eq:tDs_dg_st}
    \qquad&
  \tA_s^{\dg}
  &=c\tB_s(0)\rd_s+c\tB_x(0)\rd_{x'}+c\tB_y(0)\rd_{y'} ,
\end{alignat}
where we omit the transverse arguments $\xv_\perp'=(x',y')$ of the functions for brevity,
\begin{align}
  \tEv(0)=\tEv(\xv_\perp',0) ,
    \qquad
  \tBv(0)=\tBv(\xv_\perp',0) .
\end{align}
$\tilde{\bm S}^{\dg}$ and $\tilde{\bm T}^{\dg}$ are operators
with respect to $\xv'$, which have the current components in $s>0$,
\begin{align}
  \tSv^{\dg}(\xv')=ik\beta\tJv(\xv')+\tJ_0(\xv')\nablav' ,
   \qquad
  \tTv^{\dg}(\xv')=\tJv(\xv')\times\nablav' ,
   \qquad
  \nablav'
  =\ev_x\rd_{x'}+\ev_y\rd_{y'}+\ev_s\rd_{s'} .
   \label{eq:tST_dg_st}
\end{align}
We can replace $\rd_{s'}$ by $-\rd_s$ in Eq.(\ref{eq:tST_dg_st})
since $\tSv^{\dg}$ and $\tTv^{\dg}$ act only on $\Phi(s-s')$ in
Eqs.(\ref{eq:tEx_st_DS}-\ref{eq:tBs_st_AT}).

We show the differential expressions of the integrands of the separated form,
which are involved in Eqs.(\ref{eq:tEx_st_DS}-\ref{eq:tBs_st_AT}).
Omitting the transverse arguments $(\xv_{\perp},\xv_{\perp}')$ for brevity,
the integrands involved in the expression of $\tEv$ are given as
\begin{align}
  \tD_x^{\dg}\Phi_{-}^{+}(s)
  &=\sum_{n=1}^{\infty}\cY_{+}^n\sum_{\ell=0}^\infty\bTh_{\ell}^n
    \bigg\{
       \bdelta_{0}^{\ell}\sum_{m=0}^{\ell-1}\mfp_x(s)
      +\sum_{m=\ell}^{\infty}\bmfp_x(s)
    \bigg\} ,
   \label{eq:Dx_dg_Phi}
  \\
  \tD_y^{\dg}\Phi_{+}^{-}(s)
  &=\sum_{n=0}^{\infty}\sum_{\ell=0}^{\infty}\bTh_{\ell}^n
    \bigg\{
       \bdelta_{01}^{\ell}\sum_{m=1}^{\ell-1}\mfp_y(s)
      +\sum_{m=\ell'}^{\infty}\bmfp_y(s)
    \bigg\} ,
  \\
  \tD_s^{\dg}\Phi_{+}^{+}(s)
  &=\sum_{n=1}^{\infty}\sum_{\ell=0}^{\infty}\bTh_{\ell}^n
    \bigg\{
       \bdelta_{01}^{\ell}\sum_{m=1}^{\ell-1}\mfp_s(s)
      +\sum_{m=\ell'}^{\infty}\bmfp_s(s)
    \bigg\} ,
   \label{eq:Ds_dg_Phi}
\end{align}
where $\ell'$ is the index defined in Eq.(\ref{eq:lprm}),
\begin{align}
  \ell'
  =\ell+\delta_0^\ell .
\end{align}
The summands of Eqs.(\ref{eq:Dx_dg_Phi}-\ref{eq:Ds_dg_Phi}) are given as follows,
\begin{align}
  \mfp_x(s)
  &=\Big[
       \tE_x(0)\cos(k_s^{mn}s)
      +\frac{\sin(k_s^{mn}s)}{k_s^{mn}}\{c\tB_y(0)ik\beta -\tE_s(0)\rd_{x'}\}
    \Big]
    \cX_{-}^m ,
   \\
  \bmfp_x(s)
  &=\frac{e^{ik_s^{mn}s}}{2}
    \Big[
       \tE_x(0)
      +\frac{1}{ik_s^{mn}}\{c\tB_y(0)ik\beta -\tE_s(0)\rd_{x'}\}
    \Big]
    \cX_{-}^m ,
   \\
  \mfp_y(s)
  &=\Big[
       \tE_y(0)\cos(k_s^{mn}s)
      -\frac{\sin(k_s^{mn}s)}{k_s^{mn}}\{\tE_s(0)\rd_{y'}+c\tB_x(0)ik\beta\}
    \Big]
    \cX_{+}^m\cY_{-}^n ,
   \\
  \bmfp_y(s)
  &=\frac{e^{ik_s^{mn}s}}{2}
    \Big[
       \tE_y(0)
      -\frac{1}{ik_s^{mn}}\{\tE_s(0)\rd_{y'}+c\tB_x(0)ik\beta\}
    \Big]
    \cX_{+}^m\cY_{-}^n ,
   \\
  \mfp_s(s)
  &=\Big[
       \tE_s(0)\cos(k_s^{mn}s)
      +\frac{\sin(k_s^{mn}s)}{k_s^{mn}}\{\tE_x(0)\rd_{x'}+\tE_y(0)\rd_{y'}\}
    \Big]
    \cX_{+}^m\cY_{+}^n ,
   \\
  \bmfp_s(s)
  &=\frac{e^{ik_s^{mn}s}}{2}
    \Big[
       \tE_s(0)
      +\frac{1}{ik_s^{mn}}\{\tE_x(0)\rd_{x'}+\tE_y(0)\rd_{y'}\}
    \Big]
    \cX_{+}^m\cY_{+}^n .
\end{align}
Similarly, the integrands involved in the expression of $\tBv$ are given as
\begin{align}
  \tA_x^{\dg}\Phi_{+}^{-}(s)
  &=\sum_{n=0}^{\infty}\cY_{-}^n\sum_{\ell=0}^{\infty}\bTh_{\ell}^n
    \bigg\{
       \bdelta_{01}^{\ell}\sum_{m=1}^{\ell-1}\mfq_x(s)
      +\sum_{m=\ell'}^{\infty}\bmfq_x(s)
    \bigg\} ,
  \\
  \tA_y^{\dg}\Phi_{-}^{+}(s)
  &=\sum_{n=1}^{\infty}\sum_{\ell=0}^\infty \bTh_{\ell}^n
    \bigg\{
       \bdelta_{0}^{\ell}\sum_{m=0}^{\ell-1}\mfq_y(s)
      +\sum_{m=\ell}^{\infty}\bmfq_y(s)
    \bigg\} ,
  \\
  \tA_s^{\dg}\Phi_{-}^{-}(\vsig)
  &=\sum_{n=0}^{\infty}\sum_{\ell=0}^\infty \bTh_{\ell}^n
    \bigg\{
       \bdelta_{0}^{\ell}\sum_{m=0}^{\ell-1}\mfq_s(s)
      +\sum_{m=\ell}^{\infty}\bmfq_s(s)
    \bigg\} ,
\end{align}
where
\begin{align}
  \mfq_x(s)
  &=\Big[
       c\tB_x(0)\cos(k_s^{mn}s)
      -\frac{\sin(k_s^{mn}s)}{k_s^{mn}}\{c\tB_s(0)\rd_{x'}+\tE_y(0)ik\beta\}
    \Big]
    \cX_{+}^m ,
   \\
  \bmfq_x(s)
  &=\frac{e^{ik_s^{mn}s}}{2}
    \Big[
       c\tB_x(0)
      -\frac{1}{ik_s^{mn}}\{c\tB_s(0)\rd_{x'}+\tE_y(0)ik\beta\}
    \Big]
    \cX_{+}^m ,
   \\
  \mfq_y(s)
  &=\Big[
       c\tB_y(0)\cos(k_s^{mn}s)
       +\frac{\sin(k_s^{mn}s)}{k_s^{mn}}\{\tE_x(0)ik\beta-c\tB_s(0)\rd_{y'}\}
    \Big]
    \cX_{-}^m\cY_{+}^n ,
   \\
  \bmfq_y(s)
  &=\frac{e^{ik_s^{mn}s}}{2}
    \Big[
       c\tB_y(0)
       +\frac{1}{ik_s^{mn}}\{\tE_x(0)ik\beta-c\tB_s(0)\rd_{y'}\}
    \Big]
    \cX_{-}^m\cY_{+}^n ,
   \\
  \mfq_s(s)
  &=\Big[
       c\tB_s(0)\cos(k_s^{mn}s)
      +\frac{\sin(k_s^{mn}s)}{k_s^{mn}}\{c\tB_x(0)\rd_{x'}+c\tB_y(0)\rd_{y'}\}
    \Big]
    \cX_{-}^m\cY_{-}^n ,
   \\
  \bmfq_s(s)
  &=\frac{e^{ik_s^{mn}s}}{2}
    \Big[
       c\tB_s(0)
      +\frac{1}{ik_s^{mn}}\{c\tB_x(0)\rd_{x'}+c\tB_y(0)\rd_{y'}\}
    \Big]
    \cX_{-}^m\cY_{-}^n .
\end{align}
The integrands involving the source terms in the expression of $\tEv$ are given as
\begin{align}
  \tS_x^{\dg}(s')\Phi_{-}^{+}(\vsig)
  &=\sum_{n=1}^{\infty}\cY_{+}^n\sum_{\ell=0}^\infty \bTh_{\ell}^n
    \bigg\{
       \bdelta_{0}^{\ell}\theta(\vsig)\sum_{m=0}^{\ell-1}\mfm_x(\vsig,s')
      +\sum_{m=\ell}^{\infty}\bmfm_x(\vsig,s')
    \bigg\} ,
   \label{eq:tSx_dg_xv_prm}
  \\
  \tS_y^{\dg}(s')\Phi_{+}^{-}(\vsig)
  &=\sum_{n=0}^{\infty}\sum_{\ell=0}^{\infty}\bTh_{\ell}^n
    \bigg\{
       \bdelta_{01}^{\ell}\theta(\vsig)\sum_{m=1}^{\ell-1}\mfm_y(\vsig,s')
      +\sum_{m=\ell'}^{\infty}\bmfm_y(\vsig,s')
    \bigg\} ,
  \\
  \tS_s^{\dg}(s')\Phi_{+}^{+}(\vsig)
  &=\sum_{n=1}^{\infty}\cY_{+}^n\sum_{\ell=0}^{\infty}\bTh_{\ell}^n
    \bigg\{
       \bdelta_{01}^{\ell}\theta(\vsig)\sum_{m=1}^{\ell-1}\mfm_s(\vsig,s')
      +\sum_{m=\ell'}^{\infty}\bmfm_s(\vsig,s')
    \bigg\} .
\end{align}
$\vsig$ represents only $s-s'$ (\ie, $\vsig\ne s$) in
Eqs.(\ref{eq:tSx_dg_xv_prm}-\ref{eq:bmfns_str}).
The summands are given as
\begin{alignat}{2}
  \bigg\{{\mfm_x(\vsig,s') \atop \bmfm_x(\vsig,s')}\bigg\}
  &=\frac{1}{k_s^{mn}}\bigg\{{\sin(k_s^{mn}\vsig) \atop e^{ik_s^{mn}|\vsig|}/(2i)}\bigg\}
    \{\tJ_0(s')\rd_{x'}+ik\beta\tJ_x(s')\}\cX_{-}^m ,
   \label{eq:mfmx_str}
   \\
  \bigg\{{\mfm_y(\vsig,s') \atop \bmfm_y(\vsig,s')}\bigg\}
  &=\frac{1}{k_s^{mn}}\bigg\{{\sin(k_s^{mn}\vsig) \atop e^{ik_s^{mn}|\vsig|}/(2i) }\bigg\}
    \{\tJ_0(s')\rd_{y'}+ik\beta\tJ_y(s')\}\cX_{+}^m\cY_{-}^n ,
\end{alignat}
\begin{alignat}{2}
  \mfm_s(\vsig,s')
  &=\Big\{
       \tJ_s(s')\frac{ik\beta}{k_s^{mn}}\sin(k_s^{mn}\vsig)
      -\tJ_0(s')\cos(k_s^{mn}\vsig)
    \Big\}
    \cX_{+}^m ,
   \\
  \bmfm_s(\vsig,s')
  &=\frac{e^{ik_s^{mn}|\vsig|}}{2}
    \Big\{
       \tJ_s(s')\frac{k\beta}{k_s^{mn}}
      -\tJ_0(s')\sgn(\vsig)
    \Big\}
    \cX_{+}^m .
   \label{eq:mfms_str}
\end{alignat}
The integrands involving the source terms in the expression of $\tBv$ are given as
\begin{align}
  \tT_x^{\dg}(s')\Phi_{+}^{-}(\vsig)
  &=\sum_{n=0}^{\infty}\sum_{\ell=0}^{\infty}\bTh_{\ell}^n
    \bigg\{
       \bdelta_{01}^{\ell}\theta(\vsig)\sum_{m=1}^{\ell-1}\mfn_x(\vsig,s')
      +\sum_{m=\ell'}^{\infty}\bmfn_x(\vsig,s')
    \bigg\} ,
   \\
  \tT_y^{\dg}(s')\Phi_{-}^{+}(\vsig)
  &=\sum_{n=1}^{\infty}\cY_{+}^n\sum_{\ell=0}^\infty \bTh_{\ell}^n
    \bigg\{
       \bdelta_{0}^{\ell}\theta(\vsig)\sum_{m=0}^{\ell-1}\mfn_y(\vsig,s')
      +\sum_{m=\ell}^{\infty}\bmfn_y(\vsig,s')
    \bigg\} ,
   \\
  \tT_s^{\dg}(s')\Phi_{-}^{-}(\vsig)
  &=\sum_{n=0}^{\infty}\sum_{\ell=0}^\infty \bTh_{\ell}^n
    \bigg\{
       \bdelta_{0}^{\ell}\theta(\vsig)\sum_{m=0}^{\ell-1}\mfn_s(\vsig,s')
      +\sum_{m=\ell}^{\infty}\bmfn_s(\vsig,s')
    \bigg\} .
\end{align}
These summands are given as
\begin{align}
  \mfn_x(\vsig,s')
  &=-\Big\{
       \tJ_y(s')\cos(k_s^{mn}\vsig)
      +\tJ_s(s')\sin(k_s^{mn}\vsig)\frac{\rd_{y'}}{k_s^{mn}}
    \Big\}
    \cX_{+}^m\cY_{-}^n ,
   \label{eq:mfnx_str}
   \\
  \bmfn_x(\vsig,s')
  &=-\frac{e^{ik_s^{mn}|\vsig|}}{2}
    \Big\{
       \tJ_y(s')\sgn(\vsig)
      +\tJ_s(s')\frac{\rd_{y'}}{ik_s^{mn}}
    \Big\}
    \cX_{+}^m\cY_{-}^n ,
   \\
  \mfn_y(\vsig,s')
  &=\Big\{
       \tJ_s(s')\sin(k_s^{mn}\vsig)\frac{\rd_{x'}}{k_s^{mn}}
       +\tJ_x(s')\cos(k_s^{mn}\vsig)
    \Big\}
    \cX_{-}^m ,
   \\
  \bmfn_y(\vsig,s')
  &=\frac{e^{ik_s^{mn}|\vsig|}}{2}
    \Big\{
       \tJ_s(s')\frac{\rd_{x'}}{ik_s^{mn}}
       +\tJ_x(s')\sgn(\vsig)
    \Big\}
    \cX_{-}^m ,
\end{align}
\begin{align}
  \bigg\{{ \mfn_s(\vsig,s') \atop \bmfn_s(\vsig,s')}\bigg\}
  &=\frac{1}{k_s^{mn}}\bigg\{{ \sin(k_s^{mn}\vsig) \atop e^{ik_s^{mn}|\vsig|}/(2i) }\bigg\}
    \{\tJ_x(s')\rd_{y'}-\tJ_y(s')\rd_{x'}\}\cX_{-}^m\cY_{-}^n .
   \label{eq:bmfns_str}
\end{align}
For brevity, we do not indicate the transverse arguments $\xv_{\perp}'=(x',y')$ of
the current $\tJ=(\tJ_0,\tJv)$ involved in Eqs.(\ref{eq:mfmx_str}-\ref{eq:mfms_str}) and
(\ref{eq:mfnx_str}-\ref{eq:bmfns_str}), \ie,
\begin{align}
  \tJ_0(s')
  =\tJ_0(\xv') ,
    \qquad
  \tJv(s')
  =\tJv(\xv') .
\end{align}
%

\subsection{Complete form of the fields in a straight pipe}

We rewrite Eqs.(\ref{eq:cGp_st}-\ref{eq:cGm_st}) in order to create
the complete sum over all the horizontal eigenmodes of the straight pipe,
similar to Eq.(\ref{eq:cGpm_cCpm_bcCpm}) which is for the curved pipe,
\begin{align}
  \cG_{\pm}^n(x,x',\vsig)
  =\cC_{\pm}^n(x,x',\vsig)-\bcC_{\pm}^n(x,x',\vsig)
   ,\qquad
  \vsig=\{s-s',s\} ,
  \label{eq:cGpm_cCpm_st}
\end{align}
where $\cC_{\pm}^n$ denotes the Fourier coefficients of the Green functions of the fields of
the complete form in the straight rectangular pipe.
$\bcC_{\pm}^n$ is the difference of $\cG_{\pm}^n$ from $\cC_{\pm}^n$,
\begin{alignat}{2}
  \cC_{+}^n(x,x',\vsig)
  &=\btheta(\vsig)
   \sum_{m=1}^{\infty}\cX_{+}^m(x,x')\frac{\sin(k_s^{mn}\vsig)}{k_s^{mn}} ,
    \qquad&
  \bcC_{+}^n(x,x',\vsig)
  &=-\sum_{\ell=0}^{\infty}\bTh_{\ell}^n\sum_{m=\ell'}^{\infty}
   \cX_{+}^m(x,x')\frac{e^{-ik_s^{mn}\vsig}}{2ik_s^{mn}} ,
  \label{eq:cCp_str}
   \\
  \cC_{-}^n(x,x',\vsig)
  &=\btheta(\vsig)
   \sum_{m=0}^{\infty}\cX_{-}^m(x,x')\frac{\sin(k_s^{mn}\vsig)}{k_s^{mn}} ,
    \qquad&
  \bcC_{-}^n(x,x',\vsig)
  &=-\sum_{\ell=0}^{\infty}\bTh_{\ell}^n
   \sum_{m=\ell}^{\infty}\cX_{-}^m(x,x')\frac{e^{-ik_s^{mn}\vsig}}{2ik_s^{mn}} .
  \label{eq:cCm_str}
\end{alignat}
$\ell'=\ell+\delta_0^\ell$ as in Eq.(\ref{eq:lprm}).
$\btheta(\vsig)$ is the extended step function given by Eq.(\ref{eq:bstep}).
In deriving the expressions of $\cC_{\pm}^n$, similar to Eqs.(\ref{eq:sum_Xmn}),
we used the following identities which hold for an arbitrary summand $d_m$,
\begin{align}
  \sum_{m=1}^{\infty}d_m
  &=\sum_{\ell=0}^{\infty}\bTh_{\ell}^n
   \bigg\{
     \bdelta_{01}^{\ell}\sum_{m=1}^{\ell-1}d_m
    +\sum_{m=\ell'}^{\infty}d_m
   \bigg\}
  =\sum_{\ell=0}^{\infty}
   \bigg\{
     \bdelta_{0}^{\ell}\bTh_{\ell+1}^n\sum_{m=1}^{\ell}d_m
    +\bTh_{\ell}^n\sum_{m=\ell'}^{\infty}d_m
   \bigg\} ,
  \\
  \sum_{m=0}^{\infty}d_m
  &=\sum_{\ell=0}^\infty \bTh_{\ell}^n
    \bigg\{\bdelta_{0}^{\ell}\sum_{m=0}^{\ell-1}d_m+\sum_{m=\ell}^{\infty}d_m\bigg\} .
\end{align}
Both $\cC_{\pm}^n$ and $\bcC_{\pm}^n$ diverge exponentially for $s\to\infty$
since $k_s^{mn}\in i\mathbb{R}$ for the mode $(m,n)$ such that
$|k|<\hk_m^n$ as seen from Eq.(\ref{eq:ks_mn}).
$\cC_{\pm}^n$ and $\bcC_{\pm}^n$ each satisfy the following wave equations,
\begin{alignat}{2}
  \{\vec{\rd}^2+(k\beta)^2\}\cC_{\pm}^n(x,x',\vsig)
  &=\delta(x-x')\bdelta(\vsig) ,
    \qquad
  \{\vec{\rd}^2+(k\beta)^2\}\bcC_{\pm}^n(x,x',\vsig)
  =0 .
  \label{eq:bNb_cCpm_st}
\end{alignat}
$\vec{\rd}^2$ and $\bdelta(\vsig)$ are given by Eqs.(\ref{eq:we_xprim_drift}) and
(\ref{eq:bdelta}).
$\rd_s\cC_{\pm}^n(\vsig)$ at $\vsig=0$ is gotten using Eq.(\ref{eq:delta_Tbe}),
\begin{align}
  [\rd_s\cC_{\pm}^n(x,x',\vsig)]_{\vsig=0}=\delta(x-x')
  \qquad \text{for~ $\vsig=\{s-s',s\}$} .
  \label{eq:rdsik_cCpm_s0_st}
\end{align}
We use Eq.(\ref{eq:rdsik_cCpm_s0_st}) in verifying the fact that
the solution of the field reproduces its initial value at $s=0$.
According to the second equations of (\ref{eq:we_cX}),
$\cC_{\pm}^n$ and $\bcC_{\pm}^n$ satisfy
\begin{align}
  \rd_{x}\cC_{\pm}^n
  +\rd_{x'}\cC_{\mp}^n
  =0,
   \qquad
  \rd_{x}\bcC_{\pm}^n
  +\rd_{x'}\bcC_{\mp}^n
  =0 .
    \label{lim_cC}
\end{align}
We can also get the first equation of (\ref{lim_cC}) from
the second equations of (\ref{eq:cUVC}-\ref{eq:cUcVcC}) in the limit of $\rho\to\infty$.

Substituting Eq.(\ref{eq:cGpm_cCpm_st}) into Eqs.(\ref{eq:cEy_strt}-\ref{eq:cBy_strt}) and
(\ref{eq:sols_even_strt}-\ref{eq:sols_odd_strt}), the terms involving $\bcC_{\pm}^n$ cancel
out between the term of the initial field at $s=0$ and the source term in $s>0$ through
the wave equation (\ref{eq:we_xprim_drift}).
Then we get the Fourier coefficients of the fields of the complete form in
the straight rectangular pipe,
\begin{align}
  \bigg\{{\cE_x^n(x,s) \atop c\cB_{y,s}^n(x,s)}\bigg\}
  &=
   \int_{x_a}^{x_b}dx'
   \bigg[
     \bigg\{{\cD_x^n(x') \atop \cA_{y,s}^n(x')}\bigg\}\cC_{-}^n(x,x',s)
           +Z_0\int_0^\infty ds'\bigg\{{\cS_x^n(x',s') \atop \cT_{y,s}^n(x',s')}
     \bigg\}\cC_{-}^n(x,x',s-s')
   \bigg] ,
   \label{eq:cEx_str_cmpl}
  \\
  \bigg\{{c\cB_x^n(x,s) \atop \cE_{y,s}^n(x,s)}\bigg\}
  &=
   \int_{x_a}^{x_b}dx'
   \bigg[
     \bigg\{{\cA_x^n(x') \atop \cD_{y,s}^n(x')}\bigg\}\cC_{+}^n(x,x',s)
           +Z_0\int_0^\infty ds'\bigg\{{\cT_x^n(x',s') \atop \cS_{y,s}^n(x',s')}
     \bigg\}\cC_{+}^n(x,x',s-s')
   \bigg] .
   \label{eq:cBx_str_cmpl}
\end{align}
Next we find the complete form of the fields in the frequency domain from
Eqs.(\ref{eq:cEx_str_cmpl}-\ref{eq:cBx_str_cmpl}).
We first rewrite $(\vec{\cD}^n,\vec{\cA}^n)$ and $(\vec{\cS}^n,\vec{\cT}^n)$ into
the integral expressions with respect to $y'$, which involve $(\tDv,\tAv)$ and $(\tSv,\tTv)$
as shown in Eqs.(\ref{eq:Fp_coeff}-\ref{eq:Fm_coeff}).
Then we get the complete form of the fields in the straight pipe,
\begin{alignat}{2}
  \bigg\{{\tEv(\xv) \atop c\tBv(\xv)}\bigg\}
  &=\int_{x_a}^{x_b}dx'\int_{-h/2}^{h/2}dy'
    \bigg\{{ \hUv(\xv,\xv_\perp') \atop \hVv(\xv,\xv_\perp') }\bigg\}
    \qquad &&\text{(scalar expr.)}
   \label{eq:EB_str_cmp_scl}
   \\
  &=\int_{x_a}^{x_b}dx'\int_{-h/2}^{h/2}dy'
    \bigg\{{ \hUv^{\dg}(\xv,\xv_\perp') \atop \hVv^{\dg}(\xv,\xv_\perp') }\bigg\}
    \qquad &&\text{(differential expr.)} .
   \label{eq:EB_str_cmp}
\end{alignat}
Similar to Eqs.(\ref{eq:EB_str_sep_scl}-\ref{eq:EB_str_sep}),
we rewrite Eq.(\ref{eq:EB_str_cmp_scl}) into Eq.(\ref{eq:EB_str_cmp}) by integrating
the terms involving the derivatives of the current components and the initial fields
by parts with respect to $x'$, $y'$ or $s'$.
$(\hUv,\hVv)$ and $(\hUv^{\dg},\hVv^{\dg})$ are the integrands of
the scalar expression and the differential expression of the complete form of the fields,
\begin{alignat}{2}
  \hUv
  &=\hU_x\ev_x+\hU_y\ev_y+\hU_s\ev_s ,
    \qquad&
  \hUv^{\dg}
  &=\hU_x^{\dg}\ev_x+\hU_y^{\dg}\ev_y+\hU_s^{\dg}\ev_s ,
   \\
  \hVv
  &=\hV_x\ev_x+\hV_y\ev_y+\hV_s\ev_s ,
    \qquad&
  \hVv^{\dg}
  &=\hV_x^{\dg}\ev_x+\hV_y^{\dg}\ev_y+\hV_s^{\dg}\ev_s .
\end{alignat}
The components of $\hUv$ and $\hVv$ are given as follows,
\begin{align}
  \bigg\{{\hU_x(\xv,\xv_\perp') \atop \hV_y(\xv,\xv_\perp')}\bigg\}
  &=
       \bigg\{{\tD_x(\xv_\perp') \atop \tA_y(\xv_\perp')}\bigg\}
       \Gam_{-}^{+}(\xv_\perp,\xv_\perp',s)
      +Z_0\int_0^{\infty}ds'\bigg\{{\tS_x(\xv') \atop \tT_y(\xv')}\bigg\}
       \Gam_{-}^{+}(\xv_\perp,\xv_\perp',s-s') ,
  \label{eq:tEx_tBy_cmpl_st}
   \\
  \bigg\{{\hU_y(\xv,\xv_\perp') \atop \hV_x(\xv,\xv_\perp')}\bigg\}
  &=
       \bigg\{{\tD_y(\xv_\perp') \atop \tA_x(\xv_\perp')}\bigg\}
       \Gam_{+}^{-}(\xv_\perp,\xv_\perp',s)
      +Z_0\int_0^{\infty}ds'\bigg\{{\tS_y(\xv') \atop \tT_x(\xv')}\bigg\}
       \Gam_{+}^{-}(\xv_\perp,\xv_\perp',s-s') ,
  \label{eq:tEy_tBx_cmpl_st}
   \\
  \hU_s(\xv,\xv_\perp')
  &=
       \tD_s(\xv_\perp')\Gam_{+}^{+}(\xv_\perp,\xv_\perp',s)
      +Z_0\int_0^{\infty}ds'\tS_s(\xv')\Gam_{+}^{+}(\xv_\perp,\xv_\perp',s-s') ,
  \label{eq:tEs_cmpl_st}
   \\
  \hV_s(\xv,\xv_\perp')
  &=
       \tA_s(\xv_\perp')\Gam_{-}^{-}(\xv_\perp,\xv_\perp',s)
      +Z_0\int_0^{\infty}ds'\tT_s(\xv')\Gam_{-}^{-}(\xv_\perp,\xv_\perp',s-s') .
  \label{eq:tBs_cmpl_st}
\end{align}
$(\tDv,\tAv)$ and $(\tSv,\tTv)$ are given by
Eqs.(\ref{eq:tDv_tDA_str}) and (\ref{eq:src_E_str}-\ref{eq:src_B_str}).
On the other hand, $\hUv^{\dg}$ and $\hVv^{\dg}$ consist of
$(\tDv^{\dg},\tAv^{\dg})$ and $(\tSv^{\dg},\tTv^{\dg})$ which are the operators
given by Eqs.(\ref{eq:tDx_dg_st}-\ref{eq:tDs_dg_st}) and (\ref{eq:tST_dg_st}),
\begin{align}
  \bigg\{{\hU_x^{\dg}(\xv,\xv_{\perp}) \atop \hV_y^{\dg}(\xv,\xv_{\perp})}\bigg\}
  &=
       \bigg\{{\tD_x^{\dg}(\xv_\perp') \atop \tA_y^{\dg}(\xv_\perp')}\bigg\}
       \Gam_{-}^{+}(\xv_\perp,\xv_\perp',s)
      -Z_0\int_0^{\infty}ds'\bigg\{{\tS_x^{\dg}(\xv') \atop \tT_y^{\dg}(\xv')}\bigg\}
       \Gam_{-}^{+}(\xv_\perp,\xv_\perp',s-s') ,
  \label{eq:tEx_st_DS_cmpl}
  \\
  \bigg\{{\hU_y^{\dg}(\xv,\xv_{\perp}) \atop \hV_x^{\dg}(\xv,\xv_{\perp})}\bigg\}
  &=
       \bigg\{{\tD_y^{\dg}(\xv_\perp') \atop \tA_x^{\dg}(\xv_\perp')}\bigg\}
       \Gam_{+}^{-}(\xv_\perp,\xv_\perp',s)
      -Z_0\int_0^{\infty}ds'\bigg\{{\tS_y^{\dg}(\xv') \atop \tT_x^{\dg}(\xv')}\bigg\}
       \Gam_{+}^{-}(\xv_\perp,\xv_\perp',s-s') ,
  \label{eq:tEy_st_DS_cmpl}
   \\
  \hU_s^{\dg}(\xv,\xv_{\perp})
  &=
       \tD_s^{\dg}(\xv_\perp')\Gam_{+}^{+}(\xv_\perp,\xv_\perp',s)
      -Z_0\int_0^{\infty}ds'\tS_s^{\dg}(\xv')\Gam_{+}^{+}(\xv_\perp,\xv_\perp',s-s') ,
  \label{eq:tEs_st_DS_cmpl}
  \\
  \hV_s^{\dg}(\xv,\xv_{\perp})
  &=
       \tA_s^{\dg}(\xv_\perp')\Gam_{-}^{-}(\xv_\perp,\xv_\perp',s)
      -Z_0\int_0^{\infty}ds'\tT_s^{\dg}(\xv')\Gam_{-}^{-}(\xv_\perp,\xv_\perp',s-s') .
  \label{eq:tBs_st_AT_cmpl}
\end{align}
In the differential expressions of the fields given by Eq.(\ref{eq:EB_str_cmp}),
the operators $(\tDv^{\dg},\tAv^{\dg})$ and $(\tSv^{\dg},\tTv^{\dg})$ act on
$\Gam$ $[=(\Gam_{+}^{+},\Gam_{-}^{+},\Gam_{+}^{-},\Gam_{-}^{-})]$
which represents the Green functions of the complete form in the frequency domain,
\begin{align}
  \Gam_{\pm}^{+}(\xv_\perp,\xv_\perp',\vsig)
  =\sum_{n=1}^{\infty}\cC_{\pm}^n(x,x',\vsig)\cY_{+}^n(y,y') ,
   \qquad
  \Gam_{\pm}^{-}(\xv_\perp,\xv_\perp',\vsig)
  =\sum_{n=0}^{\infty}\cC_{\pm}^n(x,x',\vsig)\cY_{-}^n(y,y') .
   \label{eq:Gam_pm_pm}
\end{align}
The signs of $\Gam_{\pm}$ and $\Gam^{\pm}$ correspond to
those of $\cC_{\pm}^n$ and $\cY_{\pm}^n$ respectively.
That is, similar to $\Phi$ given by Eqs.(\ref{eq:Phi}), the signs of
the subscript and superscript of $\Gam_{\pm}^{\pm}$ denote respectively
the horizontal and vertical parities of the fields in the straight pipe.
$\Gam$ satisfies the following wave equation for $\vsig=\{s-s',s\}$,
\begin{align}
  \{\nablav^2+(k\beta)^2\}\Gam(\xv_\perp,\xv_\perp',\vsig)
  &=\delta(x-x')\delta(y-y')\bdelta(\vsig) .
\end{align}
According to Eqs.(\ref{eq:we_cX}) and (\ref{eq:cYpm_relations}),
$\Gam$ satisfies the following relations,
\begin{align}
  \rd_x\Gam_{-}^{\pm}+\rd_{x'}\Gam_{+}^{\pm}
  =\rd_x\Gam_{+}^{\pm}+\rd_{x'}\Gam_{-}^{\pm}
  =0 ,
    \qquad
  \rd_y\Gam_{\pm}^{-}+\rd_{y'}\Gam_{\pm}^{+}
  =\rd_y\Gam_{\pm}^{+}+\rd_{y'}\Gam_{\pm}^{-}
  =0 .
\end{align}
According to Eq.(\ref{eq:rdsik_cCpm_s0_st}), $\rd_s\Gam(\vsig)$ becomes
the transverse $\delta$-function at $\vsig=0$ for $\vsig=\{s-s',s\}$,
\begin{align}
  [\rd_s\Gam(\xv_\perp,\xv_\perp',\vsig)]_{\vsig=0}
  =\delta(x-x')\delta(y-y') .
  \label{eq:rdsik_tGam_p_s0_st}
\end{align}

Let us summarize the expressions of the fields in the frequency domain.
In this appendix we found four expressions of the fields in
the straight rectangular pipe as shown in Eqs.(\ref{eq:EB_str_sep_scl}-\ref{eq:EB_str_sep}) 
and (\ref{eq:EB_str_cmp_scl}-\ref{eq:EB_str_cmp}),
\begin{alignat}{3}
  & \quad\text{(Scalar expressions)}
  &&\quad\text{(Differential expressions)}
   \nonumber
   \\
  \bigg\{{\tEv \atop c\tBv}\bigg\}
  &=\int_{x_a}^{x_b}dx'\int_{-h/2}^{h/2}dy'
    \bigg\{{\tUv \atop \tVv}\bigg\}
  &&=\int_{x_a}^{x_b}dx'\int_{-h/2}^{h/2}dy'
    \bigg\{{\tUv^{\dg} \atop \tVv^{\dg}}\bigg\}
    \qquad&&\text{(Separated forms)}
   \\
  &=\int_{x_a}^{x_b}dx'\int_{-h/2}^{h/2}dy'
    \bigg\{{ \hUv \atop \hVv }\bigg\}
  &&=\int_{x_a}^{x_b}dx'\int_{-h/2}^{h/2}dy'
    \bigg\{{ \hUv^{\dg} \atop \hVv^{\dg} }\bigg\}
    \qquad&&\text{(Complete forms)} .
\end{alignat}
These integrands are given by the following equations,
\begin{align}
  \begin{array}{rll}
                           & \text{Scalar expressions},
                           & \text{Differential expressions}
    \\
    \text{Separated forms}:& (\tUv,\tVv)~
                             \text{in Eqs.(\ref{eq:tEx_tBy_st}-\ref{eq:tBs_st})} ,
                              \quad
                           & (\tUv^{\dg},\tVv^{\dg})~
                             \text{in Eqs.(\ref{eq:tEx_st_DS}-\ref{eq:tBs_st_AT})} ,
    \\
    \text{Complete forms}: & (\hUv,\hVv)~
                             \text{in Eqs.(\ref{eq:tEx_tBy_cmpl_st}-\ref{eq:tBs_cmpl_st})} ,
                               \quad
                           & (\hUv^{\dg},\hVv^{\dg})~
                             \text{in Eqs.(\ref{eq:tEx_st_DS_cmpl}-\ref{eq:tBs_st_AT_cmpl})}.
  \end{array}
  \label{eq:4express_str}
\end{align}
The corresponding four expressions of the fields in the curved pipe are
summarized in Eq.(\ref{eq:4express}).

\subsection{Transient fields created by a rigid beam in a straight pipe}
\label{sec:rigid_trans_str}

We consider the transient field created by a rigid bunch moving at
a constant speed $v\,(=c\beta)$ in a straight rectangular pipe.
The initial field at $s=0$ is arbitrary.
The bunch can have a nonzero dimension in both the longitudinal and transverse directions.
Assuming a rigid bunch, since $\tilde{\bm S}(s')e^{-iks'}$ and $\tilde{\bm T}(s')e^{-iks'}$ 
do not depend on $s'$, we can take them out of the integrals with respect to $s'$ in
Eqs.(\ref{eq:tEx_tBy_st}-\ref{eq:tBs_st}),
\begin{align}
  \bigg\{{\tE_x(\xv) \atop c\tB_y(\xv)}\bigg\}
  &=\int_{x_a}^{x_b}dx'\int_{-h/2}^{h/2}dy'
   \bigg[
      \bigg\{{\tD_x(\xv_\perp') \atop \tA_y(\xv_\perp')}\bigg\}
      \Phi_{-}^{+}(\xv_\perp,\xv_\perp',s)
     +Z_0\bigg\{{\tS_x(\xv') \atop \tT_y(\xv')}\bigg\}e^{-iks'}
      \tI_{-}^{+}(\xv_\perp,\xv_\perp',s)
   \bigg] ,
   \label{eq:tEx_tBy_rgd_str}
   \\
  \bigg\{{\tE_y(\xv) \atop c\tB_x(\xv)}\bigg\}
  &=\int_{x_a}^{x_b}dx'\int_{-h/2}^{h/2}dy'
   \bigg[
      \bigg\{{\tD_y(\xv_\perp') \atop \tA_x(\xv_\perp')}\bigg\}
      \Phi_{+}^{-}(\xv_\perp,\xv_\perp',s)
     +Z_0\bigg\{{\tS_y(\xv') \atop \tT_x(\xv')}\bigg\}e^{-iks'}
      \tI_{+}^{-}(\xv_\perp,\xv_\perp',s)
   \bigg] .
\end{align}
The longitudinal components of the transient fields created by the rigid bunch are given as
\begin{align}
  \tE_s(\xv)
  &=\int_{x_a}^{x_b}dx'\int_{-h/2}^{h/2}dy'
    \Big[
       \tD_s(\xv_\perp')\Phi_{+}^{+}(\xv_\perp,\xv_\perp',s)
      +Z_0\tS_s(\xv')e^{-iks'}
       \tI_{+}^{+}(\xv_\perp,\xv_\perp',s)
    \Big] ,
   \\
  c\tB_s(\xv)
  &=\int_{x_a}^{x_b}dx'\int_{-h/2}^{h/2}dy'
    \Big[
       \tA_s(\xv_\perp')\Phi_{-}^{-}(\xv_\perp,\xv_\perp',s)
      +Z_0\tT_s(\xv')e^{-iks'}
       \tI_{-}^{-}(\xv_\perp,\xv_\perp',s)
    \Big] .
   \label{eq:tBs_rgd_str}
\end{align}
Eqs.(\ref{eq:tEx_tBy_rgd_str}-\ref{eq:tBs_rgd_str}) are the scalar expressions of
the separated form.
$\tI_{\pm}^{\pm}$ is given as follows,
\begin{align}
  \tI_{\pm}^{+}(\xv_\perp,\xv_\perp',s)
  &=\int_{0}^{\infty}ds'\Phi_{\pm}^{+}(\xv_\perp,\xv_\perp',s-s')e^{iks'}
   =\sum_{n=1}^{\infty}\cY_{+}^n(y,y')\cI_{\pm}^n(x,x',s) ,
  \\
  \tI_{\pm}^{-}(\xv_\perp,\xv_\perp',s)
  &=\int_{0}^{\infty}ds'\Phi_{\pm}^{-}(\xv_\perp,\xv_\perp',s-s')e^{iks'}
   =\sum_{n=0}^{\infty}\cY_{-}^n(y,y')\cI_{\pm}^n(x,x',s) ,
\end{align}
where
\begin{align}
  \cI_{\pm}^n(x,x',s)
  &=\int_{0}^{\infty}ds'\cG_{\pm}^n(x,x',s-s')e^{iks'} .
   \label{eq:tIpm_apdx}
\end{align}
Substituting Eqs.(\ref{eq:cGp_st}-\ref{eq:cGm_st}) into Eq.(\ref{eq:tIpm_apdx}),
we calculate the integral with respect to $s'$,
\begin{align}
  \cI_{+}^n(x,x',s)
  &=\sum_{\ell=0}^{\infty}\bTh_{\ell}^n
    \bigg[
       \bdelta_{01}^{\ell}\sum_{m=1}^{\ell-1}\frac{\cX_{+}^m(x,x')}{k_s^{mn}}L(s,k_s^{mn})
     +\sum_{m=\ell'}^{\infty}\frac{\cX_{+}^m(x,x')}{2ik_s^{mn}}\bL(s,k_s^{mn})
    \bigg] ,
  \label{eq:cIp_str_rgd}
  \\
  \cI_{-}^n(x,x',s)
  &=\sum_{\ell=0}^\infty \bTh_{\ell}^n
    \bigg[
      \bdelta_{0}^{\ell}\sum_{m=0}^{\ell-1}\frac{\cX_{-}^m(x,x')}{k_s^{mn}}L(s,k_s^{mn})
      +\sum_{m=\ell}^{\infty}\frac{\cX_{-}^m(x,x')}{2ik_s^{mn}}\bL(s,k_s^{mn})
    \bigg] .
  \label{eq:cIm_str_rgd}
\end{align}
$L$ and $\bL$ are functions of $s$, which correspond respectively to
the oscillatory mode and the damped mode of the fields in the longitudinal direction,
\begin{align}
  L(s,k_s)
  &=\int_0^{s}ds'e^{iks'}\sin[k_s(s-s')]
   =\frac{1}{2}
    \bigg(
           \frac{e^{iks}-e^{-ik_ss}}{k+k_s}
          -\frac{e^{iks}-e^{ik_ss}}{k-k_s}
    \bigg)
  \\
   &=\frac{e^{iks}}{2}
    \bigg\{
           \frac{1-e^{-i(k_s+k)s}}{k_s+k}
          +\frac{1-e^{i(k_s-k)s}}{k_s-k}
    \bigg\}
   \\
  &=\frac{ik\sin(k_ss)+k_s\{\cos(k_ss)-e^{iks}\}}{k^2-k_s^2} ,
   \label{eq:L}
   \\
  \bL(s,k_s)
  &=\int_0^{\infty}\! ds'e^{iks'}e^{ik_s|s-s'|}
   =i\bigg(\frac{e^{iks}-e^{ik_ss}}{k_s-k} +\frac{e^{iks}}{k_s+k}\bigg)
   \\
  &=i\frac{(k+k_s)e^{ik_ss}-2k_se^{iks}}{k^2-k_s^2} .
   \label{eq:bL}
\end{align}
From the dispersion relation (\ref{eq:ks_mn}), the denominator of Eqs.(\ref{eq:L}) and
(\ref{eq:bL}) at the pole $k_s=k_s^{mn}$ is given as
\begin{align}
  k^2-(k_s^{mn})^2
  =(k/\gamma)^2+(k_x^m)^2+(k_y^n)^2 ,
    \qquad
  \gam
  =(1-\beta^2)^{-1/2} .
\end{align}
$k_x^m$ and $k_y^n$ are the transverse eigen-wavenumbers of the straight rectangular pipe,
given by Eqs.(\ref{eq:hkmn}).

Similar to the scalar expression (\ref{eq:tEx_tBy_rgd_str}-\ref{eq:tBs_rgd_str}),
we get the differential expression of the transient fields created by the rigid bunch
in the straight rectangular pipe,
\begin{align}
  \bigg\{{\tE_x(\xv) \atop c\tB_y(\xv)}\bigg\}
  &=\int_{x_a}^{x_b}dx'\int_{-h/2}^{h/2}dy'
    \bigg[
       \bigg\{{\tD_x^{\dg}(\xv_\perp') \atop \tA_y^{\dg}(\xv_\perp')}\bigg\}
       \Phi_{-}^{+}(\xv_\perp,\xv_\perp',s)
      -Z_0\bigg\{{\tS_x^{\dg}(\xv') \atop \tT_y^{\dg}(\xv')}\bigg\}e^{-iks'}
       \tI_{-}^{+}(\xv_\perp,\xv_\perp',s)
    \bigg] ,
   \label{eq:ExBy_dg_rgd_str}
  \\
  \bigg\{{\tE_y(\xv) \atop c\tB_x(\xv)}\bigg\}
  &=\int_{x_a}^{x_b}dx'\int_{-h/2}^{h/2}dy'
    \bigg[
       \bigg\{{\tD_y^{\dg}(\xv_\perp') \atop \tA_x^{\dg}(\xv_\perp')}\bigg\}
       \Phi_{+}^{-}(\xv_\perp,\xv_\perp',s)
      -Z_0\bigg\{{\tS_y^{\dg}(\xv') \atop \tT_x^{\dg}(\xv')}\bigg\}e^{-iks'}
       \tI_{+}^{-}(\xv_\perp,\xv_\perp',s)
    \bigg] .
\end{align}
The longitudinal components of the transient fields created by the rigid bunch are
given as follows,
\begin{align}
  \tE_s(\xv)
  &=\int_{x_a}^{x_b}dx'\int_{-h/2}^{h/2}dy'
    \Big[
       \tD_s^{\dg}(\xv_\perp')\Phi_{+}^{+}(\xv_\perp,\xv_\perp',s)
      -Z_0\tS_s^{\dg}(\xv')e^{-iks'}\tI_{+}^{+}(\xv_\perp,\xv_\perp',s)
    \Big] ,
  \\
  c\tB_s(\xv)
  &=\int_{x_a}^{x_b}dx'\int_{-h/2}^{h/2}dy'
    \Big[
       \tA_s^{\dg}(\xv_\perp')\Phi_{-}^{-}(\xv_\perp,\xv_\perp',s)
      -Z_0\tT_s^{\dg}(\xv')e^{-iks'}\tI_{-}^{-}(\xv_\perp,\xv_\perp',s)
    \Big] .
   \label{eq:Bs_dg_rgd_str}
\end{align}
%

\subsection{Time domain fields in a straight pipe}
\label{sec:TD_straight}

We try to find the expression of the field in the time domain.
Fourier transforming Eq.(\ref{eq:EB_str_cmp_scl}) through Eq.(\ref{eq:Fourier_trans}),
we get the scalar expression of the complete form in the time domain,
\begin{align}
  \bigg\{{\Ev(\xv,t) \atop c\Bv(\xv,t)}\bigg\}
  &=\int_{x_a}^{x_b}dx'\int_{-h/2}^{h/2}dy'\int_{-\infty}^{\infty}dt'
  \bigg\{{\Uv(\xv,t;\xv_\perp',t') \atop \Vv(\xv,t;\xv_\perp',t')}\bigg\} .
\end{align}
$\Uv$ and $\Vv$ are given as
\begin{align}
  \Uv=U_x\ev_x+U_y\ev_y+U_s\ev_s ,
    \qquad
  \Vv=V_x\ev_x+V_y\ev_y+V_s\ev_s ,
\end{align}
where
\begin{align}
  \bigg\{{U_x \atop V_y}\bigg\}
  &=\bigg\{{D_x(\xv_\perp',t') \atop A_y(\xv_\perp',t')}\bigg\}
    \bGam_{-}^{+}(\xv_\perp,\xv_\perp',s,\tau)
   +Z_0\int_0^{\infty}ds'\bigg\{{S_x(\xv',t') \atop T_y(\xv',t')}\bigg\}
    \bGam_{-}^{+}(\xv_\perp,\xv_\perp',s-s',\tau) ,
  \label{eq:Ex_By_cmpl_st}
  \\
  \bigg\{{U_y \atop V_x}\bigg\}
  &=\bigg\{{D_y(\xv_\perp',t') \atop A_x(\xv_\perp',t')}\bigg\}
    \bGam_{+}^{-}(\xv_\perp,\xv_\perp',s,\tau)
   +Z_0\int_0^{\infty}ds'\bigg\{{S_y(\xv',t') \atop T_x(\xv',t')}\bigg\}
    \bGam_{+}^{-}(\xv_\perp,\xv_\perp',s-s',\tau) ,
  \label{eq:Ey_Bx_cmpl_st}
  \\
  U_s
  &=D_s(\xv_\perp',t')\bGam_{+}^{+}(\xv_\perp,\xv_\perp',s,\tau)
    +Z_0\int_0^{\infty}ds'S_s(\xv',t')\bGam_{+}^{+}(\xv_\perp,\xv_\perp',s-s',\tau) ,
  \label{eq:Es_cmpl_st}
  \\
  V_s
  &=A_s(\xv_\perp',t')\bGam_{-}^{-}(\xv_\perp,\xv_\perp',s,\tau)
    +Z_0\int_0^{\infty}ds'T_s(\xv',t')\bGam_{-}^{-}(\xv_\perp,\xv_\perp',s-s',\tau) .
  \label{eq:Bs_cmpl_st}
\end{align}
$t'$ is the dummy variable for $t$.
$\tau$ is defined as
\begin{align}
  \tau=t-t' .
\end{align}
$(\Dv,\Av)$ and $(\Sv,\Tv)$ are given as follows,
\begin{alignat}{2}
  \Dv(\xv_\perp',t')
  &=[\rd_{s'}\Ev(\xv',t')+\Ev(\xv',t')\rd_s]_{s'=+0}
   ,\qquad&
  \Sv(\xv',t')
  &=\nablav'J_0(\xv',t')+\frac{\rd_{t'}}{c}\Jv(\xv',t') ,
   \\
  \Av(\xv_\perp',t')
  &=[\rd_{s'}c\Bv(\xv',t')+c\Bv(\xv',t')\rd_s]_{s'=+0}
   ,\qquad&
  \Tv(\xv',t')
  &=-\nablav'\times\Jv(\xv',t') .
\end{alignat}
$\nablav'$ is given by Eq.(\ref{eq:tST_dg_st}) which is the nabla with respect to $\xv'$.
$\bGam=(\bGam_{+}^{+},\bGam_{-}^{+},\bGam_{+}^{-},\bGam_{-}^{-})$ denotes
the Green functions of the straight rectangular pipe in the time domain,
\begin{align}
  \bGam_{\pm}^{+}(\xv_\perp,\xv_\perp',\vsig,\tau)
  &=\int_{-\infty}^{\infty}\frac{d\omg}{2\pi}e^{-i\omg\tau}
    \Gam_{\pm}^{+}(\xv_\perp,\xv_\perp',\vsig,\omg)
   =\sum_{n=1}^{\infty}\cY_{+}^n(y,y')L_{\pm}^n(x,x',\vsig,\tau) ,
  \label{eq:Gam_cmpl_st}
   \\
  \bGam_{\pm}^{-}(\xv_\perp,\xv_\perp',\vsig,\tau)
  &=\int_{-\infty}^{\infty}\frac{d\omg}{2\pi}e^{-i\omg\tau}
    \Gam_{\pm}^{-}(\xv_\perp,\xv_\perp',\vsig,\omg)
   =\sum_{n=0}^{\infty}\cY_{-}^n(y,y')L_{\pm}^n(x,x',\vsig,\tau) .
\end{align}
$L_{\pm}^n$ is given as
\begin{align}
  L_{\pm}^n(x,x',\vsig,\tau)
  &=\int_{-\infty}^{\infty}\frac{d\omg}{2\pi}e^{-i\omg\tau}\cC_{\pm}^n(x,x',\vsig,\omg)
   =\sum_{m=0,1}^{\infty}\cX_{\pm}^m(x,x')T_m^n(\vsig,\tau) ,
   \label{eq:Lpm_TD}
\end{align}
where the lower limits of the sum $m=0$ and 1 correspond to $\cX_{+}^m$ and $\cX_{-}^m$
respectively.
$T_m^n$ is given as
\begin{align}
  T_m^n(\vsig,\tau)
  &=\btheta(\vsig)\int_{-\infty}^{\infty}\frac{d\omg}{2\pi}e^{-i\omg\tau}
    \frac{\sin(k_s^{mn}\vsig)}{k_s^{mn}} ,
   \qquad
  ck_s^{mn}=\{\omg^2-(\omg_m^n)^2\}^{1/2} .
   \label{eq:Tmn_def}
\end{align}
$k_s^{mn}$ and $\btheta(\vsig)$ are given by Eq.(\ref{eq:ks_mn}) and Eq.(\ref{eq:bstep}).
$\omg_m^n$ is the cutoff frequency of the straight pipe, which corresponds to
the cutoff wavenumber $\hk_m^n\in\mathbb{R}^+$ given by Eq.(\ref{eq:hkmn}),
\begin{align}
   (\omg_m^n/c)^2=(\hk_m^n\beta)^2=(k_x^m)^2+(k_y^n)^2 .
\end{align}
The integrand in Eq.(\ref{eq:Tmn_def}) has no poles in the $\omg$-plane.
We integrate $T_m^n$ by parts,
\begin{align}
  T_m^n(\vsig,\tau)
  =-\frac{c^2}{2\pi\vsig}\int_{-\infty}^{\infty}d\omg
    \frac{1+i\omg\tau}{\omg^2} e^{-i\omg\tau}C_m^n(\vsig,\omg) ,
   \qquad
  C_m^n(\vsig,\omg)
  =\cos\Big[\frac{\vsig}{c}\{\omg^2-(\omg_m^n)^2\}^{1/2}\Big] .
  \label{eq:Tmn}
\end{align}

\clearpage

\section{Fourier coefficients of the radial and longitudinal fields}
\label{sec:HL_field_sdom}

In section \ref{sec:XS_field_sdom} we found the expressions of the Fourier coefficients of 
the radial and longitudinal components of the fields $\cE_{x,s}^n$ and $\cB_{x,s}^n$ by 
inverting their Laplace transform $\mfE_{x,s}^n$ and $\mfB_{x,s}^n$.
Instead of this way, in this appendix we will find $\cE_{x,s}^n$ and $\cB_{x,s}^n$ from
the Fourier coefficients of the vertical components $\cE_y^n$ and $\cB_y^n$ given by
Eq.(\ref{eq:cEBy_sol}) through Eqs.(\ref{eq:cBx}-\ref{eq:cBs}) in order to ensure that
$\cE_{x,s}^n$ and $\cB_{x,s}^n$ given by Eqs.(\ref{eq:cExs_cBxs_sprt}) are correct.

In order to find the expressions of $\cE_{x,s}^n$ and $\cB_{x,s}^n$ from
$\cE_y^n$ and $\cB_y^n$,
we first rewrite $\cE_y^n$ and $\cB_y^n$ so that they each have the initial values of
the radial and longitudinal components of the electric field $\cE_{x,s}^n$ only or
the magnetic field $\cB_{x,s}^n$ only, instead of the vertical components
$\cE_y^n$ and $\cB_y^n$.
Multiplying $ik\beta$ or $(-1)^nk_y^n$ with $\cE_y^n$ and $\cB_y^n$ given by
Eq.(\ref{eq:cEBy_sol}),
we rewrite them as follows using Eqs.(\ref{eq:Gauss_cF}-\ref{eq:Ampere_s_cF})
which are Maxwell equations for the Fourier coefficients of the fields,
\begin{align}
  -\,\bigg\{{(-1)^nk_y^n \atop ik\beta}\bigg\}\cE_y^n(r,s)
  &=\int_{r_a}^{r_b}dr'
    \Big[
      \cL_{+}^n(r,r',s)
     +Z_0\frac{r'}{\rho}\int_0^{\infty}ds'\cW_{+}^n(r,r',s-s',s')
    \Big]
   +Z_0\bigg\{{\cJ_0^n(r,s) \atop -\cJ_y^n(r,s)}\bigg\} ,
  \label{eq:ikb_monkyn_cEy}
  \\
  \bigg\{{(-1)^nk_y^n \atop ik\beta}\bigg\}c\cB_y^n(r,s)
  &=\int_{r_a}^{r_b}dr'
    \Big[
      \cL_{-}^n(r,r',s)
     +Z_0\frac{r'}{\rho}\int_0^{\infty}ds'\cW_{-}^n(r,r',s-s',s')
    \Big] .
  \label{eq:ikb_monkyn_cBy}
\end{align}
We can also get Eqs.(\ref{eq:ikb_monkyn_cEy}-\ref{eq:ikb_monkyn_cBy}) by
the inverse Laplace transform of Eqs.(\ref{eq:mfEy_rwrtn}-\ref{eq:mfBy_rwrtn}).
$\cL_{\pm}^n$ and $\cW_{\pm}^n$ are given as
\begin{align}
  \cL_{+}^n(r,r',s)
  &=
    \bigg\{{\cD_x^n(r')  \atop \cA_s^n(r')}\bigg\}\rd_{r'}\cG_{+}^n(r,r',s)
   +\bigg\{{-\cD_s^n(r') \atop \cA_x^n(r')}\bigg\}\frac{\rho}{r'}\rd_s\cG_{+}^n(r,r',s) ,
   \label{eq:cLpn}
   \\
  \cL_{-}^n(r,r',s)
  &=
    \bigg\{{\cA_x^n(r')  \atop \cD_s^n(r')}\bigg\}\rd_{r'}\cG_{-}^n(r,r',s)
   +\bigg\{{-\cA_s^n(r') \atop \cD_x^n(r')}\bigg\}\frac{\rho}{r'}\rd_s\cG_{-}^n(r,r',s) ,
   \\
  \cW_{+}^n(r,r',\vsig,s')
  &=
    \bigg\{{\cS_x^n(r',s')  \atop \cT_s^n(r',s')}\bigg\}\rd_{r'}\cG_{+}^n(r,r',\vsig)
   +\bigg\{{-\cS_s^n(r',s') \atop \cT_x^n(r',s')}\bigg\}
    \frac{\rho}{r'}\rd_s\cG_{+}^n(r,r',\vsig) ,
   \\
  \cW_{-}^n(r,r',\vsig,s')
  &=
    \bigg\{{\cT_x^n(r',s')  \atop \cS_s^n(r',s')}\bigg\}\rd_{r'}\cG_{-}^n(r,r',\vsig)
   +\bigg\{{-\cT_s^n(r',s') \atop \cS_x^n(r',s')}\bigg\}
    \frac{\rho}{r'}\rd_s\cG_{-}^n(r,r',\vsig) .
   \label{eq:cWmn}
\end{align}
The upper and lower quantities in the braces of
Eqs.(\ref{eq:ikb_monkyn_cEy}-\ref{eq:cWmn}) correspond.
Hereinafter in this appendix, we omit the radial arguments $(r,r')$ of
$\cG_{\pm}^n(r,r',\vsig)$ for clarity to show the longitudinal arguments
$\vsig=\{s-s',s\}$.
$\cD_{x,s}^n$ and $\cA_{x,s}^n$ are the longitudinal operators
(\ref{eq:bcEx_init}-\ref{eq:bcBs_init})
which have the initial values of $\cE_{x,s}^n$ and $\cB_{x,s}^n$ at $s=0$.
$\cS_{x,s}^n$ and $\cT_{x,s}^n$ are the source terms in the wave equations for
$\cE_{x,s}^n$ and $\cB_{x,s}^n$, given by Eqs.(\ref{eq:src_cSx_cTx}-\ref{eq:src_cSs_cTs}).

In order to derive the expressions of $\cE_{x,s}^n$ from $\cE_y^n$ and $\cB_y^n$
through Eqs.(\ref{eq:cBx}-\ref{eq:cBs}),
\begin{align}
  \cE_x^n
   &=\frac{-1}{(k_r^n)^2}
     \{ik\beta(\brd_sc\cB_y^n+Z_0\cJ_x^n)+(-1)^nk_y^n\rd_r\cE_y^n\} ,
   \\
  \cE_s^n
  &=\frac{1}{(k_r^n)^2}
    \{ik\beta(\rd_rc\cB_y^n-Z_0\cJ_s^n)-(-1)^nk_y^n\brd_s\cE_y^n\} ,
\end{align}
we differentiate the upper equation of (\ref{eq:ikb_monkyn_cEy}) and
the lower equation of (\ref{eq:ikb_monkyn_cBy}) with respect to $s$,
\begin{align}
  (-1)^nk_y^n\brd_s\cE_y^n(r,s)
  &=
   -ik\beta Z_0\cJ_s^n(r,s)
   -\int_{r_a}^{r_b}dr'
    \Big\{
      \hcD_{+}^n(r,r',s)
     +Z_0\frac{r'}{\rho}\int_0^{\infty}ds'\hcS_{+}^n(r,r',s-s',s')
    \Big\} ,
  \label{eq:kyn_rdsik_cEy}
  \\
  ik\beta\brd_sc\cB_y^n(r,s)
  &=
  \frac{Z_0}{r}\cS_x^n(r,s)
  -\int_{r_a}^{r_b}dr'
   \Big\{
      \hcD_{-}^n(r,r',s)
     +Z_0\frac{r'}{\rho}\int_0^{\infty}ds'\hcS_{-}^n(r,r',s-s',s')
   \Big\} ,
  \label{eq:ikb_rdsik_cBy}
\end{align}
where
\begin{align}
  \hcD_{+}^n(r,r',s)
  &=\cD_s^n(r')(k_r^n)^2\breve{\cG}_{+}^n(s)
   +\cD_x^n(r')\brd_s\rd_{r'}\cG_{+}^n(s) ,
   \\
  \hcD_{-}^n(r,r',s)
  &=\cD_x^n(r')(k_r^n)^2\breve{\cG}_{-}^n(s)
   -\cD_s^n(r')\brd_s\rd_{r'}\cG_{-}^n(s) ,
   \\
  \hcS_{+}^n(r,r',\vsig,s')
  &=\cS_s^n(r',s')(k_r^n)^2\breve{\cG}_{+}^n(\vsig)
   +\cS_x^n(r',s')\brd_s\rd_{r'}\cG_{+}^n(\vsig) ,
   \\
  \hcS_{-}^n(r,r',\vsig,s')
  &=\cS_x^n(r',s')(k_r^n)^2\breve{\cG}_{-}^n(\vsig)
   -\cS_s^n(r',s')\brd_s\rd_{r'}\cG_{-}^n(\vsig) .
\end{align}
$\breve{\cG}_{\pm}^n$ is given as
\begin{align}
  \breve{\cG}_{\pm}^n(\vsig)
  &=\frac{\hr'}{\hr}(\brd_{\hr'}\rd_{\hr'}+1)\cG_{\pm}^n(\vsig)
   =\frac{\rho}{\hr}
    \Big\{\delta(\hr-\hr')\bar{\delta}(\vsig)-\frac{\rho}{\hr'}\rd_s^2\cG_{\pm}^n(\vsig)
    \Big\} .
\end{align}
$\bdelta(\vsig)$ is the extended $\delta$-function of $\vsig=\{s-s',s\}$,
defined in Eq.(\ref{eq:bdelta}).

Next, we differentiate the upper equation of (\ref{eq:ikb_monkyn_cEy}) and
the lower equation of (\ref{eq:ikb_monkyn_cBy}) with respect to $r$,
\begin{align}
  (-1)^nk_y^n\rd_r\cE_y^n(r,s)
  &=
   -Z_0\rd_r\cJ_0^n(r,s)
   +\int_{r_a}^{r_b}dr'
    \Big\{
      \bcD_{+}^n(r,r',s)
      +Z_0\frac{r'}{\rho}\int_0^{\infty}ds'\bcS_{+}^n(r,r',\vsig,s')
    \Big\} ,
  \label{eq:kyn_rdr_cEy}
  \\
  ik\beta\rd_rc\cB_y^n(r,s)
  &=\int_{r_a}^{r_b}dr'
    \Big\{
      \bcD_{-}^n(r,r',s)
     +Z_0\frac{r'}{\rho}\int_0^{\infty}ds'\bcS_{-}^n(r,r',\vsig,s')
    \Big\} ,
  \label{eq:ikb_rdr_cBy}
\end{align}
where
\begin{align}
  \bcD_{+}^n(r,r',s)
  &=\cD_s^n(r')\frac{\rho}{r'}\rd_s\rd_r\cG_{+}^n(s)
   -\cD_x^n(r')\rd_r\rd_{r'}\cG_{+}^n(s) ,
   \\
  \bcD_{-}^n(r,r',s)
  &=\cD_s^n(r')\rd_r\rd_{r'}\cG_{-}^n(s)
   +\cD_x^n(r')\frac{\rho}{r'}\rd_s\rd_r\cG_{-}^n(s) ,
   \\
  \bcS_{+}^n(r,r',\vsig,s')
  &=\cS_s^n(r',s')\frac{\rho}{r'}\rd_s\rd_r\cG_{+}^n(\vsig)
   -\cS_x^n(r',s')\rd_r\rd_{r'}\cG_{+}^n(\vsig) ,
   \\
  \bcS_{-}^n(r,r',\vsig,s')
  &=\cS_s^n(r',s')\rd_r\rd_{r'}\cG_{-}^n(\vsig)
   +\cS_x^n(r',s')\frac{\rho}{r'}\rd_s\rd_r\cG_{-}^n(\vsig) .
\end{align}
Substituting Eqs.(\ref{eq:kyn_rdsik_cEy}-\ref{eq:ikb_rdsik_cBy}) and
(\ref{eq:kyn_rdr_cEy}-\ref{eq:ikb_rdr_cBy}) into Eqs.(\ref{eq:cBx}-\ref{eq:cBs}),
we rewrite them as follows,
\begin{align}
  \cE_{x,s}^n(r,s)
  &=\int_{r_a}^{r_b}dr'
    \Big\{\cP_{x,s}^n(r,r',s)
         +Z_0\int_0^{\infty}ds'\cM_{x,s}^n(r,r',s-s',s')
    \Big\} .
  \label{eq:cExs}
\end{align}
$\cP_{x,s}^n$ and $\cM_{x,s}^n$ respectively involve the initial values and
the source terms of $\cE_{x,s}^n$,
\begin{align}
  \cP_x^n(s)
  &=\cD_x^n(r')\big\{\breve{\cG}_{-}^n(s)+\rd_{\hr}\rd_{\hr'}\cG_{+}^n(s)\big\}
   -\cD_s^n(r')\rho\rd_s
    \bigg\{\frac{\rd_{\hr'}}{\hr}\cG_{-}^n(s)+\frac{\rd_{\hr}}{\hr'}\cG_{+}^n(s)\bigg\} ,
   \label{eq:cPx}
  \\
  \cP_s^n(s)
  &=\cD_s^n(r')\big\{\breve{\cG}_{+}^n(s)+\rd_{\hr}\rd_{\hr'}\cG_{-}^n(s)\big\}
   +\cD_x^n(r')\rho\rd_s
    \bigg\{\frac{\rd_{\hr'}}{\hr}\cG_{+}^n(s)+\frac{\rd_{\hr}}{\hr'}\cG_{-}^n(s)\bigg\} ,
  \\
  \cM_x^n(\vsig,s')
  &=
    \frac{r'}{\rho}\cS_x^n(r',s')
    \big\{\breve{\cG}_{-}^n(\vsig)+\rd_{\hr}\rd_{\hr'}\cG_{+}^n(\vsig)\big\}
   -\cS_s^n(r',s')r'\rd_s
    \bigg\{\frac{\rd_{\hr'}}{\hr}\cG_{-}^n(\vsig)+\frac{\rd_{\hr}}{\hr'}\cG_{+}^n(\vsig)
    \bigg\} ,
  \\
  \cM_s^n(\vsig,s')
  &=
    \frac{r'}{\rho}\cS_s^n(r',s')
    \big\{\breve{\cG}_{+}^n(\vsig)+\rd_{\hr}\rd_{\hr'}\cG_{-}^n(\vsig)\big\}
   +\cS_x^n(r',s')r'\rd_s
    \bigg\{\frac{\rd_{\hr'}}{\hr}\cG_{+}^n(\vsig)+\frac{\rd_{\hr}}{\hr'}\cG_{-}^n(\vsig)
    \bigg\} .
   \label{eq:cMs}
\end{align}

Through Eqs.(\ref{eq:cBx}) and (\ref{eq:cBs}), we can get $\cB_{x,s}^n$ from
the lower equation of (\ref{eq:ikb_monkyn_cEy}) and the upper equation of
(\ref{eq:ikb_monkyn_cBy}) which have $\cA_{x,s}^n$ and $\cT_{x,s}^n$.
Omitting the intermediate calculations, let us show the expressions of $\cB_{x,s}^n$
which are rearranged in a similar manner as Eq.(\ref{eq:cExs}),
\begin{align}
  c\cB_{x,s}^n(r,s)
  &=\int_{r_a}^{r_b}dr'
    \Big\{\cQ_{x,s}^n(r,r',s)
         +Z_0\int_0^{\infty}ds'\cN_{x,s}^n(r,r',s-s',s')
    \Big\} .
  \label{eq:cBxs}
\end{align}
$\cQ_{x,s}^n$ and $\cN_{x,s}^n$ involve respectively
the initial values and the source terms of $\cB_{x,s}^n$,
\begin{align}
  \cQ_x^n(s)
  &=\cA_x^n(r')\big\{\breve{\cG}_{+}^n(s)+\rd_{\hr}\rd_{\hr'}\cG_{-}^n(s)\big\}
   -\cA_s^n(r')\rho\rd_s
    \bigg\{\frac{\rd_{\hr'}}{\hr}\cG_{-}^n(s)+\frac{\rd_{\hr}}{\hr'}\cG_{+}^n(s)\bigg\} ,
  \\
  \cQ_s^n(s)
  &=\cA_s^n(r')\big\{\breve{\cG}_{-}^n(s)+\rd_{\hr}\rd_{\hr'}\cG_{+}^n(s)\big\}
   +\cA_x^n(r')\rho\rd_s
    \bigg\{\frac{\rd_{\hr'}}{\hr}\cG_{-}^n(s)+\frac{\rd_{\hr}}{\hr'}\cG_{+}^n(s)\bigg\} ,
  \\
  \cN_x^n(\vsig,s')
  &=
    \frac{r'}{\rho}\cT_x^n(r',s')
    \big\{\breve{\cG}_{+}^n(\vsig)+\rd_{\hr}\rd_{\hr'}\cG_{-}^n(\vsig)\big\}
   -\cT_s^n(r',s')r'\rd_s
    \bigg\{\frac{\rd_{\hr'}}{\hr}\cG_{+}^n(\vsig)+\frac{\rd_{\hr}}{\hr'}\cG_{-}^n(\vsig)
    \bigg\} ,
  \\
  \cN_s^n(\vsig,s')
  &=
    \frac{r'}{\rho}\cT_s^n(r',s')
    \big\{\breve{\cG}_{-}^n(\vsig)+\rd_{\hr}\rd_{\hr'}\cG_{+}^n(\vsig)\big\}
   +\cT_x^n(r',s')r'\rd_s
    \bigg\{\frac{\rd_{\hr'}}{\hr}\cG_{-}^n(\vsig)+\frac{\rd_{\hr}}{\hr'}\cG_{+}^n(\vsig)
    \bigg\} .
   \label{eq:cNs}
\end{align}
Eqs.(\ref{eq:cExs}) and (\ref{eq:cBxs}) agree with Eq.(\ref{eq:cExs_cBxs_sprt})
in which we use the Green functions given by Eqs.(\ref{eq:cHpm}),
\begin{align}
  \cH_{\pm}^n(\vsig)
  =\breve{\cG}_{\pm}^n(\vsig)+\rd_{\hr}\rd_{\hr'}\cG_{\mp}^n(\vsig) ,
    \qquad
  \cK_{\pm}^n(\vsig)
  =\rho\rd_s
   \bigg\{\frac{\rd_{\hr'}}{\hr}\cG_{\pm}^n(\vsig)+\frac{\rd_{\hr}}{\hr'}\cG_{\mp}^n(\vsig)
   \bigg\} .
  \label{eq:HK_pm_again}
\end{align}
$\cK_{\pm}^n$ is the Fourier coefficient of the Green functions which entangle
the radial and longitudinal components of the fields.
Eqs.(\ref{eq:HK_pm_again}) are the inverse Laplace transforms of Eqs.(\ref{eq:mfHKpm}).
We showed that we can get the expressions of $\cE_{x,s}^n$ and $\cB_{x,s}^n$ from
$\cE_y^n$ and $\cB_y^n$ through Eqs.(\ref{eq:cBx}-\ref{eq:cBs})
without calculating the Bromwich integral (\ref{eq:cHK}) which is the inverse Laplace 
transform of $\mfE_{x,s}^n$ and $\mfB_{x,s}^n$ with respect to $\nu$.
However, it requires a lot more work to rewrite Eq.(\ref{eq:cEBy_sol}) into
Eqs.(\ref{eq:ikb_monkyn_cEy}-\ref{eq:ikb_monkyn_cBy}) through
Eqs.(\ref{eq:Gauss_cF}-\ref{eq:Ampere_s_cF}).

\clearpage

\section{Radial and longitudinal fields in the Laplace domain}
\label{sec:we_xs}

Eqs.(\ref{eq:mfBsEx}-\ref{eq:mfBxEs}) are the expressions of $\mfE_{x,s}^n$ and
$\mfB_{x,s}^n$ which are the radial and longitudinal components of
the Laplace domain fields in the bending section.
We got them from the expressions of the vertical components
$\mfE_y^n$ and $\mfB_y^n$ given by Eqs.(\ref{eq:mfEBy_solution})
through Eqs.(\ref{eq:mfBx_EyBy}-\ref{eq:mfBs_EyBy}).
Besides this way, we can get $\mfE_{x,s}^n$ and $\mfB_{x,s}^n$ by solving
the wave equations (\ref{eq:we_tExsBxs}) for them using the eigenfunctions of the operator.
We will demonstrate it in this appendix to ensure Eqs.(\ref{eq:mfBsEx}-\ref{eq:mfBxEs})
analytically.
We first rewrite Eqs.(\ref{eq:we_tExsBxs}) using $r$ defined in Eq.(\ref{eq:r}),
\begin{align}
  \nablav_\vdash^2\bigg({\tE_x \atop \tE_s}\bigg)
  +\frac{2}{r}\brd_s\bigg({-\tE_s \atop +\tE_x}\bigg)
  =Z_0\bigg({\tS_x \atop \tS_s}\bigg) ,
   \qquad
  \nablav_\vdash^2\bigg({c\tB_x \atop c\tB_s}\bigg)
  +\frac{2}{r}\brd_s\bigg({-c\tB_s \atop +c\tB_x}\bigg)
  =Z_0\bigg({\tT_x \atop \tT_s}\bigg) .
  \label{eq:we_xs_apdx}
\end{align}
$\nablav_{\vdash}^2$ and $\brd_s$ are given by Eqs.(\ref{eq:brdxs}) and (\ref{eq:operator}).
The source terms $\tS_{x,s}$ and $\tT_{x,s}$ are given by Eqs.(\ref{eq:tSv}-\ref{eq:tTv}),
\begin{alignat}{2}
  \tS_x(\xv)&=\rd_{r}\tJ_0(\xv)-ik\beta\tJ_x(\xv) ,
    \qquad&
  \tT_x(\xv)&=\brd_{s}\tJ_y(\xv)-\rd_{y}\tJ_s(\xv) ,
  \\
  \tS_s(\xv)&=\brd_{s}\tJ_0(\xv)-ik\beta\tJ_s(\xv) ,
    \qquad&
  \tT_s(\xv)&=\rd_{y}\tJ_x(\xv)-\rd_{r}\tJ_y(\xv) .
\end{alignat}
We expand the fields and current in their vertical oscillation modes through
Eqs.(\ref{eq:Four_Exp_plus}-\ref{eq:Four_Exp_minus}).
The $n$th Fourier coefficients of the radial and longitudinal components of the fields
satisfy the wave equations (\ref{eq:we_cFxs}),
\begin{align}
  \rd_{\vdash}^2\bigg({\cE_x^n \atop \cE_s^n}\bigg)
  +\frac{2}{r}\brd_s\bigg({-\cE_s^n \atop +\cE_x^n}\bigg)
  =Z_0\bigg({\cS_x^n \atop \cS_s^n}\bigg) ,
   \qquad
  \rd_{\vdash}^2\bigg({c\cB_x^n \atop c\cB_s^n}\bigg)
  +\frac{2}{r}\brd_s\bigg({-c\cB_s^n \atop +c\cB_x^n}\bigg)
  =Z_0\bigg({\cT_x^n \atop \cT_s^n}\bigg) .
  \label{eq:we_cExBx}
\end{align}
$\rd_{\vdash}^2$ is the operator given by Eq.(\ref{eq:wen_bend}).
$\cS_{x,s}^n$ and $\cT_{x,s}^n$ are the $n$th Fourier coefficients of
$\tS_{x,s}$ and $\tT_{x,s}$,
\begin{alignat}{2}
  \cS_x^n(r,s)&=\rd_r\cJ_0^n(r,s)-ik\beta\cJ_x^n(r,s) ,
   \qquad&
  \cT_x^n(r,s)&=\brd_s\cJ_y^n(r,s)-(-1)^nk_y^n\cJ_s^n(r,s) ,
  \label{eq:src_cSx_cTx}
  \\
  \cS_s^n(r,s)&=\brd_s\cJ_0^n(r,s)-ik\beta\cJ_s^n(r,s) ,
   \qquad&
  \cT_s^n(r,s)&=(-1)^nk_y^n\cJ_x^n(r,s)-\rd_r\cJ_y^n(r,s) .
  \label{eq:src_cSs_cTs}
\end{alignat}
%

\subsection{Eigenfunctions and general solutions of the wave equations}

The radial and longitudinal components of the fields are coupled in
the wave equations (\ref{eq:we_cExBx}).
In order to disentangle these field variables,
we define the eigenfunctions of the operator of Eqs.(\ref{eq:we_cExBx}),
\begin{align}
  \cE_{\pm}^n=\frac{\cE_x^n\pm i\cE_s^n}{2},
   \qquad
  \cB_{\pm}^n=\frac{\cB_x^n\pm i\cB_s^n}{2};
   \qquad
  \cS_{\pm}^n=\frac{\cS_x^n\pm i\cS_s^n}{2},
   \qquad
  \cT_{\pm}^n=\frac{\cT_x^n\pm i\cT_s^n}{2} .
  \label{eq:cEBpm_def}
\end{align}
Rewriting Eqs.(\ref{eq:we_cExBx}) using Eqs.(\ref{eq:cEBpm_def}),
the field variables are disentangled in the wave equations,
\begin{align}
  \bigg[\brd_{r}\rd_{r}+(k_r^n)^2+\frac{(\rho\rd_s+i)^2}{r^2}\bigg]
  \bigg\{{\cE_{\pm}^n(r,s) \atop c\cB_{\pm}^n(r,s) }\bigg\}
  =Z_0\bigg\{{\cS_{\pm}^n(r,s) \atop \cT_{\pm}^n(r,s) }\bigg\} .
  \label{eq:we_xs_FC}
\end{align}
$\brd_r$ is given by Eq.(\ref{eq:rs_opera}).
Using the Laplace transform ($\cL$) defined in Eq.(\ref{eq:Laplace}),
we solve Eq.(\ref{eq:we_xs_FC}) as an initial value problem with respect to $s$,
similar to Eq.(\ref{eq:wenu}) for $\cE_y^n$ and $\cB_y^n$ in section \ref{sec:we_in_LD},
\begin{alignat}{2}
  \mfE_{\pm}^n(r,\nu)
  &=\cL[\cE_{\pm}^n(r,s)]
   =\frac{\mfE_x^n\pm i\mfE_s^n}{2} ,
    \qquad&
  \mfS_{\pm}^n(r,\nu)
  &=\cL[\cS_{\pm}^n(r,s)]
   =\frac{\mfS_x^n\pm i\mfS_s^n}{2} ,
  \label{eq:mfEpm}
   \\
  \mfB_{\pm}^n(r,\nu)
  &=\cL[\cB_{\pm}^n(r,s)]
   =\frac{\mfB_x^n\pm i\mfB_s^n}{2} ,
    \qquad&
  \mfT_{\pm}^n(r,\nu)
  &=\cL[\cT_{\pm}^n(r,s)]
   =\frac{\mfT_x^n\pm i\mfT_s^n}{2} .
  \label{eq:mfBpm}
\end{alignat}
$\nu$ is the Laplace variable defined in Eq.(\ref{eq:u_nu}) instead of $\kap$
for convenience.
$(\mfE_{\pm}^n,\mfB_{\pm}^n)$ and $(\mfS_{\pm}^n,\mfT_{\pm}^n)$ are
the Laplace transform of the eigenfunctions (\ref{eq:cEBpm_def}).
$\mfE_{\pm}^n$ and $\mfB_{\pm}^n$ satisfy the following wave equation,
\begin{align}
  \bigg[\brd_{\hr}\rd_{\hr}+1-\frac{(\nu\pm1)^2}{\hr^2}\bigg]
  \bigg\{{\mfE_{\pm}^n(r,\nu) \atop c\mfB_{\pm}^n(r,\nu) }\bigg\}
  =\frac{1}{k_r^n\hr}
    \bigg\{{\bmfD_{\pm}^n(r,\nu) \atop \bmfA_{\pm}^n(r,\nu) }\bigg\} .
  \label{eq:we_xs_LD}
\end{align}
$\hr$ and $\brd_{\hr}$ are given by Eqs.(\ref{eq:krn}) and (\ref{eq:rd_hr}).
Eq.(\ref{eq:we_xs_LD}) is an inhomogeneous Bessel differential equation of
the order $\nu\pm1$.
$\bmfD_\pm^n$ and $\bmfA_\pm^n$ are the driving terms of $\mfE_{\pm}^n$ and $\mfB_{\pm}^n$,
\begin{alignat}{2}
  \bmfD_{\pm}^n(r,\nu)
  &=\mfD_{\pm}^n(r,\nu)
   +Z_0r\mfS_{\pm}^n(r,\nu) ,
    \qquad&
  \mfD_{\pm}^n(r,\nu)
  &=\frac{1}{r}[\{i(\nu\pm2)+\rho\rd_{s'}\}\cE_{\pm}^n(r,s')]_{s'=+0} ,
  \label{eq:mfUpm}
  \\
  \bmfA_{\pm}^n(r,\nu)
  &=\mfA_{\pm}^n(r,\nu)
   +Z_0r\mfT_{\pm}^n(r,\nu) ,
    \qquad&
  \mfA_{\pm}^n(r,\nu)
  &=\frac{1}{r}[\{i(\nu\pm2)+\rho\rd_{s'}\}c\cB_{\pm}^n(r,s')]_{s'=+0} .
  \label{eq:mfVpm}
\end{alignat}
$(\mfS_{\pm}^n,\mfT_{\pm}^n)$ and $(\mfS_{x,s}^n,\mfT_{x,s}^n)$
are the Laplace transform of the source terms
$(\cS_{\pm}^n,\cT_{\pm}^n)$ and $(\cS_{x,s}^n,\cT_{x,s}^n)$,
\begin{alignat}{2}
  \mfS_x^n(r,\nu)
  &=\rd_{r}\mfJ_0^n(r,\nu)-ik\beta\mfJ_x^n(r,\nu) ,
   \\
  \mfS_s^n(r,\nu)
  &=\frac{1}{r}[i\nu\mfJ_0^n(r,\nu)-\cJ_0^n(r,0)]-ik\beta\mfJ_s^n(r,\nu) ,
   \\
  \mfT_x^n(r,\nu)
  &=\frac{1}{r}[i\nu\mfJ_y^n(r,\nu)-\cJ_y^n(r,0)]-(-1)^nk_y^n\mfJ_s^n(r,\nu) ,
   \\
  \mfT_s^n(r,\nu)
  &=(-1)^nk_y^n\mfJ_x^n(r,\nu)-\rd_r\mfJ_y^n(r,\nu) .
\end{alignat}
The current in the Laplace domain $\mfJ^n=(\mfJ_0^n,\mfJ_{x,y,s}^n)$ is
defined in Eq.(\ref{eq:J_LD}).

The field variables $\mfE_{\pm}^n$ and $\mfB_{\pm}^n$ are the eigenfunctions 
which are not entangled in the wave equation (\ref{eq:we_xs_LD})
for the radial and longitudinal components of the electric and magnetic fields.
Conversely, however, $\mfE_{\pm}^n$ and $\mfB_{\pm}^n$ each are entangled in
the boundary conditions on the sidewalls of the curved pipe,
\begin{align}
  \brd_r\mfE_x^n=\brd_r(c\mfB_s^n)=0
   ,\qquad
  \mfE_s^n=c\mfB_x^n=0
  \qquad \text{at}~r=r_a,r_b .
  \label{eq:BC_ExEs_BxBs}
\end{align}
We solve Eq.(\ref{eq:we_xs_LD}) using the variation of parameters.
Choosing $J_{\nu\pm1}$ and $Y_{\nu\pm1}$ as a pair of the fundamental solutions of 
Eq.(\ref{eq:we_xs_LD}), the general solutions are given as their linear combinations,
\begin{align}
  \mfE_\pm^n(r)
  =a_\pm(r) J_{\nu\pm1}(\hr) +b_\pm(r) Y_{\nu\pm1}(\hr) ,
    \qquad
  c\mfB_\pm^n(r)
  =p_\pm(r) J_{\nu\pm1}(\hr) +q_\pm(r) Y_{\nu\pm1}(\hr) .
   \label{eq:mfEB_gnrl}
\end{align}
The coefficients $(a_\pm,b_\pm)$ and $(p_\pm,q_\pm)$ are functions of $r$ and $\nu$
(we omit the argument $\nu$ for brevity excluding appendix \ref{sec:rearrange}).
Substituting Eqs.(\ref{eq:mfEB_gnrl}) into Eq.(\ref{eq:we_xs_LD}),
we get these coefficients by indefinite integration,
\begin{align}
  \bigg\{{ a_{\pm}(r) \atop p_{\pm}(r)}\bigg\}
  &=\bigg\{{ a_{\pm}^0 \atop p_{\pm}^0}\bigg\}
    +\frac{\pi}{2}\int_{r}^{r_b}dr'
     \bigg\{{ \bmfD_{\pm}^n(r') \atop \bmfA_{\pm}^n(r')}\bigg\}
     Y_{\nu\pm1}(\hr') ,
   \label{eq:ap_pm}
   \\
  \bigg\{{ b_{\pm}(r) \atop  q_{\pm}(r)}\bigg\}
  &=\bigg\{{ b_{\pm}^0 \atop  q_{\pm}^0}\bigg\}
   +\frac{\pi}{2}\int_{r_a}^{r}dr'
    \bigg\{{ \bmfD_{\pm}^n(r') \atop  \bmfA_{\pm}^n(r')}\bigg\}
    J_{\nu\pm1}(\hr') .
   \label{eq:bq_pm}
\end{align}
$r'$ is a dummy for the radial variable, and $\hr'=k_r^nr'$ similar to Eq.(\ref{eq:krn}).
$(a_{\pm}^0,b_{\pm}^0)$ and $(p_{\pm}^0,q_{\pm}^0)$ are the constants of integration,
which will be determined by imposing Eqs.(\ref{eq:BC_ExEs_BxBs}).
From Eqs.(\ref{eq:cEBpm_def}), we have
\begin{align}
  \mfE_x^n=\mfE_{+}^n+\mfE_{-}^n ,
   \qquad
  i\mfE_s^n=\mfE_{+}^n-\mfE_{-}^n ,
   \qquad
  \mfB_x^n=\mfB_{+}^n+\mfB_{-}^n ,
   \qquad
  i\mfB_s^n=\mfB_{+}^n-\mfB_{-}^n .
\end{align}
We rewrite $J_{\nu\pm1}$ and $Y_{\nu\pm1}$, which are involved in
Eqs.(\ref{eq:mfEB_gnrl}-\ref{eq:bq_pm}), in terms of $(J_{\nu},Y_{\nu})$ and
$(J_{\nu}',Y_{\nu}')$ using the recurrence relations of the Bessel functions,
given by 9.1.27 in \cite{abramo_stegun} and summarized in
appendix \ref{sec:tnu_recurr} (p.\pageref{sec:tnu_recurr}).
Then we get the general solutions of the radial and longitudinal components of the fields 
in the Laplace domain before imposing the boundary conditions (\ref{eq:BC_ExEs_BxBs}),
\begin{align}
  \bigg\{{ \mfE_x^n \atop i\mfE_s^n }\bigg\}
  &=\bigg\{{ a_{+}+a_{-}  \atop a_{+}-a_{-} }\bigg\}
    \frac{\nu}{\hr}J_{\nu}(\hr)
   +\bigg\{{ b_{+}+b_{-}  \atop b_{+}-b_{-} }\bigg\}
    \frac{\nu}{\hr}Y_{\nu}(\hr)
   -\bigg\{{ a_{+}-a_{-} \atop a_{+}+a_{-} }\bigg\}
    J_{\nu}'(\hr)
   -\bigg\{{ b_{+}-b_{-} \atop b_{+}+b_{-} }\bigg\}
    Y_{\nu}'(\hr) ,
  \label{eq:2mfEx_ab_pm}
  \\
  \bigg\{{ c\mfB_x^n \atop ic\mfB_s^n }\bigg\}
  &=\bigg\{{ p_{+}+p_{-}  \atop p_{+}-p_{-} }\bigg\}\frac{\nu}{\hr}J_{\nu}(\hr)
   +\bigg\{{ q_{+}+q_{-}  \atop q_{+}-q_{-} }\bigg\}\frac{\nu}{\hr}Y_{\nu}(\hr)
   -\bigg\{{ p_{+}-p_{-}  \atop p_{+}+p_{-} }\bigg\}J_{\nu}'(\hr)
   -\bigg\{{ q_{+}-q_{-}  \atop q_{+}+q_{-} }\bigg\}Y_{\nu}'(\hr) .
  \label{eq:2mfBx_pq_pm}
\end{align}
The quantities in the braces on the R.H.S. of
Eqs.(\ref{eq:2mfEx_ab_pm}-\ref{eq:2mfBx_pq_pm}) are given as follows,
\begin{alignat}{2}
  a_{+}(r)\pm a_{-}(r)
  &=\bar{a}_{\pm}^0+\bar{a}_{\pm}(r) ,
    \qquad&
  \bar{a}_\pm^0
  &=a_{+}^0\pm a_{-}^0 ,
   \label{eq:bara_pm}
   \\
  b_{+}(r)\pm b_{-}(r)
  &=\bar{b}_{\pm}^0+\bar{b}_{\pm}(r) ,
    \qquad&
  \bar{b}_\pm^0
  &=b_{+}^0\pm b_{-}^0 ,
   \label{eq:barb_pm}
   \\
  p_{+}(r)\pm p_{-}(r)
  &=\bar{p}_{\pm}^0+\bar{p}_{\pm}(r) ,
    \qquad&
  \bar{p}_\pm^0
  &=p_{+}^0\pm p_{-}^0 ,
  \label{eq:barp_pm}
   \\
  q_{+}(r)\pm q_{-}(r)
  &=\bar{q}_{\pm}^0+\bar{q}_{\pm}(r) ,
    \qquad&
  \bar{q}_\pm^0
  &=q_{+}^0\pm q_{-}^0 ,
  \label{eq:barq_pm}
\end{alignat}
where we rewrote the constants of integration ($a_\pm^0$, $b_\pm^0$, $p_\pm^0$, $q_\pm^0$) 
into ($\bar{a}_\pm^0$, $\bar{b}_\pm^0$, $\bar{p}_\pm^0$, $\bar{q}_\pm^0$)
for convenience in later calculations.
These unknown constants are determined in order that
Eqs.(\ref{eq:2mfEx_ab_pm}-\ref{eq:2mfBx_pq_pm}) satisfy Eqs.(\ref{eq:BC_ExEs_BxBs}).
We will find $(\bar{a}_\pm^0,\bar{b}_\pm^0)$ and $(\bar{p}_\pm^0,\bar{q}_\pm^0)$
respectively in appendices \ref{sec:mfExs} and \ref{sec:mfBxs}.
$(\bar{a}_{\pm},\bar{b}_{\pm})$ and $(\bar{p}_{\pm},\bar{q}_{\pm})$ are given as
\begin{align}
  \bigg\{{ \bar{a}_{\pm}(r) \atop \bar{p}_{\pm}(r) }\bigg\}
  &=\frac{\pi}{2}\int_{r}^{r_b}dr'
   \bigg[
     \bigg\{{ \bmfD_{+}^n(r')\pm\bmfD_{-}^n(r')
       \atop  \bmfA_{+}^n(r')\pm\bmfA_{-}^n(r')
     }\bigg\}
     \frac{\nu}{\hr'}Y_{\nu}(\hr')
    -\bigg\{{ \bmfD_{+}^n(r')\mp\bmfD_{-}^n(r')
       \atop  \bmfA_{+}^n(r')\mp\bmfA_{-}^n(r')
     }\bigg\}
     Y_{\nu}'(\hr')
   \bigg] ,
  \label{eq:bar_ap}
   \\
  \bigg\{{ \bar{b}_{\pm}(r) \atop \bar{q}_{\pm}(r) }\bigg\}
  &=\frac{\pi}{2}\int_{r_a}^{r}dr'
   \bigg[
     \bigg\{{ \bmfD_{+}^n(r')\pm\bmfD_{-}^n(r')
        \atop \bmfA_{+}^n(r')\pm\bmfA_{-}^n(r')
     }\bigg\}
     \frac{\nu}{\hr'}J_{\nu}(\hr')
    -\bigg\{{ \bmfD_{+}^n(r')\mp\bmfD_{-}^n(r')
       \atop  \bmfA_{+}^n(r')\mp\bmfA_{-}^n(r')
     }\bigg\}
     J_{\nu}'(\hr')
   \bigg] .
  \label{eq:bar_bq}
\end{align}
%

\subsection{Radial and longitudinal components of the electric field}
\label{sec:mfExs}

We impose the boundary conditions (\ref{eq:BC_ExEs_BxBs}) for $r=r_a$ and $r_b$
to the general solution of $\mfE_s^n$ given by Eq.(\ref{eq:2mfEx_ab_pm}),
\begin{align}
  \frac{\nu}{\hr_a}
  \{\bar{a}_{-}^0J_{\nu}(\hr_a)+\bar{b}_{-}^0Y_{\nu}(\hr_a)
       +\bar{a}_{-}(r_a)J_{\nu}(\hr_a)
  \}
  &-\{\bar{a}_{+}^0J_{\nu}'(\hr_a)+\bar{b}_{+}^0Y_{\nu}'(\hr_a)
         +\bar{a}_{+}(r_a)J_{\nu}'(\hr_a)
    \}
    =0 ,
  \label{eq:BC_Es0_a}
  \\
  \frac{\nu}{\hr_b}
  \{\bar{a}_{-}^0J_{\nu}(\hr_b)+\bar{b}_{-}^0Y_{\nu}(\hr_b)
       +\bar{b}_{-}(r_b)Y_{\nu}(\hr_b)
  \}
  &-\{\bar{a}_{+}^0J_{\nu}'(\hr_b)+\bar{b}_{+}^0Y_{\nu}'(\hr_b)
         +\bar{b}_{+}(r_b)Y_{\nu}'(\hr_b)
    \}
    =0 .
  \label{eq:BC_Es0_b}
\end{align}
Similarly, imposing Eq.(\ref{eq:BC_ExEs_BxBs}) to the general solution of
$\mfE_x^n$ given by Eq.(\ref{eq:2mfEx_ab_pm}), we get
\begin{align}
  \bigg(1-\frac{\nu^2}{\hr_a^2}\bigg)
  \{\bar{a}_{-}^0J_{\nu}(\hr_a)+\bar{b}_{-}^0Y_{\nu}(\hr_a)
       +\bar{a}_{-}(r_a)J_{\nu}(\hr_a)
  \}
  &+\frac{\nu}{\hr_a}
    \{\bar{a}_{+}^0J_{\nu}'(\hr_a)+\bar{b}_{+}^0Y_{\nu}'(\hr_a)
        +\bar{a}_{+}(r_a)J_{\nu}'(\hr_a)
    \}
    =0 ,
  \label{eq:BC_Ex0_a}
  \\
  \bigg(1-\frac{\nu^2}{\hr_b^2}\bigg)
  \{\bar{a}_{-}^0J_{\nu}(\hr_b)+\bar{b}_{-}^0Y_{\nu}(\hr_b)
       +\bar{b}_{-}(r_b)Y_{\nu}(\hr_b)
  \}
  &+\frac{\nu}{\hr_b}
    \{\bar{a}_{+}^0J_{\nu}'(\hr_b)+\bar{b}_{+}^0Y_{\nu}'(\hr_b)
        +\bar{b}_{+}(r_b)Y_{\nu}'(\hr_b)
    \}
   =0 .
  \label{eq:BC_Ex0_b}
\end{align}
Eqs.(\ref{eq:BC_Es0_a}-\ref{eq:BC_Ex0_b}) are linear equations with respect to
the four unknowns $\bar{a}_{\pm}^0$ and $\bar{b}_{\pm}^0$.
Multiplying $\nu/\hr_a$ with Eq.(\ref{eq:BC_Es0_a}), $\nu/\hr_b$ with Eq.(\ref{eq:BC_Es0_b}),
we add the former and the latter respectively to
Eq.(\ref{eq:BC_Ex0_a}) and Eq.(\ref{eq:BC_Ex0_b}),
\begin{align}
  \bar{a}_{-}^0J_{\nu}(\hr_b)+\bar{b}_{-}^0Y_{\nu}(\hr_b)
  =-\bar{b}_{-}(r_b)Y_{\nu}(\hr_b) ,
   \qquad
  \bar{a}_{-}^0J_{\nu}(\hr_a)+\bar{b}_{-}^0Y_{\nu}(\hr_a)
  =-\bar{a}_{-}(r_a)J_{\nu}(\hr_a) .
   \label{eq:ba_bb_mns}
\end{align}
It is easy to solve Eqs.(\ref{eq:ba_bb_mns}) with respect to
$\bar{a}_{-}^0$ and $\bar{b}_{-}^0$, \ie,
there is no need to deal with the $4\times4$ matrix,
\begin{align}
  \bar{a}_{-}^0
  =\frac{Y_{\nu}(\hr_b)}{p_{\nu}(\hr_b,\hr_a)}
   \{\bar{a}_{-}(r_a)J_{\nu}(\hr_a)-\bar{b}_{-}(r_b)Y_{\nu}(\hr_a)\} ,
   \qquad
  \bar{b}_{-}^0
  =\frac{J_{\nu}(\hr_a)}{p_{\nu}(\hr_b,\hr_a)}
   \{\bar{b}_{-}(r_b)Y_{\nu}(\hr_b)-\bar{a}_{-}(r_a)J_{\nu}(\hr_b)\} .
  \label{eq:ab_mns0}
\end{align}
Substituting Eqs.(\ref{eq:ab_mns0}) into Eqs.(\ref{eq:BC_Es0_a}-\ref{eq:BC_Es0_b}),
we solve them with respect to $\bar{a}_{+}^0$ and $\bar{b}_{+}^0$,
\begin{align}
  \bar{a}_{+}^0
  =\frac{Y_{\nu}'(\hr_b)}{s_{\nu}(\hr_b,\hr_a)}
   \{\bar{a}_{+}(r_a)J_{\nu}'(\hr_a)-\bar{b}_{+}(r_b)Y_{\nu}'(\hr_a)\} ,
   \qquad
  \bar{b}_{+}^0
  =\frac{J_{\nu}'(\hr_a)}{s_{\nu}(\hr_b,\hr_a)}
   \{\bar{b}_{+}(r_b)Y_{\nu}'(\hr_b)-\bar{a}_{+}(r_a)J_{\nu}'(\hr_b)\} .
  \label{eq:ab_pls0}
\end{align}
Into Eqs.(\ref{eq:ab_mns0}-\ref{eq:ab_pls0})
we substitute $\bar{a}_{\pm}(r_a)$ and $\bar{b}_{\pm}(r_b)$ 
given by Eqs.(\ref{eq:bar_ap}-\ref{eq:bar_bq}) for $r=r_a$ and $r_b$,
\begin{align}
  \bigg\{{\bar{a}_{+}^0 \atop \bar{b}_{+}^0}\bigg\}
  &=\frac{\pi/2}{s_{\nu}(\hr_b,\hr_a)}
    \bigg\{{Y_{\nu}'(\hr_b) \atop J_{\nu}'(\hr_a)}\bigg\}\int_{r_a}^{r_b}dr'
    \bigg[i\bmfD_s^n(r')\bigg\{{s_{\nu}(\hr',\hr_a) \atop s_{\nu}(\hr_b,\hr')}\bigg\}
         -\bmfD_x^n(r')\frac{\nu}{\hr'}
          \bigg\{{q_{\nu}(\hr',\hr_a) \atop r_{\nu}(\hr_b,\hr')}\bigg\}
    \bigg] ,
  \label{eq:bar_a_p0}
  \\
  \bigg\{{ \bar{a}_{-}^0 \atop \bar{b}_{-}^0}\bigg\}
  &=\frac{\pi/2}{p_{\nu}(\hr_b,\hr_a)}
    \bigg\{{Y_{\nu}(\hr_b) \atop J_{\nu}(\hr_a)}\bigg\} \int_{r_a}^{r_b}dr'
    \bigg[\bmfD_x^n(r')\bigg\{{r_{\nu}'(\hr',\hr_a) \atop q_{\nu}(\hr_b,\hr')}\bigg\}
         -i\bmfD_s^n(r')\frac{\nu}{\hr'}
          \bigg\{{p_{\nu}(\hr',\hr_a) \atop p_{\nu}(\hr_b,\hr')}\bigg\}
    \bigg] .
  \label{eq:bar_a_m0}
\end{align}
$\bmfD_{x,s}^n$ is the driving term of $\mfE_{x,s}^n$, given by Eqs.(\ref{eq:bmfDA_xs}).
$\bmfD_{x,s}^n$ is related to $\bmfD_{\pm}^n$ given by Eqs.(\ref{eq:mfUpm}),
\begin{align}
  \bmfD_x^n(r')
  =\bmfD_{+}^n(r')+\bmfD_{-}^n(r') ,
    \qquad
  i\bmfD_s^n(r')
  =\bmfD_{+}^n(r')-\bmfD_{-}^n(r') .
\end{align}
Substituting Eqs.(\ref{eq:bar_a_p0}-\ref{eq:bar_a_m0}) into the first equations of
(\ref{eq:bara_pm}-\ref{eq:barb_pm}), we rewrite them using
Eqs.(\ref{eq:Cnu_pnu_ba}-\ref{eq:Cnu_prm_snu_ba}),
\begin{align}
  a_{+}+a_{-}
  &=\frac{Y_{\nu}'(\hr_b)}{s_{\nu}(\hr_b,\hr_a)}\int_{r_a}^{r}dr'\frac{\pi}{2}
    \Big\{
       i\bmfD_s^n(r')s_{\nu}(\hr',\hr_a)
       -\bmfD_x^n(r')\frac{\nu}{\hr'}q_{\nu}(\hr',\hr_a)
    \Big\}
  \nonumber\\&\quad
   -\frac{Y_{\nu}'(\hr_a)}{s_{\nu}(\hr_b,\hr_a)}\int_{r}^{r_b}dr'\frac{\pi}{2}
    \Big\{i\bmfD_s^n(r')s_{\nu}(\hr_b,\hr')
          -\bmfD_x^n(r')\frac{\nu}{\hr'}r_{\nu}(\hr_b,\hr')
    \Big\} ,
   \label{eq:ap_pls_am}
  \\
  a_{+}-a_{-}
  &=\frac{Y_{\nu}(\hr_b)}{p_{\nu}(\hr_b,\hr_a)}\int_{r_a}^{r}dr'\frac{\pi}{2}
    \Big\{
       \bmfD_x^n(r')r_{\nu}(\hr',\hr_a)
     -i\bmfD_s^n(r')\frac{\nu}{\hr'}p_{\nu}(\hr',\hr_a)
    \Big\}
  \nonumber\\&\quad
   -\frac{Y_{\nu}(\hr_a)}{p_{\nu}(\hr_b,\hr_a)}\int_{r}^{r_b}dr'\frac{\pi}{2}
    \Big\{\bmfD_x^n(r')q_{\nu}(\hr_b,\hr')
        -i\bmfD_s^n(r')\frac{\nu}{\hr'}p_{\nu}(\hr_b,\hr')
    \Big\} ,
   \\
  b_{+}+b_{-}
  &=\frac{J_{\nu}'(\hr_a)}{s_{\nu}(\hr_b,\hr_a)}\int_{r}^{r_b}dr'\frac{\pi}{2}
    \Big\{i\bmfD_s^n(r')s_{\nu}(\hr_b,\hr')
          -\bmfD_x^n(r')\frac{\nu}{\hr'}r_{\nu}(\hr_b,\hr')
    \Big\}
  \nonumber\\&\quad
   -\frac{J_{\nu}'(\hr_b)}{s_{\nu}(\hr_b,\hr_a)}\int_{r_a}^{r}dr'\frac{\pi}{2}
    \Big\{i\bmfD_s^n(r')s_{\nu}(\hr',\hr_a)
          -\bmfD_x^n(r')\frac{\nu}{\hr'}q_{\nu}(\hr',\hr_a)
    \Big\} ,
  \\
  b_{+}-b_{-}
  &=\frac{J_{\nu}(\hr_a)}{p_{\nu}(\hr_b,\hr_a)}\int_{r}^{r_b}dr'\frac{\pi}{2}
    \Big\{\bmfD_x^n(r')q_{\nu}(\hr_b,\hr')
        -i\bmfD_s^n(r')\frac{\nu}{\hr'}p_{\nu}(\hr_b,\hr')
    \Big\}
  \nonumber\\&\quad
   -\frac{J_{\nu}(\hr_b)}{p_{\nu}(\hr_b,\hr_a)}\int_{r_a}^{r}dr'\frac{\pi}{2}
    \Big\{\bmfD_x^n(r')r_{\nu}(\hr',\hr_a)
        -i\bmfD_s^n(r')\frac{\nu}{\hr'}p_{\nu}(\hr',\hr_a)
    \Big\} .
   \label{eq:bp_mns_bm}
\end{align}
Substituting Eqs.(\ref{eq:ap_pls_am}-\ref{eq:bp_mns_bm}) into
Eqs.(\ref{eq:2mfEx_ab_pm}-\ref{eq:2mfBx_pq_pm}), and rearranging them using
the cross products of the Bessel functions (\ref{eq:CP_pq}-\ref{eq:CP_sr}),
we get the exact expressions of the radial and longitudinal components of
the transient electric field of synchrotron radiation in the Laplace domain,
\begin{align}
  \mfE_x^n(r,\nu)
  &=\frac{r_{\nu}(\hr_b,\hr)}{s_{\nu}(\hr_b,\hr_a)}
    \int_{r_a}^{r}dr'\frac{\pi\nu}{2\hr}
    \Big\{\bmfD_x^n(r')\frac{\nu}{\hr'}q_{\nu}(\hr',\hr_a)
         -i\bmfD_s^n(r')s_{\nu}(\hr',\hr_a)\Big\}
  \nonumber\\&\quad
   +\frac{q_{\nu}(\hr,\hr_a)}{s_{\nu}(\hr_b,\hr_a)}
    \int_{r}^{r_b}dr'\frac{\pi\nu}{2\hr}
    \Big\{\bmfD_x^n(r')\frac{\nu}{\hr'}r_{\nu}(\hr_b,\hr')
         -i\bmfD_s^n(r')s_{\nu}(\hr_b,\hr')\Big\}
  \nonumber\\&\quad
   +\frac{q_{\nu}(\hr_b,\hr)}{p_{\nu}(\hr_b,\hr_a)}
    \int_{r_a}^{r}dr'\frac{\pi}{2}
    \Big\{\bmfD_x^n(r')r_{\nu}(\hr',\hr_a)
         -i\bmfD_s^n(r')\frac{\nu}{\hr'}p_{\nu}(\hr',\hr_a)
    \Big\}
  \nonumber\\&\quad
   +\frac{r_{\nu}(\hr,\hr_a)}{p_{\nu}(\hr_b,\hr_a)}
    \int_{r}^{r_b}dr'\frac{\pi}{2}
    \Big\{\bmfD_x^n(r')q_{\nu}(\hr_b,\hr')
         -i\bmfD_s^n(r')\frac{\nu}{\hr'}p_{\nu}(\hr_b,\hr')
    \Big\} ,
  \label{eq:mfEx_0}
   \\
  \mfE_s^n(r,\nu)
  &=\frac{p_{\nu}(\hr_b,\hr)}{p_{\nu}(\hr_b,\hr_a)}
    \int_{r_a}^{r}dr'\frac{\pi\nu}{2\hr}
    \Big\{\bmfD_s^n(r')\frac{\nu}{\hr'}p_{\nu}(\hr',\hr_a)
         +i\bmfD_x^n(r')r_{\nu}(\hr',\hr_a)\Big\}
  \nonumber\\&\quad
   +\frac{p_{\nu}(\hr,\hr_a)}{p_{\nu}(\hr_b,\hr_a)}
    \int_{r}^{r_b}dr'\frac{\pi\nu}{2\hr}
    \Big\{\bmfD_s^n(r')\frac{\nu}{\hr'}p_{\nu}(\hr_b,\hr')
         +i\bmfD_x^n(r')q_{\nu}(\hr_b,\hr')\Big\}
  \nonumber\\&\quad
   +\frac{s_{\nu}(\hr_b,\hr)}{s_{\nu}(\hr_b,\hr_a)}
    \int_{r_a}^{r}dr'\frac{\pi}{2}
    \Big\{\bmfD_s^n(r')s_{\nu}(\hr',\hr_a)
        +i\bmfD_x^n(r')\frac{\nu}{\hr'}q_{\nu}(\hr',\hr_a)
    \Big\}
  \nonumber\\&\quad
   +\frac{s_{\nu}(\hr,\hr_a)}{s_{\nu}(\hr_b,\hr_a)}
    \int_{r}^{r_b}dr'\frac{\pi}{2}
    \Big\{\bmfD_s^n(r')s_{\nu}(\hr_b,\hr')
        +i\bmfD_x^n(r')\frac{\nu}{\hr'}r_{\nu}(\hr_b,\hr')
    \Big\} .
  \label{eq:mfEs_0}
\end{align}
Using the Green functions, furthermore, we will rearrange
Eqs.(\ref{eq:mfEx_0}-\ref{eq:mfEs_0}) in section \ref{sec:rearrange}.

\subsection{Radial and longitudinal components of the magnetic field}
\label{sec:mfBxs}

Similar to finding the expressions of the radial and longitudinal components of
the electric field in appendix \ref{sec:mfExs}, we find the radial and longitudinal 
components of the magnetic field in the Laplace domain.
The rest work is to find the constants of integration $\bar{p}_{\pm}^0$ and $\bar{q}_{\pm}^0$
which are involved in Eq.(\ref{eq:2mfBx_pq_pm}).
Imposing the boundary conditions (\ref{eq:BC_ExEs_BxBs}) for $r=r_a$ and $r_b$
to the general solution of $\mfB_x^n$ given by the upper equation of
(\ref{eq:2mfBx_pq_pm}), we get
\begin{align}
  \frac{\nu}{\hr_a}
  \{\bar{p}_{+}^0J_{\nu}(\hr_a)+\bar{q}_{+}^0Y_{\nu}(\hr_a)
       +\bar{p}_{+}(\hr_a)J_{\nu}(\hr_a)
  \}
  &-\{\bar{p}_{-}^0J_{\nu}'(\hr_a)+\bar{q}_{-}^0Y_{\nu}'(\hr_a)
        +\bar{p}_{-}(\hr_a)J_{\nu}'(\hr_a)
    \}
   =0 ,
  \label{eq:BC_Bx0_a}
  \\
  \frac{\nu}{\hr_b}
  \{\bar{p}_{+}^0J_{\nu}(\hr_b)+\bar{q}_{+}^0Y_{\nu}(\hr_b)
       +\bar{q}_{+}(\hr_b)Y_{\nu}(\hr_b)
  \}
  &-\{\bar{p}_{-}^0J_{\nu}'(\hr_b)+\bar{q}_{-}^0Y_{\nu}'(\hr_b)
        +\bar{q}_{-}(\hr_b)Y_{\nu}'(\hr_b)
    \}
   =0 .
  \label{eq:BC_Bx0_b}
\end{align}
$\bar{p}_{\pm}(\hr_a)$ and $\bar{q}_{\pm}(\hr_b)$ are gotten
from Eqs.(\ref{eq:bar_ap}-\ref{eq:bar_bq}) for $r=r_a$ and $r_b$.
Imposing the boundary conditions (\ref{eq:BC_ExEs_BxBs}) for $r=r_a$ and $r_b$ to
$\mfB_s^n$ given by the lower equation of (\ref{eq:2mfBx_pq_pm}), we get
\begin{align}
  \bigg(1-\frac{\nu^2}{\hr_a^2}\bigg)
  \{\bar{p}_{+}^0J_{\nu}(\hr_a)+\bar{q}_{+}^0Y_{\nu}(\hr_a)
       +\bar{p}_{+}(\hr_a)J_{\nu}(\hr_a)
  \}
  &+\frac{\nu}{\hr_a}
   \{\bar{p}_{-}^0J_{\nu}'(\hr_a)+\bar{q}_{-}^0Y_{\nu}'(\hr_a)
       +\bar{p}_{-}(\hr_a)J_{\nu}'(\hr_a)
   \}
   =0 ,
  \label{eq:BC_rdr_rBs0_a}
  \\
  \bigg(1-\frac{\nu^2}{\hr_b^2}\bigg)
  \{\bar{p}_{+}^0J_{\nu}(\hr_b)+\bar{q}_{+}^0Y_{\nu}(\hr_b)
       +\bar{q}_{+}(\hr_b)Y_{\nu}(\hr_b)
  \}
  &+\frac{\nu}{\hr_b}
    \{\bar{p}_{-}^0J_{\nu}'(\hr_b)+\bar{q}_{-}^0Y_{\nu}'(\hr_b)
        +\bar{q}_{-}(\hr_b)Y_{\nu}'(\hr_b)
    \}
   =0 .
  \label{eq:BC_rdr_rBs0_b}
\end{align}
We solve Eqs.(\ref{eq:BC_Bx0_a}-\ref{eq:BC_rdr_rBs0_b}) with respect to the four unknowns
$\bar{p}_{\pm}^0$ and $\bar{q}_{\pm}^0$ without dealing with the $4\times4$ matrix.
Multiplying $\nu/\hr_a$ with Eq.(\ref{eq:BC_Bx0_a}), $\nu/\hr_b$ with Eq.(\ref{eq:BC_Bx0_b}),
we add the former and the latter respectively to Eq.(\ref{eq:BC_rdr_rBs0_a}) and
Eq.(\ref{eq:BC_rdr_rBs0_b}),
\begin{align}
  \bar{p}_{+}^0J_{\nu}(\hr_a)+\bar{q}_{+}^0Y_{\nu}(\hr_a)=-\bar{p}_{+}(\hr_a)J_{\nu}(\hr_a),
    \qquad
  \bar{p}_{+}^0J_{\nu}(\hr_b)+\bar{q}_{+}^0Y_{\nu}(\hr_b)=-\bar{q}_{+}(\hr_b)Y_{\nu}(\hr_b) .
   \label{eq:bp_bq_pls}
\end{align}
Then we solve Eqs.(\ref{eq:bp_bq_pls}) with respect to $\bar{p}_{+}^0$ and $\bar{q}_{+}^0$,
\begin{align}
  \bar{p}_{+}^0
  =\frac{Y_{\nu}(\hr_b)}{p_{\nu}(\hr_b,\hr_a)}
   \{\bar{p}_{+}(\hr_a)J_{\nu}(\hr_a)-\bar{q}_{+}(\hr_b)Y_{\nu}(\hr_a)\} ,
   \qquad
  \bar{q}_{+}^0
  =\frac{J_{\nu}(\hr_a)}{p_{\nu}(\hr_b,\hr_a)}
   \{\bar{q}_{+}(\hr_b)Y_{\nu}(\hr_b)-\bar{p}_{+}(\hr_a)J_{\nu}(\hr_b)\} .
  \label{eq:pq_pls0_sol}
\end{align}
Substituting Eqs.(\ref{eq:pq_pls0_sol}) into Eqs.(\ref{eq:BC_Bx0_a}-\ref{eq:BC_Bx0_b}),
we solve them with respect to $\bar{p}_{-}^0$ and $\bar{q}_{-}^0$,
\begin{align}
  \bar{p}_{-}^0
  =\frac{Y_{\nu}'(\hr_b)}{s_{\nu}(\hr_b,\hr_a)}
   \{\bar{p}_{-}(\hr_a)J_{\nu}'(\hr_a)-\bar{q}_{-}(\hr_b)Y_{\nu}'(\hr_a)\} ,
   \qquad
  \bar{q}_{-}^0
  =\frac{J_{\nu}'(\hr_a)}{s_{\nu}(\hr_b,\hr_a)}
   \{\bar{q}_{-}(\hr_b)Y_{\nu}'(\hr_b)-\bar{p}_{-}(\hr_a)J_{\nu}'(\hr_b)\} .
  \label{eq:pq_mns0_sol}
\end{align}
Eqs.(\ref{eq:pq_pls0_sol}-\ref{eq:pq_mns0_sol}) are the constants of integration
involved in the general solutions of $\mfB_{x,s}^n$ given by Eq.(\ref{eq:2mfBx_pq_pm}).

Into Eqs.(\ref{eq:pq_pls0_sol}-\ref{eq:pq_mns0_sol}) we substitute $\bar{p}_{\pm}(\hr_a)$
and $\bar{q}_{\pm}(\hr_b)$ given by Eqs.(\ref{eq:bar_ap}-\ref{eq:bar_bq}) for
$r=r_a$ and $r_b$,
\begin{align}
  \bar{p}_{+}^0
  &=\frac{Y_{\nu}(\hr_b)}{p_{\nu}(\hr_b,\hr_a)}\int_{r_a}^{r_b}dr'\frac{\pi}{2}
    \Big\{i\bmfA_s^n(r')r_{\nu}(\hr',\hr_a)
         -\bmfA_x^n(r')\frac{\nu}{\hr'}p_{\nu}(\hr',\hr_a)\Big\} ,
  \label{barp_pls0_sol}
  \\
  \bar{p}_{-}^0
  &=\frac{Y_{\nu}'(\hr_b)}{s_{\nu}(\hr_b,\hr_a)}\int_{r_a}^{r_b}dr'\frac{\pi}{2}
    \Big\{\bmfA_x^n(r')s_{\nu}(\hr',\hr_a)
         -i\bmfA_s^n(r')\frac{\nu}{\hr'}q_{\nu}(\hr',\hr_a)\Big\} ,
  \label{barp_mns0_sol}
  \\
  \bar{q}_{+}^0
  &=\frac{J_{\nu}(\hr_a)}{p_{\nu}(\hr_b,\hr_a)}\int_{r_a}^{r_b}dr'\frac{\pi}{2}
    \Big\{i\bmfA_s^n(r')q_{\nu}(\hr_b,\hr')
         -\bmfA_x^n(r')\frac{\nu}{\hr'}p_{\nu}(\hr_b,\hr')\Big\} ,
  \label{barq_pls0_sol}
  \\
  \bar{q}_{-}^0
  &=\frac{J_{\nu}'(\hr_a)}{s_{\nu}(\hr_b,\hr_a)}
    \int_{r_a}^{r_b}dr'\frac{\pi}{2}
    \Big\{\bmfA_x^n(r')s_{\nu}(\hr_b,\hr')
         -i\bmfA_s^n(r')\frac{\nu}{\hr'}r_{\nu}(\hr_b,\hr')\Big\} .
  \label{barq_mns0_sol}
\end{align}
$\bmfA_{x,s}^n$ is given by Eqs.(\ref{eq:bmfDA_xs}) which are the driving terms of
$\mfB_{x,s}^n$.
$\bmfA_{x,s}^n$ is related to $\bmfA_{\pm}^n$ given by Eqs.(\ref{eq:mfVpm}),
\begin{align}
  \bmfA_x^n(r')
  =\bmfA_{+}^n(r')+\bmfA_{-}^n(r') ,
    \qquad
  i\bmfA_s^n(r')
  =\bmfA_{+}^n(r')-\bmfA_{-}^n(r') .
\end{align}
Substituting Eqs.(\ref{barp_pls0_sol}-\ref{barp_mns0_sol}) into $p_{+}\pm p_{-}$
given by Eq.(\ref{eq:barp_pm}), we rewrite them using
Eqs.(\ref{eq:Cnu_pnu_ba}-\ref{eq:Cnu_prm_snu_ba}),
\begin{align}
  p_{+}+p_{-}
  &=\frac{Y_{\nu}(\hr_b)}{p_{\nu}(\hr_b,\hr_a)}\int_{r_a}^{r}dr'\frac{\pi}{2}
    \Big\{
       i\bmfA_s^n(r')r_{\nu}(\hr',\hr_a)
      -\bmfA_x^n(r')\frac{\nu}{\hr'}p_{\nu}(\hr',\hr_a)
    \Big\}
  \nonumber\\&\quad
   -\frac{Y_{\nu}(\hr_a)}{p_{\nu}(\hr_b,\hr_a)}\int_{r}^{r_b}dr'\frac{\pi}{2}
    \Big\{
       i\bmfA_s^n(r')q_{\nu}(\hr_b,\hr')
      -\bmfA_x^n(r')\frac{\nu}{\hr'}p_{\nu}(\hr_b,\hr')
    \Big\} ,
   \label{eq:pp_pls_pm}
  \\
  p_{+}-p_{-}
  &=\frac{Y_{\nu}'(\hr_b)}{s_{\nu}(\hr_b,\hr_a)}\int_{r_a}^{r}dr'\frac{\pi}{2}
    \Big\{
       \bmfA_x^n(r')s_{\nu}(\hr',\hr_a)
      -i\bmfA_s^n(r')\frac{\nu}{\hr'}q_{\nu}(\hr',\hr_a)
    \Big\}
  \nonumber\\&\quad
   -\frac{Y_{\nu}'(\hr_a)}{s_{\nu}(\hr_b,\hr_a)}\int_{r}^{r_b}dr'\frac{\pi}{2}
    \Big\{
       \bmfA_x^n(r')s_{\nu}(\hr_b,\hr')
      -i\bmfA_s^n(r')\frac{\nu}{\hr'}r_{\nu}(\hr_b,\hr')
    \Big\} .
\end{align}
Similarly, substituting Eqs.(\ref{barq_pls0_sol}-\ref{barq_mns0_sol}) into $q_{+}\pm q_{-}$
given by Eq.(\ref{eq:barq_pm}), we rewrite them as follows,
\begin{align}
  q_{+}+q_{-}
  &=\frac{J_{\nu}(\hr_a)}{p_{\nu}(\hr_b,\hr_a)}\int_{r}^{r_b}dr'\frac{\pi}{2}
    \Big\{
       i\bmfA_s^n(r')q_{\nu}(\hr_b,\hr')
      -\bmfA_x^n(r')\frac{\nu}{\hr'}p_{\nu}(\hr_b,\hr')
    \Big\}
  \nonumber\\&\quad
   -\frac{J_{\nu}(\hr_b)}{p_{\nu}(\hr_b,\hr_a)}\int_{r_a}^{r}dr'\frac{\pi}{2}
    \Big\{
       i\bmfA_s^n(r')r_{\nu}(\hr',\hr_a)
      -\bmfA_x^n(r')\frac{\nu}{\hr'}p_{\nu}(\hr',\hr_a)
    \Big\} ,
  \\
  q_{+}-q_{-}
  &=\frac{J_{\nu}'(\hr_a)}{s_{\nu}(\hr_b,\hr_a)}\int_{r}^{r_b}dr'\frac{\pi}{2}
    \Big\{
       \bmfA_x^n(r')s_{\nu}(\hr_b,\hr')
      -i\bmfA_s^n(r')\frac{\nu}{\hr'}r_{\nu}(\hr_b,\hr')
    \Big\}
  \nonumber\\&\quad
   -\frac{J_{\nu}'(\hr_b)}{s_{\nu}(\hr_b,\hr_a)}\int_{r_a}^{r}dr'\frac{\pi}{2}
    \Big\{
       \bmfA_x^n(r')s_{\nu}(\hr',\hr_a)
      -i\bmfA_s^n(r')\frac{\nu}{\hr'}q_{\nu}(\hr',\hr_a)
    \Big\} .
   \label{eq:qp_mns_qm}
\end{align}
From Eq.(\ref{eq:2mfBx_pq_pm}) and Eqs.(\ref{eq:pp_pls_pm}-\ref{eq:qp_mns_qm}),
we get the exact expressions of the radial and longitudinal components of
the transient magnetic field of synchrotron radiation in the constant bend in
the Laplace domain,
\begin{align}
  c\mfB_x^n(r,\nu)
  &=\frac{p_{\nu}(\hr_b,\hr)}{p_{\nu}(\hr_b,\hr_a)}
    \int_{r_a}^{r}dr'\frac{\pi\nu}{2\hr}
    \Big\{\bmfA_x^n(r')\frac{\nu}{\hr'}p_{\nu}(\hr',\hr_a)
        -i\bmfA_s^n(r')r_{\nu}(\hr',\hr_a)
    \Big\}
  \nonumber\\&\quad
   +\frac{p_{\nu}(\hr,\hr_a)}{p_{\nu}(\hr_b,\hr_a)}
    \int_{r}^{r_b}dr'\frac{\pi\nu}{2\hr}
    \Big\{\bmfA_x^n(r')\frac{\nu}{\hr'}p_{\nu}(\hr_b,\hr')
        -i\bmfA_s^n(r')q_{\nu}(\hr_b,\hr')
    \Big\}
  \nonumber\\&\quad
   +\frac{s_{\nu}(\hr_b,\hr)}{s_{\nu}(\hr_b,\hr_a)}
    \int_{r_a}^{r}dr'\frac{\pi}{2}
    \Big\{
       \bmfA_x^n(r')s_{\nu}(\hr',\hr_a)
      -i\bmfA_s^n(r')\frac{\nu}{\hr'}q_{\nu}(\hr',\hr_a)
    \Big\}
  \nonumber\\&\quad
   +\frac{s_{\nu}(\hr,\hr_a)}{s_{\nu}(\hr_b,\hr_a)}
    \int_{r}^{r_b}dr'\frac{\pi}{2}
    \Big\{
       \bmfA_x^n(r')s_{\nu}(\hr_b,\hr')
      -i\bmfA_s^n(r')\frac{\nu}{\hr'}r_{\nu}(\hr_b,\hr')
    \Big\} ,
   \label{eq:mfBx_0}
   \\
  c\mfB_s^n(r,\nu)
  &=\frac{r_{\nu}(\hr_b,\hr)}{s_{\nu}(\hr_b,\hr_a)}
    \int_{r_a}^{r}dr'\frac{\pi\nu}{2\hr}
    \Big\{\bmfA_s^n(r')\frac{\nu}{\hr'}q_{\nu}(\hr',\hr_a)
        +i\bmfA_x^n(r')s_{\nu}(\hr',\hr_a)
    \Big\}
  \nonumber\\&\quad
   +\frac{q_{\nu}(\hr,\hr_a)}{s_{\nu}(\hr_b,\hr_a)}
    \int_{r}^{r_b}dr'\frac{\pi\nu}{2\hr}
    \Big\{\bmfA_s^n(r')\frac{\nu}{\hr'}r_{\nu}(\hr_b,\hr')
        +i\bmfA_x^n(r')s_{\nu}(\hr_b,\hr')
    \Big\}
  \nonumber\\&\quad
   +\frac{q_{\nu}(\hr_b,\hr)}{p_{\nu}(\hr_b,\hr_a)}
    \int_{r_a}^{r}dr'\frac{\pi}{2}
    \Big\{
       \bmfA_s^n(r')r_{\nu}(\hr',\hr_a)
      +i\bmfA_x^n(r')\frac{\nu}{\hr'}p_{\nu}(\hr',\hr_a)
    \Big\}
  \nonumber\\&\quad
   +\frac{r_{\nu}(\hr,\hr_a)}{p_{\nu}(\hr_b,\hr_a)}
    \int_{r}^{r_b}dr'\frac{\pi}{2}
    \Big\{
       \bmfA_s^n(r')q_{\nu}(\hr_b,\hr')
      +i\bmfA_x^n(r')\frac{\nu}{\hr'}p_{\nu}(\hr_b,\hr')
    \Big\} .
   \label{eq:mfBs_0}
\end{align}
%

\subsection{Rearrangement of the expressions of the fields in the Laplace domain}
\label{sec:rearrange}

In this appendix we found the expressions of $\mfE_{x,s}^n$ and $\mfB_{x,s}^n$
given by Eqs.(\ref{eq:mfEx_0}-\ref{eq:mfEs_0}) and (\ref{eq:mfBx_0}-\ref{eq:mfBs_0}).
In order to ensure their correctness analytically, we rewrite them using $\mfG_{\pm}^n$
and $\bmfG_{\pm}^n$ which are the Green functions for the vertical components of the fields 
in the Laplace domain, given by Eqs.(\ref{eq:mfGe}-\ref{eq:mfGb}) and
(\ref{eq:bmfG_pls}-\ref{eq:bmfG_mns}),
\begin{align}
  \mfG_{+}^n(\hr,\hr',\nu)
  &=\frac{\pi}{2}\rho
    \bigg\{
        \theta(\hr-\hr')\frac{p_{\nu}(\hr_b,\hr)p_{\nu}(\hr',\hr_a)}{p_{\nu}(\hr_b,\hr_a)}
       +\theta(\hr'-\hr)\frac{p_{\nu}(\hr_b,\hr')p_{\nu}(\hr,\hr_a)}{p_{\nu}(\hr_b,\hr_a)}
    \bigg\} ,
  \label{eq:mfG_pls_def}
   \\
  \mfG_{-}^n(\hr,\hr',\nu)
  &=\frac{\pi}{2}\rho
    \bigg\{
        \theta(\hr-\hr')\frac{r_{\nu}(\hr_b,\hr)q_{\nu}(\hr',\hr_a)}{s_{\nu}(\hr_b,\hr_a)}
       +\theta(\hr'-\hr)\frac{r_{\nu}(\hr_b,\hr')q_{\nu}(\hr,\hr_a)}{s_{\nu}(\hr_b,\hr_a)}
    \bigg\} ,
  \label{eq:mfG_mns_def}
   \\
  \bmfG_{+}^n(\hr,\hr',\nu)
  &=\frac{\pi}{2}\rho
    \bigg\{\theta(\hr-\hr')\frac{q_{\nu}(\hr_b,\hr)r_{\nu}(\hr',\hr_a)}{p_{\nu}(\hr_b,\hr_a)}
          +\theta(\hr'-\hr)\frac{q_{\nu}(\hr_b,\hr')r_{\nu}(\hr,\hr_a)}{p_{\nu}(\hr_b,\hr_a)}
    \bigg\} ,
  \label{eq:bmfG_pls_apndx}
   \\
  \bmfG_{-}^n(\hr,\hr',\nu)
  &=\frac{\pi}{2}\rho
    \bigg\{
       \theta(\hr-\hr')\frac{s_{\nu}(\hr_b,\hr)s_{\nu}(\hr',\hr_a)}{s_{\nu}(\hr_b,\hr_a)}
      +\theta(\hr'-\hr)\frac{s_{\nu}(\hr_b,\hr')s_{\nu}(\hr,\hr_a)}{s_{\nu}(\hr_b,\hr_a)}
    \bigg\} .
  \label{eq:bmfG_mns_apndx}
\end{align}
$\theta$ is the Heaviside step function.
The derivatives of $\mfG_{\pm}^n$ with respect to $\hr$ or $\hr'$ are given as follows,
\begin{align}
  \rd_{\hr}\mfG_{+}^n(\hr,\hr',\nu)
  &=\frac{\pi}{2}\rho
    \bigg\{
       \theta(\hr-\hr')\frac{q_{\nu}(\hr_b,\hr)p_{\nu}(\hr',\hr_a)}{p_{\nu}(\hr_b,\hr_a)}
      +\theta(\hr'-\hr)\frac{p_{\nu}(\hr_b,\hr')r_{\nu}(\hr,\hr_a)}{p_{\nu}(\hr_b,\hr_a)}
    \bigg\} ,
  \label{eq:rdr_mfGp_fin}
   \\
  \rd_{\hr}\mfG_{-}^n(\hr,\hr',\nu)
  &=\frac{\pi}{2}\rho
    \bigg\{
       \theta(\hr-\hr')\frac{s_{\nu}(\hr_b,\hr)q_{\nu}(\hr',\hr_a)}{s_{\nu}(\hr_b,\hr_a)}
      +\theta(\hr'-\hr)\frac{r_{\nu}(\hr_b,\hr')s_{\nu}(\hr,\hr_a)}{s_{\nu}(\hr_b,\hr_a)}
    \bigg\} ,
  \label{eq:rdr_mfGm_fin}
   \\
  \rd_{\hr'}\mfG_{+}^n(\hr,\hr',\nu)
  &=\frac{\pi}{2}\rho
    \bigg\{
       \theta(\hr-\hr')\frac{p_{\nu}(\hr_b,\hr)r_{\nu}(\hr',\hr_a)}{p_{\nu}(\hr_b,\hr_a)}
      +\theta(\hr'-\hr)\frac{q_{\nu}(\hr_b,\hr')p_{\nu}(\hr,\hr_a)}{p_{\nu}(\hr_b,\hr_a)}
    \bigg\} ,
  \label{eq:rdrp_mfGp_fin}
   \\
  \rd_{\hr'}\mfG_{-}^n(\hr,\hr',\nu)
  &=\frac{\pi}{2}\rho
    \bigg\{
       \theta(\hr-\hr')\frac{r_{\nu}(\hr_b,\hr)s_{\nu}(\hr',\hr_a)}{s_{\nu}(\hr_b,\hr_a)}
      +\theta(\hr'-\hr)\frac{s_{\nu}(\hr_b,\hr')q_{\nu}(\hr,\hr_a)}{s_{\nu}(\hr_b,\hr_a)}
    \bigg\} .
  \label{eq:rdrp_mfGm_fin}
\end{align}
We rewrite Eqs.(\ref{eq:mfEx_0}-\ref{eq:mfEs_0}) using
Eqs.(\ref{eq:mfG_pls_def}-\ref{eq:rdrp_mfGm_fin}),
\begin{align}
  \mfE_x^n(r,\nu)
  &=\int_{r_a}^{r_b}\frac{dr'}{\rho}
    \Big\{\bmfD_x^n(r',\nu)
           \Big(\bmfG_{+}^n+\frac{\nu^2}{\hr\hr'}\mfG_{-}^n\Big)
         -i\bmfD_s^n(r',\nu)
           \Big(\frac{\nu}{\hr}\rd_{\hr'}\mfG_{-}^n
               +\frac{\nu}{\hr'}\rd_{\hr}\mfG_{+}^n\Big)
    \Big\} ,
   \label{eq:mfEx_apdx}
  \\
  \mfE_s^n(r,\nu)
  &=\int_{r_a}^{r_b}\frac{dr'}{\rho}
    \Big\{\bmfD_s^n(r',\nu)
          \Big(\bmfG_{-}^n+\frac{\nu^2}{\hr\hr'}\mfG_{+}^n\Big)
          +i\bmfD_x^n(r',\nu)
           \Big(\frac{\nu}{\hr}\rd_{\hr'}\mfG_{+}^n
                +\frac{\nu}{\hr'}\rd_{\hr}\mfG_{-}^n\Big)
    \Big\} .
   \label{eq:mfEs_apdx}
\end{align}
Eqs.(\ref{eq:mfEx_apdx}-\ref{eq:mfEs_apdx}) are the radial and longitudinal
components of the electric field in the Laplace domain.
They agree with Eqs.(\ref{eq:mfBsEx}-\ref{eq:mfBxEs}) 
which we got from the solutions of the vertical components of the fields
$\mfE_y^n$ and $\mfB_y^n$ through Eqs.(\ref{eq:mfBx_EyBy}-\ref{eq:mfBs_EyBy}).
Similar to Eqs.(\ref{eq:mfEx_apdx}-\ref{eq:mfEs_apdx}), we rewrite
Eqs.(\ref{eq:mfBx_0}-\ref{eq:mfBs_0}) using
Eqs.(\ref{eq:mfG_pls_def}-\ref{eq:rdrp_mfGm_fin}),
\begin{align}
  c\mfB_x^n(r,\nu)
  &=\int_{r_a}^{r_b}\frac{dr'}{\rho}
    \Big\{\bmfA_x^n(r',\nu)
         \Big(\bmfG_{-}^n+\frac{\nu^2}{\hr\hr'}\mfG_{+}^n\Big)
        -i\bmfA_s^n(r',\nu)
         \Big(\frac{\nu}{\hr}\rd_{\hr'}\mfG_{+}^n
              +\frac{\nu}{\hr'}\rd_{\hr}\mfG_{-}^n\Big)
    \Big\} ,
  \label{eq:mfBx_from_we}
  \\
  c\mfB_s^n(r,\nu)
  &=\int_{r_a}^{r_b}\frac{dr'}{\rho}
    \Big\{\bmfA_s^n(r',\nu)
          \Big(\bmfG_{+}^n+\frac{\nu^2}{\hr\hr'}\mfG_{-}^n\Big)
        +i\bmfA_x^n(r',\nu)
         \Big(\frac{\nu}{\hr}\rd_{\hr'}\mfG_{-}^n
              +\frac{\nu}{\hr'}\rd_{\hr}\mfG_{+}^n\Big)
    \Big\} .
  \label{eq:mfBs_from_we}
\end{align}
Eqs.(\ref{eq:mfBx_from_we}-\ref{eq:mfBs_from_we}) agree with
Eqs.(\ref{eq:mfBsEx}-\ref{eq:mfBxEs}) which are gotten from
$\mfE_y^n$ and $\mfB_y^n$ through Eqs.(\ref{eq:mfBx_EyBy}) and (\ref{eq:mfBs_EyBy}).
We showed that we can find the expressions of the radial and longitudinal fields in
the Laplace domain, $\mfE_{x,s}^n$ and $\mfB_{x,s}^n$, by solving the wave equations
(\ref{eq:we_xs_apdx}) in terms of the eigenfunctions of the operator.
However, it requires somewhat complicated calculations to disentangle
the field variables in the wave equations.

Eqs.(\ref{eq:mfBsEx}-\ref{eq:mfBxEs}) are given using Eqs.(\ref{eq:mfHpm}-\ref{eq:mfKpm})
which are the Green functions $\mfH_{\pm}^n$ and the coupling Green functions
$\mfK_{\pm}^n$ of the radial and longitudinal components of the transient fields in 
the Laplace domain,
\begin{align}
  \mfH_{\pm}^n
  &=\frac{\nu^2}{\hr\hr'}\mfG_{\pm}^n
   +\bmfG_{\mp}^n ,
    \qquad
  \mfK_{\pm}^n
  =i\nu
   \bigg(
     \frac{\rd_{\hr'}}{\hr}\mfG_{\pm}^n
    +\frac{\rd_{\hr}}{\hr'}\mfG_{\mp}^n
   \bigg) .
   \label{eq:mfHKpm}
\end{align}
According to Eqs.(\ref{eq:lim_rho_pnu}-\ref{eq:lim_rho_snu}),
$\mfH_{\pm}^n$, $\mfK_{\pm}^n$ and $\bmfG_{\pm}^n$ go to the following limits
for $\rho\to\infty$,
\begin{align}
  \lim_{\rho\to\infty}\mfH_{\pm}^n(r,r',\nu)
  &=\lim_{\rho\to\infty}\mfG_{\pm}^n(r,r',\nu)
   =\mfG_{\pm}^n(x,x',k_s) ,
   \label{eq:lim_mfH}
   \\
  \lim_{\rho\to\infty}\mfK_{\pm}^n(r,r',\nu)
  &=0 ,
    \qquad
  \lim_{\rho\to\infty}\bmfG_{\pm}^n(r,r',\nu)
  =\frac{k_x^2}{k_x^2+k_s^2}\mfG_{\mp}^n(x,x',k_s) .
   \label{eq:lim_mfK}
\end{align}
$\mfG_{\pm}^n(x,x',k_s)$ is given by Eq.(\ref{eq:mfGbe_strt})
which denotes the Green functions in a straight rectangular pipe.
$k_s$ and $k_x$ are the longitudinal and horizontal wavenumbers of the field in 
the straight section, given by Eq.(\ref{eq:def_kx}).

\clearpage

\section{Bessel functions and their cross products}
\label{sec:bessel}

We review various (asymptotic) expressions of the Bessel functions
which we use in the present study.
In our view by putting aside the rigorous definition,
we intuitively understand the Bessel functions of the first and second kind,
$J_{\nu}$ and $Y_{\nu}$, as those which are, so to speak,
the cosine and sine functions in a uniformly curved space.
Therefore $J_{\nu}$ and $Y_{\nu}$ are suited to describing standing waves in
a round structure such as a vibration of a drumhead.
On the other hand, the Hankel functions $H_\nu^{(1,2)}=J_{\nu}\pm iY_{\nu}$ correspond to
the complex notation of the trigonometric functions $e^{\pm ik_xx}$
which are suited to describing traveling waves propagating in free space.

The Bessel function $C_\nu(z)$ is defined as that which satisfies
the Bessel differential equation,
\begin{align}
  \bigg(\brd_z\rd_z+1-\frac{\nu^2}{z^2}\bigg)C_\nu(z)=0 ,
    \qquad
  \brd_z=\rd_z+\frac{1}{z}
   \quad(\nu\in\mathbb{C},~z\in\mathbb{C}) .
   \label{eq:BDE_C}
\end{align}
The order $\nu$ and argument $z$ are both complex variables in general,
$C_\nu(z):\mathbb{C}\times\mathbb{C}\to\mathbb{C}$.
$C_\nu$ has a transition point at $\nu=z$.
If $\nu/z\in\mathbb{R}$, $C_\nu(z)$ is oscillatory and exponential
respectively for $\nu/z<1$ and $\nu/z>1$.
The ascending series representation of $J_{\nu}$ is given by 9.1.2 in \cite{abramo_stegun},
which is an absolutely convergent series,
\begin{align}
  J_{\nu}(z)
  =\sum_{k=0}^{\infty}\frac{(-1)^k(z/2)^{\nu+2k}}{k!\Gam(\nu+k+1)}
   ,\qquad
  Y_{\nu}(z)
  =\frac{J_{\nu}(z)\cos(\pi\nu)-J_{-\nu}(z)}{\sin(\pi\nu)}
   ,\qquad
  q_{\nu}(z,z)
  =\frac{2}{\pi z} .
   \label{eq:JY}
\end{align}
$Y_{\nu}$ is given by 9.1.10 in \cite{abramo_stegun}.
$q_\nu(z,z)$ is the Wronskian of $J_{\nu}(z)$ and $Y_{\nu}(z)$,
given by 9.1.16 in \cite{abramo_stegun}.
The ascending series of $Y_\nu$ for integer order $\nu=n\in\mathbb{Z}_0^{+}$ is
given by 9.1.11 in \cite{abramo_stegun},
\begin{align}
  Y_n(z)
  &=\frac{1}{\pi}
   \{
    2\log(z/2)J_n(z)
   -(1-\delta_n^0)A_n(z)-B_n(z)
   \} ,
   \qquad
  Y_0'(z)
  =-Y_1(z) ,
    \label{eq:Yn}
   \\
  Y_n'(z)
  &=\frac{1}{\pi}
   \{
    (2/z)J_n(z)+2\log(z/2)J_n'(z)
   -(1-\delta_n^0)A_n'(z)-B_n'(z)
   \} .
    \label{eq:Ynp}
\end{align}
The prime ($'$) denotes the derivative with respect to the argument.
$\delta_n^0$ is the Kronecker delta (\ref{eq:lprm}).
$A_n$ and $B_n$ are respectively defined for $n\geq1$ and $n\geq0$ as follows,
\begin{alignat}{2}
  A_n(z)
  &=\sum_{k=0}^{n-1}\frac{(n-k-1)!}{k!}(z/2)^{2k-n} ,
    \qquad&
  A_n'(z)
  &=\frac{1}{z}\sum_{k=0}^{n-1}\frac{(n-k-1)!}{k!}(2k-n)(z/2)^{2k-n} ,
    \label{eq:Anz}
   \\
  B_n(z)
  &=\sum_{k=0}^{\infty}\frac{(-1)^k\omg_{nk}}{k!(n+k)!}(z/2)^{2k+n} ,
    \qquad&
  B_n'(z)
  &=\frac{1}{z}\sum_{k=0}^{\infty}\frac{(-1)^k\omg_{nk}}{k!(n+k)!}(2k+n)(z/2)^{2k+n} .
    \label{eq:Bnz}
\end{alignat}
$\omg_{nk}$ consists of the digamma function $\psi$ of integer argument,
given by 6.3.1-6.3.2 in \cite{abramo_stegun} and Eq.(\ref{eq:digamma}),
\begin{align}
  \omg_{nk}
  &=\psi(k+1)+\psi(n+k+1) .
    \label{eq:omg_nk}
\end{align}
We can approximate Eqs.(\ref{eq:Yn}-\ref{eq:Ynp})
for large $n\in\mathbb{N}$ and small $|z|$ as follows,
\begin{align}
  Y_n(z)
  \simeq
  -\frac{1+O(\veps_n^2/n)}{\pi}A_n(z)
   ,\qquad
  Y_n'(z)
  \simeq
  -\frac{1+O(\veps_n^2)}{\pi}A_n'(z)
   ,\qquad
  \veps_n
  =\frac{(z/2)^n}{(n-1)!} .
    \label{eq:Yn_Ynp_asympt}
\end{align}
The derivatives of $J_{\nu}$ and $Y_{\nu}$ with respect to $\nu$
at $\nu=n\in\mathbb{Z}_0^{+}$ are given by 9.1.66-9.1.67 (p.362) in \cite{abramo_stegun},
\begin{align}
  \bigg[\rd_{\nu}\bigg\{{ J_{\nu}(z) \atop Y_{\nu}(z) }\bigg\}\bigg]_{\nu=n}
  &=
      \frac{\pi}{2}\bigg\{{ Y_{n}(z) \atop -J_{n}(z) }\bigg\}
     +\frac{1-\delta_n^0}{2}
      \sum_{k=0}^{n-1}\frac{n!(z/2)^{k-n}}{k!(n-k)}\bigg\{{ J_k(z) \atop Y_k(z) }\bigg\} ,
    \label{eq:dJYn}
   \\
  \bigg[\rd_{\nu}\bigg\{{ J_{\nu}'(z) \atop Y_{\nu}'(z) }\bigg\}\bigg]_{\nu=n}
  &=
      \frac{\pi}{2}\bigg\{{ Y_{n}'(z) \atop -J_{n}'(z) }\bigg\}
     +\frac{1-\delta_n^0}{2}\sum_{k=0}^{n-1} \frac{n!(z/2)^{k-n}}{k!(n-k)}
      \bigg(\frac{k-n}{z}+\rd_z\bigg)
      \bigg\{{ J_k(z) \atop Y_k(z) }\bigg\} .
    \label{eq:dJYpn}
\end{align}
The ascending series of $\rd_{\nu}J_{\nu}$ is given by 9.1.64 in \cite{abramo_stegun},
\begin{align}
  \rd_{\nu}J_{\nu}(z)
  &=J_{\nu}(z)\log(z/2)
   -\sum_{k=0}^{\infty}\frac{(-1)^k}{k!}(z/2)^{\nu+2k}
    \frac{\psi(\nu+k+1)}{\Gam(\nu+k+1)} ,
    \label{eq:dJnu}
   \\
  \rd_{\nu}J_{\nu}'(z)
  &=J_{\nu}'(z)\log(z/2)
   +\frac{1}{z}J_{\nu}(z)
   -\frac{1}{z}\sum_{k=0}^{\infty}\frac{(-1)^k}{k!}(\nu+2k)(z/2)^{\nu+2k}
    \frac{\psi(\nu+k+1)}{\Gam(\nu+k+1)} .
    \label{eq:dJnup}
\end{align}

We introduce the cross products of $J_{\nu}$ and $Y_{\nu}$,
defined in 9.1.32 of \cite{abramo_stegun},
\begin{alignat}{2}
  p_{\nu}(b,a)
  &=J_{\nu}(b)Y_{\nu}(a)-Y_{\nu}(b)J_{\nu}(a) ,
    \qquad&
  q_\nu(b,a)
  &=J_{\nu}(b)Y_{\nu}'(a)-Y_{\nu}(b)J_{\nu}'(a) ,
  \label{eq:pq_nu}
  \\
  s_\nu(b,a)
  &=J_{\nu}'(b)Y_{\nu}'(a)-Y_{\nu}'(b)J_{\nu}'(a) ,
    \qquad&
  r_\nu(b,a)
  &=J_{\nu}'(b)Y_{\nu}(a)-Y_{\nu}'(b)J_{\nu}(a) .
  \label{eq:sr_nu}
\end{alignat}
The arguments $a$ and $b$ are complex variables in general.
We rewrite Eqs.(\ref{eq:pq_nu}-\ref{eq:sr_nu}) using $J_{-\nu}$ instead of $Y_{\nu}$,
\begin{alignat}{2}
  p_{\nu}(b,a)
  &=\frac{J_{-\nu}(b)J_{\nu}(a)-J_{\nu}(b)J_{-\nu}(a)}{\sin(\pi\nu)} ,
    \qquad&
  q_{\nu}(b,a)
  &=\frac{J_{-\nu}(b)J_{\nu}'(a)-J_{\nu}(b)J_{-\nu}'(a)}{\sin(\pi\nu)} ,
  \label{eq:pq_Jpm}
  \\
  s_{\nu}(b,a)
  &=\frac{J_{-\nu}'(b)J_{\nu}'(a)-J_{\nu}'(b)J_{-\nu}'(a)}{\sin(\pi\nu)} ,
    \qquad&
  r_{\nu}(b,a)
  &=\frac{J_{-\nu}'(b)J_{\nu}(a)-J_{\nu}'(b)J_{-\nu}(a)}{\sin(\pi\nu)} .
  \label{eq:sr_Jpm}
\end{alignat}
We substitute the ascending series of $J_{\pm\nu}$, given by the first equation of
(\ref{eq:JY}), into Eqs.(\ref{eq:pq_Jpm}-\ref{eq:sr_Jpm}),
\begin{align}
  \bigg\{{ p_\nu(b,a) \atop bas_\nu(b,a) }\bigg\}\sin(\pi\nu)
  &=\sum_{j,k=0}^{\infty}
    \bigg\{{ \pi_{\nu}^{jk} \atop \sig_{\nu}^{jk} }\bigg\}
    (b^{2k-\nu}a^{2j+\nu}-b^{2j+\nu}a^{2k-\nu})
  \label{eq:pnu_series}
  \\
  &=\sum_{j,k=0}^{\infty}b^{2j}a^{2k}
    \bigg[
       \bigg\{{ \pi_{-\nu}^{jk} \atop \sig_{-\nu}^{jk} }\bigg\}
       \bigg(\frac{b}{a}\bigg)^{-\nu}
      -\bigg\{{ \pi_{\nu}^{jk} \atop \sig_{\nu}^{jk} }\bigg\}
       \bigg(\frac{b}{a}\bigg)^{\nu}
    \bigg] ,
  \label{eq:pnu_series_pm}
   \\
  \bigg\{{ aq_\nu(b,a) \atop br_\nu(b,a) }\bigg\}\sin(\pi\nu)
  &=\sum_{j,k=0}^{\infty}
    \bigg[
       \bigg\{{ \xi_{\nu}^{jk} \atop \eta_{\nu}^{jk} }\bigg\}
       b^{2k-\nu}a^{2j+\nu}
      -\bigg\{{ \eta_{\nu}^{jk} \atop \xi_{\nu}^{jk} }\bigg\}
       b^{2j+\nu}a^{2k-\nu}
    \bigg]
  \label{eq:qr_nu_series}
   \\
  &=\sum_{j,k=0}^{\infty}b^{2j}a^{2k}
    \bigg[
      \bigg\{{ \eta_{-\nu}^{jk} \atop \xi_{-\nu}^{jk} }\bigg\}
      \bigg(\frac{b}{a}\bigg)^{-\nu}
     -\bigg\{{ \eta_{\nu}^{jk} \atop \xi_{\nu}^{jk} }\bigg\}
      \bigg(\frac{b}{a}\bigg)^{\nu}
    \bigg] ,
  \label{eq:qr_pm_series}
\end{align}
where the coefficients are given as follows,
\begin{alignat}{2}
  \pi_{\nu}^{jk}
  &=\frac{(-1/4)^{j+k}}{j!k!\Gamma(j+\nu+1)\Gamma(k-\nu+1)} ,
     \qquad&
  \xi_{\nu}^{jk}
  &=(2j+\nu)\pi_{\nu}^{jk} ,
   \label{eq:pi}
   \\
  \sig_{\nu}^{jk}
  &=(2j+\nu)(2k-\nu)\pi_{\nu}^{jk} ,
     \qquad&
  \eta_{\nu}^{jk}
  &=(2k-\nu)\pi_{\nu}^{jk} .
   \label{eq:sigma}
\end{alignat}
They have the following symmetries with respect to the order $\nu$ and the exchange of
the indices $(j,k)$,
\begin{alignat}{4}
  \pi_{-\nu}^{jk}
  &=\pi_{\nu}^{kj}
   ,\qquad&
  \pi_{\nu^{\ast}}^{jk}
  =(\pi_{\nu}^{jk})^{\ast} ;
    \qquad\quad&
  \xi_{-\nu}^{jk}
  &=\eta_{\nu}^{kj}
   ,\qquad&
  \xi_{\nu^{\ast}}^{jk}
  =(\xi_{\nu}^{jk})^{\ast} ,
   \\
  \sig_{-\nu}^{jk}
  &=\sig_{\nu}^{kj}
   ,\qquad&
  \sig_{\nu^{\ast}}^{jk}
  =(\sig_{\nu}^{jk})^{\ast} ;
    \qquad\quad&
  \eta_{-\nu}^{jk}
  &=\xi_{\nu}^{kj}
   ,\qquad&
  \eta_{\nu^{\ast}}^{jk}
  =(\eta_{\nu}^{jk})^{\ast} ,
\end{alignat}
where the asterisk $(\ast)$ denotes the complex conjugate.
The cross products, represented as $t_{\nu}$ for brevity, have the following symmetries
with respect to the order and arguments,
\begin{align}
  &
  t_{\nu}(b^{\ast},a^{\ast})
  =t_{\nu^{\ast}}(b,a)
  =t_\nu^{\ast}(b,a) ,
    \qquad
  t_{-\nu}(b,a)
  =t_\nu(b,a) ,
    \qquad
  t_{\nu}^{\pm}(-b,-a)
  =\pm t_{\nu}^{\pm}(b,a) ,
  \label{eq:cp_symm}
   \\
  &
  \text{where}\quad
  t_{\nu}
  =\{p_\nu,s_\nu; q_\nu,r_\nu\}
  =\{t_{\nu}^{+};t_{\nu}^{-}\} ,
    \qquad
  t_{\nu}^{+}
  =\{p_{\nu},s_{\nu}\}
   \quad\text{and}\quad
  t_{\nu}^{-}
  =\{q_{\nu},r_{\nu}\} .
  \label{eq:cp_symm2}
\end{align}

Eqs.(\ref{eq:pnu_series}) and (\ref{eq:qr_nu_series}) are rewritten as
\begin{align}
  \bigg\{{ p_\nu(b,a) \atop bas_\nu(b,a) }\bigg\}
  &=-\frac{2}{\pi}\sum_{n=0}^{\infty}\bigg(-\frac{ba}{4}\bigg)^n
    \sum_{\ell=0}^{n}
    \bigg\{{ 1 \atop n^2-\kap_{n\ell}^2 }\bigg\}
    \frac{\sinh[\kap_{n\ell}\log(b/a)]}{\ell!(n-\ell)!\lam_{\ell,n-\ell}} ,
    \label{ps_nu}
   \\
  \bigg\{{ aq_\nu(b,a) \atop -br_\nu(b,a) }\bigg\}
  &=\frac{2}{\pi}\sum_{n=0}^{\infty}\bigg(\!-\frac{ba}{4}\bigg)^n
    \sum_{\ell=0}^{n}
    \frac{\kap_{n\ell}\cosh[\kap_{n\ell}\log(b/a)] \mp n\sinh[\kap_{n\ell}\log(b/a)]}
         {\ell!(n-\ell)!\lam_{\ell,n-\ell}} .
    \label{aq_br_nu}
\end{align}
The double sign $(-,+)$ in Eq.(\ref{aq_br_nu}) corresponds to $(q_\nu,r_\nu)$.
$\kap_{n\ell}$ and $\lam_{\ell,m}$ are given as
\begin{align}
  \kap_{n\ell}
  =\nu+2\ell-n
   ,\qquad
  \lam_{\ell,m}
   =\bigg\{\prod_{j=0}^{\ell}(j+\nu)\bigg\}
    \bigg\{\delta_0^m+(1-\delta_0^m)\prod_{j'=1}^{m}(j'-\nu)\bigg\}
   ,\qquad
  \bigg[\frac{\lam_{\ell,m}}{\nu}\bigg]_{\nu=0}
  =\ell!m! .
\end{align}
When $\nu=0$, Eqs.(\ref{ps_nu}-\ref{aq_br_nu}) are given as follows,
\begin{align}
  \bigg[
  \begin{array}{cc}
     p_0(b,a) & aq_0(b,a)\\
   bas_0(b,a) & br_0(b,a)
  \end{array}
  \bigg]
  &=-\frac{2}{\pi}\sum_{j,k=0}^{\infty}b^{2j}a^{2k}
  \bigg[
  \begin{array}{cc}
    \pi_{0}^{jk}\Psi_{jk}                       & \eta_{0}^{jk}\Psi_{jk}-\pi_{0}^{jk} \\
    \sig_{0}^{jk}\Psi_{jk} +2(k-j)\pi_{0}^{jk}~ & \xi_{0}^{jk}\Psi_{jk}+\pi_{0}^{jk}
  \end{array}
  \bigg]
   \label{eq:pqrs0_series}
   \\
  &=
   -\frac{2}{\pi}\sum_{n=0}^{\infty}\bigg(-\frac{ba}{4}\bigg)^n\sum_{\ell=0}^{n}
    \frac{A_{n\ell}\cosh u_{n\ell}+B_{n\ell}\sinh u_{n\ell}}{\{\ell!(n-\ell)!\}^2} .
    \label{cp0}
\end{align}
$\Psi_{jk}$ is given as
\begin{alignat}{2}
  \Psi_{jk}
  &=\log(b/a)+\psi(k+1)-\psi(j+1) ,
    \qquad
  \psi(j+1)
  =-\gam_{\rm e}
   +(1-\delta_j^0)\sum_{m=1}^{j}\frac{1}{m}
   \quad
   (j\in\mathbb{Z}_0^{+}) .
   \label{eq:digamma}
\end{alignat}
$\psi$ is the digamma function given by 6.3.1-6.3.2 in \cite{abramo_stegun}.
$\gam_{\rm e}=0.5772156649\cdots$ is Euler's constant.
$A_{n\ell}$ and $B_{n\ell}$ in Eq.(\ref{cp0}) represent
$A_{n\ell}=(A_{n\ell}^p,A_{n\ell}^q,A_{n\ell}^r,A_{n\ell}^s)$ and
$B_{n\ell}=(B_{n\ell}^p,B_{n\ell}^q,B_{n\ell}^r,B_{n\ell}^s)$ 
which correspond to $t_0=(p_0,q_0,r_0,s_0)$ on the L.H.S. of Eq.(\ref{eq:pqrs0_series}),
\begin{alignat}{2}
  A_{n\ell}^p
  &=\log(b/a) ,
    \qquad&
  B_{n\ell}^p
  &=\bpsi_{n\ell} ,
   \\
  A_{n\ell}^q
  &=n\log(b/a)-(2\ell-n)\bpsi_{n\ell}-1 ,
    \qquad&
  B_{n\ell}^q
  &=n\bpsi_{n\ell}-(2\ell-n)\log(b/a) ,
   \\
  A_{n\ell}^r
  &=n\log(b/a)+(2\ell-n)\bpsi_{n\ell}+1 ,
    \qquad&
  B_{n\ell}^r
  &=n\bpsi_{n\ell}+(2\ell-n)\log(b/a) ,
   \\
  A_{n\ell}^s
  &=4\ell(n-\ell)\log(b/a) ,
    \qquad&
  B_{n\ell}^s
  &=2\{2\ell(n-\ell)\bpsi_{n\ell} -(2\ell-n)\} .
\end{alignat}
$u_{n\ell}$ in Eq.(\ref{cp0}) and $\bpsi_{n\ell}$ are given as
\begin{align}
  u_{n\ell}
  &=(2\ell-n)\log(b/a) ,
    \qquad
  \bpsi_{n\ell}
  =\psi(n-\ell+1)-\psi(\ell+1) .
\end{align}
In the limit of $k_r^n\to0$, Eqs.(\ref{ps_nu}-\ref{aq_br_nu}) for $b=\hr$ and $a=\hr'$
converge as
\begin{alignat}{2}
  \lim_{k_r^n\to0} 
  \bigg\{{-\nu p_\nu(\hr,\hr') \atop (\hr\hr'/\nu)s_\nu(\hr,\hr')}\bigg\}
  &=\frac{2}{\pi}\sinh l_{\nu} ,
    \qquad&
  \lim_{k_r^n\to0}\bigg\{{ \hr'q_\nu(\hr,\hr') \atop -\hr r_\nu(\hr,\hr')}\bigg\}
  &=\frac{2}{\pi}\cosh l_{\nu} ,
    \label{lim_kr0_aq_br_bas}
   \\
  \lim_{k_r^n\to0} 
  \bigg\{{\bnu p_{i\bnu}(\hr,\hr') \atop (\hr\hr'/\bnu)s_{i\bnu}(\hr,\hr')}\bigg\}
  &=-\frac{2}{\pi}\sin\bar{l}_{\bnu} ,
    \qquad&
  \lim_{k_r^n\to0}\bigg\{{ \hr'q_{i\bnu}(\hr,\hr') \atop -\hr r_{i\bnu}(\hr,\hr')}\bigg\}
  &=\frac{2}{\pi}\cos\bar{l}_{\bnu} .
\end{alignat}
$\hr$ and $\hr'$ are the dimensionless radii normalized by the radial wavenumber $k_r^n$
given by Eq.(\ref{eq:krn}).
$l_{\nu}$ and $\bar{l}_{\bnu}$ are functions of $r$ and $r'$, which do not depend on $k_r^n$,
\begin{align}
  \hr=k_r^nr
   ,\qquad
  \hr'=k_r^nr'
   ,\qquad
  l_{\nu}
  =\nu\log(r/r')
  =i\bar{l}_{\bnu}
   ,\qquad
  \bar{l}_{\bnu}
  =\bnu\log(r/r') .
   \label{eq:hr_lnu}
\end{align}
When $\nu\ne0$, $s_\nu(\hr,\hr')$ diverges in the limit of $k_r^n\to0$
since $s_\nu(\hr,\hr')\propto(\hr\hr')^{-1}\propto(k_r^n)^{-2}\to\infty$.
On the other hand, $s_0(\hr,\hr')$ converges in the limit of $k_r^n\to0$,
because $s_0(\hr,\hr')=p_1(\hr,\hr')$, and the limit of $\nu\to0$ commutes with
the limit of $k_r^n\to0$,
\begin{alignat}{2}
  &
  \lim_{\nu\to0}\lim_{k_r^n\to0}\bigg\{{ p_\nu(\hr,\hr') \atop s_\nu(\hr,\hr') }\bigg\}
   =\lim_{k_r^n\to0}\bigg\{{ p_0(\hr,\hr') \atop s_0(\hr,\hr') }\bigg\}
   =-\frac{2}{\pi}\bigg\{{ \log(r/r') \atop \sinh[\log(r/r')] }\bigg\} ,
   \label{eq:lim_kr0_p0s0}
   \\
  &
  \lim_{\nu\to0}\lim_{k_r^n\to0}
  \bigg\{{ \hr'q_\nu(\hr,\hr') \atop -\hr r_\nu(\hr,\hr') }\bigg\}
   =\lim_{k_r^n\to0}\bigg\{{ \hr'q_0(\hr,\hr') \atop -\hr r_0(\hr,\hr') }\bigg\}
   =\frac{2}{\pi} .
   \label{eq:lim_kr0_aq0_br0}
\end{alignat}
According to Eq.(\ref{ps_nu}) and the first equation of (\ref{lim_kr0_aq_br_bas}),
the limit of $k_r^n\to0$ commutes with the operator $\rd_{\nu}$ for
$p_\nu(\hr,\hr')$ and $\hr\hr's_\nu(\hr,\hr')$, \ie,
\begin{align}
  &
  \lim_{k_r^n\to0}\rd_{\nu}p_\nu(\hr,\hr')
   =\rd_{\nu}\lim_{k_r^n\to0}p_\nu(\hr,\hr')
   =\frac{2}{\pi\nu^2}(\sinh l_{\nu}-l_{\nu}\cosh l_{\nu})
   =\frac{2i}{\pi\bnu^2}(\bar{l}_{\bnu}\cos\bar{l}_{\bnu}-\sin\bar{l}_{\bnu}) ,
  \label{eq:lim_kr0_rd_pnu}
   \\
  &
  \lim_{k_r\to0}\rd_{\nu}\{\hr\hr's_\nu(\hr,\hr')\}
   =\rd_{\nu}\lim_{k_r\to0} \hr\hr's_\nu(\hr,\hr')
   =\frac{2}{\pi}(\sinh l_{\nu}+l_{\nu}\cosh l_{\nu})
   =i\frac{2}{\pi}(\bar{l}_{\bnu}\cos\bar{l}_{\bnu}+\sin\bar{l}_{\bnu}) .
  \label{eq:lim_kr0_rd_snu}
\end{align}
For $k_r^n\to0$, the second derivatives of $p_\nu(\hr,\hr')$ and
$\hr\hr's_\nu(\hr,\hr')$ with respect to $\nu$ go to the following limits,
\begin{align}
  &
  \lim_{k_r^n\to0}\rd_{\nu}^2p_\nu(\hr,\hr')
  =
  -\frac{2 l_{\nu}^3}{\pi\nu^3}
    \bigg\{
      \frac{\sinh l_{\nu}}{l_{\nu}}
     +\frac{2}{l_{\nu}^2}
      \bigg(\frac{\sinh l_{\nu}}{l_{\nu}}-\cosh l_{\nu}\bigg)
    \bigg\} ,
  \label{eq:lim_kr0_rd2_pnu}
    \\
  &
  \lim_{k_r\to0}\rd_{\nu}^2\{\hr\hr's_\nu(\hr,\hr')\}
  =
    \frac{2l_{\nu}}{\pi\nu}
    (2\cosh l_{\nu}+l_{\nu}\sinh l_{\nu}) .
  \label{eq:lim_kr0_rd2_snu}
\end{align}
Taking the limit of $\nu\to0$ for Eqs.(\ref{eq:lim_kr0_rd2_pnu}-\ref{eq:lim_kr0_rd2_snu}), 
we get
\begin{align}
  \lim_{k_r^n\to0}\big[\rd_{\nu}^2p_\nu(\hr,\hr')\big]_{\nu=0}
  &=
   -\frac{2}{3\pi}\{\log(r/r')\}^3 ,
    \qquad
  \lim_{k_r\to0}\big[\rd_{\nu}^2\{\hr\hr's_\nu(\hr,\hr')\}\big]_{\nu=0}
  =\frac{4}{\pi}\log(r/r') .
  \label{eq:kr0_nu0_rd2_ps_nu}
\end{align}
Thus, in the limit of $k_r^n\to0$, the second derivatives of $p_\nu(\hr,\hr')$ and
$\hr\hr's_\nu(\hr,\hr')$ with respect to $\nu$ are not zero at $\nu=0$ unless $r=r'$.

\subsection{Asymptotic expressions of the Bessel functions for large order}
\label{sec:AE_bignu}

$J_\nu(z)$ and $Y_\nu(z)$ have several asymptotic expressions,
depending on the treatment of the order $\nu$ and argument $z$.
For $\Re\nu\to+\infty$ with $z$ fixed, $J_\nu(z)$ and $Y_\nu(z)$ behave as
those in 9.3.1 of \cite{abramo_stegun},
\begin{align}
  J_\nu(z)\simeq
  \frac{1}{(2\pi\nu)^{1/2}}\bigg(\frac{ez}{2\nu}\bigg)^{\!\nu} ,
    \qquad
  Y_\nu(z)\simeq
  -\bigg(\frac{2}{\pi\nu}\bigg)^{1/2}\bigg(\frac{ez}{2\nu}\bigg)^{\!-\nu}
  \qquad(\Re\nu\to+\infty).
   \label{eq:JY_nuinf}
\end{align}
From Eqs.(\ref{eq:JY_nuinf}), we get the asymptotic expressions of the cross products
$t_{\nu}(\hr,\hr')=\{p_{\nu},q_{\nu},r_{\nu},s_{\nu}\}$ for $|\Re\nu|\to\infty$ with
the arguments $\hr\,(=k_r^nr)$ and $\hr'\,(=k_r^nr')$ fixed,
\begin{align}
  \bigg\{{ -p_\nu(\hr,\hr') \atop \hr\hr' s_\nu(\hr,\hr') }\bigg\}
  &\simeq
    \bigg\{{ \nu^{-1} \atop \nu }\bigg\} \frac{2}{\pi}\sinh[\nu\log(r/r')] ,
    \qquad
  \bigg\{{ \hr' q_\nu(\hr,\hr') \atop -\hr r_\nu(\hr,\hr') }\bigg\}
  \simeq \frac{2}{\pi}\cosh[\nu\log(r/r')] .
   \label{eq:pnu_asymp}
\end{align}
The R.H.S. of Eqs.(\ref{eq:pnu_asymp}) do not depend on $k_r^n$.
According to Eqs.(\ref{eq:pimu_asymp_mu_inf}-\ref{eq:simu_asymp_mu_inf}),
Eqs.(\ref{eq:pnu_asymp}) also hold with respect to the purely imaginary order
$\nu=i\bnu\in i\mathbb{R}$ for $|\bnu|\to\infty$ with $\hr$ and $\hr'$ fixed.

\subsection{Asymptotic expressions of the Bessel functions for large argument}
\label{sec:hankel}

According to 9.2.5-9.2.6 and 9.2.11-9.2.12 in \cite{abramo_stegun},
we can expand $J_{\nu}(z)$, $Y_{\nu}(z)$, $J_{\nu}'(z)$ and $Y_{\nu}'(z)$
with respect to $z^{-1}$ for $|z|\gg1$ and $|\arg z|<\pi$ with the order $\nu$ fixed,
\begin{alignat}{2}
  J_{\nu}(z)
  &\simeq
    A(\hP_{\nu}\cos\chi_{\nu}-\hQ_{\nu}\sin\chi_{\nu}),
    \qquad&
  J_{\nu}'(z)
  &\simeq
   -A(\hR_{\nu}\sin\chi_{\nu}+\hS_{\nu}\cos\chi_{\nu}),
  \label{eq:J_hankel}
  \\
  Y_{\nu}(z)
  &\simeq
    A(\hP_{\nu}\sin\chi_{\nu}+\hQ_{\nu}\cos\chi_{\nu}),
    \qquad&
  Y_{\nu}'(z)
  &\simeq
    A(\hR_{\nu}\cos\chi_{\nu}-\hS_{\nu}\sin\chi_{\nu}) .
  \label{eq:Y_hankel}
\end{alignat}
Eqs.(\ref{eq:J_hankel}-\ref{eq:Y_hankel}) are referred to as Hankel's asymptotic expansions.
$A$ and $\chi_{\nu}$ are given as
\begin{align}
  A
  =\bigg(\frac{2}{\pi z}\bigg)^{1/2}
   ,\qquad
  \chi_{\nu}
  =z-\frac{\pi}{2}\bigg(\nu+\frac{1}{2}\bigg)
   \qquad
  (z,\nu\in\mathbb{C}) .
   \label{eq:Achi}
\end{align}
$(\hP_{\nu},\hQ_{\nu})$ and $(\hR_{\nu},\hS_{\nu})$ are ``$(P,Q)$'' and ``$(R,S)$''
given by 9.2.9-9.2.10 and 9.2.15-9.2.16 in \cite{abramo_stegun},
\begin{align}
  \bigg\{{ \hP_{\nu}(z) \atop \hR_{\nu}(z)}\bigg\}
  =\sum_{j=0}^{\infty}\frac{(-1)^j}{(2j)!(8z)^{2j}}
    \bigg\{{ \hat{a}_{2j}(\kap) \atop \hat{b}_{2j}(\kap) }\bigg\}
   ,\qquad
  \bigg\{{ \hQ_{\nu}(z) \atop \hS_{\nu}(z) }\bigg\}
  =\sum_{j=0}^{\infty}\frac{(-1)^j}{(2j+1)!(8z)^{2j+1}}
   \bigg\{{ \hat{a}_{2j+1}(\kap) \atop \hat{b}_{2j+1}(\kap) }\bigg\} ,
   \label{eq:Pnu_coe}
\end{align}
where
\begin{align}
  \kap=4\nu^2 .
\end{align}
The coefficients $\hat{a}_{\ell}$ and $\hat{b}_{\ell}$ for $\ell\in\mathbb{N}$
are functions of $\kap$,
\begin{alignat}{4}
  \hat{a}_0
  &=\hat{a}_1'
   =1
    ,\qquad&
  \hat{a}_{\ell}(\kap)
  &=\prod_{k=1}^{\ell}\alp_{k}(\kap)
    ,\qquad&
  \hat{a}_{\ell}'
  &=\rd_{\kap}\hat{a}_{\ell}
   =\tau_{\ell}\hat{a}_{\ell}
    ,\qquad&
  \alp_{k}(\kap)
  &=\kap-(2k-1)^2 ,
    \label{eq:ahat}
    \\
  \hat{b}_0
  &=\hat{b}_1'
   =1
    ,\qquad&
  \hat{b}_{\ell}(\kap)
  &=\beta_{\ell}(\kap)\hat{a}_{\ell-1}(\kap)
    ,\qquad&
  \hat{b}_{\ell}'
  &=\rd_{\kap}\hat{b}_{\ell}
   =\sig_{\ell}\hat{a}_{\ell-1}
    ,\qquad&
  \beta_{\ell}(\kap)
  &=\kap+4{\ell}^2-1 ,
    \label{eq:bhat}
\end{alignat}
where the prime $(')$ denotes the derivative with respect to $\kap$.
$\tau_{\ell}$ and $\sig_{\ell}$ are given as
\begin{align}
  \tau_0
  =0
   ,\qquad
  \tau_{\ell}
  =\sum_{k=1}^{\ell}\frac{1}{\alp_k} ,
    \qquad
  \sig_1
  =1
   ,\qquad
  \sig_{\ell}
  =1+\beta_{\ell}\tau_{\ell-1}
   \qquad
  (\ell\in\mathbb{N}) .
    \label{eq:tau_sig}
\end{align}
$\hat{a}_{\ell}$ and $\hat{b}_{\ell}$ satisfy the following recurrence relations,
\begin{align}
  \frac{\hat{a}_{\ell+1}}{\hat{a}_{\ell}}
  =\alp_{\ell+1}
   ,\qquad
  \frac{\hat{a}_{\ell+1}}{\hat{a}_{\ell-1}}
  =\alp_{\ell+1}\alp_{\ell}
   ,\qquad
  \frac{\hat{b}_{\ell+1}}{\hat{b}_{\ell}}
  =\alp_{\ell}\frac{\beta_{\ell+1}}{\beta_{\ell}}
   ,\qquad
  \frac{\hat{b}_{\ell+1}}{\hat{b}_{\ell-1}}
  =\alp_{\ell-1}\alp_{\ell}\frac{\beta_{\ell+1}}{\beta_{\ell-1}} .
\end{align}
$(\hat{a}_j,\hat{b}_j)$ and $(\hat{a}_j',\hat{b}_j')$ for $j=1$ and 2 are given as follows,
\begin{alignat}{4}
  \hat{a}_1
  &=\alp_1 ,
    \qquad&
  \hat{a}_1'
  &=\rd_{\kap}\alp_1
   =1 ,
    \qquad&
  \hat{a}_2
  &=\alp_1\alp_2 ,
    \qquad&
  \hat{a}_2'
  &=\alp_1+\alp_2 ,
    \\
  \hat{b}_1
  &=\beta_1 ,
    \qquad&
  \hat{b}_1'
  &=\rd_{\kap}\beta_1
   =1 ,
    \qquad&
  \hat{b}_2
  &=\alp_1\beta_2 ,
    \qquad&
  \hat{b}_2'
  &=\alp_1+\beta_2 .
\end{alignat}
From Eqs.(\ref{eq:J_hankel}-\ref{eq:Y_hankel}), we get the asymptotic expressions of
the cross products (\ref{eq:pq_nu}-\ref{eq:sr_nu}) for large arguments
$|z|\gg1$ and $|z'|\gg1$ with $\nu$ fixed,
\begin{align}
  \frac{\pi}{2}(zz')^{1/2}
  \bigg\{{ p_{\nu}(z,z') \atop s_{\nu}(z,z') }\bigg\}
  &\simeq
    \bigg\{{ Y_{+}(z,z') \atop Y_{-}(z,z') }\bigg\}\cos(z-z')
   -\bigg\{{ X_{+}(z,z') \atop X_{-}(z,z') }\bigg\}\sin(z-z') ,
   \label{eq:pnu_hankel}
   \\
  \frac{\pi}{2}(zz')^{1/2}\bigg\{{ q_\nu(z,z') \atop r_\nu(z,z') }\bigg\}
  &\simeq
    \bigg\{{ N(z,z') \atop -N(z',z) }\bigg\}\cos(z-z')
   +\bigg\{{ M(z,z') \atop M(z',z) }\bigg\}\sin(z-z') ,
   \label{eq:qr_hankel}
\end{align}
where
\begin{alignat}{2}
  X_{+}(z,z')
  &=\hP_{\nu}(z)\hP_{\nu}(z')+\hQ_{\nu}(z)\hQ_{\nu}(z') ,
   \qquad&
  Y_{+}(z,z')
  &=\hP_{\nu}(z)\hQ_{\nu}(z')-\hQ_{\nu}(z)\hP_{\nu}(z') ,
   \label{eq:XY_pls}
   \\
  X_{-}(z,z')
  &=\hR_{\nu}(z)\hR_{\nu}(z')+\hS_{\nu}(z)\hS_{\nu}(z') ,
   \qquad&
  Y_{-}(z,z')
  &=\hR_{\nu}(z)\hS_{\nu}(z')-\hS_{\nu}(z)\hR_{\nu}(z') ,
   \label{eq:XY_mns}
   \\
  M(z,z')
  &=\hP_{\nu}(z)\hS_{\nu}(z')-\hQ_{\nu}(z)\hR_{\nu}(z') ,
   \qquad&
  N(z,z')
  &=\hP_{\nu}(z)\hR_{\nu}(z')+\hQ_{\nu}(z)\hS_{\nu}(z') .
   \label{eq:MN_mns}
\end{alignat}
We use Eq.(\ref{eq:pnu_hankel}) in Eq.(\ref{eq:ps_hankel})
to find the poles of the fields in the range $|\nu|\leq O(1)$.
Then we find the asymptotic expressions of the cutoff wavenumbers of the curved pipe,
implicitly defined in Eqs.(\ref{eq:kmn}-\ref{eq:bkmn}).

The derivative of Eq.(\ref{eq:pnu_hankel}) with respect to the order $\nu$ is gotten as
\begin{align}
  \frac{\pi}{2}(zz')^{1/2}
  \rd_{\nu}\bigg\{{ p_{\nu}(z,z') \atop s_{\nu}(z,z') }\bigg\}
  &\simeq
    \bigg\{{ Y_{+}'(z,z') \atop Y_{-}'(z,z') }\bigg\}\cos(z-z')
   -\bigg\{{ X_{+}'(z,z') \atop X_{-}'(z,z') }\bigg\}\sin(z-z') .
   \label{eq:psnu_prm_hankel}
\end{align}
$X_{\pm}'$ and $Y_{\pm}'$ are the derivatives of $X_{\pm}$ and $Y_{\pm}$ with respect to
$\nu$,
\begin{align}
  X_{\pm}'
  =\rd_{\nu}X_{\pm},
   \qquad
  Y_{\pm}'
  =\rd_{\nu}Y_{\pm} .
\end{align}
They are given as follows,
\begin{align}
  X_{+}'(z,z')
  &=
    \hP_{\nu}'(z)\hP_{\nu}(z')
   +\hP_{\nu}(z)\hP_{\nu}'(z')
   +\hQ_{\nu}'(z)\hQ_{\nu}(z')
   +\hQ_{\nu}(z)\hQ_{\nu}'(z') ,
   \\
  X_{-}'(z,z')
  &=
    \hR_{\nu}'(z)\hR_{\nu}(z')
   +\hR_{\nu}(z)\hR_{\nu}'(z')
   +\hS_{\nu}'(z)\hS_{\nu}(z')
   +\hS_{\nu}(z)\hS_{\nu}'(z') ,
   \\
  Y_{+}'(z,z')
  &=
    \hP_{\nu}'(z)\hQ_{\nu}(z')
   +\hP_{\nu}(z)\hQ_{\nu}'(z')
   -\hQ_{\nu}'(z)\hP_{\nu}(z')
   -\hQ_{\nu}(z)\hP_{\nu}'(z') ,
   \\
  Y_{-}'(z,z')
  &=
    \hR_{\nu}'(z)\hS_{\nu}(z')
   +\hR_{\nu}(z)\hS_{\nu}'(z')
   -\hS_{\nu}'(z)\hR_{\nu}(z')
   -\hS_{\nu}(z)\hR_{\nu}'(z') .
\end{align}
$\hP_{\nu}',\hQ_{\nu}',\hR_{\nu}'$ and $\hS_{\nu}'$ are the derivatives of
$\hP_{\nu},\hQ_{\nu},\hR_{\nu}$ and $\hS_{\nu}$ with respect to $\nu$,
\begin{align}
  \bigg\{{ \hP_{\nu}'(z) \atop \hR_{\nu}'(z)}\bigg\}
  &=\rd_{\nu}\bigg\{{ \hP_{\nu}(z) \atop \hR_{\nu}(z)}\bigg\}
   =\sum_{j=1}^{\infty}\frac{8\nu(-1)^j}{(2j)!(8z)^{2j}}
    \bigg\{{ \hat{a}_{2j}'(\kap) \atop \hat{b}_{2j}'(\kap) }\bigg\} ,
   \label{eq:PR_prm}
   \\
  \bigg\{{ \hQ_{\nu}'(z) \atop \hS_{\nu}'(z) }\bigg\}
  &=\rd_{\nu}\bigg\{{ \hQ_{\nu}(z) \atop \hS_{\nu}(z) }\bigg\}
   =\sum_{j=0}^{\infty}\frac{8\nu(-1)^j}{(2j+1)!(8z)^{2j+1}}
    \bigg\{{ \hat{a}_{2j+1}'(\kap) \atop \hat{b}_{2j+1}'(\kap) }\bigg\} .
   \label{eq:QS_prm}
\end{align}
$\hat{a}_{\ell}'$ and $\hat{b}_{\ell}'$ are the derivatives of
$\hat{a}_{\ell}$ and $\hat{b}_{\ell}$ with respect to the argument $\kap\,(\,=4\nu^2)$ as
shown in Eqs.(\ref{eq:ahat}-\ref{eq:bhat}).
The numerical method to compute Eqs.(\ref{eq:pnu_hankel}-\ref{eq:qr_hankel}) and
(\ref{eq:psnu_prm_hankel}) is shown in appendix \ref{sec:hankel_numerical}
(p.\pageref{sec:hankel_numerical}).

According to 9.7.1-9.7.4 in \cite{abramo_stegun},
we can expand the modified Bessel functions $I_{\nu}$ and $K_{\nu}$ for large $|z|$,
\begin{alignat}{3}
  \bigg\{{ I_{\nu}(z) \atop I_{\nu}'(z) }\bigg\}
  &=\frac{e^z}{(2\pi z)^{1/2}}
    \bigg\{{ \cI_0(z) \atop \cI_1(z) }\bigg\}
   ,\qquad&
  \rd_{\nu}\bigg\{{ I_{\nu}(z) \atop I_{\nu}'(z) }\bigg\}
  &=\frac{8\nu e^z}{(2\pi z)^{1/2}}
    \bigg\{{ \bcI_0(z) \atop \bcI_1(z) }\bigg\} ,
    \label{eq:Inu_HAE}
   \\
  \bigg\{{ K_{\nu}(z) \atop -K_{\nu}'(z) }\bigg\}
  &=\frac{e^{-z}}{(2z/\pi)^{1/2}}
    \bigg\{{ \cK_0(z) \atop \cK_1(z) }\bigg\}
   ,\qquad&
  \rd_{\nu}\bigg\{{ K_{\nu}(z) \atop -K_{\nu}'(z) }\bigg\}
  &=\frac{8\nu e^{-z}}{(2z/\pi)^{1/2}}
    \bigg\{{ \bcK_0(z) \atop \bcK_1(z) }\bigg\} ,
    \label{eq:Knu_HAE}
\end{alignat}
where $\nu$ is fixed.
Eqs.(\ref{eq:Inu_HAE}-\ref{eq:Knu_HAE}) consist of the following series,
\begin{alignat}{3}
  \bigg\{{ \cI_0(z) \atop \cI_1(z) }\bigg\}
  &=
    \sum_{\ell=0}^{\infty}\frac{(-1)^{\ell}}{\ell!(8z)^{\ell}}
    \bigg\{{ \hat{a}_{\ell}(\kap) \atop \hat{b}_{\ell}(\kap) }\bigg\}
   ,\qquad&
  \bigg\{{ \bcI_0(z) \atop \bcI_1(z) }\bigg\}
  &=
    \sum_{\ell=1}^{\infty}\frac{(-1)^{\ell}}{\ell!(8z)^{\ell}}
    \bigg\{{ \hat{a}_{\ell}'(\kap) \atop \hat{b}_{\ell}'(\kap) }\bigg\}
   \qquad&&
   (|\arg z|<\pi/2) ,
    \label{eq:cI_HAE}
   \\
  \bigg\{{ \cK_0(z) \atop \cK_1(z) }\bigg\}
  &=
    \sum_{\ell=0}^{\infty}\frac{1}{\ell!(8z)^{\ell}}
    \bigg\{{ \hat{a}_{\ell}(\kap) \atop \hat{b}_{\ell}(\kap) }\bigg\}
   ,\qquad&
  \bigg\{{ \bcK_0(z) \atop \bcK_1(z) }\bigg\}
  &=
    \sum_{\ell=1}^{\infty}\frac{1}{\ell!(8z)^{\ell}}
    \bigg\{{ \hat{a}_{\ell}'(\kap) \atop \hat{b}_{\ell}'(\kap) }\bigg\}
   \qquad&&
   (|\arg z|<3\pi/2) .
\end{alignat}
The condition that the series $\cI_{0}$ and $\cI_{1}$ converge is approximately given as
\begin{align}
  \bigg|\frac{\kap-1}{8z}\bigg|
  \lesssim 1
   \qquad\Lra\qquad
  \nu
  \lesssim (2z+1/4)^{1/2} .
\end{align}
We substitute Eqs.(\ref{eq:Inu_HAE}-\ref{eq:Knu_HAE}) into
the cross products (\ref{eq:pnu_IK}-\ref{eq:snu_IK})  of purely imaginary arguments,
\begin{align}
  w_c p_{\nu}(iz_b,iz_a)
  &=
    e_{+}\cI_0(z_b)\cK_0(z_a)
   -e_{-}\cK_0(z_b)\cI_0(z_a) ,
   \\
  w_c s_{\nu}(iz_b,iz_a)
  &=
    e_{+}\cI_1(z_b)\cK_1(z_a)
   -e_{-}\cK_1(z_b)\cI_1(z_a) ,
   \\
  -iw_c q_{\nu}(iz_b,iz_a)
  &=
    e_{+}\cI_0(z_b)\cK_1(z_a)
   +e_{-}\cK_0(z_b)\cI_1(z_a) ,
   \\
  iw_c r_{\nu}(iz_b,iz_a)
  &=
    e_{+}\cI_1(z_b)\cK_0(z_a)
   +e_{-}\cK_1(z_b)\cI_0(z_a) ,
\end{align}
where
\begin{align}
  e_{\pm}
  =e^{\pm(z_b-z_a)} ,
    \qquad
  w_c
  =-\pi(z_bz_a)^{1/2} ,
    \qquad
  \bw_c
  =\frac{w_c}{8\nu}
   \qquad
  (z_{a,b}\in\mathbb{R}) .
\end{align}
The derivative of the cross products with respect to the order $\nu$ is given as follows,
\begin{align}
  \bw_c\rd_{\nu}p_{\nu}(iz_b,iz_a)
  &=
    e_{+}\{\bcI_0(z_b)\cK_0(z_a)+\cI_0(z_b)\bcK_0(z_a)\}
   -e_{-}\{\bcK_0(z_b)\cI_0(z_a)+\cK_0(z_b)\bcI_0(z_a)\} ,
   \\
  \bw_c\rd_{\nu}s_{\nu}(iz_b,iz_a)
  &=
    e_{+}\{\bcI_1(z_b)\cK_1(z_a)+\cI_1(z_b)\bcK_1(z_a)\}
   -e_{-}\{\bcK_1(z_b)\cI_1(z_a)+\cK_1(z_b)\bcI_1(z_a)\} ,
   \\
  -i\bw_c\rd_{\nu}q_{\nu}(iz_b,iz_a)
  &=
    e_{+}\{\bcI_0(z_b)\cK_1(z_a)+\cI_0(z_b)\bcK_1(z_a)\}
   +e_{-}\{\bcK_0(z_b)\cI_1(z_a)+\cK_0(z_b)\bcI_1(z_a)\} ,
   \\
  i\bw_c\rd_{\nu}r_{\nu}(iz_b,iz_a)
  &=
    e_{+}\{\bcI_1(z_b)\cK_0(z_a)+\cI_1(z_b)\bcK_0(z_a)\}
   +e_{-}\{\bcK_1(z_b)\cI_0(z_a)+\cK_1(z_b)\bcI_0(z_a)\} .
\end{align}
When $z_b-z_a\gg1$, the Green functions, given by Eqs.(\ref{eq:mfGe}-\ref{eq:mfGb}) and
(\ref{eq:bmfG_pls}-\ref{eq:bmfG_mns}), behave as
\begin{align}
  d_{\rho}\mfG_{+}^n(r,r',\nu)
  &\simeq
     \theta(r-r')\bar{e}_{-}
     \frac{\cL_p(z_b,z)\cL_p(z',z_a)}{\cL_p(z_b,z_a)}
    +\theta(r'-r)\bar{e}_{+}
     \frac{\cL_p(z_b,z')\cL_p(z,z_a)}{\cL_p(z_b,z_a)} ,
   \\
  d_{\rho}\mfG_{-}^n(r,r',\nu)
  &\simeq
     \theta(r-r')\bar{e}_{-}
     \frac{\cL_r(z_b,z)\cL_q(z',z_a)}{\cL_s(z_b,z_a)}
    +\theta(r'-r)\bar{e}_{+}
     \frac{\cL_r(z_b,z')\cL_q(z,z_a)}{\cL_s(z_b,z_a)} ,
   \\
  d_{\rho}\bmfG_{+}^n(r,r',\nu)
  &\simeq
     \theta(r-r')\bar{e}_{-}
     \frac{\cL_q(z_b,z)\cL_r(z',z_a)}{\cL_p(z_b,z_a)}
    +\theta(r'-r)\bar{e}_{+}
     \frac{\cL_q(z_b,z')\cL_r(z,z_a)}{\cL_p(z_b,z_a)} ,
   \\
  d_{\rho}\bmfG_{-}^n(r,r',\nu)
  &\simeq
     \theta(r-r')\bar{e}_{-}
     \frac{\cL_s(z_b,z)\cL_s(z',z_a)}{\cL_s(z_b,z_a)}
    +\theta(r'-r)\bar{e}_{+}
     \frac{\cL_s(z_b,z')\cL_s(z,z_a)}{\cL_s(z_b,z_a)} ,
\end{align}
where
\begin{alignat}{2}
  \cL_p(z,z')
  &=\cI_0(z)\cK_0(z') ,
    \qquad&
  \cL_q(z,z')
  &=\cI_0(z)\cK_1(z') ,
   \\
  \cL_s(z,z')
  &=\cI_1(z)\cK_1(z') ,
    \qquad&
  \cL_r(z,z')
  &=\cI_1(z)\cK_0(z') .
\end{alignat}
$z$ and $z'$ are real for $k_r^n=i\bk_r^n\in i\mathbb{R}$ ($\bk_r^n\in\mathbb{R}$),
\begin{align}
  d_{\rho}
  =-\frac{2}{\rho}(zz')^{1/2} ,
    \qquad
  \bar{e}_{\pm}
  =e^{\pm(z-z')} ,
    \qquad
  z=\bk_r^nr ,
    \qquad
  z'=\bk_r^nr' ,
    \qquad
  z_{a,b}=\bk_r^nr_{a,b} .
\end{align}
%

\subsection{Uniform asymptotic expansion of the Bessel functions $J_{\nu}$ and $Y_{\nu}$}
\label{sec:UAE_JY_nu}

For $\nu\gg1$ ($\nu\in\mathbb{R}$), the uniform asymptotic expansions of
$(J_{\nu},Y_{\nu})$ and $(J_{\nu}',Y_{\nu}')$ are given by 9.3.35-9.3.36 and 9.3.43-9.3.44
in \cite{abramo_stegun}, which were shown by Olver \cite{olver_0},
\begin{align}
   \bigg\{{ J_{\nu}(\nu z) \atop -Y_{\nu}(\nu z) }\bigg\}
  &=\frac{2^{1/2}}{f(z)}
    \bigg[
       \frac{1}{\nu^{1/3}}\bigg\{{ \Ai(u) \atop \Bi(u) }\bigg\}
       \sum_{j=0}^{\infty}\frac{a_j(\zeta)}{\nu^{2j}}
      +\frac{1}{\nu^{5/3}}\bigg\{{ \Ai'(u) \atop \Bi'(u) }\bigg\}
       \sum_{j=0}^{\infty}\frac{b_j(\zeta)}{\nu^{2j}}
    \bigg] ,
  \label{eq:JY_uae}
   \\
  \bigg\{{ -J_{\nu}'(\nu z) \atop Y_{\nu}'(\nu z) }\bigg\}
  &=\frac{2^{1/2}f(z)}{z}
    \bigg[
      \frac{1}{\nu^{2/3}}\bigg\{{ \Ai'(u) \atop \Bi'(u) }\bigg\}
      \sum_{j=0}^{\infty}\frac{d_j(\zeta)}{\nu^{2j}}
     +\frac{1}{\nu^{4/3}}\bigg\{{ \Ai(u) \atop \Bi(u) }\bigg\}
      \sum_{j=0}^{\infty}\frac{c_j(\zeta)}{\nu^{2j}}
    \bigg] .
  \label{eq:dJY_uae}
\end{align}
Eqs.(\ref{eq:JY_uae}-\ref{eq:dJY_uae}) hold for $|\arg z|\leq\pi-\veps~(\veps>0)$.
$u$ and $f$ are given by
\begin{align}
  &
  u
  =\nu^{2/3}\zeta ,
    \qquad
  f(z)
  =\bigg(\frac{\eps_z}{\zeta}\bigg)^{1/4}
  =\psi^{-1/6} ,
    \qquad
  \eps_z
  =1-z^2 ,
    \qquad
  f^2\rd_{\zeta}
  =\nu\rd_{\nu} ,
   \label{eq:f_z_zeta}
   \\
  &
  \rd_z\zeta
  =-\frac{f^2}{z} ,
    \qquad
  \rd_{\nu}\zeta
  =\frac{f^2}{\nu} ,
    \qquad
  \rd_zf
  =\frac{f}{2z}\bigg(\frac{f^2}{2\zeta}-\frac{z^2}{\eps_z}\bigg) ,
    \qquad
  \frac{d\nu}{\nu}+\frac{dz}{z}
  =0 .
\end{align}
$z$ differs from that in appendices \ref{sec:AE_bignu}-\ref{sec:hankel}.
$\zeta$ is given by 9.3.38-9.3.39 in \cite{abramo_stegun},
\begin{alignat}{2}
  \eta_{+}(z)
  &=\frac{2}{3}\zeta^{3/2}
   =\log\bigg(\frac{1+\eps_z^{1/2}}{z}\bigg)-\eps_z^{1/2}
  \qquad&& (z<1,~\zeta>0) ,
  \label{eq:zeta_p}
   \\
  \eta_{-}(z)
  &=\frac{2}{3}(-\zeta)^{3/2}
  =(-\eps_z)^{1/2}-\cos^{-1}\frac{1}{z}
  \qquad&& (z>1,~\zeta<0) .
  \label{eq:zeta_m}
\end{alignat}
$\zeta=0$ at $z=1$ which is the transition point of the Bessel functions.
$\psi$ is expanded around $z=1$, according to Eqs.(5.22) and (5.24) in \cite{warnock_morton},
\begin{align}
  \psi
  &=\bigg(\frac{\zeta}{\eps_z}\bigg)^{3/2}
  =\frac{3}{2}\sum_{j=0}^{\infty}\frac{\eps_z^j}{2j+3} ,
    \qquad
  \lim_{z\to1}\psi
  =\frac{1}{2} ,
    \qquad
  \lim_{z\to1}\frac{\eps_z}{\zeta}
  =2^{2/3} ,
    \qquad
  \lim_{z\to1}\zeta
  =0 .
   \label{eq:psi}
\end{align}
$(2\psi)^{\pm2/3}$ and $(2\psi)^2$ are expanded around $z=1$,
\begin{align}
  (2\psi)^{2/3}
  &=1
   +\frac{2}{5}\eps_z
   +\frac{43}{5^2\cd7}\eps_z^2
   +\frac{2^3\cd173}{5^3\cd7\cd9}\eps_z^3
   +O(\eps_z^4) ,
  \label{eq:Sig}
   \\
  (2\psi)^{-2/3}
  &=1
   -\frac{2}{5}\eps_z
   -\frac{3}{5\cd7}\eps_z^2
   -\frac{2^2\cd17}{5^2\cd7\cd9}\eps_z^3
   +O(\eps_z^4) ,
  \label{rcprc_psi23}
\end{align}
and
\begin{align}
  &
  (2\psi)^2
  =\sum_{j=0}^{\infty}\alp_j\eps_z^j ,
    \qquad
  \alp_j
  =\sum_{k=0}^{j}\frac{3^2}{(2k+3)\{2(j-k)+3\}} :
    \qquad
   \alp_0=1 ,
     \qquad
   \alp_1=\frac{2\cd3}{5} ,
   \\
  &\qquad
   \alp_2=\frac{3\cd71}{5^2\cd7} ,
    \qquad
   \alp_3=\frac{2^2\cd31}{3\cd5\cd7} ,
    \qquad
   \alp_4=\frac{17\cd179}{5\cd7^2\cd11} ,
    \qquad
   \alp_5=\frac{2\cd2689}{5\cd7\cd11\cd13} .
\end{align}
Conversely, $z$ and $\eps_z$ have the following series representations with respect to
$\zeta$, according to Eq.(4.13) in \cite{olver_0} and p.421 in \cite{olver},
\begin{align}
  z
  &=
    1
   -\bzeta
   +\frac{3}{10}\bzeta^2
   +\frac{1}{350}\bzeta^3
   -\frac{479}{63000}\bzeta^4
   -\frac{20231}{8085000}\bzeta^5
   +O(\bzeta^6) ,
   \\
  \frac{\eps_z}{2\bzeta}
  &=
    1
   -\frac{4}{5}\bzeta
   +\frac{52}{175}\bzeta^2
   -\frac{272}{7875}\bzeta^3
   -\frac{18064}{3031875}\bzeta^4
   +O(\bzeta^5),
    \qquad
  \bzeta
  =\frac{\zeta}{2^{1/3}} .
\end{align}
Using Eqs.(\ref{eq:JY_uae}-\ref{eq:dJY_uae}), we expand
Eqs.(\ref{eq:pq_nu}-\ref{eq:sr_nu}) in the uniform asymptotic series,
\begin{alignat}{2}
  p_{\nu}(\nu z,\nu z')
  &=-\frac{2}{\nu^{2/3}f(z)f(z')}
     \sum_{j,l=0}^{\infty}\frac{A_{\nu}^{jl}(\zeta,\zeta')}{\nu^{2(j+l)}} ,
    \qquad
  &q_{\nu}(\nu z,\nu z')
  &=\frac{2f(z')}{\nu z'f(z)}
    \sum_{j,l=0}^{\infty}\frac{B_\nu^{jl}(\zeta,\zeta')}{\nu^{2(j+l)}} ,
  \label{eq:pnu_uae}
   \\
  s_{\nu}(\nu z,\nu z')
  &=-\frac{2f(z)f(z')}{\nu^{4/3}zz'}
    \sum_{j,l=0}^{\infty}\frac{D_{\nu}^{jl}(\zeta,\zeta')}{\nu^{2(j+l)}} ,
    \qquad
  &r_{\nu}(\nu z,\nu z')
  &=\frac{2f(z)}{\nu zf(z')}
    \sum_{j,l=0}^{\infty}\frac{C_{\nu}^{jl}(\zeta,\zeta')}{\nu^{2(j+l)}} .
  \label{eq:snu_uae}
\end{alignat}
$z'$ and $\zeta'$ are the variables which obey
Eqs.(\ref{eq:zeta_p}-\ref{eq:zeta_m}), similar to $z$ and $\zeta$.
$A_{\nu}^{jl}$, $B_{\nu}^{jl}$, $C_{\nu}^{jl}$ and $D_{\nu}^{jl}$
are given as
\begin{align}
  A_{\nu}^{jl}(\zeta,\zeta')
  &=a_j(\zeta)a_l(\zeta')\mfp(u,u')
   +\frac{a_j(\zeta)b_l(\zeta')}{\nu^{4/3}}\mfq(u,u')
   +\frac{b_j(\zeta)a_l(\zeta')}{\nu^{4/3}}\mfr(u,u')
   +\frac{b_j(\zeta)b_l(\zeta')}{\nu^{8/3}}\mfs(u,u') ,
   \label{eq:Pnu_jl}
   \\
  B_\nu^{jl}(\zeta,\zeta')
  &=a_j(\zeta)d_l(\zeta')\mfq(u,u')
   +\frac{a_j(\zeta)c_l(\zeta')}{\nu^{2/3}}\mfp(u,u')
   +\frac{b_j(\zeta)d_l(\zeta')}{\nu^{4/3}}\mfs(u,u')
   +\frac{b_j(\zeta)c_l(\zeta')}{\nu^2}\mfr(u,u') ,
   \\
  C_{\nu}^{jl}(\zeta,\zeta')
  &=d_j(\zeta)a_l(\zeta')\mfr(u,u')
   +\frac{c_j(\zeta)a_l(\zeta')}{\nu^{2/3}}\mfp(u,u') 
   +\frac{d_j(\zeta)b_l(\zeta')}{\nu^{4/3}}\mfs(u,u')
   +\frac{c_j(\zeta)b_l(\zeta')}{\nu^2}\mfq(u,u') ,
   \\
  D_{\nu}^{jl}(\zeta,\zeta')
  &=d_j(\zeta)d_l(\zeta')\mfs(u,u')
   +\frac{c_j(\zeta)d_l(\zeta')}{\nu^{2/3}}\mfq(u,u')
   +\frac{d_j(\zeta)c_l(\zeta')}{\nu^{2/3}}\mfr(u,u')
   +\frac{c_j(\zeta)c_l(\zeta')}{\nu^{4/3}}\mfp(u,u') .
  \label{eq:Snu_jl}
\end{align}
$\mfp$, $\mfq$, $\mfr$ and $\mfs$ are the cross products of the Airy functions
$\Ai$ and $\Bi$:
\begin{alignat}{2}
  \mfp(u,u')&=\Ai(u)\Bi(u')-\Bi(u)\Ai(u') ,
    \qquad&
  \mfq(u,u')&=\Ai(u)\Bi'(u')-\Bi(u)\Ai'(u') ,
  \label{eq:cp_airy_p}
  \\
  \mfs(u,u')&=\Ai'(u)\Bi'(u')-\Bi'(u)\Ai'(u') ,
    \qquad&
  \mfr(u,u')&=\Ai'(u)\Bi(u')-\Bi'(u)\Ai(u') .
  \label{eq:cp_airy_sr}
\end{alignat}
$\Ai'$ and $\Bi'$ are the derivatives of $\Ai$ and $\Bi$ with respect to the argument,
\begin{align}
  u=\nu^{2/3}\zeta ,
    \qquad
  u'=\nu^{2/3}\zeta' .
  \label{eq:u_up}
\end{align}
The coefficients $(a_j,b_j)$ and $(d_j,c_j)$ are the functions of $\zeta$,
given by 9.3.40 and 9.3.46 in \cite{abramo_stegun},
\begin{alignat}{3}
  a_0
  &=1
   ,\qquad&
  a_j(\zeta)
  &=\sum_{s=0}^{2j}\frac{\kap_s}{\zeta^{3s/2}}u_{2j-s}(t)
  ,\qquad&
  b_j(\zeta)
  &=-\zeta^{-1/2}\sum_{s=0}^{2j+1}
    \frac{\lambda_s}{\zeta^{3s/2}}u_{2j-s+1}(t) ,
  \label{eq:aj_bj}
   \\
  d_0
  &=1
   ,\qquad&
  d_j(\zeta)
  &=\sum_{s=0}^{2j}\frac{\lambda_s}{\zeta^{3s/2}}v_{2j-s}(t)
  ,\qquad&
  c_j(\zeta)
  &=-\zeta^{1/2}\sum_{s=0}^{2j+1}\frac{\kap_s}{\zeta^{3s/2}}v_{2j-s+1}(t) ,
  \label{eq:cj_dj}
\end{alignat}
where $t$ denotes the following variable which is a function of $\eps_z$ given by
Eq.(\ref{eq:f_z_zeta}),
\begin{align}
  t=\eps_z^{-1/2}
   \qquad\Lra\qquad
   z
   =\frac{(t^2-1)^{1/2}}{t}
   ,\qquad
  \frac{dt}{dz}
  =z\eps_z^{-3/2}
  =t^2(t^2-1)^{1/2} .
   \label{eq:t}
\end{align}
According to Eqs.(6.10) in \cite{olver_0}, Eqs.(\ref{eq:aj_bj}-\ref{eq:cj_dj}) have
the following relations,
\begin{alignat}{2}
  d_j(\zeta)
  &=a_j(\zeta)+(1-\delta_j^0)\{\chi(\zeta)+\rd_{\zeta}\}b_{j-1}(\zeta) ,
    \qquad&
  c_j(\zeta)
  &=\zeta b_j(\zeta)
  +\{\chi(\zeta)+\rd_{\zeta}\}a_j(\zeta) ,
   \label{eq:djcj_ajbj}
   \\
  d_j(\zeta)
  &=a_j(\zeta)
   +\frac{1-\delta_j^0}{\zeta}\{\chi(\zeta)-\rd_{\zeta}\}c_{j-1}(\zeta) ,
    \qquad&
  c_j(\zeta)
  &=\zeta b_j(\zeta)
   +\{\chi(\zeta)-\rd_{\zeta}\}d_j(\zeta) ,
   \label{eq:djcj_ajbj_1}
\end{alignat}
where $j\in\mathbb{Z}_0^{+}$.
$\delta_j^0$ is the Kronecker delta.
From Eqs.(\ref{eq:djcj_ajbj}-\ref{eq:djcj_ajbj_1}),
\begin{align}
  \{\chi(\zeta)-\rd_{\zeta}\}d_j(\zeta)
  =\{\chi(\zeta)+\rd_{\zeta}\}a_j(\zeta) ,
    \qquad
  \{\chi(\zeta)-\rd_{\zeta}\}c_j(\zeta)  
  =\zeta\{\chi(\zeta)+\rd_{\zeta}\}b_j(\zeta) .
\end{align}
$\chi$ is an odd function given by Eqs.(6.11) in \cite{olver_0},
\begin{align}
  \chi(\zeta)
  &=-\chi(-\zeta)
  =\nu\rd_{\nu}\bigg(\frac{1}{2e_z}\bigg)
  =\frac{1}{4}\bigg(\frac{1}{\zeta}+\frac{2z}{\eps_z}\cd\frac{dz}{d\zeta}\bigg)
  =\frac{1-2z^2\psi}{4\zeta}
  =
    \frac{3\eps_z}{2\zeta}\sum_{j=0}^{\infty}
    \frac{\eps_z^j}{(2j+3)(2j+5)} ,
   \label{eq:M}
   \\
  \chi(0)
  &=\frac{1}{5\cd2^{1/3}}
   ,\qquad
  \frac{dz}{d\zeta}
  =-\frac{z}{e_z}
   ,\qquad
  e_z
  =f^2
  =\nu\rd_{\nu}\zeta
  =\bigg(\frac{\eps_z}{\zeta}\bigg)^{1/2}
  =\frac{1}{\psi^{1/3}}
   ,\qquad
  \eps_z
  =1-z^2 .
\end{align}
The coefficients $\kap_s$ (``$\mu_s$'' in \cite{abramo_stegun}) and $\lambda_s$ in
Eqs.(\ref{eq:aj_bj}-\ref{eq:cj_dj}) are given by 9.3.41 in \cite{abramo_stegun},
\begin{align}
  &
  \bigg({\kap_0 \atop \lam_0}\bigg)=1
  ,\qquad
  \kap_s
  =-\frac{6s+1}{6s-1}\lam_s
  ,\qquad
  \lam_s
  =\frac{1}{s!}\prod_{\ell=1}^{2s}\frac{2(s+\ell)-1}{12}
   \quad(s\geq1):
  \label{eq:kap_lam}
   \\
  &
  \bigg({\kap_1 \atop \lam_1}\bigg)=\frac{1}{2^4\cd3}\bigg({-7 \atop 5}\bigg)
  ,\qquad
  \bigg({\kap_2 \atop \lam_2}\bigg)
  =\frac{5\cd7}{2^9\cd3^2}\bigg({-13 \atop 11}\bigg)
  ,\qquad
  \bigg({\kap_3 \atop \lam_3}\bigg)
  =\frac{5\cd7\cd11\cd13}{2^{13}\cd3^4}\bigg({-19 \atop 17}\bigg) .
\end{align}
$u_j$ and $v_j$, involved in Eqs.(\ref{eq:aj_bj}-\ref{eq:cj_dj}), are defined in
9.3.10 and 9.3.14 of \cite{abramo_stegun},
\begin{alignat}{3}
  u_0&=1 ,
   \qquad&
  u_{j+1}(t)
  &=\frac{1}{2}t^2(1-t^2)\rd_tu_j(t)+\frac{1}{8}\int_{0}^{t}(1-5t'^2)u_j(t')dt' ,
  \label{eq:uj}
   \\
  v_0&=1 ,
   \qquad&
  v_{j+1}(t)
  &=u_{j+1}(t)+t(t^2-1)\bigg(\frac{1}{2}+t\rd_t\bigg)u_{j}(t) .
  \label{eq:vj}
\end{alignat}
They are given by the following finite power series of $t$,
\begin{align}
  u_{j}(t)
   =\sum_{k=0}^{j}\mfa_k^{(j)}t^{2k+j} ,
    \qquad
  v_{j}(t)
   =\sum_{k=0}^{j}\mfb_k^{(j)}t^{2k+j} ,
    \qquad
  \mfa_0^{(0)}
  =\mfb_0^{(0)}
  =1 .
  \label{eq:uv}
\end{align}
The coefficient $\mfa_k^{(j)}$ is given by the following recurrence relations,
\begin{align}
  \mfa_0^{(j+1)}
  &=\frac{1}{8}f_0^{(j)}\mfa_0^{(j)}
   ,\qquad
  \mfa_{j+1}^{(j+1)}
  =-\frac{1}{8}g_j^{(j)}\mfa_j^{(j)}
   \qquad
   (j\in\mathbb{Z}_0^{+}) ,
    \label{eq:akj_rcrr}
   \\
  \mfa_{k+1}^{(j+1)}
  &=\frac{1}{8}
    \big\{
      \mfa_{k+1}^{(j)}f_{k+1}^{(j)}
     -\mfa_k^{(j)}g_k^{(j)}
    \big\}
   \qquad
   (0\leq k\leq j-1,~j\in\mathbb{N}) .
    \label{eq:akj_rcrr_0}
\end{align}
The coefficient $\mfb_k^{(j)}$ is given from $\mfa_k^{(j)}$ as follows,
\begin{align}
  \mfb_0^{(j+1)}
  &=
    \mfa_0^{(j+1)}
   -\frac{2j+1}{2}\mfa_0^{(j)}
   =\frac{\mfa_0^{(j)}}{8}
    \big\{
      f_0^{(j)}
     -4(2j+1)
    \big\}
   =-\frac{(2j+1)(2j+3)}{8(j+1)}\mfa_0^{(j)}
   \quad
  (j\in\mathbb{Z}_0^{+}) ,
   \\
  \mfb_{j+1}^{(j+1)}
  &=
    \mfa_{j+1}^{(j+1)}
   +\frac{6j+1}{2}\mfa_j^{(j)}
   =\frac{\mfa_j^{(j)}}{8}
    \big\{
      4(6j+1)
     -g_j^{(j)}
    \big\}
   =\frac{(6j+1)(6j+7)}{24(j+1)}\mfa_j^{(j)}
   \quad
  (j\in\mathbb{Z}_0^{+}) ,
   \\
  \mfb_{k+1}^{(j+1)}
  &=
     \mfa_{k+1}^{(j+1)}
    +\frac{4k+2j+1}{2}\mfa_k^{(j)}
    -\frac{4k+2j+5}{2}\mfa_{k+1}^{(j)}
   \qquad
  (0\leq k\leq j-1,~j\in\mathbb{N})
   \\
  &=
    \frac{\mfa_k^{(j)}}{8}
    \big\{
      4(4k+2j+1)
     -g_k^{(j)}
    \big\}
   -\frac{\mfa_{k+1}^{(j)}}{8}
    \big\{
      4(4k+2j+5)
     -f_{k+1}^{(j)}
    \big\}
    \label{eq:bkj_rcrr}
   \\
  &=\frac{4k+2j+7}{8(2k+j+3)}
    \big\{
      (4k+2j+1)\mfa_k^{(j)}
     -(4k+2j+5)\mfa_{k+1}^{(j)}
    \big\} .
\end{align}
$f_k^{(j)}$ and $g_k^{(j)}$ in Eqs.(\ref{eq:akj_rcrr}-\ref{eq:bkj_rcrr}) are given as
\begin{align}
  f_k^{(j)}
  &=\frac{(2m+1)^2}{m+1} ,
    \qquad
  g_k^{(j)}
  =\frac{(2m+1)(2m+5)}{m+3} ,
    \quad\text{where}\quad
  m
  =2k+j .
\end{align}
The explicit expression of $u_j$ is given by 9.3.9 in \cite{abramo_stegun} up to $j=4$,
\begin{align}
  u_1(t)
  &=-\frac{t}{2^3}\bigg(\frac{5}{3}t^2-1\bigg)
   ,\qquad
  u_2(t)
  =\frac{7}{2^7}t^2\bigg(\frac{5\cd11}{9}t^4-\frac{2\cd11}{3}t^2+\frac{9}{7}\bigg) ,
   \label{u1_u2}
   \\
  u_3(t)
  &=-\frac{13}{2^{10}}t^3
    \bigg(
       \frac{5\cd7\cd11\cd17}{3^4}t^6
      -\frac{7\cd11\cd17}{9}t^4
      +\frac{3^3\cd13}{5}t^2
      -\frac{3\cd5^2}{13}
    \bigg) ,
   \\
  u_4(t)
  &=\frac{11}{2^{15}}t^4
   \bigg\{
     7\cd13\cd17\cd19
     \bigg(
       \frac{5\cd23}{3^5}t^8
      -\frac{2^2\cd23}{3^4}t^6
      +\frac{2\cd53}{7\cd17}t^4
     \bigg)
    -\frac{2^2\cd8803}{5}t^2
    +\frac{3\cd5^2\cd7^2}{11}
   \bigg\} ,
   \\
  u_5(t)
  &=-\frac{7\cd11\cd13\cd17}{2^{18}}t^5\hu_5(t)
    ,\qquad
  u_6(t)
  =\frac{1062347}{2^{21}}t^6\hu_6(t)
   ,\qquad
  u_7(t)
  =-\frac{1062347}{2^{24}}t^7\hu_7(t) ,
\end{align}
where
\begin{align}
  \hu_5(t)
  &=\frac{5\cd19}{3^2}
    \bigg\{
      \frac{5\cd23\cd29}{3^4}(t^{10}-3t^8)
     +2\cd67t^6
    \bigg\}
    -\frac{2\cd23\cd641}{3^2\cd5}t^4
    +\frac{23\cd47\cd283}{5\cd7^2\cd11}t^2
    -\frac{3^5\cd5\cd7}{11\cd13\cd17} ,
   \\
  \hu_6(t)
  &=
    \frac{1101275}{729}t^{10}
    \bigg(\frac{5}{18}t^2-1\bigg)
    +\frac{t^4}{6}
     \bigg(
       12555 t^4
      -\frac{124292}{15} t^2
      +\frac{14924597}{5775} 
     \bigg)
    \nonumber\\&\quad
    -\frac{45}{1062347}
     \bigg(
       \frac{8642131}{7} t^2
      -\frac{53361}{2}
     \bigg) ,
   \\
  \hu_7(t)
  &=
    \frac{1101275}{729}t^{10}
    \bigg(
      \frac{7585}{378} t^4
     -\frac{1517}{18} t^2
     +\frac{69597}{490}
    \bigg)
   -\frac{t^6}{18}
    \bigg(\frac{9939721}{3} t^2 -\frac{16590407}{11}\bigg)
    \nonumber\\&\quad
   -\frac{3118855661}{164450} t^4
   +\frac{1493441525}{874874} t^2
   -\frac{405405}{14858} .
\end{align}
Similarly, $v_j$ is given by 9.3.13 in \cite{abramo_stegun} up to $j=3$,
\begin{align}
  v_1(t)
  &=\frac{t}{2^3}\bigg(\frac{7}{3}t^2-3\bigg)
   ,\qquad
   v_2(t)
  =-\frac{5}{2^7}t^2
   \bigg(\frac{7\cd13}{9}t^4 -\frac{6\cd11}{5}t^2 +3\bigg) ,
   \\
  v_3(t)
  &=\frac{11}{2^{10}}t^3
    \bigg(
       \frac{5\cd7\cd13\cd19}{3^4}t^6
      -\frac{5\cd7\cd17}{3}t^4
      +\frac{3\cd13^2}{5}t^2
      -\frac{3\cd5\cd7}{11}
    \bigg) ,
   \\
  v_4(t)
  &=-\frac{13}{2^{15}}t^4
    \bigg\{
      \frac{7\cd11\cd17\cd19}{3}
      \bigg(
        \frac{5^3}{3^4}t^8
       -\frac{2^2\cd7\cd23}{3^2\cd19}t^6
       +\frac{2\cd53}{5\cd7}t^4
      \bigg)
     -\frac{2^2\cd8803}{5}t^2
     +\frac{3^3\cd5^2\cd7}{13}
    \bigg\} ,
   \\
  v_5(t)
  &=\frac{7\cd11\cd17\cd23}{2^{18}}t^5\hv_5(t)
   ,\qquad
  v_6(t)
  =-\frac{13\cd17}{2^{21}}t^6\hv_6(t)
   ,\qquad
  v_7(t)
  =\frac{1062347}{2^{24}}t^7\hv_7(t) ,
   \label{v5}
\end{align}
where
\begin{align}
  \hv_5(t)
  &=
    \frac{13\cd19}{3^2}
    \bigg(
       \frac{5^2\cd31}{3^4}t^{10}
      -29t^8
      +\frac{2\cd5\cd67}{3\cd7}t^6
      -\frac{2\cd641}{5\cd17}t^4
    \bigg)
    +\frac{3\cd47\cd283}{7^2\cd11}t^2
    -\frac{3^3\cd5\cd7}{17\cd23} ,
   \\
  \hv_6(t)
  &=
    \frac{4879105}{9} t^6
    \bigg(
       \frac{5735}{1458}t^6
      -\frac{385}{27} t^4
      +\frac{279}{14} t^2
      -\frac{386}{29} 
    \bigg)
   +t^2
    \bigg(
      \frac{343265731}{150}t^2
     -\frac{25926393}{91}
    \bigg)
   +\frac{218295}{34} ,
   \\
  \hv_7(t)
  &=
   \frac{1101275}{729}t^{10}
   \bigg(
     \frac{7955}{378} t^4
    -\frac{533}{6} t^2
    +\frac{2109}{14} 
   \bigg)
   -\frac{t^6}{18}
    \bigg(
       \frac{10625219}{3}t^2
      -\frac{447940989}{275}
    \bigg)
    \nonumber\\&\quad
   -\frac{1}{1062347}
    \bigg(
      \frac{23169978705569}{1050} t^4
     -\frac{28375388975}{14} t^2
     +\frac{66891825}{2}
    \bigg) .
\end{align}
$b_0$ is given by 9.3.42 in \cite{abramo_stegun} and Eq.(6.7) in \cite{olver_0},
\begin{align}
  b_0(\zeta)
  &=\frac{1}{8\zeta^2}
    \bigg\{
        \psi(z)\bigg(\frac{5}{3}-\eps_z\bigg)
       -\frac{5}{6}
    \bigg\}
   =\frac{\eps_z^2}{4\zeta^2}\sum_{j=0}^{\infty}\frac{(j+1)\eps_z^j}{(2j+5)(2j+7)} ,
   \\
  c_0(\zeta)
  &=\frac{1}{8\zeta}
    \bigg\{
       \frac{7}{6}
      -\psi(z)\bigg(\frac{7}{3}-3\eps_z\bigg)
    \bigg\}
   =\frac{\eps_z}{4\zeta}\sum_{j=0}^{\infty}\frac{(j+6)\eps_z^j}{(2j+3)(2j+5)} ,
   \label{eq:c0_series}
\end{align}
where $\eps_z=1-z^2=t^{-2}$ as in Eq.(\ref{eq:t}).
$\psi$ is given by Eq.(\ref{eq:psi}).
$a_1$ and $d_1$ are given by
\begin{align}
  a_1(\zeta)
  &=u_2(t)
   +\frac{\kap_1}{\zeta^{3/2}}u_1(t)
   +\frac{\kap_2}{\zeta^{3}} ,
    \qquad
  d_1(\zeta)
  =v_2(t)
   +\frac{\lam_1}{\zeta^{3/2}}v_1(t)
   +\frac{\lam_2}{\zeta^3} .
   \label{eq:a1_d1}
\end{align}
Substituting $(\kap_s,\lam_s)$ and $(u_j,v_j)$ into Eqs.(\ref{eq:a1_d1}), we rearrange them,
\begin{align}
  a_1(\zeta)
  &=\frac{7}{2^7}
   \bigg\{
     -\frac{5\cd13}{6^2\zeta^3}
     +\frac{t}{3\zeta^{3/2}}\bigg(\frac{5}{3}t^2-1\bigg)
     +t^2\bigg(\frac{5\cd11}{9}t^4-\frac{2\cd11}{3}t^2+\frac{9}{7}\bigg)
   \bigg\}
   \\
  &=\frac{7}{2^7\zeta^3}
   \bigg\{
     -\frac{5\cd13}{6^2}
     +\frac{\psi(z)}{3}\bigg(\frac{5}{3}-\eps_z\bigg)
     +\psi^2(z)\bigg(\frac{5\cd11}{9}-\frac{2\cd11}{3}\eps_z+\frac{9}{7}\eps_z^2\bigg)
   \bigg\} ,
   \\
  d_1(\zeta)
  &=\frac{5}{2^7}
    \bigg\{
       \frac{7\cd11}{6^2\zeta^3}
      +\frac{t}{\zeta^{3/2}}\bigg(\frac{7}{9}t^2-1\bigg)
      -t^2\bigg(\frac{7\cd13}{9}t^4-\frac{6\cd11}{5}t^2+3\bigg)
    \bigg\}
   \\
  &=\frac{5}{2^7\zeta^3}
    \bigg\{
       \frac{7\cd11}{6^2}
      +\psi(z)\bigg(\frac{7}{9}-\eps_z\bigg)
      -\psi^2(z)\bigg(\frac{7\cd13}{9}-\frac{6\cd11}{5}\eps_z+3\eps_z^2\bigg)
    \bigg\} .
\end{align}
We describe the numerical method to compute the coefficients $(a_j,b_j,d_j,c_j)$
in appendix \ref{sec:uae_coeff}.
It is not easy to compute the values of these coefficients with a good accuracy
because of the following reasons.
At first, they have terms which diverge at $\zeta=0$ ($z=1$) even though the value of
the coefficient is finite, \ie, the divergent terms cancel out in them.
Second, a digit loss can happen in computing $u_{\ell}$ and $v_{\ell}$ given by
Eqs.(\ref{eq:uv}).

\subsection{Identities involving the cross products of the Bessel functions}

We show some useful identities which involve the Bessel functions and their cross products
(\ref{eq:pq_nu}-\ref{eq:sr_nu}),
\begin{align}
  C_{\nu}(\hr)p_{\nu}(\hr_b,\hr_a)
  &=C_{\nu}(\hr_a)p_{\nu}(\hr_b,\hr)+C_{\nu}(\hr_b)p_{\nu}(\hr,\hr_a) ,
  \label{eq:Cnu_pnu_ba}
  \\
  C_{\nu}(\hr)s_{\nu}(\hr_b,\hr_a)
  &=C_{\nu}'(\hr_a)r_{\nu}(\hr_b,\hr)+C_{\nu}'(\hr_b)q_{\nu}(\hr,\hr_a) ,
  \label{eq:Cnu_snu_ba}
  \\
  C_{\nu}'(\hr)p_{\nu}(\hr_b,\hr_a)
  &=C_{\nu}(\hr_a)q_{\nu}(\hr_b,\hr)+C_{\nu}(\hr_b)r_{\nu}(\hr,\hr_a) ,
  \label{eq:Cnu_prm_pnu_ba}
  \\
  C_{\nu}'(\hr)s_{\nu}(\hr_b,\hr_a)
  &=C_{\nu}'(\hr_a)s_{\nu}(\hr_b,\hr)+C_{\nu}'(\hr_b)s_{\nu}(\hr,\hr_a) ,
  \label{eq:Cnu_prm_snu_ba}
\end{align}
where $C_{\nu}$ represents $J_{\nu}$ and $Y_{\nu}$, and
$C_{\nu}'(\hr)=\rd_{\hr}C_{\nu}(\hr)$.
We use Eqs.(\ref{eq:Cnu_pnu_ba}-\ref{eq:Cnu_prm_snu_ba}) in rewriting the coefficients
given by Eqs.(\ref{eq:bara_pm}-\ref{eq:barq_pm}) in appendix \ref{sec:we_xs}.
From Eqs.(\ref{eq:Cnu_pnu_ba}-\ref{eq:Cnu_prm_snu_ba}), we get
\begin{align}
  p_\nu(\hr,\hr')
  =\frac{p_\nu(\hr_b,\hr')p_\nu(\hr,\hr_a)-p_\nu(\hr_b,\hr)p_\nu(\hr',\hr_a)}
        {p_\nu(\hr_b,\hr_a)}
  =\frac{r_\nu(\hr_b,\hr')q_\nu(\hr,\hr_a)-r_\nu(\hr_b,\hr)q_\nu(\hr',\hr_a)}
        {s_\nu(\hr_b,\hr_a)} ,
  \label{eq:pppp_rqrq}
   \\
  q_\nu(\hr,\hr')
  =\frac{q_\nu(\hr_b,\hr')p_\nu(\hr,\hr_a)-p_\nu(\hr_b,\hr)r_\nu(\hr',\hr_a)}
        {p_\nu(\hr_b,\hr_a)}
  =\frac{s_\nu(\hr_b,\hr')q_\nu(\hr,\hr_a)-r_\nu(\hr_b,\hr)s_\nu(\hr',\hr_a)}
        {s_\nu(\hr_b,\hr_a)} ,
  \label{eq:pqqp_rssq}
   \\
  r_\nu(\hr,\hr')
  =\frac{p_\nu(\hr_b,\hr')r_\nu(\hr,\hr_a)-q_\nu(\hr_b,\hr)p_\nu(\hr',\hr_a)}
       {p_\nu(\hr_b,\hr_a)}
  =\frac{r_\nu(\hr_b,\hr')s_\nu(\hr,\hr_a)-s_\nu(\hr_b,\hr)q_\nu(\hr',\hr_a)}
        {s_\nu(\hr_b,\hr_a)} ,
  \label{eq:qppr_sqrs_0212}
   \\
  s_\nu(\hr,\hr')
  =\frac{q_\nu(\hr_b,\hr')r_\nu(\hr,\hr_a)-q_\nu(\hr_b,\hr)r_\nu(\hr',\hr_a)}
        {p_\nu(\hr_b,\hr_a)}
  =\frac{s_\nu(\hr_b,\hr')s_\nu(\hr,\hr_a)-s_\nu(\hr_b,\hr)s_\nu(\hr',\hr_a)}
        {s_\nu(\hr_b,\hr_a)} .
  \label{eq:qrqr_ssss}
\end{align}
As described in Eqs.(\ref{eq:pnu_poles}-\ref{eq:snu_poles}),
$p_{\nu}(\hr_b,\hr_a)$ and $s_{\nu}(\hr_b,\hr_a)$ become zero at
$\nu=\nu_m^n$ and $\mu_m^n$ respectively.
When $\nu=\nu_m^n$ or $\mu_m^n$,
the following relations hold for $\forall\hr$ and $\forall\hr'$,
according to Eqs.(\ref{eq:pppp_rqrq}) and (\ref{eq:qrqr_ssss}), 
\begin{alignat}{4}
  p_\nu(\hr_b,\hr')p_\nu(\hr,\hr_a)
  &=p_\nu(\hr_b,\hr)p_\nu(\hr',\hr_a)
   ,\qquad&
  q_\nu(\hr_b,\hr')r_\nu(\hr,\hr_a)
  &=q_\nu(\hr_b,\hr)r_\nu(\hr',\hr_a)
   &&\qquad
  (\nu=\nu_m^n) ,
   \label{eq:nu_mn_rels}
   \\
  s_\nu(\hr_b,\hr')s_\nu(\hr,\hr_a)
  &=s_\nu(\hr_b,\hr)s_\nu(\hr',\hr_a)
   ,\qquad&
  r_\nu(\hr_b,\hr')q_\nu(\hr,\hr_a)
  &=r_\nu(\hr_b,\hr)q_\nu(\hr',\hr_a)
   &&\qquad
  (\nu=\mu_m^n) .
   \label{eq:mu_mn_rels}
\end{alignat}
$q_{\nu}(\hr_b,\hr_a)$ and $r_{\nu}(\hr_b,\hr_a)$ for $\nu=\nu_m^n$ and $\mu_m^n$ are
rewritten as follows,
\begin{alignat}{4}
  \frac{\pi\hr_a}{2}q_{\nu}(\hr_b,\hr_a)
  &=\frac{J_{\nu}(\hr_b)}{J_{\nu}(\hr_a)}
   =\frac{Y_{\nu}(\hr_b)}{Y_{\nu}(\hr_a)}
   =\frac{F_{\nu}(\hr_b)}{F_{\nu}(\hr_a)}
   =\frac{G_{\nu}(\hr_b)}{G_{\nu}(\hr_a)}
   &&\qquad
  (\nu=\nu_m^n) ,
   \label{eq:qr_p0}
   \\
  -\frac{\pi\hr_b}{2}r_{\nu}(\hr_b,\hr_a)
  &=\frac{J_{\nu}(\hr_a)}{J_{\nu}(\hr_b)}
   =\frac{Y_{\nu}(\hr_a)}{Y_{\nu}(\hr_b)}
   =\frac{F_{\nu}(\hr_a)}{F_{\nu}(\hr_b)}
   =\frac{G_{\nu}(\hr_a)}{G_{\nu}(\hr_b)}
   &&\qquad
  (\nu=\nu_m^n) ,
   \\
  \frac{\pi\hr_b}{2}q_{\nu}(\hr_b,\hr_a)
  &=\frac{J_{\nu}'(\hr_a)}{J_{\nu}'(\hr_b)}
   =\frac{Y_{\nu}'(\hr_a)}{Y_{\nu}'(\hr_b)}
   =\frac{F_{\nu}'(\hr_a)}{F_{\nu}'(\hr_b)}
   =\frac{G_{\nu}'(\hr_a)}{G_{\nu}'(\hr_b)}
   &&\qquad
  (\nu=\mu_m^n) ,
   \\
  -\frac{\pi\hr_a}{2}r_{\nu}(\hr_b,\hr_a)
  &=\frac{J_{\nu}'(\hr_b)}{J_{\nu}'(\hr_a)}
   =\frac{Y_{\nu}'(\hr_b)}{Y_{\nu}'(\hr_a)}
   =\frac{F_{\nu}'(\hr_b)}{F_{\nu}'(\hr_a)}
   =\frac{G_{\nu}'(\hr_b)}{G_{\nu}'(\hr_a)}
   &&\qquad
  (\nu=\mu_m^n) .
   \label{eq:qr_s0}
\end{alignat}
$F_{\nu}$ and $G_{\nu}$ are the Bessel functions defined in Eq.(\ref{eq:FG_JY}).
We use Eqs.(\ref{eq:nu_mn_rels}-\ref{eq:qr_s0}) in rewriting the radial eigenfunctions
$\cR_{\pm}^{mn}$ given by Eqs.(\ref{eq:Ven}-\ref{eq:mfRm}) and
(\ref{eq:cRp_ba}-\ref{eq:cRm_ba}).
According to Eqs.(\ref{eq:qr_p0}-\ref{eq:qr_s0}),
the following relation holds at the poles,
\begin{align}
  \hr_b\hr_aq_{\nu}(\hr_b,\hr_a)r_{\nu}(\hr_b,\hr_a)
  =-(2/\pi)^2
   \qquad\text{where}\quad
  \nu=(\nu_m^n,\mu_m^n) .
    \label{eq:qr_relat}
\end{align}
We can also get Eq.(\ref{eq:qr_relat}) from Eqs.(\ref{eq:nu_mn_rels}-\ref{eq:mu_mn_rels})
and the Wronskian given by the last equation of (\ref{eq:JY}).

We get the following relations from Eqs.(\ref{eq:Cnu_snu_ba}-\ref{eq:Cnu_prm_pnu_ba}),
\begin{align}
  s_{\nu}(\hr_b,\hr_a)\frac{J_{\nu}(\hr_a)}{J_{\nu}'(\hr_a)}
  -p_{\nu}(\hr_b,\hr_a)\frac{J_{\nu}'(\hr_b)}{J_{\nu}(\hr_b)}
  &=\frac{2}{\pi}
    \bigg\{
       \frac{1}{\hr_a}\frac{J_{\nu}'(\hr_b)}{J_{\nu}'(\hr_a)}
      -\frac{1}{\hr_b}\frac{J_{\nu}(\hr_a)}{J_{\nu}(\hr_b)}
    \bigg\} ,
   \label{eq:fml_ps_b}
   \\
  s_{\nu}(\hr_b,\hr_a)\frac{J_{\nu}(\hr_b)}{J_{\nu}'(\hr_b)}
  -p_{\nu}(\hr_b,\hr_a)\frac{J_{\nu}'(\hr_a)}{J_{\nu}(\hr_a)}
  &=\frac{2}{\pi}
    \bigg\{
       \frac{1}{\hr_a}\frac{J_{\nu}(\hr_b)}{J_{\nu}(\hr_a)}
      -\frac{1}{\hr_b}\frac{J_{\nu}'(\hr_a)}{J_{\nu}'(\hr_b)}
    \bigg\} .
   \label{eq:fml_ps_a}
\end{align}
From Eqs.(\ref{eq:fml_ps_b}-\ref{eq:fml_ps_a}), we get
\begin{align}
  s_{\nu}(\hr_b,\hr_a)
  =
   \frac{J_{\nu}'(\hr_b)J_{\nu}'(\hr_a)}{J_{\nu}(\hr_b)J_{\nu}(\hr_a)}p_{\nu}(\hr_b,\hr_a)
  +\frac{2}{\pi}
   \bigg\{
      \frac{1}{\hr_a}\frac{J_{\nu}'(\hr_b)}{J_{\nu}(\hr_a)}
     -\frac{1}{\hr_b}\frac{J_{\nu}'(\hr_a)}{J_{\nu}(\hr_b)}
   \bigg\} .
   \label{eq:pnu_snu_id}
\end{align}
When $\nu=\mu_m^n$, Eq.(\ref{eq:pnu_snu_id}) becomes
\begin{align}
  [p_{\nu}(\hr_b,\hr_a)]_{\nu=\mu_m^n}
  =\frac{2}{\pi}
   \bigg[
      \frac{1}{\hr_b}\frac{J_{\nu}(\hr_a)}{J_{\nu}'(\hr_b)}
     -\frac{1}{\hr_a}\frac{J_{\nu}(\hr_b)}{J_{\nu}'(\hr_a)}
   \bigg]_{\nu=\mu_m^n} .
\end{align}
Moreover, since both $\mu_m^n/\rho$ and $\nu_m^n/\rho$ go to $k_s^{mn}$ in
the limit of $\rho\to\infty$ as shown in Eq.(\ref{eq:lim_nu}),
$p_{\nu}(\hr_b,\hr_a)$ at $\nu=\mu_m^n$ goes to zero in this limit,
\begin{align}
  [\hr_bJ_{\nu}'(\hr_b)J_{\nu}(\hr_b)]_{\nu=\mu_m^n}
  \simeq[\hr_aJ_{\nu}'(\hr_a)J_{\nu}(\hr_a)]_{\nu=\mu_m^n}
   \qquad (\nu\gg1) .
\end{align}
%

\subsection{Recurrence relations of the cross products of the Bessel functions}
\label{sec:tnu_recurr}

According to 9.1.27 (p.361) in \cite{abramo_stegun},
the Bessel functions satisfy the following recurrence relations,
\begin{alignat}{2}
  \frac{2\nu}{\hr}C_{\nu}
  &=C_{\nu-1}+C_{\nu+1} ,
    \qquad&
  C_{\nu}'
  &=C_{\nu-1}-\frac{\nu}{\hr}C_{\nu} ,
   \\
  2C_{\nu}'
  &=C_{\nu-1}-C_{\nu+1} ,
    \qquad&
  C_{\nu}'
  &=-C_{\nu+1}+\frac{\nu}{\hr}C_{\nu} ,
\end{alignat}
where $C_{\nu}(\hr)$ represents an arbitrary Bessel function.
According to 9.1.33 and 9.1.34 in \cite{abramo_stegun},
the cross products $t_{\nu}(\hr_b,\hr_a)=\{p_{\nu},q_{\nu},r_{\nu},s_{\nu}\}$
satisfy the following recurrence relations with respect to the order,
\begin{alignat}{2}
  p_{\nu-1}+p_{\nu+1}
  &=
   2s_{\nu}+\frac{2\nu^2}{\hr_b\hr_a}p_{\nu} ,
    \qquad&
  q_{\nu+1}
  +r_{\nu}
  &=
    \frac{\nu}{\hr_b}p_{\nu}
   -\frac{\nu+1}{\hr_a}p_{\nu+1} ,
   \label{eq:pp_qr_rec}
   \\
  p_{\nu-1}-p_{\nu+1}
  &=
   \frac{2\nu}{\hr_b}q_{\nu}+\frac{2\nu}{\hr_a}r_{\nu} ,
    \qquad&
  r_{\nu+1}
  +q_{\nu}
  &=
    \frac{\nu}{\hr_a}p_{\nu}
   -\frac{\nu+1}{\hr_b}p_{\nu+1} ,
\end{alignat}
%
%
\begin{align}
  p_{\nu}s_{\nu}-q_{\nu}r_{\nu}
  &=\frac{4}{\pi^2\hr_b\hr_a} .
   \label{eq:pq_rec}
\end{align}
All the cross products $t_{\nu}$ and $t_{\nu\pm1}$ have the arguments
$(\hr_b,\hr_a)$ in Eqs.(\ref{eq:pp_qr_rec}-\ref{eq:s_mns1_rec}).
We can rewrite $t_{\nu\pm1}$ in terms of $t_{\nu}$ as follows,
according to Eqs.(\ref{eq:pp_qr_rec}-\ref{eq:pq_rec}),
\begin{align}
  p_{\nu\pm1}
  &=
    s_{\nu}
   \mp\frac{\nu}{\hr_b}q_{\nu}
   \mp\frac{\nu}{\hr_a}r_{\nu}
   +\frac{\nu^2}{\hr_b\hr_a}p_{\nu} ,
   \\
  q_{\nu\pm1}
  &=
   -\frac{1\pm\nu}{\hr_a}\Big(s_{\nu}\mp\frac{\nu}{\hr_b}q_{\nu}\Big)
   -\bigg\{1-\frac{\nu(\nu\pm1)}{\hr_a^2}\bigg\}\Big(r_{\nu}\mp\frac{\nu}{\hr_b}p_{\nu}\Big),
   \\
  r_{\nu\pm1}
  &=
   -\frac{1\pm\nu}{\hr_b}\Big(s_{\nu}\mp\frac{\nu}{\hr_a}r_{\nu}\Big)
   -\bigg\{1-\frac{\nu(\nu\pm1)}{\hr_b^2}\bigg\}\Big(q_{\nu}\mp\frac{\nu}{\hr_a}p_{\nu}\Big),
   \\
  s_{\nu\pm1}
  &=
    \bigg\{1-\frac{\nu(\nu\pm1)}{\hr_b^2}\bigg\}
    \bigg\{1-\frac{\nu(\nu\pm1)}{\hr_a^2}\bigg\}p_{\nu}
   +\frac{1\pm\nu}{\hr_a}\bigg\{1-\frac{\nu(\nu\pm1)}{\hr_b^2}\bigg\}q_{\nu}
   \nonumber\\&\quad
   +\frac{1\pm\nu}{\hr_b}\bigg\{1-\frac{\nu(\nu\pm1)}{\hr_a^2}\bigg\}r_{\nu}
   +\frac{(\nu\pm1)^2}{\hr_b\hr_a}s_{\nu} .
   \label{eq:s_mns1_rec}
\end{align}
%


\section{Cross products of the modified Bessel functions}

\subsection{Modified Bessel functions of complex order}

The modified Bessel function $I_{\nu}$ has the ascending series representation
given by 9.6.10 in \cite{abramo_stegun},
\begin{align}
  I_{\nu}(z)
  =\frac{J_{\nu}(c_{\nu}z)}{c_{\nu}}
  =\sum_{j=0}^{\infty}\frac{(z/2)^{\nu+2j}}{j!\Gam(\nu+j+1)},
   \qquad
  c_{\nu}
  =
  \bigg\{
  \begin{array}{ll}
    e^{i\pi\nu/2}   & (-\pi<\arg z\leq\pi/2)  \\
    e^{-3i\pi\nu/2} & (-\pi/2<\arg z\leq\pi)
  \end{array} .
   \label{eq:Inu}
\end{align}
In general, $I_{i\bnu}(x)$ is complex for $\bnu\in\mathbb{R}$ and $x\in\mathbb{R}^{+}$.
In order to describe the field of synchrotron radiation using only the real Bessel functions,
instead of $I_\nu(z)$, we introduce $L_\nu(z)$ given by Eq.(2.2) in \cite{dunster},
\begin{align}
  L_\nu(z)=i\pi\frac{I_{-\nu}(z)+I_\nu(z)}{2\sin(\pi\nu)}
   \qquad
  (\nu\ne0),
   \qquad
  K_\nu(z)=\pi\frac{I_{-\nu}(z)-I_\nu(z)}{2\sin(\pi\nu)} ,
   \label{eq:def_KL_nu}
\end{align}
where $\nu\in\mathbb{C}$ and $z\in\mathbb{C}$ in general,
however, $L_\nu$ is not defined at $\nu=0$.
$L_{i\bnu}(x)$ is real for $\bnu\in\mathbb{R}$ and $x\in\mathbb{R}^{+}$,
similar to $K_{i\bnu}(x)\in\mathbb{R}$.
$K_\nu$ and $L_\nu$ are linearly independent of each other.
$I_{-\nu}\ne I_{\nu}$ unless $\nu=n\in\mathbb{Z}$.
On the other hand, $K_\nu$ and $L_\nu$ are respectively even and odd with respect to $\nu$, 
\ie,
\begin{align}
  K_{-\nu}(z)=K_{\nu}(z),
   \qquad
  L_{-\nu}(z)=-L_{\nu}(z) .
\end{align}
Their Wronskian is given by Eq.(2.10) in \cite{dunster},
\begin{align}
  W[K_\nu(z),L_\nu(z)]=\frac{i\pi}{z\sin(\pi\nu)}
   \qquad
  (\nu, z\in\mathbb{C},~\nu\ne0) .
  \label{eq:W_KL}
\end{align}
Eq.(\ref{eq:W_KL}) diverges at $\nu=0$ since $L_\nu$ diverges.
So, instead of $L_\nu$, we consider another modified Bessel function
which is linearly independent of $K_\nu$.
For example, the Bessel function $\bI_{\bnu}$ given by \S 10.45 of \cite{nist} is defined
without the factor $i\pi/\sin(\pi\nu)$ which $L_\nu$ has as in Eq.(\ref{eq:def_KL_nu}),
\begin{align}
  \bI_{\bnu}(x)
  =\Re I_{i\bnu}(x)
   \qquad
  (\bnu\in\mathbb{R},~x\in\mathbb{R}^{+}) .
\end{align}
Extending the order and argument of $\bI_{\bnu}(x)$ to complex variables,
we define $\bL_{\nu}(z)$ as
\begin{align}
  \bL_{\nu}(z)
  =\frac{I_{-\nu}(z)+I_\nu(z)}{2},
   \qquad
  W[K_\nu(z),\bL_\nu(z)]
  =W[K_\nu(z),I_\nu(z)]
  =\frac{1}{z}
   \quad
  (\nu, z\in\mathbb{C}) .
  \label{eq:W_KI}
\end{align}
$\bL_{\nu}$ is defined in full $\mathbb{C}$ including $\nu=0$ unlike $L_{\nu}$.
But $\bL_{\nu}$ does not have a symmetric expression with $K_{\nu}$.

The cross products $t_{\nu}(\hr,\hr')=\{p_{\nu},q_{\nu},r_{\nu},s_{\nu}\}$ are defined in 
Eqs.(\ref{eq:CP_pq}-\ref{eq:CP_sr}) using $J_{\nu}(\hr)$ and $Y_{\nu}(\hr)$.
If the arguments are purely imaginary $\hr=i\br\in i\mathbb{R}$,
$t_{\nu}(i\bbar,i\abar)$ is given by Eqs.(\ref{eq:CP_P_mod}-\ref{eq:CP_S_mod}) using
$I_{\nu}(\br)$ and $K_{\nu}(\br)$,
\begin{alignat}{2}
  -\frac{\pi}{2}p_\nu(i\bbar,i\abar)
  &=I_{\nu}(\bbar)K_{\nu}(\abar)
   -K_{\nu}(\bbar)I_{\nu}(\abar),
   \qquad&
  -i\frac{\pi}{2}q_\nu(i\bbar,i\abar)
  &=I_{\nu}(\bbar)K_{\nu}'(\abar)
   -K_{\nu}(\bbar)I_{\nu}'(\abar) ,
   \label{eq:pnu_IK}
   \\
  \frac{\pi}{2}s_\nu(i\bbar,i\abar)
  &=I_{\nu}'(\bbar)K_{\nu}'(\abar)
   -K_{\nu}'(\bbar)I_{\nu}'(\abar),
   \qquad&
  -i\frac{\pi}{2}r_\nu(i\bbar,i\abar)
  &=I_{\nu}'(\bbar)K_{\nu}(\abar)
   -K_{\nu}'(\bbar)I_{\nu}(\abar) .
   \label{eq:snu_IK}
\end{alignat}
We rewrite Eqs.(\ref{eq:pnu_IK}-\ref{eq:snu_IK}) in terms of $I_{\pm\nu}$
through $K_{\nu}$ given by Eq.(\ref{eq:def_KL_nu}),
\begin{alignat}{2}
  p_{\nu}(i\bbar,i\abar)
  &=\frac{
    I_{-\nu}(\bbar)I_{\nu}(\abar)
   -I_{\nu}(\bbar)I_{-\nu}(\abar)
    }{\sin(\pi\nu)},
    \qquad&
  iq_{\nu}(i\bbar,i\abar)
  &=\frac{
      I_{-\nu}(\bbar)I_{\nu}'(\abar)
     -I_{\nu}(\bbar)I_{-\nu}'(\abar)
    }{\sin(\pi\nu)} ,
    \label{eq:pnu_Ipm}
   \\
  -s_{\nu}(i\bbar,i\abar)
  &=\frac{
      I_{-\nu}'(\bbar)I_{\nu}'(\abar)
     -I_{\nu}'(\bbar)I_{-\nu}'(\abar)
    }{\sin(\pi\nu)},
    \qquad&
  ir_{\nu}(i\bbar,i\abar)
  &=\frac{
    I_{-\nu}'(\bbar)I_{\nu}(\abar)
   -I_{\nu}'(\bbar)I_{-\nu}(\abar)
    }{\sin(\pi\nu)} .
    \label{eq:snu_Ipm}
\end{alignat}
When $\nu=n\in\mathbb{Z}$, we cannot use Eqs.(\ref{eq:pnu_Ipm}-\ref{eq:snu_Ipm})
since $I_{-n}=I_n$.
We use Eqs.(\ref{eq:pnu_IK}-\ref{eq:snu_IK}) in calculating
$t_n$ and $[\rd_{\nu}t_{\nu}]_{\nu=n}$.
The derivatives of Eqs.(\ref{eq:pnu_IK}-\ref{eq:snu_IK}) with respect to $\nu$ are
given as follows,
\begin{align}
  -\frac{\pi}{2}\rd_{\nu}p_{\nu}(i\bbar,i\abar)
  &=\rd_{\nu}I_{\nu}(\bbar)K_{\nu}(\abar)
   -\rd_{\nu}K_{\nu}(\bbar)I_{\nu}(\abar)
   +I_{\nu}(\bbar)\rd_{\nu}K_{\nu}(\abar)
   -K_{\nu}(\bbar)\rd_{\nu}I_{\nu}(\abar) ,
   \\
  \frac{\pi}{2}\rd_{\nu}s_{\nu}(i\bbar,i\abar)
  &=\rd_{\nu}I_{\nu}'(\bbar)K_{\nu}'(\abar)
   -\rd_{\nu}K_{\nu}'(\bbar)I_{\nu}'(\abar)
   +I_{\nu}'(\bbar)\rd_{\nu}K_{\nu}'(\abar)
   -K_{\nu}'(\bbar)\rd_{\nu}I_{\nu}'(\abar) ,
   \\
  -i\frac{\pi}{2}\rd_{\nu}q_{\nu}(i\bbar,i\abar)
  &=\rd_{\nu}I_{\nu}(\bbar)K_{\nu}'(\abar)
   -\rd_{\nu}K_{\nu}(\bbar)I_{\nu}'(\abar)
   +I_{\nu}(\bbar)\rd_{\nu}K_{\nu}'(\abar)
   -K_{\nu}(\bbar)\rd_{\nu}I_{\nu}'(\abar) ,
   \\
  -i\frac{\pi}{2}\rd_{\nu}r_{\nu}(i\bbar,i\abar)
  &=\rd_{\nu}I_{\nu}'(\bbar)K_{\nu}(\abar)
   -\rd_{\nu}K_{\nu}'(\bbar)I_{\nu}(\abar)
   +I_{\nu}'(\bbar)\rd_{\nu}K_{\nu}(\abar)
   -K_{\nu}'(\bbar)\rd_{\nu}I_{\nu}(\abar) .
\end{align}
We rewrite Eqs.(\ref{eq:pnu_IK}-\ref{eq:snu_IK}) using $K_\nu$ and $L_\nu$,
\begin{alignat}{2}
  P_\nu(\bbar,\abar)
  &=\frac{\sin(\pi\nu)}{i\pi}\{L_\nu(\bbar)K_\nu(\abar)-K_\nu(\bbar)L_\nu(\abar)\}
  &&=-\frac{\pi}{2}p_\nu(i\bbar,i\abar) ,
  \label{eq:Pnu_ba}
  \\
  Q_\nu(\bbar,\abar)
  &=\frac{\sin(\pi\nu)}{i\pi}\{L_\nu(\bbar)K_\nu'(\abar)-K_\nu(\bbar)L_\nu'(\abar)\}
  &&=-i\frac{\pi}{2}q_\nu(i\bbar,i\abar) ,
  \\
  R_\nu(\bbar,\abar)
  &=\frac{\sin(\pi\nu)}{i\pi}\{L_\nu'(\bbar)K_\nu(\abar)-K_\nu'(\bbar)L_\nu(\abar)\}
  &&=-i\frac{\pi}{2}r_\nu(i\bbar,i\abar) ,
  \\
  S_\nu(\bbar,\abar)
  &=\frac{\sin(\pi\nu)}{i\pi}\{L_\nu'(\bbar)K_\nu'(\abar)-K_\nu'(\bbar)L_\nu'(\abar)\}
  &&=\frac{\pi}{2}s_\nu(i\bbar,i\abar) ,
  \label{eq:Snu_ba}
\end{alignat}
where $\nu\in\mathbb{C}$ and $\abar,\bbar\in\mathbb{C}$ in general.
Eqs.(\ref{eq:Pnu_ba}-\ref{eq:Snu_ba}) are also defined at $\nu=0$
even though $L_{\nu}$ itself is not defined at $\nu=0$.
As in Eqs.(\ref{eq:mfEBy_solution}) and (\ref{eq:mfBxEs}),
the radiation field in the Laplace domain is given in terms of the cross products of
the Bessel functions which are defined $\forall\nu\in\mathbb{C}$ including $\nu=0$.
Since the expressions of the fields do not have a single $L_{\nu}$ 
which is not involved in the cross products,
the singularity of $L_{\nu}$ at $\nu=0$ causes no problem in describing the field of
synchrotron radiation.

The uniform asymptotic expansion of $I_{\nu}(\br)$ and $K_{\nu}(\br)$ for $\nu\to\infty$ is
given by 9.7.7-9.7.10 in \cite{abramo_stegun},
\begin{align}
  \bigg\{{ I_{\nu}(\nu\bz) \atop I_{\nu}'(\nu\bz) }\bigg\}
  &=\frac{e^{\nu\xi}}{(2\pi\nu)^{1/2}}
    \bigg\{{ U_{+}(p)p^{1/2} \atop V_{+}(p)/(\bz p^{1/2}) }\bigg\} ,
    \qquad
  \bigg\{{ K_{\nu}(\nu\bz) \atop K_{\nu}'(\nu\bz) }\bigg\}
  =\Big(\frac{\pi}{2\nu}\Big)^{1/2}
    e^{-\nu\xi} \bigg\{{ U_{-}(p)p^{1/2} \atop -V_{-}(p)/(\bz p^{1/2}) }\bigg\} ,
    \label{eq:IKnu_uae}
\end{align}
where $|\arg\bz|<\pi/2$, \ie, $\Re\bz>0$.
$p$ and $\xi$ are functions of $\bz$ $(=\br/\nu)$,
\begin{align}
  p
  =(1+\bz^2)^{-1/2} ,
   \qquad
  \xi
  =\frac{1}{p}
  +\log\bigg(\frac{\bz}{1+1/p}\bigg),
   \qquad
  pd\xi
  =\frac{d\bz}{\bz}
  =-\frac{d\nu}{\nu} .
    \label{eq:p_xi}
\end{align}
$U_{\pm}$, $V_{\pm}$ and their derivatives with respect to $\nu$ are given as
\begin{align}
  \bigg\{{ U_{\pm}(p) \atop V_{\pm}(p) }\bigg\}
  =\sum_{j=0}^{\infty}\frac{(\pm1)^j}{\nu^j}
    \bigg\{{ u_j(p) \atop v_j(p) }\bigg\} ,
    \qquad
  \rd_{\nu}\bigg\{{ U_{\pm}(p) \atop V_{\pm}(p) }\bigg\}
  =\sum_{j=0}^{\infty}\frac{(\pm1)^j}{\nu^{j+1}}
    \bigg\{{ \chu_j(p) \atop \chv_j(p) }\bigg\} .
    \label{eq:UV_pm}
\end{align}
$u_j$ and $v_j$ are given by Eqs.(\ref{eq:uj}-\ref{eq:vj}).
$\chu_j$ and $\chv_j$ are given as follows,
\begin{alignat}{2}
   \bigg\{{ u_j(p) \atop v_j(p) }\bigg\}
  &=\sum_{k=0}^{j}  \bigg\{{ \mfa_k^{(j)} \atop \mfb_k^{(j)} }\bigg\} p^{2k+j} ,
    \qquad&
    \bigg\{{ \chu_j(p) \atop \chv_j(p) }\bigg\}
  &=\sum_{k=0}^{j}(2k\bz^2-j)
      \bigg\{{ \mfa_k^{(j)} \atop \mfb_k^{(j)} }\bigg\} p^{2(k+1)+j} .
    \label{eq:ch_uv}
\end{alignat}
The coefficients $\mfa_k^{(j)}$ and $\mfb_k^{(j)}$ are given by
Eqs.(\ref{eq:akj_rcrr}-\ref{eq:bkj_rcrr}).

\subsection{Modified Bessel functions of integer order}

We substitute $\nu=n\in\mathbb{Z}_0^{+}$ into Eq.(\ref{eq:Inu}) and its derivative
with respect to the argument $z$,
\begin{align}
  I_n(z)
  =\sum_{k=0}^{\infty}\frac{(z/2)^{n+2k}}{k!(n+k)!} ,
    \qquad
  I_n'(z)
  =\frac{1}{z}\sum_{k=0}^{\infty}\frac{n+2k}{k!(n+k)!}(z/2)^{n+2k}
    \quad
   (n\in\mathbb{Z}_0^{+}) .
   \label{eq:In}
\end{align}
The ascending series of $K_n$ is given by 9.6.11 (p.375) in \cite{abramo_stegun},
\begin{align}
  K_n(z)
  &=
    (-1)^{n+1}I_n(z)\log(z/2)
   +\frac{1}{2}\{(1-\delta_n^0)\bA_n(z)+(-1)^n\bB_n(z)\}
   \quad(n\in\mathbb{Z}_0^{+}) ,
   \label{eq:Kn}
   \\
  K_n'(z)
  &=
    (-1)^{n+1}\bigg\{I_n'(z)\log(z/2)+\frac{I_n(z)}{z}\bigg\}
   +\frac{1}{2}\{(1-\delta_n^0)\bA_n'(z)+(-1)^n\bB_n'(z)\} .
   \label{eq:Knp}
\end{align}
$\delta_n^0$ is the Kronecker delta.
$\bA_n$ and $\bB_n$ are defined respectively for $n\in\mathbb{N}$ and $\mathbb{Z}_0^{+}$
as follows,
\begin{align}
  \bA_n(z)
  &=\sum_{k=0}^{n-1}\frac{(n-k-1)!}{(-1)^k k!}(z/2)^{2k-n} ,
    \qquad
  \bB_n(z)
  =\sum_{k=0}^{\infty}\frac{\omg_{nk}}{k!(n+k)!}(z/2)^{2k+n} ,
   \\
  \bA_n'(z)
  &=\rd_z\bA_n(z)
   =\frac{1}{z}\sum_{k=0}^{n-1}\frac{(n-k-1)!(2k-n)}{(-1)^k k!}(z/2)^{2k-n}
   \qquad(n\in\mathbb{N}) ,
   \\
  \bB_n'(z)
  &=\rd_z\bB_n(z)
   =\frac{1}{z}\sum_{k=0}^{\infty}\omg_{nk}\frac{2k+n}{k!(n+k)!}(z/2)^{2k+n}
   \qquad(n\in\mathbb{Z}_0^{+}) .
\end{align}
According to 9.6.6 and 9.6.27 in \cite{abramo_stegun},
\begin{align}
  I_{-n}(z)
  =I_n(z) ,
    \qquad
  K_{-\nu}(z)
  =K_{\nu}(z) ,
    \qquad
  I_0'(z)
  =I_1(z) ,
    \qquad
  K_0'(z)
  =-K_1(z) .
\end{align}
The ascending series of the derivative of $I_{\nu}$ with respect to the order $\nu$ is 
given by 9.6.42 in \cite{abramo_stegun},
\begin{align}
  \rd_{\nu}I_{\nu}(z)
  &=I_{\nu}(z)\log(z/2)
   -\sum_{k=0}^{\infty}\frac{\psi(\nu+k+1)}{k!\Gam(\nu+k+1)}(z/2)^{\nu+2k} ,
    \label{eq:dInu}
   \\
  [\rd_{\nu}I_{\nu}(z)]_{\nu=n}
  &=I_n(z)\log(z/2)
   -\sum_{k=0}^{\infty}\frac{\psi(n+k+1)}{k!(n+k)!}(z/2)^{n+2k} .
    \label{eq:dIn_In}
\end{align}
According to 9.6.44-9.6.46 in \cite{abramo_stegun},
\begin{align}
  (-1)^n[\rd_{\nu}I_{\nu}(z)]_{\nu=n}
  &=-K_n(z)
   +\frac{n!}{2}\sum_{k=0}^{n-1}(-1)^k\frac{(z/2)^{k-n}}{(n-k)k!}I_k(z) ,
    \label{eq:dIn}
   \\
  [\rd_{\nu}K_{\nu}(z)]_{\nu=n}
  &=\frac{n!}{2}\sum_{k=0}^{n-1}\frac{(z/2)^{k-n}}{(n-k)k!}K_k(z) .
    \label{eq:dKn}
\end{align}
Differentiating Eqs.(\ref{eq:dInu}-\ref{eq:dKn}) with respect to the argument $z$, we get
\begin{align}
  \rd_{\nu}I_{\nu}'(z)
  &=I_{\nu}'(z)\log(z/2)
   +\frac{I_{\nu}(z)}{z}
   -\frac{1}{z}\sum_{k=0}^{\infty}
    \frac{(\nu+2k)\psi(\nu+k+1)}{k!\Gam(\nu+k+1)}(z/2)^{\nu+2k} ,
   \\
  [\rd_{\nu}I_{\nu}'(z)]_{\nu=n}
  &=I_n'(z)\log(z/2)
   +\frac{I_n(z)}{z}
   -\frac{1}{z}\sum_{k=0}^{\infty}\frac{(n+2k)\psi(n+k+1)}{k!(n+k)!}(z/2)^{n+2k} ,
    \label{eq:dInp_Inp}
\end{align}
and
\begin{align}
  (-1)^n[\rd_{\nu}I_{\nu}'(z)]_{\nu=n}
  &=-K_n'(z)
   +\frac{n!}{2z}\sum_{k=0}^{n-1}(-1)^k\frac{(z/2)^{k-n}}{k!(n-k)}
    \{zI_k'(z)+(k-n)I_k(z)\} ,
    \label{eq:dInp}
   \\
  [\rd_{\nu}K_{\nu}'(z)]_{\nu=n}
  &=\frac{n!}{2}\sum_{k=0}^{n-1}\frac{(z/2)^{k-n}}{k!(n-k)}
    \bigg(\rd_z+\frac{k-n}{z}\bigg)K_k(z) .
    \label{eq:dKnp}
\end{align}
When $\nu=0$,
\begin{alignat}{2}
  [\rd_{\nu}I_{\nu}(z)]_{\nu=0}
  &=-K_0(z) ,
    \qquad&
  [\rd_{\nu}K_{\nu}(z)]_{\nu=0}
  &=0 ,
    \label{eq:dI0_dK0}
   \\
  [\rd_{\nu}I_{\nu}'(z)]_{\nu=0}
  &=-K_0'(z) ,
    \qquad&
  [\rd_{\nu}K_{\nu}'(z)]_{\nu=0}
  &=0 .
\end{alignat}
$\rd_{\nu}(I_{\nu},I_{\nu}')$ for $\nu=n$ has two expressions.
But Eqs.(\ref{eq:dIn}) and (\ref{eq:dInp}) are not suited in the numerical calculation, 
because they involve $K_n$ and $K_n'$ which tend to increase for $n\to\infty$ roughly as 
follows,
\begin{alignat}{3}
  K_n(z)
  &\propto \frac{(n-1)!}{(z/2)^{n}} ,
    \qquad&
  K_n'(z)
  &\propto \frac{n!}{(z/2)^{n-1}}
   \qquad&
   (n\to\infty) ,
   \\
  I_n(z)
  &\propto \frac{(z/2)^{n}}{n!} ,
    \qquad&
  I_n'(z)
  &\propto \frac{(z/2)^{n-1}}{(n-1)!}
   \qquad&
   (n\to\infty) .
   \label{eq:In_ninf}
\end{alignat}
On the contrary, $I_n$ and $I_n'$ tend to decrease for $n\to\infty$
roughly as shown in Eq.(\ref{eq:In_ninf}).
This means that there is a large cancellation between the first and second terms on
the R.H.S. of Eqs.(\ref{eq:dIn}) and (\ref{eq:dInp}), which causes a large digit loss.
Therefore, in calculating the numerical values of $\rd_{\nu}(I_{\nu},I_{\nu}')$ for
$\nu=n\in\mathbb{Z}_0^{+}$, we should use Eqs.(\ref{eq:dIn_In}) and (\ref{eq:dInp_Inp})
instead of Eqs.(\ref{eq:dIn}) and (\ref{eq:dInp}).

\clearpage

\section{Bessel functions of purely imaginary order}
\label{sec:dunster}

The radial eigenfunctions of the curved pipe (\ref{eq:Ven}-\ref{eq:mfRm}) consist of
the cross products of the Bessel functions given by Eqs.(\ref{eq:CP_pq}-\ref{eq:CP_sr}).
They can have a purely imaginary order
since the fields in the Laplace domain have poles on the imaginary axis of
the $\nu$-plane as described in section \ref{sec:pole}.
We review Dunster's study on the Bessel functions of purely imaginary order and
their asymptotic expansions \cite{dunster}.

\subsection{Bessel functions $F_{\nu}$ and $G_{\nu}$}
\label{sec:bessel_FG}

When $J_{\nu}(z)$ and $Y_{\nu}(z)$ have a purely imaginary order
$\nu=i\bnu~(\bnu\in\mathbb{R})$ and a real argument $z=x\in\mathbb{R}^{+}$,
$J_{i\bnu}(x)$ and $Y_{i\bnu}(x)$ are complex functions in general.
But the cross products $t_{\nu}(z_1,z_2)=\{p_{\nu},q_{\nu},r_{\nu},s_{\nu}\}$
are real for $\nu=i\bnu\in i\mathbb{R}$
and $z_{1,2}=x_{1,2}\in\mathbb{R}^{+}$ since the imaginary parts of
$J_{i\bnu}(x_{1,2})$, $Y_{i\bnu}(x_{1,2})$ and their derivatives with respect to
$x_{1,2}$ cancel out in $t_{i\bnu}(x_1,x_2)$.
In order to describe $t_{i\bnu}(x_1,x_2)$ using only real functions,
we introduce a pair of linearly independent Bessel functions
$F_{\nu}(z)$ and $G_{\nu}(z)$ given by Eqs.(3.2-3.4) in \cite{dunster},
\begin{align}
  \bigg[
  \begin{array}{c}
   \! F_{\nu}(z) \! \\
   \! G_{\nu}(z) \!
  \end{array}
  \bigg]
  &=
  \bigg[
  \begin{array}{cc}
   \cos\tht & -\sin\tht\\
   \sin\tht & \cos\tht
  \end{array}
  \bigg]
  \bigg[
  \begin{array}{c}
   \! J_{\nu}(z) \!  \\
   \! Y_{\nu}(z) \! 
  \end{array}
  \bigg] ,
    \qquad
  \tht
  =\frac{\pi\nu}{2}
   \quad
  (\nu\in\mathbb{C},~z\in\mathbb{C}) .
   \label{eq:FG_JY}
\end{align}
Using the Hankel functions $H_\nu^{(1,2)}=J_{\nu}\pm iY_{\nu}$ or $J_{\pm\nu}$,
we rewrite Eq.(\ref{eq:FG_JY}),
\begin{alignat}{3}
  F_{\nu}(z)
  &=\frac{1}{2}\big\{H_\nu^{(1)}(z)e^{i\tht}+H_\nu^{(2)}(z)e^{-i\tht}\big\}  
  &&=\frac{J_\nu(z)+J_{-\nu}(z)}{2\cos\tht} ,
    \qquad&
  F_0(z)
  &=J_0(z) ,
   \label{eq:Fnu}
   \\
  G_{\nu}(z)
  &=\frac{1}{2i}\big\{H_\nu^{(1)}(z)e^{i\tht}-H_\nu^{(2)}(z)e^{-i\tht}\big\}
  &&=\frac{J_\nu(z)-J_{-\nu}(z)}{2\sin\tht} ,
    \qquad&
  G_0(z)
  &=Y_0(z) .
   \label{eq:Gnu}
\end{alignat}
$F_{\nu}$ and $G_{\nu}$ are even with respect to $\nu$.
The Wronskian of $F_\nu$ and $G_\nu$ is given by Eq.(3.10) in \cite{dunster},
which is equal to Eq.(\ref{eq:JY}),
\begin{align}
  W[F_\nu(z),G_\nu(z)]
  =W[J_\nu(z),Y_\nu(z)]
  =\frac{2}{\pi z} .
\end{align}
Using $F_\nu$ and $G_\nu$ instead of $J_\nu$ and $Y_\nu$, we rewrite
the cross products $t_{\nu}$ defined in Eqs.(\ref{eq:CP_pq}-\ref{eq:CP_sr}),
\begin{alignat}{2}
  p_{\nu}(b,a)
  &=F_{\nu}(b)G_{\nu}(a)-G_{\nu}(b)F_{\nu}(a) ,
    \qquad&
  q_{\nu}(b,a)
  &=F_{\nu}(b)G_{\nu}'(a)-G_{\nu}(b)F_{\nu}'(a) ,
  \label{eq:CP_pq_appendix}
  \\
  s_{\nu}(b,a)
  &=F_{\nu}'(b)G_{\nu}'(a)-G_{\nu}'(b)F_{\nu}'(a) ,
    \qquad&
  r_{\nu}(b,a)
  &=F_{\nu}'(b)G_{\nu}(a)-G_{\nu}'(b)F_{\nu}(a) ,
  \label{eq:CP_sr_appendix}
\end{alignat}
where the order $\nu$ and the arguments $a$ and $b$ are complex in general.
The prime $(')$ denotes the derivative with respect to the argument.
In investigating the pole structure of the fields in the Laplace domain,
it is enough to consider the arguments as either $a,b\in\bbR_{0}^{+}$ for
$|k\beta|\geq k_y^n$ or $a,b\in i\bbR^{+}$ for $|k\beta|<k_y^n$
since $k_r^n\in\mathbb{A}$ as in Eq.(\ref{eq:krn}).
In appendix \ref{sec:dunster_mod_Bess} we will discuss the cross products
of purely imaginary order and arguments.

We consider $F_{i\bnu}(x)$ and $G_{i\bnu}(x)$ which have a purely imaginary order
$\nu=i\bnu$ ($\bnu\in\mathbb{R}$) and a nonzero positive real argument $x\in\bbR^{+}$.
They can be given using $J_{i\bnu}(x)$ and $Y_{i\bnu}(x)$ as follows,
\begin{align}
  F_{i\bnu}(x)
  =\frac{\Re J_{i\bnu}(x)}{\cosh(\pi\bnu/2)}
  \in\mathbb{R} ,
    \qquad
  G_{i\bnu}(x)
  =\frac{\Re Y_{i\bnu}(x)}{\cosh(\pi\bnu/2)}
  \in\mathbb{R}
   \qquad
   (\bnu\in\bbR,~x\in\bbR^{+}) .
   \label{eq:FG_ibnu_x}
\end{align}
Therefore the cross products (\ref{eq:CP_pq_appendix}-\ref{eq:CP_sr_appendix}) are real
for $\nu=i\bnu\in i\mathbb{R}$ and $a,b\in\mathbb{R}^{+}$.
The asymptotic series of $F_{i\bnu}(x)$ and $G_{i\bnu}(x)$ for $\bnu\to\infty$
with $x$ fixed are given by Eqs.(3.8-3.9) in \cite{dunster},
\begin{alignat}{2}
  F_{i\bnu}(x)
  &=a_{\bnu}
    \sum_{j=0}^\infty
    \frac{(-x^2/4)^j}{j! h_{\bnu,j}}\cos\vphi_j ,
    \qquad&
  F_{i\bnu}'(x)
  &=\frac{a_{\bnu}}{x}\sum_{j=0}^{\infty}
    \frac{(-x^2/4)^j}{j!h_{\bnu,j}}
    (2j\cos\vphi_j-\bnu\sin\vphi_j) ,
   \label{eq:Fimu_x}
  \\
  G_{i\bnu}(x)
  &=b_{\bnu}
    \sum_{j=0}^\infty
    \frac{(-x^2/4)^j}{j!h_{\bnu,j}}\sin\vphi_j ,
    \qquad&
  G_{i\bnu}'(x)
  &=\frac{b_{\bnu}}{x}\sum_{j=0}^{\infty}
    \frac{(-x^2/4)^j}{j!h_{\bnu,j}}
    (2j\sin\vphi_j+\bnu\cos\vphi_j) ,
   \label{eq:Gimu_x}
\end{alignat}
where
\begin{align}
  a_{\bnu}
  =\Big\{\frac{2}{\pi}\bnu\tanh\Big(\frac{\pi}{2}\bnu\Big)\Big\}^{1/2} ,
    \qquad
  b_{\bnu}
  =\Big\{\frac{2}{\pi}\bnu\coth\Big(\frac{\pi}{2}\bnu\Big)\Big\}^{1/2} ,
    \qquad
  h_{\bnu,j}
  =\prod_{l=0}^{j}(\bnu^2+l^2)^{1/2} .
  \label{eq:phij_hj}
\end{align}
The phase $\vphi_j$ is a function of $x$ and $\bnu$,
\begin{align}
  \vphi_j
  =\bnu\log(x/2)-\phi_{\bnu,j} ,
    \qquad
  \phi_{\bnu,j}
  =\arg\Gamma(j+1+i\bnu)
   \qquad
  (j\in\mathbb{Z}_0^{+}) .
   \label{eq:phij}
\end{align}
According to the description below Eq.(2.7) in \cite{dunster},
for each $j$ we define the branch of $\phi_{\bnu,j}$ so that $\phi_{\bnu,j}$
is continuous for $0<\bnu<\infty$, with $\lim_{\bnu\to0}\phi_{\bnu,j}=0$.
Using 6.1.27 in \cite{abramo_stegun}, $\phi_{\bnu,j}$ is given as
\begin{align}
  \phi_{\bnu,j}
  &=\bnu\psi(j+1)
   +\sum_{\ell=1}^{\infty}(c_{j\ell}-\tan^{-1}c_{j\ell}) ,
    \qquad
  c_{j\ell}
  =\frac{\bnu}{j+\ell}
   \qquad
  (\bnu>0) .
    \label{arg_Gamj_190620}
\end{align}
$\psi$ is the digamma function of integer argument (\ref{eq:digamma}).
$\phi_{\bnu,j}$ goes to 0 in the limit of $\bnu\to0$ for
$\forall j\in\mathbb{Z}_0^{+}$.
On the other hand, for large $\bnu$, since the sum in Eq.(\ref{arg_Gamj_190620}) converges 
algebraically (\ie, slowly) with respect to $\ell$, we rewrite the second equation of
(\ref{eq:phij}) as follows,
\begin{align}
  \phi_{\bnu,j}
  &=\arg\Gam(i\bnu)
   +t_{\bnu,j}
   ,\qquad
  t_{\bnu,j}
   =\sum_{\ell=0}^{j}\arg(\ell+i\bnu)
   =\frac{\pi}{2}
   +(1-\delta_j^0)\sum_{\ell=1}^{j}\tan^{-1}(\bnu/\ell) .
\end{align}
According to 6.1.44 (p.257) in \cite{abramo_stegun}, $\arg\Gamma(i\bnu)$ has
the following asymptotic expression for $\bnu\gg1$,
\begin{align}
  &
  \arg\Gamma(i\bnu)
  \simeq
   \bnu(\log\bnu-1)-\frac{\pi}{4}
   -\sum_{n=1}^{\infty}\frac{\bB_n}{\bnu^{2n-1}}
  \qquad(\bnu\to\infty)
   ,\qquad
  \bB_n
  =\frac{(-1)^{n-1}B_{2n}}{2n(2n-1)} ,
\end{align}
where $B_{2n}$ is the Bernoulli number given by Table 23.2 (p.810) in \cite{abramo_stegun}.
$\bB_n$ is given as follows,
\begin{align}
  \bB_1
  &=\frac{1}{12}
   ,\qquad
  \bB_2
  =\frac{1}{360}
   ,\qquad
  \bB_3
  =\frac{1}{1260}
   ,\qquad
  \bB_4
  =\frac{1}{1680}
   ,\qquad
  \bB_5
  =\frac{1}{1188}
   ,\qquad
  \bB_6
  =\frac{691}{360360}
   ,
   \\
  \bB_7
  &=\frac{1}{156}
   ,\qquad
  \bB_8
  =\frac{3617}{122400}
   ,\qquad
  \bB_9
  =\frac{43867}{244188}
   ,\qquad
  \bB_{10}
  =\frac{174611}{125400}
   ,\qquad
  \bB_{11}
  =\frac{77683}{5796}
   ,
   \\
  \bB_{12}
  &=\frac{236364091}{1506960}
   ,\qquad
  \bB_{13}
  =\frac{657931}{300}
   ,\qquad
  \bB_{14}
  =\frac{3392780147}{93960}
   ,\qquad
  \bB_{15}
  =\frac{1723168255201}{2492028}
   ,
   \\
  \bB_{16}
  &=\frac{7709321041217}{505920}
   ,\qquad
  \bB_{17}
  =\frac{151628697551}{396}
   ,\qquad
  \bB_{18}
  =\frac{26315271553053477373}{2418179400}
   ,
   \\
  \bB_{19}
  &=\frac{154210205991661}{444}
   ,\quad
  \bB_{20}
  =\frac{261082718496449122051}{21106800}
   ,\quad
  \bB_{21}
  =\frac{1520097643918070802691}{3109932}
   ,
   \\
  \bB_{22}
  &=\frac{2530297234481911294093}{118680}
   ,\qquad
  \bB_{23}
  =\frac{25932657025822267968607}{25380}
   .
\end{align}
For example, $\bB_n/\bnu^{2n-1}=10^{-16}$ for $n=11$ and $\bnu\approx 6.54$,
or for $n=30$ and $\bnu\approx 6.21$.

We substitute Eqs.(\ref{eq:Fimu_x}-\ref{eq:Gimu_x}) into
Eqs.(\ref{eq:CP_pq_appendix}-\ref{eq:CP_sr_appendix}) for
$\nu=i\bnu\in i\mathbb{R}$ and $a,b\in\mathbb{R}^{+}$,
\begin{align}
  \bigg[
  \begin{array}{cc}
    p_{i\bnu}(b,a) & -aq_{i\bnu}(b,a) \\
    abs_{i\bnu}(b,a) & br_{i\bnu}(b,a)
  \end{array}
  \bigg]
  &=
   -\frac{2\bnu}{\pi}\sum_{j,k=0}^\infty
    \frac{(-b^2/4)^j}{j! h_{\bnu,j}}
    \frac{(-a^2/4)^k}{k!h_{\bnu,k}}
  \bigg[
  \begin{array}{cc}
    T_{jk}^p & T_{jk}^q \\
    T_{jk}^s & T_{jk}^r
  \end{array}
  \bigg] ,
    \label{eq:t_ibnu_x}
\end{align}
\begin{align}
  T_{jk}^p
  &=\sin\vphi_{jk} ,
    \qquad
  T_{jk}^s
  =(\bnu^2+4jk)\sin\vphi_{jk}
  -2\bnu(j-k)\cos\vphi_{jk} ,
   \\
  T_{jk}^q
  &=\bnu\cos\vphi_{jk}-2k\sin\vphi_{jk} ,
    \qquad
  T_{jk}^r
  =\bnu\cos\vphi_{jk} +2j\sin\vphi_{jk} .
\end{align}
The phase $\vphi_{jk}$ is given as
\begin{align}
  \vphi_{jk}
  &=\bnu\log(b/a)
   -\alp_{jk}
   \quad
  (\bnu,b,a\in\mathbb{R}^{+})
   ,\qquad
  \alp_{jk}
  =\arg\bigg[\frac{\Gamma(1+j+i\bnu)}{\Gamma(1+k+i\bnu)}\bigg] .
\end{align}
From Eq.(\ref{arg_Gamj_190620}), $\alp_{jk}$ is given as
\begin{align}
  \alp_{jj}
  =0,
   \qquad
  \alp_{jk}
  =\sum_{\ell=k+1}^{j}\tan^{-1}(\bnu/\ell)
   \quad
  (j>k),
   \qquad
  \alp_{jk}
  =-\sum_{\ell=j+1}^{k}\tan^{-1}(\bnu/\ell)
   \quad
  (j<k) .
\end{align}
We examine the asymptotic behavior of Eq.(\ref{eq:t_ibnu_x}) for $\bnu\to\pm\infty$
with the arguments $a$ and $b$ fixed.
Taking only the leading order term into account in Eq.(\ref{eq:t_ibnu_x}),
we get the asymptotic expressions of the cross products $t_{i\bnu}(b,a)$ for
$\bnu\to\pm\infty$ with $b=\hr=k_r^nr$ and $a=\hr'=k_r^nr'$ fixed,
\begin{alignat}{2}
  p_{i\bnu}(\hr,\hr')
  &\sim -\frac{2}{\pi\bnu}\{\sin\tht_{\bnu}+O(\veps)\} ,
    \qquad&
  q_{i\bnu}(\hr,\hr')
  &\sim +\frac{2}{\pi\hr'}\{\cos\tht_{\bnu}+O(\veps)\} ,
  \label{eq:pimu_asymp_mu_inf}
   \\
  s_{i\bnu}(\hr,\hr')
  &\sim -\frac{2\bnu}{\pi\hr\hr'}\{\sin\tht_{\bnu}+O(\veps)\} ,
    \qquad&
  r_{i\bnu}(\hr,\hr')
  &\sim -\frac{2}{\pi\hr}\{\cos\tht_{\bnu}+O(\veps)\} .
  \label{eq:simu_asymp_mu_inf}
\end{alignat}
$\tht_{\bnu}$ is given below.
Eqs.(\ref{eq:pimu_asymp_mu_inf}-\ref{eq:simu_asymp_mu_inf}) agree with
Eqs.(\ref{lim_kr0_aq_br_bas}) and (\ref{eq:pnu_asymp}) for $\nu=i\bnu$.
\begin{align}
  \tht_{\bnu}=\bnu\log(r/r') ,
    \qquad
  O(\veps)=O(\hr^2/\bnu)=O(\hr'^2/\bnu) .
\end{align}
%

\subsection{Uniform asymptotic expansion of the Bessel functions of
imaginary order and real argument}
\label{sec:UAE_JY_inu}

The uniform asymptotic expansion of $\{J_{\nu}(\hr),Y_{\nu}(\hr)\}$ and
$\{J_{\nu}'(\hr),Y_{\nu}'(\hr)\}$ for $\nu\to\infty$ and $\hr\to\infty$ are
given by Eqs.(\ref{eq:JY_uae}-\ref{eq:dJY_uae}).
On the other hand, we use Eqs.(\ref{eq:CP_pq_appendix}-\ref{eq:CP_sr_appendix}) in expanding
the cross products of purely imaginary order $t_{i\bnu}(\hr,\hr')$
in the uniform asymptotic series.
In appendix \ref{sec:UAE_JY_inu} we use the following variables,
\begin{align}
  \nu
  =i\bnu
   ,\qquad
  \bz
  =iz
   ,\qquad
  \hr
  =\nu z
  =\bnu\bz
   \qquad
  (\bnu\in\mathbb{R}^{+}) .
\end{align}
$\bz$ depends on $\bnu$ through the argument $\hr$ which is independent of $\bnu$.
The uniform asymptotic expansions of $F_{i\bnu}(\hr)$ and $G_{i\bnu}(\hr)$
around $\bz=0$ are given by Eqs.(5.15-5.16) in \cite{dunster},
\begin{alignat}{2}
  F_{i\bnu}(\bnu\bz)
  &=\bar{f}_{\bnu}
    \big(\hU_n\cos\bphi_{\bnu}
        +\bU_n\sin\bphi_{\bnu}
        +\cE_n^{+}
    \big) ,
    \qquad&
  F_{i\bnu}'(\bnu\bz)
  &=\hat{f}_{\bnu}
    \big(
       \bV_n\cos\bphi_{\bnu}
      -\hV_n\sin\bphi_{\bnu}
      +\bcE_n^{+}
    \big) ,
   \label{eq:Fimu_muz}
  \\
  G_{i\bnu}(\bnu\bz)
  &=\bar{f}_{\bnu}
    \big(\hU_n\sin\bphi_{\bnu}
        -\bU_n\cos\bphi_{\bnu}
        +\cE_n^{-}
    \big) ,
    \qquad&
  G_{i\bnu}'(\bnu\bz)
  &=\hat{f}_{\bnu}
    \big(
       \bV_n\sin\bphi_{\bnu}
      +\hV_n\cos\bphi_{\bnu}
      +\bcE_n^{-}
    \big) ,
   \label{eq:Gimu_muz}
\end{alignat}
where $|\arg\bz|<\pi/2$ ($\Re\bz>0$).
$n$ in Eqs.(\ref{eq:Fimu_muz}-\ref{eq:Gimu_muz}) is the error index
which is involved in Eqs.(\ref{eq:err_p}-\ref{eq:err_m}).
The phase $\bphi_{\bnu}$ depends on $\xi$ given by Eq.(3.24) in \cite{dunster},
\begin{align}
  \bphi_{\bnu}(\bz)
  =\bnu\xi(\bz)-\frac{\pi}{4} ,
    \qquad
  \xi(\bz)
  =\frac{1}{p}+\log\bigg(\frac{\bz}{1+1/p}\bigg)
  =\frac{1}{p}+\frac{d\bphi_{\bnu}}{d\bnu}
   ,\qquad
  \lim_{\bz\to\infty}\xi
  =\infty .
  \label{eq:xi_bz}
\end{align}
$\xi$ is related to $\eta_{\pm}$ given by Eqs.(\ref{eq:zeta_p}-\ref{eq:zeta_m}),
\begin{align}
  \xi(\bz)+\eta_{+}(z)
  =i\pi/2,
   \qquad
  \eta_{+}(z)
  =i\eta_{-}(z) .
   \label{xi_eta_pm}
\end{align}
$p$ is a function of $\bz$, which is related to $t$ defined in Eq.(\ref{eq:t})
as $p(iz)=t(z)$, \ie,
\begin{align}
  p(\bz)
  =(1+\bz^2)^{-1/2} ,
    \qquad
  \bz=\frac{1}{p}(1-p^2)^{1/2} ,
    \qquad
  \frac{dp}{d\bz}
  =-\bz p^3 ,
    \qquad
  pd\xi
  =\frac{d\bz}{\bz}
  =-\frac{d\bnu}{\bnu} .
   \label{eq:p_bz}
\end{align}
$\bar{f}_{\bnu}$ and $\hat{f}_{\bnu}$ in Eqs.(\ref{eq:Fimu_muz}-\ref{eq:Gimu_muz}) are
given as
\begin{align}
  \bar{f}_{\bnu}
  =(\chi_{\bnu}p)^{1/2} ,
    \qquad
  \hat{f}_{\bnu}
  =\frac{1}{\bz}(\chi_{\bnu}/p)^{1/2} ,
    \qquad
  \chi_{\bnu}
  =\frac{2}{\pi\bnu} ,
    \qquad
  \frac{d}{d\bnu}\{\bar{f}_{\bnu},\hat{f}_{\bnu}\}
  =\frac{p^2}{2\bnu}\{-\bar{f}_{\bnu},\hat{f}_{\bnu}\} .
\end{align}
$\hU_n$, $\hV_n$, $\bU_n$ and $\bU_n$ are the following series,
\begin{align}
  \bigg\{{ \hU_n \atop \hV_n}\bigg\}
  &=\sum_{j=0}^{n}\frac{(-1)^j}{\bnu^{2j}}\bigg\{{ u_{2j}(p) \atop v_{2j}(p)}\bigg\} ,
    \qquad
  \bigg\{{ \bU_n \atop \bV_n}\bigg\}
  =\sum_{j=0}^{n-1}\frac{(-1)^j}{\bnu^{2j+1}}\bigg\{{ u_{2j+1}(p) \atop v_{2j+1}(p)}\bigg\} .
   \label{eq:Unu_Vnu}
\end{align}
$u_{\ell}$ and $v_{\ell}$ are given by Eqs.(\ref{eq:uv}).
The derivatives of Eqs.(\ref{eq:Unu_Vnu}) with respect to $\bnu$ are given as
\begin{alignat}{2}
  \frac{d}{d\bnu}\bigg\{{ \hU_n \atop \hV_n}\bigg\}
  =\sum_{j=1}^{n}\frac{(-1)^j}{\bnu^{2j+1}}
    \bigg\{{ \chu_{2j}(p) \atop \chv_{2j}(p)}\bigg\} ,
    \qquad
  \frac{d}{d\bnu}\bigg\{{ \bU_n \atop \bV_n}\bigg\}
  =\sum_{j=0}^{n-1}\frac{(-1)^j}{\bnu^{2j+2}}
    \bigg\{{ \chu_{2j+1}(p) \atop \chv_{2j+1}(p)}\bigg\} ,
   \label{eq:rdbnu_UV}
\end{alignat}
where $\chu_{\ell}$ and $\chv_{\ell}$ are given by Eq.(\ref{eq:ch_uv}),
\begin{align}
  \bigg\{{ \chu_{\ell}(p) \atop \chv_{\ell}(p) }\bigg\}
  =(\bz^2p^3\rd_{p}-\ell) \bigg\{{ u_{\ell}(p) \atop v_{\ell}(p) }\bigg\}
  =\sum_{k=0}^{\ell}(2k\bz^2-\ell)
    \bigg\{{ \mfa_k^{(\ell)} \atop \mfb_k^{(\ell)} }\bigg\}
    p^{2(k+1)+\ell} .
\end{align}
Eqs.(\ref{eq:rdbnu_UV}) are involved in the derivatives of
Eqs.(\ref{eq:Fimu_muz}-\ref{eq:Gimu_muz}) with respect to $\bnu$,
\begin{align}
  \frac{d}{d\bnu}\bigg\{{ F_{i\bnu}(\bnu\bz) \atop G_{i\bnu}(\bnu\bz) }\bigg\}
  =
   \bar{f}_{\bnu}\bigg\{{ \cF_0 \atop \cG_0 }\bigg\}
   -\frac{p^2}{2\bnu}\bigg\{{ F_{i\bnu}(\bnu\bz) \atop G_{i\bnu}(\bnu\bz) }\bigg\}
   ,\qquad
  \frac{d}{d\bnu}\bigg\{{ F_{i\bnu}'(\bnu\bz) \atop G_{i\bnu}'(\bnu\bz) }\bigg\}
  =
    \hat{f}_{\bnu}\bigg\{{ \cF_1 \atop \cG_1 }\bigg\}
   +\frac{p^2}{2\bnu}\bigg\{{ F_{i\bnu}'(\bnu\bz) \atop G_{i\bnu}'(\bnu\bz) }\bigg\} .
\end{align}
$\cF_{0,1}$ and $\cG_{0,1}$ are given below.
$\bphi_{\bnu}$ and $\rd_{\bnu}\bphi_{\bnu}\,(=d\bphi_{\bnu}/d\bnu)$ are
given by Eqs.(\ref{eq:xi_bz}).
\begin{align}
  \cF_0
  &=
    (\rd_{\bnu}\hU_n+\bU_n\rd_{\bnu}\bphi_{\bnu})\cos\bphi_{\bnu}
   +(\rd_{\bnu}\bU_n-\hU_n\rd_{\bnu}\bphi_{\bnu})\sin\bphi_{\bnu}
   +\rd_{\bnu}\cE_n^{+} ,
   \\
  \cG_0
  &=
    (\rd_{\bnu}\hU_n+\bU_n\rd_{\bnu}\bphi_{\bnu})\sin\bphi_{\bnu}
   -(\rd_{\bnu}\bU_n-\hU_n\rd_{\bnu}\bphi_{\bnu})\cos\bphi_{\bnu}
   +\rd_{\bnu}\cE_n^{-} ,
   \\
  \cF_1
  &=
    (\rd_{\bnu}\bV_n-\hV_n\rd_{\bnu}\bphi_{\bnu})\cos\bphi_{\bnu}
   -(\rd_{\bnu}\hV_n+\bV_n\rd_{\bnu}\bphi_{\bnu})\sin\bphi_{\bnu}
   +\rd_{\bnu}\bcE_n^{+} ,
   \\
  \cG_1
  &=
    (\rd_{\bnu}\bV_n-\hV_n\rd_{\bnu}\bphi_{\bnu})\sin\bphi_{\bnu}
   +(\rd_{\bnu}\hV_n+\bV_n\rd_{\bnu}\bphi_{\bnu})\cos\bphi_{\bnu}
   +\rd_{\bnu}\bcE_n^{-} .
\end{align}

$n$ is the index to denote the order of the error in
Eqs.(\ref{eq:Fimu_muz}-\ref{eq:Gimu_muz}) as shown in
Sec.4 of \cite{dunster} and Chap.10, Sec.3 of \cite{olver}.
$n$ is not so important in using Eqs.(\ref{eq:Fimu_muz}-\ref{eq:Gimu_muz}).
$\cE_n^{\pm}$ are the error terms of the $n$th order,
\begin{align}
  \cE_n^{+}(\bnu,\xi)
  &=\frac{1}{2}
    \big\{e^{-i\pi/4}\veps_{2n+1}^{(1)}(\bnu,\xi)+e^{i\pi/4}\veps_{2n+1}^{(2)}(\bnu,\xi)
    \big\} ,
   \label{eq:err_p}
  \\
  \cE_n^{-}(\bnu,\xi)
  &=\frac{1}{2i}
    \big\{e^{-i\pi/4}\veps_{2n+1}^{(1)}(\bnu,\xi)-e^{i\pi/4}\veps_{2n+1}^{(2)}(\bnu,\xi)
    \big\} .
   \label{eq:err_m}
\end{align}
$\veps_n^{(1,2)}$ is bounded as shown in Eq.(3.04) of Chap.10, p.366 in \cite{olver}.
According to the explanation described in p.1007 of \cite{dunster}, it may be enough to 
understand the order of magnitude of the error with respect to $\bnu$,
\begin{align}
  \veps_n^{(1,2)}(\bnu,\xi)=e^{-\bnu\xi}\times O(\bnu^{-n}) .
  \label{eq:err_bnd}
\end{align}
$\bcE_n^{\pm}$ in Eqs.(\ref{eq:Fimu_muz}-\ref{eq:Gimu_muz}) are the error terms of
$F_{i\bnu}'$ and $G_{i\bnu}'$,
\begin{align}
   \bigg\{{ \bcE_n^{+}(\bnu,\xi) \atop \bcE_n^{-}(\bnu,\xi) }\bigg\}
  &=\cE_n^{\ast}(p) \bigg\{{ \cos\bphi_{\bnu} \atop \sin\bphi_{\bnu} }\bigg\}
   +\frac{1}{\bnu}\Big[\rd_{\xi}+\frac{p}{2}(p^2-1)\Big]
    \bigg\{{ \cE_n^{+}(\bnu,\xi) \atop \cE_n^{-}(\bnu,\xi) }\bigg\} ,
  \label{eq:errn}
\end{align}
where
\begin{align}
  \cE_n^{\ast}(p)
  &=\frac{(-1)^n}{\bnu^{2n+1}}\{v_{2n+1}(p)-u_{2n+1}(p)\}
   =\frac{(-1)^n}{\bnu^{2n+1}}p(p^2-1)\bigg(\frac{1}{2}+p\rd_p\bigg)u_{2n}(p) .
\end{align}

$\xi$ is a function of $\bz$, given by Eq.(\ref{eq:xi_bz}).
We consider the asymptotic expansion of $\xi$ for $|\bz|\ll1$ or $|\bz|\gg1$.
When $|\bz|\gg1$, we can expand $\xi/\bz$ with respect to $\bz^{-2}$ as follows,
\begin{align}
  &
  \frac{\xi(\bz)}{\bz}
  =\sum_{j=0}^{\infty}\frac{(-1)^j\gam_j}{1-2j}\bz^{-2j}
   \qquad(|\bz|\gg1)
   ,\qquad
  \gam_j
  =-\frac{\Gam(j-1/2)}{2\pi^{1/2}j!}
  =\frac{(2j+1)!!}{2^j(1-4j^2)j!} :
   \label{eq:xit}
   \\
  &
  \gam_0
  =1
   ,\quad~~
  \gam_1
  =-\frac{1}{2}
   ,\quad~~
  \gam_2
  =-\frac{1}{2^3}
   ,\quad~~
  \gam_3
  =-\frac{1}{2^4}
   ,\quad~~
  \gam_4
  =-\frac{5}{2^7}
   ,\quad~~
  \gam_5
  =-\frac{7}{2^8} .
    \label{eq:gamj}
\end{align}
$\gam_j$ is the coefficient of the $j$th order term in
the Maclaurin series of $(1-\delta)^{1/2}$ with respect to $\delta$.
We derived Eq.(\ref{eq:xit}) as follows.
For convenience, we define $\bt$ as
\begin{align}
  \bt
  =\frac{1}{\bz}
   \qquad
  (|\bt|\ll1) ,
    \qquad
  \frac{1}{p}
  =(1+\bz^2)^{1/2}
  =\frac{1}{\bt}(1+\bt^2)^{1/2}
   \qquad
  (|p|\ll1) .
\end{align}
Then we expand $\xi$, given by Eq.(\ref{eq:xi_bz}), with respect to $\bt$,
\begin{align}
  \xi
  &=\frac{1}{p}
    -\frac{1}{2}\log(1+\bt^2)
    -\log(1+p)
   =\frac{1}{\bt}
   +\sum_{j=1}^{\infty}\frac{\bt^{2j-1}}{1-2j}
    \bigg(
      \vpi_j^1
     +\sum_{\ell=0}^{\infty}\vpi_{\ell}^j\bt^{2\ell}
    \bigg)
   =\sum_{j=0}^{\infty}\frac{(-1)^j\gam_j}{1-2j}\bt^{2j-1} ,
\end{align}
where $\vpi_{\ell}^j$ and $\gam_j$ are given as
\begin{align}
  \vpi_{\ell}^j
  =(-1)^{\ell}\frac{\Gam(\ell+j-1/2)}{\ell!\Gam(j-1/2)}
   ,\qquad
  \gam_j
   =(-1)^j\sum_{n=0}^{j}\frac{2j-1}{2n-1}\vpi_{j-n}^n
   =\sum_{n=0}^{j}(-1)^nC_n^j \frac{n!\Gam(j+1/2)}{j!\Gam(n+1/2)} .
   \label{gamj_intm}
\end{align}
$C_n^j$ is the binomial coefficient,
\begin{align}
  C_n^j
  =\frac{j!}{n!(j-n)!}
   \qquad
   (j,n\in\mathbb{Z}_0^{+}) .
\end{align}
We rewrite $\gam_j$, given by Eq.(\ref{gamj_intm}),
using the ninth formula in p.12, vol.\,II, part I, chap.1, \S 3-(ii) in \cite{iwanami},
\begin{align}
  \sum_{j=0}^{n}\frac{(-1)^jC_j^n}{j+a}
  =\frac{n!\Gam(a)}{\Gam(n+a+1)}
   \qquad\Lra\qquad
  \sum_{n=0}^{j}\frac{(-1)^nC_n^jn!\Gam(a)}{\Gam(n+a+1)}
  =\frac{1}{j+a}
    \qquad
   (a\not\in\mathbb{Z}_0^{-}) ,
   \label{iwanami_formula}
\end{align}
where we inverted the binomial sum using Eqs.(4.2.1-4.2.2) in \cite{gross}.
$a$ is an arbitrary variable which is not a negative integer or 0.
Using the second equation of (\ref{iwanami_formula}) for $a=-1/2$, we can rewrite $\gam_j$
given by Eq.(\ref{gamj_intm}) into Eq.(\ref{eq:xit}).
Thus we can expand $\xi/\bz$ with respect to $\bz^{-2}$ as shown in Eqs.(\ref{eq:xit}).

\noindent
When $|\bz|\ll1$, we can expand $\xi-\log(\bz/2)$ with respect to $\bz^2$ as follows,
\begin{align}
  &
  \xi(\bz)
  =\log(\bz/2)
   +\sum_{j=0}^{\infty}h_j\bz^{2j}
   \qquad
   (|\bz|\ll1)
   \label{xi_expd}
   ,\qquad
  h_j
  =\frac{(-1)^j\gam_j}{2j+\delta_j^0}
  =\frac{(-1)^j(2j+1)!!}{2^j(2j+\delta_j^0)(1-4j^2)j!} :
   \\
  &
  h_0
  =1
   ,\quad~~
  h_1
  =\frac{1}{2^2}
   ,\quad~~
  h_2
  =-\frac{1}{2^5}
   ,\quad~~
  h_3
  =\frac{1}{2^5\cd 3}
   ,\quad~~
  h_4
  =-\frac{5}{2^{10}}
   ,\quad~~
  h_5
  =\frac{7}{2^9\cd5} .
   \label{hj_coeff}
\end{align}

Substituting Eqs.(\ref{eq:Fimu_muz}-\ref{eq:Gimu_muz}) into
Eqs.(\ref{eq:CP_pq_appendix}-\ref{eq:CP_sr_appendix}) for
$\nu=i\bnu$ $(\bnu\in\mathbb{R}^{+})$, we rearrange them,
\begin{align}
  p_{i\bnu}(\bnu\bz,\bnu\bz')
  &\simeq
   f_{\bnu}^p(\bz,\bz')
   \{
     -\Ups_0^p(p,p')\sin\vphi
     -\Ups_1^p(p,p')\cos\vphi
     +\cE_n^p(\bnu,\xi,\xi')
   \} ,
   \label{eq:pibnu}
   \\
  q_{i\bnu}(\bnu\bz,\bnu\bz')
  &\simeq
   f_{\bnu}^q(\bz,\bz')
   \{
     +\Ups_0^q(p,p')\cos\vphi
     -\Ups_1^q(p,p')\sin\vphi
     +\cE_n^q(\bnu,\xi,\xi')
   \} ,
   \\
  r_{i\bnu}(\bnu\bz,\bnu\bz')
  &\simeq
   f_{\bnu}^r(\bz,\bz')
   \{
     -\Ups_0^r(p,p')\cos\vphi
     +\Ups_1^r(p,p')\sin\vphi
     +\cE_n^r(\bnu,\xi,\xi')
   \} ,
   \\
  s_{i\bnu}(\bnu\bz,\bnu\bz')
  &\simeq
   f_{\bnu}^s(\bz,\bz')
   \{
     -\Ups_0^s(p,p')\sin\vphi
     -\Ups_1^s(p,p')\cos\vphi
     +\cE_n^s(\bnu,\xi,\xi')
   \} .
   \label{eq:sibnu}
\end{align}
Eqs.(\ref{eq:pibnu}-\ref{eq:sibnu}) are valid around $\bz=0$.
$\bz$ and $\bz'$ can be complex if $\Re\bz>0$ and $\Re\bz'>0$.
$\vphi$ is given as
\begin{align}
  \vphi(\bz,\bz')
  =\bnu(\xi-\xi')
   ,\qquad
  \xi
  =\xi(\bz)
   ,\qquad
  \xi'
  =\xi(\bz') .
   \label{eq:vphi}
\end{align}
$\xi$ and $p$ are the functions of $\bz$, given by Eqs.(\ref{eq:xi_bz}) and (\ref{eq:p_bz}).
Similarly, $\xi'$ and $p'$ are the functions of $\bz'$,
\begin{alignat}{2}
  p
  =p(\bz)
  =(1+\bz^2)^{-1/2}
   ,\qquad
  p'
  =p(\bz')
  =(1+\bz'^2)^{-1/2} .
   \label{eq:p_pprm}
\end{alignat}
$f_{\bnu}^{p,q,r,s}(\bz,\bz')$ denotes the following functions,
\begin{align}
  f_{\bnu}^p
  =\frac{2}{\pi\bnu}(pp')^{1/2}
   ,\qquad
  f_{\bnu}^q
  =\frac{2}{\pi\bnu}\frac{(p/p')^{1/2}}{\bz'}
   ,\qquad
  f_{\bnu}^r
  =\frac{2}{\pi\bnu}\frac{(p'/p)^{1/2}}{\bz}
   ,\qquad
  f_{\bnu}^s
  =\frac{2}{\pi\bnu}\frac{(pp')^{-1/2}}{\bz\bz'} .
\end{align}
$\cE_n^{p,q,r,s}$ denotes the error terms of
the cross products in the uniform asymptotic expansion.
$n$ is the order of the error as described below Eqs.(\ref{eq:Unu_Vnu}).
It is enough to know that $\cE_n^{p,q,r,s}$ is of the order of
$\cE_n^{\pm}$ and $\bcE_n^{\pm}$ given by Eqs.(\ref{eq:err_p}-\ref{eq:err_m}) and
(\ref{eq:errn}).
$\Ups_{0,1}^{p,q,r,s}$ in Eqs.(\ref{eq:pibnu}-\ref{eq:sibnu}) is given as
\begin{alignat}{2}
  \Ups_0^p(p,p')
  &=\hU_n(p)\hU_n(p')+\bU_n(p)\bU_n(p') ,
    \qquad&
  \Ups_1^p(p,p')
  &=\hU_n(p)\bU_n(p')-\bU_n(p)\hU_n(p') ,
   \label{eq:Ups_p_0}
    \\
  \Ups_0^q(p,p')
  &=\hU_n(p)\hV_n(p')+\bU_n(p)\bV_n(p') ,
    \qquad&
  \Ups_1^q(p,p')
  &=\hU_n(p)\bV_n(p')-\bU_n(p)\hV_n(p') ,
   \\
  \Ups_0^r(p,p')
  &=\hV_n(p)\hU_n(p')+\bV_n(p)\bU_n(p') ,
    \qquad&
  \Ups_1^r(p,p')
  &=\hV_n(p)\bU_n(p')-\bV_n(p)\hU_n(p') ,
    \\
  \Ups_0^s(p,p')
  &=\hV_n(p)\hV_n(p')+\bV_n(p)\bV_n(p') ,
    \qquad&
  \Ups_1^s(p,p')
  &=\hV_n(p)\bV_n(p')-\bV_n(p)\hV_n(p') .
   \label{eq:Ups_m_0}
\end{alignat}
$\hU_n$, $\bU_n$, $\hV_n$ and $\bV_n$ are the series given by Eqs.(\ref{eq:Unu_Vnu}).
$\Ups_0^{p,s}=O(1)$ and $\Ups_1^{p,s}=O(\bz^2/\bnu)$.

We rewrite Eqs.(\ref{eq:pibnu}) and (\ref{eq:sibnu}) in order to find the asymptotic 
expressions of the zeros of $p_{\nu}(\hr_b,\hr_a)$ and $s_{\nu}(\hr_b,\hr_a)$ 
with respect to $\nu$ in appendix \ref{sec:poles}.
Substituting $p=p_b$ and $p'=p_a$ into Eqs.(\ref{eq:Ups_p_0}) and (\ref{eq:Ups_m_0}),
we rewrite them as follows,
\begin{align}
  \Ups_0^p(p_b,p_a)
  &=d_p\cos\tht_{+} ,
    \qquad
  \Ups_1^p(p_b,p_a)
  =d_p\sin\tht_{+} ,
    \qquad
  d_p^2
  =[\hU_n^2(p_b)+\bU_n^2(p_b)][\hU_n^2(p_a)+\bU_n^2(p_a)] ,
   \label{eq:Ups_p_tf}
   \\
  \Ups_0^s(p_b,p_a)
  &=d_s\cos\tht_{-} ,
    \qquad
  \Ups_1^s(p_b,p_a)
  =d_s\sin\tht_{-} ,
    \qquad
  d_s^2
  =[\hV_n^2(p_b)+\bV_n^2(p_b)][\hV_n^2(p_a)+\bV_n^2(p_a)] .
   \label{eq:Ups_s_tf}
\end{align}
Substituting $\hr_b=\bnu\bz$ and $\hr_a=\bnu\bz'$ into 
Eqs.(\ref{eq:pibnu}) and (\ref{eq:sibnu}), we rewrite them,
\begin{align}
  \frac{\pi\bnu}{2} p_{i\bnu}(\hr_b,\hr_a)
  &\simeq
   -\frac{d_p}{\vkap_0}\sin(\vphi+\tht_{+}) ,
    \qquad
  \frac{\pi\bnu}{2} s_{i\bnu}(\hr_b,\hr_a)
   \simeq
   -\frac{\vkap_0d_s}{\bz_b\bz_a}\sin(\vphi+\tht_{-}) ,
  \label{eq:pimu_sin}
\end{align}
where
\begin{align}
  \vkap_0
  =(p_bp_a)^{-1/2}
  =\{(1+\bz_b^2)(1+\bz_a^2)\}^{1/4} ,
    \qquad
  \hr_{a,b}
  =\bnu\bz_{a,b}
   \qquad
  (\bnu\in\mathbb{R}^{+}) .
\end{align}
$\vphi$ is given by Eq.(\ref{eq:vphi}) for $\bz=\bz_b$ and $\bz'=\bz_a$,
\begin{align}
  \vphi
  =\vphi(\bz_b,\bz_a)
  =\bnu\{\xi(\bz_b)-\xi(\bz_a)\} ,
    \qquad
  p_{a,b}
  =p(\bz_{a,b})
  =(1+\bz_{a,b}^2)^{-1/2} .
  \label{eq:variables}
\end{align}
$\tht_{\pm}$ in Eqs.(\ref{eq:Ups_p_tf}-\ref{eq:pimu_sin}) is given as
\begin{align}
  &
  \tht_{\pm}
  =\tan^{-1}c_{\pm}
  =\phi_{\pm}(p_a)-\phi_{\pm}(p_b)
   ,\qquad
  c_{+}
  =\frac{\Ups_1^p(p_b,p_a)}{\Ups_0^p(p_b,p_a)}
   ,\qquad
  c_{-}
  =\frac{\Ups_1^s(p_b,p_a)}{\Ups_0^s(p_b,p_a)} ,
  \label{tht}
\end{align}
where
\begin{align}
  \bigg\{{ \phi_{+}(p)  \atop \phi_{-}(p) }\bigg\}
  &=\bigg\{{ \tan^{-1}\bar{c}_{+}  \atop \tan^{-1}\bar{c}_{-} }\bigg\}
   =\sum_{k=0}^{2n-1}\frac{(-1)^k}{\bnu^{2k+1}}
    \bigg\{{  \bu_{2k+1}(p)  \atop \bv_{2k+1}(p) }\bigg\} ,
    \qquad
  \bar{c}_{+}
  =\frac{\bU_n(p)}{\hU_n(p)} ,
    \qquad
  \bar{c}_{-}
  =\frac{\bV_n(p)}{\hV_n(p)} .
\end{align}
$\bu_{2k+1}$ and $\bv_{2k+1}$ for $k\in\{0,1,2,3\}$ are given as follows,
\begin{alignat}{2}
  \bu_1
  &=u_1 ,
    \qquad&
  \bu_3
  &=u_3+u_1(u_1^2/3-u_2) ,
   \label{eq:bu3}
   \\
  \bv_1
  &=v_1 ,
    \qquad&
  \bv_3
  &=v_3+v_1(v_1^2/3-v_2) ,
\end{alignat}
\begin{align}
  \bu_5
  &=u_5
   +u_1(u_1^4/5-u_4)
   +(u_2-u_1^2)(u_2u_1-u_3) ,
   \\
  \bv_5
  &=v_5
   +v_1(v_1^4/5-v_4)
   +(v_2-v_1^2)(v_2v_1-v_3) ,
   \label{eq:bv5}
\end{align}
\begin{align}
  \bu_7
  &=
    u_7
   +u_1(u_3^2-u_2^3+2u_4u_2-u_6)
   +u_1^2(u_5-3u_2u_3)
    \nonumber\\&\quad
   +u_1^3(2u_2^2-u_4)
   +u_1^4(u_3 -u_2u_1)
   +u_1^7/7
   +u_3(u_2^2-u_4)
   -u_2u_5 ,
   \\
  \bv_7
  &=
    v_7
   +v_1(v_3^2-v_2^3+2v_4v_2-v_6)
   +v_1^2(v_5-3v_2v_3)
    \nonumber\\&\quad
   +v_1^3(2v_2^2-v_4)
   +v_1^4(v_3 -v_2v_1)
   +v_1^7/7
   +v_3(v_2^2-v_4)
   -v_2v_5 .
   \label{eq:bv7}
\end{align}
Substituting $u_j$ and $v_j$, given by Eqs.(\ref{eq:uv}), into
Eqs.(\ref{eq:bu3}-\ref{eq:bv7}), we rewrite them as
\begin{alignat}{2}
  (\bu_1,\bv_1)
  &=\sum_{\ell=0}^{1}\frac{(\mfa_{\ell}^{+},\mfa_{\ell}^{-})}{2^3}p^{2\ell+1} ,
    \qquad&
  (\bu_3,\bv_3)
  &=\sum_{\ell=0}^{3}\frac{(\mfb_{\ell}^{+},\mfb_{\ell}^{-})}{2^7}p^{2\ell+3} ,
   \label{buv_13}
   \\
  (\bu_5,\bv_5)
  &=\sum_{\ell=0}^{5}\frac{(\mfc_{\ell}^{+},\mfc_{\ell}^{-})}{2^{10}}p^{2\ell+5} ,
    \qquad&
  (\bu_7,\bv_7)
  &=\sum_{\ell=0}^{7}\frac{(\mfd_{\ell}^{+},\mfd_{\ell}^{-})}{2^{15}}p^{2\ell+5} .
   \label{buv_57}
\end{alignat}
The coefficients of the series (\ref{buv_13}-\ref{buv_57}) are given as follows,
\begin{alignat}{4}
  \mfa_{0}^{+}
  &=1 ,
     \qquad&
  \mfa_{1}^{+}
  &=-\frac{5}{3};
     \qquad&
  \mfa_{0}^{-}
  &=-3,
     \qquad&
  \mfa_{1}^{-}
  &=\frac{7}{3} ;
   \label{mfa_pm}
   \\
  \mfb_{0}^{+}
  &=\frac{5^2}{3},
     \qquad&
  \mfb_{1}^{+}
  &=-\frac{3^2\cd59}{5},
     \qquad&
  \mfb_{2}^{+}
  &=13\cd17,
     \qquad&
  \mfb_{3}^{+}
  &=-\frac{5\cd13\cd17}{3^2} ;
   \label{mfab_pls}
   \\
  \mfb_{0}^{-}
  &=-3\cd7,
     \qquad&
  \mfb_{1}^{-}
  &=\frac{11\cd79}{5},
     \qquad&
  \mfb_{2}^{-}
  &=-3^2\cd5\cd7,
     \qquad&
  \mfb_{3}^{-}
  &=\frac{7\cd11\cd19}{3^2} ,
\end{alignat}
\begin{alignat}{3}
  \mfc_{0}^{+}
  &=\frac{29\cd37}{5},
     \qquad&
  \mfc_{1}^{+}
  &=-\frac{3^2\cd67\cd83}{7},
     \qquad&
  \mfc_{2}^{+}
  &=\frac{2\cd227\cd823}{3^2} ,
   \\
  \mfc_{3}^{+}
  &=-2\cd59\cd761,
     \qquad&
  \mfc_{4}^{+}
  &=5^2\cd3313,
     \qquad&
  \mfc_{5}^{+}
  &=-\frac{5^2\cd3313}{3} ;
   \\
  \mfc_{0}^{-}
  &=-\frac{3^2\cd211}{5},
     \qquad&
  \mfc_{1}^{-}
  &=\frac{3\cd24001}{7},
     \qquad&
  \mfc_{2}^{-}
  &=-\frac{2\cd5\cd49547}{3^2} ,
   \\
  \mfc_{3}^{-}
  &=2\cd31\cd1831,
     \qquad&
  \mfc_{4}^{-}
  &=-5\cd7\cd2897,
     \qquad&
  \mfc_{5}^{-}
  &=\frac{7\cd70753}{3\cd5} ,
   \label{mfc_mns}
\end{alignat}
\begin{alignat}{4}
  \mfd_{0}^{+}
  &=\frac{375733}{7} ,
     \qquad&
  \mfd_{1}^{+}
  &=-3407083 ,
     \qquad&
  \mfd_{2}^{+}
  &=38038059 ,
     \qquad&
  \mfd_{3}^{+}
  &=-168434443 ,
    \\
  \mfd_{4}^{+}
  &=\frac{1852155413}{5} ,
     \qquad&
  \mfd_{5}^{+}
  &=-432038457 ,
     \qquad&
  \mfd_{6}^{+}
  &=256406305 ,
     \qquad&
  \mfd_{7}^{+}
  &=-\frac{1282031525}{3\cd7} ;
   \\
  \mfd_{0}^{-}
  &=-\frac{543483}{7} ,
     \qquad&
  \mfd_{1}^{-}
  &=4451141 ,
     \qquad&
  \mfd_{2}^{-}
  &=-47257957 ,
     \qquad&
  \mfd_{3}^{-}
  &=202607973 ,
    \\
  \mfd_{4}^{-}
  &=-\frac{2177459227}{5} ,
     \qquad&
  \mfd_{5}^{-}
  &=499237207 ,
     \qquad&
  \mfd_{6}^{-}
  &=-292312335 ,
     \qquad&
  \mfd_{7}^{-}
  &=\frac{206530429}{3} .
\end{alignat}

\clearpage

\subsection{Modified Bessel functions of purely imaginary order}
\label{sec:dunster_mod_Bess}

In examining the pole structure of the fields in the Laplace domain for $|k\beta|<k_y^n$,
as shown in Eqs.(\ref{eq:CP_P_mod}-\ref{eq:CP_S_mod}),
we deal with the modified Bessel functions of purely imaginary order
$K_{i\bnu}(x)$ and $I_{i\bnu}(x)$, where $\bnu\in\bbR$ and $x\in\bbR^{+}$.
$K_{i\bnu}(x)\in\mathbb{R}$ and $I_{i\bnu}(x)\in\mathbb{C}$ in general.
Also, we must know the asymptotic behavior of the Green functions $\mfG_{\pm}^n$ for
$\nu\to\pm i\infty$ in order to examine the convergence of the Bromwich integral
given by Eq.(\ref{eq:cGeb_def}).

We consider the modified Bessel functions $L_\nu(z)$ and $K_\nu(z)$ for
$\nu=i\bnu$ $(\bnu\in\bbR)$ and $z=x\in\bbR^{+}$.
The power series representations of $L_{i\bnu}(x)$ and $K_{i\bnu}(x)$
with respect to $x$ are given by Eqs.(2.6) and (2.8) in \cite{dunster},
\begin{align}
  \bigg\{{ L_{i\bnu}(x) \atop K_{i\bnu}(x) }\bigg\}
  &=\bigg[\frac{\pi\bnu}{\sinh(\pi\bnu)}\bigg]^{\!1/2} \sum_{j=0}^\infty
    \frac{(x/2)^{2j}}{j!h_{\bnu,j}}
    \bigg\{{ \cos\vphi_{\bnu,j} \atop -\sin\vphi_{\bnu,j} }\bigg\}
   ,\qquad
  \vphi_{\bnu,j}
  =\bnu\log(x/2)-\phi_{\bnu,j} .
   \label{eq:LK_ibnu}
\end{align}
$h_{\bnu,j}$ and $\phi_{\bnu,j}$ are given by Eqs.(\ref{eq:phij_hj}-\ref{eq:phij}).
Neglecting the higher order terms with respect to $\bnu^{-1}$ in Eq.(\ref{eq:LK_ibnu}),
the asymptotic expressions of Eqs.(\ref{eq:Pnu_ba}-\ref{eq:Snu_ba})
for $\nu=i\bnu$ and $\bnu\to\pm\infty$ are given as
\begin{alignat}{2}
  \bigg\{{ p_{i\bnu}(ix,iy) \atop s_{i\bnu}(ix,iy) }\bigg\}
   &=\frac{2}{\pi}\bigg\{{ -P_{i\bnu}(x,y) \atop S_{i\bnu}(x,y) }\bigg\}
   &&\simeq \bigg\{{ -1/\bnu \atop \bnu/xy }\bigg\}
    \frac{2}{\pi}\{\sin[\bnu\log(x/y)]+O(\bnu^{-2})\} ,
   \label{eq:p_ibnu_ixy}
   \\
  \bigg\{{ q_{i\bnu}(ix,iy) \atop r_{i\bnu}(ix,iy) }\bigg\}
   &=\frac{2}{i\pi}\bigg\{{ Q_{i\bnu}(x,y) \atop R_{i\bnu}(x,y) }\bigg\}
   &&\simeq \bigg\{{ 1/y \atop -1/x }\bigg\}
     \frac{2}{i\pi}\{\cos[\bnu\log(x/y)]+O(\bnu^{-2})\} ,
   \label{eq:s_ibnu_ixy}
\end{alignat}
where $x,y\in\mathbb{R}^{+}$.
Eqs.(\ref{eq:p_ibnu_ixy}-\ref{eq:s_ibnu_ixy}) agree with
Eqs.(\ref{eq:pimu_asymp_mu_inf}-\ref{eq:simu_asymp_mu_inf})
for $\hr=ix$ and $\hr'=iy$.
The cross products of purely imaginary order $\nu=i\bnu\in i\mathbb{R}$ and arguments
are oscillatory with respect to $\bnu$.

According to Eqs.(4.8) and (4.14) in \cite{dunster}, we can expand
$K_{i\bnu}(\br)$ and $L_{i\bnu}(\br)$ in the uniform asymptotic series using
the Airy functions, similar to Eqs.(\ref{eq:JY_uae}-\ref{eq:dJY_uae})
which are those of $J_{\nu}(\hr)$ and $Y_{\nu}(\hr)$,
\begin{align}
  K_{i\bnu}(\bnu z)
  &=\pi e^{-\pi\bnu/2}
    \frac{2^{1/2}}{f(z)}
    \bigg[
      \frac{\Ai(\bu)}{\bnu^{1/3}}\sum_{j=0}^{n}\frac{a_j(-\zeta)}{\bnu^{2j}}
     +\frac{\Ai'(\bu)}{\bnu^{5/3}}\sum_{j=0}^{n-1}\frac{b_j(-\zeta)}{\bnu^{2j}}
     +\veps_{2n+1}^{(0)}(\bnu,-\zeta)
    \bigg] ,
   \label{eq:Kibnu_uae}
   \\
  L_{i\bnu}(\bnu z)
  &=\frac{\pi e^{\pi\bnu/2}}{2\sinh(\pi\bnu)}
    \cd\frac{2^{1/2}}{f(z)}
    \bigg[
      \frac{\Bi(\bu)}{\bnu^{1/3}}\sum_{j=0}^{n}\frac{a_j(-\zeta)}{\bnu^{2j}}
     +\frac{\Bi'(\bu)}{\bnu^{5/3}}\sum_{j=0}^{n-1}\frac{b_j(-\zeta)}{\bnu^{2j}}
     +\veps_{2n+1}^{(\ast)}(\bnu,-\zeta)
    \bigg] ,
   \label{eq:Libnu_uae}
   \\
  K_{i\bnu}'(\bnu z)
  &=\pi e^{-\pi\bnu/2}
    \frac{2^{1/2}f(z)}{z}
   \bigg[
    \frac{\Ai'(\bu)}{\bnu^{2/3}}
    \sum_{j=0}^{n}\frac{d_j(-\zeta)}{\bnu^{2j}}
   +\frac{\Ai(\bu)}{\bnu^{4/3}}\sum_{j=0}^{n-1}\frac{c_j(-\zeta)}{\bnu^{2j}}
   +\bveps_n^{(0)}(\bnu,-\zeta)
   \bigg] ,
   \label{eq:K_prm_uae}
   \\
  L_{i\bnu}'(\bnu z)
  &=\frac{\pi e^{\pi\bnu/2}}{2\sinh(\pi\bnu)}
    \cd\frac{2^{1/2}f(z)}{z}
    \bigg[
      \frac{\Bi'(\bu)}{\bnu^{2/3}}\sum_{j=0}^{n}\frac{d_j(-\zeta)}{\bnu^{2j}}
     +\frac{\Bi(\bu)}{\bnu^{4/3}}\sum_{j=0}^{n-1}\frac{c_j(-\zeta)}{\bnu^{2j}}
     +\bveps_n^{(\ast)}(\bnu,-\zeta)
    \bigg] ,
   \label{eq:L_prm_uae}
\end{align}
where $\bnu\in\mathbb{R}^{+}$ and $|\arg z|<\pi$.
Eq.(\ref{eq:Kibnu_uae}) is also given by Exercise\,10.6 (p.425), Chap.\,11 in \cite{olver}.
$f(z)$ is the function given by Eq.(\ref{eq:f_z_zeta}).
The argument of the Airy functions $\bu$ can be positive or negative, depending on the sign 
of $\zeta$, similar to $u$ appearing in Eqs.(\ref{eq:JY_uae}-\ref{eq:dJY_uae}),
\begin{align}
  \bu
  =-\bnu^{2/3}\zeta ,
   \qquad
  \eta_{\pm}(z)
  =\frac{2}{3}(\pm\zeta)^{3/2} .
\end{align}
$\eta_{\pm}$ is given by Eqs.(\ref{eq:zeta_p}-\ref{eq:zeta_m}).
The coefficients $(a_j,b_j,d_j,c_j)$ are given by Eqs.(\ref{eq:aj_bj}-\ref{eq:cj_dj}).
According to Eqs.(4.6) in \cite{dunster}, they have the following symmetry
with respect to the sign of $\zeta$,
\begin{alignat}{2}
  a_j(-\zeta)
  &=(-1)^ja_j(\zeta) ,
    \qquad&
  b_j(-\zeta)
  &=(-1)^jb_j(\zeta) ,
   \label{eq:aj_bj_symm}
   \\
  d_j(-\zeta)
  &=(-1)^jd_j(\zeta) ,
    \qquad&
  c_j(-\zeta)
  &=-(-1)^jc_j(\zeta) .
  \label{eq:cj_dj_symm}
\end{alignat}
Eqs.(\ref{eq:cj_dj_symm}) are gotten from Eqs.(\ref{eq:djcj_ajbj}) and (\ref{eq:aj_bj_symm}).
We substitute $\nu=i\bnu$, $\bbar=\bnu z$ and $\abar=\bnu z'$ into the cross products
(\ref{eq:Pnu_ba}-\ref{eq:Snu_ba}),
\begin{alignat}{2}
  P_{i\bnu}(\bnu z,\bnu z')
  &=\frac{\sinh(\pi\bnu)}{\pi}
    \{L_{i\bnu}(\bnu z)K_{i\bnu}(\bnu z')-K_{i\bnu}(\bnu z)L_{i\bnu}(\bnu z')\}
  &&=-\frac{\pi}{2}p_\nu(\nu z,\nu z') ,
  \label{eq:P_ibnu}
   \\
  Q_{i\bnu}(\bnu z,\bnu z')
  &=\frac{\sinh(\pi\bnu)}{\pi}
    \{L_{i\bnu}(\bnu z)K_{i\bnu}'(\bnu z')-K_{i\bnu}(\bnu z)L_{i\bnu}'(\bnu z')\}
  &&=-i\frac{\pi}{2}q_\nu(\nu z,\nu z') ,
   \\
  R_{i\bnu}(\bnu z,\bnu z')
  &=\frac{\sinh(\pi\bnu)}{\pi}
    \{L_{i\bnu}'(\bnu z)K_{i\bnu}(\bnu z')-K_{i\bnu}'(\bnu z)L_{i\bnu}(\bnu z')\}
  &&=-i\frac{\pi}{2}r_\nu(\nu z,\nu z') ,
   \\
  S_{i\bnu}(\bnu z,\bnu z')
  &=\frac{\sinh(\pi\bnu)}{\pi}
    \{L_{i\bnu}'(\bnu z)K_{i\bnu}'(\bnu z')-K_{i\bnu}'(\bnu z)L_{i\bnu}'(\bnu z')\}
  &&=\frac{\pi}{2}s_\nu(\nu z,\nu z') .
  \label{eq:S_ibnu}
\end{alignat}

In Eqs.(\ref{eq:Kibnu_uae}-\ref{eq:L_prm_uae}) $n\in\mathbb{N}$ is the index
to denote the order of the error.
If $n\gg1$, $n$ is not so important in using Eqs.(\ref{eq:Kibnu_uae}-\ref{eq:L_prm_uae}).
The error terms $\veps_{2n+1}^{(\ell)}$ for $\ell\in\{0,\pm1\}$,
which are involved in Eqs.(\ref{eq:Kibnu_uae}) and (\ref{eq:berr}),
are bounded as shown by Eq.(9.03) in p.418, Sec.\,9, Chap.\,11 in \cite{olver}.
But it is enough to know Eq.(\ref{eq:err_bnd})
which is the order of magnitude of the error.
$\veps_{2n+1}^{(\ast)}$ in Eq.(\ref{eq:Libnu_uae}) is the error term of $L_{i\bnu}$,
\begin{align}
  \veps_{2n+1}^{(\ast)}(\bnu,-\zeta)
  &=e^{-\pi i/6}\veps_{2n+1}^{(1)}(\bnu,-\zeta)
   +e^{\pi i/6}\veps_{2n+1}^{(-1)}(\bnu,-\zeta) .
   \label{eq:berr}
\end{align}
$\bveps_n^{(0)}$ and $\bveps_n^{(\ast)}$ in
Eqs.(\ref{eq:K_prm_uae}-\ref{eq:L_prm_uae}) are given as
\begin{align}
  \bigg\{{ \bveps_n^{(0)}(\bnu,-\zeta) \atop \bveps_n^{(\ast)}(\bnu,-\zeta) }\bigg\}
  &=\frac{\chi(-\zeta)-\rd_{\zeta}}{\bnu}
    \bigg\{{\veps_{2n+1}^{(0)}(\bnu,-\zeta)\atop \veps_{2n+1}^{(\ast)}(\bnu,-\zeta)}\bigg\}
   +\frac{\zeta b_n(-\zeta)+c_n(-\zeta)}{\bnu^{2n+4/3}}
    \bigg\{{ \Ai(-\bnu^{2/3}\zeta) \atop \Bi(-\bnu^{2/3}\zeta) }\bigg\} .
\end{align}
$\chi$ is the function given by Eq.(\ref{eq:M}).

\section{Airy functions}

The Wronskian of the Airy functions $\Ai$ and $\Bi$ is given by 10.4.10 in
\cite{abramo_stegun},
\begin{align}
  W[\Ai(u),\Bi(u)]
  =\mfq(u,u)
  =\frac{1}{\pi}
   \qquad
  (u\in\mathbb{C}) .
   \label{eq:wronski_AiBi}
\end{align}
$\{\mfp,\mfq,\mfr,\mfs\}$ are given by Eqs.(\ref{eq:cp_airy_p}-\ref{eq:cp_airy_sr})
which are the cross products of $\Ai$ and $\Bi$.
Similar to Eqs.(\ref{eq:qr_p0}-\ref{eq:qr_s0}), 
when $u$ and $v$ satisfy $\mfp(u,v)=0$ or $\mfs(u,v)=0$,
we can rewrite $\mfq(u,v)$ and $\mfr(u,v)$ using Eq.(\ref{eq:wronski_AiBi}),
\begin{alignat}{4}
  \mfp(u,v)
  &=0
   \qquad&\Ra&\qquad&
  \pi\mfq(u,v)
  &=\frac{\Ai(u)}{\Ai(v)}
   =\frac{\Bi(u)}{\Bi(v)}
   ,\qquad&
  \pi\mfr(u,v)
  &=-\frac{\Ai(v)}{\Ai(u)}
   =-\frac{\Bi(v)}{\Bi(u)} ,
   \\
  \mfs(u,v)
  &=0
   \qquad&\Ra&\qquad&
  \pi\mfq(u,v)
  &=\frac{\Ai'(v)}{\Ai'(u)}
   =\frac{\Bi'(v)}{\Bi'(u)}
   ,\qquad&
  \pi\mfr(u,v)
  &=-\frac{\Ai'(u)}{\Ai'(v)}
   =-\frac{\Bi'(u)}{\Bi'(v)} .
\end{alignat}
%

\subsection{Asymptotic expansion of the Airy functions for large argument}

The asymptotic series of the Airy functions for $u\to\infty$ are given by
10.4.59-10.4.67 in \cite{abramo_stegun},
\begin{alignat}{3}
  \Ai(u)
  &\simeq
  \frac{e^{-U}}{2\pi^{1/2}u^{1/4}} \sum_{j=0}^{\infty}(-1)^j\frac{C_j}{U^j}
   ,\qquad&
  \Ai'(u)
  &\simeq
  -\frac{u^{1/4}e^{-U}}{2\pi^{1/2}} \sum_{j=0}^{\infty}(-1)^j\frac{D_j}{U^j}
  \qquad&&
  (|\arg u|<\pi) ,
   \label{eq:Ai_Aip}
  \\
  \Bi(u)
  &\simeq
  \frac{e^{U}}{\pi^{1/2}u^{1/4}} \sum_{j=0}^{\infty}\frac{C_j}{U^j}
   ,\qquad&
  \Bi'(u)
  &\simeq
  \frac{u^{1/4}e^{U}}{\pi^{1/2}} \sum_{j=0}^{\infty}\frac{D_j}{U^j}
  \qquad&&
  (|\arg u|<\pi/3) .
   \label{eq:Bi_Bip}
\end{alignat}
The Airy functions for negative large argument are expanded as shown in
10.4.60-10.4.67 of \cite{abramo_stegun},
\begin{align}
  \Ai(-u)
  &\simeq
  \frac{u^{-1/4}}{\pi^{1/2}}
  \{
     \Omg_0^{+}(U)\sin(U+\pi/4)
    -\Omg_1^{+}(U)\cos(U+\pi/4)
  \} ,
  \label{eq:Ai_mns_ae}
   \\
  \Bi(-u)
  &\simeq
  \frac{u^{-1/4}}{\pi^{1/2}}
  \{
     \Omg_0^{+}(U)\cos(U+\pi/4)
    +\Omg_1^{+}(U)\sin(U+\pi/4)
  \} ,
   \\
  \Ai'(-u)
  &\simeq
  -\frac{u^{1/4}}{\pi^{1/2}}
  \{
     \Omg_0^{-}(U)\cos(U+\pi/4)
    +\Omg_1^{-}(U)\sin(U+\pi/4)
  \} ,
   \\
  \Bi'(-u)
  &\simeq
  \frac{u^{1/4}}{\pi^{1/2}}
  \{
     \Omg_0^{-}(U)\sin(U+\pi/4)
    -\Omg_1^{-}(U)\cos(U+\pi/4)
  \} ,
  \label{eq:Bi_prm_mns}
\end{align}
where $|\arg u|<2\pi/3$.
$U$ involved in Eqs.(\ref{eq:Ai_Aip}-\ref{eq:Bi_prm_mns}) is given as
\begin{align}
  U
  =\frac{2}{3}u^{3/2}
  =\nu\frac{2}{3}(\pm\zeta)^{3/2}
  =\nu\eta_{\pm}(z)
   ,\qquad
  u=\nu^{2/3}(\pm\zeta) .
   \label{eq:arg_airy}
\end{align}
$\eta_{\pm}$ and $\zeta$ are given by Eqs.(\ref{eq:zeta_p}-\ref{eq:zeta_m}).
$\Omg_{0,1}^{\pm}$ represents
\begin{align}
  \bigg\{{ \Omg_0^{+}(U) \atop \Omg_0^{-}(U)}\bigg\}
  =\sum_{j=0}^{\infty}\frac{(-1)^j}{U^{2j}}\bigg\{{ C_{2j} \atop D_{2j}}\bigg\} ,
    \qquad
  \bigg\{{ \Omg_1^{+}(U) \atop \Omg_1^{-}(U)}\bigg\}
  =\sum_{j=0}^{\infty}\frac{(-1)^j}{U^{2j+1}}\bigg\{{ C_{2j+1} \atop D_{2j+1}}\bigg\} .
  \label{alp01_beta01}
\end{align}
When $|U|\gg1$, $\Omg_0^{\pm}$ has a value close to 1.
On the other hand, $\Omg_1^{\pm}$ is of $O(U^{-1})$, \ie,
\begin{align}
  \Omg_0^{\pm}
  \approx 1+O(U^{-1})
  \gg |\Omg_1^{\pm}|
  =O(U^{-1})
   \qquad
  (|U|\gg1) .
\end{align}
The coefficients $C_s$ and $D_s$, involved in Eqs.(\ref{alp01_beta01}), are
given by 10.4.58 in \cite{abramo_stegun},
\begin{align}
  C_0
  &=D_0=1
  ,\qquad
  D_s
  =-\frac{6s+1}{6s-1}C_s
   \qquad
   (s\in\mathbb{N}),
  \label{eq:Coeff_Dj}
   \\
  C_s
  &=\frac{\Gamma(3s+1/2)}{ {54}^s s! \Gamma(s+1/2)}
   =\frac{(2s+1)(2s+3)\cdots(6s-1)}{ {6}^{3s}\cd s!} .
  \label{eq:Coeff_Cj}
\end{align}
$C_s$ and $D_s$ have the following relations to $\lam_s$ and $\kap_s$
given by Eqs.(\ref{eq:kap_lam}),
\begin{align}
  \bigg\{{ C_s \atop D_s }\bigg\}
  =\bigg(\frac{2}{3}\bigg)^s\bigg\{{ \lam_s \atop \kap_s}\bigg\}
   \qquad
   (s\in\mathbb{Z}_0^{+}) .
\end{align}
$C_s$ and $D_s$ up to $s=5$ are given as follows,
\begin{align}
  &
  \bigg({C_1 \atop D_1}\bigg)
  =\frac{1}{2^3\cdot3^2}\bigg({5 \atop -7}\bigg)
   ,\qquad
  \bigg({C_2 \atop D_2}\bigg)
  =\frac{5\cdot7}{2^7\cdot3^4}\bigg({11 \atop -13}\bigg)
   ,\qquad
  \bigg({C_3 \atop D_3}\bigg)
  =\frac{5\cdot7\cdot11\cdot13}{2^{10}\cdot3^7}\bigg({17 \atop -19}\bigg) ,
  \label{CD_012}
   \\
  &
  \bigg({C_4 \atop D_4}\bigg)
  =\frac{5\cdot7\cdot11\cdot13\cdot17\cdot19}{2^{15}\cdot3^9}
   \bigg({23 \atop -25}\bigg)
   ,\qquad
  \bigg({C_5 \atop D_5}\bigg)
  =\frac{5^2\cd7\cd11\cd13\cd17\cd19\cd23}{2^{18}\cd3^{11}}
    \bigg({29 \atop -31}\bigg) .
\end{align}
Their recurrence relations are given as
\begin{align}
  \frac{C_{s+1}}{C_s}
  =\frac{s}{2}+\frac{C_1}{s+1} ,
    \qquad
  \frac{D_{s+1}}{D_s}
  =\frac{s}{2}+\frac{D_1}{s+1}
   \qquad(s\in\mathbb{Z}_0^{+}) .
\end{align}
Therefore $C_s$ and $D_s$ for $s\geq2$ are given respectively using $C_1$ and $D_1$,
\begin{alignat}{2}
  C_s
  &=\frac{C_1}{s!}\prod_{l=2}^{s}\bigg\{\frac{l(l-1)}{2}+C_1\bigg\} ,
    \qquad
  D_s
  =\frac{D_1}{s!}\prod_{l=2}^{s}\bigg\{\frac{l(l-1)}{2}+D_1\bigg\} .
\end{alignat}

Using Eqs.(\ref{eq:Ai_mns_ae}-\ref{eq:Bi_prm_mns}), we expand the cross products of
the Airy functions, given by Eqs.(\ref{eq:cp_airy_p}-\ref{eq:cp_airy_sr}),
for $u\to\infty$ and $u'\to\infty$,
\begin{align}
  \pi(uu')^{1/4}\mfp(-u,-u')
  &=f_{+}^{(s)}(U,U')\sin\phi_u+f_{+}^{(c)}(U,U')\cos\phi_u ,
   \label{eq:mfp_mm}
   \\
  \pi(uu')^{-1/4}\mfs(-u,-u')
  &=f_{-}^{(s)}(U,U')\sin\phi_u+f_{-}^{(c)}(U,U')\cos\phi_u ,
   \label{eq:mfs_mm}
   \\
  \pi(u/u')^{1/4}\mfq(-u,-u')
  &=\{\Omg_0^{+}(U)\Omg_0^{-}(U')+\Omg_1^{+}(U)\Omg_1^{-}(U')\}\cos\phi_u
   \nonumber\\&\quad
   +\{\Omg_1^{+}(U)\Omg_0^{-}(U')-\Omg_0^{+}(U)\Omg_1^{-}(U')\}\sin\phi_u ,
   \label{eq:mfq_mm}
   \\
  -\pi(u'/u)^{1/4}\mfr(-u,-u')
  &=
    \{\Omg_0^{-}(U)\Omg_0^{+}(U')+\Omg_1^{-}(U)\Omg_1^{+}(U')\}\cos\phi_u
   \nonumber\\&\quad
   +\{\Omg_1^{-}(U)\Omg_0^{+}(U')-\Omg_0^{-}(U)\Omg_1^{+}(U')\}\sin\phi_u ,
   \label{eq:mfr_mm}
\end{align}
where
\begin{align}
  \phi_u
  =U-U' .
\end{align}
$f_{\pm}^{(s)}$ and $f_{\pm}^{(c)}$ are given as follows,
\begin{alignat}{2}
  f_{+}^{(s)}(U,U')
  &=\Omg_0^{+}(U)\Omg_0^{+}(U')+\Omg_1^{+}(U)\Omg_1^{+}(U')
  &&=\sum_{j,l=0}^{\infty}\frac{(-1)^{j+l}}{U^{2j}U'^{2l}}
    \bigg(C_{2j}C_{2l}+\frac{C_{2j+1}C_{2l+1}}{UU'}\bigg) ,
   \\
  f_{+}^{(c)}(U,U')
  &=\Omg_0^{+}(U)\Omg_1^{+}(U')-\Omg_1^{+}(U)\Omg_0^{+}(U')
  &&=\sum_{j,l=0}^{\infty}
    \frac{(-1)^{j+l}}{U^{2j}U'^{2l}}
    \bigg(\frac{C_{2j}C_{2l+1}}{U'}-\frac{C_{2j+1}C_{2l}}{U}\bigg) ,
   \\
  f_{-}^{(s)}(U,U')
  &=\Omg_0^{-}(U)\Omg_0^{-}(U')+\Omg_1^{-}(U)\Omg_1^{-}(U')
  &&=\sum_{j,l=0}^{\infty}\frac{(-1)^{j+l}}{U^{2j}U'^{2l}}
    \bigg(D_{2j}D_{2l}+\frac{D_{2j+1}D_{2l+1}}{UU'}\bigg) ,
   \\
  f_{-}^{(c)}(U,U')
  &=\Omg_0^{-}(U)\Omg_1^{-}(U')-\Omg_1^{-}(U)\Omg_0^{-}(U')
  &&=\sum_{j,l=0}^{\infty}\frac{(-1)^{j+l}}{U^{2j}U'^{2l}}
    \bigg(\frac{D_{2j}D_{2l+1}}{U'}-\frac{D_{2j+1}D_{2l}}{U}\bigg) .
\end{alignat}

\clearpage

\subsection{Taylor series of the Airy functions for small argument}

Taylor series of the Airy functions around $u=0$ are given as follows,
\begin{align}
  \bigg\{{ \Ai(u) \atop \Bi(u) }\bigg\}
  &=\bigg\{{ \Ai(0) \atop \Bi(0) }\bigg\}
    \sum_{\ell=0}^{\infty}\frac{e_{\ell}u^{3\ell}}{(3\ell)!}
   +\bigg\{{ \Ai'(0) \atop \Bi'(0) }\bigg\}
    \sum_{\ell=0}^{\infty}\frac{e_{\ell}'u^{3\ell+1}}{(3\ell+1)!} ,
  \label{eq:Ai_ts}
   \\
  \bigg\{{ \Ai'(u) \atop \Bi'(u) }\bigg\}
  &=\bigg\{{ \Ai'(0) \atop \Bi'(0) }\bigg\}
    \sum_{\ell=0}^{\infty}\frac{e_{\ell}'u^{3\ell}}{(3\ell)!}
   +\bigg\{{ \Ai(0) \atop \Bi(0) }\bigg\}
    \sum_{\ell=1}^{\infty}\frac{e_{\ell}u^{3\ell-1}}{(3\ell-1)!} .
  \label{eq:Aip_ts}
\end{align}
$e_{\ell}$ and $e_{\ell}'$ are the coefficients,
excluding $1/(3\ell)!$ and $1/(3\ell\pm1)!$ respectively,
\begin{align}
  e_{\ell}
  =3^{\ell}\frac{\Gam(\ell+1/3)}{\Gam(1/3)}
  =\prod_{j=0}^{\ell-1}(3j+1) ,
    \qquad
  e_{\ell}'
  =3^{\ell}\frac{\Gam(\ell+2/3)}{\Gam(2/3)}
  =\prod_{j=0}^{\ell-1}(3j+2) .
  \label{eq:elp}
\end{align}
The second expressions of (\ref{eq:elp}) using $\prod_{j=0}^{\ell-1}$ hold
for $\ell\in\mathbb{N}$.
$e_{\ell}$ and $e_{\ell}'$ up to $\ell=4$ are given as follows,
\begin{alignat}{5}
  e_0&=1,
   \qquad&
  e_1&=1,
   \qquad&
  e_2&=4,
   \qquad&
  e_3&=28,
   \qquad&
  e_4&=280,
   \\
  e_0'&=1,
   \qquad&
  e_1'&=2,
   \qquad&
  e_2'&=10,
   \qquad&
  e_3'&=80,
   \qquad&
  e_4'&=880.
\end{alignat}
Using Eqs.(\ref{eq:Ai_ts}-\ref{eq:Aip_ts}), the cross products of the Airy functions
(\ref{eq:cp_airy_p}-\ref{eq:cp_airy_sr}) are expanded as
\begin{align}
  \pi\mfp(u,v)
  &=\sum_{\ell=0}^{\infty}\sum_{\kap=0}^{\infty}
    \frac{e_{\ell}e_{\kap}'}{(3\ell)!(3\kap+1)!}
    (u^{3\ell}v^{3\kap+1} -u^{3\kap+1}v^{3\ell})
  \label{eq:mfp_ts}
   \\
  &=(v-u)
   +\frac{1}{3!}\Big\{uv(u^2-v^2)+\frac{1}{2}(v^4-u^4)\Big\}
   \nonumber\\&\quad
   +\frac{1}{3\cd4!}
    \Big\{(uv)^3(v-u)+\frac{2}{5}uv(u^5-v^5)+\frac{1}{7}(v^7-u^7)\Big\}
   +O(u^{10},v^{10}) ,
   \\
  \pi\mfs(u,v)
  &=\sum_{\ell=1}^{\infty}\sum_{\kap=0}^{\infty}
    \frac{e_{\ell}e_{\kap}'}{(3\ell-1)!(3\kap)!}
    (u^{3\ell-1}v^{3\kap}-u^{3\kap}v^{3\ell-1})
   \\
  &=\frac{1}{2!}(u^2-v^2)
   +\frac{1}{3!}\Big\{(uv)^2(v-u)+\frac{1}{5}(u^5-v^5)\Big\}
   \nonumber\\&\quad
   +\frac{1}{6!}
    \Big\{8(uv)^3(u^2-v^2)+5(uv)^2(v^4-u^4)+\frac{1}{2}(u^8-v^8)\Big\}
   +O(u^{11},v^{11}) ,
\end{align}
and
\begin{align}
  \pi\mfq(u,v)
  &=\sum_{\ell=0}^{\infty}\sum_{\kap=0}^{\infty}
    \frac{e_{\ell}e_{\kap}'}{(3\ell)!(3\kap)!}
    \Big(u^{3\ell}v^{3\kap} -\frac{3\ell}{3\kap+1}u^{3\kap+1}v^{3\ell-1}\Big)
  \label{eq:mfq_ts}
   \\
  &=\sum_{\kap=0}^{\infty}\frac{e_0e_{\kap}'}{(3\kap)!}v^{3\kap}
   +\sum_{\ell=1}^{\infty}\sum_{\kap=0}^{\infty}
    \frac{e_{\ell}e_{\kap}'}{(3\ell)!(3\kap)!}
    \Big(u^{3\ell}v^{3\kap} -\frac{3\ell}{3\kap+1}u^{3\kap+1}v^{3\ell-1}\Big)
   \\
  &=1
   +\bigg(\frac{v^3}{3} +\frac{u^3}{6} -\frac{uv^2}{2}\bigg)
   +\frac{v^2}{3!}\bigg\{\frac{v^4}{12}+u^3\bigg(\frac{v}{3}-\frac{u}{4}\bigg)\bigg\}
   \nonumber\\&\quad
   +\frac{v^2}{6!}
    \bigg\{
      \frac{10}{63}v^7
     +u^3\bigg(\frac{5}{3}v^4 -\frac{5}{7}u^4 +\frac{4}{3}u^3v -2uv^3\bigg)
    \bigg\}
   +O(u^{12},v^{12}) ,
   \\
  -\pi\mfr(u,v)
  &=\sum_{\ell=0}^{\infty}\sum_{\kap=0}^{\infty}
    \frac{e_{\ell}e_{\kap}'}{(3\ell)!(3\kap)!}
    \Big(u^{3\kap}v^{3\ell} -\frac{3\ell}{3\kap+1}u^{3\ell-1}v^{3\kap+1}\Big)
  \label{eq:mfr_ts}
   \\
  &=\sum_{\kap=0}^{\infty}\frac{e_0e_{\kap}'}{(3\kap)!}u^{3\kap}
   +\sum_{\ell=1}^{\infty}\sum_{\kap=0}^{\infty}
    \frac{e_{\ell}e_{\kap}'}{(3\ell)!(3\kap)!}
    \Big(u^{3\kap}v^{3\ell} -\frac{3\ell}{3\kap+1}u^{3\ell-1}v^{3\kap+1}\Big)
   \\
  &=1
   +\bigg(\frac{u^3}{3}+\frac{v^3}{6} -\frac{u^2v}{2}\bigg)
   +\frac{u^2}{3!}\bigg\{\frac{u^4}{12}+v^3\bigg(\frac{u}{3}-\frac{v}{4}\bigg)\bigg\}
   \nonumber\\&\quad
   +\frac{u^2}{6!}
    \bigg\{
      \frac{10}{63}u^7
     +v^3\bigg(\frac{5}{3}u^4 -\frac{5}{7}v^4 +\frac{4}{3}uv^3 -2u^3v\bigg)
    \bigg\}
   +O(u^{12},v^{12}) .
  \label{eq:mfr_ts9}
\end{align}
We use Eqs.(\ref{eq:mfp_ts}-\ref{eq:mfr_ts9}) in deriving the asymptotic expression 
of the zeroth pole of $\mfG_{-}^n$ in appendix \ref{sec:0th_0}.

\clearpage

\section{Zeros of the cross products of the Bessel functions with respect to the order}
\label{sec:cochran}

We review Cochran's study \cite{cochran} on the zeros of $p_\nu(z,z')$ and $s_\nu(z,z')$ in
the $\nu$-plane (order domain).
The cross products of the Bessel functions
$t_{\nu}(z,z')=\{p_{\nu},q_{\nu},r_{\nu},s_{\nu}\}$ are defined in
Eqs.(\ref{eq:CP_pq}-\ref{eq:CP_sr}).
We begin with the Bessel differential equations for $p_\mu(\xi,t\xi)$ and
$r_{\mu}(\xi,t\xi)$ with respect to $t\in\mathbb{R}^{+}$
which denotes the ratio of the second argument of the cross products to the first one,
\begin{align}
  t\xi\rd_t\{tq_\mu(\xi,t\xi)\}
  +\{(t\xi)^2-\mu^2\}p_\mu(\xi,t\xi)
  =0 ,
    \qquad
  t\xi\rd_t\{ts_{\mu}(\xi,t\xi)\}
  +\{(t\xi)^2-\mu^2\}r_{\mu}(\xi,t\xi)
  =0 .
  \label{eq:bessel_cochran}
\end{align}
$\mu,\nu\in\mathbb{C}$ in general.
$\xi\in\mathbb{A}_0^{+}=\{\mathbb{R}_0^{+},i\mathbb{R}^{+}\}$ is a parameter.
From Eqs.(\ref{eq:bessel_cochran}), we get
\begin{align}
  \int_\tau^1\frac{dt}{t\tau} \omg
  \bigg\{{ p_\mu(\xi,t\xi)p_\nu(\eta,t\eta) \atop r_\mu(\xi,t\xi)r_\nu(\eta,t\eta) }\bigg\}
  &=\eta 
    \bigg\{{ p_\mu(\xi,\tau\xi)q_\nu(\eta,\tau\eta) \atop
             r_\mu(\xi,\tau\xi)s_\nu(\eta,\tau\eta) }\bigg\}
   -\xi 
    \bigg\{{ p_\nu(\eta,\tau\eta)q_\mu(\xi,\tau\xi) \atop
             r_\nu(\eta,\tau\eta)s_\mu(\xi,\tau\xi) }\bigg\} ,
   \label{eq:watson134}
\end{align}
where
\begin{align}
  \omg
  =(\mu^2-\nu^2)-t^2(\xi^2-\eta^2) .
\end{align}
$\eta\in\mathbb{A}_0^{+}$ is also a parameter, similar to $\xi$.
We define $\tau$ as $\tau=r_a/r_b\in(0,1)$ which denotes the ratio of
the radius of the inner wall to the outer wall of the curved pipe.
We do not need to consider the case $\tau=1$ for which
Eq.(\ref{eq:watson134}) becomes a trivial identity.
For a typical accelerator, $\tau$ has a value slightly smaller than 1 as $0<1-\tau\ll1$
since the bending radius $\rho$ is usually much larger than the width of the pipe
$w=r_b-r_a$.
Eq.(\ref{eq:watson134}) is given by Eq.(1a) in \cite{cochran}
where ``$k$''$=\tau^{-1}\in(1,\infty)$ is used instead of $\tau$.
As described in p.590-591 of \cite{cochran}, if we limit the range of $\tau$ to
either $(0,1)$ or $(1,\infty)$, the current discussion does not lose the generality
since $p_\nu(z,z')$ and $s_\nu(z,z')$ are antisymmetric with respect to the exchange of
$z$ and $z'$.

\subsection{Absence of complex zero}
\label{sec:no_cmplx_zero}

We first show that the cross products $p_\nu(\xi,t\xi)$ and $s_\nu(\xi,t\xi)$ have 
no complex zero in the $\nu$-plane.
Substituting $\eta=\xi$ into Eq.(\ref{eq:watson134}), we rewrite it as follows,
\begin{align}
  \frac{\mu^2-\nu^2}{\tau\xi}\int_\tau^1 \frac{dt}{t}p_\mu(\xi,t\xi)p_\nu(\xi,t\xi)
  &=p_\mu(\xi,\tau\xi)q_\nu(\xi,\tau\xi)-p_\nu(\xi,\tau\xi)q_\mu(\xi,\tau\xi) ,
  \label{eq:watson_xyeq}
   \\
  \frac{\mu^2-\nu^2}{\tau\xi}\int_{\tau}^{1}\frac{dt}{t}r_{\mu}(\xi,t\xi)r_{\nu}(\xi,t\xi)
  &=r_{\mu}(\xi,\tau\xi)s_{\nu}(\xi,\tau\xi)-r_{\nu}(\xi,\tau\xi)s_{\mu}(\xi,\tau\xi) .
  \label{eq:watson_xyeq_s}
\end{align}
Into Eqs.(\ref{eq:watson_xyeq}-\ref{eq:watson_xyeq_s}) we substitute $\mu=\nu^{\ast}$
which is the complex conjugate of $\nu$,
\begin{align}
  \frac{(\nu^{\ast})^2-\nu^2}{\tau\xi}\int_\tau^1 \frac{dt}{t}|p_\nu(\xi,t\xi)|^2
  &=p_\nu^{\ast}(\xi,\tau\xi)q_\nu(\xi,\tau\xi)
   -p_\nu(\xi,\tau\xi)q_\nu^{\ast}(\xi,\tau\xi) ,
  \label{eq:watson134x}
   \\
  \frac{(\nu^{\ast})^2-\nu^2}{\tau\xi}
  \int_{\tau}^{1}\frac{dt}{t}|r_{\nu}(\xi,t\xi)|^2
  &=r_{\nu}^{\ast}(\xi,\tau\xi)s_{\nu}(\xi,\tau\xi)
   -r_{\nu}(\xi,\tau\xi)s_{\nu}^{\ast}(\xi,\tau\xi) ,
  \label{eq:watson134x_s}
\end{align}
where we used the second equation of (\ref{eq:cp_symm})
which means mirror symmetry of the cross products in the $\nu$-plane.
The integrals in Eqs.(\ref{eq:watson134x}-\ref{eq:watson134x_s}) cannot be zero for
$\tau\in\mathbb(0,1)$.

We define $\kap$ and $\lam$ respectively as the zeros of
$p_\nu(\xi,\tau\xi)$ and $s_\nu(\xi,\tau\xi)$ with respect to $\nu$, \ie,
\begin{alignat}{4}
  p_\kap(\xi,\tau\xi)
  &=0
    \qquad&\Lra&\qquad&
  p_{\kap^{\ast}}(\xi,\tau\xi)
  &=p_{\kap}^{\ast}(\xi,\tau\xi)
  &&=0 ,
  \label{eq:plam0}
   \\
  s_{\lam}(\xi,\tau\xi)
  &=0
   \qquad&\Lra&\qquad&
  s_{\lam^{\ast}}(\xi,\tau\xi)
  &=s_{\lam}^{\ast}(\xi,\tau\xi)
  &&=0 .
  \label{eq:slam0}
\end{alignat}
Their complex conjugate, $\kap^{\ast}$ and $\lam^{\ast}$, are also zeros of
$p_\nu$ and $s_\nu$ since they have symmetry in the $\nu$-plane
as shown in Eq.(\ref{eq:cp_symm}).
Substituting $\nu=\kap$ and $\nu=\lam$ respectively into
Eq.(\ref{eq:watson134x}) and Eq.(\ref{eq:watson134x_s}), we get
\begin{align}
  \Re\kap\Im\kap\int_{\tau}^{1}\frac{dt}{t}|p_{\kap}(\xi,t\xi)|^2=0 ,
   \qquad
  \Re\lam\Im\lam\int_{\tau}^{1}\frac{dt}{t}|r_{\lam}(\xi,t\xi)|^2=0 .
  \label{eq:watson134x_lam}
\end{align}
In the first equation of (\ref{eq:watson134x_lam}),
either $\Re\kap$ or $\Im\kap$ must be zero, 
\ie, $\kap\in\mathbb{A}=\{\mathbb{R},i\mathbb{R}\}$
since the integral cannot be zero for $\tau\in\mathbb(0,1)$.
This means that $p_{\nu}$ has no complex zero with respect to $\nu$,
\ie, $p_{\nu}$ has zeros only on the real and imaginary axes in the $\nu$-plane.
Similarly, $s_{\nu}$ has zeros only on the real and imaginary axes of the $\nu$-plane
since either $\Re\lam$ or $\Im\lam$ must be zero, \ie, $\lam\in\mathbb{A}$,
according to the second equation of (\ref{eq:watson134x_lam}).

\subsection{Simple zeros on the real and imaginary axes of the order domain}

We examine the order of the zeros of $p_{\nu}$ and $s_{\nu}$ with respect to $\nu$
(order domain).
Taking the limit of $\mu\to\nu$ in Eqs.(\ref{eq:watson_xyeq}-\ref{eq:watson_xyeq_s}),
their asymptotic limits are given by Eq.(3) in \cite{cochran},
\begin{align}
  \frac{2\nu}{\tau\xi}\int_\tau^1 \frac{dt}{t}p_\nu^2(\xi,t\xi)
  &=q_\nu(\xi,\tau\xi)\rd_\nu p_\nu(\xi,\tau\xi)
   -p_\nu(\xi,\tau\xi)\rd_\nu q_\nu(\xi,\tau\xi) ,
  \label{eq:watson134xnu}
   \\
  \frac{2\nu}{\tau\xi}\int_{\tau}^{1}\frac{dt}{t}r_{\nu}^2(\xi,t\xi)
  &=s_{\nu}(\xi,\tau\xi)\rd_{\nu}r_{\nu}(\xi,\tau\xi)
   -r_{\nu}(\xi,\tau\xi)\rd_{\nu}s_{\nu}(\xi,\tau\xi) .
  \label{eq:watson134xnu_s}
\end{align}
Into Eqs.(\ref{eq:watson134xnu}) and (\ref{eq:watson134xnu_s}), respectively,
we substitute $\nu=\kap$ and $\nu=\lam$ which satisfy Eqs.(\ref{eq:plam0}-\ref{eq:slam0}),
\begin{alignat}{2}
  \frac{2\kap}{\tau\xi}\int_\tau^1 \frac{dt}{t}p_\kap^2(\xi,t\xi)
  &=q_\kap(\xi,\tau\xi)[\rd_\nu p_\nu(\xi,\tau\xi)]_{\nu=\kap}
   \qquad
   &&(\kap\in\mathbb{A}) ,
  \label{eq:watson_lambda}
   \\
  \frac{2\lam}{\tau\xi}\int_{\tau}^{1}\frac{dt}{t}r_{\lam}^2(\xi,t\xi)
  &=-r_{\lam}(\xi,\tau\xi)[\rd_{\nu}s_{\nu}(\xi,\tau\xi)]_{\nu=\lam}
   \qquad
   &&(\lam\in\mathbb{A}) .
  \label{eq:watson_lambda_s}
\end{alignat}
$p_\kap^2(\xi,t\xi)\in\mathbb{R}_0^{+}$ since $p_\kap(\xi,t\xi)\in\mathbb{R}$
for $\xi\in\mathbb{A}_0^{+}$.
Therefore the integral in Eq.(\ref{eq:watson_lambda}) has a value in $\mathbb{R}^{+}$
for $\tau\in(0,1)$, \ie, the integral cannot be zero.
It follows that the L.H.S. of Eq.(\ref{eq:watson_lambda}) cannot be zero unless $\kap=0$.
In addition, $q_\kap(\xi,\tau\xi)$ on the R.H.S. of Eq.(\ref{eq:watson_lambda}) is not zero
since $\kap$ is a zero of $p_{\nu}(\xi,\tau\xi)$ as defined in Eq.(\ref{eq:plam0}).
$p_\kap(\xi,\tau\xi)$ and $q_\kap(\xi,\tau\xi)$ cannot be zero at the same time,
because  $W[J_\kap(\tau\xi),Y_\kap(\tau\xi)]=2/(\pi\tau\xi)$
which holds for $\forall\tau$ and $\forall\xi$.
That is, if $p_\kap=0$ and $q_\kap=0$, it contradicts with the Wronskian of
$J_{\kap}$ and $Y_{\kap}$.
The same is true for Eq.(\ref{eq:watson_lambda_s}).
Since $q_\kap(\xi,\tau\xi)\ne0$ and $r_{\lam}(\xi,\tau\xi)\ne0$ respectively
in Eqs.(\ref{eq:watson_lambda}) and (\ref{eq:watson_lambda_s}),
\begin{align}
  [\rd_\nu p_\nu(\xi,\tau\xi)]_{\nu=\kap}\ne0
  \qquad(\kap\ne0)
   ,\qquad\quad
  [\rd_\nu s_\nu(\xi,\tau\xi)]_{\nu=\lam}\ne0
  \qquad(\lam\ne0) ,
\end{align}
because the L.H.S. of Eqs.(\ref{eq:watson_lambda}-\ref{eq:watson_lambda_s}) cannot be zero
unless $\kap=0$ and $\lam=0$.
Accordingly, $\nu=\kap$ and $\nu=\lam$ are respectively simple zeros of $p_\nu$ and $s_\nu$
unless $\kap=0$ and $\lam=0$ which we will discuss in appendix \ref{sec:orgn_odr}.

\subsection{Zero of the second order at the origin of the order domain}
\label{sec:orgn_odr}

We examine the zero of $p_\nu(\xi,\tau\xi)$ and $s_\nu(\xi,\tau\xi)$ at $\nu=0$.
In general, $p_0(\xi,\tau\xi)$ is not zero for $\tau\in(0,1)$ and $\xi\in\mathbb{A}_0^{+}$.
$p_0(\xi,\tau\xi)$ becomes zero if $\tau$ and $\xi$ have a particular relation.
In terms of geometry in the $(\xi,\tau,\nu)$-space, $\nu=0$ denotes a plane
which intersects with the hypersurface $p_\nu(\xi,\tau\xi)=0$.
In our study on synchrotron radiation, for given $m$ and $n$
which are the horizontal and vertical mode numbers,
the cutoff wavenumber $k=k_m^n$, defined in Eq.(\ref{eq:kmn}), forms a subset
$\mathbb{S}_m^n$ which is the intersection of the plane $\nu_m^n=0$ and
the hypersurface $p_{\nu_m^n}(\hr_b,\hr_a)=0$ in the space $\mathbb{S}$
given by Eq.(\ref{eq:hypersurface}).
For a given $n$, there are infinite number of the hypersurfaces in $\mathbb{S}$
since $m\in\mathbb{N}$ as shown in Fig.\ref{fig:poles}.
The same is also true for $s_\nu$, \ie,
$k=\bk_m^n$ ($m\in\mathbb{Z}_0^{+}$), defined in Eq.(\ref{eq:bkmn}), 
forms a subset $\bar{\mathbb{S}}_m^n$ 
which is the intersection of $\mu_m^n=0$ and $s_{\mu_m^n}(\hr_b,\hr_a)=0$ in $\mathbb{S}$.
To the following, we assume that $(\tau,\xi)\in\mathbb{S}_m^n$ or $\bar{\mathbb{S}}_m^n$
which corresponds to $k=k_m^n$ or $\bk_m^n$ defined in Eqs.(\ref{eq:kmn}-\ref{eq:bkmn}),
\begin{align}
   \nu_m^n=0
  \quad\Ra\quad
  p_0(\xi,\tau\xi)=0 ,
   \qquad\text{or}\qquad
  \mu_m^n=0
  \quad\Ra\quad
  s_0(\xi,\tau\xi)=0 .
  \label{eq:p0_xitau_0}
\end{align}
$\nu_m^n=0$ and $\mu_m^n=0$ do not hold simultaneously unless $\rho=\infty$, \ie,
$s_0(\xi,\tau\xi)\ne0$ when $p_0(\xi,\tau\xi)=0$, and vice versa.
$\rd_\nu p_\nu(\xi,\tau\xi)=0$ at $\nu=0$ since $p_\nu(\xi,\tau\xi)$ is even
with respect to $\nu$ as shown in Eq.(\ref{eq:cp_symm}).
Therefore the zero of $p_\nu(\xi,\tau\xi)$ at $\nu=0$ is second order
if $\rd_\nu^2p_\nu(\xi,\tau\xi)\ne0$ at $\nu=0$.
The same is also true for $s_{\nu}(\xi,\tau\xi)$.

In order to show that $p_\nu(\xi,\tau\xi)$ and $s_\nu(\xi,\tau\xi)$ each have
a zero of the second order at $\nu=0$,
dividing Eqs.(\ref{eq:watson134xnu}-\ref{eq:watson134xnu_s}) by $\nu$,
we take the limit of $\nu\to0$ for them,
\begin{align}
  \frac{2}{\tau\xi}\int_\tau^1 \frac{dt}{t}p_0^2(\xi,t\xi)
  &=q_0(\xi,\tau \xi)[\rd_\nu^2p_\nu(\xi,\tau \xi)]_{\nu=0} ,
  \label{eq:watson_nu0}
   \\
  \frac{2}{\tau\xi}\int_{\tau}^{1}\frac{dt}{t}r_0^2(\xi,t\xi)
  &=-r_0(\xi,\tau\xi)[\rd_{\nu}^2s_{\nu}(\xi,\tau\xi)]_{\nu=0} ,
  \label{eq:watson_nu0_s}
\end{align}
where the integrals are not zero since $p_0(\xi,t\xi)\in\mathbb{R}$ and
$r_0(\xi,t\xi)\in\mathbb{R}$ for $t\in(0,1)$.
Therefore the L.H.S. of Eqs.(\ref{eq:watson_nu0}-\ref{eq:watson_nu0_s}) cannot be zero.
Also, $q_0(\xi,\tau\xi)$ and $r_0(\xi,\tau\xi)$ on the R.H.S. of
Eqs.(\ref{eq:watson_nu0}-\ref{eq:watson_nu0_s}) are not zero,
because we assume $p_0(\xi,\tau\xi)=0$ and $s_0(\xi,\tau\xi)=0$ respectively in
Eqs.(\ref{eq:watson_nu0}) and (\ref{eq:watson_nu0_s}).
Accordingly,
\begin{align}
  [\rd_{\nu}^2p_{\nu}(\xi,\tau\xi)]_{\nu=0}\ne0 ,
    \qquad
  [\rd_{\nu}^2s_{\nu}(\xi,\tau\xi)]_{\nu=0}\ne0 .
   \label{eq:rdnu2_non0}
\end{align}
It follows that $p_\nu(\xi,\tau\xi)$ and $s_\nu(\xi,\tau\xi)$ each have a zero of
the second order at the origin $\nu=0$ when $\tau$ and $\xi$ satisfy
either $\nu_m^n=0$ or $\mu_m^n=0$.

\subsection{Coupling of the poles and the residue at the origin}
\label{sec:analytic}

We consider the residue of a fraction function at the origin of the $\nu$-plane,
which is a pole of the second order as shown in appendix \ref{sec:orgn_odr}.
As mentioned in sections \ref{sec:pole} and \ref{sec:analytic_continuation},
we can regard the pole of the second order at $\nu=0$ as the coupling of a pair of
simple poles which are located symmetrically on the real axis of the $\nu$-plane.
That is, the residue at $\nu=0$ is equal to the sum of the residues at the pair of
the simple real poles which coupled at $\nu=0$.
We will show it using the simple example given by Eqs.(\ref{eq:phi_simple}) and
(\ref{eq:frac_func}) since it is enough to examine the pole structure of
a fraction function in the vicinity of the origin.
We consider the following integral with respect to $\nu$,
which involves a function $F(\nu)$ having a real parameter $\delta$,
\begin{align}
  I
  =\int_{-\infty}^{\infty}F(\nu,\delta)d\nu ,
    \qquad
  F(\nu,\delta)
  =\frac{f(\nu,\delta)}{\nu^2-\delta} ,
    \qquad
  f(\pm\delta^{1/2},\delta)\ne0
   \qquad
  (\delta\in\mathbb{R}).
  \label{eq:Int_smpl_dbl}
\end{align}
The integrand $F$ is a fraction function of $\nu$,
whose denominator is even with respect to $\nu$.
We assume that the numerator $f$ is an analytic function for
$\forall\nu\in\mathbb{C}$ and $\forall\delta\in\mathbb{R}$.
In addition, we assume that $f\ne0$ at the points such that $\nu^2=\delta$.
If $\delta=0$, $F$ has a pole of the second order at $\nu=0$
as described below Eq.(\ref{eq:phi_simple}).
From Eq.(\ref{eq:Res2}),
we get $R(0)$ which is the residue of $F(\nu,0)$ at $\nu=0$,
\begin{align}
  R(0)=[\rd_\nu f(\nu,0)]_{\nu=0} .
   \label{eq:res_double}
\end{align}
If $\delta>0$, $F$ has a pair of simple poles
which symmetrically locate on the real axis of the $\nu$-plane,
\begin{align}
  \nu=\pm\lam ,
    \quad\text{where}\quad
  \lam=\delta^{1/2}\in\mathbb{R}^{+}
   \qquad
  (\delta>0) .
  \label{eq:lam_alp}
\end{align}
We define $R_1(\lam)$ as the residue of $F$ at the simple pole $\nu=\lam$.
We consider $S(\lam)$ which is the sum of the residues of $F$ at $\nu=\pm\lam$, \ie,
\begin{align}
  S(\lam)
  &=R_1(\lam)+R_1(-\lam)
   =\frac{f(\lam,\delta)-f(-\lam,\delta)}{2\lam} .
   \label{eq:sum_simple_res}
\end{align}
If we take the limit of $\lam\to0$ independent of $\delta$
(\ie, simply substituting $\delta=+0$ regardless of $\lam$) in Eq.(\ref{eq:sum_simple_res}),
it seems that $S(\lam)$ tends to be the partial derivative of $f$ with respect to $\lam$
as follows,
\begin{align}
  S(0)
  &=\lim_{\lam\to0}S(\lam)
   =[\rd_\lam f(\lam,\delta)]_{\lam=0,\delta=+0} .
   \label{eq:S0_res}
\end{align}
But the poles $\pm\lam$ depend on $\delta$ as shown in Eq.(\ref{eq:lam_alp}).
Therefore, in calculating $S(0)$ from $S(\lam)$,
we must take the limits of $\lam\to0$ and $\delta\to+0$ 
simultaneously in Eq.(\ref{eq:sum_simple_res}) while keeping $\lam^2=\delta$.
Eq.(\ref{eq:res_double}) agrees with the asymptotic limit $S(0)$ given by
Eq.(\ref{eq:S0_res}), \ie,
we can get the residue at the pole of the second order from the sum of the residues at
the coupled simple real poles.
Eq.(\ref{eq:S0_res}) seems to be intuitively obvious and trivial
since we assume that $f$ is an analytic function everywhere in the $\nu$-plane
like a smooth mountain range.

We show that Eq.(\ref{eq:S0_res}) is equivalent to Eq.(\ref{eq:res_double})
in a proper way.
Since we assume that $f$ is an analytic function, we expand $f$ in Taylor series with 
respect to both $\nu$ and $\delta$ around their origins,
\begin{align}
  f(\nu,\delta)=\sum_{m,n=0}^{\infty}\frac{\nu^m\delta^n}{m!n!}
  [\rd_\nu^m\rd_\delta^n f(\nu,\delta)]_{\nu=0,\delta=0} ,
    \qquad\text{where}\quad
  \rd_{\nu}
  =\frac{\rd}{\rd\nu} ,
    \qquad
  \rd_{\delta}
  =\frac{\rd}{\rd\delta} .
  \label{eq:f_taylor}
\end{align}
Substituting Eq.(\ref{eq:f_taylor}) into Eq.(\ref{eq:sum_simple_res}),
we take the limit of $\lam\to0$ which is equivalent to $\delta\to0$,
\begin{align}
  \lim_{\lam\to0}S(\lam)
  =\lim_{\delta\to0}\sum_{j,n=0}^{\infty}\frac{\delta^{j+n}}{(2j+1)!n!}
   [\rd_\lam^{2j+1}\rd_\delta^nf(\lam,\delta)]_{\lam=0,\delta=0}
  =[\rd_\lam f(\lam,\delta)]_{\lam=0,\delta=0} .
  \label{eq:lim_Slam}
\end{align}
Thus, $S(0)$ is equal to $R(0)$ given by Eq.(\ref{eq:res_double}).
We showed the equivalence between the residues at the coupled simple poles on the real axis
and that of the second order at the origin
using a simple function given by Eq.(\ref{eq:Int_smpl_dbl}).
But Eq.(\ref{eq:S0_res}) holds for any fraction function
if the denominator is even with respect to $\nu$,
which goes to zero of the second order for $\nu\to0$.

According to the above discussion, we can regard the pole of the second order as
the pair of the simple poles on the real axis, which coupled at $\nu=0$.
Instead of this interpretation, in terms of mathematics, we can regard the pole of
the second order as the pair of the simple imaginary poles
$\nu=\pm i(-\delta)^{1/2}\in i\mathbb{R}$ ($\delta\leq 0$) which coupled at $\nu=0$.
In terms of physics, however, we cannot regard the pole of the second order at $\nu=0$
as the pair of the simple imaginary poles,
because they are partitioned into the inside and outside 
of the contours of the Bromwich integral shown in Fig.\ref{fig:contour}
(p.\pageref{fig:contour}).
That is, we cannot use Eq.(\ref{eq:lim_Slam}) for $\delta<0$,
because it does not fit the contours in the $\nu$-plane,
which determine the causality of the transient electromagnetic field.

\clearpage

\section{Asymptotic expressions of the poles for large radius}
\label{sec:poles}

$\mfG_{\pm}^n$ is given by Eqs.(\ref{eq:mfGe}-\ref{eq:mfGb}) which are
the Green functions of the vertical components of the fields
in the Laplace domain $\nu\in\mathbb{C}$.
We will find two kinds of asymptotic expressions of the poles of
$\mfG_{\pm}^n$ in the $\nu$-plane for large radius $w/\rho\ll1$.
That is, we will find the asymptotic expressions of the zeros of
the cross products of the Beseel functions $p_\nu(\hr_b,\hr_a)$ and $s_\nu(\hr_b,\hr_a)$ 
with respect to the order $\nu$ under the two sets of assumptions
(\ref{eq:cond1}-\ref{eq:cond2}).
According to \cite{cochran} which we reviewed in appendix \ref{sec:cochran} using
our notation, $p_\nu$ and $s_\nu$ each have zeros only on the real and imaginary axes of 
the $\nu$-plane as described in Eqs.(\ref{eq:pnu_poles}-\ref{eq:snu_poles}),
\begin{alignat}{4}
  p_\nu(\hr_b,\hr_a)&=0
   \quad\text{for}~~
  \mfG_{+}^n
   \qquad&\Lra&\qquad
  \nu
  =\nu_m^n
  &&\in\mathbb{A}
   \qquad&&
  (m\in\mathbb{N}) ,
  \label{eq:pnu_eq0}
  \\
  s_\nu(\hr_b,\hr_a)&=0
   \quad\text{for}~~
  \mfG_{-}^n
   \qquad&\Lra&\qquad
  \nu
  =\mu_m^n
  &&\in\mathbb{A}
   \qquad&&
  (m\in\mathbb{Z}_0^{+}) ,
  \label{eq:snu_eq0}
\end{alignat}
where $\mathbb{A}=\{\mathbb{R},i\mathbb{R}\}$.
The indices $m$ and $n$ denote respectively the horizontal (= radial) and
vertical mode numbers of the fields in the pipe which has a rectangular cross section
as shown in Fig.\ref{fig:pipe3D}.
$\hr_a$ and $\hr_b$ are the dimensionless radii of the inner and outer walls of
the curved rectangular pipe,
\begin{align}
  \hr_{a,b}
  =k_r^nr_{a,b}
  =k_r^n\rho g_{a,b}
  ,\qquad
  r_{a,b}
  =\rho+x_{a,b}
    ,\qquad
  g_{a,b}
  =\frac{r_{a,b}}{\rho} .
  \label{eq:z_hr}
\end{align}
$k_r^n$ is the radial wavenumber of the field in the pipe,
given by Eqs.(\ref{eq:krn}).
$g_a$ and $g_b$ are given by Eq.(\ref{eq:coordinates}), which are the geometric factors $g$ 
on the sidewalls $r=r_a$ and $r_b$ in a constant bend $\vrho=\rho$.

In deriving the asymptotic expressions of the poles $\nu=(\nu_m^n,\mu_m^n)$,
it is enough to consider them on the positive axes
$\nu\in\mathbb{A}_0^{+}=\{\mathbb{R}_0^{+},i\mathbb{R}^{+}\}$
since $p_{\nu}$ and $s_{\nu}$ have mirror symmetry in the $\nu$-plane
as shown in Eqs.(\ref{eq:even_symm_ps}).
$\nu/\rho\in\mathbb{C}$ denotes the longitudinal wavenumber of the field in the bend
since $\kap\,(=i\nu)$, given by Eq.(\ref{eq:u_nu}), is the original Laplace variable
with respect to $s$ as defined in Eqs.(\ref{eq:Laplace}).
Therefore $\nu/\rho$ goes to $k_s$ for $\rho\to\infty$, where $k_s\in\mathbb{C}$ is
the Laplace variable of the field in the straight section, given by Eq.(\ref{eq:def_kx}).
Eqs.(\ref{eq:pnu_eq0}-\ref{eq:snu_eq0}) are the dispersion relations of $E_y$ and $B_y$ in 
the uniformly curved rectangular pipe as discussed in section \ref{sec:pole}.
On the other hand, in the straight pipe which has the same rectangular cross section as
the curved pipe, the dispersion relation of all the components of
the electromagnetic field is given by Eq.(\ref{eq:ks_mn}),
\begin{align}
  (k\beta)^2
  =(k_x^m)^2+(k_y^n)^2+(k_s^{mn})^2 ,
   \qquad
  k\beta=\frac{\omg}{c} ,
    \qquad
  k_x^m=\frac{m\pi}{w} ,
    \qquad
  k_y^n=\frac{n\pi}{h} .
   \label{eq:dispersion}
\end{align}
$v$ $(=\beta c)$ is the speed of the reference particle, defined in Eq.(\ref{eq:z}).
$\omg$ is the frequency defined in Eqs.(\ref{eq:Fourier_trans}-\ref{eq:omg}).
$k_x^m$, $k_y^n$ and $k_s^{mn}$ are the horizontal, vertical and longitudinal wavenumbers
of the field in the straight rectangular pipe.
From a mathematical point of view,
$k_s^{mn}$ is the pole of the Laplace domain field in
the straight pipe as shown in Eq.(\ref{eq:kxm}).
Although $k_s^{mn}\in\mathbb{A}$ similar to $\nu_m^n$ and $\mu_m^n$,
it is enough to consider $k_s^{mn}$ in $\mathbb{A}_0^{+}$
because of the symmetry in the $k_s$-plane.
The dispersion relation (\ref{eq:dispersion}) is symmetric with respect to
the exchange of $x_a$ and $x_b$.
Similarly, $\nu_m^n$ and $\mu_m^n$ are symmetric with respect to the exchange of
$x_a$ and $x_b$, because $p_{\nu}(\hr_b,\hr_a)$ and $s_{\nu}(\hr_b,\hr_a)$ are 
antisymmetric with respect to the exchange of $r_a$ and $r_b$.
Excluding the whispering gallery modes which we discussed in
sections \ref{sec:wgm} and \ref{sec:iwgm},
Eqs.(\ref{eq:pnu_eq0}-\ref{eq:snu_eq0}) must tend to be Eq.(\ref{eq:dispersion}) in
the limit of $\rho\to\infty$ with the pipe width $w=r_b-r_a=x_b-x_a$ kept constant.
From Eq.(\ref{eq:dispersion}), the radial wavenumber $k_r^n$ can be given using
$k_x^m$ and $k_s^{mn}$ which are the variables of the field in the straight pipe.
It follows that $k_r^n$ has the following two expressions
including the original definition (\ref{eq:krn}),
\begin{align}
  (k_r^n)^2
  =(k\beta)^2-(k_y^n)^2
  =(k_s^{mn})^2+(k_x^m)^2 .
  \label{eq:krn_apdx}
\end{align}
$k_s^{mn}$ for $m=0$ equals $k_r^n$.
We define $m$ for $\nu_m^n$ and $\mu_m^n$ respectively as it begins with 1 and 0
in accordance with $m$ in Eqs.(\ref{eq:dispersion}).
Thus, $\mfG_{-}^n$ has the pole of the zeroth radial mode $\nu=\mu_0^n$
which is somewhat special from the higher order radial modes ($m\in\mathbb{N}$)
as seen from Eqs.(\ref{eq:lim_nurho_ksmn}) and (\ref{eq:cXm}).

For large $|\nu|$, the uniform asymptotic expansion of the Bessel functions
$C_{\nu}(\hr)=\{J_{\nu}(\hr),Y_{\nu}(\hr)\}$ and their derivatives with respect to
the argument $\rd_{\hr}C_{\nu}(\hr)=C_{\nu}'(\hr)$ are given by
Eqs.(\ref{eq:JY_uae}-\ref{eq:dJY_uae}) where the variable $z$ is not the argument of
$C_{\nu}$ but the ratio of the argument $\hr$ to the order $\nu$ as follows,
\begin{align}
  z
  =\frac{\hr}{\nu}
  =g\frac{k_r^n}{\nu/\rho} ,
    \qquad
  z_{a,b}
  =\frac{\hr_{a,b}}{\nu} ;
    \qquad
  \lim_{\rho\to\infty}z
  =\frac{k_r^n}{k_s} ,
    \qquad
  \lim_{\rho\to\infty}(z^2-1)^{1/2}
  =\frac{k_x}{k_s} .
  \label{eq:z_uae}
\end{align}
Although $z\in\mathbb{C}$ in general, it is enough to consider the domain
$z\in\mathbb{A}_0^{+}$ in considering the poles.
$z$ denotes the ratio of the radial wavenumber $k_r^n$ to the longitudinal wavenumber
$\nu/\rho$, modified by the geometric factor $g\,(=r/\rho)$ in the curved pipe.
Therefore $(z^2-1)^{1/2}$ is a quantity which goes to the ratio of $k_x$ to $k_s$ in
the limit of $\rho\to\infty$ for $\forall x$.
$k_x$ and $k_s$ are given by Eqs.(\ref{eq:def_kx}),
which are the complex variables unlike $k_x^m$ and $k_s^{mn}$.
As seen from Eqs.(\ref{eq:BDE}) and (\ref{eq:zeta_p}-\ref{eq:zeta_m}),
$z=1$ is the transition point that $C_{\nu}$ and $C_{\nu}'$ switch from
exponential functions ($z<1$) to oscillatory functions ($z>1$) with respect to $U$
given by Eq.(\ref{eq:arg_airy}).

\subsection{Poles of the real and imaginary whispering gallery modes in a curved pipe}
\label{sec:whisper}

In Eq.(\ref{eq:nu_mt}) we defined $m_a$ as the largest $m$ such that
$C_{\nu}(\hr_a)$ and $C_{\nu}'(\hr_a)$ for $\nu=(\nu_m^n,\mu_m^n)\in\mathbb{R}^{+}$ behave
as exponential functions.
The outer wall of the curved pipe dominates the poles for $m\leq m_a$,
because $J_{\nu}(\hr_a)$ and $J_{\nu}'(\hr_a)$ in
$p_{\nu}(\hr_b,\hr_a)$ and $s_{\nu}(\hr_b,\hr_a)$
are exponentially small and negligible for $Y_{\nu}(\hr_a)$ and $Y_{\nu}'(\hr_a)$
which are exponentially large, according to Eqs.(\ref{eq:Ai_Aip}-\ref{eq:Bi_Bip}).
Therefore, when $m\leq m_a$ which denotes the real whispering gallery modes,
we can approximate the first equations of (\ref{eq:pnu_eq0}-\ref{eq:snu_eq0}) as
Eq.(3.2) in \cite{cochran_zero},
\begin{align}
  J_{\nu}(\hr_b)
  \simeq0
   ,\qquad
  J_{\nu}'(\hr_b)
  \simeq0
   ,\qquad\text{where}\quad
  1\ll\hr_a<(\nu_m^n,\mu_m^n)<\hr_b
   \quad\Lra\quad
  m\leq m_a .
   \label{Jb_zero}
\end{align}
This approximation implies the assumption that the inner wall of the curved pipe hardly 
affects the field.
It is equivalent to the model that the vacuum chamber has no inner wall
like a sector of a pillbox.
Olver studied the zeros of $J_{\nu}(\hr)$ and $J_{\nu}'(\hr)$ for $\nu\in\mathbb{R}^{+}$
in section 7 (p.343) of \cite{olver_0}.
On the basis of it, Cochran found the asymptotic solutions of Eqs.(\ref{Jb_zero}) with 
respect to $\nu$, which are given by Eqs.(3.4-3.5) in \cite{cochran_zero},
\begin{align}
  \frac{\{\nu_m^n,\mu_m^n\}}{\hr_b}
  &\simeq
   1+\sum_{j=1}^{\infty}\frac{\{A_m^{(j)},B_m^{(j)}\}}{\hr_b^{2j/3}}
   \qquad
   [ m\leq m_a,~ m\in(\mathbb{N},\mathbb{Z}_0^{+}) ] .
   \label{cochran_sol}
\end{align}
$A_m^{(j)}$ and $B_m^{(j)}$ are the coefficients of the $j$th order terms,
which are ``$R_r$'' and ``$S_r$'' ($j=$``$r$'') given by p.516 in \cite{cochran_zero}.
For simplicity, we rearrange them using $a_m$ and $b_m$ as follows,
\begin{alignat}{3}
  A_m^{(1)}
  &=-\frac{a_m}{2},
   \qquad&
  A_m^{(2)}
  &=\frac{a_m^2}{120},
   \qquad&
  A_m^{(3)}
  &=\frac{1}{70}\bigg(\frac{a_m^3}{40}-1\bigg) ,
   \label{eq:Ajm}
   \\
  B_m^{(1)}
  &=-\frac{b_m}{2},
   \qquad&
  B_m^{(2)}
  &=\frac{1}{5}\bigg(\frac{b_m^2}{24}+\frac{1}{b_m}\bigg),
   \qquad&
  B_m^{(3)}
  &=\frac{1}{25}\bigg(\frac{b_m^3}{112}+1+\frac{1}{b_m^3}\bigg) .
   \label{eq:Bjm}
\end{alignat}
$a_m$ and $b_m$ are those from which $-2^{-2/3}$ is factored out of
``$a_s$'' and ``$a_s'$'' given by 10.4.94-10.4.95 in \cite{abramo_stegun},
p.516 in \cite{cochran_zero} and Eqs.(A20) in \cite{olver_0},
\begin{align}
  (a_m,b_m)
  &=(d_m^{\pm})^{2/3}T_{\pm}(d_m^{\pm}/2),
    \qquad
  d_m^{\pm}
  =3\pi(m\mp1/4),
    \qquad
  m\in(\mathbb{N},\mathbb{Z}_0^{+}) .
   \label{eq:am_bm}
\end{align}
$(a_m,b_m)$ corresponds to $d_m^{\pm}=(d_m^{+},d_m^{-})$ and $T_{\pm}=(T_{+},T_{-})$.
``$a_s$'' and ``$a_s'$'' in \cite{cochran_zero,olver_0} are
the ``$s$''th zeros of $\Ai$ and $\Ai'$ respectively.
Their asymptotic expressions for large ``$s$'' are given by Eqs.(A19-A20) in
\cite{olver_0}.
The radial mode index ``$s$'' in \cite{cochran_zero,olver_0} corresponds to 
ours as ``$s$''=\,$(m,m+1)$ for $(\nu_m^n,\mu_m^n)$ in the present paper
since we defined it as $m\in(\mathbb{N},\mathbb{Z}_0^{+})$ for $(a_m,b_m)$
unlike ``$s$''$\in\mathbb{N}$ for both ``$a_s$'' and ``$a_s'$''.
$T_{+}$ and $T_{-}$ are those from which $\lam^{2/3}$ is
factored out of ``$T$'' and ``$U$'' given by Eq.(A19) in \cite{olver_0},
\begin{align}
   &
  T_{\pm}(\lam)
  =1+\sum_{\ell=1}^{\infty}\frac{c_\ell^{\pm}}{\lam^{2\ell}},
   \qquad\text{where}\qquad
  (c_1^{+},c_1^{-})
  =\bigg(\frac{5}{48},\, -\frac{7}{48}\bigg),
   \label{eq:Tpm_lam}
   \\
   &
  (c_2^{+},c_2^{-})
  =\bigg(\! -\frac{5}{36},\, \frac{35}{288}\bigg),
   \qquad
  (c_3^{+},c_3^{-})
  =\bigg(\frac{77125}{82944},\, -\frac{181223}{207360}\bigg).
   \label{eq:cj_pm}
\end{align}
The relative error of Eq.(\ref{cochran_sol}) for the numerical solution of
Eqs.(\ref{eq:pnu_eq0}-\ref{eq:snu_eq0}) is plotted in Fig.\ref{fig:nu_err_4K}
(p.\pageref{fig:nu_err_4K}).
Eq.(\ref{eq:am_bm}) is accurate for $m\geq4$
unless $m$ is close to $m_a$ given by Eq.(\ref{eq:mt}).
Using Eq.(\ref{eq:am_bm}), $\nu_m^n$ and $\mu_m^n$ for small $m$ such as $m\leq3$ are not 
so accurate, because Eq.(\ref{eq:am_bm}) is an asymptotic expression for 
large $m$ which means a negative large argument of $\Ai$ and $\Ai'$.
Especially, $\mu_0^n$ has a relatively large error of $O(10^{-4})$
since the first zero of $\Ai'$ is ``$a_1'$''$\approx-1.01879$
which is not large enough to use Eq.(\ref{eq:am_bm}).
In order to get the values of $(\nu_m^n,\mu_m^n)$ for $m\leq3$ more precisely,
instead of using Eq.(\ref{eq:am_bm}), we use the following values of
$a_m$ and $b_m$ which are gotten from the numerical values of ``$a_s$'' and ``$a_s'$''
given by Table 10.13 (p.478) in \cite{abramo_stegun},
\begin{alignat}{4}
   &\qquad&
  a_1
  &=3.71151416
   ,\qquad&
  a_2
  &=6.48921524
   ,\qquad&
  a_3
  &=8.76334248 ,
   \\
  b_0
  &=1.61723303
   ,\qquad&
  b_1
  &=5.15619225
   ,\qquad&
  b_2
  &=7.65143055
   ,\qquad&
  b_3
  &=9.78364058 .
\end{alignat}

We get the asymptotic limit of $\mu_0^n$ from Eq.(\ref{cochran_sol}) for $m=0$,
\begin{align}
  \lim_{\rho\to\infty}\frac{\mu_0^n}{\rho}
  =k_r^n
   ,\qquad
  \lim_{\rho\to\infty}\frac{d}{dk_r^n}\bigg(\frac{\mu_0^n}{\rho}\bigg)^2
  =2k_r^n .
    \label{eq:lim_mu0_sol}
\end{align}
This is used in Eq.(\ref{eq:cRm_ba}) for $m=0$.
$\nu_m^n/\rho$ and $\mu_m^n/\rho$ for $m\in\mathbb{N}$, given by Eq.(\ref{cochran_sol}), 
have no asymptotic limit for $\rho\to\infty$,
\ie, $(\nu_m^n,\mu_m^n)$ for $m\leq m_a$ does not exist in the straight pipe,
because $m_a\to0$ in the limit of $\rho\to\infty$ as shown in Eq.(\ref{eq:lim_gl_g12}).
Thus, the zeroth pole is special in this regard.

Eq.(\ref{cochran_sol}) is the asymptotic solution of Eqs.(\ref{eq:pnu_eq0}-\ref{eq:snu_eq0})
with respect to $\nu\in\mathbb{R}^{+}$, which is asymmetric with respect to
the exchange of $r_b$ and $r_a$ since it does not depend on $r_a$.
Eq.(\ref{cochran_sol}) is applicable to the modes for $m\leq m_a$
which is the left side of the vertical dashed line in
Fig.\ref{fig:nu_err_4K} (p.\pageref{fig:nu_err_4K}).
The error of $\nu_m^n$ and $\mu_m^n$ increases as $m$ gets close to $m_a$
since $\nu_m^n$ and $\mu_m^n$ get closer to $\hr_a$ for $m\to m_a$.
That is, for $m\to m_a$, $J_{\nu}(\hr_a)$ and $J_{\nu}'(\hr_a)$ in
Eqs.(\ref{eq:pnu_eq0}-\ref{eq:snu_eq0}) tend to be comparable to
$Y_{\nu}(\hr_a)$ and $Y_{\nu}'(\hr_a)$,
because $C_{\nu}(\hr_a)$ and $C_{\nu}'(\hr_a)$ at $\nu=(\nu_m^n,\mu_m^n)$ make
the transition between $m_a$ and $m_a+1$ as defined in Eq.(\ref{eq:nu_mt}).
Thus, when $m$ is close or equal to $m_a$, Eqs.(\ref{Jb_zero}) are not accurate
as the approximate expressions of Eqs.(\ref{eq:pnu_eq0}-\ref{eq:snu_eq0}).
As seen from Fig.\ref{fig:nu_err_4K}, Eq.(\ref{cochran_sol}) has a relatively large error of
$O(10^{-5}\sim 10^{-4})$ at $m=m_a$, similar to the one at $m=0$.
Regardless whether $m\leq m_a$ or $m>m_a$,
it is difficult to find the asymptotic expressions of the poles
which are accurate [\eg, $O(10^{-6})$ or less] around the transition mode $m=m_a$,
because it is difficult to deal with $\Ai(\pm u_a)$ and $\Ai'(\pm u_a)$ around $z_a=1$,
where $u_a=\nu^{2/3}(\pm \zeta_a)$ given by Eq.(\ref{eq:arg_airy}) for $\zeta=\zeta_a$
(\ie, $z=z_a$).

We consider the cross products $p_{\nu}(\hr_b,\hr_a)$ and $s_{\nu}(\hr_b,\hr_a)$ for
purely imaginary $\nu$ and $k_r^n$, \ie,
\begin{align}
  \nu
  =i\bnu
  \in i\mathbb{R}^{+}
   ,\qquad
  k_r^n
  =i\bk_r^n
  \in i\mathbb{R}^{+}
   ,\qquad
  \hr_{b,a}
  =i\brr_{b,a}
  \in i\mathbb{R}^{+}
   ,\qquad
  \brr_{b,a}
  =\bk_r^nr_{b,a}
  \in \mathbb{R}^{+} .
    \label{eq:Im_nu_krn}
\end{align}
According to Eqs.(\ref{eq:P_ibnu}) and (\ref{eq:S_ibnu}),
$p_{\nu}$ and $s_{\nu}$ have zeros for Eqs.(\ref{eq:Im_nu_krn}),
\begin{alignat}{5}
  p_{\nu}(\hr_b,\hr_a)
  &=0
   \qquad&\Lra&\qquad&
  P_{i\bnu}(\brr_b,\brr_a)
  &=0
   \qquad&\Ra&\qquad&
  K_{i\bnu}(\brr_a)
  &\simeq 0 ,
    \label{eq:Kibnu_zero}
   \\
  s_{\nu}(\hr_b,\hr_a)
  &=0
   \qquad&\Lra&\qquad&
  S_{i\bnu}(\brr_b,\brr_a)
  &=0
   \qquad&\Ra&\qquad&
  K_{i\bnu}'(\brr_a)
  &\simeq 0 .
    \label{eq:Kp_ibnu_zero}
\end{alignat}
The last equations of (\ref{eq:Kibnu_zero}-\ref{eq:Kp_ibnu_zero}) do not depend on $r_b$, 
\ie, their approximate zeros are determined only by the inner wall radius $r_a$.
Using Eqs.(\ref{eq:Kibnu_uae}) and (\ref{eq:K_prm_uae})
which are the uniform asymptotic series of $K_{i\bnu}$ and $K_{i\bnu}'$,
we get the asymptotic solutions of the last equations of
(\ref{eq:Kibnu_zero}-\ref{eq:Kp_ibnu_zero}) with respect to $\bnu$,
\begin{align}
  \frac{\{\bnu_m^n,\bmu_m^n\}}{\brr_a}
  &\simeq
   1+\sum_{j=1}^{\infty}\frac{\{A_m^{(j)},B_m^{(j)}\}}{(-1)^j\brr_a^{2j/3}}
    \qquad
   [m\leq m_b,~  m\in(\mathbb{N},\mathbb{Z}_0^{+})] .
   \label{eq:imag_whisper}
\end{align}
$m_b$ is defined in Eq.(\ref{eq:mb_def}) which is
the largest radial mode number of the imaginary whispering gallery modes.
We plot Eq.(\ref{eq:imag_whisper}) on the left side of the vertical dashed line
($m=m_b$) in Fig.\ref{fig:bnu_err_bkr} (p.\pageref{fig:bnu_err_bkr})
which is for $k_r^n\in i\mathbb{R}$.
Similar to Eq.(\ref{cochran_sol}), Eq.(\ref{eq:imag_whisper}) tends to be more accurate for
larger $m$ unless $m$ is close or equal to $m_b$.

\subsection{Symmetric asymptotic solutions of the poles with respect to the sidewalls}
\label{sec:symmetric_poles}

In appendix \ref{sec:whisper} we described the real whispering gallery modes
($m\leq m_a$) for $k_r^n\in\mathbb{R}$
and the imaginary whispering gallery modes ($m\leq m_b$) for $k_r^n\in i\mathbb{R}$.
Next we will consider the poles $\nu=(\nu_m^n,\mu_m^n)$ of the normal modes:
$m>m_a$ for $k_r^n\in\mathbb{R}$ and $m>m_b$ for $k_r^n\in i\mathbb{R}$.
These poles are the solutions of Eqs.(\ref{eq:pnu_eq0}-\ref{eq:snu_eq0})
such that are symmetric with respect to the exchange of $r_b$ and $r_a$.
Since $p_{\nu}(\hr_b,\hr_a)$ and $s_{\nu}(\hr_b,\hr_a)$ are antisymmetric
with respect to the exchange of $r_b$ and $r_a$,
Eqs.(\ref{eq:pnu_eq0}-\ref{eq:snu_eq0}) do not change for this exchange.
Excluding the real and imaginary whispering gallery modes,
the solutions of Eqs.(\ref{eq:pnu_eq0}-\ref{eq:snu_eq0}) satisfy the following asymptotic 
limit for $\rho\to\infty$ with the width of the pipe $w$ ($=r_b-r_a=x_b-x_a$) fixed,
\begin{align}
  \lim_{\rho\to\infty}\frac{(\nu_m^n,\mu_m^n)}{\rho}
  =k_s^{mn}
   ,\qquad
  m
  \geq
  \left\{
  \begin{array}{l}
    m_a+1 \\
    m_b+1
  \end{array}
  \right.
   \quad\text{for}\quad
  k_r^n
  \in
  \left\{
  \begin{array}{l}
    \mathbb{R} \\
    i\mathbb{R}
  \end{array}
  \right.  .
  \label{eq:asympt_lim}
\end{align}
As seen from Eqs.(\ref{eq:dispersion}), the longitudinal wavenumber $k_s^{mn}$ is symmetric 
with respect to the exchange of $x_b$ and $x_a$ since it is related to the square of
the horizontal wavenumber $k_x^m$ which depends on the width $w$.
Since Eq.(\ref{cochran_sol}) does not depend on $r_a$,
it does not satisfy Eq.(\ref{eq:asympt_lim}) except the zeroth pole $\mu_0^n$
which is special in this regard as described around Eq.(\ref{eq:lim_nurho_ksmn}).
Eq.(\ref{eq:asympt_lim}) is also required from the condition that Eqs.(\ref{eq:we_cX}) 
must be the asymptotic limit of Eqs.(\ref{eq:BDE_cRp}-\ref{eq:BDE_cRm}) for $\rho\to\infty$,
\ie, $\cR_{\pm}^{mn}$ goes to $\cX_{\pm}^m$ in the limit of $\rho\to\infty$
as shown by Eq.(\ref{eq:lim_cRcX}).
$\cR_{\pm}^{mn}$ and $\cX_{\pm}^m$ denote
the horizontal eigenfunctions of the curved pipe and the straight pipe respectively,
which are given by Eqs.(\ref{eq:cRp_ba}-\ref{eq:cRm_ba}) and (\ref{eq:cXp}-\ref{eq:cXm}).

We introduce an approximate solution of Eq.(\ref{eq:pnu_eq0}),
presented by Horvat and Prosen \cite{horvat_prosen}.
It is given by Eq.(26) in \cite{horvat_prosen} and 
Eq.(\ref{eq:HP26}) in the present paper using our notation.
The authors got it from Eq.(25) in \cite{horvat_prosen},
which is given by Eq.(\ref{eq:HP25}) using our notation.
Eqs.(25-26) in \cite{horvat_prosen} are symmetric with respect to the exchange of
$r_b$ and $r_a$.
In addition, Eqs.(25-26) in \cite{horvat_prosen} satisfy Eq.(\ref{eq:asympt_lim}).
According to our private communication with Horvat, however, they got
Eq.(25) in \cite{horvat_prosen} empirically by their own intuition,
combining the following two kinds of asymptotic expressions of $C_\nu(\hr)$ for
(i) $\nu\to\infty$ and $\hr=\text{fixed}$,
(ii) $\nu=\text{fixed}$ and $\hr\to\infty$.
That is, Eqs.(25-26) in \cite{horvat_prosen} are their conjectures
in the absence of proper proof.
In appendix \ref{sec:asympt_numu} we will show that Eq.(26) in \cite{horvat_prosen}
is the asymptotic solution of Eq.(\ref{eq:pnu_eq0}) to the zeroth order with respect to
$\eps_r$ where
\begin{align}
  \eps_r
  =\bigg(\frac{k_r^n}{k_x^m}\bigg)^2 ,
    \qquad
  \eps_s
  =\bigg(\frac{k_s^{mn}}{k_r^n}\bigg)^2 ,
    \qquad
  \eps_x
  =\bigg(\frac{k_x^m}{k_r^n}\bigg)^2 ,
    \qquad
  \eps
  =\frac{w}{\rho} .
  \label{eq:eps}
\end{align}
On the other hand, Eq.(25) in \cite{horvat_prosen} is a pseudo asymptotic solution of
Eq.(\ref{eq:pnu_eq0}) to the zeroth order with respect to $\eps_s$, 
because, to be rigorous, it is incorrect as the asymptotic solution of
Eq.(\ref{eq:pnu_eq0}) for $\eps_s\to0$, though it is valid by accident up to $O(\eps)$ 
and the zeroth order with respect to $(k_r^n\rho)^{-2}$ as shown later in
Eqs.(\ref{bg1_gl2}).
That is, Eqs.(25-26) in \cite{horvat_prosen} are approximate solutions
which roughly hold under the following conditions,
\begin{alignat}{5}
  \text{Eq.(\ref{cochran_sol}) in App.\,\ref{sec:whisper}:}
   \quad&&
  |\eps_x|
  &\ll1
   ,\qquad&
  k_s^{mn}
  &=k_r^n(1-\eps_x)^{1/2}
   ,\qquad&
  &|\nu/\rho|\gg k_x^m ,
   \label{eq:high_freq}
   \\
  \text{Eq.(26) in \cite{horvat_prosen}\,\,=\,\,Eq.(\ref{eq:HP26})
        in App.\,\ref{sec:asympt_numu}:}
   \quad&&
  |\eps_r|&\ll1
   ,\qquad&
  k_s^{mn}
  &=ik_x^m(1-\eps_r)^{1/2}
   ,\qquad&
  &|\nu/\rho|\sim k_x^m ,
   \label{eq:cond26}
   \\
  \text{Eq.(25) in \cite{horvat_prosen}\,\,=\,\,Eq.(\ref{eq:HP25}) in App.\,\ref{sec:HP2}:}
   \quad&&
  |\eps_s|&\ll1
   ,\qquad&
  k_x^m
  &=k_r^n(1-\eps_s)^{1/2}
   ,\qquad&
  &|\nu/\rho|\ll k_x^m .
   \label{eq:cond25}
\end{alignat}
The limit of $\eps_r\to0$ is inconsistent with the limit of $\eps_s\to0$, because
the dispersion relation (\ref{eq:krn_apdx}) must hold in the limit of $\rho\to\infty$.
In appendices \ref{sec:HP} and \ref{sec:HP2} we discuss Eqs.(25-26) in
\cite{horvat_prosen} in detail while comparing with the asymptotic solutions of the poles
which we will find in appendices \ref{sec:asympt_numu} and \ref{sec:HP2}.

We use the uniform asymptotic expansion of the Bessel functions in calculating
the cross products $p_{\nu}(\hr_b,\hr_a)$ and $s_{\nu}(\hr_b,\hr_a)$.
There are two kinds of the uniform asymptotic expansion:
Olver's expansion for $J_{\nu}(\nu z)$ and $Y_{\nu}(\nu z)$ ($\nu\in\mathbb{R}^{+}$), and
Dunster's expansion for $F_{\nu}(\nu z)$ and $G_{\nu}(\nu z)$ ($\nu\in i\mathbb{R}^{+}$),
which are respectively given by Eqs.(\ref{eq:JY_uae}-\ref{eq:dJY_uae}) and
(\ref{eq:Fimu_muz}-\ref{eq:Gimu_muz}).
These expansions are described in terms of the variables $(\nu,z)$ and $(\bnu,\bz)$,
\begin{align}
  \hr
  =k_r^nr
  =\nu z
  =\bnu\bz
   \qquad\text{where}\quad
  \nu=i\bnu
   ,\qquad
  \bz=iz .
  \label{eq:nu_bnu}
\end{align}
We use $z$ and $\bz$ for $r=r_{a,b}$ as in the second equation of (\ref{eq:z_uae}).
As we will demonstrate in Eqs.(\ref{eq:bnu_sols}-\ref{eq:nu_ks}),
we can rewrite the expressions of the poles from
$\bnu$ to $\nu$ through Eq.(\ref{eq:nu_bnu}) after the derivation.
In order to find the asymptotic expressions of the poles $\nu=(\nu_m^n,\mu_m^n)$ for
$m\in\mathbb{N}$ under the assumptions (\ref{eq:cond26}-\ref{eq:cond25}),
we use Eqs.(\ref{eq:Fimu_muz}-\ref{eq:Gimu_muz}),
because these series are given in terms of the trigonometric functions,
\ie, we do not have to further expand the Airy functions involved in
Eqs.(\ref{eq:JY_uae}-\ref{eq:dJY_uae}).
In short, it is not easy to rearrange the coefficients of the asymptotic expansions 
given by Eqs.(\ref{eq:JY_uae}-\ref{eq:dJY_uae}) and 
(\ref{eq:Ai_mns_ae}-\ref{eq:Bi_prm_mns})
in finding the asymptotic solutions to a higher order with respect to $\eps_{r,s,x}$.
But it is not so troublesome to use Eqs.(\ref{eq:JY_uae}-\ref{eq:dJY_uae}) only if
we calculate the leading order term in the series to find
the asymptotic expressions or asymptotic limits of the cross products for $\rho\to\infty$
as shown later in appendices \ref{sec:asympt_numu} and \ref{sec:AL_GR}.
We will find the asymptotic expression of the zeroth pole $\mu_0^n$ later in appendix
\ref{sec:0th_0} under an assumption different from Eq.(\ref{eq:high_freq}) since it requires
a special treatment which differs from the higher order radial modes ($m\in\mathbb{N}$).

We begin the derivation of the asymptotic expressions of the poles
$\nu=i\bnu=(\nu_m^n,\mu_m^n)\in\mathbb{A}$ of the normal modes
($m\geq m_{a,b}+1$) which satisfy Eqs.(\ref{eq:pnu_eq0}-\ref{eq:snu_eq0})
under assumption (\ref{eq:cond26}) or (\ref{eq:cond25}).
$m_a$ and $m_b$ are the largest real and imaginary whispering gallery mode numbers
given by Eqs.(\ref{eq:mt}) and (\ref{eq:mb}).
According to Eqs.(\ref{eq:pimu_sin}), we can find the asymptotic expressions of the poles 
of the normal mode by solving the following equation,
\begin{align}
  \vphi
  +\tht_{\pm}
  \simeq m\pi
   \qquad\Ra\qquad
  \vphi^2+2\vphi\tht_{\pm}+\tht_{\pm}^2
  \simeq (k_x^mw)^2
   \qquad (m\geq m_{a,b}+1) .
  \label{eq:eta_ba}
\end{align}
$\vphi$ and $\tht_{\pm}$ are given by Eqs.(\ref{eq:variables}-\ref{tht}).
$\tht_{\pm}=(\tht_{+},\tht_{-})$ corresponds to $(\nu_m^n,\mu_m^n)$.
Since the derivation of the asymptotic expression of $\nu_m^n$ for $m>m_{a,b}$ is 
similar to $\mu_m^n$, for brevity, we often write similar two equations for
$\nu_m^n$ and $\mu_m^n$ (\ie, $p_{\nu}=0$ and $s_{\nu}=0$) together with
a single equation as Eqs.(\ref{eq:eta_ba}).
We got the second equation of (\ref{eq:eta_ba}) by squaring the first equation
on the basis of the following considerations:
\begin{itemize}
\item
  $p_{\nu}(\hr_b,\hr_a)$ and $s_{\nu}(\hr_b,\hr_a)$ are even
  with respect to $\nu$ as shown in Eq.(\ref{eq:cp_symm}).
\item
  $(\nu_m^n)^2$ and $(\mu_m^n)^2$ are real
  similar to $(k_s^{mn})^2\in\mathbb{R}$ in the straight pipe,
  according to Eqs.(\ref{eq:pnu_eq0}-\ref{eq:snu_eq0}).
\item
  $\nu/\rho$ denotes the longitudinal wavenumber of the field in the bending section as
  seen from Eq.(\ref{eq:ILT_nu}).
\item
  $\nu=(\nu_m^n,\mu_m^n)$ for $m>m_{a,b}$ tend to be $k_s^{mn}$
  in the limit of $\rho\to\infty$ as described in Eq.(\ref{eq:asympt_lim}).
\item
  $k_s^{mn}$ is related to $k_x^m$ through Eq.(\ref{eq:dispersion})
  which consists of their squares $(k_s^{mn})^2$ and $(k_x^m)^2$.
\end{itemize}
According to these considerations, it is easier to find the asymptotic expression of
$(\nu/\rho)^2\in\mathbb{R}$ or $(\bnu/\rho)^2\in\mathbb{R}$ than finding
the asymptotic expression of $\nu/\rho\in\mathbb{A}$ or $\bnu/\rho\in\mathbb{A}$.
From a more practical point of view, we will find the expressions of
$(\bnu/k_x^m\rho)^2$ or $(\bnu/k_r^n\rho)^2$
since their absolute values are much larger or smaller than 1
under the assumptions listed in Eqs.(\ref{eq:cond26}-\ref{eq:cond25}).
We can get the asymptotic expression of $\nu/\rho$ or $\bnu/\rho$ by expanding
the square root of the expression of $(\nu/\rho)^2$ or $(\bnu/\rho)^2$ with respect to
$\eps_{r,s}$ if necessary.

\clearpage

\subsection{Asymptotic solutions of the poles for small radial wavenumber}
\label{sec:asympt_numu}

We solve Eq.(\ref{eq:eta_ba}) with respect to $(\bnu/\rho)^2$ under the assumption
(\ref{eq:cond26}),
\begin{align}
  &
  |k_r^n|\ll k_x^m
   \quad\text{and}\quad
  w\ll\rho
   \quad~~\Lra~~\quad
  |\bnu|\gg|k_r^n\rho|
   \quad\text{and}\quad
  \eps\ll1
   \quad~~\Ra~~\quad
  |\bz_{b,a}^2|
  =|g_{b,a}^2\veps_r|
  \ll1
   \label{eq:bz_assm}
   \\
  &\Ra\quad
  \frac{\bnu}{\rho}
  =O(k_x^m)
   \quad\text{and}\quad
  |\veps_r|
  =O(\eps_r)\ll1 ,
    \qquad\text{where}\quad
  \veps_r
  =\bigg(\frac{k_r^n}{\bnu/\rho}\bigg)^2 ,
     \qquad
 \eps_r
  =\bigg(\frac{k_r^n}{k_x^m}\bigg)^2 .
   \label{eq:bz_assm_1}
\end{align}
$\veps_r^{-1}$ is the solution which we are going to find.
Although we assume $|\veps_r|\ll1$ in this derivation,
the asymptotic solution of the poles is fairly accurate in a relatively wide range
unless the radial mode number $m$ is close or equal to the whispering gallery modes
$m_{a,b}$ as shown in Fig.\ref{fig:nu_err_4K} and Fig.\ref{fig:bnu_err_bkr}
(p.\pageref{fig:nu_err_4K}).

After neglecting $\tht_{\pm}^2$ in the second equation of (\ref{eq:eta_ba}),
we divide the equation by $(\eps\bnu)^2$ for convenience in calculating
$\vphi$ and $\tht_{\pm}$ for $|\bz_{b,a}|\ll1$,
\begin{align}
   \Xi^2
  +2\Xi\hTh_{\pm}
  +O[\eps_r^2(\eps/m)^4]
  \simeq \frac{\veps_r}{\eps_r} ,
    \qquad
  \Xi
  =\frac{\xi(\bz_b)-\xi(\bz_a)}{\eps} ,
    \qquad
  \hTh_{\pm}
  =\frac{\tht_{\pm}}{\eps\bnu}
   \quad (m>m_{a,b}) .
  \label{eq:xi_sq}
\end{align}
In order to calculate $\tht_{\pm}$ for $|\bz_{b,a}|\ll1$, we expand $p^{2\ell+1}$,
which is involved in Eqs.(\ref{buv_13}-\ref{buv_57}), in Taylor series
with respect to $\bz^2$,
\begin{align}
  p^{2\ell+1}
  &=(1+\bz^2)^{-(\ell+1/2)}
   =\sum_{j=0}^{\infty}(-1)^j\frac{\Gam(j+\ell+1/2)}{j!\Gam(\ell+1/2)}\bz^{2j}
   \qquad
   (|\bz|\ll1) .
\end{align}
$p$ is given by Eq.(\ref{eq:variables}).
We expand $\hTh_{\pm}$ with respect to $\veps_r$ using Eqs.(\ref{buv_13}-\ref{buv_57}),
\begin{align}
  (\hTh_{+},\hTh_{-})
  &\simeq
   \frac{g_1\veps_r^2}{2(k_r^n\rho)^2}
   \bigg[
     (-1,1)
    +\veps_r
     \bigg\{
        \frac{(11,-13)}{4}g_2
       +\frac{(1,-1)}{(k_r^n\rho)^2}
     \bigg\}
    +O(\veps_r^2)
   \bigg] .
   \label{eq:tht_ps}
\end{align}
$g_j$ is given as follows,
\begin{align}
  g_1
  &=\frac{g_b+g_a}{2},
    \qquad
  g_2
  =\frac{g_b^2+g_a^2}{2},
    \qquad
  g_3
  =\frac{g_b^4+(g_bg_a)^2+g_a^4}{3},
   \\
  g_l
  &=\frac{\rho}{w}\log\bigg(\frac{g_b}{g_a}\bigg),
    \qquad
  g_{b,a}
  =\frac{r_{b,a}}{\rho};
    \qquad
   \lim_{\rho\to\infty}(g_{j},g_l,g_{b,a})=1 .
   \label{eq:gl}
\end{align}
$g_l$ is a geometric factor involved in Eq.(\ref{eq:xiba_2nd}).
Both $g_j$ and $g_l$ are symmetric with respect to the exchange of $r_a$ and $r_b$.
$g_j$, $g_l$ and $g_{a,b}$ all go to 1 in the limit of $\rho\to\infty$ with
$x_{a,b}$ kept constant.

We expand $\Xi$ with respect to $\bz^2$ using Eq.(\ref{xi_expd}), 
\begin{align}
  \Xi
  &=g_l
    +2g_1\sum_{j=1}^{\infty}h_j\frac{g_b^{2j}-g_a^{2j}}{g_b^2-g_a^2}\veps_r^j
  \simeq
    g_l
   +\frac{g_1}{2}\veps_r
    \Big\{
       1
      -\frac{g_2}{4}\veps_r
      +\frac{g_3}{8}\veps_r^2
      +O(\veps_r^3)
    \Big\} .
  \label{eq:xiba_2nd}
\end{align}
$h_j$ is the coefficient of the expansion (\ref{xi_expd}).
Substituting Eqs.(\ref{eq:tht_ps}) and (\ref{eq:xiba_2nd}) into the first equation of
(\ref{eq:xi_sq}), we rewrite it into the following quadratic equation
with respect to $\veps_r^{-1}$,
\begin{align}
   \frac{1}{\veps_r^2}
  -\bigg(\frac{1}{g_l^2\eps_r}-\frac{g_1}{g_l}\bigg)\frac{1}{\veps_r}
  -\frac{g_1}{g_l}
   \bigg\{
      \frac{1}{4}\bigg(g_2-\frac{g_1}{g_l}\bigg)
     \pm\frac{1}{(k_r^n\rho)^2}
   \bigg\}
  +O(\veps_r)
  &\simeq 0 ,
   \label{eq:veps_quad}
\end{align}
where the sign $(+,-)$ corresponds to $\nu=(\nu_m^n,\mu_m^n)=i(\bnu_m^n,\bmu_m^n)$.
The upper and lower signs correspond to $\nu_m^n$ and $\mu_m^n$ in what follows.
Solving Eq.(\ref{eq:veps_quad}) with respect to $\veps_r^{-1}$,
we expand it with respect to $\eps_r$ given by Eq.(\ref{eq:bz_assm_1}).
Taking the terms up to $O(\eps_r^2)$ into account, we get the asymptotic solutions of 
Eqs.(\ref{eq:pnu_eq0}-\ref{eq:snu_eq0}) with respect to $(\bnu/\rho)^2$ of
the normal mode ($m>m_{a,b}$) under the assumptions (\ref{eq:bz_assm}),
\begin{align}
  \bigg\{\frac{g_l(\bnu_m^n,\bmu_m^n)}{k_x^m\rho}\bigg\}^2
  &\simeq
    1
   -g_1g_l\bigg(\frac{k_r^n}{k_x^m}\bigg)^2
   +g_1g_l^2\bigg\{\frac{g_2g_l-g_1}{4}\pm\frac{g_l}{(k_r^n\rho)^2}\bigg\}
    \bigg(\frac{k_r^n}{k_x^m}\bigg)^4
   +O(\eps_r^3) .
   \label{eq:bnu_sols}
\end{align}
Through $\nu=i\bnu$ as described in Eq.(\ref{eq:nu_bnu}), we rewrite Eq.(\ref{eq:bnu_sols})
in accordance with the expression on the L.H.S. of Eq.(\ref{eq:asympt_lim}),
\begin{align}
  \bigg\{\frac{(\nu_m^n,\mu_m^n)}{k_s^{mn}\rho}\bigg\}^2
  &\simeq
    \frac{1}{g_l^2}
   +\frac{1-g_1g_l}{g_l^2}\bigg(\frac{k_r^n}{k_x^m}\bigg)^2
   +\bigg\{
      \frac{1-g_1g_l}{g_l^2}+g_1\frac{g_2g_l-g_1}{4}\pm\frac{g_1g_l}{(k_r^n\rho)^2}
    \bigg\}
    \bigg(\frac{k_r^n}{k_x^m}\bigg)^4
   +O(\eps_r^3) .
  \label{eq:nu_ks}
\end{align}
$\nu_m^n$ differs from $\mu_m^n$ in the terms higher order than $O(\eps_r^2)$ under
the assumptions (\ref{eq:bz_assm}).
When $m$ is large, Eq.(\ref{eq:nu_ks}) has a good accuracy even to the first order with 
respect to $\eps_r$,
\begin{align}
  \bigg\{\frac{(\nu_m^n,\mu_m^n)}{k_s^{mn}\rho}\bigg\}^2
  &\simeq
   \frac{1}{g_l^2}
   \bigg\{
     1
    +(1-g_1g_l)\bigg(\frac{k_r^n}{k_x^m}\bigg)^2
   \bigg\}
  +O(\eps_r^2)
    \qquad(m>m_{a,b}) ,
  \label{eq:nu_ks_1st}
\end{align}
where $\nu_m^n$ is equal to $\mu_m^n$ up to $O(\eps_r)$.
As shown in Fig.\ref{fig:nu_err_4K}  (p.\pageref{fig:nu_err_4K}), however,
Eqs.(\ref{eq:bnu_sols}-\ref{eq:nu_ks_1st}) are not accurate around $m=m_{\pm}$
since $|\nu/\rho|\ll O(k_x^m)$ at $m=m_{\pm}$.
Conversely, Eq.(\ref{nu2_sum}) is accurate around $m=m_{\pm}$.
We expand the geometric factors (\ref{eq:nu_ks}) up to the second order
with respect to $x_{b,a}/\rho$,
\begin{align}
  g_1g_l
  \simeq 1+\frac{\eps^2}{12}+O(\eps^3)
   ,\qquad
  \frac{g_2g_l-g_1}{4}
  \simeq \frac{\eps^2}{12}+O(\eps^3)
   ,\qquad
  \frac{1}{g_l^2}
  \simeq g_bg_a+\frac{\eps^2}{12}+O(\eps^3) .
   \label{eq:g0_gl}
\end{align}
$\eps=w/\rho$ as in Eq.(\ref{eq:eps}).
Using Eqs.(\ref{eq:g0_gl}), we expand Eq.(\ref{eq:nu_ks}) with respect to $x_{b,a}/\rho$,
\begin{align}
  \bigg\{\frac{(\nu_m^n,\mu_m^n)}{k_s^{mn}\rho}\bigg\}^2
  \simeq
    g_bg_a
   -\frac{\eps^2}{12}\bigg(\frac{k_s^{mn}}{k_x^m}\bigg)^2
   \pm\frac{1}{(k_x^m\rho)^2}\bigg(1+\frac{\eps^2}{12}\bigg)
    \bigg(\frac{k_r^n}{k_x^m}\bigg)^2
   +O(\eps_r^3,\eps_r\eps^3) .
   \label{eq:nu_eps}
\end{align}
Eq.(\ref{eq:nu_eps}) satisfies Eq.(\ref{eq:asympt_lim}) which is the asymptotic limit of
the poles of the normal mode ($m>m_{a,b}$) for $\rho\to\infty$.
On the other hand, for $\eps_r\to0$ with $\rho$ fixed, 
Eqs.(\ref{eq:bnu_sols}-\ref{eq:nu_ks}) go to the following asymptotic limit,
\begin{align}
  \lim_{\eps_r\to0}\bigg\{\frac{(\bnu_m^n,\bmu_m^n)}{k_x^m\rho}\bigg\}^2
  =\lim_{\eps_r\to0}\bigg\{\frac{(\nu_m^n,\mu_m^n)}{k_s^{mn}\rho}\bigg\}^2
  =\frac{1}{g_l^2}
    \qquad(m> m_{a,b}) .
  \label{eq:lim_nu_ks}
\end{align}
$g_l$ is given by Eq.(\ref{eq:gl}) which does not depend on the wavenumbers of the field:
$k$, $k_r^n$, $k_s^{mn}$, $k_x^m$ and $k_y^n$.
We get the derivatives of $(\nu_m^n)^2$ and $(\mu_m^n)^2$ with respect to $k_r^n$,
which are involved in Eqs.(\ref{eq:cRp_ba}-\ref{eq:cRm_ba}),
\begin{align}
  \frac{d}{dk_r^n}
  \bigg\{\frac{(\nu_m^n,\mu_m^n)}{\rho}\bigg\}^2
  &\simeq
   2g_1k_r^n
   \bigg[
     \frac{1}{g_l}
    +\frac{1}{(k_x^m\rho)^2}
     \bigg\{
        \frac{g_1-g_2g_l}{2}(k_r^n\rho)^2
       \mp g_l
     \bigg\}
    +O(\eps_r^2)
   \bigg] .
   \label{eq:rd_kr_numu2}
\end{align}

We derived Eq.(\ref{eq:nu_eps}) using Eqs.(\ref{eq:Fimu_muz}-\ref{eq:Gimu_muz})
which are Dunster's uniform asymptotic series of $F_{i\bnu}$ and $G_{i\bnu}$.
Besides this way, we can get Eq.(\ref{eq:nu_eps}) using Olver's uniform asymptotic 
series (\ref{eq:JY_uae}-\ref{eq:dJY_uae}) of $J_{\nu}$ and $Y_{\nu}$
under the assumption $m\gg1$ as shown below.
Taking only the leading order terms in the first equations of
(\ref{eq:pnu_uae}-\ref{eq:snu_uae}),
we approximate Eqs.(\ref{eq:pnu_eq0}-\ref{eq:snu_eq0}) as follows,
\begin{align}
  \mfp(-u_b,-u_a)\simeq0
   ,\qquad
  \mfs(-u_b,-u_a)\simeq0
   \qquad\Ra\qquad
  U_b-U_a
  \simeq m\pi
   ,\qquad
  U_{b,a}
  =\nu\frac{2}{3}(-\zeta_{b,a})^{3/2} .
   \label{mfp_mfs_0}
\end{align}
$\mfp$ and $\mfs$ are the cross products of the Airy functions,
given by Eqs.(\ref{eq:cp_airy_p}-\ref{eq:cp_airy_sr}).
We rewrite $U_b-U_a$ as
\begin{align}
  \frac{U_b-U_a}{\nu}
  &=\sig-\sin^{-1}\!\bigg(\frac{\sig}{z_bz_a}\bigg) ,
    \qquad
  \sig=(z_b^2-1)^{1/2}-(z_a^2-1)^{1/2} .
  \label{eq:Uba_nu}
\end{align}
Substituting Eq.(\ref{eq:Uba_nu}) into the third equation of (\ref{mfp_mfs_0}),
we expand it with respect to $\sig$,
\begin{align}
  \frac{m\pi}{\nu}
  &\simeq
   \sig\bigg(1-\frac{1}{z_bz_a}\bigg)
   \bigg\{1-\frac{\sig^2}{6(z_bz_a)^2(z_bz_a-1)} +O(\sig^4)\bigg\} 
   \qquad
  (|\sig|\ll 1) .
  \label{eq:mpi_phi}
\end{align}
Moreover, expanding the square of Eq.(\ref{eq:mpi_phi}) with respect to $x_{b,a}/\rho$
up to $O(\eps^2)$, we get
\begin{align}
  \frac{(k_x^m)^2}{\vkap_x^2}
  &\simeq
     1
     +\frac{x_b+x_a}{\rho}\frac{(\nu/\rho)^2}{\vkap_x^2}
     -\bigg\{\frac{x_b^2+x_bx_a+x_a^2}{\rho^2}
            +\frac{w^2}{\rho^2}\frac{(\nu/\rho)^2}{12\vkap_x^2}
      \bigg\}
      \frac{(\nu/\rho)^2}{\vkap_x^2}
     +O(\eps^3) ,
  \label{eq:nu_quadra}
\end{align}
where
\begin{align}
  \vkap_x^2
  =(k_r^n)^2-(\nu/\rho)^2 .
\end{align}
Solving Eq.(\ref{eq:nu_quadra}) with respect to $(\nu/\rho)^2$, we get
Eq.(\ref{eq:nu_eps}) up to $O(\eps^2)$.
Thus, we can get Eq.(\ref{eq:nu_eps}) in two different ways using
Dunster's expansion for $\nu=i\bnu$ and
Olver's expansion for $\nu\in\mathbb{R}$.
Using Olver's expansion, however, since it is harder to rearrange
the asymptotic series of the Airy functions to the higher order,
it is harder to find the asymptotic solution of the poles to the higher order
with respect to $\eps_r$ and $\eps$.
In this regard, it is easier to deal with Dunster's series than Olver's one.
If it is possible, we wish to find the uniform asymptotic series of the cross products
for $\nu\in\mathbb{C}$ in terms of the trigonometric functions.

\subsection{Approximate solution of the pole by Horvat and Prosen}
\label{sec:HP}

We discuss Eq.(26) in \cite{horvat_prosen}, which is presented as
an approximate solution of Eq.(\ref{eq:pnu_eq0}) without a proper proof.
Using our notation, Eq.(26) in \cite{horvat_prosen} is given and rewritten as follows,
\begin{align}
  (k_r^nw)^2-\{\nu_m^n\log(g_b/g_a)\}^2
  \simeq (m\pi)^2
   \qquad\Lra\qquad
  \bigg(\frac{\nu_m^n}{k_s^{mn}\rho}\bigg)^2
  \simeq \frac{1}{g_l^2}
    \qquad
  (m\in\mathbb{N}) .
  \label{eq:HP26}
\end{align}
In order to compare our asymptotic solution with Eq.(\ref{eq:HP26}),
we rearrange $\nu_m^n$ which we derived in Eq.(\ref{eq:nu_ks_1st}),
\begin{align}
  \bigg(\frac{\nu_m^n}{k_s^{mn}\rho}\bigg)^2
  \simeq
    \frac{1}{g_l^2}
   +\eps_1
   +O[(\eps\eps_r)^2] ,
     \qquad
  \eps_1
  =\frac{1-g_1g_l}{g_l^2}\eps_r
  \simeq
  \frac{\eps_r}{12}\eps^2\{1+O(\eps)\}
    \qquad(m>m_{a,b}) .
   \label{eq:nu2_1st}
\end{align}
$\eps_1$ is the first order term with respect to $\eps_r$ in the asymptotic solution
(\ref{eq:nu_ks_1st}).
Since Eq.(\ref{eq:HP26}) agrees with Eq.(\ref{eq:nu2_1st}) to leading order
with respect to $\eps_r$,
Eq.(26) in \cite{horvat_prosen} is the asymptotic solution of Eq.(\ref{eq:pnu_eq0})
to zeroth order with respect to $\eps_r$ and first order with respect to $\eps$.
In other words, Eq.(\ref{eq:HP26}) is the asymptotic limit of $(\nu_m^n/\rho)^2$ for
$\eps_r\to0$ as shown in Eq.(\ref{eq:lim_nu_ks}).
Eq.(\ref{eq:HP26}) has an error $\eps_1$ for Eq.(\ref{eq:nu_ks_1st})
which is the correct asymptotic expression of the poles as shown
analytically in appendix \ref{sec:asympt_numu} in two ways
(using Dunster's series and Olver's series) and also numerically in
Fig.\ref{fig:nu_err_4K} and Fig.\ref{fig:bnu_err_bkr} (p.\pageref{fig:nu_err_4K})
for $k_r^n\in\mathbb{R}$ and $i\mathbb{R}$.

We examine the applicability and behavior of the error in Eq.(\ref{eq:HP26}).
Eq.(\ref{eq:HP26}) is approximately correct for
$|k_r^n/k_x^m|\to0$ and $\eps=w/\rho\ll 1$.
For Eq.(\ref{eq:nu_ks}), Eq.(\ref{eq:HP26}) has a relative error
$\eps_1\simeq[k_r^nw/(k_x^m\rho)]^2/12$.
Therefore Eq.(\ref{eq:HP26}) tends to have a larger error for a larger $k_r^n$ and
smaller $k_x^m$.
Actually, Fig.2 (b) in \cite{horvat_prosen} shows the fact that Eq.(\ref{eq:HP26}) 
tends to brake down for larger $(k_r^n)^2$ and smaller $m$
(as $m$ increases, the corresponding curve appears toward the left side in Fig.2 (b) of
\cite{horvat_prosen}).
Also, Eq.(\ref{eq:HP26}) does not hold for the whispering gallery modes ($1\leq m\leq m_a$)
similar to $\nu_m^n$ given by Eq.(\ref{eq:nu_ks}).
On the contrary, for $(k_r^n)^2\to0$ and larger $m$, Fig.2 (b) in \cite{horvat_prosen}
shows that Eq.(\ref{eq:HP26}) tends to agree with the true solution which the authors got
by numerically solving Eq.(\ref{eq:pnu_eq0}).
As mentioned below Eq.(\ref{eq:nu_ks_1st}), however,
Eq.(\ref{eq:nu2_1st}) is not accurate around $m=m_{\pm}$
for which $|\nu/\rho|\ll O(k_x^m)$.
The asymptotic solution of the poles for $|\nu/\rho|\ll O(k_x^m)$, \ie, $|\eps_s|\ll 1$
will be given by Eq.(\ref{nu2_sum}) in appendix \ref{sec:HP2},
which is valid only around $m=m_{\pm}$.
Since the dispersion relation (\ref{eq:krn_apdx}) always holds in the straight pipe,
$|\eps_s|\ll 1$ is inconsistent with $|\eps_r|\ll 1$.

We would like to point out a problem in \cite{horvat_prosen},
which is about the normalization of the system in the model to calculate the field.
We can take the limit of $\eps_r=(k_r^n/k_x^m)^2\to0$ for
Eqs.(\ref{eq:bnu_sols}-\ref{eq:nu_ks}) as shown in Eq.(\ref{eq:lim_nu_ks}).
However, we must take this limit after expanding the Bessel functions
in the uniform asymptotic series for large $\rho$.
The ``radial wavenumber $k$'' in \cite{horvat_prosen} corresponds to 
our $\hr_b\,(=k_r^nr_b)$ which is the dimensionless radius of the outer wall $r_b$ 
normalized by the radial wavenumber $k_r^n$ as in Eq.(\ref{eq:z_hr}).
But since the authors fix the radius of the outer wall to 1 in \cite{horvat_prosen},
we cannot consider the limits of $\rho\to\infty$ and $k_r^n\to0$ independently in
their theory.
These two limits do not contradict with each other, because $\rho$ is a geometric parameter
which is independent of the wavenumbers of the field.
And therefore we can consider the limit of $k_r^n\rho\to\infty$ under the assumption that
the limit of $\rho\to\infty$ is faster than $k_r^n\to0$.
According to these considerations, we derived Eq.(\ref{eq:nu_ks}) which is the asymptotic
expression of the poles for $\rho\to\infty$ and $\eps_r\to0$.
Also, as shown in appendix \ref{sec:cRcX}, for $\rho\to\infty$,
the horizontal eigenfunctions of the curved pipe $\cR_{\pm}^{mn}$,
given by Eqs.(\ref{eq:Ven}-\ref{eq:mfRm}), tend to be those of
the straight pipe which has the same cross section as the curved pipe.
Unfortunately, the normalization of the whole geometry of the system by
the outer wall radius in \cite{horvat_prosen} reduces the degree of freedom
which is necessary in manipulating the parameters to expand the Bessel functions in
the uniform asymptotic series.
Due to this, the theory in \cite{horvat_prosen} does not allow us to consider the asymptotic
limit of the field and the poles in the curved pipe to those in the straight pipe
by taking the limit of $\rho\to\infty$ with $w$ kept constant.
It is not a good idea to normalize the geometry of the system in such a way.
For example, in the framework of the paraxial approximation, we normalized
the horizontal variable $x$ as $\hx=x/\ell_{\perp}$ as shown in Eq.(15) of \cite{agoh},
where $\ell_{\perp}=(2k^2/\rho)^{-1/3}$ denotes the typical transverse scale 
length (extent) of the steady field of coherent synchrotron radiation (CSR) in
a bending magnet as described in Eq.(\ref{eq:shield}).
Actually, Eq.(15) in \cite{agoh} fits the mathematical framework of
the paraxial approximation in terms of the Airy functions
which are the fundamental solutions of the CSR field in this theoretical framework.
On the other hand, in the current problem involving the Bessel functions
in the exact theory, we should normalize the geometry as shown in Eqs.(\ref{eq:z_hr}).
That is, we normalize the radius $r$ in the following two ways
using the field variable $k_r^n$ or the geometric parameter $\rho$:
$\hr=k_r^nr$ and $g=r/\rho$.
These two normalized radii are rewritten as $\hr=gk_r^n\rho$ and
$g=1+x/\rho$, where the limit of $k_r^n\to0$ does not necessarily contradict with
$\hr\to\infty$ since $\hr$ involves $\rho$ which is independent of $k_r^n$ and $x_{b,a}$ in
our model shown in section \ref{sec:frame}.

We would also like to describe the coordinate system in
\cite{horvat_prosen} where the inner wall of the curved pipe represents the reference axis.
In contrast, in the present paper we define the reference axis ($s$-axis)
independent of the sidewalls of the pipe ($r_a$ and $r_b$).
Owing to this independence, our model has geometric degrees of freedom
which allow us to consider various asymptotic limits of the field, \eg,
absence of the sidewalls in a bending section
($\rho=\text{fixed}$, $r_a\to0$, $r_b\to\infty$),
a straight pipe ($\rho\to\infty$, $x_{b,a}=\text{fixed}$),
absence of the sidewalls in a straight section
($\rho\to\infty$, $x_a\to-\infty$, $x_b\to\infty$) and so on.
We can remove the upper-lower walls by taking the limit of $h\to\infty$.
If necessary, one can extend our model of the rectangular pipe so that
the upper and lower walls of the pipe can be moved independently
as $y=(y_b,y_a)$ similar to the sidewalls $r=(r_b,r_a)$,
though we set them symmetrically as $y=\pm h/2$ in the present study
since it is not important to take into account the vertical asymmetry with respect to
the reference axis in calculating the field of CSR.
Generally, a field in a structure tends to be a certain asymptotic limit in accordance with
a geometrical limit such as $\rho\to\infty$ or $r_a\to0$.
In studying the mathematical structure of a complex field like CSR,
it is often instructive to look up a relatively simple field as a guide such as the one in
a straight pipe ($\rho\to\infty$ with $w=r_b-r_a$ kept constant) as described in
appendix \ref{sec:travel_wave}.
Besides, one can find the asymptotic conditions which the field must obey in a certain 
limit:
Eqs.(\ref{eq:dispersion}), (\ref{eq:asympt_lim}),
(\ref{eq:lim_nu}-\ref{eq:BDE_asympt}), (\ref{eq:lim_kmn}), (\ref{eq:lim_RX}),
(\ref{eq:lim_cGpm}), (\ref{eq:lim_cH_cK}), (\ref{eq:lim_Cpm}) and
(\ref{eq:lim_Gam_xy}-\ref{eq:lim_Gam_s}).
Thus, in modeling a system in physics, one should consider additional degrees of freedom
in both the field and geometry in order to consider various kinds of asymptotic limits.
Then one should normalize the system as it fits the physical and mathematical frameworks of
the theory.
In physics, a smaller number of variables and parameters does not necessarily make
the problem simpler and easier to solve.
If a model has a redundant degree of freedom, we can get rid of it after solving
the problem by substituting a value or taking a limit for the solution.

\subsection{Asymptotic solutions of the poles around the origin of the Laplace plane}
\label{sec:HP2}

We discuss Eq.(25) in \cite{horvat_prosen}, which is given as follows
using our notation,
\begin{align}
  \bigg(\frac{k_r^n}{\kap_r^m}\bigg)^2
  -\bigg(g_l\frac{\nu_m^n}{k_x^m\rho}\bigg)^2
  \simeq 1 ,
    \qquad
  [\nu_m^n]_{k_r^n=\kap_r^m}=0 .
  \label{eq:HP25}
\end{align}
$k_x^m$ and $k_r^n$ are given by Eqs.(\ref{eq:dispersion}-\ref{eq:krn_apdx}).
$\nu_m^n$ is a function of the radial wavenumber $k_r^n$.
The first equation of (\ref{eq:HP25}) involves $\kap_r^m$ (``$k_0$'' in \cite{horvat_prosen})
which denotes the zero of $\nu_m^n$ with respect to $k_r^n$
as shown in the second equation of (\ref{eq:HP25}) and Eq.(\ref{eq:kaprm_p}).
According to \cite{horvat_prosen}, the authors got Eq.(\ref{eq:HP26}) by substituting
$\kap_r^m\simeq k_x^m$ into Eq.(\ref{eq:HP25}).
But in reality, Eq.(\ref{eq:HP25}) is incompatible with Eq.(\ref{eq:HP26}),
because $\nu_m^n/\rho\to k_s^{mn}$ ($m>m_a$) in the limit of $\rho\to\infty$
as shown in Eq.(\ref{eq:asympt_lim}).
That is, Eqs.(\ref{eq:HP26}) and (\ref{eq:HP25}) are valid respectively in
the different limits $\eps_r\to0$ and  $\eps_s\to0$ which are inconsistent with each other 
as seen from Eqs.(\ref{eq:cond26}-\ref{eq:cond25}).

As mentioned below Eq.(\ref{eq:eps}),
Eq.(\ref{eq:HP25}) is a pseudo asymptotic solution of Eq.(\ref{eq:pnu_eq0}).
We will show it by solving Eq.(\ref{eq:eta_ba}) with respect to
$(\bnu/k_r^n\rho)^2$ under assumption (\ref{eq:cond25}).
As shown in Eq.(\ref{eq:hkmn}), the straight pipe has the cutoff wavenumber $k=\hk_m^n$
such that $k_s^{mn}=0$.
This is the transition point that the $(m,n)$-th longitudinal mode of
the field interchanges between propagation ($k_s^{mn}\in\mathbb{R}$) and
exponential decay ($k_s^{mn}\in i\mathbb{R}$).
The curved pipe has two cutoff wavenumbers $k=k_m^n$ and $\bk_m^n$ which are
of $\tE_y$ and $\tB_y$ as in Eqs.(\ref{eq:kmn}-\ref{eq:bkmn}).
Since $\nu/\rho$ is the longitudinal wavenumber of the field in the bend, the zeros of
$p_{\nu}(\hr_b,\hr_a)$ and $s_{\nu}(\hr_b,\hr_a)$ for small $|\nu|$ denote the dispersion 
relations of $\tE_y$ and $\tB_y$ in the curved pipe around their cutoff wavenumbers.
In order to solve Eq.(\ref{eq:eta_ba}) with respect to $\nu$
around the origin of the $\nu$-plane, we assume Eq.(\ref{eq:cond25}), \ie,
\begin{align}
  &
  \eps\ll1
    \quad\text{and}\quad
  |\veps_s|
   =O(\eps_s)
  \ll1
   \qquad\Ra\qquad
  |\bz|\gg1
    \quad\text{and}\quad
  k_r^n\rho
  =O(k_x^m\rho)
  =O(\eps^{-1})
  \gg1 ,
  \label{eq:bz_cond_big}
\end{align}
where
\begin{align}
  \eps
  =\frac{w}{\rho}
   ,\qquad
   \veps_s
   =\bigg(\frac{\bnu}{k_r^n\rho}\bigg)^{\!2}
   ,\qquad
  \eps_s
  =\bigg(\frac{k_s^{mn}}{k_r^n}\bigg)^{\!2}
   ,\qquad
  \bz
  =\frac{\hr}{\bnu}
  =g\veps_s^{-1/2}
   \qquad
  (m>m_{a,b}) .
  \label{eq:sig_bsig}
\end{align}
$\veps_s$ is the solution which we will find.
We divide the second equation of (\ref{eq:eta_ba}) by $(k_r^nw)^2$ for convenience
in calculating $\vphi$ and $\tht_{\pm}$, given by Eqs.(\ref{eq:variables}-\ref{tht}),
for $|\bz|\gg1$,
\begin{align}
   \bigg(\frac{k_x^m}{k_r^n}\bigg)^2
  &\simeq \bigg(\frac{\vphi}{k_r^nw}\bigg)^2
  +2\frac{\vphi}{k_r^nw}\cd\frac{\tht_{\pm}}{k_r^nw}
  +\bigg(\frac{\tht_{\pm}}{k_r^nw}\bigg)^2 .
  \label{zeroeq2}
\end{align}
$\vphi$ involves $\xi$ given by Eq.(\ref{eq:xi_bz}).
We expand $\vphi$ with respect to $\veps_s$
by expanding $\xi/\bz$ with respect to $\bz^{-2}$ using Eq.(\ref{eq:xit}),
\begin{align}
  \frac{\vphi}{k_r^nw}
   =\sum_{j=0}^{\infty}\gam_j\bg_j(-\veps_s)^j
   \simeq
      1
     +\frac{\bg_1}{2}\veps_s
     -\frac{\bg_2}{2^3}\veps_s^2
     +\frac{\bg_3}{2^4}\veps_s^3
     +O(\veps_s^4) .
   \label{eq:xi_ba}
\end{align}
$\gam_j$ is given by Eq.(\ref{eq:xit}-\ref{eq:gamj}).
$\bg_j$ is the following geometric factor which goes to 1 in the limit of $\rho\to\infty$,
\begin{align}
  \bg_j
  &=\frac{g_b^{2j-1}-g_a^{2j-1}}{(2j-1)(g_b-g_a)(g_bg_a)^{2j-1}} ,
    \qquad
  g_{b,a}
  =\frac{r_{b,a}}{\rho} ;
    \qquad
  \lim_{\rho\to\infty}\bg_j
  =1 .
   \label{eq:bgj}
\end{align}
$\bg_j$ and $g_{b,a}$ are equal or close to 1 when $\eps\ll1$.
We can rewrite $\bg_{j\geq2}$ using $\bg_1$ as follows,
\begin{align}
  \bg_0
  &=1 ,
    \qquad
  \bg_1
  =\frac{1}{g_bg_a}
  \simeq
    1
   -\frac{x_b+x_a}{\rho}
   +\frac{w^2+3x_bx_a}{\rho^2}
   -\frac{(x_b+x_a)(x_b^2+x_a^2)}{\rho^3}
   +O(\eps^4) ,
   \label{eq:bg1}
   \\
  \bg_2
  &=\bg_1^2\bigg(1+\frac{\eps^2\bg_1}{3}\bigg)
   ,\qquad
  \bg_3
  =
   \bg_1^3
    \bigg\{
      1
     +\eps^2\bg_1
     +\frac{(\eps^2\bg_1)^2}{5}
    \bigg\}
   ,\qquad
  \bg_4
   =\bg_1^4
    \bigg\{
      (1+\eps^2\bg_1)^2
     +\frac{(\eps^2\bg_1)^3}{7}
    \bigg\} .
\end{align}

We calculate $\tht_{\pm}$ by expanding Eq.(\ref{tht}) with respect to $\veps_s$.
$\tht_{\pm}$ involves the following functions,
\begin{alignat}{3}
  \bpi_1
  &=\bu_1(p_a)-\bu_1(p_b) ,
    \qquad&
  \bpi_3
  &=\bu_3(p_a)-\bu_3(p_b) ,
    \qquad&
  \bpi_5
  &=\bu_5(p_a)-\bu_5(p_b) ,
   \label{bpi35_sep}
   \\
  \bsig_1
  &=\bv_1(p_a)-\bv_1(p_b) ,
    \qquad&
  \bsig_3
  &=\bv_3(p_a)-\bv_3(p_b) ,
    \qquad&
  \bsig_5
  &=\bv_5(p_a)-\bv_5(p_b) .
   \label{bsig35_sep}
\end{alignat}
$\bu_{2k+1}$ and $\bv_{2k+1}$ are given by Eqs.(\ref{eq:bu3}-\ref{eq:bv5}).
In order to expand $\bpi_{2k+1}$ and $\bsig_{2k+1}$ with respect to $\veps_s$,
we first expand $\bz p^{2\ell+1}$ with respect to $\bz^{-2}$,
where $p$ represents $p_{a,b}$ given by Eq.(\ref{eq:variables}),
\begin{align}
  p^{2\ell+1}
  =\sum_{j=0}^{\infty}\frac{(-1)^j\Gam(j+\ell+1/2)}{j!\Gam(\ell+1/2)}\bz^{-\{2(j+\ell)+1\}}
   \qquad
   (|\bz|\gg1) .
   \label{pexpnd}
\end{align}
Substituting Eq.(\ref{pexpnd}) into Eqs.(\ref{buv_13}-\ref{buv_57}),
we rewrite Eqs.(\ref{bpi35_sep}-\ref{bsig35_sep}) using $\veps_s$,
\begin{align}
  \frac{(\bpi_1,\bsig_1)}{\bnu}
  &=\frac{\eps}{2^2(k_r^n\rho)}
    \sum_{j=0}^{\infty}\frac{(-1)^j}{j!}
    \sum_{\ell=0}^{1}\frac{\Gam(j+\ell+3/2)}{\Gam(\ell+1/2)}
    (\mfa_{\ell}^{+},\mfa_{\ell}^{-})\bg_{j+\ell+1}\veps_s^{j+\ell} ,
   \label{bpi_bsig_1}
   \\
  \frac{(\bpi_3,\bsig_3)}{\bnu^3}
  &=\frac{\eps}{2^6(k_r^n\rho)^3}\sum_{j=0}^{\infty}\frac{(-1)^j}{j!}
    \sum_{\ell=0}^{3}\frac{\Gam(j+\ell+5/2)}{\Gam(\ell+3/2)}
    (\mfb_{\ell}^{+},\mfb_{\ell}^{-})\bg_{j+\ell+2}\veps_s^{j+\ell} ,
   \\
  \frac{(\bpi_5,\bsig_5)}{\bnu^5}
  &=\frac{\eps}{2^9(k_r^n\rho)^5}
    \sum_{j=0}^{\infty}\frac{(-1)^j}{j!}
    \sum_{\ell=0}^{5}\frac{\Gam(j+\ell+7/2)}{\Gam(\ell+5/2)}
    (\mfc_{\ell}^{+},\mfc_{\ell}^{-})\bg_{j+\ell+3}\veps_s^{j+\ell} .
   \label{bpi_bsig_5}
\end{align}
The coefficients $\mfa_{\ell}^{\pm}$, $\mfb_{\ell}^{\pm}$ and $\mfc_{\ell}^{\pm}$ are
given by Eqs.(\ref{mfab_pls}-\ref{mfc_mns}).
Substituting Eqs.(\ref{bpi_bsig_1}-\ref{bpi_bsig_5}) into Eq.(\ref{tht}), $\tht_{\pm}$ is
given as the following series with respect to $\veps_s$ and $(k_r^n\rho)^{-2}$,
\begin{align}
  \frac{\tht_{\pm}}{k_r^nw}
  &=\frac{1}{8}\sum_{j=0}^{\infty}f_j^{\pm}\veps_s^j ,
    \qquad
  f_j^{\pm}
  =\sum_{\ell=0}^{\infty}\frac{d_{j,\ell}^{\pm}\bg_{j+\ell+1}}{(k_r^n\rho)^{2(\ell+1)}} .
   \label{eq:theta_lt}
\end{align}
$f_j^{\pm}$ and $d_{j,\ell}^{\pm}$ are the coefficients of the expansions
with respect to $\veps_s$ and $(k_r^n\rho)^{-2}$ respectively.
The coefficients of the lower order terms are given as follows,
\begin{alignat}{3}
  d_{0,0}^{+}
  &=1 ,
    \qquad&
  d_{0,1}^{+}
  &=-\frac{5^2}{2^4} ,
    \qquad&
  d_{0,2}^{+}
  &=\frac{29\cd37}{2^7} ,
   \\
  d_{1,0}^{+}
  &=-\frac{13}{2} ,
    \qquad&
  d_{1,1}^{+}
  &=\frac{1187}{2^5} ,
    \qquad&
  d_{1,2}^{+}
  &=-\frac{107609}{2^8} ,
   \\
  d_{0,0}^{-}
  &=-3 ,
    \qquad&
  d_{0,1}^{-}
  &=\frac{3^2\cd7}{2^4} ,
    \qquad&
  d_{0,2}^{-}
  &=-\frac{3^2\cd211}{2^7} ,
   \\
  d_{1,0}^{-}
  &=\frac{23}{2} ,
    \qquad&
  d_{1,1}^{-}
  &=-\frac{2053}{2^5} ,
    \qquad&
  d_{1,2}^{-}
  &=\frac{3\cd52433}{2^8} ,
\end{alignat}
\begin{alignat}{3}
  d_{2,0}^{+}
  &=\frac{5\cd23}{2^3} ,
    \qquad&
  d_{2,1}^{+}
  &=-\frac{3\cd7\cd13\cd103}{2^7} ,
    \qquad&
  d_{2,2}^{+}
  &=\frac{4858499}{2^{10}} ,
   \\
  d_{3,0}^{+}
  &=-\frac{5\cd7\cd11}{2^4} ,
    \qquad&
  d_{3,1}^{+}
  &=\frac{5\cd39719}{2^8} ,
    \qquad&
  d_{3,2}^{+}
  &=-\frac{11\cd199\cd26881}{2^{11}} ,
   \\
  d_{2,0}^{-}
  &=-\frac{5\cd37}{2^3} ,
    \qquad&
  d_{2,1}^{-}
  &=\frac{7\cd6311}{2^7} ,
    \qquad&
  d_{2,2}^{-}
  &=-\frac{5\cd223\cd5987}{2^{10}} ,
   \\
  d_{3,0}^{-}
  &=\frac{5\cd7\cd17}{2^4} ,
    \qquad&
  d_{3,1}^{-}
  &=-\frac{7\cd42611}{2^8} ,
    \qquad&
  d_{3,2}^{-}
  &=\frac{3\cd5\cd11\cd53\cd8951}{2^{11}} .
\end{alignat}
Substituting Eqs.(\ref{eq:xi_ba}) and (\ref{eq:theta_lt}) into Eq.(\ref{zeroeq2}),
we rewrite it as
\begin{align}
  \bigg(\frac{k_x^m}{k_r^n}\bigg)^2
  &\simeq \sum_{j=0}^{\infty}q_j^{\pm}\veps_s^j ,
    \qquad
  q_j^{\pm}
  =\sum_{\ell=0}^{\infty}\frac{e_{j,\ell}^{\pm}}{(k_r^n\rho)^{2\ell}} .
   \label{eq:zeroeq}
\end{align}
$q_j^{\pm}$ is the coefficient of the $j$th order term with respect to $\veps_s$.
$(q_j^{+},q_j^{-})$ corresponds to $(\nu_m^n,\mu_m^n)$.
$e_{j,\ell}^{\pm}$ denotes the $\ell$th order coefficients of the expansion of $q_j^{\pm}$
with respect to $(k_r^n\rho)^{-2}$,
\begin{align}
  e_{0,0}^{\pm}
  &=1
   ,\qquad
  e_{0,1}^{\pm}
  =\frac{\bg_1}{2^2}d_{0,0}^{\pm}
   ,\qquad
  e_{0,2}^{\pm}
  =\frac{\bg_2}{2^2}d_{0,1}^{\pm}
   +\frac{\bg_1^2}{2^6}(d_{0,0}^{\pm})^2
   ,\qquad
  e_{0,3}^{\pm}
  =\frac{\bg_3}{2^2}d_{0,2}^{\pm}
   +\frac{\bg_1\bg_2}{2^5}d_{0,0}^{\pm}d_{0,1}^{\pm} ,
   \\
  e_{1,0}^{\pm}
  &=\bg_1
   ,\qquad
  e_{1,1}^{\pm}
  =\frac{\bg_2}{2^2}d_{1,0}^{\pm}
   +\frac{\bg_1^2}{2^3}d_{0,0}^{\pm}
   ,\qquad
  e_{1,2}^{\pm}
  =\frac{\bg_3}{2^2}d_{1,1}^{\pm}
   +\frac{\bg_1\bg_2}{2^5}(d_{0,0}^{\pm}d_{1,0}^{\pm}+2^2d_{0,1}^{\pm}) ,
   \\
  e_{1,3}^{\pm}
  &=\frac{\bg_4}{2^2}d_{1,2}^{\pm}
    +\frac{\bg_1\bg_3}{2^5}(d_{0,0}^{\pm}d_{1,1}^{\pm}+2^2d_{0,2}^{\pm})
    +\frac{\bg_2^2}{2^5}d_{0,1}^{\pm}d_{1,0}^{\pm} ,
\end{align}
\begin{align}
  e_{2,0}^{\pm}
  &=\frac{\bg_1^2-\bg_2}{2^2}
   =-\frac{\eps^2\bg_1^3}{2^2\cd3}
   ,\qquad\quad
  e_{2,1}^{\pm}
  =\frac{\bg_3}{2^2}d_{2,0}^{\pm}
   +\frac{\bg_1\bg_2}{2^5}(2^2d_{1,0}^{\pm}-d_{0,0}^{\pm}) ,
   \\
  e_{2,2}^{\pm}
  &=\frac{\bg_4}{2^2}d_{2,1}^{\pm}
    +\frac{\bg_1\bg_3}{2^5}(d_{0,0}^{\pm}d_{2,0}^{\pm}+2^2d_{1,1}^{\pm})
    +\frac{\bg_2^2}{2^6}\{(d_{1,0}^{\pm})^2-2d_{0,1}^{\pm}\} ,
   \\
  e_{2,3}^{\pm}
  &=\frac{\bg_5}{2^2}d_{2,2}^{\pm}
    +\frac{\bg_1\bg_4}{2^5}(2^2d_{1,2}^{\pm}+d_{0,0}^{\pm}d_{2,1}^{\pm})
    +\frac{\bg_2\bg_3}{2^5}
     (d_{0,1}^{\pm}d_{2,0}^{\pm}+d_{1,0}^{\pm}d_{1,1}^{\pm}-d_{0,2}^{\pm}) ,
\end{align}
\begin{align}
  e_{3,0}^{\pm}
  &=\frac{\bg_3-\bg_1\bg_2}{2^3}
   =\frac{\eps^2\bg_1^4}{2^3}\bigg(\frac{2}{3}+\frac{\eps^2\bg_1}{5}\bigg)
   ,\qquad\quad
  e_{3,1}^{\pm}
  =\frac{\bg_4}{2^2}d_{3,0}^{\pm}
   +\frac{\bg_1\bg_3}{2^6}(2^3d_{2,0}^{\pm}+d_{0,0}^{\pm})
   -\frac{\bg_2^2}{2^5}d_{1,0}^{\pm} ,
   \\
  e_{3,2}^{\pm}
  &=\frac{\bg_5}{2^2}d_{3,1}^{\pm}
    +\frac{\bg_1\bg_4}{2^5}(2^2d_{2,1}^{\pm}+d_{0,0}^{\pm}d_{3,0}^{\pm})
    +\frac{\bg_2\bg_3}{2^6}(d_{0,1}^{\pm}-2d_{1,1}^{\pm}+2d_{1,0}^{\pm}d_{2,0}^{\pm}) ,
   \\
  e_{3,3}^{\pm}
  &=\frac{\bg_6}{2^2}d_{3,2}^{\pm}
    +\frac{\bg_1\bg_5}{2^5}(2^2d_{2,2}^{\pm}+d_{0,0}^{\pm}d_{3,1}^{\pm})
    +\frac{\bg_2\bg_4}{2^5}
     (d_{0,1}^{\pm}d_{3,0}^{\pm}+d_{1,0}^{\pm}d_{2,1}^{\pm}-d_{1,2}^{\pm})
    +\frac{\bg_3^2}{2^6}(d_{0,2}^{\pm}+2d_{1,1}^{\pm}d_{2,0}^{\pm}) .
\end{align}
We find the asymptotic solution of Eq.(\ref{eq:zeroeq}) with respect to $\veps_s$
up to $O(\hveps^3)$ by iteration,
\begin{align}
 -\veps_s
  &\simeq
    \hveps
   +\hveps^2\frac{q_2^{\pm}}{q_1^{\pm}}
   +\hveps^3
    \bigg\{2\bigg(\frac{q_2^{\pm}}{q_1^{\pm}}\bigg)^{\!2}-\frac{q_3^{\pm}}{q_1^{\pm}}\bigg\}
   +O(\hveps^4) ,
    \qquad
  \hveps
  =\frac{q_0^{\pm}-(k_x^m/k_r^n)^2}{q_1^{\pm}} .
   \label{eq:vepss}
\end{align}
$O(\hveps^j)$ denotes $O(\eps^{2j})$ for simplicity:
\begin{alignat}{2}
  O(\hveps)
  &=O(\eps^2)
   =O(\eps^2,\eps_s) ,
    \qquad&
  O(\hveps^3)
  &=O(\eps^6)=O(\eps^6,\eps^4\eps_s,\eps^2\eps_s^2,\eps_s^3) ,
   \\
  O(\veps^2)
  &=O(\eps^4)
   =O(\eps^4,\eps^2\eps_s,\eps_s^2) ,
    \qquad&
  O(\hveps^4)
  &=O(\eps^8)=O(\eps^8,\eps^6\eps_s,\eps^4\eps_s^2,\eps^2\eps_s^3,\eps_s^4) .
\end{alignat}
In order to find the asymptotic solution $\veps_s$ up to $O(\hveps^4)$,
we must take $(\bpi_7,\bsig_7)/\bnu^7$ into account in Eq.(\ref{tht}).

We expand the R.H.S. of Eq.(\ref{eq:vepss}) with respect to $(k_r^n\rho)^{-2}$ and
$\eps_s$ given by Eq.(\ref{eq:sig_bsig}).
Rewriting Eq.(\ref{eq:vepss}) using $\nu\,(=i\bnu)$ instead of $\bnu$,
we get the asymptotic solutions of Eqs.(\ref{eq:pnu_eq0}-\ref{eq:snu_eq0})
under the assumptions given by Eqs.(\ref{eq:bz_cond_big}),
\begin{align}
  \bigg\{\frac{(\nu_m^n,\mu_m^n)}{k_r^n\rho}\bigg\}^2
  &\simeq \sum_{j,\ell=0}^{\infty}
   \frac{(N_{\ell,j}^{+},N_{\ell,j}^{-})}{(k_r^n\rho)^{2\ell}}
   \eps_s^{j}
   \label{nu2_sum}
   \\
  &\simeq
    \bigg\{
       N_{0,1}^{\pm}\bigg(\frac{k_s^{mn}}{k_r^n}\bigg)^2
      +\frac{N_{1,0}^{\pm}}{(k_r^n\rho)^2}
    \bigg\}
   +\bigg\{
       \frac{N_{1,1}^{\pm}}{(k_r^n\rho)^2}\bigg(\frac{k_s^{mn}}{k_r^n}\bigg)^2
      +\frac{N_{2,0}^{\pm}}{(k_r^n\rho)^4}
    \bigg\}
   \nonumber\\&\quad
   +\bigg\{
      \frac{N_{1,2}^{\pm}}{(k_r^n\rho)^2}\bigg(\frac{k_s^{mn}}{k_r^n}\bigg)^4
     +\frac{N_{2,1}^{\pm}}{(k_r^n\rho)^4}\bigg(\frac{k_s^{mn}}{k_r^n}\bigg)^2
     +\frac{N_{3,0}^{\pm}}{(k_r^n\rho)^6}
    \bigg\}
   +O(\eps^8) ,
   \label{nu2_6th}
\end{align}
where the sign $(+,-)$ of $N_{\ell,j}^{\pm}$ corresponds to $(\nu_m^n,\mu_m^n)$.
$N_{\ell,j}^{\pm}$ is given as follows,
\begin{alignat}{2}
  N_{0,1}^{\pm}
  &=\frac{1}{\bg_1}
   ,\qquad
  N_{1,0}^{\pm}
  =\frac{(1,-3)}{4}
   ,\qquad\quad&
  N_{1,1}^{\pm}
  &=\frac{(3,-5)}{2}
   ,\qquad
  N_{2,0}^{-}
  =\bg_1
  =\frac{1}{g_bg_a} ,
   \\
  N_{1,2}^{\pm}
  &=\frac{1}{2^2}
    \bigg\{
       (11,-17)
      -\frac{(k_r^nw)^2}{3}
    \bigg\}
   ,\qquad\quad&
  N_{2,1}^{\pm}
  &=\frac{\bg_1}{2}
    \bigg\{
       \frac{(3,-5)}{3}(k_r^nw)^2
      +\frac{3}{2}(-7,39)
    \bigg\} ,
   \\
  N_{3,0}^{-}
  &=\bg_1^2
    \bigg\{
       (k_r^nw)^2
      -\frac{3^2\cd47}{2^4}
    \bigg\}
   ,\qquad\quad&
  N_{0,3}^{\pm}
  &=
   -\frac{\eps^2}{12}\bigg(1+\frac{2}{15}\bg_1\eps^2\bigg)
   ,\qquad
   \eps
   =\frac{w}{\rho} ,
   \\
  N_{0,0}^{\pm}
  &=N_{0,2}^{\pm}
  =N_{2,0}^{+}
  =N_{3,0}^{+}
  =0,
\end{alignat}
where the sign $(+,-)$ of $N_{\ell,j}^{\pm}$ corresponds to the values in
the parentheses as $N_{1,0}^{+}=1/4$ and $N_{1,0}^{-}=-3/4$.
$N_{0,j}^{\pm}$ for $0\leq j\leq 3$ is common between $\nu_m^n$ and $\mu_m^n$.
$N_{0,3}^{\pm}\eps_s^3=O(\eps^2\eps_s^3)=O(\eps^8)$ which is negligible in
Eq.(\ref{nu2_6th}).
Eqs.(\ref{nu2_sum}-\ref{nu2_6th}) hold for $m\geq m_{a,b}+1$ which are the normal modes
(\ie, not the whispering gallery modes).
Let us neglect the terms of $O(\eps^6)$ or $O(\eps^4)$ in Eq.(\ref{nu2_6th}),
\begin{align}
  \bigg\{\frac{(\nu_m^n,\mu_m^n)}{k_r^n\rho}\bigg\}^2
  &\simeq
    \bigg\{
      \frac{1}{\bg_1}
     +\frac{(3,-5)}{2(k_r^n\rho)^2}
    \bigg\}
    \bigg(\frac{k_s^{mn}}{k_r^n}\bigg)^2
   +\frac{(1,-3)}{4(k_r^n\rho)^2}
   +\frac{(0,3)\bg_1}{(k_r^n\rho)^4}
   +O(\eps^6)
   \label{eq:nu2}
   \\
  &\simeq
    \frac{1}{\bg_1}\bigg(\frac{k_s^{mn}}{k_r^n}\bigg)^2
   +\frac{(1,-3)}{4(k_r^n\rho)^2}
   +O(\eps^4) .
   \label{eq:nu_rho2}
\end{align}
$\bg_1$ is given by Eq.(\ref{eq:bg1}).
Eq.(\ref{eq:nu_rho2}) satisfies the asymptotic condition (\ref{eq:asympt_lim}).
We got Eq.(\ref{eq:nu_rho2}) from Eq.(\ref{eq:nu2}) just by neglecting the terms of
$O(\eps^4,\eps^2\eps_s,\eps_s^2)$ in it on purpose to compare it with that which
we will find in another way as shown below.

We got Eq.(\ref{eq:nu_rho2}) using Eqs.(\ref{eq:Fimu_muz}-\ref{eq:Gimu_muz})
which are Dunster's uniform asymptotic expansions of $F_{i\bnu}$ and $G_{i\bnu}$.
We can get Eq.(\ref{eq:nu_rho2}) without using
Eqs.(\ref{eq:Fimu_muz}-\ref{eq:Gimu_muz}).
That is, it is also possible to derive Eq.(\ref{eq:nu_rho2}) using Hankel's asymptotic 
expansions (\ref{eq:J_hankel}-\ref{eq:Y_hankel}),
because Eq.(\ref{eq:nu_rho2}) is the asymptotic solution of
Eqs.(\ref{eq:pnu_eq0}-\ref{eq:snu_eq0}) for $|\nu/k_r^n\rho|\ll1$.
Eqs.(\ref{eq:J_hankel}-\ref{eq:Y_hankel}) are the asymptotic expansions of
$J_{\nu}$ and $Y_{\nu}$ for large argument with the order $\nu$ fixed.
Assuming that $|\hr_{b,a}|\gg1$ with $\nu$ fixed, we expand the cross products
$p_{\nu}(\hr_b,\hr_a)$ and $s_{\nu}(\hr_b,\hr_a)$ with respect to $\hr_{b,a}^{-1}$
as shown in Eq.(\ref{eq:pnu_hankel}),
\begin{align}
  \bigg\{{p_{\nu}(\hr_b,\hr_a) \atop s_{\nu}(\hr_b,\hr_a)}\bigg\}
  &\simeq
   \frac{2}{\pi}\bigg(\frac{X_{\pm}^2+Y_{\pm}^2}{\hr_b\hr_a}\bigg)^{1/2}
    \sin(\vth_{\pm}-k_r^nw) ,
   \qquad
  \vth_{\pm}
  =\tan^{-1}\!\bigg(\frac{Y_{\pm}}{X_{\pm}}\bigg)
   \qquad
  (|k_r^n\rho|\gg1) ,
  \label{eq:ps_hankel}
\end{align}
where the double sign $(+,-)$ correspond to $(p_{\nu},s_{\nu})$.
$X_{\pm}$ and $Y_{\pm}$ are given by Eqs.(\ref{eq:XY_pls}-\ref{eq:XY_mns}),
\begin{alignat}{2}
  X_{+}
  &=\hP_{\nu}(\hr_b)\hP_{\nu}(\hr_a)+\hQ_{\nu}(\hr_b)\hQ_{\nu}(\hr_a) ,
   \qquad&
  Y_{+}
  &=\hP_{\nu}(\hr_b)\hQ_{\nu}(\hr_a)-\hQ_{\nu}(\hr_b)\hP_{\nu}(\hr_a) ,
   \\
  X_{-}
  &=\hR_{\nu}(\hr_b)\hR_{\nu}(\hr_a)+\hS_{\nu}(\hr_b)\hS_{\nu}(\hr_a) ,
   \qquad&
  Y_{-}
  &=\hR_{\nu}(\hr_b)\hS_{\nu}(\hr_a)-\hS_{\nu}(\hr_b)\hR_{\nu}(\hr_a) .
\end{alignat}
$\hP_{\nu},\hQ_{\nu},\hR_{\nu}$ and $\hS_{\nu}$ are given by
Eqs.(\ref{eq:Pnu_coe}).
In addition to the condition $|\hr_{b,a}|=O(k_r^n\rho)\gg1$, we assume
\begin{align}
  |\nu|\leq O(1)
   \qquad\Ra\qquad
  |\nu/k_r^n\rho|\ll1 .
   \label{eq:assm_hankel}
\end{align}
Eq.(\ref{eq:assm_hankel}) includes the case $|\nu|\ll1$.
Assumption (\ref{eq:assm_hankel}) is equivalent to 
$|\bz|\gg 1$ given by Eq.(\ref{eq:bz_cond_big}) where $(k_r^n\rho)^{-1}=O(\eps/m)$.
Eq.(\ref{eq:assm_hankel}) allows us to truncate the series (\ref{eq:Pnu_coe}) up to
the terms higher than the second order with respect to $\nu/\hr$ and $\nu^2/\hr$,
\begin{alignat}{2}
  \hP_{\nu}(\hr)
  \simeq \hR_{\nu}(\hr)
  \simeq 1+O(\nu^{\ell}/\hr^2) ,
    \qquad
  \{\hQ_{\nu}(\hr),\hS_{\nu}(\hr)\}
  \simeq\frac{4\nu^2-\{1,-3\}}{8\hr}+O(\nu^{\ell'}/\hr^3) ,
   \label{eq:PQRS_hankel}
\end{alignat}
where $\ell\in\{0,1,2\}$ and $\ell'\in\{0,1,2,3\}$ are indices which indicate
the power of $\nu$ in the error terms.

Expanding $\vth_{\pm}$ with respect to $Y_{\pm}/X_{\pm}=O(\eps^2/m)$,
we take into account only the leading order term in it,
\begin{align}
  (\vth_{+},\vth_{-})
  &\simeq
   \frac{\eps^2}{8g_bg_ak_r^nw}\{4(\nu_m^n,\mu_m^n)^2-(1,-3)\}
  +O(\eps^4) .
\end{align}
$\vth_{\pm}=O(\eps^2/m)$ similar to $\tht_{\pm}$ given by Eq.(\ref{eq:theta_lt}).
We can get the asymptotic expressions of the zeros of Eq.(\ref{eq:ps_hankel}) by solving
\begin{align}
  k_r^nw-\vth_{\pm}
  \simeq k_x^mw
   \qquad\Ra\qquad
  (k_s^{mn}w)^2
  \simeq 2k_r^nw\vth_{\pm}+O(\eps^4/m^2) ,
  \label{eq:hankel}
\end{align}
where we neglected the term $\vth_{\pm}^2=O(\eps^4/m^2)$ after squaring the first equation
since we are seeking the square of the solution $\nu^2$ to leading order with respect 
to $\eps$.
Solving the second equation of (\ref{eq:hankel}) with respect to $\nu^2$
which is involved in $\vth_{\pm}$, we get
\begin{align}
  (\nu_m^n,\mu_m^n)^2
  \simeq
    g_bg_a(k_s^{mn}\rho)^2
   +\frac{(1,-3)}{4}
   +O(\eps^2/m^2)
   \qquad
  (m>m_{a,b}) .
   \label{eq:nu2_hankel}
\end{align}
Eq.(\ref{eq:nu2_hankel}) agrees with Eq.(\ref{eq:nu_rho2}).
Thus, we can get Eq.(\ref{eq:nu_rho2}) in the following two ways using
the different asymptotic expansions:
(i) Dunster's uniform asymptotic expansions for $\nu=i\bnu$ under the assumptions
$\eps\ll1$ and $|\bnu/k_r^n\rho|\ll1$,
(ii) Hankel's asymptotic expansions for $|k_r^n\rho|\gg1$ with $\nu$ fixed 
under the assumption $|\nu/k_r^n\rho|\ll1$.

\subsection{Cutoff wavenumbers and Horvat-Prosen's approximate solutions}
\label{sec:kcutoff}

In order to examine Eq.(25) in \cite{horvat_prosen}, which is given by Eq.(\ref{eq:HP25})
using our notation, we consider $k_m^n$ and $\bk_m^n$ which are the cutoff wavenumbers of 
the curved pipe, defined implicitly by Eqs.(\ref{eq:kmn}-\ref{eq:bkmn}).
They are the transition points with respect to the wavenumber $k$
such that $\nu_m^n=0$ and $\mu_m^n=0$ respectively,
\ie, the longitudinal wavenumber of $E_y$ and $B_y$ become zero in the curved pipe.
We define $\kap_r^m$ and $\bkap_r^m$ respectively as the zeros of the poles
$\nu_m^n$ and $\mu_m^n$ with respect to the radial wavenumber $k_r^n$, \ie,
\begin{alignat}{4}
  \nu_m^n(k_r^n)
  &=0
   \qquad&\Lra&\qquad&
  (k_r^n)^2
  &=(\kap_r^m)^2
   =(k_m^n\beta)^2-(k_y^n)^2
   \qquad&&
  (m\in\mathbb{N}) ,
   \label{eq:kaprm_p}
   \\
  \mu_m^n(k_r^n)
  &=0
   \qquad&\Lra&\qquad&
  (k_r^n)^2
  &=(\bkap_r^m)^2
   =(\bk_m^n\beta)^2-(k_y^n)^2
   \qquad&&
  (m\in\mathbb{Z}_0^{+}) .
   \label{eq:kaprm_s}
\end{alignat}
We do not have the explicit and exact expressions of $\kap_r^m$ and $\bkap_r^m$ except
for $\bkap_r^0$.
We can get the expression of $\bkap_r^0$ exactly from $(k_r^n)^2=0$ since $\mu_0^n=0$
at $k_r^n=0$ as seen from Eq.(\ref{eq:nu_krn0}),
\begin{align}
  \bkap_r^0=0
   ,\qquad
  \bk_0^n=k_y^n/\beta .
\end{align}
From Eq.(\ref{eq:nu2}), we get the asymptotic expressions of $(\kap_r^m,\bkap_r^m)$ 
and $(k_m^n,\bk_m^n)$ for $m\in\mathbb{N}$ and $\eps=w/\rho\ll1$ up to $O(\eps^4)$,
\begin{align}
  \bigg\{\frac{(\kap_r^m,\bkap_r^m)}{k_x^m}\bigg\}^2
  &\simeq
     1
    -\frac{(1,-3)}{4r_br_a(k_x^m)^2}
     \bigg\{1-\frac{3}{2r_br_a(k_x^m)^2}\bigg\}
    +O(\eps^6) ,
   \label{eq:kaprm2_p}
   \\
  \bigg\{\frac{(k_m^n,\bk_m^n)}{\hk_m^n}\bigg\}^2
  &\simeq
     1
    -\frac{(1,-3)}{4r_br_a(\hk_m^n\beta)^2}
     \bigg\{1-\frac{3}{2r_br_a(k_x^m)^2}\bigg\}
    +O(\eps^6) .
  \label{eq:cutoff}
\end{align}
$\hk_m^n$ is the cutoff wavenumber of the straight pipe, given by Eq.(\ref{eq:hkmn}).
Eqs.(\ref{eq:kaprm2_p}-\ref{eq:cutoff}) have a value close to 1
since we assume $\eps\ll1$, \ie, the curvature of the pipe is not so sharp.
We rewrite Eq.(\ref{eq:nu2}) using $(\kap_r^m,\bkap_r^m)$,
\begin{align}
  \bigg\{\frac{(\nu_m^n,\mu_m^n)}{\rho}\bigg\}^2
  &\simeq
    \big\{(k_r^n)^2-(\kap_r^m,\bkap_r^m)^2\big\}
    \bigg\{\frac{1}{\bg_1}+\frac{(3,-5)}{2(k_r^n\rho)^2}\bigg\}
   +O(\eps^6)
   \qquad
  (m>m_a) .
   \label{eq:numu_kap}
\end{align}
$\bg_j$ is given by Eqs.(\ref{eq:bgj}-\ref{eq:bg1}).
Eq.(\ref{eq:numu_kap}) is the asymptotic expression of the poles of the normal mode under
the assumption $|\nu/\rho|\ll|k_r^n|=O(k_x^m)$ as described in Eqs.(\ref{eq:bz_cond_big}).
In order to compare Eq.(\ref{eq:numu_kap}) with Eq.(25) in \cite{horvat_prosen},
we rewrite Eq.(\ref{eq:numu_kap}) using Eq.(\ref{eq:kaprm2_p}),
\begin{align}
  \bigg\{\frac{k_r^n}{(\kap_r^m,\bkap_r^m)}\bigg\}^2
  -\bg_1'\bigg\{\frac{(\nu_m^n,\mu_m^n)}{k_x^m\rho}\bigg\}^2
  &\simeq 1+O(\eps^6)
   \qquad
  (m>m_a) .
  \label{eq:my25_nu}
\end{align}
Eq.(\ref{eq:my25_nu}) is the correct asymptotic relation between $(\nu_m^n,\mu_m^n)$ and
$(\kap_r^m,\bkap_r^m)$ for $|\eps_s|\ll1$ as we assumed in Eq.(\ref{eq:bz_cond_big}).
$\bg_1'$ depends on not only the geometric parameters ($\rho$ and $r_{b,a}$)
but also the radial wavenumber $k_r^n$,
\begin{align}
  \bg_1'
  &\simeq
   \bg_1
    \bigg\{
       1
      +\frac{(-5,7)}{4(k_r^n\rho)^2}\bg_1
      +O(\eps^4,\eps^2\eps_s,\eps_s^2)
     \bigg\} .
   \label{bg1p_gl2}
\end{align}
$\bg_1'$ represents two functions depending on the coefficients $(-5,7)$
which correspond to $(\nu_m^n,\mu_m^n)$ and $(\kap_r^m,\bkap_r^m)$.
$\bg_1$ and $g_l$ are the geometric factors given by Eqs.(\ref{eq:bg1}) and (\ref{eq:gl}).

Let us discuss Eq.(\ref{eq:HP25}) which is the conjecture given by
Eq.(25) in \cite{horvat_prosen}.
Although Eq.(\ref{eq:HP25}) looks similar to Eq.(\ref{eq:my25_nu}),
it slightly differs from Eq.(\ref{eq:my25_nu}) since $\bg_1'\ne g_l^2$ as seen from
\begin{align}
  \bg_1
  \simeq
   g_l^2
   +\frac{w^2}{12\rho^2}
    \bigg(
      1
     -2\frac{x_b+x_a}{\rho}
    \bigg)
   +O(\eps^4) .
   \label{bg1_gl2}
\end{align}
Although $\bg_1$ is equal to $g_l^2$ up to $O(\eps)$,
they differ in the terms higher order than $O(\eps^2)$.
Therefore, to be strict, Eq.(\ref{eq:HP25}) is incorrect, \ie,
Eq.(25) in \cite{horvat_prosen} is a pseudo asymptotic solution of
Eq.(\ref{eq:pnu_eq0}) for $|\eps_s|\ll1$ and $\eps\ll1$.
The difference of Eq.(\ref{eq:HP25}) from Eq.(\ref{eq:my25_nu}) is of
$O[\eps_s\eps^2,\eps_s(k_r^n\rho)^{-2}]$
which is the first order with respect to $\eps_s$.
Accordingly, Eq.(25) of \cite{horvat_prosen} is the approximate solution of
Eq.(\ref{eq:pnu_eq0}) to the zeroth order with respect to $\eps_s$ and
the first order with respect to $\eps$.
As already seen in appendix \ref{sec:asympt_numu},
the logarithmic factor $g_l$ is involved in Eq.(\ref{eq:nu_ks})
which is the asymptotic expression of the poles for $|\eps_r|\ll1$, \ie,
$|\nu/\rho|\gg|k_r^n|$.

According to the discussion in appendices \ref{sec:asympt_numu}-\ref{sec:HP2},
we conclude that Eq.(25) in \cite{horvat_prosen} is
a different kind of approximate solution from Eq.(26) in \cite{horvat_prosen},
because we found the former for $|\nu/\rho|\ll O(k_r^n)$, \ie, $k_x^m=O(k_r^n)$
unlike the latter which is gotten assuming $|\nu/\rho|\gg O(k_r^n)$, \ie, $k_x^m\gg k_r^n$
as described in Eqs.(\ref{eq:cond26}-\ref{eq:cond25}).
Thus, Eqs.(\ref{eq:HP26}) is inconsistent with Eq.(\ref{eq:HP25}), \ie, they are 
approximate expressions which do not hold simultaneously under a common condition.
Therefore it is incorrect to derive Eq.(\ref{eq:HP26}) from Eq.(\ref{eq:HP25})
by substituting $k_x^m\simeq\kap_r^m$ into the latter.
One may wonder why Eq.(26) in \cite{horvat_prosen} is gotten from 
Eq.(25) in \cite{horvat_prosen} by substituting $k_x^m\simeq \kap_r^m$ into it.
This is because of the absence of consideration on the error terms in
the asymptotic solutions.
It is true that we derived Eqs.(\ref{eq:nu_ks}) and (\ref{nu2_6th}) under
different assumptions with respect to the variables of the field,
\ie, $|\eps_r|\ll1$ or $|\eps_s|\ll1$ as listed in Eqs.(\ref{eq:cond26}-\ref{eq:cond25}),
but Eqs.(\ref{eq:nu_ks}) and (\ref{nu2_6th}) have common terms up to $O(\eps)$ under
the assumption $\eps=w/\rho\ll1$ as shown in Eq.(\ref{nu_rho_ks}),
\begin{align}
  \bigg\{\frac{(\nu_m^n,\mu_m^n)}{k_s^{mn}\rho}\bigg\}^2
  \simeq 
  1+\frac{x_b+x_a}{\rho}+O(\eps^2)
   \qquad
  (m>m_a) ,
   \label{nu_rho_ks_1}
\end{align}
where we do not indicate the order of the error with respect to $\eps_{r,s}$ on purpose
to get Eq.(26) in \cite{horvat_prosen} from Eq.(25) in \cite{horvat_prosen} in a wrong way.
That is, the obscurity of the error terms with respect to $\eps_{r,s}$ in
Eq.(\ref{nu_rho_ks_1}) allows us to confuse Eq.(25) with Eq.(26) in \cite{horvat_prosen}.
Let us demonstrate the wrong derivation from now.
Since Eq.(\ref{nu_rho_ks_1}) has an accuracy of $O(\eps)$ at most,
we can rewrite it using either $\bg_1$ or $g_l$ as follows,
\begin{align}
  &
  \bigg\{\frac{(\nu_m^n,\mu_m^n)}{k_s^{mn}\rho}\bigg\}^2
  \simeq \frac{1}{\bg_1}+O(\eps^2)
   \simeq \frac{1}{g_l^2}+O(\eps^2)
   \label{nu_rho_ks_2}
   \\
  \Ra\quad~~&
  \bigg(\frac{k_r^n}{k_x^m}\bigg)^2
  -\bg_1\bigg\{\frac{(\nu_m^n,\mu_m^n)}{k_x^m\rho}\bigg\}^2
  \simeq 1+O[\eps^2(k_s^{mn}/k_x^m)^2] .
   \label{nu_rho_ks_3}
\end{align}
In Eqs.(\ref{nu_rho_ks_2}-\ref{nu_rho_ks_3}) we can replace
$\bg_1$ directly by $g_l^2$ without using Eq.(\ref{bg1_gl2})
since Eqs.(\ref{nu_rho_ks_2}-\ref{nu_rho_ks_3}) have an error of $O(\eps^2)$.
In accordance with it, neglecting the higher order terms $O(\eps^2)$ in
Eq.(\ref{eq:kaprm2_p}), we get
\begin{align}
  (\kap_r^m,\bkap_r^m)
  \simeq k_x^m+O(\eps^2)
   \qquad
  (\eps\ll1,~ |\eps_s|\ll1) .
  \label{kap_bkap}
\end{align}
If we do not assume $|\eps_s|\ll1$, we cannot replace $k_x^m$ with $\kap_r^m$ in
Eq.(\ref{nu_rho_ks_3}) since Eq.(\ref{eq:kaprm2_p}) is the asymptotic expression
for $|\eps_s|\ll1$.
Substituting Eq.(\ref{kap_bkap}) only into the first term on the L.H.S. of
Eq.(\ref{nu_rho_ks_3}), it seems to become equivalent to Eq.(\ref{eq:my25_nu}) up to
the zeroth order with respect to $\eps_s$ and the first order with respect to $\eps$.
However, it is arbitrary and strange to substitute Eq.(\ref{kap_bkap}) only into
the first term on the L.H.S. of Eq.(\ref{nu_rho_ks_3})
despite the fact that the second term also involves $k_x^m$.
Thus, we can get Eq.(26) from Eq.(25) in \cite{horvat_prosen}
in an incorrect way which does not properly deal with the error terms with respect to
$\eps_{r,s}$ and $\eps^2$, violating Eq.(\ref{eq:krn_apdx}) which is
the dispersion relation of the field in the straight rectangular pipe.
According to the above considerations, Eqs.(25-26) in \cite{horvat_prosen} are
pseudo-equivalent to each other up to the first order with respect to $\eps$ and
the zeroth order with respect to $\eps_r$ and $\eps_s$.
They are not equivalent since the condition $|\eps_r|\ll1$ contradicts with $|\eps_s|\ll1$ 
according to Eq.(\ref{eq:krn_apdx}).

\subsection{Pole of the zeroth radial mode}
\label{sec:0th_0}

The Green function $\mfG_{-}^n$ has the zeroth poles
$\nu=\pm\mu_0^n$ in the Laplace plane unlike $\mfG_{+}^n$.
That is, the cross product $s_{\nu}(\hr_b,\hr_a)$ has the zeros of the zeroth radial mode
($m=0$) with respect to $\nu$.
We will find the asymptotic expression of $\mu_0^n$ which is symmetric with
respect to the exchange of the inner and outer walls of the curved pipe.
It differs from Eqs.(\ref{cochran_sol}) and (\ref{eq:imag_whisper}) for $m=0$,
which are the real and imaginary whispering gallery modes of the curved pipe.
We find the expression of $\mu_0^n$ under the following assumptions,
\begin{align}
  \nu^{-1}=O(\eps)
   ,\qquad
  (k_r^n\rho)^{-1}=O(\eps)
   ,\qquad
  k_r^nw=O(1)
   ,\qquad
  \eps
  =w/\rho
  \ll1 .
   \label{eq:eps_def}
\end{align}
As described below Eq.(\ref{eq:eta_ba}), for facility, we find the following $\tau_0$ as
the asymptotic solution of Eq.(\ref{eq:snu_eq0}) with respect to $\nu$ instead of
the pole $\nu=\mu_0^n\in\mathbb{A}$ itself,
\begin{align}
  \tau_0
  =\bigg(\frac{\mu_0^n}{k_r^n\rho}\bigg)^2
   \in\mathbb{R} ,
    \qquad
  k_r^n
  =k_s^{0n} .
  \label{eq:snu0_m0}
\end{align}
When $m=0$, the radial wavenumber $k_r^n$ equals $k_s^{mn}$
which is the longitudinal wavenumber of the field in the straight pipe,
given by Eqs.(\ref{eq:ks_mn}) and (\ref{eq:krn_apdx}).
$\tau_0$ is a quantity of $O(1)$ under the assumptions (\ref{eq:eps_def}) since $\mu_0^n$ 
satisfies the asymptotic condition (\ref{eq:lim_nurho_ksmn}) in the limit of $\rho\to\infty$.
For brevity, we write $\mu_0^n$ simply as $\nu$ in the derivation excluding the final
expression of $\tau_0$ which will be shown later in Eq.(\ref{eq:tau0_2nd}).

Assuming Eqs.(\ref{eq:eps_def}), we expand $s_{\nu}(\hr_b,\hr_a)$ in
the uniform asymptotic series which involves the variables
$z$, $\zeta$ and $\eps_z\,(=1-z^2)$ for both $r=r_a$ and $r_b$:
\begin{alignat}{3}
  z_b
  &=g_b\tau_0^{-1/2}
   >1
   ,\qquad &&
  \zeta_b<0
   ,\qquad &&
  \eps_b
  =1-z_b^2
  <0 ,
  \label{eq:zb}
   \\
  z_a
  &=g_a\tau_0^{-1/2}
   <1
   ,\qquad &&
  \zeta_a>0
   ,\qquad &&
  \eps_a
  =1-z_a^2
  >0 .
  \label{eq:za}
\end{alignat}
$g_{b,a}$ is given by Eq.(\ref{eq:gl}).
In finding the asymptotic expression of $\mu_0^n$ which is symmetric with respect to
the exchange of $r_a$ and $r_b$ like a normal mode,
we assume $z_a<1<z_b$ similar to the whispering gallery modes.
The zeroth mode differs from the higher order modes in this regard.
We expand $s_{\nu}(\hr_b,\hr_a)$ in the uniform asymptotic series
given by Eq.(\ref{eq:snu_uae}),
\begin{align}
  s_{\nu}(\hr_b,\hr_a)
  &\simeq
  -\frac{2\nu}{\hr_b\hr_a}\bigg(\frac{-\eps_b\eps_a}{u_bu_a}\bigg)^{1/4}
   \{D_{\nu}^{00}+\delta_1+O(\eps^{14/3})\} ,
    \qquad
  \delta_1
  =\frac{D_{\nu}^{01}+D_{\nu}^{10}}{\nu^2}
  =O(\eps^{8/3}) .
  \label{eq:snu_Snu}
\end{align}
$u_a$ and $-u_b$ are the variables of the Airy functions
which are involved in Eqs.(\ref{eq:JY_uae}-\ref{eq:dJY_uae}),
\begin{alignat}{3}
  u_b
  &=\nu^{2/3}(-\zeta_b)
  &&=(\nu/2)^{2/3}(-\eps_b)\{2\psi(z_b)\}^{2/3}
  &&>0 ,
  \label{eq:ub_m0}
   \\
  u_a
  &=\nu^{2/3}(+\zeta_a)
  &&=(\nu/2)^{2/3}(+\eps_a)\{2\psi(z_a)\}^{2/3}
  &&>0 .
  \label{eq:ua_m0}
\end{alignat}
$\psi(z)$ is given by Eq.(\ref{eq:psi}).
Since $\eps_{a,b}=O(\eps)$ under the assumptions (\ref{eq:eps_def}),
we can expand $\{2\psi(z)\}^{2/3}$ with respect to $\eps_z$ using Eq.(\ref{eq:Sig}).
$D_{\nu}^{jl}$ is the function given by Eq.(\ref{eq:Snu_jl}),
\begin{align}
  D_{\nu}^{00}(\zeta_b,\zeta_a)
  &=
    \mfs(-u_b,u_a)
    +\frac{c_0(\zeta_a)}{\nu^{2/3}}\mfr(-u_b,u_a)
    +\frac{c_0(\zeta_b)}{\nu^{2/3}}\mfq(-u_b,u_a)
    +\frac{c_0(\zeta_a)c_0(\zeta_b)}{\nu^{4/3}}\mfp(-u_b,u_a) .
  \label{eq:S00}
\end{align}
$\mfp$, $\mfq$, $\mfr$ and $\mfs$ are the cross products of $\Ai$ and $\Bi$,
given by Eqs.(\ref{eq:cp_airy_p}-\ref{eq:cp_airy_sr}).
$c_0$ is given by Eq.(\ref{eq:c0_series}),
\begin{align}
  c_0(\zeta_{b,a})
  &=\frac{\{\psi(z_{b,a})\}^{-2/3}}{4}
    \sum_{j=0}^{\infty}\frac{(j+6)\eps_{b,a}^j}{(2j+3)(2j+5)}
  \label{eq:c0_zeta}
   \\
  &\simeq
    \frac{2^{2/3}}{10}
    \Big\{
      1
     +\frac{1}{10}\eps_{b,a}
     +\frac{2}{63}\eps_{b,a}^2
     +O(\eps_{b,a}^3)
    \Big\} .
\end{align}
$c_0$ is a quantity of the zeroth order with respect to $\eps$ and $\eps_z$.
In deriving the asymptotic expression of $\tau_0$ for $\eps=w/\rho\ll1$, it is instructive
to keep in mind the following orders of the variables with respect to $\eps$ under 
the assumptions (\ref{eq:eps_def}),
\begin{align}
  \nu
  =O(\eps^{-1})
   ,\qquad
  z_{b,a}=O(1)
   ,\qquad
  \zeta_{b,a}=O(\eps)
   ,\qquad
  \eps_{b,a}=O(\eps)
   ,\qquad
  u_{b,a}=O(\eps^{1/3}) .
   \label{eq:order}
\end{align}
In appendix \ref{sec:0th_0}, since we assume $u_{b,a}=O(\eps^{1/3})$
which goes to zero for $\eps\to0$,
we expand the Airy functions in Taylor series around $u_{b,a}=0$
as shown in Eqs.(\ref{eq:mfp_ts}-\ref{eq:mfr_ts}),
\begin{align}
  \pi\mfs(-u_b,u_a)
  &\simeq
   \frac{1}{2}(u_b^2-u_a^2)
   +\frac{1}{6}\Big\{(u_bu_a)^2(u_a+u_b)-\frac{1}{5}(u_b^5+u_a^5)\Big\}
   +O(u_{a,b}^8) ,
  \label{eq:mfs_7th}
   \\
  \pi\mfp(-u_b,u_a)
  &\simeq
   (u_a+u_b)+\frac{1}{12}(u_a-u_b)(u_b+u_a)^3 +O(u_{a,b}^7) ,
  \label{eq:mfp_6th}
   \\
  \pi\mfq(-u_b,u_a)
  &\simeq
   1+\bigg(\frac{u_a^3}{3}-\frac{u_b^3}{6}+\frac{u_bu_a^2}{2}\bigg)+O(u_{a,b}^6) ,
   \label{eq:mfq_8th}
   \\
  \pi\mfr(-u_b,u_a)
  &\simeq
   -1+\bigg(\frac{u_b^3}{3}-\frac{u_a^3}{6}+\frac{u_b^2u_a}{2}\bigg)+O(u_{a,b}^6) .
   \label{eq:mfr_8th}
\end{align}
According to Eqs.(\ref{eq:order}) and (\ref{eq:mfs_7th}-\ref{eq:mfp_6th}),
\begin{align}
  \mfs(-u_b,u_a)=O(\eps^{2/3})
   ,\qquad
  \mfp(-u_b,u_a)=O(\eps^{1/3}) .
\end{align}
The last term in Eq.(\ref{eq:S00}) is $O(\eps^{5/3})$.
On the other hand, $\mfq(-u_b,u_a)$ and $\mfr(-u_b,u_a)$ are quantities of $O(1)$
for $u_{b,a}\to0$.
But since their leading order terms cancel out in Eq.(\ref{eq:S00}),
the sum of the second and third terms of Eq.(\ref{eq:S00}) is $O(\eps^{5/3})$
which is the same order as the last term involving $\mfp(-u_b,u_a)$.
Therefore, in Eq.(\ref{eq:S00}), the first term $\mfs(-u_b,u_a)$ is dominant
over the remaining part by $O(\eps)$.

According to the above consideration, by neglecting $\delta_1=O(\eps^{8/3})$ in
the braces of Eq.(\ref{eq:snu_Snu}), we find the asymptotic expression of $\tau_0$
with the accuracy of $O(\eps^2)$,
\begin{align}
  s_{\nu}(\hr_b,\hr_a)=0
   \qquad\Ra\qquad
  D_{\nu}^{00}+O(\eps^{8/3})
  \simeq 0 .
  \label{eq:S00_0}
\end{align}
We find the asymptotic expression of $\tau_0$ to $O(\eps^2)$ by solving
the second equation of (\ref{eq:S00_0}) iteratively with respect to $\eps$.
We first find $\tau_0$ to $O(\eps)$.
Since $\mfs(-u_b,u_a)$ is the leading order term in Eq.(\ref{eq:S00}),
it is enough to solve the following equation to find the asymptotic expression of
$\tau_0$ to $O(\eps)$,
\begin{align}
  \mfs(-u_b,u_a)+O(\eps^{5/3})\simeq 0
   \quad~~\Ra~~\quad
  u_b^2-u_a^2+O(\eps^{5/3})\simeq 0
   \quad~~\Ra~~\quad
  z_b^2+z_a^2
  \simeq
   2+O(\eps^2) .
   \label{zb2_za2_2}
\end{align}
From the last equation of (\ref{zb2_za2_2}), we get the asymptotic solution to $O(\eps)$,
\begin{align}
  \tau_0
  \simeq g_2+O(\eps^2)
   ,\qquad\text{where}\quad
  g_2
  =\frac{g_b^2+g_a^2}{2}
   ,\qquad
  g_{b,a}
  =\frac{r_{b,a}}{\rho} .
  \label{tau2}
\end{align}
For facility in deriving the asymptotic solution to $O(\eps^2)$,
we calculate the reciprocal of Eq.(\ref{tau2}),
\begin{align}
  \tau_0^{-1}
  \simeq g_2^{-1}+t_2+O(\eps^3) ,
    \qquad
  t_2
  =O(\eps^2) .
  \label{eq:rcpr}
\end{align}
$t_2$ is an unknown which we will find to the leading order.
By the way, $p_{\nu}(\hr_b,\hr_a)$ cannot be zero under the assumptions (\ref{eq:eps_def}), 
because $u_{b,a}>0$ as shown in Eqs.(\ref{eq:ub_m0}-\ref{eq:ua_m0}),
and therefore $u_b+u_a$ cannot be zero,
which is the leading order term of Eq.(\ref{eq:mfp_6th}).
Accordingly, $p_{\nu}(\hr_b,\hr_a)$ has no zeros of the zeroth radial mode.

We find $t_2$ by solving Eq.(\ref{eq:S00_0}).
We multiply $\pi\nu^{2/3}$ with the second equation of (\ref{eq:S00_0}) for convenience,
\begin{align}
   K_s
  +(K_q+K_r)
  +K_p
  +O(\eps^2)
  \simeq 0 ,
   \label{eq:Keq}
\end{align}
where
\begin{alignat}{2}
  K_s
  &=\nu^{2/3}\pi\mfs(-u_b,u_a) ,
    \qquad&
  K_q
  &=c_0(\zeta_b)\pi\mfq(-u_b,u_a) ,
  \label{eq:Ks_Kq}
   \\
  K_p
  &=c_0(\zeta_b)c_0(\zeta_a)\frac{\pi\mfp(-u_b,u_a)}{\nu^{2/3}} ,
    \qquad&
  K_r
  &=c_0(\zeta_a)\pi\mfr(-u_b,u_a) .
  \label{eq:Kp_Kr}
\end{alignat}
We expand each term involved in Eq.(\ref{eq:Keq}) with respect to $\eps_{a,b}$ using
Eqs.(\ref{eq:ub_m0}-\ref{eq:ua_m0}) and (\ref{eq:mfs_7th}-\ref{eq:mfr_8th}),
\begin{align}
  \frac{K_s}{\eps_b-\eps_a}
  &\simeq
    \frac{1}{2}\bigg(\frac{\nu^2}{2^{4/3}}\bigg)
    \Big\{
      (\eps_b+\eps_a)
     +\frac{4}{5}(\eps_b^2+\eps_a\eps_b+\eps_a^2)
    \Big\}
    \nonumber\\&\quad
   +\frac{1}{30}\bigg(\frac{\nu^4}{2^{10/3}}\bigg)
    (\eps_a^4+\eps_a^3\eps_b-4\eps_a^2\eps_b^2+\eps_a\eps_b^3+\eps_b^4)
   +O(\eps) ,
  \label{eq:Ks_eps}
\end{align}
\begin{align}
  \frac{K_q+K_r}{\eps_b-\eps_a}
  &\simeq
   \frac{2^{2/3}}{20}
   \Big\{
     \frac{1}{5}
    -\frac{\nu^2}{12}(\eps_b-\eps_a)^2
   \Big\}
  +O(\eps) ,
    \qquad
  \frac{K_p}{\eps_b-\eps_a}
  \simeq
   -\frac{2^{2/3}}{100}
   +O(\eps) .
  \label{eq:K_pqr_eps}
\end{align}
Substituting Eqs.(\ref{eq:Ks_eps}-\ref{eq:K_pqr_eps}) into Eq.(\ref{eq:Keq}),
we rewrite it as follows,
\begin{align}
    \frac{(k_r^nw)^2}{\nu^2}
    \Big\{
      (\eps_b+\eps_a)
     +\frac{\veps_2}{30}
    \Big\}
   +\frac{(k_r^nw)^2}{60}\veps_4
   +O(\eps^5)
  &\simeq 0 ,
   \label{eq:K2K4_eq}
\end{align}
where
\begin{align}
  \frac{(k_r^nw)^2}{\nu^2}
  &=\frac{\eps^2}{g_2}
    \{1+t_2+O(\eps^3)\} ,
\end{align}
and
\begin{align}
  \veps_2
  =23\eps_b^2+26\eps_a\eps_b+23\eps_a^2 ,
    \qquad
  \veps_4
  =\eps_b^4+\eps_b^3\eps_a-4\eps_b^2\eps_a^2+\eps_b\eps_a^3+\eps_a^4
   .
   \label{eq:K2K4}
\end{align}
Substituting Eq.(\ref{eq:rcpr}) into the last equations of (\ref{eq:zb}-\ref{eq:za}),
we calculate $\eps_{b,a}$ up to $O(\eps^2)$,
\begin{align}
  \eps_b
  &=
   -\frac{g_1}{g_2}\eps
   -t_2
   +O(\eps^3) ,
    \qquad
  \eps_a
  =
    \frac{g_1}{g_2}\eps
   -t_2
   +O(\eps^3) .
    \label{eq:eps_ba}
\end{align}
The geometric factors are given as follows,
\begin{align}
  g_1
  =\frac{g_b+g_a}{2}
   ,\qquad
  g_2
  =\frac{g_b^2+g_a^2}{2}
   ,\qquad
  g_{b,a}
  =\frac{r_{b,a}}{\rho}
   ,\qquad
  \eps
  =\frac{w}{\rho} .
\end{align}
Solving Eq.(\ref{eq:K2K4_eq}) with respect to $t_2$ up to $O(\eps^2)$
which is its leading order, we get
\begin{align}
  t_2
  &\simeq
    \frac{\eps^2}{3}
    \bigg\{
      1
     -\frac{(k_r^nw)^2}{10}
    \bigg\}
   +O(\eps^3) .
   \label{t2_fin}
\end{align}
Substituting Eq.(\ref{t2_fin}) into Eq.(\ref{eq:rcpr}), and calculating the reciprocal
up to $O(\eps^2)$, we get the asymptotic expression of $\tau_0$
under the assumptions (\ref{eq:eps_def}),
\begin{align}
  \bigg(\frac{\mu_0^n}{k_r^n\rho}\bigg)^2
  \simeq
    \frac{r_br_a}{\rho^2}
   +\frac{w^2}{6\rho^2}\bigg\{1+\frac{(k_r^nw)^2}{5}\bigg\}
   +O(\eps^3) .
  \label{eq:tau0_2nd}
\end{align}
$w~(=r_b-r_a)$ is the width of the pipe.
Eq.(\ref{eq:tau0_2nd}) is symmetric with respect to the exchange of $r_a$ and $r_b$.
The square root of Eq.(\ref{eq:tau0_2nd}) up to $O(\eps^2)$ is gotten as follows,
\begin{align}
  \frac{\mu_0^n}{k_r^n\rho}
  &\simeq
    \frac{r_b+r_a}{2\rho}
   +\frac{w^2}{12\rho^2}\bigg\{\frac{(k_r^nw)^2}{5}-\frac{1}{2}\bigg\}
   +O(\eps^3) .
    \label{eq:mu0n}
\end{align}
Eqs.(\ref{eq:tau0_2nd}-\ref{eq:mu0n}) satisfy the asymptotic condition given by
Eqs.(\ref{eq:lim_nurho_ksmn}) and (\ref{eq:asympt_lim}),
\begin{align}
  \lim_{\rho\to\infty}\frac{\mu_0^n}{\rho}
  &=k_r^n
   =k_s^{0n} .
    \label{eq:asymp_mu0}
\end{align}
Also, the zeroth poles of the whispering gallery modes $\nu=\mu_0^n$,
which are given by Eqs.(\ref{cochran_sol}) and (\ref{eq:imag_whisper}) for $m=0$,
satisfy Eq.(\ref{eq:asymp_mu0}) since $k_s^{0n}$ is equal to $k_r^n$
which does not depend on $w$.
On the other hand, $k_s^{mn}$ for $m\geq1$ depends on $w$.
In this regard, the zeroth radial mode $m=0$ is special compared to
the higher order modes ($m\geq1$) as described around
Eqs.(\ref{eq:lim_nurho_ksmn}) and (\ref{eq:asympt_lim}).
We get the following limit from Eq.(\ref{eq:tau0_2nd}),
\begin{align}
  \lim_{\rho\to\infty}\frac{d}{dk_r^n}\bigg(\frac{\mu_0^n}{\rho}\bigg)^2
  &=2k_r^n .
    \label{eq:lim_mu0sq}
\end{align}
We use Eqs.(\ref{eq:tau0_2nd}-\ref{eq:mu0n}) in appendix \ref{sec:R0n_AE} to show
Eq.(\ref{eq:lim_cRm_m0}) which is the asymptotic limit of the zeroth order radial 
eigenfunction of the curved pipe for $\rho\to\infty$.

By the way, in order to find the asymptotic expression of $\mu_0^n$ up to $O(\eps^3)$,
we must take into account the terms of $O(u^8)$, $O(u^4)$ and $O(u^6)$ respectively in
Eqs.(\ref{eq:mfs_7th}), (\ref{eq:mfp_6th}) and (\ref{eq:mfq_8th}-\ref{eq:mfr_8th})
which are involved in $D_{\nu}^{00}$.
In addition, we must take into account the leading order term of $\delta_1=O(\eps^{8/3})$
given by Eq.(\ref{eq:snu_Snu}).

\subsection{Number of the real poles in the Laplace plane}
\label{sec:repoles}

As described in Eqs.(\ref{eq:ps_poles}), the cross products $p_{\nu}(\hr_b,\hr_a)$
and $s_{\nu}(\hr_b,\hr_a)$ each have a finite number of zeros on the real axis of
the $\nu$-plane.
We will find the expressions of $m_{+}$ and $m_{-}+1$ which are the number of their zeros on
the positive real axis in the $\nu$-plane as defined in Eqs.(\ref{eq:ps_poles}).
The origin $\nu=0$ is the transition point that the poles $(\nu_m^n,\mu_m^n)$ interchange 
between the real and purely imaginary numbers as explained using
a simple function in Eq.(\ref{eq:phi_simple}).

According to Eqs.(\ref{eq:nu_ks_1st}) and (\ref{eq:nu_rho2}), the square of the poles
$(\nu_m^n)^2$ and $(\mu_m^n)^2$ are real.
$(\nu_m^n)^2$ and $(\mu_m^n)^2$ decrease monotonically on
the real $\nu$-axis as $m$ increases in the range $m\leq m_{\pm}$
if the other variables and parameters are fixed.
It implies that the number of the real poles is finite.
On the other hand, for $m>m_{\pm}$, $|\nu_m^n|$ and $|\mu_m^n|$ increase monotonically
on the imaginary $\nu$-axis as $m$ increases.
The number of the zeros on the imaginary axis is infinite
since $m$ takes an integer to infinity.
We first consider the zeroth pole $\mu_0^n$ for a given $n$,
which has the largest value among $\mu_m^n$ for $0\leq m\leq m_{-}$
on the real axis as Fig.\ref{fig:poles} in p.\pageref{fig:poles} shows.
According to Eqs.(\ref{eq:mu0n}) and (\ref{cochran_sol}) for $m=0$,
whether $\mu_0^n$ is real or purely imaginary is determined by 
the relation between $k$ and $k_y^n$, \ie,
\begin{align}
  |k\beta|\geq k_y^n
   \quad\Ra\quad
  \mu_0^n\in\mathbb{R} ;
    \qquad\quad
  |k\beta|<k_y^n
   \quad\Ra\quad
  \mu_0^n\in i\mathbb{R} .
\end{align}
It follows that, when $|k\beta|<k_y^n$, all the poles $\mu_m^n$ for
$\forall m\in\mathbb{Z}_0^{+}$ are purely imaginary, \ie,
$s_{\nu}(\hr_b,\hr_a)$ has no zeros on the real axis for $|k\beta|<k_y^n$.

In what follows in appendix \ref{sec:repoles}, we assume $|k\beta|>k_y^n$, \ie,
$k_r^n\in\mathbb{R}$ in considering the number of the real poles.
In Eqs.(\ref{eq:ps_poles}) we defined $m_{\pm}$ which correspond to
the number of the zeros of $p_{\nu}(\hr_b,\hr_a)$ and $s_{\nu}(\hr_b,\hr_a)$ on
the real axis of the $\nu$-plane.
Substituting $\nu=0$ into Eq.(\ref{eq:nu_rho2}) and solving it with respect to $m$, we get
\begin{align}
  (m_{+},m_{-})
  =\bigg\lfloor
   \frac{k_r^n}{k_x^1}
   \Big\{
     1+\frac{(1,-3)}{2^3\hr_b\hr_a}
      +\frac{(-25,63)}{2^7(\hr_b\hr_a)^2}
      +O(\hr_{b,a}^{-6})
   \Big\}
   \bigg\rfloor
    \qquad
  ( |k\beta|>k_y^n,~ k_r^n\rho\gg1 ) .
   \label{eq:m_pm}
\end{align}
$\lfloor \chi \rfloor$ is the floor function which denotes
the largest integer less than $\chi\in\mathbb{R}^{+}$.
$k_x^1\,(=\pi/w)$ is given by Eq.(\ref{eq:kxm}) for $m=1$,
which is the fundamental horizontal wavenumber of the straight rectangular pipe.
In Eq.(\ref{eq:m_pm}) the coefficients $(1,-3)$ and $(-25,63)$ correspond to
$m_{\pm}=(m_{+},m_{-})$ and $\nu=(\nu_m^n,\mu_m^n)$.
$m_{+}$ given by Eq.(\ref{eq:m_pm}) agrees with Eq.(19) in \cite{horvat_prosen} to
$O(\hr_{b,a}^{-2})$.
In the numerical calculation of the field, we can use Eq.(\ref{eq:m_pm}) to find
the number of the real poles if $\hr_{a,b}\gg1$.
In the limit of $\rho\to\infty$, $m_{\pm}$ goes to $m_0$ given by Eq.(\ref{eq:Re_poles_st})
which is the number of the real poles of the Laplace domain field in
the straight rectangular pipe.
Since we assume $\hr_{a,b}\gg1$ in deriving Eq.(\ref{eq:m_pm}),
$m_{\pm}$ is equal to $m_0$ in most cases under this assumption.

\clearpage

\subsection{Problems in the uniform asymptotic expansion of the Bessel functions}
\label{sec:uae_problem}

In appendices \ref{sec:symmetric_poles}-\ref{sec:HP2}
we derived the asymptotic expressions of the poles of the field in the Laplace domain
which is the order plane of the Bessel functions.
We describe several mathematical problems in our derivation,
which are about the uniform asymptotic series of
the cross products of the Bessel functions in the following points:
(i) the range of applicability with respect to the order and argument,
(ii) the rearrangement of the coefficients of the uniform asymptotic series.
In addition to these problems, a digit loss occurs due to cancellation in calculating
the coefficients of the uniform asymptotic series.
(i) might be related to symmetries of the cross products with respect to the order and 
arguments as shown in Eqs.(\ref{eq:cp_symm}-\ref{eq:cp_symm2}).
We explain these problems below.

According to \cite{dunster}, the uniform asymptotic expansions
(\ref{eq:Fimu_muz}-\ref{eq:Gimu_muz}) are valid for $\bnu\in\mathbb{R}^{+}$ around $\bz=0$.
In finding the asymptotic expressions of the poles in the Laplace domain, however,
we used Eqs.(\ref{eq:Fimu_muz}-\ref{eq:Gimu_muz}) not only for $|\bz|\ll1$
but also for $|\bz|\gg1$ as described in Eq.(\ref{eq:bz_cond_big}).
Under the assumption $|\bz|\gg1$, we derived Eq.(\ref{eq:nu_rho2}) in which we rewrote
the order from $\bnu$ to $\nu$ through the relation $\nu=i\bnu$ as shown in
Eq.(\ref{eq:nu_bnu}).
Then we showed that Eq.(\ref{eq:nu_rho2}) agrees with Eq.(\ref{eq:nu2_hankel})
at least to the first order with respect to $\eps^2$ and $\eps_s$.
In addition to this analytical verification, we ensured that
Eq.(\ref{nu2_6th}) agrees with the numerical solution using Newton's method
especially around $m=m_{\pm}$ which means $|\nu|=0$, \ie, $|\bz|\to\infty$
as shown in Fig.\ref{fig:nu_err_4K} (p.\pageref{fig:nu_err_4K}).
These analytical and numerical agreements imply that we can apply
Eqs.(\ref{eq:Fimu_muz}-\ref{eq:Gimu_muz}) to the cross products of the Bessel functions
$F_{i\bnu}(\bnu\bz)$ and $G_{i\bnu}(\bnu\bz)$
not only around $\bz=0$ but also for $|\bz|\gg1$.
Besides, we guess that it may be possible to use Eqs.(\ref{eq:Fimu_muz}-\ref{eq:Gimu_muz}) 
for not only $\bnu\in\mathbb{R}^{+}$ but also $\bnu\in i\mathbb{R}^{+}$, \ie,
$\nu\in\mathbb{R}^{+}$ if we expand the cross products of the Bessel functions in
the uniform asymptotic series.
Actually we changed the variable from $\bnu$ to $\nu$ in deriving 
Eq.(\ref{eq:nu_ks}) from Eq.(\ref{eq:bnu_sols}).
The range of applicability is one of the problems on the uniform asymptotic series of
the Bessel functions, which we have not figured out at the present time.

The cross product $p_{\nu}(\hr_b,\hr_a)$ has a mathematical structure
(symmetry, singularity, zeros, asymptotic behavior for small/large order or arguments)
which is not the same as that of $F_{\nu}(\hr)$ and $G_{\nu}(\hr)$,
because $p_{\nu}(\hr_b,\hr_a)$ has symmetries with respect to the order and arguments
as shown in Eqs.(\ref{eq:cp_symm}).
A single Bessel function does not have such symmetries.
We guess that it may be possible to expand $p_{i\bnu}(\bnu\bz_b,\bnu\bz_a)$
in the uniform asymptotic series also for $|\bz_{b,a}|\gg1$ as we described in
the previous paragraph.
If this is true, the same may also be true for $s_{\nu}(\hr_b,\hr_a)$,
$F_{\nu}'(\hr)$ and $G_{\nu}'(\hr)$.
We think that the ranges of $\nu$ and $z_{b,a}$, in which we can expand
the cross products of the Bessel functions in the uniform asymptotic series,
might differ from those for a single Bessel function.
But, at present, we do not have the tools to show it.
Thus, our discussion on the uniform asymptotic expansions of the Bessel functions in 
appendices \ref{sec:asympt_numu} and \ref{sec:HP2} has a mathematical ambiguity about
the range of applicability with respect to $\nu$ and $z_{b,a}$.

As described in appendices \ref{sec:uae_coeff}-\ref{sec:uae_KL},
in computing the coefficients of the uniform asymptotic series of the Bessel functions,
some significant digits are lost in several points because of the cancellation among
the terms involved in the summations.
Therefore it is not easy to compute the cross products of the Bessel functions precisely in
a numerical calculation.
In our numerical calculation, usually, the numerical accuracy of
the uniform asymptotic series is about $O(10^{-12}\sim 10^{-13})$ or so.
But in the worst case in our numerical calculation of the cross products of
the Bessel functions,
the numerical accuracy goes down to single precision due to the digit loss,
depending on the values of the order and arguments of the cross products.
In addition to the numerical accuracy, there is a problem that it takes time to compute
the coefficients of the uniform asymptotic series, because they consist of multiple sums
as seen from Eqs.(\ref{eq:aj_bj_rw}-\ref{eq:dj_cj_rw}) and (\ref{eq:taj}-\ref{eq:tcj}).
We think that the uniform asymptotic series of the cross products
$p_{\nu}(\hr_b,\hr_a)$ and $s_{\nu}(\hr_b,\hr_a)$ may have simpler expressions than
Eqs.(\ref{eq:pnu_uae}-\ref{eq:snu_uae}) and (\ref{eq:pibnu}-\ref{eq:sibnu})
which are gotten respectively from Eqs.(\ref{eq:JY_uae}-\ref{eq:dJY_uae}) and
(\ref{eq:Fimu_muz}-\ref{eq:Gimu_muz}).
That is, we guess that we may be able to reformulate the uniform asymptotic series of
the cross products of the Bessel functions for $\nu\in\mathbb{C}$ and
$(z_b,z_a)\in\mathbb{C}^2$ in a more elegant manner than that of each Bessel function,
because the cross products have symmetries with respect to
the order and arguments as shown by Eqs.(\ref{eq:cp_symm}).
Since our purpose in the present paper is to find the exact expression of a transient field 
of (coherent) synchrotron radiation which is shielded by a curved rectangular pipe,
we have not yet gone through with a profound study on the fundamental structure of
the cross products of the Bessel functions in the space $(\nu,\hr_b,\hr_a)\in\mathbb{C}^3$.

\clearpage
\section{Green functions and radial eigenfunctions for large radius}
\label{sec:AL_GR}

$\mfG_{\pm}^n$ and $\cR_{\pm}^{mn}$ are given by
Eqs.(\ref{eq:mfGe}-\ref{eq:mfGb}) and (\ref{eq:cRp_ba}-\ref{eq:cRm_ba})
which are the Green functions in the Laplace domain and
the radial eigenfunctions of the curved rectangular pipe.
We will find their asymptotic limits for $\rho\to\infty$ to ensure
their expressions by showing how they tend to be those of the straight rectangular pipe.

\subsection{Green functions in the Laplace domain in the limit of large radius}
\label{sec:mfGpm_AL}

We show that $\mfG_{\pm}^n(\nu)$ in the curved pipe goes to $\mfG_{\pm}^n(k_s)$ in
the limit of $\rho\to\infty$.
$\mfG_{\pm}^n(k_s)$ is the Green function of the straight pipe in
the Laplace domain ($k_s$-domain), given by Eq.(\ref{eq:mfGbe_strt}).
We find the asymptotic limit of the cross products of the Bessel functions for
$\rho\to\infty$ using the uniform asymptotic expansion (\ref{eq:JY_uae}-\ref{eq:dJY_uae}).
In expanding the cross products $t_{\nu}(\hr,\hr')=\{p_{\nu},q_{\nu},r_{\nu},s_{\nu}\}$,
we use the following variables:
\begin{alignat}{8}
  \hr
  &=k_r^nr
   ,\qquad&
  z
  &=\frac{\hr}{\nu}
   =\frac{k_r^n\rho}{\nu}g
  &&>1
   ,\qquad&
  u
  &=\nu^{2/3}(-\zeta)
  &&>0
   ,\qquad&
  U
  &=\frac{2}{3}u^{3/2}
  &&=\nu\eta_{-}(z) ,
   \label{zuU}
   \\
  \hr'
  &=k_r^nr'
   ,\qquad&
  z'
  &=\frac{\hr'}{\nu}
   =\frac{k_r^n\rho}{\nu}g'
  &&>1
   ,\qquad&
  u'
  &=\nu^{2/3}(-\zeta')
  &&>0
   ,\qquad&
  U'
  &=\frac{2}{3}u'^{3/2}
  &&=\nu\eta_{-}(z') .
   \label{zuU_prm}
\end{alignat}
Similar to $z_{a,b}$ in Eqs.(\ref{mfp_mfs_0}-\ref{eq:Uba_nu}), we assume that
both $z$ and $z'$ are larger than 1 since the straight pipe does not have
the whispering gallery modes (\ref{eq:nu_mt}) and (\ref{eq:mb_def}).
The Airy functions, involved in Eqs.(\ref{eq:pnu_uae}-\ref{eq:snu_uae}),
have the argument $-u$ or $-u'$.
$g$ and $g'$ are the geometric factors defined in Eq.(\ref{eq:coordinates}).
$\eta_{-}$ is given by Eq.(\ref{eq:zeta_m}).
In finding the asymptotic limit of $\mfG_{\pm}^n(\nu)$ for $\rho\to\infty$
while keeping $\nu/\rho$ constant as in Eq.(\ref{eq:lim_nu}), it is enough to consider
the leading order terms of Eqs.(\ref{eq:pnu_uae}-\ref{eq:snu_uae}),
\begin{alignat}{2}
  p_{\nu}(\hr,\hr')
  &\simeq -\frac{2}{\nu}\bigg\{\frac{uu'}{(z^2-1)(z'^2-1)}\bigg\}^{1/4}\mfp(-u,-u') ,
  \label{eq:pnu_lt}
   \\
  s_{\nu}(\hr,\hr')
  &\simeq -\frac{2\nu}{\hr\hr'}\bigg\{\frac{(z^2-1)(z'^2-1)}{uu'}\bigg\}^{1/4}\mfs(-u,-u') ,
  \label{eq:snu_lt}
\end{alignat}
\begin{alignat}{2}
  q_{\nu}(\hr,\hr')
  \simeq \frac{2}{\hr'}\bigg\{\frac{u(z'^2-1)}{u'(z^2-1)}\bigg\}^{1/4}\mfq(-u,-u') ,
    \qquad
  r_{\nu}(\hr,\hr')
  \simeq \frac{2}{\hr}\bigg\{\frac{u'(z^2-1)}{u(z'^2-1)}\bigg\}^{1/4}\mfr(-u,-u') .
  \label{eq:qr_nu_lt}
\end{alignat}
Taking only the leading order terms in Eqs.(\ref{eq:mfp_mm}-\ref{eq:mfr_mm}),
we get the asymptotic expressions of the cross products of the Airy functions for
negative large argument,
\begin{alignat}{2}
  \pi\mfp(-u,-u')
  &\simeq \nu^{-1/3}(\zeta\zeta')^{-1/4}\sin\phi ,
    \qquad&
  \pi\mfq(-u,-u')
  &\simeq (\zeta/\zeta')^{-1/4}\cos\phi ,
   \label{eq:mfpq_0}
   \\
  \pi\mfs(-u,-u')
  &\simeq \nu^{1/3}(\zeta\zeta')^{1/4}\sin\phi ,
    \qquad&
  \pi\mfr(-u,-u')
  &\simeq -(\zeta/\zeta')^{1/4}\cos\phi .
   \label{eq:mfsr_0}
\end{alignat}
The phase $\phi$ is given and rewritten as follows,
\begin{align}
  \phi(z,z')
  =U-U'
  =\nu\bigg\{\sig-\sin^{-1}\!\bigg(\frac{\sig}{zz'}\bigg)\bigg\} ,
    \qquad
  \sig
  =(z^2-1)^{1/2}-(z'^2-1)^{1/2} .
   \label{eq:UUp}
\end{align}
From Eqs.(\ref{eq:pnu_lt}-\ref{eq:qr_nu_lt}) and (\ref{eq:mfpq_0}-\ref{eq:mfsr_0}), we get
\begin{align}
  p_{\nu}(\nu z,\nu z')
  &\simeq
   -\frac{2\mfp(-u,-u')}{\nu^{2/3}}
    \bigg\{\frac{\zeta\zeta'}{(z^2-1)(z'^2-1)}\bigg\}^{1/4}
   \simeq
   -\frac{2\sin\phi}{\pi\nu}\{(z^2-1)(z'^2-1)\}^{-1/4} ,
  \label{eq:pnu_sin}
   \\
  s_{\nu}(\nu z,\nu z')
  &\simeq
   -\frac{2\mfs(-u,-u')}{zz'\nu^{4/3}}
    \bigg\{\frac{(z^2-1)(z'^2-1)}{\zeta\zeta'}\bigg\}^{1/4}
   \simeq
   -\frac{2\sin\phi}{\pi\nu zz'}\{(z^2-1)(z'^2-1)\}^{1/4} ,
  \label{eq:snu_sin}
\end{align}
\begin{alignat}{2}
  q_{\nu}(\nu z,\nu z')
  &\simeq
    \frac{2\mfq(-u,-u')}{\nu z'}
    \bigg\{\frac{\zeta(z'^2-1)}{\zeta'(z^2-1)}\bigg\}^{1/4}
  &&\simeq
    \frac{2\cos\phi}{\pi\nu z'}\bigg(\frac{z^2-1}{z'^2-1}\bigg)^{-1/4} ,
  \label{eq:qnu_cos}
   \\
  r_{\nu}(\nu z,\nu z')
  &\simeq
    \frac{2\mfr(-u,-u')}{\nu z}
    \bigg\{\frac{\zeta'(z^2-1)}{\zeta(z'^2-1)}\bigg\}^{1/4}
  &&\simeq
    -\frac{2\cos\phi}{\pi\nu z}\bigg(\frac{z^2-1}{z'^2-1}\bigg)^{1/4} .
  \label{eq:rnu_cos}
\end{alignat}

We consider the asymptotic limit of Eqs.(\ref{eq:pnu_sin}-\ref{eq:rnu_cos})
for $\rho\to\infty$.
In this limit, the following $\kap_x$, which depends on $x$ through $z$, goes to $k_x$
for $\forall x$,
\begin{align}
  &
  \kap_x
  =\frac{\nu}{\rho}(z^2-1)^{1/2}
  =\{(k_r^ng)^2-(\nu/\rho)^2\}^{1/2} ,
    \qquad
  \lim_{\rho\to\infty}\kap_x
  =k_x
  =\{(k_r^n)^2-k_s^2\}^{1/2} ,
   \label{eq:kapx}
\end{align}
where
\begin{align}
  (k_r^n)^2
  =k_x^2+k_s^2
  =(k\beta)^2-(k_y^n)^2 .
   \label{eq:kx_ks_krn}
\end{align}
$k_s$ is the Laplace variable defined in Eqs.(\ref{eq:Lap_trans_kap}),
which is related to $k_x$ through Eq.(\ref{eq:kx_ks_krn}).
$k_x$ and $k_s$ are the complex horizontal and longitudinal wavenumbers of
the Laplace domain field in the straight section.
They differ from $k_x^m\in\mathbb{R}_0^{+}$ and $k_s^{mn}\in\mathbb{A}$
which are the eigenmodes of the straight rectangular pipe, given by Eq.(\ref{eq:ks_mn}).
For $\rho\to\infty$, the phase $\phi$ behaves as
\begin{alignat}{2}
  &
  \phi
  \simeq\nu\sig\bigg(1-\frac{1}{zz'}\bigg)
  \simeq \vkap_x(x-x') ,
    \qquad&&
  \lim_{\rho\to\infty}\phi
  =\frac{k_x^2}{(k_r^n)^2}\lim_{\rho\to\infty}\nu\sig
  =k_x(x-x') ,
  \label{eq:U_mns_Up}
   \\
  &
  \vkap_x
  =[\kap_x]_{x=0}
  =\{(k_r^n)^2-(\nu/\rho)^2\}^{1/2} ,
    \qquad&&
  \lim_{\rho\to\infty}\vkap_x
  =k_x ,
    \qquad
  \lim_{\rho\to\infty}\nu/\rho
  =k_s .
  \label{eq:vkapx}
\end{alignat}
$\nu/\rho$ is the complex longitudinal wavenumber of the field in the bending section,
which goes to $k_s$ in the limit of $\rho\to\infty$ as shown in Eq.(\ref{eq:lim_nu}).
$\vkap_x$ is a complex wavenumber of the field in the bend,
which goes to $k_x$ in the limit of $\rho\to\infty$.
Accordingly, the uniform asymptotic limit of the cross products of the Bessel functions
$t_{\nu}(\hr,\hr')=\{p_{\nu},q_{\nu},r_{\nu},s_{\nu}\}$ for
$\rho\to\infty$ and $\nu\to\infty$ keeping $\nu/\rho$ constant is given as follows,
\begin{alignat}{2}
  \lim_{\rho\to\infty}\rho p_{\nu}(\hr,\hr')
  &=\frac{2}{\pi k_x}p(\hx,\hx') ,
     \qquad&
  p(\hx,\hx')
  &=-\sin(\hx-\hx') ,
   \label{eq:lim_rho_pnu}
   \\
  \lim_{\rho\to\infty}\rho q_{\nu}(\hr,\hr')
  &=\frac{2}{\pi k_r^n}q(\hx,\hx') ,
     \qquad&
  q(\hx,\hx')
  &=+\cos(\hx-\hx') ,
   \\
  \lim_{\rho\to\infty}\rho r_{\nu}(\hr,\hr')
  &=\frac{2}{\pi k_r^n}r(\hx,\hx') ,
     \qquad&
  r(\hx,\hx')
  &=-\cos(\hx-\hx') ,
   \\
  \lim_{\rho\to\infty}\rho s_{\nu}(\hr,\hr')
  &=\frac{2k_x}{\pi(k_r^n)^2}s(\hx,\hx') ,
     \qquad&
  s(\hx,\hx')
  &=-\sin(\hx-\hx') ,
   \label{eq:lim_rho_snu}
\end{alignat}
where $t(\hx,\hx')=\{p,q,r,s\}$ denotes the cross products of the trigonometric functions
$\cos\hx$ and $\sin\hx$, defined in Eqs.(\ref{eq:CP_p_st}-\ref{eq:CP_s_st}).
$\hx$ and $\hx'$ are the dimensionless horizontal variables normalized by $k_x$,
\begin{align}
  \hx=k_xx ,
    \qquad
  \hx'=k_xx' ;
    \qquad
  r=\rho+x ,
    \qquad
  r'=\rho+x' .
\end{align}
In the Laplace domain, the Green functions for $E_y$ and $B_y$ in the curved pipe are
given by Eqs.(\ref{eq:mfGe}-\ref{eq:mfGb}),
\begin{align}
  \mfG_{+}^n(r,r',\nu)
  &=\frac{\pi}{2}
    \bigg\{\theta(r-r')
         \frac{\rho p_\nu(\hr_b,\hr)\rho p_\nu(\hr',\hr_a)}{\rho p_\nu(\hr_b,\hr_a)}
        +\theta(r'-r)
         \frac{\rho p_\nu(\hr_b,\hr')\rho p_\nu(\hr,\hr_a)}{\rho p_\nu(\hr_b,\hr_a)}
    \bigg\} ,
   \\
  \mfG_{-}^n(r,r',\nu)
  &=\frac{\pi}{2}
    \bigg\{\theta(r-r')
         \frac{\rho r_\nu(\hr_b,\hr)\rho q_\nu(\hr',\hr_a)}{\rho s_\nu(\hr_b,\hr_a)}
        +\theta(r'-r)
         \frac{\rho r_\nu(\hr_b,\hr')\rho q_\nu(\hr,\hr_a)}{\rho s_\nu(\hr_b,\hr_a)}
    \bigg\} .
\end{align}
According to Eqs.(\ref{eq:lim_rho_pnu}-\ref{eq:lim_rho_snu}),
$\mfG_{\pm}^n(\nu)$ goes to the following asymptotic limit for $\rho\to\infty$,
\begin{align}
  \lim_{\rho\to\infty}\mfG_{\pm}^n(r,r',\nu)
  &=\mp\frac{\Xi_{\pm}(x,x',k_x)+\Xi_{\pm}(x',x,k_x)}{k_x\sin[k_x(x_b-x_a)]} ,
  \label{eq:lim_mfGp_uae}
\end{align}
where
\begin{align}
  \Xi_{+}(x,x',k_x)
  &=\theta(x-x')\sin[k_x(x_b-x)]\sin[k_x(x'-x_a)] ,
   \\
  \Xi_{-}(x,x',k_x)
  &=\theta(x-x')\cos[k_x(x_b-x)]\cos[k_x(x'-x_a)] .
\end{align}
The R.H.S. of Eq.(\ref{eq:lim_mfGp_uae}) agrees with Eq.(\ref{eq:mfGbe_strt})
which is the Green functions in the $k_s$-domain
(Laplace domain in the straight section),
\begin{align}
  \lim_{\rho\to\infty}\mfG_{\pm}^n(r,r',\nu)=\mfG_{\pm}^n(x,x',k_s) .
  \label{eq:lim_mfGpm}
\end{align}
Thus, in the limit of $\rho\to\infty$, the Green functions in
the curved pipe go to those in the straight pipe.

\subsection{Radial eigenfunctions in the limit of large radius}
\label{sec:cRcX}

The radial eigenfunctions of the curved pipe $\cR_{\pm}^{mn}$ are given by
Eqs.(\ref{eq:cRp_ba}-\ref{eq:cRm_ba}).
We will show that $\cR_{\pm}^{mn}$ of the normal mode ($m\geq m_{a,b}+1$) and
the zeroth mode ($m=0$) goes to $\cX_{\pm}^m$ in the limit of
$\rho\to\infty$ with the width of the pipe $w$ kept constant.
$\cX_{\pm}^m$ is given by Eqs.(\ref{eq:cXp}-\ref{eq:cXm})
which are the horizontal eigenfunctions of the straight pipe
having the same rectangular cross section as the curved pipe.
For convenience, we first rewrite Eqs.(\ref{eq:cRp_ba}-\ref{eq:cRm_ba}) as follows,
\begin{align}
  \cR_{+}^{mn}(r,r')
  &=k_r^n\frac{d}{dk_r^n}\bigg(\frac{\pi\nu_m^n}{2\rho}\bigg)^{\!2}
    \bigg[
      \frac{\rho p_{\nu}(\hr_b,\hr)\cdot \rho p_{\nu}(\hr',\hr_a)}{u_{\nu}(\hr_b,\hr_a)}
    \bigg]_{\nu=\nu_m^n} ,
  \label{eq:cRp_ba_rw}
  \\
  \cR_{-}^{mn}(r,r')
  &=k_r^n\frac{d}{dk_r^n}\bigg(\frac{\pi\mu_m^n}{2\rho}\bigg)^{\!2}
    \bigg[
     \frac{\rho r_{\nu}(\hr_b,\hr)\cdot \rho q_{\nu}(\hr',\hr_a)}{v_{\nu}(\hr_b,\hr_a)}
    \bigg]_{\nu=\mu_m^n} ,
  \label{eq:cRm_ba_rw}
\end{align}
where $m\geq m_{a,b}+1$ which means the normal mode as described in
sections \ref{sec:wgm} and \ref{sec:iwgm}.
In addition to the normal mode, we deal with the zeroth radial mode $m=0$ of
Eq.(\ref{eq:cRm_ba_rw}).
$u_{\nu}$ and $v_{\nu}$ are given as
\begin{align}
  u_{\nu}(\hr_b,\hr_a)
  =\rho
   \bigg\{\frac{J_{\nu}(\hr_b)}{J_{\nu}(\hr_a)}-\frac{J_{\nu}(\hr_a)}{J_{\nu}(\hr_b)}\bigg\}
   ,\qquad
  v_{\nu}(\hr_b,\hr_a)
  =\rho
   \bigg\{
     \bigg(1-\frac{\nu^2}{\hr_a^2}\bigg)\frac{J_{\nu}'(\hr_b)}{J_{\nu}'(\hr_a)}
    -\bigg(1-\frac{\nu^2}{\hr_b^2}\bigg)\frac{J_{\nu}'(\hr_a)}{J_{\nu}'(\hr_b)}
   \bigg\} .
  \label{eq:uv_nu}
\end{align}
From Eqs.(\ref{eq:rd_kr_numu2}), (\ref{eq:nu_rho2}) and (\ref{eq:lim_mu0sq}), we get
\begin{align}
  \lim_{\rho\to\infty}
  \frac{d}{dk_r^n}\bigg\{\frac{(\nu_m^n,\mu_m^n)}{\rho}\bigg\}^2
  =2k_r^n .
  \label{eq:lim_rd_kr_nu2}
\end{align}

We expand the numerators of Eqs.(\ref{eq:cRp_ba_rw}-\ref{eq:cRm_ba_rw})
using Eqs.(\ref{eq:pnu_sin}-\ref{eq:rnu_cos}),
\begin{alignat}{2}
  \rho p_{\nu}(\hr_b,\hr)\cdot \rho p_{\nu}(\hr',\hr_a)
  &\simeq
   \bigg(\frac{2}{\pi}\bigg)^2
   \frac{\sin(U_b-U)\sin(U'-U_a)}{(\nu/\rho)^2\{(z_b^2-1)(z^2-1)(z'^2-1)(z_a^2-1)\}^{1/4}} ,
  \label{eq:num_p}
   \\
  \rho r_{\nu}(\hr_b,\hr)\cdot \rho q_{\nu}(\hr',\hr_a)
  &\simeq
   -\bigg(\frac{2}{\pi}\bigg)^2\frac{\cos(U_b-U)\cos(U'-U_a)}{(k_r^n)^2g_bg_a}
    \bigg\{\frac{(z_b^2-1)(z_a^2-1)}{(z^2-1)(z'^2-1)}\bigg\}^{1/4} .
  \label{eq:num_m}
\end{alignat}
$(z,z')$ and $(U,U')$ are given by Eqs.(\ref{zuU}-\ref{zuU_prm}).
Similar to them, we define $z_{b,a}$ and $U_{b,a}$ as follows,
\begin{alignat}{7}
  z_b
  &=\frac{\hr_b}{\nu}
   =\frac{k_r^n\rho}{\nu}g_b
  &&>1
   ,\qquad&&
  \zeta_b<0
   ,\qquad&
  u_b
  &=\nu^{2/3}(-\zeta_b)
  &&>0
   ,\qquad&
  U_b
  &=\frac{2}{3}u_b^{3/2}
  &&=\nu\eta_{-}(z_b) ,
   \label{zuUb}
   \\
  z_a
  &=\frac{\hr_a}{\nu}
   =\frac{k_r^n\rho}{\nu}g_a
  &&>1
   ,\qquad&&
  \zeta_a<0
   ,\qquad&
  u_a
  &=\nu^{2/3}(-\zeta_a)
  &&>0
   ,\qquad&
  U_a
  &=\frac{2}{3}u_a^{3/2}
  &&=\nu\eta_{-}(z_a) .
   \label{zuUa}
\end{alignat}
According to Eqs.(\ref{eq:krn_apdx}) and (\ref{eq:asympt_lim}),
$\kap_x$ given by Eq.(\ref{eq:kapx}) for $\nu=(\nu_m^n,\mu_m^n)$ of the normal mode
goes to $k_x^m$ in the limit of $\rho\to\infty$, \ie,
\begin{align}
  \lim_{\rho\to\infty}[\kap_x]_{\nu=(\nu_m^n,\mu_m^n)}
  &=\{(k_r^n)^2-(k_s^{mn})^2\}^{1/2}
   =k_x^m .
  \label{eq:kap_mn}
\end{align}
We calculate the phases involved in Eqs.(\ref{eq:num_p}-\ref{eq:num_m})
using $\phi$ given by Eqs.(\ref{eq:UUp}).
According to Eq.(\ref{eq:U_mns_Up}) for $k_x=k_x^m$, the phases for $\nu=(\nu_m^n,\mu_m^n)$ 
go to the following limits in the limit of $\rho\to\infty$, 
\begin{align}
  \lim_{\rho\to\infty}[U_b-U]_{\nu=(\nu_m^n,\mu_m^n)}=k_x^m(x_b-x) ,
    \qquad
  \lim_{\rho\to\infty}[U'-U_a]_{\nu=(\nu_m^n,\mu_m^n)}=k_x^m(x'-x_a) .
\end{align}
Then the asymptotic limits of Eqs.(\ref{eq:num_p}-\ref{eq:num_m}) for
$\nu=(\nu_m^n,\mu_m^n)$ and $\rho\to\infty$ are gotten as
\begin{align}
  \lim_{\rho\to\infty}
  \big[\rho p_{\nu}(\hr_b,\hr)\cdot \rho p_{\nu}(\hr',\hr_a)\big]_{\nu=\nu_m^n}
  &=+\bigg(\frac{2}{\pi k_x^m}\bigg)^2\sin[k_x^m(x_b-x)]\sin[k_x^m(x'-x_a)] ,
   \label{lim_num_p}
   \\
  \lim_{\rho\to\infty}
  \big[\rho r_{\nu}(\hr_b,\hr)\cdot \rho q_{\nu}(\hr',\hr_a)\big]_{\nu=\mu_m^n}
  &=-\bigg(\frac{2}{\pi k_r^n}\bigg)^2\cos[k_x^m(x_b-x)]\cos[k_x^m(x'-x_a)] .
   \label{lim_num_m}
\end{align}

Next, we find the asymptotic limit of $u_{\nu}$ and $v_{\nu}$ 
respectively for $\nu=\nu_m^n$ and $\mu_m^n$ $(m>m_{a,b})$
in the limit of $\rho\to\infty$.
We consider $v_{\nu}$ for $\nu=\mu_0^n$ in appendix \ref{sec:R0n_AE}
since $m=0$ is special for $v_{\nu}$.
Taking the leading order term in Eqs.(\ref{eq:JY_uae}-\ref{eq:dJY_uae}),
we expand $J_{\nu}$ and $J_{\nu}'$ for $\nu=\nu_m^n$ and $\mu_m^n$,
\begin{align}
  [J_{\nu}(\nu z)]_{\nu=\nu_m^n}
  &\simeq
   \bigg[\frac{2^{1/2}u^{1/4}}{(\kap_x\rho)^{1/2}}\Ai(-u)\bigg]_{\nu=\nu_m^n}
   \simeq
   \bigg[\bigg(\frac{2}{\pi\kap_x\rho}\bigg)^{1/2}\sin(U+\pi/4)\bigg]_{\nu=\nu_m^n} ,
  \label{eq:Jnu_sin}
   \\
  [J_{\nu}'(\nu z)]_{\nu=\mu_m^n}
  &\simeq
   -\bigg[\frac{2^{1/2}(\kap_x\rho)^{1/2}}{\nu zu^{1/4}}\Ai'(-u)
    \bigg]_{\nu=\mu_m^n}
   \simeq
   \bigg[\bigg(\frac{2\kap_x\rho}{\pi}\bigg)^{1/2}\frac{\cos(U+\pi/4)}{\nu z}
   \bigg]_{\nu=\mu_m^n} .
  \label{eq:Jpnu_sin}
\end{align}
$\kap_x$ is a function of $x$ as in Eq.(\ref{eq:kapx}).
We define $\kap_a$ and $\kap_b$ as $\kap_x$ for $x=x_a$ and $x_b$.
According to Eq.(\ref{eq:kap_mn}), $\kap_a$ and $\kap_b$ at the poles of the normal mode
go to $k_x^m$ in the limit of $\rho\to\infty$,
\begin{align}
  \kap_{a,b}
   =\frac{\nu}{\rho}(z_{a,b}^2-1)^{1/2} ,
    \qquad
  \lim_{\rho\to\infty}[\kap_{a,b}]_{\nu=(\nu_m^n,\mu_m^n)}
  =k_x^m
   \qquad
  (m>m_{a,b}) .
\end{align}
According to Eqs.(\ref{eq:Jnu_sin}-\ref{eq:Jpnu_sin}) and (\ref{eq:U_mns_Up}),
$J_{\nu}(\hr_b)/J_{\nu}(\hr_a)$ and $J_{\nu}'(\hr_b)/J_{\nu}'(\hr_a)$ respectively for
$\nu=\nu_m^n$ and $\mu_m^n$ ($m>m_{a,b}$) behave as follows for $\rho\to\infty$,
\begin{align}
  \bigg[\frac{J_{\nu}(\hr_b)}{J_{\nu}(\hr_a)}\bigg]_{\nu=\nu_m^n}
  &\simeq
  \bigg[\bigg(\frac{\kap_a}{\kap_b}\bigg)^{1/2}\frac{\sin(U_b+\pi/4)}{\sin(U_a+\pi/4)}
  \bigg]_{\nu=\nu_m^n}
  \simeq
   (-1)^m\bigg[\bigg(\frac{\kap_a}{\kap_b}\bigg)^{1/2}\bigg]_{\nu=\nu_m^n} ,
   \\
  \bigg[\frac{J_{\nu}'(\hr_b)}{J_{\nu}'(\hr_a)}\bigg]_{\nu=\mu_m^n}
  &\simeq
  \bigg[\frac{g_a}{g_b}\bigg(\frac{\kap_b}{\kap_a}\bigg)^{1/2}
        \frac{\cos(U_b+\pi/4)}{\cos(U_a+\pi/4)}
  \bigg]_{\nu=\mu_m^n}
  \simeq
   (-1)^m\frac{g_a}{g_b}\bigg[\bigg(\frac{\kap_b}{\kap_a}\bigg)^{1/2}\bigg]_{\nu=\mu_m^n} .
\end{align}
Therefore, for $\rho\to\infty$, the asymptotic expressions of $u_{\nu}$ and $v_{\nu}$
respectively at $\nu=\nu_m^n$ and $\mu_m^n$ $(m>m_{a,b})$ are gotten as
\begin{align}
  [u_{\nu}(\hr_b,\hr_a)]_{\nu=\nu_m^n}
  &\simeq
   -(-1)^m\big[\rho(\kap_b-\kap_a)(\kap_a\kap_b)^{-1/2}\big]_{\nu=\nu_m^n} ,
   \label{eq:u_nu_kap}
   \\
  [v_{\nu}(\hr_b,\hr_a)]_{\nu=\mu_m^n}
  &\simeq
   -\frac{(-1)^m}{(k_r^n)^2g_ag_b}
    \big[\rho(\kap_b-\kap_a)(\kap_a\kap_b)^{1/2}\big]_{\nu=\mu_m^n} .
   \label{eq:v_nu_kap}
\end{align}
In the limit of $\rho\to\infty$, $\rho(\kap_b-\kap_a)$
at $\nu=(\nu_m^n,\mu_m^n)$ goes to the following limit,
\begin{align}
   \lim_{\rho\to\infty}[\rho(\kap_b-\kap_a)]_{\nu=(\nu_m^n,\mu_m^n)}
   =w\frac{(k_r^n)^2}{k_x^m}
   \qquad
  (m>m_{a,b}) .
   \label{lim_kap_ba}
\end{align}
Eq.(\ref{lim_kap_ba}) does not hold for $m\leq m_{a,b}$ including $m=0$.
From Eqs.(\ref{eq:u_nu_kap}-\ref{lim_kap_ba}), we get the asymptotic limits of $u_{\nu}$
and $v_{\nu}$ at the poles $\nu=\nu_m^n$ and $\mu_m^n$ in the limit of $\rho\to\infty$,
\begin{alignat}{2}
  \lim_{\rho\to\infty}[u_{\nu}(\hr_b,\hr_a)]_{\nu=\nu_m^n}
  =-(-1)^mw\bigg(\frac{k_r^n}{k_x^m}\bigg)^2
   ,\qquad
  \lim_{\rho\to\infty}[v_{\nu}(\hr_b,\hr_a)]_{\nu=\mu_m^n}
  =-(-1)^mw .
  \label{eq:lim_unu_vnu}
\end{alignat}
Substituting Eqs.(\ref{lim_num_p}-\ref{lim_num_m}) and Eqs.(\ref{eq:lim_unu_vnu}) into
Eqs.(\ref{eq:cRp_ba_rw}-\ref{eq:cRm_ba_rw}), we get the asymptotic limit of the radial 
eigenfunctions of the normal mode ($m\geq m_{a,b}+1$) in the limit of $\rho\to\infty$,
\begin{align}
  \lim_{\rho\to\infty}\cR_{+}^{mn}(r,r')
  &=-\frac{2}{w}(-1)^m\sin[k_x^m(x_b-x)]\sin[k_x^m(x'-x_a)] ,
   \label{eq:lim_cRp}
   \\
  \lim_{\rho\to\infty}\cR_{-}^{mn}(r,r')
  &=+\frac{2}{w}(-1)^m\cos[k_x^m(x_b-x)]\cos[k_x^m(x'-x_a)] .
   \label{eq:lim_cRm}
\end{align}
As shown in Eqs.(\ref{eq:lim_gl_g12}) and (\ref{eq:lim_mb}), the real and imaginary 
whispering gallery modes vanish ($m_{a,b}\to0$) in the limit of $\rho\to\infty$.
It follows that Eqs.(\ref{eq:lim_cRp}-\ref{eq:lim_cRm}) hold for $\forall m\in\mathbb{N}$,
\begin{align}
  \lim_{\rho\to\infty}\cR_{\pm}^{mn}(r,r')
  &=\cX_{\pm}^m(x,x')
   \qquad
  (m\in\mathbb{N}) .
   \label{eq:lim_cR_cX}
\end{align}
$\cX_{\pm}^m$ is given by Eqs.(\ref{eq:mfXp_new}-\ref{eq:mfXm_new})
which are the horizontal eigenfunctions of the straight rectangular pipe.

\subsection{Zeroth radial eigenfunction in the limit of large radius}
\label{sec:R0n_AE}

We show that $\cR_{-}^{0n}$ goes to $\cX_{-}^0$ in the limit of $\rho\to\infty$.
$\cR_{-}^{0n}$ and $\cX_{-}^0$ are the zeroth horizontal eigenfunctions of the curved and 
straight pipes, respectively given by Eqs.(\ref{eq:cRm_ba_rw}) and (\ref{eq:cXm}) for $m=0$,
\begin{alignat}{2}
  \cR_{-}^{0n}
  &=k_r^n\frac{d}{dk_r^n}\bigg(\frac{\pi\mu_0^n}{2\rho}\bigg)^{\!2}
    \bigg[
     \frac{\rho r_{\nu}(\hr_b,\hr)\cdot \rho q_{\nu}(\hr',\hr_a)}{v_{\nu}(\hr_b,\hr_a)}
    \bigg]_{\nu=\mu_0^n} ,
    \qquad
  \frac{\mu_0^n}{k_r^n\rho}
  \simeq 1+\frac{x_b+x_a}{2\rho}+O(\eps^2) .
  \label{eq:cRm_m0}
\end{alignat}
The derivative of $(\mu_0^n/\rho)^2$ with respect to $k_r^n$ goes to
Eq.(\ref{eq:lim_mu0sq}) in the limit of $\rho\to\infty$.
We find the asymptotic limit of $q_{\nu}$ and $r_{\nu}$ for $\nu=\mu_0^n$ and
$\rho\to\infty$.
Assuming $z_a<1<z_b$ similar to the whispering gallery,
we expand $q_{\nu}(\hr,\hr_a)$ and $r_{\nu}(\hr_b,\hr)$ in the uniform asymptotic series,
\begin{align}
  q_{\nu}(\hr,\hr_a)
  \simeq \frac{2f(z_a)}{\hr_af(z)}\mfq(u,u_a) ,
    \qquad
  r_{\nu}(\hr_b,\hr)
  \simeq \frac{2f(z_b)}{\hr_bf(z)}\mfr(-u_b,u) .
  \label{eq:qr_uae_lt}
\end{align}
$f(z)$ is given by Eq.(\ref{eq:f_z_zeta}).
The arguments of the Airy functions are $-u_b$ and $u_a$ similar to
Eqs.(\ref{eq:ub_m0}-\ref{eq:ua_m0}) since we deal with the zeroth pole $\nu=\mu_0^n$
under the assumption $z_a<1<z_b$, even though it is not a whispering gallery mode
which exists only in the curved pipe,
\begin{alignat}{2}
  u_b
  &=(\nu/2)^{2/3}(z_b^2-1)\{2\psi(z_b)\}^{2/3} ,
    \qquad&
  \hr_b
  &=\nu z_b ,
   \\
  u_a
  &=(\nu/2)^{2/3}(1-z_a^2)\{2\psi(z_a)\}^{2/3} ,
    \qquad&
  \hr_a
  &=\nu z_a .
\end{alignat}
From Eqs.(\ref{eq:qr_uae_lt}) for $\nu=\mu_0^n$, we calculate the numerator of
Eq.(\ref{eq:cRm_m0}) in the limit of $\rho\to\infty$,
\begin{align}
  \lim_{\rho\to\infty}
  [\rho r_{\nu}(\hr_b,\hr) \cd \rho q_{\nu}(\hr',\hr_a)]_{\nu=\mu_0^n}
   =-\bigg(\frac{2}{\pi k_r^n}\bigg)^2
     \lim_{\rho\to\infty}\bigg[\frac{f(z_b)f(z_a)}{f(z)f(z')}\bigg]_{\nu=\mu_0^n}
   =-\bigg(\frac{2}{\pi k_r^n}\bigg)^2 .
  \label{eq:lim_rq_m0}
\end{align}
We got Eq.(\ref{eq:lim_rq_m0}) by expanding $2^{1/6}/f(z)$ for $\nu=\mu_0^n$
in a power series with respect to $\eps_z$ and $x_{b,a}/\rho$
which is involved in $\mu_0^n$ through $z\,(=\hr/\nu)$ for $\nu=\mu_0^n$,
\begin{align}
  &
  \bigg[\frac{2^{1/6}}{f(z)}\bigg]_{\nu=\mu_0^n}
   \simeq \Big[1+\frac{\eps_z}{10}+O\big(\eps_z^2)\Big]_{\nu=\mu_0^n}
   \simeq 1+\frac{x_b+x_a-2x}{10\rho}+O(\eps^2) ,
   \\
  &
  \Ra\qquad
  \lim_{\rho\to\infty}[f(z)]_{\nu=\mu_0^n}
  =2^{1/6}
   \quad\text{for}~
  \forall x
   \qquad
  (\eps_z=1-z^2,~ \eps=w/\rho) .
\end{align}

The denominator of Eq.(\ref{eq:cRm_m0}), $v_{\nu}(\hr_b,\hr_a)$, is given by
Eq.(\ref{eq:uv_nu}).
When $\nu=\mu_0^n$, it is given as
\begin{align}
  [v_{\nu}(\hr_b,\hr_a)]_{\nu=\mu_0^n}
  &=\bigg[
      W_a\frac{J_{\nu}'(\hr_b)}{J_{\nu}'(\hr_a)}
     -W_b\frac{J_{\nu}'(\hr_a)}{J_{\nu}'(\hr_b)}
    \bigg]_{\nu=\mu_0^n}
   ,\qquad
  W_{a,b}
  =\rho\bigg\{1-\Big(\frac{\mu_0^n}{\hr_{a,b}}\Big)^2\bigg\} .
   \label{eq:v_mu0}
\end{align}
Expanding $W_{a,b}$ with respect to $x_{b,a}/\rho$, we take the limit of $\rho\to\infty$
while keeping $w$ constant,
\begin{align}
  W_b
  \simeq -W_a
  \simeq  w\{1+O(\eps)\}
   \qquad\Ra\qquad
  \lim_{\rho\to\infty}W_b=-\lim_{\rho\to\infty}W_a=w .
  \label{eq:Wba}
\end{align}
We find the asymptotic limit of $J_{\nu}'(\hr_b)/J_{\nu}'(\hr_a)$ for $\nu=\mu_0^n$
in the limit of $\rho\to\infty$.
Expanding $J_{\nu}'(\hr_{b,a})$ using Eq.(\ref{eq:dJY_uae}),  
we take the limit of $\rho\to\infty$ for the leading order term,
\begin{align}
  \frac{J_{\nu}'(\hr_b)}{J_{\nu}'(\hr_a)}
  &\simeq
   \frac{z_af(z_b)}{z_bf(z_a)}\cd\frac{\Ai'(-u_b)}{\Ai'(u_a)}
   \qquad\Ra\qquad
  \lim_{\rho\to\infty}
  \bigg[\frac{J_{\nu}'(\hr_b)}{J_{\nu}'(\hr_a)}\bigg]_{\nu=\mu_0^n}=1 ,
  \label{eq:limJJ}
\end{align}
where we used Eq.(\ref{eq:Aip_ts}) which is the Taylor series of $\Ai'(u)$ around $u=0$.
According to Eqs.(\ref{eq:Wba}-\ref{eq:limJJ}), the denominator
$v_{\mu_0^n}(\hr_b,\hr_a)$ goes to the following asymptotic limit for $\rho\to\infty$,
\begin{align}
  \lim_{\rho\to\infty}[v_{\nu}(\hr_b,\hr_a)]_{\nu=\mu_0^n}
  &=-2w .
   \label{eq:lim_rho_vnu_m0}
\end{align}
Eq.(\ref{eq:lim_rho_vnu_m0}) differs from the second equation of (\ref{eq:lim_unu_vnu}) for 
$m=0$.
Eq.(\ref{eq:lim_unu_vnu}) does not hold for $m=0$ since we assume $1<z_a<z_b$ in
deriving Eq.(\ref{eq:lim_unu_vnu}).
Thus, the zeroth horizontal mode is special from the higher order modes ($m\in\mathbb{N}$).
Eqs.(\ref{eq:lim_unu_vnu}) and (\ref{eq:lim_rho_vnu_m0}) are summarized using
the Kronecker delta $\delta_0^m$ as follows,
\begin{alignat}{2}
  \lim_{\rho\to\infty}[v_{\nu}(\hr_b,\hr_a)]_{\nu=\mu_m^n}
  =-\{(-1)^m+\delta_0^m\}w
   \qquad
  (m\in\mathbb{Z}_0^{+}).
  \label{eq:lim_vnu_m0}
\end{alignat}
From Eqs.(\ref{eq:lim_rd_kr_nu2}), (\ref{eq:lim_rq_m0}) and (\ref{eq:lim_rho_vnu_m0}),
we get the asymptotic limit of the zeroth radial eigenfunction for $\rho\to\infty$,
\begin{alignat}{2}
  \lim_{\rho\to\infty}\cR_{-}^{0n}
   =\frac{\pi^2}{4}\cd2(k_r^n)^2\cd\frac{-4}{(\pi k_r^n)^2}\cd\frac{-1}{2w}
   =\frac{1}{w}
   =\cX_{-}^0 .
  \label{eq:lim_cRm_m0}
\end{alignat}
$\cX_{-}^0$ is given by Eq.(\ref{eq:cXm}) for $m=0$,
which is the zeroth horizontal eigenfunction of the straight rectangular pipe.
At last, let us summarize Eqs.(\ref{eq:lim_cR_cX}) and (\ref{eq:lim_cRm_m0}),
\begin{align}
  \lim_{\rho\to\infty}\cR_{\pm}^{mn}(r,r')
  &=\cX_{\pm}^m(x,x') ,
   \label{eq:lim_cRcX}
\end{align}
where $m\in(\mathbb{N},\mathbb{Z}_0^{+})$ for $(+,-)$.
Thus, in the limit of $\rho\to\infty$, the horizontal eigenfunctions of the curved pipe
go to those of the straight pipe
that has the same rectangular cross section as the curved pipe.
We showed Eq.(\ref{eq:lim_cRcX}) by taking into account the symmetry between $r_a$ and $r_b$
in deriving the asymptotic expressions of the poles $\nu=(\nu_m^n,\mu_m^n)$
in appendix \ref{sec:poles}.
As described around Eq.(\ref{eq:asympt_lim}), if we use Cochran's asymptotic solutions
given by Eqs.(3.4-3.5) in \cite{cochran_zero}, we cannot derive Eq.(\ref{eq:lim_cRcX}),
because Cochran's asymptotic solutions are the poles of the whispering gallery modes of
the curved pipe.
In deriving Eq.(\ref{eq:lim_cRcX}) only for the zeroth horizontal mode ($m=0$),
however, we must assume $z_a<1<z_b$ similar to the whispering gallery modes
even though the zeroth pole satisfies the asymptotic condition (\ref{eq:lim_nurho_ksmn}).

\clearpage

\section{Orthogonality of the cross products of the Bessel functions with respect to
the order}
\label{sec:orthogonality}

The cross products of the Bessel functions are given by Eqs.(\ref{eq:CP_pq}-\ref{eq:CP_sr}).
We find two orthogonality relations of the cross products
with respect to the radial mode number $m$ of the order $\nu$
at the poles $\nu=(\nu_m^n,\mu_m^n)$ given by Eq.(\ref{eq:ps_poles}).
One is the orthogonality of $p_{\nu}(\hr,\hr')$ at $\nu=\nu_m^n$,
the other is that of $q_{\nu}(\hr,\hr')$ at $\nu=\mu_m^n$,
which are shown respectively in appendices \ref{sec:ortho_pp} and \ref{sec:ortho_snu}.
From these orthogonality relations, we find the expressions of the $\delta$-function
in terms of the radial eigenfunctions $\cR_{\pm}^{mn}$ given by
Eqs.(\ref{eq:Ven}-\ref{eq:mfRm}).
We use these expressions of the radial $\delta$-functions in verifying
the exact solutions of the fields in appendix \ref{sec:verify}.

\subsection{Orthogonality of the cross products of the vertical electric field}
\label{sec:ortho_pp}

We find the orthogonality relation of the cross products
$p_{\nu}(\hr_b,\hr)$ and $p_{\nu}(\hr,\hr_a)$
which are involved in the radial eigenfunction $\cR_{+}^{mn}$ given by Eq.(\ref{eq:mfRp}).
The dimensionless radii $\hr$ and $\hr_{a,b}$ are given by Eqs.(\ref{eq:krn}).
We first consider the Bessel differential equations for
$p_\nu(\hr_b,\hr)$ and $p_\mu(\hr,\hr_a)$ with respect to $\hr$,
\begin{align}
  \brd_{\hr}q_\nu(\hr_b,\hr)
  =\bigg(\frac{\nu^2}{\hr^2}-1\bigg)p_\nu(\hr_b,\hr) ,
    \qquad
  \brd_{\hr}r_\mu(\hr,\hr_a)
  =\bigg(\frac{\mu^2}{\hr^2}-1\bigg)p_\mu(\hr,\hr_a) ,
  \label{eq:BDE_pp}
\end{align}
where $\nu,\mu\in\mathbb{C}$ in general.
$\brd_{\hr}$ is the operator given by Eq.(\ref{eq:rd_hr}),
\begin{align}
  \brd_{\hr}
  =\rd_{\hr}+\hr^{-1} .
\end{align}
Making a cross product by multiplying $p_\mu(\hr,\hr_a)$ and $p_\nu(\hr_b,\hr)$ with
Eqs.(\ref{eq:BDE_pp}), we integrate it from $\hr_a$ to $\hr_b$,
\begin{align}
  (\mu^2-\nu^2)\int_{\hr_a}^{\hr_b}\frac{d\hr}{\hr}p_\nu(\hr_b,\hr)p_\mu(\hr,\hr_a)
  =\int_{\hr_a}^{\hr_b}d\hr p_\nu(\hr_b,\hr)\rd_{\hr}\{\hr r_\mu(\hr,\hr_a)\}
  -\int_{\hr_a}^{\hr_b}d\hr p_\mu(\hr,\hr_a)\rd_{\hr}\{\hr q_\nu(\hr_b,\hr)\} .
  \label{eq:BDE_munu}
\end{align}
Further integrating the terms on the R.H.S. of Eq.(\ref{eq:BDE_munu}) by parts, we get
\begin{align}
  -\frac{\pi}{2}\int_{\hr_a}^{\hr_b}\frac{d\hr}{\hr}p_\nu(\hr_b,\hr)p_\mu(\hr,\hr_a)
  =\frac{p_\nu(\hr_b,\hr_a)-p_\mu(\hr_b,\hr_a)}{\nu^2-\mu^2} .
  \label{eq:int_pmu_pnu}
\end{align}

Here we suppose that both $\mu$ and $\nu$ are zeros of $p_{\lam}(\hr_b,\hr_a)$
with respect to $\lam$ as shown in Eq.(\ref{eq:pnu_poles}),
\begin{align}
  p_\nu(\hr_b,\hr_a)
  =p_\mu(\hr_b,\hr_a)
  =0
   \qquad\Lra\qquad
  \nu
  =\nu_m^n
  \in\mathbb{A}
    ,\qquad
  \mu
  =\nu_{m'}^n
  \in\mathbb{A}
   \qquad
  (m,m'\in\bbN) .
   \label{eq:pnu_zero}
\end{align}
$m$ is the radial mode number.
The superscript $n$ is the vertical mode number defined in Eq.(\ref{eq:kyn}).
$n$ is also involved in $\hr$ and $\hr_{a,b}$ through the radial wavenumber $k_r^n$
given by Eq.(\ref{eq:krn}) though we omit it for brevity.
We assume that $n$ is given and fixed in this appendix
($n$ can be omitted as $\nu_m$ and $\nu_{m'}$ for brevity since $n$ is not important in
this appendix).
We substitute Eqs.(\ref{eq:pnu_zero}) into Eq.(\ref{eq:int_pmu_pnu}).
When $m\ne m'$, since $\mu\ne\nu$, the R.H.S. of Eq.(\ref{eq:int_pmu_pnu}) is zero,
\begin{align}
  \int_{\hr_a}^{\hr_b}\frac{d\hr}{\hr}p_{\nu_m^n}(\hr_b,\hr)p_{\nu_{m'}^n}(\hr,\hr_a)
  =0 
   \qquad
  (m\ne m') .
  \label{eq:int_pmu_pnu_zero}
\end{align}
On the contrary, when $m=m'$, we take the limit of $\mu\to\nu$ for Eq.(\ref{eq:int_pmu_pnu}),
\begin{align}
  \int_{\hr_a}^{\hr_b}\frac{d\hr}{\hr}
  [p_\nu(\hr_b,\hr)p_\nu(\hr,\hr_a)]_{\nu=\nu_m^n}
  =-\Pi_{\nu_m^n}
   \qquad
  (m\in\mathbb{N})
   ,\qquad
  \Pi_{\nu}
  =\frac{\rd_\nu p_\nu(\hr_b,\hr_a)}{\pi\nu} .
  \label{eq:int_pnu_pnu}
\end{align}
Eqs.(\ref{eq:int_pmu_pnu_zero}-\ref{eq:int_pnu_pnu}) represent the orthogonality relation of
$p_{\nu}(\hr_b,\hr)$ and $p_{\nu}(\hr,\hr_a)$ for $\nu=\nu_m^n$
with respect to $m$ with a weight of $\hr^{-1}$,
\begin{align}
  \int_{\hr_a}^{\hr_b}\frac{d\hr}{\hr}
  p_{\nu_{m}^n}(\hr_b,\hr) p_{\nu_{m'}^n}(\hr,\hr_a)
  =-\delta_{m'}^m \Pi_{\nu_m^n}
   \qquad
  (m,m'\in\mathbb{N}) .
  \label{eq:ortho_pnu_pnu}
\end{align}
$\delta_{m'}^m$ is the Kronecker delta.
In the limit of $\rho\to\infty$,
Eq.(\ref{eq:ortho_pnu_pnu}) goes to Eq.(\ref{eq:sin_ortho_strt})
which is the orthogonality relation of the cross products of
the trigonometric functions involved in $E_y$ in a straight rectangular pipe.
Normalizing $p_{\nu}(\hr,\hr')$ by $\hr\Pi_{\nu}^{1/2}$, we can rewrite
Eq.(\ref{eq:ortho_pnu_pnu}) into the orthonormality of
$\bar{p}_{\nu_m^n}(\hr,\hr_{b,a})$ with a weight of $\hr$ which corresponds to
the Jacobian of the polar coordinates $(\hr,s/\rho)$ in the constant bending section,
\begin{align}
  \int_{\hr_a}^{\hr_b}
  \bar{p}_{\nu_{m}^n}(\hr,\hr_b)\bar{p}_{\nu_{m'}^n}(\hr,\hr_a) \hr d\hr
  =\delta_{m'}^m
   \qquad
  (m,m'\in\mathbb{N})
   ,\qquad
  \bar{p}_{\nu}(\hr,\hr')
  =\frac{p_{\nu}(\hr,\hr')}{\hr \Pi_{\nu}^{1/2}} .
  \label{eq:orthonorm_pnu}
\end{align}
$\Pi_{\nu}$ has $\rd_{\nu}p_\nu(\hr_b,\hr_a)$ which is not zero at $\nu=\nu_m^n$ in general.
As shown in appendix \ref{sec:orgn_odr},
$\rd_{\nu}p_\nu(\hr_b,\hr_a)$ has a zero of the first order at the origin $\nu=\nu_m^n=0$
since $p_\nu(\hr_b,\hr_a)$ has a zero of the second order at $\nu=0$
when $|k|=k_m^n$ which is the cutoff wavenumber of the vertical electric field $E_y$
in the curved pipe as defined in Eq.(\ref{eq:kmn}).
For $\nu\to0$, $\Pi_{\nu}$ goes to the following asymptotic limit,
\begin{align}
  \lim_{\nu\to0}\Pi_{\nu}
  =\frac{1}{\pi}[\rd_\nu^2 p_\nu(\hr_b,\hr_a)]_{\nu=0}
  \ne 0
   ,\qquad\text{where}\quad
  \nu_m^n=0
   \quad\Lra\quad
  |k|=k_m^n 
   \quad
   (m\in\mathbb{N}).
  \label{eq:int_pnu_pnu_nu0}
\end{align}
$\Pi_{0}$ is not zero as seen from the first equation of (\ref{eq:rdnu2_non0}).
According to Eq.(\ref{eq:ortho_pnu_pnu}), the radial $\delta$-function is given in
terms of the cross products of the Bessel functions as follows,
\begin{align}
  \delta(\hr-\hr')
  =-\frac{\pi}{\hr}\sum_{m=1}^{\infty}
  \bigg[\nu\frac{p_\nu(\hr_b,\hr)p_\nu(\hr',\hr_a)}{\rd_\nu p_\nu(\hr_b,\hr_a)}
  \bigg]_{\nu=\nu_m^n}
  =-\frac{\pi}{\hr}\sum_{m=1}^{\infty}
  \bigg[\nu\frac{p_\nu(\hr_b,\hr')p_\nu(\hr,\hr_a)}{\rd_\nu p_\nu(\hr_b,\hr_a)}
  \bigg]_{\nu=\nu_m^n} ,
  \label{eq:delta_pp}
\end{align}
where we used the first equation of (\ref{eq:nu_mn_rels}) which holds at
the poles $\nu=\nu_m^n$.
We rewrite Eq.(\ref{eq:delta_pp}) using the radial eigenfunction $\cR_{+}^{mn}$
given by Eq.(\ref{eq:mfRp}),
\begin{align}
  \frac{r}{\rho}\delta(r-r')=\sum_{m=1}^{\infty}\cR_{+}^{mn}(r,r') .
  \label{eq:delta_Ve}
\end{align}
Thus, the complete sum over all the radial eigenmodes of the curved pipe forms
the radial $\delta$-function.

\subsection{Orthogonality of the cross products of the vertical magnetic field}
\label{sec:ortho_snu}

Next, we find the orthogonality relation of the cross products of the Bessel functions
$r_{\nu}(\hr_b,\hr)$ and $q_{\nu}(\hr,\hr_a)$ which are involved in
the radial eigenfunction $\cR_{-}^{mn}$ given by Eq.(\ref{eq:mfRm}).
Then we find the expression of the radial $\delta$-function which consists of
the complete sum of $\cR_{-}^{mn}$ over all the radial eigenmodes,
similar to Eq.(\ref{eq:delta_Ve}).
The derivation is basically similar to appendix \ref{sec:ortho_pp}
except that $\cR_{-}^{mn}$ has the zeroth radial mode ($m=0$)
unlike $\cR_{+}^{mn}$ which is defined for $m\in\mathbb{N}$.

We first consider the Bessel differential equations for
the cross products $r_\nu(\hr_b,\hr)$ and $q_\mu(\hr,\hr_a)$ with respect to $\hr$,
\begin{align}
  \brd_{\hr}s_\nu(\hr_b,\hr)
  =\bigg(\frac{\nu^2}{\hr^2}-1\bigg)r_\nu(\hr_b,\hr) ,
    \qquad
  \brd_{\hr}s_\mu(\hr,\hr_a)
  =\bigg(\frac{\mu^2}{\hr^2}-1\bigg)q_\mu(\hr,\hr_a) .
  \label{eq:BDE_qr}
\end{align}
From Eqs.(\ref{eq:BDE_qr}), we get the following equation in a similar way
to deriving Eqs.(\ref{eq:BDE_pp}-\ref{eq:int_pmu_pnu}),
\begin{align}
  -\frac{\pi}{2}\int_{\hr_a}^{\hr_b}\frac{d\hr}{\hr}r_\nu(\hr_b,\hr)q_\mu(\hr,\hr_a)
  =\frac{s_\nu(\hr_b,\hr_a)-s_\mu(\hr_b,\hr_a)}{\nu^2-\mu^2} .
  \label{eq:int_qmu_rnu}
\end{align}
We suppose that $\mu$ and $\nu$ are zeros of $s_{\lam}(\hr_b,\hr_a)$
with respect to the order $\lam$ as shown in Eq.(\ref{eq:snu_poles}),
\begin{align}
  s_{\mu}(\hr_b,\hr_a)=s_{\nu}(\hr_b,\hr_a)=0
   \qquad\Lra\qquad
  \mu=\mu_m^n
  \in\mathbb{A}
   ,\qquad
  \nu=\mu_{m'}^n
  \in\mathbb{A}
   \qquad
  (m,m'\in\mathbb{Z}_0^{+}) .
  \label{eq:munu_zeros}
\end{align}
When $m\ne m'$, since $\mu\ne\nu$, the R.H.S. of Eq.(\ref{eq:int_qmu_rnu}) is zero,
\begin{align}
  \int_{\hr_a}^{\hr_b}\frac{d\hr}{\hr}r_{\mu_m^n}(\hr_b,\hr)q_{\mu_{m'}^n}(\hr,\hr_a)
  =0
   \qquad
  (m\ne m') .
  \label{eq:int_qnu_rnu_0}
\end{align}
On the contrary, when $m=m'$, \ie, in the limit of $\mu\to\nu$,
Eq.(\ref{eq:int_qmu_rnu}) becomes
\begin{align}
  \int_{\hr_a}^{\hr_b}\frac{d\hr}{\hr}
  [r_\nu(\hr_b,\hr)q_\nu(\hr,\hr_a)]_{\nu=\mu_m^n}
  =-\Sig_{\mu_m^n}
   \qquad
  (m\in\mathbb{Z}_0^{+})
   ,\qquad
  \Sig_{\nu}
  =\frac{\rd_{\nu}s_\nu(\hr_b,\hr_a)}{\pi\nu} .
  \label{eq:int_qnu_rnu}
\end{align}
Eqs.(\ref{eq:int_qnu_rnu_0}-\ref{eq:int_qnu_rnu}) represent the orthogonality relation of
$r_{\nu}(\hr_b,\hr)$ and $q_{\nu}(\hr,\hr_a)$ at $\nu=\mu_m^n$
with respect to $m$ with a weight of $\hr^{-1}$,
\begin{align}
  \int_{\hr_a}^{\hr_b}\frac{d\hr}{\hr}
  r_{\mu_{m}^n}(\hr_b,\hr) q_{\mu_{m'}^n}(\hr,\hr_a)
  =-\delta_{m'}^m \Sig_{\mu_m^n}
   \qquad
  (m,m'\in\mathbb{Z}_0^{+}) .
  \label{eq:ortho_rmu_qmu}
\end{align}
In the limit of $\rho\to\infty$,
Eq.(\ref{eq:ortho_rmu_qmu}) goes to Eq.(\ref{eq:cos_ortho_strt})
which is the orthogonality relation of the cross products of the trigonometric functions
involved in the vertical magnetic field $B_y$ in the straight rectangular pipe.
Normalizing $q_{\nu}(\hr,\hr')$ by $\hr\Sig_{\nu}^{1/2}$ in Eq.(\ref{eq:ortho_rmu_qmu}),
we get the orthonormality relation of $\bar{q}_{\mu_m^n}(\hr,\hr_{a,b})$ with respect to $m$ 
with a weight of $\hr$,
\begin{align}
  \int_{\hr_a}^{\hr_b}
  \bar{q}_{\mu_{m}^n}(\hr,\hr_b) \bar{q}_{\mu_{m'}^n}(\hr,\hr_a) \hr d\hr
  =\delta_{m'}^m
   \qquad
  (m,m'\in\mathbb{Z}_0^{+})
   ,\qquad
  \bar{q}_{\nu}(\hr,\hr')
  =\frac{q_{\nu}(\hr,\hr')}{\hr\Sig_{\nu}^{1/2}} .
  \label{eq:orthonorm_qmu}
\end{align}
$\Sig_{\nu}$ involves $\rd_{\nu}s_\nu(\hr_b,\hr_a)$
which is, in general, not zero at $\nu=\mu_m^n$.
As shown in appendix \ref{sec:orgn_odr}, 
$\rd_\nu s_\nu(\hr_b,\hr_a)$ for $\nu=\mu_m^n$ becomes zero of the first order at
the origin $\mu_m^n=0$ since $s_\nu(\hr_b,\hr_a)$ has the pole of the second order at
$\nu=\mu_m^n=0$ when $|k|=\bk_m^n$ which is the cutoff wavenumber of $B_y$ in
the curved rectangular pipe as described in Eq.(\ref{eq:bkmn}).
Therefore, in the limit of $\nu\to0$,
\begin{align}
  \lim_{\nu\to0}\Sig_{\nu}
  =\frac{1}{\pi}[\rd_\nu^2 s_\nu(\hr_b,\hr_a)]_{\nu=0}
  \ne 0
   ,\qquad\text{where}\quad
  \mu_m^n=0
   \quad\Lra\quad
  |k|=\bk_m^n .
   \label{eq:int_rnu_qnu_nu0}
\end{align}
$\Sig_{0}$ is not zero as seen from the second equation of (\ref{eq:rdnu2_non0}).
According to Eq.(\ref{eq:ortho_rmu_qmu}), the radial $\delta$-function is given
in terms of the cross products of the Bessel functions for $\nu=\mu_m^n$,
\begin{align}
  \delta(\hr-\hr')
  &=-\frac{\pi}{\hr}\sum_{m=0}^{\infty}
    \bigg[
       \nu\frac{r_\nu(\hr_b,\hr)q_\nu(\hr',\hr_a)}{\rd_\nu s_\nu(\hr_b,\hr_a)}
    \bigg]_{\nu=\mu_m^n}
  =-\frac{\pi}{\hr}\sum_{m=0}^{\infty}
    \bigg[
       \nu\frac{r_\nu(\hr_b,\hr')q_\nu(\hr,\hr_a)}{\rd_\nu s_\nu(\hr_b,\hr_a)}
    \bigg]_{\nu=\mu_m^n} ,
   \label{eq:deltar_snu}
\end{align}
where we used Eq.(\ref{eq:mu_mn_rels}) which holds at $\nu=\mu_m^n$.
From Eq.(\ref{eq:deltar_snu}), we get the complete sum over all the radial eigenfunctions
$\cR_{-}^{mn}$ given by Eq.(\ref{eq:cRm_ba}),
\begin{align}
  \frac{r}{\rho}\delta(r-r')
  &=\sum_{m=0}^{\infty}\cR_{-}^{mn}(r,r') .
  \label{eq:delta_Vb}
\end{align}

Using Eq.(\ref{eq:rdelr}) and the Bessel differential equations
(\ref{eq:BDE_cRp}-\ref{eq:BDE_cRm}) for $\cR_{\pm}^{mn}$,
we can rewrite Eqs.(\ref{eq:delta_Ve}) and (\ref{eq:delta_Vb}) as follows,
\begin{align}
  \frac{r}{\rho}\delta(r-r')
  &=\sum_{m=1}^{\infty}
    \bigg\{\frac{(\nu_m^n)^2}{\hr\hr'}+\rd_{\hr}\rd_{\hr'}\bigg\}\cR_{+}^{mn}
   =\sum_{m=0}^{\infty}
    \bigg\{\frac{(\mu_m^n)^2}{\hr\hr'}+\rd_{\hr}\rd_{\hr'}\bigg\}\cR_{-}^{mn} .
   \label{eq:delr2}
\end{align}
Eqs.(\ref{eq:delr2}) are related to Eqs.(\ref{eq:mfHpm}) and (\ref{eq:bmfG_pm_delta})
which are the Green functions of the radial and longitudinal components of the fields in 
the Laplace domain.
Eqs.(\ref{eq:delta_Ve}) and (\ref{eq:delta_Vb}) go to Eq.(\ref{eq:delta_Tbe}) in
the limit of $\rho\to\infty$, according to Eq.(\ref{eq:lim_cRcX}) and
the last equation of (\ref{eq:lim_gl_g12}),
\begin{align}
  \lim_{\rho\to\infty}\frac{r}{\rho}\delta(r-r')
  &=\sum_{m=1,0}^{\infty}\lim_{\rho\to\infty}\cR_{\pm}^{mn}(r,r')
  =\sum_{m=1,0}^{\infty}\cX_{\pm}^m(x,x')
  =\delta(x-x') ,
\end{align}
where the lower limit of the sums $m=(1,0)$ corresponds to the sign $(+,-)$ of
$\cR_{\pm}^{mn}$ and $\cX_{\pm}^m$ as those in Eqs.(\ref{eq:delr2}).
$\cX_{\pm}^m$ is given by Eqs.(\ref{eq:mfXp_new}-\ref{eq:mfXm_new})
which are the horizontal eigenfunctions of the straight rectangular pipe.
The horizontal variables $r$ and $x$ are related through Eq.(\ref{eq:r}).

\clearpage

\section{Differential expressions of the fields}
\label{sec:opera_express}

We found four expressions of the exact solution of the field as summarized in
Eq.(\ref{eq:4express}).
In this appendix we explicitly write out the differential expressions of
the separated and complete forms
since we use them in the analytical and numerical calculations
in appendices \ref{sec:verify} and \ref{sec:impedance}.
We often omit the transverse arguments $(\xv_\perp,\xv_\perp')$ of the functions
for brevity, indicating only the longitudinal arguments $(s,s')$ as follows,
\begin{align}
  \cR_{\pm}^{mn}
  &=\cR_{\pm}^{mn}(r,r')
   ,\qquad
  \cY_{\pm}^n
  =\cY_{\pm}^n(y,y')
   ,\qquad
  \tJ_0(s')
  =\tJ_0(\xv')
   ,\qquad
  \tJv(s')
  =\tJv(\xv') ,
   \label{RYJ_abbrev}
   \\
  \tEv(s')
  &=\tEv(\xv')
   ,\qquad
  \tBv(s')
  =\tBv(\xv')
   ,\qquad
  \tEv(0)
  =\tEv(\xv_\perp',0)
   ,\qquad
  \tBv(0)
  =\tBv(\xv_\perp',0) .
   \label{abbrev}
\end{align}
In this appendix $\vsig$ represents only $s-s'$ (\ie, $\vsig\ne s$)
instead of the one defined originally in Eq.(\ref{eq:cGeb_def}).

\subsection{Differential expression of the separated form}

The differential expression of the electromagnetic field in the separated form is
given by Eq.(\ref{eq:tEBv_dg}),
\begin{align}
  \bigg\{{\tEv(\xv) \atop c\tBv(\xv)}\bigg\}
  &=\int_{r_a}^{r_b}dr'\int_{-h/2}^{h/2}dy'
    \bigg[
       \bigg\{{\tPv^{\dg}(\xv,\xv_\perp') \atop
               \tQv^{\dg}(\xv,\xv_\perp')}
       \bigg\}
      -Z_0\int_{0}^{\infty}ds'
       \bigg\{{\tMv^{\dg}(\xv_{\perp},s-s',\xv') \atop \tNv^{\dg}(\xv_{\perp},s-s',\xv')}
       \bigg\}
    \bigg] .
  \label{eq:tEBv_dg_app}
\end{align}
$(\tPv^{\dg},\tQv^{\dg})$ and $(\tMv^{\dg},\tNv^{\dg})$ are given by 
Eqs.(\ref{eq:tPy_dg_cG}-\ref{eq:tQy_dg_cG}) and (\ref{eq:tMy_dg_rw}-\ref{eq:tNs_dg_rw}).
Substituting Eqs.(\ref{eq:cGp_sep}-\ref{eq:cGm_sep}) into the first equations of
(\ref{eq:tPy_dg_cG}-\ref{eq:tQy_dg_cG}), we rearrange them,
\begin{align}
  \tP_y^{\dg}(s)
  &=\sum_{n=0}^{\infty}\sum_{\ell=0}^{\infty}\Th_{+}^{n\ell}
    \bigg[
        \bdelta_{01}^{\ell}\sum_{m=1}^{\ell-1}\cR_{+}^{mn}\mfp_y(s)
       +\sum_{m=\ell'}^{\infty}\cR_{+}^{mn}\bmfp_y(s)
    \bigg] ,
  \label{eq:tPy_dg}
  \\
  \tQ_y^{\dg}(s)
  &=\sum_{n=1}^{\infty}\sum_{\ell=0}^{\infty}\Th_{-}^{n\ell}
    \bigg[\bdelta_{0}^{\ell}\sum_{m=0}^{\ell-1}\cR_{-}^{mn}\mfq_y(s)
         +\sum_{m=\ell}^{\infty}\cR_{-}^{mn}\bmfq_y(s)
    \bigg] .
  \label{eq:tQy_dg}
\end{align}
$\ell'$, $\bdelta_{01}^{\ell}$ and $\bdelta_{0}^{\ell}$ are given by 
Eqs.(\ref{eq:lprm}-\ref{eq:d01_d0}).
The integrands of Eqs.(\ref{eq:tPy_dg}-\ref{eq:tQy_dg}) are given as follows,
\begin{align}
  \mfp_y(s)
  &=\bigg[
       \frac{\rho}{r'}\tE_y(0)\cos\!\bigg(\frac{\hnu_m^ns}{\rho}\bigg)
      -\frac{\rho}{\hnu_m^n}\sin\!\bigg(\frac{\hnu_m^ns}{\rho}\bigg)
       \big\{\tE_s(0)\rd_{y'}+c\tB_x(0)ik\beta\big\}
    \bigg]
    \cY_{-}^n ,
  \label{eq:mfpy}
  \\
  \bmfp_y(s)
  &=\frac{e^{-\cnu_m^ns/\rho}}{2}
    \bigg[\frac{\rho}{r'}\tE_y(0)
         +\frac{\rho}{\cnu_m^n}\big\{\tE_s(0)\rd_{y'}+c\tB_x(0)ik\beta\big\}
    \bigg]
    \cY_{-}^n ,
  \\
  \mfq_y(s)
  &=\bigg[
       \frac{\rho}{r'}\cos\!\bigg(\frac{\hmu_m^ns}{\rho}\bigg)c\tB_y(0)
      -\frac{\rho}{\hmu_m^n}\sin\!\bigg(\frac{\hmu_m^ns}{\rho}\bigg)
       \big\{c\tB_s(0)\rd_{y'}-\tE_x(0)ik\beta\big\}
    \bigg]\cY_{+}^n ,
  \\
  \bmfq_y(s)
  &=\frac{e^{-\cmu_m^ns/\rho}}{2}
    \bigg[
       \frac{\rho}{r'}c\tB_y(0)
      +\frac{\rho}{\cmu_m^n}\big\{c\tB_s(0)\rd_{y'}-\tE_x(0)ik\beta\big\}
    \bigg]
    \cY_{+}^n .
\end{align}
$(\hnu_m^n,\hmu_m^n)$ and $(\cnu_m^n,\cmu_m^n)$ are the real and imaginary poles
defined in Eqs.(\ref{eq:ps_poles}).
Substituting Eqs.(\ref{eq:cGp_sep}-\ref{eq:cGm_sep}) and (\ref{eq:cHp}-\ref{eq:cKm}) into
the second equation of (\ref{eq:tPy_dg_cG}), we rearrange them,
\begin{align}
  \tP_x^{\dg}(s)
  &=\sum_{n=1}^{\infty}\sum_{\ell=0}^{\infty}
    \Bigg[
      \Th_{-}^{n\ell}
      \bigg\{
         \bdelta_{0}^{\ell}\sum_{m=0}^{\ell-1}\mfp_x^{-}(s)
        +\sum_{m=\ell}^{\infty}\bmfp_x^{-}(s)
      \bigg\}
     +\Th_{+}^{n\ell}
      \bigg\{
         \bdelta_{01}^{\ell}\sum_{m=1}^{\ell-1}\mfp_x^{+}(s)
        +\sum_{m=\ell'}^{\infty}\bmfp_x^{+}(s)
      \bigg\}
    \Bigg] ,
  \\
  \tP_s^{\dg}(s)
  &=\sum_{n=1}^{\infty}\sum_{\ell=0}^{\infty}
    \Bigg[
       \Th_{+}^{n\ell}
       \bigg\{
          \bdelta_{01}^{\ell}\sum_{m=1}^{\ell-1}\mfp_s^{+}(s)
         +\sum_{m=\ell'}^{\infty}\bmfp_s^{+}(s)
       \bigg\}
      +\Th_{-}^{n\ell}
       \bigg\{
          \bdelta_{0}^{\ell}\sum_{m=0}^{\ell-1}\mfp_s^{-}(s)
         +\sum_{m=\ell}^{\infty}\bmfp_s^{-}(s)
       \bigg\}
    \Bigg] ,
\end{align}
where
\begin{align}
  \mfp_x^{-}(s)
  &=\bigg[
       \cos\!\bigg(\frac{\hmu_m^ns}{\rho}\bigg)
       \big\{\tE_x(0)k_r^n -\tE_y(0)\rd_{y'}\rd_{\hr'}\big\}
      +\frac{\hmu_m^n}{\hr'}\sin\!\bigg(\frac{\hmu_m^ns}{\rho}\bigg)c\tB_y(0)ik\beta
    \bigg]
    \frac{\rho}{\hr}\cR_{-}^{mn}\cY_{+}^n ,
   \label{eq:mfpxm}
  \\
  \bmfp_x^{-}(s)
  &=\frac{e^{-\cmu_m^ns/\rho}}{2}
    \bigg[
       \tE_x(0)k_r^n -\tE_y(0)\rd_{y'}\rd_{\hr'}
      +c\tB_y(0)ik\beta\frac{\cmu_m^n}{\hr'}
    \bigg]
    \frac{\rho}{\hr}\cR_{-}^{mn}\cY_{+}^n ,
  \\
  \mfp_x^{+}(s)
  &=\bigg[
       \frac{\rho}{\hnu_m^n}\sin\!\bigg(\frac{\hnu_m^ns}{\rho}\bigg)
       \big\{\tE_s(0)k_r^n+c\tB_y(0)ik\beta\rd_{\hr'}\big\}
      -\frac{\rho}{\hr'}\cos\!\bigg(\frac{\hnu_m^ns}{\rho}\bigg)\tE_y(0)\rd_{y'}
    \bigg]
    \rd_{\hr}\cR_{+}^{mn}\cY_{+}^n ,
  \\
  \bmfp_x^{+}(s)
  &=-\frac{e^{-\cnu_m^ns/\rho}}{2}
    \bigg[
       \tE_y(0)\frac{\rho}{\hr'}\rd_{y'}
      +\frac{\rho}{\cnu_m^n}\big\{c\tB_y(0)ik\beta\rd_{\hr'}+\tE_s(0)k_r^n\big\}
    \bigg]
    \rd_{\hr}\cR_{+}^{mn}\cY_{+}^n ,
\end{align}
\begin{align}
  \mfp_s^{+}(s)
  &=\bigg[
       \cos\!\bigg(\frac{\hnu_m^ns}{\rho}\bigg)
       \big\{\tE_s(0)k_r^n+c\tB_y(0)ik\beta\rd_{\hr'}\big\}
      +\frac{\hnu_m^n}{\hr'}\sin\!\bigg(\frac{\hnu_m^ns}{\rho}\bigg)\tE_y(0)\rd_{y'}
    \bigg]
    \frac{\rho}{\hr}\cR_{+}^{mn}\cY_{+}^n ,
  \\
  \bmfp_s^{+}(s)
  &=\frac{e^{-\cnu_m^ns/\rho}}{2}
    \bigg[
       \tE_s(0)k_r^n+c\tB_y(0)ik\beta\rd_{\hr'}
      +\tE_y(0)\rd_{y'}\frac{\cnu_m^n}{\hr'}
    \bigg]
    \frac{\rho}{\hr}\cR_{+}^{mn}\cY_{+}^n ,
  \\
  \mfp_s^{-}(s)
  &=\bigg[
       \frac{\rho}{\hr'}\cos\!\bigg(\frac{\hmu_m^ns}{\rho}\bigg)c\tB_y(0)ik\beta
      +\frac{\rho}{\hmu_m^n}\sin\!\bigg(\frac{\hmu_m^ns}{\rho}\bigg)
       \big\{\tE_y(0)\rd_{\hr'}\rd_{y'}-\tE_x(0)k_r^n\big\}
    \bigg]
    \rd_{\hr}\cR_{-}^{mn}\cY_{+}^n ,
  \\
  \bmfp_s^{-}(s)
  &=\frac{e^{-\cmu_m^ns/\rho}}{2}
    \bigg[
       c\tB_y(0)ik\beta\frac{\rho}{\hr'}
      -\frac{\rho}{\cmu_m^n}\big\{\tE_y(0)\rd_{\hr'}\rd_{y'}-\tE_x(0)k_r^n\big\}
    \bigg]
    \rd_{\hr}\cR_{-}^{mn}\cY_{+}^n .
\end{align}
Similarly, substituting Eqs.(\ref{eq:cGp_sep}-\ref{eq:cGm_sep}) and
(\ref{eq:cHp}-\ref{eq:cKm}) into the second equation of (\ref{eq:tQy_dg_cG}),
we rearrange them,
\begin{align}
  \tQ_x^{\dg}(s)
  &=\sum_{n=0}^{\infty}\sum_{\ell=0}^{\infty}
    \Bigg[
       \Th_{+}^{n\ell}
       \bigg\{
          \bdelta_{01}^{\ell}\sum_{m=1}^{\ell-1}\mfq_x^{+}(s)
         +\sum_{m=\ell'}^{\infty}\bmfq_x^{+}(s)
       \bigg\}
      +\Th_{-}^{n\ell}
       \bigg\{
          \bdelta_{0}^{\ell}\sum_{m=0}^{\ell-1}\mfq_x^{-}(s)
         +\sum_{m=\ell}^{\infty}\bmfq_x^{-}(s)
       \bigg\}
    \Bigg] ,
  \\
  \tQ_s^{\dg}(s)
  &=\sum_{n=0}^{\infty}\sum_{\ell=0}^{\infty}
    \Bigg[
      \Th_{-}^{n\ell}
      \bigg\{
         \bdelta_{0}^{\ell}\sum_{m=0}^{\ell-1}\mfq_s^{-}(s)
        +\sum_{m=\ell}^{\infty}\bmfq_s^{-}(s)
      \bigg\}
     +\Th_{+}^{n\ell}
      \bigg\{
         \bdelta_{01}^{\ell}\sum_{m=1}^{\ell-1}\mfq_s^{+}(s)
        +\sum_{m=\ell'}^{\infty}\bmfq_s^{+}(s)
      \bigg\}
    \Bigg] ,
\end{align}
where
\begin{align}
  \mfq_x^{+}(s)
  &=\bigg[
       \cos\!\bigg(\frac{\hnu_m^ns}{\rho}\bigg)
       \big\{c\tB_x(0)k_r^n-c\tB_y(0)\rd_{y'}\rd_{\hr'}\big\}
      -\frac{\hnu_m^n}{\hr'}\sin\!\bigg(\frac{\hnu_m^ns}{\rho}\bigg)\tE_y(0)ik\beta
    \bigg]
    \frac{\rho}{\hr}\cR_{+}^{mn}\cY_{-}^n ,
   \\
  \bmfq_x^{+}(s)
  &=\frac{e^{-\cnu_m^ns/\rho}}{2}
    \bigg[
       c\tB_x(0)k_r^n
      -c\tB_y(0)\rd_{y'}\rd_{\hr'}
      -\tE_y(0)ik\beta\frac{\cnu_m^n}{\hr'}
    \bigg]
    \frac{\rho}{\hr}\cR_{+}^{mn}\cY_{-}^n ,
   \\
  \mfq_x^{-}(s)
  &=\bigg[
       \frac{\rho}{\hmu_m^n}\sin\!\bigg(\frac{\hmu_m^ns}{\rho}\bigg)
       \big\{c\tB_s(0)k_r^n-\tE_y(0)ik\beta\rd_{\hr'}\big\}
      -\frac{\rho}{\hr'}\cos\!\bigg(\frac{\hmu_m^ns}{\rho}\bigg)c\tB_y(0)\rd_{y'}
    \bigg]
    \rd_{\hr}\cR_{-}^{mn}\cY_{-}^n ,
   \\
  \bmfq_x^{-}(s)
  &=\frac{e^{-\cmu_m^ns/\rho}}{2}
    \bigg[
       \frac{\rho}{\cmu_m^n}\big\{\tE_y(0)ik\beta\rd_{\hr'}-c\tB_s(0)k_r^n\big\}
      -\frac{\rho}{\hr'}c\tB_y(0)\rd_{y'}
    \bigg]
    \rd_{\hr}\cR_{-}^{mn}\cY_{-}^n ,
   \\
   \nonumber
   \\
  \mfq_s^{-}(s)
  &=\bigg[
        \frac{\hmu_m^n}{\hr'}\sin\!\bigg(\frac{\hmu_m^ns}{\rho}\bigg)c\tB_y(0)\rd_{y'}
       +\cos\!\bigg(\frac{\hmu_m^ns}{\rho}\bigg)
        \big\{c\tB_s(0)k_r^n-\tE_y(0)ik\beta\rd_{\hr'}\big\}
    \bigg]
    \frac{\rho}{\hr}\cR_{-}^{mn}\cY_{-}^n ,
   \\
  \bmfq_s^{-}(s)
  &=\frac{e^{-\cmu_m^ns/\rho}}{2}
    \bigg[
        c\tB_s(0)k_r^n
       +c\tB_y(0)\frac{\cmu_m^n}{\hr'}\rd_{y'}
       -\tE_y(0)ik\beta\rd_{\hr'}
    \bigg]
    \frac{\rho}{\hr}\cR_{-}^{mn}\cY_{-}^n ,
   \\
  \mfq_s^{+}(s)
  &=\bigg[
       \frac{\rho}{\hnu_m^n}\sin\!\bigg(\frac{\hnu_m^ns}{\rho}\bigg)
       \big\{c\tB_y(0)\rd_{\hr'}\rd_{y'}-c\tB_x(0)k_r^n\big\}
      -\frac{\rho}{\hr'}\cos\!\bigg(\frac{\hnu_m^ns}{\rho}\bigg)\tE_y(0)ik\beta
    \bigg]
    \rd_{\hr}\cR_{+}^{mn}\cY_{-}^n ,
   \\
  \bmfq_s^{+}(s)
  &=\frac{e^{-\cnu_m^ns/\rho}}{2}
    \bigg[
       \frac{\rho}{\cnu_m^n}\big\{c\tB_x(0)k_r^n-c\tB_y(0)\rd_{\hr'}\rd_{y'}\big\}
       -\tE_y(0)ik\beta\frac{\rho}{\hr'}
    \bigg]
    \rd_{\hr}\cR_{+}^{mn}\cY_{-}^n .
   \label{eq:bmfq_sp}
\end{align}
We omit the transverse arguments $\xv_{\perp}'=(r',y')$ of the functions.
$\tEv(0)$ and $\tBv(0)$ are the initial fields at $s'=0$ as described in Eqs.(\ref{abbrev}).

We rearrange $\tMv^{\dg}$ and $\tNv^{\dg}$ which involve the source terms of
the fields in Eq.(\ref{eq:tEBv_dg_app}).
Substituting the Green functions of the separated form
(\ref{eq:cGp_sep}-\ref{eq:cGm_sep}) into the first equations of
(\ref{eq:tMy_dg_rw}-\ref{eq:tNy_dg_rw}), we rearrange them,
\begin{align}
  \tM_y^{\dg}(\vsig,s')
  &=\frac{r'}{\rho}\sum_{n=0}^{\infty}
    \big\{\tJ_0(s')\rd_{y'}+\tJ_y(s')ik\beta\big\}\cY_{-}^n
    \sum_{\ell=0}^{\infty}\Th_{+}^{n\ell}
    \bigg\{
      \bdelta_{01}^{\ell}\theta(\vsig)\sum_{m=1}^{\ell-1}\mfm_y(\vsig)
     +\sum_{m=\ell'}^{\infty}\bmfm_y(\vsig)
    \bigg\} ,
   \label{tMy_dg_app}
   \\
  \tN_y^{\dg}(\vsig,s')
  &=\sum_{n=1}^{\infty}\cY_{+}^n\sum_{\ell=0}^{\infty}\Th_{-}^{n\ell}
    \bigg\{
       \bdelta_{0}^{\ell}\theta(\vsig)\sum_{m=0}^{\ell-1}\mfn_y(\vsig,s')
      +\sum_{m=\ell}^{\infty}\bmfn_y(\vsig,s')
    \bigg\} .
   \label{tNy_dg_app}
\end{align}
$\vsig$ represents $s-s'$ ($\vsig\ne s$) in this appendix.
The summands of Eqs.(\ref{tMy_dg_app}-\ref{tNy_dg_app}) are given as follows,
\begin{alignat}{2}
  \mfm_y
  &=\frac{\rho}{\hnu_m^n}\cR_{+}^{mn}\sin\!\bigg(\frac{\hnu_m^n\vsig}{\rho}\bigg) ,
    \qquad
  &\mfn_y
  &=\bigg\{
       \frac{\hr'}{\hmu_m^n}\sin\!\bigg(\frac{\hmu_m^n\vsig}{\rho}\bigg)\tJ_s(s')\rd_{\hr'}
      +\cos\!\bigg(\frac{\hmu_m^n\vsig}{\rho}\bigg)\tJ_x(s')
    \bigg\}\cR_{-}^{mn} ,
  \\
  \bmfm_y
  &=-\frac{\rho}{2\cnu_m^n}\cR_{+}^{mn}e^{-\cnu_m^n|\vsig|/\rho} ,
    \qquad
  &\bmfn_y
  &=\frac{e^{-\cmu_m^n|\vsig|/\rho}}{2}
    \bigg\{\tJ_x(s')\sgn(\vsig)-\tJ_s(s')\frac{\hr'}{\cmu_m^n}\rd_{\hr'}\bigg\}
    \cR_{-}^{mn} .
\end{alignat}
Substituting Eqs.(\ref{eq:cGp_sep}-\ref{eq:cGm_sep}) and (\ref{eq:cHp}-\ref{eq:cKm}) into 
the second equation of (\ref{eq:tMy_dg_rw}), we rearrange them,
\begin{align}
  \tM_x^{\dg}(\vsig,s')
  &=\sum_{n=1}^{\infty}\cY_{+}^n\sum_{\ell=0}^{\infty}
    \Bigg[
      \Th_{-}^{n\ell}
      \bigg\{
        \bdelta_{0}^{\ell}\theta(\vsig)\sum_{m=0}^{\ell-1}\mfm_x^{-}
       +\sum_{m=\ell}^{\infty}\bmfm_x^{-}
      \bigg\}
     +\Th_{+}^{n\ell}
      \bigg\{
        \bdelta_{01}^{\ell}\theta(\vsig)\sum_{m=1}^{\ell-1}\mfm_x^{+}
       +\sum_{m=\ell'}^{\infty}\bmfm_x^{+}
      \bigg\}
    \Bigg] ,
  \\
  \tM_s^{\dg}(\vsig,s')
  &=\sum_{n=1}^{\infty}\cY_{+}^n\sum_{\ell=0}^{\infty}
    \Bigg[
       \Th_{+}^{n\ell}
       \bigg\{
          \bdelta_{01}^{\ell}\theta(\vsig)\sum_{m=1}^{\ell-1}\mfm_s^{+}
         +\sum_{m=\ell'}^{\infty}\bmfm_s^{+}
       \bigg\}
      +\Th_{-}^{n\ell}
       \bigg\{
          \bdelta_{0}^{\ell}\theta(\vsig)\sum_{m=0}^{\ell-1}\mfm_s^{-}
         +\sum_{m=\ell}^{\infty}  \bmfm_s^{-}
       \bigg\}
    \Bigg] ,
\end{align}
where
\begin{align}
  \mfm_x^{-}(\vsig,s')
  &=-\frac{ik\beta}{k_r^n}
    \bigg[
       \cos\!\bigg(\frac{\hmu_m^n\vsig}{\rho}\bigg)\tJ_s(s')\frac{\hr'}{\rho}\rd_{\hr'}
      -\frac{\hmu_m^n}{\rho}\sin\!\bigg(\frac{\hmu_m^n\vsig}{\rho}\bigg)\tJ_x(s')
    \bigg]
    \frac{\rho}{\hr}\cR_{-}^{mn} ,
  \\
  \bmfm_x^{-}(\vsig,s')
  &=-\frac{ik\beta}{k_r^n}\frac{e^{-\cmu_m^n|\vsig|/\rho}}{2}
    \bigg[
       \tJ_s(s')\sgn(\vsig)\frac{\hr'}{\rho}\rd_{\hr'}
      -\tJ_x(s')\frac{\cmu_m^n}{\rho}
    \bigg]
    \frac{\rho}{\hr}\cR_{-}^{mn} ,
  \\
  \mfm_x^{+}(\vsig,s')
  &=-\bigg[
      \tJ_s(s')\frac{ik\beta}{k_r^n}\cos\!\bigg(\frac{\hnu_m^n\vsig}{\rho}\bigg)
     +\frac{\hr'}{\hnu_m^n}\sin\!\bigg(\frac{\hnu_m^n\vsig}{\rho}\bigg)
      \Big\{\tJ_0(s')-\tJ_x(s')\frac{ik\beta}{k_r^n}\rd_{\hr'}\Big\}
    \bigg]
    \rd_{\hr}\cR_{+}^{mn} ,
  \\
  \bmfm_x^{+}(\vsig,s')
  &=\frac{e^{-\cnu_m^n|\vsig|/\rho}}{2}
    \bigg[
       \frac{\hr'}{\cnu_m^n}\Big\{\tJ_0(s')-\tJ_x(s')\frac{ik\beta}{k_r^n}\rd_{\hr'}\Big\}
      -\tJ_s(s')\sgn(\vsig)\frac{ik\beta}{k_r^n}
    \bigg]
    \rd_{\hr}\cR_{+}^{mn} ,
  \\
  \nonumber
  \\
  \mfm_s^{+}(\vsig,s')
  &=\bigg[
       \sin\!\bigg(\frac{\hnu_m^n\vsig}{\rho}\bigg)
       \tJ_s(s')\frac{ik\beta}{k_r^n}\frac{\hnu_m^n}{\rho}
      +\frac{\hr'}{\rho}\cos\!\bigg(\frac{\hnu_m^n\vsig}{\rho}\bigg)
       \Big\{\tJ_x(s')\frac{ik\beta}{k_r^n}\rd_{\hr'}-\tJ_0(s')\Big\}
    \bigg]
    \frac{\rho}{\hr}\cR_{+}^{mn} ,
  \label{eq:mfmsp}
  \\
  \bmfm_s^{+}(\vsig,s')
  &=\frac{e^{-\cnu_m^n|\vsig|/\rho}}{2}
    \bigg[
       \tJ_s(s')\frac{ik\beta}{k_r^n}\frac{\cnu_m^n}{\rho}
      +\sgn(\vsig)\frac{\hr'}{\rho}
       \Big\{\tJ_x(s')\frac{ik\beta}{k_r^n}\rd_{\hr'}-\tJ_0(s')\Big\}
    \bigg]
    \frac{\rho}{\hr}\cR_{+}^{mn} ,
  \\
  \mfm_s^{-}(\vsig,s')
  &=\frac{ik\beta}{k_r^n}
    \bigg[
       \sin\!\bigg(\frac{\hmu_m^n\vsig}{\rho}\bigg)\tJ_s(s')\frac{\hr'}{\hmu_m^n}\rd_{\hr'}
      +\cos\!\bigg(\frac{\hmu_m^n\vsig}{\rho}\bigg)\tJ_x(s')
    \bigg]
    \rd_{\hr}\cR_{-}^{mn} ,
  \label{eq:mfmsm}
  \\
  \bmfm_s^{-}(\vsig,s')
  &=-\frac{ik\beta}{k_r^n}\frac{e^{-\cmu_m^n|\vsig|/\rho}}{2}
    \bigg[\tJ_s(s')\frac{\hr'}{\cmu_m^n}\rd_{\hr'}-\sgn(\vsig)\tJ_x(s')\bigg]
    \rd_{\hr}\cR_{-}^{mn} .
\end{align}
Similarly, substituting Eqs.(\ref{eq:cGp_sep}-\ref{eq:cGm_sep}) and
(\ref{eq:cHp}-\ref{eq:cKm}) into the second equation of (\ref{eq:tNy_dg_rw}),
we rearrange them,
\begin{align}
  \tN_x^{\dg}(\vsig,s')
  &=\sum_{n=0}^{\infty}\sum_{\ell=0}^{\infty}
    \Bigg[
       \Th_{+}^{n\ell}
       \bigg\{
          \bdelta_{01}^{\ell}\theta(\vsig)\sum_{m=1}^{\ell-1}\mfn_x^{+}
         +\sum_{m=\ell'}^{\infty}\bmfn_x^{+}
       \bigg\}
      +\Th_{-}^{n\ell}
       \bigg\{\bdelta_{0}^{\ell}\theta(\vsig)\sum_{m=0}^{\ell-1} \mfn_x^{-}
             +\sum_{m=\ell}^{\infty} \bmfn_x^{-}
       \bigg\}
    \Bigg] ,
   \\
  \tN_s^{\dg}(\vsig,s')
  &=\sum_{n=0}^{\infty}\sum_{\ell=0}^{\infty}
    \Bigg[
      \Th_{-}^{n\ell}
      \bigg\{
        \bdelta_{0}^{\ell}\theta(\vsig)\sum_{m=0}^{\ell-1}\mfn_s^{-}
       +\sum_{m=\ell}^{\infty}\bmfn_s^{-}
      \bigg\}
     +\Th_{+}^{n\ell}
      \bigg\{
        \bdelta_{01}^{\ell}\theta(\vsig)\sum_{m=1}^{\ell-1}\mfn_s^{+}
       +\sum_{m=\ell'}^{\infty}\bmfn_s^{+}
      \bigg\}
    \Bigg] ,
\end{align}
where
\begin{align}
  \mfn_x^{+}(\vsig,s')
  &=-\bigg[
       \sin\!\bigg(\frac{\hnu_m^n\vsig}{\rho}\bigg)
       \tJ_s(s')\frac{\hnu_m^n}{\rho}\frac{\rd_{y'}}{k_r^n}
      +\frac{\hr'}{\rho}\cos\!\bigg(\frac{\hnu_m^n\vsig}{\rho}\bigg)
       \Big\{\tJ_x(s')\rd_{\hr'}\frac{\rd_{y'}}{k_r^n}+\tJ_y(s')\Big\}
    \bigg]
    \frac{\rho}{\hr}\cR_{+}^{mn}\cY_{-}^n ,
  \label{eq:mfnxp}
  \\
  \bmfn_x^{+}(\vsig,s')
  &=-\frac{e^{-\cnu_m^n|\vsig|/\rho}}{2}
    \bigg[
       \tJ_s(s')\frac{\cnu_m^n}{\rho}\frac{\rd_{y'}}{k_r^n}
      +\sgn(\vsig)\frac{\hr'}{\rho}
       \Big\{\tJ_x(s')\rd_{\hr'}\frac{\rd_{y'}}{k_r^n}+\tJ_y(s')\Big\}
    \bigg]
    \frac{\rho}{\hr}\cR_{+}^{mn}\cY_{-}^n ,
  \\
  \mfn_x^{-}(\vsig,s')
  &=-\bigg[
       \sin\!\bigg(\frac{\hmu_m^n\vsig}{\rho}\bigg)\tJ_s(s')\frac{\hr'}{\hmu_m^n}\rd_{\hr'}
      +\cos\!\bigg(\frac{\hmu_m^n\vsig}{\rho}\bigg)\tJ_x(s')
    \bigg]
    \rd_{\hr}\cR_{-}^{mn}\frac{\rd_{y'}}{k_r^n}\cY_{-}^n ,
  \label{eq:mfnxm}
  \\
  \bmfn_x^{-}(\vsig,s')
  &=-\frac{e^{-\cmu_m^n|\vsig|/\rho}}{2}
    \bigg[\tJ_x(s')\sgn(\vsig)-\tJ_s(s')\frac{\hr'}{\cmu_m^n}\rd_{\hr'}\bigg]
    \rd_{\hr}\cR_{-}^{mn}\frac{\rd_{y'}}{k_r^n}\cY_{-}^n ,
  \\
  \nonumber
  \\
  \mfn_s^{-}(\vsig,s')
  &=-\bigg[
       \cos\!\bigg(\frac{\hmu_m^n\vsig}{\rho}\bigg)\tJ_s(s')\frac{\hr'}{\rho}\rd_{\hr'}
      -\sin\!\bigg(\frac{\hmu_m^n\vsig}{\rho}\bigg)\tJ_x(s')\frac{\hmu_m^n}{\rho}
    \bigg]
    \frac{\rho}{\hr}\cR_{-}^{mn}\frac{\rd_{y'}}{k_r^n}\cY_{-}^n ,
  \label{eq:mfnsm}
  \\
  \bmfn_s^{-}(\vsig,s')
  &=-\frac{e^{-\cmu_m^n|\vsig|/\rho}}{2}
    \bigg[\tJ_s(s')\sgn(\vsig)\frac{\hr'}{\rho}\rd_{\hr'}-\tJ_x(s')\frac{\cmu_m^n}{\rho}
    \bigg]
    \frac{\rho}{\hr}\cR_{-}^{mn}\frac{\rd_{y'}}{k_r^n}\cY_{-}^n ,
  \\
  \mfn_s^{+}(\vsig,s')
  &=-\bigg[
       \cos\!\bigg(\frac{\hnu_m^n\vsig}{\rho}\bigg)\tJ_s(s')\frac{\rd_{y'}}{k_r^n}
      -\frac{\hr'}{\hnu_m^n}\sin\!\bigg(\frac{\hnu_m^n\vsig}{\rho}\bigg)
       \Big\{\tJ_x(s')\rd_{\hr'}\frac{\rd_{y'}}{k_r^n}+\tJ_y(s')\Big\}
    \bigg]
    \rd_{\hr}\cR_{+}^{mn}\cY_{-}^n ,
  \label{eq:mfnsp}
  \\
  \bmfn_s^{+}(\vsig,s')
  &=-\frac{e^{-\cnu_m^n|\vsig|/\rho}}{2}
    \bigg[
       \tJ_s(s')\sgn(\vsig)\frac{\rd_{y'}}{k_r^n}
      +\frac{\hr'}{\cnu_m^n}\Big\{\tJ_x(s')\rd_{\hr'}\frac{\rd_{y'}}{k_r^n}+\tJ_y(s')\Big\}
    \bigg]
    \rd_{\hr}\cR_{+}^{mn}\cY_{-}^n .
\end{align}

\clearpage

\subsection{Differential expression of the complete form}

The differential expression of the electromagnetic field in the complete form is given by
Eqs.(\ref{eq:tEBv_cmpl_dg}),
\begin{align}
  \bigg\{{\tEv(\xv) \atop c\tBv(\xv)}\bigg\}
  &=\int_{r_a}^{r_b}dr'\int_{-h/2}^{h/2}dy'
    \bigg[
       \bigg\{{\hPv^{\dg}(\xv,\xv_\perp') \atop
               \hQv^{\dg}(\xv,\xv_\perp')}
       \bigg\}
      -Z_0\int_{0}^{\infty}ds'
       \bigg\{{\hMv^{\dg}(\xv_{\perp},s-s',\xv') \atop \hNv^{\dg}(\xv_{\perp},s-s',\xv')}
       \bigg\}
    \bigg] .
  \label{eq:tEBv_cmpl_dg_app}
\end{align}
$(\hPv^{\dg},\hQv^{\dg})$ and $(\hMv^{\dg},\hNv^{\dg})$ are given by
Eqs.(\ref{eq:hPv_hMv_dg}-\ref{eq:hQv_hNv_dg}), which involve respectively
the initial fields at $s=0$ and the source terms in $s>0$.
Substituting Eqs.(\ref{eq:cCp}-\ref{eq:cCm}) into the first equations of
(\ref{eq:hPy_dg_cCp}-\ref{eq:hQy_dg_cCm}), we rearrange them,
\begin{align}
  \hP_y^{\dg}(s)
  &=\sum_{n=0}^{\infty}\sum_{m=1}^{\infty}\cR_{+}^{mn}
    \bigg[
       \tE_y(0)\frac{\rho}{r'}\cos\!\bigg(\frac{\nu_m^ns}{\rho}\bigg)
      -\frac{\rho}{\nu_m^n}\sin\!\bigg(\frac{\nu_m^ns}{\rho}\bigg)
       \big\{\tE_s(0)\rd_{y'}+c\tB_x(0)ik\beta\big\}
    \bigg]
    \cY_{-}^n ,
  \label{eq:hPy_dg}
  \\
  \hQ_y^{\dg}(s)
  &=\sum_{n=1}^{\infty}\sum_{m=0}^{\infty}\cR_{-}^{mn}
    \bigg[
       c\tB_y(0)\frac{\rho}{r'}\cos\!\bigg(\frac{\mu_m^ns}{\rho}\bigg)
      -\frac{\rho}{\mu_m^n}\sin\!\bigg(\frac{\mu_m^ns}{\rho}\bigg)
       \big\{c\tB_s(0)\rd_{y'}-\tE_x(0)ik\beta\big\}
    \bigg]
    \cY_{+}^n .
  \label{eq:hQy_dg}
\end{align}
Both $\nu_m^n$ and $\mu_m^n$ represent the real and imaginary poles, depending on $m$ as 
shown in Eqs.(\ref{eq:ps_poles}).
Substituting Eqs.(\ref{eq:cCp}-\ref{eq:cCm}) and (\ref{eq:cUp}-\ref{eq:cVm}) into
the second equations of (\ref{eq:hPy_dg_cCp}-\ref{eq:hQy_dg_cCm}), we rearrange them,
\begin{alignat}{2}
  \hP_{x,s}^{\dg}(s)
  &=\sum_{n=1}^{\infty}\sum_{m=0}^{\infty}
    \big\{(1-\delta_0^m)\hmfp_{x,s}^{+}(s)+\hmfp_{x,s}^{-}(s)\big\} ,
   \qquad
  &\hQ_{x,s}^{\dg}(s)
  &=\sum_{n=0}^{\infty}\sum_{m=0}^{\infty}
    \big\{(1-\delta_0^m)\hmfq_{x,s}^{+}(s)+\hmfq_{x,s}^{-}(s)\big\} .
  \label{eq:hPxs_dg}
\end{alignat}
The double sign of $\mathfrak{p}_{x,s}^{\pm}$ and $\mathfrak{q}_{x,s}^{\pm}$ 
corresponds to that of $\cR_{\pm}^{mn}$,
\begin{align}
  \hmfp_x^{+}(s)
  &=\bigg[
       \frac{\rho}{\nu_m^n}\sin\!\bigg(\frac{\nu_m^ns}{\rho}\bigg)
       \big\{\tE_s(0)k_r^n+c\tB_y(0)ik\beta\rd_{\hr'}\big\}
      -\frac{\rho}{\hr'}\cos\!\bigg(\frac{\nu_m^ns}{\rho}\bigg)\tE_y(0)\rd_{y'}
    \bigg]
    \rd_{\hr}\cR_{+}^{mn}\cY_{+}^n ,
  \\
  \hmfp_x^{-}(s)
  &=\bigg[
       \cos\!\bigg(\frac{\mu_m^ns}{\rho}\bigg)
       \big\{\tE_x(0)k_r^n-\tE_y(0)\rd_{\hr'}\rd_{y'}\big\}
      +\frac{\mu_m^n}{\hr'}\sin\!\bigg(\frac{\mu_m^ns}{\rho}\bigg)c\tB_y(0)ik\beta
    \bigg]\frac{\rho}{\hr}\cR_{-}^{mn}\cY_{+}^n ,
  \\
  \hmfp_s^{+}(s)
   &=\bigg[
       \cos\!\bigg(\frac{\nu_m^ns}{\rho}\bigg)
       \big\{\tE_s(0)k_r^n+c\tB_y(0)ik\beta\rd_{\hr'}\big\}
      +\frac{\nu_m^n}{\hr'}\sin\!\bigg(\frac{\nu_m^ns}{\rho}\bigg)\tE_y(0)\rd_{y'}
    \bigg]\frac{\rho}{\hr}\cR_{+}^{mn}\cY_{+}^n ,
  \\
  \hmfp_s^{-}(s)
   &=\bigg[
        \frac{\rho}{\mu_m^n}\sin\!\bigg(\frac{\mu_m^ns}{\rho}\bigg)
        \big\{\tE_y(0)\rd_{\hr'}\rd_{y'}-\tE_x(0)k_r^n\big\}
       +\frac{\rho}{\hr'}\cos\!\bigg(\frac{\mu_m^ns}{\rho}\bigg)c\tB_y(0)ik\beta
    \bigg]\rd_{\hr}\cR_{-}^{mn}\cY_{+}^n ,
   \\
   \nonumber
   \\
  \hmfq_x^{+}(s)
  &=\bigg[
       \cos\!\bigg(\frac{\nu_m^ns}{\rho}\bigg)
       \big\{c\tB_x(0)k_r^n-c\tB_y(0)\rd_{\hr'}\rd_{y'}\big\}
      -\frac{\nu_m^n}{\hr'}\sin\!\bigg(\frac{\nu_m^ns}{\rho}\bigg)\tE_y(0)ik\beta
    \bigg]\frac{\rho}{\hr}\cR_{+}^{mn}\cY_{-}^n ,
  \label{eq:hmfq_x_p}
  \\
  \hmfq_x^{-}(s)
  &=\bigg[
       \frac{\rho}{\mu_m^n}\sin\!\bigg(\frac{\mu_m^ns}{\rho}\bigg)
       \big\{c\tB_s(0)k_r^n-\tE_y(0)ik\beta\rd_{\hr'}\big\}
      -\frac{\rho}{\hr'}\cos\!\bigg(\frac{\mu_m^ns}{\rho}\bigg)c\tB_y(0)\rd_{y'}
    \bigg]\rd_{\hr}\cR_{-}^{mn}\cY_{-}^n ,
  \\
  \hmfq_s^{+}(s)
  &=\bigg[
       \frac{\rho}{\nu_m^n}\sin\!\bigg(\frac{\nu_m^ns}{\rho}\bigg)
       \big\{c\tB_y(0)\rd_{\hr'}\rd_{y'}-c\tB_x(0)k_r^n\big\}
      -\frac{\rho}{\hr'}\cos\!\bigg(\frac{\nu_m^ns}{\rho}\bigg)\tE_y(0)ik\beta
    \bigg]\rd_{\hr}\cR_{+}^{mn}\cY_{-}^n ,
  \\
  \hmfq_s^{-}(s)
  &=\bigg[
       \cos\!\bigg(\frac{\mu_m^ns}{\rho}\bigg)
       \big\{c\tB_s(0)k_r^n-\tE_y(0)ik\beta\rd_{\hr'}\big\}
      +\frac{\mu_m^n}{\hr'}\sin\!\bigg(\frac{\mu_m^ns}{\rho}\bigg)c\tB_y(0)\rd_{y'}
    \bigg]\frac{\rho}{\hr}\cR_{-}^{mn}\cY_{-}^n .
  \label{eq:hmfq_s_m}
\end{align}
$\tEv(0)$ and $\tBv(0)$ are the initial values of the fields at the entrance of the bend,
given by Eqs.(\ref{abbrev}).

Substituting Eqs.(\ref{eq:cCp}-\ref{eq:cCm}) into the first equations of
(\ref{eq:hMy_cCp}-\ref{eq:hNy_cCm}), we rearrange them,
\begin{align}
  \hM_y^{\dg}(\vsig,s')
  &=\theta(\vsig)\big\{\tJ_0(s')\rd_{y'}+ik\beta\tJ_y(s')\big\}
    \sum_{n=0}^{\infty}\cY_{-}^n\sum_{m=1}^{\infty}
    \frac{r'}{\nu_m^n}\sin\!\bigg(\frac{\nu_m^n\vsig}{\rho}\bigg)\cR_{+}^{mn} ,
  \label{eq:hMy_dg}
  \\
  \hN_y^{\dg}(\vsig,s')
  &=\theta(\vsig)\sum_{n=1}^{\infty}\cY_{+}^n\sum_{m=0}^{\infty}
    \bigg[
       \sin\!\bigg(\frac{\mu_m^n\vsig}{\rho}\bigg)\tJ_s(s')\frac{\hr'}{\mu_m^n}\rd_{\hr'}
      +\cos\!\bigg(\frac{\mu_m^n\vsig}{\rho}\bigg)\tJ_x(s')
    \bigg]\cR_{-}^{mn} .
  \label{eq:hNy_dg}
\end{align}
Substituting Eqs.(\ref{eq:cCp}-\ref{eq:cCm}) and (\ref{eq:cUp}-\ref{eq:cVm})
into the second equations of (\ref{eq:hMy_cCp}-\ref{eq:hNy_cCm}), we rearrange them,
\begin{align}
  \hM_{x,s}^{\dg}(\vsig,s')
  &=\theta(\vsig)\sum_{n=1}^{\infty}\cY_{+}^n\sum_{m=0}^{\infty}
    \big\{(1-\delta_0^m)\hmfm_{x,s}^{+}(\vsig,s')+\hmfm_{x,s}^{-}(\vsig,s')\big\} ,
  \label{eq:hMxs_dg}
   \\
  \hN_{x,s}^{\dg}(\vsig,s')
  &=\theta(\vsig)\sum_{n=0}^{\infty}\sum_{m=0}^{\infty}
    \big\{(1-\delta_0^m)\hmfn_{x,s}^{+}(\vsig,s')+\hmfn_{x,s}^{-}(\vsig,s')\big\} .
  \label{eq:hNxs_dg}
\end{align}
$\mathfrak{m}_{x,s}^{\pm}$ and $\hmfn_{x,s}^{\pm}$ are given as
\begin{align}
  \hmfm_x^{+}(\vsig,s')
  &=-\bigg[
        \cos\!\bigg(\frac{\nu_m^n\vsig}{\rho}\bigg)\tJ_s(s')\frac{ik\beta}{k_r^n}
       +\frac{\hr'}{\nu_m^n}\sin\!\bigg(\frac{\nu_m^n\vsig}{\rho}\bigg)
        \Big\{\tJ_0(s')-\tJ_x(s')\frac{ik\beta}{k_r^n}\rd_{\hr'}\Big\}
   \bigg]
   \rd_{\hr}\cR_{+}^{mn} ,
   \\
  \hmfm_x^{-}(\vsig,s')
  &=-\frac{ik\beta}{k_r^n}\bigg[
       \cos\!\bigg(\frac{\mu_m^n\vsig}{\rho}\bigg)\tJ_s(s')\frac{\hr'}{\rho}\rd_{\hr'}
      -\frac{\mu_m^n}{\rho}\sin\!\bigg(\frac{\mu_m^n\vsig}{\rho}\bigg)\tJ_x(s')
    \bigg]\frac{\rho}{\hr}\cR_{-}^{mn} ,
   \\
  \hmfm_s^{+}(\vsig,s')
  &=\bigg[
      \frac{\nu_m^n}{\rho}\sin\!\bigg(\frac{\nu_m^n\vsig}{\rho}\bigg)
      \tJ_s(s')\frac{ik\beta}{k_r^n}
     +\frac{\hr'}{\rho}\cos\!\bigg(\frac{\nu_m^n\vsig}{\rho}\bigg)
      \Big\{\tJ_x(s')\frac{ik\beta}{k_r^n}\rd_{\hr'}-\tJ_0(s')\Big\}
    \bigg]\frac{\rho}{\hr}\cR_{+}^{mn} ,
   \\
  \hmfm_s^{-}(\vsig,s')
  &=\frac{ik\beta}{k_r^n}\bigg[
      \frac{\hr'}{\mu_m^n}\sin\!\bigg(\frac{\mu_m^n\vsig}{\rho}\bigg)\tJ_s(s')\rd_{\hr'}
     +\cos\!\bigg(\frac{\mu_m^n\vsig}{\rho}\bigg)\tJ_x(s')
    \bigg]\rd_{\hr}\cR_{-}^{mn} ,
\end{align}
and
\begin{align}
  \hmfn_x^{+}(\vsig,s')
  &=-\bigg[
       \frac{\nu_m^n}{\rho}\sin\!\bigg(\frac{\nu_m^n\vsig}{\rho}\bigg)
       \tJ_s(s')\frac{\rd_{y'}}{k_r^n}
      +\frac{\hr'}{\rho}\cos\!\bigg(\frac{\nu_m^n\vsig}{\rho}\bigg)
       \Big\{\tJ_x(s')\rd_{\hr'}\frac{\rd_{y'}}{k_r^n}+\tJ_y(s')\Big\}
    \bigg]\frac{\rho}{\hr}\cR_{+}^{mn}\cY_{-}^n ,
  \label{eq:hmfn_x_p}
  \\
  \hmfn_x^{-}(\vsig,s')
  &=-\bigg[
        \sin\!\bigg(\frac{\mu_m^n\vsig}{\rho}\bigg)\tJ_s(s')\frac{\hr'}{\mu_m^n}\rd_{\hr'}
       +\cos\!\bigg(\frac{\mu_m^n\vsig}{\rho}\bigg)\tJ_x(s')
    \bigg]\rd_{\hr}\cR_{-}^{mn}\frac{\rd_{y'}}{k_r^n}\cY_{-}^n ,
  \\
  \hmfn_s^{+}(\vsig,s')
  &=-\bigg[
       \cos\!\bigg(\frac{\nu_m^n\vsig}{\rho}\bigg)\tJ_s(s')\frac{\rd_{y'}}{k_r^n}
      -\frac{\hr'}{\nu_m^n}\sin\!\bigg(\frac{\nu_m^n\vsig}{\rho}\bigg)
       \Big\{\tJ_x(s')\rd_{\hr'}\frac{\rd_{y'}}{k_r^n}+\tJ_y(s')\Big\}
   \bigg]\rd_{\hr}\cR_{+}^{mn}\cY_{-}^n ,
  \\
  \hmfn_s^{-}(\vsig,s')
  &=-\bigg[
       \frac{\hr'}{\rho}\cos\!\bigg(\frac{\mu_m^n\vsig}{\rho}\bigg)\tJ_s(s')\rd_{\hr'}
      -\frac{\mu_m^n}{\rho}\sin\!\bigg(\frac{\mu_m^n\vsig}{\rho}\bigg)\tJ_x(s')
    \bigg]\frac{\rho}{\hr}\cR_{-}^{mn}\frac{\rd_{y'}}{k_r^n}\cY_{-}^n .
  \label{eq:hmfn_s_m}
\end{align}
The current components $\tJ=(\tJ_0,\tJv)$ have the arguments $\xv'=(r',y',s')$ and $k$ 
as shown in Eqs.(\ref{RYJ_abbrev}).

\clearpage

\section{Steady fields of coherent synchrotron radiation in a bend}
\label{sec:steady_field}

The transient fields of CSR in the Laplace domain $(\mfE^n,\mfB^n)$ are given by
Eqs.(\ref{eq:mfEBy_solution}) and (\ref{eq:mfBsEx}-\ref{eq:mfBxEs}).
They have the source terms in the Laplace domain $(\mfS^n,\mfT^n)$
which are arbitrary under the assumptions listed in section \ref{sec:assumption}.
Assuming a rigid bunch in this appendix, we find the expressions of
the longitudinal components of the steady fields of CSR.
Also, we rederive Eq.(3.9) in \cite{warnock_morton} in a different way,
which is the longitudinal impedance of steady CSR in a toroidal chamber.
In appendix \ref{sec:impedance} we need the expressions of the steady field and
impedance in rewriting the expression of the longitudinal electric field in
a transient state and the longitudinal impedance
such that are usable in the numerical calculation.

In terms of collective effect of a bunch of particles,
one of the most important effects by the field of CSR is
an inhomogeneous variation of the energy distribution of the bunch.
In estimating the effect of CSR to the emitting bunch itself,
it is important to calculate the longitudinal electric field $E_s$
regardless whether assuming a rigid bunch.
We can assume a rigid bunch moving in a bending section
if the bunch has no horizontal extent and no energy spread.
In a bend of radius $\rho$, a steady field of CSR with a wavenumber $k$ has
a transverse spread $\ell_{\perp}$ around the bunch as described in Eq.(\ref{eq:shield}).
Therefore, in calculating $E_s$,
if the transverse beam size $\sig_{\perp}$ is much smaller than $\ell_{\perp}$,
\ie, if $2\sig_{\perp}^3\ll \rho\sig_z^2$ as shown in Eq.(11) of \cite{derbenev},
then we can assume an infinitely thin bunch
since $E_s$ hardly depends on $\sig_{\perp}$ under this condition.
This condition is also applicable to the vertical Lorentz force $F_y$
which the particles receive from their own CSR field in a horizontal bend.
If $\sig_{\perp}=0$, however, the horizontal Lorentz force $F_x$ in the bend may diverge at 
the transverse beam position.

\subsection{Vertical fields in a steady state or transient state}

Assuming a rigid bunch which moves along the $s$-axis in a bend of radius $\rho$,
the current $\mfJ^n$ and the source terms $\mfS^n$ and $\mfT^n$ in the Laplace domain are 
given using their Fourier coefficients $(\cJ^n;\cS^n,\cT^n)$ as follows,
\begin{align}
  \mfF^n(r,\nu)
  =\frac{\cF^n(r,s)e^{-iks}}{i(\nu-k\rho)} ,
    \qquad
  \mfF^n
  =(\mfJ^n;\mfS^n,\mfT^n) ,
    \qquad
  \cF^n
  =(\cJ^n;\cS^n,\cT^n) .
  \label{eq:J_thin2}
\end{align}
$\cJ^n$ represents $(\cJ_0^n,\cJ_{x,y,s}^n)$.
$\cS^n$ and $\cT^n$ denote $\cS_{x,y,s}^n$ and $\cT_{x,y,s}^n$ given by
Eqs.(\ref{eq:cSx_cTx}-\ref{eq:cSs_cTs}).
Assuming a rigid bunch, $\cS^n(s)e^{-iks}$ and $\cT^n(s)e^{-iks}$ do not depend on $s$
since we define the Fourier transform with respect to $t$ (not $z$) in
Eqs.(\ref{eq:Fourier_trans}).
That is, $e^{iks}$ is not factored out of the time domain quantities in
Eqs.(\ref{eq:Fourier_trans}) which differ from Eqs.(\ref{eq:FT_zk_9}) where
the Fourier transform is defined with respect to $z$ defined in Eq.(\ref{eq:z}).
Similarly, $\cE^ne^{-iks}$ and $\cB^ne^{-iks}$ do not depend on $s$
if the fields are in a steady state which means the asymptotic limit of $s\to\infty$ in
the semi-infinite bend.
Using the final value theorem of the Laplace transform for $\mfE_y^n$ and $\mfB_y^n$,
we get the Fourier coefficients of the vertical components of the steady fields in the bend,
\begin{alignat}{2}
  \lim_{s\to\infty}\cE_y^n(r,s)e^{-iks}
  &=\lim_{\nu\to k\rho}i(\nu-k\rho)\mfE_y^n(r,\nu)
  &&=Z_0\int_{r_a}^{r_b}dr'\frac{r'}{\rho}\mfG_{+}^n(r,r',k\rho)
     \cS_y^n(r',s')e^{-iks'} ,
  \label{eq:cEy_steady}
  \\
  \lim_{s\to\infty}c\cB_y^n(r,s)e^{-iks}
  &=\lim_{\nu\to k\rho}i(\nu-k\rho)c\mfB_y^n(r,\nu)
  &&=Z_0\int_{r_a}^{r_b}dr'\frac{r'}{\rho}\mfG_{-}^n(r,r',k\rho)
     \cT_y^n(r',s')e^{-iks'} .
  \label{eq:cBy_steady}
\end{alignat}
$\mfG_{\pm}^n$ is given by Eqs.(\ref{eq:mfGe}-\ref{eq:mfGb}) which are the Green functions 
of the vertical fields in the Laplace domain.

Assuming a rigid bunch in Eq.(\ref{eq:cEBy_sol}), the Fourier coefficients of
the vertical components of the transient fields in the separated form are given as
\begin{align}
  \cE_y^n(r,s)
  &=\int_{r_a}^{r_b}dr'
    \Big[
      \cD_y^n(r')\cG_{+}^n(r,r',s)+Z_0\frac{r'}{\rho}\cS_y^n(r',s')e^{-iks'}\cI_{+}^n(r,r',s)
    \Big] ,
  \\
  c\cB_y^n(r,s)
  &=\int_{r_a}^{r_b}dr'
    \Big[
      \cA_y^n(r')\cG_{-}^n(r,r',s)+Z_0\frac{r'}{\rho}\cT_y^n(r',s')e^{-iks'}\cI_{-}^n(r,r',s)
    \Big] ,
\end{align}
where $\cS_y^n(s')e^{-iks'}$ and $\cT_y^n(s')e^{-iks'}$ do not depend on $s'$
since we assume a rigid bunch.
$\cI_{\pm}^n$ denotes
\begin{align}
  \cI_{\pm}^n(r,r',s)
  &=\int_0^{\infty}ds'\cG_{\pm}^n(r,r',s-s')e^{iks'} .
   \label{eq:cIpm}
\end{align}
Substituting Eqs.(\ref{eq:cGp_sep}-\ref{eq:cGm_sep}) into Eq.(\ref{eq:cIpm}),
we calculate the $s'$-integral,
\begin{align}
  \cI_{+}^n(r,r',s)
  &=\sum_{\ell=0}^{\infty}\Th_{+}^{n\ell}
   \bigg[
    \bdelta_{01}^{\ell}\sum_{m=1}^{\ell-1}
    \cR_{+}^{mn}(r,r')\frac{\rho}{\hnu_m^n}L(s,\hnu_m^n)
   -\sum_{m=\ell'}^{\infty}
    \cR_{+}^{mn}(r,r')\frac{\rho}{\cnu_m^n}\bL(s,\cnu_m^n)
   \bigg] ,
   \\
  \cI_{-}^n(r,r',s)
  &=\sum_{\ell=0}^{\infty}\Th_{-}^{n\ell}
   \bigg[
     \bdelta_{0}^{\ell}\sum_{m=0}^{\ell-1}\cR_{-}^{mn}(r,r')
     \frac{\rho}{\hmu_m^n}L(s,\hmu_m^n)
     -\sum_{m=\ell}^{\infty}\cR_{-}^{mn}(r,r')\frac{\rho}{\cmu_m^n}\bL(s,\cmu_m^n)
   \bigg] ,
\end{align}
where $\ell'$ $(=\ell+\delta_0^\ell)$ is given by Eq.(\ref{eq:lprm}).
$\cR_{\pm}^{mn}$ denotes the radial eigenfunctions of the curved pipe,
given by Eqs.(\ref{eq:Ven}-\ref{eq:mfRm}).
$L$ and $\bL$ are the following functions of $s$,
\begin{align}
  L(s,\hnu)
  &=\int_0^{s}ds'e^{iks'}\sin[\hnu(s-s')/\rho]
  =\frac{1}{2}
    \bigg(
           \frac{e^{iks}-e^{-i\hnu s/\rho}}{\hnu/\rho+k}
          +\frac{e^{iks}-e^{i\hnu s/\rho}}{\hnu/\rho-k}
    \bigg) ,
   \\
  \bL(s,\cnu)
  &=\int_0^{\infty}ds'e^{iks'}e^{-\cnu|s-s'|/\rho}
  =\frac{1}{k^2+(\cnu/\rho)^2}
   \bigg\{(\cnu/\rho)e^{iks}+\frac{ik-\cnu/\rho}{2}e^{-\cnu s/\rho}\bigg\} .
\end{align}

Next, we consider a steady field of CSR created by a rigid bunch in a semi-infinite bend.
Since the steady field in the frequency domain is proportional to $e^{iks}$ as seen from
Eqs.(\ref{eq:Fourier_trans}), $\cE_y^n(s)e^{-iks}$ and $\cB_y^n(s)e^{-iks}$ do not depend 
on $s$ in the limit of $s\to\infty$,
\begin{align}
  &
  \lim_{s\to\infty}
  \bigg\{{\cE_y^n(r,s) \atop c\cB_y^n(r,s)}\bigg\}e^{-iks}
  =\int_{r_a}^{r_b}dr' \bigg\{{O_{+}^n(r,s) \atop O_{-}^n(r,s)}\bigg\} ,
\end{align}
where
\begin{align}
  O_{+}^n(r,s)
  &=
     \cD_y^n(r')\lim_{s\to\infty}\cG_{+}^n(r,r',s)e^{-iks}
    +Z_0\frac{r'}{\rho}\cS_y^n(r',s')e^{-iks'}\lim_{s\to\infty}\cI_{+}^n(r,r',s)e^{-iks} ,
  \label{eq:Op_sinf}
  \\
  O_{-}^n(r,s)
  &=
    \cA_y^n(r')\lim_{s\to\infty}\cG_{-}^n(r,r',s)e^{-iks}
    +Z_0\frac{r'}{\rho}\cT_y^n(r',s')e^{-iks'}\lim_{s\to\infty}\cI_{-}^n(r,r',s)e^{-iks} .
  \label{eq:Om_sinf}
\end{align}
Here we put an infinitesimal imaginary part into the wavenumber $k$
so that $e^{-iks}$ damps for $s\to\infty$,
\begin{align}
  k=\lim_{\eps\to+0}(k-i\eps)
  \qquad\Ra\qquad
  e^{-iks}\to e^{-i(k-i\eps)s}\to0
   \quad\text{for}~~s\to\infty
   \qquad(\eps=+0) .
  \label{eq:exp_miks}
\end{align}
The sign of the infinitesimal offset ($-i\eps$) from the real $k$-axis is opposite to
$k=k+i\eps~(\eps=+0)$ which is used in Eq.(136) of \cite{agoh}.
This is because, in the present paper, we are considering the field in the frequency domain 
with respect to $t$ and $\omg$ where $\omg=kv$ as defined in
Eqs.(\ref{eq:Fourier_trans}-\ref{eq:omg}),
\begin{align}
  F(t)\propto \tF(\omg)e^{-i\omg t}
  ,\qquad
  e^{-i\omg t}
  \to e^{-i\omg t}e^{-\eps't}\to0
   \quad\text{for}~~t\to\infty
  \qquad(\eps'=\eps v=+0) .
\end{align}
On the other hand, we employed the Fourier transform with respect to $\{z,k\}$ in
\cite{agoh}.
That is, in the present paper, we define $k$ with the opposite sign to that in \cite{agoh}.
Therefore, according to Eq.(\ref{eq:omg}), we must consider the infinitesimal factor
so that $e^{-i\omg t}$ damps for $t\to\infty$.
The limit of $s\to\infty$ is equivalent to $t\to\infty$ in considering the field around
the reference particle defined in Eq.(\ref{eq:z}).
According to the treatment (\ref{eq:exp_miks}), we get
\begin{align}
  \lim_{s\to\infty}\cG_{\pm}^n(r,r',s)e^{-iks}=0 .
   \label{eq:lim_cGpm_exp}
\end{align}
Eq.(\ref{eq:lim_cGpm_exp}) is obvious since the first terms of
Eqs.(\ref{eq:Op_sinf}-\ref{eq:Om_sinf}) involve the initial fields at $s=0$,
which are unrelated to the steady fields in the limit of $s\to\infty$.
$\cI_{\pm}^n(s)e^{-iks}$ converges in the limit of $s\to\infty$ as follows,
\begin{align}
  \lim_{s\to\infty}\cI_{+}^n(r,r',s)e^{-iks}
  =\sum_{m=1}^{\infty}\frac{\cR_{+}^{mn}(r,r')}{(\nu_m^n/\rho)^2-k^2} ,
   \qquad
  \lim_{s\to\infty}\cI_{-}^n(r,r',s)e^{-iks}
  =\sum_{m=0}^{\infty}\frac{\cR_{-}^{mn}(r,r')}{(\mu_m^n/\rho)^2-k^2} .
   \label{eq:lim_cIpm_e}
\end{align}
We used Eqs.(\ref{eq:sum_Xmn}) in deriving Eqs.(\ref{eq:lim_cIpm_e}).
$\nu_m^n$ and $\mu_m^n$ are respectively the zeros of $p_\nu(\hr_b,\hr_a)$ and
$s_\nu(\hr_b,\hr_a)$ with respect to $\nu$ as shown in
Eqs.(\ref{eq:pnu_poles}-\ref{eq:snu_poles}).
From Eqs.(\ref{eq:lim_cIpm_e}), we get the Fourier coefficients of
the vertical components of the steady fields created by the rigid bunch,
\begin{align}
  \lim_{s\to\infty}\cE_y^n(r,s)e^{-iks}
  &=Z_0\int_{r_a}^{r_b}dr'\cS_y^n(r',s')e^{-iks'}\frac{r'}{\rho}
    \sum_{m=1}^{\infty}\frac{ \cR_{+}^{mn}(r,r')}{(\nu_m^n/\rho)^2-k^2} ,
  \label{eq:cEy_rgd_sinf}
  \\
  \lim_{s\to\infty}c\cB_y^n(r,s)e^{-iks}
  &=Z_0\int_{r_a}^{r_b}dr'\cT_y^n(r',s')e^{-iks'}\frac{r'}{\rho}
    \sum_{m=0}^{\infty}\frac{\cR_{-}^{mn}(r,r')}{(\mu_m^n/\rho)^2-k^2} .
  \label{eq:cBy_rgd_sinf}
\end{align}
From Eqs.(\ref{eq:cEy_steady}-\ref{eq:cBy_steady}) and
(\ref{eq:cEy_rgd_sinf}-\ref{eq:cBy_rgd_sinf}),
$\mfG_{\pm}^n(\nu)$ for $\nu=k\rho$ can be given using $\cR_{\pm}^{mn}$ as follows,
\begin{align}
  \mfG_{+}^n(r,r',k\rho)
  &=\sum_{m=1}^{\infty}\frac{\cR_{+}^{mn}(r,r')}{(\nu_m^n/\rho)^2-k^2} ,
   \qquad
  \mfG_{-}^n(r,r',k\rho)
  =\sum_{m=0}^{\infty}\frac{\cR_{-}^{mn}(r,r')}{(\mu_m^n/\rho)^2-k^2} .
  \label{eq:mfGpm_mfRpm}
\end{align}
Eqs.(\ref{eq:mfGpm_mfRpm}) correspond to Eqs.(\ref{eq:mfGpm_cXpm}) in the straight pipe.

\noindent
According to Eqs.(\ref{eq:delta_Ve}), (\ref{eq:delta_Vb}) and (\ref{eq:mfGpm_mfRpm}),
the Green functions $\bmfG_{\pm}^n(\nu)$ for $\nu=k\rho$,
given by Eq.(\ref{eq:bmfG_pm_delta}), can be expressed using $\cR_{\pm}^{mn}$ as follows,
\begin{align}
  \bmfG_{+}^n(r,r',k\rho)
  &=\sum_{m=1}^{\infty}
    \bigg\{
      \frac{\rd_{\hr}\rd_{\hr'}}{(\nu_m^n/\rho)^2-k^2}
     +\frac{\rho^2}{\hr^2}
    \bigg\}
    \cR_{+}^{mn}(r,r') ,
   \\
  \bmfG_{-}^n(r,r',k\rho)
  &=\sum_{m=0}^{\infty}
    \bigg\{
      \frac{\rd_{\hr}\rd_{\hr'}}{(\mu_m^n/\rho)^2-k^2}
     +\frac{\rho^2}{\hr^2}
    \bigg\}
    \cR_{-}^{mn}(r,r') ,
\end{align}
where we can replace $\rho^2/\hr^2$ in the braces by $\rho^2/\hr'^2$ or $\rho^2/\hr\hr'$
since it is a factor multiplied with $\delta(\hr-\hr')$.

\subsection{Steady fields created by a thin bunch in a bending section}
\label{sec:steady_thin}

In considering a steady CSR in a bend,
we assume a thin bunch in the absence of radial extent.
It is a rigid bunch whose distribution has no correlation between the radial and 
longitudinal directions.
We assume that the rigid and thin bunch is moving along the $s$-axis at a constant speed
equal to the reference particle $v$ $(=c\beta)$ defined in Eq.(\ref{eq:z}).
Out of the six components of the electromagnetic field, the three components
$\tB_x$ and $\tE_{y,s}$ are continuous with respect to $r$ even for the thin bunch
having no radial extent.
On the other hand, the rest three $\tE_x$ and $\tB_{y,s}$ are discontinuous
at $r=\rho$ which is the radial position of the thin bunch.
This is one of the reasons why it is difficult to calculate the horizontal Lorentz force of 
CSR in the horizontal bend.

We define $\psi_y(y)$ and $\lambda(s,t)$ respectively as the vertical and longitudinal
charge distributions of the bunch, which are both normalized to unity as follows,
\begin{align}
  \int_{-h/2}^{h/2}\psi_y(y)dy=1
   ,\qquad
  \int_{-\infty}^{\infty}\lambda(s,t)dz=1
   ,\qquad
  \tlam(s,k)
  =\int_{-\infty}^{\infty}\lambda(s,t)e^{i\omg t}dt .
   \label{eq:lam}
\end{align}
$\tlam$ is the line density of the bunch in the frequency domain with respect to
$\{t,\omg\}$.
Even if the bunch is rigid, $\tlam$ depends on $s$ as $\tlam\propto e^{iks}$
since $\tlam$ is the Fourier transform of $\lam$ with respect to $t$ (not $z$).
As described in section \ref{sec:assumption}, we assume $\psi_y(\pm h/2)=0$, \ie,
there is no charge on the upper-lower walls of the perfectly conducting rectangular pipe.
Let $\tpsi_y^n$ be the $n$th Fourier coefficient of $\psi_y$.
From Eq.(\ref{eq:Fp_coeff}),
\begin{align}
  \tpsi_y^n
  &=\bigg\{{ \hat{\psi}_y^{2p-1} \atop \check{\psi}_y^{2p}}\bigg\}
  =\int_{-h/2}^{h/2}\frac{dy}{h/2}\psi_y(y)
   \bigg\{{ \cos(k_y^{2p-1}y) \atop \sin(k_y^{2p}y)}\bigg\} ,
    \qquad
  n=\bigg\{{ 2p-1 \atop 2p }\bigg\}
    \in\mathbb{N}
    \quad
   (p\in\mathbb{N}) .
   \label{eq:tpsi}
\end{align}
$k_y^n$ is the vertical wavenumber (\ref{eq:kyn}).
If the charge distribution of the bunch is symmetric with respect to the median horizontal 
plane, $\psi_y$ has no Fourier coefficients of even order ($n=2p$), \ie,
\begin{align}
  \text{if}\quad
  \psi_y(y)
  &=\psi_y(-y)
   \qquad\Ra\qquad
  \tpsi_y^{2p}
  =0 .
   \label{eq:tpsi_2p_eq0}
\end{align}
For example, if the bunch has the following vertical distributions which are symmetric with
respect to $y=0$, the Fourier coefficients of the even modes $\tpsi_y^{2p}$ are all zero
as shown in Eq.(\ref{eq:tpsi_2p_eq0}):
(A) no vertical extent ($\delta$-function of $y$) or
(B) a Gaussian distribution having a spread $\sig_y\,(\ll h)$
or (C) a rectangular distribution in $y=[-u_y,u_y]$ which has a height $2u_y$
like a vertical ribbon as used in Eq.(2.49) of \cite{warnock_morton}.
For these vertical distributions, $\tpsi_y^n$ of the odd modes ($n=2p-1$) are given as
follows,
\begin{alignat}{5}
  &\text{(A)}
   \quad&
  \psi_y(y)
  &=\delta(y)
   \qquad&\Ra&\qquad&
  \tpsi_y^{2p-1}
  &=\frac{2}{h} ,
   \label{eq:psiy_delta}
   \\
  &\text{(B)}
   \quad&
  \psi_y(y)
  &=\frac{e^{-(y/\sig_y)^2/2}}{\sig_y(2\pi)^{1/2}}
   \qquad&\Ra&\qquad&
  \tpsi_y^{2p-1}
  &\simeq \frac{2}{h}e^{-(k_y^{2p-1}\sig_y)^2/2} ,
   \label{eq:tpsi_gaussian}
   \\
  &\text{(C)}
   \quad&
  \psi_y(y)
  &=\frac{\theta(y+u_y)-\theta(y-u_y)}{2u_y}
   \qquad&\Ra&\qquad&
  \tpsi_y^{2p-1}
  &=\frac{2}{h}\cd\frac{\sin(k_y^{2p-1}u_y)}{k_y^{2p-1}u_y} .
\end{alignat}
$\tpsi_y^{2p-1}$ in Eq.(\ref{eq:tpsi_gaussian}) is an approximate expression since
we extended the limit of integration from $y=\pm h/2$ to $\pm\infty$ in Eq.(\ref{eq:tpsi}),
assuming $\sig_y\ll h/2$ and hence $\psi_y(\pm h/2)\simeq0$ approximately.
For an arbitrary $\psi_y$ which satisfies the condition $\psi_y(\pm h/2)=0$,
the Fourier coefficients of the current are given as
\begin{align}
  \cJ_0^n(r,s)=cq\tlam(s)\tpsi_y^n\delta(r-\rho)
   ,\qquad
  \vec{\cJ}^n(r,s)=\beta\cJ_0^n(r,s)\ev_s
   ,\qquad
  \tlam_0(k)
  =\tlam(k,0)
  =\tlam(k,s)e^{-iks} .
  \label{eq:J_thin}
\end{align}
$q$ is the total charge of the bunch.
$\tlam_0$ is the bunch spectrum which is a function of $k$ only.
$\tlam_0$ does not depend on $s$ under the assumption of a rigid bunch.
For example, if the bunch has a rigid Gaussian distribution $\lam$ with a spread
$\sig_z$ in the longitudinal direction, $\tlam_0$ is given as follows,
\begin{align}
  \lam(s,t)
  &=\frac{e^{-(z/\sig_z)^2/2}}{\sig_z(2\pi)^{1/2}}
   =\frac{v}{2\pi}\int_{-\infty}^{\infty}\tlam_0(k)e^{ikz}dk
   \qquad\Lra\qquad
  \tlam_0(k)
  =\frac{1}{v}e^{-(k\sig_z)^2/2} .
  \label{eq:tlam0}
\end{align}

Substituting Eqs.(\ref{eq:J_thin}) into Eqs.(\ref{eq:cEy_steady}-\ref{eq:cBy_steady}) 
through Eqs.(\ref{eq:cSy_cTy}), we get the Fourier coefficients of
the vertical components of the steady fields in the limit of $s\to\infty$,
\begin{align}
  \lim_{s\to\infty}\bigg\{ {\cE_y^n(r,s) \atop c\cB_y^n(r,s)}\bigg\}e^{-iks}
  =\frac{q\tlam_0}{\eps_0}\tpsi_y^n
   \bigg\{{
      (-1)^nk_y^n\mfG_{+}^n(r,r',k\rho) \atop -\beta\rd_{r'}\mfG_{-}^n(r,r',k\rho)
   }\bigg\}_{r'=\rho} .
  \label{eq:lim_cEycBy}
\end{align}
$\cB_y^n$ is discontinuous at $r=\rho$ which is the radial beam position.
Substituting Eq.(\ref{eq:lim_cEycBy}) into Eq.(\ref{eq:cBx}),
we get the Fourier coefficient of the longitudinal component of the steady electric field
in the limit of $s\to\infty$,
\begin{align}
  \lim_{s\to\infty}\cE_s^n(r,s)e^{-iks}
  &=-ik\frac{q\tlam_0}{\eps_0}\tpsi_y^n \mfG_s^n(r,\rho,k\rho) ,
  \label{eq:lim_cEs_thin}
\end{align}
where
\begin{align}
  \mfG_s^n(r,r',\nu)
  &=\beta^2\bmfG_{-}^n(r,r',\nu)
   +\bigg(\frac{k_y^n}{k_r^n}\bigg)^{\!2}\frac{\rho}{r}\mfG_{+}^n(r,r',\nu) ,
   \label{eq:mfGs}
   \\
  \mfG_s^n(r,r,\nu)
  &=\frac{\pi}{2}\rho
    \bigg\{
      \beta^2\frac{s_\nu(\hr_b,\hr)s_\nu(\hr,\hr_a)}{s_\nu(\hr_b,\hr_a)}
     +\bigg(\frac{k_y^n}{k_r^n}\bigg)^{\!2}
      \frac{p_\nu(\hr_b,\hr)p_\nu(\hr,\hr_a)}{p_\nu(\hr_b,\hr_a)}
    \bigg\} .
\end{align}
The Green functions $\mfG_{+}^n$ and $\bmfG_{-}^n$ are given by Eqs.(\ref{eq:mfGe})
and (\ref{eq:bmfG_mns}).
From Eq.(\ref{eq:lim_cEs_thin}), we get the frequency domain representation of
the longitudinal electric field in the steady state in the semi-infinite bend,
\begin{align}
  \lim_{s\to\infty}\tE_s(\xv)e^{-iks}
  &=-2ik\frac{q\tlam_0}{\eps_0h}\sum_{n=1}^{\infty}\hcY_{+}^n(y) \mfG_s^n(r,\rho,k\rho) .
   \label{eq:tEs_std}
\end{align}
$\hcY_{+}^n$ is the following dimensionless function of $y$,
\begin{align}
  \hcY_{+}^n(y)
  &=\frac{h}{2}\int_{-h/2}^{h/2}dy'\psi_y(y')\cY_{+}^n(y,y')
   =\frac{h}{2}\tpsi_y^n
    \bigg\{{ \cos(k_y^ny) \atop \sin(k_y^ny)}\bigg\} ,
   \qquad
  n=\bigg\{{ 2p-1 \atop 2p }\bigg\}
     \quad
   (p\in\mathbb{N}) .
  \label{eq:hcYp}
\end{align}
$\cY_{+}^n$ is the vertical eigenfunction given by Eq.(\ref{eq:cYp}).
For example, if the bunch has no vertical spread as in Eq.(\ref{eq:psiy_delta}),
$\hcY_{+}^n$ is given as follows,
\begin{align}
  \psi_y(y)=\delta(y)
   \qquad\Ra\qquad
  \hcY_{+}^{2p-1}(y)=\cos(k_y^{2p-1}y)
   ,\qquad
  \hcY_{+}^{2p}=0 .
\end{align}
%

\subsection{Longitudinal impedance and power spectrum of synchrotron radiation}

We consider Eq.(\ref{eq:tEs_std}) for $\xv_{\perp}=(r,y)=(\rho,0)$ which is
the transverse position of the rigid and thin bunch given by Eq.(\ref{eq:J_thin}).
We define $\cZ$ as the longitudinal impedance of the steady field per
unit length along the $s$-axis at the transverse beam position,
\begin{align}
  \cZ(k)
  \equiv\frac{-1}{qv\tlam_0}\lim_{s\to\infty}\tE_s(\rho,0,s)e^{-iks}
   \quad [\Omega/{\rm m}] .
  \label{eq:cZ_def}
\end{align}
From Eq.(\ref{eq:tEs_std}), $\cZ$ in the curved rectangular pipe is gotten as
\begin{align}
  \frac{\cZ(k)}{Z_0}
   &=\frac{\pi}{2}ik\rho\sum_{p=1}^{\infty}\tpsi_y^{2p-1}
     \bigg[
       \beta\frac{s_\nu(\hr_b,\hrho)s_\nu(\hrho,\hr_a)}{s_\nu(\hr_b,\hr_a)}
      +\frac{1}{\beta}\bigg(\frac{k_y^{2p-1}}{k_r^{2p-1}}\bigg)^{\!2}
       \frac{p_\nu(\hr_b,\hrho)p_\nu(\hrho,\hr_a)}{p_\nu(\hr_b,\hr_a)}
     \bigg]_{\nu=k\rho}
   \quad \big[{\rm m}^{-1}\big] .
  \label{eq:cZ_thin}
\end{align}
The Fourier coefficient $\tpsi_y^n$ is given by Eq.(\ref{eq:tpsi}).
$\hr_{b,a}$ and $\hrho$ are the dimensionless radii normalized by $k_r^{2p-1}$ as
defined in Eq.(\ref{eq:krn}).

Let us compare Eq.(\ref{eq:cZ_thin}) with Eq.(3.9) in \cite{warnock_morton},
which is the longitudinal impedance of synchrotron radiation per revolution in
a rectangular toroidal chamber.
In \cite{warnock_morton} the impedance is defined averaged over the vertical dimension
unlike Eq.(\ref{eq:cZ_def}).
In Eq.(3.9) of \cite{warnock_morton} ``$n$'' is an integer which denotes the harmonic 
number of the field for the circumference of the circular orbit of the beam, \ie,
``$n$''$=k\rho$, similar to the quantum number of Bohr's quantum condition in an atom.
Eq.(\ref{eq:cZ_thin}) is equivalent to Eq.(3.9) in \cite{warnock_morton} per unit length
if we replace ``$n$'' in \cite{warnock_morton} by $\nu$ in Eq.(\ref{eq:cZ_thin}).
$\nu$ takes any real value in Eq.(\ref{eq:cZ_thin})
since our chamber model is not a toroid, \ie, we do not impose the periodic boundary 
condition to the field with respect to $s$ as described in Eq.(\ref{eq:krho}).
Eq.(\ref{eq:cZ_thin}) is purely imaginary,
because the field is confined in the perfectly conducting beam pipe.
That is, our model is a closed system in which the energy of the electromagnetic field is 
not dissipated.

Similar to Eq.(136) in \cite{agoh}, however, by taking an infinitesimal damping of
the field into account through the wavenumber $k$,
we get the expression of the real impedance which is concealed in the poles
of the field on the real axis of the $k$-plane,
\begin{align}
  k=\lim_{\eps\to+0}(k-i\eps)
   \qquad\Ra\qquad
  \frac{1}{k-k_0-i\eps}={\cal P}\frac{1}{k-k_0}+i\pi\delta(k-k_0) ,
  \label{eq:torus_kmn_resona}
\end{align}
where the sign of $i\eps$ is opposite to that of Eq.(136) in \cite{agoh},
because, in the present paper, we define $k$ with the opposite sign to that in \cite{agoh}.
${\cal P}$ is the principal value of the integral, which denotes the imaginary impedance,
\begin{align}
  {\cal P}\cZ=\Im\cZ .
\end{align}
On the other hand, the second term involving $\delta(k-k_0)$ in
Eq.(\ref{eq:torus_kmn_resona}) corresponds to the real part of the impedance.
In Eq.(\ref{eq:torus_kmn_resona}) $k_0\in\bbR$ represents $\vkap_m^n$ and $\vbkap_m^n$
which are the zeros of the cross products $p_{\nu}(\hr_b,\hr_a)$ and $s_{\nu}(\hr_b,\hr_a)$ 
for $\nu=k\rho$ with respect to $k$, \ie,
\begin{alignat}{3}
  p_{k\rho}(k_r^nr_b,k_r^nr_a)
  &=0
   \qquad&\Lra&\qquad&
  k&=\vkap_m^n ,
   \label{eq:p_krho}
   \\
  s_{k\rho}(k_r^nr_b,k_r^nr_a)
  &=0
   \qquad&\Lra&\qquad&
  k&=\vbkap_m^n .
   \label{eq:s_krho}
\end{alignat}
Eqs.(\ref{eq:p_krho}-\ref{eq:s_krho}) correspond to Eqs.(92) in \cite{agoh}.
The approximate solution of $(\vkap_m^n,\vbkap_m^n)$ is given by Eqs.(\ref{eq:vkapm_approx})
and (\ref{eq:ivkapm_approx}).
By the treatment (\ref{eq:torus_kmn_resona}), we get the real part 
of the impedance of the steady synchrotron radiation emitted in
the perfectly conducting rectangular pipe,
\begin{align}
  &
  \frac{\Re\cZ(k)}{Z_0}
  =\frac{\pi^2}{2}k\rho\sum_{p=1}^{\infty}\tpsi_y^{2p-1}\!\!
    \sum_{\substack{m=-\infty\\(m\ne0)}}^{\infty}\!
    \big\{\Delta_{-}^{2p-1}\delta(k-\vbkap_m^{2p-1})
         +\Delta_{+}^{2p-1}\delta(k-\vkap_m^{2p-1})\big\} ,
  \label{eq:Re_cZ_thin}
   \\
  &~
  \Delta_{-}^n
  =\beta\frac{s_{k\rho}(\hr_b,\hrho)s_{k\rho}(\hrho,\hr_a)}{\rd_ks_{k\rho}(\hr_b,\hr_a)} ,
    \qquad
  \Delta_{+}^n
  =\frac{1}{\beta}\bigg(\frac{k_y^n}{k_r^n}\bigg)^{\!2}
    \frac{p_{k\rho}(\hr_b,\hrho)p_{k\rho}(\hrho,\hr_a)}{\rd_kp_{k\rho}(\hr_b,\hr_a)} .
\end{align}
The denominators of $\Delta_{\pm}^n$ are given by
\begin{align}
  \rd_ks_{k\rho}(\hr_b,\hr_a)
  &=\bigg[\rho\rd_\nu s_\nu(\hr_b,\hr_a)
          +\frac{2k\beta^2}{\pi(k_r^n)^2}
           \bigg\{\bigg(\frac{\nu^2}{\hr_b^2}-1\bigg)\frac{J_\nu'(\hr_a)}{J_\nu'(\hr_b)}
                 -\bigg(\frac{\nu^2}{\hr_a^2}-1\bigg)\frac{J_\nu'(\hr_b)}{J_\nu'(\hr_a)}
           \bigg\}
    \bigg]_{\nu=k\rho} ,
  \\
  \rd_kp_{k\rho}(\hr_b,\hr_a)
  &=\bigg[\rho\rd_\nu p_\nu(\hr_b,\hr_a)+\frac{2k\beta^2}{\pi(k_r^n)^2}
          \bigg\{\frac{J_\nu(\hr_b)}{J_\nu(\hr_a)}-\frac{J_\nu(\hr_a)}{J_\nu(\hr_b)}\bigg\}
    \bigg]_{\nu=k\rho} .
\end{align}
In the paraxial approximation, by using the uniform asymptotic expansion of
the Bessel functions, Eqs.(\ref{eq:cZ_thin}) and (\ref{eq:Re_cZ_thin}) respectively
tend to be Eqs.(68) and (139) in \cite{agoh}.

We assume a rigid and thin bunch (\ref{eq:J_thin})
which is moving on a circular orbit of radius $\rho$.
The power of synchrotron radiation emitted by this bunch per unit frequency
per revolution is given as
\begin{align}
  \frac{dP}{d\omega}
  =2\beta\rho e^2\{N+N(N-1)|\tlam_0(k)|^2\}\Re\cZ(k)
   ,\qquad
  \omg=kv 
   \qquad
  (v=c\beta).
   \label{eq:dP_dw}
\end{align}
The expression of $dP/d\omega$ given by Eq.(206) in \cite{agoh} is wrong.
$N$ is the number of the particles forming the bunch ($q=Ne$).
$\tlam_0$ is the bunch spectrum given by Eq.(\ref{eq:J_thin}).
In order to find the exact expression of the power spectrum of
synchrotron radiation from a single particle which is moving on the $s$-axis in
the perfectly conducting rectangular pipe,
let us assume a point charge having the elementary charge $e$
as the source current of the radiation field,
\begin{align}
  N=1
   ,\qquad
  \tlam_0=1
   ,\qquad
  \tpsi_y^{2p-1}=2/h .
   \label{eq:N1}
\end{align}
Substituting Eqs.(\ref{eq:Re_cZ_thin}) and (\ref{eq:N1}) into Eq.(\ref{eq:dP_dw}), we get
\begin{align}
  \frac{dP}{d\omega}
  &=Z_0(\pi e\rho)^2\frac{k}{h}\sum_{p=1}^{\infty}
    \sum_{\substack{m=-\infty\\(m\ne0)}}^{\infty}\!
    \big\{\Delta_{-}^{2p-1}\delta(k-\vbkap_m^{2p-1})
         +\Delta_{+}^{2p-1}\delta(k-\vkap_m^{2p-1})
    \big\} .
  \label{eq:pwr_spctr_point}
\end{align}
This spectrum is discrete since the field resonates in the uniformly curved rectangular pipe.
In reality, however, the resonance may not happen at very high frequencies
since the walls of an actual pipe are not perfectly flat but rough.
If we consider the power spectrum in a range much higher than the shielding threshold of 
the pipe, we should take the limit of $x_a\to-\rho$, $x_b\to\infty$ and $h\to\infty$ for
Eq.(\ref{eq:pwr_spctr_point}).
Then the sums in Eq.(\ref{eq:pwr_spctr_point}) with respect to $p$ and $m$ become integrals.


\clearpage

\section{Analytical verification of the expressions of the fields}
\label{sec:verify}

In the present study we found the expressions of all the components of
the fields of synchrotron radiation in the frequency domain as functions of $\xv$ and $k$.
The coordinate variables in the bending section are given as
\begin{align}
  \xv=(r,y,s) ,
    \qquad
  \xv_{\perp}=(r,y) ,
    \qquad
  \xv'=(r',y',s') ,
    \qquad
  \xv_{\perp}'=(r',y') .
\end{align}
In this appendix we analytically verify that the solutions of the fields
satisfy the wave equations and Maxwell equations.
Also, we show that these solutions satisfy the initial condition at
the entrance of the bend ($s=0$).

\subsection{Initial field and its longitudinal derivative at the entrance of the bend}
\label{sec:initial_field}

We first show that the expressions of the fields satisfy the initial condition at $s=0$.
Substituting $s=0$ into Eq.(\ref{eq:tEBv_cmpl})
which is the scalar expression of the fields in the complete form,
we rewrite the integrands $(\hPv,\hQv)$ and $(\hMv,\hNv)$ 
for $s=0$ using Eqs.(\ref{eq:tGamy_tLamy_vsig0}) and
(\ref{eq:GamLam_vsig0}-\ref{eq:rds_GamLam_vsig0}),
\begin{align}
  \hPv(\xv_{\perp},0,\xv_{\perp}')
  &=\frac{\rho}{r'}
    \big[
      \ev_x\tE_x(\xv')\rd_s\Gam_x(s)
     +\ev_y\tE_y(\xv')\rd_s\Gam_y(s)
     +\ev_s\tE_s(\xv')\rd_s\Gam_s(s)
    \big]_{s=s'=0}
   \\                                            
  &=\delta(r-r')\delta(y-y')\tEv(\xv_{\perp}',0) ,
   \\
  \hQv(\xv_{\perp},0,\xv_{\perp}')
  &=\frac{\rho}{r'}
    \big[
      \ev_xc\tB_x(\xv')\rd_s\Lam_x(s)
     +\ev_yc\tB_y(\xv')\rd_s\Lam_y(s)
     +\ev_sc\tB_s(\xv')\rd_s\Lam_s(s)
    \big]_{s=s'=0}                    
   \\
  &=\delta(r-r')\delta(y-y')c\tBv(\xv_{\perp}',0) ,
\end{align}
and
\begin{align}
  [\hMv(\xv_{\perp},s-s',\xv')]_{s=0}=0 ,
    \qquad
  [\hNv(\xv_{\perp},s-s',\xv')]_{s=0}=0 .
\end{align}
Then we calculate the integrals with respect to $r'$ and $y'$ in Eq.(\ref{eq:tEBv_cmpl})
for $s=0$,
\begin{align}
  \bigg\{{\tEv(\xv_{\perp},0) \atop c\tBv(\xv_{\perp},0)}\bigg\}
  &=\int_{r_a}^{r_b}dr'\int_{-h/2}^{h/2}dy'
    \bigg\{{\hPv(\xv_{\perp},0,\xv_{\perp}') \atop \hQv(\xv_{\perp},0,\xv_{\perp}')}\bigg\}
   \label{eq:tEBv_ini}
   \\
  &=\int_{r_a}^{r_b}dr'\int_{-h/2}^{h/2}dy'
    \bigg\{{\tEv(\xv_{\perp}',0) \atop c\tBv(\xv_{\perp}',0)}\bigg\}
    \delta(r-r')\delta(y-y')
   \\
  &=\bigg\{{\tEv(\xv_{\perp},0) \atop c\tBv(\xv_{\perp},0)}\bigg\} .
\end{align}
Thus, the initial values of the fields at the entrance of the bend are reproduced by
substituting $s=0$ into the solution of the fields of the complete form (\ref{eq:tEBv_cmpl}).

Similarly, we will show that the solutions of the fields in the bend,
given by Eq.(\ref{eq:tEBv_cmpl}),
reproduce the longitudinal derivative of the initial fields at the entrance of the bend.
Differentiating the components of Eq.(\ref{eq:tEBv_cmpl}) with respect to $s$,
we substitute $s=+0$ into them (where we should not differentiate
$\tEv$ and $\tBv$ as vectors since $\ev_x$ and $\ev_s$ depend on $s$).
Then we integrate them with respect to $r'$ and $y'$,
\begin{align}
  \bigg[
  \rd_s\bigg\{{\tE_{x,y,s}(\xv) \atop c\tB_{x,y,s}(\xv)}\bigg\}
  \bigg]_{s=+0}
  &=\int_{r_a}^{r_b}dr'\int_{-h/2}^{h/2}dy'
    \bigg[
       \rd_s\bigg\{{\hP_{x,y,s}(\xv,\xv_{\perp}') \atop\hQ_{x,y,s}(\xv,\xv_{\perp}')}\bigg\}
    \bigg]_{s=+0}
   \label{eq:rds_tEB_vf}
   \\
  &=\int_{r_a}^{r_b}dr'\int_{-h/2}^{h/2}dy'
    \delta(r-r')\delta(y-y')
    \bigg[\rd_{s'}
       \bigg\{{\tE_{x,y,s}(\xv') \atop
               c\tB_{x,y,s}(\xv')}
       \bigg\}
    \bigg]_{s'=+0}
   \label{eq:rds_tEB_vf_0}
   \\
  &=\bigg[\rd_{s'}
       \bigg\{{\tE_{x,y,s}(\xv_{\perp},s') \atop
               c\tB_{x,y,s}(\xv_{\perp},s')}
       \bigg\}
    \bigg]_{s'=+0} .
\end{align}
Thus, Eq.(\ref{eq:tEBv_cmpl}) reproduces the longitudinal derivative of
the initial field at $s=0$.
We used Eqs.(\ref{eq:tGamy_tLamy_vsig0}-\ref{eq:rds2_Gamy_s0}) and
(\ref{eq:GamLam_vsig0}-\ref{eq:rds2_GL}) in rewriting
Eq.(\ref{eq:rds_tEB_vf}) into Eq.(\ref{eq:rds_tEB_vf_0}),
\begin{align}
  [\rd_s\hP_{x,y,s}(\xv,\xv_{\perp}')]_{s=+0}
  &=\delta(r-r')\delta(y-y')[\rd_{s'}\tE_{x,y,s}(\xv')]_{s'=+0} ,
   \\
  [\rd_s\hQ_{x,y,s}(\xv,\xv_{\perp}')]_{s=+0}
  &=\delta(r-r')\delta(y-y')[\rd_{s'}c\tB_{x,y,s}(\xv')]_{s'=+0} ,
   \label{eq:rds_hQxys}
\end{align}
and
\begin{align}
  [\rd_s\hM_{x,y,s}(\xv_{\perp},s-s',\xv')]_{s=0}=0 ,
    \qquad
  [\rd_s\hN_{x,y,s}(\xv_{\perp},s-s',\xv')]_{s=0}=0 .
\end{align}
%

\subsection{Verification of the expressions of the fields through the wave equations}
\label{sec:EyBy_verification}

We show that the expressions of the fields (\ref{eq:tEBv_cmpl}) satisfy
the wave equations given by Eqs.(\ref{eq:we_tEy_pm}) and
(\ref{eq:we_tBy_pm}-\ref{eq:we_tExsBxs}).
Eq.(\ref{eq:tEBv_cmpl}) involves $(\hPv,\hQv)$ and $(\hMv,\hNv)$
given by Eqs.(\ref{eq:hPv_hMv}-\ref{eq:hQv_hNv}).
In appendix \ref{sec:EyBy_verification}, for brevity,
we omit the transverse arguments $(\xv_\perp,\xv_\perp')$ of the functions
except for the $\delta$-functions.
Acting the operator $\nablav_{\rm v}^2$, given by Eq.(\ref{eq:vopera}),
on the vertical components of Eq.(\ref{eq:tEBv_cmpl}),
we rewrite them using Eqs.(\ref{eq:we_hPQMN_y}),
\begin{align}
  \nablav_{\rm v}^2\bigg\{{\tE_y(s) \atop c\tB_y(s)}\bigg\}
  &=\int_{r_a}^{r_b}dr'\int_{-h/2}^{h/2}dy'
    \bigg[
       \nablav_{\rm v}^2\bigg\{{\hP_y(s) \atop \hQ_y(s)}\bigg\}
      +Z_0\int_{0}^{\infty}ds'
       \nablav_{\rm v}^2
       \bigg\{{\hM_y(s-s',s') \atop \hN_y(s-s',s')}\bigg\}
    \bigg]
  \label{nab2_tEBy_cmpl}
   \\
  &=Z_0\int_{r_a}^{r_b}dr'\int_{-h/2}^{h/2}dy'\int_{0}^{\infty}ds'
    \bigg\{{\tS_y(s') \atop \tT_y(s')}\bigg\}\delta(r-r')\delta(y-y')\delta(s-s')
   \\
  &=Z_0\bigg\{{\tS_y(s) \atop \tT_y(s)}\bigg\} .
   \label{eq:src_y}
\end{align}
Eq.(\ref{eq:src_y}) agrees with the second equations of (\ref{eq:we_tEy_pm}) and
(\ref{eq:we_tBy_pm}) which are the wave equations for $\tE_y$ and $\tB_y$.

Next, we show that the radial and longitudinal components of Eq.(\ref{eq:tEBv_cmpl})
satisfy the wave equations (\ref{eq:we_tExsBxs}).
Acting the operators $\nablav_{\vdash}^2$ and $(2/r)\brd_s$, given by Eqs.(\ref{eq:vopera}), 
on the radial and longitudinal components of Eq.(\ref{eq:tEBv_cmpl}),
we calculate the L.H.S. of the radial components of Eqs.(\ref{eq:we_tExsBxs}),
\begin{align}
  &
  \nablav_{\vdash}^2\bigg\{{\tE_x(s) \atop c\tB_x(s)}\bigg\}
 -\frac{2}{r}\brd_s\bigg\{{\tE_s(s) \atop c\tB_s(s)}\bigg\}
  \\&\quad
  =\int_{r_a}^{r_b}dr'\int_{-h/2}^{h/2}dy'
    \bigg[
       \nablav_{\vdash}^2\bigg\{{\hP_x(s) \atop \hQ_x(s)}\bigg\}
      -\frac{2}{r}\brd_s\bigg\{{\hP_s(s) \atop \hQ_s(s)}\bigg\}
    \bigg]
   \nonumber\\&\qquad
   +Z_0\int_{r_a}^{r_b}dr'\int_{-h/2}^{h/2}dy'\int_{0}^{\infty}ds'
    \bigg[
       \nablav_{\vdash}^2\bigg\{{\hM_x(s-s',s') \atop \hN_x(s-s',s')}\bigg\}
      -\frac{2}{r}\brd_s\bigg\{{\hM_s(s-s',s') \atop \hN_s(s-s',s')}\bigg\}
    \bigg]
  \label{nab2_tEBx_cmpl}
   \\
  &\quad
   =Z_0\int_{r_a}^{r_b}dr'\int_{-h/2}^{h/2}dy'\int_{0}^{\infty}ds'
    \bigg\{{\tS_x(s') \atop \tT_x(s')}\bigg\}\delta(r-r')\delta(y-y')\delta(s-s')
   \label{eq:int_delta_x}
   \\
  &\quad
   =Z_0\bigg\{{\tS_x(s) \atop \tT_x(s)}\bigg\} .
\end{align}
Similarly, we calculate the L.H.S. of the longitudinal components of
Eqs.(\ref{eq:we_tExsBxs}) using Eq.(\ref{eq:tEBv_cmpl}),
\begin{align}
  &
  \nablav_{\vdash}^2\bigg\{{\tE_s(s) \atop c\tB_s(s)}\bigg\}
 +\frac{2}{r}\brd_s\bigg\{{\tE_x(s) \atop c\tB_x(s)}\bigg\}
  \\&\quad
  =\int_{r_a}^{r_b}dr'\int_{-h/2}^{h/2}dy'
    \bigg[
       \nablav_{\vdash}^2\bigg\{{\hP_s(s) \atop \hQ_s(s)}\bigg\}
      +\frac{2}{r}\brd_s\bigg\{{\hP_x(s) \atop \hQ_x(s)}\bigg\}
    \bigg]
   \nonumber\\&\qquad
   +Z_0\int_{r_a}^{r_b}dr'\int_{-h/2}^{h/2}dy'\int_{0}^{\infty}ds'
    \bigg[
       \nablav_{\vdash}^2\bigg\{{\hM_s(s-s',s') \atop \hN_s(s-s',s')}\bigg\}
      +\frac{2}{r}\brd_s\bigg\{{\hM_x(s-s',s') \atop \hN_x(s-s',s')}\bigg\}
    \bigg]
  \label{nab2_tEBs_cmpl}
   \\
  &\quad
   =Z_0\int_{r_a}^{r_b}dr'\int_{-h/2}^{h/2}dy'\int_{0}^{\infty}ds'
    \bigg\{{\tS_s(s') \atop \tT_s(s')}\bigg\}\delta(r-r')\delta(y-y')\delta(s-s')
   \label{eq:int_delta_s}
  \\
  &\quad
   =Z_0\bigg\{{\tS_s(s) \atop \tT_s(s)}\bigg\} .
\end{align}
Thus, all the components of Eq.(\ref{eq:tEBv_cmpl}) satisfy the wave equations
(\ref{eq:we_tEy_pm}) and (\ref{eq:we_tBy_pm}-\ref{eq:we_tExsBxs}) in the frequency domain.
We used Eqs.(\ref{eq:we_hPxs}-\ref{eq:we_hMNs})
in rewriting the integrands in Eqs.(\ref{nab2_tEBx_cmpl}) and (\ref{nab2_tEBs_cmpl}).

\subsection{Verification of the radial and longitudinal fields through the wave equations}
\label{sec:EBxs_verification}

We show that $\tE_{x,s}$, given by Eq.(\ref{eq:tEBv_cmpl}), satisfies the wave equations
(\ref{eq:we_tExsBxs}) without using Eqs.(\ref{eq:we_hPQMN_y}-\ref{eq:we_hMNs}), \ie,
without calculating the integrals which involve the $\delta$-functions as in
Eqs.(\ref{eq:int_delta_x}) and (\ref{eq:int_delta_s}).
Opposite to the way of the proof in appendix \ref{sec:EyBy_verification},
we reproduce $\tE_{x,s}$ by rewriting the R.H.S. of Eq.(\ref{eq:tEBv_cmpl}) 
using the source terms $\tS_{x,s}$ involved in Eq.(\ref{eq:we_tExsBxs}).
We rewrite the source terms of $\tE_{x,s}$ in Eq.(\ref{eq:we_tExsBxs})
using $\xv'$ instead of $\xv$,
\begin{align}
  Z_0\bigg\{{ \tS_x(\xv') \atop \tS_s(\xv')}\bigg\}
  &=\nablav_{\vdash}'^2
    \bigg\{{ \tE_x(\xv') \atop \tE_s(\xv')}\bigg\}
   +\frac{2\rho}{r'^2}\rd_{s'}\bigg\{{ -\tE_s(\xv') \atop +\tE_x(\xv')}\bigg\}
   ,\qquad
  \nablav_{\vdash}'^2
  =\rd_{r'}\brd_{r'}+\rd_{y'}^2+(k\beta)^2+\frac{\rho^2}{r'^2}\rd_{s'}^2 .
  \label{eq:tSxs_tExs_xvp}
\end{align}
We substitute the source terms (\ref{eq:tSxs_tExs_xvp}) into the integrands of $\tE_x(\xv)$,
involved in Eq.(\ref{eq:tEBv_cmpl}),
\begin{align}
  \tE_x(\xv)
  &=K_x^0
   +\int_{r_a}^{r_b}dr'\int_{-h/2}^{h/2}dy'
    \{\tD_x(\xv_\perp')\Gam_x(s)-\tD_s(\xv_\perp')\Gam_x^{\ast}(s)\} ,
  \label{eq:tEx_K}
   \\
  K_x^0
  &=K_x^x-K_x^s -\bK_x^s-\bK_x^x .
\end{align}
For brevity, we omit the transverse arguments $(\xv_{\perp},\xv_{\perp}')$ of
the Green functions ($\Gam_{x,s}$, $\Gam_{x,s}^{\ast}$) given by Eqs.(\ref{eq:tGam_xs}).
The driving terms $\tD_{x,s}$ are given by Eqs.(\ref{eq:bEx}) and (\ref{eq:bEs}).
$K_x^{x,s}$ and $\bK_x^{s,x}$ denote the following integrals,
\begin{align}
  \bigg\{{K_x^x \atop K_x^s}\bigg\}
  &=\int_{r_a}^{r_b}dr'\int_{-h/2}^{h/2}dy'\int_{0}^{\infty}ds'
    \Gam_x(s-s')
    \bigg\{{(r'/\rho)\nablav_{\vdash}'^2\tE_x(\xv') \atop (2/r')\rd_{s'}\tE_s(\xv')}\bigg\} ,
  \label{eq:Kxx}
  \\
  \bigg\{{\bK_x^s \atop \bK_x^x}\bigg\}
  &=\int_{r_a}^{r_b}dr'\int_{-h/2}^{h/2}dy'\int_{0}^{\infty}ds'
    \Gam_x^{\ast}(s-s')
    \bigg\{{(r'/\rho)\nablav_{\vdash}'^2\tE_s(\xv') \atop (2/r')\rd_{s'}\tE_x(\xv')}\bigg\} .
  \label{eq:bKxs}
\end{align}
In Eqs.(\ref{eq:Kxx}-\ref{eq:bKxs}) we integrate the derivatives of $\tE_{x,s}(\xv')$
by parts with respect to $r'$, $y'$ or $s'$,
\begin{align}
  K_x^x
  &=\tE_x(\xv)
   +\int_{r_a}^{r_b}dr'\int_{-h/2}^{h/2}dy'\int_{0}^{\infty}\frac{ds'}{\rho}
    2\tE_x(\xv')\frac{\rho}{r'}\rd_s\Gam_x^{\ast}(s-s')
  \nonumber\\&\quad
   -\int_{r_a}^{r_b}\frac{dr'}{r'}\int_{-h/2}^{h/2}dy'
    \big[\rho\rd_{s'}\tE_x(\xv')+\tE_x(\xv')\rho\rd_s\big]_{s'=+0}\Gam_x(s) ,
   \\
  K_x^s
  &=\int_{r_a}^{r_b}dr'\int_{-h/2}^{h/2}dy'\int_{0}^{\infty}\frac{ds'}{\rho}
    2\tE_s(\xv')\frac{\rho}{r'}\rd_s\Gam_x(s-s')
   -\int_{r_a}^{r_b}\frac{dr'}{r'}\int_{-h}^{h}dy'2\tE_s(\xv_\perp',0)\Gam_x(s) ,
   \\
  \bK_x^s
  &=-\int_{r_a}^{r_b}dr'\int_{-h/2}^{h/2}dy'\int_{0}^{\infty}\frac{ds'}{\rho}2\tE_s(\xv')
     \frac{\rho}{r'}\rd_s\Gam_x(s-s')
  \nonumber\\&\quad
    -\int_{r_a}^{r_b}\frac{dr'}{r'}\int_{-h/2}^{h/2}dy'
     \big[\rho\rd_{s'}\tE_s(\xv')+\tE_s(\xv')\rho\rd_s\big]_{s'=+0}\Gam_x^{\ast}(s),
   \\
  \bK_x^x
  &=\int_{r_a}^{r_b}dr'\int_{-h/2}^{h/2}dy'\int_{0}^{\infty}\frac{ds'}{\rho}
    2\tE_x(\xv')\frac{\rho}{r'}\rd_s\Gam_x^{\ast}(s-s')
   -\int_{r_a}^{r_b}\frac{dr'}{r'}\int_{-h}^{h}dy'2\tE_x(\xv_\perp',0)\Gam_x^{\ast}(s) .
\end{align}
Then $K_x^0$ is rewritten as
\begin{align}
  K_x^0
  &=\tE_x(\xv)
   -\int_{r_a}^{r_b}dr'\int_{-h/2}^{h/2}dy'
    \Big[
        \frac{\rho}{r'}\{\tE_x(\xv')\rd_s+\rd_{s'}\tE_x(\xv')\}
       -\frac{2}{r'}\tE_s(\xv')
    \Big]_{s'=+0}
    \Gam_x(s)
  \nonumber\\&\quad
   +\int_{r_a}^{r_b}dr'\int_{-h/2}^{h/2}dy'
    \Big[
       \frac{\rho}{r'}\{\tE_s(\xv')\rd_s+\rd_{s'}\tE_s(\xv')\}
      +\frac{2}{r'}\tE_x(\xv')
    \Big]_{s'=+0}
    \Gam_x^{\ast}(s)
   \\
  &=\tE_x(\xv)
   -\int_{r_a}^{r_b}dr'\int_{-h/2}^{h/2}dy'
    \{\tD_x(\xv_{\perp}')\Gam_x(s) -\tD_s(\xv_{\perp}')\Gam_x^{\ast}(s)\} .
  \label{eq:K4}
\end{align}
Substituting Eq.(\ref{eq:K4}) into Eq.(\ref{eq:tEx_K}), we rewrite it as follows,
\begin{align}
  \tE_x(\xv)
  &=\tE_x(\xv)
   -\int_{r_a}^{r_b}dr'\int_{-h/2}^{h/2}dy'
    \{\tD_x(\xv_\perp')\Gam_x(s)-\tD_s(\xv_\perp')\Gam_x^{\ast}(s)\}
  \nonumber\\&\quad
   +\int_{r_a}^{r_b}dr'\int_{-h/2}^{h/2}dy'
    \{\tD_x(\xv_\perp')\Gam_x(s)-\tD_s(\xv_\perp')\Gam_x^{\ast}(s)\}
  \\
  &=\tE_x(\xv) .
\end{align}
Thus, the expression of $\tE_x(\xv)$ reproduces itself through the wave equations
(\ref{eq:we_tExsBxs}) for $\tE_x(\xv')$ and $\tE_s(\xv')$.
It follows that $\tE_x(\xv)$, given by Eq.(\ref{eq:tEBv_cmpl}), is the exact solution of
the wave equation (\ref{eq:we_tExsBxs}).

In a similar way to the above verification for $\tE_x$,
we will show that $\tE_s$ satisfies the wave equation (\ref{eq:we_tExsBxs}).
Substituting the source terms (\ref{eq:tSxs_tExs_xvp}) into $\tE_s(\xv)$
given by Eq.(\ref{eq:tEBv_cmpl}), we rewrite it as follows,
\begin{align}
  \tE_s(\xv)
  &=K_s^0
   +\int_{r_a}^{r_b}dr'\int_{-h/2}^{h/2}dy'
    \{\tD_s(\xv_\perp')\Gam_s(s)+\tD_x(\xv_\perp')\Gam_s^{\ast}(s)\} ,
  \label{eq:tEs_K}
   \\
  K_s^0
  &=K_s^s+K_s^x+\bK_s^x-\bK_s^s ,
\end{align}
where
\begin{align}
  \bigg\{{K_s^s \atop K_s^x}\bigg\}
  &=\int_{r_a}^{r_b}dr'\int_{-h/2}^{h/2}dy'\int_{0}^{\infty}ds'
    \,\Gam_s(s-s')
    \bigg\{{(r'/\rho)\nablav_{\vdash}'^2\tE_s(\xv') \atop (2/r')\rd_{s'}\tE_x(\xv')}\bigg\} ,
  \label{eq:Kss}
   \\
  \bigg\{{\bK_s^x \atop \bK_s^s}\bigg\}
  &=\int_{r_a}^{r_b}dr'\int_{-h/2}^{h/2}dy'\int_{0}^{\infty}ds'
    \,\Gam_s^{\ast}(s-s')
    \bigg\{{(r'/\rho)\nablav_{\vdash}'^2\tE_x(\xv') \atop (2/r')\rd_{s'}\tE_s(\xv')}\bigg\} .
  \label{eq:bKss}
\end{align}
In Eqs.(\ref{eq:Kss}-\ref{eq:bKss}) we integrate the derivatives of $\tE_{x,s}(\xv')$
by parts with respect to $r'$, $y'$ or $s'$,
\begin{align}
  K_s^s
  &=\tE_s(\xv)
   +\int_{r_a}^{r_b}dr'\int_{-h/2}^{h/2}dy'\int_{0}^{\infty}\frac{ds'}{\rho}
    2\tE_s(\xv')\frac{\rho}{r'}\rd_s\Gam_s^{\ast}(s-s')
  \nonumber\\&\quad
   -\int_{r_a}^{r_b}\frac{dr'}{r'}\int_{-h/2}^{h/2}dy'
    \big[\rho\rd_{s'}\tE_s(\xv')+\tE_s(\xv')\rho\rd_s\big]_{s'=+0}\Gam_s(s) ,
  \label{eq:Kss_0}
   \\
  K_s^x
  &=\int_{r_a}^{r_b}dr'\int_{-h/2}^{h/2}dy'\int_{0}^{\infty}\frac{ds'}{\rho}2\tE_x(\xv')
    \frac{\rho}{r'}\rd_s\Gam_s(s-s')
   -\int_{r_a}^{r_b}\frac{dr'}{r'}\int_{-h/2}^{h/2}dy'2\tE_x(\xv_\perp',0)\Gam_s(s) ,
   \\
  \bK_s^x
  &=-\int_{r_a}^{r_b}dr'\int_{-h/2}^{h/2}dy'\int_{0}^{\infty}\frac{ds'}{\rho}
     2\tE_x(\xv')\frac{\rho}{r'}\rd_s\Gam_s(s-s')
  \nonumber\\&\quad
   -\int_{r_a}^{r_b}\frac{dr'}{r'}\int_{-h/2}^{h/2}dy'
    \big[\rho\rd_{s'}\tE_x(\xv')+\tE_x(\xv')\rho\rd_s\big]_{s'=+0}\Gam_s^{\ast}(s) ,
  \label{eq:bKsx_0}
   \\
  \bK_s^s
  &=\int_{r_a}^{r_b}dr'\int_{-h/2}^{h/2}dy'\int_{0}^{\infty}\frac{ds'}{\rho}2\tE_s(\xv')
    \frac{\rho}{r'}\rd_s\Gam_s^{\ast}(s-s')
   -\int_{r_a}^{r_b}\frac{dr'}{r'}\int_{-h/2}^{h/2}dy'2\tE_s(\xv_\perp',0)\Gam_s^{\ast}(s) .
\end{align}
Then $K_s^0$ on the R.H.S. of Eq.(\ref{eq:tEs_K}) is rewritten as follows,
\begin{align}
  K_s^0
  &=\tE_s(\xv)
   -\int_{r_a}^{r_b}dr'\int_{-h/2}^{h/2}dy'
    \Big[
      \frac{\rho}{r'}\{\tE_s(\xv')\rd_s+\rd_{s'}\tE_s(\xv')\}
     +\frac{2}{r'}\tE_x(\xv')
    \Big]_{s'=+0}\Gam_s(s)
  \nonumber\\&\quad
   -\int_{r_a}^{r_b}dr'\int_{-h/2}^{h/2}dy'
    \Big[
       \frac{\rho}{r'}\{\tE_x(\xv')\rd_s+\rd_{s'}\tE_x(\xv')\}
      -\frac{2}{r'}\tE_s(\xv')
    \Big]_{s'=+0}\Gam_s^{\ast}(s)
  \\
  &=\tE_s(\xv)
   -\int_{r_a}^{r_b}dr'\int_{-h/2}^{h/2}dy'
    \{\tD_s(\xv_\perp')\Gam_s(s)+\tD_x(\xv_\perp')\Gam_s^{\ast}(s)\} .
  \label{eq:Ks0}
\end{align}
Substituting Eq.(\ref{eq:Ks0}) into Eq.(\ref{eq:tEs_K}), we rewrite it as
\begin{align}
  \tE_s(\xv)
  &=\tE_s(\xv)
   -\int_{r_a}^{r_b}dr'\int_{-h/2}^{h/2}dy'
    \{\tD_s(\xv_\perp')\Gam_s(s)+\tD_x(\xv_\perp')\Gam_s^{\ast}(s)\}
  \nonumber\\&\quad
   +\int_{r_a}^{r_b}dr'\int_{-h/2}^{h/2}dy'
    \{\tD_s(\xv_\perp')\Gam_s(s)+\tD_x(\xv_\perp')\Gam_s^{\ast}(s)\}
  \\
  &=\tE_s(\xv) .
\end{align}
We showed that $\tE_s(\xv)$ reproduces itself through the wave equations for
$\tE_s(\xv')$ and $\tE_x(\xv')$.
It follows that $\tE_s(\xv)$, given by Eq.(\ref{eq:tEBv_cmpl}), is the exact solution of
the wave equation (\ref{eq:we_tExsBxs}).
Similar to the above verification for the solutions of $\tE_{x,s}$,
we can show that the radial and longitudinal components of the magnetic field,
$\tB_{x,s}$ given by Eqs.(\ref{eq:tEBv_cmpl}), satisfy their wave equations
(\ref{eq:we_tExsBxs}), though we omit the proof in the present paper.

\subsection{Verification of the electric field through Gauss's law}
\label{sec:verify_gauss}

We will analytically show that the expression of the electric field in the bending section
satisfies Gauss's law in the frequency domain.
We use Eq.(\ref{eq:tEBv_cmpl_dg}) which is the differential expression of the complete 
form since we already verified the scalar expressions of the complete form in appendices 
\ref{sec:initial_field}, \ref{sec:EyBy_verification} and \ref{sec:EBxs_verification}.
Gauss's law for the electric field in the frequency domain is given by
Eq.(\ref{eq:Gauss_law_FD2}),
\begin{align}
  \nablav\cdot\Ev&=c\mu_0J_0
   \qquad\Lra\qquad
  \brd_r\tE_x+\rd_y\tE_y+\brd_s\tE_s=Z_0\tJ_0 .
  \label{eq:Gauss_law_appdx}
\end{align}
The operators $\brd_r$ and $\brd_s$ are given by Eqs.(\ref{eq:rdv2}).
We substitute Eq.(\ref{eq:tEBv_cmpl_dg}) into Eq.(\ref{eq:Gauss_law_appdx}), 
\begin{align}
  I_0=Z_0(I_1+\tJ_0) ,
  \label{eq:tEv_gauss}
\end{align}
where
\begin{align}
  I_0=\int_{r_a}^{r_b}dr'\int_{-h/2}^{h/2}dy'\hF_0(s) ,
    \qquad
  I_1=\int_{r_a}^{r_b}dr'\int_{-h/2}^{h/2}dy'\int_0^{\infty}ds'\hF_1(s-s') .
  \label{eq:I1_I2}
\end{align}
$\hF_0$ and $\hF_1$ are given as
\begin{align}
  \hF_0(s)
  =\nablav\cd\hPv^{\dg}
  =\brd_r\hat{P}_x^{\dg}+\brd_s\hat{P}_s^{\dg}
   +\rd_y\hat{P}_y^{\dg} ,
    \qquad
  \hF_1(\vsig)
  =\nablav\cd\hMv^{\dg}
  =\brd_r\hat{M}_x^{\dg}+\brd_s\hat{M}_s^{\dg}
   +\rd_y\hat{M}_y^{\dg} .
\end{align}
We substitute Eqs.(\ref{eq:hPxs_dg}) and (\ref{eq:hMxs_dg}) respectively into
$\hF_0$ and $\hF_1$,
\begin{align}
  \hF_0(s)
  &=\sum_{n=1}^{\infty}\sum_{m=0}^{\infty}
    \big\{
       (1-\delta_0^m)\big(\brd_r\hmfp_x^{+}+\brd_s\hmfp_s^{+}\big)
      +\big(\brd_r\hmfp_x^{-}+\brd_s\hmfp_s^{-}\big)
    \big\}
   +\rd_y\hat{P}_y^{\dg} ,
   \label{eq:hF0}
   \\
  \hF_1(\vsig)
  &=\theta(\vsig)\sum_{n=1}^{\infty}\cY_{+}^n\sum_{m=0}^{\infty}
    \big\{
       (1-\delta_0^m)\big(\brd_r\hmfm_x^{+}+\brd_s\hmfm_s^{+}\big)
      +\big(\brd_r\hmfm_x^{-}+\brd_s\hmfm_s^{-}\big)
    \big\}
   +\rd_y\hat{M}_y^{\dg}
    \nonumber\\&\quad
   -\tJ_0(\xv')\delta(r-r')\delta(y-y')\delta(\vsig) .
   \label{eq:hF1}
\end{align}
$\hat{P}_y^{\dg}$ and $\hat{M}_y^{\dg}$ are given by Eqs.(\ref{eq:hPy_dg}) and
(\ref{eq:hMy_dg}).
The terms involved in Eq.(\ref{eq:hF0}) are given as follows,
\begin{align}
  \brd_{r}\hmfp_{x}^{+}
  &=\Big[
       \frac{\rho}{r'}\cos\!\Big(\frac{\nu_m^ns}{\rho}\Big)\tE_y(0)\rd_{y'}
      -\frac{\rho}{\nu_m^n}\sin\!\Big(\frac{\nu_m^ns}{\rho}\Big)
       \{\tE_s(0)(k_r^n)^2+c\tB_y(0)ik\beta\rd_{r'}\}
    \Big]
    \bigg\{1-\frac{(\nu_m^n)^2}{\hr^2}\bigg\}\cR_{+}^{mn}\cY_{+}^n ,
  \label{eq:rdr_rhmfpx_p}
   \\
  \brd_{r}\hmfp_{x}^{-}
  &=\Big[
       \cos\!\Big(\frac{\mu_m^ns}{\rho}\Big)
       \{\tE_x(0)k_r^n-\tE_y(0)\rd_{\hr'}\rd_{y'}\}
      +\frac{\mu_m^n}{\hr'}\sin\!\Big(\frac{\mu_m^ns}{\rho}\Big)c\tB_y(0)ik\beta
    \Big]\frac{\rho}{r}\rd_{\hr}\cR_{-}^{mn}\cY_{+}^n ,
   \\
  \rd_s\hmfp_s^{+}
   &=\Big[
       \frac{\nu_m^n}{\hr'}\cos\!\Big(\frac{\nu_m^ns}{\rho}\Big)\tE_y(0)\rd_{y'}
      -\sin\!\Big(\frac{\nu_m^ns}{\rho}\Big)
       \{\tE_s(0)k_r^n+c\tB_y(0)ik\beta\rd_{\hr'}\}
     \Big]\frac{\nu_m^n}{\hr}\cR_{+}^{mn}\cY_{+}^n ,
   \\
  \rd_s\hmfp_s^{-}
   &=\Big[
        \cos\!\Big(\frac{\mu_m^ns}{\rho}\Big)
        \{\tE_y(0)\rd_{\hr'}\rd_{y'}-\tE_x(0)k_r^n\}
       -\frac{\mu_m^n}{\hr'}\sin\!\Big(\frac{\mu_m^ns}{\rho}\Big)c\tB_y(0)ik\beta
     \Big]\rd_{\hr}\cR_{-}^{mn}\cY_{+}^n .
  \label{eq:rds_hmfps_m}
\end{align}
$\tEv(0)$ and $\tBv(0)$ are the initial fields at $s'=0$ as in Eqs.(\ref{abbrev}).
In deriving Eq.(\ref{eq:rdr_rhmfpx_p}), we used Eq.(\ref{eq:BC_mfRp})
which is the boundary condition for $\cR_{+}^{mn}$.
According to Eqs.(\ref{eq:rdr_rhmfpx_p}-\ref{eq:rds_hmfps_m}),
Eq.(\ref{eq:hF0}) is rewritten as
\begin{align}
  \hF_0
  &=ik\beta\sum_{n=1}^{\infty}\sum_{m=1}^{\infty}
    \frac{\rho}{\nu_m^n}\sin\!\Big(\frac{\nu_m^ns}{\rho}\Big)
    \{c\tB_x(0)\rd_{y'}-c\tB_y(0)\rd_{r'}+ik\beta\tE_s(0)\}\cR_{+}^{mn}\cY_{+}^n .
   \label{eq:hF0_intm}
\end{align}
$I_0$ involves $\rd_{r'}\cR_{+}^{mn}$ and $\rd_{y'}\cY_{+}^n$ through Eq.(\ref{eq:hF0_intm}).
Integrating these terms by parts, we rewrite $I_0$ as
\begin{align}
  I_0
  &=\int_{r_a}^{r_b}dr'\int_{-h/2}^{h/2}dy'ik\beta\sum_{n=1}^{\infty}\sum_{m=1}^{\infty}
    \frac{\rho}{\nu_m^n}\sin\!\Big(\frac{\nu_m^ns}{\rho}\Big)\cR_{+}^{mn}\cY_{+}^n
    \{\rd_{r'}c\tB_y(0)-\rd_{y'}c\tB_x(0)+ik\beta\tE_s(0)\}
   \label{I1_intm}
  \\
  &=ik\beta\int_{r_a}^{r_b}dr'\int_{-h/2}^{h/2}dy'Z_0\tJ_s(0)
    \sum_{n=1}^{\infty}\sum_{m=1}^{\infty}\cR_{+}^{mn}\cY_{+}^n
    \frac{\rho}{\nu_m^n}\sin\!\Big(\frac{\nu_m^ns}{\rho}\Big) ,
   \label{I1_fin}
\end{align}
where $\tJ_s(0)=\tJ_s(\xv_{\perp}',0)$ denotes the initial value of $\tJ_s(\xv')$ at $s'=0$.
In rewriting Eq.(\ref{I1_intm}) into Eq.(\ref{I1_fin}), we used Eq.(\ref{eq:Ampere_s_FD2}) 
in terms of $\xv'=(r',y',s')$ at $s'=0$, which is the longitudinal component of Ampere's law,
\begin{align}
  \rd_{r'}c\tB_y(\xv_{\perp}',0)-\rd_{y'}c\tB_x(\xv_{\perp}',0)+ik\beta\tE_s(\xv_{\perp}',0)
  =Z_0\tJ_s(\xv_{\perp}',0) .
\end{align}

Substituting Eq.(\ref{eq:hF1}) into Eq.(\ref{eq:I1_I2}), we rewrite $I_1$ 
using Eqs.(\ref{eq:hMy_dg}) and (\ref{eq:hMxs_dg}),
\begin{align}
  I_1
  &=\int_{r_a}^{r_b}dr'\frac{r'}{\rho}\int_{-h/2}^{h/2}dy'\int_0^{s}ds'
    \sum_{n=1}^{\infty}\cY_{+}^n\sum_{m=1}^{\infty}
    \frac{\rho}{\nu_m^n}\sin\!\Big(\frac{\nu_m^n\vsig}{\rho}\Big)
    \{\tJ_0(\xv')(k\beta)^2-ik\beta\tJ_x(\xv')\rd_{r'}\}\cR_{+}^{mn}
  \nonumber\\&\quad
   +\int_{r_a}^{r_b}dr'\int_{-h/2}^{h/2}dy'\int_0^{s}ds'
    ik\beta\tJ_s(\xv')\sum_{n=1}^{\infty}\cY_{+}^n\sum_{m=1}^{\infty}
    \cos\!\Big(\frac{\nu_m^n\vsig}{\rho}\Big)\cR_{+}^{mn}
  \nonumber\\&\quad
   -\int_{r_a}^{r_b}\!dr'\frac{r'}{\rho}\int_{-h/2}^{h/2}dy'\int_0^{s}ds'
    ik\beta\tJ_y(\xv')
    \sum_{n=0}^{\infty}\rd_{y'}\cY_{+}^n\sum_{m=1}^{\infty}
    \sin\!\Big(\frac{\nu_m^n\vsig}{\rho}\Big)\frac{\rho}{\nu_m^n}\cR_{+}^{mn}
  \nonumber\\&\quad
   -\int_{r_a}^{r_b}dr'\int_{-h/2}^{h/2}dy'\int_0^{\infty}ds'
    \tJ_0(\xv')\delta(\vsig)\delta(r-r')\delta(y-y') .
  \label{eq:I2_int}
\end{align}
In Eq.(\ref{eq:I2_int}) we integrate the terms which involve
$\rd_{r'}\cR_{+}^{mn}$ or $\rd_{y'}\cY_{+}^n$ by parts with respect to $r'$ or $y'$,
\begin{align}
  I_1
  &=\int_{r_a}^{r_b}dr'\frac{r'}{\rho}\int_{-h/2}^{h/2}dy'\int_0^{s}ds'
    C(\xv')ik\beta
    \sum_{n=1}^{\infty}\cY_{+}^n\sum_{m=1}^{\infty}\cR_{+}^{mn}
    \frac{\rho}{\nu_m^n}\sin\!\Big(\frac{\nu_m^n\vsig}{\rho}\Big)
  \nonumber\\&\quad
   +\int_{r_a}^{r_b}\!dr'\int_{-h/2}^{h/2}\!dy'
    \tJ_s(0)ik\beta\sum_{n=1}^{\infty}\cY_{+}^n\sum_{m=1}^{\infty}\cR_{+}^{mn}
    \frac{\rho}{\nu_m^n}\sin\!\Big(\frac{\nu_m^ns}{\rho}\Big)
   -\tJ_0(\xv) ,
  \label{eq:I2_intm}
\end{align}
where $C(\xv')$ is zero, according to the equation of continuity in the frequency domain,
given by Eq.(\ref{eq:eq_cont_fd}),
\begin{align}
  C(\xv')
  =\brd_{r'}\tJ_x(\xv')+\frac{\rho}{r'}\rd_{s'}\tJ_s(\xv')
   +\rd_{y'}\tJ_y(\xv')-ik\beta\tJ_0(\xv')
  =0 .
\end{align}
Therefore Eq.(\ref{eq:I2_intm}) is rewritten as
\begin{align}
  I_1
  &=
    ik\beta\int_{r_a}^{r_b}\!dr'\int_{-h/2}^{h/2}\!dy'
    \tJ_s(0)\sum_{n=1}^{\infty}\cY_{+}^n\sum_{m=1}^{\infty}\cR_{+}^{mn}
    \frac{\rho}{\nu_m^n}\sin\!\Big(\frac{\nu_m^ns}{\rho}\Big)
   -\tJ_0(\xv)
   \\
  &=I_0/Z_0-\tJ_0(\xv) .
   \label{I2_fin}
\end{align}
Eq.(\ref{I2_fin}) is equivalent to Eq.(\ref{eq:tEv_gauss}).
It follows that the electric field $\tEv$ given by Eq.(\ref{eq:tEBv_cmpl_dg}) satisfies
Eq.(\ref{eq:Gauss_law_appdx}) which is Gauss's law in the frequency domain.
In this proof we used the equation of continuity for the current $(\tJ_0,\tJv)$ and
the longitudinal component of Ampere's law which involves $\tE_s$ and $\tB_{x,y}$.
That is, we assume that the initial values of $\tE_s$ and $\tB_{x,y}$ at $s=0$ satisfy
Ampere's law.

\subsection{Verification of the magnetic field through the no-monopole law}
\label{sec:verf_nomonopole}

We will show that $\tBv$, given by Eq.(\ref{eq:tEBv_cmpl_dg}), satisfies
Eq.(\ref{eq:No_monopole_FD2}) which means the absence of magnetic monopole,
\begin{align}
  \nablav\cdot\Bv=0
  \qquad\Lra\qquad
  K_0+\rd_yc\tB_y=0
   ,\qquad\text{where}\quad
  K_0
  =\brd_r c\tB_x+\brd_sc\tB_s .
  \label{eq:No_monopole_appdx}
\end{align}
We substitute Eq.(\ref{eq:tEBv_cmpl_dg}) into $K_0$,
\begin{align}
  K_0
  &=\int_{r_a}^{r_b}dr'\int_{-h/2}^{h/2}dy'
    \Big\{\bF_0(s) -Z_0\int_{0}^{\infty}ds'\bF_1(s-s')\Big\} ,
  \label{eq:hK0}
\end{align}
where
\begin{alignat}{3}
  \bF_0(s)
  &=\brd_r\hQ_{x}^{\dg}
   +\brd_s\hQ_{s}^{\dg}
  &&=\sum_{n=0}^{\infty}\sum_{m=0}^{\infty}
     \{(1-\delta_0^m)\mff_0^{+}+\mff_0^{-}\}
   ,\qquad&
  \mff_0^{\pm}
  &=\brd_r\hmfq_{x}^{\pm}+\brd_s\hmfq_{s}^{\pm} ,
   \\
  \bF_1(\vsig)
  &=\brd_r\hN_{x}^{\dg}
   +\brd_s\hN_{s}^{\dg}
  &&=\frac{\theta(\vsig)}{\rho}\sum_{n=0}^{\infty}\sum_{m=0}^{\infty}
     \{(1-\delta_0^m)\mff_1^{+}+\mff_1^{-}\}
   ,\qquad&
  \mff_1^{\pm}
  &=\brd_{r}\hmfn_x^{\pm}+\brd_s\hmfn_s^{\pm} .
\end{alignat}
$\hQ_{x,s}^{\dg}$ and $\hN_{x,s}^{\dg}$ are given by
Eqs.(\ref{eq:hPxs_dg}) and (\ref{eq:hNxs_dg}).
We substitute Eqs.(\ref{eq:hmfq_x_p}-\ref{eq:hmfq_s_m}) and
(\ref{eq:hmfn_x_p}-\ref{eq:hmfn_s_m}) into $\mff_{0,1}^{\pm}$,
\begin{align}
  \mff_0^{+}
  &=0
  ,\qquad
  \mff_1^{+}
  =0 ,
   \\
  \mff_0^{-}
  &=
    \Big[
       \frac{\rho}{\mu_m^n}\sin\!\Big(\frac{\mu_m^ns}{\rho}\Big)
       \{\tE_y(0)ik\beta\rd_{r'}-c\tB_s(0)(k_r^n)^2\}
      +\frac{\rho}{r'}\cos\!\Big(\frac{\mu_m^ns}{\rho}\Big)
       c\tB_y(0)\rd_{y'}
    \Big]\cR_{-}^{mn}\cY_{-}^n ,
   \\
  \mff_1^{-}
  &=\rho
    \Big[
      \tJ_s\sin\!\Big(\frac{\mu_m^n\vsig}{\rho}\Big)\frac{\hr'\rd_{\hr'}}{\mu_m^n}
     +\tJ_x\cos\!\Big(\frac{\mu_m^n\vsig}{\rho}\Big)
    \Big]\cR_{-}^{mn}\rd_{y'}\cY_{-}^n .
\end{align}
Then Eq.(\ref{eq:hK0}) is rewritten as
\begin{align}
  K_0
  &=\int_{r_a}^{r_b}dr'\int_{-h/2}^{h/2}dy'
    \bigg\{
    \sum_{n=0}^{\infty}\sum_{m=0}^{\infty}U_m^n\cR_{-}^{mn}\cY_{-}^n
   -Z_0\int_{0}^{s}ds'
    \sum_{n=0}^{\infty}\rd_{y'}\cY_{-}^n\sum_{m=0}^{\infty}
    \bU_m^n
    \cR_{-}^{mn}
    \bigg\} ,
   \label{eq:nomonopole_1st}
\end{align}
where $U_m^n$ and $\bU_m^n$ are operators which have $\rd_{y'}$ and $\rd_{\hr'}$,
\begin{align}
  U_m^n
  &=
       \frac{\rho}{r'}\cos\!\Big(\frac{\mu_m^ns}{\rho}\Big)
       c\tB_y(0)\rd_{y'}
      +\frac{\rho}{\mu_m^n}\sin\!\Big(\frac{\mu_m^ns}{\rho}\Big)
       \{ik\beta\tE_y(0)\rd_{r'}-c\tB_s(0)(k_r^n)^2\} ,
   \label{eq:Umn}
   \\
  \bU_m^n
  &=
        \tJ_s\sin\!\Big(\frac{\mu_m^n\vsig}{\rho}\Big)\frac{\hr'\rd_{\hr'}}{\mu_m^n}
       +\tJ_x\cos\!\Big(\frac{\mu_m^n\vsig}{\rho}\Big) .
\end{align}

Next, we calculate $\rd_yc\tB_y$ which is the second term on the R.H.S. of
the second equation of (\ref{eq:No_monopole_appdx}),
\begin{align}
  \rd_yc\tB_y
  &=K_1
   -Z_0K_2 ,
  \label{eq:rdy_tBy_intm}
\end{align}
where
\begin{align}
  K_1
  =\int_{r_a}^{r_b}dr'\int_{-h/2}^{h/2}dy'\rd_y\hQ_{y}^{\dg} ,
    \qquad
  K_2
  =\int_{r_a}^{r_b}dr'\int_{-h/2}^{h/2}dy'\int_{0}^{\infty}ds'\rd_y\hN_{y}^{\dg} .
  \label{eq:K1_K2}
\end{align}
We calculate $\rd_y\hQ_y^{\dg}$
using Eq.(\ref{eq:hQy_dg}) and the second equation of (\ref{eq:cYpm_relations}),
\begin{align}
  &
  \rd_y\hQ_y^{\dg}
  =-\sum_{n=1}^{\infty}\sum_{m=0}^{\infty}V_m^n\cR_{-}^{mn}\cY_{-}^n ,
   \label{eq:rdy_hQy}
   \\
  &
  V_m^n
  =\frac{\rho}{r'}\cos\!\Big(\frac{\mu_m^ns}{\rho}\Big)c\tB_y(0)\rd_{y'}
   +\frac{\rho}{\mu_m^n}\sin\!\Big(\frac{\mu_m^ns}{\rho}\Big)
    \{(k_y^n)^2c\tB_s(0)+ik\beta\tE_x(0)\rd_{y'}\} .
   \label{eq:Vmn}
\end{align}
$K_1$ involves $\tE_x(0)\rd_{y'}\cY_{-}^n$ since $V_m^n$ is an operator which has $\rd_{y'}$.
In the first equation of (\ref{eq:K1_K2}) we integrate the term
$\tE_x(0)\rd_{y'}\cY_{-}^n$ by parts with respect to $y'$ and rewrite it as follows,
\begin{align}
  I_y
  &=\int_{-h/2}^{h/2}dy'\tE_x(0)\rd_{y'}\cY_{-}^n
   =-\int_{-h/2}^{h/2}dy'\cY_{-}^n\rd_{y'}\tE_x(0)
   =\int_{-h/2}^{h/2}dy' \{ik\beta c\tB_s(0)-\rd_{r'}\tE_y(0)\}\cY_{-}^n ,
  \label{eq:Iy_intm}
\end{align}
where we used the longitudinal component of Faraday's law (\ref{eq:Faraday_s_FD2}) with
respect to $\xv'$ at $s'=0$, \ie,
\begin{align}
  \rd_{y'}\tE_x(\xv_{\perp}',0)
  =\rd_{r'}\tE_y(\xv_{\perp}',0)-ik\beta c\tB_s(\xv_{\perp}',0) .
  \label{eq:faraday_s_appdx}
\end{align}
According to Eq.(\ref{eq:Iy_intm}),
$K_1$ is rewritten using $U_m^n$ given by Eq.(\ref{eq:Umn}),
\begin{align}
  K_1
  &=-\int_{r_a}^{r_b}dr'\int_{-h/2}^{h/2}dy'
    \sum_{n=1}^{\infty}\sum_{m=0}^{\infty}U_m^n\cR_{-}^{mn}\cY_{-}^n .
   \label{eq:K1}
\end{align}
We rewrite $K_2$ given by Eq.(\ref{eq:K1_K2}).
Differentiating Eq.(\ref{eq:hNy_dg}) with respect to $y$,
we rewrite it using Eq.(\ref{eq:cYpm_relations}),
\begin{align}
  \rd_y\hN_y^{\dg}
  &=-\theta(\vsig)
    \sum_{n=1}^{\infty}\rd_{y'}\cY_{-}^n\sum_{m=0}^{\infty}
    \bU_m^n
    \cR_{-}^{mn} .
   \label{eq:rdy_hNy_dg}
\end{align}
Substituting Eqs.(\ref{eq:K1}-\ref{eq:rdy_hNy_dg}) into Eq.(\ref{eq:rdy_tBy_intm}), we get
\begin{align}
  \rd_yc\tB_y
  &=Z_0\int_{r_a}^{r_b}dr'\int_{-h/2}^{h/2}dy'\int_{0}^{s}ds'
    \sum_{n=1}^{\infty}\rd_{y'}\cY_{-}^n\sum_{m=0}^{\infty}
    \bU_m^n
   \cR_{-}^{mn}
  \nonumber\\&\quad
  -\int_{r_a}^{r_b}dr'\int_{-h/2}^{h/2}dy'\sum_{n=1}^{\infty}\sum_{m=0}^{\infty}
   U_m^n\cR_{-}^{mn}\cY_{-}^n
   \\
  &=-K_0 .
   \label{eq:nomonopole_2nd}
\end{align}
Eq.(\ref{eq:nomonopole_2nd}) is equivalent to Eq.(\ref{eq:No_monopole_appdx}).
It follows that the expression of the magnetic field $\tBv$
given by Eq.(\ref{eq:tEBv_cmpl_dg})
satisfies the law of absence of magnetic monopole in the frequency domain.
In this proof we used Faraday's law for $\tE_{x,y}$ and $\tB_s$ at $s'=0$,
given by Eq.(\ref{eq:faraday_s_appdx}), \ie,
we assume that the initial fields are correctly given.

\subsection{Verification of the expressions of the fields through Ampere's law}

We will show that the expressions of $\tB_{x,s}$ and $\tE_y$, given by
Eq.(\ref{eq:tEBv_cmpl_dg}), satisfy Eq.(\ref{eq:Ampere_y_FD2})
which is the vertical component of Ampere's law in the frequency domain,
\begin{align}
  \brd_sc\tB_x-\brd_r c\tB_s
  =Z_0\tJ_y-ik\beta\tE_y .
  \label{eq:ampere_y_appdx}
\end{align}
We substitute $\tB_{x,s}$, given by Eq.(\ref{eq:tEBv_cmpl_dg}),
into the L.H.S. of Eq.(\ref{eq:ampere_y_appdx}),
\begin{align}
  \brd_sc\tB_x-\brd_rc\tB_s
  =\int_{r_a}^{r_b}dr'\int_{-h/2}^{h/2}dy'
   \Big\{\hG_0(s)+Z_0\int_{0}^{\infty}ds'\hG_1(s-s')\Big\} .
  \label{eq:LHS_ampere_y}
\end{align}
$\hG_0$ and $\hG_1$ are given as
\begin{alignat}{3}
  \hG_0(s)
  &=\brd_s\hQ_{x}^{\dg}-\brd_{r}\hQ_{s}^{\dg}
  &&=\sum_{n=0}^{\infty}\sum_{m=0}^{\infty}
     \{(1-\delta_0^m)\mfg_0^{+}+\mfg_0^{-}\} ,
  \label{eq:G0_appdx}
   \\
  \hG_1(\vsig)
  &=\brd_{r}\hN_{s}^{\dg}-\brd_s\hN_{x}^{\dg}
  &&=\sum_{n=0}^{\infty}
    \bigg[
       \theta(\vsig)\sum_{m=0}^{\infty}
       \{(1-\delta_0^m)\mfg_1^{+}+\mfg_1^{-}\}
      +\hat{\cG}_1^n
    \bigg] ,
   \label{eq:G1_intm}
\end{alignat}
where
\begin{align}
  \mfg_0^{\pm}
  =\brd_s\hmfq_x^{\pm}-\brd_{r}\hmfq_s^{\pm} ,
    \qquad
  \mfg_1^{\pm}
  =\brd_{r}\hmfn_s^{\pm}-\brd_s\hmfn_x^{\pm} .
\end{align}
$\hQ_{x,s}^{\dg}$ and $\hN_{x,s}^{\dg}$ are
given by Eqs.(\ref{eq:hPxs_dg}) and (\ref{eq:hNxs_dg}).
$\hat{\cG}_1^n$ is given and rewritten as
\begin{align}
  \hat{\cG}_1^n
  &=-\frac{\rho}{r}\delta(\vsig)\sum_{m=0}^{\infty}
     \{(1-\delta_0^m)\hmfn_{x}^{+}+\hmfn_{x}^{-}\}
   =\tJ_y(\xv')\delta(r-r')\delta(s-s')\cY_{-}^n(y,y') ,
   \label{eq:Sigmax}
\end{align}
where we used Eqs.(\ref{eq:hmfn_x_p}-\ref{eq:hmfn_s_m}) and Eq.(\ref{eq:rdel}).
From Eqs.(\ref{eq:hmfq_x_p}-\ref{eq:hmfq_s_m}) and (\ref{eq:hmfn_x_p}-\ref{eq:hmfn_s_m}), 
we get
\begin{align}
  \mfg_0^{+}
  &=\Big[
       \frac{\rho}{\nu_m^n}\sin\!\Big(\frac{\nu_m^ns}{\rho}\Big)
       \{c\tB_y(0)\rd_{r'}\rd_{y'}-c\tB_x(0)(k_r^n)^2\}
      -\tE_y(0)ik\beta\frac{\rho}{r'}\cos\!\Big(\frac{\nu_m^ns}{\rho}\Big)
    \Big]\cR_{+}^{mn}\cY_{-}^n ,
   \\
  \mfg_1^{+}
  &=\Big[
       \tJ_s(\xv')\cos\!\Big(\frac{\nu_m^n\vsig}{\rho}\Big)\rd_{y'}
      -\sin\!\Big(\frac{\nu_m^n\vsig}{\rho}\Big)
       \frac{\hr'}{\nu_m^n}\{\tJ_x(\xv')\rd_{\hr'}\rd_{y'}+k_r^n\tJ_y(\xv')\}
    \Big]\cR_{+}^{mn}\cY_{-}^n ,
   \\
  \mfg_0^{-}
  &=0 
    ,\qquad
  \mfg_1^{-}
  =0 .
\end{align}
Using Eq.(\ref{eq:delta_yyp}), $\hG_1$ is rewritten as follows,
\begin{align}
  \hG_1
  &=\theta(\vsig)\sum_{n=0}^{\infty}\sum_{m=1}^{\infty}
    \Big[
       \tJ_s(\xv')\cos\!\Big(\frac{\nu_m^n\vsig}{\rho}\Big)\rd_{y'}
      -\frac{\hr'}{\nu_m^n}\sin\!\Big(\frac{\nu_m^n\vsig}{\rho}\Big)
       \{\tJ_x(\xv')\rd_{\hr'}\rd_{y'}+k_r^n\tJ_y(\xv')\}
    \Big]\cR_{+}^{mn}\cY_{-}^n
   \nonumber\\&\quad
   +\tJ_y(\xv')\delta(r-r')\delta(y-y')\delta(s-s') .
  \label{eq:G1_appdx}
\end{align}
Substituting $\hG_{0,1}$ into Eq.(\ref{eq:LHS_ampere_y}), we get
\begin{align}
  &
  c(\brd_s\tB_x-\brd_r\tB_s)
  =Z_0\tJ_y(\xv)
   +\sum_{n=0}^{\infty}\sum_{m=1}^{\infty}\int_{r_a}^{r_b}dr'\int_{-h/2}^{h/2}\!dy'
    \Big\{
       \cW_0^{mn}(s)
      +Z_0\int_{0}^{s}ds'\cW_1^{mn}(s-s')
    \Big\} ,
  \label{eq:LHS_amp_y_intm}
   \\
  &
  \cW_0^{mn}(s)
  =\Big[
      \frac{\rho}{\nu_m^n}\sin\!\Big(\frac{\nu_m^ns}{\rho}\Big)
      \{c\tB_y(0)\rd_{r'}\rd_{y'}-c\tB_x(0)(k_r^n)^2\}
     -\frac{\rho}{r'}\cos\!\Big(\frac{\nu_m^ns}{\rho}\Big)ik\beta\tE_y(0)
   \Big]\cR_{+}^{mn}\cY_{-}^n ,
   \\
  &
  \cW_1^{mn}(\vsig)
  =\Big[
       \tJ_s(\xv')\cos\!\Big(\frac{\nu_m^n\vsig}{\rho}\Big)\rd_{y'}
      -\frac{r'}{\nu_m^n}\sin\!\Big(\frac{\nu_m^n\vsig}{\rho}\Big)
       \{\tJ_x(\xv')\rd_{r'}\rd_{y'}+\tJ_y(\xv')(k_r^n)^2\}
    \Big]\cR_{+}^{mn}\cY_{-}^n .
\end{align}
We integrate $\rd_{r'}\cR_{+}^{mn}$ in $\cW_1^{mn}$ by parts, which is involved in
the $r'$-integral given by Eq.(\ref{eq:LHS_amp_y_intm}),
\begin{align}
  \int_{r_a}^{r_b}dr'r'\tJ_x\rd_{r'}\cR_{+}^{mn}
   =-\int_{r_a}^{r_b}dr'\cR_{+}^{mn}\rd_{r'}(r'\tJ_x)
   =\int_{r_a}^{r_b}dr'\cR_{+}^{mn}
    r'\Big(\rd_{y'}\tJ_y-ik\beta\tJ_0+\frac{\rho}{r'}\rd_{s'}\tJ_s\Big) ,
  \label{eq:Irp_appdx}
\end{align}
where we used Eq.(\ref{eq:eq_cont_fd}).
According to Eq.(\ref{eq:Irp_appdx}), Eq.(\ref{eq:LHS_amp_y_intm}) is rewritten as
\begin{align}
  &
  c(\brd_s\tB_x-\brd_r\tB_s)
  =Z_0\tJ_y(\xv)
   +\sum_{n=0}^{\infty}\sum_{m=1}^{\infty}\int_{r_a}^{r_b}dr'\int_{-h/2}^{h/2}dy'
    \Big\{
       \cW_0^{mn}(s)
      +Z_0\int_{0}^{s}ds'\bcW_1^{mn}(s-s')
    \Big\} ,
  \label{eq:LHS_amp_y_intm2}
\end{align}
where
\begin{align}
  &
  \bcW_1^{mn}(\vsig)
  =\Big[
      \tJ_s\cos\!\Big(\frac{\nu_m^n\vsig}{\rho}\Big)\rd_{y'}
      -\frac{r'}{\nu_m^n}\sin\!\Big(\frac{\nu_m^n\vsig}{\rho}\Big)\bT_1
   \Big]\cR_{+}^{mn}\cY_{-}^n .
   \label{bW1_mn}
   \\
  &
  \bT_1
   =\Big(\rd_{y'}\tJ_y-ik\beta\tJ_0+\frac{\rho}{r'}\rd_{s'}\tJ_s\Big)\rd_{y'}
   +(k_r^n)^2\tJ_y .
\end{align}
$\bcW_1^{mn}$ has the terms $(\rd_{s'}\tJ_s)\sin(\nu_m^n\vsig/\rho)$ and
$(\rd_{y'}\tJ_y)(\rd_{y'}\cY_{-}^n)$ which are involved in the integrals
(\ref{eq:LHS_amp_y_intm2}).
We integrate these terms by parts with respect to $s'$ and $y'$ respectively,
\begin{align}
  \int_{0}^{s}ds'
  \sin\!\Big(\frac{\nu_m^n\vsig}{\rho}\Big)\rd_{s'}\tJ_s(s')
  &=
   -\tJ_s(0)\sin\!\Big(\frac{\nu_m^ns}{\rho}\Big)
   +\int_{0}^{s}ds'\tJ_s(s')\frac{\nu_m^n}{\rho}\cos\!\Big(\frac{\nu_m^n\vsig}{\rho}\Big) ,
   \label{eq:Is_prm}
   \\
  \int_{-h/2}^{h/2}dy'(\rd_{y'}\tJ_y)(\rd_{y'}\cY_{-}^n)
  &=(k_y^n)^2\int_{-h/2}^{h/2}\tJ_y\cY_{-}^ndy' ,
   \label{eq:Iy_prm}
\end{align}
where $\vsig=s-s'$ and $\tJ_s(0)=\tJ_s(\xv_\perp',0)$.
According to Eqs.(\ref{eq:Is_prm}-\ref{eq:Iy_prm}),
Eq.(\ref{eq:LHS_amp_y_intm2}) is rewritten as
\begin{align}
  c(\brd_s\tB_x-\brd_r\tB_s)
  &=Z_0\tJ_y
   +\int_{r_a}^{r_b}dr'\int_{-h/2}^{h/2}dy'\sum_{n=0}^{\infty}\sum_{m=1}^{\infty}
    \Big\{
       \cL_m^n(s)
      +Z_0ik\beta\int_{0}^{\infty}\cM_m^n(s-s')ds'
    \Big\} ,
  \label{eq:LHS_amp_y_ZZ}
\end{align}
where
\begin{align}
  \cL_m^n(s)
  &=\Big\{
       \frac{\rho}{\nu_m^n}\sin\!\Big(\frac{\nu_m^ns}{\rho}\Big)H^n
      -\frac{\rho}{r'}\cos\!\Big(\frac{\nu_m^ns}{\rho}\Big)ik\beta\tE_y(0)
    \Big\}
    \cR_{+}^{mn}\cY_{-}^n ,
   \\
  H^n
  &=\{Z_0\tJ_s(0)+c\tB_y(0)\rd_{r'}\}\rd_{y'}
   -c\tB_x(0)(k_r^n)^2 ,
  \label{eq:H0_def}
    \\
  \cM_m^n(\vsig)
  &=\theta(\vsig)\frac{r'}{\nu_m^n}\sin\!\Big(\frac{\nu_m^n\vsig}{\rho}\Big)
    \{ik\beta\tJ_y(s')+\tJ_0(s')\rd_{y'}\}\cR_{+}^{mn}\cY_{-}^n .
\end{align}
$H^n$ is an operator which has $\rd_{r'}$ and $\rd_{y'}$.
We rewrite Eq.(\ref{eq:H0_def}) using Eq.(\ref{eq:Ampere_s_FD2}) for $s'=0$,
\begin{align}
  Z_0\tJ_s(\xv_{\perp}',0)
  =\rd_{r'}c\tB_y(\xv_{\perp}',0)-\rd_{y'}c\tB_x(\xv_{\perp}',0)
  +ik\beta\tE_s(\xv_{\perp}',0) .
   \label{eq:ampere_s_app}
\end{align}
Eq.(\ref{eq:ampere_s_app}) is the longitudinal component of Ampere's law.
We rewrite Eq.(\ref{eq:H0_def}) using Eq.(\ref{eq:ampere_s_app}),
\begin{align}
  H^n
  &=ik\beta\tE_s(0)\rd_{y'}
   +\{\rd_{r'}c\tB_y(0)\}\rd_{y'}
   +c\tB_y(0)\rd_{r'}\rd_{y'}
   -\{\rd_{y'}c\tB_x(0)\}\rd_{y'}
   -c\tB_x(0)(k_r^n)^2 ,
  \label{eq:H0_intm}
\end{align}
where the initial values of the fields are functions of $\xv_{\perp}'$,
similar to those in Eqs.(\ref{eq:ampere_s_app}),
\begin{align}
  \rd_{r'}c\tB_y(0)=\rd_{r'}c\tB_y(\xv_{\perp}',0) ,
    \qquad
  \rd_{y'}c\tB_x(0)=\rd_{y'}c\tB_x(\xv_{\perp}',0) .
\end{align}
We integrate these terms by parts with respect to $r'$ and $y'$ in
Eq.(\ref{eq:LHS_amp_y_ZZ}) which involves $H^n\cR_{+}^{mn}\cY_{-}^n$,
\begin{align}
  \int_{r_a}^{r_b}dr'\int_{-h/2}^{h/2}dy'H^n\cR_{+}^{mn}\cY_{-}^n
   =ik\beta\int_{r_a}^{r_b}dr'\int_{-h/2}^{h/2}dy'
    \{\tE_s(0)\rd_{y'}+c\tB_x(0)ik\beta\}\cR_{+}^{mn}\cY_{-}^n .
\end{align}
Then Eq.(\ref{eq:LHS_amp_y_ZZ}) is rewritten as follows,
\begin{align}
  c(\brd_s\tB_x-\brd_r\tB_s)
  &=Z_0\tJ_y
   -ik\beta\int_{r_a}^{r_b}dr'\int_{-h/2}^{h/2}dy'
    \sum_{n=0}^{\infty}\sum_{m=1}^{\infty}
    \Big\{
       \cP_m^n(s)
      -Z_0\int_{0}^{\infty}ds'\cM_m^n(s-s')
    \Big\} .
  \label{eq:LHS_amp_y_barH}
\end{align}
$\cP_m^n$ is given as
\begin{align}
  \cP_m^n(s)
  &=
    \Big[
       \frac{\rho}{r'}\cos\!\Big(\frac{\nu_m^ns}{\rho}\Big)\tE_y(0)
      -\frac{\rho}{\nu_m^n}\sin\!\Big(\frac{\nu_m^ns}{\rho}\Big)
       \{\tE_s(0)\rd_{y'}+c\tB_x(0)ik\beta\}
    \Big]\cR_{+}^{mn}\cY_{-}^n .
\end{align}
$\cP_m^n$ and $\cM_m^n$ are the summands of $\hP_y^{\dg}$ and $\hM_y^{\dg}$
as seen from Eqs.(\ref{eq:hPy_dg}) and (\ref{eq:hMy_dg}),
\begin{align}
  \hP_y^{\dg}(s)
  =\sum_{n=0}^{\infty}\sum_{m=1}^{\infty}\cP_m^n(s) ,
    \qquad
  \hM_y^{\dg}(\vsig)
  =\sum_{n=0}^{\infty}\sum_{m=1}^{\infty}\cM_m^n(\vsig) .
   \label{eq:hPMy}
\end{align}
We rewrite Eq.(\ref{eq:LHS_amp_y_barH}) using Eqs.(\ref{eq:hPMy}),
\begin{align}
  c(\brd_s\tB_x-\brd_r\tB_s)
  &=Z_0\tJ_y
   -ik\beta\int_{r_a}^{r_b}dr'\int_{-h/2}^{h/2}dy'
    \Big\{
       \hP_y^{\dg}(s)
      -\int_{0}^{\infty}ds'\hM_y^{\dg}(s-s')
    \Big\}
   \\
  &=Z_0\tJ_y-ik\beta\tE_y ,
   \label{eq:ampy}
\end{align}
where we used the expression of $\tE_y$ given by Eq.(\ref{eq:tEBv_cmpl_dg}).
Eq.(\ref{eq:ampy}) agrees with the R.H.S. of Eq.(\ref{eq:ampere_y_appdx}), \ie,
the expressions of $\tB_{x,s}$ and $\tE_y$ given by Eq.(\ref{eq:tEBv_cmpl_dg})
satisfy the vertical component of Ampere's law.
In this proof we used the longitudinal component of Ampere's law at $s=0$ in
Eq.(\ref{eq:ampere_s_app}).
That is, we assume that the initial values of $\tE_s$ and $\tB_{x,y}$ at the entrance of
the bending section ($s=0$) satisfy Ampere's law.
In addition, in Eq.(\ref{eq:Irp_appdx}) we used Eq.(\ref{eq:eq_cont_fd})
which is the equation of continuity for the current in the frequency domain.

\clearpage

\section{Longitudinal electric field and impedance of a rigid bunch}
\label{sec:impedance}

We have four expressions of the transient field of CSR as described in
Eq.(\ref{eq:4express}), however, we cannot use these expressions in the numerical calculation
as they are, because they include the space charge field of the current which is singular.
We will show the expressions of $\tE_s$ and $Z$ which are usable in
the numerical calculation.
$\tE_s$ and $Z$ are the longitudinal electric field in the frequency domain
and the impedance of a transient CSR.
In order to find numerically computable expressions of $\tE_s$ and $Z$,
we rewrite Eq.(\ref{eq:tEBv_dg}) which is the differential expression of the field in
the separated form.
Eq.(\ref{eq:tEBv_dg}) has no terms which diverge exponentially for $s\to\infty$.
In this appendix we assume that the beam current is a rigid and horizontally thin bunch 
moving along the $s$-axis at a constant speed $v$ $(=c\beta)$ as
shown in Eqs.(\ref{eq:thin_bunch}).
The vertical and longitudinal charge distributions of the bunch are arbitrary
under the assumptions listed in section \ref{sec:assumption}.

We can consider several expressions of $\tE_s$ in a transient state, depending on
whether separating the initial field at $s=0$ or the steady field in the limit of
$s\to\infty$, which includes the space charge field in the perfectly conducting pipe
as discussed in appendices \ref{sec:space_charge} and \ref{sec:steady_thin}.
We will show the following expressions of $\tE_s$ and $Z$ in
the curved pipe (CP) or the straight pipe (SP) which follows the curved one.
\begin{alignat}{3}
  &
  \text{Appendix:}
   \qquad&&
  \text{Field/Impedance}
   ,\qquad&&
  \text{Separation,}
   \nonumber
   \\
  &
  \text{\ref{sec:Es_bend} (CP):}
   \qquad&&
  \tE_s
   ,\qquad&&
  \text{None,}
   \label{app_Es_1}
   \\
  &
  \text{\ref{sec:Es_bend} (CP):}
   \qquad&&
  \tE_s=\tE_s^0+\Del\tE_s
   ,\qquad&&
  \text{Initial field $+$ rest,}
   \label{app_Es_2}
   \\
  &
  \text{\ref{sec:Es_bend_std} (CP):}
   \qquad&&
  \tE_s=\tE_s^{\infty}+\Del\brE_s
   ,\qquad&&
  \text{Steady field $+$ rest,}
   \label{app_Es_3}
   \\
  &
  \text{\ref{sec:Es_str} (SP):}
   \qquad&&
  \tE_s
   ,\qquad&&
  \text{None,}
   \label{app_Es_4}
   \\
  &
  \text{\ref{sec:Es_str} (SP):}
   \qquad&&
  \tE_s=\tE_s^{\infty}+\Del\tE_s
   ,\qquad&&
  \text{Steady field $+$ rest,}
   \label{app_Es_5}
   \\
   \nonumber
   \\
  &
  \text{\ref{sec:impedance_trans} (CP):}
   \qquad&&
  Z
   ,\qquad&&
  \text{None,}
   \label{app_Z_1}
   \\
  &
  \text{\ref{sec:impedance_trans} (CP):}
   \qquad&&
  Z=Z_{i}+\Del Z
   ,\qquad&&
  \text{Initial field $+$ rest,}
   \label{app_Z_2}
   \\
  &
  \text{\ref{sec:Z_mns_Zstd} (CP):}
   \qquad&&
  Z=Z_{s}+\Del \breve{Z}
   ,\qquad&&
  \text{Steady field $+$ rest,}
   \label{app_Z_3}
   \\
  &
  \text{\ref{sec:impedance_str} (SP):}
   \qquad&&
  \bZ
   ,\qquad&&
  \text{None,}
   \label{app_Z_4}
   \\
  &
  \text{\ref{sec:impedance_str} (SP):}
   \qquad&&
  \bZ=\bZ_{s}+\Del\bZ
   ,\qquad&&
  \text{Steady field $+$ rest.}
   \label{app_Z_5}
\end{alignat}
We define the origin of the $s$-axis ($s=0$) at the entrance of
the curved pipe and the straight pipe respectively in
Eqs.(\ref{app_Es_1}-\ref{app_Es_3}) and (\ref{app_Es_4}-\ref{app_Es_5}).
$\tE_s^{\infty}$ in Eqs.(\ref{app_Es_3}) and (\ref{app_Es_5}) denotes $\tE_s$ in
the steady state, \ie, $\tE_s^{\infty}$ is the final field in the limit of $s\to\infty$
under the assumption that the pipe extends semi-infinitely in
the bending section or straight section.
In calculating $\tE_s^{\infty}$, we must take the limit of $s\to\infty$
as described in Eqs.(\ref{eq:tEs_init}) and (\ref{eq:tEs_std})
since we use the Fourier transform with respect to $t$ (not $z$) in the present paper,
\begin{align}
  \tE_s^{\infty}(\xv)
  =e^{iks}\lim_{s\to\infty}\tE_s(\xv)e^{-iks} ,
    \qquad
  \xv=(x,y,s) .
   \label{Es_inf}
\end{align}
$\tE_s^{\infty}(s)e^{-iks}$ does not depend on $s$.
$\Del\brE_s$ and $\Del\tE_s$ in Eqs.(\ref{app_Es_3}) and (\ref{app_Es_5}) are
the differences of $\tE_s$ from $\tE_s^{\infty}$ in the curved pipe and straight pipe,
respectively.
$\tE_s^{\infty}$ in Eq.(\ref{app_Es_3}) is the steady field in the curved pipe,
which consists of the steady radiation and the space charge field in the curved pipe
as described in appendix \ref{sec:steady_thin}.
They are usually inseparable unless we consider a special case such as
Eq.(87) in \cite{agoh}, which is the asymptotic expression of the longitudinal impedance
of a steady field of CSR for $k\to0$.
On the other hand, $\tE_s^{\infty}$ in Eq.(\ref{app_Es_5}) is the space charge field
which satisfies the boundary condition of the perfectly conducting straight rectangular pipe
as shown in appendix \ref{sec:space_charge}.
We use Eqs.(\ref{eq:tEs_init}) as $\tE_s^0e^{-iks}$
which is the initial field at
the entrance of the bend in Eqs.(\ref{app_Es_1}-\ref{app_Es_3}), \ie,
we assume that the initial field before entering the bend is
the space charge field in the straight pipe.
In short, we use the expression of $\tE_s^{\infty}$ in
Eq.(\ref{app_Es_5}) as $\tE_s^0$ in Eq.(\ref{app_Es_1}-\ref{app_Es_3}).
The expressions of $Z$ given by Eqs.(\ref{app_Z_1}-\ref{app_Z_5}) correspond to 
Eqs.(\ref{app_Es_1}-\ref{app_Es_5}) in this order, \ie, we deal with the rest part of $Z$, 
separating the part coming from the initial or steady field.

We derived Eq.(\ref{app_Es_2}) to remove the space charge field from $\tE_s$ in the bend.
In the numerical calculation, however, it is difficult to calculate the value of
$\Del\tE_s$ in Eq.(\ref{app_Es_2}) since it has sums which do not converge
as explained in appendix \ref{sec:Es_bend}.
Similarly, although we can separate $\tE_s^0$ from $\tE_s$ in
the straight pipe following the bend, it is unusable in the numerical calculation.
Among Eqs.(\ref{app_Es_1}-\ref{app_Z_5}), only Eqs.(\ref{app_Es_3}) and (\ref{app_Es_5}) 
are usable in the numerical calculation.
Nevertheless, we also show the unusable expressions to explain the reason.
Although we cannot use Eqs.(\ref{app_Es_1}) and (\ref{app_Es_4}) themselves in
the numerical calculation, they are the intermediate expressions to get
Eqs.(\ref{app_Es_3}) and (\ref{app_Es_5}).
We cannot use Eqs.(\ref{app_Z_1}-\ref{app_Z_5}) in the numerical calculation, \ie,
we found no expression of $Z$ which is numerically computable.
In section \ref{sec:discussion} we computed the value of $Z$ in the bending magnet by
integrating $\tE_s$, given by Eq.(\ref{app_Es_3}), numerically with respect to $s$.

\subsection{Longitudinal electric field in a bend}
\label{sec:Es_bend}

We show the expressions of $\tE_s$ described in Eqs.(\ref{app_Es_1}-\ref{app_Es_2})
which are unusable in the numerical calculation.
Substituting Eqs.(\ref{eq:thin_bunch}) into $\tE_s$ given by Eq.(\ref{eq:tEBv_dg}),
and integrating it with respect to $r'$ and $s'$, we get the expression of
the transient $\tE_s$ created by a rigid bunch moving in the bending section,
\begin{align}
  \bigg\{{ \tE_s(\xv) \atop \Del\tE_s(\xv)}\bigg\}
  &=\frac{q\tlam_0}{\eps_0h}
    \sum_{n=1}^{\infty}\bigg\{{ \hcE_s^n(x,s) \atop \Del\hcE_s^n(x,s)}\bigg\}\hcY_{+}^n(y) ,
   \qquad
  \tlam_0(k)
  =\tlam(k,s)e^{-iks} .
   \label{eq:tEs_thn}
\end{align}
$\tlam_0$ is the spectrum of the rigid bunch as in Eq.(\ref{eq:ribbon_str}).
$\lam$ does not have to be a Gaussian distribution with respect to $z$
if $\tlam_0$ is computable.
For convenience, we use both $x$ and $r$ $(=\rho+x)$ as the horizontal variable in the bend
though it is enough to use either one.
Similarly, we use both $x'$ and $r'$ $(=\rho+x')$.
$\hcY_{+}^n$ is given by Eq.(\ref{eq:hcYp}).
$\Del\tE_s$ is the difference of $\tE_s$ from the initial $\tE_s$ at $s=0$
as described in Eq.(\ref{app_Es_2}),
\begin{align}
  \tE_s(\xv)e^{-iks}
  &=\tE_s(\xv_\perp,0)+\Del\tE_s(\xv)e^{-iks} ,
    \qquad
  \hcE_s^n(x,s)e^{-iks}
  =\hcE_s^n(x,0)+\Del\hcE_s^n(x,s)e^{-iks}.
   \label{eq:tEs_tEs0}
\end{align}
$\hcE_s^n$ and $\Del\hcE_s^n$ are the Fourier coefficients of
$\tE_s$ and $\Del\tE_s$ with respect to $y$ as in section \ref{sec:ymode},
\begin{align}
  \bigg\{{ \hcE_s^n(x,s) \atop \Del\hcE_s^n(x,s)}\bigg\}
  &=
    \sum_{\ell=0}^{\infty}
    \bigg[
      \int_{x_a}^{x_b}\bigg\{{ K_{\ell}^n(x,x',s) \atop \Del K_{\ell}^n(x,x',s)}\bigg\}dx'
     -\bigg\{{ L_{\ell}^n(x,s) \atop \Del L_{\ell}^n(x,s)}\bigg\}
    \bigg] .
   \label{eq:tEs_KL_thn}
\end{align}
Using Eq.(\ref{eq:tEs_init}) as $\tE_s(\xv_\perp,0)$ $(=\tE_s^0e^{-iks})$ in
Eq.(\ref{eq:tEs_tEs0}), $\hcE_s^n$ at $s=0$ is given by Eq.(\ref{eq:cZ0_straight}),
\begin{align}
  \hcE_s^n(x,0)
  =\ccE_s^n(x,0)
  =\ccE_s^n(x,s)e^{-iks}
  =-\frac{2ik}{\gamma^2}\cd\frac{\mfX_{+}^n(x)}{\bk_x^n} ,
    \qquad
  (\bk_x^n)^2
  =(k_y^n)^2+(k/\gamma)^2 .
   \label{eq:ccEs_th}
\end{align}
$\gamma$ is the Lorentz factor.
$\ccE_s^n$ is the Fourier coefficient of the steady $\tE_s$ in the straight rectangular pipe
which is perfectly conducting.
That is, $\ccE_s^n$ is the Fourier coefficient of the longitudinal space charge field
of the thin bunch in the straight pipe.
$\mfX_{+}^n$ is given by Eq.(\ref{eq:mfHp}).
As in Eq.(\ref{eq:bkx}), $i\bk_x^n$ is the horizontal wavenumber of the space charge 
field in the straight pipe.
$K_{\ell}^n$ and $L_{\ell}^n$ are the following functions which have respectively
the initial value of $\tE_s$ at $s=0$ and the source term of $\tE_s$ in $s>0$,
\begin{align}
  K_{\ell}^n(x,x',s)
  &=
    \Th_{+}^{n\ell}
    \bigg\{
       \bdelta_{01}^{\ell}\sum_{m=1}^{\ell-1}\kap_{+}^{mn}
      +\sum_{m=\ell'}^{\infty}\bkap_{+}^{mn}
    \bigg\}
   +\Th_{-}^{n\ell}
    \bigg\{
       \bdelta_{0}^{\ell}\sum_{m=0}^{\ell-1}\kap_{-}^{mn}
      +\sum_{m=\ell}^{\infty}\bkap_{-}^{mn}
    \bigg\} ,
   \label{eq:Kln}
   \\
  L_{\ell}^n(x,s)
  &=\Th_{+}^{n\ell}
    \bigg\{
       \bdelta_{01}^{\ell}\sum_{m=1}^{\ell-1}\lm_{+}^{mn}
      +\sum_{m=\ell'}^{\infty}\blam_{+}^{mn}
    \bigg\}
   +\Th_{-}^{n\ell}
    \bigg\{
       \bdelta_{0}^{\ell}\sum_{m=0}^{\ell-1}\lm_{-}^{mn}
      +\sum_{m=\ell}^{\infty}\blam_{-}^{mn}
    \bigg\} ,
   \label{eq:Lln}
   \\
  \Del K_{\ell}^n(x,x',s)
  &=K_{\ell}^n(x,x',s)-K_{\ell}^n(x,x',0)e^{iks}
   \\
  &=
    \Th_{+}^{n\ell}
    \bigg\{
       \bdelta_{01}^{\ell}\sum_{m=1}^{\ell-1}
       \Del\kap_{+}^{mn}
      +\sum_{m=\ell'}^{\infty}
       \Del\bkap_{+}^{mn}
    \bigg\}
   +\Th_{-}^{n\ell}
    \bigg\{
       \bdelta_{0}^{\ell}\sum_{m=0}^{\ell-1}
       \Del\kap_{-}^{mn}
      +\sum_{m=\ell}^{\infty}
       \Del\bkap_{-}^{mn}
    \bigg\} ,
   \label{eq:Del_Kln}
   \\
  \Del L_{\ell}^n(x,s)
  &=L_{\ell}^n(x,s)-L_{\ell}^n(x,0)e^{iks}
   \\
  &=
    \Th_{+}^{n\ell}
    \bigg\{
       \bdelta_{01}^{\ell}\sum_{m=1}^{\ell-1}
       \lm_{+}^{mn}
      +\sum_{m=\ell'}^{\infty}
       \Del\blam_{+}^{mn}
    \bigg\}
   +\Th_{-}^{n\ell}
    \bigg\{
       \bdelta_{0}^{\ell}\sum_{m=0}^{\ell-1}
       \lm_{-}^{mn}
      +\sum_{m=\ell}^{\infty}
       \Del\blam_{-}^{mn}
    \bigg\} .
   \label{eq:Del_Lln}
\end{align}
$\Th_{\pm}^{n\ell}$, $\ell'$, $\bdelta_{01}^{\ell}$ and $\bdelta_{0}^{\ell}$ are given by
Eqs.(\ref{eq:Th_def}-\ref{eq:bTh_def}) and (\ref{eq:lprm}-\ref{eq:d01_d0}).
The summands of $K_{\ell}^n$ and $\Del K_{\ell}^n$ are given as
\begin{align}
  \kap_{+}^{mn}
  &=i
    \big\{
       e^{i\hnu_m^ns/\rho} \Omg_{+}^n(\hnu_m^n)
      +e^{-i\hnu_m^ns/\rho} \Omg_{+}^n(-\hnu_m^n)
    \big\}
    \frac{\rho}{\hr}\cR_{+}^{mn}(r,r') ,
  \label{eq:kap_p}
   \\
  \kap_{-}^{mn}
  &=i
    \big\{
      e^{i\hmu_m^ns/\rho} \Omg_{-}^n(\hmu_m^n)
     +e^{-i\hmu_m^ns/\rho} \Omg_{-}^n(-\hmu_m^n)
    \big\}
    \rd_{\hr}\cR_{-}^{mn}(r,r') ,
  \label{eq:kap_m}
   \\
  \Del\kap_{+}^{mn}
  &=i
    \big\{
       e^{i\hnu_m^ns/\rho} \Omg_{+}^n(\hnu_m^n)
      +e^{-i\hnu_m^ns/\rho} \Omg_{+}^n(-\hnu_m^n)
      -2e^{iks}\Omg_{+}^n(0)
    \big\}
    \frac{\rho}{\hr}\cR_{+}^{mn}(r,r') ,
   \label{eq:Del_kap_p}
   \\
  \Del\kap_{-}^{mn}
  &=i
    \big\{
      e^{i\hmu_m^ns/\rho} \Omg_{-}^n(\hmu_m^n)
     +e^{-i\hmu_m^ns/\rho} \Omg_{-}^n(-\hmu_m^n)
     -2e^{iks}\Omg_{-}^n(\infty)
    \big\}
    \rd_{\hr}\cR_{-}^{mn}(r,r') ,
   \label{eq:Del_kap_m}
\end{align}
\begin{alignat}{2}
  \bkap_{+}^{mn}
  &=ie^{-\cnu_m^ns/\rho}\Omg_{+}^n(i\cnu_m^n)\frac{\rho}{\hr}\cR_{+}^{mn}(r,r') ,
   \qquad&
  \Del\bkap_{+}^{mn}
  &=i(e^{-\cnu_m^ns/\rho}-e^{iks})\Omg_{+}^n(i\cnu_m^n)\frac{\rho}{\hr}\cR_{+}^{mn}(r,r') ,
  \label{eq:bkap_Delbkap_p}
   \\
  \bkap_{-}^{mn}
  &=ie^{-\cmu_m^ns/\rho}\Omg_{-}^n(i\cmu_m^n)\rd_{\hr}\cR_{-}^{mn}(r,r') ,
   \qquad&
  \Del\bkap_{-}^{mn}
  &=i(e^{-\cmu_m^ns/\rho}-e^{iks})\Omg_{-}^n(i\cmu_m^n)\rd_{\hr}\cR_{-}^{mn}(r,r') .
  \label{eq:bkap_Delbkap_m}
\end{alignat}
$\nu_m^n=(\hnu_m^n,i\cnu_m^n)$ and $\mu_m^n=(\hmu_m^n,i\cmu_m^n)$ are
the poles of the fields in the Laplace domain as shown in Eqs.(\ref{eq:ps_poles}).
$\Omg_{\pm}^n$ denotes the following operators with respect to $\hr'$ $(=k_r^nr')$,
\begin{align}
  \Omg_{+}^n(\nu)
  &=
    \frac{\mfX_{+}^n(x')}{\bk_x^n}
    \Big\{\frac{\nu}{\hr'}(k_y^n)^2-\frac{k}{\gamma^2}k_r^n\Big\}
   +k\beta^2\mfX_{-}^n(x')\rd_{\hr'} ,
   \label{eq:Omg_p}
   \\
  \Omg_{-}^n(\nu)
  &=
    \mfX_{-}^n(x')\Big(\frac{\rho}{\hr'}k\beta^2+\frac{\rho}{\nu}k_r^n\Big)
   +\frac{\rho}{\nu}(k_y^n)^2\frac{\mfX_{+}^n(x')}{\bk_x^n}\rd_{\hr'} .
   \label{eq:Omg_m}
\end{align}
$\Omg_{\pm}^n$ acts on the radial eigenfunctions $\cR_{\pm}^{mn}$ given by
Eqs.(\ref{eq:Ven}-\ref{eq:mfRm}) and (\ref{eq:cRp_ba}-\ref{eq:cRm_ba}).
$\rd_{\hr}\cR_{\pm}^{mn}$ and $\rd_{\hr'}\cR_{\pm}^{mn}$ are given by
Eqs.(\ref{eq:Dr_cRp}-\ref{eq:D2_mfRm}).
$\mfX_{\pm}^n$ is given by Eqs.(\ref{eq:mfHp}-\ref{eq:mfHm}) which are
the horizontal functions of the space charge field in the straight rectangular pipe.
The summands of $L_{\ell}^n$ and $\Del L_{\ell}^n$ are given as
\begin{align}
  \lm_{+}^{mn}
  &=i
    \bigg\{
       \iota(\hnu_m^n)\frac{e^{iks}-e^{i\hnu_m^ns/\rho}}{k-\hnu_m^n/\rho}
      +\iota(-\hnu_m^n)\frac{e^{iks}-e^{-i\hnu_m^ns/\rho}}{k+\hnu_m^n/\rho}
    \bigg\}
    \frac{\rho}{r}\cR_{+}^{mn}(r,\rho) ,
   \label{eq:lam_p}
   \\
  \blam_{+}^{mn}
  &=i
    \bigg\{
       \iota(i\cnu_m^n)\frac{e^{iks}-e^{-\cnu_m^ns/\rho}}{k-i\cnu_m^n/\rho}
      +\iota(-i\cnu_m^n)\frac{e^{iks}}{k+i\cnu_m^n/\rho}
     \bigg\}
     \frac{\rho}{r}\cR_{+}^{mn}(r,\rho) ,
   \label{eq:blam_p}
  \\
  \lm_{-}^{mn}
  &=i
    \bigg\{
       \frac{e^{iks}-e^{-i\hmu_m^ns/\rho}}{k+\hmu_m^n/\rho}
      -\frac{e^{iks}-e^{i\hmu_m^ns/\rho}}{k-\hmu_m^n/\rho}
    \bigg\}
    \frac{k\beta^2}{\hmu_m^n/\rho}[\rd_{\hr'}\rd_{\hr}\cR_{-}^{mn}(r,r')]_{r'=\rho} ,
   \label{eq:lam_m}
  \\
  \blam_{-}^{mn}
  &=i
    \bigg\{
       \frac{e^{iks}}{k+i\cmu_m^n/\rho}
      -\frac{e^{iks}-e^{-\cmu_m^ns/\rho}}{k-i\cmu_m^n/\rho}
    \bigg\}
    \frac{k\beta^2}{i\cmu_m^n/\rho}[\rd_{\hr'}\rd_{\hr}\cR_{-}^{mn}(r,r')]_{r'=\rho}
    ,
   \label{eq:blam_m}
  \\
  \Del\blam_{+}^{mn}
  &=i
    \bigg\{
       \iota(i\cnu_m^n)\frac{e^{iks}-e^{-\cnu_m^ns/\rho}}{k-i\cnu_m^n/\rho}
    \bigg\}
    \frac{\rho}{r}\cR_{+}^{mn}(r,\rho) ,
   \label{eq:Dblam_p}
   \\
  \Del\blam_{-}^{mn}
  &=i
    \bigg\{
      -\frac{e^{iks}-e^{-\cmu_m^ns/\rho}}{k-i\cmu_m^n/\rho}
    \bigg\}
    \frac{k\beta^2}{i\cmu_m^n/\rho}[\rd_{\hr'}\rd_{\hr}\cR_{-}^{mn}(r,r')]_{r'=\rho} .
   \label{eq:Dblam_m}
\end{align}
$\rd_{\hr'}\rd_{\hr}\cR_{\pm}^{mn}$ is given by
Eqs.(\ref{eq:bU_pls}-\ref{eq:rdr_rdrp_cRm_2}).
$\iota$ is given as
\begin{align}
  \iota(\nu)
  =1-\frac{\nu}{\rho}\cd\frac{k\beta^2}{(k_r^n)^2} .
  \label{eq:iota}
\end{align}
We cannot use Eqs.(\ref{eq:tEs_thn}-\ref{eq:tEs_tEs0}) to compute the values of $\tE_s$ in 
the numerical calculation, because the infinite series with respect to $m$,
which are involved in Eqs.(\ref{eq:Kln}-\ref{eq:Del_Lln}), do not converge exponentially.
We must separate the steady field in the curved pipe from $\tE_s$
as shown in appendix \ref{sec:Es_bend_std}.

\subsection{Longitudinal electric field in a bend, separating the steady field}
\label{sec:Es_bend_std}

We show the expression of $\tE_s$ in the bend, described in Eq.(\ref{app_Es_3})
which is computable.
We define $\brE_s$ as the steady $\tE_s$ in the bend, \ie, $\brE_s=\tE_s^{\infty}$
which includes the space charge field of the current in the curved pipe.
We define $\Del\brE_s$ as the difference of $\tE_s$ from $\brE_s$, 
\begin{align}
  \tE_s(\xv)
  =\brE_s(\xv)+\Del\brE_s(\xv) ,
    \qquad
  \Del\brE_s(\xv)
  &=\frac{q\tlam_0}{\eps_0h}\sum_{n=1}^{\infty}\Del\brcE_s^n(x,s)\hcY_{+}^n(y) .
   \label{eq:tEs_std_sep}
\end{align}
$\tE_s$ is given by the upper equation of (\ref{eq:tEs_thn}).
From Eq.(\ref{Es_inf}),
\begin{align}
  \brE_s(\xv)
  =e^{iks}\lim_{s\to\infty}\tE_s(\xv)e^{-iks}
  =\frac{q\tlam_0}{\eps_0h}\sum_{n=1}^{\infty}\brcE_s^n(x,s)\hcY_{+}^n(y) .
   \label{brEs_app}
\end{align}
$\brcE_s^n$ and $\Del\brcE_s^n$ are respectively the $n$th Fourier coefficients of
$\brE_s$ and $\Del\brE_s$, excluding the factor $q\tlam_0/(\eps_0h)$,
\begin{align}
  \brcE_s^n(x,s)
  =e^{iks}\lim_{s\to\infty}\hcE_s^n(x,s)e^{-iks} ,
    \qquad
  \hcE_s^n(x,s)
  =\brcE_s^n(x,s)+\Del\brcE_s^n(x,s) .
   \label{hcEs_brcEs}
\end{align}
We get $\brcE_s^n(x,s)$ from Eq.(\ref{eq:tEs_std}) which involves $\mfG_{s}^n$ given by
Eq.(\ref{eq:mfGs}),
\begin{align}
  \brcE_s^n(x,0)
  =\brcE_s^n(x,s)e^{-iks}
  =-2ik\mfG_{s}^n(r,\rho,k\rho) .
   \label{eq:cEs_inf}
\end{align}
As shown in Eq.(\ref{eq:bmfG_pm_delta}), $\bmfG_{-}^n$ has the radial $\delta$-function
which denotes the space charge in the bend.
$\brcE_s^n(x,0)$ includes the space charge field in the bend,
which is indistinguishable from the radiation field in Eq.(\ref{eq:cEs_inf}).

From Eqs.(\ref{eq:bmfG_pm_delta}), (\ref{eq:rdel}), (\ref{eq:sum_Xmn}) and
(\ref{eq:mfGpm_mfRpm}), we can get another expression for Eq.(\ref{eq:cEs_inf}),
\begin{align}
   &
  \mfG_{s}^n(r,\rho,k\rho)
  =\frac{i}{2k}\brcE_s^n(x,s)e^{-iks}
   =\sum_{\ell=0}^{\infty}\delta \brL_{\ell}^n(r),
   \label{eq:ik2_hcEs}
   \\
   &
  \delta \brL_{\ell}^n(r)
  =
    \Th_{+}^{n\ell}
    \bigg\{
       \bdelta_{01}^{\ell}\sum_{m=1}^{\ell-1}\del\Lam_{+}^{mn}
      +\sum_{m=\ell'}^{\infty}\del\bLam_{+}^{mn}
    \bigg\}
   +\Th_{-}^{n\ell}
    \bigg\{
       \bdelta_{0}^{\ell}\sum_{m=0}^{\ell-1}\del\Lam_{-}^{mn}
      +\sum_{m=\ell}^{\infty}\del\bLam_{-}^{mn}
    \bigg\} ,
   \label{eq:delta_Leln}
\end{align}
where
\begin{alignat}{2}
  \del\Lam_{+}^{mn}
  &=
    \frac{1}{(k_r^n)^2}
    \bigg\{\beta^2-\frac{(k_y^n)^2}{k^2-(\hnu_m^n/\rho)^2}\bigg\}
    \frac{\rho}{r}\cR_{+}^{mn}(r,\rho)
   ,\qquad&
  \del\Lam_{-}^{mn}
  &=-\beta^2\frac{[\rd_{\hr'}\rd_{\hr}\cR_{-}^{mn}(r,r')]_{r'=\rho}}{k^2-(\hmu_m^n/\rho)^2} ,
   \label{eq:delta_Lam}
   \\
  \del\bLam_{+}^{mn}
  &=
    \frac{1}{(k_r^n)^2}
    \bigg\{\beta^2-\frac{(k_y^n)^2}{k^2+(\cnu_m^n/\rho)^2}\bigg\}
    \frac{\rho}{r}\cR_{+}^{mn}(r,\rho)
   ,\qquad&
  \del\bLam_{-}^{mn}
  &=-\beta^2\frac{[\rd_{\hr'}\rd_{\hr}\cR_{-}^{mn}(r,r')]_{r'=\rho}}{k^2+(\cmu_m^n/\rho)^2} .
   \label{eq:delta_bLam}
\end{alignat}
Eq.(\ref{eq:delta_Leln}) cannot be used in the numerical calculation
since the infinite series involving $\del\bLam_{\pm}^{mn}$ with respect to $m$ in
Eq.(\ref{eq:delta_Leln}) do not converge exponentially.
These infinite series are the parts of the radial $\delta$-function
involved in Eqs.(\ref{eq:bmfG_pm_delta}) and (\ref{eq:rdel}).
Instead of the last expression of Eq.(\ref{eq:ik2_hcEs}), we use Eq.(\ref{eq:cEs_inf})
in computing $\brcE_s^n$.
We use Eq.(\ref{eq:ik2_hcEs}) to verify the fact that
$\Del\brcE_s^n$ given by Eq.(\ref{eq:Del_brcEs}) is correct.
That is, we can analytically show that Eq.(\ref{eq:Del_brcEs}) agrees with the expression 
of $\Del\brcE_s^n$ which is derived from Eq.(\ref{eq:delta_Leln}) through
Eqs.(\ref{eq:dL_L_DL}).

We consider $\Del\brcE_s^n$ defined in Eq.(\ref{hcEs_brcEs}).
We got $\brcE_s^n$, given by Eq.(\ref{eq:cEs_inf}), by
using the final value theorem in Eqs.(\ref{eq:cEy_steady}-\ref{eq:cBy_steady}).
Instead of this way, we can get $\brcE_s^n$ by taking the limit of $s\to\infty$ 
for $\hcE_s^ne^{-iks}$ using Eq.(\ref{eq:tEs_KL_thn}),
where we must take into account an infinitesimal damping in the field through $k$
as shown in Eq.(\ref{eq:torus_kmn_resona}).
Using Eq.(\ref{eq:torus_kmn_resona}) to Eqs.(\ref{hcEs_brcEs}),
we get $\Del\brcE_s^n$ as follows without using Eq.(\ref{eq:ik2_hcEs}),
\begin{align}
  \Del\brcE_s^n(x,s)
   =
    \sum_{\ell=0}^{\infty}
    \bigg\{
       \int_{x_a}^{x_b}K_{\ell}^n(x,x',s)dx'
      -\Del\brL_{\ell}^n(x,s)
    \bigg\} .
   \label{eq:Del_brcEs}
\end{align}
$K_{\ell}^n$ is given by Eq.(\ref{eq:Kln}),
which involves the initial value of $\tE_s$ at the entrance of the bend.
On the other hand, $\Del\brL_{\ell}^n$ involves the source term of the field in the bend,
\begin{align}
  \Del\brL_{\ell}^n(x,s)
  &=
       \Th_{+}^{n\ell}
       \bigg\{
          \bdelta_{01}^{\ell}\sum_{m=1}^{\ell-1}
          \Del\Lam_{+}^{mn}
         +\sum_{m=\ell'}^{\infty}
          \Del\bLam_{+}^{mn}
       \bigg\}
      +\Th_{-}^{n\ell}
       \bigg\{
          \bdelta_{0}^{\ell}\sum_{m=0}^{\ell-1}
          \Del\Lam_{-}^{mn}
         +\sum_{m=\ell}^{\infty}
          \Del\bLam_{-}^{mn}
       \bigg\} ,
   \label{eq:Del_brLln}
\end{align}
where
\begin{align}
  \Del\Lam_{+}^{mn}(r,s)
  &=-i
    \bigg\{
       \iota(\hnu_m^n)\frac{e^{i\hnu_m^ns/\rho}}{k-\hnu_m^n/\rho}
      +\iota(-\hnu_m^n)\frac{e^{-i\hnu_m^ns/\rho}}{k+\hnu_m^n/\rho}
    \bigg\}
    \frac{\rho}{r}\cR_{+}^{mn}(r,\rho) ,
   \label{eq:Del_Lam_p}
   \\
  \Del\bLam_{+}^{mn}(r,s)
  &=-i
    \iota(i\cnu_m^n)\frac{e^{-\cnu_m^ns/\rho}}{k-i\cnu_m^n/\rho}
    \cd\frac{\rho}{r}\cR_{+}^{mn}(r,\rho) ,
   \label{eq:Del_bLam_p}
   \\
  \Del\Lam_{-}^{mn}(r,s)
  &=i
    \bigg(
       \frac{e^{i\hmu_m^ns/\rho}}{k-\hmu_m^n/\rho}
      -\frac{e^{-i\hmu_m^ns/\rho}}{k+\hmu_m^n/\rho}
    \bigg)
    \frac{k\beta^2}{\hmu_m^n/\rho}[\rd_{\hr'}\rd_{\hr}\cR_{-}^{mn}(r,r')]_{r'=\rho} ,
   \\
  \Del\bLam_{-}^{mn}(r,s)
  &=i
    \frac{e^{-\cmu_m^ns/\rho}}{k-i\cmu_m^n/\rho}\cd
    \frac{k\beta^2}{i\cmu_m^n/\rho}[\rd_{\hr'}\rd_{\hr}\cR_{-}^{mn}(r,r')]_{r'=\rho} .
   \label{eq:Del_bLam_m}
\end{align}
$\iota(\nu)$ is given by Eq.(\ref{eq:iota}),
\begin{align}
  \iota(\nu)
  =1-\alp\frac{(k\beta)^2}{(k_r^n)^2}
   ,\qquad
  \alp
  =\frac{\nu}{k\rho}
   ,\qquad
  \frac{\iota(\nu)}{1-\alp}
  +\frac{\iota(-\nu)}{1+\alp}
  =\frac{2}{(k_r^n)^2}\bigg\{(k\beta)^2-\frac{(k_y^n)^2}{1-\alp^2}\bigg\} .
\end{align}
$\Del\bLam_{\pm}^{mn}$ has the exponential damping factors
$e^{-\cnu_m^ns/\rho}$ and $e^{-\cmu_m^ns/\rho}$ unlike $\del\bLam_{\pm}^{mn}$ 
given by Eqs.(\ref{eq:delta_bLam}).
By this, Eq.(\ref{eq:Del_brLln}) can be used in the numerical calculation
since the infinite series involving $\Del\bLam_{\pm}^{mn}$ with respect to $m$ converge
exponentially.
Eqs.(\ref{eq:delta_Leln}-\ref{eq:delta_bLam}) are related to
Eqs.(\ref{eq:Del_brLln}-\ref{eq:Del_bLam_m}) as follows,
\begin{align}
  f_k\del\brL_{\ell}^n
  =L_{\ell}^n-\Del\brL_{\ell}^n
   ;\qquad
  f_k\del\Lam_{\pm}^{mn}
  =\lm_{\pm}^{mn}-\Del\Lam_{\pm}^{mn}
   ,\qquad
  f_k\del\bLam_{\pm}^{mn}
  =\blam_{\pm}^{mn}-\Del\bLam_{\pm}^{mn} ,
    \label{eq:dL_L_DL}
\end{align}
where $f_k=2ike^{iks}$.
$L_{\ell}^n$ and its summands $\lm_{\pm}^{mn}$ and $\blam_{\pm}^{mn}$ are given by
Eq.(\ref{eq:Lln}) and Eqs.(\ref{eq:lam_p}-\ref{eq:blam_m}).

\clearpage

\subsection{Longitudinal electric field in a straight rectangular pipe}
\label{sec:Es_str}

We show the expressions of $\tE_s$ which propagates in a straight rectangular pipe
together with the rigid and thin bunch (\ref{eq:thin_bunch}).
It is a transient field in a perfectly conducting straight pipe,
which tends to be the space charge field for $s\to\infty$.
As described in Eqs.(\ref{app_Es_4}-\ref{app_Es_5}),
we find the two expressions of $\tE_s$ in the straight pipe,
depending on whether we separate the space charge field from $\tE_s$.
In appendix \ref{sec:Es_str} we define the origin of the $s$-axis ($s=0$)
at the entrance of the straight rectangular pipe
unlike appendices \ref{sec:Es_bend} and \ref{sec:Es_bend_std}.

Substituting Eqs.(\ref{eq:thin_bunch}) into $\tE_s$ given by Eq.(\ref{eq:EB_str_sep}),
we integrate it with respect to $x'$ and $s'$,
\begin{align}
  \bigg\{{ \tE_s(\xv) \atop \Del\tE_s(\xv)}\bigg\}
  &=\frac{q\tlam_0}{\eps_0h}\sum_{n=1}^{\infty}
    \bigg\{{ \bcE_s^n(x,s) \atop \Del\bcE_s^n(x,s)}\bigg\} \hcY_{+}^n(y) ,
    \qquad
  \tE_s(\xv)
  =\chE_s(\xv)+\Del\tE_s(\xv) .
  \label{eq:tEs_st_th}
\end{align}
$\Del\tE_s$ is the difference of $\tE_s$ from $\chE_s$ which is the space charge field
given by Eq.(\ref{eq:tEs_init}),
\begin{align}
  \chE_s(\xv)
   =\frac{q\tlam_0}{\eps_0h}\sum_{n=1}^{\infty}\ccE_s^n(x,s)\hcY_{+}^n(y) ,
    \qquad
  \bcE_s^n(x,s)
  =\ccE_s^n(x,s)+\Del\bcE_s^n(x,s) .
   \label{eq:bcEs_th_SC}
\end{align}
$\chE_s$ denotes $\tE_s^{\infty}$ in Eq.(\ref{app_Es_5}).
$\bcE_s^n$ and $\Del\bcE_s^n$ are respectively the Fourier coefficients of 
$\tE_s$ and $\Del\tE_s$ in the straight pipe.
$\ccE_s^n$ is the Fourier coefficient of $\chE_s$, given by Eq.(\ref{eq:ccEs_th}).
$\hcY_{+}^n$ is given by Eq.(\ref{eq:hcYp}).

We find the two expressions of $\bcE_s^n$, which correspond to
the first and second equations of (\ref{eq:tEs_st_th}).
It depends on whether we separate the space charge field $\ccE_s^n$ from $\bcE_s^n$.
Assuming that the initial field at the entrance of the straight pipe is
the field at the exit of the bend, $\bcE_s^n$ and $\Del\bcE_s^n$ are given as
\begin{align}
  \bigg\{{ \bcE_s^n(x,s) \atop \Del\bcE_s^n(x,s)}\bigg\}
  &=\sum_{\ell=0}^{\infty}
    \bigg[
       \int_{x_a}^{x_b}\bK_{\ell}^n(x,x',s,s_0)dx'
      -\bigg\{{ \bL_{\ell}^n(x,s) \atop \Del\bL_{\ell}^n(x,s)}\bigg\}
    \bigg] .
   \label{eq:bcEs_th}
\end{align}
$\bK_{\ell}^n$ has the initial value of the field at the entrance of the straight pipe.
$s_0$ is the length of the bending magnet along the $s$-axis.
The sum of $\bK_{\ell}^n$ over all $\ell$ is given as
\begin{align}
  \sum_{\ell=0}^{\infty}\bK_{\ell}^n(x,x',s,s')
  &=(\rd_s+\rd_{s'})\{\cG_{+}^n(x,x',s)\hcE_s^n(x',s')\} .
\end{align}
$\hcE_s^n$ is the Fourier coefficient of $\tE_s$ in the bend,
given by Eq.(\ref{eq:tEs_KL_thn}).
$\cG_{+}^n(x,x',s)$ is the Green function in the straight pipe,
given by Eq.(\ref{eq:cGp_st}).
$\bL_{\ell}^n$ and $\Del\bL_{\ell}^n$ are functions
which vanish in the limit of $\gam\to\infty$,
\begin{align}
  \bL_{\ell}^n(x,s)
  &=\bTh_{\ell}^n
    \bigg\{
       \bdelta_{01}^\ell\sum_{m=1}^{\ell-1}
       \lm_m^n(x,s)
      +\sum_{m=\ell'}^{\infty}
       \blam_m^n(x,s)
    \bigg\} ,
   \label{eq:bLnl}
   \\
  \Del\bL_{\ell}^n(x,s)
  &=\bTh_{\ell}^n
    \bigg\{
    \bdelta_{01}^\ell\sum_{m=1}^{\ell-1}
    \Del\lm_m^n(x,s)
   +\sum_{m=\ell'}^{\infty}
    \Del\blam_m^n(x,s)
    \bigg\} .
  \label{eq:Del_bLln}
\end{align}
$\bTh_{\ell}^n$ is the window function given by Eq.(\ref{eq:def_Th_ell}).
The summands of Eqs.(\ref{eq:bLnl}-\ref{eq:Del_bLln}) are given as follows,
\begin{align}
  \lm_m^n(x,s)
  &=\frac{ik}{\gamma^2k_s^{mn}}\cX_{+}^m(x,0)
    \bigg(
       \frac{e^{iks}-e^{ik_s^{mn}s}}{k-k_s^{mn}}
      -\frac{e^{iks}-e^{-ik_s^{mn}s}}{k+k_s^{mn}}
    \bigg) ,
   \label{eq:blam}
   \\
  \blam_m^n(x,s)
  &=\frac{ik}{\gamma^2k_s^{mn}}\cX_{+}^m(x,0)
    \bigg(
       \frac{e^{iks}-e^{ik_s^{mn}s}}{k-k_s^{mn}}
      -\frac{e^{iks}}{k+k_s^{mn}}
    \bigg) ,
   \\
  \Del\lm_m^n(x,s)
  &=
    -\frac{ik}{\gamma^2k_s^{mn}}\cX_{+}^m(x,0)
    \bigg(
       \frac{e^{ik_s^{mn}s}}{k-k_s^{mn}}
      -\frac{e^{-ik_s^{mn}s}}{k+k_s^{mn}}
    \bigg) ,
   \label{eq:Del_lam}
   \\
  \Del\blam_m^n(x,s)
  &=-\frac{ik}{\gamma^2k_s^{mn}}\cX_{+}^m(x,0)
    \frac{e^{ik_s^{mn}s}}{k-k_s^{mn}} .
   \label{eq:Del_blam}
\end{align}
$k_s^{mn}$ and $\cX_{+}^m$ are given by Eqs.(\ref{eq:ks_mn}) and (\ref{eq:cXp}).
The infinite series involving $\blam_m^n$ with respect to $m$ does not converge
in the numerical calculation of Eq.(\ref{eq:bLnl}).
Instead, Eq.(\ref{eq:Del_bLln}) is computable since $\Del\blam_m^n$ has
the exponential damping factor $e^{ik_s^{mn}s}=e^{-\bk_s^{mn}s}$
($\bk_s^{mn}\in\mathbb{R}^{+}$) for $m\geq \ell'$, 
similar to $\Del\bLam_{\pm}^{mn}$ in Eq.(\ref{eq:Del_brLln}).

As shown in appendix \ref{sec:impedance_str}, in calculating the longitudinal impedance
in the semi-infinite straight pipe from $s=0$ to $\infty$,
we must separate the space charge field as in the second equation of
(\ref{eq:tEs_st_th}), because the integral involving the space charge term with
respect to $s$ diverges unless $\gamma=\infty$.

\subsection{Longitudinal impedance of a transient field of radiation in a bend}
\label{sec:impedance_trans}

We find the expressions of the longitudinal impedance $Z$ in the bend, described in
Eqs.(\ref{app_Z_1}-\ref{app_Z_2}) which are both unusable in the numerical calculation.
In this appendix we define $Z$ as a function of $(\xv_{\perp},s_0,k)$,
assuming a rigid and thin bunch given by Eq.(\ref{eq:thin_bunch}),
\begin{align}
  Z(\xv_{\perp},s_0,k)
  =\frac{-1}{vq\tlam_0}\int_0^{s_0}\tE_s(\xv)e^{-iks}ds .
  \label{eq:Z_def}
\end{align}
We define the origin of the $s$-axis ($s=0$) at the entrance of the bend,
similar to appendices \ref{sec:Es_bend} and \ref{sec:Es_bend_std}.
$s_0$ is the length of the bend along the $s$-axis.
We substitute $\tE_s$ given by Eq.(\ref{eq:tEs_thn}) or (\ref{eq:tEs_tEs0}) into
Eq.(\ref{eq:Z_def}),
\begin{align}
  Z(\xv_{\perp},s_0,k)
  &=\frac{Z_0}{h\beta}\sum_{n=1}^{\infty} \hcY_{+}^n(y) \zeta_b^n(x,s_0) ,
    \qquad
  \zeta_b^n(x,s_0)
  =-\int_0^{s_0}\hcE_s^n(s)e^{-iks}ds .
   \label{eq:Zk}
\end{align}
$\hcY_{+}^n$ is given by Eq.(\ref{eq:hcYp}).
$\zeta_b^n$ is the Fourier coefficient of $Z$ in the bend,
of which $Z_0/(h\beta)$ is factored out,
\begin{align}
  \zeta_b^n(x,s_0)
  &=\sum_{\ell=0}^{\infty}
    \Big\{
       U_{\ell}^n(x,s_0)
      -\int_{x_a}^{x_b}V_{\ell}^n(x,x',s_0)dx'
    \Big\}
   \label{eq:zetan}
   \\
  &=
   -s_0\hcE_s^n(x,0)
   +\sum_{\ell=0}^{\infty}
    \Big\{
       \Del U_{\ell}^n(x,s_0)
      -\int_{x_a}^{x_b}\Del V_{\ell}^n(x,x',s_0)dx'
    \Big\} .
   \label{eq:zetan0}
\end{align}
Eqs.(\ref{eq:zetan}-\ref{eq:zetan0}) correspond to Eqs.(\ref{eq:tEs_thn}-\ref{eq:tEs_tEs0}),
\ie, Eq.(\ref{eq:zetan0}) is the expression in which
the initial value of $\tE_s$ at $s=0$ is separated.
The functions of $s_0$ involved in Eqs.(\ref{eq:zetan}-\ref{eq:zetan0}) are given as
\begin{align}
  (U_{\ell}^n,V_{\ell}^n,\Del U_{\ell}^n,\Del V_{\ell}^n)
  &=\int_{0}^{s_0}ds (L_{\ell}^n,K_{\ell}^n,\Del L_{\ell}^n,\Del K_{\ell}^n)e^{-iks} .
   \label{eq:UV_LK}
\end{align}
In Eqs.(\ref{eq:UV_LK}) and (\ref{eq:u_lam}-\ref{eq:v_kap}),
the quantities in the parentheses on the left and right hand sides correspond,
\begin{align}
  U_{\ell}^n
  &=
    \Th_{+}^{n\ell}
    \bigg\{
       \bdelta_{01}^{\ell}\sum_{m=1}^{\ell-1}
       u_{+}^{mn}
      +\sum_{m=\ell'}^{\infty}
       \bu_{+}^{mn}
    \bigg\}
   +\Th_{-}^{n\ell}
    \bigg\{
       \bdelta_{0}^{\ell}\sum_{m=0}^{\ell-1}
       u_{-}^{mn}
      +\sum_{m=\ell}^{\infty}
       \bu_{-}^{mn}
    \bigg\} ,
   \label{eq:Uln}
   \\
  V_{\ell}^n
  &=
    \Th_{+}^{n\ell}
    \bigg\{
       \bdelta_{01}^{\ell}\sum_{m=1}^{\ell-1}
       v_{+}^{mn}
      +\sum_{m=\ell'}^{\infty}
       \bv_{+}^{mn}
    \bigg\}
   +\Th_{-}^{n\ell}
    \bigg\{
       \bdelta_{0}^{\ell}\sum_{m=0}^{\ell-1}
       v_{-}^{mn}
      +\sum_{m=\ell}^{\infty}
       \bv_{-}^{mn}
    \bigg\} ,
   \label{eq:Vln}
   \\
  \Del U_{\ell}^n
  &=
    \Th_{+}^{n\ell}
    \bigg\{
       \bdelta_{01}^{\ell}\sum_{m=1}^{\ell-1}
       u_{+}^{mn}
      +\sum_{m=\ell'}^{\infty}
       \Del\bu_{+}^{mn}
    \bigg\}
   +\Th_{-}^{n\ell}
    \bigg\{
       \bdelta_{0}^{\ell}\sum_{m=0}^{\ell-1}
       u_{-}^{mn}
      +\sum_{m=\ell}^{\infty}
       \Del\bu_{-}^{mn}
    \bigg\} ,
   \label{eq:DelUln}
   \\
  \Del V_{\ell}^n
  &=
    \Th_{+}^{n\ell}
    \bigg\{
       \bdelta_{01}^{\ell}\sum_{m=1}^{\ell-1}
       \Del v_{+}^{mn}
      +\sum_{m=\ell'}^{\infty}
       \Del\bv_{+}^{mn}
    \bigg\}
   +\Th_{-}^{n\ell}
    \bigg\{
       \bdelta_{0}^{\ell}\sum_{m=0}^{\ell-1}
       \Del v_{-}^{mn}
      +\sum_{m=\ell}^{\infty}
       \Del\bv_{-}^{mn}
    \bigg\} .
   \label{eq:DelVln}
\end{align}
The summands of Eqs.(\ref{eq:Uln}-\ref{eq:DelVln}) are given from
Eqs.(\ref{eq:lam_p}-\ref{eq:Dblam_m}) and (\ref{eq:kap_p}-\ref{eq:bkap_Delbkap_m}),
\begin{align}
  (u_{\pm}^{mn},\bu_{\pm}^{mn},\Del\bu_{\pm}^{mn})
  &=\int_{0}^{s_0}ds (\lm_{\pm}^{mn},\blam_{\pm}^{mn},\Del\blam_{\pm}^{mn})
    e^{-iks} ,
   \label{eq:u_lam}
   \\
  (v_{\pm}^{mn},\bv_{\pm}^{mn},\Del v_{\pm}^{mn},\Del\bv_{\pm}^{mn})
  &=\int_{0}^{s_0}ds
    (\kap_{\pm}^{mn},\bkap_{\pm}^{mn},\Del\kap_{\pm}^{mn},\Del\bkap_{\pm}^{mn})
    e^{-iks} .
   \label{eq:v_kap}
\end{align}
Calculating the integrals (\ref{eq:u_lam}-\ref{eq:v_kap}) with respect to $s$, we get
\begin{align}
  u_{+}^{mn}
  &=\bigg\{
       2iks_0\frac{1-(\beta\hnu_m^n/k_r^n\rho)^2}{k^2-(\hnu_m^n/\rho)^2}
      -\iota(\hnu_m^n)\frac{\phi(\hnu_m^n)}{k-\hnu_m^n/\rho}
      -\iota(-\hnu_m^n)\frac{\phi(-\hnu_m^n)}{k+\hnu_m^n/\rho}
    \bigg\}
    \frac{\rho}{r}\cR_{+}^{mn}(r,\rho) ,
   \label{eq:u_p}
   \\
  u_{-}^{mn}
  &=
    \bigg\{
       \frac{\rho}{\hmu_m^n}
       \bigg[
          \frac{\phi(\hmu_m^n)}{k-\hmu_m^n/\rho}
         -\frac{\phi(-\hmu_m^n)}{k+\hmu_m^n/\rho}
       \bigg]
      -\frac{2is_0}{k^2-(\hmu_m^n/\rho)^2}
    \bigg\}
    k\beta^2[\rd_{\hr'}\rd_{\hr}\cR_{-}^{mn}(r,r')]_{r'=\rho} ,
   \label{eq:u_m}
   \\
  \bu_{+}^{mn}
  &=  
    \bigg\{
       2iks_0\frac{1+(\beta\cnu_m^n/k_r^n\rho)^2}{k^2+(\cnu_m^n/\rho)^2}
      -\iota(i\cnu_m^n)\frac{\phi(i\cnu_m^n)}{k-i\cnu_m^n/\rho}
    \bigg\}
    \frac{\rho}{r}\cR_{+}^{mn}(r,\rho) ,
   \label{eq:bu_p}
   \\
  \bu_{-}^{mn}
  &=
    \bigg\{
       \frac{\rho}{i\cmu_m^n}\cd\frac{\phi(i\cmu_m^n)}{k-i\cmu_m^n/\rho}
      -\frac{2is_0}{k^2+(\cmu_m^n/\rho)^2}
    \bigg\}
    k\beta^2[\rd_{\hr'}\rd_{\hr}\cR_{-}^{mn}(r,r')]_{r'=\rho} ,
   \label{eq:bu_m}
\end{align}
and
\begin{align}
  \Del\bu_{+}^{mn}
  &=
    \frac{\iota(i\cnu_m^n)}{k-i\cnu_m^n/\rho}
    \{\phi(k\rho)  -\phi(i\cnu_m^n)\}
    \frac{\rho}{r}\cR_{+}^{mn}(r,\rho) ,
   \label{eq:Del_bu_p}
   \\
  \Del\bu_{-}^{mn}
  &=
    \frac{-1}{k-i\cmu_m^n/\rho}
    \{\phi(k\rho) -\phi(i\cmu_m^n)\}
    \frac{k\beta^2}{i\cmu_m^n/\rho}
    [\rd_{\hr'}\rd_{\hr}\cR_{-}^{mn}(r,r')]_{r'=\rho} .
   \label{eq:Del_bu_m}
\end{align}
$\iota(\nu)$ is the function given by Eq.(\ref{eq:iota}).
$\phi(\nu)$ is given as follows,
\begin{align}
  \phi(\nu)
  =\frac{1-e^{-i(k-\nu/\rho)s_0}}{k-\nu/\rho}
   ,\qquad
  \lim_{\nu\to k\rho}\frac{\phi}{is_0}
  =\lim_{s_0\to0}\frac{\phi}{is_0}
  =1
   ,\qquad
  \lim_{s_0\to\infty}\phi
  =\frac{1}{k-\nu/\rho}
   \quad
  (k\ne\nu/\rho) .
  \label{eq:phi_nu}
\end{align}
In taking the limit of $s_0\to\infty$ for $\phi$ as in the last equation of
(\ref{eq:phi_nu}), we take into account an infinitesimal damping of the field
as shown in Eq.(\ref{eq:torus_kmn_resona}).
The summands of Eqs.(\ref{eq:Vln}) and (\ref{eq:DelVln}) are given as
\begin{align}
  v_{+}^{mn}
  &=
    \{
       \phi(\hnu_m^n)\Omg_{+}^n(\hnu_m^n)
      +\phi(-\hnu_m^n)\Omg_{+}^n(-\hnu_m^n)
    \}
    \frac{\rho}{\hr}\cR_{+}^{mn} ,
  \label{eq:v_p}
   \\
  v_{-}^{mn}
  &=
    \{
       \phi(\hmu_m^n)\Omg_{-}^n(\hmu_m^n)
      +\phi(-\hmu_m^n)\Omg_{-}^n(-\hmu_m^n)
    \}
    \rd_{\hr}\cR_{-}^{mn} ,
  \label{eq:v_m}
   \\
  \Del v_{+}^{mn}
  &=
    \{
       \phi(\hnu_m^n)\Omg_{+}^n(\hnu_m^n)
      +\phi(-\hnu_m^n)\Omg_{+}^n(-\hnu_m^n)
      -2\phi(k\rho)\Omg_{+}^n(0)
    \}
    \frac{\rho}{\hr}\cR_{+}^{mn} ,
   \label{eq:Del_v_p}
   \\
  \Del v_{-}^{mn}
  &=
    \{
       \phi(\hmu_m^n)\Omg_{-}^n(\hmu_m^n)
      +\phi(-\hmu_m^n)\Omg_{-}^n(-\hmu_m^n)
      -2\phi(k\rho)\Omg_{-}^n(\infty)
    \}
    \rd_{\hr}\cR_{-}^{mn} ,
   \label{eq:Del_v_m}
\end{align}
\begin{alignat}{2}
  \bv_{+}^{mn}
  &=\phi(i\cnu_m^n)\Omg_{+}^n(i\cnu_m^n)\frac{\rho}{\hr}\cR_{+}^{mn}
   ,\qquad &&
  \Del\bv_{+}^{mn}
  =\{\phi(i\cnu_m^n)-\phi(k\rho)\}\Omg_{+}^n(i\cnu_m^n)\frac{\rho}{\hr}\cR_{+}^{mn} ,
  \label{eq:bv_p}
   \\
  \bv_{-}^{mn}
  &=\phi(i\cmu_m^n)\Omg_{-}^n(i\cmu_m^n)\rd_{\hr}\cR_{-}^{mn}
   ,\qquad &&
  \Del\bv_{-}^{mn}
  =\{\phi(i\cmu_m^n)-\phi(k\rho)\}\Omg_{-}^n(i\cmu_m^n)\rd_{\hr}\cR_{-}^{mn} .
  \label{eq:bv_m}
\end{alignat}
We omit the radial arguments of $\cR_{\pm}^{mn}(r,r')$ in Eqs.(\ref{eq:v_p}-\ref{eq:bv_m}).
$\Omg_{\pm}^n$ is the operator given by Eqs.(\ref{eq:Omg_p}-\ref{eq:Omg_m}).
Eqs.(\ref{eq:zetan}-\ref{eq:zetan0}) cannot be used in the numerical calculation,
because the infinite series with respect to $m$ involved in
Eqs.(\ref{eq:Uln}-\ref{eq:DelVln}) do not converge exponentially.

\subsection{Longitudinal impedance of a field in a bend, separating the steady field}
\label{sec:Z_mns_Zstd}

As described in Eq.(\ref{app_Z_3}), we find another expression of $Z$
by separating the part involving $\brE_s$ which is the steady $\tE_s$ in the curved pipe,
given by Eq.(\ref{brEs_app}).
From Eqs.(\ref{hcEs_brcEs}) and (\ref{eq:Zk}), we get
\begin{align}
  \zeta_b^n(x,s_0)
  &=
   -s_0\brcE_s^n(x,0)
   +\sum_{\ell=0}^{\infty}
    \Big\{
       \Del\brW_{\ell}^n(x,s_0)
      -\int_{x_a}^{x_b}V_{\ell}^n(x,x',s_0)dx'
    \Big\} .
   \label{zetab_app}
\end{align}
$\brcE_s^n(x,0)$ and $V_{\ell}^n$ are given by Eqs.(\ref{eq:cEs_inf}) and (\ref{eq:Vln}).
$\Del\brW_{\ell}^n$ is given as
\begin{align}
  \Del\brW_{\ell}^n(x,s_0)
  &=\int_0^{s_0}\Del\brL_{\ell}^n(x,s)e^{-iks}ds
   \\
  &=
       \Th_{+}^{n\ell}
       \bigg\{
          \bdelta_{01}^{\ell}\sum_{m=1}^{\ell-1}
          \Del w_{+}^{mn}
         +\sum_{m=\ell'}^{\infty}
          \Del\bw_{+}^{mn}
       \bigg\}
      +\Th_{-}^{n\ell}
       \bigg\{
          \bdelta_{0}^{\ell}\sum_{m=0}^{\ell-1}
          \Del w_{-}^{mn}
         +\sum_{m=\ell}^{\infty}
          \Del\bw_{-}^{mn}
       \bigg\} .
   \label{eq:Del_brTln}
\end{align}
$\Th_{\pm}^{n\ell}$ and $\ell'$ are given by Eqs.(\ref{eq:Th_def}-\ref{eq:bTh_def})
and (\ref{eq:lprm}).
The summands of Eq.(\ref{eq:Del_brTln}) are given by
\begin{align}
  \Del w_{+}^{mn}(r,s_0)
  &=-
    \bigg\{
       \iota(\hnu_m^n)\frac{\phi(\hnu_m^n)}{k-\hnu_m^n/\rho}
      +\iota(-\hnu_m^n)\frac{\phi(-\hnu_m^n)}{k+\hnu_m^n/\rho}
    \bigg\}
    \frac{\rho}{r}\cR_{+}^{mn}(r,\rho) ,
   \\
  \Del\bw_{+}^{mn}(r,s_0)
  &=-\iota(i\cnu_m^n)\frac{\phi(i\cnu_m^n)}{k-i\cnu_m^n/\rho}
    \cd\frac{\rho}{r}\cR_{+}^{mn}(r,\rho) ,
   \\
  \Del w_{-}^{mn}(r,s_0)
  &=
    \bigg\{
       \frac{\phi(\hmu_m^n)}{k-\hmu_m^n/\rho}
      -\frac{\phi(-\hmu_m^n)}{k+\hmu_m^n/\rho}
    \bigg\}
    \frac{k\beta^2}{\hmu_m^n/\rho}[\rd_{\hr'}\rd_{\hr}\cR_{-}^{mn}(r,r')]_{r'=\rho} ,
   \\
  \Del\bw_{-}^{mn}(r,s_0)
  &=
    \frac{\phi(i\cmu_m^n)}{k-i\cmu_m^n/\rho}
    \cd\frac{k\beta^2}{i\cmu_m^n/\rho}
    [\rd_{\hr'}\rd_{\hr}\cR_{-}^{mn}(r,r')]_{r'=\rho} .
\end{align}
Eq.(\ref{zetab_app}) cannot be used in the numerical calculation
since the infinite series involving $\Del\bw_{\pm}^{mn}$ with respect to $m$ in
Eqs.(\ref{eq:Del_brTln}) do not converge exponentially.
At present, we have found no expression of $Z$ usable to compute the value,
excluding the numerical integration of $\tE_s$ as in Eq.(\ref{eq:Z_def}).

\subsection{Longitudinal impedance of a transient field in a straight pipe}
\label{sec:impedance_str}

We define the origin of the $s$-axis ($s=0$) at the entrance of the straight pipe,
similar to appendix \ref{sec:Es_str}.
We define the longitudinal impedance $\bZ$ using $\tE_s$
in the perfectly conducting straight pipe for a length $s_1$ as
\begin{align}
  \bZ(\xv_\perp,s_1,k)
  =\frac{-1}{vq\tlam_0}\int_{0}^{s_1}\tE_s(\xv)e^{-iks}ds .
  \label{eq:Zs_redef}
\end{align}
We assume a rigid bunch (\ref{eq:thin_bunch}) which loses no energy.
If the Lorentz factor $\gam$ is finite, $\bZ$ diverges in the limit of $s_1\to\infty$
since $\tE_s$ includes the space charge field which is proportional to $\gamma^{-2}$
as seen from Eqs.(\ref{eq:blam}-\ref{eq:Del_blam}).
If $\gamma=\infty$, $\bZ$ converges even for $s_1\to\infty$ since we assume that
the beam pipe is perfectly conducting, \ie, the longitudinal electric field goes to zero
in the limit of $s\to\infty$ if $\gamma=\infty$.
Substituting $\tE_s$, given by Eq.(\ref{eq:tEs_st_th}), into Eq.(\ref{eq:Zs_redef}),
and calculating the integral with respect to $s$ analytically, we get
\begin{align}
  \bZ(\xv_\perp,s_1,k)
  &=\frac{Z_0}{h\beta}\sum_{n=1}^{\infty}\bzeta^n(x,s_1)\hcY_{+}^n(y) .
  \label{eq:bZ_st}
\end{align}
$\bzeta^n$ is the Fourier coefficient of $\bZ$, excluding the factor $Z_0/(h\beta)$,
\begin{align}
  \bzeta^n(x,s_1)
  &=
    \sum_{\ell=0}^{\infty}
    \Big\{
       \Del\bU_{\ell}^n(x,s_1)
      -\int_{x_a}^{x_b}\bV_{\ell}^n(x,x',s_0,s_1)dx'
    \Big\}
   -s_1\ccE_s^n(x,0) .
   \label{eq:zeta_s}
\end{align}
$\ccE_s^n$ is the Fourier coefficient of the space charge field in
the straight rectangular pipe, given by Eq.(\ref{eq:ccEs_th}).
As already described, if $\gamma$ is finite, since $\ccE_s^n\ne0$,
$s_1\ccE_s^n$ diverges in the limit of $s_1\to\infty$.
Conversely, $\ccE_s^n$ tends to be zero in the ultrarelativistic limit $\gamma\to\infty$.
$\Del\bU_{\ell}^n$ and $\bV_{\ell}^n$ are given as
\begin{align}
  \Del\bU_{\ell}^n(x,s_1)
  &=\bTh_{\ell}^n
    \bigg\{
    \bdelta_{01}^\ell\sum_{m=1}^{\ell-1}
    \Del u_m^n(x,s_1)
   +\sum_{m=\ell'}^{\infty}
    \Del\bu_m^n(x,s_1)
    \bigg\} ,
   \label{eq:DU_appU6}
   \\
  \bV_{\ell}^n(x,x',s',s_1)
  &=
    \bTh_{\ell}^n
    \bigg\{
       \bdelta_{01}^\ell\sum_{m=1}^{\ell-1}v_m^n(x,x',s',s_1)
      +\sum_{m=\ell'}^{\infty}\bv_m^n(x,x',s',s_1)
    \bigg\} ,
   \label{eq:V_appU6}
\end{align}
where
\begin{align}
  \Del u_m^n(x,s_1)
  &=-\frac{k}{\gamma^2k_s^{mn}}\cX_{+}^m(x,0)
    \bigg\{
       \frac{\bphi(k_s^{mn})}{k-k_s^{mn}}
      -\frac{\bphi(-k_s^{mn})}{k+k_s^{mn}}
    \bigg\} ,
   \label{eq:Du_mn_appT}
   \\
  \Del\bu_m^n(x,s_1)
  &=-\frac{k}{\gamma^2k_s^{mn}}\cX_{+}^m(x,0)
    \frac{\bphi(k_s^{mn})}{k-k_s^{mn}} ,
   \label{eq:Dbu_mn_appT}
\end{align}
\begin{align}
  v_m^n(x,x',s',s_1)
  &=\frac{1}{2i}
       \cX_{+}^m(x,x')
       \Big\{
          \bphi(k_s^{mn})\Big(1+\frac{\rd_{s'}}{ik_s^{mn}}\Big)
         +\bphi(-k_s^{mn})\Big(1-\frac{\rd_{s'}}{ik_s^{mn}}\Big)
       \Big\}
       \hcE_s^n(x',s') ,
   \label{eq:vmn_appT}
   \\
  \bv_m^n(x,x',s',s_1)
  &=\frac{1}{2i}
       \cX_{+}^m(x,x')\bphi(k_s^{mn})
       \Big(1+\frac{\rd_{s'}}{ik_s^{mn}}\Big)\hcE_s^n(x',s') .
   \label{eq:bvmn_appT}
\end{align}
$k_s^{mn}$, $\bTh_{\ell}^n$ and $\cX_{+}^m$ are given by Eq.(\ref{eq:ks_mn}),
(\ref{eq:def_Th_ell}) and (\ref{eq:cXp}).
$\hcE_s^n$ is the Fourier coefficient of $\tE_s$ in the curved pipe,
given by Eq.(\ref{eq:tEs_tEs0}).
$\bphi$ is the following function which corresponds to $\phi$ given by Eq.(\ref{eq:phi_nu}),
\begin{align}
  \bphi(k_s)
  &=\frac{1-e^{-i(k-k_s)s_1}}{k-k_s} ,
    \qquad
  \lim_{s_1\to\infty}\bphi(k_s)
  =\frac{1}{k-k_s}
   \qquad
  (k\to k-i\eps,~k\ne k_s) .
   \label{eq:k_eps_damp_1}
\end{align}
Eqs.(\ref{eq:Du_mn_appT}-\ref{eq:bvmn_appT}) converge in the limit of $s_1\to\infty$
by using Eq.(\ref{eq:torus_kmn_resona}) as shown in Eq.(\ref{eq:k_eps_damp_1}).
On the other hand, Eq.(\ref{eq:zeta_s}) has a numerical problem, \ie,
Eqs.(\ref{eq:DU_appU6}-\ref{eq:V_appU6}) cannot be used in the numerical calculation, 
because the infinite series involving $\Del\bu_m^n$ and $\bv_m^n$
with respect to the horizontal mode number $m$ do not converge exponentially,
similar to the problem of the numerical calculation of $Z$ in
appendices \ref{sec:impedance_trans} and \ref{sec:Z_mns_Zstd}.
In order to calculate the value of $\bZ$, we must separate the term $1/(k-i\bk_s^{mn})$,
which is involved in $\bphi(i\bk_s^{mn})$, from the infinite series with respect to $m$,
and we must rewrite the series into another expression having no series with
respect to $m$.
Instead, we wonder if we cannot separate $\bZ(0)$ from $\bZ(k)$ as
$\bZ(k)=\bZ(0)+\Del\bZ_1(k)$, and rewrite $\bZ(0)$ into a computable expression.
This is because $\bZ(0)$ includes the space charge field which has a series 
representation of the $\delta$-function with respect to the horizontal mode number.

\clearpage

\section{Numerical methods of the Bessel functions}

We describe the numerical methods to calculate a field of CSR
using the exact solution found in the present study.
We need to calculate $t_{\nu}=\{p_{\nu},q_{\nu},r_{\nu},s_{\nu}\}$ which denotes
the cross products of the Bessel functions, given by Eqs.(\ref{eq:CP_pq}-\ref{eq:CP_sr}).
Also, we need to calculate the derivatives of $p_{\nu}$ and $s_{\nu}$ with respect to
the order $\nu$, which are involved in the radial eigenfunctions
(\ref{eq:Ven}-\ref{eq:mfRm}).
$t_{\nu}$ has several representations:
the ascending series, Hankel's asymptotic series,
Olver's and Dunster's uniform asymptotic series,
which are given by Eqs.(\ref{eq:pnu_series}-\ref{eq:qr_pm_series}),
(\ref{eq:pnu_hankel}-\ref{eq:qr_hankel}), (\ref{eq:pnu_uae}-\ref{eq:snu_uae}) and
(\ref{eq:pibnu}-\ref{eq:sibnu}).
We use these representations properly,
depending on the order and arguments of $t_{\nu}(b,a)$.
In this appendix, for convenience, we sometimes use the same symbol in
a different meaning from the one used before in this paper,
\eg, $F_j$ in Eq.(\ref{eq:Fj_Cj}) is not $F_{\nu}$ in Eq.(\ref{eq:FG_JY}).

\subsection{Ascending series of the cross products of the Bessel functions}
\label{sec:ASR_app}

Using the ascending series of $J_{\nu}$ given by Eq.(\ref{eq:JY}), we rearrange
the cross products (\ref{eq:pq_Jpm}-\ref{eq:sr_Jpm}) for $\nu\not\in\mathbb{Z}$,
\begin{align}
  p_{\nu}(b,a)
  &=-\frac{1}{\pi\nu}
    \{
      (b/a)^{\nu}S(\nu,b)S(-\nu,a)
     -(a/b)^{\nu}S(\nu,a)S(-\nu,b)
    \} ,
    \label{eq:pnu_SS}
   \\
  q_{\nu}(b,a)
  &=\frac{1}{a\pi}
    \{
      (b/a)^{\nu}S(\nu,b)T(-\nu,a)
     +(a/b)^{\nu}S(-\nu,b)T(\nu,a)
    \} ,
   \\
  r_{\nu}(b,a)
  &=-\frac{1}{b\pi}
     \{
       (b/a)^{\nu}T(\nu,b)S(-\nu,a)
      +(a/b)^{\nu}T(-\nu,b)S(\nu,a)
     \} ,
   \\
  s_{\nu}(b,a)
  &=\frac{\nu}{ba\pi}
    \{
      (b/a)^{\nu}T(\nu,b)T(-\nu,a)
     -(a/b)^{\nu}T(\nu,a)T(-\nu,b)
    \} .
    \label{eq:snu_TT}
\end{align}
$S$ and $T$ are given as follows,
\begin{alignat}{4}
  S(\nu,z)
  &=\sum_{j=0}^{\infty}F_j(\nu,z) ,
    \qquad&
  F_0
  &=1 ,
    \qquad&
  \frac{F_j}{F_{j-1}}
  &=C_j &
  &=-\frac{(z/2)^2}{j(\nu+j)} ,
   \label{eq:Fj_Cj}
   \\
  T(\nu,z)
  &=\sum_{j=0}^{\infty}G_j(\nu,z) ,
    \qquad&
  G_0
  &=1 ,
    \qquad&
  \frac{G_j}{G_{j-1}}
  &=D_j &
  &=\frac{\nu+2j}{\nu+2(j-1)}C_j ,
   \label{eq:Gj_Dj}
\end{alignat}
where $C_j$ and $D_j$ are not $C_s$ and $D_s$ given by Eqs.(\ref{eq:Coeff_Cj}).
$F_j$ and $G_j$ are given as
\begin{align}
  (-1)^jF_j(\nu,z)
  &=f_j(\nu,z)
   =\frac{(z/2)^{2j}}{j!\Pi_j(\nu)} ;
    \qquad
  \Pi_{0}
  =1
   ,\qquad
  \Pi_{j}(\nu)
  =\prod_{\ell=1}^{j}(\nu+\ell)
   \quad~(j\in\mathbb{N}) ,
    \label{eq:Fjfj}
   \\
  (-1)^jG_j(\nu,z)
  &=g_j(\nu,z)
  =\Big(1+\frac{2j}{\nu}\Big)f_j(\nu,z)
   \quad
   (\nu\ne0) ,
    \qquad
  f_0=g_0
  =F_0=G_0
  =1 .
    \label{eq:Gjgj}
\end{align}
Since $\Pi_{j}(-n)=0$ for $j\geq n\in\mathbb{N}$,
$S(\nu,z)$ and $T(\nu,z)$ diverge at $\nu=-n$.
Therefore Eqs.(\ref{eq:pnu_SS}-\ref{eq:snu_TT}) are invalid for $\nu\in\mathbb{Z}$.
We differentiate Eqs.(\ref{eq:pnu_SS}) and (\ref{eq:snu_TT}) with respect to
the order $\nu$ $(\not\in\mathbb{Z})$,
\begin{alignat}{2}
  \rd_{\nu}p_{\nu}(b,a)
  &=\frac{1}{\pi\nu}\{(b/a)^{\nu}H_p(b,a)-(a/b)^{\nu}H_p(a,b)\} ,
   \\
  \rd_{\nu}s_{\nu}(b,a)
  &=\frac{\nu}{ba\pi}\{(b/a)^{\nu}H_s(b,a)-(a/b)^{\nu}H_s(a,b)\} ,
\end{alignat}
\begin{align}
  H_p(b,a)
  &=\{1/\nu-\log(b/a)\}S(\nu,b)S(-\nu,a)
   -\{S'(\nu,b)S(-\nu,a)-S(\nu,b)S'(-\nu,a)\} ,
   \\
  H_s(b,a)
  &=\{1/\nu+\log(b/a)\}T(\nu,b)T(-\nu,a)
   +\{T'(\nu,b)T(-\nu,a)-T(\nu,b)T'(-\nu,a)\} .
\end{align}
$S'$ and $T'$ are the derivatives of $S$ and $T$ with respect to $\nu$,
\begin{alignat}{3}
  S'(\nu,z)
  &=\rd_{\nu}S(\nu,z)
   =\sum_{j=1}^{\infty}F_j'(\nu,z) ,
    \qquad&
  F_1'
  &=\frac{(z/2)^2}{(\nu+1)^2} ,
   \\
  T'(\nu,z)
  &=\rd_{\nu}T(\nu,z)
   =\sum_{j=1}^{\infty}G_j'(\nu,z) ,
    \qquad&
  G_1'
  &=\Big(1+\frac{4}{\nu}+\frac{2}{\nu^2}\Big)F_1' .
\end{alignat}
$F_j'$ and $G_j'$ are the derivatives of $F_j$ and $G_j$ with respect to $\nu$,
\begin{alignat}{2}
  F_j'(\nu,z)
  &=\rd_{\nu}F_j(\nu,z)
   =-\Sig_j(\nu)F_j(\nu,z) ,
    \qquad&
  \Sig_j(\nu)
  &=\sum_{\ell=1}^{j}\frac{1}{\nu+\ell}
   \quad~(j\in\mathbb{N}) ,
   \\
  G_j'(\nu,z)
  &=\rd_{\nu}G_j(\nu,z)
   =-\bSig_j(\nu)F_j(\nu,z) ,
    \qquad&
  \bSig_j(\nu)
  &=\Big(1+\frac{2j}{\nu}\Big)\Sig_j(\nu)+\frac{2j}{\nu^2} .
\end{alignat}
$\Sig_j(\nu)$ can be zero when $\nu=-(n+1/2)$ where $n\in\mathbb{N}$.
$\rd_{\nu}(J_{\nu},J_{\nu}')$ is given as
\begin{align}
  \rd_{\nu}J_{\nu}(z)
  =\frac{(z/2)^{\nu}}{\Gam(\nu+1)}\hat{D}_{\nu}(z)S(\nu,z) ,
    \qquad
  \rd_{\nu}J_{\nu}'(z)
  =\frac{(z/2)^{\nu}}{z\Gam(\nu+1)}\{1+\nu\hat{D}_{\nu}(z)\}T(\nu,z) .
\end{align}
$\hat{D}_{\nu}$ denotes the following operator with respect to $\nu$,
\begin{align}
  \hat{D}_{\nu}(z)
  &=\log(z/2)-\psi(\nu+1)+\rd_{\nu} ,
   \label{eq:Dnu_z}
\end{align}
where $\psi$ is the digamma function given by 6.3.1 in \cite{abramo_stegun}.

When the order is a nonzero integer, \ie, $\nu=n\in\mathbb{N}$,
we calculate Eqs.(\ref{eq:Anz}-\ref{eq:Bnz}) as follows,
\begin{align}
  \frac{A_n(z)}{w_n(z)}
  &=\sum_{k=0}^{n-1}\balp_k^{(n)}
   ,\qquad
  \frac{A_n'(z)}{w_n(z)}
  =\frac{1}{z}\sum_{k=0}^{n-1}(2k-n)\balp_k^{(n)}
   ,\qquad
  w_n(z)
  =\frac{(n-1)!}{(z/2)^{n}} ,
   \label{eq:An_balp}
   \\
  B_n(z)
  &=\frac{1}{nw_n(z)}\sum_{k=0}^{\infty}(-1)^k\omg_{nk}\bbeta_k^{(n)}
   ,\qquad
  B_n'(z)
  =\frac{1}{n zw_n(z)}
    \sum_{k=0}^{\infty}(2k+n)(-1)^k\omg_{nk}\bbeta_k^{(n)} .
\end{align}
$\omg_{nk}$ is given by Eq.(\ref{eq:omg_nk}).
$\balp_k^{(n)}$ and $\bbeta_k^{(n)}$ are given as follows,
\begin{alignat}{4}
  \balp_k^{(n)}
  &=\frac{(n-k-1)!}{k!(n-1)!}(z/2)^{2k}
   ,\qquad&
  \balp_0^{(n)}
  &=1
   ,\qquad&
  \frac{\balp_k^{(n)}}{\balp_{k-1}^{(n)}}
  &=\frac{(z/2)^2}{k(n-k)}
   \qquad&&
  (1\leq k\leq n-1) ,
    \label{eq:alp_kn}
   \\
  \bbeta_k^{(n)}
  &=\frac{n!}{k!(n+k)!}(z/2)^{2k}
   ,\qquad&
  \bbeta_0^{(n)}
  &=1
   ,\qquad&
  \frac{\bbeta_k^{(n)}}{\bbeta_{k-1}^{(n)}}
  &=\frac{(z/2)^2}{k(n+k)}
   \qquad&&
  (k\geq 1) .
    \label{eq:beta_kn}
\end{alignat}
Using Eqs.(\ref{eq:Yn_Ynp_asympt}), the asymptotic expressions of
the cross products for a large integer order $n$ are given as
\begin{align}
  -\pi p_n(b,a)
  &\simeq
   \frac{1}{n}
   \{
     (b/a)^n S(n,b)U_n(a)
    -(a/b)^n U_n(b)S(n,a)
   \} ,
   \\
  -\pi q_n(b,a)
  &\simeq
   \frac{1}{a}
   \Big\{
    \frac{1}{n}(b/a)^{n}S(n,b)V_n(a)
   -(a/b)^nU_n(b)T(n,a)
   \Big\} ,
   \\
  -\pi r_n(b,a)
  &\simeq
    \frac{1}{b}
    \Big\{
      (b/a)^nT(n,b)U_n(a)
     -\frac{1}{n}(a/b)^{n}V_n(b)S(n,a)
    \Big\} ,
   \\
  -\pi s_n(b,a)
  &\simeq
    \frac{1}{ba}
    \{
      (b/a)^nT(n,b)V_n(a)
     -(a/b)^nV_n(b)T(n,a)
    \} ,
\end{align}
where $U_n$ and $V_n$ are the series involved in Eqs.(\ref{eq:An_balp}),
\begin{align}
  U_n(z)
  =\sum_{k=0}^{n-1}\balp_k^{(n)}
   ,\qquad
  V_n(z)
  =\sum_{k=0}^{n-1}(2k-n)\balp_k^{(n)} .
\end{align}
%

\subsection{Derivatives of the ascending series of $J_{\nu}$ and $Y_{\nu}$ with
respect to the integer order}

It takes time to compute Eqs.(\ref{eq:dJYn}-\ref{eq:dJYpn}) numerically
since they have a series involving $J_k$ or $Y_k$ with respect to $k\in\mathbb{Z}_0^{+}$.
Therefore, when $\nu=n\in\mathbb{N}$, we calculate $\rd_{\nu}(J_{\nu},J_{\nu}')$ using
Eqs.(\ref{eq:dJnu}-\ref{eq:dJnup}),
\begin{align}
  w_n(z)[\rd_{\nu}J_{\nu}(z)]_{\nu=n}
  &=E_n(z)\log(z/2)
   -\sum_{k=0}^{\infty}(-1)^k \lam_k^n \psi(n+k+1) ,
   \\
  w_n(z)[\rd_{\nu}J_{\nu}'(z)]_{\nu=n}
  &=\bE_n(z)\log(z/2)
   +\frac{1}{z}
    \bigg\{
      E_n(z)
     -\sum_{k=0}^{\infty}(-1)^k \gam_k^n \psi(n+k+1)
    \bigg\} .
\end{align}
$\psi$ is the digamma function (\ref{eq:digamma}) for integer argument.
$\lam_k^n$ and $\gam_k^n$ are given as
\begin{alignat}{3}
  \lam_k^n(z)
  &=\frac{(n-1)!}{k!(n+k)!}(z/2)^{2k}
   ,\qquad&
  \lam_0^n
  &=\frac{1}{n}
   ,\qquad&
  \frac{\lam_k^n}{\lam_{k-1}^n}
  &=\frac{(z/2)^2}{k(n+k)} ,
   \\
  \gam_k^n(z)
  &=(n+2k)\lam_k^n(z)
   ,\qquad&
  \gam_0^n
  &=1
   ,\qquad&
  \frac{\gam_k^n}{\gam_{k-1}^n}
  &=\frac{(z/2)^2}{k(n+k)}\cd\frac{n+2k}{n+2(k-1)} .
\end{alignat}
Using $w_n$ given by Eq.(\ref{eq:An_balp}),
we define $(E_n,\bE_n)$ and $(X_n,\bX_n)$ for $n\in\mathbb{N}$ as follows,
\begin{alignat}{2}
  J_n(z)
  &=\frac{E_n(z)}{w_n(z)}
   ,\qquad&
  Y_n(z)
  &=w_n(z)X_n(z) ,
   \\
  J_n'(z)
  &=\frac{\bE_n(z)}{w_n(z)}
   ,\qquad&
  Y_n'(z)
  &=w_n(z)\bX_n(z) .
\end{alignat}
We compute $\rd_{\nu}(Y_{\nu},Y_{\nu}')$ for $\nu=n$ using
the lower equations of (\ref{eq:dJYn}-\ref{eq:dJYpn}),
\begin{alignat}{2}
  [\rd_{\nu}Y_{\nu}(z)]_{\nu=n}
  &=\frac{w_n(z)}{2}
    \Big\{
     F_n(z)
     -\frac{\pi}{w_n(z)} J_{n}(z)
    \Big\}
   ,\qquad&
  F_n(z)
  &=\sum_{k=0}^{n-1}c_k^nX_k(z) ,
   \\
  [\rd_{\nu}Y_{\nu}'(z)]_{\nu=n}
  &=
   \frac{w_n(z)}{2}
   \Big\{
      \bF_n(z)
     -\frac{\pi}{w_n(z)} J_n'(z)
   \Big\}
   ,\qquad&
  \bF_n(z)
  &=\sum_{k=0}^{n-1}c_k^n\Big\{\bX_k(z)-\frac{n-k}{z}X_k(z)\Big\} ,
\end{alignat}
where
\begin{align}
  w_n(z)
  &=\frac{(n-1)!}{(z/2)^{n}}
   \quad
  (n\in\mathbb{N})
   ,\qquad
  c_k^n
  =\frac{n}{(k+\delta_k^0)(n-k)}
   ,\qquad
  c_0^n
  =1 .
    \label{eq:wn_ckn}
\end{align}
$\rd_{\nu}(J_{\nu},Y_{\nu})$ for $\nu=0$ is given by 9.1.68 in \cite{abramo_stegun},
\begin{alignat}{2}
  [\rd_{\nu}J_{\nu}(z)]_{\nu=0}
  &=(\pi/2)Y_0(z) ,
    \qquad&
  [\rd_{\nu}Y_{\nu}(z)]_{\nu=0}
  &=-(\pi/2)J_0(z) ,
   \\
  [\rd_{\nu}J_{\nu}'(z)]_{\nu=0}
  &=(\pi/2)Y_0'(z) ,
    \qquad&
  [\rd_{\nu}Y_{\nu}'(z)]_{\nu=0}
  &=-(\pi/2)J_0'(z) .
\end{alignat}
%

\subsection{Ascending series of the cross products of the modified Bessel functions}

When the argument is purely imaginary $z=i\bz$ ($\bz\in\mathbb{R}$),
we use $I_{\nu}(\bz)$ given by Eq.(\ref{eq:Inu}),
\begin{alignat}{3}
  I_{\nu}(\bz)
  &=\frac{(\bz/2)^{\nu}}{\Gam(\nu+1)}\bS(\nu,\bz)
   ,\qquad&
  I_{-\nu}(\bz)
  &=\frac{(\bz/2)^{-\nu}}{\Gam(1-\nu)}\bS(-\nu,\bz)
   ,\qquad&
  \bS(\nu,\bz)
  &=\sum_{j=0}^{\infty}f_j(\nu,\bz) ,
   \label{eq:I_bS}
   \\
  I_{\nu}'(\bz)
  &=\frac{\nu(\bz/2)^{\nu}}{z\Gam(\nu+1)}\bT(\nu,\bz)
   ,\qquad&
  I_{-\nu}'(\bz)
  &=\frac{-\nu(\bz/2)^{-\nu}}{z\Gam(1-\nu)}\bT(-\nu,\bz)
   ,\qquad&
  \bT(\nu,\bz)
  &=\sum_{j=0}^{\infty}g_j(\nu,\bz) .
   \label{eq:Ip_bT}
\end{alignat}
$f_j$ and $g_j$ are given by Eqs.(\ref{eq:Fjfj}-\ref{eq:Gjgj}).
Using $\hat{D}_{\nu}$ given by Eq.(\ref{eq:Dnu_z}), $\rd_{\nu}(I_{\nu},I_{\nu}')$ is given as
\begin{align}
  \rd_{\nu}I_{\nu}(\bz)
  &=\frac{(\bz/2)^{\nu}}{\Gam(\nu+1)}
    \hat{D}_{\nu}(\bz)\bS(\nu,\bz)
   ,\qquad
  \rd_{\nu}I_{\nu}'(\bz)
  =\frac{(\bz/2)^{\nu}}{\bz\Gam(\nu+1)}
    \{1+\nu\hat{D}_{\nu}(\bz)\}\bT(\nu,\bz)
   \qquad
   (\nu\not\in\mathbb{Z}^{-}) .
\end{align}
From Eqs.(\ref{eq:pnu_Ipm}-\ref{eq:snu_Ipm}) and (\ref{eq:I_bS}-\ref{eq:Ip_bT}), we get
\begin{align}
  p_{\nu}(i\bbar,i\abar)
  &=-\frac{1}{\pi\nu}
     \{
      (\bbar/\abar)^{\nu}\bS(\nu,\bbar)\bS(-\nu,\abar)
     -(\abar/\bbar)^{\nu}\bS(\nu,\abar)\bS(-\nu,\bbar)
     \} ,
    \label{eq:pnu_bS}
   \\
  iq_{\nu}(i\bbar,i\abar)
  &=\frac{1}{\abar\pi}
    \{
      (\bbar/\abar)^{\nu}\bS(\nu,\bbar)\bT(-\nu,\abar)
     +(\abar/\bbar)^{\nu}\bS(-\nu,\bbar)\bT(\nu,\abar)
    \} ,
   \\
  ir_{\nu}(i\bbar,i\abar)
  &=-\frac{1}{\bbar\pi}
     \{
       (\bbar/\abar)^{\nu}\bT(\nu,\bbar)\bS(-\nu,\abar)
      +(\abar/\bbar)^{\nu}\bT(-\nu,\bbar)\bS(\nu,\abar)
     \} ,
   \\
  s_{\nu}(i\bbar,i\abar)
  &=-\frac{\nu}{\bbar\abar\pi}
     \{
      (\bbar/\abar)^{\nu}\bT(\nu,\bbar)\bT(-\nu,\abar)
     -(\abar/\bbar)^{\nu}\bT(\nu,\abar)\bT(-\nu,\bbar)
     \} .
    \label{eq:snu_bT}
\end{align}
$\nu\ne n\in\mathbb{Z}$ in Eqs.(\ref{eq:pnu_bS}-\ref{eq:dsnu_Ipm}) since $I_{-n}=I_n$.
$\rd_{\nu}(p_{\nu},s_{\nu})$ for purely imaginary arguments is given as
\begin{alignat}{2}
  \rd_{\nu}p_{\nu}(i\bbar,i\abar)
  &=\frac{1}{\pi\nu}
   \{(\bbar/\abar)^{\nu}\bH_p(\bbar,\abar)-(\abar/\bbar)^{\nu}\bH_p(\bbar,\abar)\} ,
    \label{eq:dpnu_Ipm}
   \\
  \rd_{\nu}s_{\nu}(i\bbar,i\abar)
  &=-\frac{\nu}{\bbar\abar\pi}
   \{(\bbar/\abar)^{\nu}\bH_s(\bbar,\abar)-(\abar/\bbar)^{\nu}\bH_s(\bbar,\abar)\} ,
    \label{eq:dsnu_Ipm}
\end{alignat}
where
\begin{align}
  \bH_p(\bbar,\abar)
  &=
    \{1/\nu-\log(\bbar/\abar)\}\bS(\nu,\bbar)\bS(-\nu,\abar)
   -\{\bS'(\nu,\bbar)\bS(-\nu,\abar)-\bS(\nu,\bbar)\bS'(-\nu,\abar)\} ,
   \\
  \bH_s(\bbar,\abar)
  &=
     \{1/\nu+\log(\bbar/\abar)\}\bT(\nu,\bbar)\bT(-\nu,\abar)
    +\{\bT'(\nu,\bbar)\bT(-\nu,\abar)-\bT(\nu,\bbar)\bT'(-\nu,\abar)\} .
\end{align}
$(\bS',\bT')$ and $(f_j',g_j')$ are the derivatives of $(\bS,\bT)$ and $(f_j,g_j)$ with 
respect to the order $\nu$ which is non-negative integer, \ie, $\nu\ne-n$ ($n\in\mathbb{N}$),
\begin{alignat}{3}
  \bS'(\nu,\bz)
  &=\rd_{\nu}\bS(\nu,\bz)
   =\sum_{j=1}^{\infty}f_j'(\nu,\bz)
   ,\qquad&
  f_1'(\nu,\bz)
  &=\rd_{\nu}f_1(\nu,\bz)
   =\frac{(\bz/2)^2}{(\nu+1)^2}
   \quad~
  (\nu\ne-n) ,
   \\
  \bT'(\nu,\bz)
  &=\rd_{\nu}\bT(\nu,\bz)
   =\sum_{j=1}^{\infty}g_j'(\nu,\bz)
   ,\qquad&
  g_1'(\nu,\bz)
  &=\rd_{\nu}g_1(\nu,\bz)
   =\Big(1+\frac{4}{\nu}+\frac{2}{\nu^2}\Big)f_1'(\nu,\bz) .
\end{alignat}
%

\subsection{Modified Bessel function of integer order}

We cannot use Eqs.(\ref{eq:pnu_bS}-\ref{eq:snu_bT}) for $\nu=n\in\mathbb{N}$
since $f_j(-n,\bz)=\infty$.
When $\nu=n$, we use Eqs.(\ref{eq:pnu_IK}-\ref{eq:snu_IK}) which involve
$(I_n,K_n)$ and $(I_n',K_n')$ given by Eqs.(\ref{eq:In}-\ref{eq:Knp}).
In order to avoid underflow and overflow in calculating the modified Bessel functions for 
large $n$, we factorize $w_n$ from them,
\begin{alignat}{4}
  I_n(\bz)
  &=\frac{H_n(\bz)}{w_n(\bz)} ,
    \qquad&
  K_n(\bz)
  &=w_n(\bz)W_n(\bz)
   \quad&&
  (n\in\mathbb{N})
   ,\qquad&
  K_0(\bz)
  &=W_0(\bz) ,
    \label{eq:HW}
   \\
  I_n'(\bz)
  &=\frac{\bH_n(\bz)}{w_n(\bz)} ,
    \qquad&
  K_n'(\bz)
  &=w_n(\bz)\bW_n(\bz)
   \quad&&
  (n\in\mathbb{N})
   ,\qquad&
  K_0'(\bz)
  &=\bW_0(\bz) ,
    \label{eq:HWp}
\end{alignat}
where
\begin{align}
  H_n(\bz)
  =\frac{1}{n}\sum_{k=0}^{\infty}\bbeta_k^{(n)}(\bz) ,
    \qquad
  \bH_n(\bz)
  =\frac{1}{n\bz}\sum_{k=0}^{\infty}(n+2k)\bbeta_k^{(n)}(\bz) .
\end{align}
$w_n$ is given by Eq.(\ref{eq:wn_ckn}), however, the argument is not $z$ but $\bz$,
\begin{align}
  w_n(\bz)
  &=\frac{(n-1)!}{(\bz/2)^n}
   \quad
  (n\in\mathbb{N})
   ,\qquad
  w_0
  =1 .
\end{align}
$w_n$ diverges in the limit of $n\to\infty$ with $\bz$ fixed.
$W_n$ and $\bW_n$ for $n\in\mathbb{N}$ are given as
\begin{align}
  W_n(\bz)
  &=
    (-1)^{n+1}\frac{I_n(\bz)}{w_n(\bz)}\log(\bz/2)
   +\frac{1}{2}
    \bigg\{(1-\delta_n^0)\frac{\bA_n(\bz)}{w_n(\bz)}+(-1)^n\frac{\bB_n(\bz)}{w_n(\bz)}
    \bigg\} ,
   \\
  \bW_n(\bz)
  &=
    (-1)^{n+1}\bigg\{\frac{I_n'(\bz)}{w_n(\bz)}\log(z/2)+\frac{I_n(\bz)}{\bz w_n(\bz)}\bigg\}
   +\frac{1}{2}
    \bigg\{(1-\delta_n^0)\frac{\bA_n'(\bz)}{w_n(\bz)}+(-1)^n\frac{\bB_n'(\bz)}{w_n(\bz)}
    \bigg\} ,
\end{align}
where
\begin{alignat}{3}
  \frac{\bA_n(\bz)}{w_n(\bz)}
  &=\sum_{k=0}^{n-1}(-1)^k\balp_k^{(n)}(\bz)
   ,\qquad&
  \frac{\bA_n'(\bz)}{w_n(\bz)}
  &=\frac{1}{\bz}\sum_{k=0}^{n-1}(-1)^k(2k-n)\balp_k^{(n)}(\bz) ,
    \label{eq:bAn_wn}
   \\
  \bB_n(\bz)
  &=\frac{1}{nw_n(\bz)}\sum_{k=0}^{\infty}\omg_{nk}\bbeta_k^{(n)}(\bz)
   ,\qquad&
  \bB_n'(\bz)
  &=\frac{1}{nw_n(\bz)\bz}\sum_{k=0}^{\infty}\omg_{nk}(2k+n)\bbeta_k^{(n)}(\bz) .
    \label{eq:bBn_wn}
\end{alignat}
$\omg_{nk}$ is given by Eq.(\ref{eq:omg_nk}).
$\balp_k^{(n)}$ and $\bbeta_k^{(n)}$ are given by
Eqs.(\ref{eq:alp_kn}-\ref{eq:beta_kn}), however, the argument is not $z$ but $\bz$,
\begin{alignat}{4}
  \balp_k^{(n)}(\bz)
  &=\frac{(n-k-1)!}{k!(n-1)!}(\bz/2)^{2k}
   ,\qquad&
  \balp_0^{(n)}
  &=1
   ,\qquad&
  \frac{\balp_k^{(n)}}{\balp_{k-1}^{(n)}}
  &=\frac{(\bz/2)^2}{k(n-k)}
   \qquad&&
  (1\leq k\leq n-1) ,
    \label{eq:balp_kn}
   \\
  \bbeta_k^{(n)}(\bz)
  &=\frac{n!}{k!(n+k)!}(\bz/2)^{2k}
   ,\qquad&
  \bbeta_0^{(n)}
  &=1
   ,\qquad&
  \frac{\bbeta_k^{(n)}}{\bbeta_{k-1}^{(n)}}
  &=\frac{(\bz/2)^2}{k(n+k)}
   \qquad&&
  (k\geq 1) .
    \label{eq:bbeta_kn}
\end{alignat}
Similar to Eqs.(\ref{eq:HW}-\ref{eq:HWp}),
we factorize $w_n$ from $\rd_{\nu}(I_{\nu},I_{\nu}')$ and $\rd_{\nu}(K_{\nu},K_{\nu}')$ for
$\nu=n$, given by Eqs.(\ref{eq:dIn_In}), (\ref{eq:dKn}) and (\ref{eq:dInp}-\ref{eq:dKnp}),
to avoid over/underflow in computing $\rd_{\nu}(p_{\nu},s_{\nu})$ for $\nu=n\in\mathbb{N}$,
\begin{alignat}{2}
  [\rd_{\nu}I_{\nu}(\bz)]_{\nu=n}
  &=\frac{N_n(\bz)}{w_n(\bz)}
   ,\qquad&
  [\rd_{\nu}K_{\nu}(\bz)]_{\nu=n}
  &=w_n(\bz)Z_n(\bz) ,
   \\
  [\rd_{\nu}I_{\nu}'(\bz)]_{\nu=n}
  &=\frac{\bN_n(\bz)}{w_n(\bz)}
   ,\qquad&
  [\rd_{\nu}K_{\nu}'(\bz)]_{\nu=n}
  &=w_n(\bz)\bZ_n(\bz) .
\end{alignat}
$Z_n$ and $\bZ_n$ are the following finite series
which involve $c_k^n$ given by Eq.(\ref{eq:wn_ckn}),
\begin{align}
  N_n(\bz)
  &=H_n(\bz)\log(\bz/2)
   -\frac{1}{n}\sum_{k=0}^{\infty}\bbeta_k^{(n)}\psi(n+k+1) ,
   \\
  \bN_n(\bz)
  &=\bH_n(\bz)\log(\bz/2)
   +\frac{1}{\bz}
    \bigg\{H_n(\bz)-\frac{1}{n}\sum_{k=0}^{\infty}(n+2k)\bbeta_k^{(n)}\psi(n+k+1)\bigg\} ,
   \\
  Z_n(\bz)
  &=\frac{1}{2}\sum_{k=0}^{n-1}c_k^nW_k(\bz)
   ,\qquad
  \bZ_n(\bz)
  =\frac{1}{2}\sum_{k=0}^{n-1}c_k^n
   \Big\{\bW_k(\bz)-\frac{n-k}{\bz}W_k(\bz)\Big\} .
\end{align}
%

\subsection{Hankel's asymptotic expansion of the cross products}
\label{sec:hankel_numerical}

We consider the numerical calculation of the cross products of the Bessel functions
using Hankel's asymptotic expansion (\ref{eq:J_hankel}-\ref{eq:Y_hankel}).
We rewrite Eqs.(\ref{eq:Pnu_coe}) and (\ref{eq:PR_prm}-\ref{eq:QS_prm}) as follows,
\begin{alignat}{3}
  \bigg\{{ \hP_{\nu} \atop \hR_{\nu}}\bigg\}
  &=\sum_{j=0}^{\infty}\bigg\{{ \mff_j \atop \mfg_j }\bigg\}
   ,\qquad&
  \bigg\{{ \hP_{\nu}' \atop \hR_{\nu}'}\bigg\}
  =\rd_{\nu}\bigg\{{ \hP_{\nu} \atop \hR_{\nu}}\bigg\}
  =\sum_{j=1}^{\infty}
  \bigg\{{ \mff_j' \atop \mfg_j' }\bigg\}
   ,\qquad&
  \bigg\{{ \mff_j' \atop \mfg_j' }\bigg\}
  &=\rd_{\nu}\bigg\{{ \mff_j \atop \mfg_j }\bigg\} ,
   \label{eq:PR_hankel}
   \\
  \bigg\{{ \hQ_{\nu} \atop \hS_{\nu} }\bigg\}
  &=\sum_{j=0}^{\infty}\bigg\{{ \bmff_j \atop \bmfg_j }\bigg\}
   ,\qquad&
  \bigg\{{ \hQ_{\nu}' \atop \hS_{\nu}' }\bigg\}
  =\rd_{\nu}\bigg\{{ \hQ_{\nu} \atop \hS_{\nu}}\bigg\}
  =\sum_{j=0}^{\infty}
  \bigg\{{ \bmff_j' \atop \bmfg_j' }\bigg\}
   ,\qquad&
  \bigg\{{ \bmff_j' \atop \bmfg_j' }\bigg\}
  &=\rd_{\nu}\bigg\{{ \bmff_j \atop \bmfg_j }\bigg\} ,
   \label{eq:QS_hankel}
\end{alignat}
where
\begin{alignat}{4}
  \mff_j
  &=\frac{(-1)^j\hat{a}_{2j}(\kap)}{(2j)!(8z)^{2j}}
   ,\qquad&
  \bmff_j
  &=\frac{(-1)^j\hat{a}_{2j+1}(\kap)}{(2j+1)!(8z)^{2j+1}}
   ;\qquad&
  \mff_0
  &=1
   ,\qquad&
  \bmff_0
  &=\frac{4\nu^2-1}{8z} ,
   \label{eq:fj_f0}
   \\
  \mfg_j
  &=\frac{(-1)^j\hat{b}_{2j}(\kap)}{(2j)!(8z)^{2j}}
   ,\qquad&
  \bmfg_j
  &=\frac{(-1)^j\hat{b}_{2j+1}(\kap)}{(2j+1)!(8z)^{2j+1}}
   ;\qquad&
  \mfg_0
  &=1
   ,\qquad&
  \bmfg_0
  &=\frac{4\nu^2+3}{8z} .
   \label{eq:gj_g0}
\end{alignat}
$(\mff_j',\mfg_j')$ and $(\bmff_j',\bmfg_j')$ are the derivatives of
$(\mff_j,\mfg_j)$ and $(\bmff_j,\bmfg_j)$ with respect to $\nu$,
\begin{alignat}{4}
  \mff_j'
  &=\frac{(-1)^j8\nu\hat{a}_{2j}'(\kap)}{(2j)!(8z)^{2j}}
   ,\qquad&
  \bmff_j'
  &=\frac{(-1)^j8\nu\hat{a}_{2j+1}'(\kap)}{(2j+1)!(8z)^{2j+1}}
   ;\qquad\quad&
  \mff_1'
  &=-\frac{\nu(4\nu^2-5)}{8z^2}
   ,\qquad&
  \bmff_0'
  &=\frac{\nu}{z} ,
    \label{eq:fj_bfj}
   \\
  \mfg_j'
  &=\frac{(-1)^j8\nu\hat{b}_{2j}'(\kap)}{(2j)!(8z)^{2j}}
   ,\qquad&
  \bmfg_j'
  &=\frac{(-1)^j8\nu\hat{b}_{2j+1}'(\kap)}{(2j+1)!(8z)^{2j+1}}
   ;\qquad\quad&
  \mfg_1'
  &=-\frac{\nu(4\nu^2+7)}{8z^2}
   ,\qquad&
  \bmfg_0'
  &=\frac{\nu}{z} .
    \label{eq:gj_bgj}
\end{alignat}
$\hat{a}_{\ell}$ and $\hat{b}_{\ell}$ are the functions of $\kap\,(=4\nu^2)$,
given by Eqs.(\ref{eq:ahat}-\ref{eq:bhat}).
When $\nu=1/2$, $\bmff_0=0$ and $\hQ_{\nu}=0$.
$\mff_1'=0$ when $\nu=5^{1/2}/2$.
We compute Eqs.(\ref{eq:fj_f0}-\ref{eq:gj_g0}) by recurrence,
\begin{alignat}{4}
  \frac{\mff_{j}}{\mff_{j-1}}
  &=-\frac{\alp_{2j-1}\alp_{2j}}{2j(2j-1)(8z)^2}
   =c_j
   ,\qquad&
  \frac{\mfg_{j}}{\mfg_{j-1}}
  &=d_j\frac{\beta_{2j}}{\beta_{2j-2}}
   ,\qquad&
  d_j
  &=-\frac{\alp_{2j-2}\alp_{2j-1}}{2j(2j-1)(8z)^2}
   \quad&&
  (j\geq1) ,
    \label{eq:fgj_cdj}
   \\  
  \frac{\bmff_{j}}{\bmff_{j-1}}
  &=-\frac{\alp_{2j}\alp_{2j+1}}{2j(2j+1)(8z)^2}
   =\bar{c}_j
   ,\qquad&
  \frac{\bmfg_{j}}{\bmfg_{j-1}}
  &=\bar{d}_j\frac{\beta_{2j+1}}{\beta_{2j-1}}  
   ,\qquad&
  \bar{d}_j
  &=-\frac{\alp_{2j-1}\alp_{2j}}{2j(2j+1)(8z)^2}
   \quad&&
  (j\geq1) .
    \label{eq:fgj_cdj_bar}
\end{alignat}
$c_j$ and $d_j$ are not the ones given by Eqs.(\ref{eq:cj_dj}).
We compute $\mfg_1$ using the first equation of (\ref{eq:gj_g0}) as follows
since the second equation of (\ref{eq:fgj_cdj}) is invalid for $j=1$,
\ie, $\beta_0$ does not exist,
\begin{align}
  \mfg_1
  =-\frac{\alp_1\beta_2}{2(8z)^2} .
\end{align}
$\alp_{\ell}$ and $\beta_{\ell}$ are given by Eqs.(\ref{eq:ahat}-\ref{eq:bhat}),
\begin{align}
  \alp_{\ell}
  =\kap-(2\ell-1)^2
    ,\qquad
  \beta_{\ell}
  =\kap+4{\ell}^2-1
   \qquad
  (\ell\in\mathbb{N})
   ,\qquad
  \kap
  =(2\nu)^2 .
\end{align}
When $\alp_n\ne0$ for $n\leq\ell$ ($\ell=2j$ or $2j+1$), the recurrence relations of
Eqs.(\ref{eq:fj_bfj}-\ref{eq:gj_bgj}) are given as follows,
\begin{alignat}{6}
  \frac{\mff_j'}{\mff_{j-1}'}
  &=c_j\frac{\tau_{2j}}{\tau_{2j-2}}
   ,\qquad&
  \frac{\mfg_j'}{\mfg_{j-1}'}
  &=d_j\frac{\sig_{2j}}{\sig_{2j-2}}
   \quad&&
  (j\geq2)
   \qquad&\text{when}&~~&
  \alp_n
  &\ne0
   \quad&&
  (1\leq n\leq 2j) ,
    \label{eq:fj_gj_rat}
   \\
  \frac{\bmff_j'}{\bmff_{j-1}'}
  &=\bar{c}_j\frac{\tau_{2j+1}}{\tau_{2j-1}}
   ,\qquad&
  \frac{\bmfg_j'}{\bmfg_{j-1}'}
  &=\bar{d}_j\frac{\sig_{2j+1}}{\sig_{2j-1}}
   \quad&&
  (j\geq1)
   \qquad&\text{when}&~~&
  \alp_n
  &\ne0
   \quad&&
  (1\leq n\leq 2j+1) .
    \label{eq:bfj_bgj_rat}
\end{alignat}
$\tau_{\ell}$ and $\sig_{\ell}$ are given by Eqs.(\ref{eq:tau_sig}).
$\tau_{\ell}$ and $\tau_{\ell-2}$ diverge when $\alp_n=0$ for $n\leq\ell-2$.
$\alp_n$ can be zero when $\nu$ is half integer.
When $\nu=n-1/2$, the recurrence relations of $\tau_{\ell}$ and $\sig_{\ell}$,
which are involved in Eqs.(\ref{eq:fj_gj_rat}-\ref{eq:bfj_bgj_rat}), are given as
\begin{align}
  \frac{\tau_{\ell}}{\tau_{\ell-2}}
  &=1
   \quad
  (1\leq n\leq \ell-2)
   ,\qquad
  \frac{\sig_{\ell}}{\sig_{\ell-2}}
  =\frac{\beta_{\ell}}{\beta_{\ell-2}}
   \quad
  (1\leq n\leq \ell-3)
   ,\qquad
  n
  =\nu+\frac{1}{2}
  \in\mathbb{N}.
    \label{eq:fp_rat_alp0}
\end{align}
If $\alp_n=0$ only for $n=\ell-1$ or $\ell$, since $\tau_{\ell}=\infty$ and
$\tau_{\ell-2}\ne\infty$, we use the following relations,
\begin{alignat}{4}
  \frac{\hat{a}_{\ell}'}{\hat{a}_{\ell-2}'}
  &=\frac{\alp_{\ell-1}+\alp_{\ell}}{\tau_{\ell-2}}
   \qquad&\text{when}&~~&
  \alp_{\ell-1}
  &=0
   ~~\text{or}~~&&
  \alp_{\ell}
  =0 ,
   \\
  \frac{\hat{b}_{\ell}'}{\hat{b}_{\ell-2}'}
  &=(\alp_{\ell-2}+\alp_{\ell-1})\frac{\beta_{\ell}}{\sig_{\ell-2}}
   \qquad&\text{when}&~~&
  \alp_{\ell-2}
  &=0
   ~~\text{or}~~&&
  \alp_{\ell-1}
  =0 .
\end{alignat}
For convenience in the numerical calculation,
we define $(\mfF_j',\mfG_j')$ and $(\bmfF_j',\bmfG_j')$ as follows,
\begin{align}
  (\mfF_j',\mfG_j')
  =\frac{(\mff_j',\mfg_j')}{8\nu}
   \quad
  (j\in\mathbb{N})
   ,\qquad
  (\bmfF_j',\bmfG_j')
  =\frac{(\bmff_j',\bmfg_j')}{8\nu}
   \quad
  (j\in\mathbb{Z}_0^{+}) .
\end{align}
That is,
\begin{alignat}{4}
  \mfF_j'
  &=\cC_j\hat{a}_{2j}' ,
    \qquad&
  \bmfF_j'
  &=\bcC_j\hat{a}_{2j+1}' ;
    \qquad&
  \mfF_1'
  &=\cC_1\hat{a}_{2}' ,
    \qquad&
  \bmfF_0'
  &=\bcC_0\hat{a}_{1}'
   =\bcC_0 ,
   \\
  \mfG_j'
  &=\cC_j\hat{b}_{2j}' ,
    \qquad&
  \bmfG_j'
  &=\bcC_j\hat{b}_{2j+1}' ;
    \qquad&
  \mfG_1'
  &=\cC_1\hat{b}_{2}' ,
    \qquad&
  \bmfG_0'
  &=\bcC_0\hat{b}_{1}'
   =\bcC_0 ,
\end{alignat}
where
\begin{alignat}{3}
  \frac{\cC_j}{\cC_{j-1}}
  &=-\frac{1}{2j(2j-1)(8z)^2}
   \quad
  (j\geq2)
   ,\qquad&
  \cC_j
  &=\frac{(-1)^j}{(2j)!(8z)^{2j}}
   ,\qquad&
  \cC_1
  &=-\frac{1}{2(8z)^2} ,
   \\
  \frac{\bcC_j}{\bcC_{j-1}}
  &=-\frac{1}{2j(2j+1)(8z)^2}
   \quad
  (j\geq1)
   ,\qquad&
  \bcC_j
  &=\frac{(-1)^j}{(2j+1)!(8z)^{2j+1}}
   ,\qquad&
  \bcC_0
  &=\frac{1}{8z} .
\end{alignat}
%

\subsection{Condition of convergence of Hankel's asymptotic expansion}

There are conditions with respect to the order $\nu$ and argument $z$
such that Hankel's asymptotic expansions of the Bessel functions
$C_{\nu}(z)$ and $C_{\nu}'(z)$ are valid.
The eight asymptotic series (\ref{eq:PR_hankel}-\ref{eq:QS_hankel}) each converge 
if the absolute values of the following quantities are smaller than 1,
\begin{alignat}{4}
  \frac{\mff_1}{\mff_0}
  &=-\frac{\hat{a}_2}{2(8z)^2}
   ,\qquad&
  \frac{\bmff_1}{\bmff_0}
  &=-\frac{\hat{a}_3/\hat{a}_1}{6(8z)^2}
   ,\qquad&
  \frac{\mff_2'}{\mff_1'}
  &=-\frac{\hat{a}_4'/\hat{a}_2'}{12(8z)^2}
   ,\qquad&
  \frac{\bmff_1'}{\bmff_0'}
  &=-\frac{\hat{a}_3'/\hat{a}_1'}{6(8z)^2} ,
   \label{eq:Hcond1}
   \\
  \frac{\mfg_1}{\mfg_0}
  &=-\frac{\hat{b}_2}{2(8z)^2}
   ,\qquad&
  \frac{\bmfg_1}{\bmfg_0}
  &=-\frac{\hat{b}_3/\hat{b}_1}{6(8z)^2}
   ,\qquad&
  \frac{\mfg_2'}{\mfg_1'}
  &=-\frac{\hat{b}_4'/\hat{b}_2'}{12(8z)^2}
   ,\qquad&
  \frac{\bmfg_1'}{\bmfg_0'}
  &=-\frac{\hat{b}_3'/\hat{b}_1'}{6(8z)^2} .
   \label{eq:Hcond2}
\end{alignat}
For example, when $\nu,z\in\mathbb{R}$, the series $\hP_{\nu}$ converges under
the following condition,
\begin{align}
  \bigg|\frac{\mff_1}{\mff_0}\bigg|
  =\bigg|\frac{(\kap-1)(\kap-9)}{2(8z)^2}\bigg|
  <1
   \qquad\Lra\qquad
  \kap
  =(2\nu)^2
  <4(8z^2+1)^{1/2}+5 .
   \label{eq:c1_cond}
\end{align}
When $|z|\gg1$, Eq.(\ref{eq:c1_cond}) is approximately given as
\begin{align}
  \nu^2\lesssim 8^{1/2}z
   \qquad
  (\nu,z\in\mathbb{R}) .
\end{align}
Eq.(\ref{eq:c1_cond}) may be the severest condition among
Eqs.(\ref{eq:Hcond1}-\ref{eq:Hcond2}).
That is, if $\nu$ and $z$ satisfy Eq.(\ref{eq:c1_cond}), all the series given by
Eqs.(\ref{eq:PR_hankel}-\ref{eq:QS_hankel}) converge as far as we have examined.

In numerically calculating the Bessel functions for $|z|\gg1$, we use Eq.(\ref{eq:c1_cond}) 
as the border of switching the representation from Hankel's asymptotic series to
the uniform asymptotic series in the $(z,\nu)$-plane.
When $\nu$ is large enough, \eg, $\nu>16$, however, the uniform asymptotic series is valid
for $\forall z$ in the numerical calculation.
Therefore Eq.(\ref{eq:c1_cond}) has the meaning of the border roughly for
$|z|<90$ or 100 in practice.
On the contrary, when $\nu<16$ and $|z|$ is not so large, we must use
the ascending series of the Bessel functions in computing the cross products.
Although the ascending series is absolutely convergent,
more significant digits are lost in the numerical calculation for larger $|z|$.

\clearpage

\subsection{Coefficients of the uniform asymptotic expansion of the Bessel functions}
\label{sec:uae_coeff}

The coefficients of the uniform asymptotic expansion $(a_j,b_j,d_j,c_j)$ are functions of
$\zeta$ as in Eqs.(\ref{eq:aj_bj}-\ref{eq:cj_dj}),
\begin{align}
  \zeta
  =\pm\bigg(\frac{3}{2}\eta_{\pm}\bigg)^{2/3}
   ,\qquad
  e_z
  =\psi^{-1/3}
  =\bigg(\frac{\eps_z}{\zeta}\bigg)^{1/2}
   ,\qquad
  \lim_{z\to1}\zeta
  =0
   ,\qquad
  \lim_{z\to1}e_z
  =2^{1/3} .
    \label{eq:zeta_ez}
\end{align}
$\psi$ is given by Eq.(\ref{eq:psi}) which is not the digamma function.
For a given $z$, we get $\zeta$ through $\eta_{\pm}$ given by
Eqs.(\ref{eq:zeta_p}-\ref{eq:zeta_m}), depending on the sign of $\zeta$.
$e_z$ is real for $\forall z\in\mathbb{R}$ since $\zeta$ has the same sign as
$\eps_z$ given below.
The coefficients $(a_j,b_j,d_j,c_j)$ involve $u_{\ell}$ or $v_{\ell}$ given by
Eqs.(\ref{eq:uj}-\ref{eq:vj}),
\begin{align}
  \bigg\{{ u_{\ell}(t) \atop v_{\ell}(t) }\bigg\}
  &=\sum_{k=0}^{\ell} \bigg\{{ \mfa_k^{(\ell)} \atop \mfb_k^{(\ell)} }\bigg\} t^{2k+\ell} ,
   \qquad
  t
  =\eps_z^{-1/2} ,
    \qquad
  \eps_z
  =1-z^2 ,
    \qquad
  z
  =\frac{\hr}{\nu} .
    \label{eq:uv_app}
\end{align}
$z$ is not the argument of the Bessel function $C_{\nu}(\hr)$ in
appendices \ref{sec:uae_coeff}-\ref{sec:uae_KL}.
$z=1$ is the transition point of $C_{\nu}(\nu z)$.
When $z>1$ ($\zeta<0$), $t$ and $\zeta^{1/2}$ are purely imaginary.
The coefficients $(a_j,b_j,d_j,c_j)$ are real for $\forall z\in\mathbb{R}$, because,
when $\zeta<0$, the imaginary units coming out of $t$ and $\zeta^{1/2}$ cancel out in
Eqs.(\ref{eq:aj_bj}-\ref{eq:cj_dj}).
Substituting Eq.(\ref{eq:uv_app}) into Eqs.(\ref{eq:aj_bj}-\ref{eq:cj_dj}),
we rewrite them as follows in order to calculate them for $\forall z\in\mathbb{R}$
using only real variables, \ie, without using complex variables in the numerical code,
\begin{alignat}{2}
  a_j(\zeta)
  &=\sum_{s=0}^{2j}\kap_s\frac{e_z^s}{\zeta^s}\bA_s^{(j)} ,
    \qquad&
  b_j(\zeta)
  &=-\sum_{s=0}^{2j+1}\lambda_s\frac{e_z^{s+1}}{\zeta^s}\bB_s^{(j)} ,
   \label{eq:aj_bj_rw}
   \\
  d_j(\zeta)
  &=\sum_{s=0}^{2j}\lambda_s\frac{e_z^s}{\zeta^s}\bD_s^{(j)} ,
    \qquad&
  c_j(\zeta)
  &=-\sum_{s=0}^{2j+1}\kap_s\frac{e_z^{s+1}}{\zeta^{s-1}}\bC_s^{(j)} ,
   \label{eq:dj_cj_rw}
\end{alignat}
where
\begin{alignat}{2}
  \bA_s^{(j)}
  &=\sum_{k=0}^{2j-s}\frac{\mfa_k^{(2j-s)}}{\eps_z^{j+k}} ,
    \qquad&
  \bB_s^{(j)}
  &=\sum_{k=0}^{2j-s+1}\frac{\mfa_k^{(2j-s+1)}}{\eps_z^{j+k+1}}
   =\bA_{s+1}^{(j+1)} ,
    \label{eq:AB_bar}
   \\
  \bD_s^{(j)}
  &=\sum_{k=0}^{2j-s} \frac{\mfb_k^{(2j-s)}}{\eps_z^{j+k}} ,
    \qquad&
  \bC_s^{(j)}
  &=\sum_{k=0}^{2j-s+1} \frac{\mfb_k^{(2j-s+1)}}{\eps_z^{j+k+1}}
   =\bD_{s+1}^{(j+1)} .
    \label{eq:DC_bar}
\end{alignat}
Eqs.(\ref{eq:aj_bj_rw}-\ref{eq:dj_cj_rw}) have terms which diverge at $z=1$.
Since these infinities occuring at $z=1$ cancel out in
Eqs.(\ref{eq:aj_bj_rw}-\ref{eq:dj_cj_rw}), $(a_j,b_j,d_j,c_j)$ are finite at $z=1$.
When $z$ is equal or close to 1, due to the divergent terms,
we cannot use Eqs.(\ref{eq:aj_bj_rw}-\ref{eq:dj_cj_rw})
in the numerical calculation as they are.
So we must separate $V_j^{a,b,d,c}$ $(=0)$ in Eqs.(\ref{eq:aj_bj_rw}-\ref{eq:dj_cj_rw}), 
which denotes the sums of the divergent terms at $z=1$,
\begin{alignat}{2}
  \frac{a_j(\zeta)}{e_z^{6j}}
  &=V_j^a
   +\sum_{\ell=0}^{\infty}\ta_{\ell}^{(j)}\eps_z^{\ell} ,
    \qquad&
  \ta_{\ell}^{(j)}
  &=
    \sum_{s=0}^{2j}
    \frac{\kap_{2j-s}}{2^{s}}
    \sum_{k=0}^{s} \mfa_{k}^{(s)} p_{3j+\ell+k-s}^{(s)} ,
    \label{eq:taj}
   \\
  -\frac{b_j(\zeta)}{e_z^{6j+4}}
  &=V_j^b
   +\sum_{\ell=0}^{\infty} \tb_{\ell}^{(j)}\eps_z^{\ell} ,
    \qquad&
  \tb_{\ell}^{(j)}
  &=\sum_{s=0}^{2j+1}\frac{\lam_{2j+1-s}}{2^{s}}
    \sum_{k=0}^{s} \mfa_{k}^{(s)} p_{3j+\ell+k-s+2}^{(s)} ,
   \\
  \frac{d_j(\zeta)}{e_z^{6j}}
  &=V_j^d
   +\sum_{\ell=0}^{\infty}\td_{\ell}^{(j)}\eps_z^{\ell} ,
    \qquad&
  \td_{\ell}^{(j)}
  &=
    \sum_{s=0}^{2j} \frac{\lam_{2j-s}}{2^{s}}
    \sum_{k=0}^{s} \mfb_{k}^{(s)}p_{3j+\ell+k-s}^{(s)} ,
   \\
  -\frac{c_j(\zeta)}{e_z^{6j+2}}
  &=V_j^c
   +\sum_{\ell=0}^{\infty}
    \tc_{\ell}^{(j)}\eps_z^{\ell} ,
    \qquad&
  \tc_{\ell}^{(j)}
  &=\sum_{s=0}^{2j+1} \frac{\kap_{2j+1-s}}{2^s}
    \sum_{k=0}^{s} \mfb_{k}^{(s)}p_{3j+\ell+k-s+1}^{(s)} ,
    \label{eq:tcj}
\end{alignat}
where
\begin{align}
  e_z^{6j}
  =\psi^{-2j}
   ,\qquad
  e_z^{6j+4}
   =\psi^{-2(j+2/3)}
   ,\qquad
  e_z^{6j+2}
   =\psi^{-2(j+1/3)} .
    \label{eq:ez6j}
\end{align}
$p_{\ell}^{(s)}$ is the coefficient of the $\ell$th order term in the power series of
$(2\psi)^{s}$ with respect to $\eps_z$,
\begin{align}
  (2\psi)^{s}
  =\sum_{\ell=0}^{\infty}p_{\ell}^{(s)}\eps_z^{\ell}
   \quad
   (s\in\mathbb{Z}_0^{+})
   ,\qquad
  p_{0}^{(s)}
  =1
   ,\qquad
  p_{\ell}^{(s)}
  =\frac{3}{\ell}\sum_{k=0}^{\ell-1}
   \frac{s(\ell-k)-k}{2(\ell-k)+3}p_k^{(s)}
   \quad
   (\ell\in\mathbb{N}) .
    \label{eq:2psi_pwr}
\end{align}
$V_j^{a,b,d,c}$ denotes the sums of the divergent terms which cancel out and vanish,
\begin{alignat}{3}
  V_j^a
  &=\sum_{\ell=1}^{3j}\cA_{\ell}^{(j)}\eps_z^{-\ell}
   =0
   ,\qquad&
  \cA_{\ell}^{(j)}
  &=\bigg\{
    \begin{array}{ll}
      \tcA_{\ell}^{(j)} & (1\leq\ell\leq j) \\
      \bcA_{\ell}^{(j)} & (j\leq\ell\leq 3j)
    \end{array}
   ,\qquad&
  \tcA_{j}^{(j)}
  =\bcA_{j}^{(j)} ,
   \\
  V_j^b
  &=\sum_{\ell=1}^{3j+2}\cB_{\ell}^{(j)}\eps_z^{-\ell}
   =0
   ,\qquad&
  \cB_{\ell}^{(j)}
  &=\bigg\{
    \begin{array}{ll}
      \tcB_{\ell}^{(j)} & (1\leq\ell\leq j+1) \\
      \bcB_{\ell}^{(j)} & (j+1\leq\ell\leq 3j+2)
    \end{array}
   ,\qquad&
  \tcB_{j+1}^{(j)}
  =\bcB_{j+1}^{(j)} ,
   \\
  V_j^d
  &=\sum_{\ell=1}^{3j}\cD_{\ell}^{(j)}\eps_z^{-\ell}
   =0
   ,\qquad&
  \cD_{\ell}^{(j)}
  &=\bigg\{
    \begin{array}{ll}
      \tcD_{\ell}^{(j)} & (1\leq\ell\leq j) \\
      \bcD_{\ell}^{(j)} & (j\leq\ell\leq 3j)
    \end{array}
   ,\qquad&
  \tcD_{j}^{(j)}
  =\bcD_{j}^{(j)} ,
   \\
  V_j^c
  &=\sum_{\ell=1}^{3j+1}\cC_{\ell}^{(j)}\eps_z^{-\ell}
   =0
   ,\qquad&
  \cC_{\ell}^{(j)}
  &=\bigg\{
    \begin{array}{ll}
      \tcC_{\ell}^{(j)} & (1\leq\ell\leq j) \\
      \bcC_{\ell}^{(j)} & (j\leq\ell\leq 3j+1)
    \end{array}
   ,\qquad&
  \tcC_{j}^{(j)}
  =\bcC_{j}^{(j)} .
\end{alignat}
$V_j^{a,b,d,c}=0$ since all the coefficients $\cA_{\ell}^{(j)}$, $\cB_{\ell}^{(j)}$,
$\cD_{\ell}^{(j)}$ and $\cC_{\ell}^{(j)}$ are zero, \ie,
\begin{alignat}{3}
  \tcA_{\ell}^{(j)}
  &=\sum_{k=0}^{2j} \alp_{jk}^{(\ell)}
   =0
   ,\qquad&
  \bcA_{\ell}^{(j)}
  &=\sum_{k=0}^{3j-\ell} \alp_{jk}^{(\ell)}
   =0
   ,\qquad&
  \alp_{jk}^{(\ell)}
  &=\sum_{s=k}^{2j}
    \bar{\alp}_{jk\ell}^{(s)} ,
    \label{eq:tcA_bcA}
   \\
  \tcB_{\ell}^{(j)}
  &=\sum_{k=0}^{2j+1} \beta_{jk}^{(\ell)}
   =0
   ,\qquad&
  \bcB_{\ell}^{(j)}
  &=\sum_{k=0}^{3j+2-\ell} \beta_{jk}^{(\ell)}
   =0
   ,\qquad&
  \beta_{jk}^{(\ell)}
  &=\sum_{s=k}^{2j+1}
    \bar{\beta}_{jk\ell}^{(s)} ,
   \\
  \tcD_{\ell}^{(j)}
  &=\sum_{k=0}^{2j} \delta_{jk}^{(\ell)}
   =0
   ,\qquad&
  \bcD_{\ell}^{(j)}
  &=\sum_{k=0}^{3j-\ell} \delta_{jk}^{(\ell)}
   =0
   ,\qquad&
  \delta_{jk}^{(\ell)}
  &=\sum_{s=k}^{2j}
    \bar{\delta}_{jk\ell}^{(s)} ,
   \\
  \tcC_{\ell}^{(j)}
  &=\sum_{k=0}^{2j+1}\gam_{jk}^{(\ell)}
   =0
   ,\qquad&
  \bcC_{\ell}^{(j)}
  &=\sum_{k=0}^{3j+1-\ell}\gam_{jk}^{(\ell)}
   =0
   ,\qquad&
  \gam_{jk}^{(\ell)}
  &=\sum_{s=k}^{2j+1}
    \bar{\gam}_{jk\ell}^{(s)} ,
    \label{eq:tcC_bcC}
\end{alignat}
where
\begin{alignat}{2}
  \bar{\alp}_{jk\ell}^{(s)}
  &=\frac{\kap_{2j-s}}{2^s}
   \mfa_{s-k}^{(s)} p_{3j-\ell-k}^{(s)} ,
    \qquad&
  \bar{\beta}_{jk\ell}^{(s)}
  &=\frac{\lam_{2j+1-s}}{2^s}
    \mfa_{s-k}^{(s)} p_{3j-\ell-k+2}^{(s)} ,
   \\
  \bar{\delta}_{jk\ell}^{(s)}
  &=\frac{\lam_{2j-s}}{2^s}
    \mfb_{s-k}^{(s)} p_{3j-\ell-k}^{(s)} ,
    \qquad&
  \bar{\gam}_{jk\ell}^{(s)}
  &=\frac{\kap_{2j+1-s}}{2^s}
    \mfb_{s-k}^{(s)} p_{3j-\ell-k+1}^{(s)} .
\end{alignat}
We change the order of the sums with respect to $k$ and $s$ in
Eqs.(\ref{eq:tcA_bcA}-\ref{eq:tcC_bcC}) for convenience in the numerical verification
that Eqs.(\ref{eq:tcA_bcA}-\ref{eq:tcC_bcC}) are zero up to a certain precision.
That is, when $1\leq\ell\leq j~(+1)$,
\begin{alignat}{2}
  \tcA_{\ell}^{(j)}
  =
    \sum_{s=0}^{2j}
    \sum_{k=0}^{s} \bar{\alp}_{jk\ell}^{(s)}
   ,\qquad
  \tcB_{\ell}^{(j)}
  =
    \sum_{s=0}^{2j+1}
    \sum_{k=0}^{s} \bar{\beta}_{jk\ell}^{(s)}
   ,\qquad
  \tcD_{\ell}^{(j)}
  =
    \sum_{s=0}^{2j}
    \sum_{k=0}^{s} \bar{\delta}_{jk\ell}^{(s)}
   ,\qquad
  \tcC_{\ell}^{(j)}
  =
    \sum_{s=0}^{2j+1}
    \sum_{k=0}^{s} \bar{\gam}_{jk\ell}^{(s)} .
\end{alignat}
When $j+1\leq\ell\leq 3j~(+1~\text{or}~2)$, we change the order of $k$ and $s$ as follows,
\begin{align}
  \bcA_{\ell}^{(j)}
  &=
    \sum_{s=0}^{3j-\ell-1}\sum_{k=0}^{s} \bar{\alp}_{jk\ell}^{(s)}
   +\sum_{s=2j}^{3j-\ell}\sum_{k=0}^{3j-\ell} \bar{\alp}_{jk\ell}^{(s)}
   =
    \sum_{s=0}^{3j-\ell}\sum_{k=0}^{s} \bar{\alp}_{jk\ell}^{(s)}
   +\sum_{s=2j}^{3j-\ell+1}\sum_{k=0}^{3j-\ell} \bar{\alp}_{jk\ell}^{(s)} ,
   \\
  \bcB_{\ell}^{(j)}
  &
   =\sum_{s=0}^{3j-\ell+1}
    \sum_{k=0}^{s} \bar{\beta}_{jk\ell}^{(s)}
   +\sum_{s=2j+1}^{3j-\ell+2}
    \sum_{k=0}^{3j-\ell+2} \bar{\beta}_{jk\ell}^{(s)}
   =\sum_{s=0}^{3j-\ell+2}
    \sum_{k=0}^{s} \bar{\beta}_{jk\ell}^{(s)}
   +\sum_{s=2j+1}^{3j-\ell+3}
    \sum_{k=0}^{3j-\ell+2} \bar{\beta}_{jk\ell}^{(s)} ,
   \\
  \bcD_{\ell}^{(j)}
  &=
    \sum_{s=0}^{3j-\ell-1}\sum_{k=0}^{s} \bar{\delta}_{jk\ell}^{(s)}
   +\sum_{s=2j}^{3j-\ell}\sum_{k=0}^{3j-\ell} \bar{\delta}_{jk\ell}^{(s)}
   =
    \sum_{s=0}^{3j-\ell}\sum_{k=0}^{s} \bar{\delta}_{jk\ell}^{(s)}
   +\sum_{s=2j}^{3j-\ell+1}\sum_{k=0}^{3j-\ell} \bar{\delta}_{jk\ell}^{(s)} ,
   \\
  \bcC_{\ell}^{(j)}
  &
   =\sum_{s=0}^{3j-\ell}
    \sum_{k=0}^{s} \bar{\gam}_{jk\ell}^{(s)}
   +\sum_{s=2j+1}^{3j-\ell+1}
    \sum_{k=0}^{3j-\ell+1} \bar{\gam}_{jk\ell}^{(s)}
   =\sum_{s=0}^{3j-\ell+1}
    \sum_{k=0}^{s} \bar{\gam}_{jk\ell}^{(s)}
   +\sum_{s=2j+1}^{3j-\ell+2}
    \sum_{k=0}^{3j-\ell+1} \bar{\gam}_{jk\ell}^{(s)} .
\end{align}
The vanishing coefficients given by Eqs.(\ref{eq:tcA_bcA}-\ref{eq:tcC_bcC}) tend to have
a larger error for larger $j$ because of digit loss in the numerical summation.
By this, the application of Eqs.(\ref{eq:taj}-\ref{eq:tcj}) is limited in
$j\leq6$ in double precision.
We need quadruple precision to compute Eqs.(\ref{eq:taj}-\ref{eq:tcj}) for a higher accuracy.

\subsection{Uniform asymptotic expansion of the cross products of $J_{\nu}$ and $Y_{\nu}$}
\label{sec:JY_uae}

$\rd_{\nu}p_{\nu}$ and $\rd_{\nu}s_{\nu}$ in terms of $J_{\nu}$ and $Y_{\nu}$ are given as
\begin{align}
  \rd_{\nu}p_{\nu}(b,a)
  &=\{\rd_{\nu}J_{\nu}(b)\}Y_{\nu}(a)
   +J_{\nu}(b)\rd_{\nu}Y_{\nu}(a)
   -\{\rd_{\nu}Y_{\nu}(b)\}J_{\nu}(a)
   -Y_{\nu}(b)\rd_{\nu}J_{\nu}(a) ,
  \\
  \rd_{\nu}s_\nu(b,a)
  &=\{\rd_{\nu}J_{\nu}'(b)\}Y_{\nu}'(a)
   +J_{\nu}'(b)\rd_{\nu}Y_{\nu}'(a)
   -\{\rd_{\nu}Y_{\nu}'(b)\}J_{\nu}'(a)
   -Y_{\nu}'(b)\rd_{\nu}J_{\nu}'(a) .
\end{align}
The uniform asymptotic expansion of $J_{\nu}$ and $Y_{\nu}$ for $\nu,z\in\mathbb{R}$ is
given by Eqs.(\ref{eq:JY_uae}-\ref{eq:dJY_uae}),
\begin{align}
  \bigg\{{ J_{\nu}(\nu z) \atop -Y_{\nu}(\nu z) }\bigg\}
  &=F_0
    \bigg[
      \frac{H_a}{\nu^{1/3}}\bigg\{{ \Ai(u) \atop \Bi(u) }\bigg\}
     +\frac{H_b}{\nu^{5/3}}\bigg\{{ \Ai'(u) \atop \Bi'(u) }\bigg\}
    \bigg] ,
   \label{eq:JY_uae_num}
   \\
  \bigg\{{ -J_{\nu}'(\nu z) \atop Y_{\nu}'(\nu z) }\bigg\}
  &=F_1
    \bigg[
      \frac{H_d}{\nu^{2/3}}\bigg\{{ \Ai'(u) \atop \Bi'(u) }\bigg\}
     +\frac{H_c}{\nu^{4/3}}\bigg\{{ \Ai(u) \atop \Bi(u) }\bigg\}
    \bigg] ,
   \\
  \rd_{\nu}\bigg\{{ J_{\nu}(\nu z) \atop -Y_{\nu}(\nu z) }\bigg\}
  &=F_0
    \bigg[
      \frac{E_a}{\nu^{1/3}}\bigg\{{ \Ai(u) \atop \Bi(u) }\bigg\}
     +\frac{E_b}{\nu^{5/3}}\bigg\{{ \Ai'(u) \atop \Bi'(u) }\bigg\}
    \bigg] ,
   \\
  \rd_{\nu}\bigg\{{ -J_{\nu}'(\nu z) \atop Y_{\nu}'(\nu z) }\bigg\}
  &=F_1
    \bigg[
      \frac{E_d}{\nu^{2/3}}\bigg\{{ \Ai'(u) \atop \Bi'(u) }\bigg\}
     +\frac{E_c}{\nu^{4/3}}\bigg\{{ \Ai(u) \atop \Bi(u) }\bigg\}
    \bigg] .
   \label{eq:dJYp_uae_num}
\end{align}
$u$ is given by Eq.(\ref{eq:f_z_zeta}),
\begin{align}
  u
  =\nu^{2/3}\zeta
   ,\qquad
  \rd_{\nu}u
  =\frac{q}{\nu^{1/3}}
   ,\qquad
  u\rd_{\nu}u
  =\nu^{1/3}\zeta q
   ,\qquad
  q
  =e_z+\frac{2}{3}\zeta
   ,\qquad
  \chi
  =\frac{1-x}{2e_z} .
   \label{eq:uq}
\end{align}
$\chi$ is given by Eq.(\ref{eq:M}).
$F_{0}$ and $F_{1}$ in Eqs.(\ref{eq:JY_uae_num}-\ref{eq:dJYp_uae_num}) are given as
\begin{alignat}{4}
  F_0
  &=\frac{2^{1/2}}{f}
   ,\qquad&
  \rd_zF_0
  &=-F_0\frac{1-x}{2z}
   ,\qquad&
  \rd_{\nu}F_0
  &=F_0\frac{1-x}{2\nu}
   ,\qquad&
  \rd_{\zeta}F_0
  &=F_0\chi ,
    \label{eq:F0_app}
   \\
  F_1
  &=\frac{2^{1/2}f}{z}
   ,\qquad&
  \rd_zF_1
  &=-F_1\frac{1+x}{2z}
   ,\qquad&
  \rd_{\nu}F_1
  &=F_1\frac{1+x}{2\nu}
   ,\qquad&
  \rd_{\zeta}F_1
  &=F_1(e_z^{-1}-\chi) .
    \label{eq:F1_app}
\end{alignat}
$f$ is given by Eq.(\ref{eq:f_z_zeta}).
$x$ is a function of $z$ given below,
\begin{align}
  &
  f
  =\bigg(\frac{\eps_z}{\zeta}\bigg)^{1/4}
   ,\qquad
  \eps_z
  =1-z^2
    ,\qquad
  \rd_{\nu}z
  =-\frac{z}{\nu}
   ,\qquad
  \rd_z\zeta
  =-\frac{e_z}{z}
    ,\qquad
  e_z
  =f^2
  =\nu\rd_{\nu}\zeta ,
  \label{eq:fz_app}
   \\
  &
  \rd_zf
  =\frac{f}{2z}(1-x)
    ,\qquad
  \rd_{\nu}f
  =\frac{f}{2\nu}(x-1)
    ,\qquad
  x\eps_z
  =1-\frac{e_z^3}{2}
    ,\qquad
  (x-1)\eps_z
  =z^2-\frac{e_z^3}{2} .
  \label{eq:dfz_app}
\end{align}
$H_{a,b,d,c}$ and $E_{a,b,d,c}$ are given as
\begin{alignat}{3}
  H_a
  &=\sum_{j=0}^{\infty}\frac{a_j(\zeta)}{\nu^{2j}}
   ,\qquad&
  a_j(\zeta)
  &=\sum_{s=0}^{2j}\kap_s\frac{e_z^s}{\zeta^s}\bA_s^{(j)}
   ,\qquad&
  \bA_s^{(j)}
  &=\sum_{k=0}^{2j-s}\frac{\mfa_k^{(2j-s)}}{\eps_z^{k+j}} ,
    \label{eq:Ha}
   \\
  H_b
  &=\sum_{j=0}^{\infty}\frac{b_j(\zeta)}{\nu^{2j}}
   ,\qquad&
  b_j(\zeta)
  &=-\sum_{s=0}^{2j+1}\lam_s\frac{e_z^{s+1}}{\zeta^s}\bB_s^{(j)}
   ,\qquad&
  \bB_s^{(j)}
  &=\sum_{k=0}^{2j-s+1}\frac{\mfa_k^{(2j-s+1)}}{\eps_z^{k+j+1}} ,
   \\
  H_d
  &=\sum_{j=0}^{\infty}\frac{d_j(\zeta)}{\nu^{2j}}
   ,\qquad&
  d_j(\zeta)
  &=\sum_{s=0}^{2j}\lam_s\frac{e_z^s}{\zeta^s}\bD_s^{(j)}
   ,\qquad&
  \bD_s^{(j)}
  &=\sum_{k=0}^{2j-s} \frac{\mfb_k^{(2j-s)}}{\eps_z^{k+j}} ,
   \\
  H_c
  &=\sum_{j=0}^{\infty}\frac{c_j(\zeta)}{\nu^{2j}}
   ,\qquad&
  c_j(\zeta)
  &=-\sum_{s=0}^{2j+1}\kap_s\frac{e_z^{s+1}}{\zeta^{s-1}}\bC_s^{(j)}
   ,\qquad&
  \bC_s^{(j)}
  &=\sum_{k=0}^{2j-s+1} \frac{\mfb_k^{(2j-s+1)}}{\eps_z^{k+j+1}} ,
    \label{eq:Hc}
\end{alignat}
and
\begin{alignat}{2}
  E_a
  &=M_a+\zeta\frac{q}{\nu}H_b
   ,\qquad&
  M_a
  &=
    \rd_{\nu}H_a
   +\bigg(\frac{1-x}{2}-\frac{1}{3}\bigg)\frac{H_a}{\nu} ,
   \\
  E_b
  &=M_b+\nu qH_a
   ,\qquad&
  M_b
  &=
    \rd_{\nu}H_b
   +\bigg(\frac{1-x}{2}-\frac{5}{3}\bigg)\frac{H_b}{\nu} ,
   \\
  E_d
  &=M_d
   +\frac{q}{\nu}H_c
   ,\qquad&
  M_d
  &=
    \rd_{\nu}H_d
   +\bigg(\frac{1+x}{2}-\frac{2}{3}\bigg)\frac{H_d}{\nu} ,
   \\
  E_c
  &=M_c
   +\zeta\nu q H_d
   ,\qquad&
  M_c
  &=
    \rd_{\nu}H_c
   +\bigg(\frac{1+x}{2}-\frac{4}{3}\bigg)\frac{H_c}{\nu} .
\end{alignat}
We cannot use Eqs.(\ref{eq:Ha}-\ref{eq:Hc}) for $z=1$ since $\eps_z$ becomes zero.
$M_{a,b,d,c}$ is given as follows,
\begin{alignat}{3}
  M_a
  &=\sum_{j=0}^{\infty}\frac{\ha_j}{\nu^{2j+1}}
   ,\qquad&
  \ha_j
  &=\sum_{s=0}^{2j}\kap_s\frac{e_z^s}{\zeta^s}\hA_s^{(j)}
   ,\qquad&
  \hA_s^{(j)}
  &=\sum_{k=0}^{2j-s}\frac{\mfa_k^{(2j-s)}}{\eps_z^{k+j+1}}
    \Phi_{s,k}^{(j)} ,
    \label{eq:Ma}
   \\
  M_b
  &=\sum_{j=0}^{\infty}\frac{\hb_j}{\nu^{2j+1}}
   ,\qquad&
  \hb_j
  &=-\sum_{s=0}^{2j+1}\lam_s\frac{e_z^{s+1}}{\zeta^s}\hB_s^{(j)}
   ,\qquad&
  \hB_s^{(j)}
  &=\sum_{k=0}^{2j-s+1}\frac{\mfa_k^{(2j-s+1)}}{\eps_z^{k+j+2}}
    \Psi_{s,k}^{(j+1)} ,
    \\
 M_d
  &=\sum_{j=0}^{\infty}\frac{\hd_j}{\nu^{2j+1}}
   ,\qquad&
  \hd_j
  &=\sum_{s=0}^{2j}\lam_s\frac{e_z^s}{\zeta^s}\hD_s^{(j)}
   ,\qquad&
  \hD_s^{(j)}
  &=\sum_{k=0}^{2j-s}\frac{\mfb_k^{(2j-s)}}{\eps_z^{k+j+1}}
    \Psi_{s,k}^{(j)} ,
   \\
  M_c
  &=\sum_{j=0}^{\infty}\frac{\hc_j}{\nu^{2j+1}}
   ,\qquad&
  \hc_j
  &=-\sum_{s=0}^{2j+1}\kap_s\frac{e_z^{s+1}}{\zeta^{s-1}}\hC_s^{(j)}
   ,\qquad&
  \hC_s^{(j)}
  &=\sum_{k=0}^{2j-s+1}\frac{\mfb_k^{(2j-s+1)}}{\eps_z^{k+j+2}}
    \Phi_{s,k}^{(j)} .
    \label{eq:Mc}
\end{alignat}
$\mfa_k^{(\ell)}$ and $\mfb_k^{(\ell)}$ are the coefficients of the series
$u_{\ell}$ and $v_{\ell}$ given by Eq.(\ref{eq:uv_app}).
$\Phi_{s,k}^{(j)}$ and $\Psi_{s,k}^{(j)}$ are given as
\begin{align}
  \{\Phi_{s,k}^{(j)},\Psi_{s,k}^{(j)}\}
  &=\Delta_{s,k}^{(j)}\mp\frac{\vpi+2}{6}
   ,\qquad
  \Delta_{s,k}^{(j)}
  =s\vpi
   -2(j+kz^2)
   ,\qquad
  \vpi
  =z^2-\frac{3}{2\psi} ,
    \label{eq:Phi_Psi}
\end{align}
where the double sign $\{-,+\}$ corresponds to $\{\Phi_{s,k}^{(j)},\Psi_{s,k}^{(j)}\}$.
The coefficients $(\ha_j,\hb_j,\hd_j,\hc_j)$, given by the second equations of
(\ref{eq:Ma}-\ref{eq:Mc}), are gotten from
\begin{alignat}{2}
  \ha_j
  &=
      \bigg(\frac{1-x}{2}-\frac{1}{3}-2j\bigg)a_j
     +\nu\rd_{\nu}a_j ,
    \qquad&
  \hb_j
  &=
      \bigg(\frac{1-x}{2}-\frac{5}{3}-2j\bigg)b_j
     +\nu\rd_{\nu}b_j ,
    \label{eq:haj_hbj}
   \\
  \hd_j
  &=
      \bigg(\frac{1+x}{2}-\frac{2}{3}-2j\bigg)d_j
     +\nu\rd_{\nu}d_j ,
    \qquad&
  \hc_j
  &=
      \bigg(\frac{1+x}{2}-\frac{4}{3}-2j\bigg)c_j
     +\nu\rd_{\nu}c_j .
    \label{eq:hdj_hcj}
\end{alignat}
$x$ is given by Eqs.(\ref{eq:uq}) and (\ref{eq:dfz_app}).
$x$ is related to $\chi$ and $e_z$ given by Eqs.(\ref{eq:M}) and (\ref{eq:zeta_ez}),
\begin{align}
  \frac{1-x}{2}
  =e_z\chi
   ,\qquad
  \frac{1+x}{2}
  =1-e_z\chi .
   \label{eq:xchi}
\end{align}
$(\ha_j,\hb_j,\hd_j,\hc_j)$ are related to $(a_j,b_j,d_j,c_j)$ through
Eqs.(\ref{eq:djcj_ajbj}-\ref{eq:djcj_ajbj_1}),
\begin{alignat}{2}
  \ha_j
  &=
    e_z(c_j-\zeta b_j)
   -(2j+1/3)a_j
   ,\qquad&
  \hb_j
  &=
    e_z(d_{j+1}-a_{j+1})
   -\{2(j+1)-1/3\}b_j ,
   \label{eq:ha_hb}
   \\
  \hd_j
  &=
    e_z(\zeta b_j-c_j)
   -(2j-1/3)d_j
   ,\qquad&
  \hc_j
  &=
    \zeta e_z(a_{j+1}-d_{j+1})
   -(2j+1/3)c_j .
   \label{eq:hd_hc}
\end{alignat}
From Eqs.(\ref{eq:ha_hb}-\ref{eq:hd_hc}), we get
\begin{align}
  -(\ha_j+\hd_j)
  &=
    (2j+1/3)a_j
   +(2j-1/3)d_j ,
   \\
  -(\zeta\hb_j+\hc_j)
  &=
    \{2(j+1)-1/3\}\zeta b_j
   +(2j+1/3)c_j .
   \label{eq:hb_hc}
\end{align}
We can use Eqs.(\ref{eq:ha_hb}-\ref{eq:hb_hc}) to evaluate the numerical accuracy of
$(a_j,b_j,d_j,c_j)$ and $(\ha_j,\hb_j,\hd_j,\hc_j)$ given by
Eqs.(\ref{eq:taj}-\ref{eq:tcj}), (\ref{eq:Ha}-\ref{eq:Mc}) and
(\ref{haj_hbj_191128}-\ref{hdj_hcj_191128}) in
which many digits are lost due to cancelation.

In computing $(a_j,b_j,d_j,c_j)$ around $z=1$,
since $\zeta=\eps_z=0$ at $z=1$, we use Eqs.(\ref{eq:taj}-\ref{eq:tcj})
instead of the second equations of (\ref{eq:Ha}-\ref{eq:Hc}).
Similarly, in computing $(\ha_j,\hb_j,\hd_j,\hc_j)$ around $z=1$, we use
the following expressions instead of the second equations of (\ref{eq:Ma}-\ref{eq:Mc}),
\begin{alignat}{2}
  \frac{\ha_j}{e_z^{6j}}
  &=
    \sum_{\ell=0}^{\infty}\eps_z^{\ell}
    \big\{
      u_a^{(j)}\ta_{\ell}^{(j)}
     +2(\ell+1)\ta_{\ell+1}^{(j)}
    \big\} ,
    \qquad&
  \frac{\hb_j}{e_z^{6j+4}}
  &=
   -\sum_{\ell=0}^{\infty}\eps_z^{\ell}
    \big\{
      u_b^{(j+1)}\tb_{\ell}^{(j)}
     +2(\ell+1)\tb_{\ell+1}^{(j)}
    \big\} ,
    \label{haj_hbj_191128}
   \\
  \frac{\hd_j}{e_z^{6j}}
  &=
    \sum_{\ell=0}^{\infty}\eps_z^{\ell}
    \big\{
      u_d^{(j)}\td_{\ell}^{(j)}
     +2(\ell+1)\td_{\ell+1}^{(j)}
    \big\} ,
    \qquad&
  \frac{\hc_j}{e_z^{6j+2}}
  &=
   -\sum_{\ell=0}^{\infty}\eps_z^{\ell}
    \big\{
      u_c^{(j)}\tc_{\ell}^{(j)}
     +2(\ell+1)\tc_{\ell+1}^{(j)}
    \big\} .
    \label{hdj_hcj_191128}
\end{alignat}
$\ta_{\ell}^{(j)}$, $\tb_{\ell}^{(j)}$, $\td_{\ell}^{(j)}$ and
$\tc_{\ell}^{(j)}$ are given by Eqs.(\ref{eq:taj}-\ref{eq:tcj}).
$u_{a,b,d,c}^{(j)}$ is given as
\begin{alignat}{3}
  u_a^{(j)}
  &=\bu_{\ell}^{(j)}
   -w_1
   ,\qquad&
  u_d^{(j)}
  &=\bu_{\ell}^{(j)}
   +w_1
   ,\qquad&
  w_1
  &=\frac{1}{2}w_0+\frac{1}{3} ,
   \\
  u_b^{(j)}
  &=\bu_{\ell}^{(j)}
   -w_2
   ,\qquad&
  u_c^{(j)}
  &=\bu_{\ell}^{(j)}
   +w_2
   ,\qquad&
  w_2
  &=\frac{5}{2}w_0-\frac{1}{3} ,
\end{alignat}
where
\begin{align}
  \bu_{\ell}^{(j)}
  =2j(3w_0-1)-2\ell
   ,\qquad
  w_0
  =-2e_z\chi
  =\frac{1}{\eps_z}
   \bigg(z^2-\frac{1}{2\psi}\bigg)
  =\frac{\psi_1}{\psi}-1 .
\end{align}
$\psi$ and $\chi$ are given by Eqs.(\ref{eq:psi}) and (\ref{eq:M}).
$\psi$ and $\psi_1$ are given as
\begin{align}
  \psi
  =\frac{3}{2}\sum_{n=0}^{\infty}\frac{\eps_z^n}{2n+3}
   ,\qquad
  \psi_1
  =(1-2e_z\chi)\psi
  =\frac{2\psi-1}{2\eps_z}
  =\frac{3}{2}\sum_{n=0}^{\infty}\frac{\eps_z^n}{2n+5} .
\end{align}
When $z=1$, since $\eps_z=0$, $\psi$, $\psi_1$ and $w_0$ go to the following values,
\begin{align}
  \lim_{z\to1}\psi
  =\frac{1}{2}
   ,\qquad
  \lim_{z\to1}\psi_1
  =\frac{3}{10}
   ,\qquad
  \lim_{z\to1}w_0
  =-\frac{2}{5} .
\end{align}
%


\subsection{Uniform asymptotic expansion of $J_{\nu}$ and $Y_{\nu}$ avoiding overflow}

In the numerical calculation,
the Airy functions $(\Ai,\Bi)$ and their derivatives $(\Ai',\Bi')$ can underflow or 
overflow for positive large argument as seen from their asymptotic expressions
(\ref{eq:Ai_Aip}-\ref{eq:Bi_Bip}).
In the cross products $t_{\nu}=\{p_{\nu},q_{\nu},r_{\nu},s_{\nu}\}$ given by
Eqs.(\ref{eq:CP_pq}-\ref{eq:CP_sr}), however, the exponential functions partially cancel.
In order to avoid the over/underflow in the numerical calculation of $t_{\nu}$,
we rewrite Eqs.(\ref{eq:CP_pq}-\ref{eq:CP_sr}) as follows,
\begin{alignat}{2}
  p_{\nu}(b,a)
  &=e^{\tau}\cJ_{\nu}(b)\cY_{\nu}(a)
   -e^{-\tau}\cY_{\nu}(b)\cJ_{\nu}(a)
   ,\qquad&
  q_{\nu}(b,a)
  &=e^{\tau}\cJ_{\nu}(b)\bcY_{\nu}(a)
   -e^{-\tau}\cY_{\nu}(b)\bcJ_{\nu}(a) ,
   \\
  s_\nu(b,a)
  &=e^{\tau}\bcJ_{\nu}(b)\bcY_{\nu}(a)
   -e^{-\tau}\bcY_{\nu}(b)\bcJ_{\nu}(a)
   ,\qquad&
  r_{\nu}(b,a)
  &=e^{\tau}\bcJ_{\nu}(b)\cY_{\nu}(a)
   -e^{-\tau}\bcY_{\nu}(b)\cJ_{\nu}(a) ,
\end{alignat}
where
\begin{align}
  \tau
  =U_a-U_b
   ,\qquad
  U_{a,b}
  =\frac{2}{3}u_{a,b}^{3/2} .
\end{align}
$u_{a,b}$ is $u$ given by Eq.(\ref{eq:u_up}) for $\hr=a,b$.
$(\cJ_{\nu},\cY_{\nu})$ and $(\bcJ_{\nu},\bcY_{\nu})$ are defined as follows,
\begin{alignat}{2}
  J_{\nu}(\hr)
  &=e^{-U}\cJ_{\nu}(\hr)
   ,\qquad&
  \cJ_{\nu}(\hr)
  &=
    \frac{2^{1/2}}{f(z)}
    \bigg[
       \frac{A_0(U)}{\nu^{1/3}}
       \sum_{j=0}^{\infty}\frac{a_j(\zeta)}{\nu^{2j}}
      +\frac{A_1(U)}{\nu^{5/3}}
       \sum_{j=0}^{\infty}\frac{b_j(\zeta)}{\nu^{2j}}
    \bigg] ,
  \\
  Y_{\nu}(\hr)
  &=e^{U}\cY_{\nu}(\hr)
   ,\qquad&
 \cY_{\nu}(\hr)
  &=-
    \frac{2^{1/2}}{f(z)}
    \bigg[
       \frac{B_0(U)}{\nu^{1/3}}
       \sum_{j=0}^{\infty}\frac{a_j(\zeta)}{\nu^{2j}}
      +\frac{B_1(U)}{\nu^{5/3}}
       \sum_{j=0}^{\infty}\frac{b_j(\zeta)}{\nu^{2j}}
    \bigg] ,
   \\
  J_{\nu}'(\hr)
  &=e^{-U}\bcJ_{\nu}(\hr)
   ,\qquad&
  \bcJ_{\nu}(\hr)
  &=-
    \frac{2^{1/2}f(z)}{z}
    \bigg[
      \frac{A_1(U)}{\nu^{2/3}}
      \sum_{j=0}^{\infty}\frac{d_j(\zeta)}{\nu^{2j}}
     +\frac{A_0(U)}{\nu^{4/3}}
      \sum_{j=0}^{\infty}\frac{c_j(\zeta)}{\nu^{2j}}
    \bigg] ,
   \\
  Y_{\nu}'(\hr)
  &=e^{U}\bcY_{\nu}(\hr)
   ,\qquad&
  \bcY_{\nu}(\hr)
  &=
    \frac{2^{1/2}f(z)}{z}
    \bigg[
      \frac{B_1(U)}{\nu^{2/3}}
      \sum_{j=0}^{\infty}\frac{d_j(\zeta)}{\nu^{2j}}
     +\frac{B_0(U)}{\nu^{4/3}}
      \sum_{j=0}^{\infty}\frac{c_j(\zeta)}{\nu^{2j}}
    \bigg] .
\end{alignat}
$A_{0,1}$ and $B_{0,1}$ are given as
\begin{alignat}{3}
  A_0(U)
  &=
   \frac{1}{2\pi^{1/2}u^{1/4}}\sum_{j=0}^{\infty}(-1)^j\frac{C_j}{U^j} ,
    \qquad&
  A_1(U)
  &=
   -\frac{u^{1/4}}{2\pi^{1/2}}\sum_{j=0}^{\infty}(-1)^j\frac{D_j}{U^j} ,
   \label{eq:A01}
  \\
  B_0(U)
  &=
   \frac{1}{\pi^{1/2}u^{1/4}}\sum_{j=0}^{\infty}\frac{C_j}{U^j} ,
    \qquad&
  B_1(U)
  &=
   \frac{u^{1/4}}{\pi^{1/2}}\sum_{j=0}^{\infty}\frac{D_j}{U^j} .
   \label{eq:B01}
\end{alignat}
They are the parts of the asymptotic expressions of the Airy functions for large $u$,
factoring $e^{\pm U}$ out,
\begin{align}
  \Ai(u)
  \simeq
  e^{-U} A_0(U)
   ,\qquad
  \Ai'(u)
  \simeq
  e^{-U} A_1(U)
   ,\qquad
  \Bi(u)
  \simeq
   e^{U}B_0(U)
   ,\qquad
  \Bi'(u)
  \simeq
   e^{U}B_1(U) .
\end{align}
%

\subsection{Uniform asymptotic expansion of $I_{\nu}$ and $K_{\nu}$ of real order}

The uniform asymptotic expansions of the modified Bessel functions
$I_{\nu}(\br)$ and $K_{\nu}(\br)$ are given by Eqs.(\ref{eq:IKnu_uae}).
We substitute them into the cross products given by Eqs.(\ref{eq:pnu_IK}-\ref{eq:snu_IK}),
\begin{alignat}{2}
  -p_{\nu}(i\br_b,i\br_a)
  &=c_1\alp_b\alp_a
    (
      U_{+}^bU_{-}^a e^{\tht}
     -U_{-}^bU_{+}^a e^{-\tht}
    ) ,
   \\
  iq_{\nu}(i\br_b,i\br_a)
  &=c_1\alp_b\beta_a
    (
      U_{+}^bV_{-}^a e^{\tht}
     +U_{-}^bV_{+}^a e^{-\tht}
    ) ,
   \\
  -ir_{\nu}(i\br_b,i\br_a)
  &=c_1\beta_b\alp_a
    (
      V_{+}^bU_{-}^a e^{\tht}
     +V_{-}^bU_{+}^a e^{-\tht}
    ) ,
   \\
  -s_{\nu}(i\br_b,i\br_a)
  &=c_1\beta_b\beta_a
    (
      V_{+}^bV_{-}^a e^{\tht}
     -V_{-}^bV_{+}^a e^{-\tht}
    ) ,
\end{alignat}
where
\begin{align}
  &
  c_1
  =\frac{1}{\pi\nu}
   ,\qquad
  \br_{a,b}
  =\nu\bz_{a,b}
   ,\qquad
  \alp_{a,b}
  =p_{a,b}^{1/2}
   ,\qquad
  \beta_{a,b}
  =(\bz_{a,b}\alp_{a,b})^{-1} ,
   \\
  &
  p_{a,b}
  =(1+\bz_{a,b}^2)^{-1/2}
   ,\qquad
  \tht
  =\nu(\xi_b-\xi_a)
   ,\qquad
  \xi_{a,b}
  =\xi(\bz_{a,b}) .
\end{align}
$\xi(\bz)$ is given by Eqs.(\ref{eq:p_xi}) and (\ref{eq:xi_bz}).
$U_{\pm}^{a,b}$ and $V_{\pm}^{a,b}$ are given by Eqs.(\ref{eq:UV_pm}) for $p=p_{a,b}$,
\begin{alignat}{2}
  U_{\pm}^{a,b}
  =U_{\pm}(p_{a,b})
   ,\qquad
  V_{\pm}^{a,b}
  =V_{\pm}(p_{a,b}) .
\end{alignat}
The derivatives of the cross products with respect to $\nu$ are given as follows,
\begin{align}
  \rd_{\nu}p_{\nu}(i\br_b,i\br_a)
  &=c_1\alp_b\alp_a
    \{\cU_{\nu}(\bz_b,\bz_a)e^{\tht}-\cU_{\nu}(\bz_a,\bz_b)e^{-\tht}\} ,
   \\
  -\rd_{\nu}s_{\nu}(i\br_b,i\br_a)
  &=c_1\beta_b\beta_a
    \{\cV_{\nu}(\bz_b,\bz_a)e^{\tht}-\cV_{\nu}(\bz_a,\bz_b)e^{-\tht}\} ,
   \\
  i\rd_{\nu}q_{\nu}(i\br_b,i\br_a)
  &=c_1\alp_b\beta_a
    \{
    \cW_{\nu}^{+}(\bz_b,\bz_a)e^{\tht}
   +\cW_{\nu}^{-}(\bz_b,\bz_a)e^{-\tht}
    \} ,
   \\
  -i\rd_{\nu}r_{\nu}(i\br_b,i\br_a)
  &=c_1\beta_b\alp_a
    \{
    \cW_{\nu}^{-}(\bz_a,\bz_b)e^{\tht}
   +\cW_{\nu}^{+}(\bz_a,\bz_b)e^{-\tht}
    \} ,
\end{align}
where
\begin{align}
  \cU_{\nu}(\bz_b,\bz_a)
  &=
    (d_{\nu}-\rd_{\nu}\tht)U_{+}^bU_{-}^a
   -(U_{-}^a\rd_{\nu}U_{+}^b+U_{+}^b\rd_{\nu}U_{-}^a) ,
   \\
  \cV_{\nu}(\bz_b,\bz_a)
  &=
    (d_{\nu}+\rd_{\nu}\tht)V_{+}^bV_{-}^a
   +(V_{-}^a\rd_{\nu}V_{+}^b+V_{+}^b\rd_{\nu}V_{-}^a) ,
   \\
  \cW_{\nu}^{+}(\bz_b,\bz_a)
  &=
      (\bar{d}_{\nu}+\rd_{\nu}\tht)U_{+}^bV_{-}^a
     +(V_{-}^a\rd_{\nu}U_{+}^b+U_{+}^b\rd_{\nu}V_{-}^a) ,
    \\
  \cW_{\nu}^{-}(\bz_b,\bz_a)
  &=
      (\bar{d}_{\nu}-\rd_{\nu}\tht)U_{-}^bV_{+}^a
     +(V_{+}^a\rd_{\nu}U_{-}^b+U_{-}^b\rd_{\nu}V_{+}^a) .
\end{align}
$\rd_{\nu}U_{\pm}$ and $\rd_{\nu}V_{\pm}$ are given by Eqs.(\ref{eq:UV_pm}).
$d_{\nu}$, $\bar{d}_{\nu}$ and $\rd_{\nu}\tht$ are given as
\begin{align}
  d_{\nu}
  &=\frac{p_b^2+p_a^2}{2\nu}
   ,\qquad
  \bar{d}_{\nu}
  =\frac{p_a^2-p_b^2}{2\nu}
   ,\qquad
  \rd_{\nu}\tht
  =\frac{\tht}{\nu}-(p_b^{-1}-p_a^{-1})
  =\log\bigg(\frac{\bz_b}{\bz_a}\cd\frac{1+p_a^{-1}}{1+p_b^{-1}}\bigg) .
\end{align}
%

\subsection{Uniform asymptotic expansion of $K_{i\bnu}$ and $L_{i\bnu}$}
\label{sec:uae_KL}

We rewrite Eqs.(\ref{eq:Pnu_ba}-\ref{eq:Snu_ba}) which are the cross products of
the modified Bessel functions $K_{i\bnu}(\br)$ and $\bL_{i\bnu}(\br)$ having
purely imaginary order,
\begin{alignat}{2}
  P_{i\bnu}(\br,\br')
  &=\bL_{i\bnu}(\br)K_{i\bnu}(\br')-K_{i\bnu}(\br)\bL_{i\bnu}(\br')
  &&=-\frac{\pi}{2}p_\nu(\hr,\hr') ,
   \label{eq:Pibnu_pnu}
   \\
  Q_{i\bnu}(\br,\br')
  &=\bL_{i\bnu}(\br)K_{i\bnu}'(\br')-K_{i\bnu}(\br)\bL_{i\bnu}'(\br')
  &&=-i\frac{\pi}{2}q_\nu(\hr,\hr') ,
   \\
  R_{i\bnu}(\br,\br')
  &=\bL_{i\bnu}'(\br)K_{i\bnu}(\br')-K_{i\bnu}'(\br)\bL_{i\bnu}(\br')
  &&=-i\frac{\pi}{2}r_\nu(\hr,\hr') ,
   \\
  S_{i\bnu}(\br,\br')
  &=\bL_{i\bnu}'(\br)K_{i\bnu}'(\br')-K_{i\bnu}'(\br)\bL_{i\bnu}'(\br')
  &&=\frac{\pi}{2}s_\nu(\hr,\hr') .
   \label{eq:Sibnu_snu}
\end{alignat}
The argument $\br$ is independent of the order $\nu$,
\begin{align}
  \br
  &=\bnu z ,
   \qquad
  \hr
  =i\br
  =i\bnu z
  =\nu z
   \qquad\Lra\qquad
  z
  =\frac{\hr}{\nu}
  =\frac{\br}{\bnu} ,
    \label{eq:br_z}
   \\
  \nu
  &=i\bnu
   \quad
  (\bnu\in\mathbb{R})
   ,\qquad
  \rd_{\nu}
  =-i\rd_{\bnu}
   \qquad\Lra\qquad
  i\rd_{\nu}
  =\rd_{\bnu} .
    \label{eq:nu_bnu_app}
\end{align}
The uniform asymptotic series of $K_{i\bnu}$ and $\bL_{i\bnu}$ are given as follows,
\begin{align}
  \bigg\{{ K_{i\bnu}(\bnu z) \atop \bL_{i\bnu}(\bnu z) }\bigg\}
  &= \bigg\{{ e_K \atop e_L }\bigg\}
    F_0(z)
    \bigg[
      \frac{S_a}{\bnu^{1/3}}
      \bigg\{{ \Ai(\bu) \atop \Bi(\bu) }\bigg\}
     +\frac{S_b}{\bnu^{5/3}}
      \bigg\{{ \Ai'(\bu) \atop \Bi'(\bu) }\bigg\}
     +\veps_{2n+1}^{(0,*)}
    \bigg] ,
   \label{eq:KLb_uae}
   \\
  \bigg\{{ K_{i\bnu}'(\bnu z) \atop \bL_{i\bnu}'(\bnu z) }\bigg\}
  &= \bigg\{{ e_K \atop e_L }\bigg\}
    F_1(z)
   \bigg[
      \frac{S_d}{\bnu^{2/3}}
      \bigg\{{ \Ai'(\bu) \atop \Bi'(\bu) }\bigg\}
     +\frac{S_c}{\bnu^{4/3}}
      \bigg\{{ \Ai(\bu) \atop \Bi(\bu) }\bigg\}
     +\bveps_n^{(0,*)}
   \bigg] .
   \label{eq:KLb_prm_uae}
\end{align}
$F_0$ and $F_1$ are given by Eqs.(\ref{eq:F0_app}-\ref{eq:F1_app}).
$e_K$ and $e_L$ are given as
\begin{align}
  e_K
  =\pi e^{-\pi\bnu/2} ,
    \qquad
  e_L
  =\frac{e^{\pi\bnu/2}}{2} .
   \label{eq:eKL}
\end{align}
$\zeta$ is related to $z$ through Eqs.(\ref{eq:zeta_p}-\ref{eq:zeta_m}).
$\bu$ is given as
\begin{align}
  \bu
  &=-\bnu^{2/3}\zeta
   ,\qquad
  \rd_{\bnu}\bu
  =-\frac{q}{\bnu^{1/3}}
   ,\qquad
  \bu\rd_{\bnu}\bu
  =\bnu^{1/3}\zeta q
   ,\qquad
  q
  =e_z+\frac{2}{3}\zeta .
    \label{eq:bu_q}
\end{align}
$S_{a,b,d,c}$ is given as follows,
\begin{align}
   \bigg\{{ S_a \atop S_d }\bigg\}
  &=\sum_{j=0}^{n}\frac{(-1)^j}{\bnu^{2j}}
     \bigg\{{ a_j(\zeta) \atop d_j(\zeta) }\bigg\}
   ,\qquad
   \bigg\{{ S_b \atop S_c }\bigg\}
  =\sum_{j=0}^{n-1}\frac{(-1)^j}{\bnu^{2j}}
     \bigg\{{ b_j(\zeta) \atop -c_j(\zeta) }\bigg\} .
    \label{eq:Sab}
\end{align}
The coefficients $(a_j,b_j,d_j,c_j)$ are defined in Eqs.(\ref{eq:aj_bj}-\ref{eq:cj_dj})
and given by Eqs.(\ref{eq:aj_bj_rw}-\ref{eq:dj_cj_rw}).
The series (\ref{eq:Sab}) have the upper limit
$n\leq32$ in our numerical code at present.
It is limited by the finite number of the coefficients $u_n$ and $v_n$
given by Eqs.(\ref{eq:uj}).
We can increase $n$ in Eqs.(\ref{eq:Sab}) if we find $u_n$ and $v_n$ up to a higher order.
Increasing $n$, however, the accuracy of Eqs.(\ref{eq:KLb_uae}-\ref{eq:KLb_prm_uae}) does not
necessarily get better since many significant digits are lost due to the cancelation among
the terms in Eqs.(\ref{eq:AB_bar}-\ref{eq:DC_bar}).

In order to rearrange the expressions of $\rd_{\nu}p_{\nu}$ and $\rd_{\nu}s_{\nu}$ for
$\nu=i\bnu$, we rewrite Eqs.(\ref{eq:Pibnu_pnu}) and (\ref{eq:Sibnu_snu}),
\begin{alignat}{2}
  p_\nu(\hr,\hr')
  &=
    \hK_{i\bnu}(\br)\hL_{i\bnu}(\br')
   -\hL_{i\bnu}(\br)\hK_{i\bnu}(\br') ,
  \label{eq:pnu_hKL}
   \\
  iq_\nu(\hr,\hr')
  &=
    \hK_{i\bnu}(\br)\hL_{i\bnu}'(\br')
   -\hL_{i\bnu}(\br)\hK_{i\bnu}'(\br') ,
   \\
  ir_\nu(\hr,\hr')
  &=
    \hK_{i\bnu}'(\br)\hL_{i\bnu}(\br')
   -\hL_{i\bnu}'(\br)\hK_{i\bnu}(\br') ,
   \\
  -s_\nu(\hr,\hr')
  &=
    \hK_{i\bnu}'(\br)\hL_{i\bnu}'(\br')
   -\hL_{i\bnu}'(\br)\hK_{i\bnu}'(\br') .
  \label{eq:snu_hKL}
\end{alignat}
Then we differentiate Eqs.(\ref{eq:pnu_hKL}) and (\ref{eq:snu_hKL}) with respect to $\nu$,
\begin{align}
  i\rd_{\nu}p_\nu(\hr,\hr')
  &=
     \rd_{\bnu}\hK_{i\bnu}(\br)\hL_{i\bnu}(\br')
    +\hK_{i\bnu}(\br)\rd_{\bnu}\hL_{i\bnu}(\br')
    -\rd_{\bnu}\hL_{i\bnu}(\br)\hK_{i\bnu}(\br')
    -\hL_{i\bnu}(\br)\rd_{\bnu}\hK_{i\bnu}(\br') ,
  \label{eq:dpnu_hKL}
   \\
  -i\rd_{\nu}s_\nu(\hr,\hr')
  &=
     \rd_{\bnu}\hK_{i\bnu}'(\br)\hL_{i\bnu}'(\br')
    +\hK_{i\bnu}'(\br)\rd_{\bnu}\hL_{i\bnu}'(\br')
    -\rd_{\bnu}\hL_{i\bnu}'(\br)\hK_{i\bnu}'(\br')
    -\hL_{i\bnu}'(\br)\rd_{\bnu}\hK_{i\bnu}'(\br') .
  \label{eq:dsnu_hKL}
\end{align}
$(\hK_{i\bnu},\hK_{i\bnu}')$ and $(\hL_{i\bnu},\hL_{i\bnu}')$ do not have
the exponential factors $e_K$ and $e_L$ given by Eqs.(\ref{eq:eKL}),
\begin{alignat}{2}
  \bigg\{{ \hK_{i\bnu}(\br) \atop \hL_{i\bnu}(\br) }\bigg\}
  &=
    F_0
    \bigg[
      \frac{S_a}{\bnu^{1/3}}
      \bigg\{{ \Ai(\bu) \atop \Bi(\bu) }\bigg\}
     +\frac{S_b}{\bnu^{5/3}}
      \bigg\{{ \Ai'(\bu) \atop \Bi'(\bu) }\bigg\}
     +\veps_{2n+1}^{(0,*)}
    \bigg] ,
    \label{eq:hKL}
   \\
  \bigg\{{ \hK_{i\bnu}'(\br) \atop \hL_{i\bnu}'(\br) }\bigg\}
  &=
    F_1
    \bigg[
      \frac{S_d}{\bnu^{2/3}}
      \bigg\{{ \Ai'(\bu) \atop \Bi'(\bu) }\bigg\}
     +\frac{S_c}{\bnu^{4/3}}
      \bigg\{{ \Ai(\bu) \atop \Bi(\bu) }\bigg\}
     +\bveps_n^{(0,*)}
    \bigg] .
    \label{eq:hKLp}
\end{alignat}
The derivatives of Eqs.(\ref{eq:hKL}-\ref{eq:hKLp}) with respect to $\bnu$ are given as
\begin{alignat}{2}
  \rd_{\bnu}\bigg\{{ \hK_{i\bnu}(\br) \atop \hL_{i\bnu}(\br) }\bigg\}
  &=F_0
    \bigg[
    \frac{Z_a}{\bnu^{1/3}}
    \bigg\{{ \Ai(\bu) \atop \Bi(\bu) }\bigg\}
   +\frac{Z_b}{\bnu^{5/3}}
    \bigg\{{ \Ai'(\bu) \atop \Bi'(\bu) }\bigg\}
    \bigg] ,
    \label{eq:dhKL}
   \\
  \rd_{\bnu}\bigg\{{ \hK_{i\bnu}'(\br) \atop \hL_{i\bnu}'(\br) }\bigg\}
  &=F_1
    \bigg[
    \frac{Z_d}{\bnu^{2/3}}
    \bigg\{{ \Ai'(\bu) \atop \Bi'(\bu) }\bigg\}
   +\frac{Z_c}{\bnu^{4/3}}
    \bigg\{{ \Ai(\bu) \atop \Bi(\bu) }\bigg\}
    \bigg] .
    \label{eq:dhKLp}
\end{alignat}
$F_0$ and $F_1$ are given by Eqs.(\ref{eq:F0_app}-\ref{eq:F1_app}).
$Z_{a,b,d,c}$ is given as follows,
\begin{alignat}{2}
  Z_a
  &=
    \bZ_a
   +\zeta\frac{q}{\bnu}S_b
   ,\qquad&
  Z_b
  &=
    \bZ_b
   -\bnu qS_a ,
   \\
  Z_d
  &=
    \bZ_d
   -\frac{q}{\bnu}S_c
   ,\qquad&
  Z_c
  &=
    \bZ_c
   +\zeta\bnu qS_d .
\end{alignat}
$\bu$ and $q$ are given by Eqs.(\ref{eq:bu_q}).
$\bZ_{a,b,d,c}$ denotes the following series,
\begin{alignat}{2}
   \bigg\{{ \bZ_a \atop \bZ_d }\bigg\}
  &=\sum_{j=0}^{n}\frac{(-1)^j}{\bnu^{2j+1}}
     \bigg\{{ \ha_j \atop \hd_j }\bigg\} ,
    \qquad&
   \bigg\{{ \bZ_b \atop \bZ_c }\bigg\}
  &=\sum_{j=0}^{n-1}\frac{(-1)^j}{\bnu^{2j+1}}
     \bigg\{{ \hb_j \atop -\hc_j }\bigg\} .
\end{alignat}
The coefficients $(\ha_j,\hb_j,\hd_j,\hc_j)$ are given by the second equations of
(\ref{eq:Ma}-\ref{eq:Mc}).
$\hc_j$ has the different symmetry from the others with respect to the change of the sign 
of $\zeta$ as seen from Eqs.(\ref{eq:aj_bj_symm}-\ref{eq:cj_dj_symm}) and (\ref{eq:Sab}).

When $\bu\gg1$, in order to avoid the overflow and underflow,
we rewrite Eqs.(\ref{eq:pnu_hKL}-\ref{eq:snu_hKL})
by factoring $e^{\pm \bU}$ out of the Airy functions and their derivatives,
\begin{align}
  p_\nu(\hr,\hr')
  &=
    e^{s}\cK_{i\bnu}(\br)\cL_{i\bnu}(\br')
   -e^{-s}\cL_{i\bnu}(\br)\cK_{i\bnu}(\br') ,
   \label{eq:pnu_ckcL}
   \\
  iq_\nu(\hr,\hr')
  &=
    e^{s}\cK_{i\bnu}(\br)\bcL_{i\bnu}(\br')
   -e^{-s}\cL_{i\bnu}(\br)\bcK_{i\bnu}(\br') ,
   \\
  ir_\nu(\hr,\hr')
  &=
    e^{s}\bcK_{i\bnu}(\br)\cL_{i\bnu}(\br')
   -e^{-s}\bcL_{i\bnu}(\br)\cK_{i\bnu}(\br') ,
   \\
  -s_\nu(\hr,\hr')
  &=
    e^{s}\bcK_{i\bnu}(\br)\bcL_{i\bnu}(\br')
   -e^{-s}\bcL_{i\bnu}(\br)\bcK_{i\bnu}(\br') ,
\end{align}
and Eqs.(\ref{eq:dpnu_hKL}-\ref{eq:dsnu_hKL}),
\begin{align}
  i\rd_{\nu}p_{\nu}(\hr,\hr')
  &=
     \rd_{\bnu}\hK_{i\bnu}(\br)\hL_{i\bnu}(\br')
    +\hK_{i\bnu}(\br)\rd_{\bnu}\hL_{i\bnu}(\br')
    -\rd_{\bnu}\hL_{i\bnu}(\br)\hK_{i\bnu}(\br')
    -\hL_{i\bnu}(\br)\rd_{\bnu}\hK_{i\bnu}(\br') ,
   \\
  -i\rd_{\nu}s_{\nu}(\hr,\hr')
  &=
     \rd_{\bnu}\hK_{i\bnu}'(\br)\hL_{i\bnu}'(\br')
    +\hK_{i\bnu}'(\br)\rd_{\bnu}\hL_{i\bnu}'(\br')
    -\rd_{\bnu}\hL_{i\bnu}'(\br)\hK_{i\bnu}'(\br')
    -\hL_{i\bnu}'(\br)\rd_{\bnu}\hK_{i\bnu}'(\br') ,
\end{align}
where
\begin{align}
  \bigg\{{ \cK_{i\bnu}(\br) \atop \cL_{i\bnu}(\br) }\bigg\}
  &=
    F_0
    \bigg[
      \frac{S_a}{\bnu^{1/3}}
      \bigg\{{ A_0(\bU) \atop B_0(\bU) }\bigg\}
     +\frac{S_b}{\bnu^{5/3}}
      \bigg\{{ A_1(\bU) \atop B_1(\bU) }\bigg\}
    \bigg] ,
    \label{eq:cK_cL}
   \\
  \bigg\{{ \bcK_{i\bnu}(\br) \atop \bcL_{i\bnu}(\br) }\bigg\}
  &=
    F_1
    \bigg[
      \frac{S_d}{\bnu^{2/3}}
      \bigg\{{ A_1(\bU) \atop B_1(\bU) }\bigg\}
     +\frac{S_c}{\bnu^{4/3}}
      \bigg\{{ A_0(\bU) \atop B_0(\bU) }\bigg\}
    \bigg] ,
   \\
  \bigg\{{ \cM_{i\bnu}(\br) \atop \cN_{i\bnu}(\br) }\bigg\}
  &=
    F_0
    \bigg[
    \frac{Z_a}{\bnu^{1/3}}
    \bigg\{{ A_0(\bU) \atop B_0(\bU) }\bigg\}
   +\frac{Z_b}{\bnu^{5/3}}
    \bigg\{{ A_1(\bU) \atop B_1(\bU) }\bigg\}
    \bigg] ,
   \\
  \bigg\{{ \bcM_{i\bnu}(\br) \atop \bcN_{i\bnu}(\br) }\bigg\}
  &=
    F_1
    \bigg[
    \frac{Z_d}{\bnu^{2/3}}
    \bigg\{{ A_1(\bU) \atop B_1(\bU) }\bigg\}
   +\frac{Z_c}{\bnu^{4/3}}
    \bigg\{{ A_0(\bU) \atop B_0(\bU) }\bigg\}
    \bigg] .
    \label{eq:bcM_bcN}
\end{align}
The variables involved in Eqs.(\ref{eq:pnu_ckcL}-\ref{eq:bcM_bcN}) are given as follows,
\begin{align}
  s
  =\bU'-\bU ,
    \qquad
  (\bU,\bU')
  =\frac{2}{3}(\bu,\bu')^{3/2} ,
    \qquad
  (\bu,\bu')
  =-\bnu^{2/3}(\zeta,\zeta')
  >0 .
\end{align}
$(\zeta,\zeta')$ is related to $(\br,\br')$ and $(z,z')$ through
Eq.(\ref{eq:br_z}) and Eq.(\ref{eq:zeta_m}).
$A_{0,1}$ and $B_{0,1}$ are the series given by Eqs.(\ref{eq:A01}-\ref{eq:B01}).
Eqs.(\ref{eq:cK_cL}-\ref{eq:bcM_bcN}) are the rest parts of
Eqs.(\ref{eq:hKL}-\ref{eq:hKLp}) and (\ref{eq:dhKL}-\ref{eq:dhKLp}) of which
$e^{\pm\bU}$ is factored out,
\begin{alignat}{2}
  \bigg\{{ \hK_{i\bnu}(\br) \atop \hK_{i\bnu}'(\br) }\bigg\}
  &=e^{-\bU}
    \bigg\{{ \cK_{i\bnu}(\br) \atop \bcK_{i\bnu}(\br) }\bigg\}
   ,\qquad&
  \rd_{\bnu}\bigg\{{ \hK_{i\bnu}(\br) \atop \hK_{i\bnu}'(\br) }\bigg\}
  &=e^{-\bU}
    \bigg\{{ \cM_{i\bnu}(\br) \atop \bcM_{i\bnu}(\br) }\bigg\} ,
   \\
  \bigg\{{ \hL_{i\bnu}(\br) \atop \hL_{i\bnu}'(\br) }\bigg\}
  &=e^{\bU}
    \bigg\{{ \cL_{i\bnu}(\br) \atop \bcL_{i\bnu}(\br) }\bigg\}
   ,\qquad&
  \rd_{\bnu}\bigg\{{ \hL_{i\bnu}(\br) \atop \hL_{i\bnu}'(\br) }\bigg\}
  &=e^{\bU}
    \bigg\{{ \cN_{i\bnu}(\br) \atop \bcN_{i\bnu}(\br) }\bigg\} .
\end{alignat}
%


\section*{Acknowledgments}

I would like to thank Martin Horvat for having the discussion on
the zeros of the cross products of the Bessel functions.
I thank Etienne Forest for the English proofreading of the present paper
and his help in computing the coefficients of the uniform asymptotic expansion.
I want to appreciate my late sister Yukie Koizumi and
her daughter Mai Koizumi for their encouragement.
Also, I deeply appreciate the following people:
my parents Takashi Agoh and Sumiko Agoh,
my uncle and aunt Mitsumasa Kurome and Junko Kurome,
and my grandmothers Yasuko Agoh and Ikuko Kurome.
I would not be able to complete this work without their great support and encouragement
over many years.

\clearpage




\begin{thebibliography}{99}
\bibitem{schwinger} J. Schwinger, {\it On Radiation by Electrons in a Betatron}
   (unpublished), transcribed by M. A. Furman, \\ LBNL-39088, CBP Note-179, UC-414.
%
\bibitem{saldin_schneidmiller_yurkov} E. L. Saldin, E. A. Schneidmiller and M. V. Yurkov,
  Nucl. Inst. Meth. A {\bf398}, p.373-394 (1997).
%
\bibitem{warnock} R. L. Warnock, SLAC-PUB-5375 (1990).
%
\bibitem{agoh} T. Agoh,  Phys. Rev. ST-AB {\bf 12}, 094402 (2009).
%
\bibitem{warnock_morton} R. L. Warnock and P. Morton, SLAC-PUB-4562 (1988).
%
\bibitem{agoh_yokoya} T. Agoh and K. Yokoya, Phys. Rev. ST-AB {\bf 7}, 054403 (2004).
%
\bibitem{stupakov_kotelnikov} G. V. Stupakov and I. A. Kotelnikov, Phys. Rev. ST-AB
  {\bf 12}, 104401 (2009).
%
\bibitem{stupakov_kotelnikov_0} G. V. Stupakov and I. A. Kotelnikov, Phys. Rev. ST-AB
  {\bf 6}, 034401 (2003).
%
\bibitem{rice} S. O. Rice, Bell System Technical Journal, {\bf 27}, 2, p.305-349 (1948).
%
\bibitem{dunster} T. M. Dunster, SIAM J. Math. Anal. {\bf 21}, 4, p.995 (1990).
%
\bibitem{yokoya} K. Yokoya, KEK International 85-7 (1985).
%
\bibitem{abramo_stegun} M. Abramowitz and I. Stegun,
  {\it Handbook of Mathematical Functions}, Dover, New York (1972).
%
\bibitem{agoh_ws} T. Agoh, {\it Error of the paraxial approximation} (unpublished),
    Mini-Workshop on CSR, Nov. 8, 2010, \\ KEK, Tsukuba, Japan.
%
\bibitem{agoh_yokoya1}
T. Agoh and K. Yokoya, MOP15004, Proc. 3rd Asian Particle Accelerator Conf. (APAC'04),
Mar. 2004, Gyeongju, Korea.
%
\bibitem{yokoya_rw}
K. Yokoya, Particle Accelerators, vol.41, p.221-248 (1993).
%
\bibitem{agoh_phD}
T. Agoh, {\it Dynamics of Coherent Synchrotron Radiation by Paraxial Approximation},
Ph.D. thesis, \\ University of Tokyo (2005).
%
\bibitem{ng_warnock}
K. Y. Ng and R. Warnock, Phys. Rev. D {\bf 40}, 231 (1989).
%
\bibitem{cochran} J. A. Cochran, J. Soc. Indust. Appl. Math. {\bf12}, 3, p.580-587 (1964).
%
\bibitem{cochran_zero} J. A. Cochran, Quart. Journ. Mech. and Applied Math.,
{\bf 19}, 4, p.511-522 (1966).
%
\bibitem{nist} National Institute of Standards and Technology,
               NIST Digital Library of Mathematical Functions, \\
               \S 10.45 {\it Functions of Imaginary Order}.
               ~ http://dlmf.nist.gov/10.45
%
\bibitem{horvat_prosen}
  M. Horvat and T. Prosen, J. Phys. A: Math. Theor., {\bf 40}, p.6349-6379 (2007).
%
\bibitem{olver}
  F. J. W. Olver, {\it Asymptotics and special functions}, Academic Press, New York (1974).
%
\bibitem{olver_0}
  F. J. W. Olver, Philos. Trans. Roy. Soc. London Ser. A {\bf 247}, p.328-368  (1954).
%
\bibitem{iwanami}
  S. Morishige, K. Udagawa and S. Hitotsumatsu, Mathematical formulas II,
  Iwanami, Tokyo (1981).
%
\bibitem{gross}
  J. L. Gross,  {\it Combinatorial Methods with Computer Applications},
  Chapman and Hall/CRC, Boca Raton,  2007
%
\bibitem{derbenev}
  Ya. S. Derbenev, J. Rossbach, E. L. Saldin and V. D. Shiltsev,
  TESLA FEL-Report 1995-05 (1995).
\end{thebibliography}

\end{document}